\documentclass[reprint,natbib209,
 amsmath,amssymb,
 aps,aa
floatfix,
]{revtex4-1}
\usepackage{graphicx}
\usepackage{dcolumn}
\usepackage{bm}
\usepackage{hyperref}

\begin{document}

\preprint{Fermi Excess}

\title{An alternative Explanation for the Fermi GeV Gamma-Ray Excess
}

\author{Wim de Boer}
\email{Wim.de.Boer@kit.edu}
\author{Iris Gebauer}%
 \email{Iris.Gebauer@kit.edu}
\author{Alexander Neumann}%
\email{alexander.neumann2@student.kit.edu}
\affiliation{Dept. of Phys., Karlsruhe Inst. for Technology KIT, Karlsruhe, Germany
}%

\author{Peter L. Biermann}
\email{plbiermann@mpifr-bonn.mpg.de}
\affiliation{MPI for Radioastronomy, Bonn, Germany
}%
\affiliation{Dept. of Phys., Karlsruhe Inst. for Technology KIT, Karlsruhe, Germany
}%
\affiliation{Dept. of Phys. \& Astr., Univ. of Alabama, Tuscaloosa, AL, USA}
\affiliation{Dept. of Phys. \& Astron., Univ. Bonn, Bonn, Germany}

\date{\today}

\begin{abstract}
  
The  "GeV-excess" of the diffuse gamma-rays in the halo is studied with a template fit based on  energy spectra   for each possible process of gamma-ray emission. Such a fit allows to determine the background and signal simultaneously, so the Galactic Disk  can be included in the analysis. 
We find evidence that the "excess", characterised by a gamma-ray spectrum peaking at 2 GeV, is much stronger  in  Molecular Clouds   in the disk  than the "GeV-excess" observed up to now in the halo.  The possible reason why the emissivity of Molecular Clouds peaks at 2 GeV are the energy losses and magnetic cutoffs inside MCs, thus depleting the low energy part of the CR spectra and shifting the maximum of the gamma-ray spectra to higher energies. This peaking of the emissivity in Molecular Clouds at 2 GeV was clearly observed from the spectrum of the Central Molecular Zone, which dominates the emission in the inner few degrees of the Galactic Centre.

 Although the spectrum of the Central Molecular Zone peaks at 2 GeV, it cannot be responsible for the ''GeV-excess'' observed in the halo, since the latitude extension of the Zone is below $|b|<0.5^\circ$. However, lines-of-sight into the halo cross Molecular Clouds in the disk, so the emissivity of clouds in the disk will be observed in the halo as an apparent "GeV-excess". 
The fact, that this "GeV-excess" has the same morphology in the disk {\it and} in the halo as the column density of Molecular Clouds, as traced by the CO map from the Planck satellite resembling an NFW-like latitude profile,  and the fact the MCs have an emissivity peaking at 2 GeV shows that the "GeV-excess" originates from Molecular Clouds in the disk, not from a process surrounding the Galactic Centre.

\end{abstract}

\maketitle

\onecolumngrid
\section{Introduction}\label{intro}
An apparent "GeV-excess" of diffuse gamma-rays in the data from the Fermi-LAT satellite around energies of 2 GeV towards the Galactic Centre (GC) has been  studied by many  groups\cite{Goodenough:2009gk,Hooper:2010mq,Boyarsky:2010dr,Morselli:2010ty,Vitale:2011zz,Wharton:2011dv,Hooper:2012sr,YusefZadeh:2012nh,
Abazajian:2012pn,Hooper:2013rwa,Mirabal:2013rba,Huang:2013pda,Huang:2013apa,Gordon:2013vta,Macias:2013vya,Daylan:2014rsa,
Macias:2014sta,Lee:2014mza,Abazajian:2014fta,Abazajian:2014hsa,Calore:2014xka,Calore:2014nla,Cholis:2014lta,
Bartels:2015aea,Lacroix:2015wfx,TheFermi-LAT:2015kwa,Lee:2015fea,Cholis:2015dea,Hooper:2015jlu,2015arXivf05310D,Carlson:2016iis,Choquette:2016xsw,Yang:2016duy}. The Galactic Centre Excess (GCE)   is usually assumed to originate from the GC with the most exciting interpretations being the contributions from dark matter (DM) annihilation\cite{Daylan:2014rsa}  and/or unresolved sources, like millisecond pulsars, see e.g. Refs.\cite{FaucherGiguere:2009df,Lee:2014mza,Lee:2015fea,Bartels:2015aea,Hooper:2015jlu} and references therein. 
Up to now people compared the data with interstellar emissivity models (IEM) in order to find an excess. IEMs are provided  by propagation models, like Galprop\cite{Moskalenko:1998id,Vladimirov:2010aq} or Dragon\cite{Evoli:2008dv} or the diffuse emission model from Fermi\cite{diffuse}. Unfortunately, the IEMs suffer from large uncertainties. E.g. the source and gas distributions used in propagation models are poorly known. Also the diffuse emission model is not optimal\cite{caveat}, since the model has smoothed spatial emissivities, so it is optimized to search for excesses from point sources, not for extended excesses, like a signal from dark matter (DM) annihilation.

Here we follow a different approach: instead of determining the background emissivity from IEMs, we use a template fit from a linear combination of the energy spectra for all processes to the data, which can be applied to  cones  with a high spatial resolution. A bad fit inside a cone indicated the need for one or more additional contributions. The templates are obtained with a data-driven method, as discussed in detail in Appendix \ref{B}. A data-driven method circumvents the uncertainties from the IEMs, which are especially large in the GD. Hence, a template fit allows to include  the Galactic Disc (GD) into the analysis, as will be shown later.

  In addition to the standard background templates for $\pi^0$ production, inverse Compton scattering (IC) and Bremsstrahlung (BR) one needs a template describing the energy spectrum of the Fermi Bubbles, as determined by the Fermi Collaboration\cite{Fermi-LAT:2014sfa}. The Bubbles have a hard spectrum corresponding to the gamma-ray spectrum from a  proton spectrum  falling with rigidity $R$ as approximately $1/R^{2.1}$.

After adding this template  the fit did not only find the Fermi Bubbles in the halo and its extension into the GD, but found fluxes from this hard template  in the GD as well, especially in the star forming regions, like the Galactic Bar and tangent points of the spiral arms\cite{2014ApJ...794L..17D}. The star forming regions in the GD can be traced by the 1.809 MeV line from $^{26}$Al, a radioactive isotope that is synthesized in sources\cite{Prantzos1996}.
The correlation betwee the flux from the $1/R^{2.1}$  template and the  $^{26}$Al flux was interpreted as the first clear evidence\cite{2014ApJ...794L..17D} for the predicted Source Cosmic Rays (SCRs)\cite{2000ApJ...540..923B}, which are CRs confined inside sources during the acceleration.   The enhanced gas density and enhanced CR density inside the shockwaves of SNRs provide  the ideal conditions for $\pi^0$ production from the hard $1/R^{2.1}$ proton spectrum predicted for diffuse shockwave acceleration\cite{Hillas:2005cs,Biermann:2010qn}.  The fact that the Bubbles have exactly this same hard spectrum in the halo as in the star forming regions suggests that the Bubbles are outflows from the GC, as discussed previously\cite{2014ApJ...794L..17D} and in  Appendix \ref{C}.

The inclusion of the Bubble spectrum in the template fit improved considerably the fit inside the GD, but it did not describe the GCE, which is  characterised by a shift of the energy spectrum (weighted by $E^2$ for each energy bin) from 0.7 GeV for the usual $\pi^0$ production dominated background to 2 GeV. This can be either explained by a new source  with a gamma-ray spectrum peaking at 2 GeV or a source providing a depletion of gamma-rays below 2 GeV.  New sources with a spectrum peaking at 2 GeV are millisecond pulsars (MSPs)\cite{FaucherGiguere:2009df,Lee:2014mza,Lee:2015fea,Bartels:2015aea,Hooper:2015jlu} or DM annihilation\cite{Daylan:2014rsa}. Sources with a depletion below 2 GeV could be e.g. molecular clouds (MCs) with a magnetic cutoff. Such a cutoff is well known from CRs entering the Earth magnetic field  near the magnetic equator: particles below typically 20 GV do not reach the Earth, but are repelled into outer space  by the Lorentz force\cite{Herbst:2013hr}. The magnetic cutoff is proportional to the magnetic moment. Although the magnetic field near the Earth (0.5 G) is an order of magnitude higher than the typical magnetic fields in dense MCs\cite{2012ARA&A..50...29C}, the much larger sizes of MCs - or its  substructure of filaments and cloudlets\cite{2000prpl.conf...97W} - yield magnetic moments easily of the same order of magnitude, so similar magnetic cutoffs can be expected.  

Experimental evidence that the gamma-ray spectrum from MCs has its maximum shifted to 2 GeV comes from the Central Molecular Zone (CMZ), a dense MC in the GC with a total mass of $5\cdot10^7 ~M_\odot$ in the tiny solid angle limited by $-1.5^\circ<l<2^\circ$ and $|b|<0.5^\circ$\cite{Tsuboi:1999,Jones:2011bv}. For comparison, the CMZ has an order of magnitude higher mass than  the supermassive black hole  SGR A$^*$ in the GC.  The emissivity from this solid angle is dominated by the CMZ, as will be shown later by the template fit and the CMZ exhibits directly the spectrum peaking at 2 GeV, the hallmark of the GCE. 

The CMZ cannot be responsible for the GCE observed in the halo, since the latitude extension of the CMZ is below $|b|<0.5^\circ$. However, lines-of-sight into the halo cross other MCs in the GD, so an "excess" from MCs can be observed in the halo.  The latitude distribution of such an "excess" will be proportional to the column density of MCs, which happens to resemble an NFW profile\cite{Navarro:1996gj} or more generalized DM profiles\cite{Coe:2010xg}. The column density of MCs can be traced by the rotation lines of the CO molecule\cite{1988ApJ...324..248B} and its latitude distribution can be obtained from the precise all-sky CO map, which has been provided by the Planck  satellite\cite{ThePlanck:2013dge} and is publicly available\cite{Planck}.

It is the purpose of the present paper to study the MC scenario as a possible explanation of the GCE by performing a multifrequency, mutimessenger analysis of the morphology of the fluxes from the SCR and MCR templates with the fluxes from the $^{26}$Al  and  CO lines.

\section{Analysis  }\label{anal}
 We have analysed the diffuse gamma-rays in the energy range between 0.1 and 100 GeV using the diffuse class of the public  P7REP\_SOURCE\_V15 data collected from August, 2008 till July 2014 (72 months) by the Fermi Space Telescope\cite{Atwood:2009ez}. The data were analysed  with the recommended selections for the diffuse class using the  Fermi Science Tools (FST)  software\cite{FST}, as detailed in Appendix \ref{A}. The point sources  from the second Fermi point source catalogue\cite{Fermi-LAT:2011iqa} were subtracted using the {\it gtsrc} routine in the FST. 
The sky maps were binned in longitude and latitude in 0.5x0.5$^\circ$ bins, which were combined to form a total of 797 cones covering the whole sky. In and around the GD the cones were  one degree in latitude with a longitude size adapted to the structures, like the CMZ and the Fermi Bubbles. In the halo the cone size increased. The cone sizes and fit results for each of the 797 cones have been  given in Appendix \ref{C}. 

The gamma-ray flux is proportional to the product of the CR densities, the "target densities" (gas or gamma-rays in the interstellar radiation field (ISRF)) and the cross sections. A template fit combines the product of these three factors into a single normalisation factor  for each gamma-ray component $k$, thus eliminating the need to know them individually.
The total flux in a given direction can be described by a linear combination of the various processes with known energy templates:
 \begin{eqnarray} |\Phi_{tot}>&=&n_1|\Phi_{PCR}>\  +\  n_2|\Phi_{BR}>\   +\  n_3|\Phi_{IC}> +\nonumber \\ &&  n_4|SCR>\  +\ n_5|MCR>\  +\ n_6|\Phi_{iso}>, \label{e1}\end{eqnarray} where the normalisation factors $n_i$ determine the fraction of the total flux for a given process:  PCR from the $\pi_0$ production by propagated CRs, BR from Bremsstrahlung, IC from inverse Compton, SCR from the $\pi_0$ production by SCRs, MCR from the $\pi_0$ production  inside MCs and "iso" for the isotropic background. The factors $n_i$, and hence the flux of each process, can be found from a $\chi^2$ fit, which adjusts  the templates to best describe the data.  Details on the test statistic  have been provided in Appendix \ref{A}.The spectrum of a each cone  has 21 energy bins  with only $n_i\le 6$  free parameters, so the fit is strongly constrained. Furthermore, the templates for each process have quite a different shape,  which allows a  determination of the flux for each process in each direction.   

\begin{figure}
\centering
\includegraphics[width=0.46\textwidth,height=0.46\textwidth,clip]{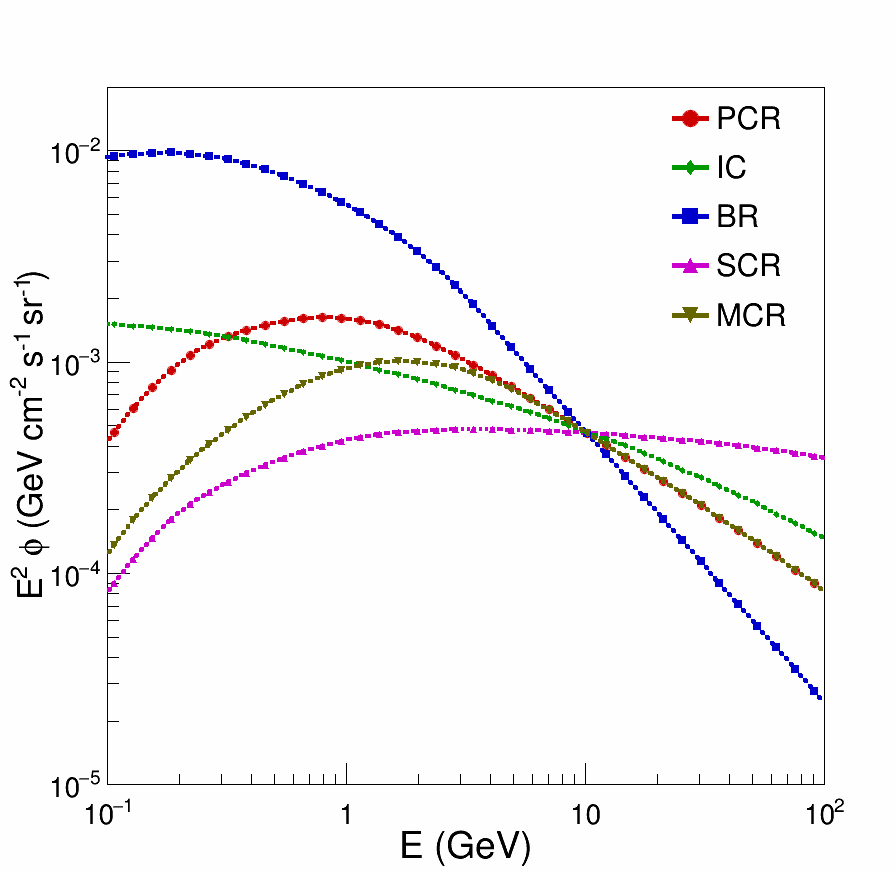}
\includegraphics[width=0.46\textwidth,height=0.46\textwidth,clip]{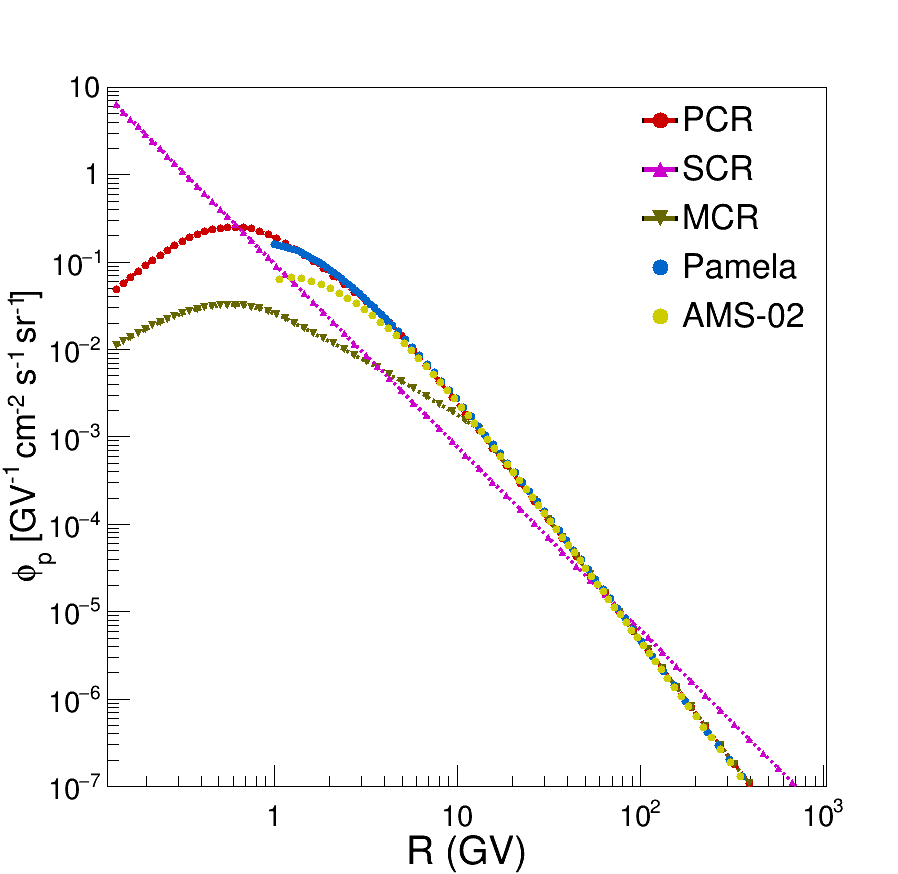}\\
(a) \hspace*{0.46\textwidth}(b)\\
\caption[]{(a)  Diffuse gamma-ray spectral templates for   Bremsstrahlung,  inverse Compton scattering, PCRs, SCRs and MCRs. The templates are normalised around 10 GeV. 
(b)Nucleon injection spectra for the gamma-ray template from $\pi^0$ production by PCRs, SCRs and MCRs. For comparison, the data from AMS-02\cite{Aguilar:2015ooa}  and Pamela\cite{Adriani:2011cu} are shown as well. The spectra are normalised  at 70 GV.
  \label{f1}}
\end{figure}

 The energy templates for the various processes can be obtained from the data by a data-driven method. E.g.  the spectum  for the MCR template can be inferred from the CMZ and the SCR template from the Fermi Bubble. Details of extracting each of the templates by a data-driven method are described in Appendix \ref{B}. The   resulting gamma-ray templates  are shown in Fig. \ref{f1}(a). 
 The  $\pi^0$ production templates  are proportional to the CR spectra, which are shown in Fig. \ref{f1}(b). The SCR template corresponds to the $1/R^{2.1}$ spectrum, which is valid for the Fermi Bubbles as well, as demonstrated in Appendix \ref{C}.  

The MCR spectrum corresponds to a nucleon spectrum with a break below 14 GV to simulate the effect of a magnetic cutoff and energy losses, as required by the spectrum from the CMZ. Some lower density MCs needed a slightly lower break,  as discussed in Appendix \ref{B}. 

The PCR component, as determined from the gamma-ray spectra, is close to the Pamela data, which are shown in Fig. \ref{f1}(b) as well. The difference between the PCR spectra and the locally observed data is caused by the solar wind, which suppresses the flux of particles below 20 GV. The MCR template also requires a suppression of low energy CRs.  Hence, the break in the MCR template and solar modulation are highly correlated. The GCE is affected in a similar way and the solar modulation is the single, most important parameter for the size of the GCE, as discussed  in Appendix \ref{B}. This fact was not realised in previous studies on the systematic errors of the GCE\cite{Calore:2014xka}.

The isotropic component (ISO) arises from  interacting hadrons misidentified as gamma-rays and the extragalactic background. The  isotropic component  was given by the Fermi Science Group\cite{FST}, but a more precise determination of both, the spectrum and the absolute flux of the isotropic background, is given in  Appendix \ref{B}. Hence, the value of $n_6$ in Eq. \ref{e1}, which is the same for all sky directions, is fixed. Finally, 
there are only 5 free parameters, namely  $n_1$, $n_2$, $n_3$, $n_4$ and $n_5$, with 21 data points in the binned gamma-ray spectrum for each cone. 
\begin{figure}
\centering
\includegraphics[width=0.45\textwidth,height=0.446\textwidth,clip]{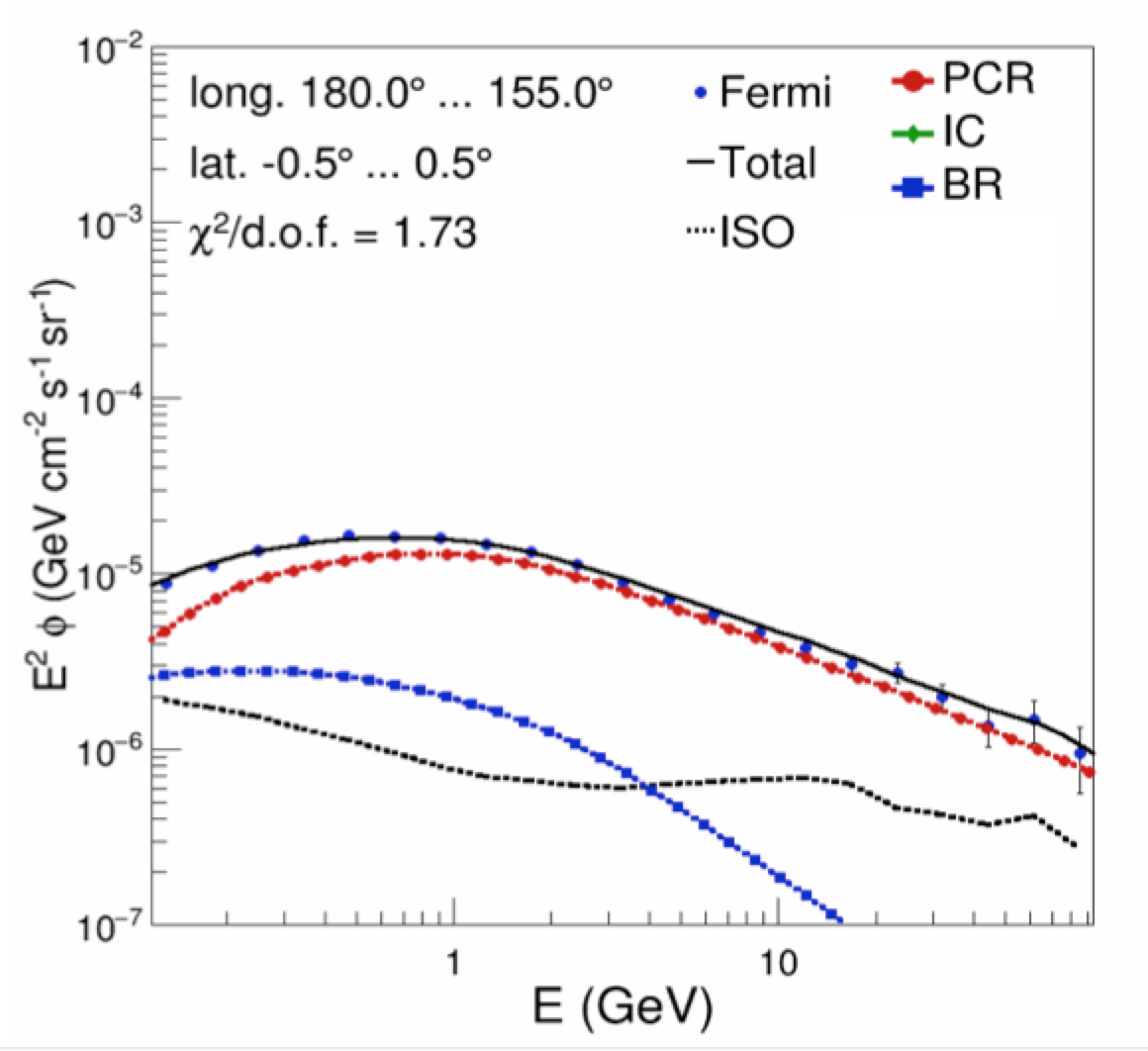}\hspace*{0.5mm}
\includegraphics[width=0.46\textwidth,height=0.446\textwidth,clip]{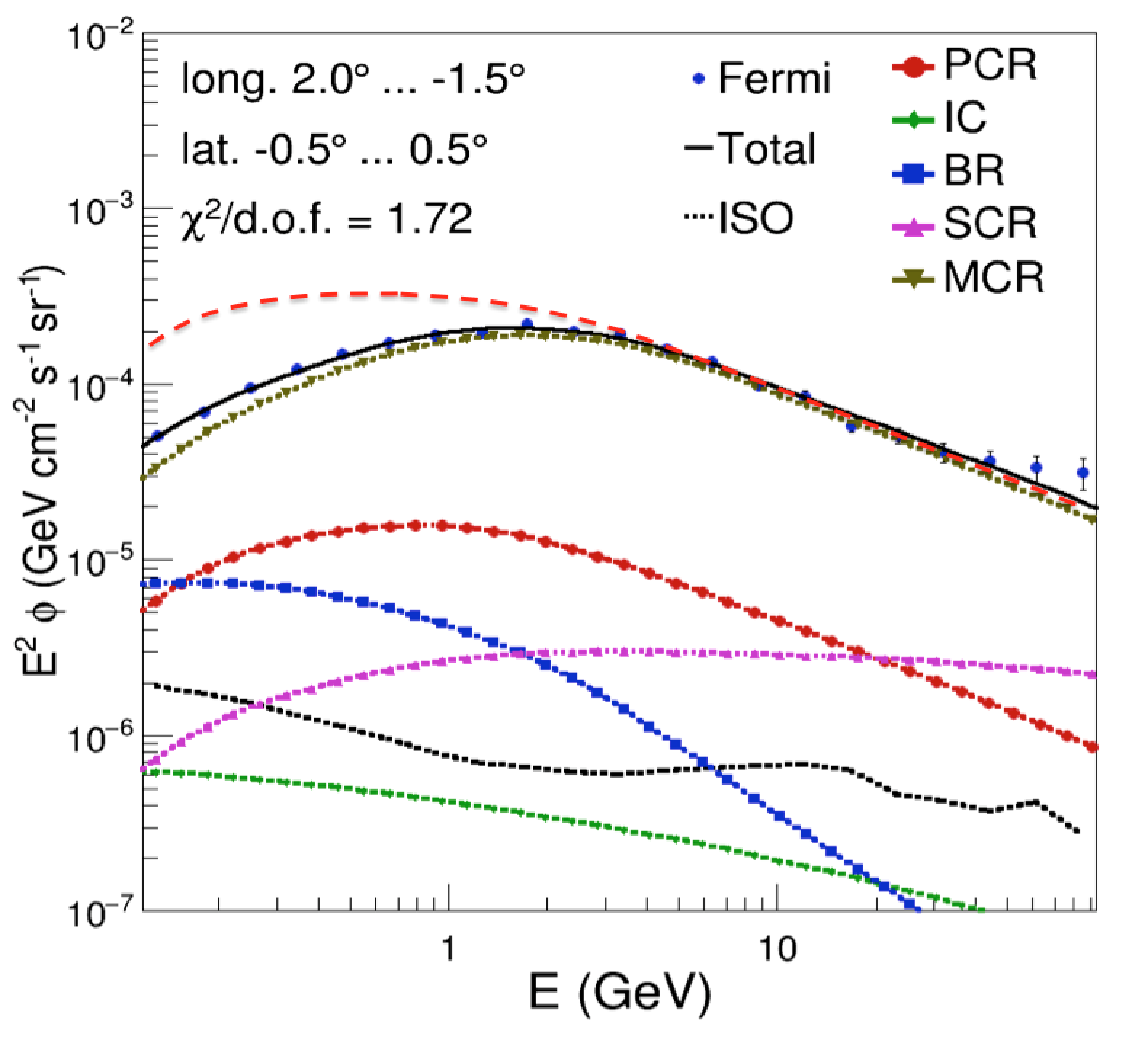}\\
(a) \hspace*{0.46\textwidth}(b)\\
\includegraphics[width=0.474\textwidth,height=0.46\textwidth,clip]{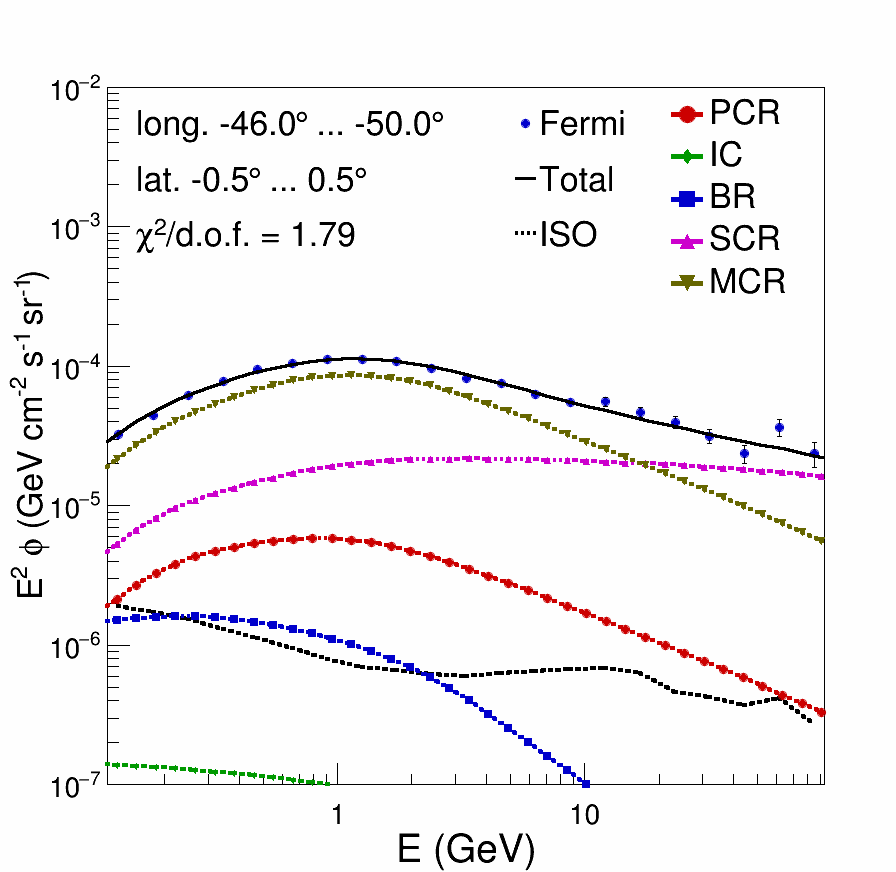}
\includegraphics[width=0.474\textwidth,height=0.46\textwidth,clip]{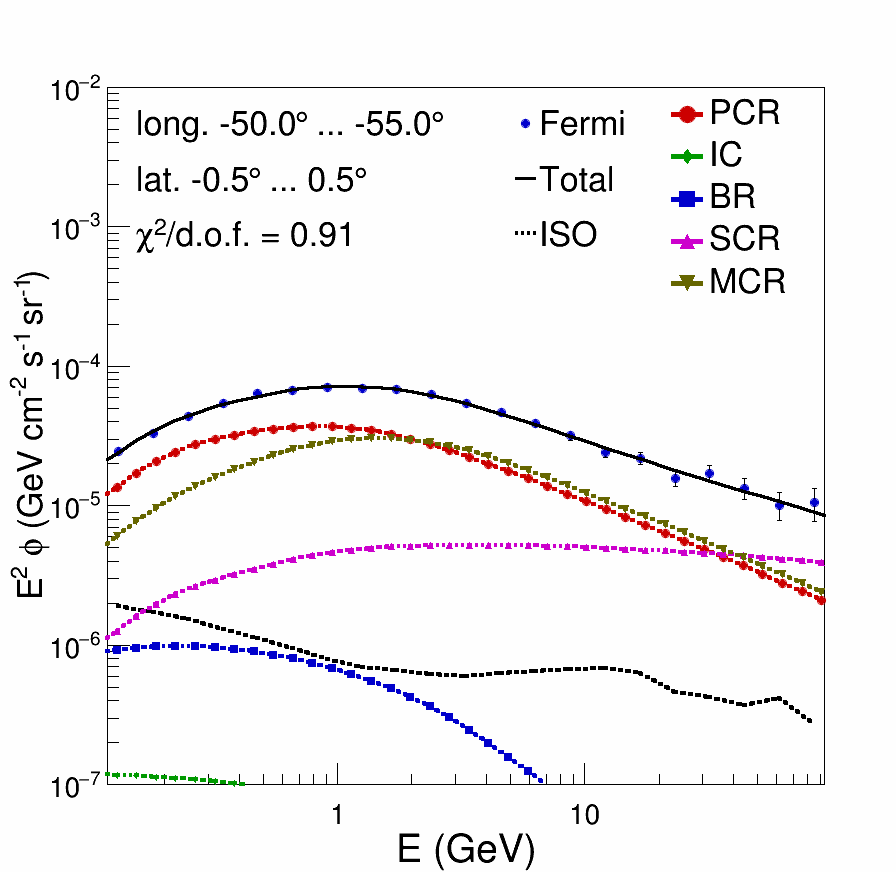}\\
(c) \hspace*{0.46\textwidth}(d)
\caption[]{Template fits towards the direction opposite of the GC  (a) and the GC (b). The cone size in (b) corresponds to the CMZ.
Fits towards the tangent point of the nearest spiral arm (c) and  one of its neighbouring cones (d).
 \label{f2}
}
\end{figure}
\begin{figure}[]\centering
\begin{minipage}[]{0.45\textwidth}\centering
\includegraphics[width=0.9\textwidth,height=0.6\textwidth,clip]{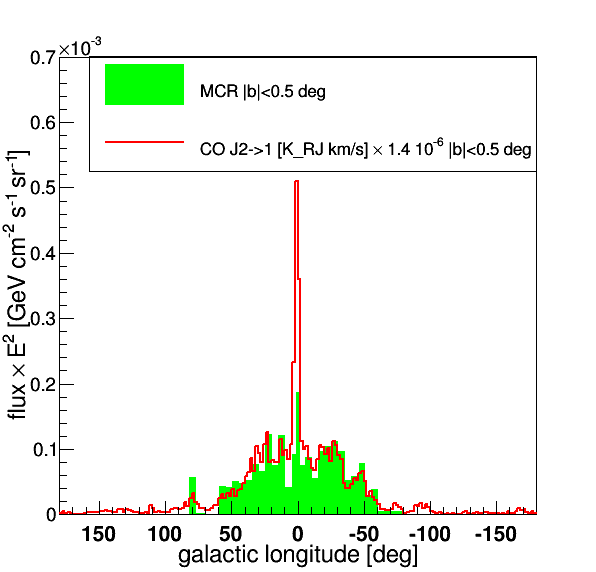}\\
\center{(a)}\\
\includegraphics[width=0.9\textwidth,height=0.6\textwidth,clip]{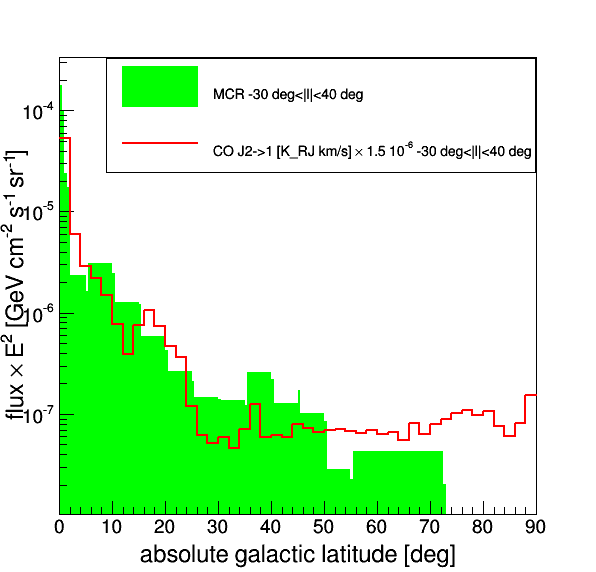}\\
\center{(b)}
\end{minipage}
\begin{minipage}[]{0.45\textwidth}\centering\vspace*{5mm}
\includegraphics[width=0.9\textwidth,height=1.30\textwidth,clip]{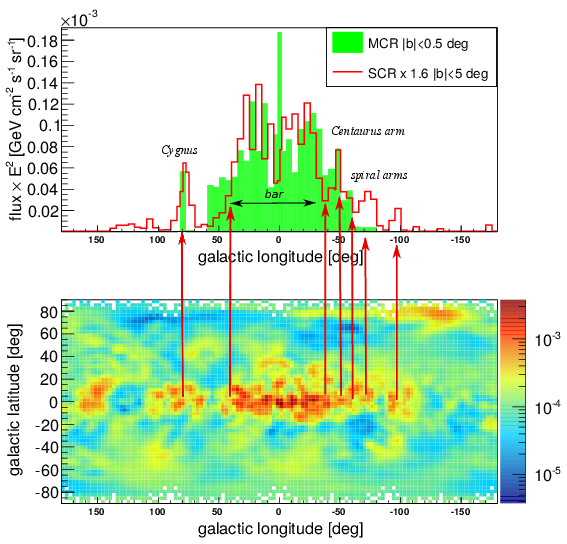}\\
\center{(c)}
\end{minipage}
	\caption[]{Longitude (a) and latitude (b) distribution  of the MCR fluxes icompared to  the CO distributions from the Planck satellite (red line)\cite{Planck}.  (c): a direct comparison of the
 longitude distribution of the SCR and MCR fluxes  (upper panel), which have the same morphology, as expected, since they are both connected to MCs,  A comparison of the SCR and MCR fluxes
  with the $^{26}$Al sky map\cite{Bouchet:2015rxa,Spi}  (lower panel).  The  $^{26}$Al flux given in units of $\rm cm^{-2}s^{-1}sr^{-1}$. 
}
\label{f3}
\end{figure}
\begin{figure}[]
\centering
\includegraphics[width=0.46\textwidth,height=0.46\textwidth,clip]{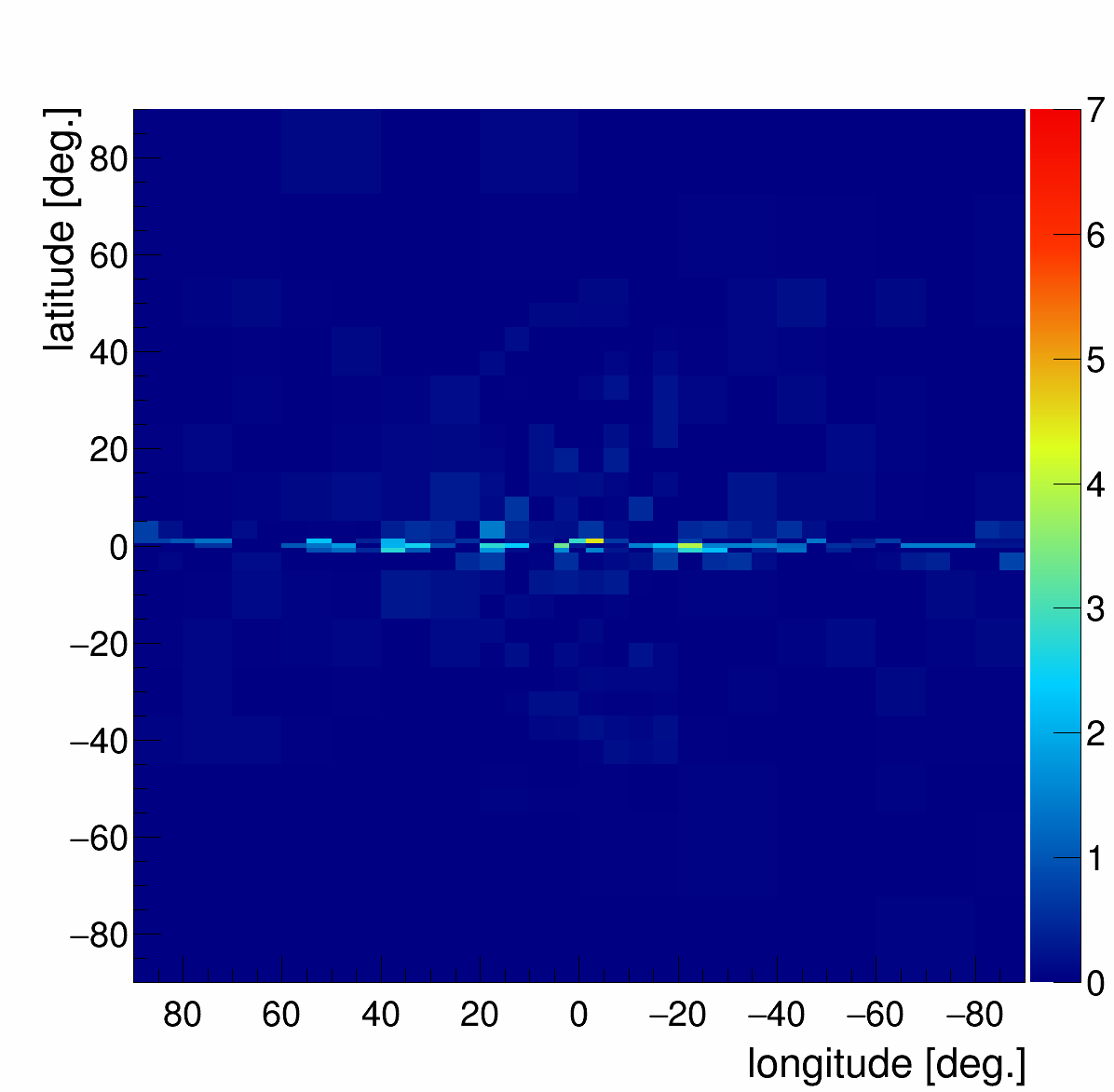}
\includegraphics[width=0.47\textwidth,height=0.44\textwidth,clip]{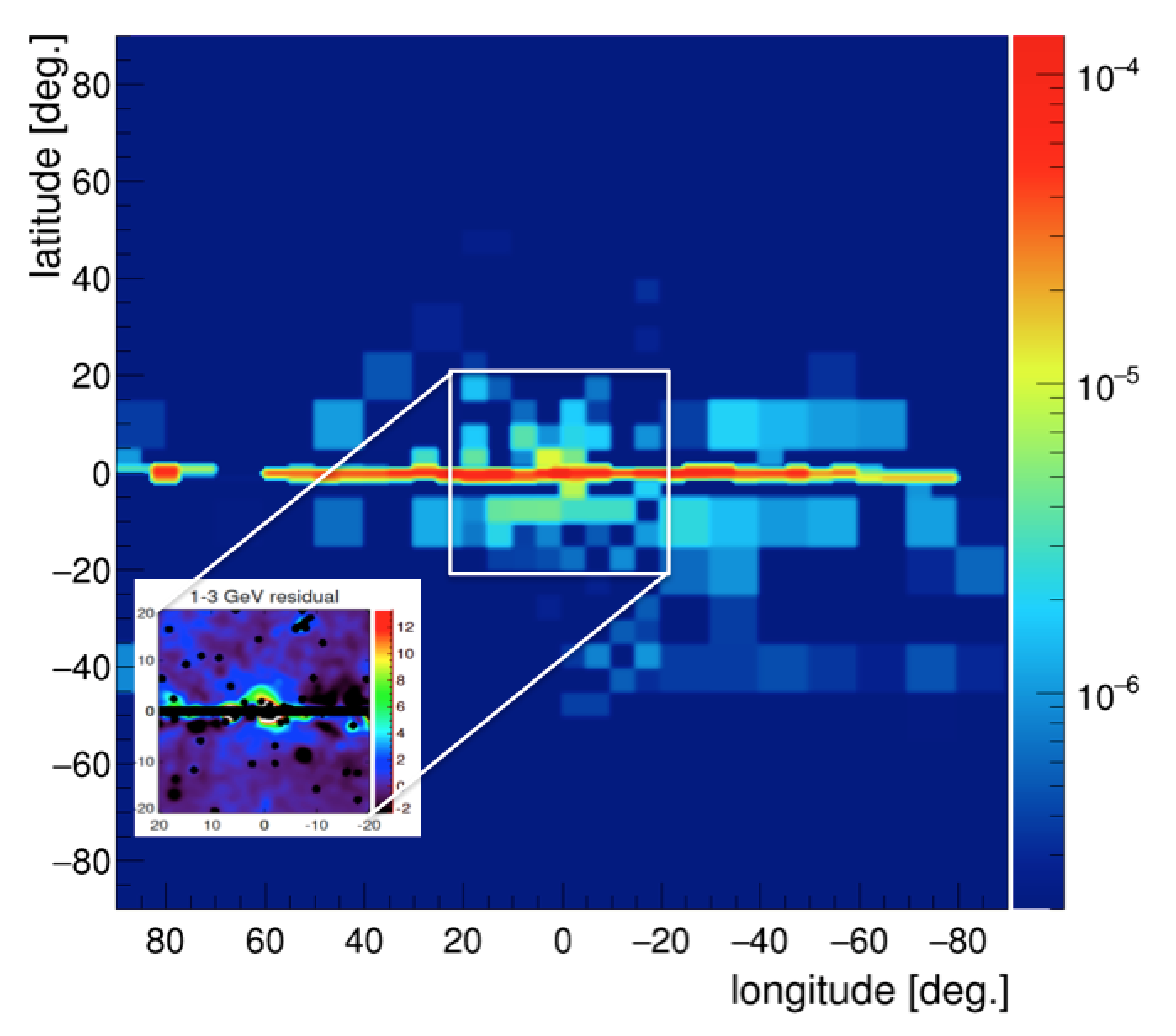}
\hspace*{0.04\textwidth}(a)\hspace*{0.46\textwidth} (b)\\ \vspace*{-1mm}
\caption[]{(a): Absolute values of the residuals  between the fit and the gamma-ray data for the energy bin at 2.3 GeV. The color coding indicates the flux difference between the fit and the data in units of $10^{-6}$ $\rm GeV cm^{-1} s^{-1} sr^{-1}$.  (b): Sky map of the MCR fluxes in $\rm GeVcm^{-1}s^{-1} sr^{-1}$. The insert shows the  GCE from Ref. \cite{Daylan:2014rsa}. Here the region $|b|<2^\circ$ is masked, so the strong GCE in the GD is not seen. 
\label{f4}
}
\end{figure}

Examples of  template fits are shown in Fig. \ref{f2} for selected cones in the GD.  Figs. \ref{f2}(a) and (b) show fits to the direction opposite to the GC (called Galactic Anticentre (GA)) and to the GC, respectively.  The maximum in the energy spectrum, multiplied by $E^2$, is shifted from 0.7 GeV in  Fig. \ref{f2}(a) to 2 GeV in Fig. \ref{f2}(b). The size of the GC cone was selected to be the size of the CMZ. The  spectrum in  Fig. \ref{f2}(a) (GA) is well described by a linear combination of the  known background templates from  PCR, IC, BR and the isotropic component (ISO). The IC contribution is not visible on this scale.  Interestingly, the slope of the spectrum above a few GeV is the same in the GC and GA, as shown by the (red) dashed line in Fig. \ref{f2}(b). This dashed line  corresponds to the shape of the GA spectrum in Fig. \ref{f2}(a). The same slope at high energies is expected in both directions, if they both originate from $\pi^0$ production uotside regions, where energy losses and magnetic cutoffs affect the spectrum. 

The data towards the GC can only be described, if one includes the SCR and MCR templates. The latter is close to the total flux (black line), so it dominates the flux in Fig. \ref{f2}(b), as expected from the high density of the CMZ.  Moving the CMZ cone in Fig. \ref{f2}(b)   as a sliding window with a fixed size in all directions reduces the flux from the MCR template, which  indicates that the MCR template in this cone is directly connected to the CMZ. 

 As a second example  the template fit towards the nearby tangent point of the Scutum spiral arm and one of its neighbouring cones are shown in Figs.  \ref{f2}(c) and (d), respectively. One observes again that towards the tangent point the spectrum is dominated by the MCR component, while in the neighbouring cone the MCR and PCR contributions are of siimilar strength. 
Apparently, the total flux of gemma-ray emission inside and outside the arm does not change, only the spectrum changes,
since the gas inside the spiral arm has more MCs than outside, as expected from the higher star formation rate.

In order to see if this shift of the maximum is obtained everywhere, where there are MCs, one can look for a spatial correlation between the CO maps and the MCR template. This is shown in Figs. \ref{f3}(a) and (b) for the longitude and latitude distributions. The red line shows the distributions obtained  from the publicly available Planck data\cite{Planck}, while the green area shows the flux from the MCR template, as obtained from the fit. The agreement in morphology between the two points to a strong correlation, which is not expected to be exact, since the MCR flux is determined by the gas density convolved with the CR density, while the CO maps are proportional to the gas density only.  

The SCR flux is expected from the $1/R^{2.1}$ spectrum in sources. Since the sources are typically embedded inside MCs one expects a similar morphology of the MCR and SCR fluxes. This is indeed the case, as shown in  Fig. \ref{f3}(c). The bar region and spiral arms are the dominant regions of sources and MCs.  
Note that the SCR template naturally describes the hard spectrum of gamma-rays  in the central Galaxy and the "normal" spectrum from PCRs in the opposite direction, as was noticed recently\cite{Yang:2016jda}.
The SCR component is driven by CRs inside sources, while the MCR component is driven by propagated CRs. This might explain the difference in intensity in the central bin  in the top panel of  Fig. \ref{f3}(c) between SCR and MCR fluxes: the SCR density is  reduced by driving the outflow into the Bubbles\cite{Everett:2007dw,Breitschwerdt:2008na}, thus reducing the density of SCRs, which are confined to sources, stronger than the density of MCRs distributed over the whole cloud region.
The SCR (MCR) fluxes in the top panel of Fig. \ref{f3}(c) are integrated over a latitude range of $|b|<6^\circ(0.5^\circ)$, respectively. The larger latitude range for the SCR component is just to increase the statistics of the SCR fluxes, since the sources can have outflows towards higher latitudes, as suggested by the broad latitude distribution of $^{26}$Al in the bottom panel of Fig. \ref{f3}(c).  A comparison of the latitude and longitude distributions  of the $^{26}$Al  and  SCR fluxes  is  shown in Appendix \ref{C}.

After including the contributions from the MCR and SCR templates a good fit is obtained for the whole gamma-ray sky, as demonstrated in Fig. \ref{f4}(a), which shows the absolute difference between the fit and the data at an energy of 2.3 GeV. The residuals for other energy bins are shown in  Appendix \ref{C}. The small regions in the GD with non-zero residuals correspond to differences between the data and the fit at the per cent level.

Finally,  the skymap of the MCR fluxes is shown in Fig. \ref{f4}(b). The insert shows the GCE, as  given in Ref. \cite{Daylan:2014rsa}. Here the region for $|b|<2^\circ$ was masked because of the large uncertainty associated with their spatial templates inside the GD, so the strong "excess" in the GD is not observed.
Outside the GD the morphology of both is similar: the MCR flux and the GCE both extend into the halo around the GC and have a similar  flux of a few times $\rm GeVcm^{-1}s^{-1} sr^{-1}$ around a latitude of 5$^\circ$. However, the template fit shows  some clumpiness, as expected from the discrete nature of the MCs or its filamentary substructure. In Ref. \cite{Lee:2015fea} some deviation from smooth templates has been observed in this region, which was interpreted as evidence for unresolved point sources, a feature  used to support the MSP interpretation of the GCE. But the nature of the clumpiness is unknown and could be related to MCs as well\cite{Lee:2014mza}.
\section{Summary}
In summary,  we performed an energy template fit with
 two new contributions: (i) the SCR template  corresponding to the hard spectrum from CRs inside the sources during the acceleration, which describes the Bubbles and the high energy tail above the power law from normal $\pi^0$ production in the GD; (ii) the MCR template, which describes the emissivity from MCs, characterised by a maximum in the gamma-ray spectrum at 2 GeV, the hallmark of the GCE. The MCR template was obtained  directly from the data, especially the dense CMZ in the inner few degrees of the GC. As it happens, the MC column density  is  steeply falling with decreasing latitude, as shown by its CO tracer, which leads to  a morphology of the "excess" in MCs similar to a DM profile.

Our multifrequency, multimessenger full-sky analysis demonstrates a fourfold correlation  between  the morphologies from  the MCR  and  SCR fluxes and the tracers of MCs, namely the 1.8 MeV line of $^{26}$Al (a tracer of sources inside MCs) and the CO rotation lines in the radiofrequency range (a direct tracer of MCs), from which we conclude   that the so-called GCE is related to the propagation inside MCs, in which case the GCE is not really an "excess", but a depletion of gamma-ray fluxes  below 2 GeV inside MCs.

‎

\acknowledgments
Support from the Deutsche Forschungsgemeinschaft  (DFG, Grant BO 1604/3-1)  is warmly  acknowledged. We thank Roland Crocker, Francesca Calore, Daniele Gaggero  and Christoph Weniger for helpful discussions. We are grateful to the Fermi scientists, engineers and technicians for collecting the Fermi data and the Fermi Science Support Center for providing the software and strong support for  guest investigators.
\appendix
\section{Data selection and Analysis Details}\label{A}
We have analysed the diffuse gamma-rays in the energy range between 0.1 and 100 GeV using the diffuse class of the public  P7REP\_SOURCE\_V15 data collected from August, 2008 till July 2014 (72 months) by the Fermi Space Telescope\cite{Atwood:2009ez}. The data were analysed  with the recommended selections for the diffuse class using the  Fermi Science Tools (FST)  software\cite{FST}. This included the higher zenith angle cut of 100$^\circ$ to eliminate Earth limb events. Gamma-rays converted in the front  and  back end of the detector were included. The residual hadronic background was included in the isotropic template, as discussed in Appendix \ref{B}.
The sky maps were binned in longitude and latitude in 0.5x0.5$^\circ$ bins, which could later be combined at will. The point sources  from the second Fermi point source catalogue\cite{Fermi-LAT:2011iqa} have been subtracted using the {\it gtsrc} routine in the FST. 

As test statistic we use the  $\chi^2$ function defined as
 \begin{equation} \chi^2=\sum_{i=1}^{N} \sum_{j=1}^{21}\left[\frac{\langle data(i,j)- \sum_{k=1}^{5} n(i,k) \times tem(i,j,k)\rangle^2}{\sigma(i,j)^2}\right], \label{e2}\end{equation}
where the sum is taken over the N=797 cones   in different sky directions $i$, $data(i,j)$ represents the total Fermi flux in direction $i$ for energy bin $j$, $tem(i,j,k)$ the template contribution with normalisation $n(i,k)$ for template $k$  and  $\sigma(i,j)$ is the total error on $data(i,j)$, obtained by adding the statistical and systematic errors  in quadrature.  Typically, latitude (longitude) steps of 1$^\circ$  (5$^\circ$) were taken near the GD with larger steps in the halo. The precise cone sizes  can be read off from the template fits for each cone in Appendix \ref{C}. The recommended systematic errors  in the Fermi Software on the total flux  are 10\% for gamma-ray energies below 100 MeV, 5\% at 562 MeV, and 20\% above 10 GeV. We used  a linear interpolation for energies in between. 
 We rescaled  these errors by a factor  0.25 at all energies to obtain 
$\chi^2/d.o.f. \approx 1$. This rescaling did not affect the main results. The systematic errors between the bins are correlated, which implies that all data points are allowed to move simultaneously up or down by an amount given by the correlated part of the systematic error. However,  a template fit with free normalisations for each template allows to move the fit up and down as well, which compensates a common shift in the data. So adding a correlated error in the data can slightly change the overall flux, but hardly affects the relative contributions of the various templates, as was verified by explicitly adding a covariance matrix to Eq. \ref{e2} with a common positive correlation between all bins, which was varied between 10\% and 70\% of the total systematic error. We did not vary the size of the correlation as function of energy, since this effect is expected to be small after rescaling the systematic errors by 0.25, which leads to maximum systematic errors of 5\%.

\section{Determination of Templates}\label{B}
\subsection{The PCR, BR and IC Templates}
A first order estimate of the standard background gamma-ray templates can be obtained from propagation models, for which we used Dragon\cite{Evoli:2008dv}. We checked that the Galprop code yields the same results. Most propagation models have been optimized for charged CRs and the gamma-ray emission is calculated after a good description of the charged cosmic ray distributions has been found. This is not optimal, because  the locally observed CR spectra differ from   interstellar spectra by the effect from the solar wind, which modulates the low energy part of the spectra (below about 20 GV) with a solar cycle of 11 years.
This modulation can be described in a simple spherical symmetric  force field approximation\cite{Gleeson:1968zza} with a single parameter, the solar modulation potential  $\Phi$, which is time dependent and can be different for different particles, since it is a complicated function of the magnetic field in the solar cycle, the strength of the solar wind and the gas density, which determines the amount of energy losses.
\begin{table}[]
\centering                                                                                          
    \begin{tabular}{ | l | l | l | l |} 
    \hline
    Parameter & Unit & Value & Description \\ \hline
    $\alpha_0$ & 1 & 1.43 & Nucleon injection spectral index below break point\\ \hline
    $\alpha_1$ & 1 & 2.16 & Nucleon injection spectral index above break point\\ \hline
    $\rho$ & GV & 4.62 & Nucleon injection break point\\ \hline
    $\alpha_{el,0}$ & 1 & 1.60 & Electron spectral injection index below break point\\ \hline
    $\alpha_{el,1}$ & 1 & 2.54 & Electron spectral injection index above break point \\ \hline
    $\rho_{el}$ & GV & 4 & Electron injection break point\\ \hline
    $L$ & kpc & 7.4 & Halo height \\ \hline
    $D_0$ & $10^{28}\text{cm}^2$/s & 4.24 & Diffusion coefficient \\ \hline
    $\delta$ & 1 & 0.63 & Diffusion rigidiy index\\ \hline
    $\eta$ & 1 & -0.53 & Diffusion velocity index\\ \hline
    $v_{\alpha}$ & km/s & 2.23 & Alfv\'{e}n velocity \\ \hline
	$v_0$ & km/s & 3.71 & Convection base velocity \\ \hline
	$f_b=v_b/v_0$ & 1 & 0.21 & Convection velocity break parameter\\ \hline
	$d V_c/d z$ & km/s/kpc & 0 & Convection gradient\\ \hline
	$\alpha_r$ & 1 & 1.29 & Convection radial source index \\ \hline
	$z_k$ & kpc & 0.11 & Convection gradient break point\\ 
    \hline
    \end{tabular}
\caption[]{Dragon configuration parameters for an optimized background model including gamma-rays. From Ref. \cite{Kunz}.}
\label{T1} 
\end{table}
\begin{figure}[]
\centering
\includegraphics[width=0.45\textwidth,height=0.35\textwidth,clip]{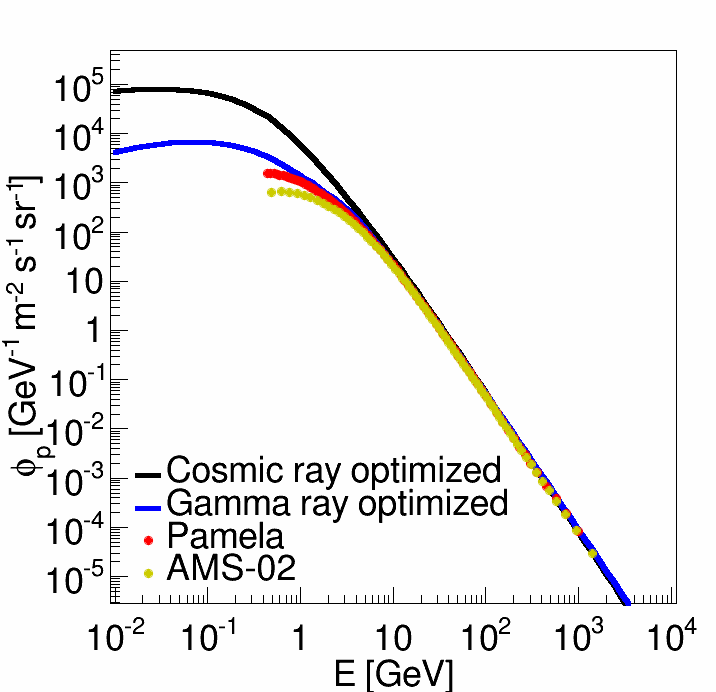}
\includegraphics[width=0.45\textwidth,height=0.35\textwidth,clip]{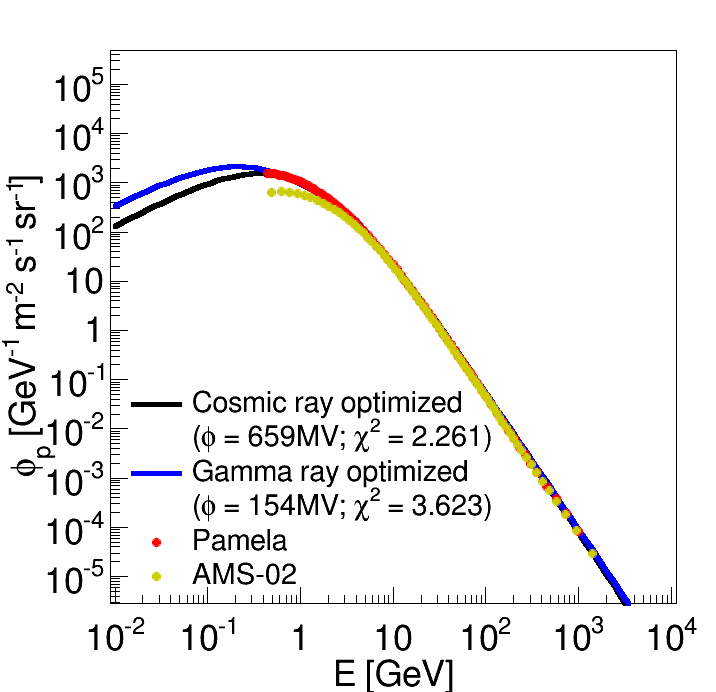}\\
(a) \hspace*{0.4\textwidth}(b)\\
\includegraphics[width=0.45\textwidth,height=0.35\textwidth,clip]{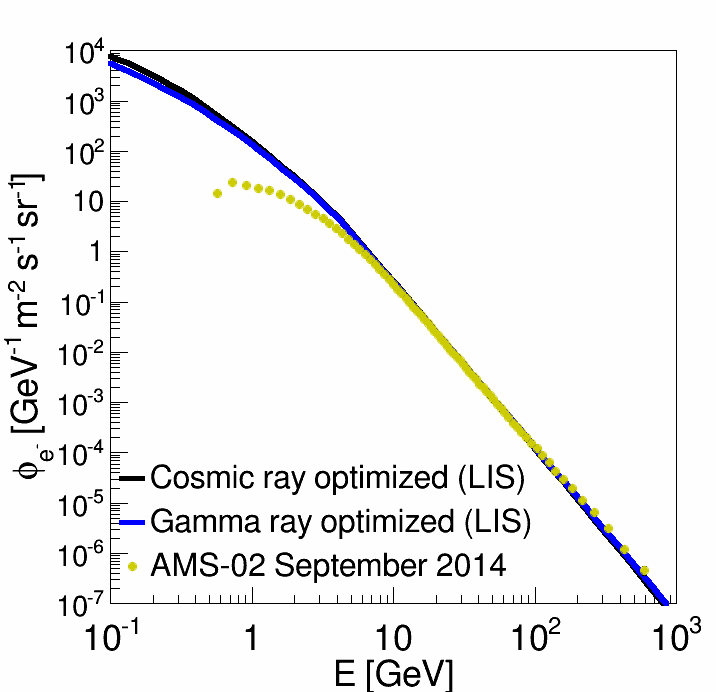}
\includegraphics[width=0.45\textwidth,height=0.35\textwidth,clip]{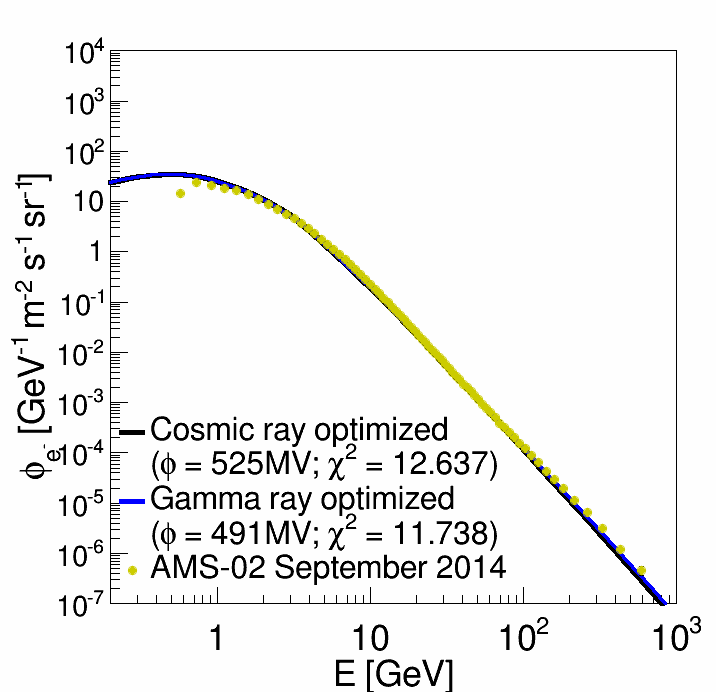}\\
(c) \hspace*{0.4\textwidth}(d)
	\caption[]{(a) The interstellar proton  spectrum for propagation models fitted to charged CRs and gamma-rays simultaneously (curve Gamma ray optimized) and fitted to CRs only (curve Cosmic ray optimized). Note that determining the interstellar spectrum from CRs only is ambiguous, since it depends on the injection spectrum. A break in the injection spectrum  at low energies is fully correlated with the SM potential. Here no break in the injection spectrum of CRs was assumed. (b) The modulated proton spectra in comparison with the Pamela data. The solar modulation parameters have been indicated.
Bottom: same as  top row, but now for electrons in comparison with AMS-02 data.
}
\label{F1}
\end{figure}

Sometimes the SM potential is determined by a comparison of the locally observed spectrum with the CR spectrum of the Voyager I satellite, which left the solar system. However,  it is still inside the bow shock, so it does not really measure the interstellar spectrum. One finds typically potentials between 285 and 650 MV, if compared with the Pamela data\cite{Corti:2015bqi,Webber:2016vgr}. These simple models are not accurate, since during cycles with minimum solar activity also particle drift has to be taken into account\cite{DiFelice:2016oec}. 

The interstellar proton spectrum can also be deduced from the gamma-ray spectra in regions with dominant $\pi^0$ production and excluding the GC and Fermi Bubbles.
This leads to  smaller values of the SM potential in comparison with the values obtained from a comparison with the Voyager I data, namely 160 MV for the Pamela proton spectrum. This value was obtained from a detailed tuning of the Dragon parameters to the locally observed CR spectra and gamma-spectra using an extensive Markov Chain Monte Carlo\cite{Kunz}. The obtained parameters are given in Table \ref{T1}.
The modulated (=locally observed) and unmodulated (=interstellar) spectra for protons and electrons are shown in Fig. \ref{F1}. 
The curves are either optimized for CRs only of include in addition the gamma-rays in the fit. For CRs (gamma-ray) optimized a SM potential of 659 (154) MV is needed to describe the Pamela data, as indicated in Fig. \ref{F1}(b) for the modulated data. These different potentials correspond to an order of magnitude differene in the unmodulated (=interstellar)  spectra  for protons below 1 GV, as demonstrated in Fig. \ref{F1}(a). As can be seen from Fig.  \ref{f1}(b)  an order of magnitude suppression below 1 GV is also the difference between the protons for the PCR and MCR template.
Hence, the solar modulation is highly correlated with the break in the MCR spectrum, since both suppress protons at low energies. And it is the suppression of protons at low energies, which shifts the gamma-ray spectrum to higher energies, the hallmark of the GCE. 

The electron spectra in Figs. \ref{F1}(c) and (d) have been compared with the  AMS-02 data, which were taken later in the 11 year solar cycle, so  a larger solar modulation parameter is needed, as indicated in Fig.   \ref{F1}(d). 
\begin{figure}[]
\centering
\includegraphics[width=0.3\textwidth,height=0.27\textwidth,clip]{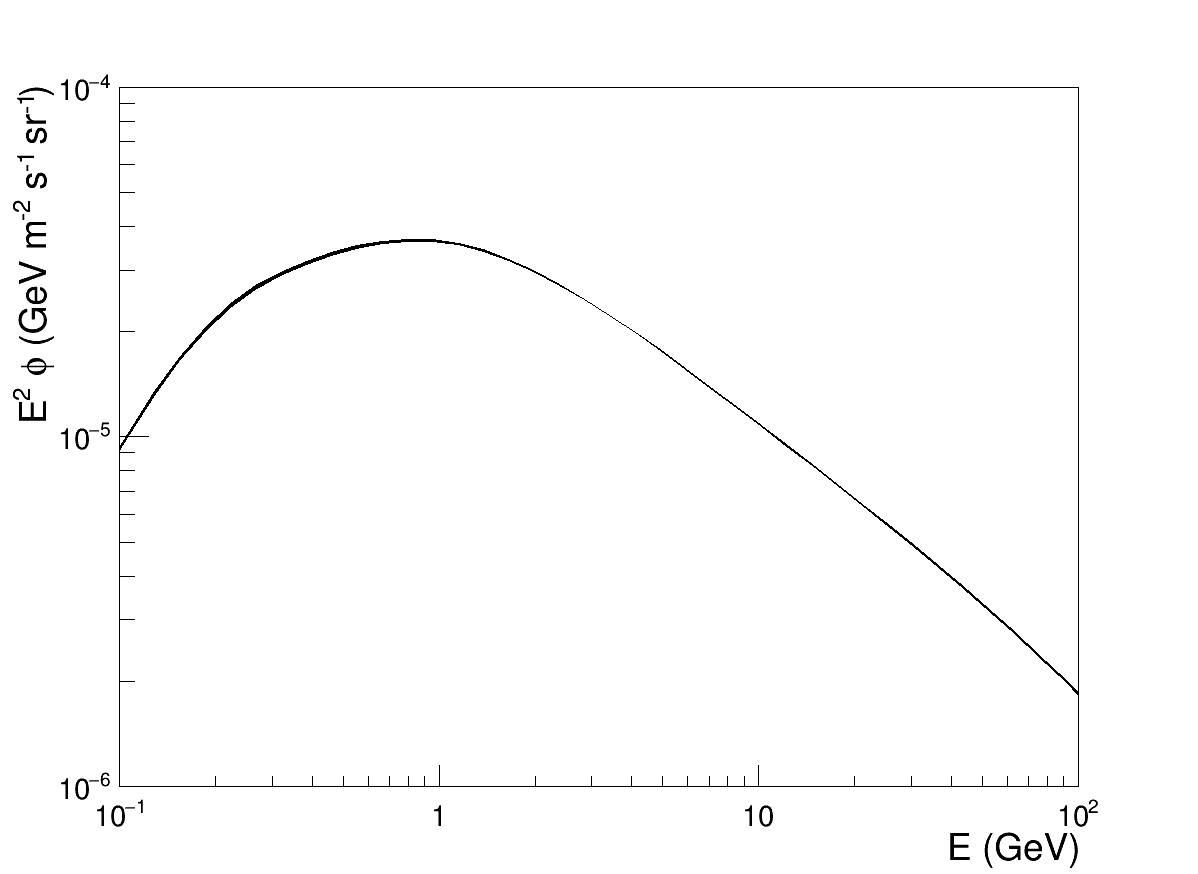}
\includegraphics[width=0.3\textwidth,height=0.27\textwidth,clip]{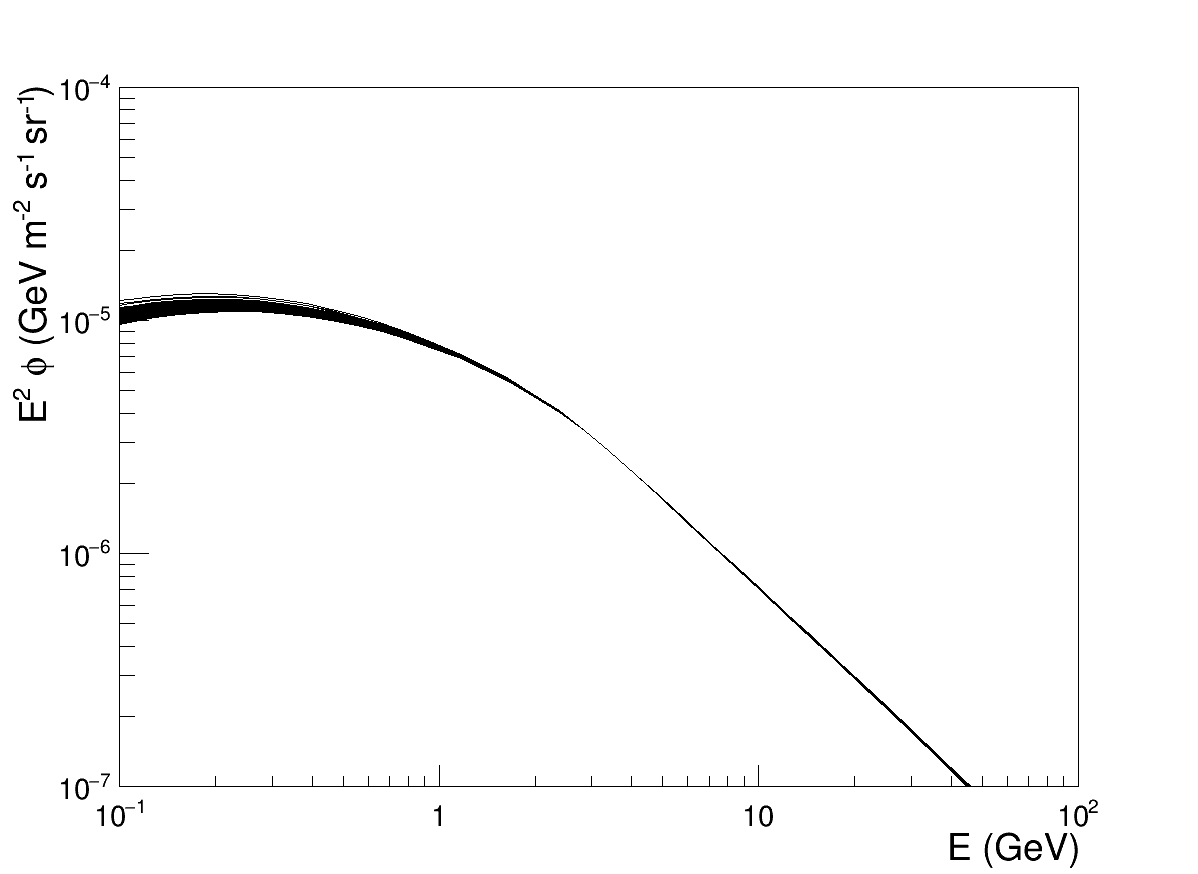}
\includegraphics[width=0.3\textwidth,height=0.27\textwidth,clip]{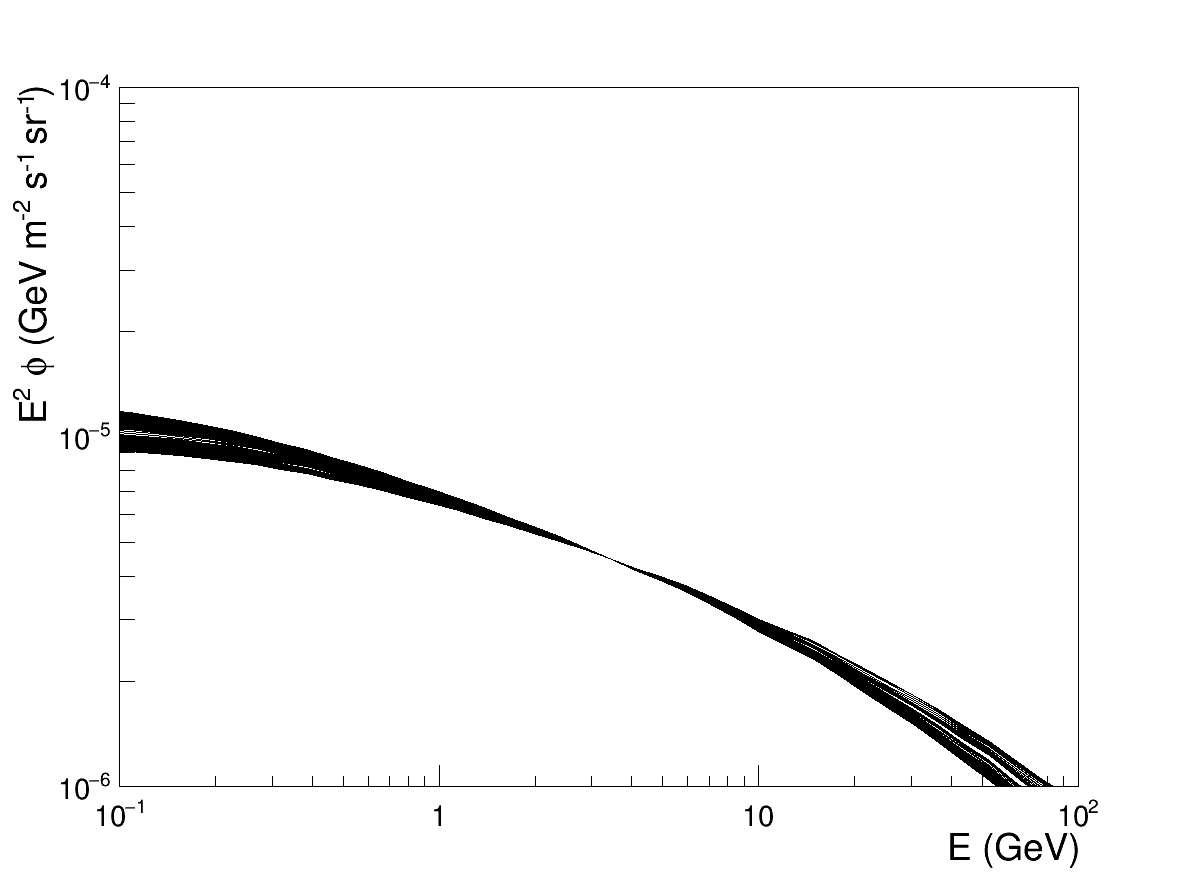}
	\caption[]{From left to right: superimposed templates in all skydirections for PCR, BR and IC contributions.
}
\label{F2}
\end{figure}
\begin{figure}[]
\centering
\includegraphics[width=0.45\textwidth,height=0.35\textwidth,clip]{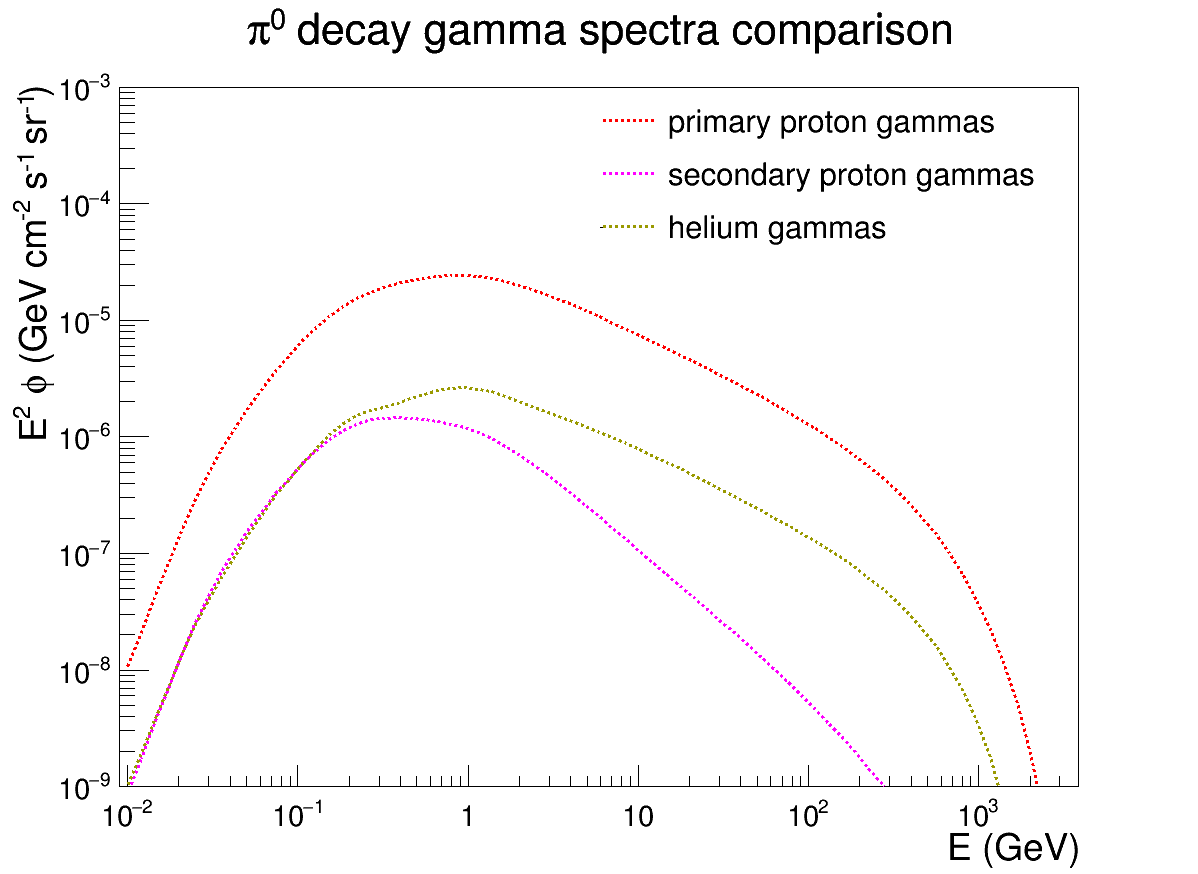}
\includegraphics[width=0.45\textwidth,height=0.355\textwidth,clip]{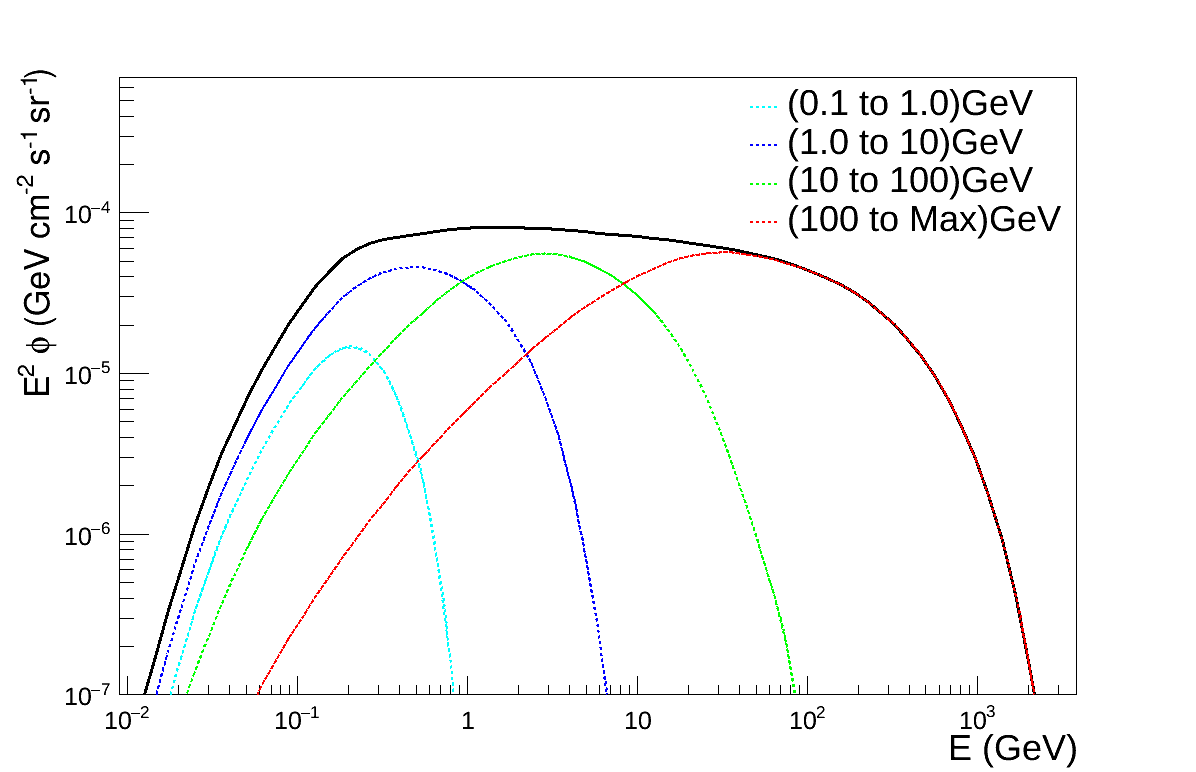}\\
(a) \hspace*{0.46\textwidth}(b)
	\caption[]{(a): A comparison of the gamma-ray spectra from $\pi^0$ production by primary protons, secondary protons produced by collisions and helium nuclei. (b): Gamma-ray spectra from various slices of proton spectra.
If one suppresses protons in the spectrum below 10 GeV the maximum of the spectrum (multiplied by $E^2$) shifts to 2 GeV, as needed to describe the gamma-ray spectrum from MCs.
}
\label{F3}
\end{figure}

With the optimized transport parameters of Dragon fitting the CR data and gamma-ray data outside the central GD and Bubbles one can calculate the gamma-ray templates for the processes included in Dragon, i.e. the  PCR-, BR- and IC-template.
In Fig. 2(a) of the paper these templates are shown for the direction towards the GC. However, the templates depend on the sky direction. In Fig. \ref{F2} the templates in various directions are superimposed on each other. For the PCR template the differences between cones are negligible, as expected from the small energy losses for nuclei. The electron spectrum is affected by the energy losses in the magnetic field, the ISRF and the gas. This leads to a small direction dependence of the the BR template (middle panel of Fig. \ref{F2}). For the IC template the direction dependence  is largest, because in addition to the direction dependence of the electron spectrum, also this ISRF, which is composed among others out of dust, star light and cosmic microwave background, is varying, as shown in the right panel of Fig. \ref{F2}.  All direction dependences of the templates were taking into account in the template fit. The contributions of IC and BR are small, so if the direction dependence of these templates is ignored, similar results are obtained. For the $\pi^0$ production one has also to take the contributions from secondary protons and helium nuclei into account. As can be seen from Fig. \ref{F3}(a) these contributions are at the 10\% level; secondary protons lead to a somewhat softer gamma-ray spectrum than primary protons, while helium leads to a somewhat harder spectrum. These effects  were taken into account.
\begin{figure}[]
\centering
\includegraphics[width=0.45\textwidth,height=0.35\textwidth,clip]{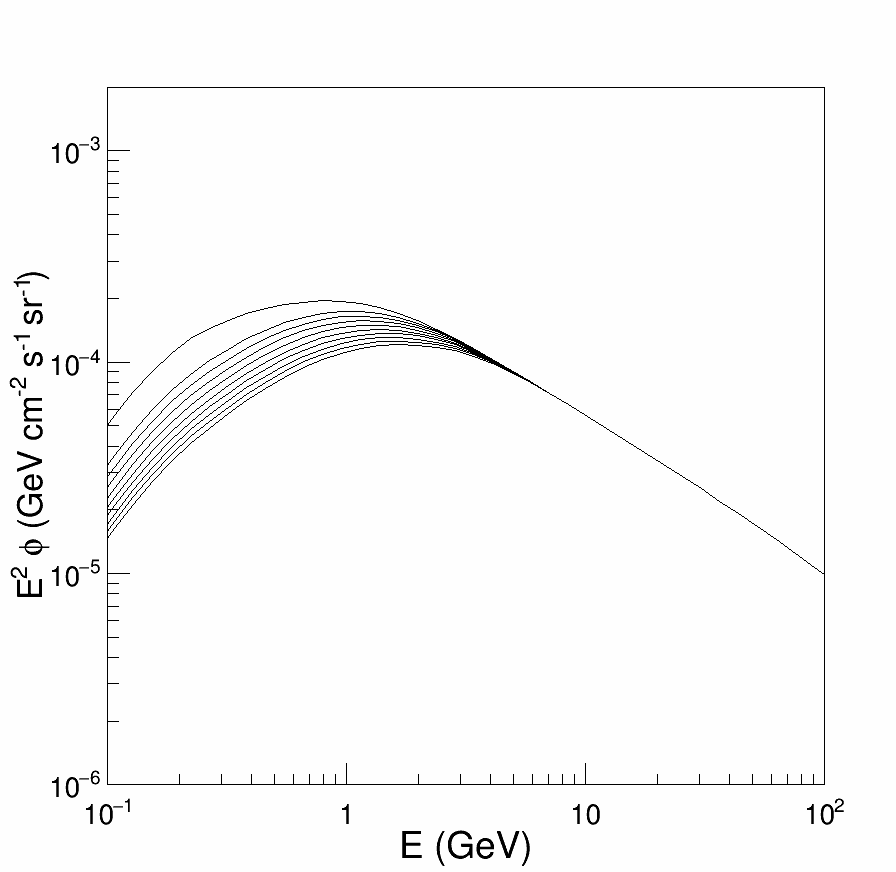}
\includegraphics[width=0.45\textwidth,height=0.35\textwidth,clip]{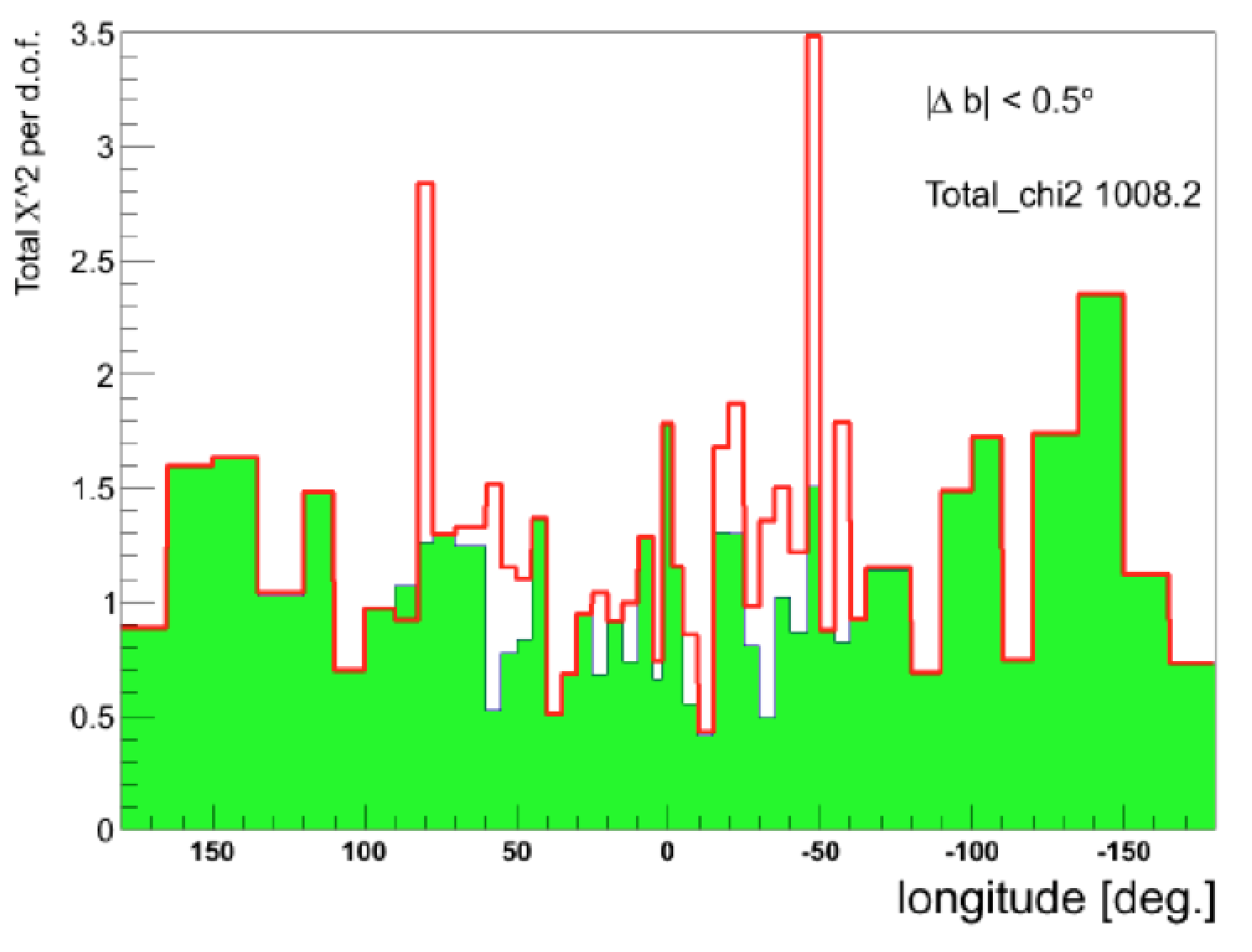}\\
(a) \hspace*{0.46\textwidth}(b)
	\caption[]{(a): A comparison of the gamma-ray spectra from $\pi^0$ production inside MCs for  breaks between 14 and 6 GV in steps of 1 GV, where the highest break point yields the maximum in the spectrum at the highest energy. (b): $\chi^2$ values for a constant break of 14 GV (red) and for varying breaks (green histogram).
}
\label{F4}
\end{figure}
\begin{figure}[]
\centering
\includegraphics[width=0.45\textwidth,height=0.35\textwidth,clip]{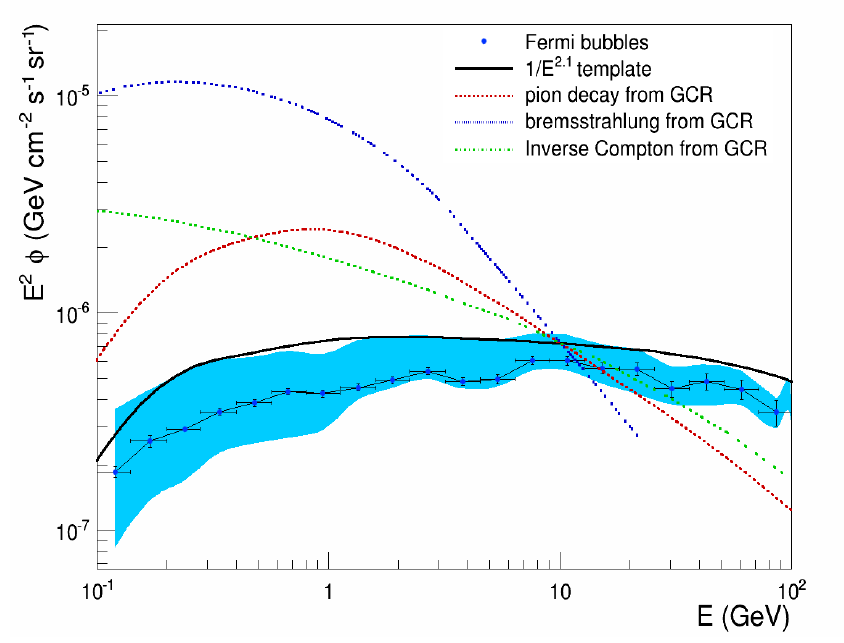}
\caption[]{A compilation of the various templates. The SCR template (black line) agrees  well with the Bubble template (blue band), as measured by the Fermi Collaboration\cite{Fermi-LAT:2014sfa}.
}
\label{F5}
\end{figure}

\subsection{The MCR Template}
As observed from the direction of the CMZ in the GC the gamma-ray spectrum has its maximum shifted to 2 GeV.
This can be obtained by suppressing the proton spectrum below about 10 GV, as can be deduced of Fig. \ref{F3}(b). Unfortunately, a suppression below 10 GV is strongly correlated with the SM, as discussed above. Hence, it is important to determine first the SM from regions with small contributions from MCs and then fit the template from MCs, e.g. by fitting the region towards the CMZ. 

A  suppression of a proton spectrum below a certain rigidity can be done by introducing a break at that rigidity
with  different slopes below and above the break. As shown in Fig. \ref{f2}(b) the high energy data towards the GC and in the opposite direction show a similar slope at high energies. In the first direction the PCR template dominates, in the second direction the MCR spectrum. Therefore, the spectral index of the MCR spectrum above the break was fixed to the slope needed for the  injection spectrum of the PCR template above the break (2.16, see Table \ref{T1}). Then we are left with only 2 parameters: the position of the break and the slope below the break. The  proton spectra were then generated for a range of breaks (6-14 GV) and a range of slopes (0.6-1.3) and propagated in the Dragon program described in the previous section. The resulting $\pi^0$ spectra and the corresponding gamma-ray templates were then generated with the Dragon code and used as a template in the fit. The templates for the different  break points are shown in Fig. \ref{F4}(a). The slope below the break point depends on the geometry of the magnetic field, i.e. the solid angle of the magnetic pole region, where there is no cutoff and the solid angle of the magnetic equator region.  It turns out that the slope is similar for all MC regions, suggesting a similar magnetic field morphology   for all MCs.

 For the CMZ the best fit was obtained for an MCR template with  a cutoff of in the injection spectrum at 14 GV and a slope of the injection spectra of  -0.7 (-2.16) below (above) the break. This template was applied to the whole sky and yields a good $\chi^2$ in most regions of the GD, as shown by the red line in Fig. \ref{F4}(b). Only in the Cygnus region at $l=80^\circ$ and the tangent point of the nearest spiral arm at $l=-50^\circ$ the $\chi^2$ is worse, as shown by the peaks in the red line  in Fig. \ref{F4}(b). Here a lower break is needed. By offering the fit each of the templates in Fig. \ref{F4}(a) the fit can decide on the best break, which results in the $\chi^2$ distribution given by the green histogram in Fig. \ref{F4}(b). 
 In regions, where the fit improves, the density of the MC is presumably lower, leading to a lower magnetic field with a somewhat lower cutoff. The magnetic field varies only with the square of the density, which is presumably the reason, why the break  in the rather narrow range of 6 to 14 GV yields a good fit.

\subsection{The SCR Template}
The Fermi Collaboration has determined the energy spectrum in the Bubbles\cite{Fermi-LAT:2014sfa}. We found that
this spectrum is well described by $\pi^0$ production from  a proton spectrum with a $1/R^{2.1}$ rigidity dependence, as can be seen from Fig. \ref{F5}. The latter spectrum is the spectrum expected for SCRs,  so the same template can be used for the SCR flux and the Bubbles, as discussed previously\cite{2014ApJ...794L..17D}. The fit also does not need templates for what is called the Loop I structure, since this is recognised by the usual background templates, mainly the PCR template.  
\begin{figure}[]
\centering
\includegraphics[width=0.45\textwidth,height=0.35\textwidth,clip]{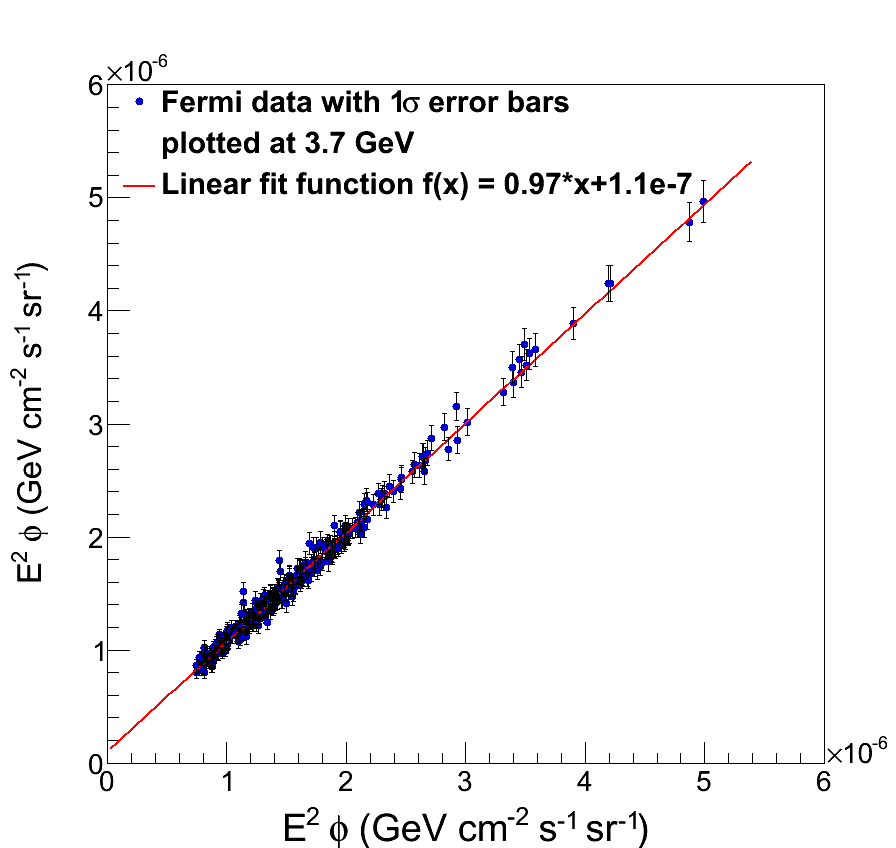}
\includegraphics[width=0.45\textwidth,height=0.35\textwidth,clip]{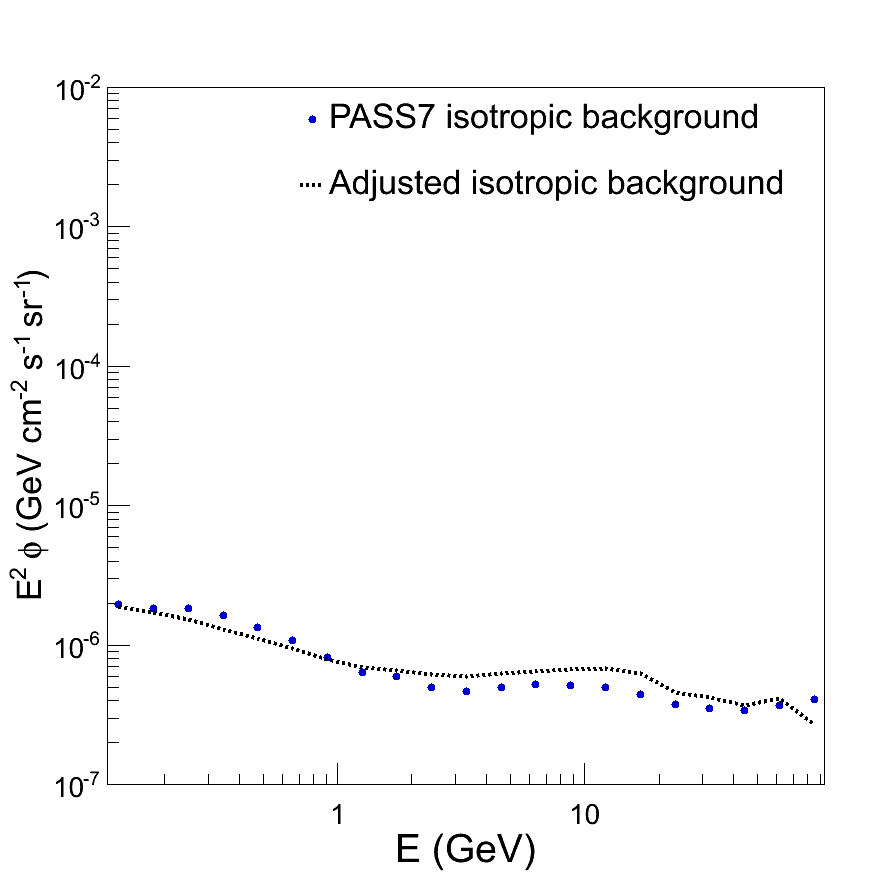}\\
(a) \hspace*{0.46\textwidth}(b)
	\caption[]{(a): The observed data versus the fitted data in various sky regions (i.e. various fluxes) for a given energy. The offset of a linear fit at the vertical axis represents the isotropic component in the data, which moves the data in all bins upwards by this amount. The data is different from the fit in all bins, if the isotropic template does not correspond to the isotropic component in the data. (b) A comparison of the isotropic template used in our analysis and the isotropic template given in the Fermi software.
}
\label{F6}
\end{figure}
\subsection{The isotropic Template}
The isotropic template represents the contribution from the extragalactic background and hadron misidentification. Its spectral shape and absolute normalisation are provided within the Fermi software\cite{FST}.  The isotropic template was redetermined for our analysis in the following way. We fit the data in regions outside the Bubbles and GD {\it without}  the isotropic template in the fit. If one plots  the observed flux versus the fitted flux in the various cones in a certain energy bin, one expects a linear relation with an offset in the observed flux given by the isotropic component, if one extrapolates the fitted flux to zero.  An example of such a fit is shown in Fig. \ref{F6}(a) for an energy bin between 3.7-5.2 GeV. This offset can be determined for each energy bin, which yields a first order  energy template of the isotropic component. One can iterate the procedure by entering this  first order template into the fit and look for deviations from the first order template. Typically, after a few iterations a stable template is obtained. The resulting template has  deviations from the Fermi template of the order of 20\%, as shown in Fig. \ref{F6}(b), but  these deviations improved the fit significantly in practically all sky regions given the small errors in the Fermi data.
\begin{figure}[]
\centering
\includegraphics[width=0.9\textwidth,height=0.45\textwidth,clip]{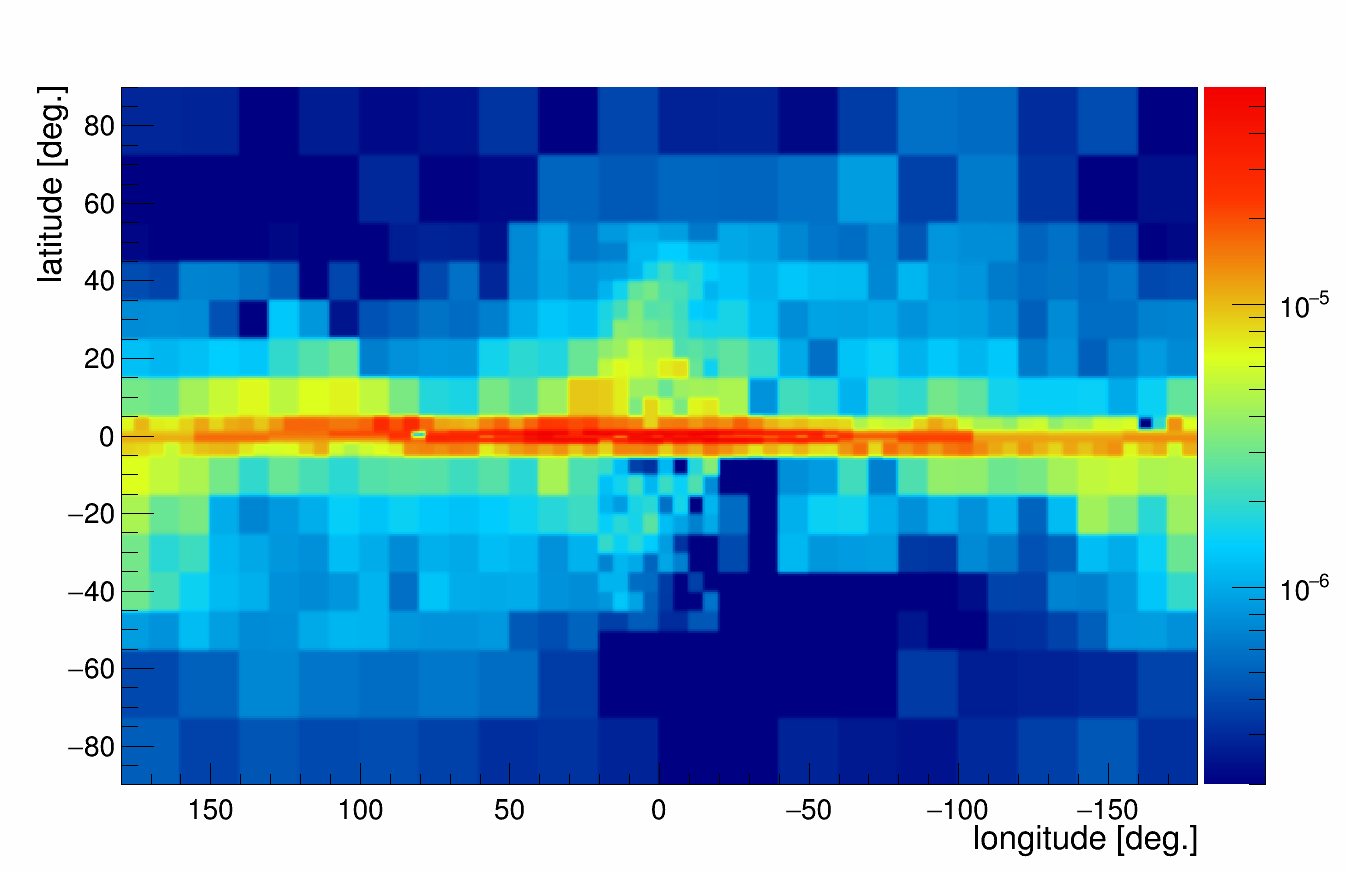}\\
\center{(a)}\\
\includegraphics[width=0.4\textwidth,height=0.4\textwidth,clip]{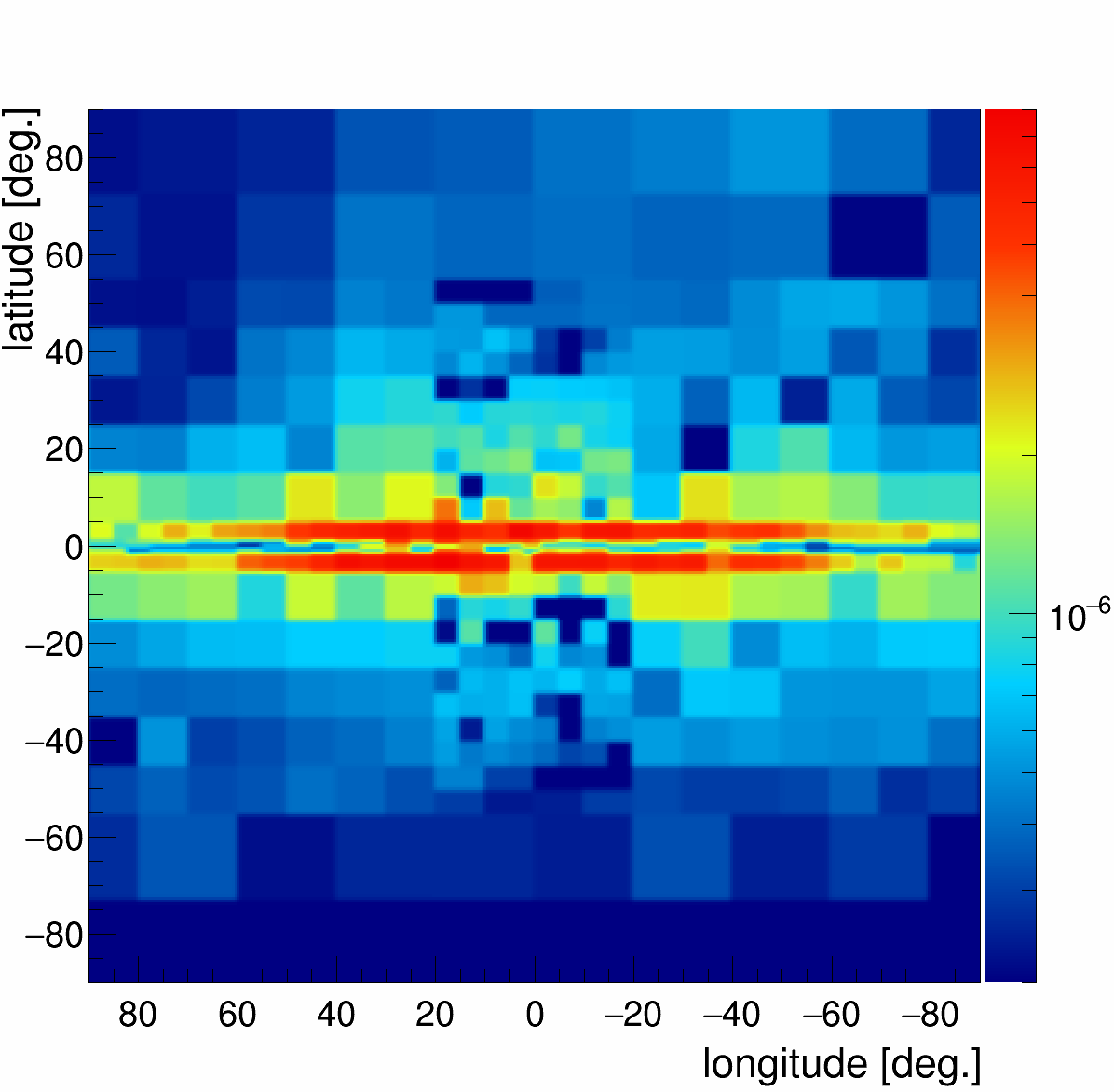}
\includegraphics[width=0.4\textwidth,height=0.4\textwidth,clip]{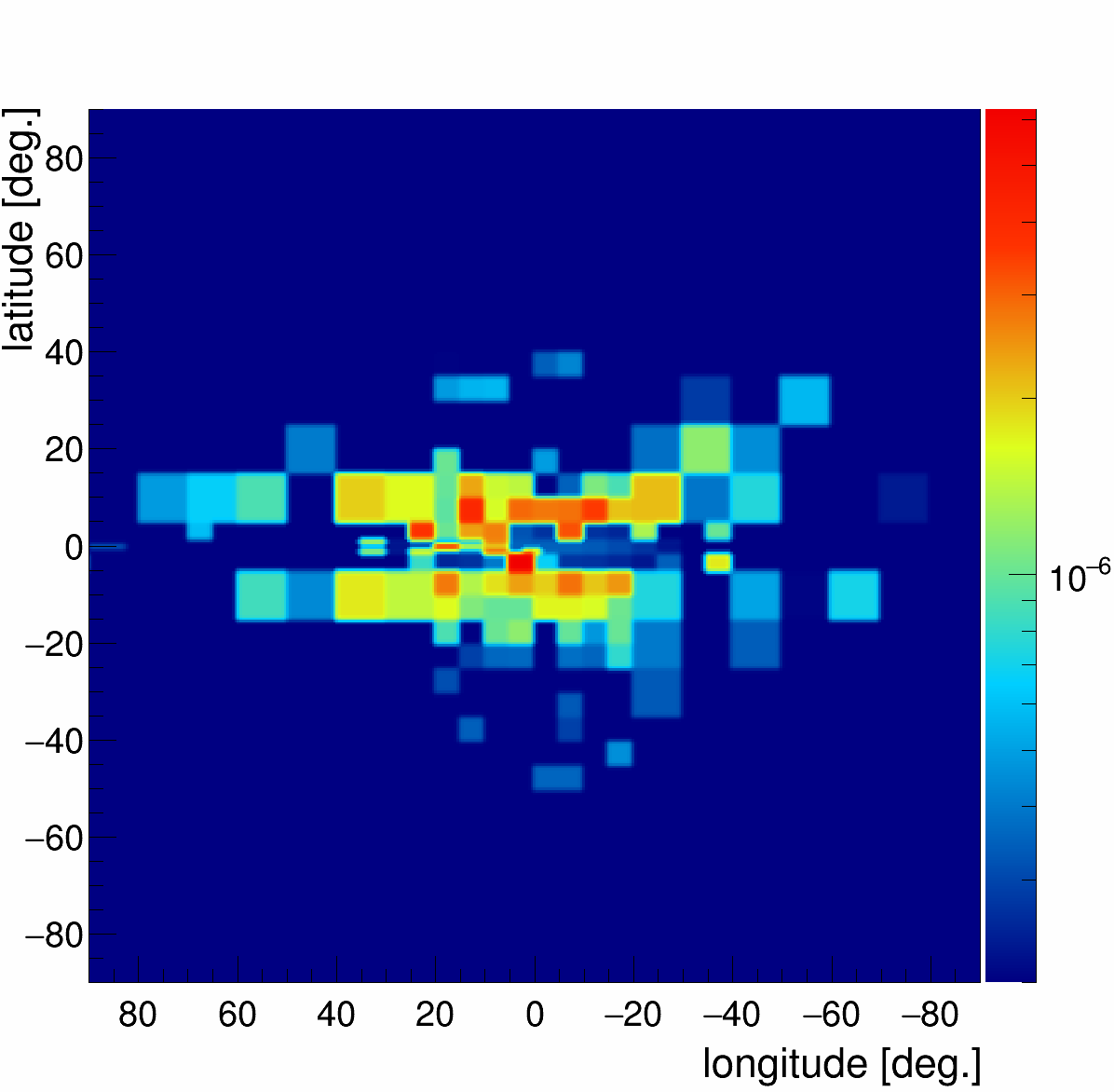}\\
(b)\hspace*{0.46\textwidth}(c)\\
\includegraphics[width=0.4\textwidth,height=0.4\textwidth,clip]{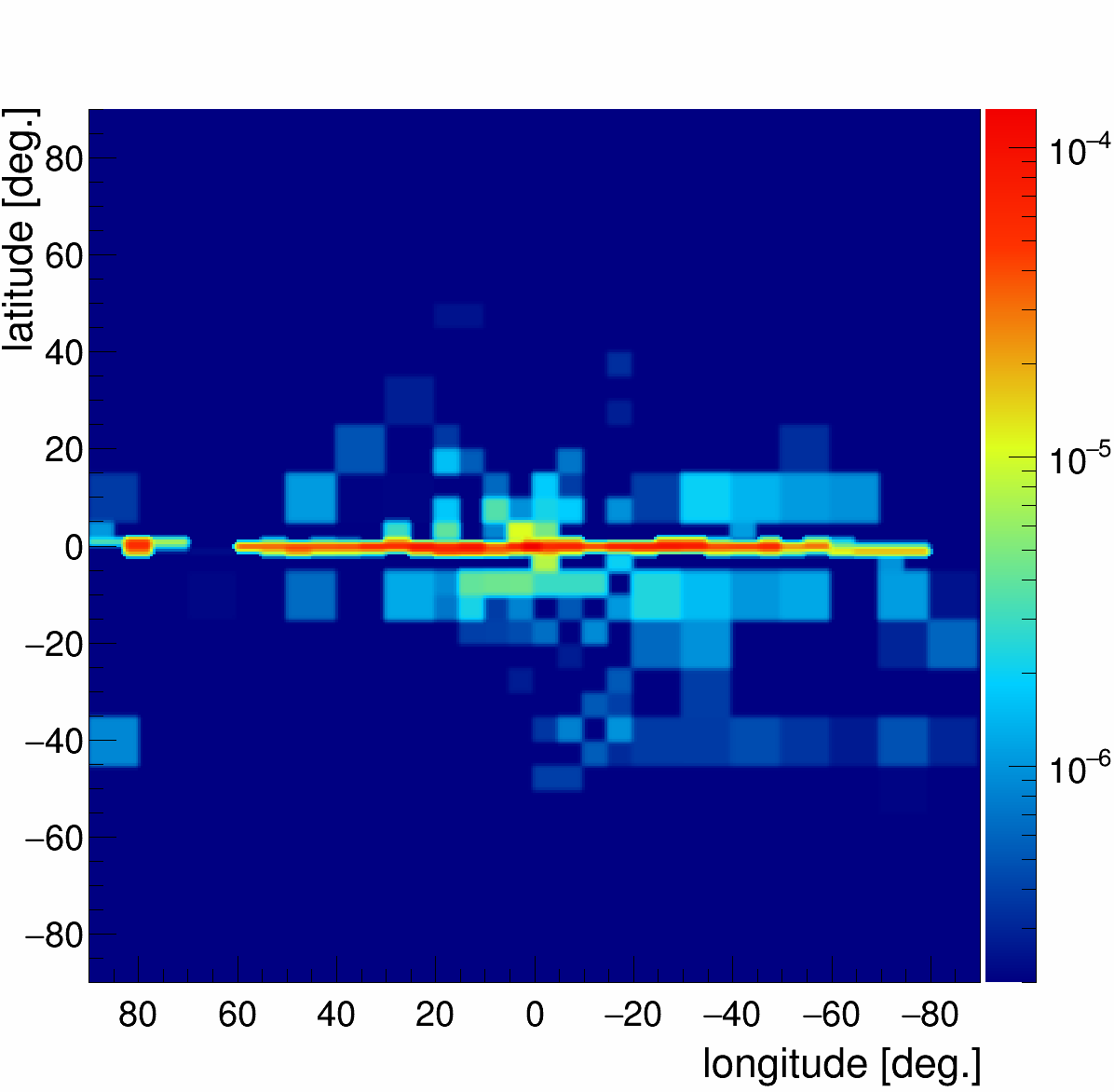}
\includegraphics[width=0.4\textwidth,height=0.4\textwidth,clip]{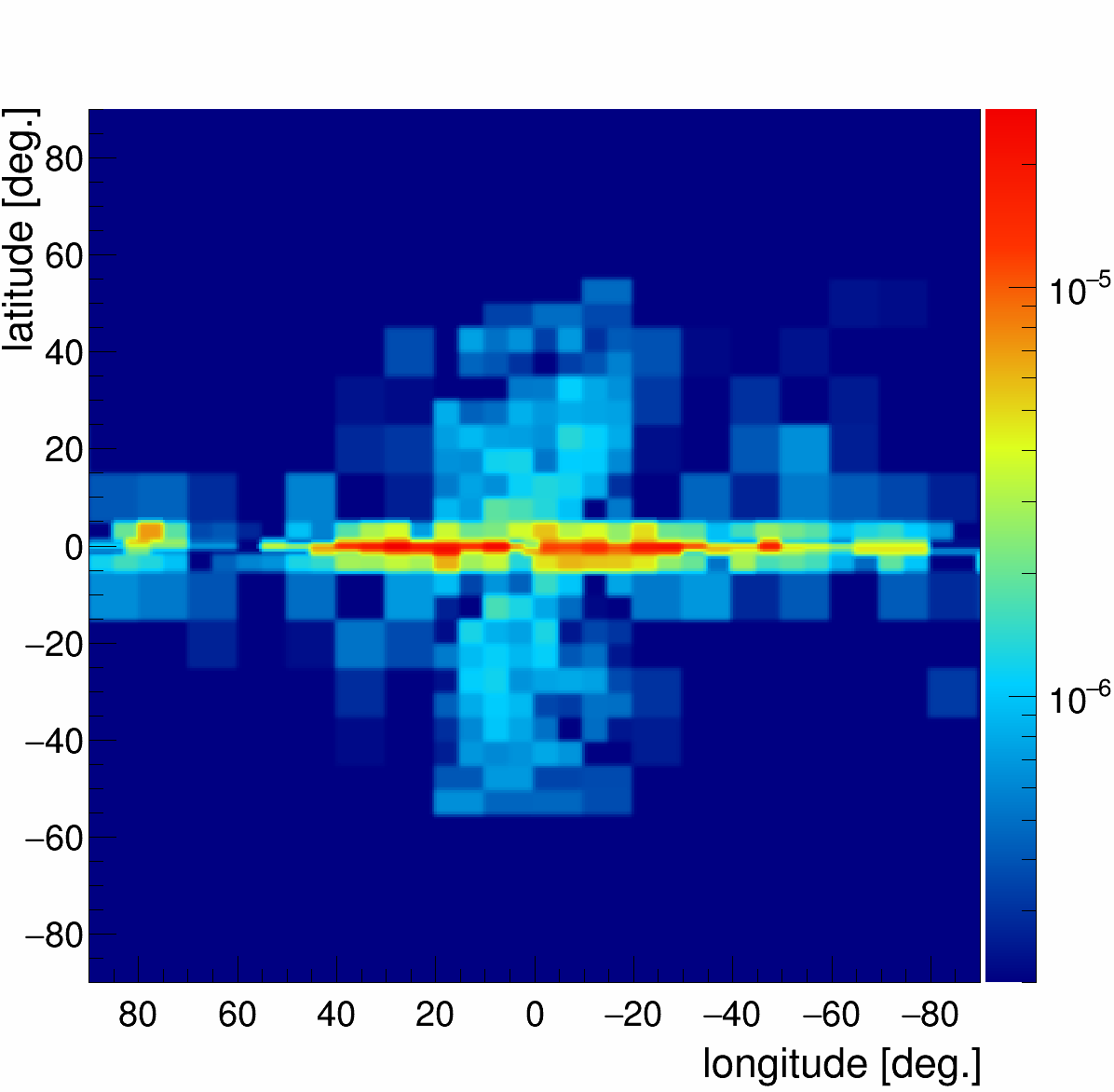}\\
(d) \hspace*{0.46\textwidth}(e)
	\caption[]{Sky maps of the fluxes at 2.3 GeV from the following templates: PCR (a);  BR (b):  IC (c);  MCR (d) and SCR (e). The scale is in units of $\rm GeV ~cm^{-1} s^{-1} sr^{-1}$. The SCR template shows  clearly the contributions in the halo (Fermi Bubbles) and in the GD (sources).}
	\label{F7}
\end{figure}
\begin{figure}[]\centering
\centering
\includegraphics[width=0.45\textwidth,height=0.35\textwidth,clip]{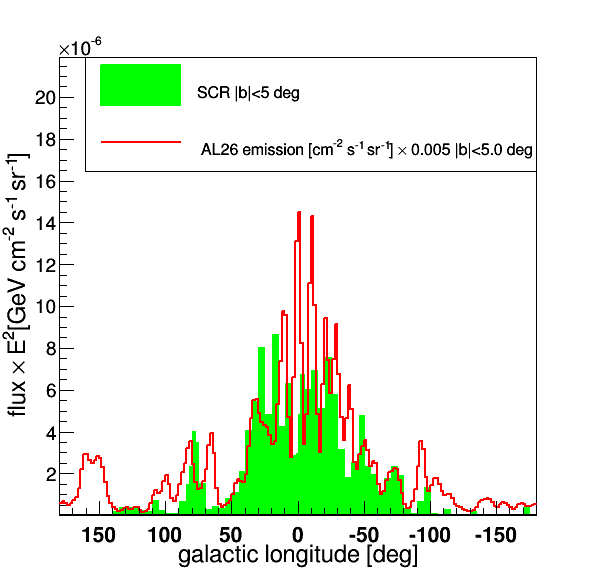}
\includegraphics[width=0.45\textwidth,height=0.35\textwidth,clip]{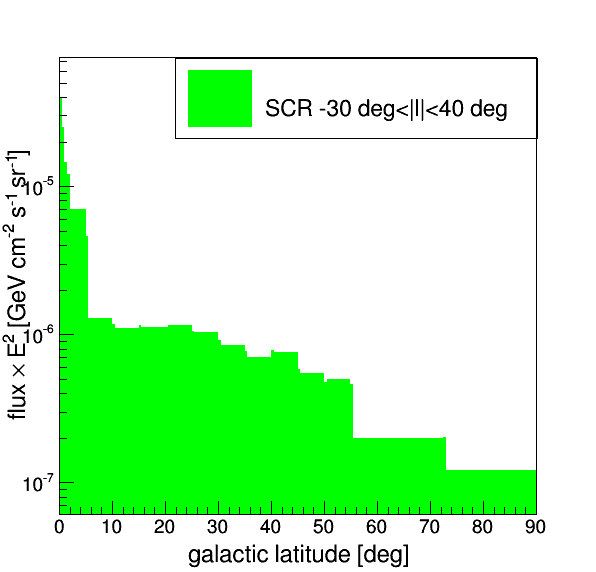}\\
(a) \hspace*{0.46\textwidth}(b)
	\caption[]{Longitude (a) and absolute latitude (b) distribution  of the SCR fluxes (green). In the longitude distribution  the $^{26}$Al  flux is shown as well (red line), which traces the sources. The long tail in latitude, till $|b|=55^\circ$, originates from the Fermi Bubbles, which are not traced by he $^{26}$Al  flux, so the  $^{26}$Al  flux is not shown.
}
\label{F8}
\end{figure}

\section{Details on Fit Results}\label{C}
The templates discussed in the previous section were used in all sky regions. The fit is supposed to find if the expected backgrounds from the PCR, BR, IC and ISO templates fit the data or if the maximum of the spectrum is shifted (a feature recognised by the MCR template) or if the data has a high energy excess above the expectations from the known backgrounds (a feature recognised by the SCR template). As shown in the paper the fluxes of the MCR and SCR templates are strongly  correlated in space in the GD, namely only at positions of MCs, as traced by the CO sky map.  The SCR fit is not only found in MCs, where the sources are, but also in the halo in the form of Fermi Bubbles, which suggests that they are  connected, e.g. because the  CR pressure inside the CMZ zone is high enough to overcome the magnetic pressure and gravity, which results in an outflow of a plasma from the CMZ with  the CRs either trapped inside the plasma or being reaccelerated in the shock wave of the plasma (or both). Of course, the same spectrum does not prove that the Bubbles and the sources are connected, since there are other processes to produce high energy gamma-rays. However, it is difficult to have other processes, which have the spectrum originating from $\pi^0$ production by CRs with a $1/E^{2.1}$ spectrum from SCRs.
 
In Fig. \ref{F7} we show the complete sky maps of every template at an energy of 2.3 GeV, as obtained from the normalisation factors in the fitted flux in Eq. \ref{e2}. One observes  the Fermi Bubbles  and its transition into the disk  in Fig. \ref{F7}(e) and the intense contributions of the MCR and SCR components in the GD in  Figs. \ref{F7}(d) and (e). Note the much narrower MCR contribution in the GD, because this contribution is only found  in the MCs. The sources, and hence the SCR contribution, can have outflows into the GD, like the Fermi Bubbles.

Fig. \ref{F8}  shows the correlation between the SCR distributions in longitude and latitude in comparison with the $^{26}$Al distributions.  The correlation is not perfect, since  $^{26}$Al is synthesized by proton capture of $^{25}$Mg, so it only happens in heavy, magnesium rich stars\cite{Prantzos1996}. The gamma-ray emission from SCRs happens inside all sources, as well as in the Fermi Bubbles, which are apparent in the step-like decrease in the latitude distribution at $|b|\approx 55^\circ$ in Fig. \ref{F8}(b). In addition, one observes that the latitude distribution of $^{26}$Al is not as strongly peaked in the GC as the SCR emission, which is expected, since the SCRs are confined to the sources, while the $^{26}$Al radioactive element with a lifetime of the order of $10^6$ years can propagate  before emitting the 1.809 MeV line\cite{Prantzos1996}.

\begin{figure}[]
\centering
\includegraphics[width=0.45\textwidth,height=0.35\textwidth,clip]{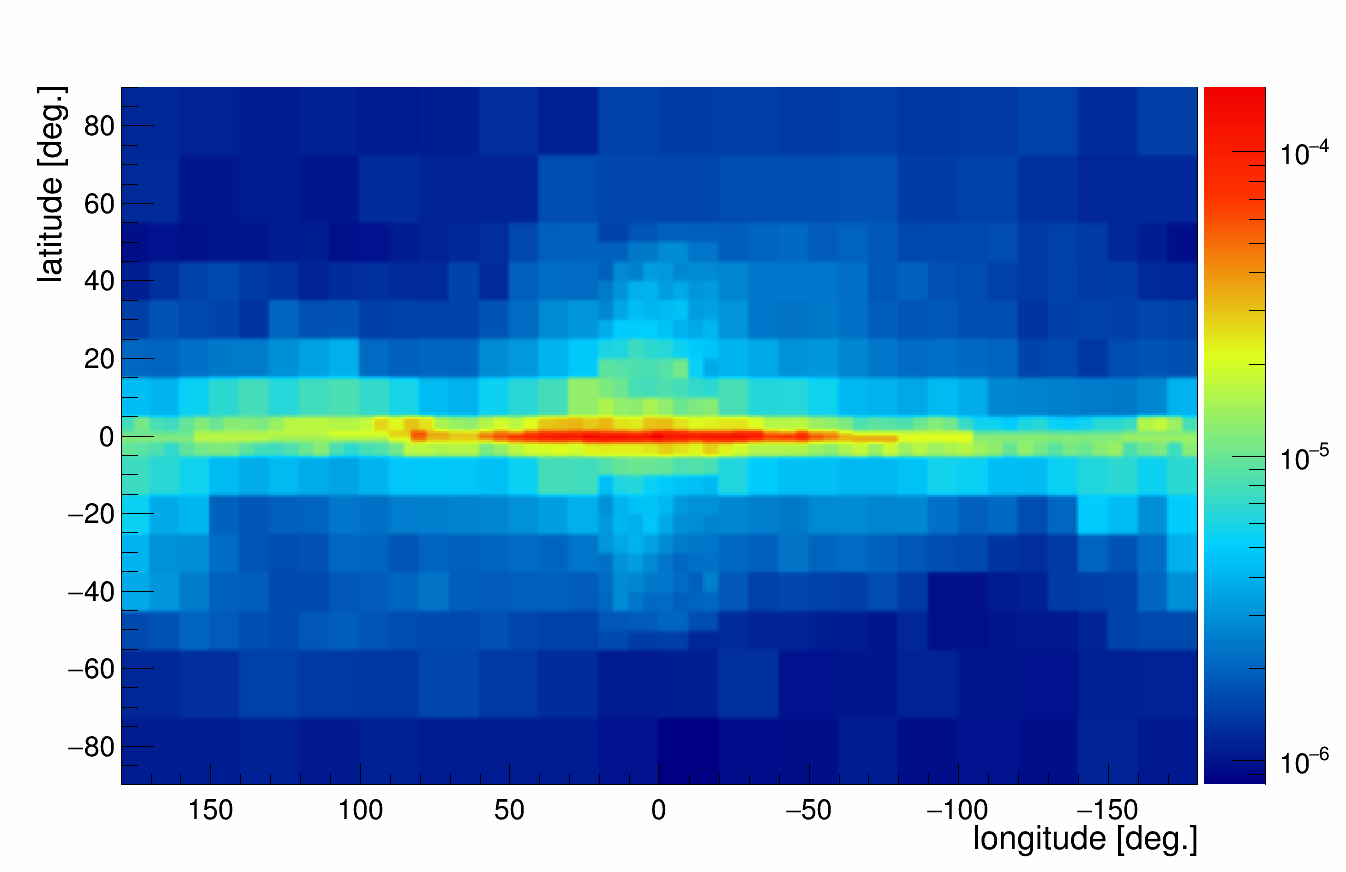}
\includegraphics[width=0.45\textwidth,height=0.35\textwidth,clip]{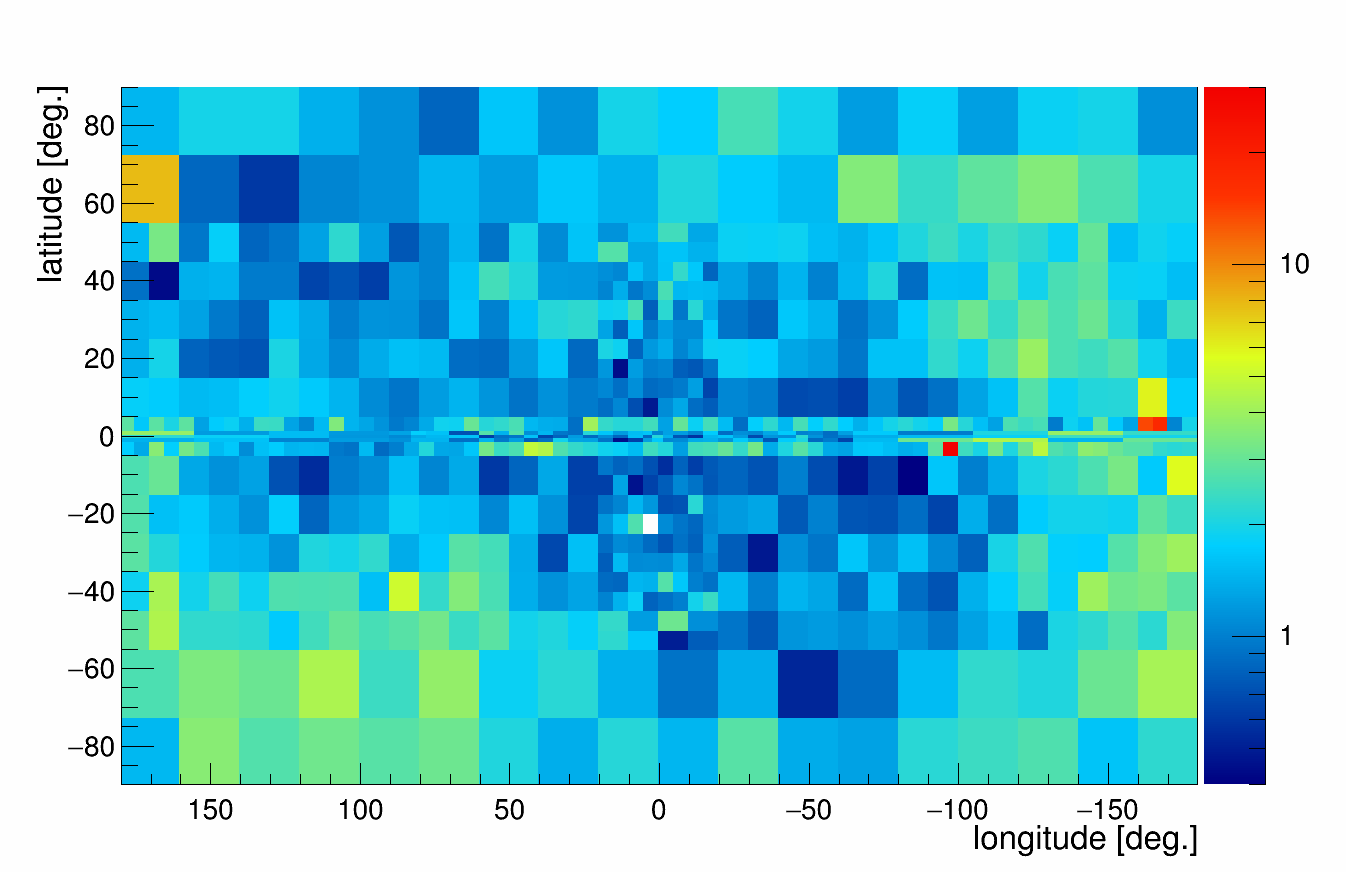}\\
(a) \hspace*{0.46\textwidth}(b)
	\caption[]{ (a) Sum of the template fluxes and (b) $\chi^2/d.o.f.$ distribution in  the l,b-plane.
}
\label{F9}
\end{figure}

\begin{figure}[]
\centering
\includegraphics[width=0.24\textwidth,height=0.24\textwidth,clip]{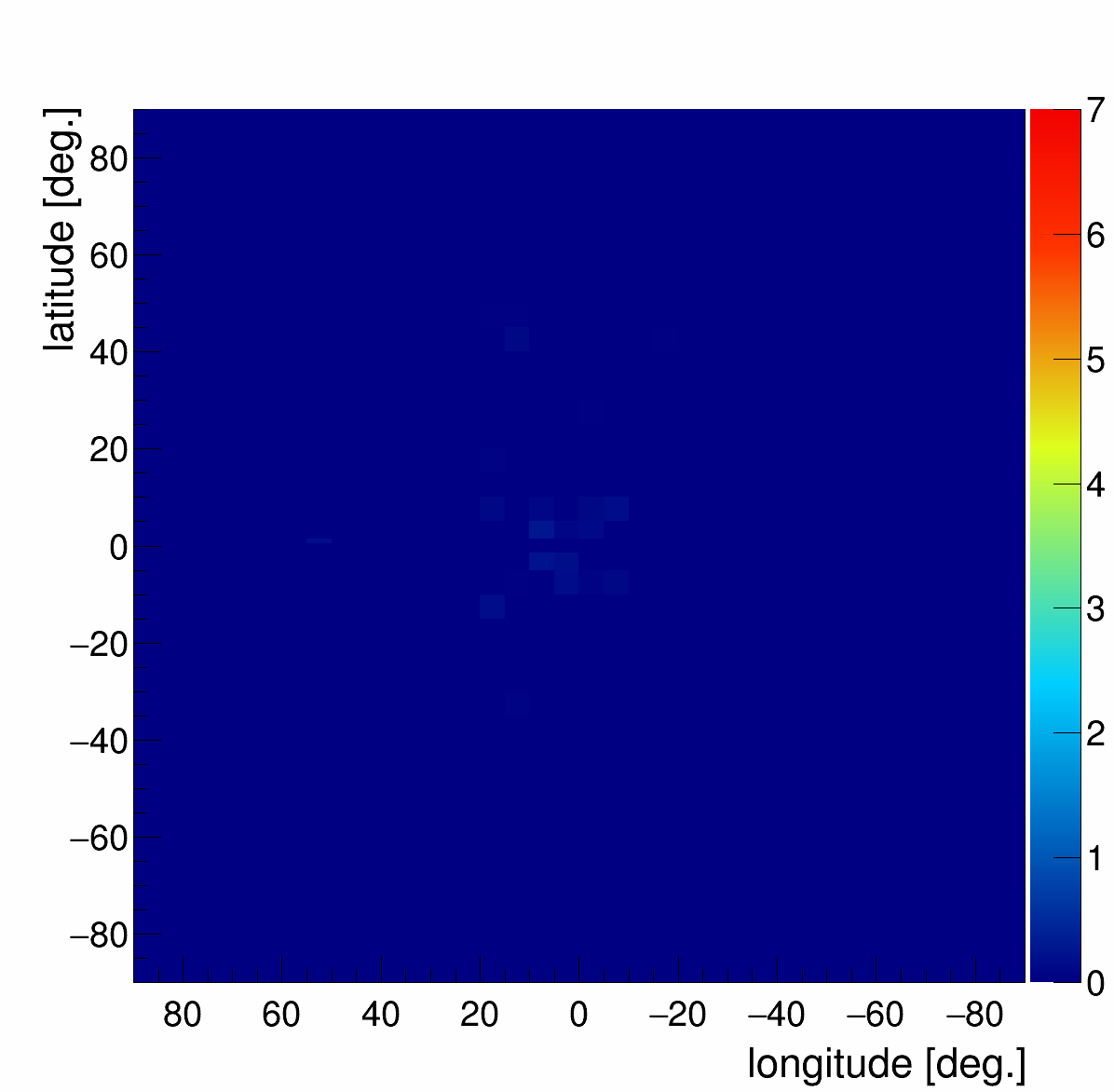}
\includegraphics[width=0.24\textwidth,height=0.24\textwidth,clip]{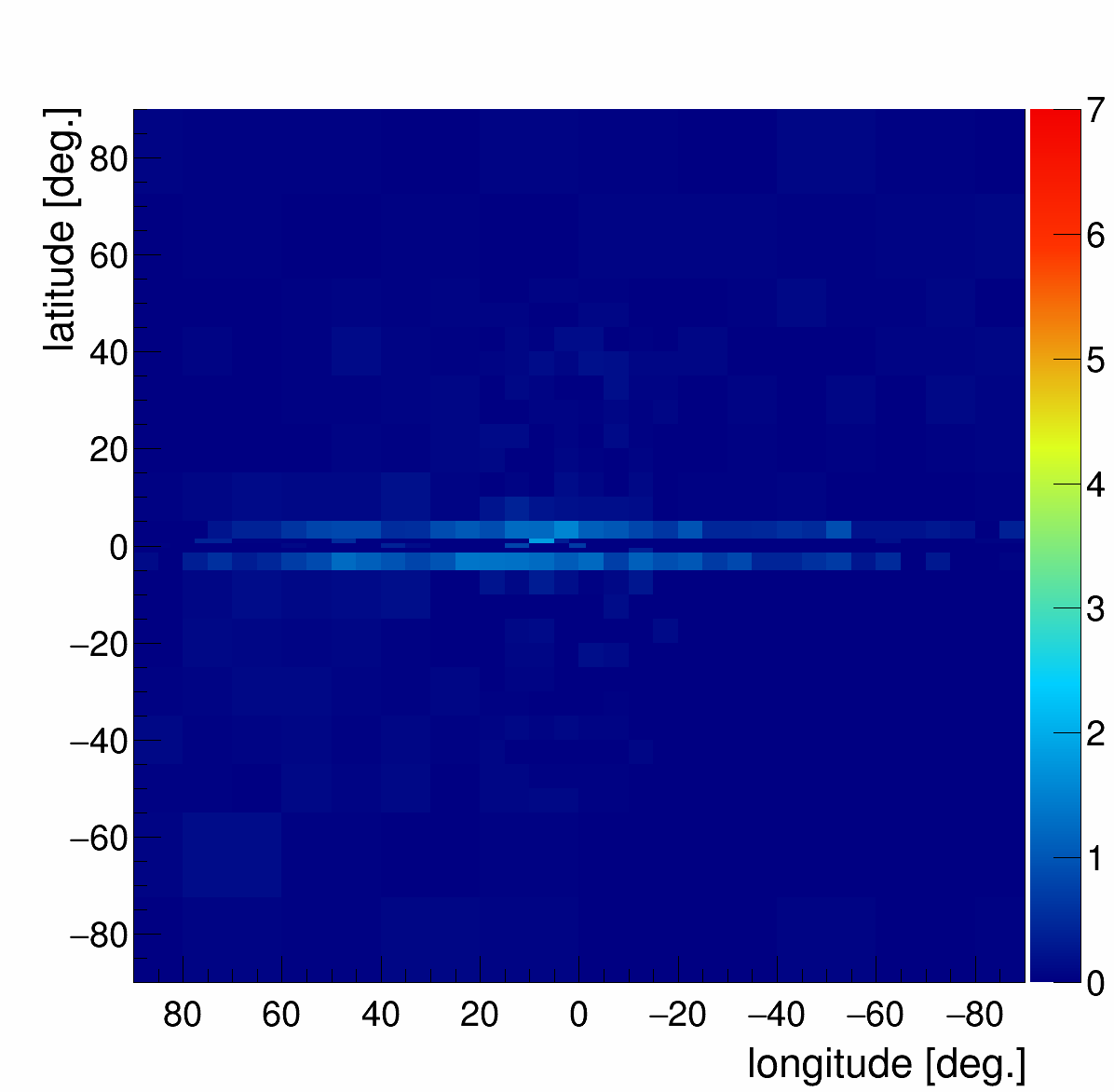}
\includegraphics[width=0.24\textwidth,height=0.24\textwidth,clip]{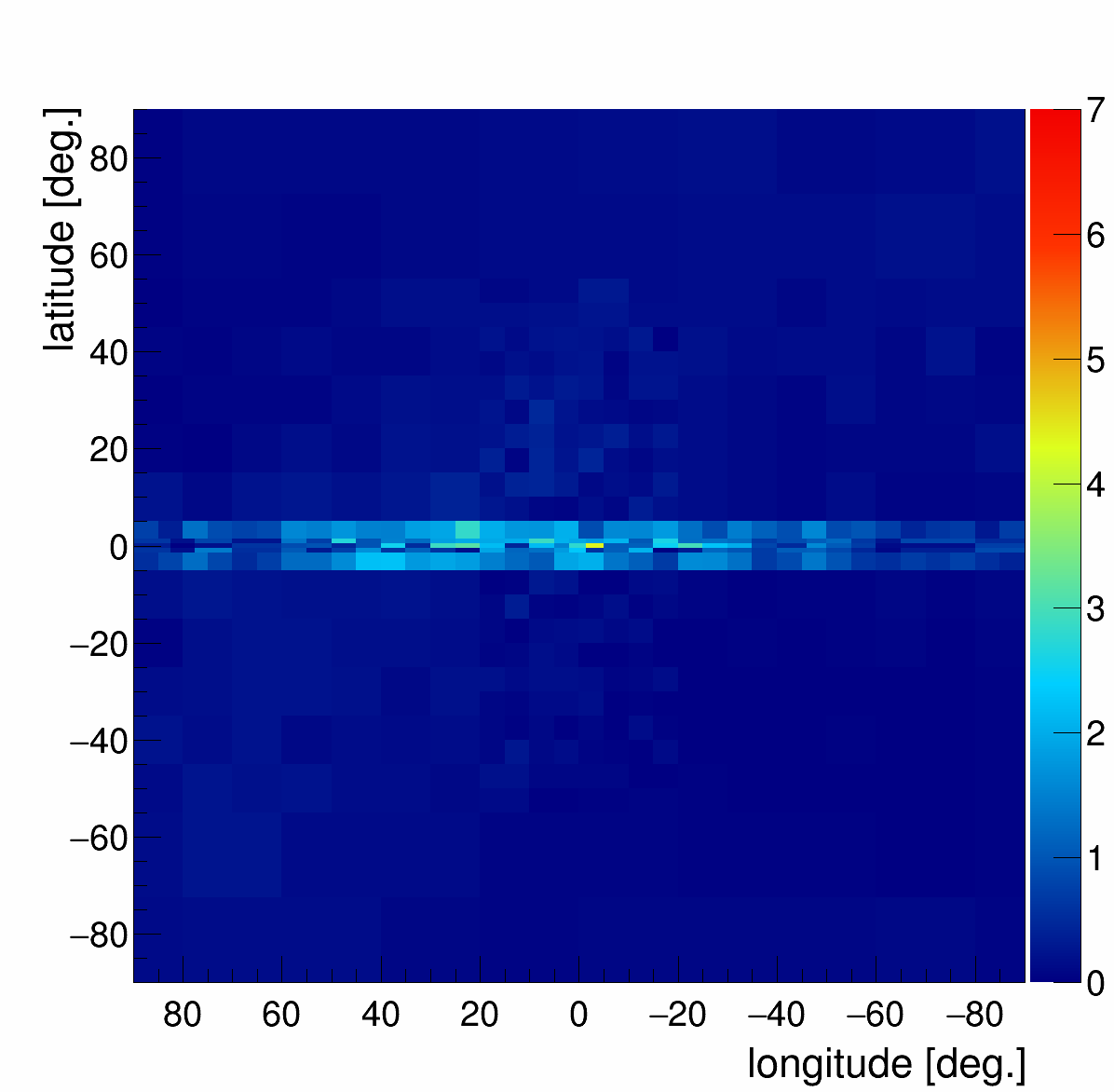}
\includegraphics[width=0.24\textwidth,height=0.24\textwidth,clip]{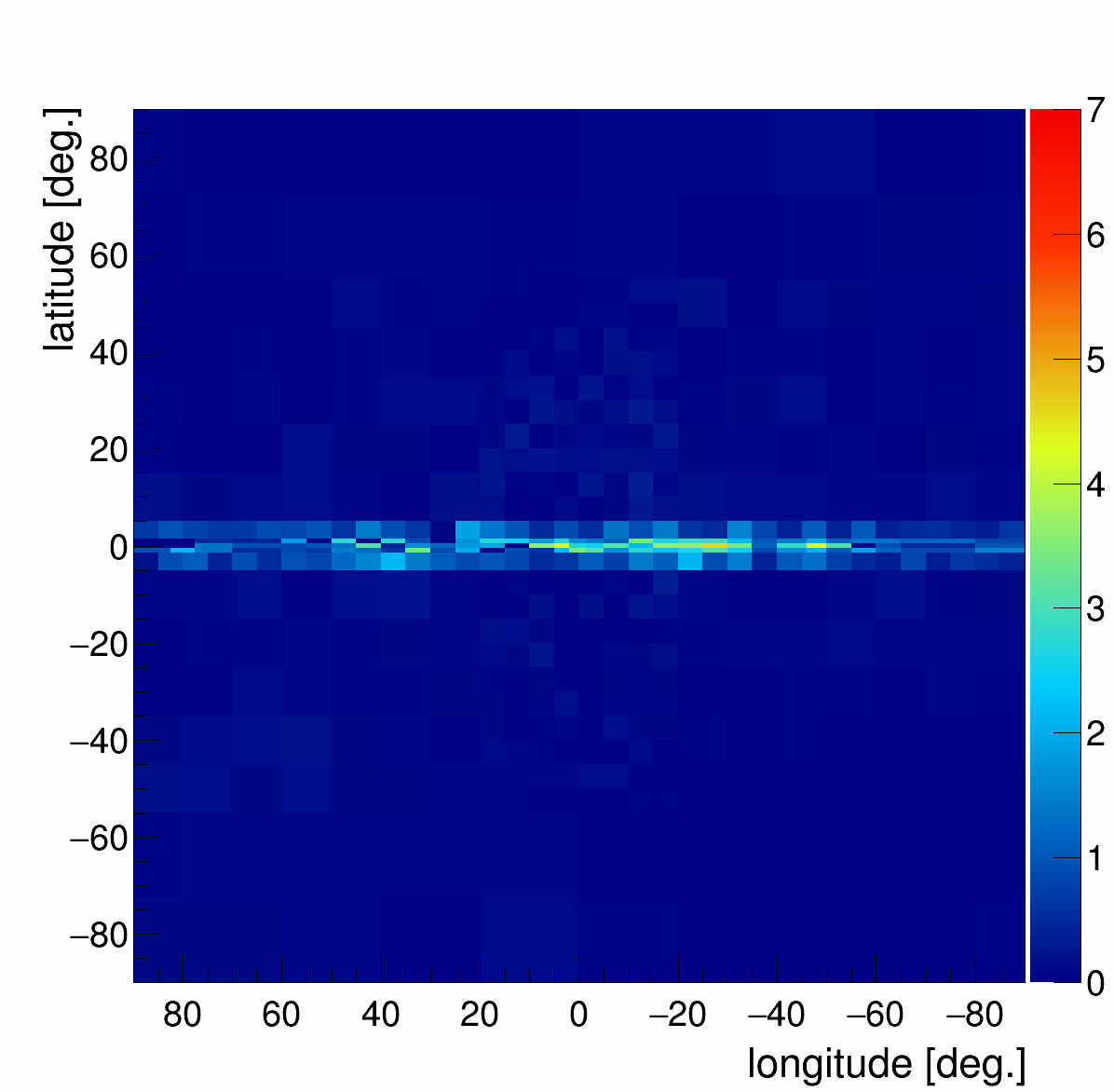}
\includegraphics[width=0.24\textwidth,height=0.24\textwidth,clip]{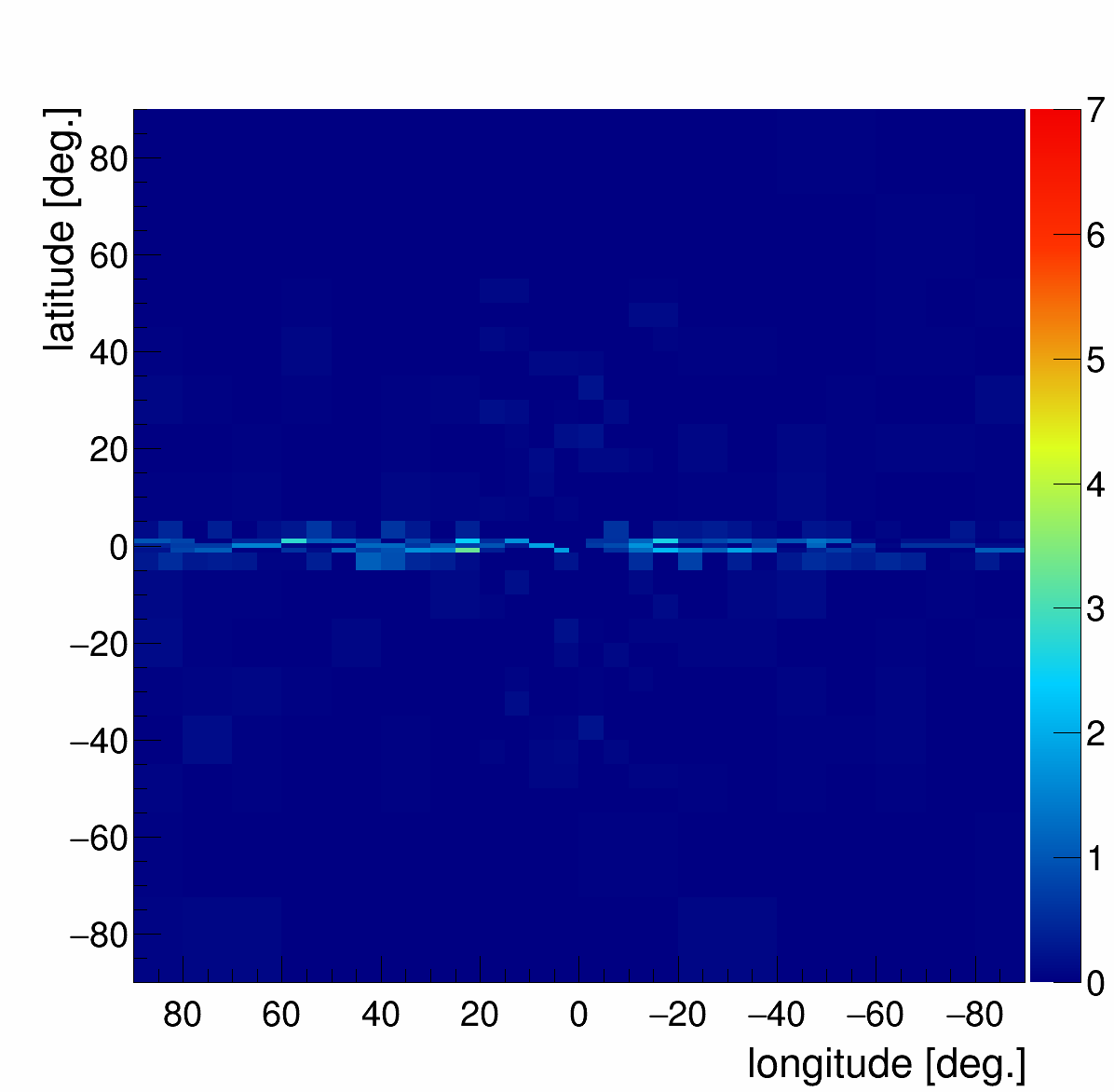}
\includegraphics[width=0.24\textwidth,height=0.24\textwidth,clip]{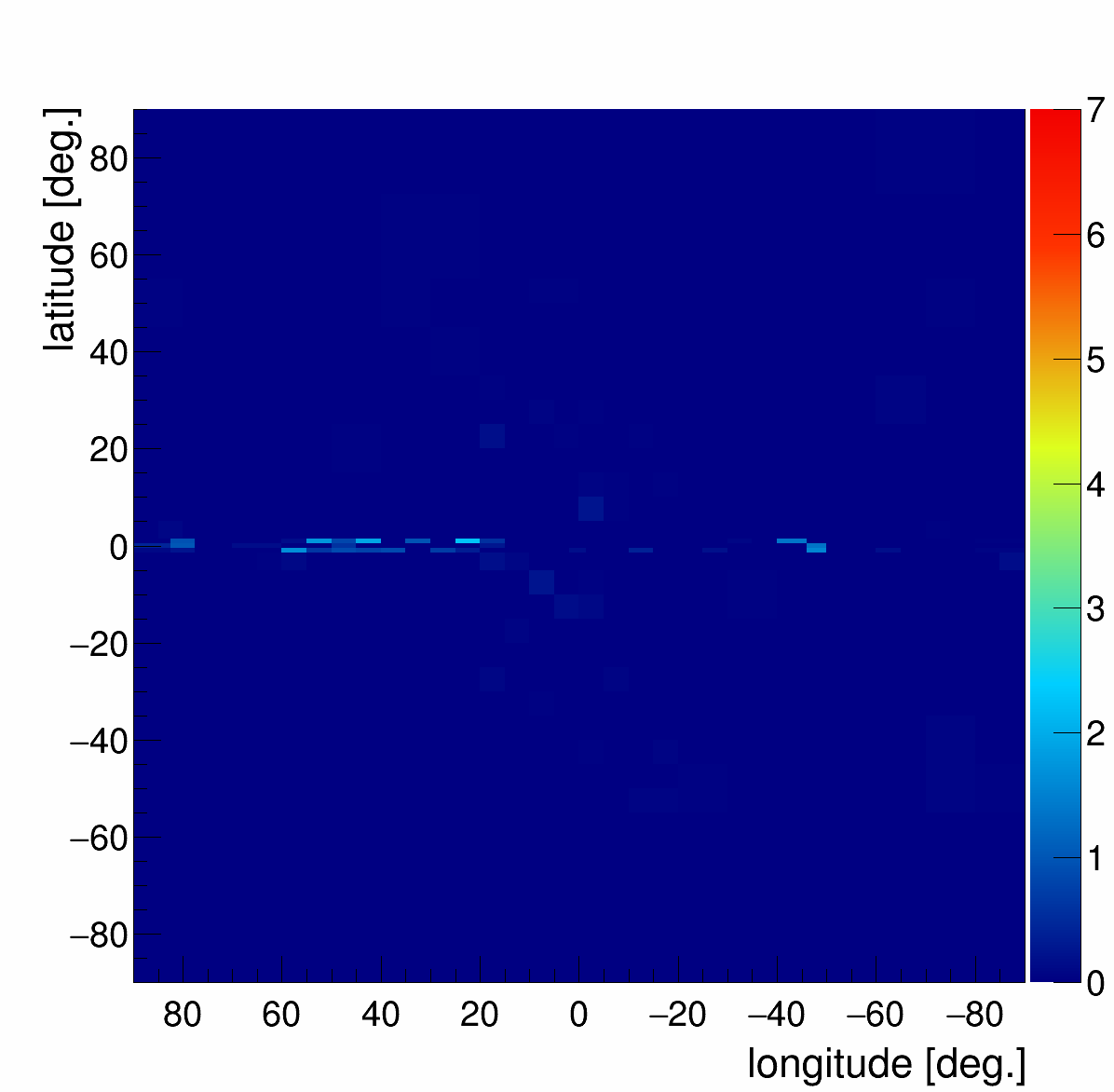}
\includegraphics[width=0.24\textwidth,height=0.24\textwidth,clip]{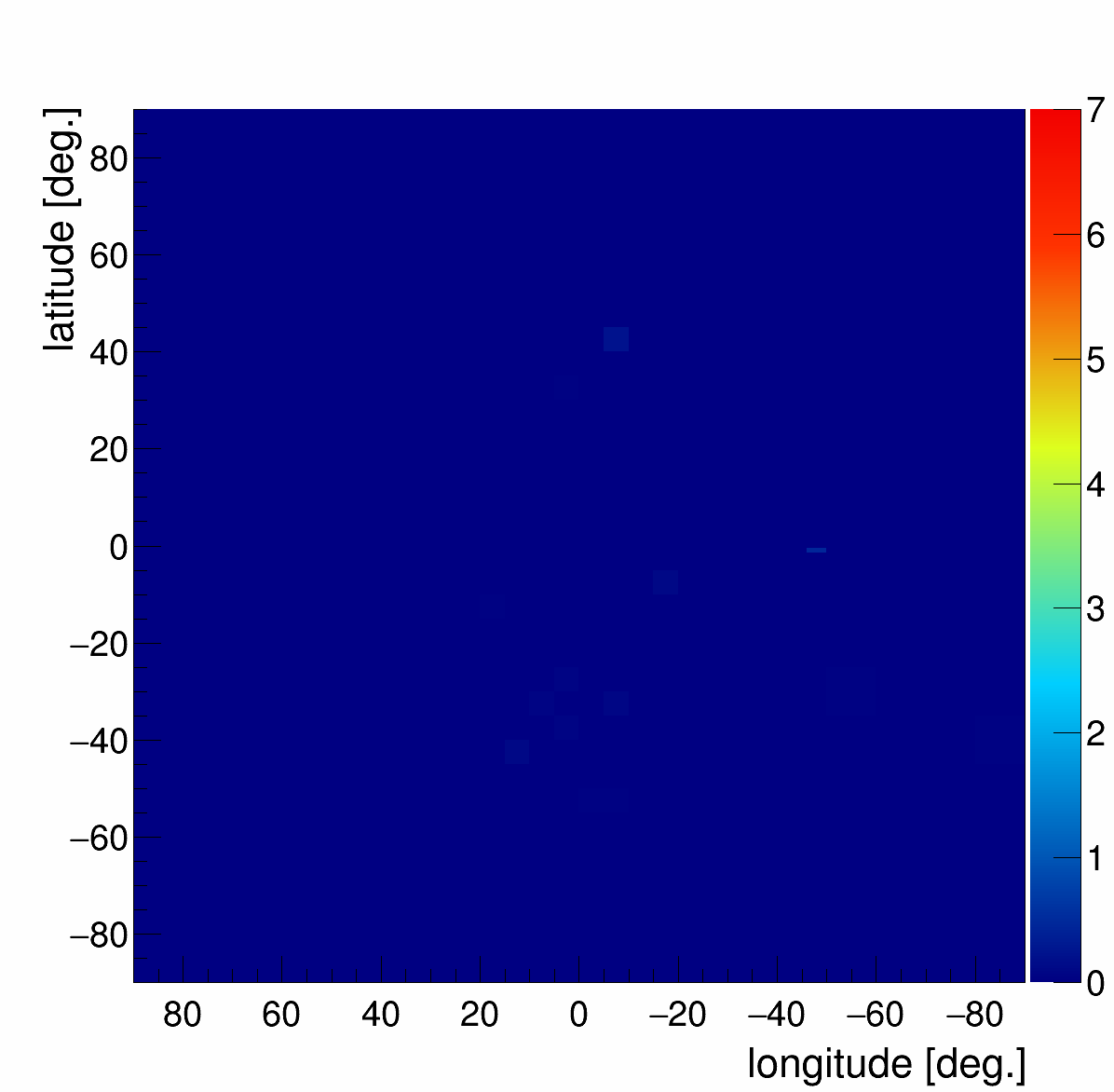}
\includegraphics[width=0.24\textwidth,height=0.24\textwidth,clip]{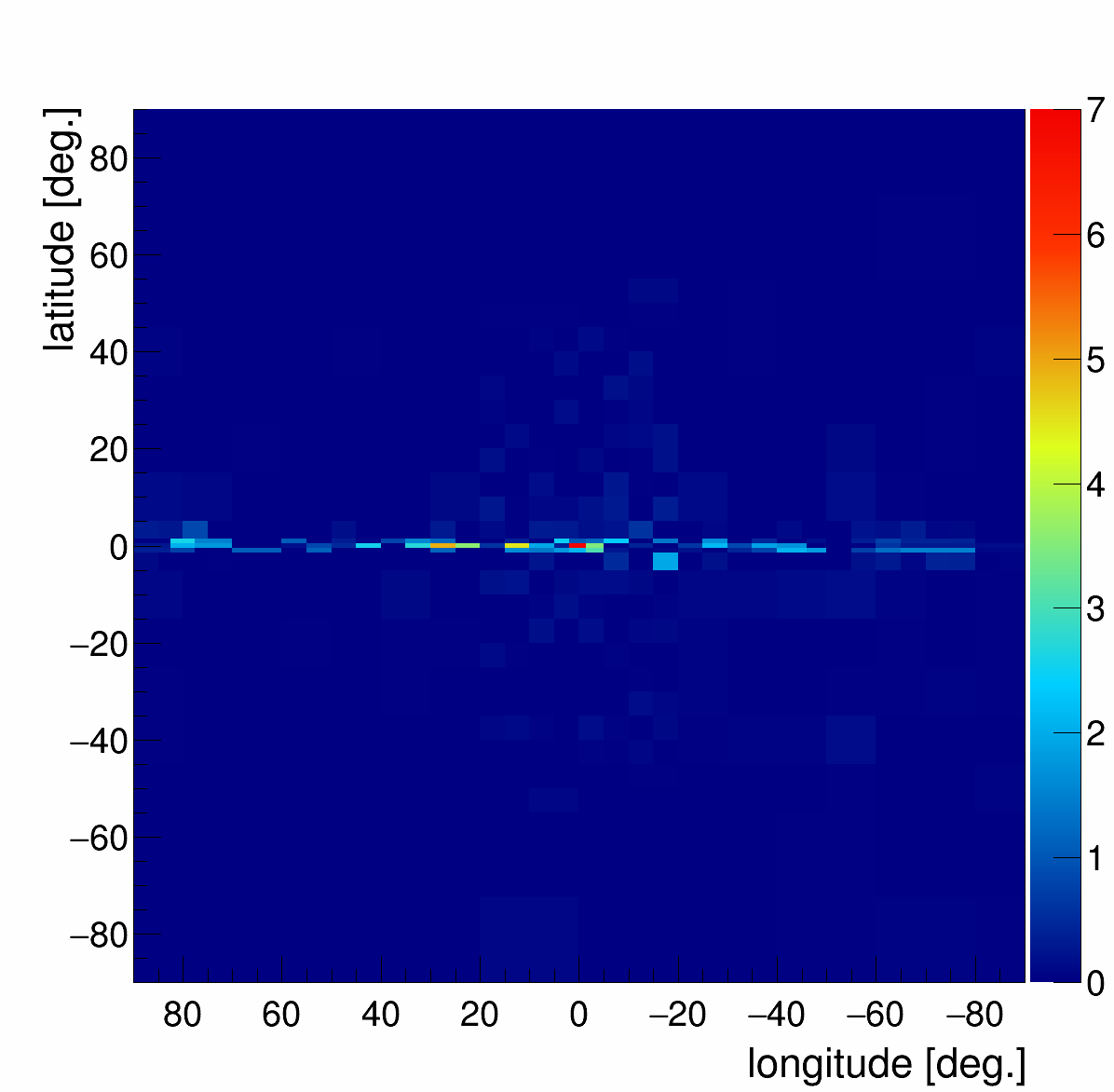}
\includegraphics[width=0.24\textwidth,height=0.24\textwidth,clip]{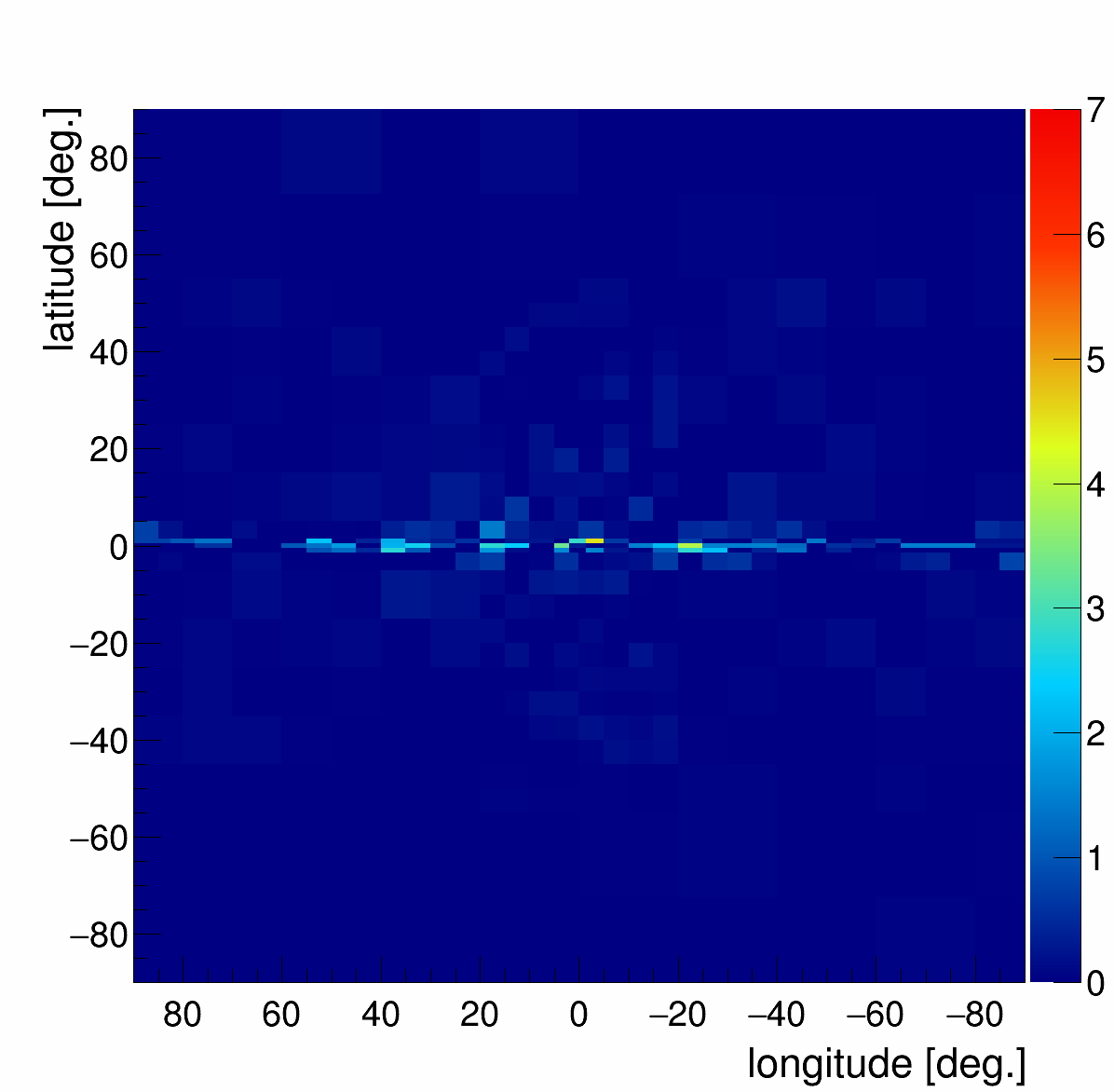}
\includegraphics[width=0.24\textwidth,height=0.24\textwidth,clip]{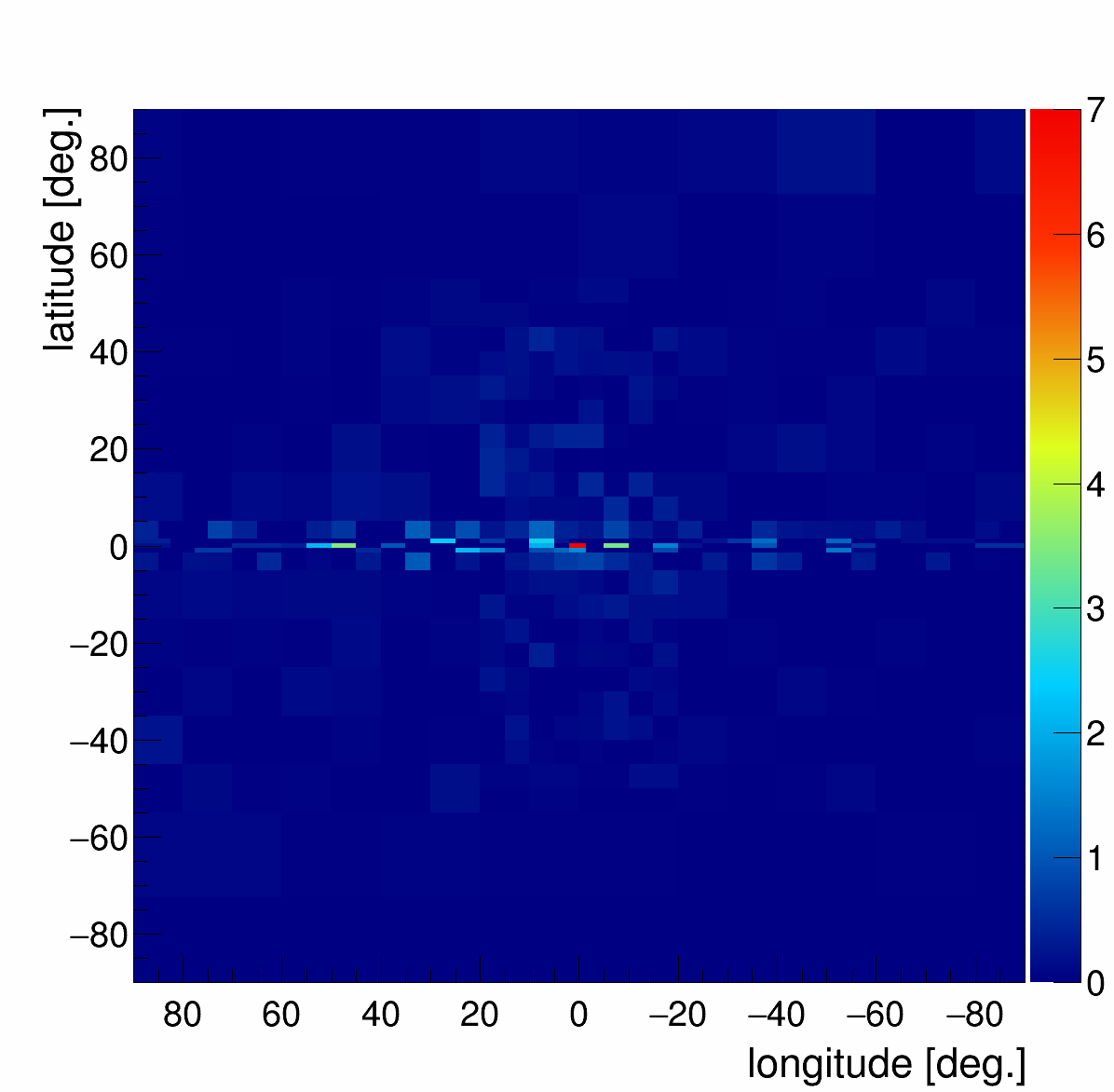}
\includegraphics[width=0.24\textwidth,height=0.24\textwidth,clip]{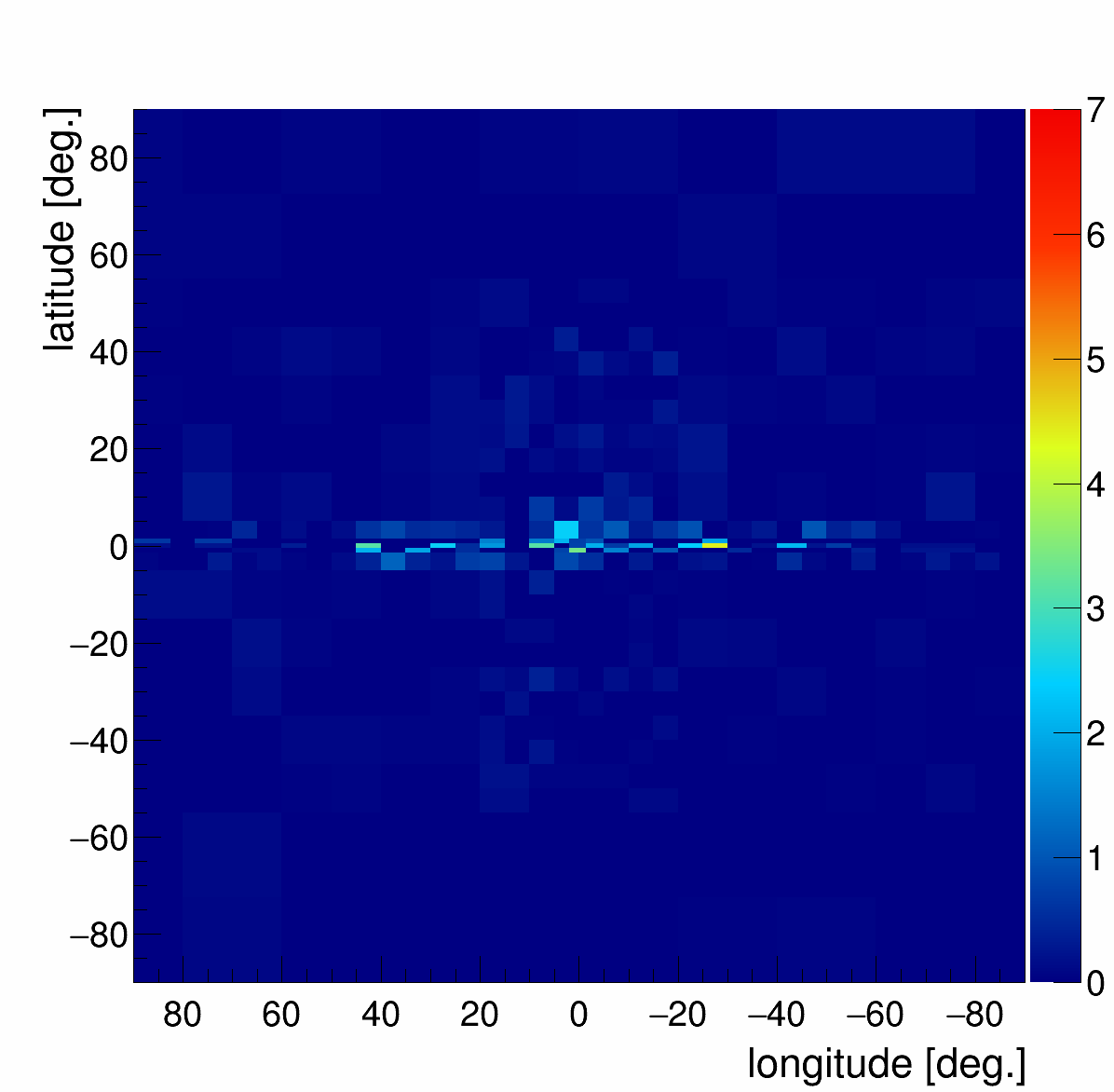}
\includegraphics[width=0.24\textwidth,height=0.24\textwidth,clip]{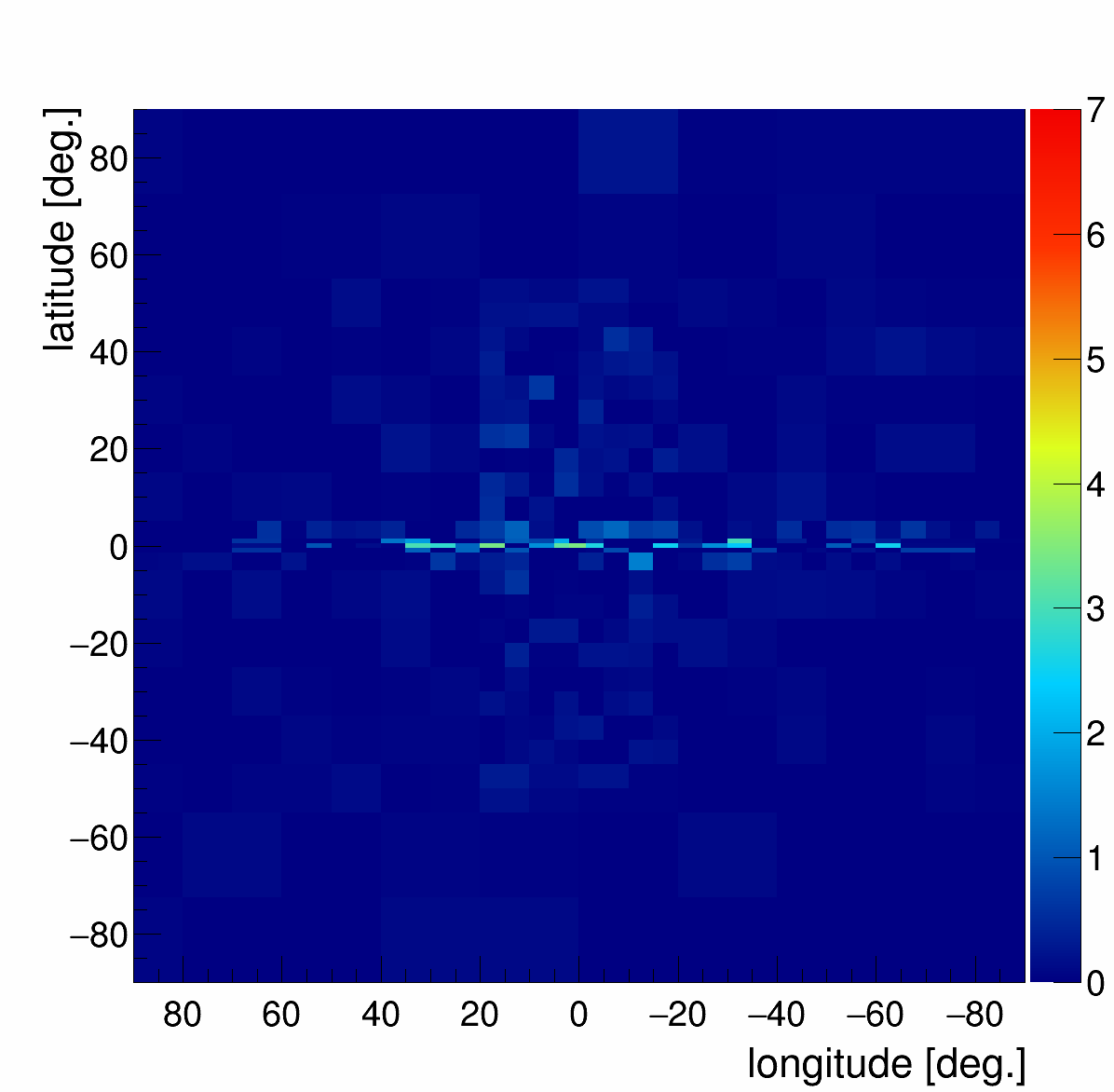}
\includegraphics[width=0.24\textwidth,height=0.24\textwidth,clip]{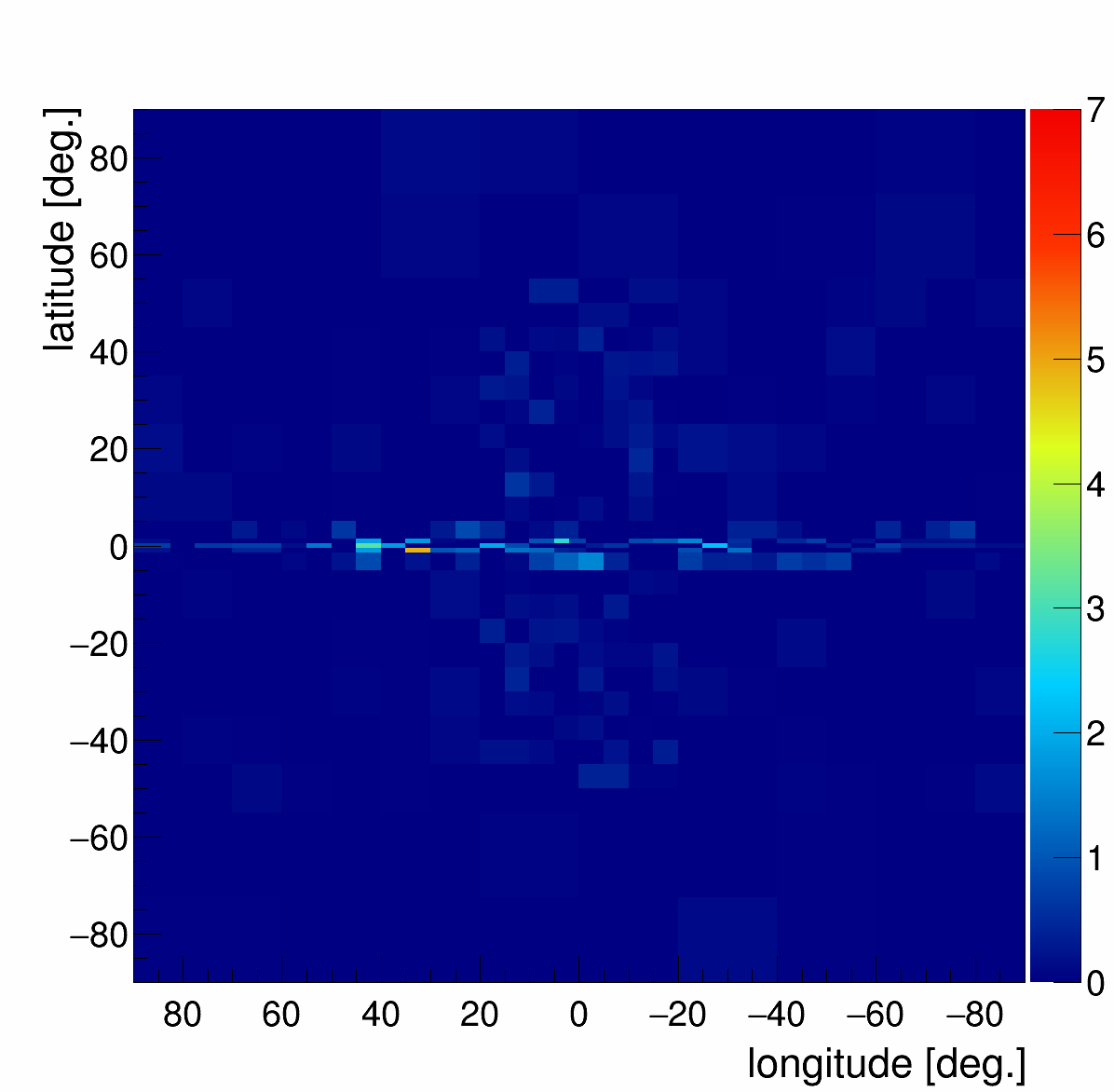}
\includegraphics[width=0.24\textwidth,height=0.24\textwidth,clip]{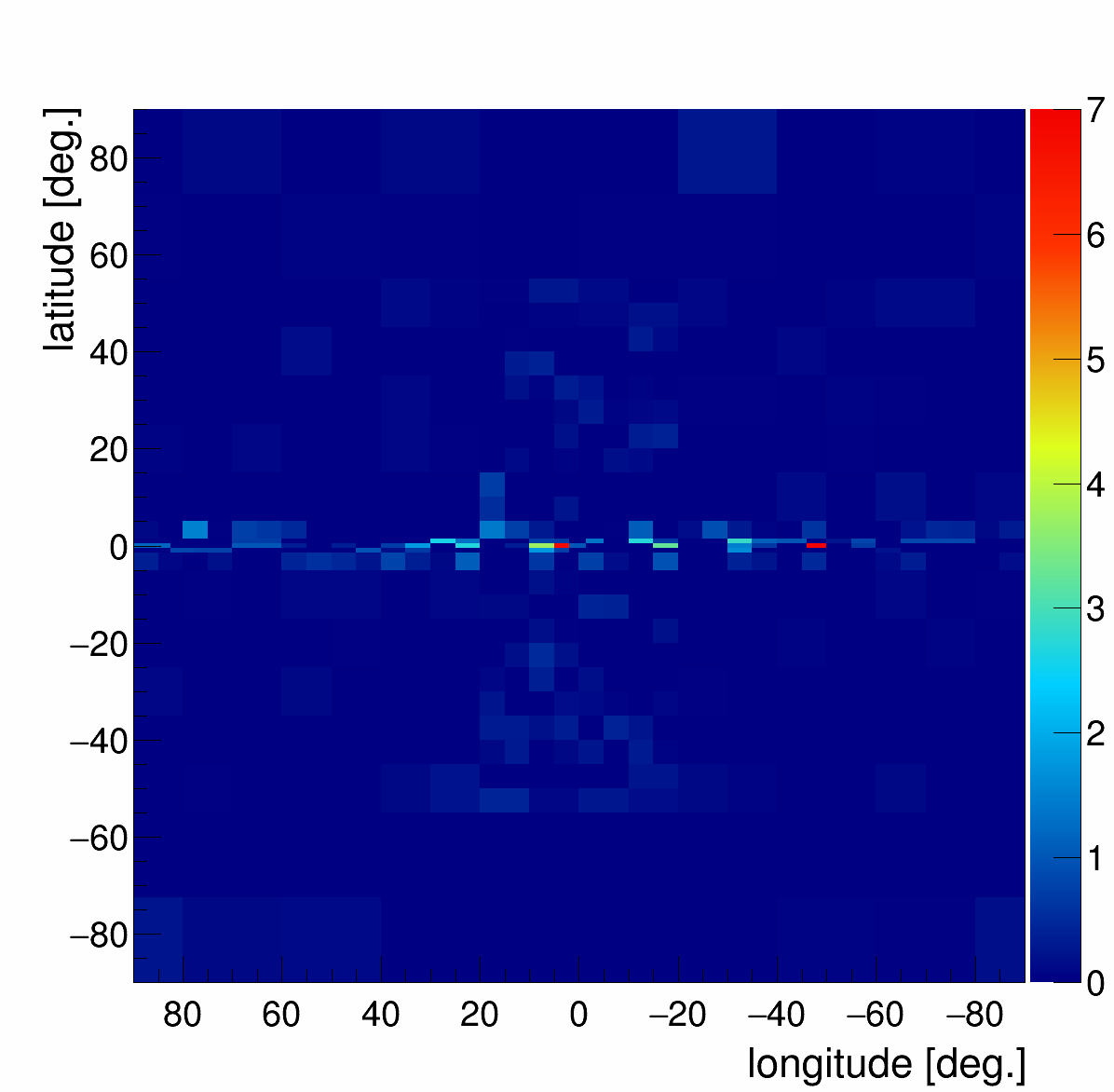}
\includegraphics[width=0.24\textwidth,height=0.24\textwidth,clip]{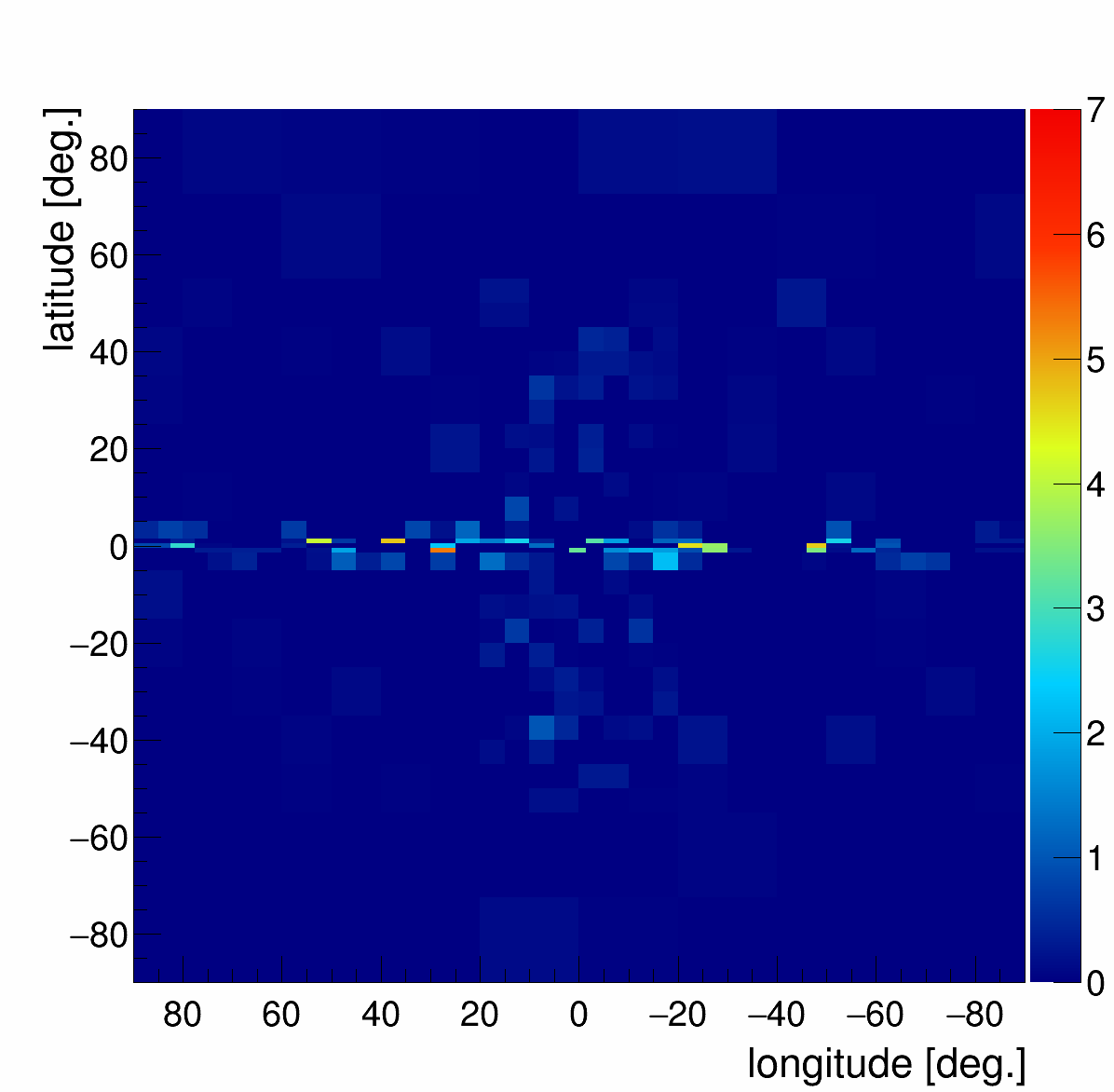}
\includegraphics[width=0.24\textwidth,height=0.24\textwidth,clip]{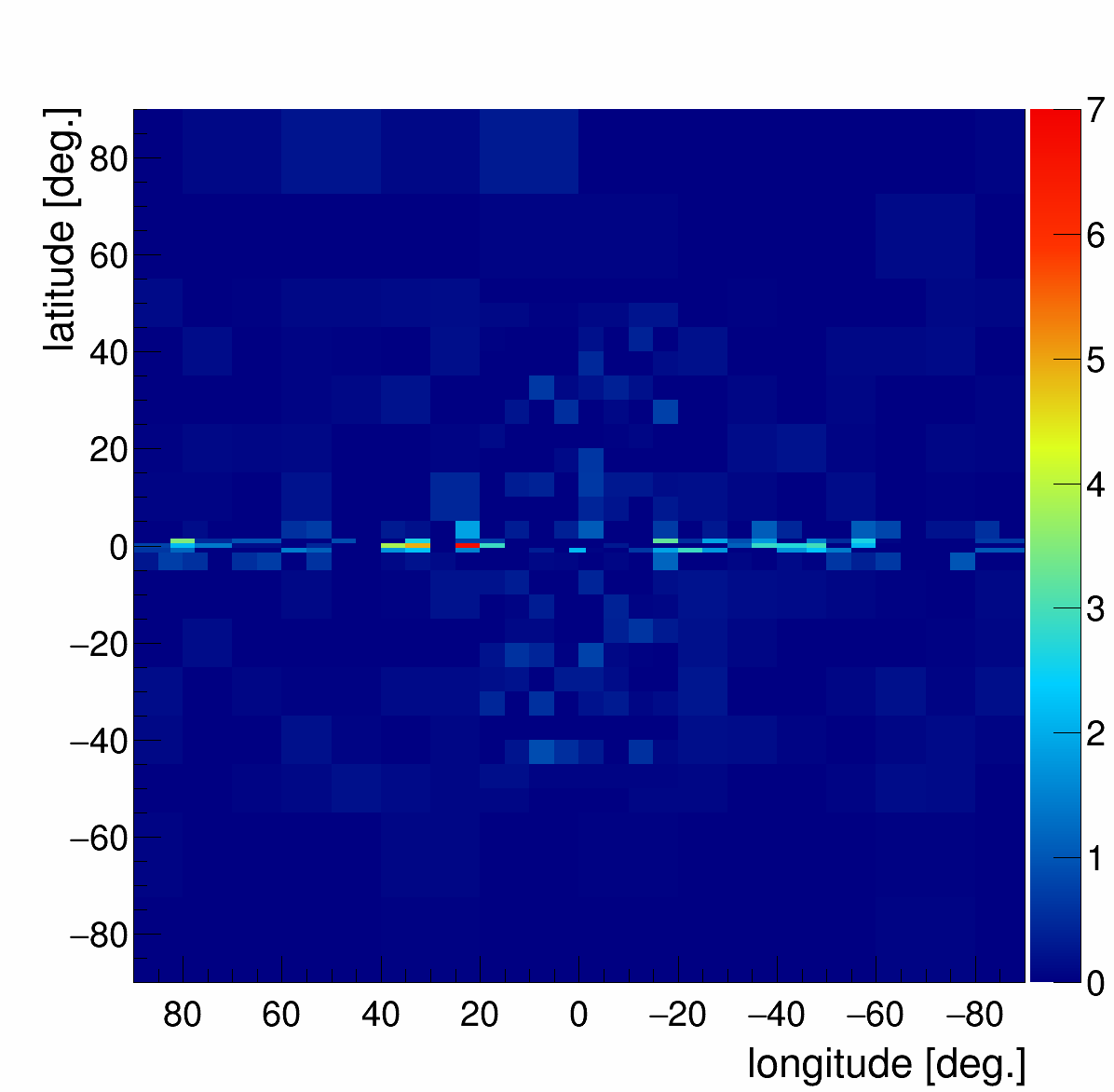}
\includegraphics[width=0.24\textwidth,height=0.24\textwidth,clip]{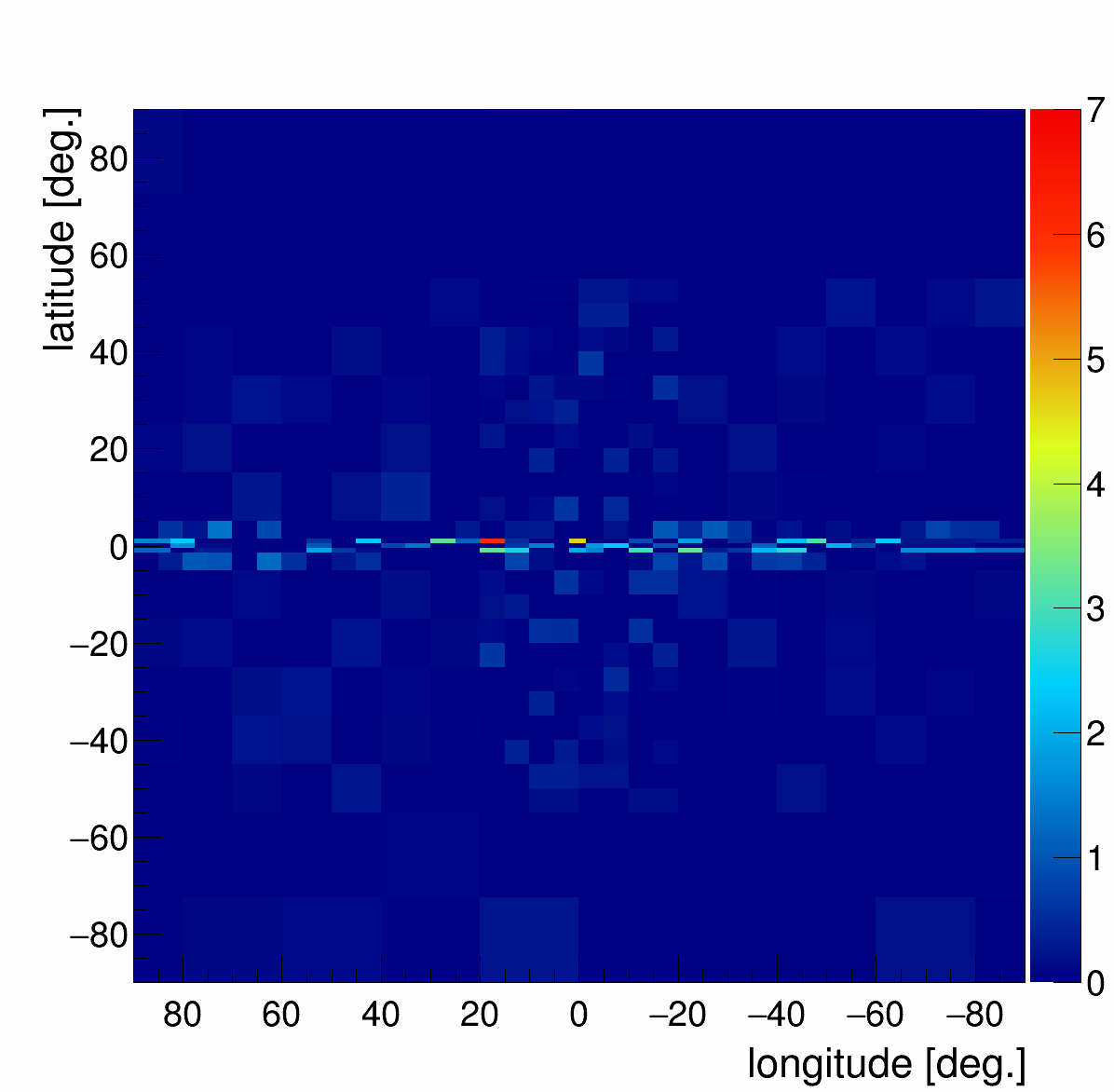}
\includegraphics[width=0.24\textwidth,height=0.24\textwidth,clip]{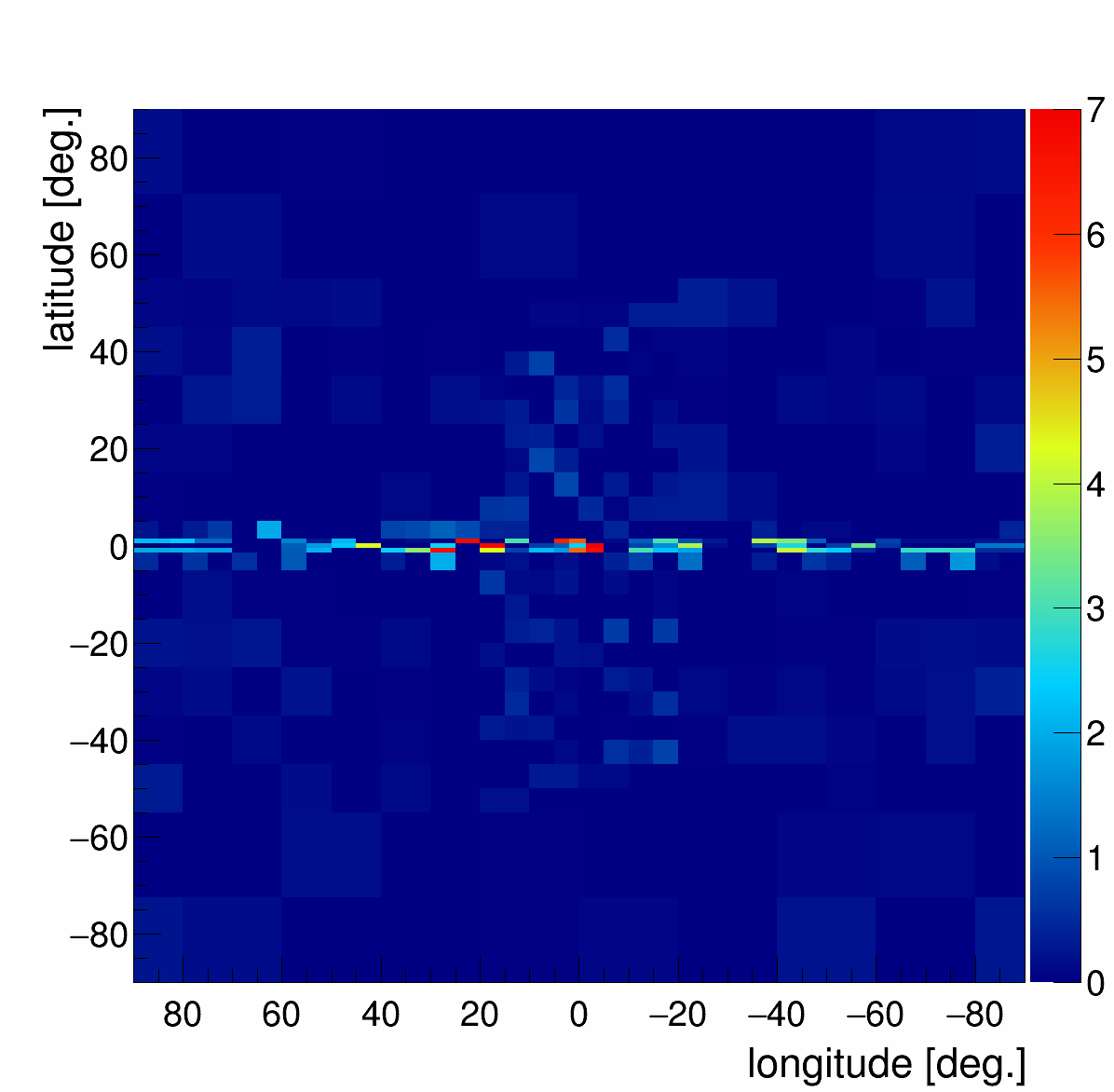}
\includegraphics[width=0.24\textwidth,height=0.24\textwidth,clip]{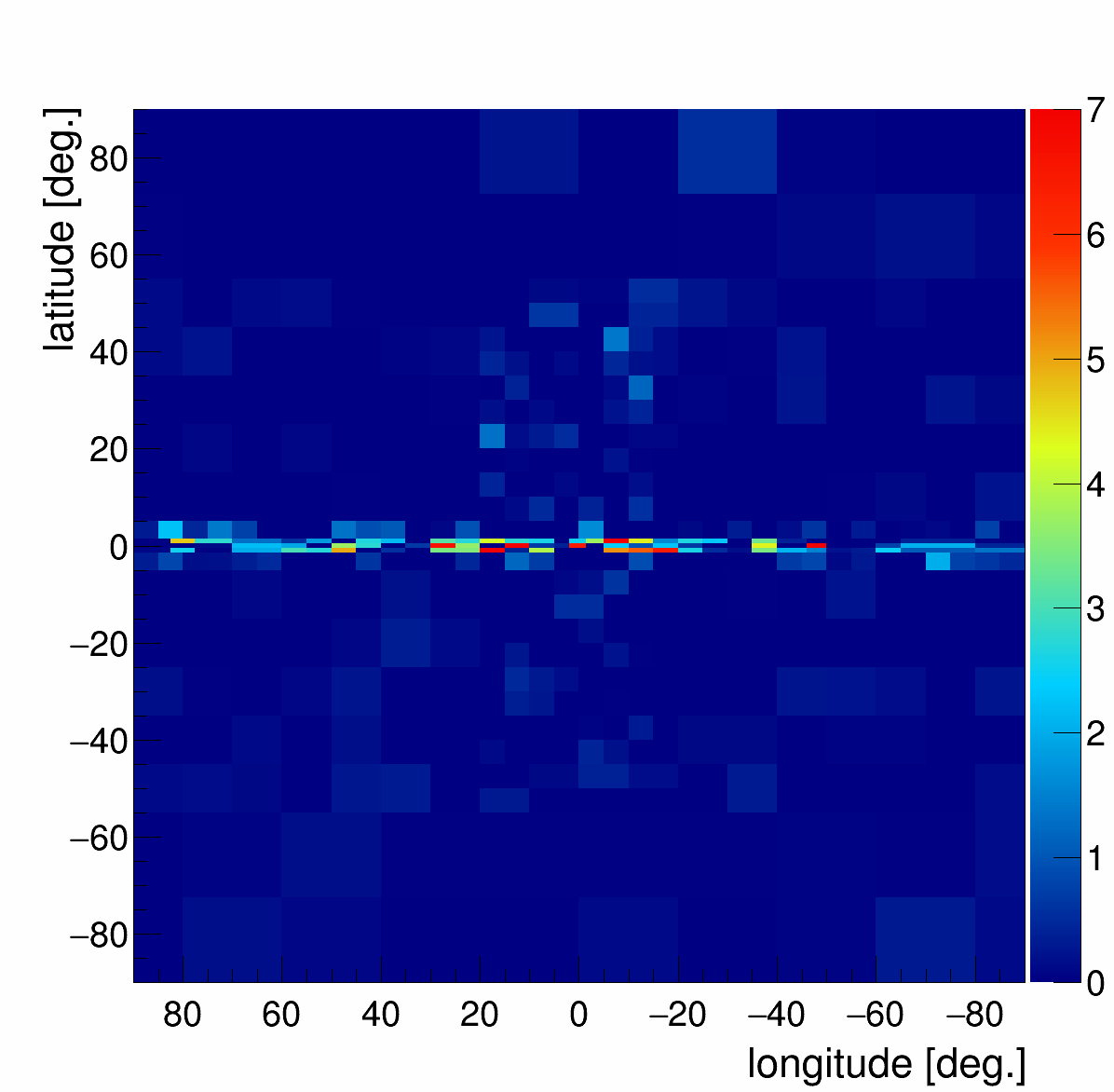}
\includegraphics[width=0.24\textwidth,height=0.24\textwidth,clip]{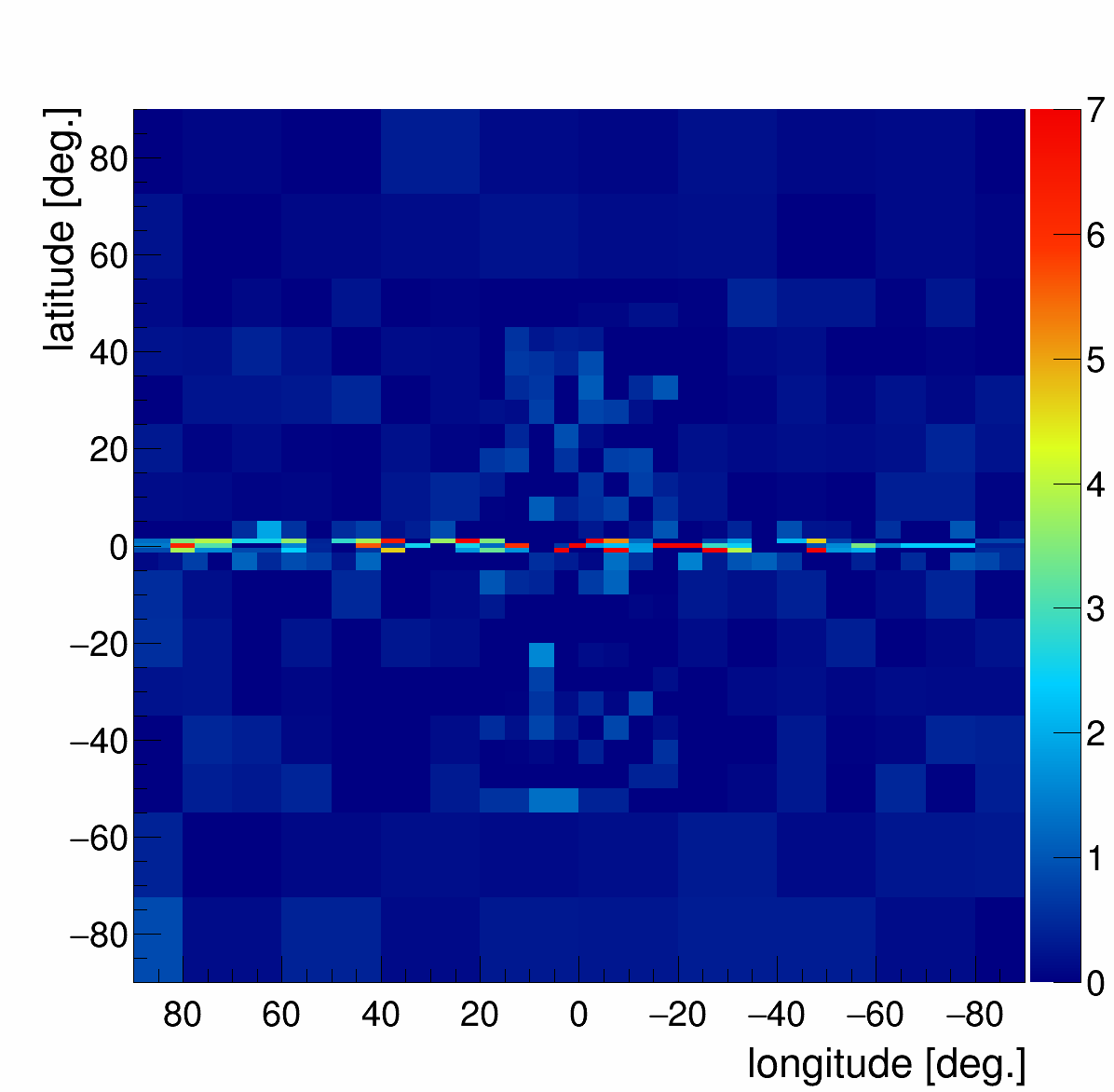}
\caption[]{ Residuals for the energy bins  (from left to right in each row starting at the top  and indicating the endpoint of each bin): 0.19, 0.27, 0.37, 0.52, 0.72, 1.00, 1.39, 1.93, 2.68, 3.73, 5.18, 7.20, 10.00, 13.89, 19.31, 26.83, 37.28, 51.79, 71.97, 100,00  GeV, respectively. The first bin from 0.1-0.14 GeV was not plotted for space reasons.The scale is in units of $10^{-6}$ $\rm GeV ~cm^{-1} s^{-1} sr^{-1}$. Note that at the highest energies the errors are typically 5\%, so the few red spots  are still typically only a few $\sigma$.
}
\label{F10}
\end{figure}

The flux of the sum of the templates is shown in Fig. \ref{F9}(a), while the overall $\chi^2/d.o.f.$ for every subcone is given in Fig. \ref{F9}(b).  
Note that the dominating systematic errors in the Fermi data were rescaled at all energies by a factor 0.25 in order to get $\chi^2/d.o.f. \approx 1$. This did not affect the  relative contributions of the fluxes in Fig. \ref{F7}.
The residuals, defined as the absolute difference between the data and the fit, are practically zero in the halo and small in the GD, as shown in Fig. \ref{F10} for different energies. 

\begin{figure}
\centering
\includegraphics[width=0.16\textwidth,height=0.16\textwidth,clip]{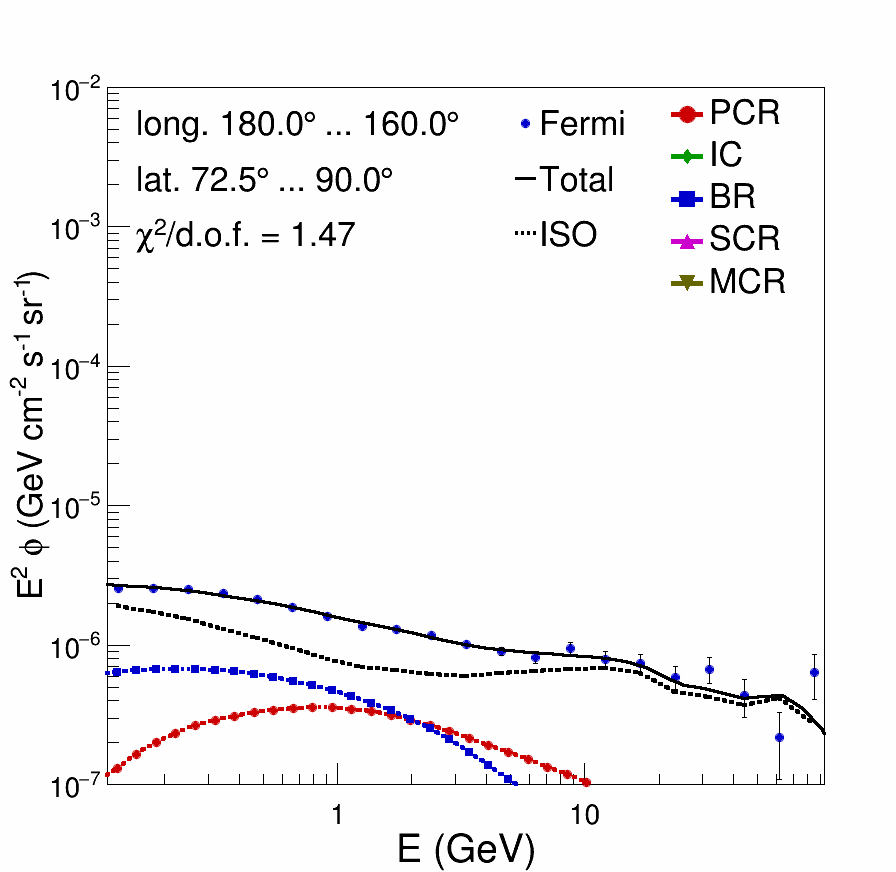}
\includegraphics[width=0.16\textwidth,height=0.16\textwidth,clip]{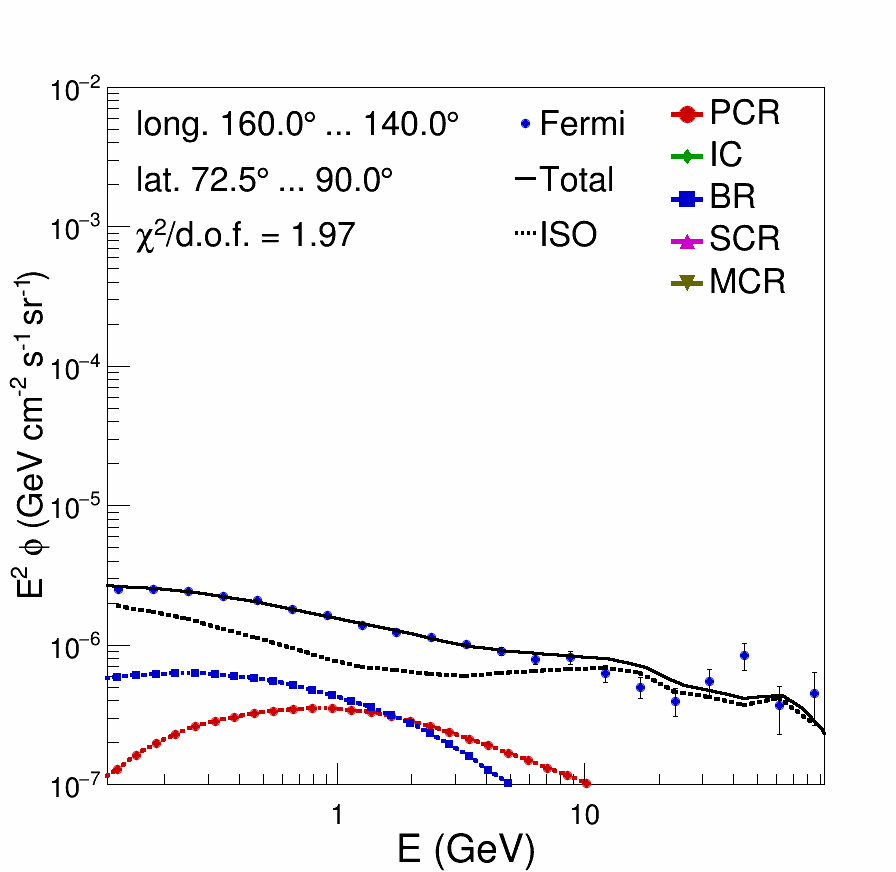}
\includegraphics[width=0.16\textwidth,height=0.16\textwidth,clip]{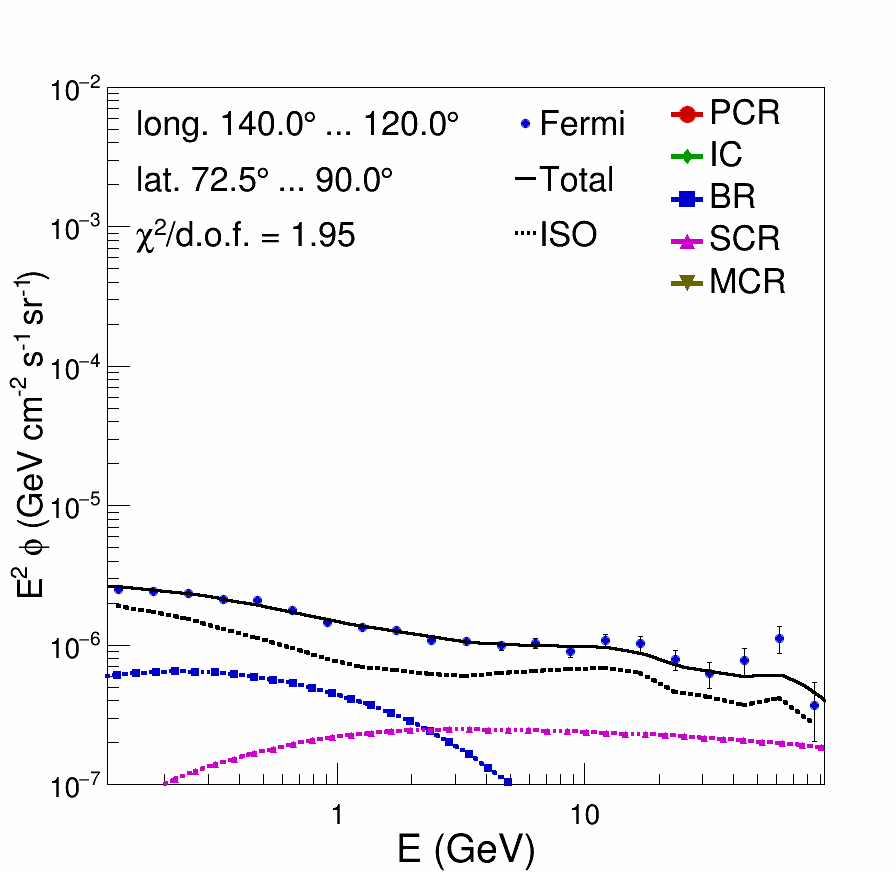}
\includegraphics[width=0.16\textwidth,height=0.16\textwidth,clip]{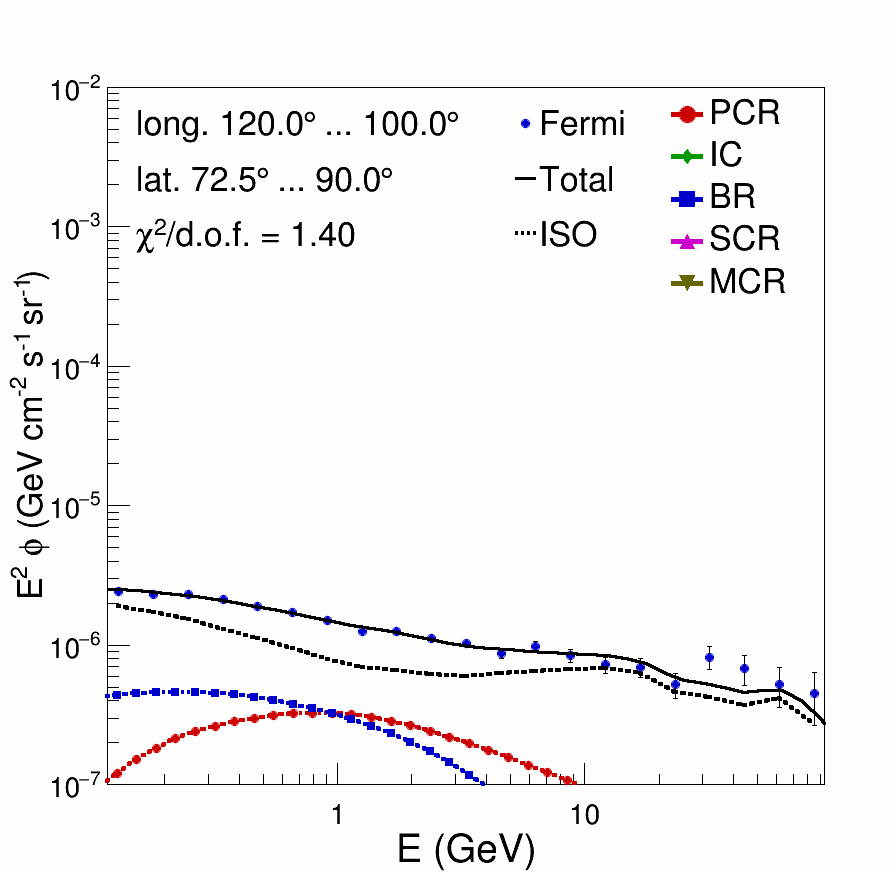}
\includegraphics[width=0.16\textwidth,height=0.16\textwidth,clip]{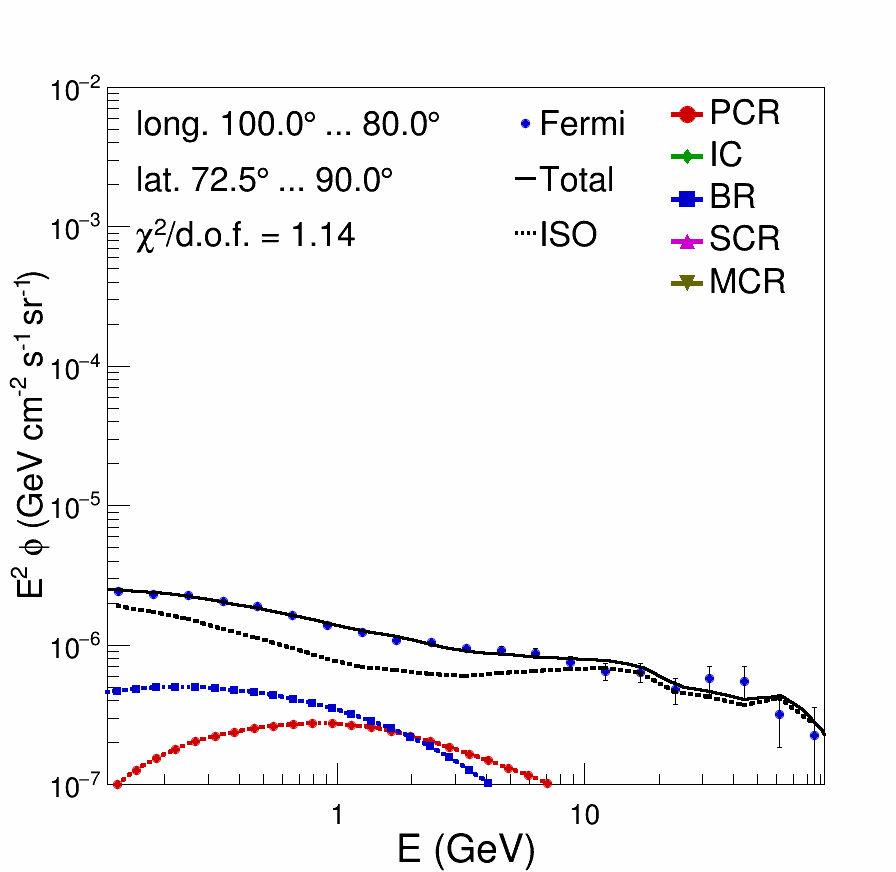}
\includegraphics[width=0.16\textwidth,height=0.16\textwidth,clip]{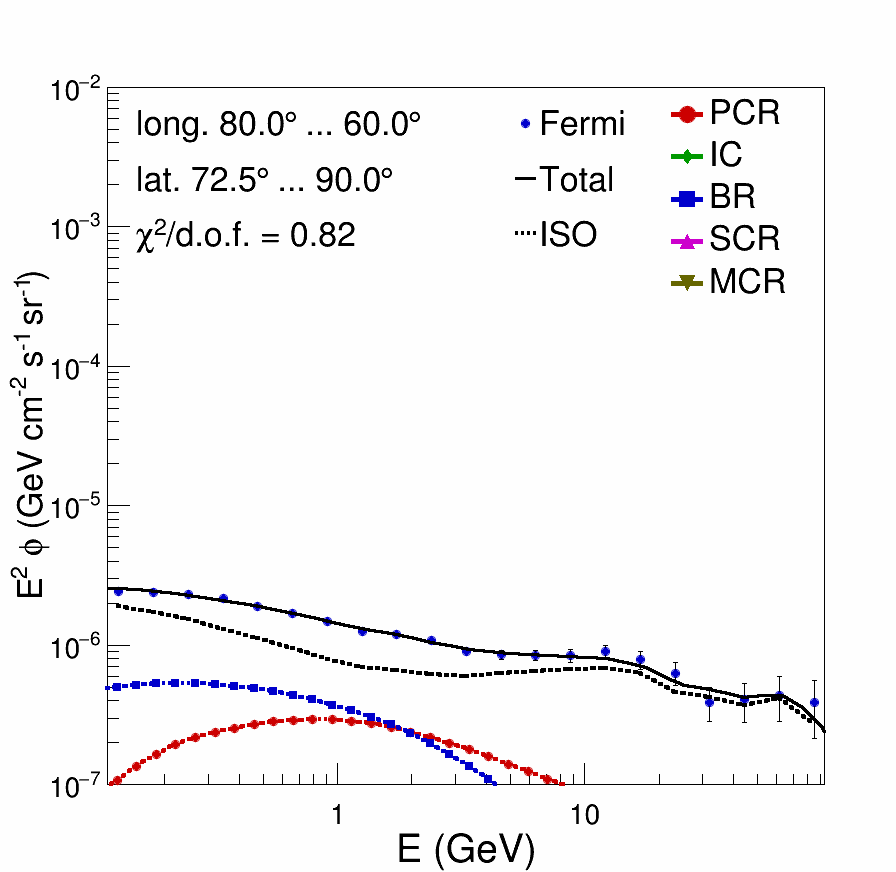}
\includegraphics[width=0.16\textwidth,height=0.16\textwidth,clip]{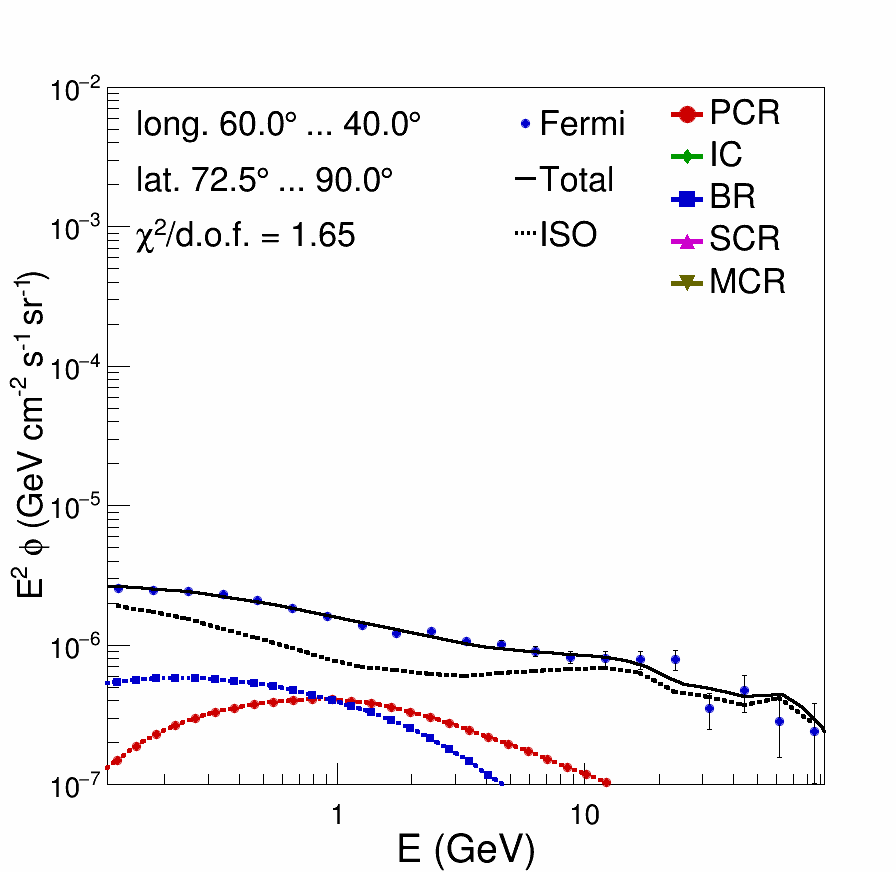}
\includegraphics[width=0.16\textwidth,height=0.16\textwidth,clip]{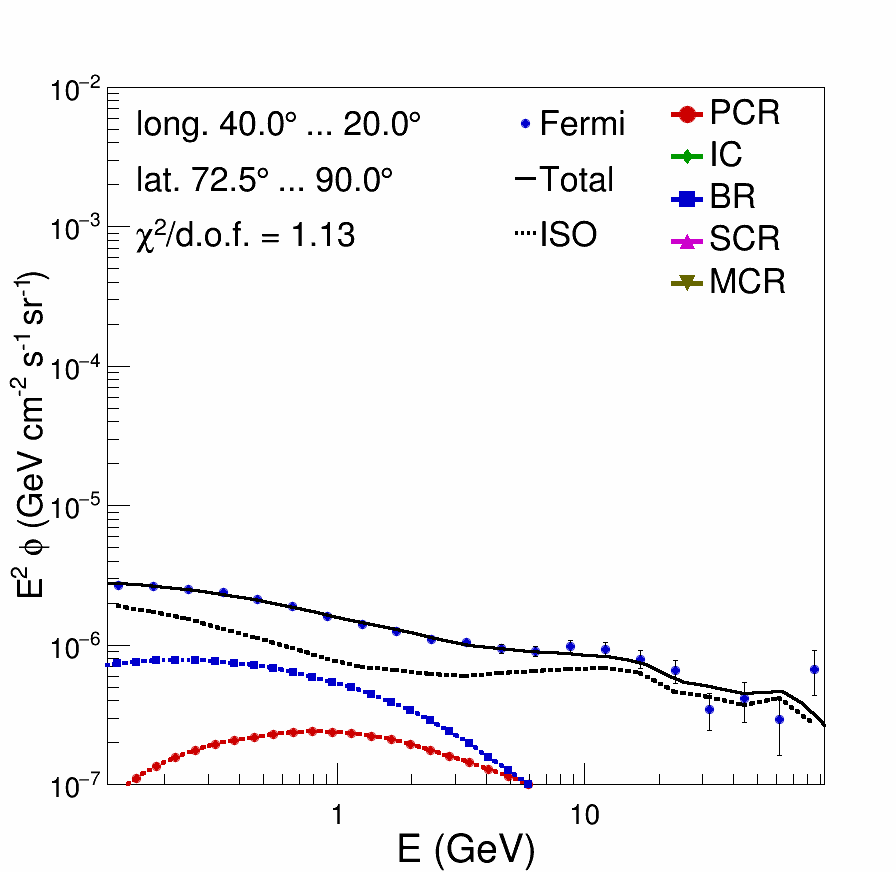}
\includegraphics[width=0.16\textwidth,height=0.16\textwidth,clip]{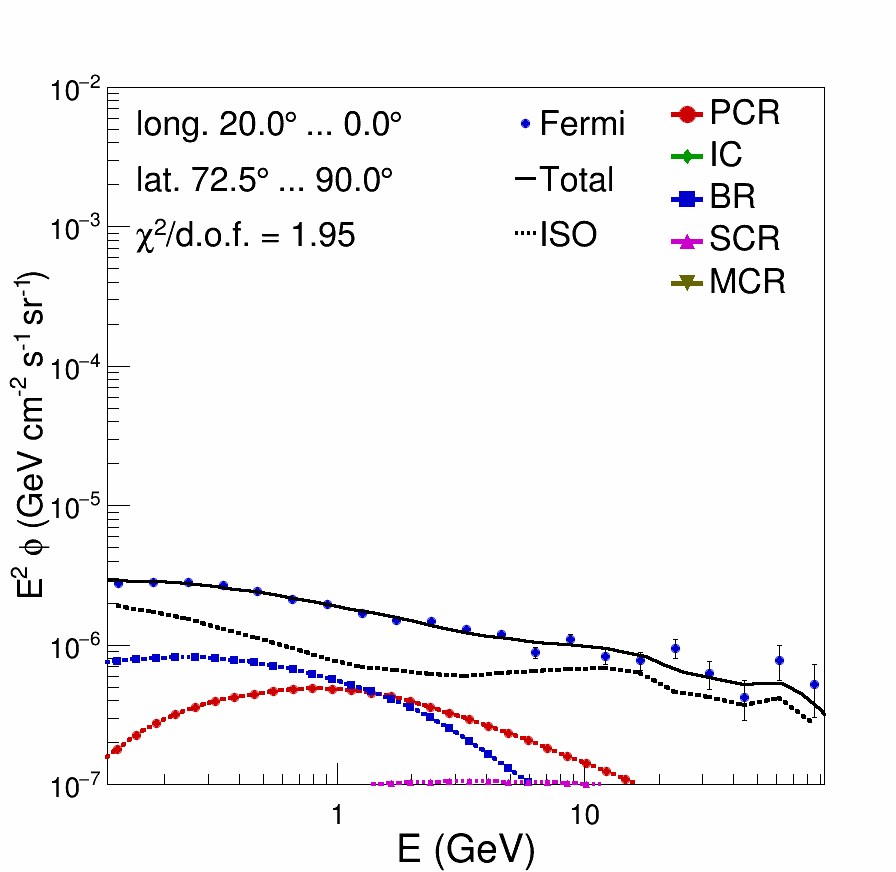}
\includegraphics[width=0.16\textwidth,height=0.16\textwidth,clip]{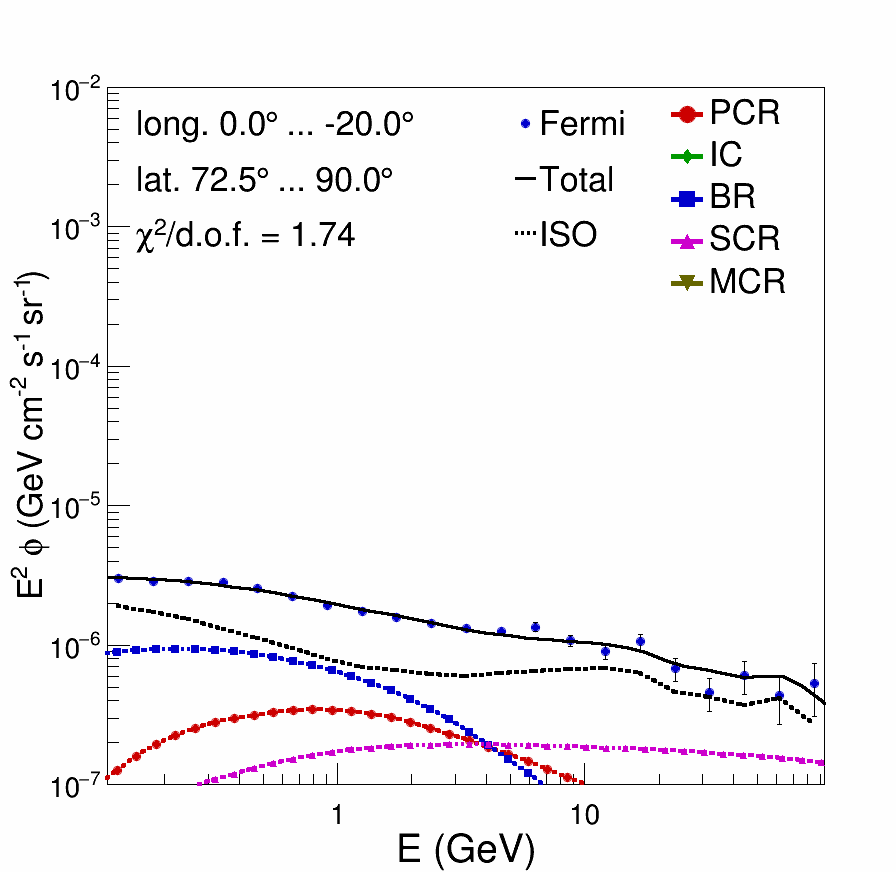}
\includegraphics[width=0.16\textwidth,height=0.16\textwidth,clip]{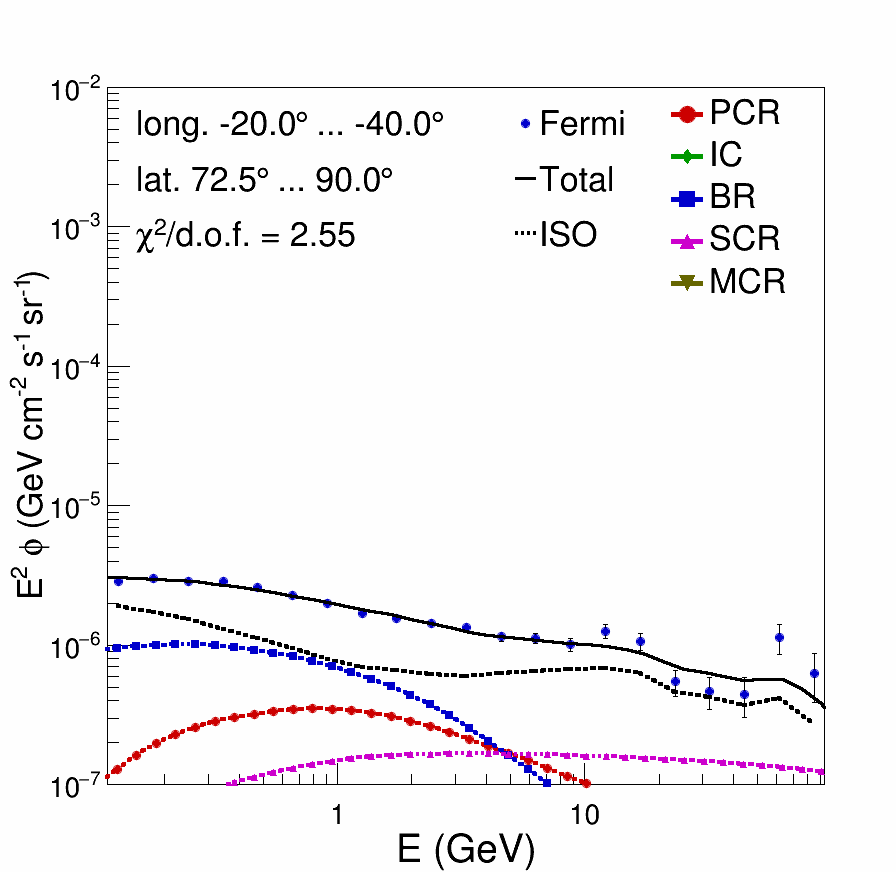}
\includegraphics[width=0.16\textwidth,height=0.16\textwidth,clip]{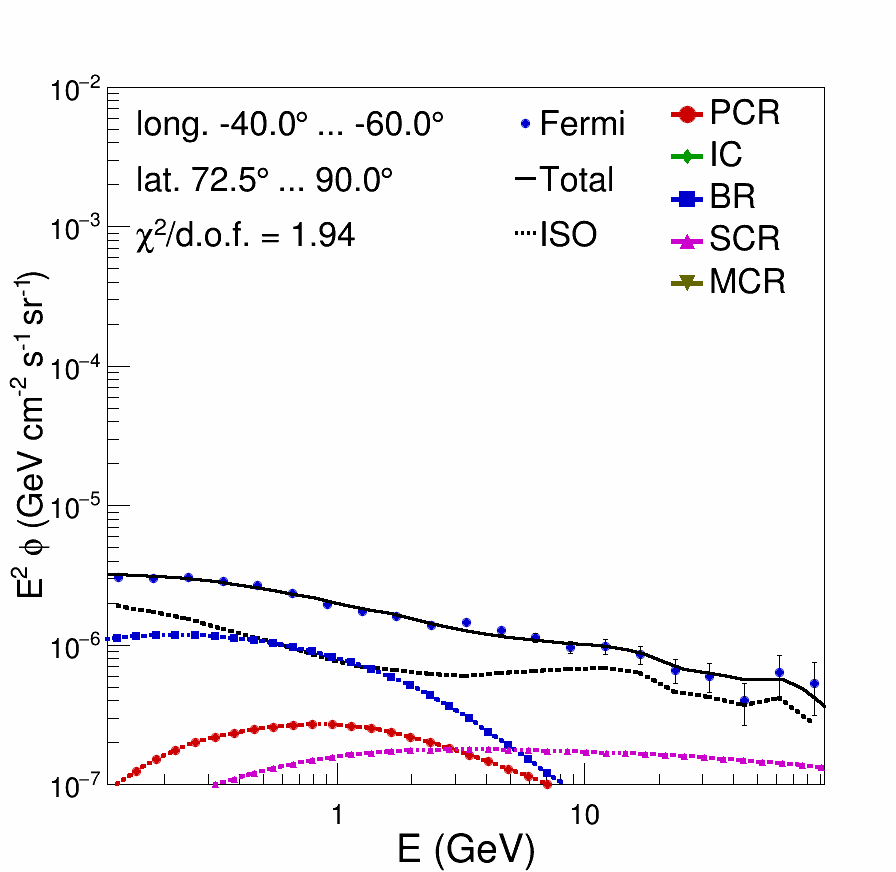}
\includegraphics[width=0.16\textwidth,height=0.16\textwidth,clip]{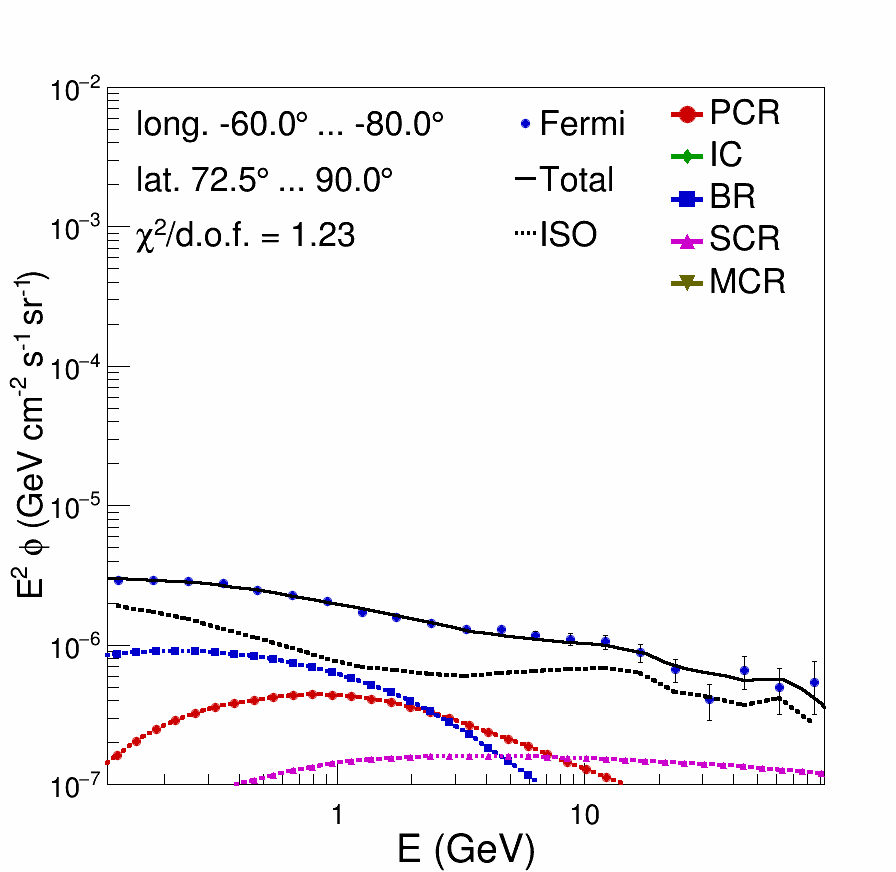}
\includegraphics[width=0.16\textwidth,height=0.16\textwidth,clip]{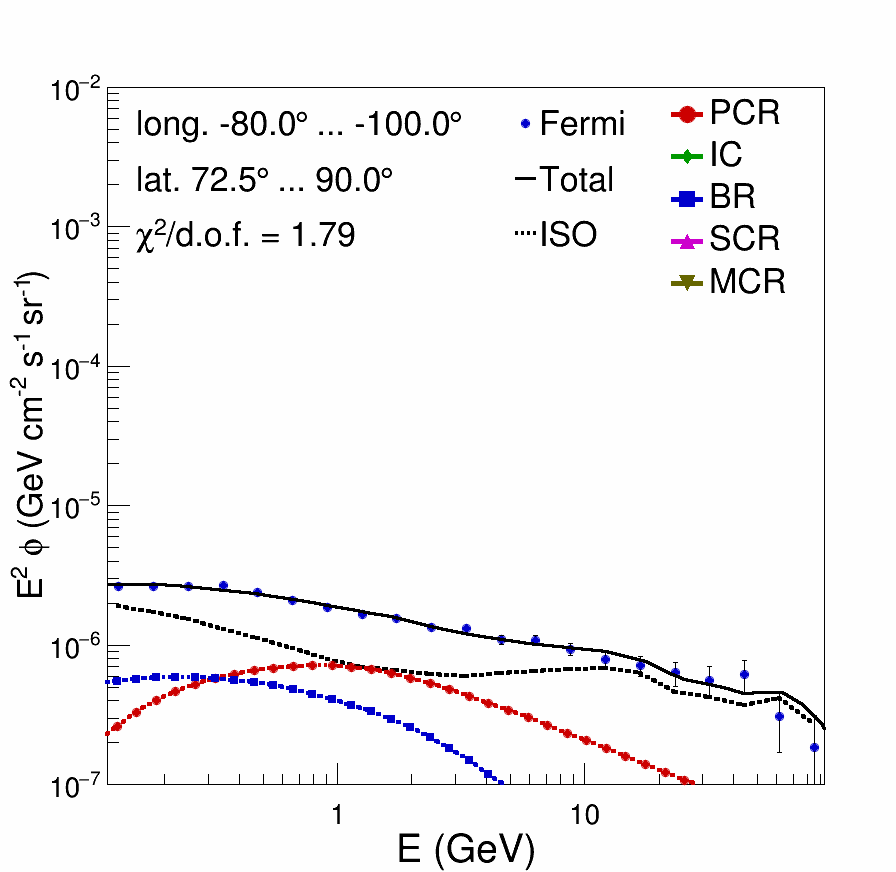}
\includegraphics[width=0.16\textwidth,height=0.16\textwidth,clip]{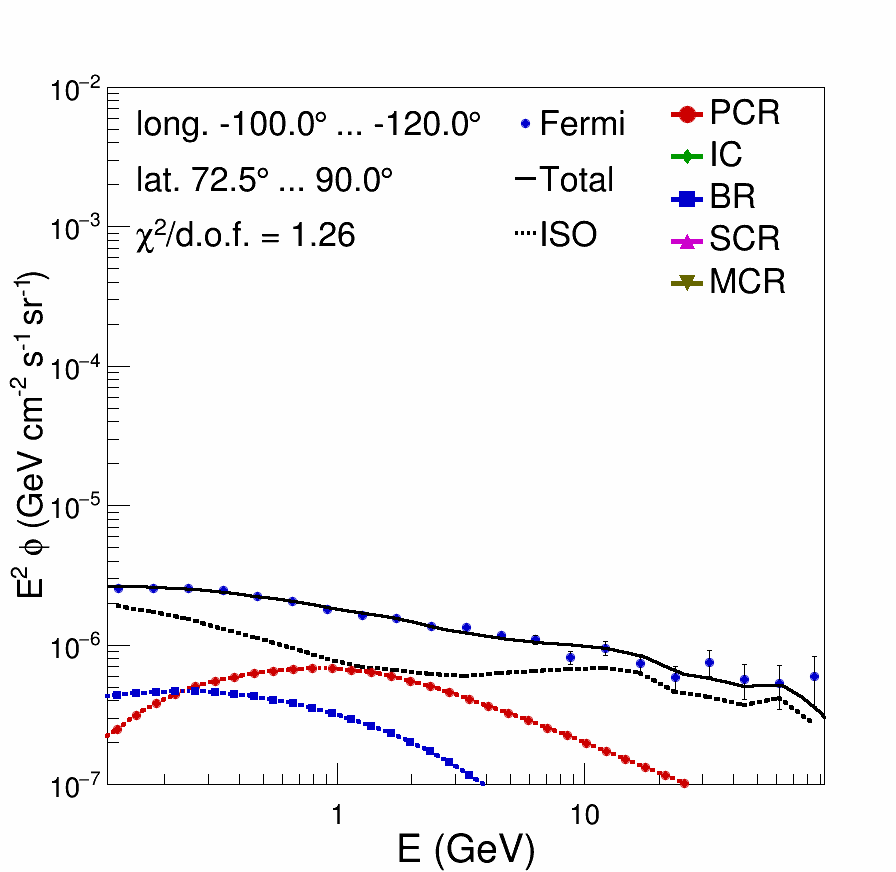}
\includegraphics[width=0.16\textwidth,height=0.16\textwidth,clip]{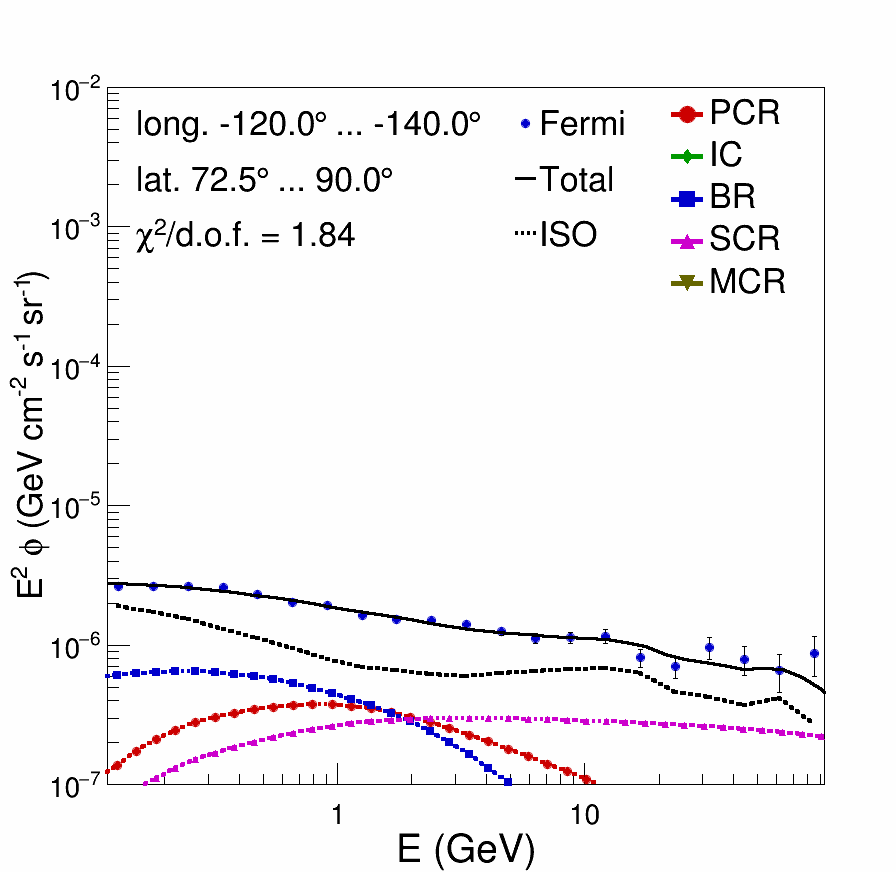}
\includegraphics[width=0.16\textwidth,height=0.16\textwidth,clip]{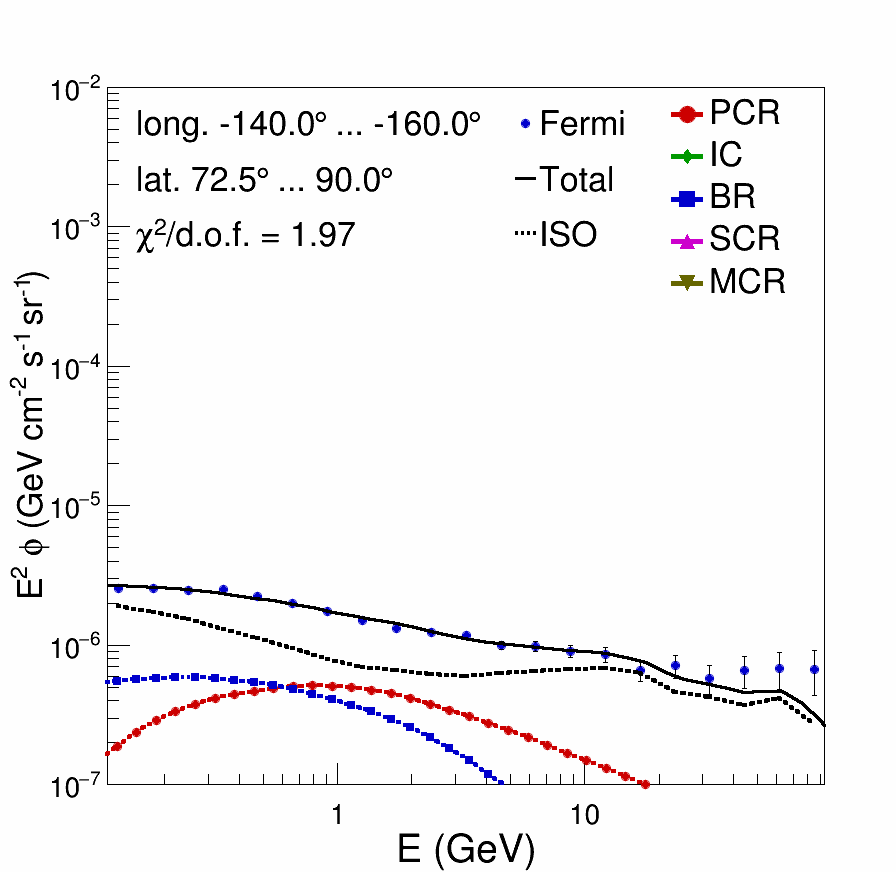}
\includegraphics[width=0.16\textwidth,height=0.16\textwidth,clip]{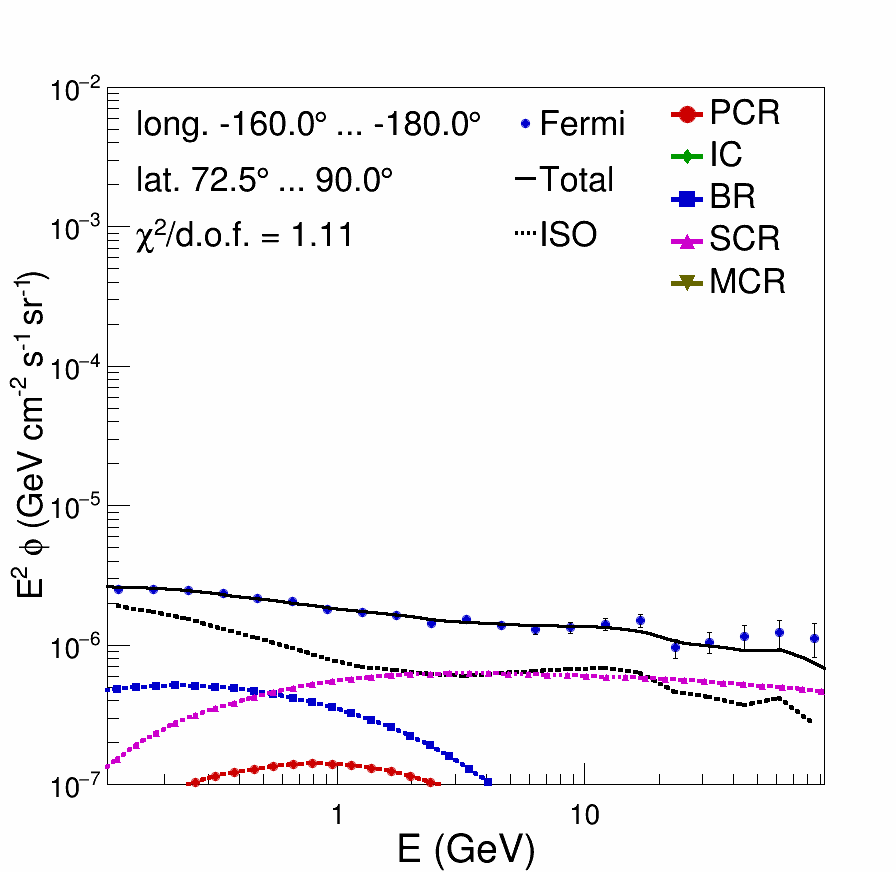}    
\caption[]{Template fits for latitudes  with $72.5^\circ<b<90.0^\circ$ and longitudes decreasing from 180$^\circ$ to -180$^\circ$.} 
\label{F11}
\end{figure}

The template fits   are shown  in Figs. \ref{F11} till \ref{F31} for each of the 797 cones separately.

\begin{figure}
\includegraphics[width=0.16\textwidth,height=0.16\textwidth,clip]{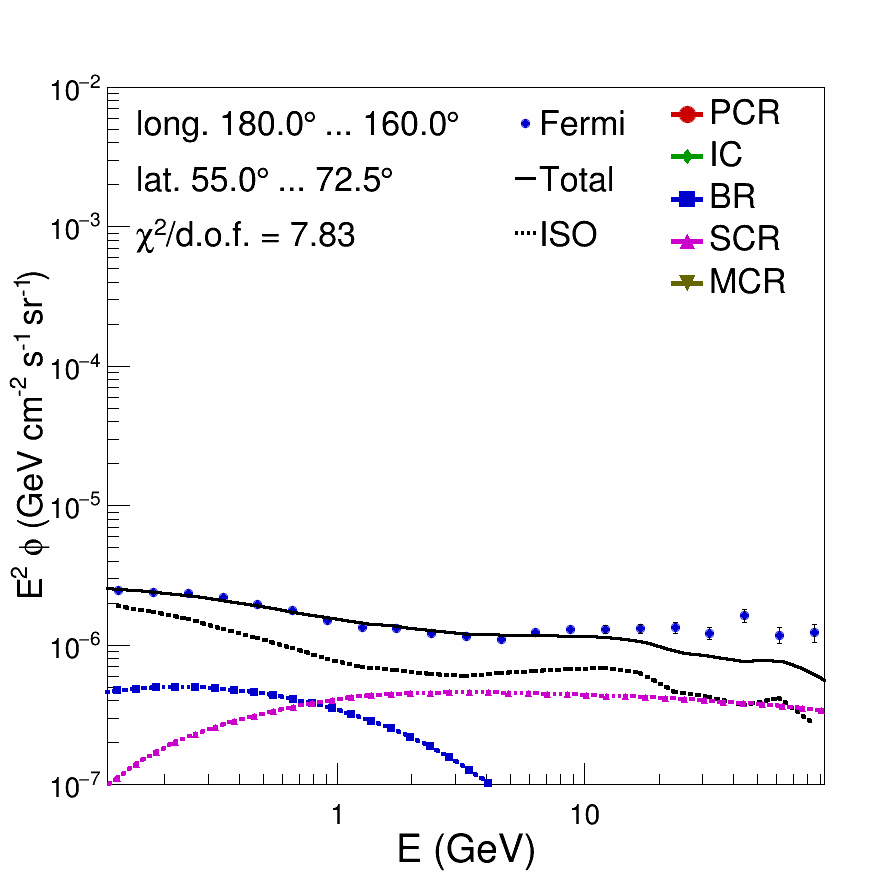}
\includegraphics[width=0.16\textwidth,height=0.16\textwidth,clip]{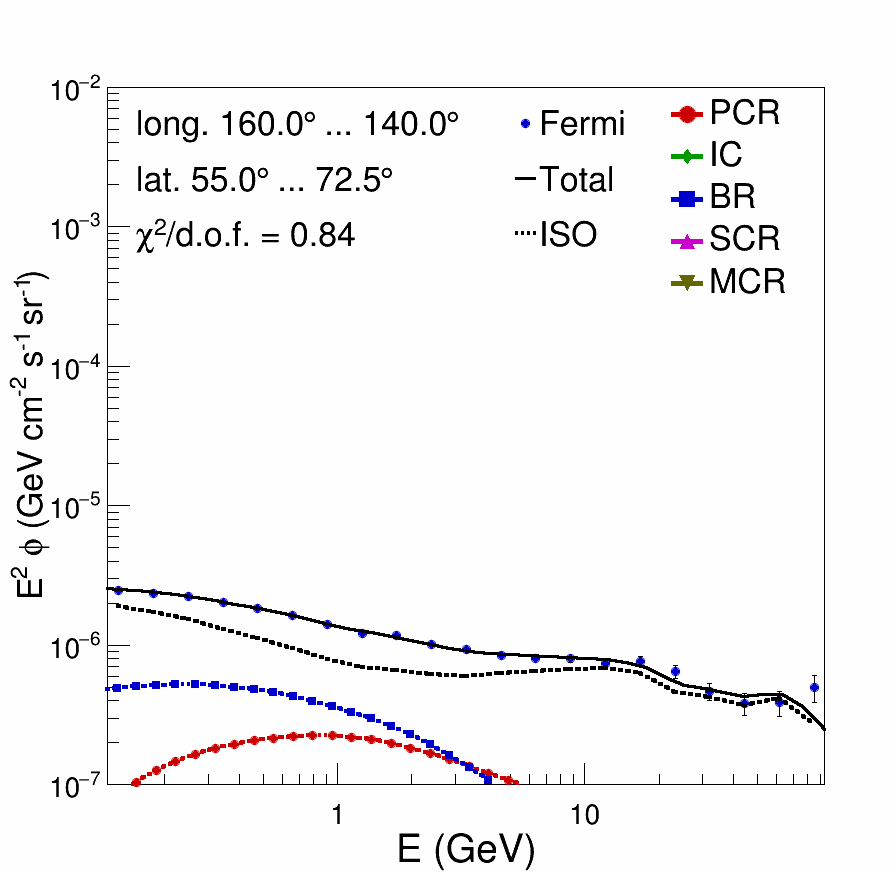}
\includegraphics[width=0.16\textwidth,height=0.16\textwidth,clip]{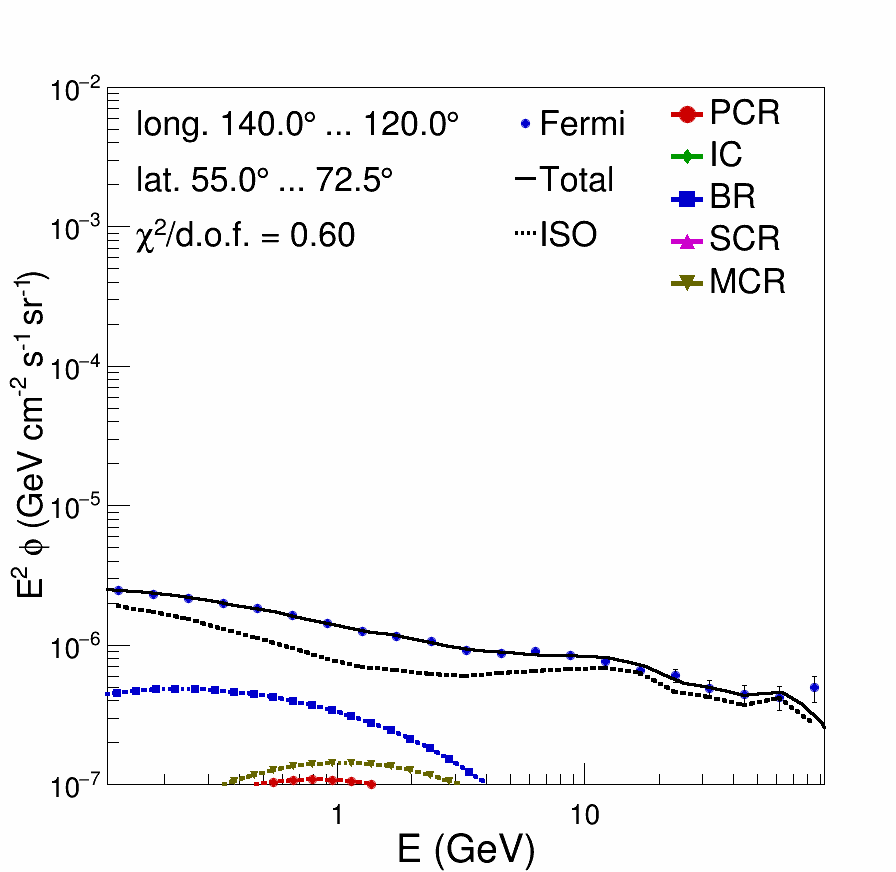}
\includegraphics[width=0.16\textwidth,height=0.16\textwidth,clip]{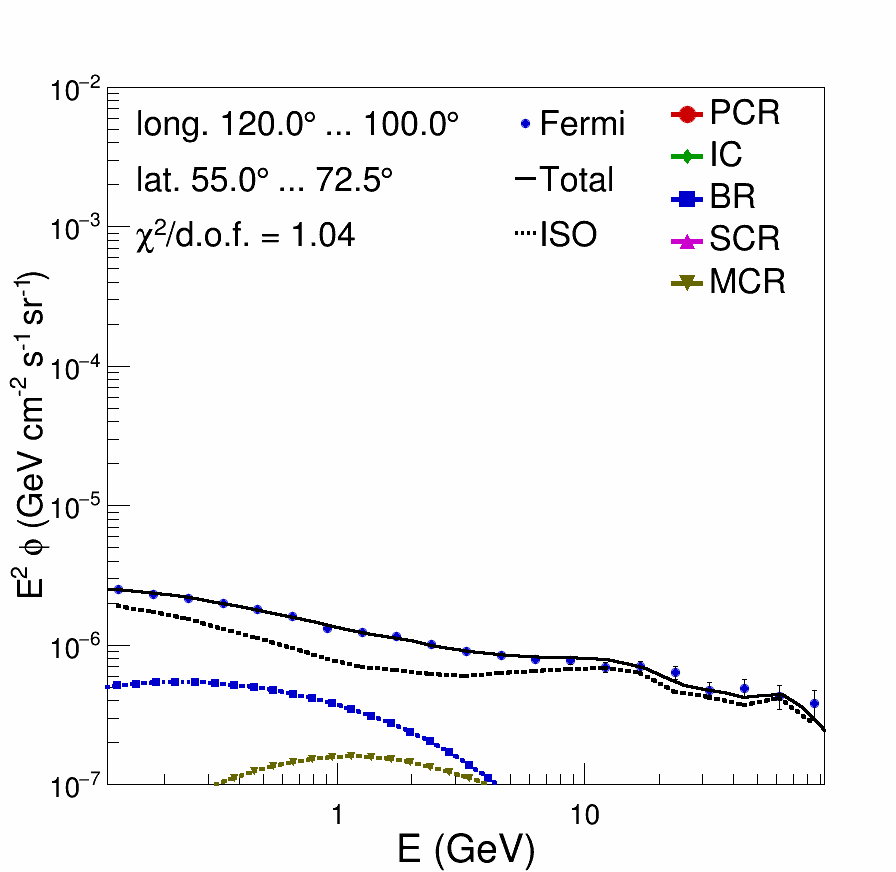}
\includegraphics[width=0.16\textwidth,height=0.16\textwidth,clip]{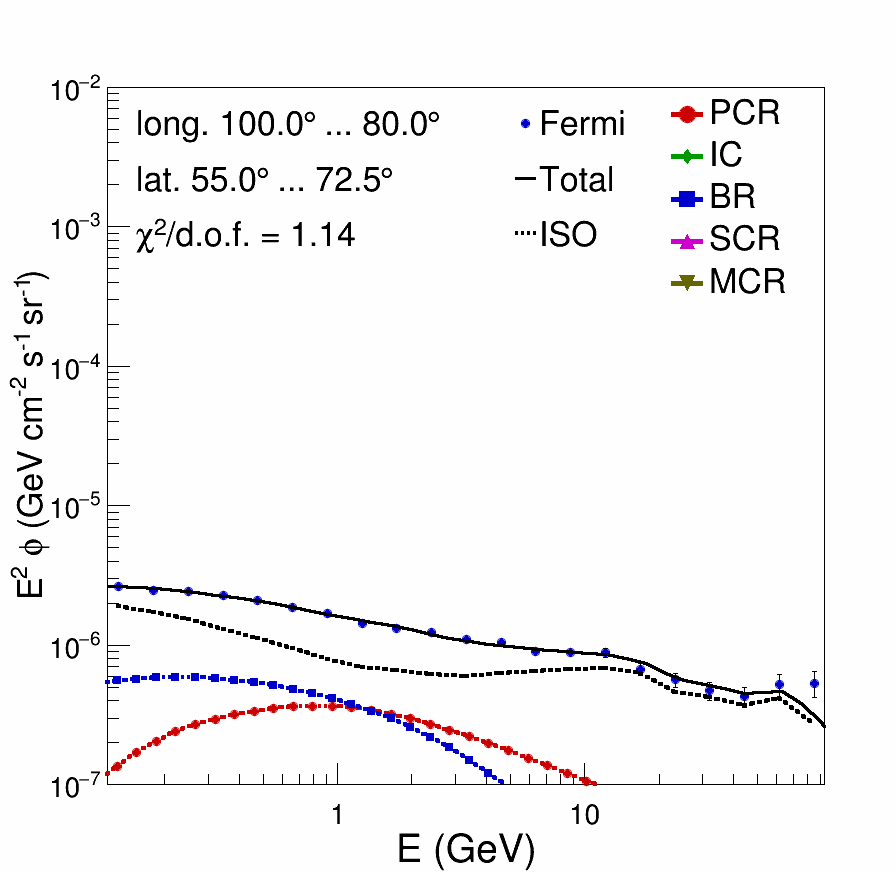}
\includegraphics[width=0.16\textwidth,height=0.16\textwidth,clip]{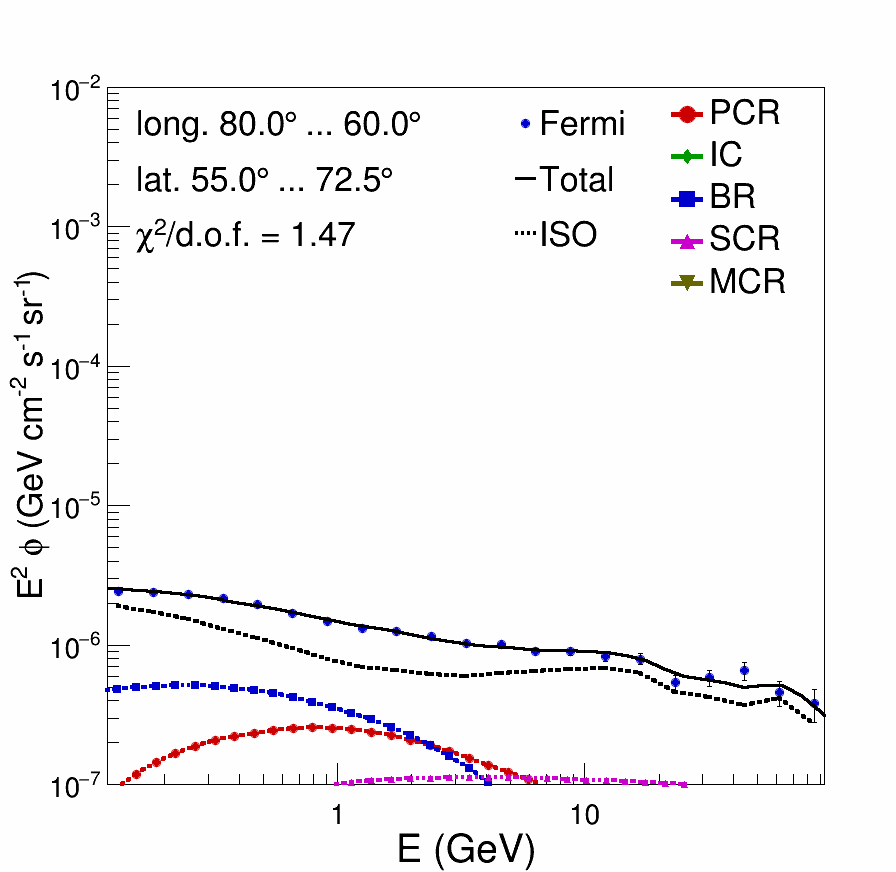}
\includegraphics[width=0.16\textwidth,height=0.16\textwidth,clip]{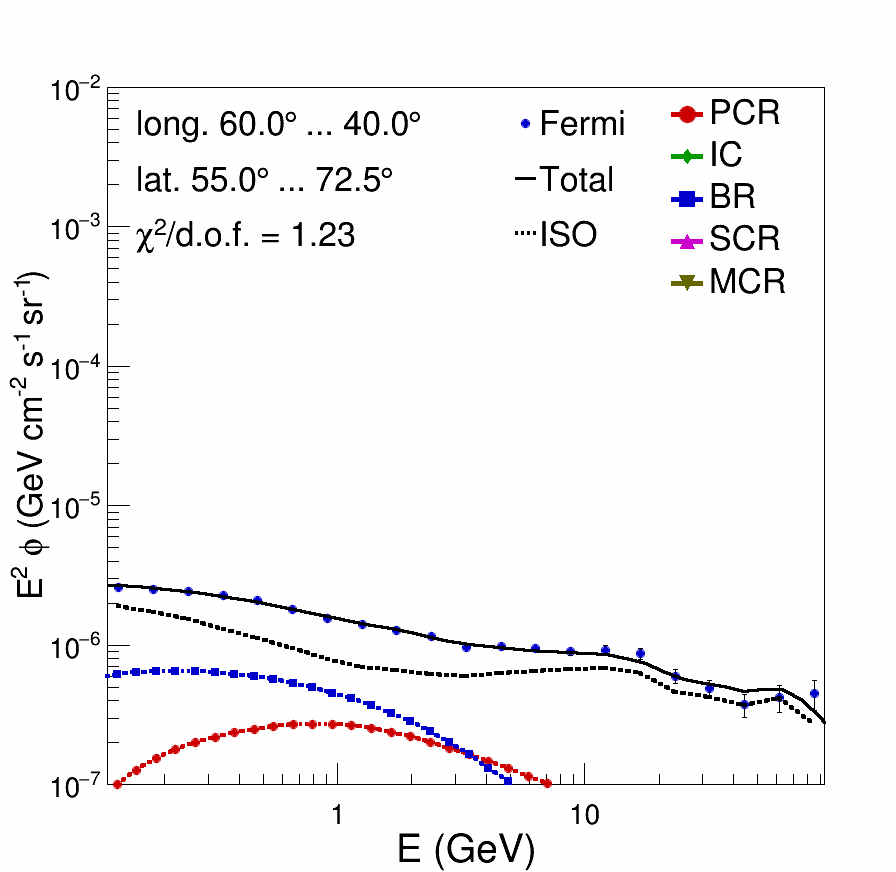}
\includegraphics[width=0.16\textwidth,height=0.16\textwidth,clip]{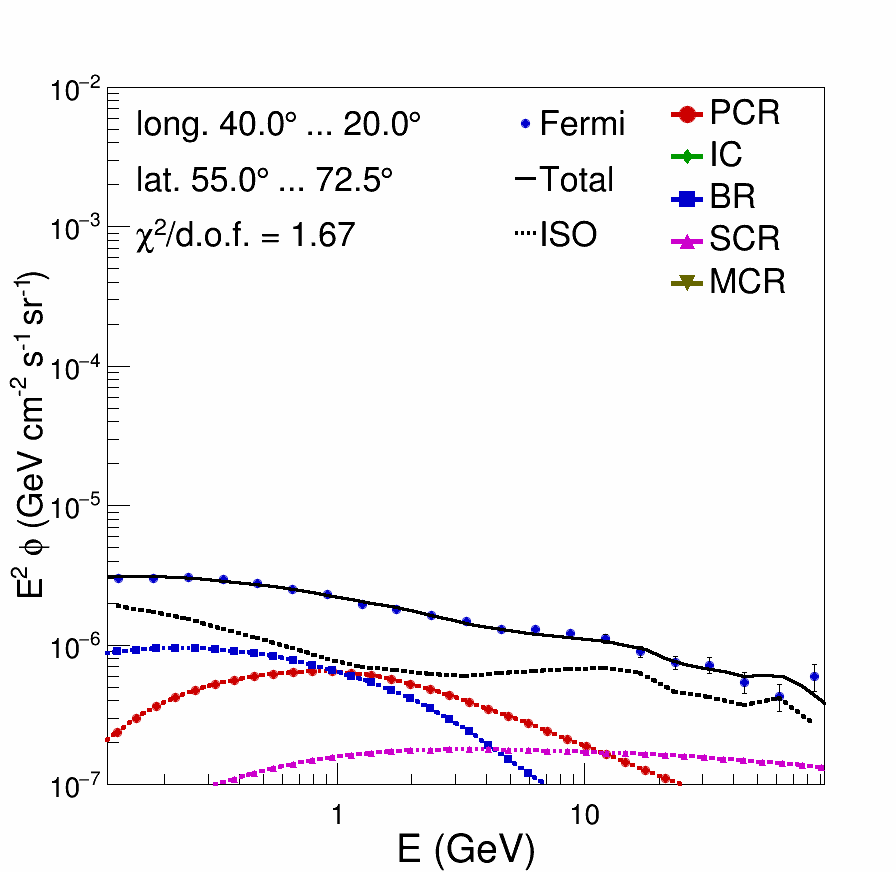}
\includegraphics[width=0.16\textwidth,height=0.16\textwidth,clip]{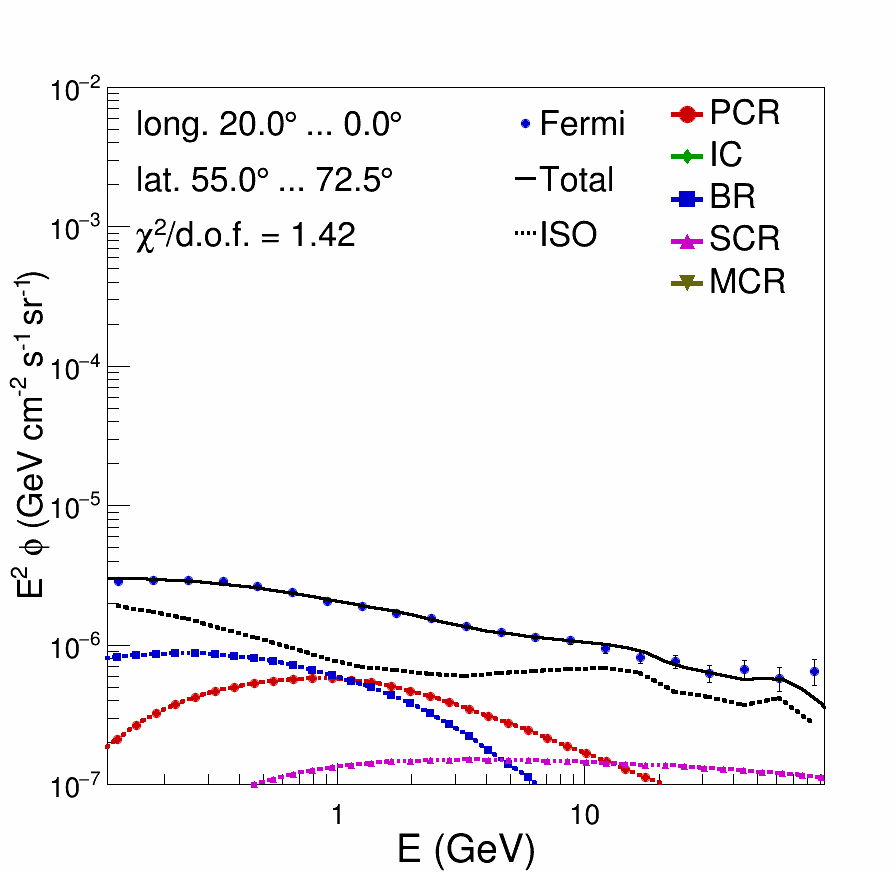}
\includegraphics[width=0.16\textwidth,height=0.16\textwidth,clip]{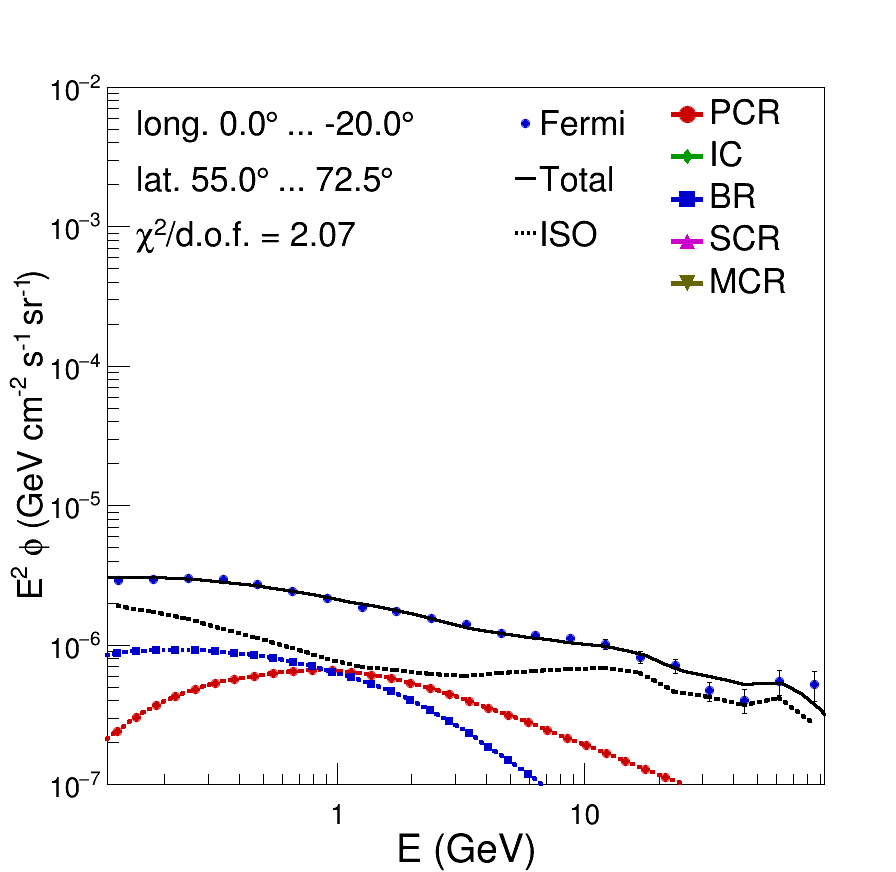}
\includegraphics[width=0.16\textwidth,height=0.16\textwidth,clip]{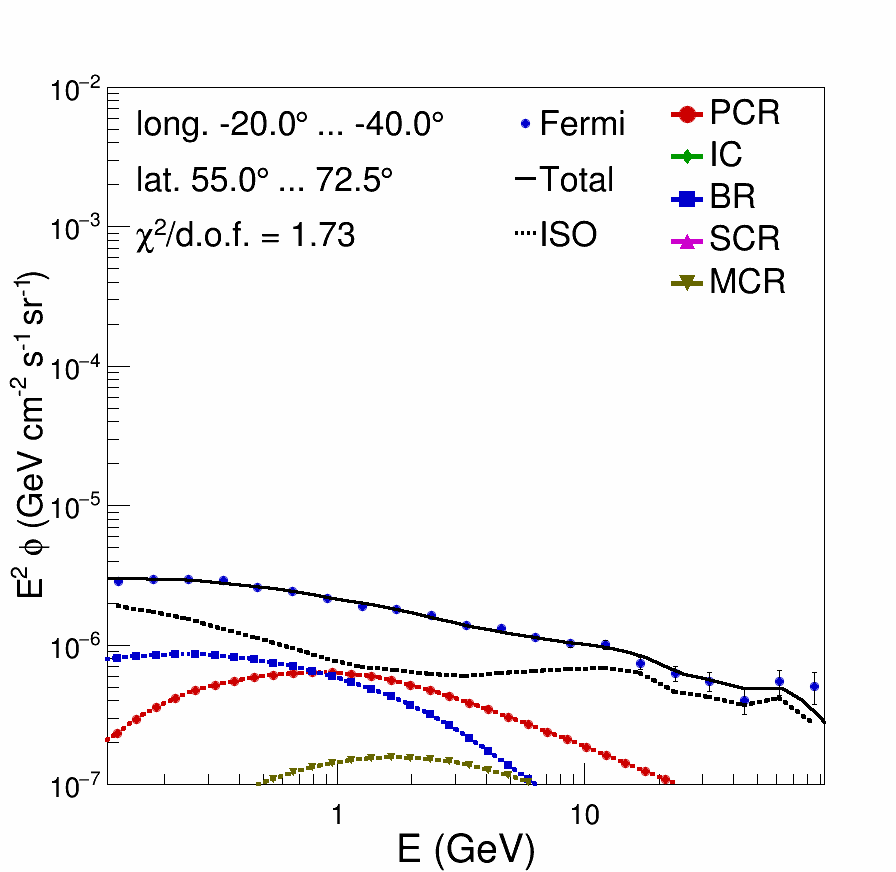}
\includegraphics[width=0.16\textwidth,height=0.16\textwidth,clip]{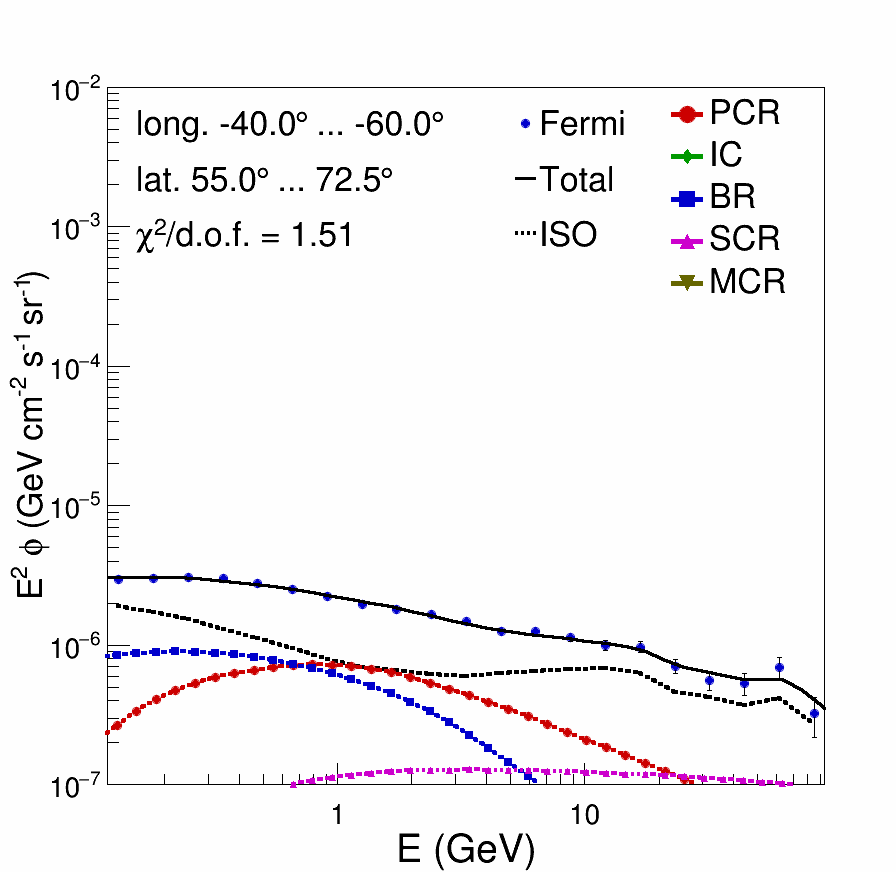}
\includegraphics[width=0.16\textwidth,height=0.16\textwidth,clip]{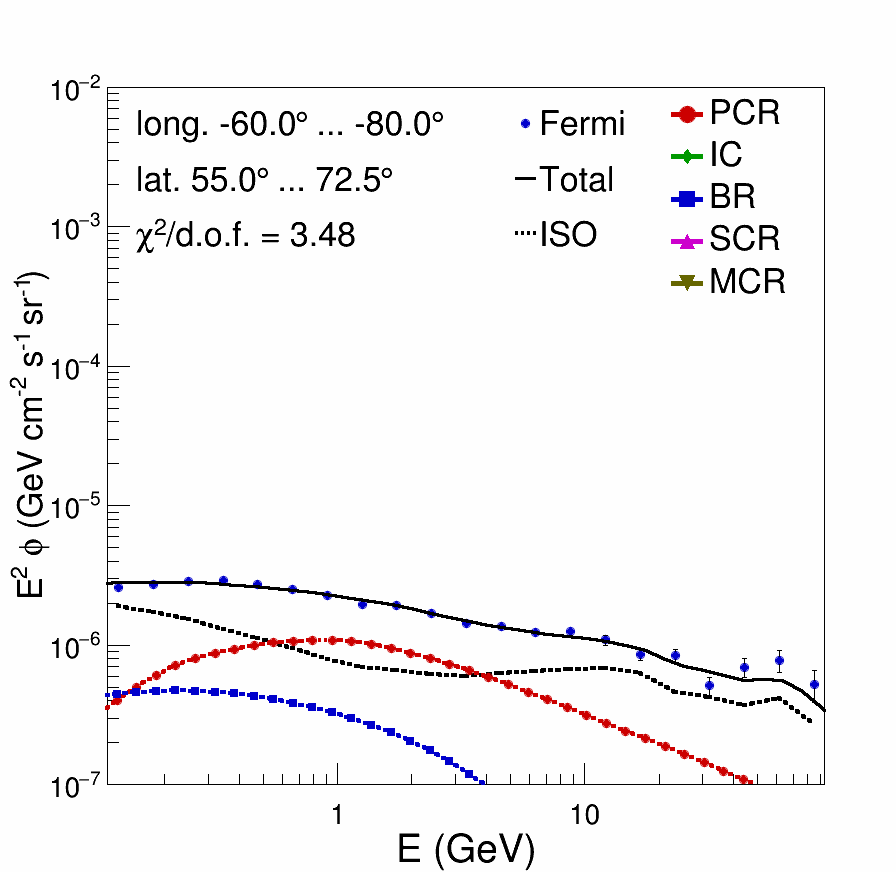}
\includegraphics[width=0.16\textwidth,height=0.16\textwidth,clip]{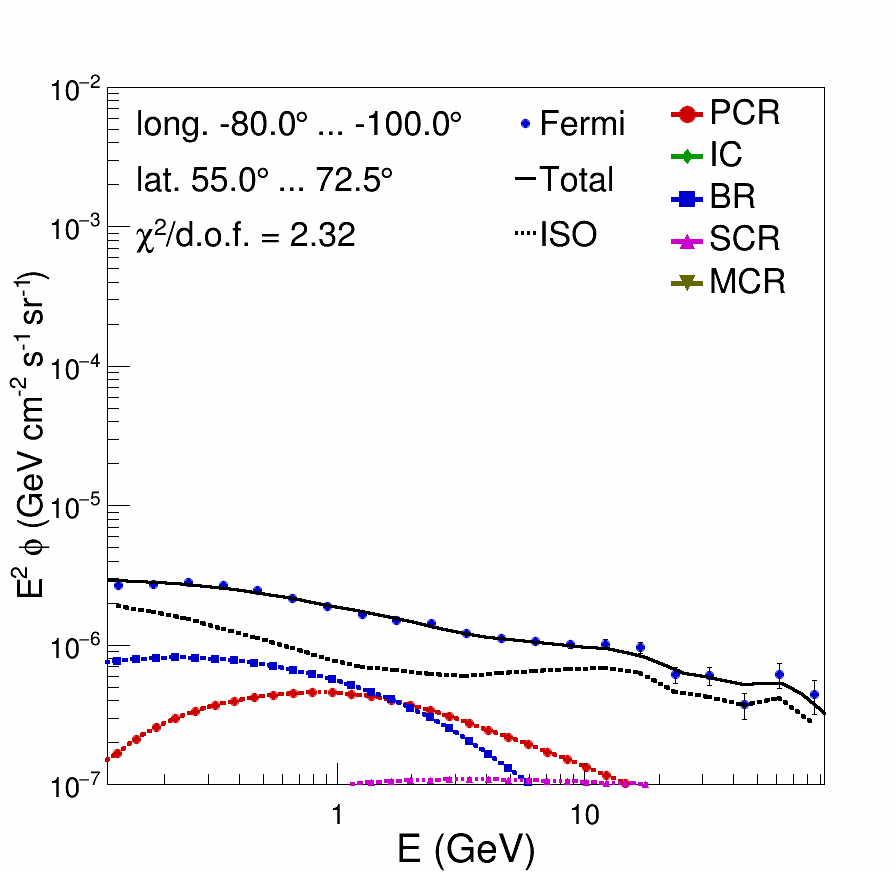}
\includegraphics[width=0.16\textwidth,height=0.16\textwidth,clip]{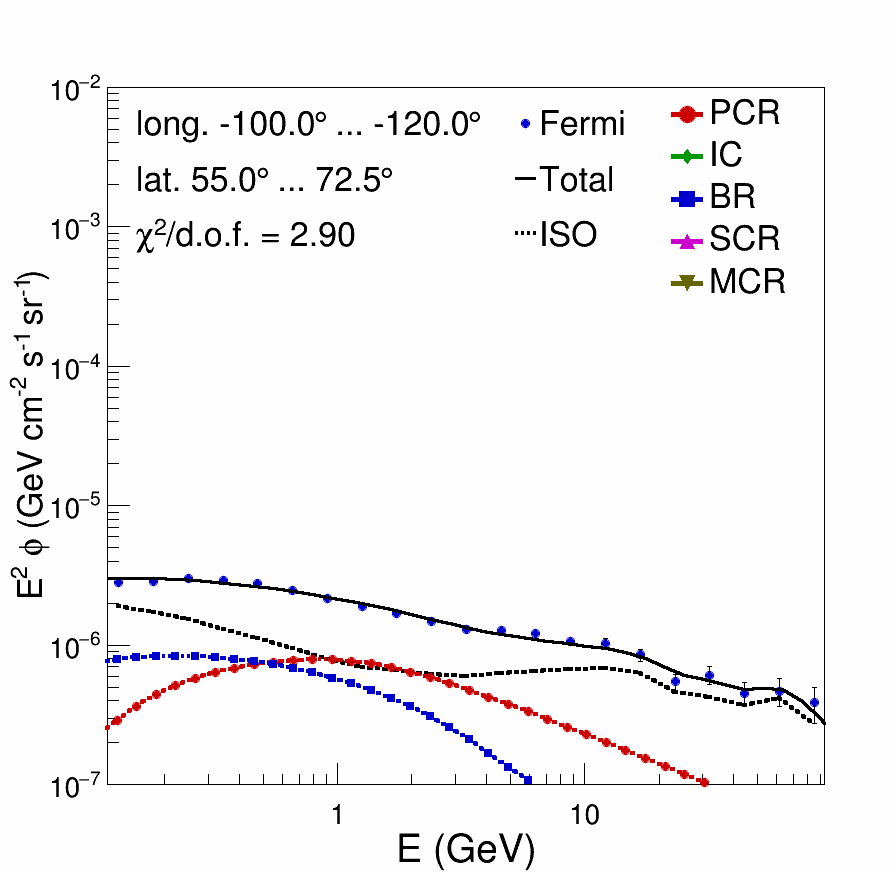}
\includegraphics[width=0.16\textwidth,height=0.16\textwidth,clip]{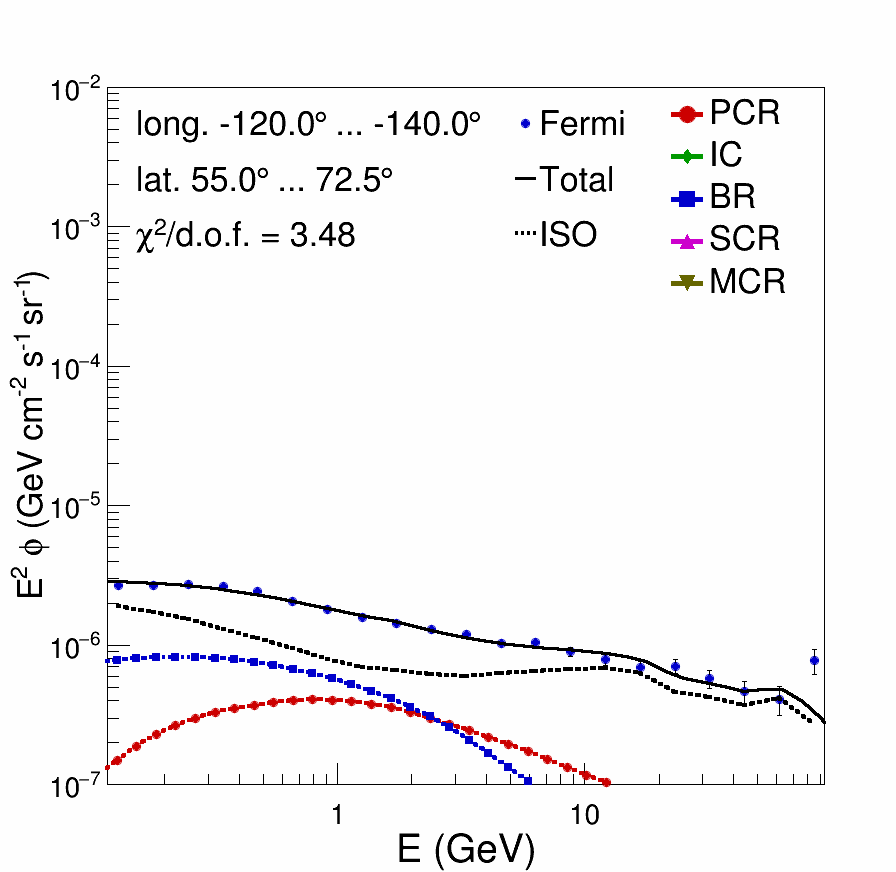}
\includegraphics[width=0.16\textwidth,height=0.16\textwidth,clip]{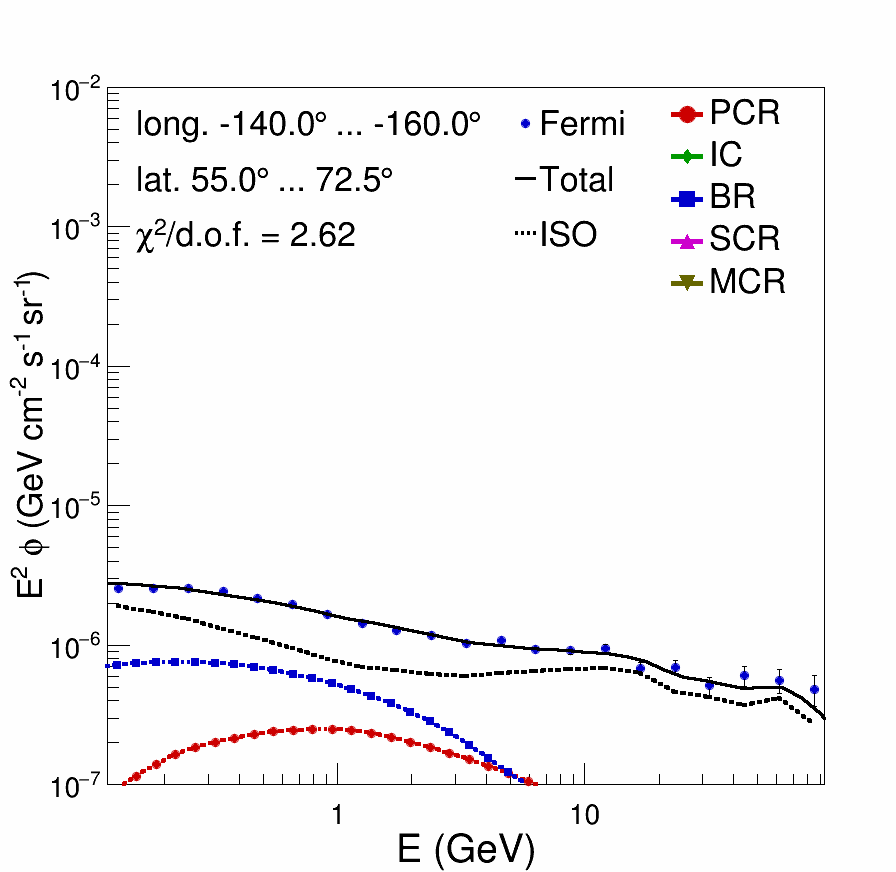}
\includegraphics[width=0.16\textwidth,height=0.16\textwidth,clip]{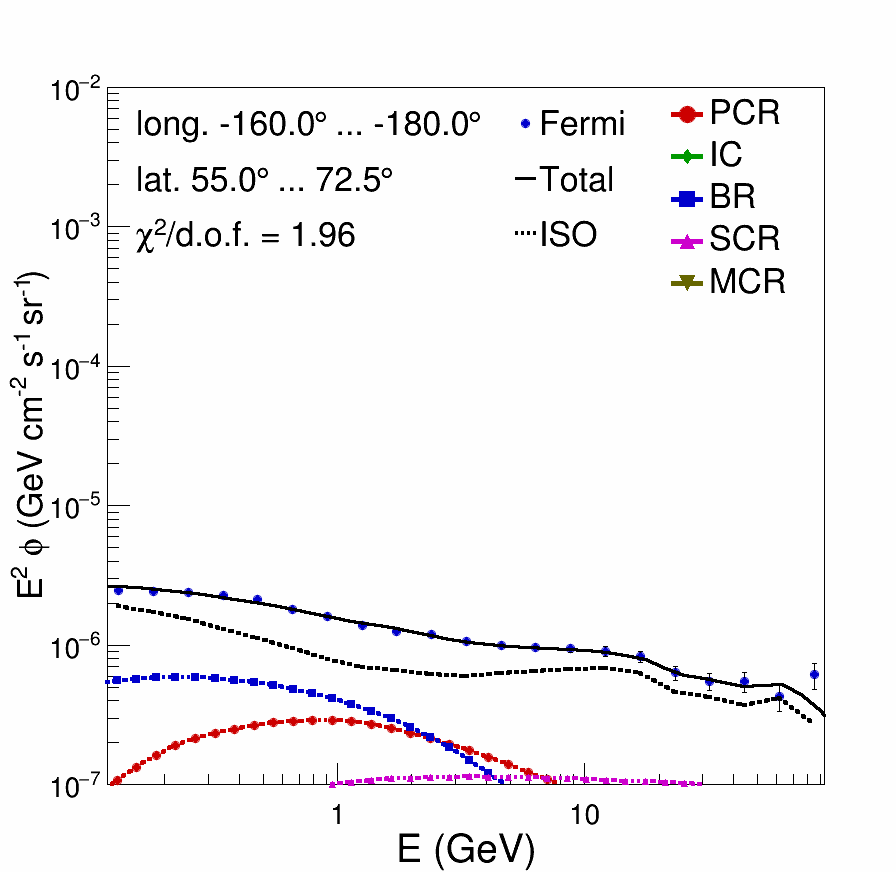}
\caption[]{Template fits for latitudes  with $55.0^\circ<b<72.5^\circ$ and longitudes decreasing from 180$^\circ$ to -180$^\circ$.} 
\label{F12}
\end{figure}
\begin{figure}
\centering
\includegraphics[width=0.16\textwidth,height=0.16\textwidth,clip]{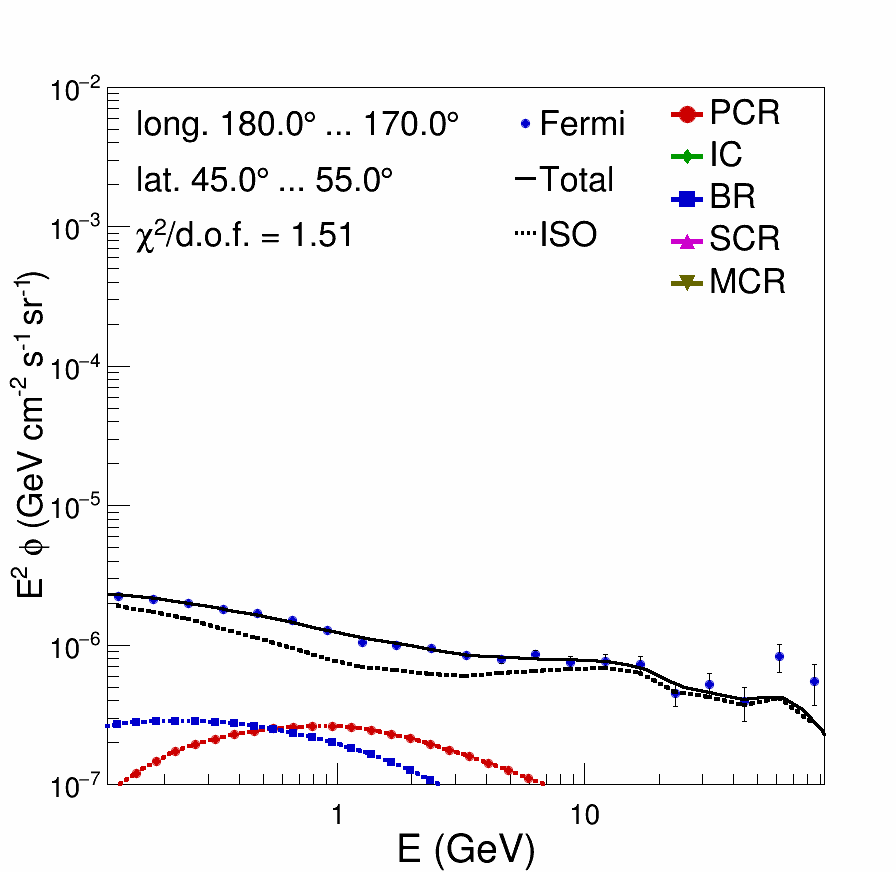}
\includegraphics[width=0.16\textwidth,height=0.16\textwidth,clip]{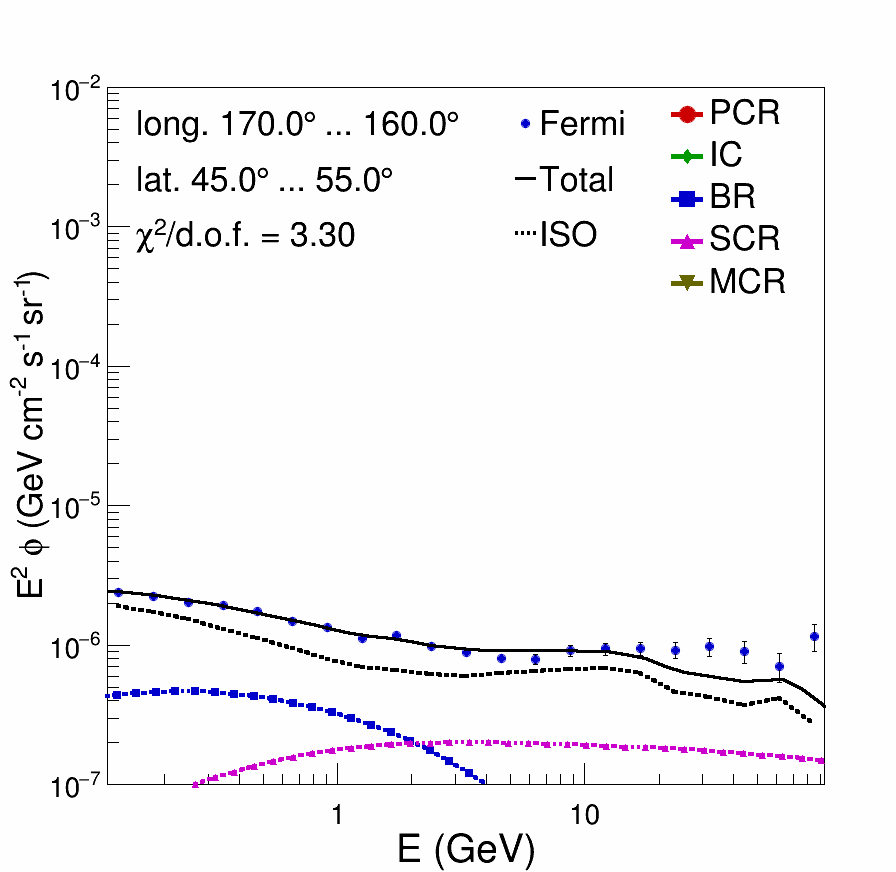}
\includegraphics[width=0.16\textwidth,height=0.16\textwidth,clip]{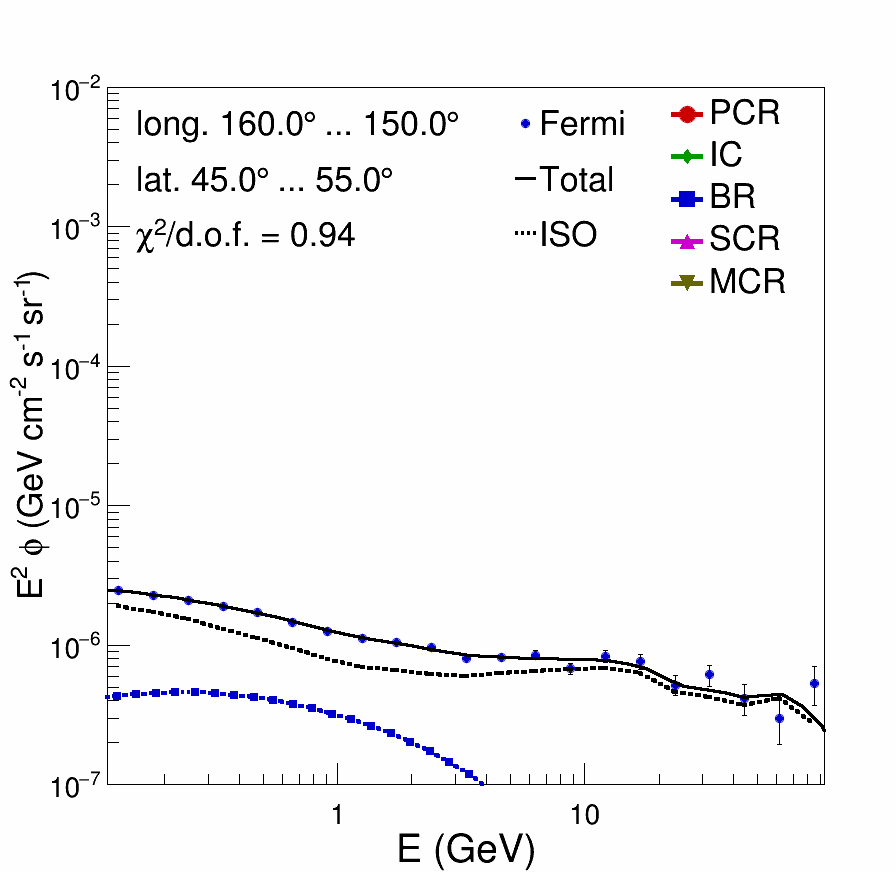}
\includegraphics[width=0.16\textwidth,height=0.16\textwidth,clip]{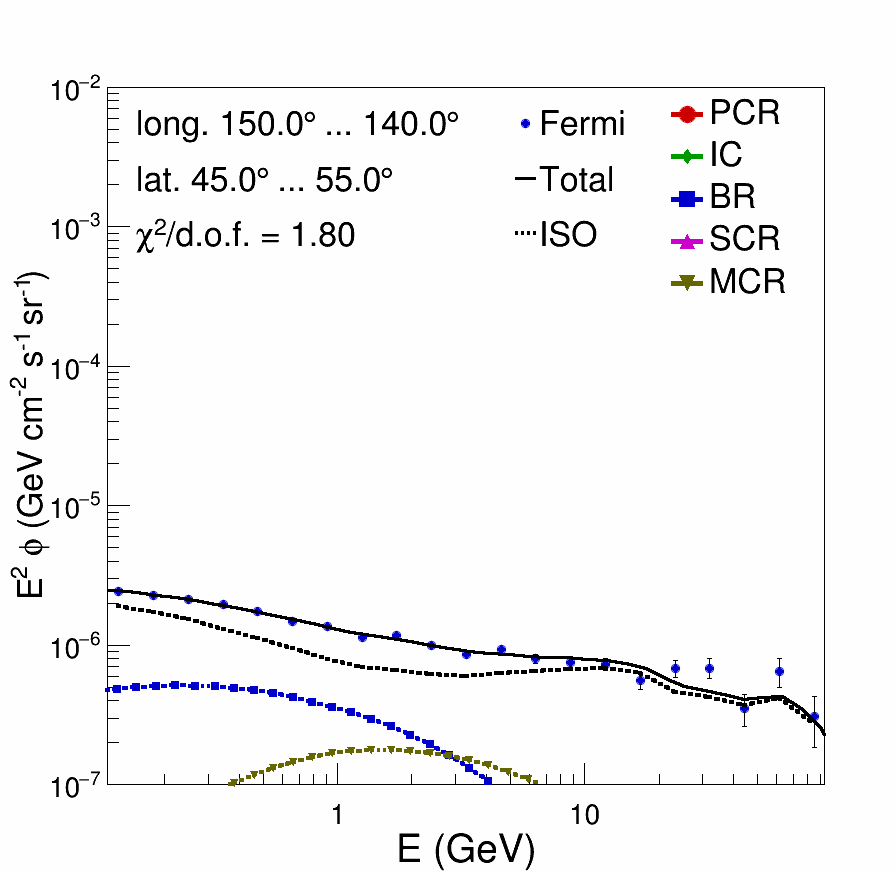}
\includegraphics[width=0.16\textwidth,height=0.16\textwidth,clip]{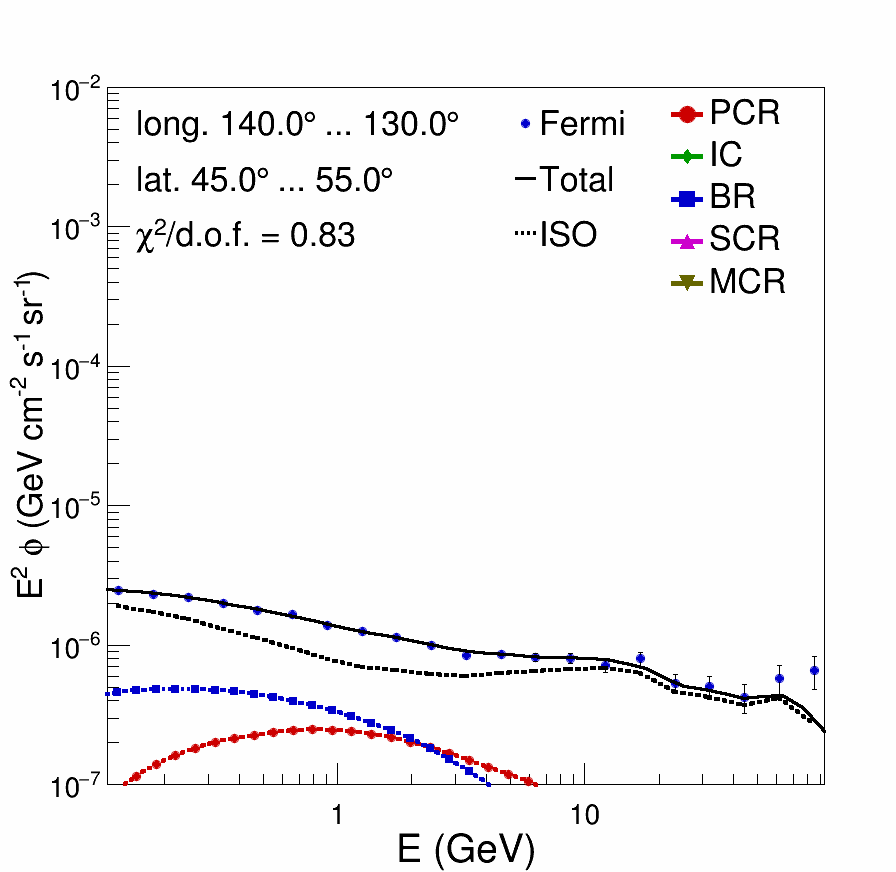}
\includegraphics[width=0.16\textwidth,height=0.16\textwidth,clip]{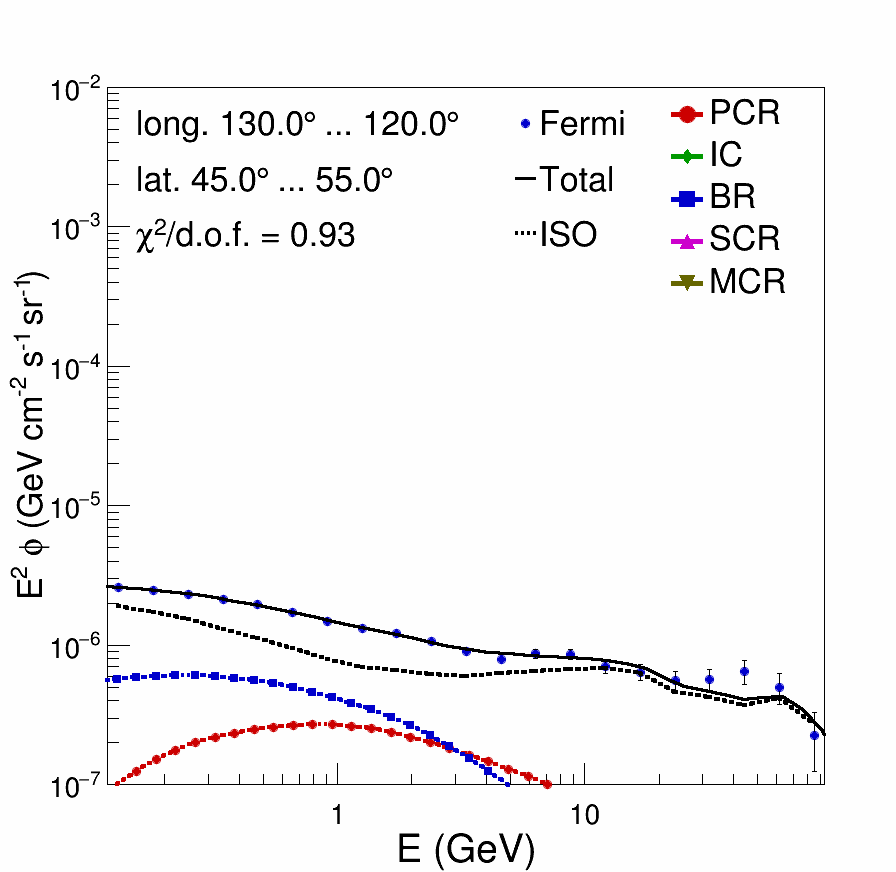}
\includegraphics[width=0.16\textwidth,height=0.16\textwidth,clip]{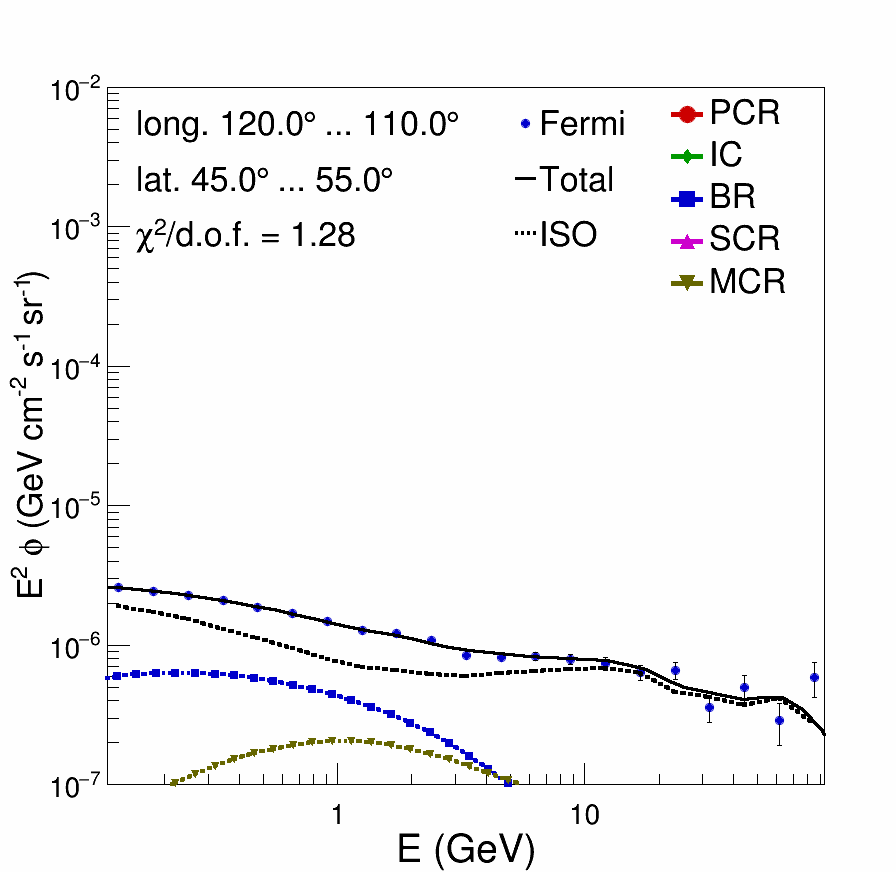}
\includegraphics[width=0.16\textwidth,height=0.16\textwidth,clip]{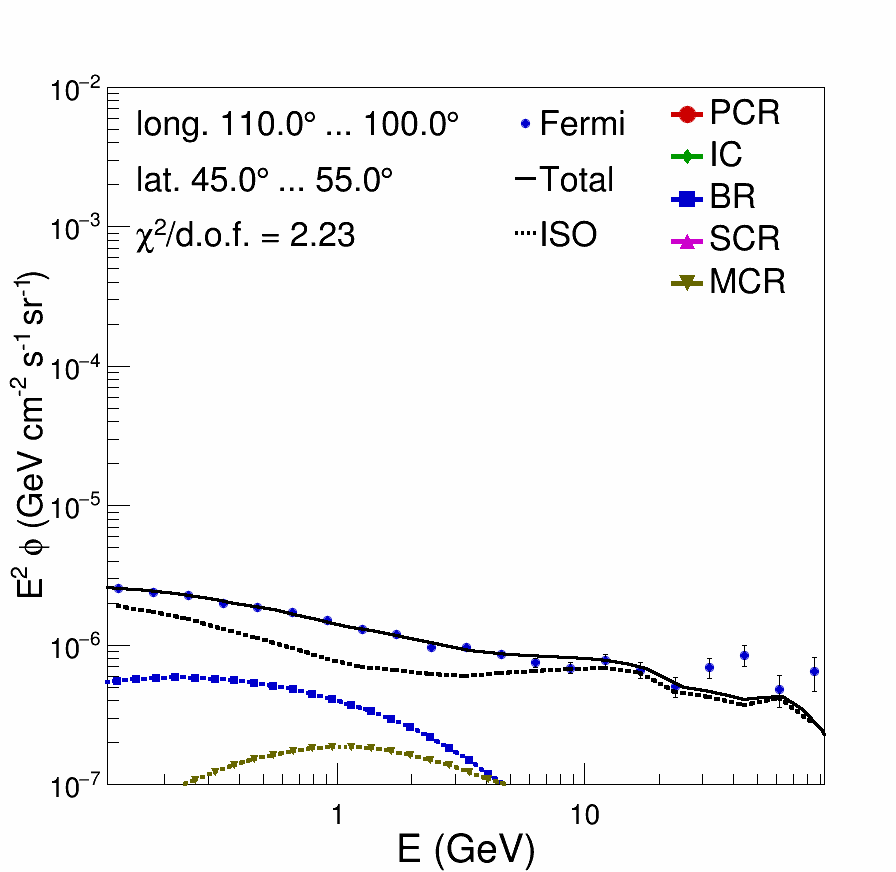}
\includegraphics[width=0.16\textwidth,height=0.16\textwidth,clip]{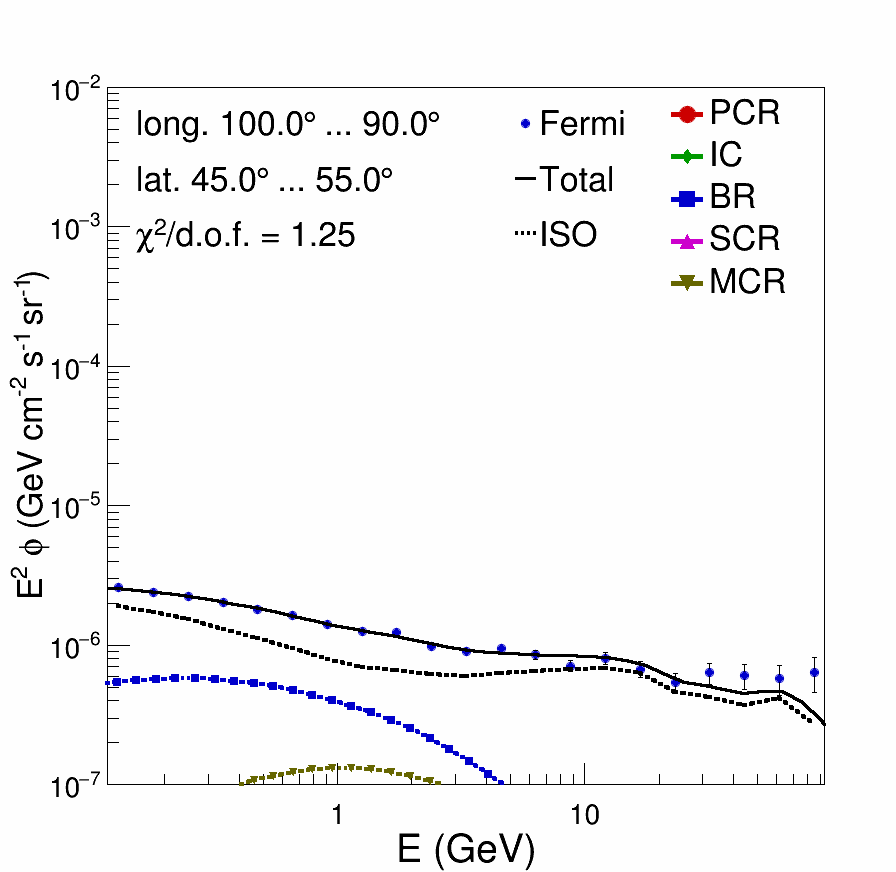}
\includegraphics[width=0.16\textwidth,height=0.16\textwidth,clip]{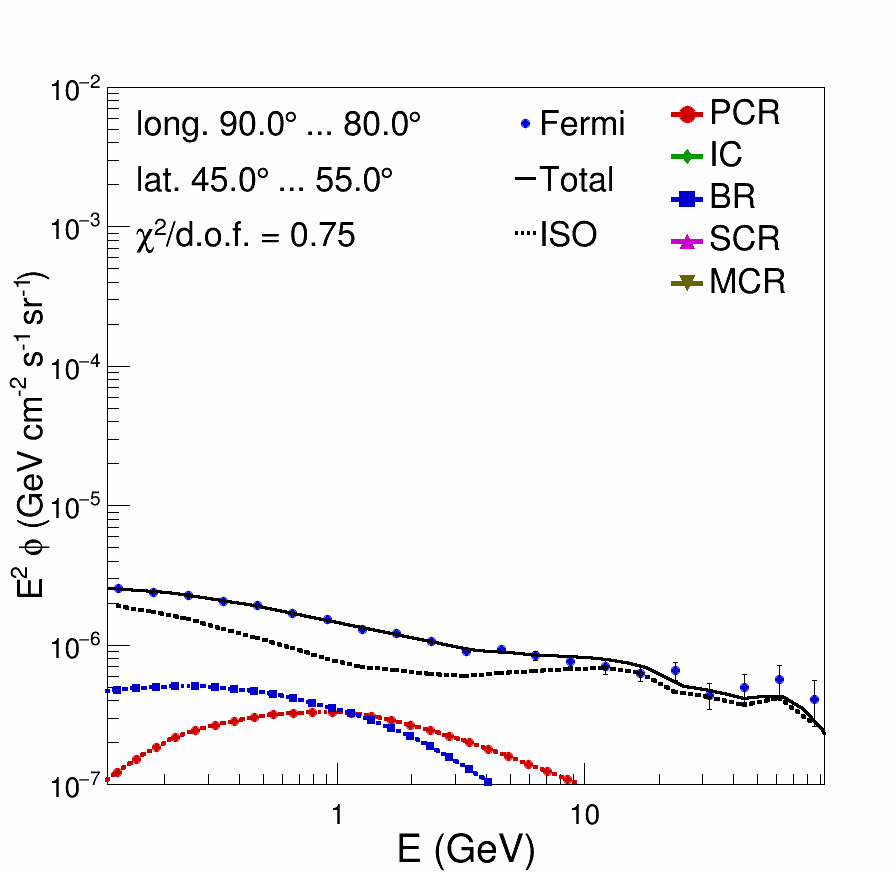}
\includegraphics[width=0.16\textwidth,height=0.16\textwidth,clip]{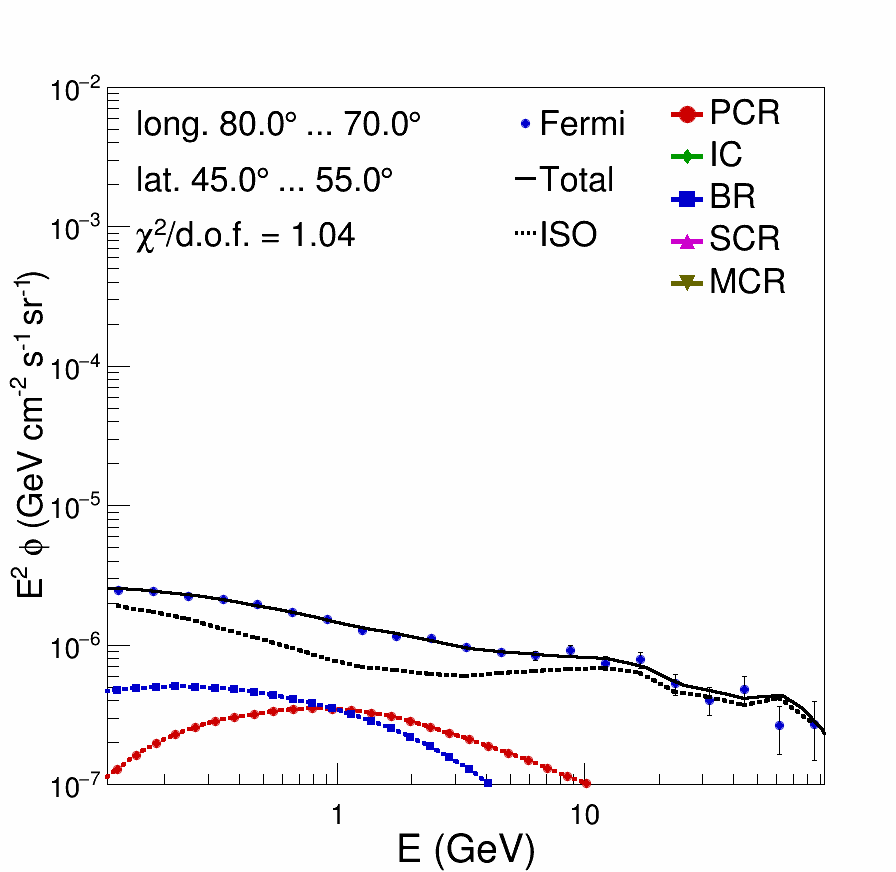}
\includegraphics[width=0.16\textwidth,height=0.16\textwidth,clip]{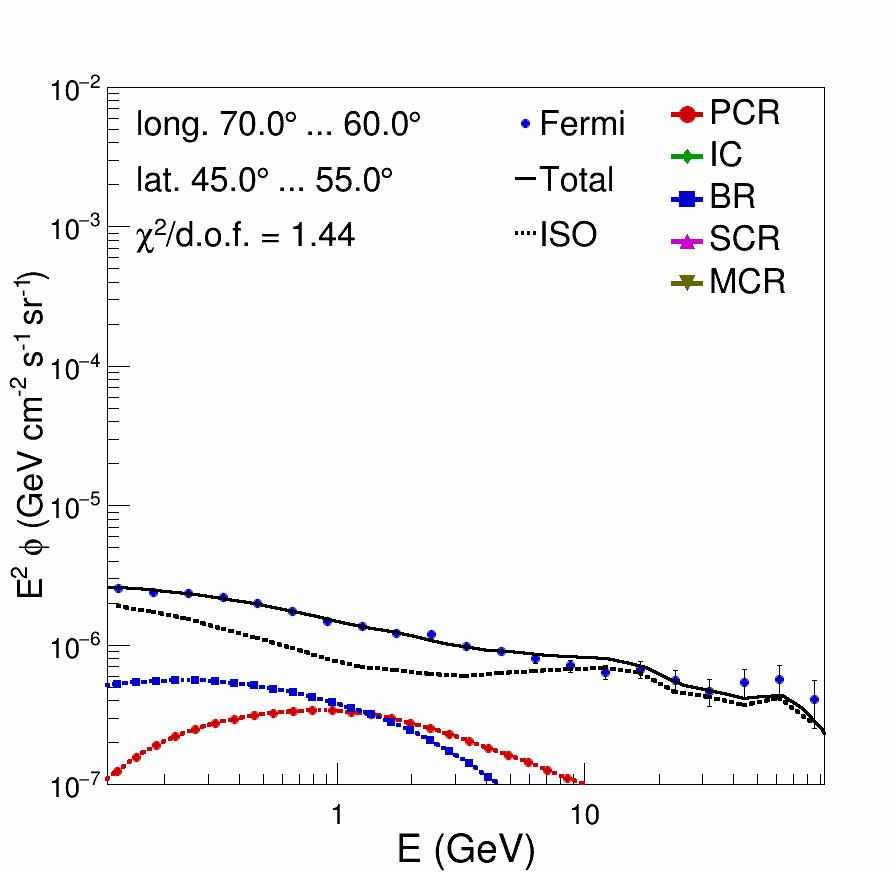}
\includegraphics[width=0.16\textwidth,height=0.16\textwidth,clip]{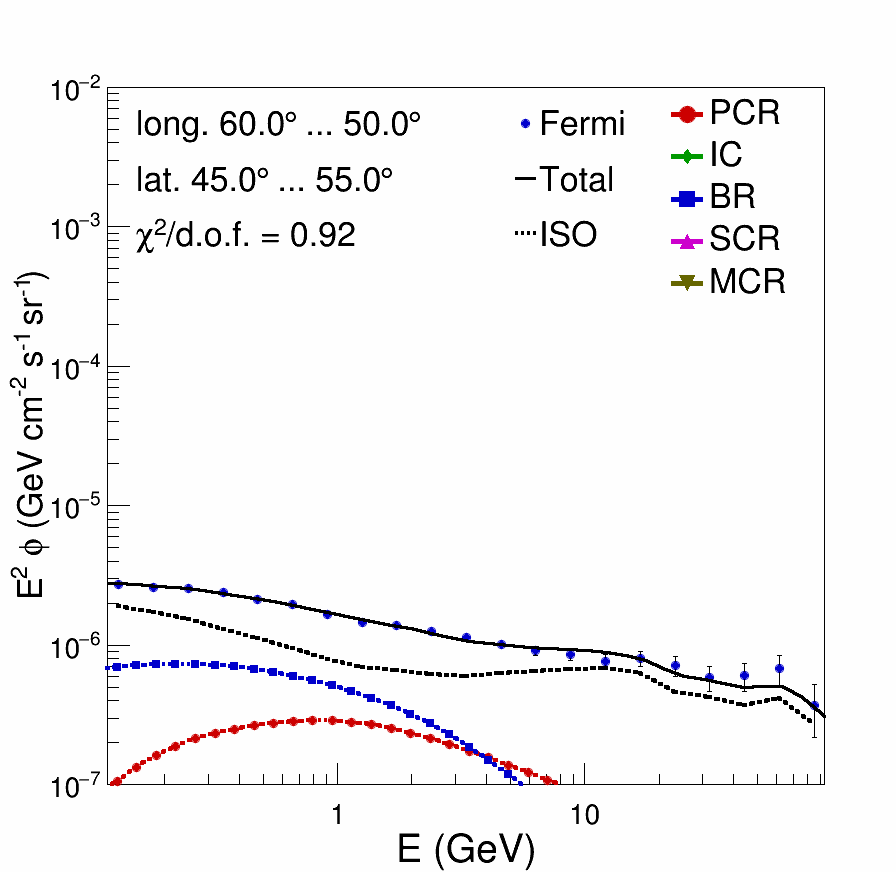}
\includegraphics[width=0.16\textwidth,height=0.16\textwidth,clip]{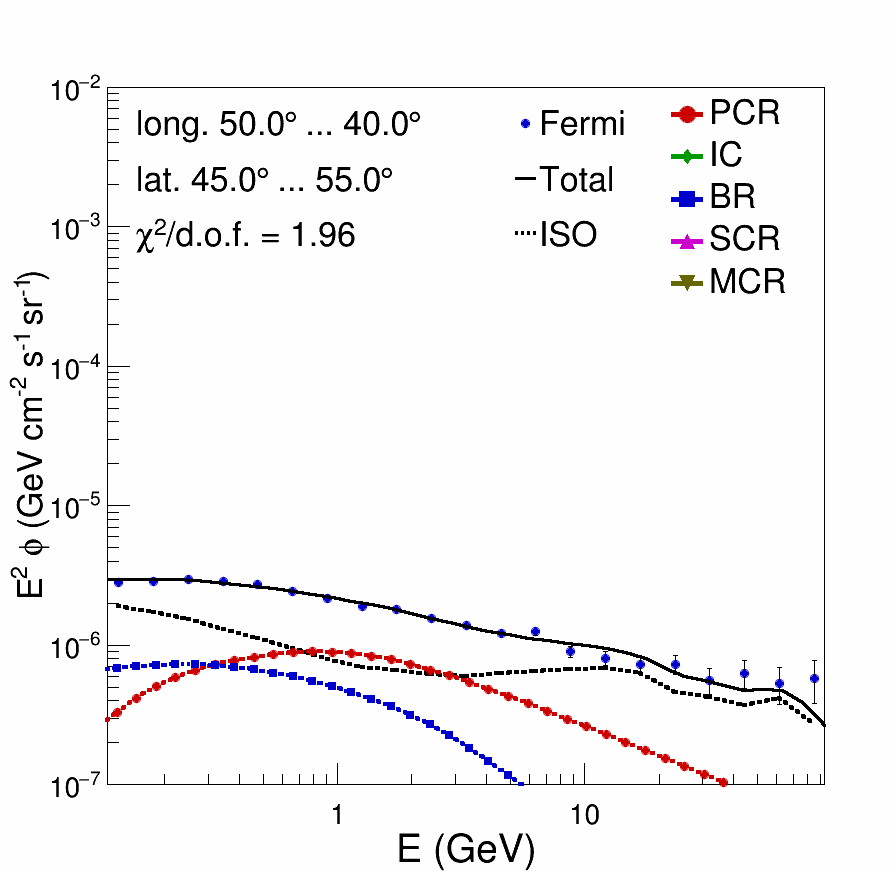}
\includegraphics[width=0.16\textwidth,height=0.16\textwidth,clip]{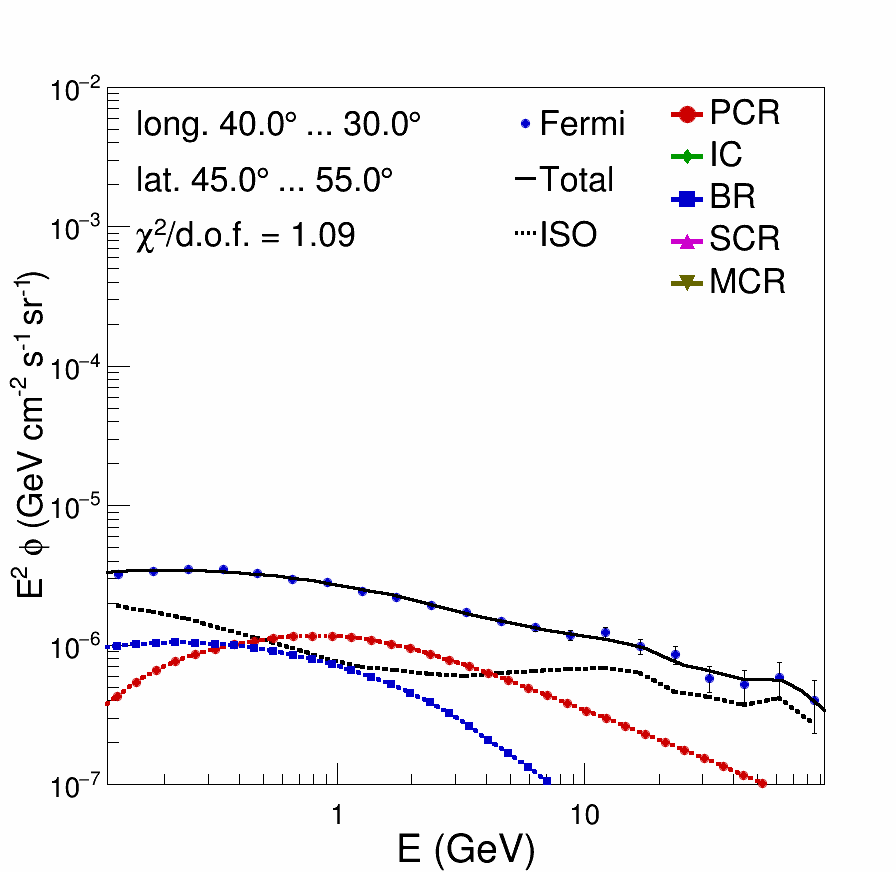}
\includegraphics[width=0.16\textwidth,height=0.16\textwidth,clip]{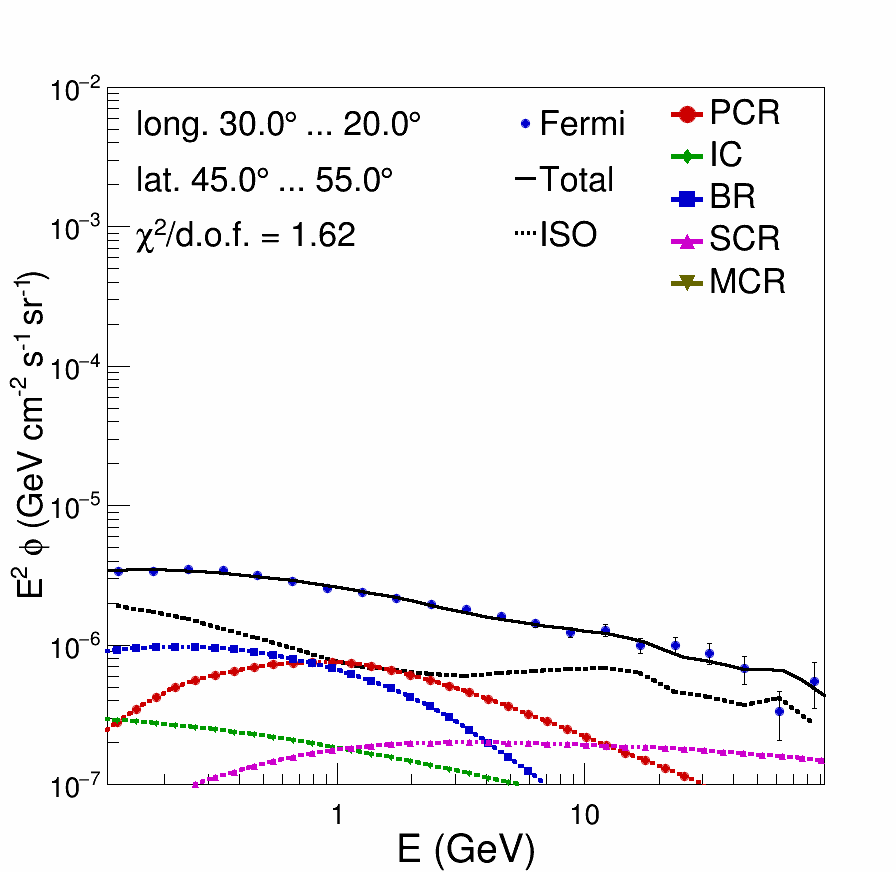}
\includegraphics[width=0.16\textwidth,height=0.16\textwidth,clip]{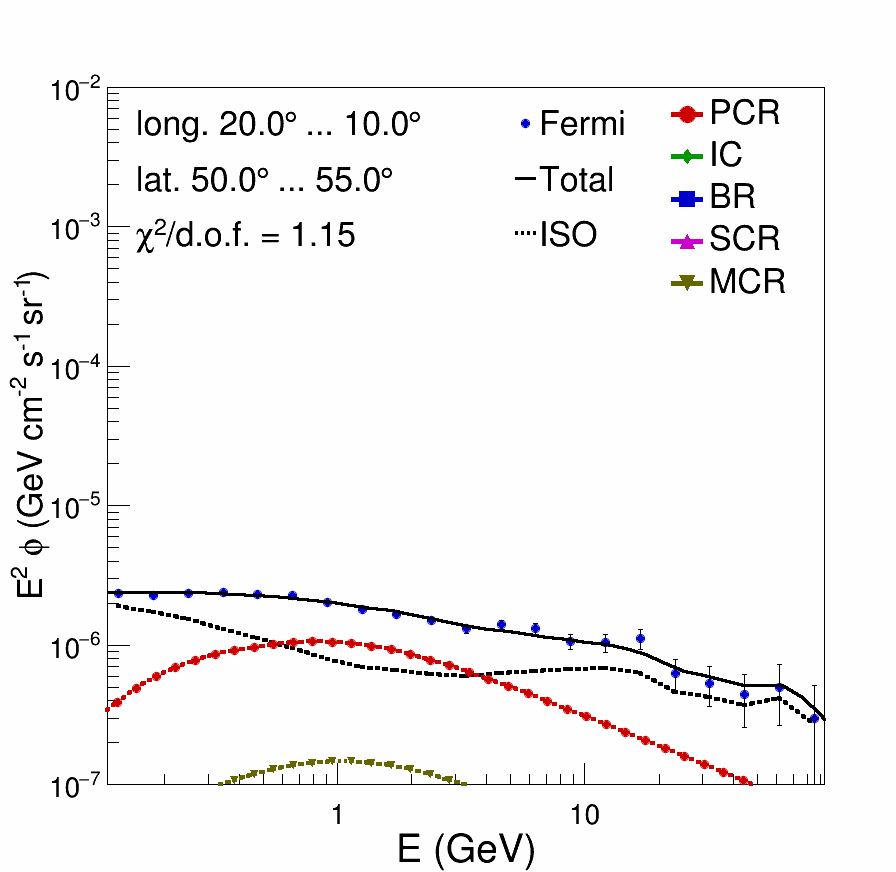}
\includegraphics[width=0.16\textwidth,height=0.16\textwidth,clip]{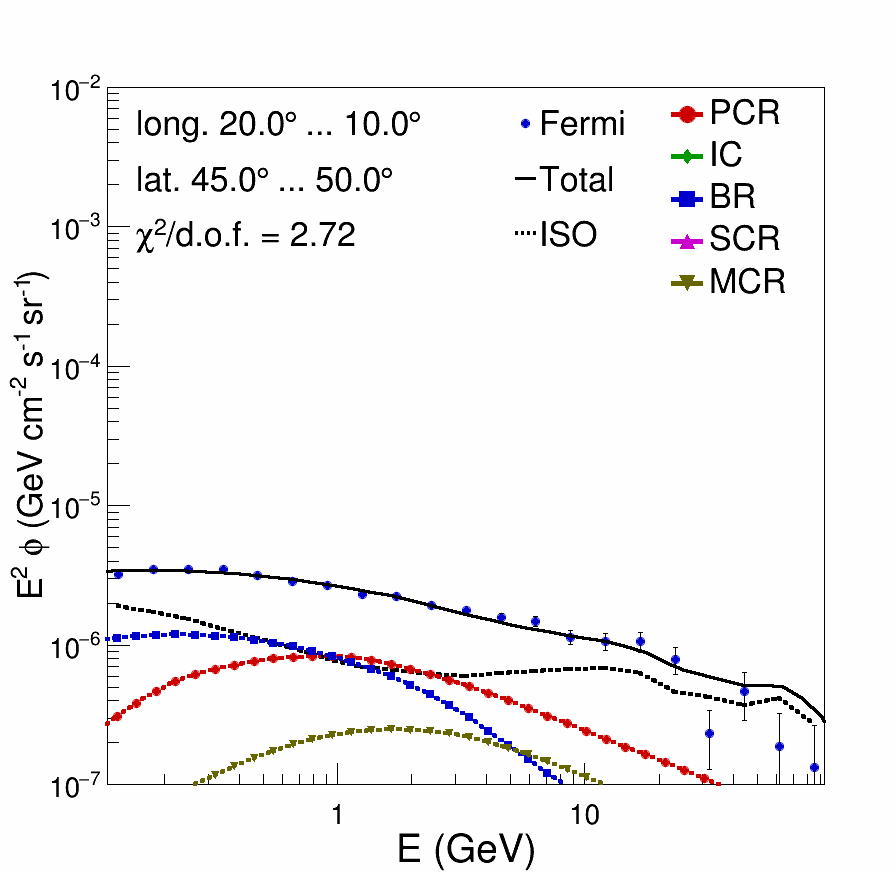}
\includegraphics[width=0.16\textwidth,height=0.16\textwidth,clip]{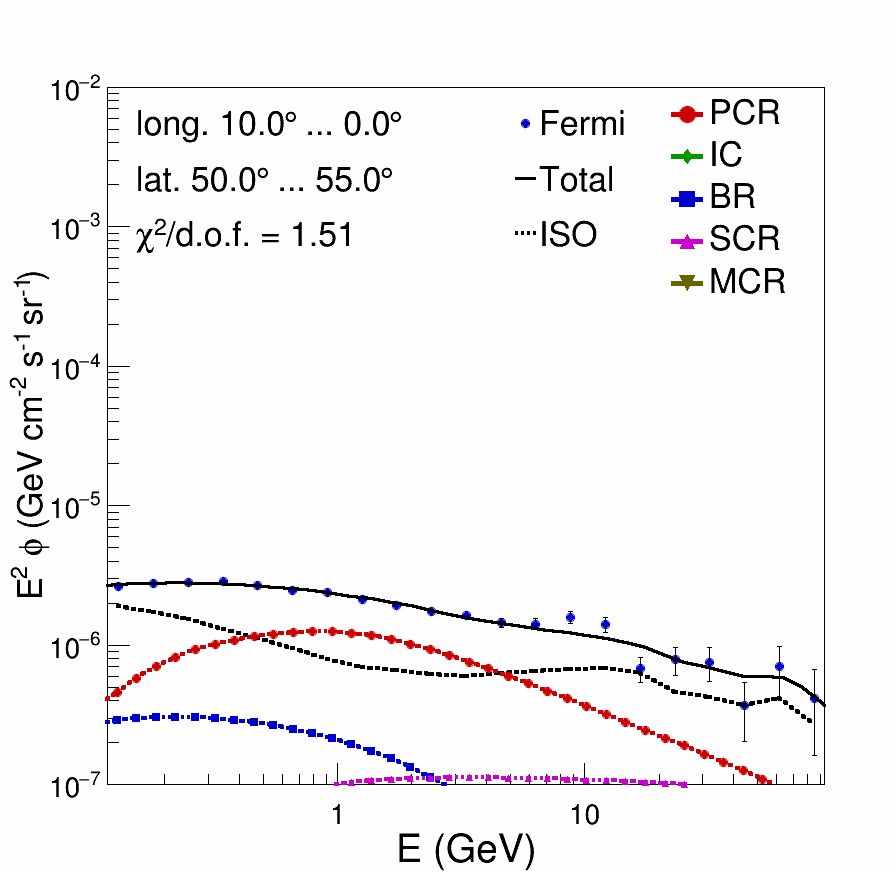}
\includegraphics[width=0.16\textwidth,height=0.16\textwidth,clip]{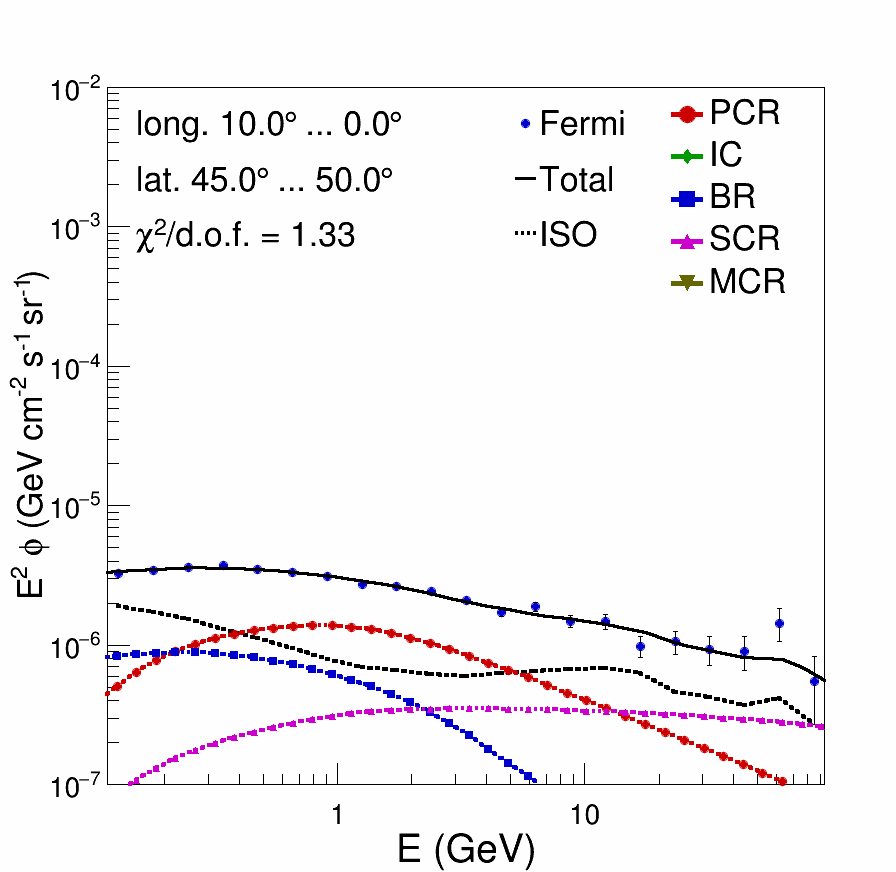}
\includegraphics[width=0.16\textwidth,height=0.16\textwidth,clip]{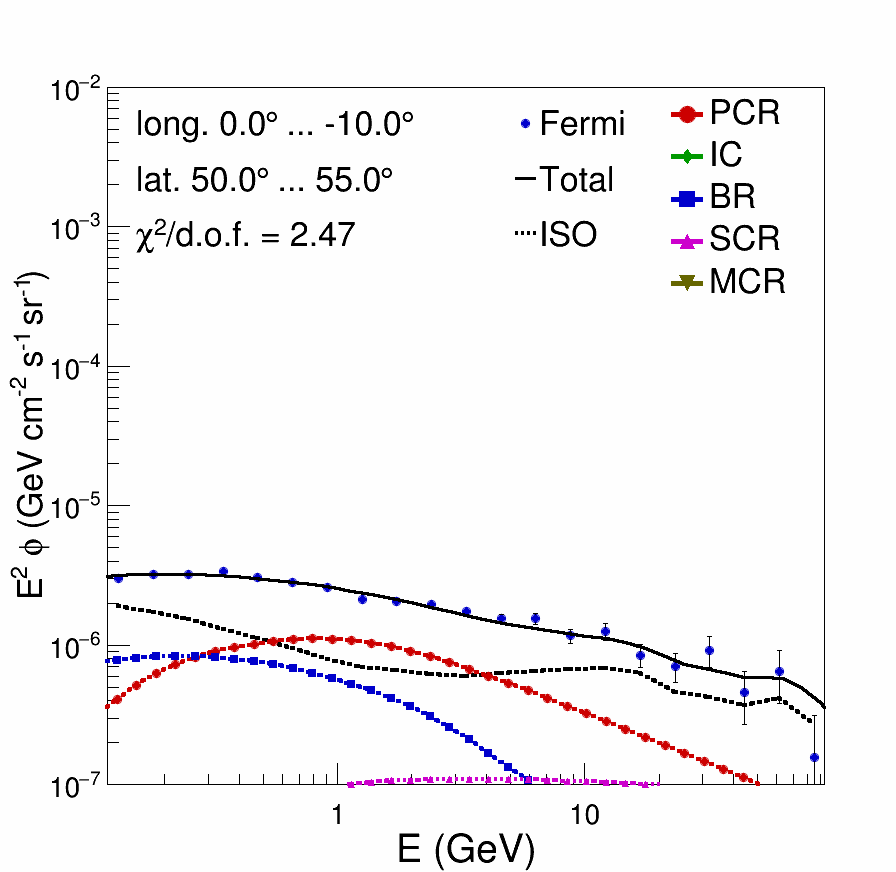}
\includegraphics[width=0.16\textwidth,height=0.16\textwidth,clip]{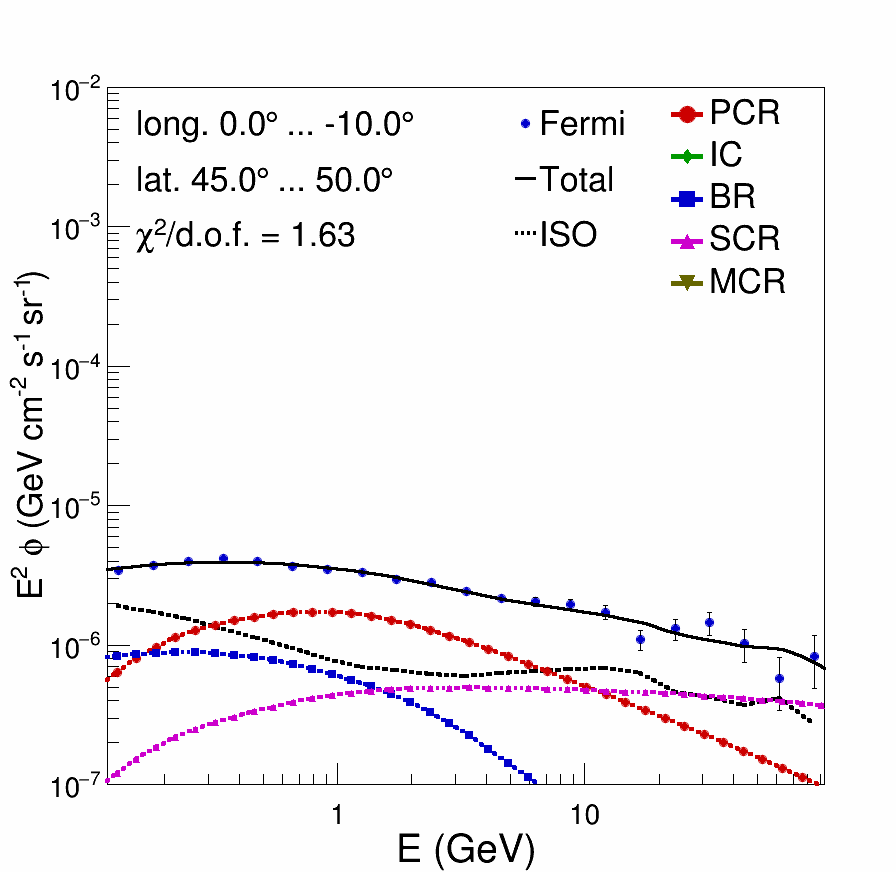}
\includegraphics[width=0.16\textwidth,height=0.16\textwidth,clip]{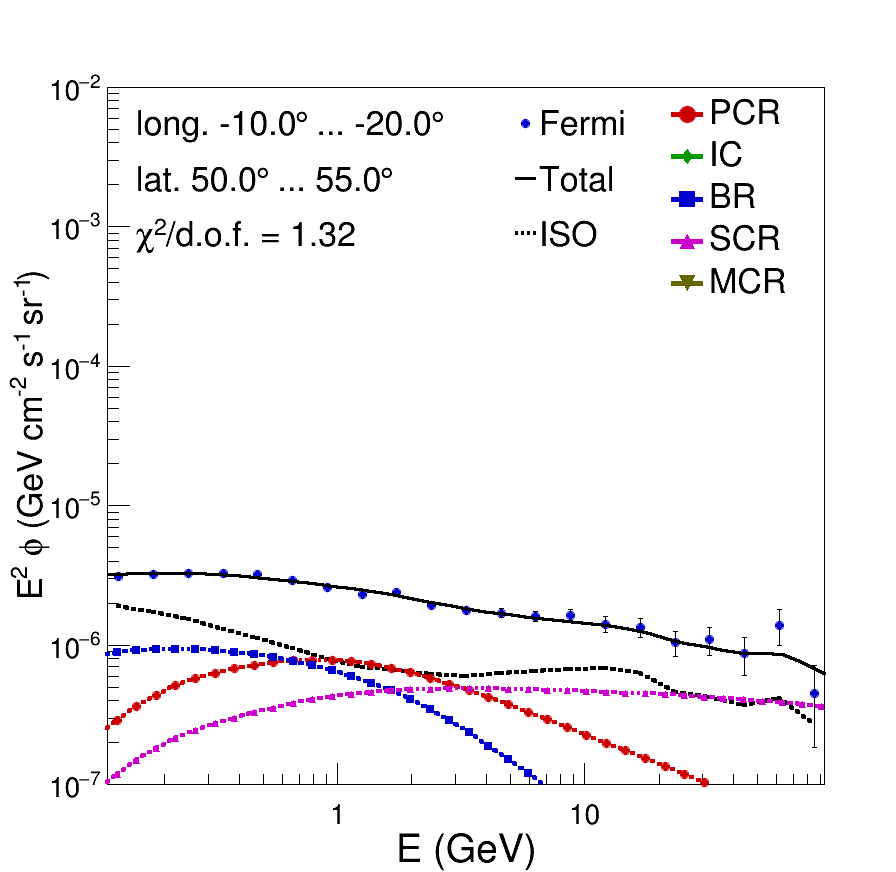}
\includegraphics[width=0.16\textwidth,height=0.16\textwidth,clip]{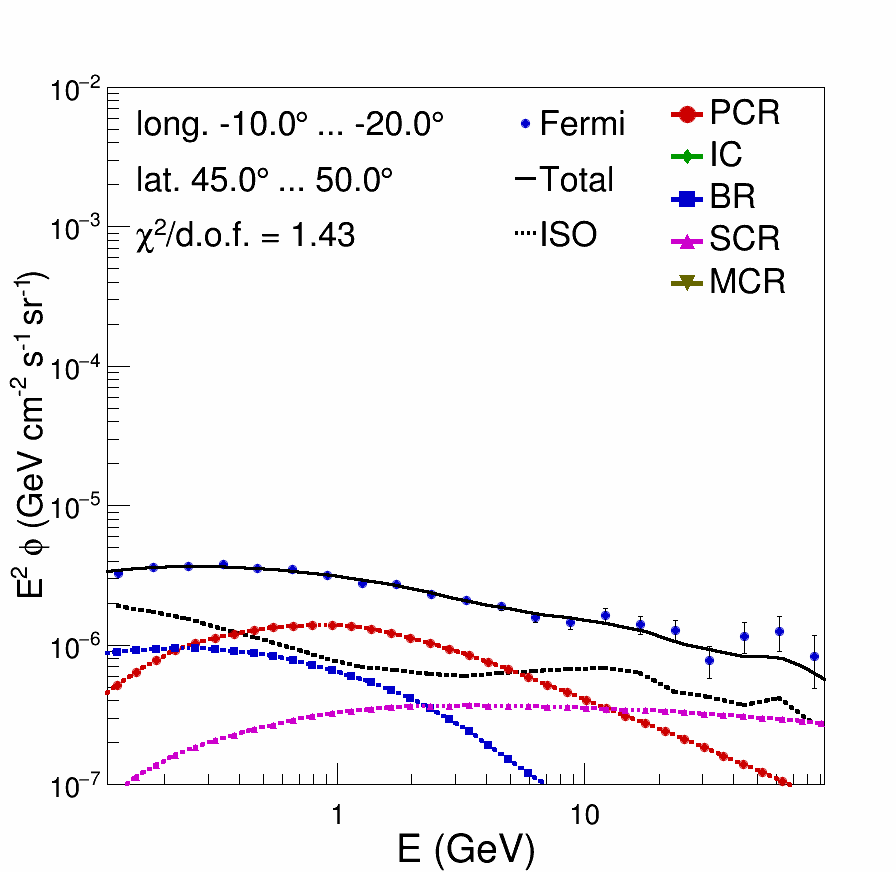}
\includegraphics[width=0.16\textwidth,height=0.16\textwidth,clip]{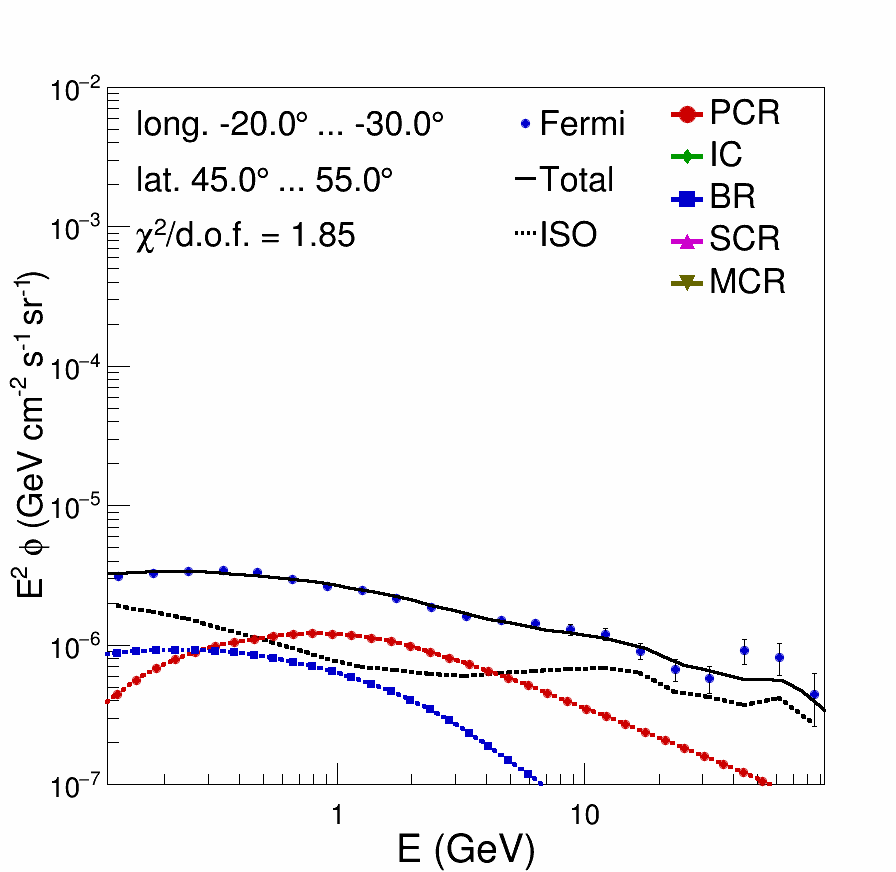}
\includegraphics[width=0.16\textwidth,height=0.16\textwidth,clip]{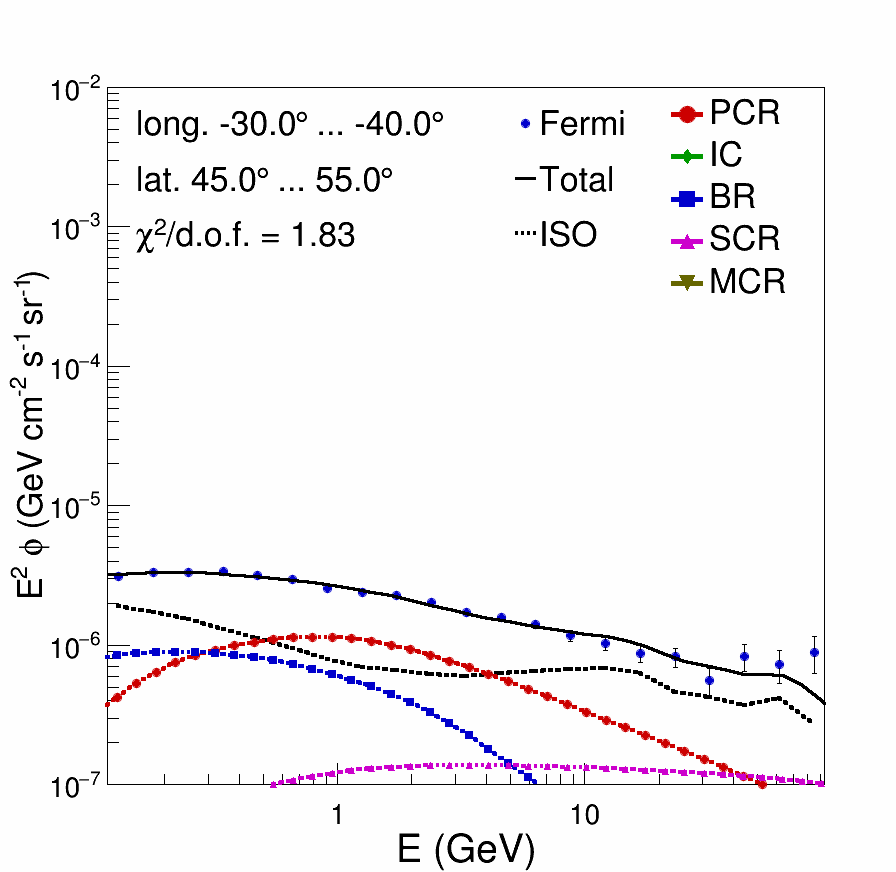}
\includegraphics[width=0.16\textwidth,height=0.16\textwidth,clip]{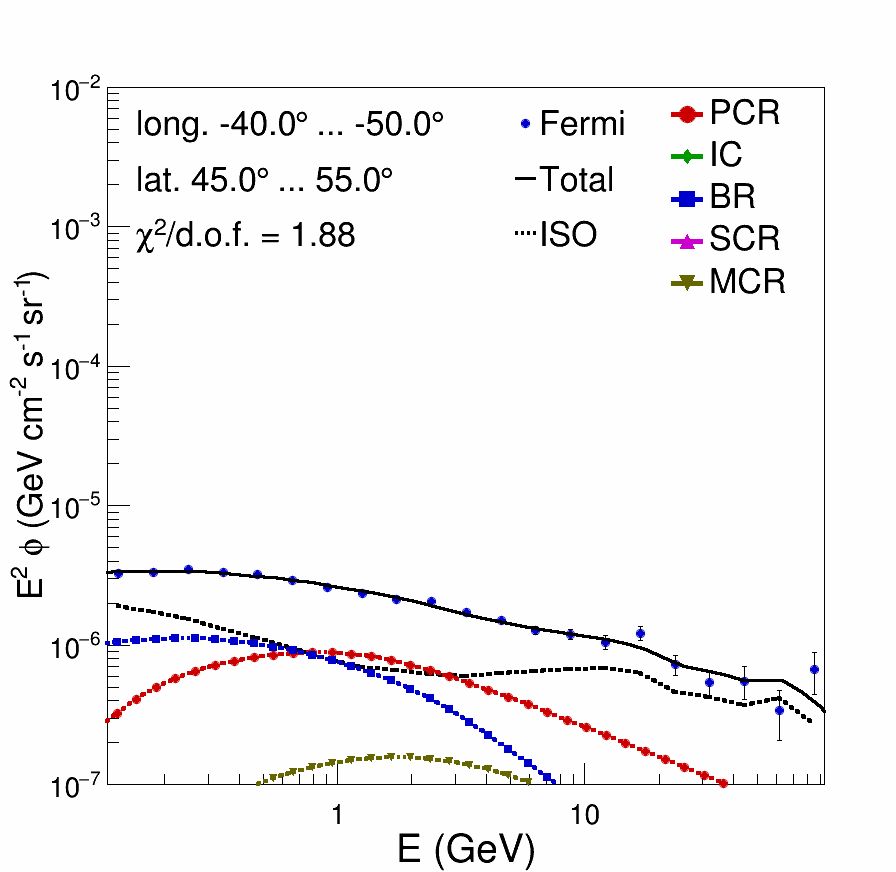}
\includegraphics[width=0.16\textwidth,height=0.16\textwidth,clip]{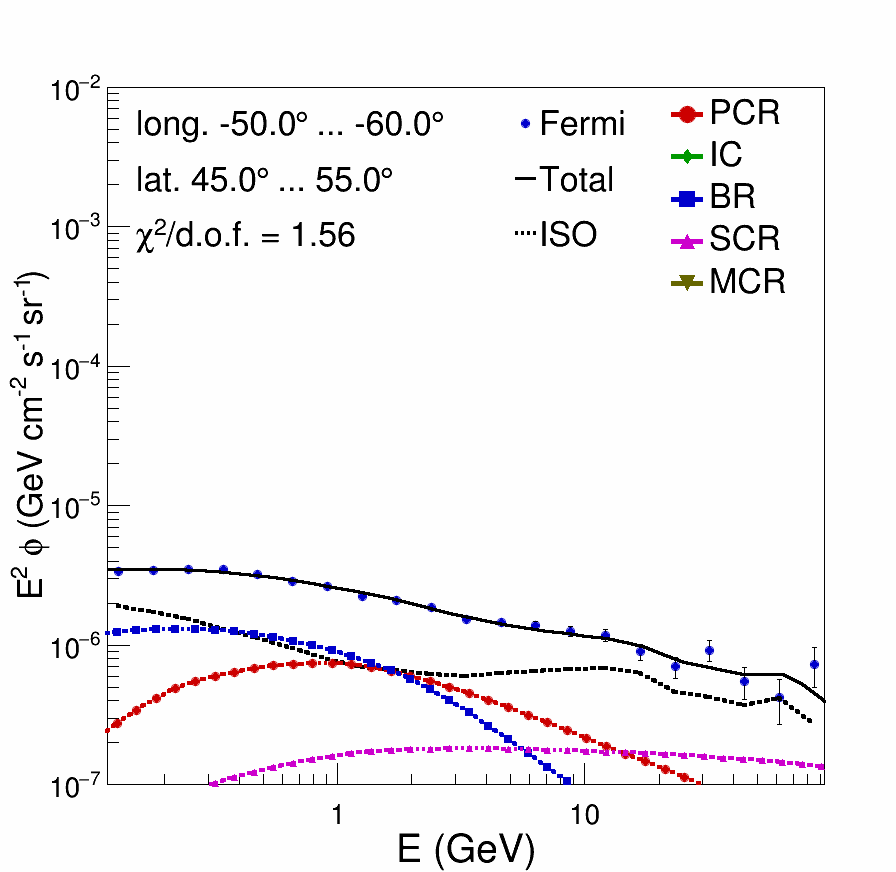}
\includegraphics[width=0.16\textwidth,height=0.16\textwidth,clip]{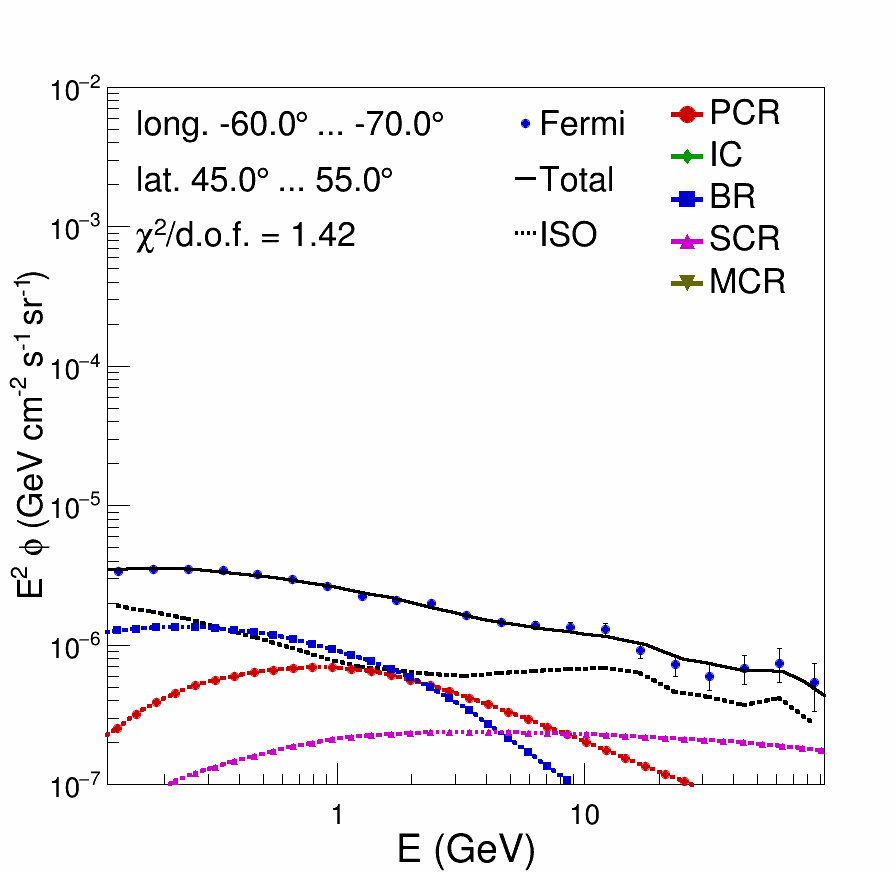}
\includegraphics[width=0.16\textwidth,height=0.16\textwidth,clip]{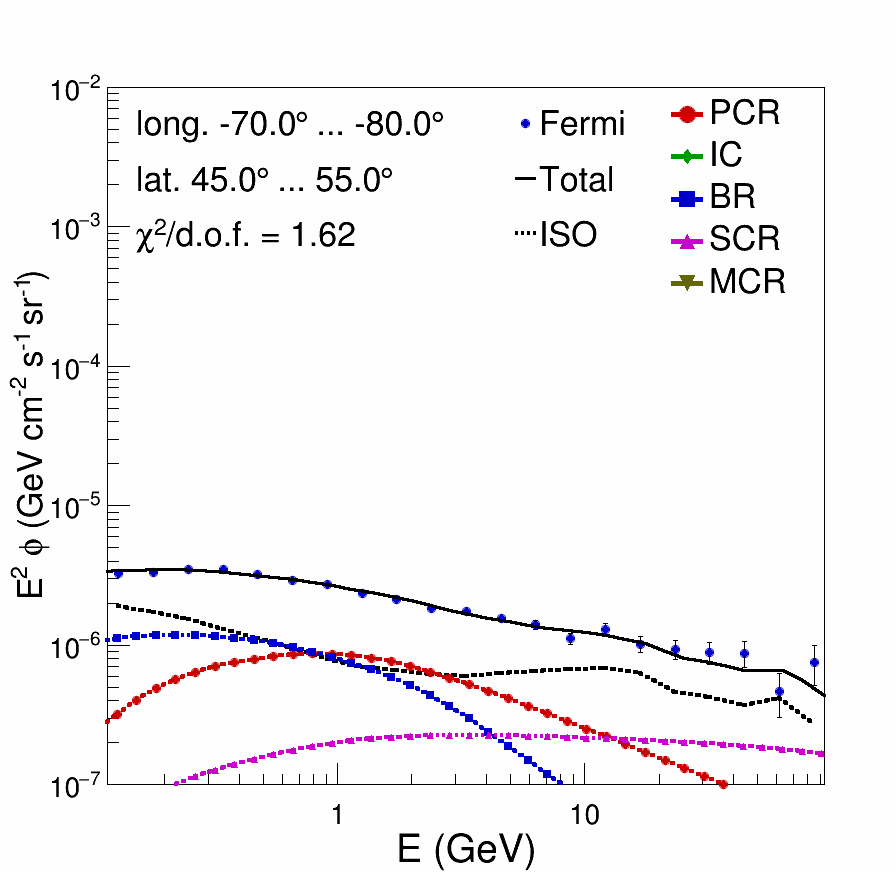}
\includegraphics[width=0.16\textwidth,height=0.16\textwidth,clip]{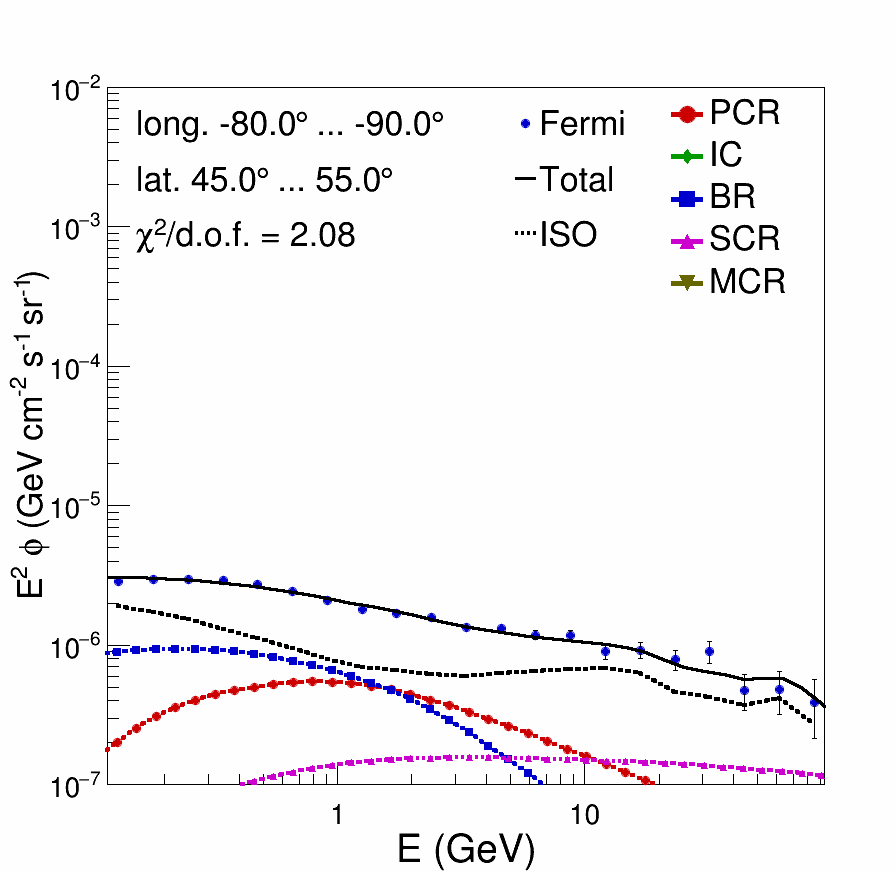}
\includegraphics[width=0.16\textwidth,height=0.16\textwidth,clip]{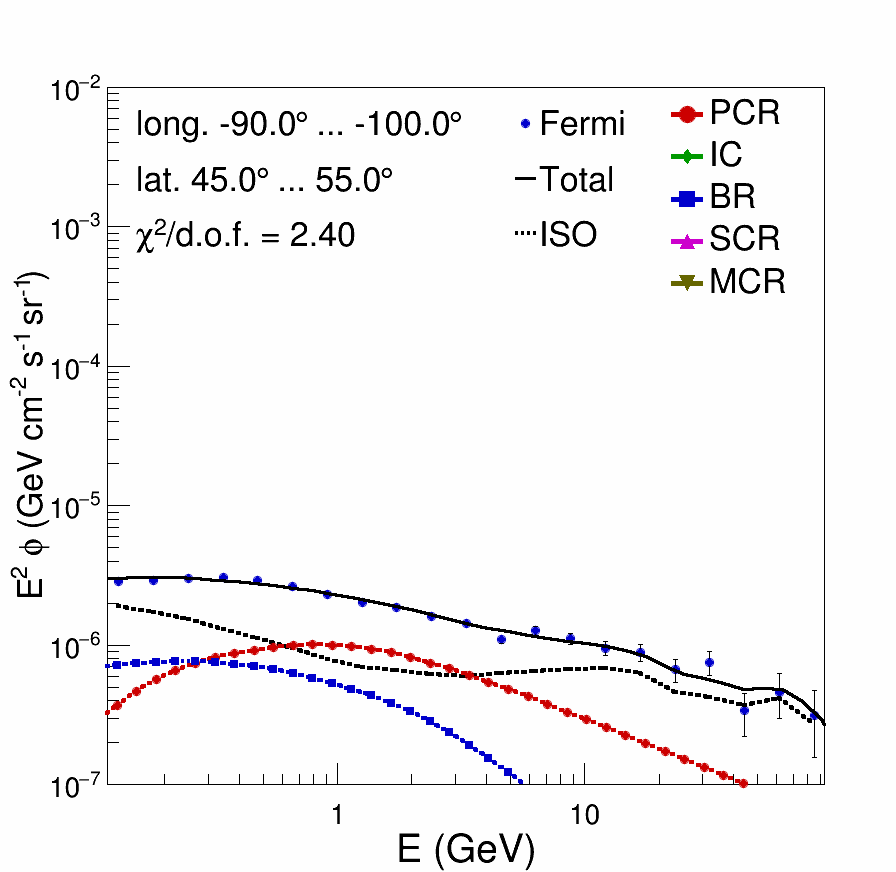}
\includegraphics[width=0.16\textwidth,height=0.16\textwidth,clip]{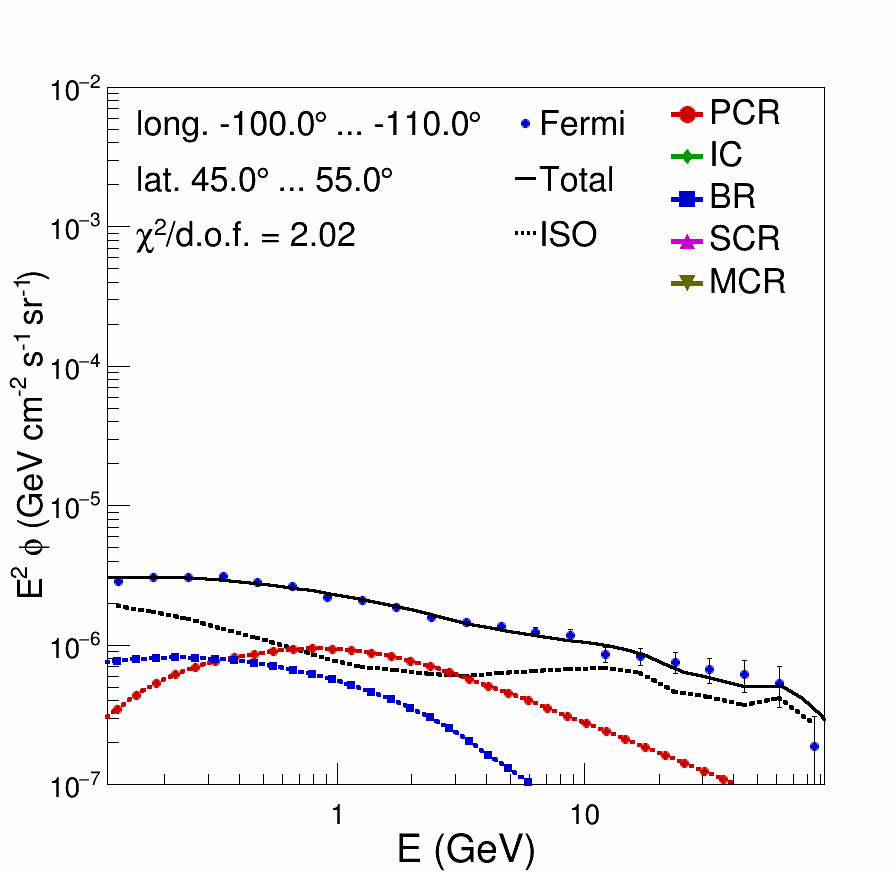}
\includegraphics[width=0.16\textwidth,height=0.16\textwidth,clip]{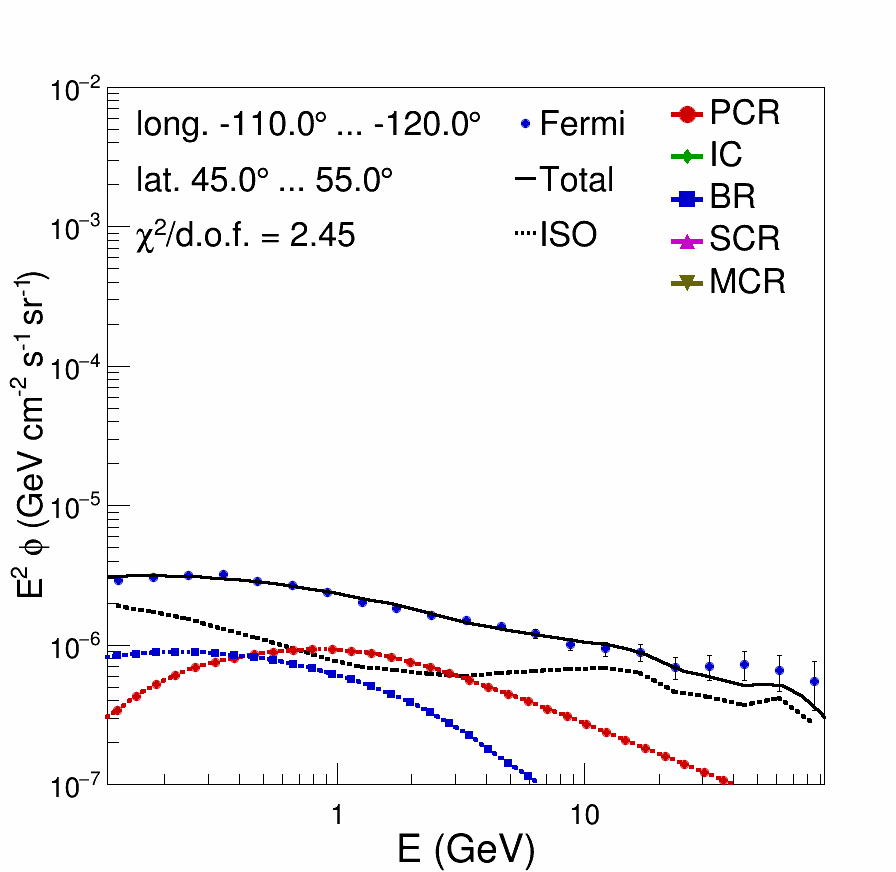}
\includegraphics[width=0.16\textwidth,height=0.16\textwidth,clip]{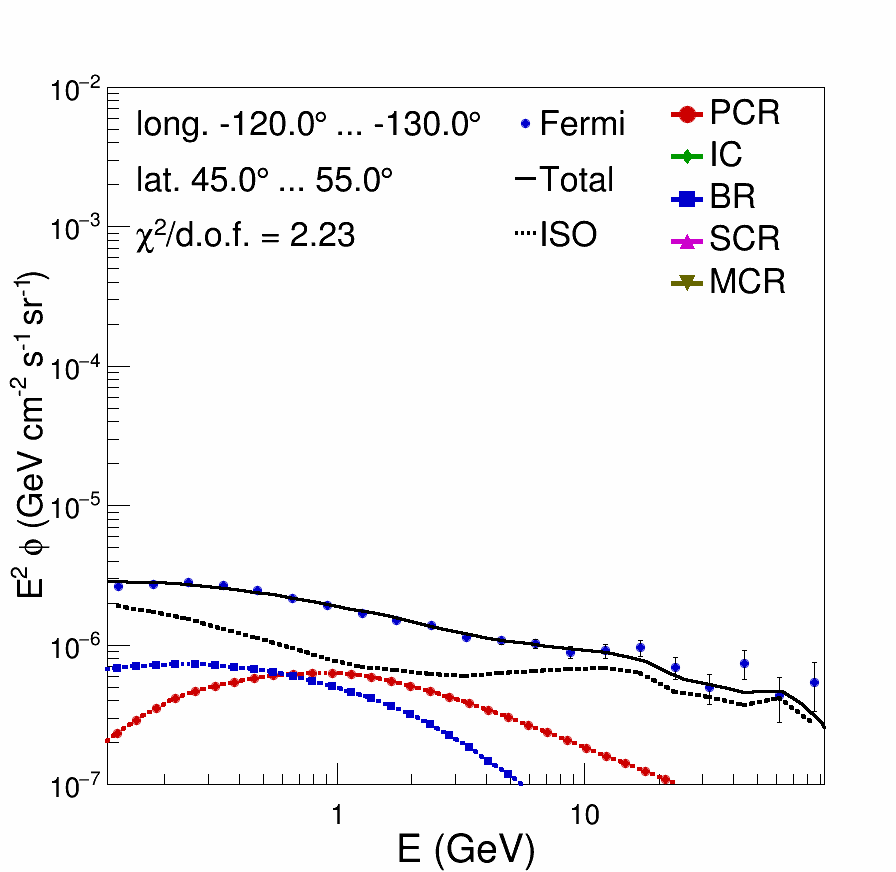}
\includegraphics[width=0.16\textwidth,height=0.16\textwidth,clip]{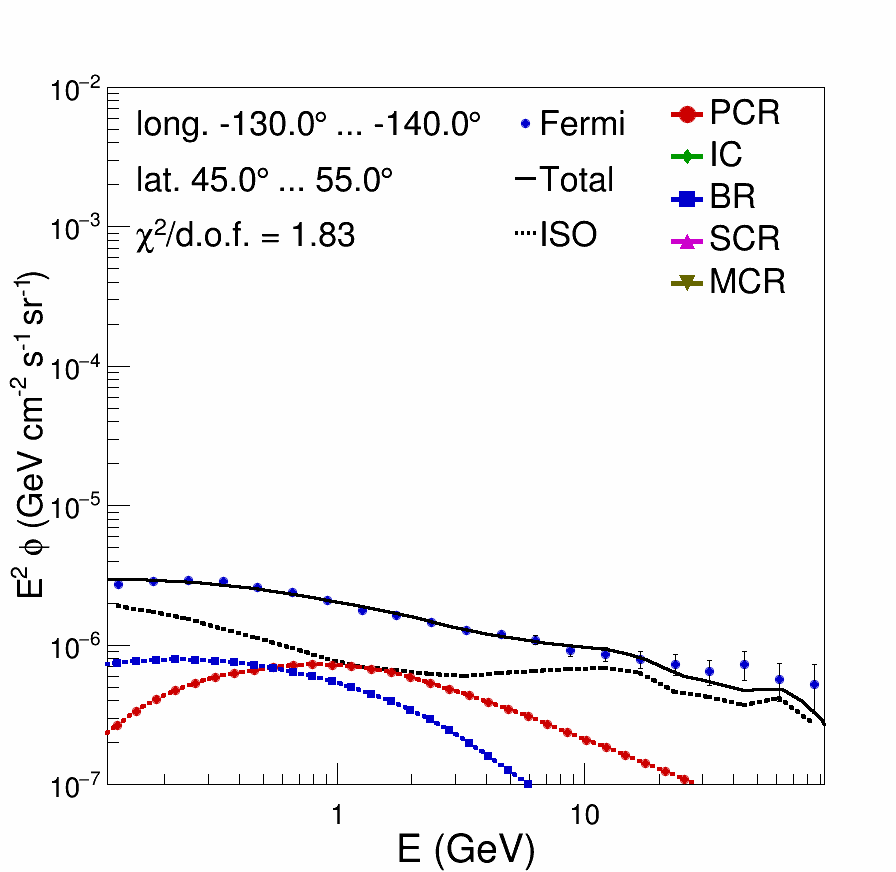}
\includegraphics[width=0.16\textwidth,height=0.16\textwidth,clip]{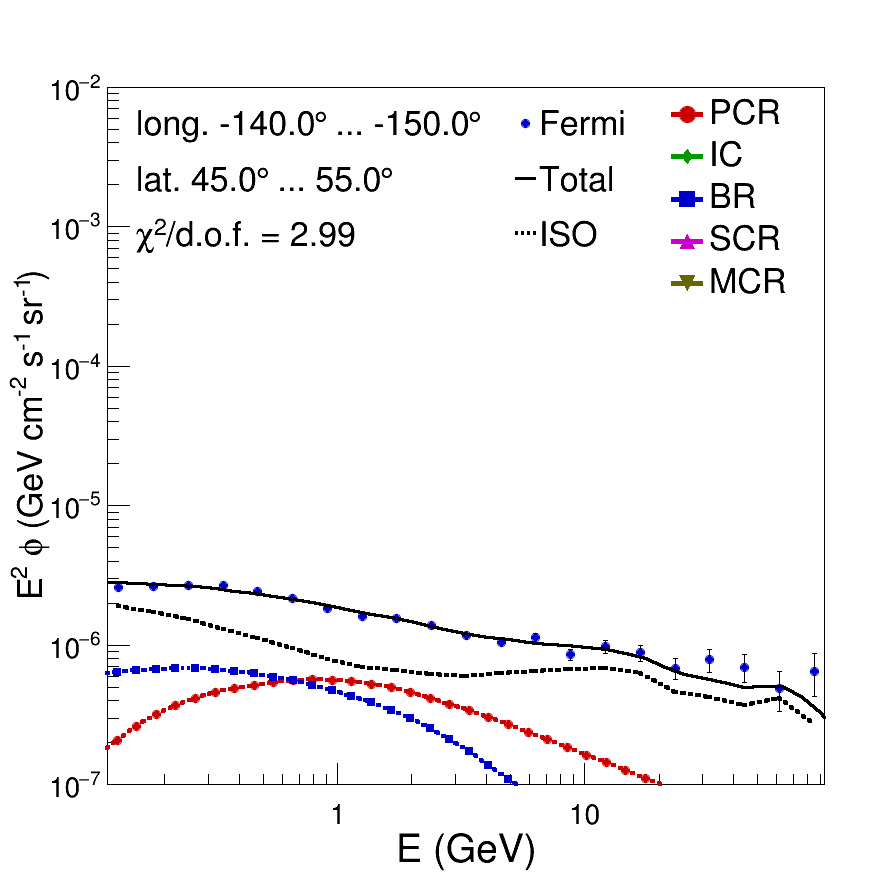}
\includegraphics[width=0.16\textwidth,height=0.16\textwidth,clip]{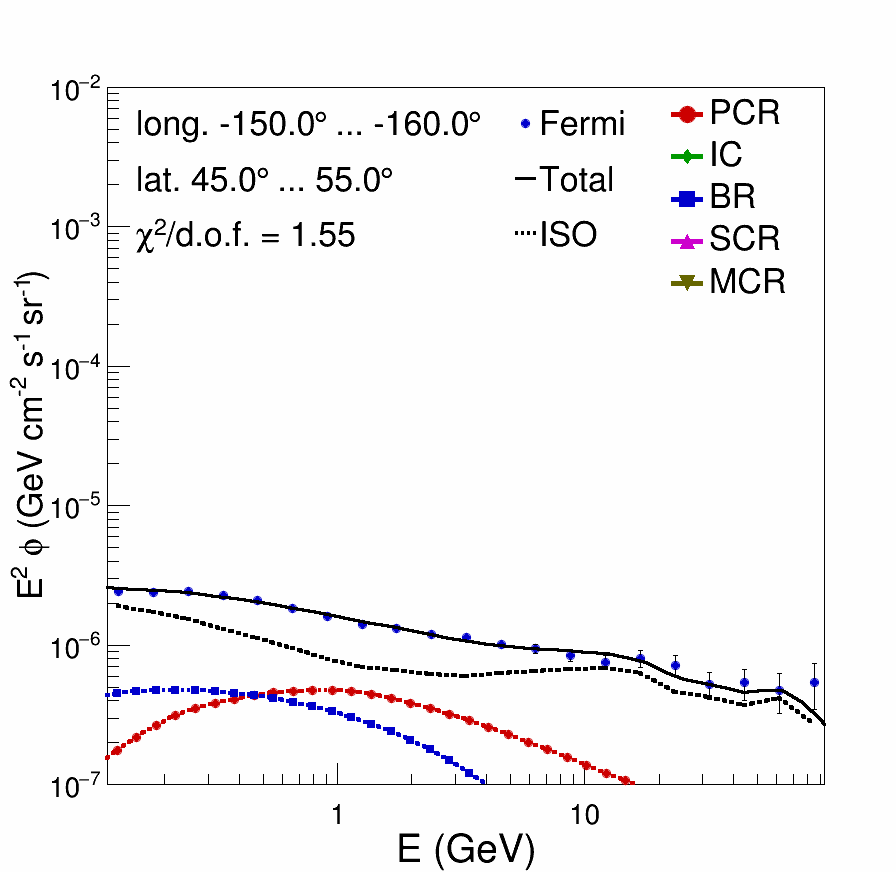}
\includegraphics[width=0.16\textwidth,height=0.16\textwidth,clip]{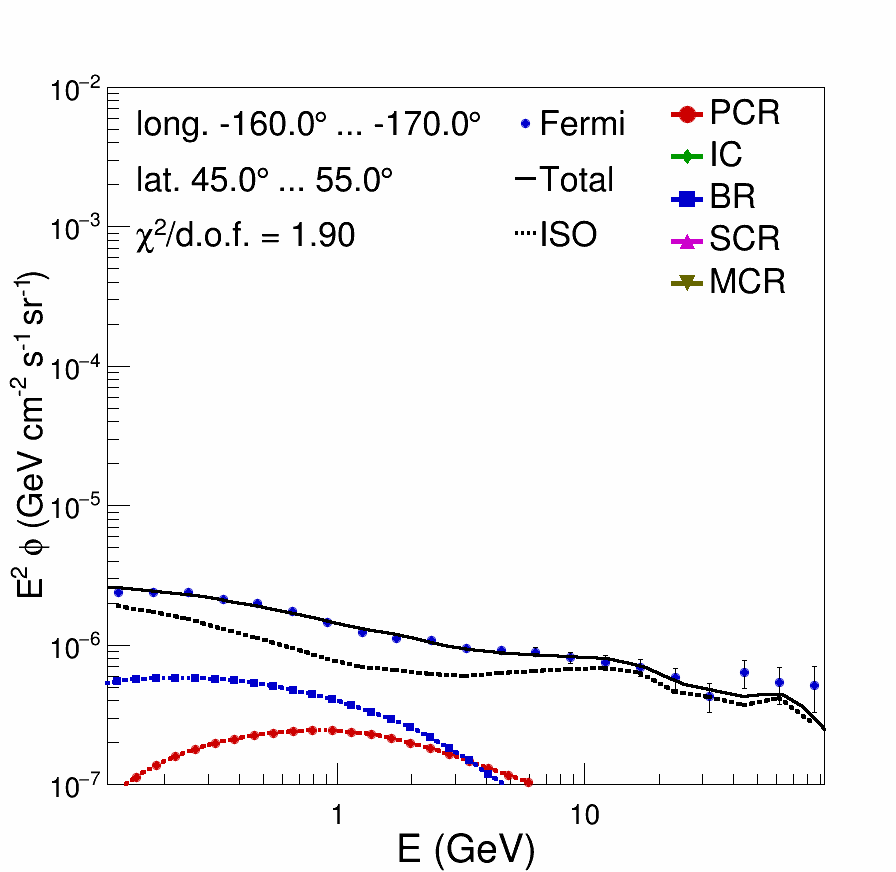}
\includegraphics[width=0.16\textwidth,height=0.16\textwidth,clip]{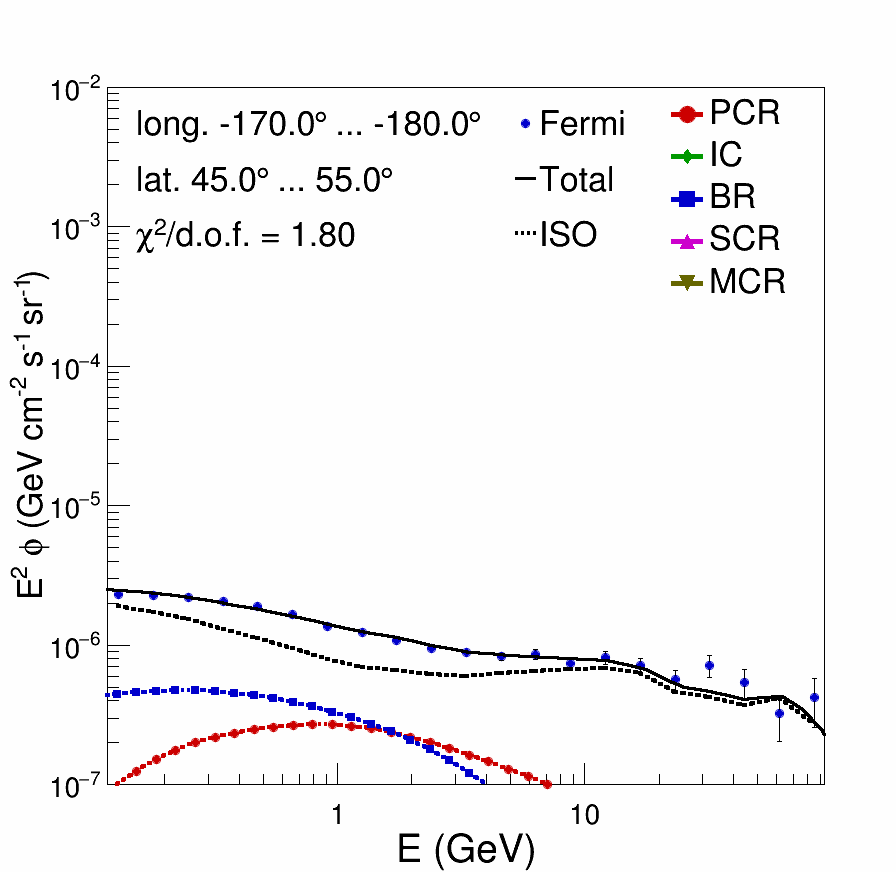}
\caption[]{Template fits for latitudes  with $45.0^\circ<b<55.0^\circ$ and longitudes decreasing from 180$^\circ$ to -180$^\circ$. \label{F13}
}
\end{figure}
\begin{figure}
\centering
\includegraphics[width=0.16\textwidth,height=0.16\textwidth,clip]{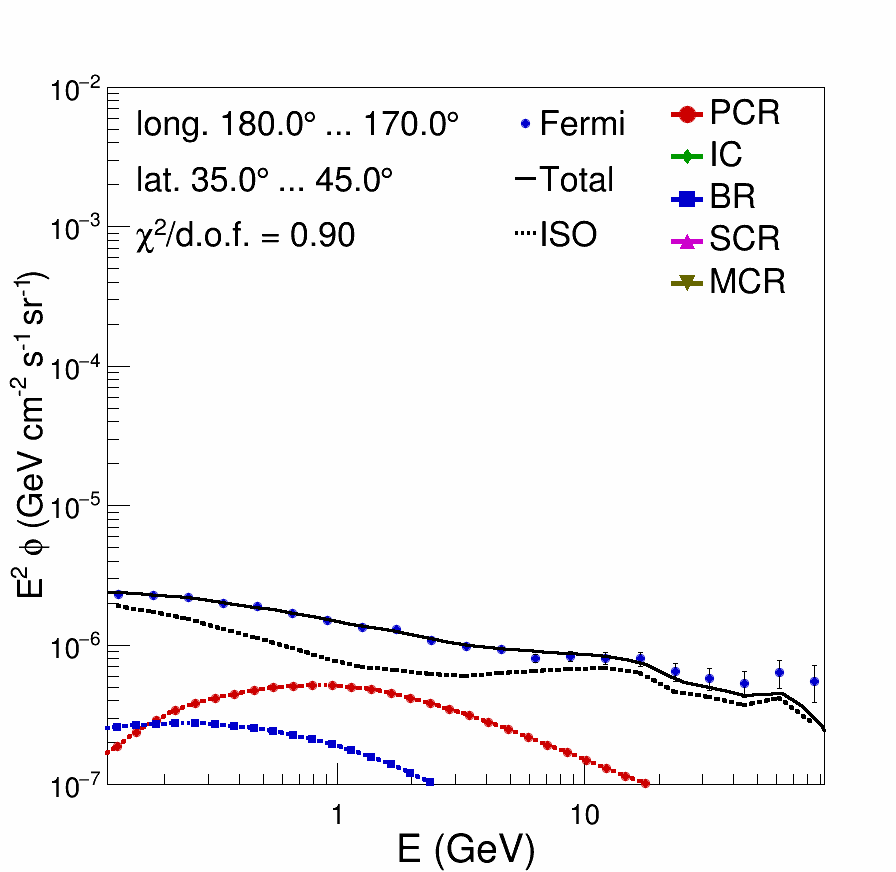}
\includegraphics[width=0.16\textwidth,height=0.16\textwidth,clip]{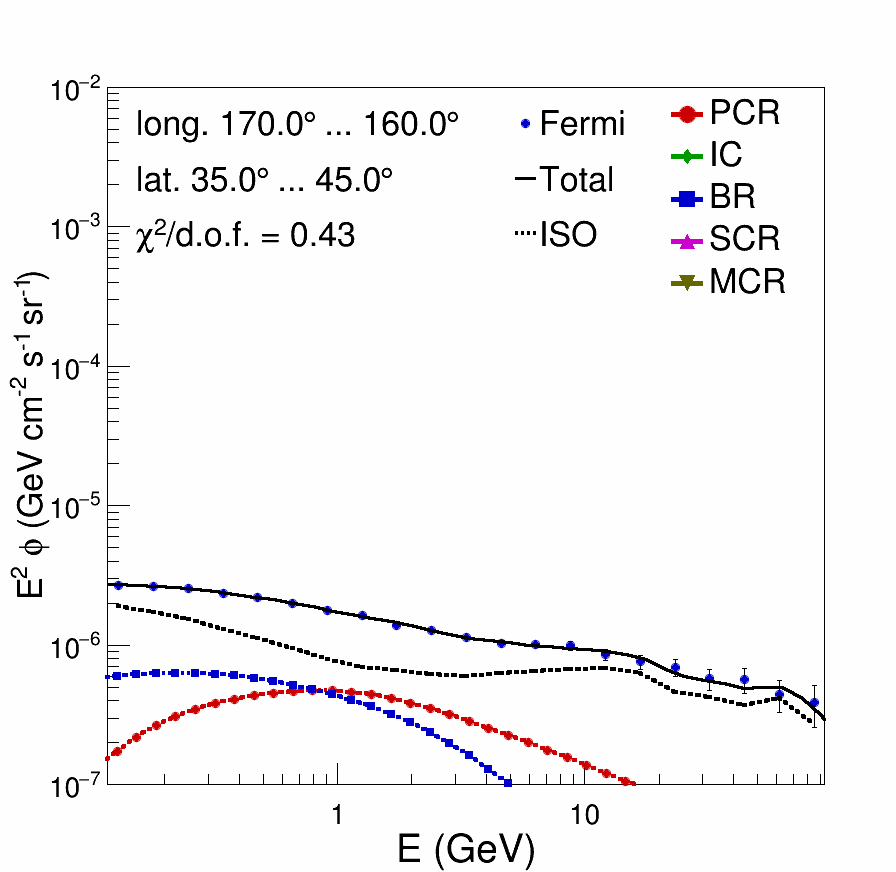}
\includegraphics[width=0.16\textwidth,height=0.16\textwidth,clip]{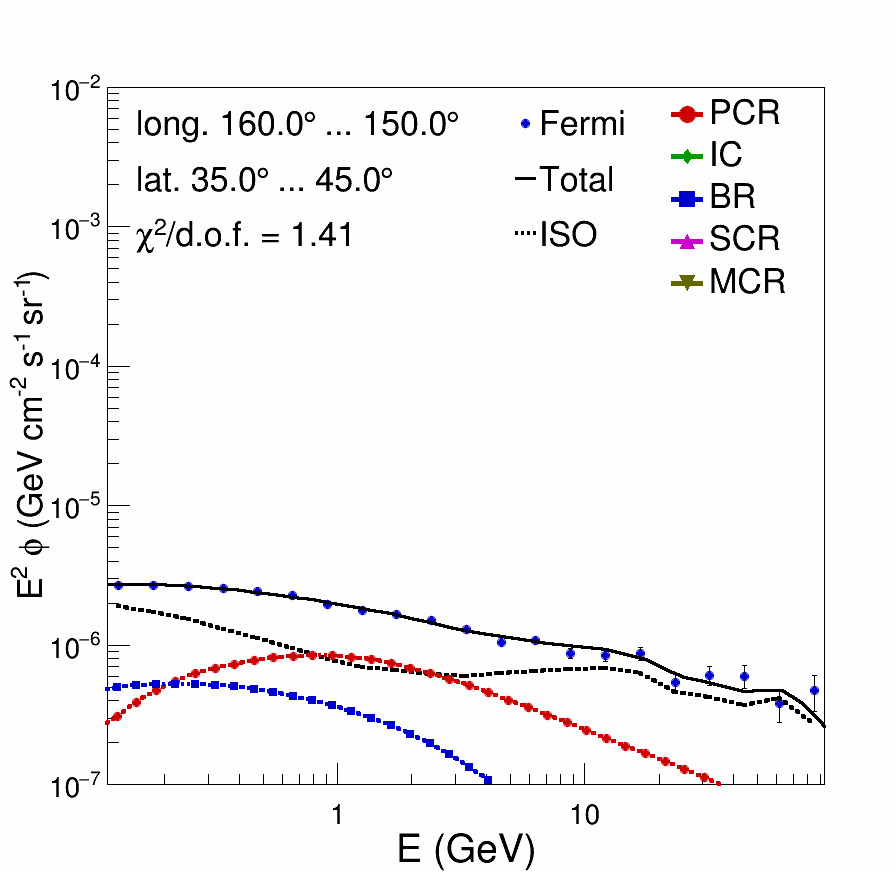}
\includegraphics[width=0.16\textwidth,height=0.16\textwidth,clip]{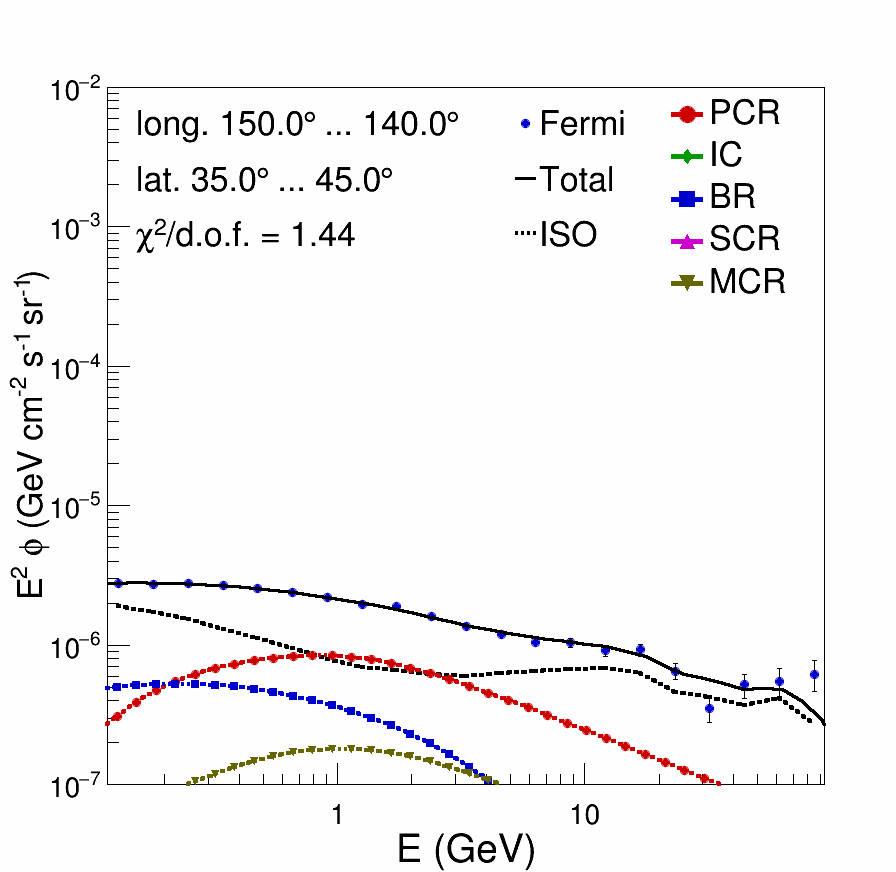}
\includegraphics[width=0.16\textwidth,height=0.16\textwidth,clip]{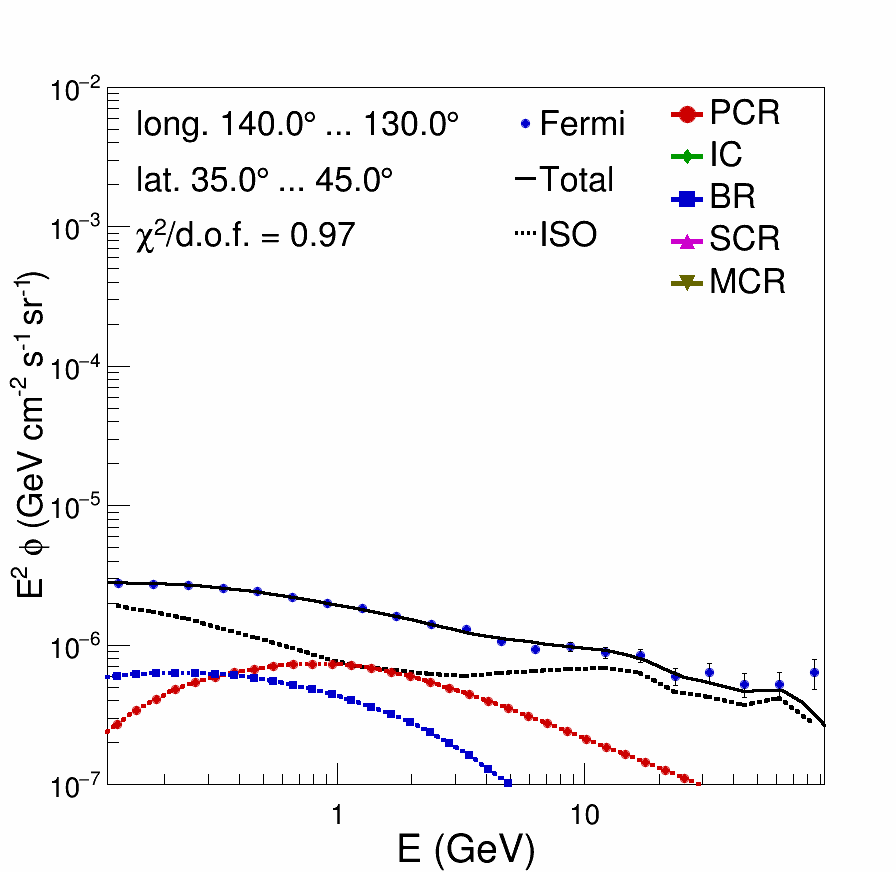}
\includegraphics[width=0.16\textwidth,height=0.16\textwidth,clip]{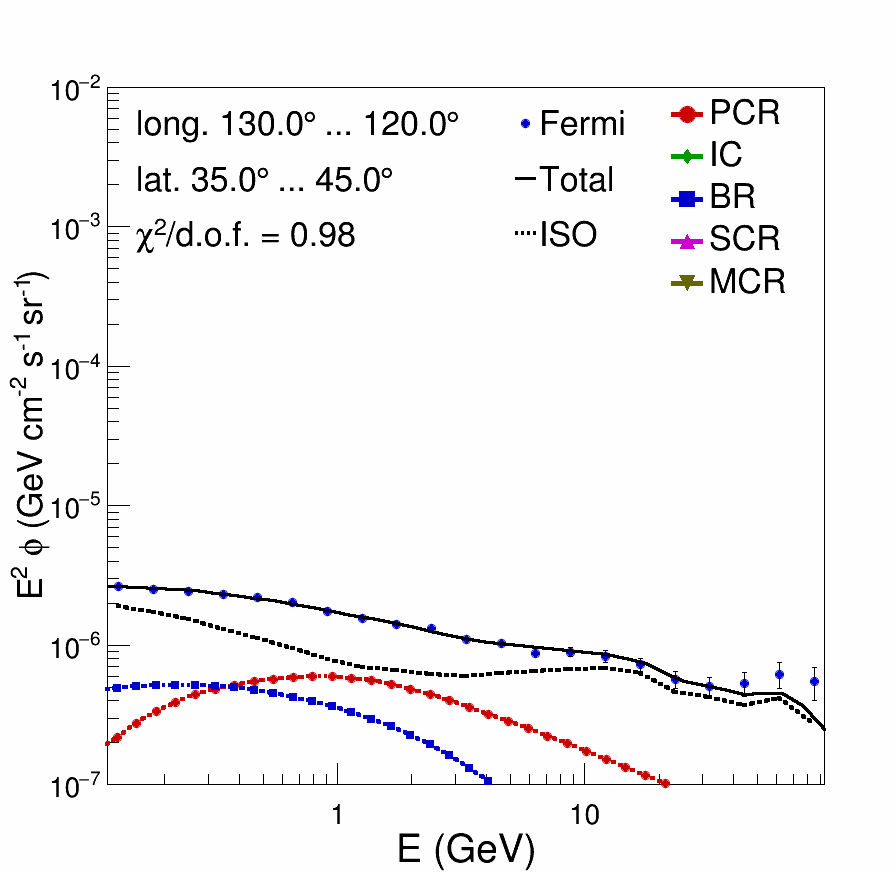}
\includegraphics[width=0.16\textwidth,height=0.16\textwidth,clip]{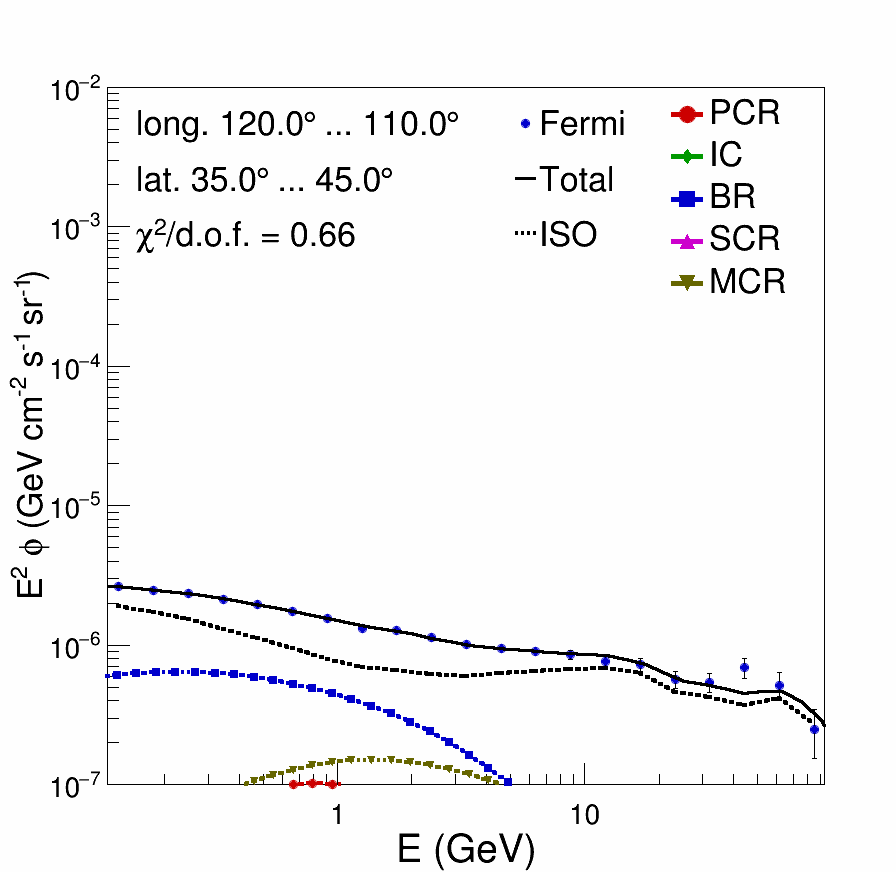}
\includegraphics[width=0.16\textwidth,height=0.16\textwidth,clip]{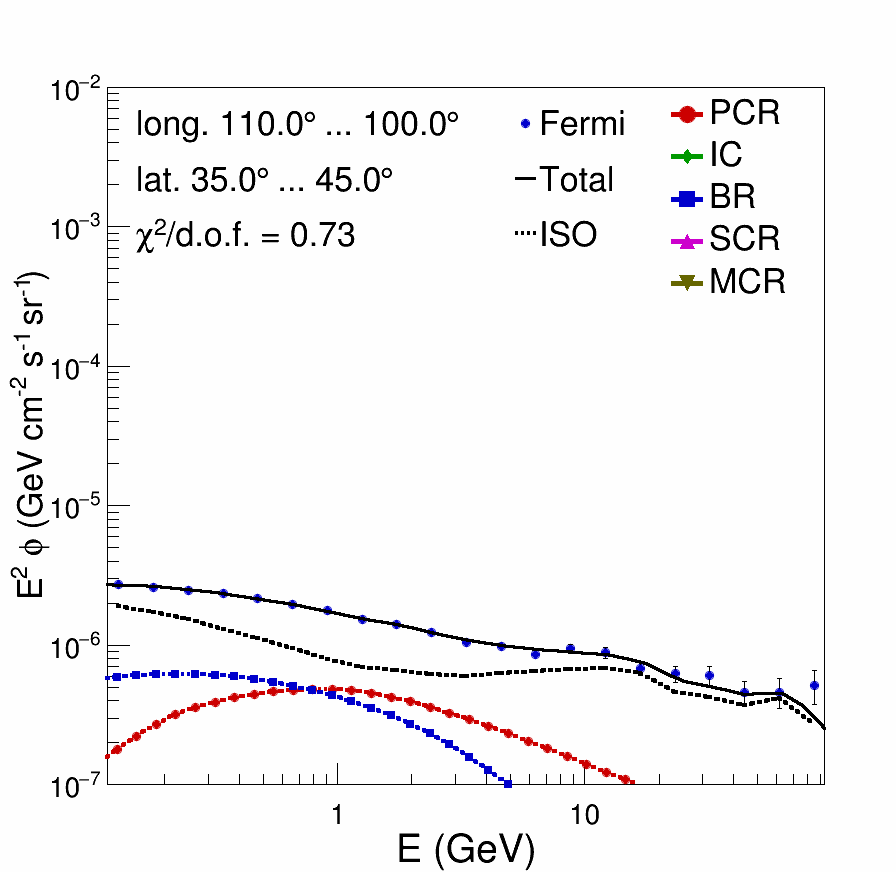}
\includegraphics[width=0.16\textwidth,height=0.16\textwidth,clip]{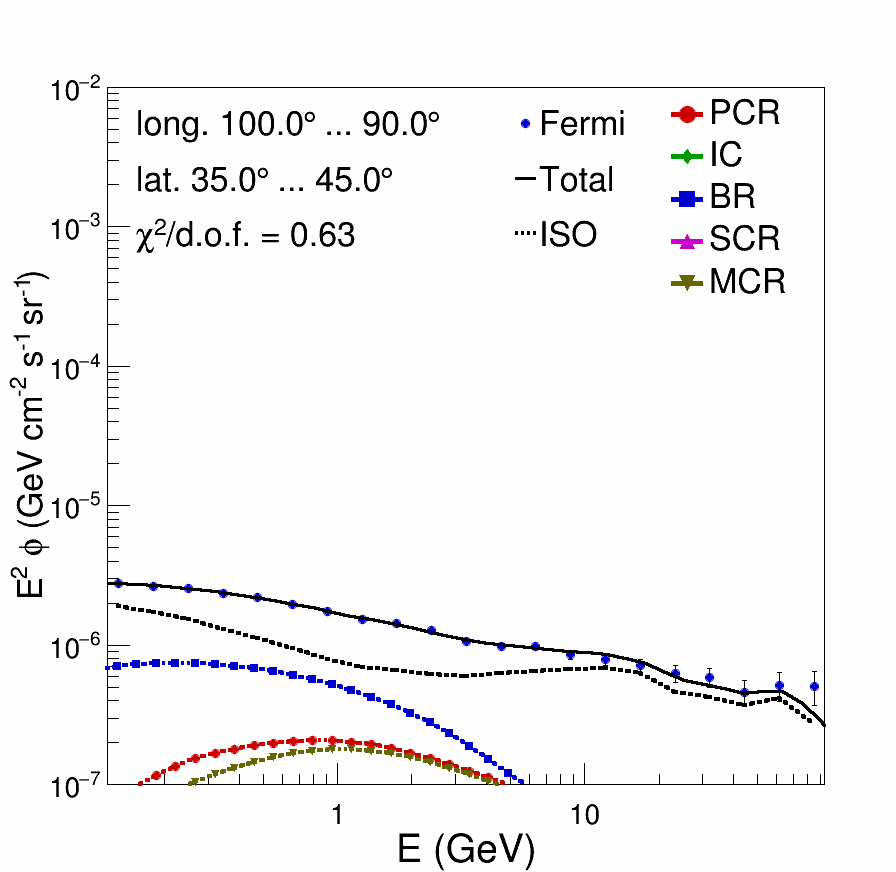}
\includegraphics[width=0.16\textwidth,height=0.16\textwidth,clip]{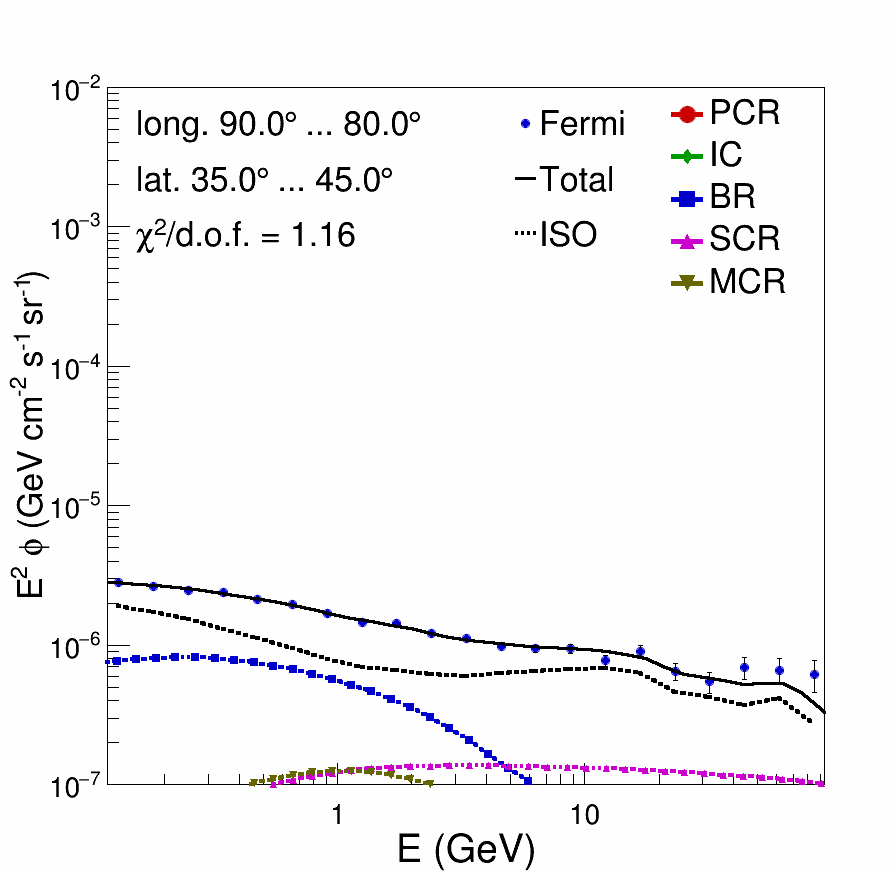}
\includegraphics[width=0.16\textwidth,height=0.16\textwidth,clip]{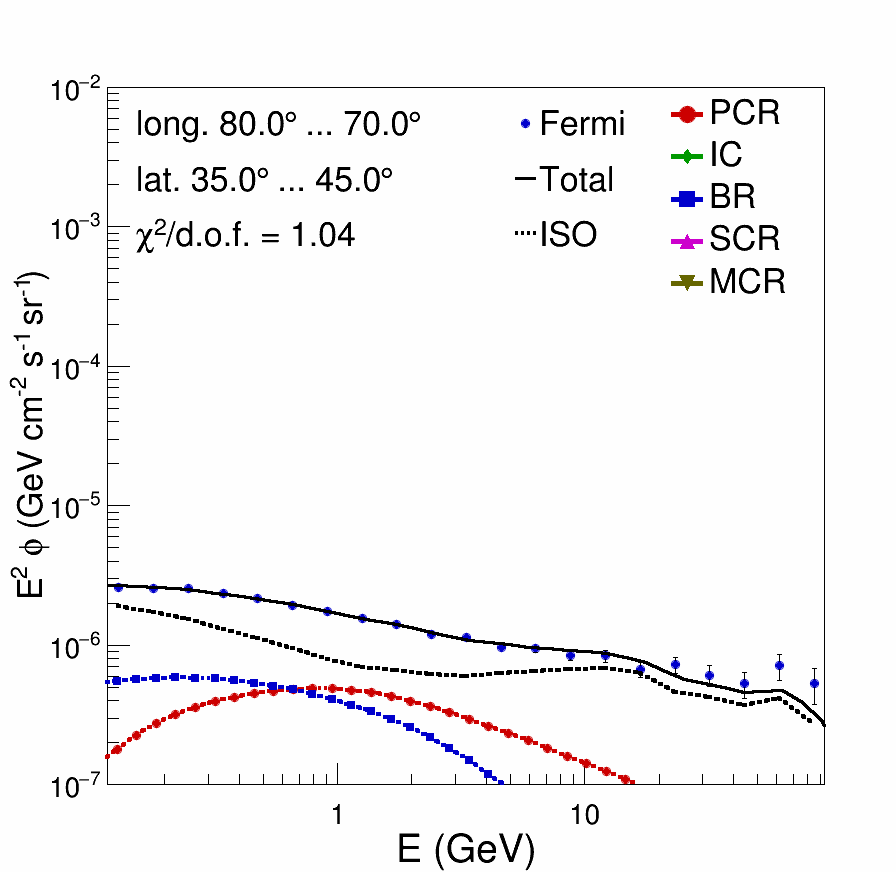}
\includegraphics[width=0.16\textwidth,height=0.16\textwidth,clip]{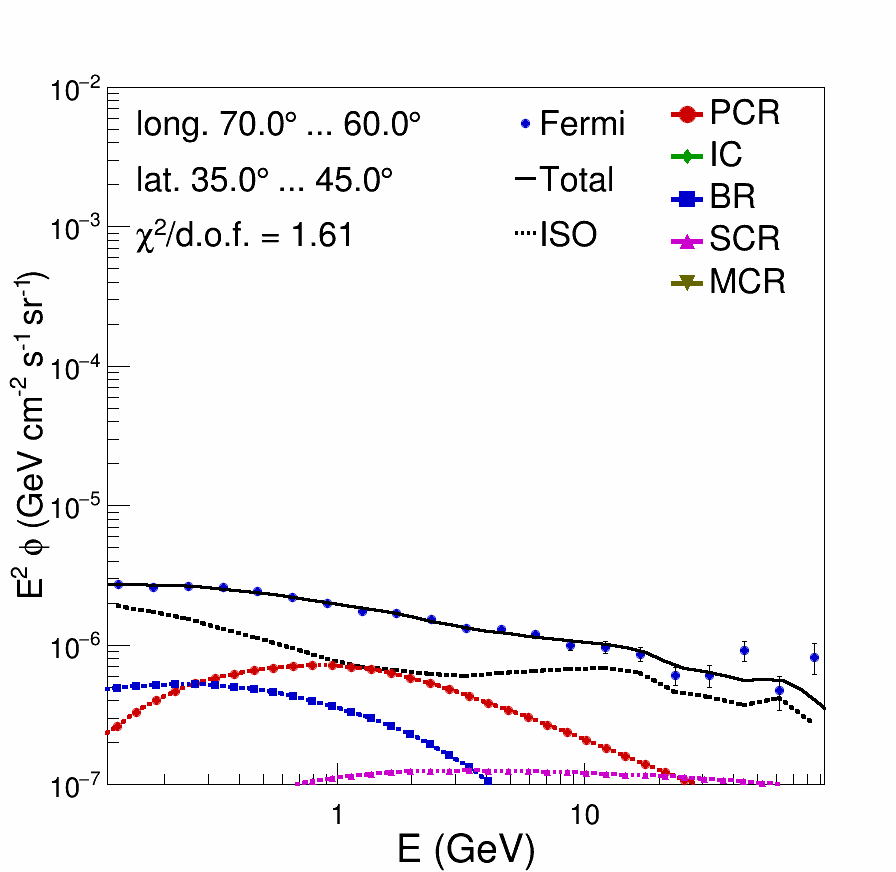}
\includegraphics[width=0.16\textwidth,height=0.16\textwidth,clip]{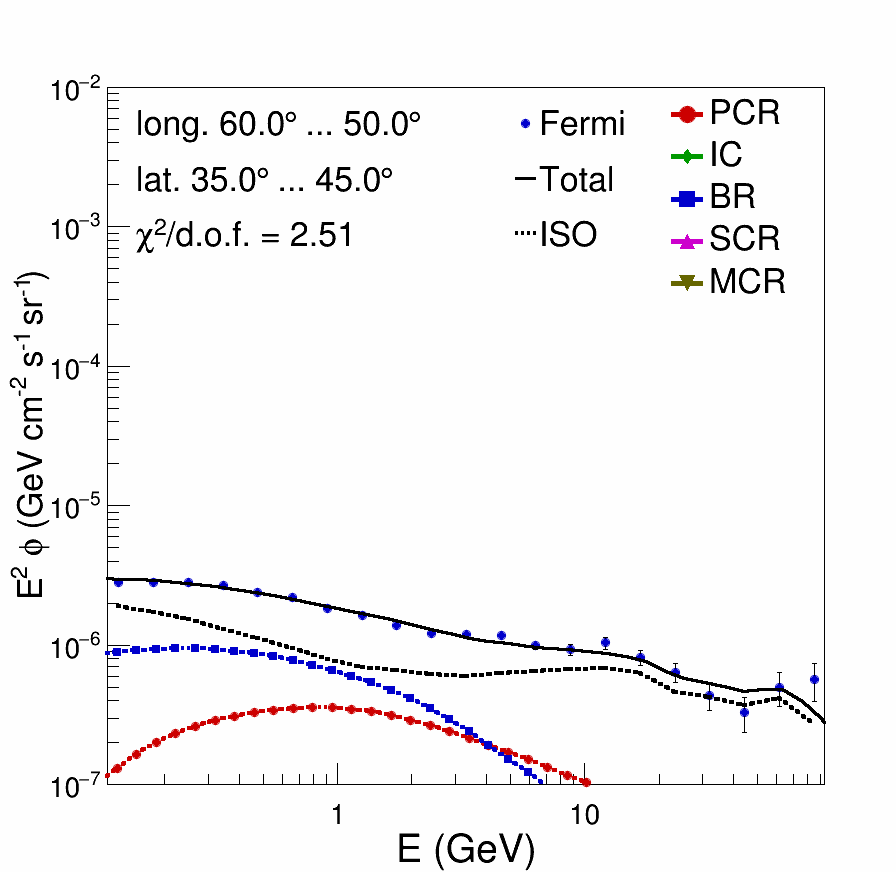}
\includegraphics[width=0.16\textwidth,height=0.16\textwidth,clip]{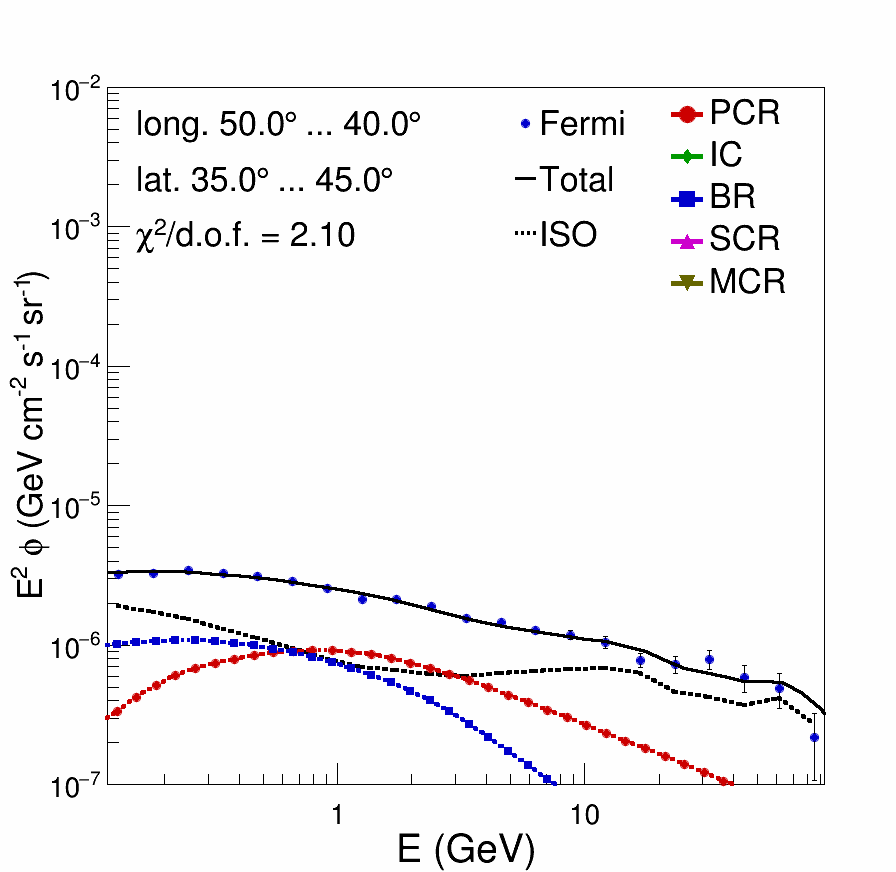}
\includegraphics[width=0.16\textwidth,height=0.16\textwidth,clip]{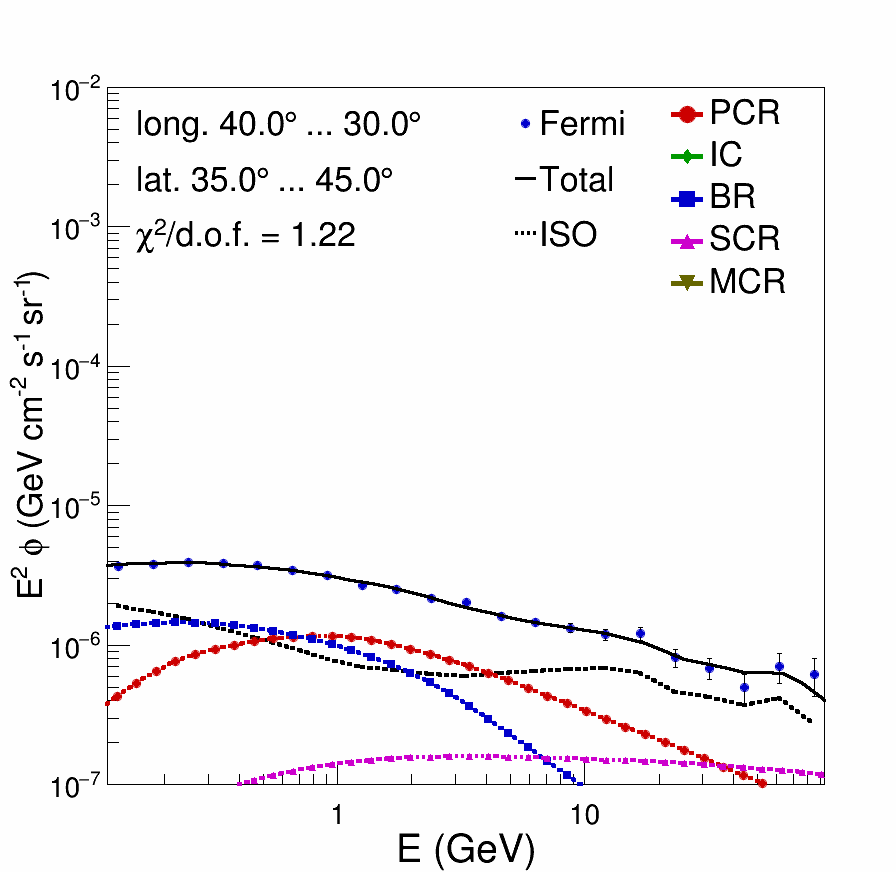}
\includegraphics[width=0.16\textwidth,height=0.16\textwidth,clip]{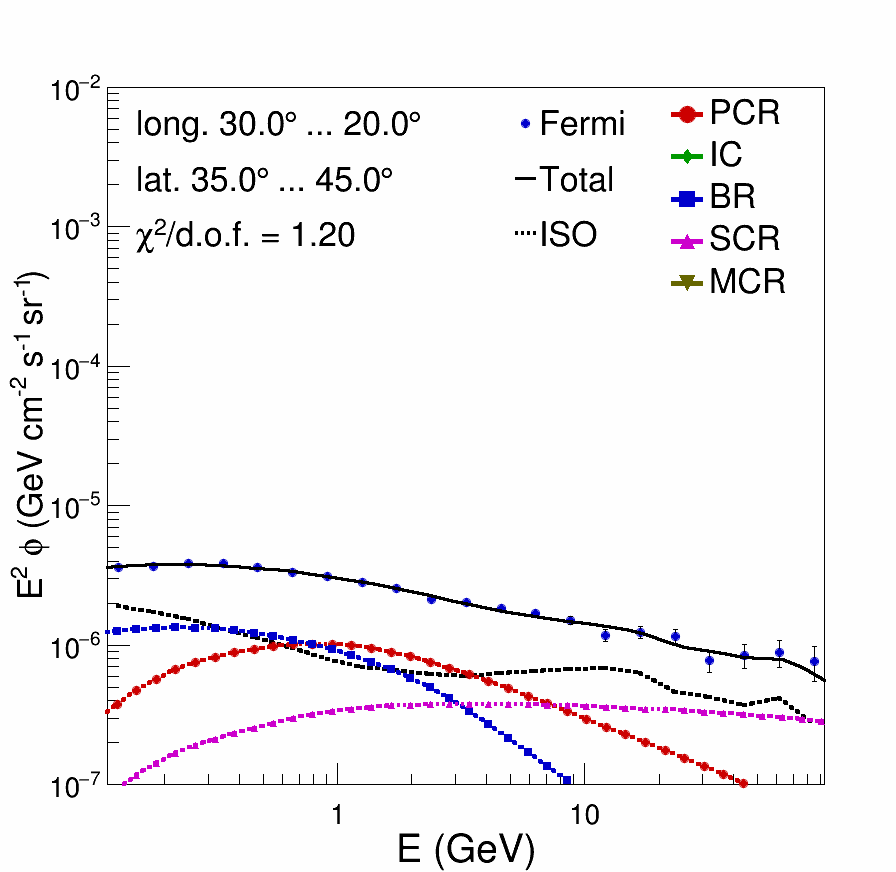}
\includegraphics[width=0.16\textwidth,height=0.16\textwidth,clip]{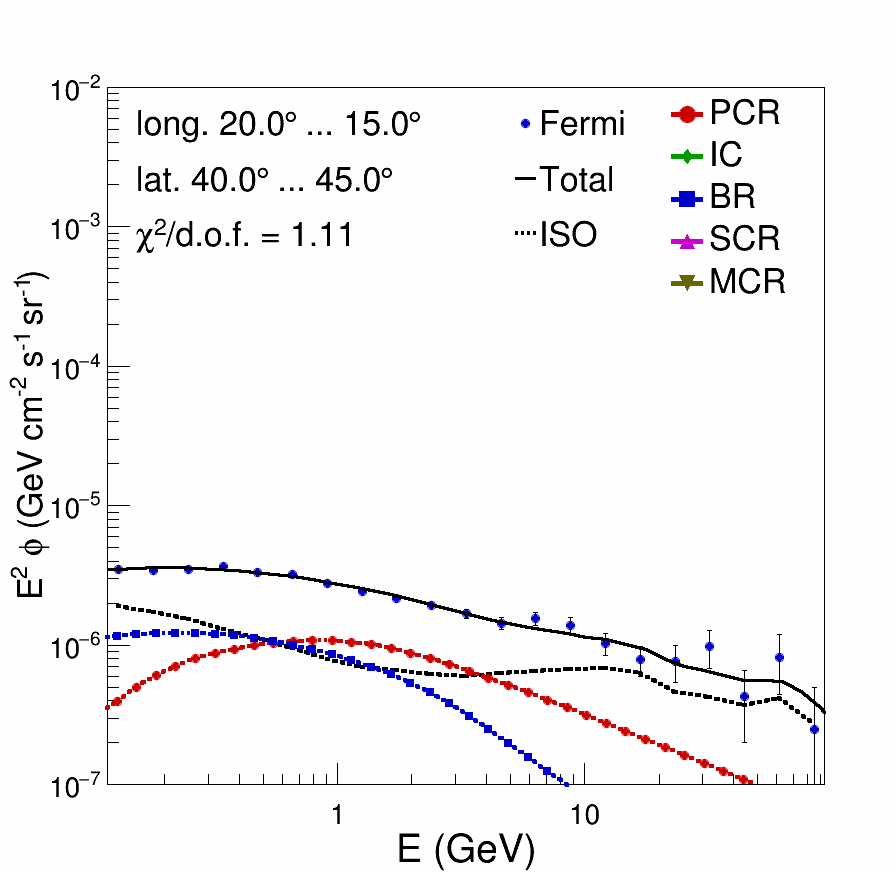}
\includegraphics[width=0.16\textwidth,height=0.16\textwidth,clip]{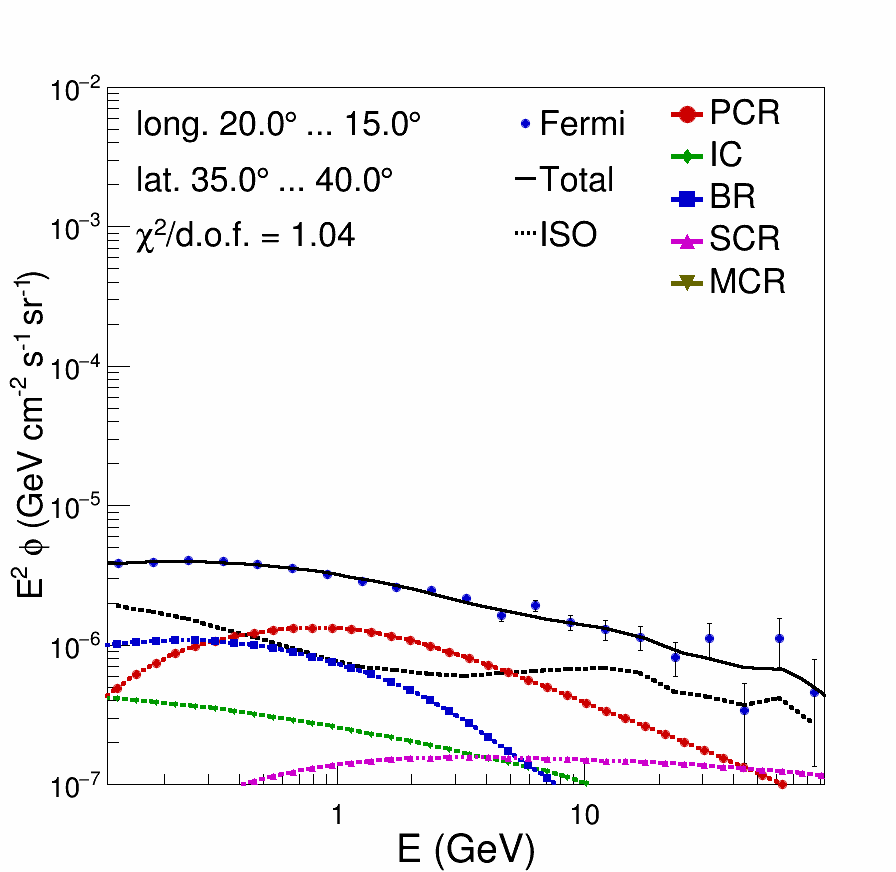}
\includegraphics[width=0.16\textwidth,height=0.16\textwidth,clip]{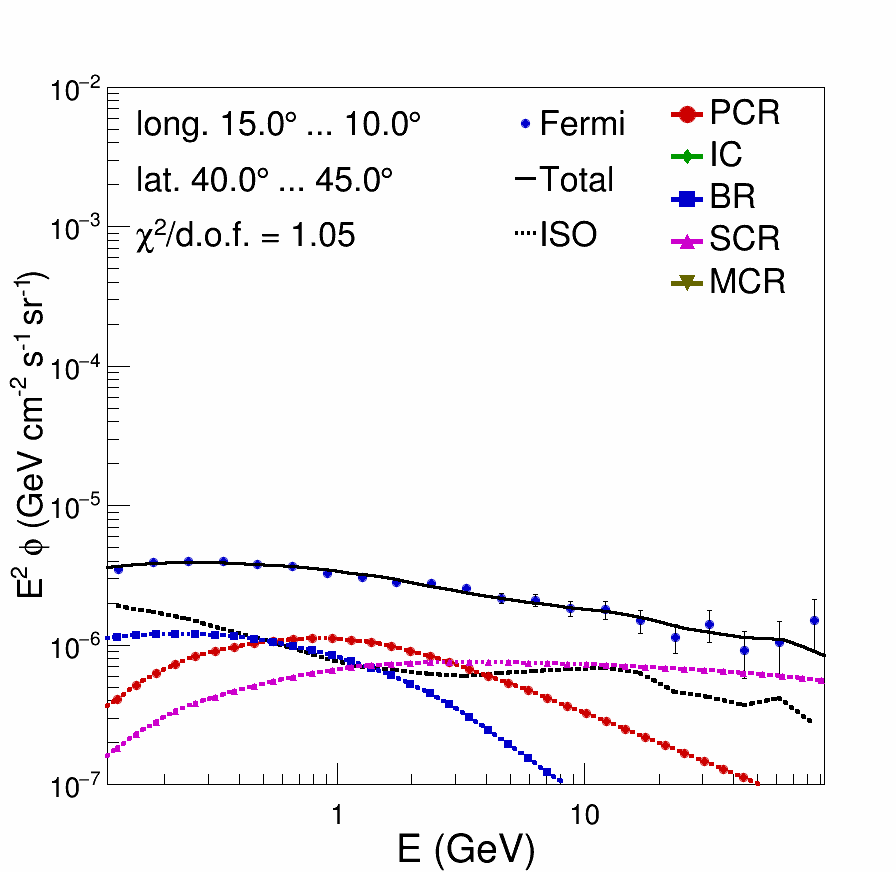}
\includegraphics[width=0.16\textwidth,height=0.16\textwidth,clip]{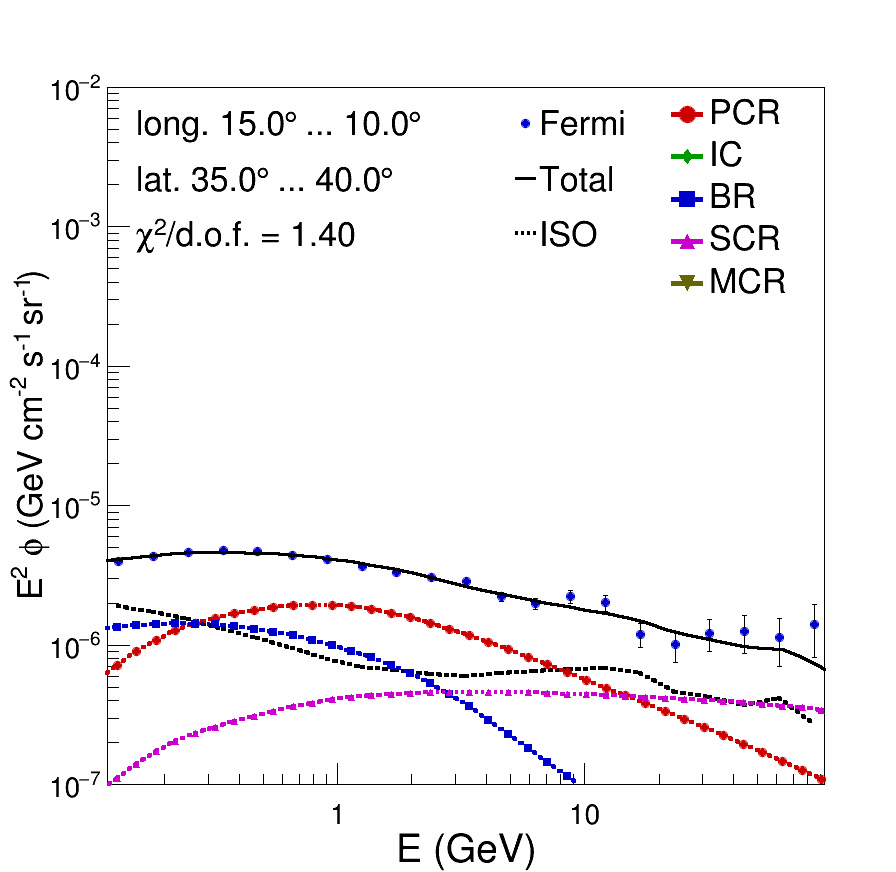}
\includegraphics[width=0.16\textwidth,height=0.16\textwidth,clip]{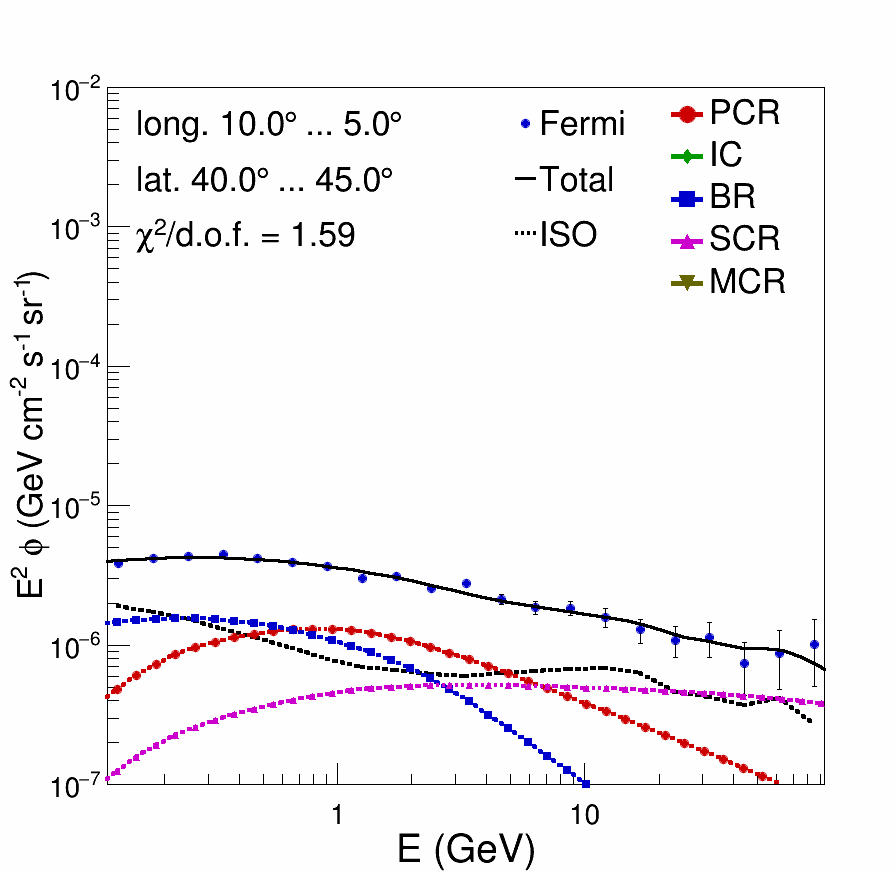}
\includegraphics[width=0.16\textwidth,height=0.16\textwidth,clip]{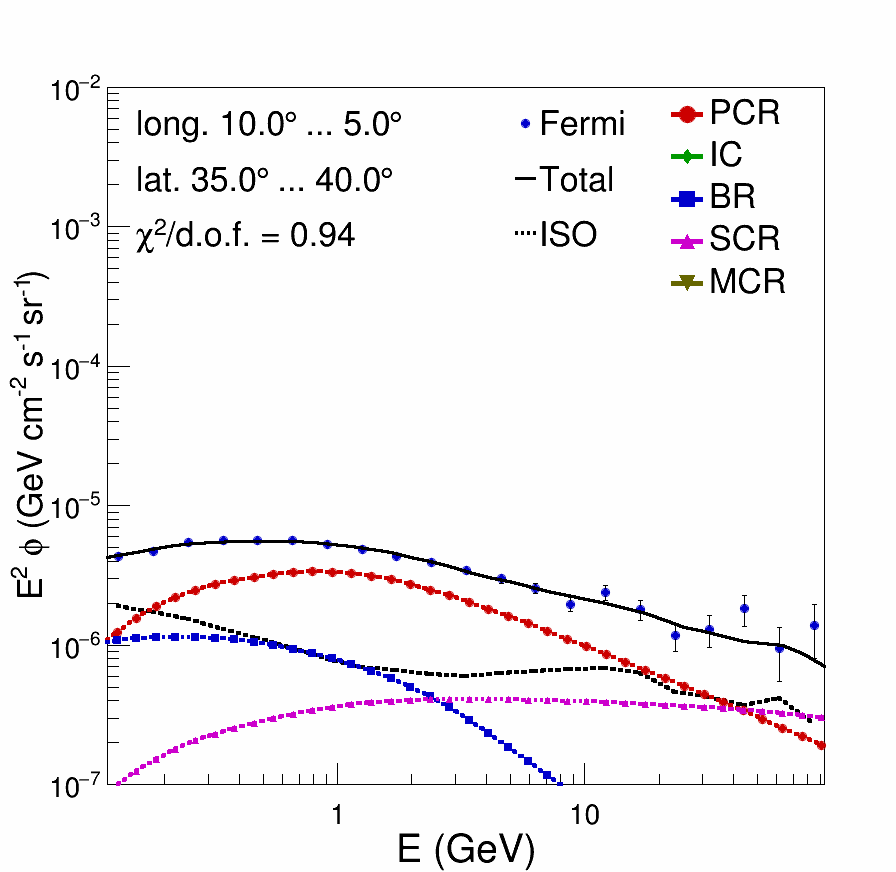}
\includegraphics[width=0.16\textwidth,height=0.16\textwidth,clip]{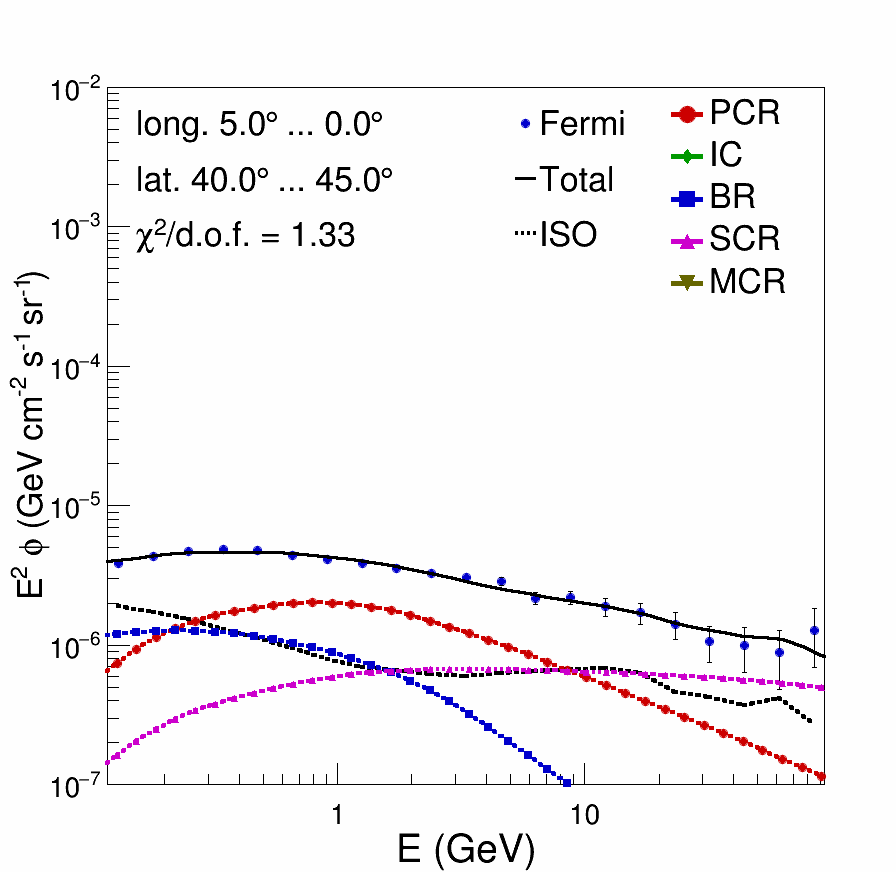}
\includegraphics[width=0.16\textwidth,height=0.16\textwidth,clip]{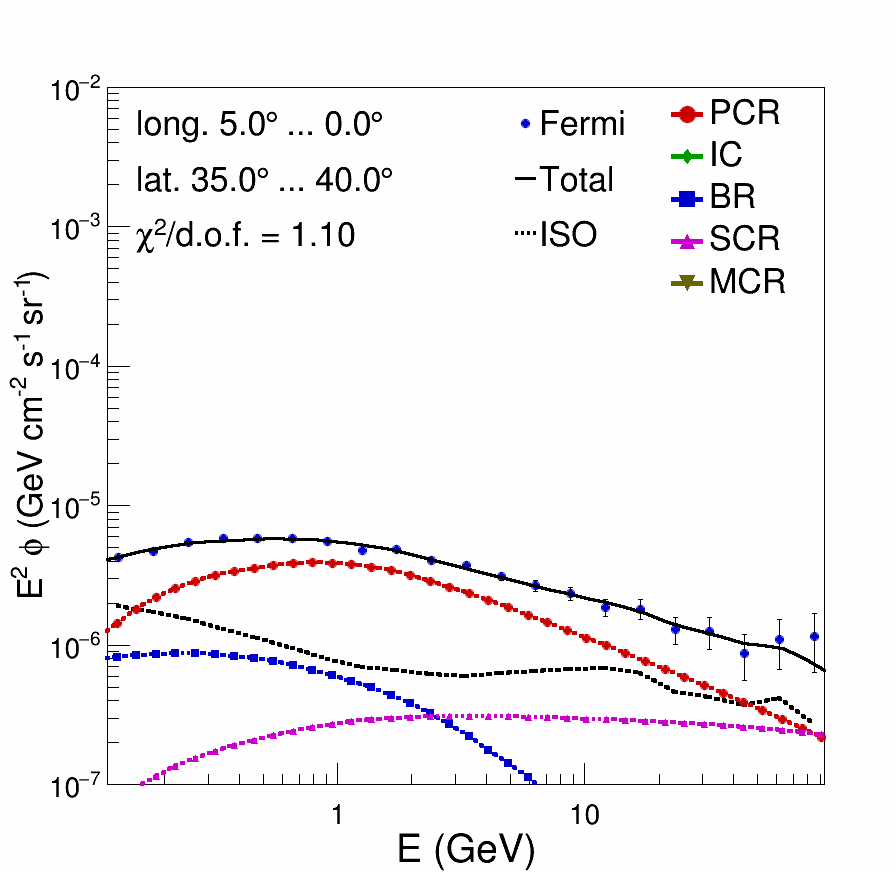}
\includegraphics[width=0.16\textwidth,height=0.16\textwidth,clip]{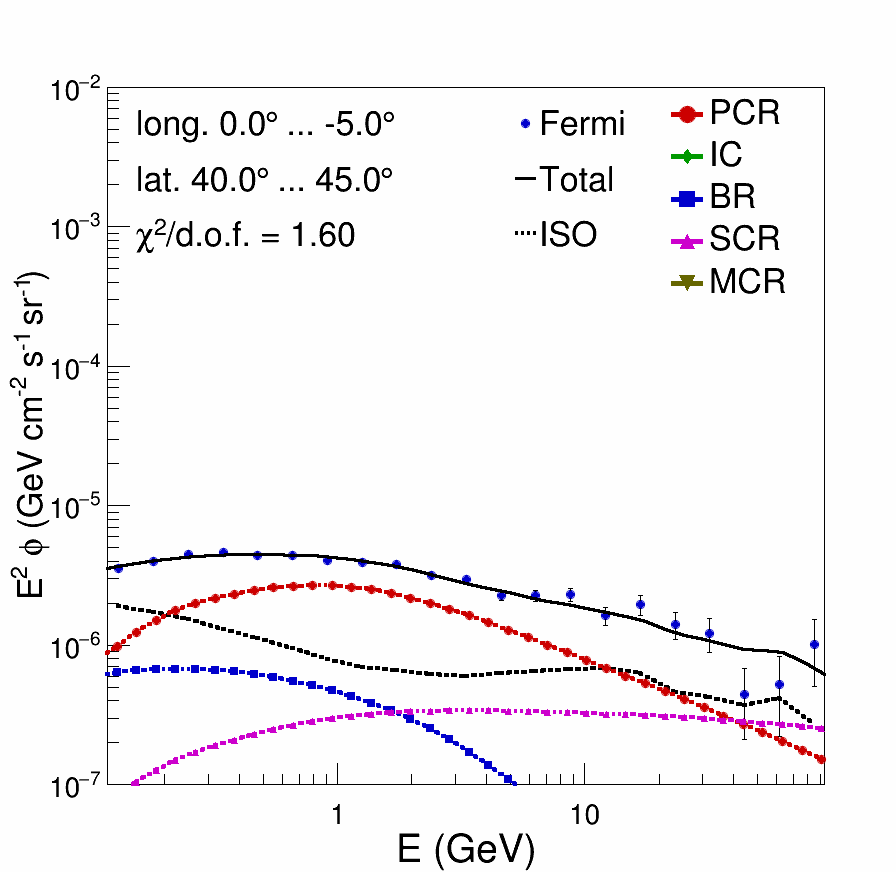}
\includegraphics[width=0.16\textwidth,height=0.16\textwidth,clip]{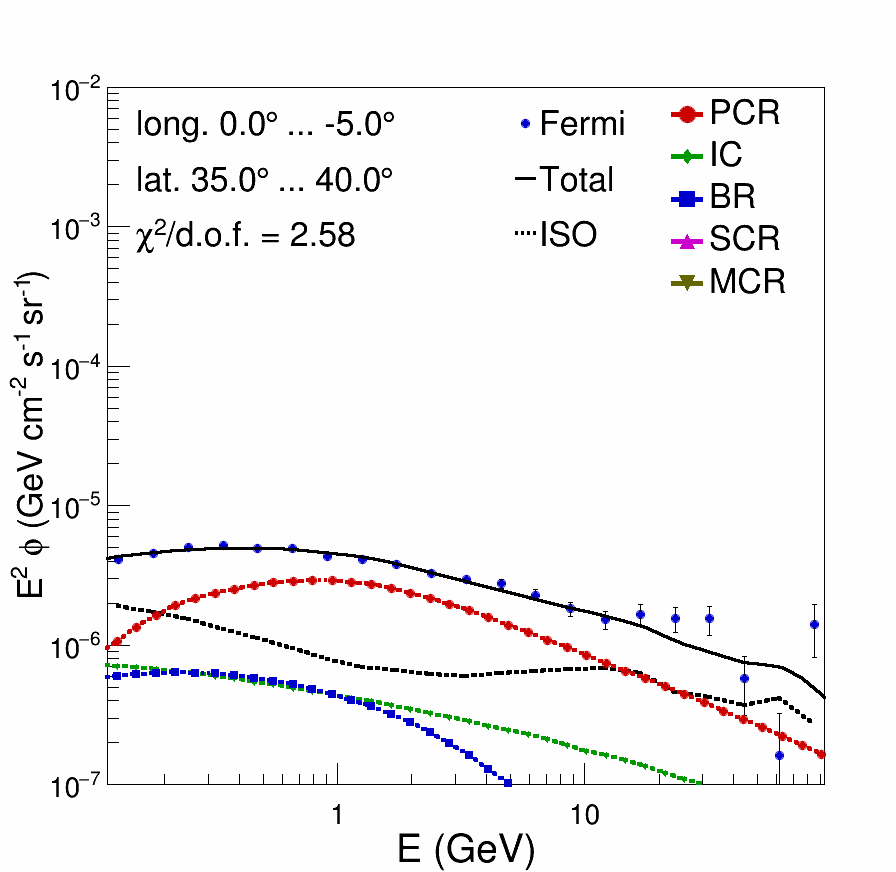}
\includegraphics[width=0.16\textwidth,height=0.16\textwidth,clip]{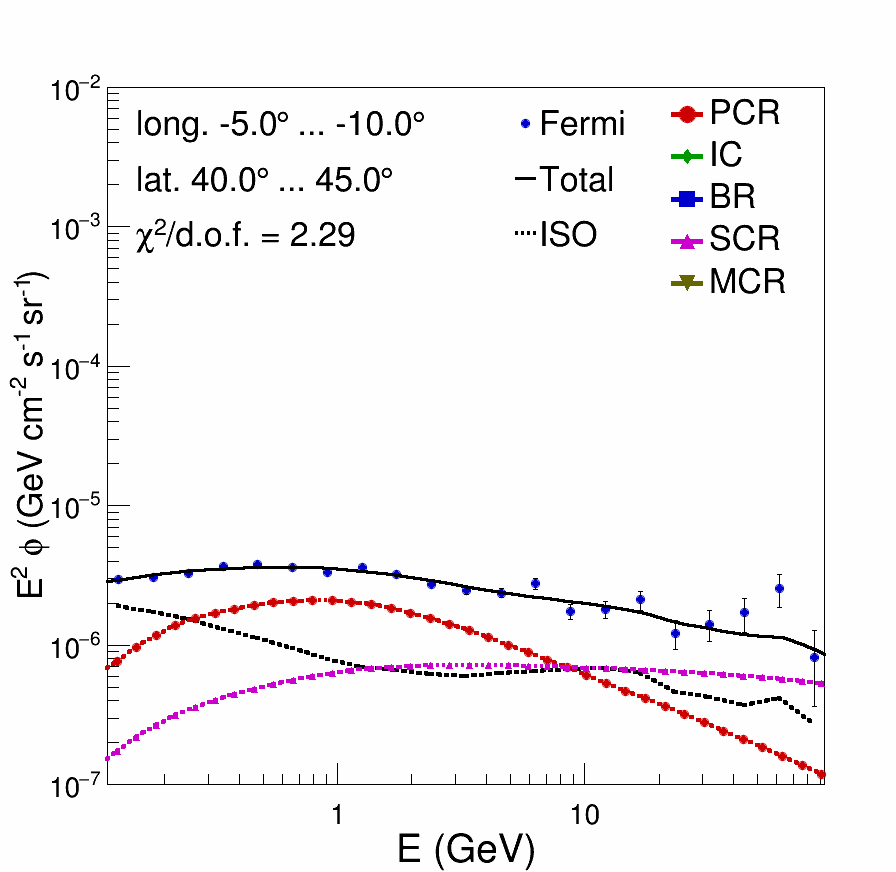}
\includegraphics[width=0.16\textwidth,height=0.16\textwidth,clip]{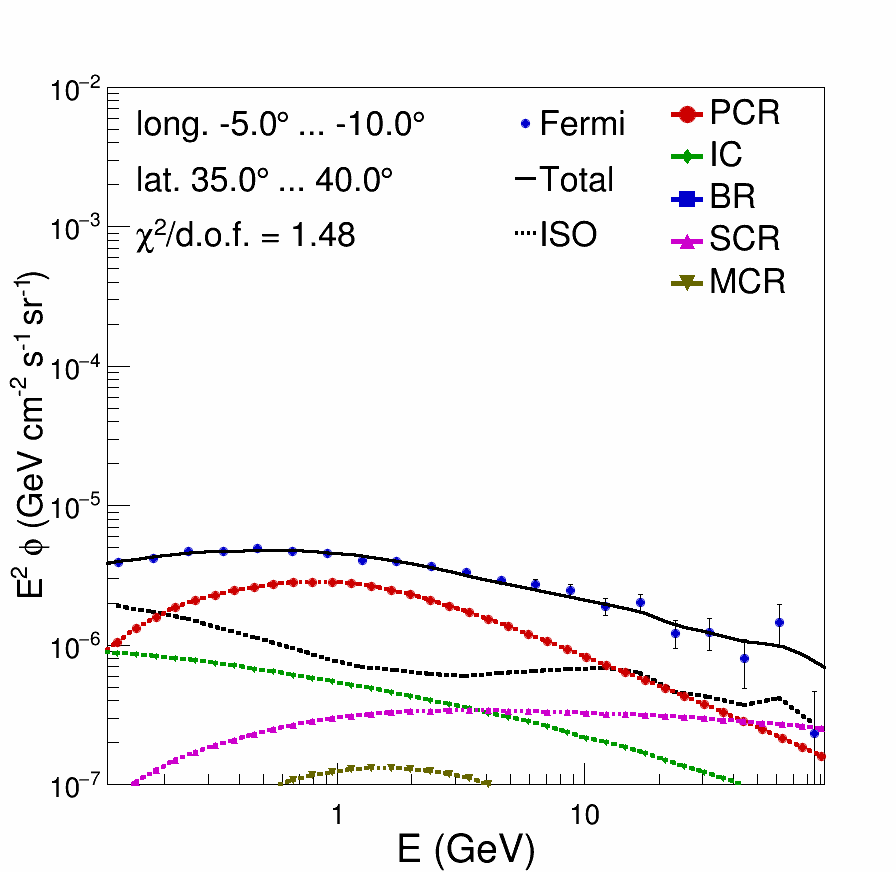}
\includegraphics[width=0.16\textwidth,height=0.16\textwidth,clip]{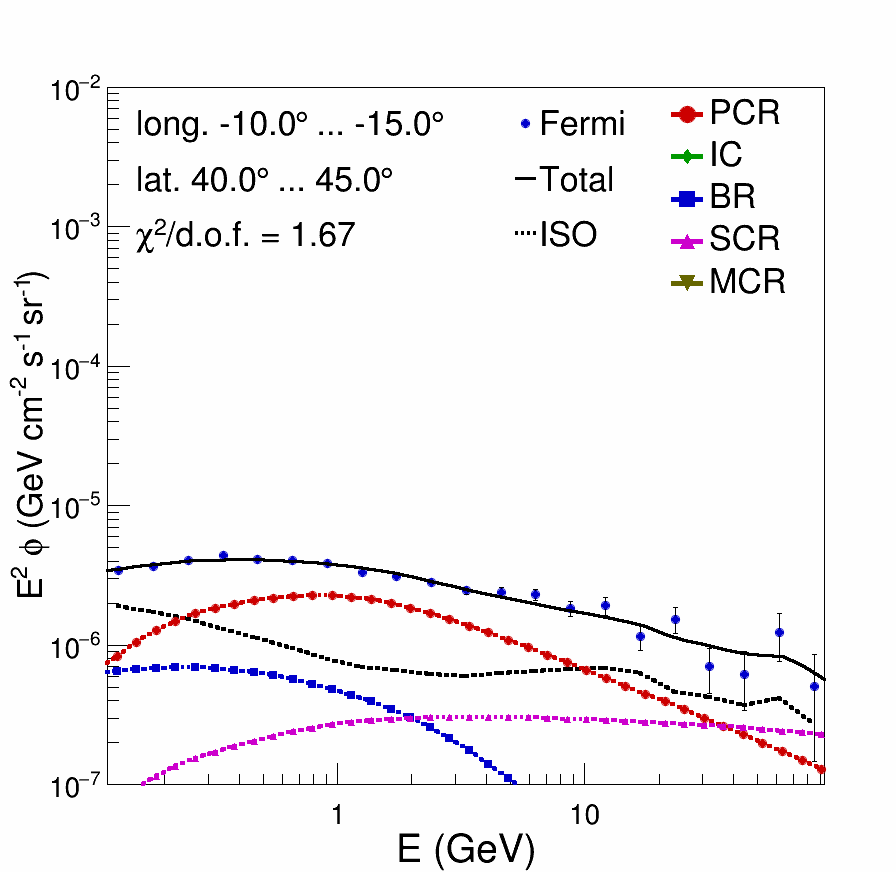}
\includegraphics[width=0.16\textwidth,height=0.16\textwidth,clip]{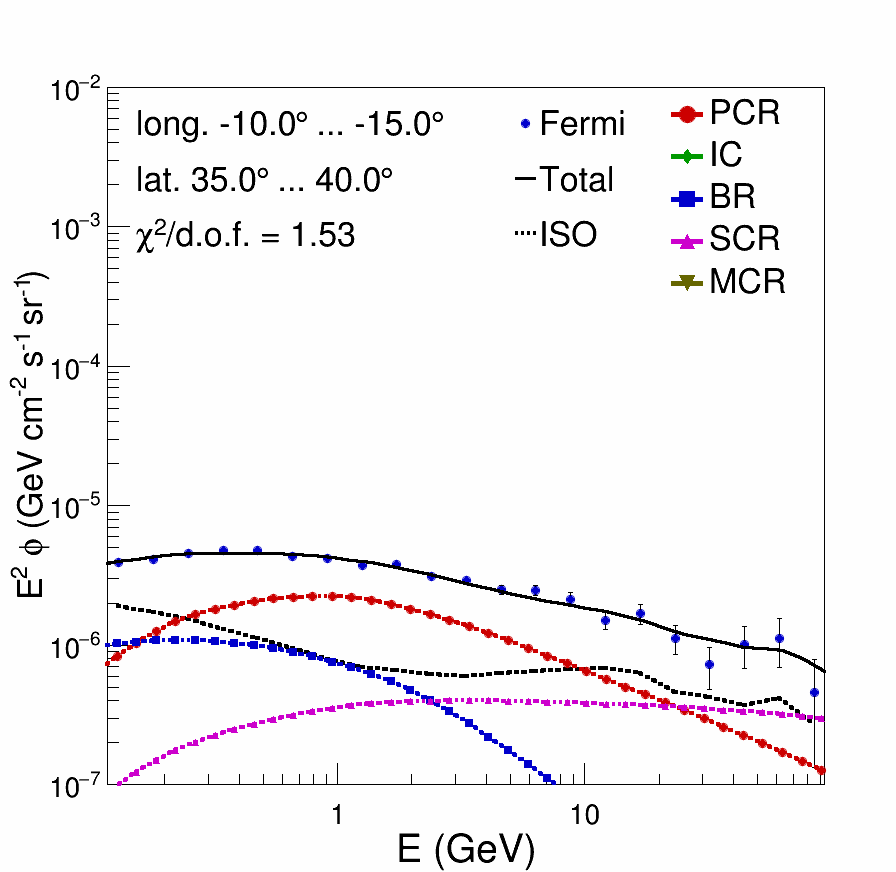}
\includegraphics[width=0.16\textwidth,height=0.16\textwidth,clip]{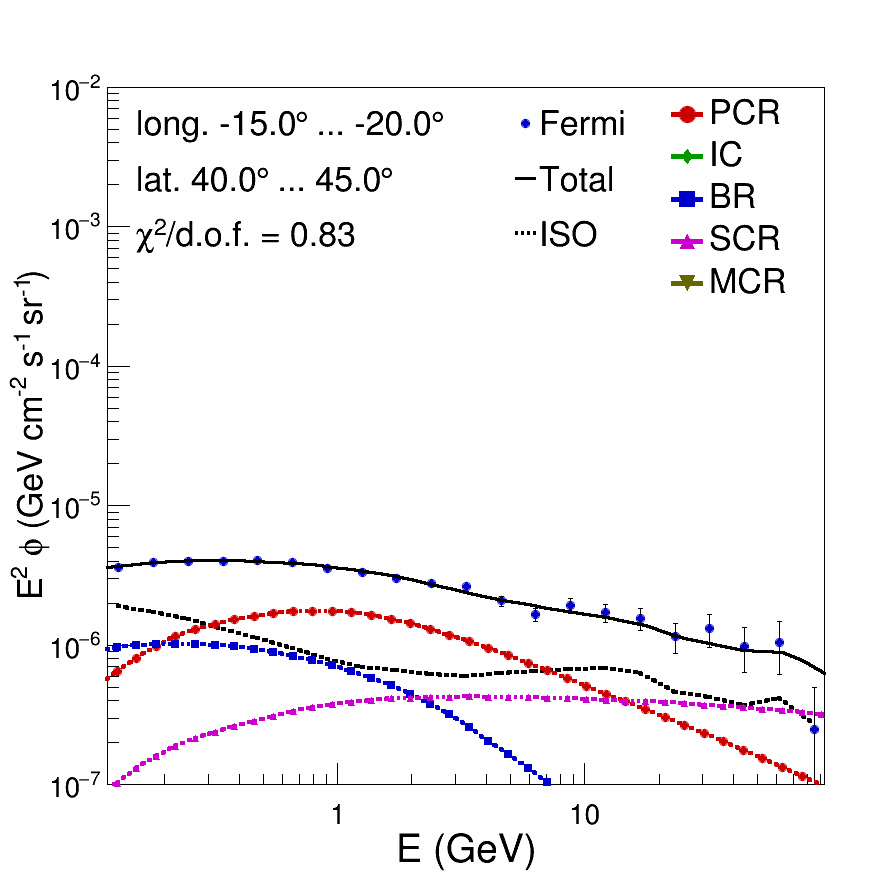}
\includegraphics[width=0.16\textwidth,height=0.16\textwidth,clip]{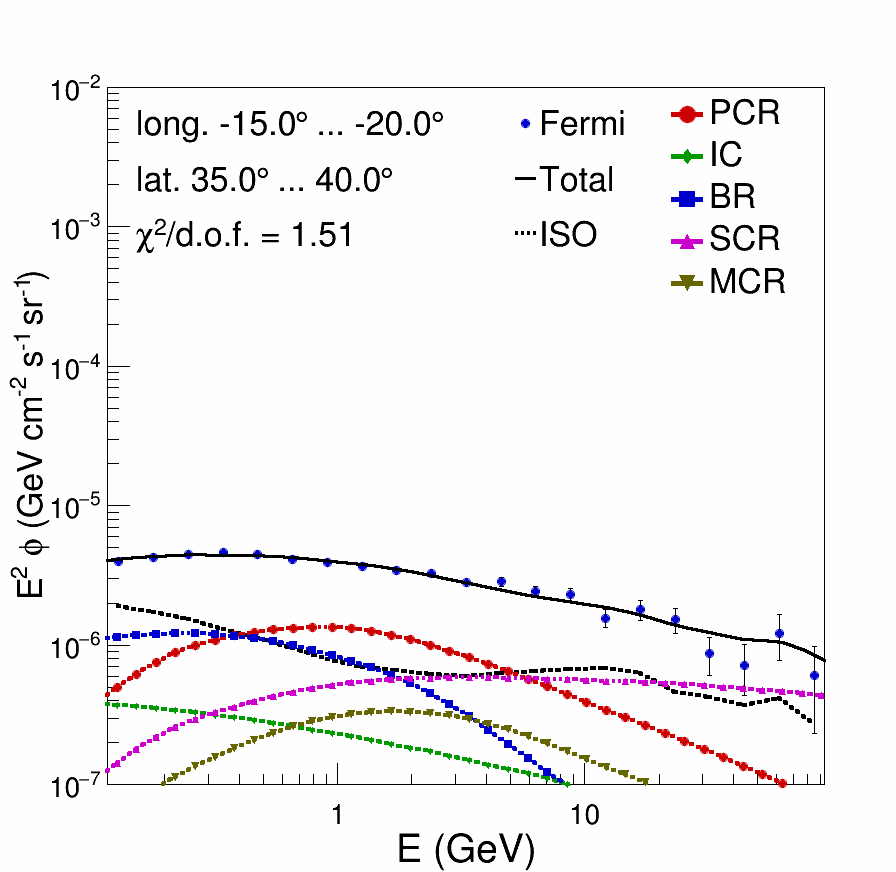}
\includegraphics[width=0.16\textwidth,height=0.16\textwidth,clip]{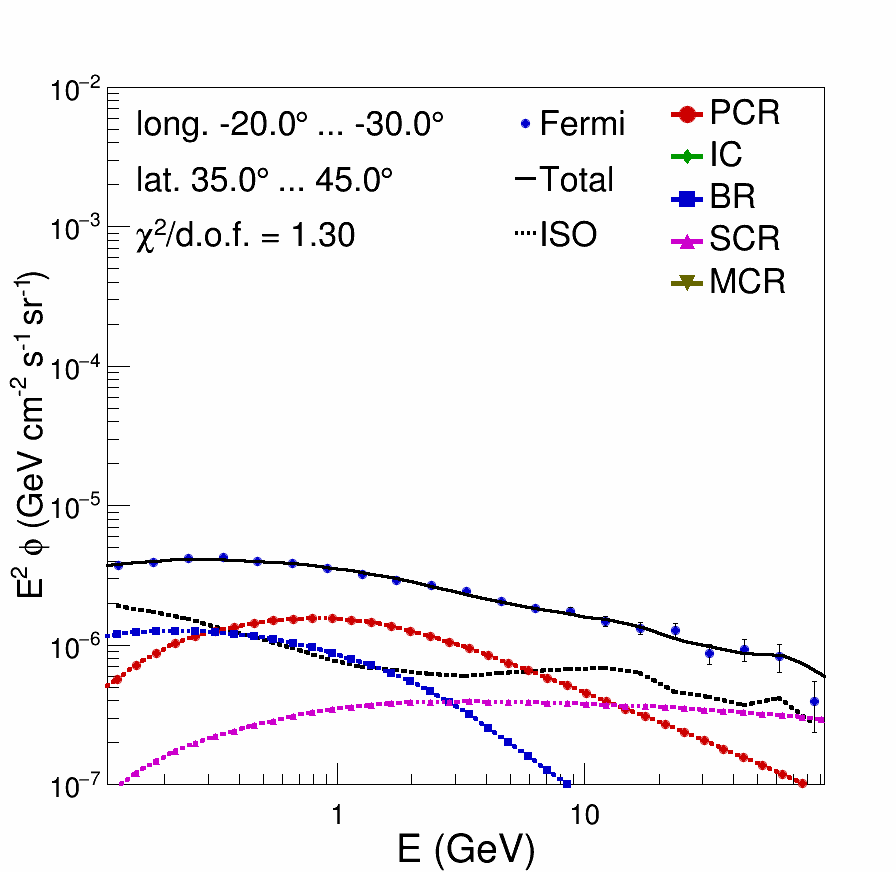}
\includegraphics[width=0.16\textwidth,height=0.16\textwidth,clip]{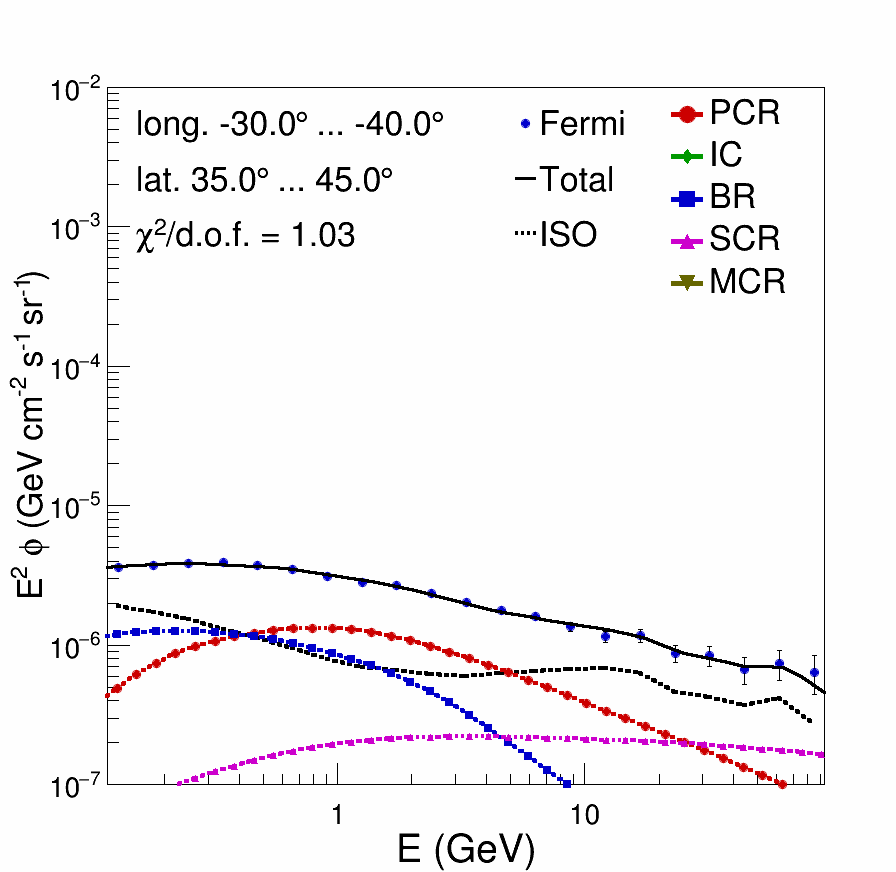}
\includegraphics[width=0.16\textwidth,height=0.16\textwidth,clip]{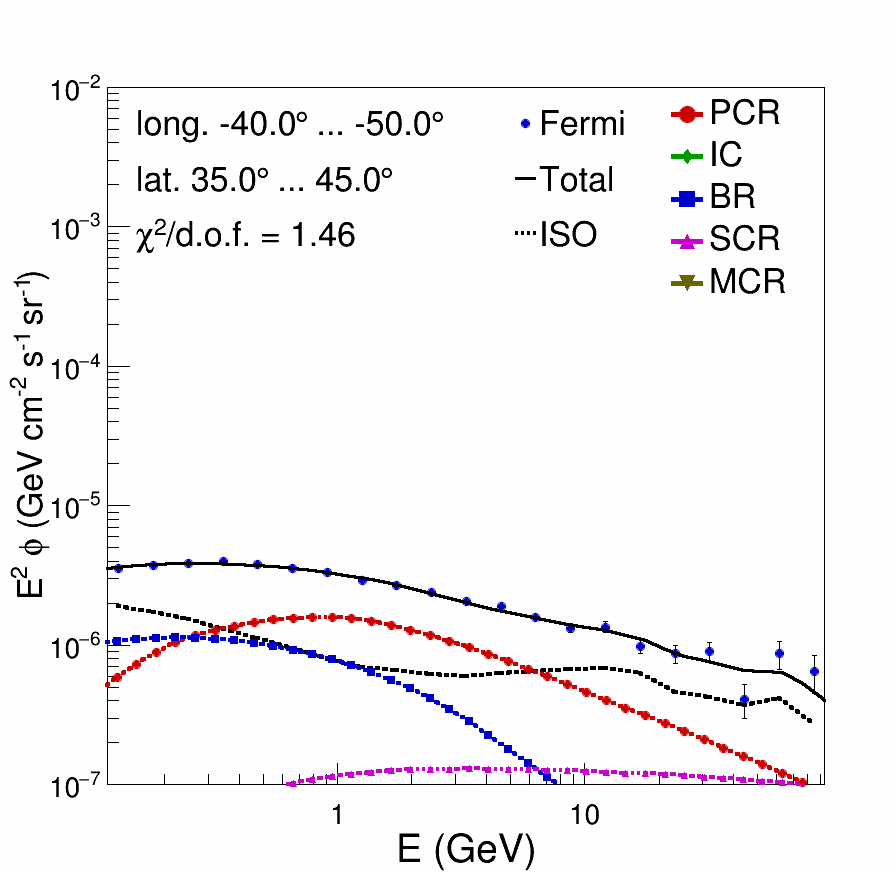}
\includegraphics[width=0.16\textwidth,height=0.16\textwidth,clip]{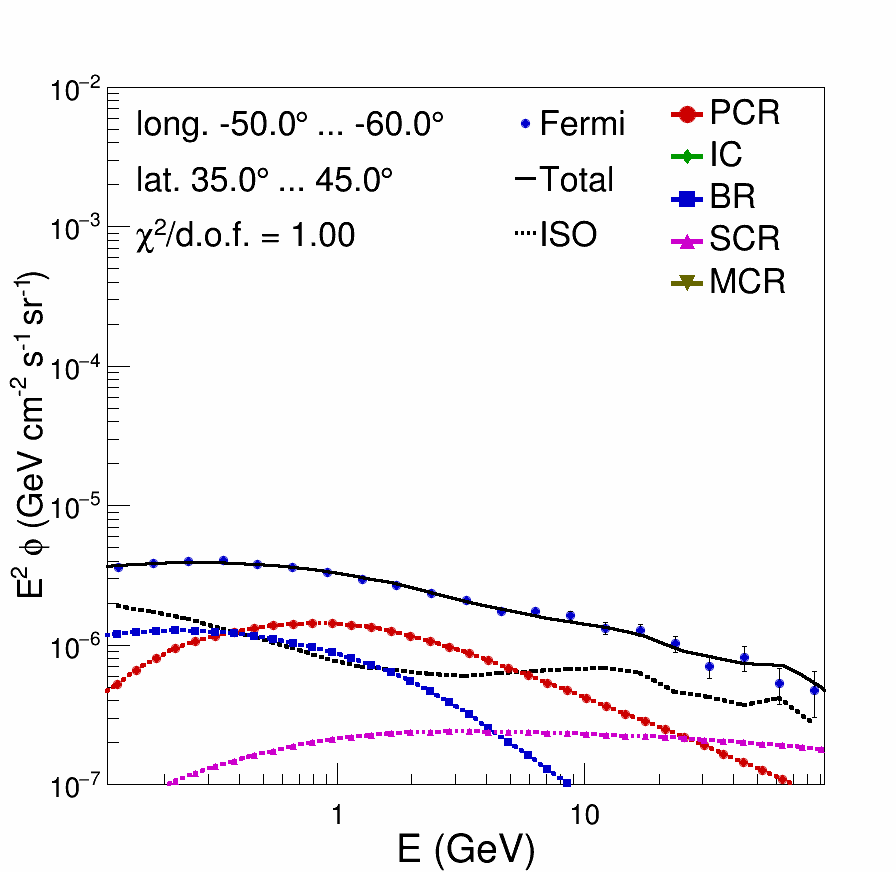}
\includegraphics[width=0.16\textwidth,height=0.16\textwidth,clip]{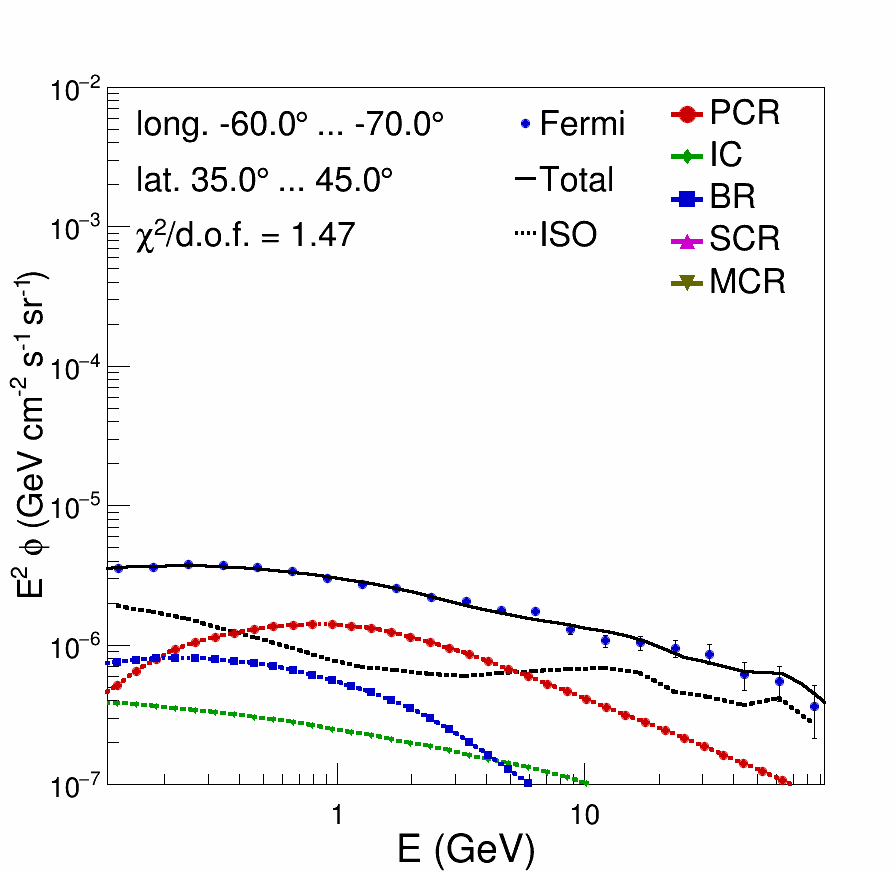}
\includegraphics[width=0.16\textwidth,height=0.16\textwidth,clip]{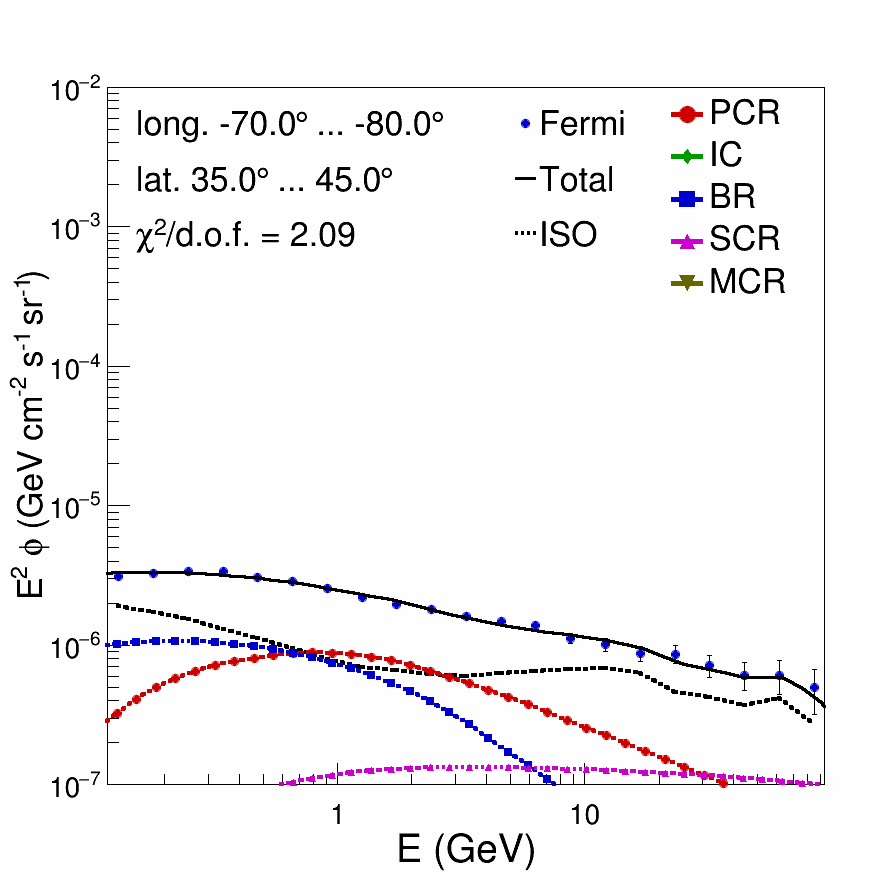}
\includegraphics[width=0.16\textwidth,height=0.16\textwidth,clip]{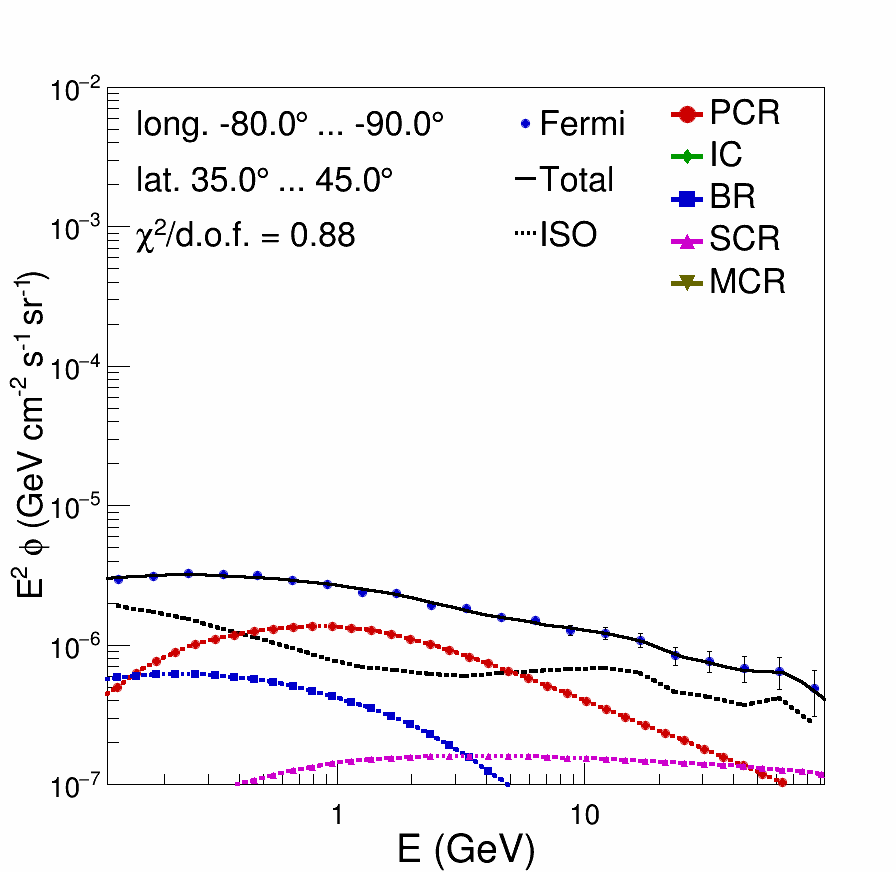}
\includegraphics[width=0.16\textwidth,height=0.16\textwidth,clip]{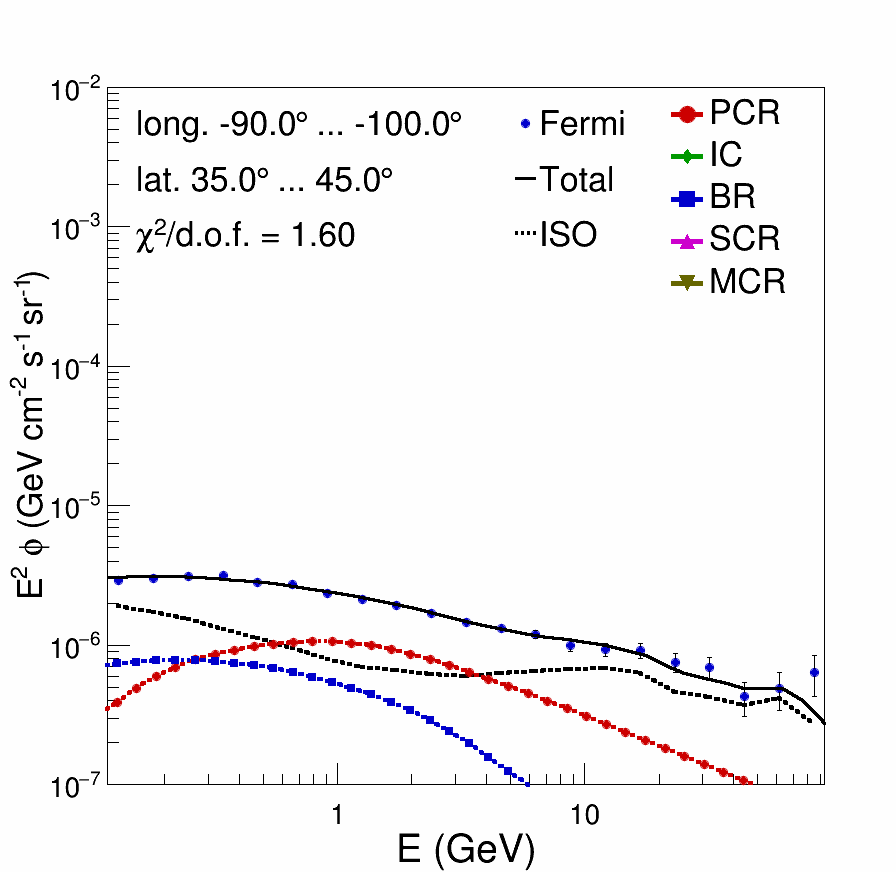}
\includegraphics[width=0.16\textwidth,height=0.16\textwidth,clip]{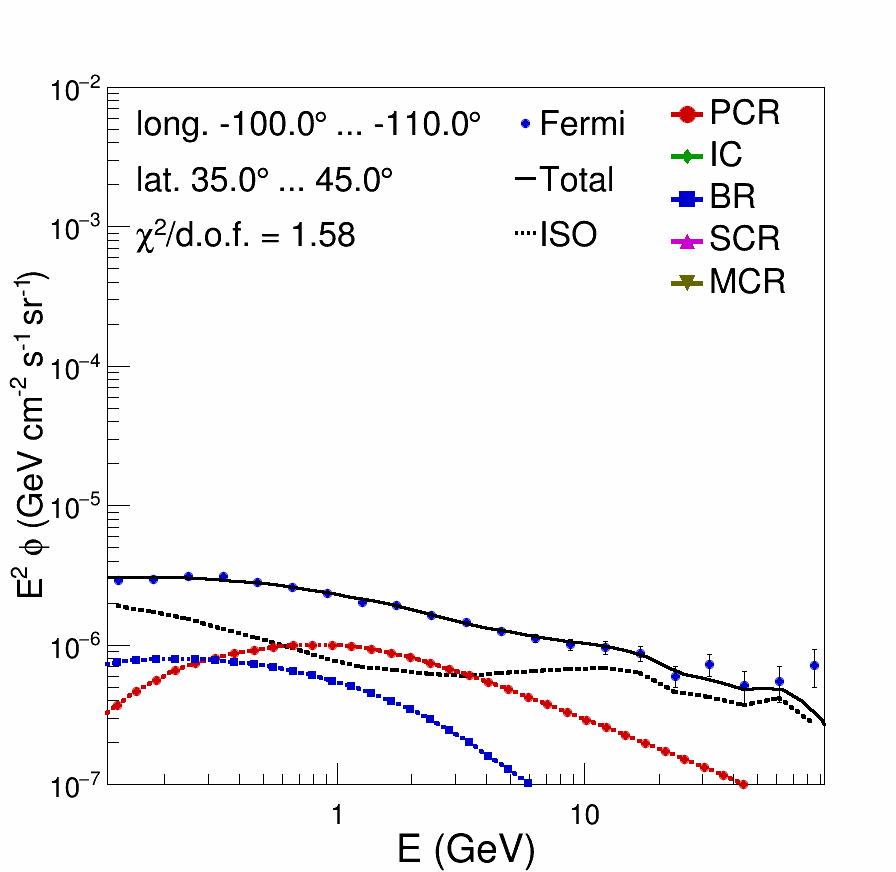}
\includegraphics[width=0.16\textwidth,height=0.16\textwidth,clip]{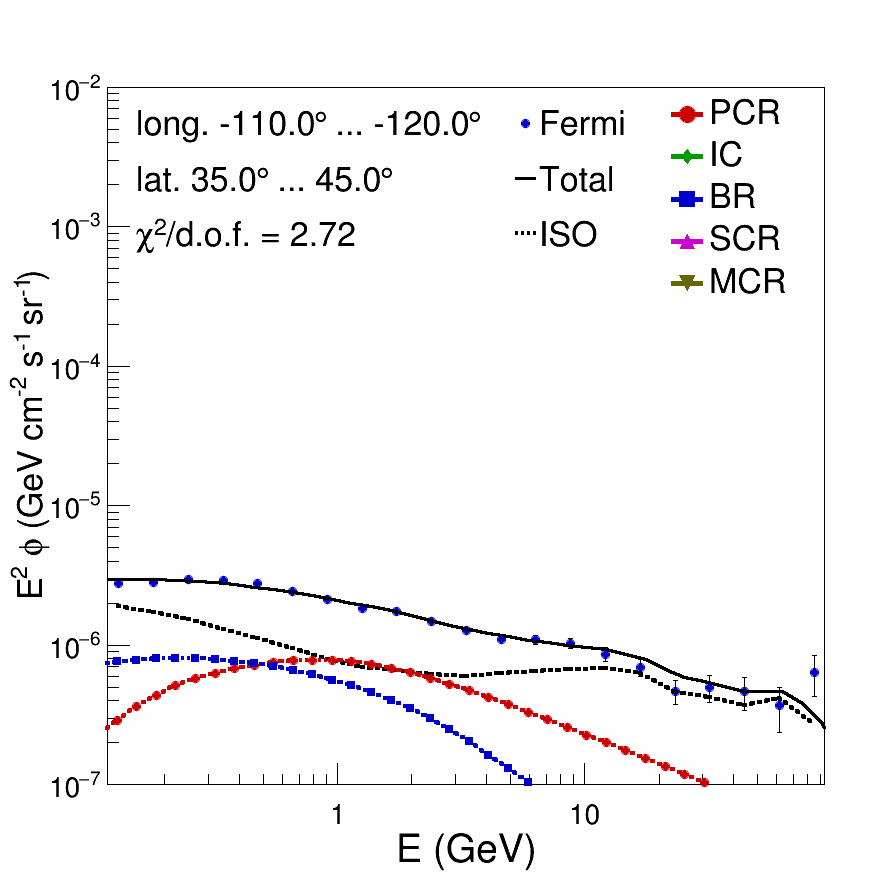}
\includegraphics[width=0.16\textwidth,height=0.16\textwidth,clip]{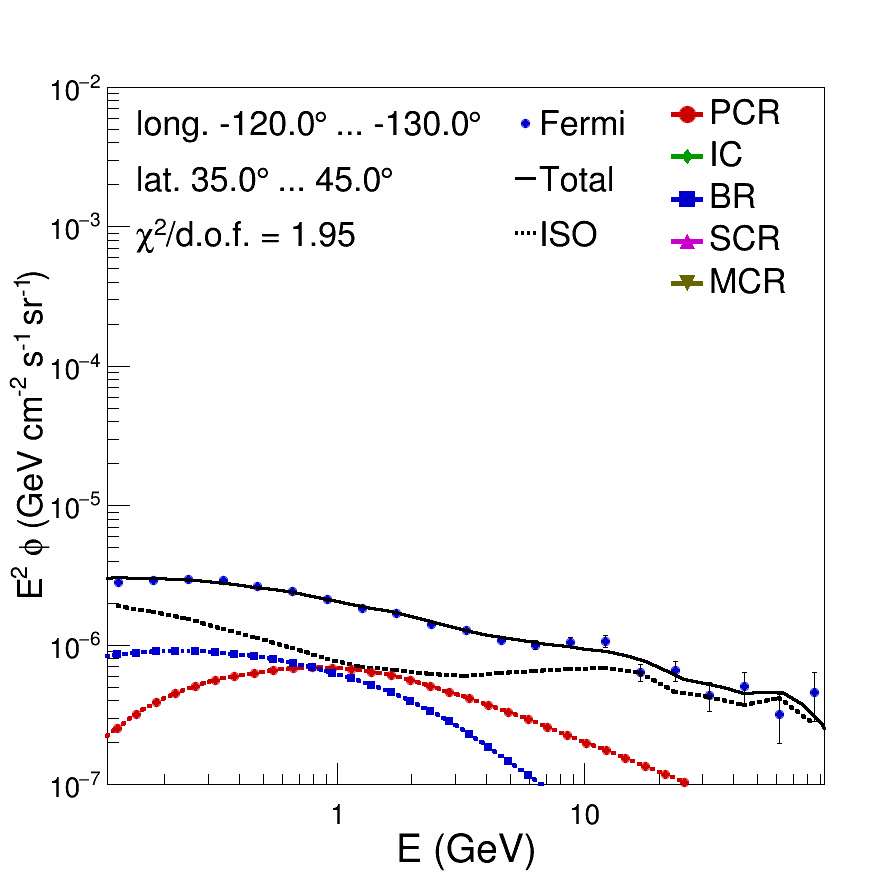}
\includegraphics[width=0.16\textwidth,height=0.16\textwidth,clip]{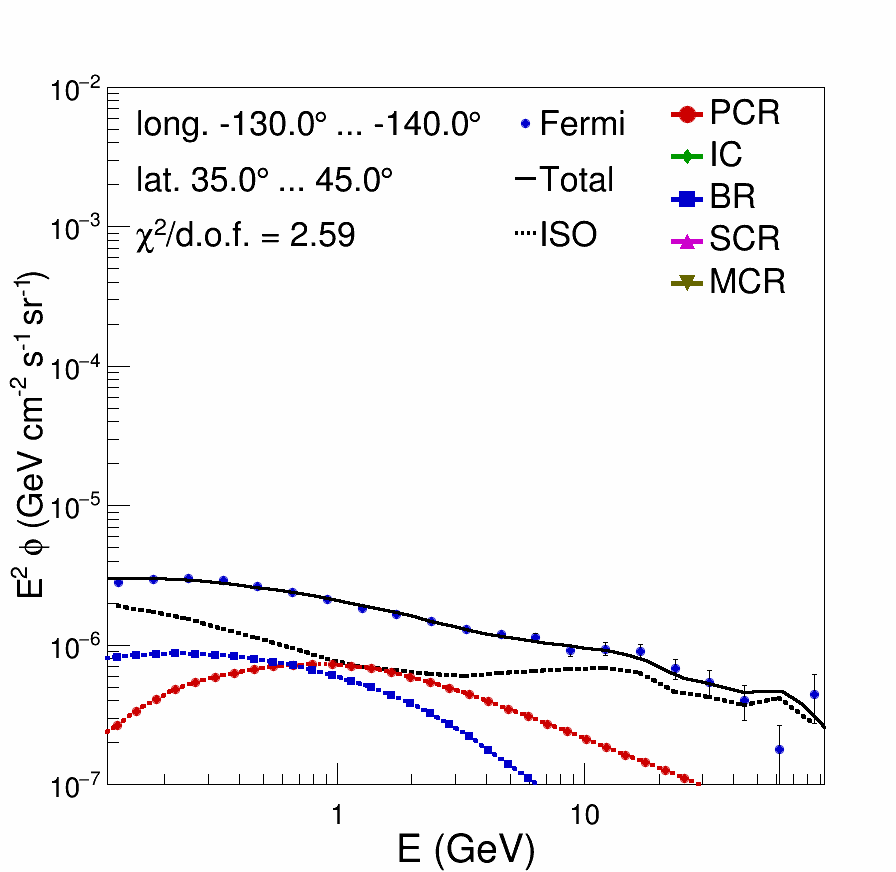}
\includegraphics[width=0.16\textwidth,height=0.16\textwidth,clip]{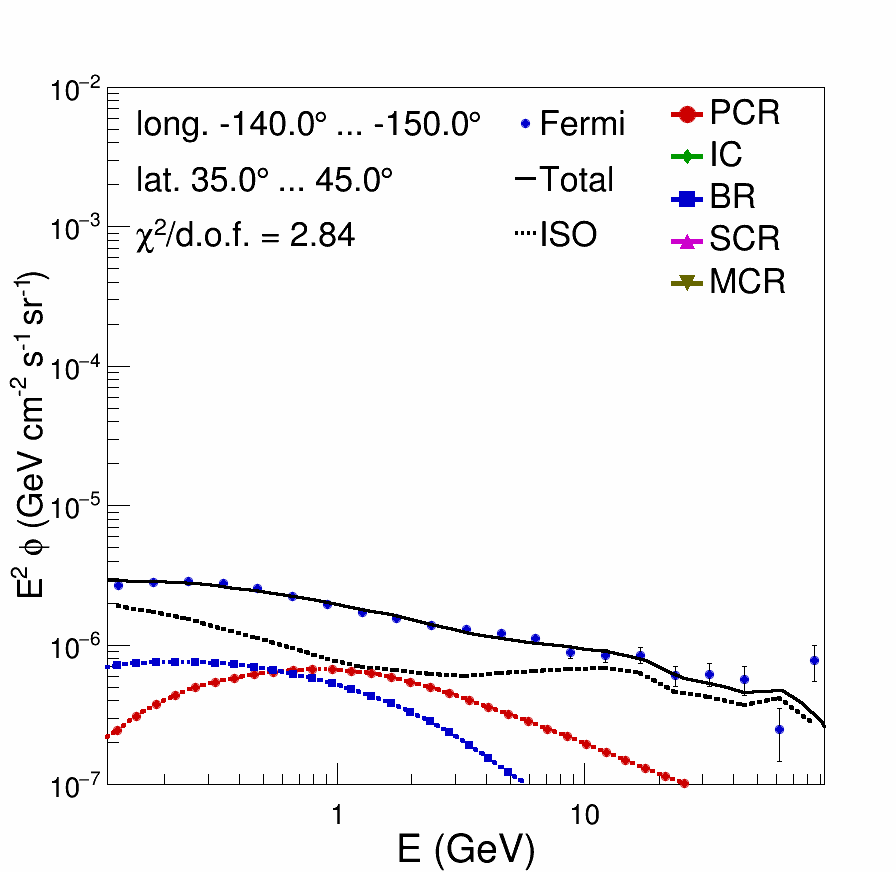}
\includegraphics[width=0.16\textwidth,height=0.16\textwidth,clip]{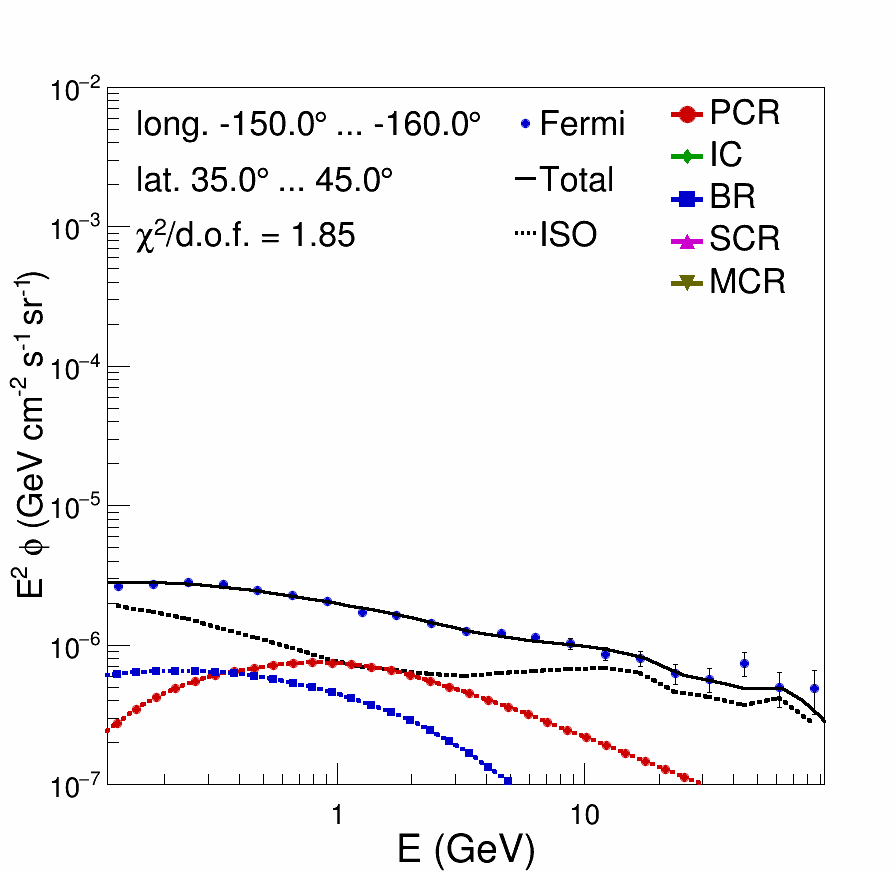}
\includegraphics[width=0.16\textwidth,height=0.16\textwidth,clip]{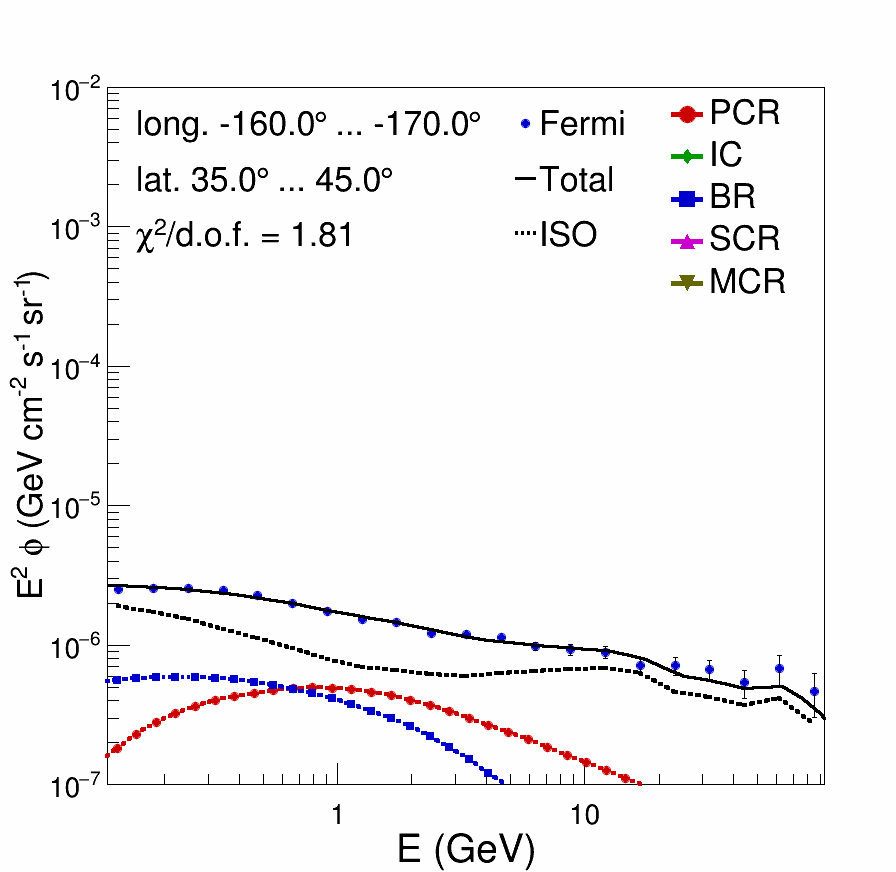}
\includegraphics[width=0.16\textwidth,height=0.16\textwidth,clip]{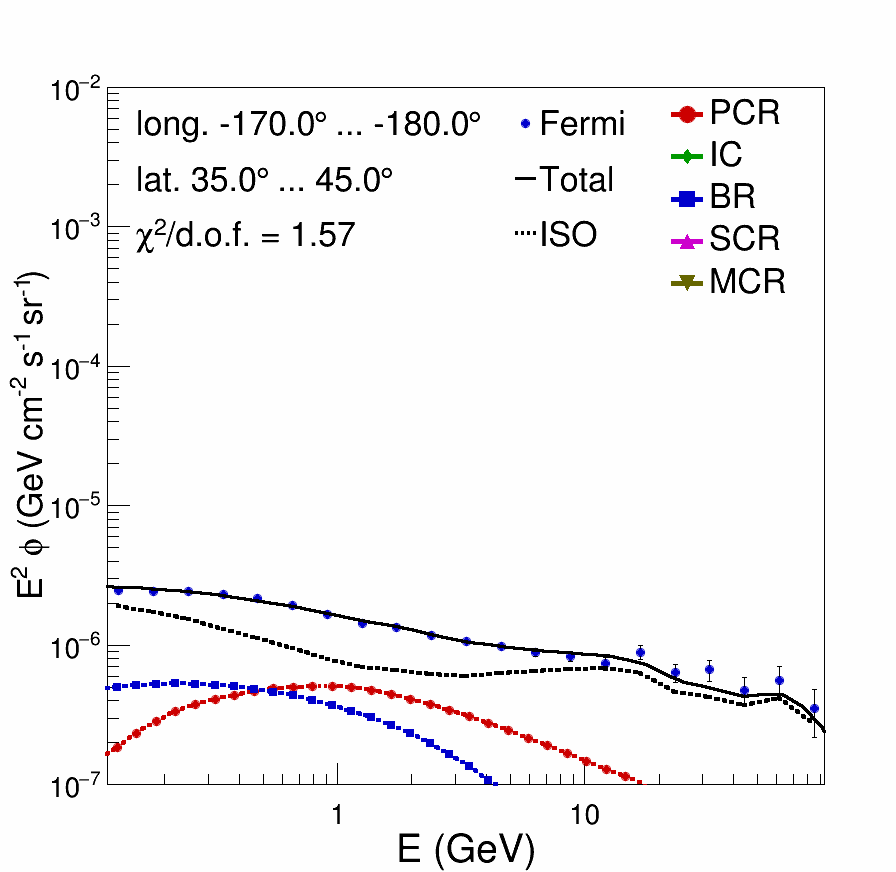}
\caption[]{Template fits for latitudes  with $35.0^\circ<b<45.0^\circ$ and longitudes decreasing from 180$^\circ$ to -180$^\circ$. \label{F14}
}
\end{figure}
\clearpage
\begin{figure}
\centering
\includegraphics[width=0.16\textwidth,height=0.16\textwidth,clip]{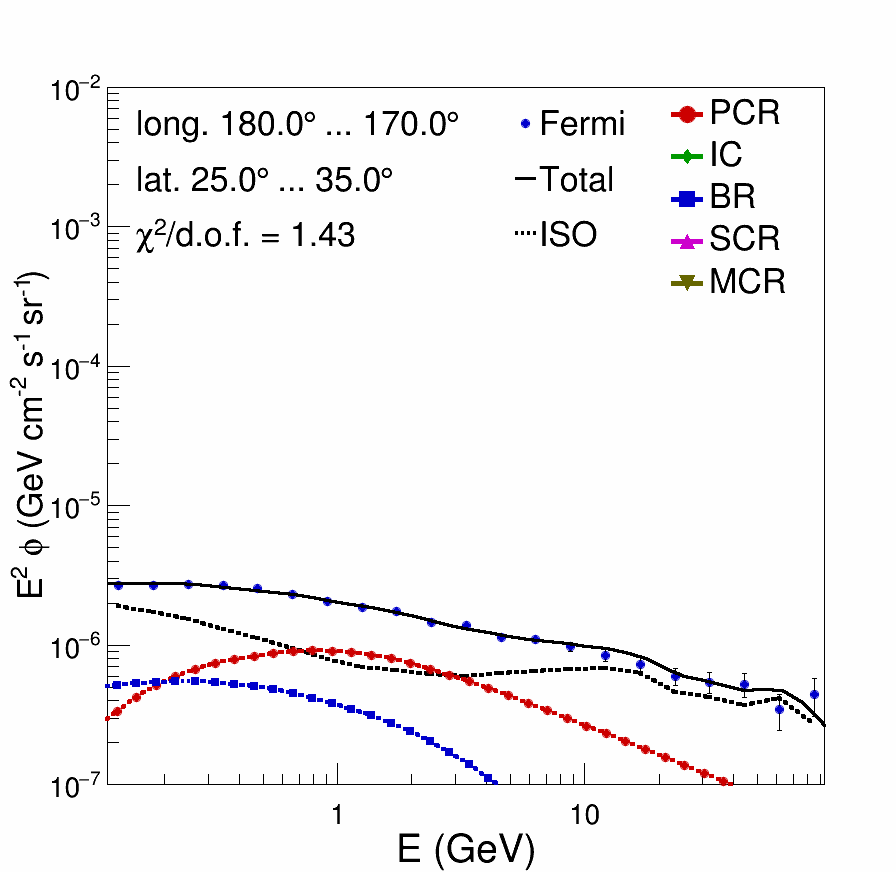}
\includegraphics[width=0.16\textwidth,height=0.16\textwidth,clip]{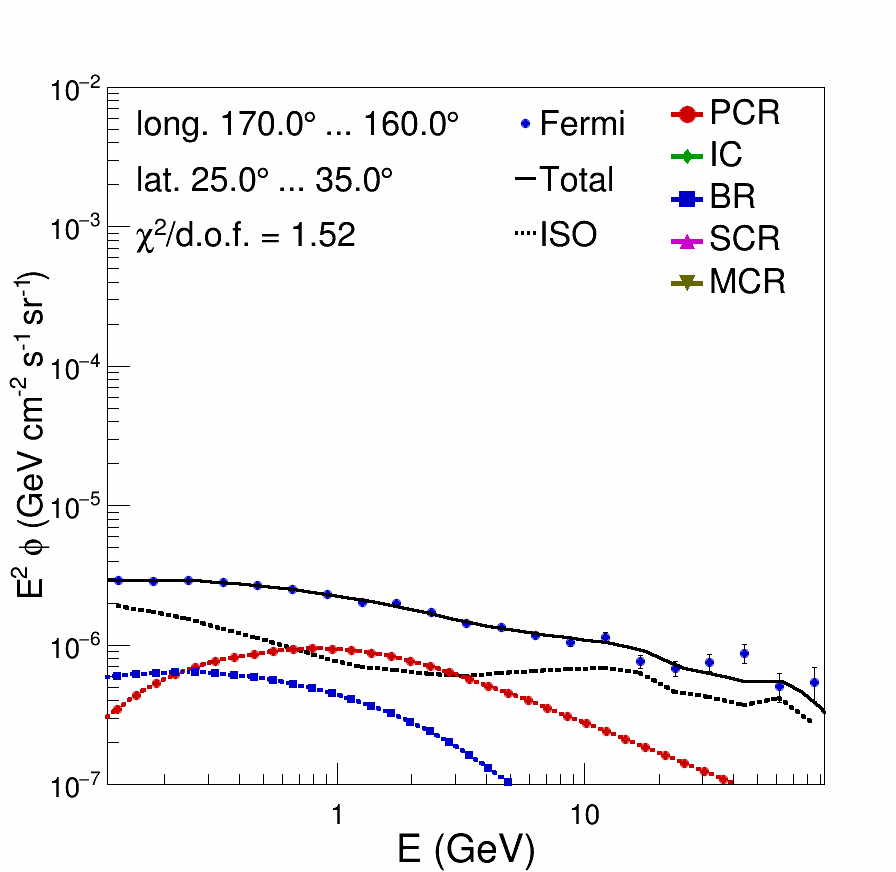}
\includegraphics[width=0.16\textwidth,height=0.16\textwidth,clip]{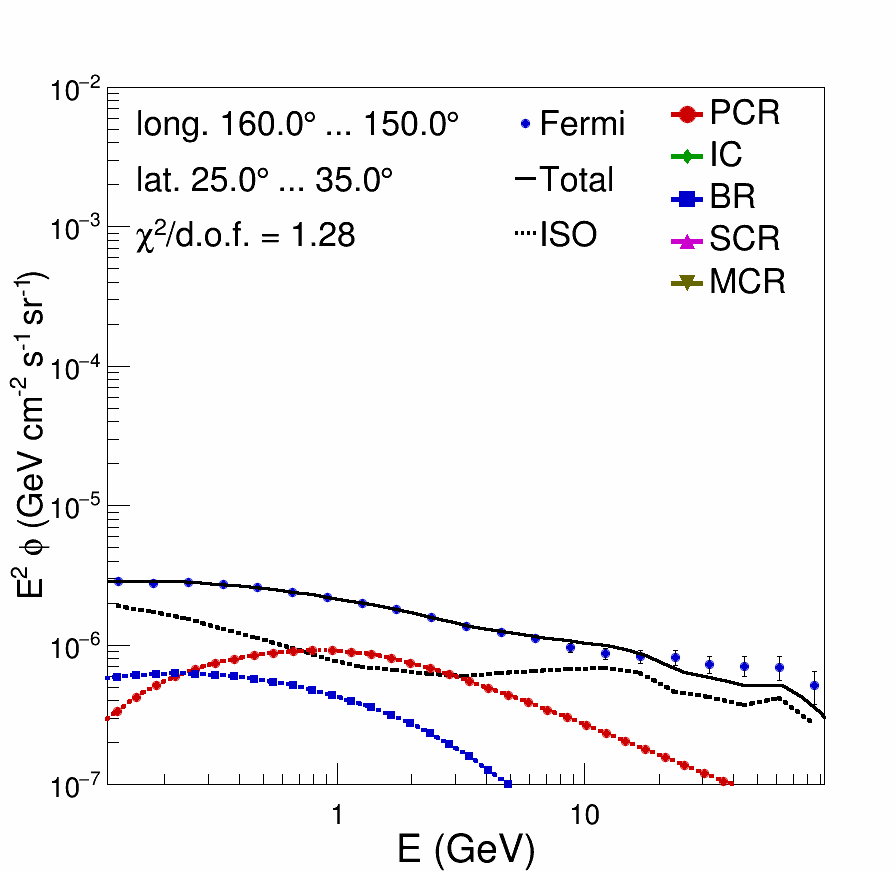}
\includegraphics[width=0.16\textwidth,height=0.16\textwidth,clip]{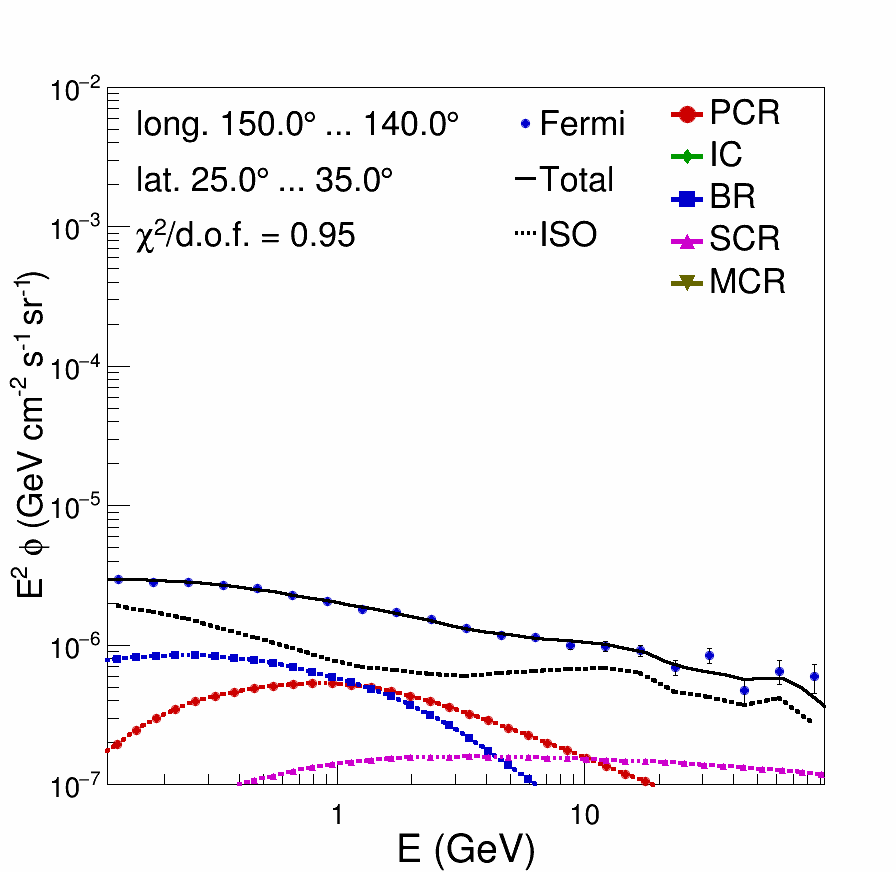}
\includegraphics[width=0.16\textwidth,height=0.16\textwidth,clip]{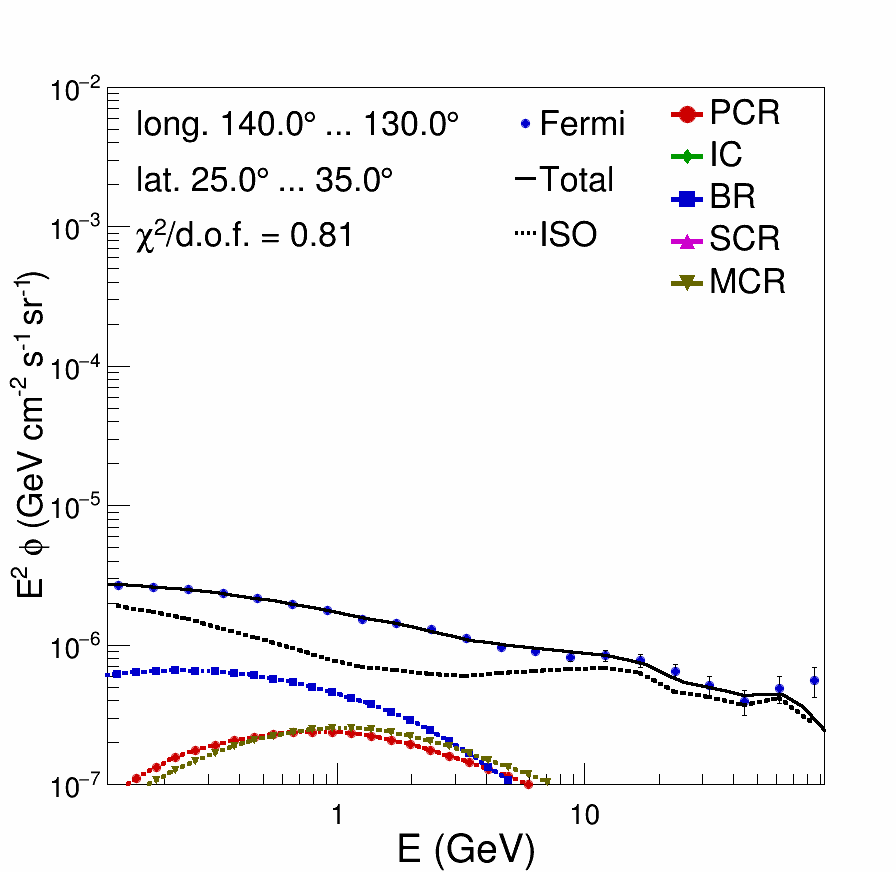}
\includegraphics[width=0.16\textwidth,height=0.16\textwidth,clip]{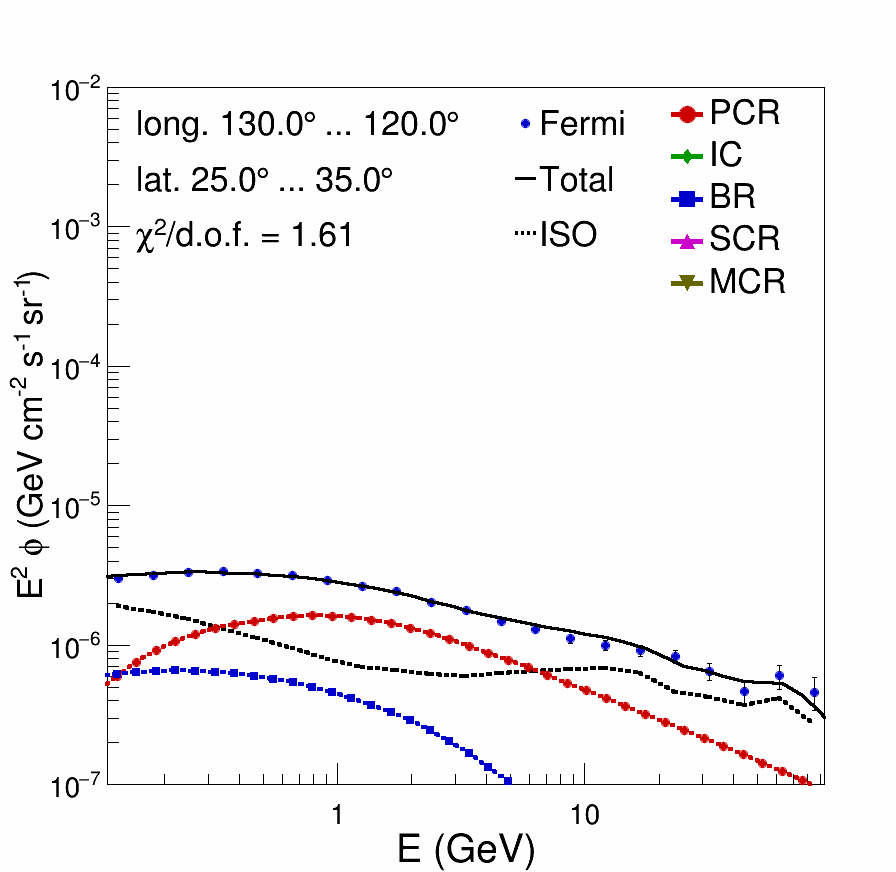}
\includegraphics[width=0.16\textwidth,height=0.16\textwidth,clip]{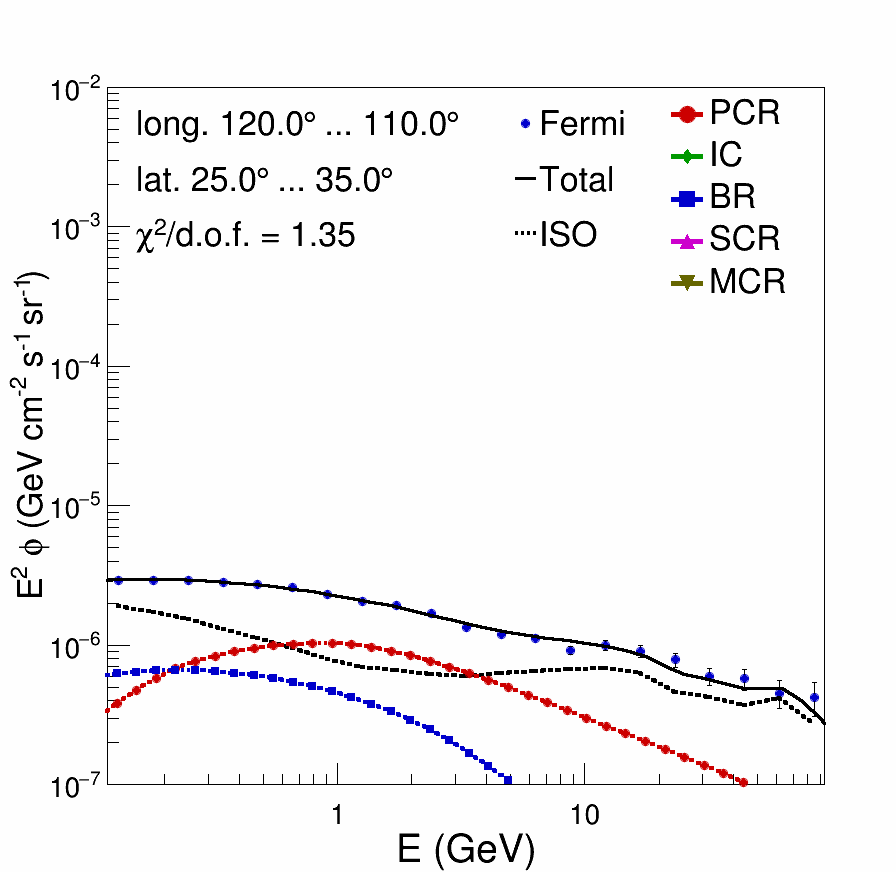}
\includegraphics[width=0.16\textwidth,height=0.16\textwidth,clip]{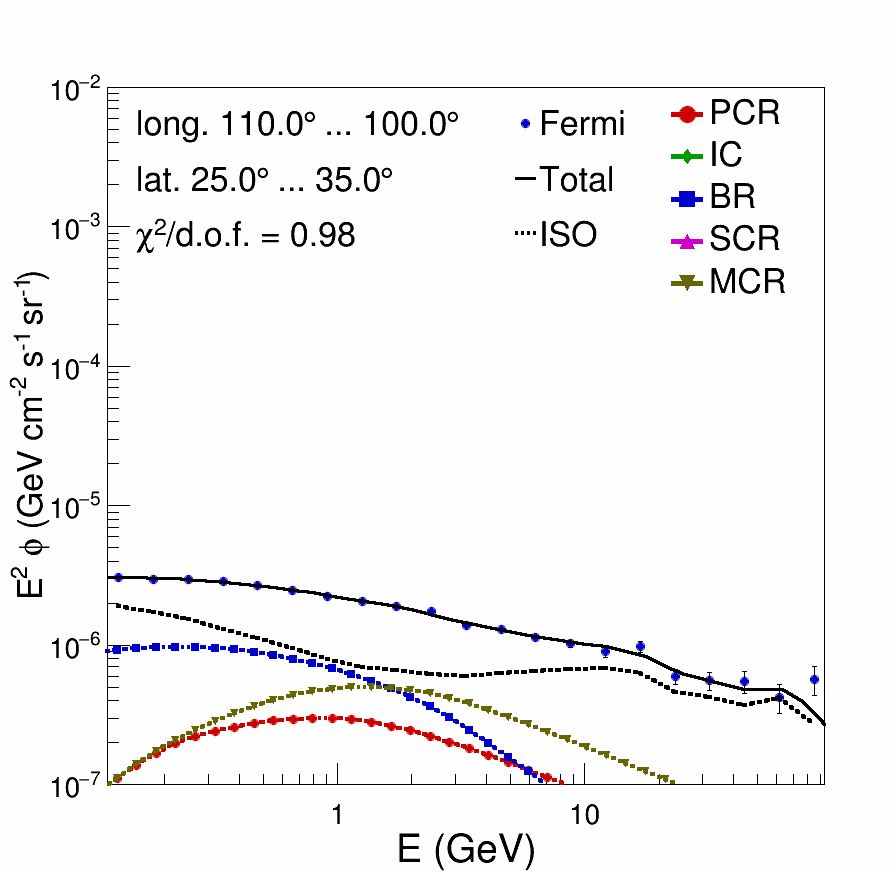}
\includegraphics[width=0.16\textwidth,height=0.16\textwidth,clip]{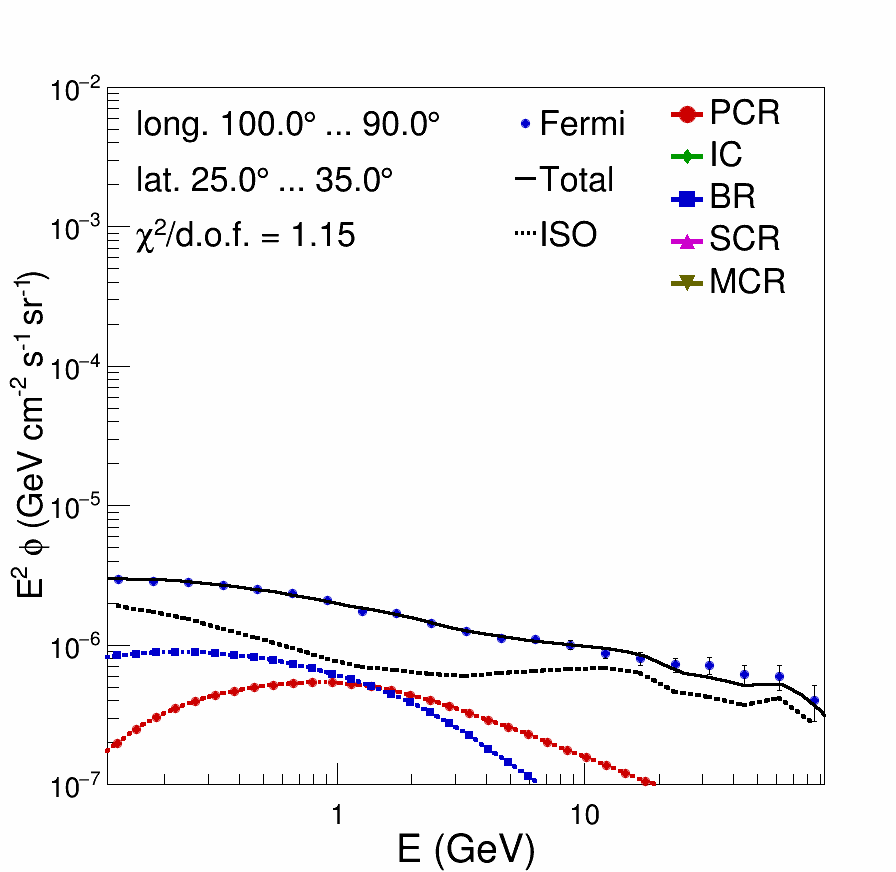}
\includegraphics[width=0.16\textwidth,height=0.16\textwidth,clip]{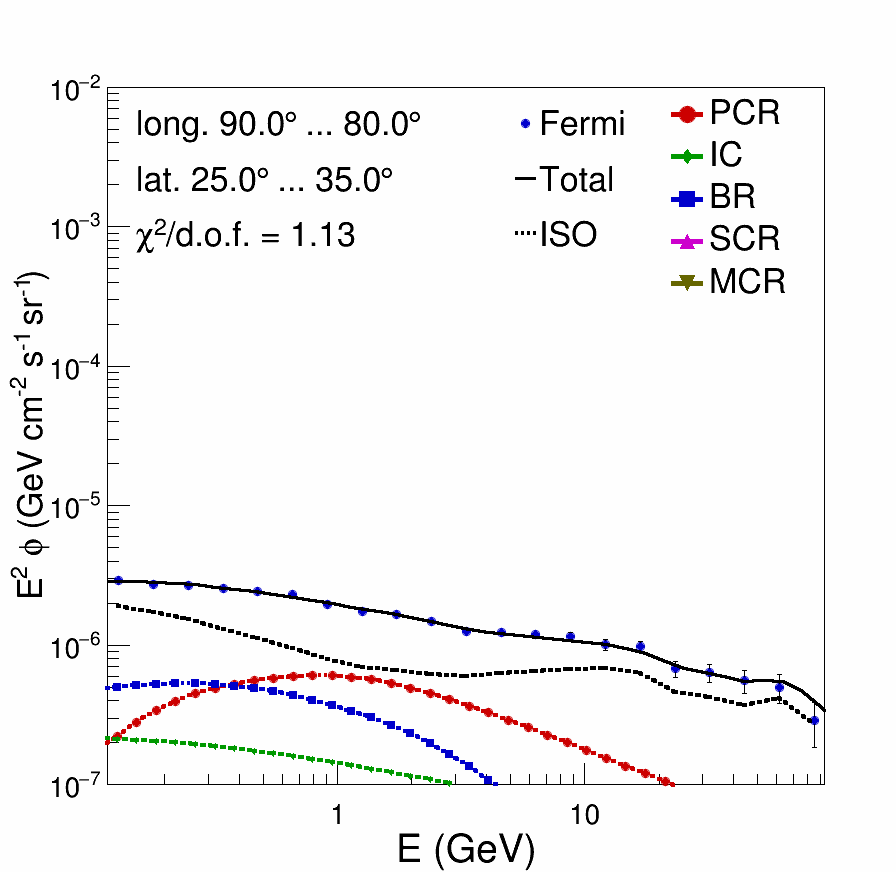}
\includegraphics[width=0.16\textwidth,height=0.16\textwidth,clip]{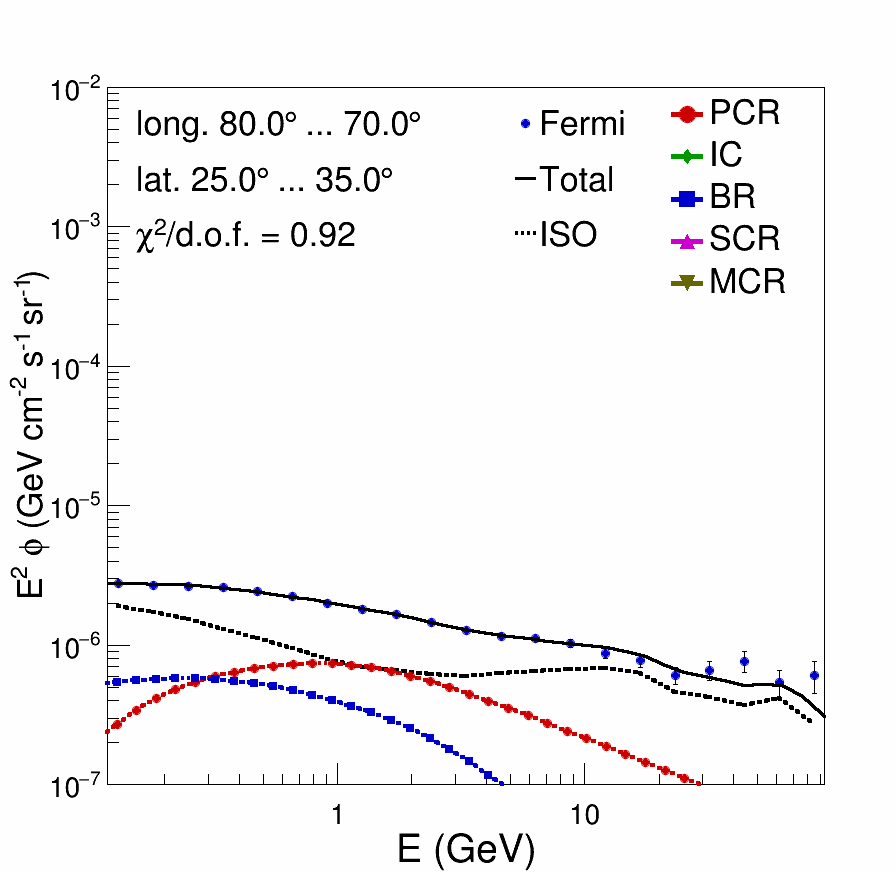}
\includegraphics[width=0.16\textwidth,height=0.16\textwidth,clip]{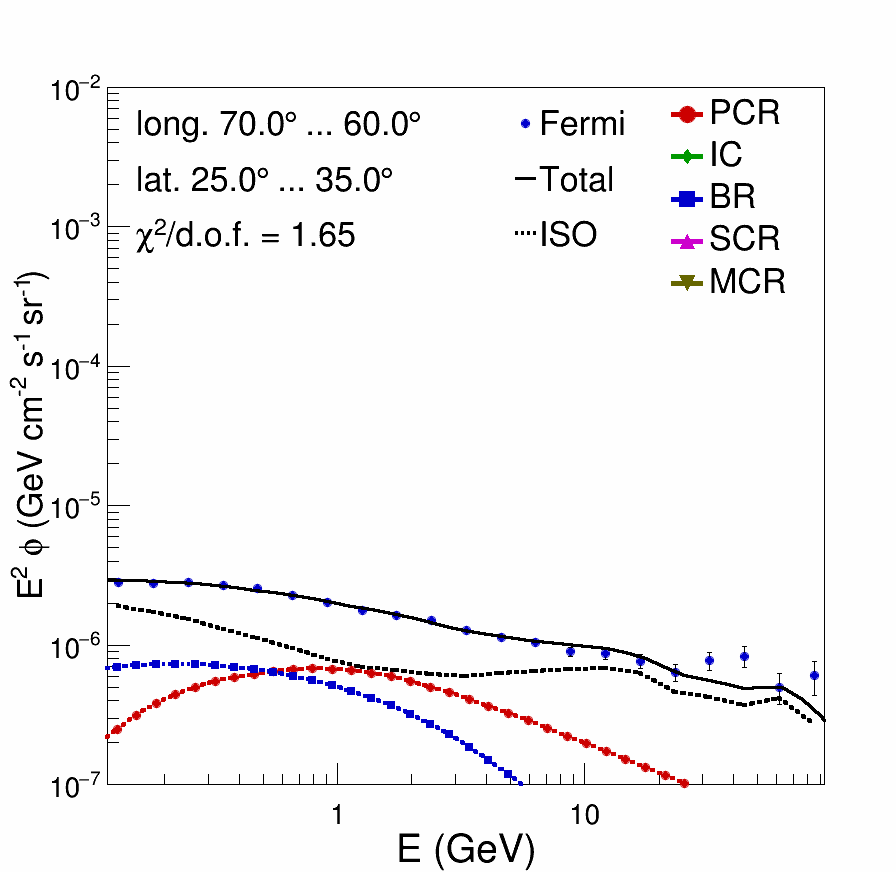}
\includegraphics[width=0.16\textwidth,height=0.16\textwidth,clip]{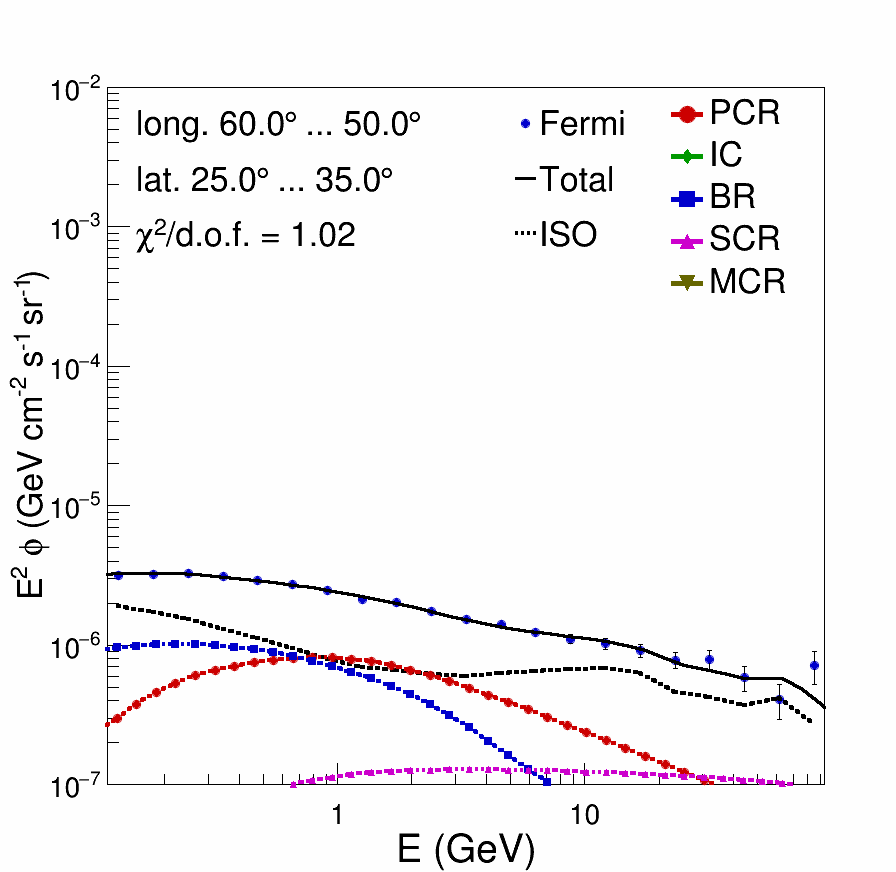}
\includegraphics[width=0.16\textwidth,height=0.16\textwidth,clip]{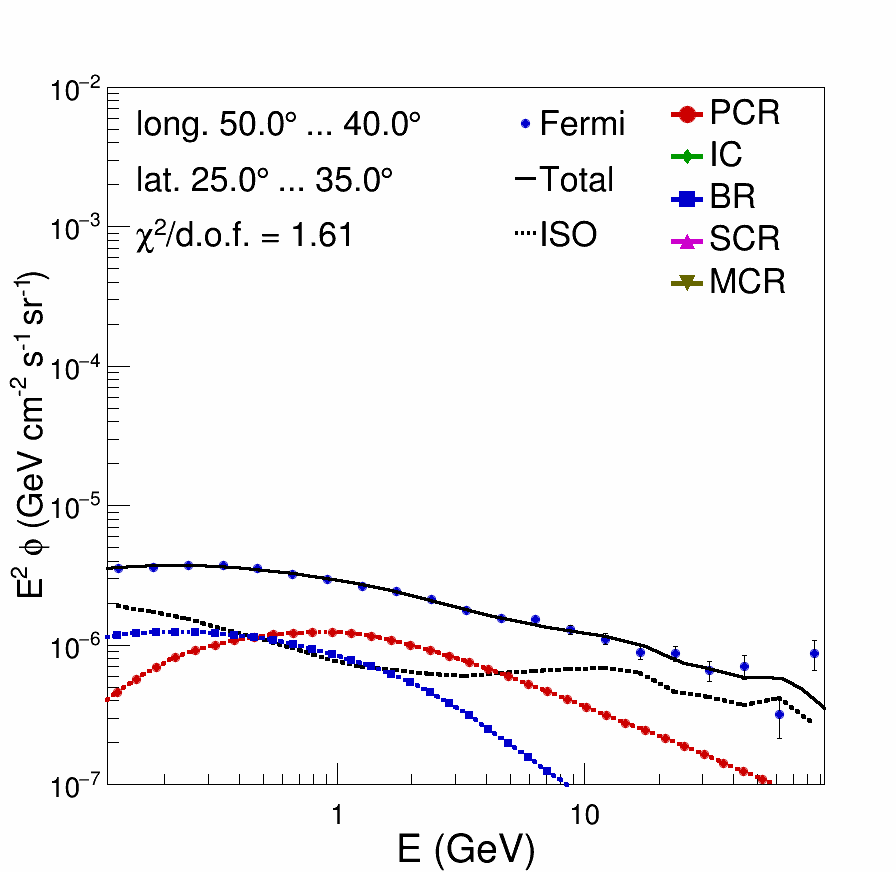}
\includegraphics[width=0.16\textwidth,height=0.16\textwidth,clip]{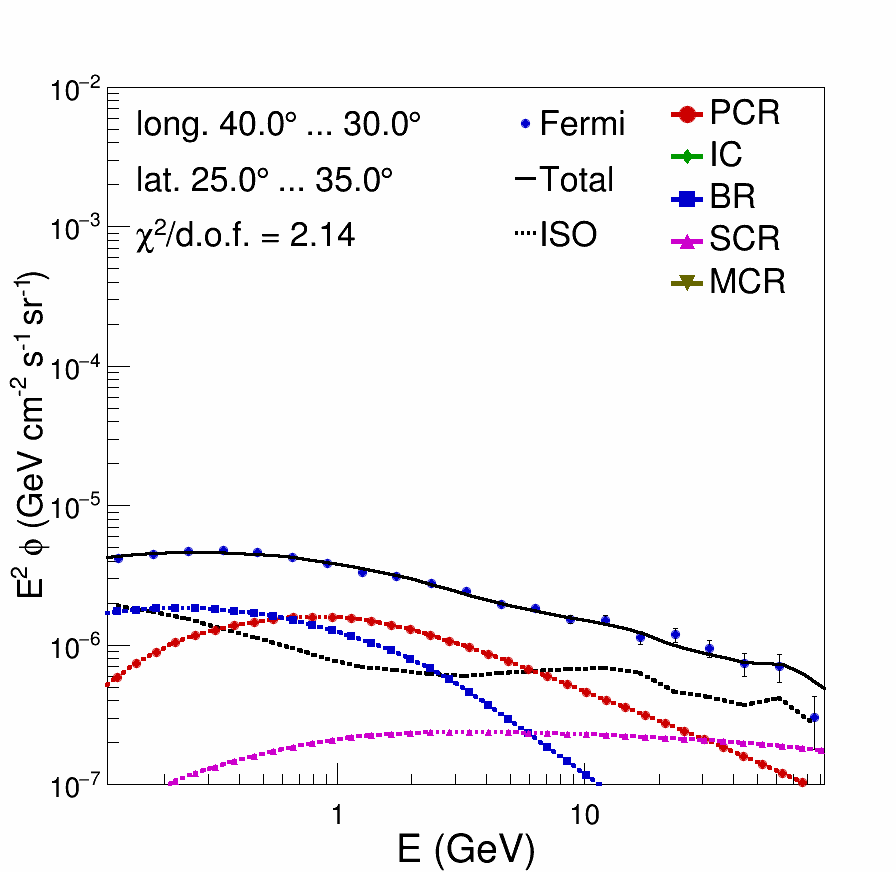}
\includegraphics[width=0.16\textwidth,height=0.16\textwidth,clip]{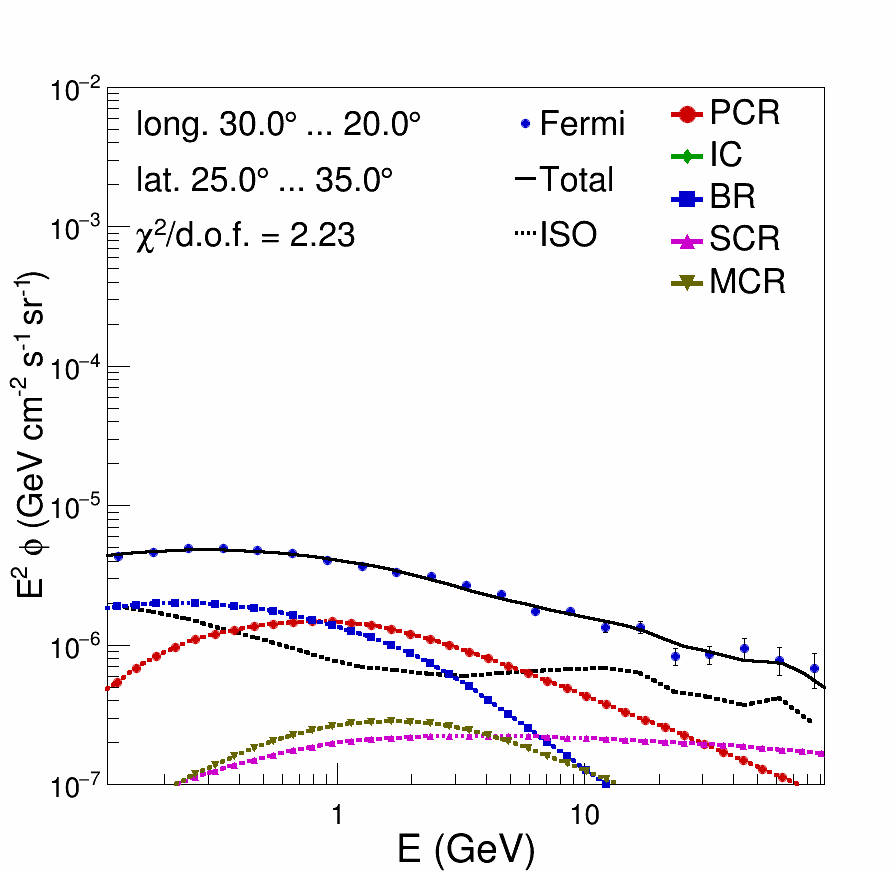}
\includegraphics[width=0.16\textwidth,height=0.16\textwidth,clip]{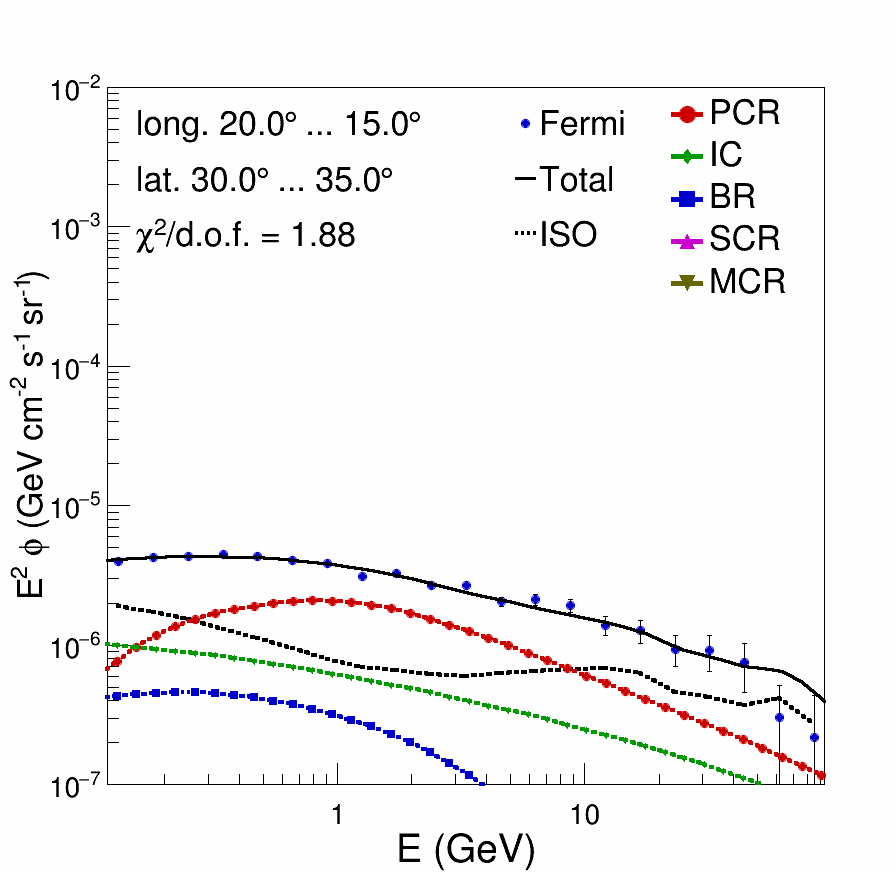}
\includegraphics[width=0.16\textwidth,height=0.16\textwidth,clip]{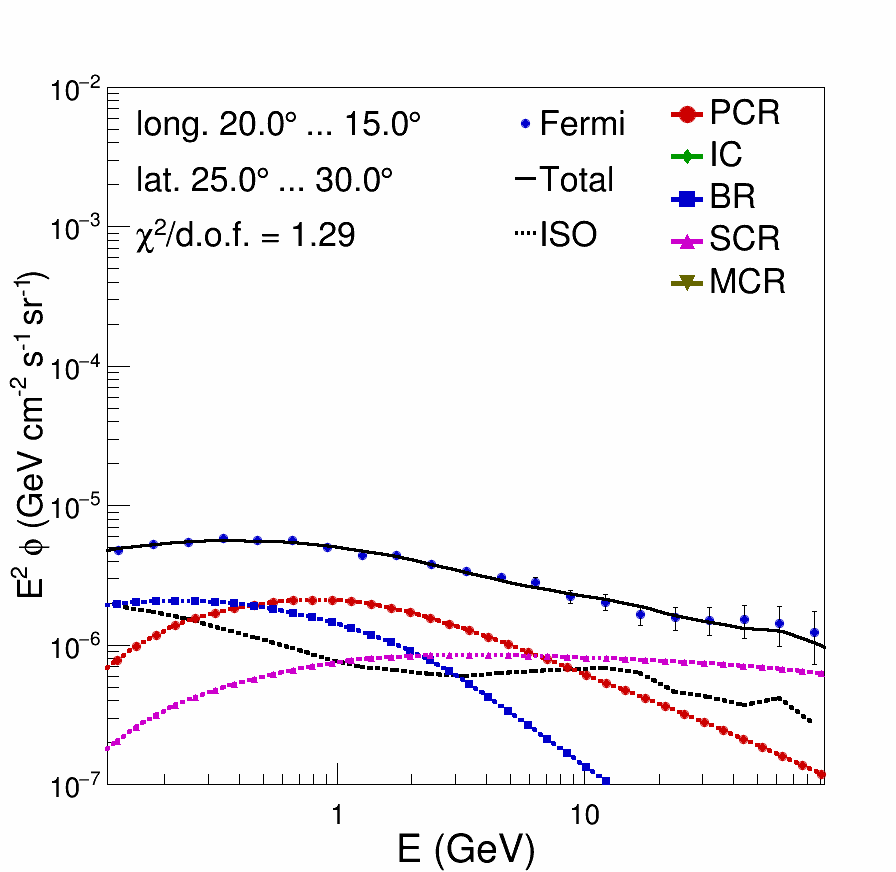}
\includegraphics[width=0.16\textwidth,height=0.16\textwidth,clip]{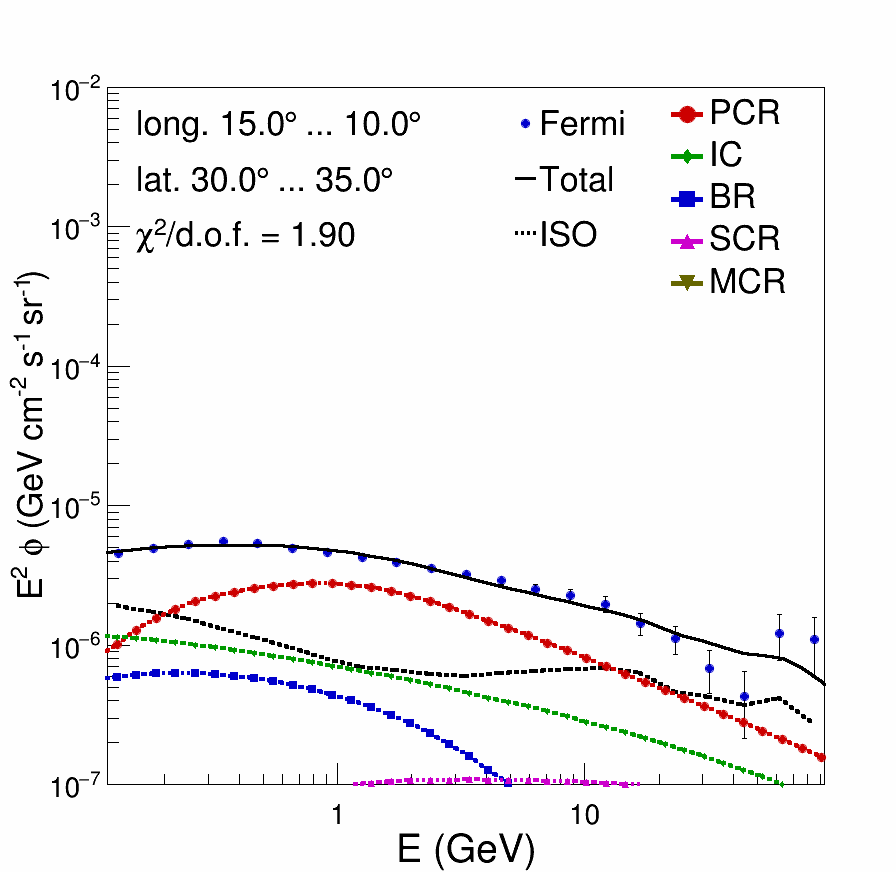}
\includegraphics[width=0.16\textwidth,height=0.16\textwidth,clip]{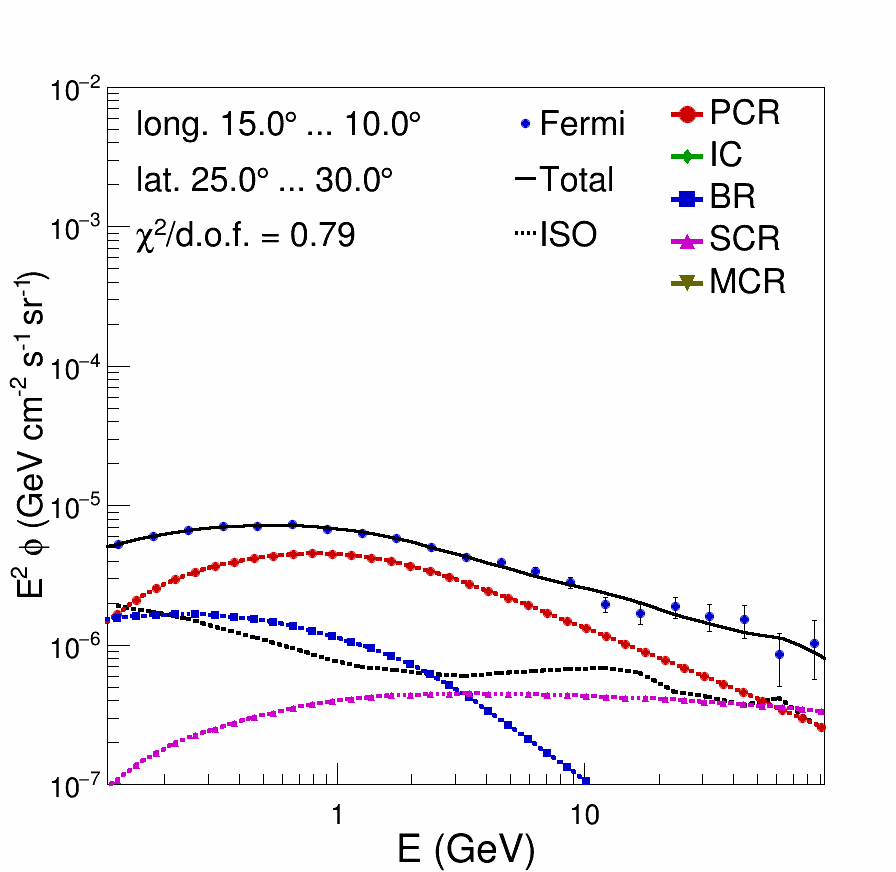}
\includegraphics[width=0.16\textwidth,height=0.16\textwidth,clip]{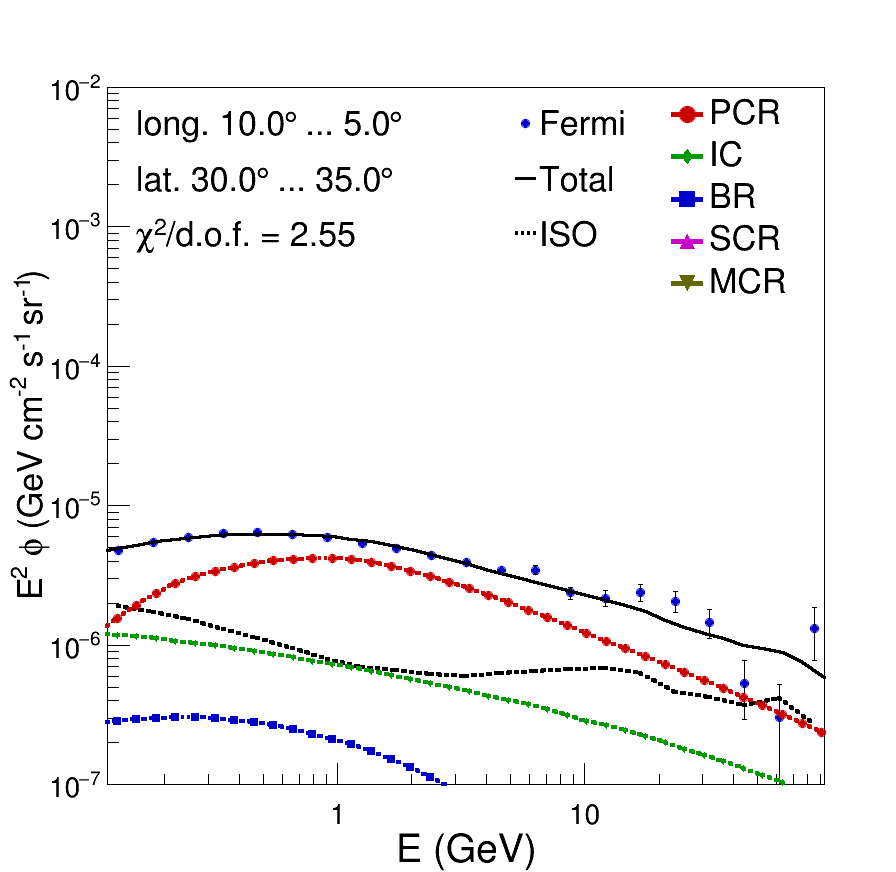}
\includegraphics[width=0.16\textwidth,height=0.16\textwidth,clip]{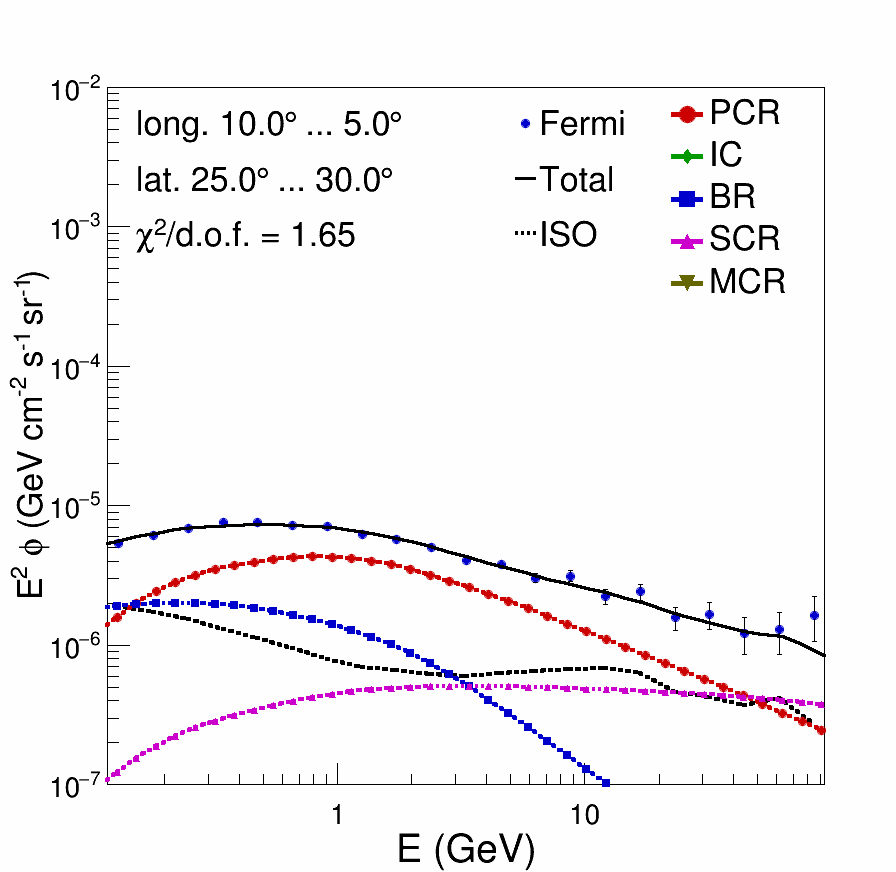}
\includegraphics[width=0.16\textwidth,height=0.16\textwidth,clip]{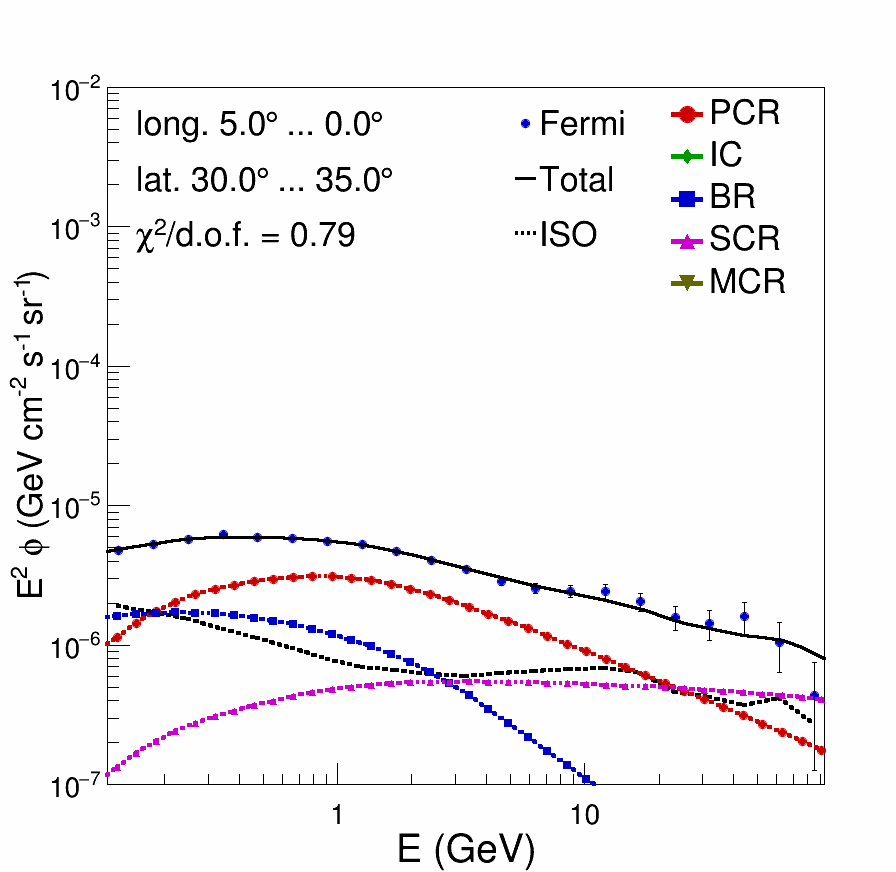}
\includegraphics[width=0.16\textwidth,height=0.16\textwidth,clip]{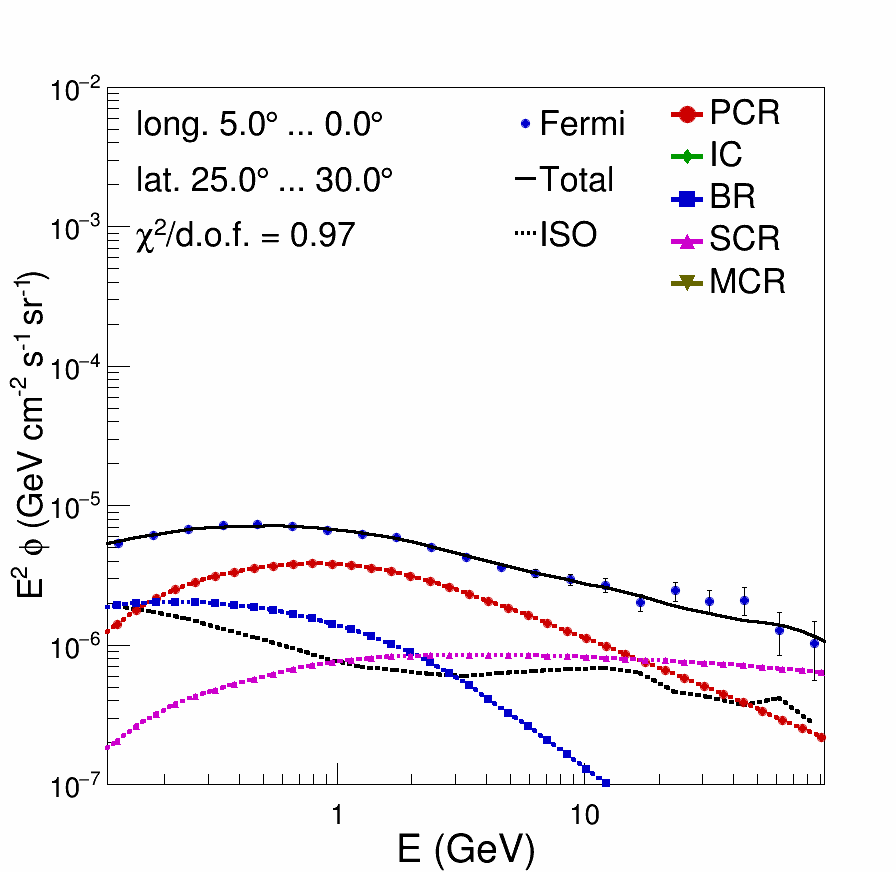}
\includegraphics[width=0.16\textwidth,height=0.16\textwidth,clip]{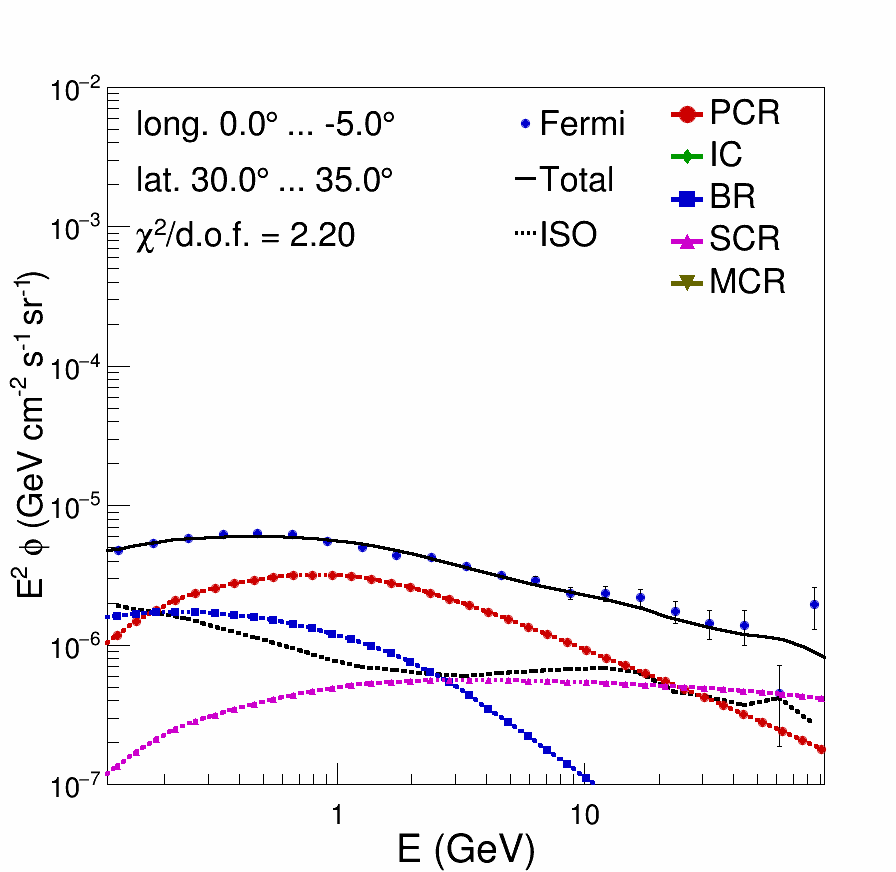}
\includegraphics[width=0.16\textwidth,height=0.16\textwidth,clip]{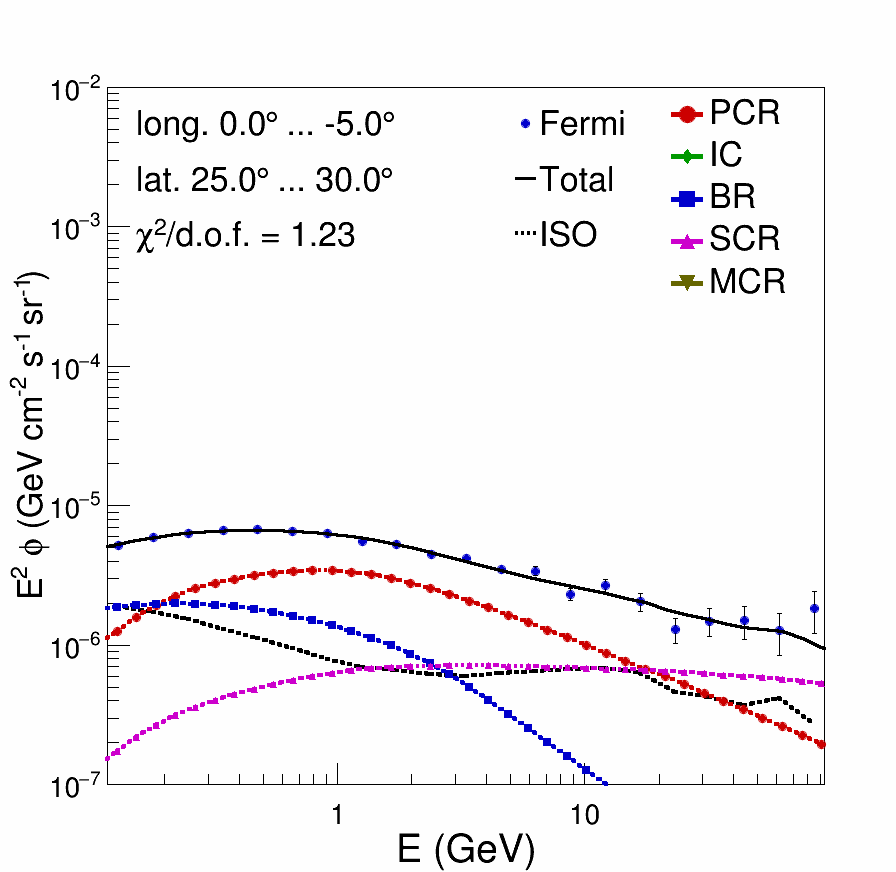}
\includegraphics[width=0.16\textwidth,height=0.16\textwidth,clip]{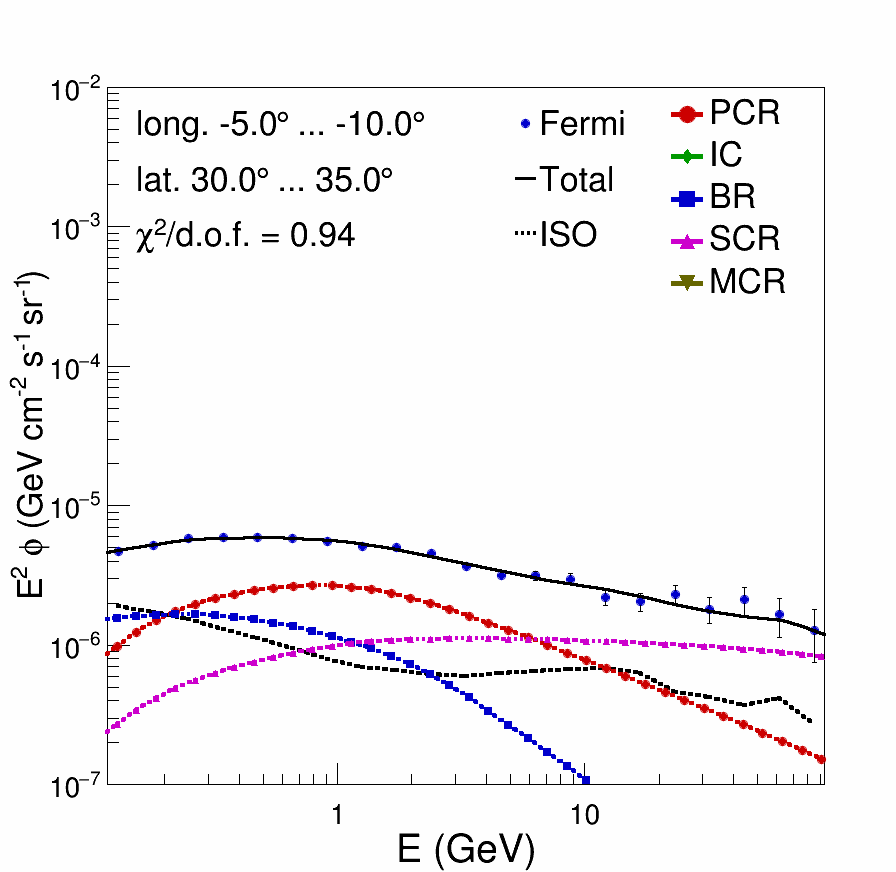}
\includegraphics[width=0.16\textwidth,height=0.16\textwidth,clip]{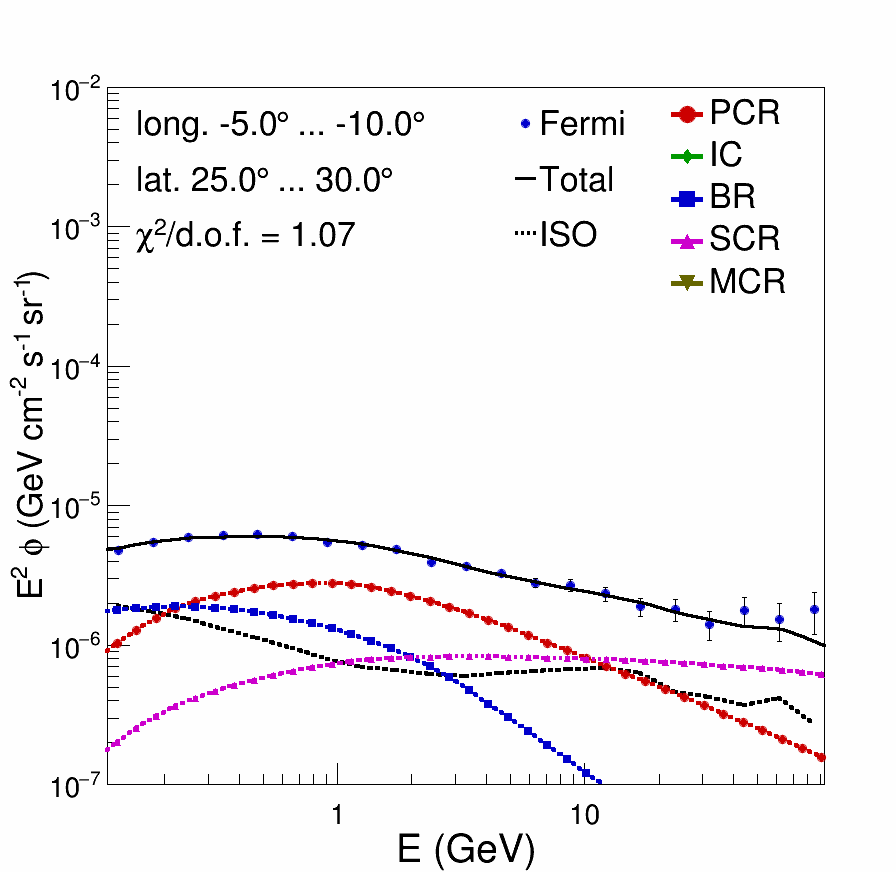}
\includegraphics[width=0.16\textwidth,height=0.16\textwidth,clip]{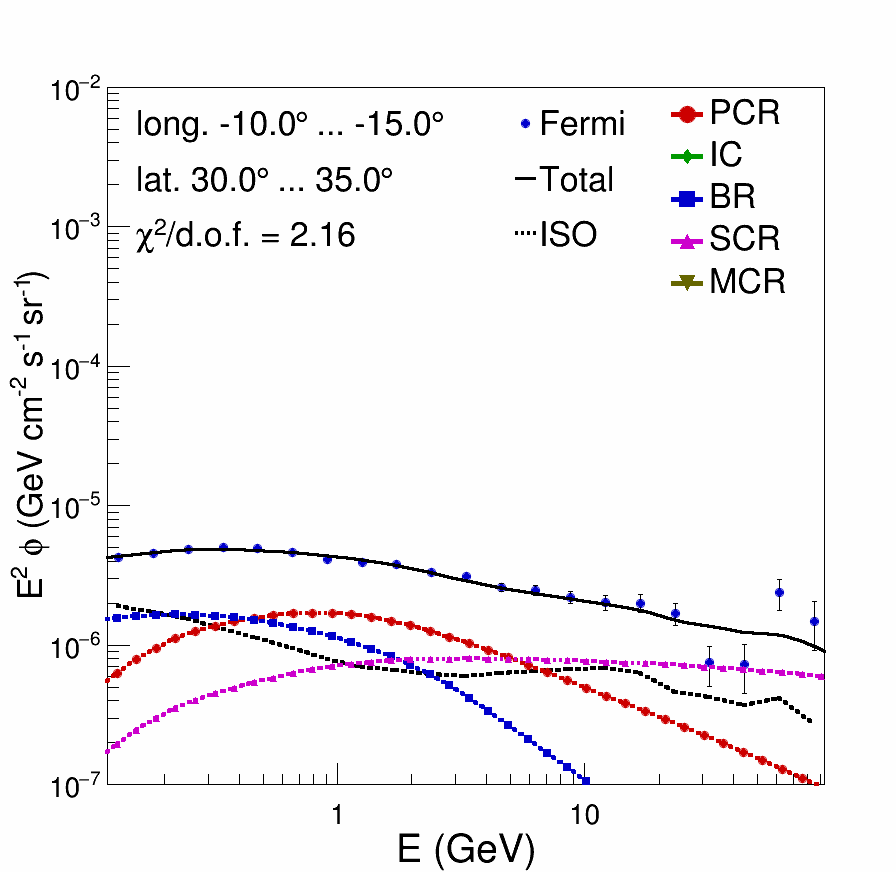}
\includegraphics[width=0.16\textwidth,height=0.16\textwidth,clip]{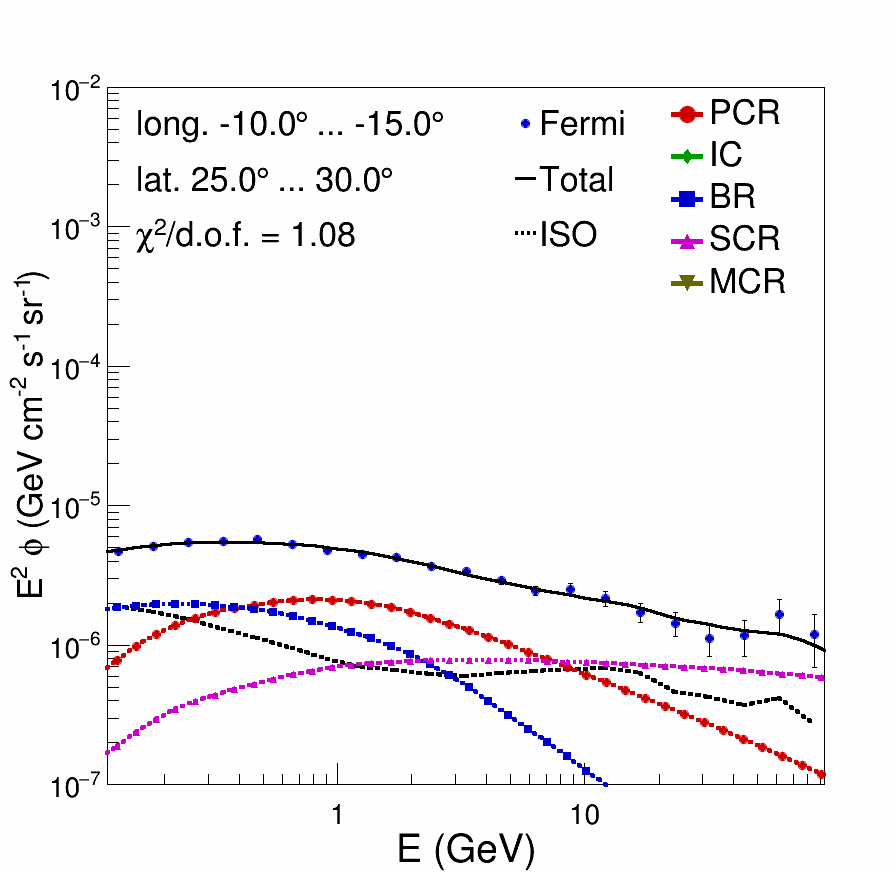}
\includegraphics[width=0.16\textwidth,height=0.16\textwidth,clip]{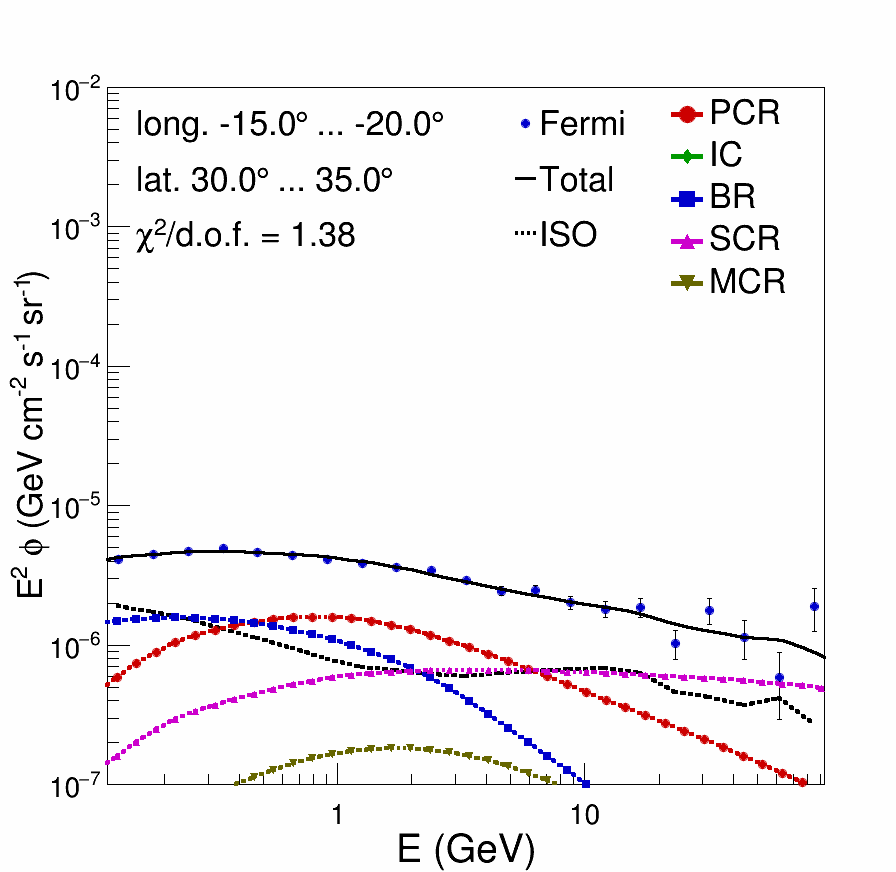}
\includegraphics[width=0.16\textwidth,height=0.16\textwidth,clip]{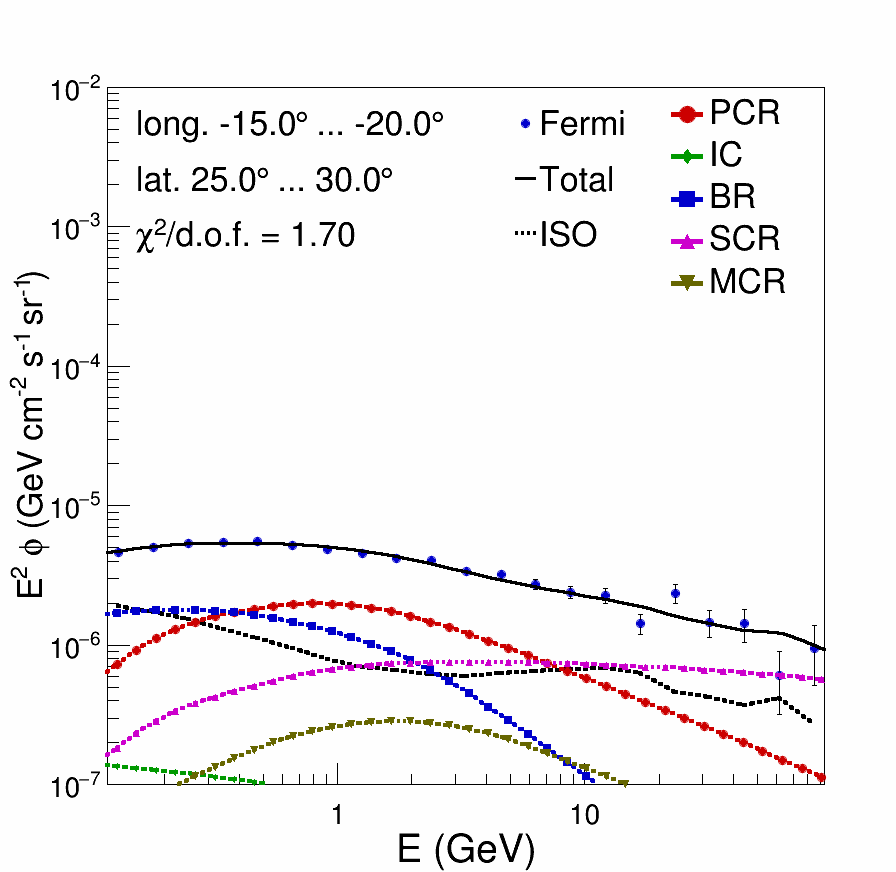}
\includegraphics[width=0.16\textwidth,height=0.16\textwidth,clip]{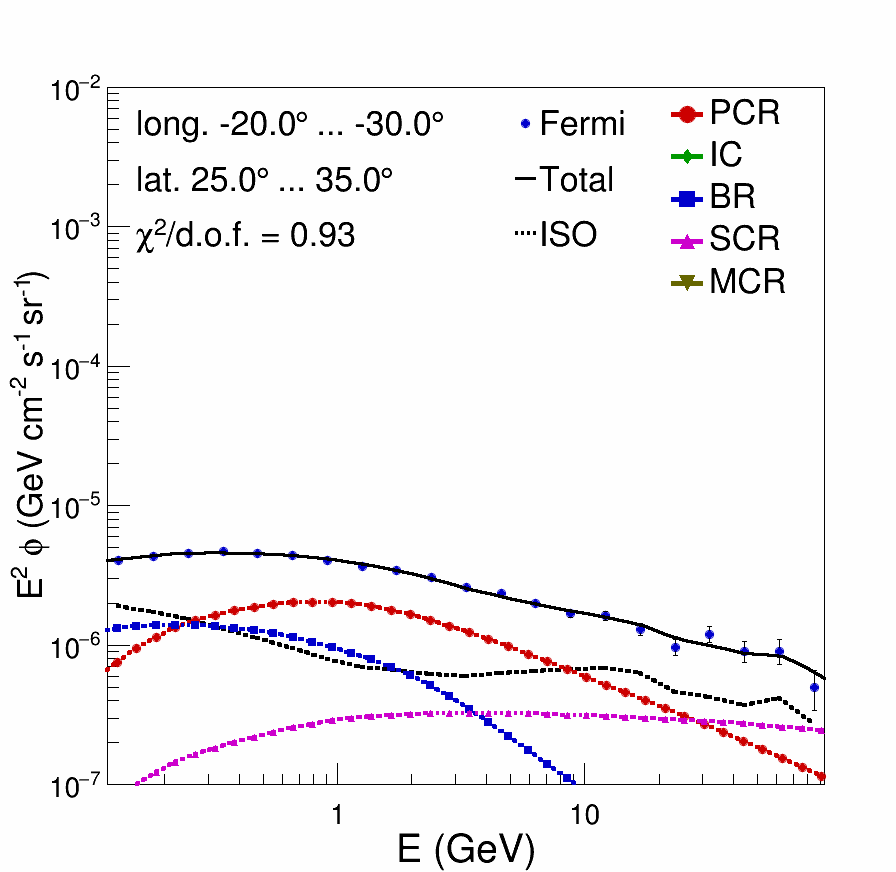}
\includegraphics[width=0.16\textwidth,height=0.16\textwidth,clip]{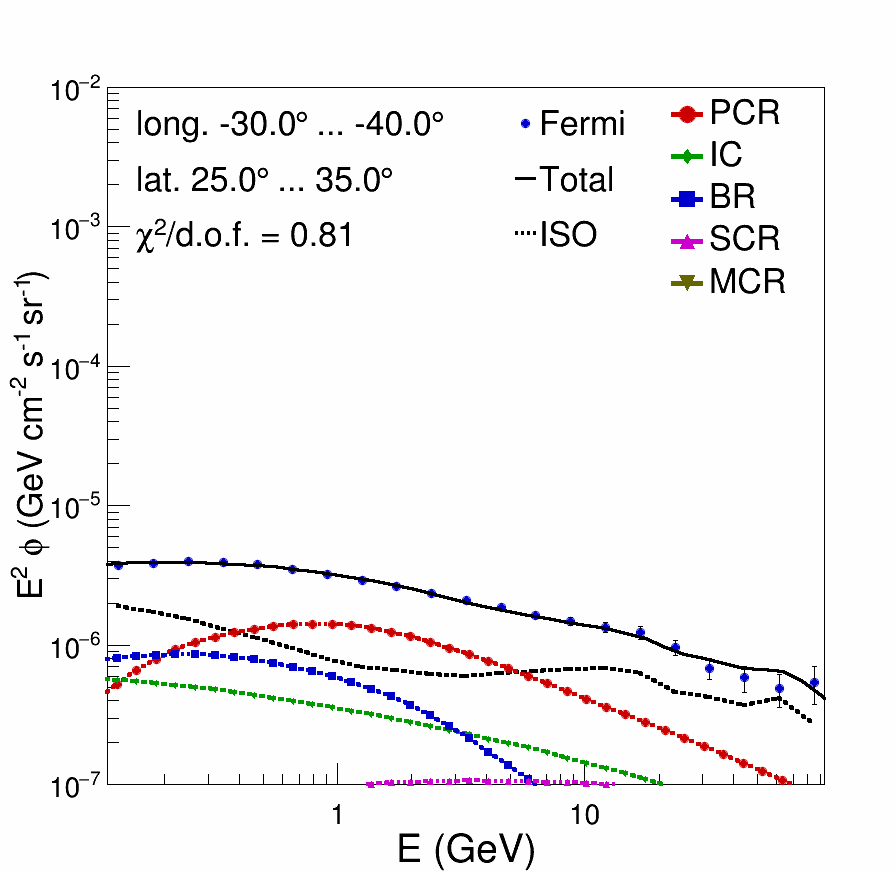}
\includegraphics[width=0.16\textwidth,height=0.16\textwidth,clip]{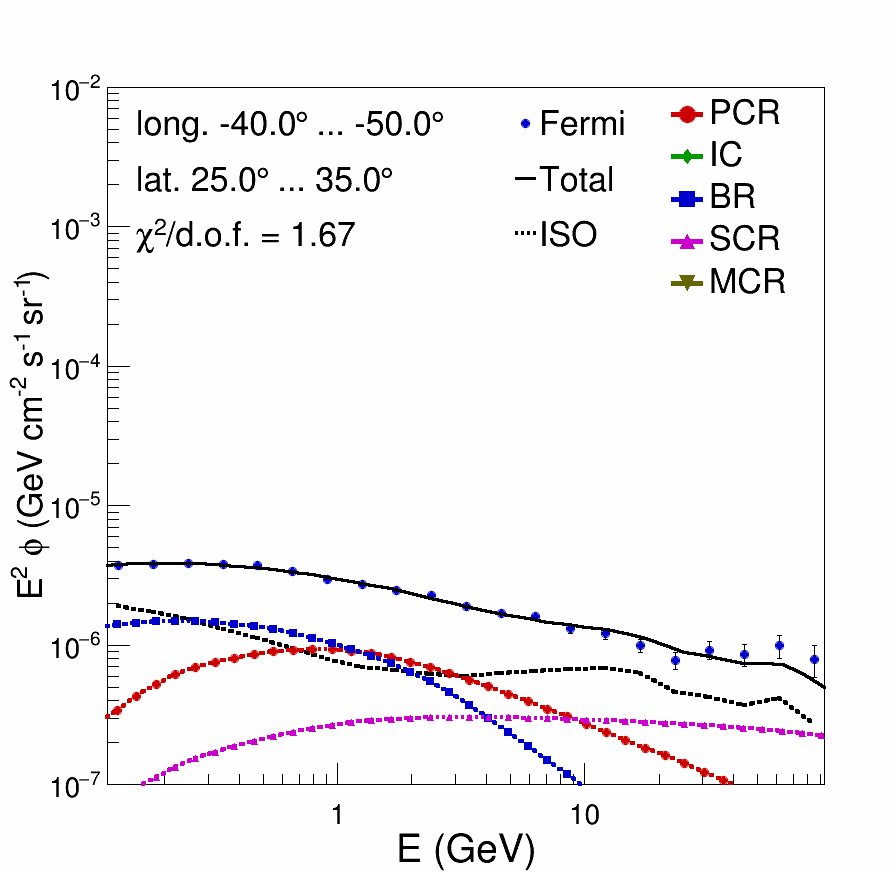}
\includegraphics[width=0.16\textwidth,height=0.16\textwidth,clip]{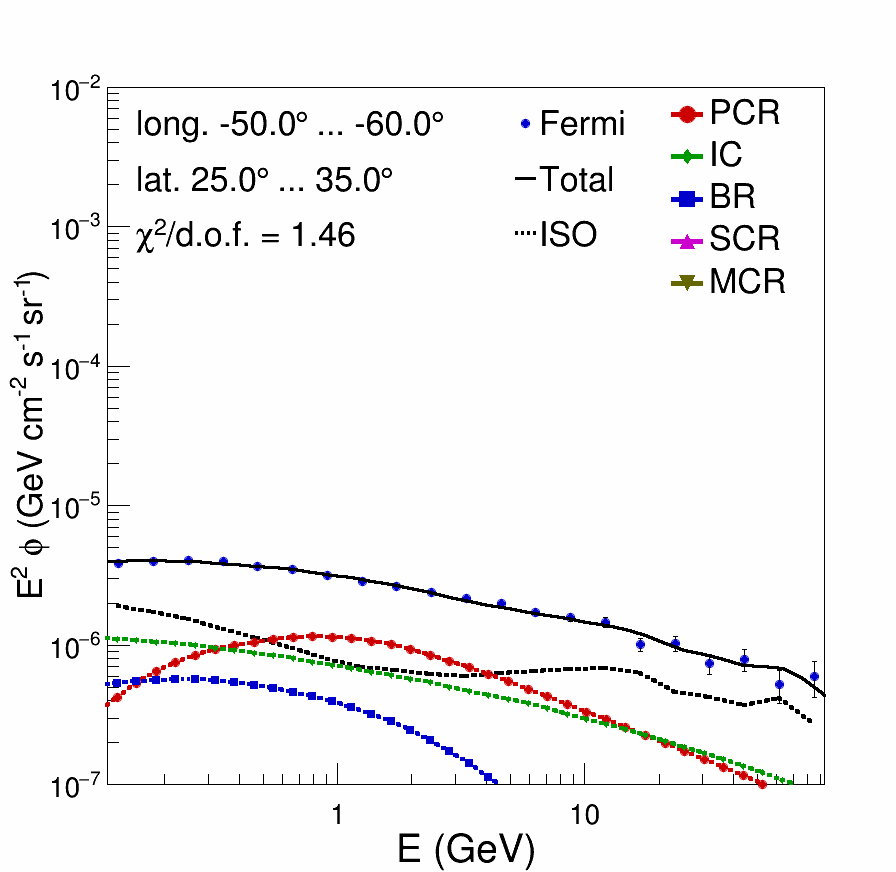}
\includegraphics[width=0.16\textwidth,height=0.16\textwidth,clip]{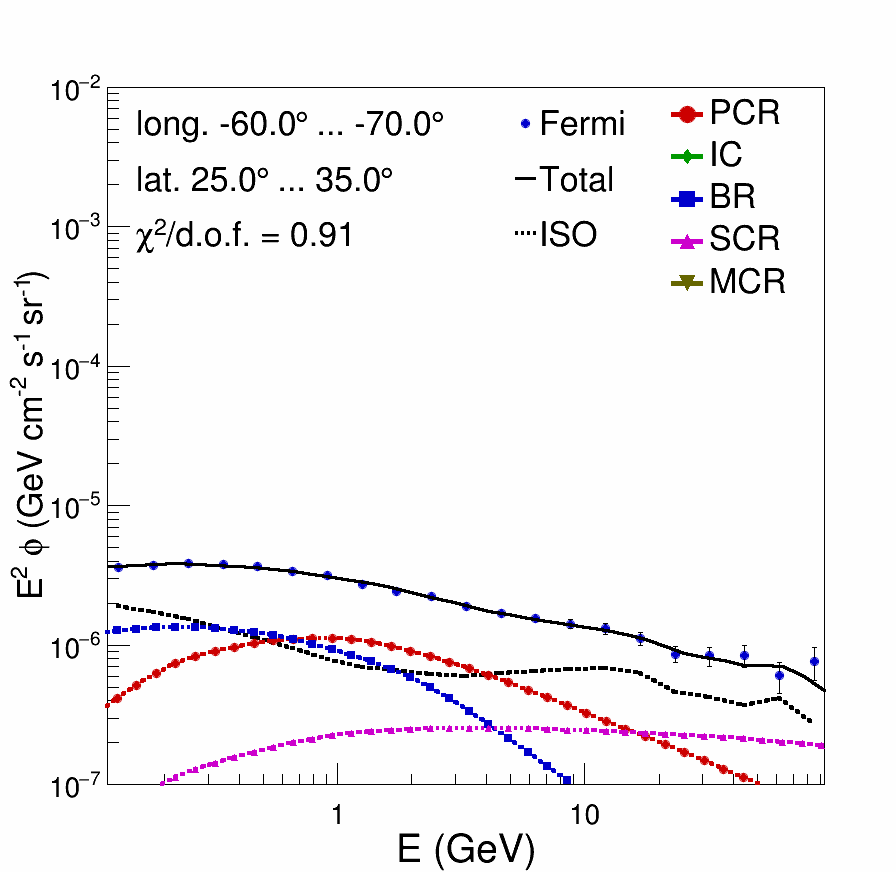}
\includegraphics[width=0.16\textwidth,height=0.16\textwidth,clip]{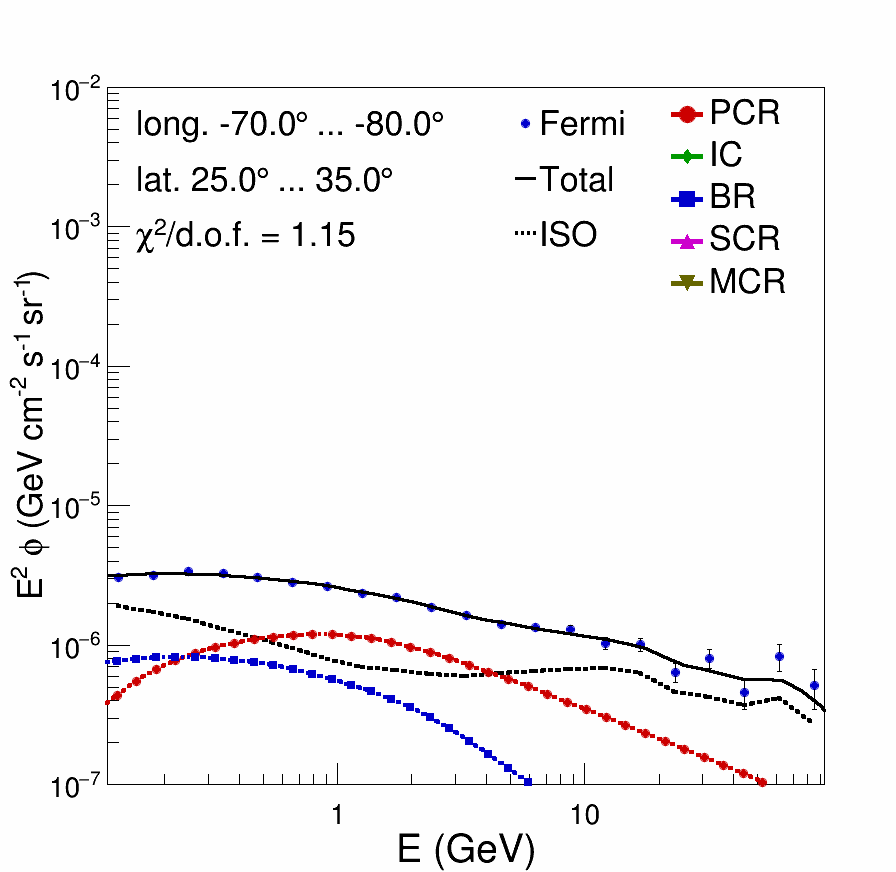}
\includegraphics[width=0.16\textwidth,height=0.16\textwidth,clip]{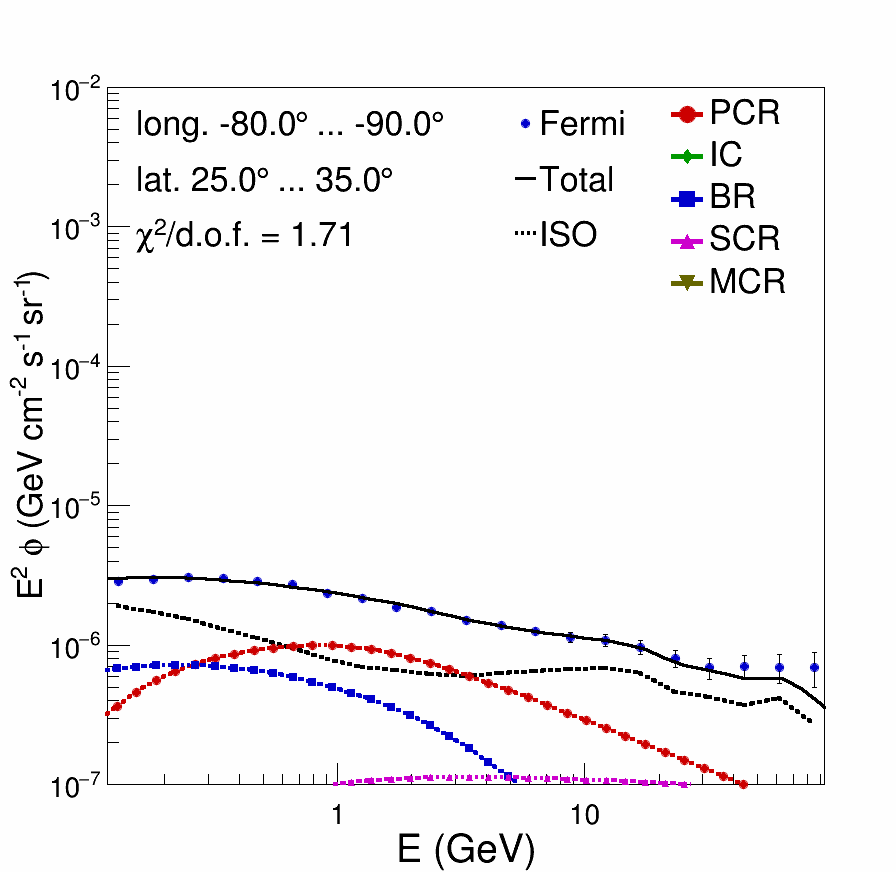}
\includegraphics[width=0.16\textwidth,height=0.16\textwidth,clip]{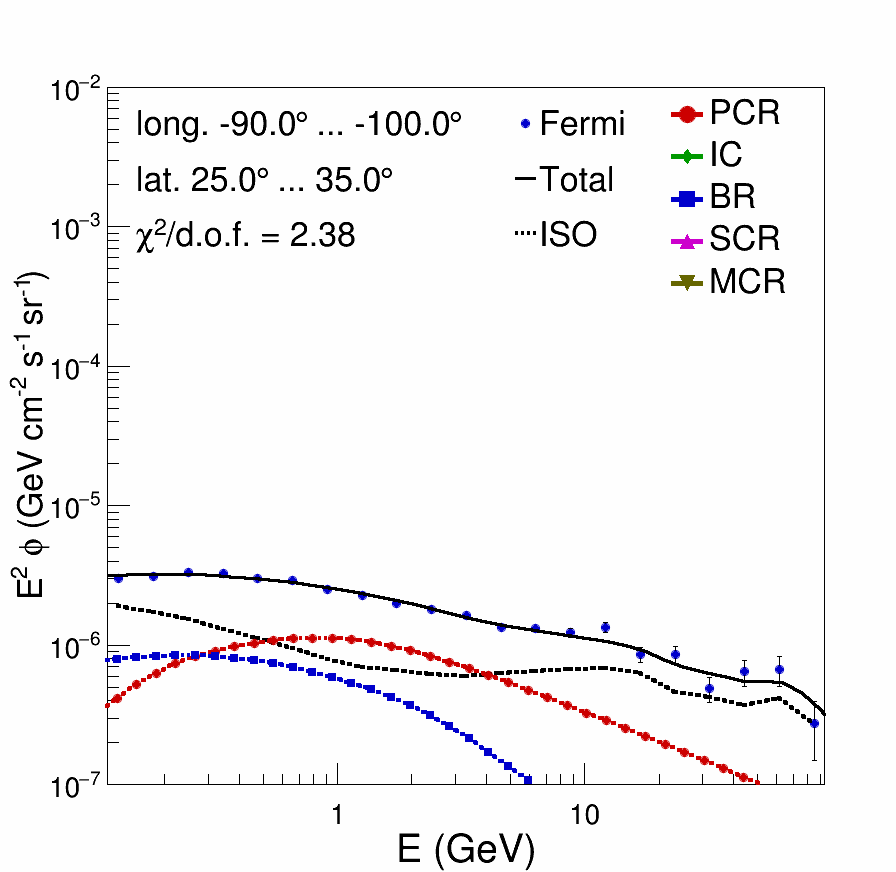}
\includegraphics[width=0.16\textwidth,height=0.16\textwidth,clip]{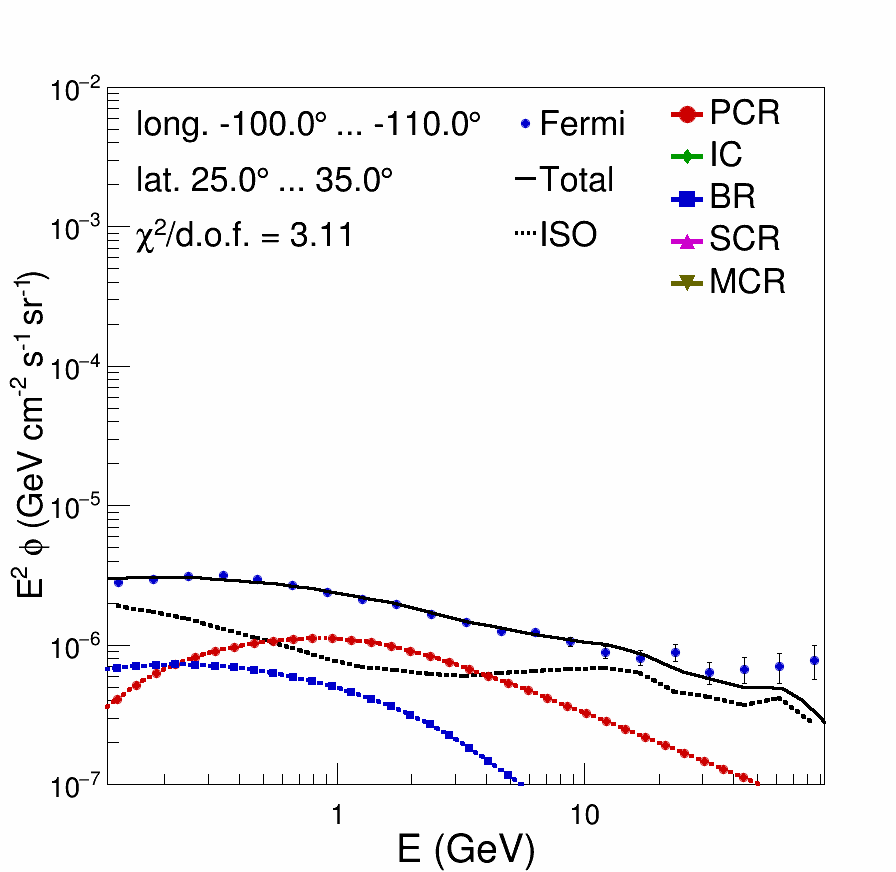}
\includegraphics[width=0.16\textwidth,height=0.16\textwidth,clip]{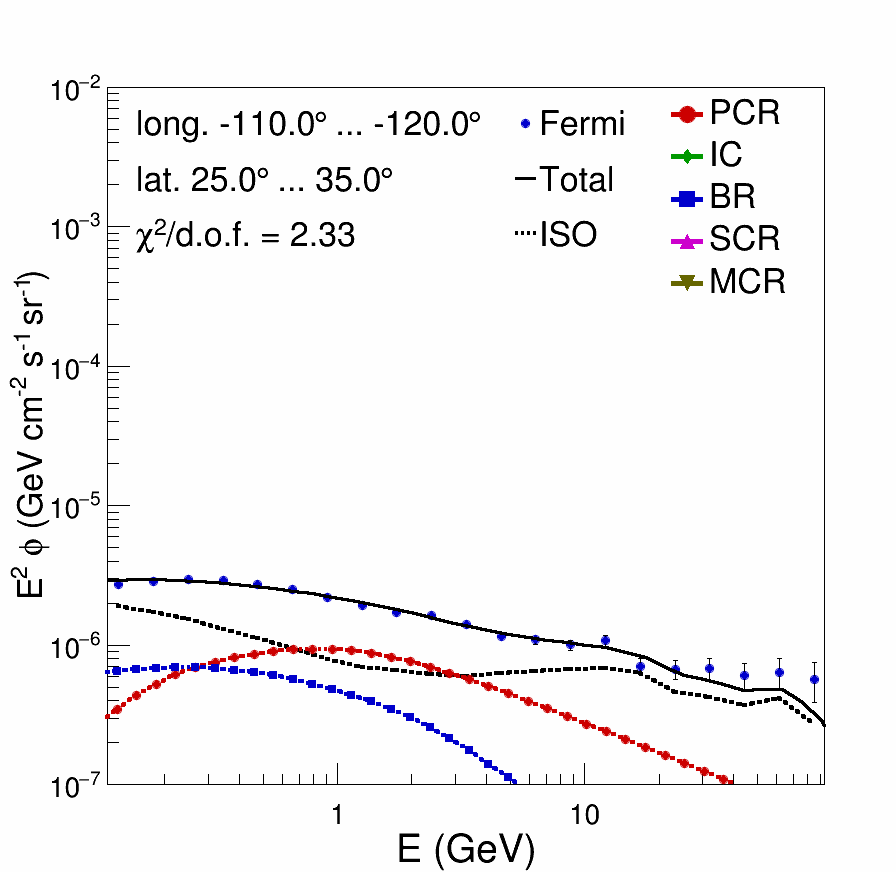}
\includegraphics[width=0.16\textwidth,height=0.16\textwidth,clip]{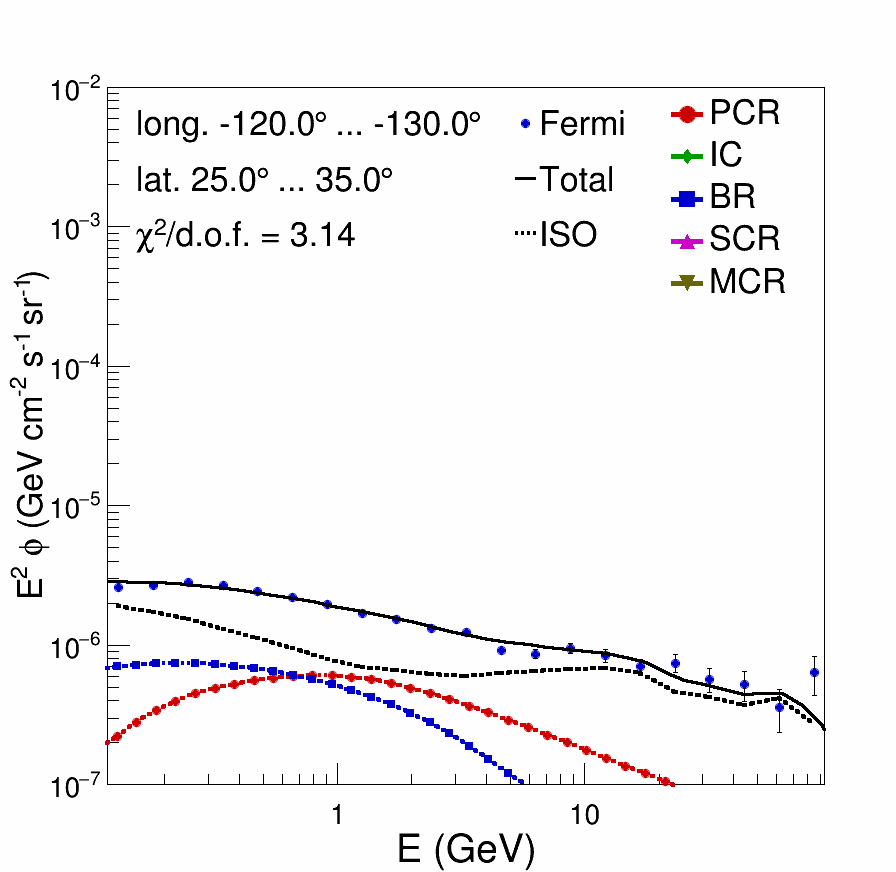}
\includegraphics[width=0.16\textwidth,height=0.16\textwidth,clip]{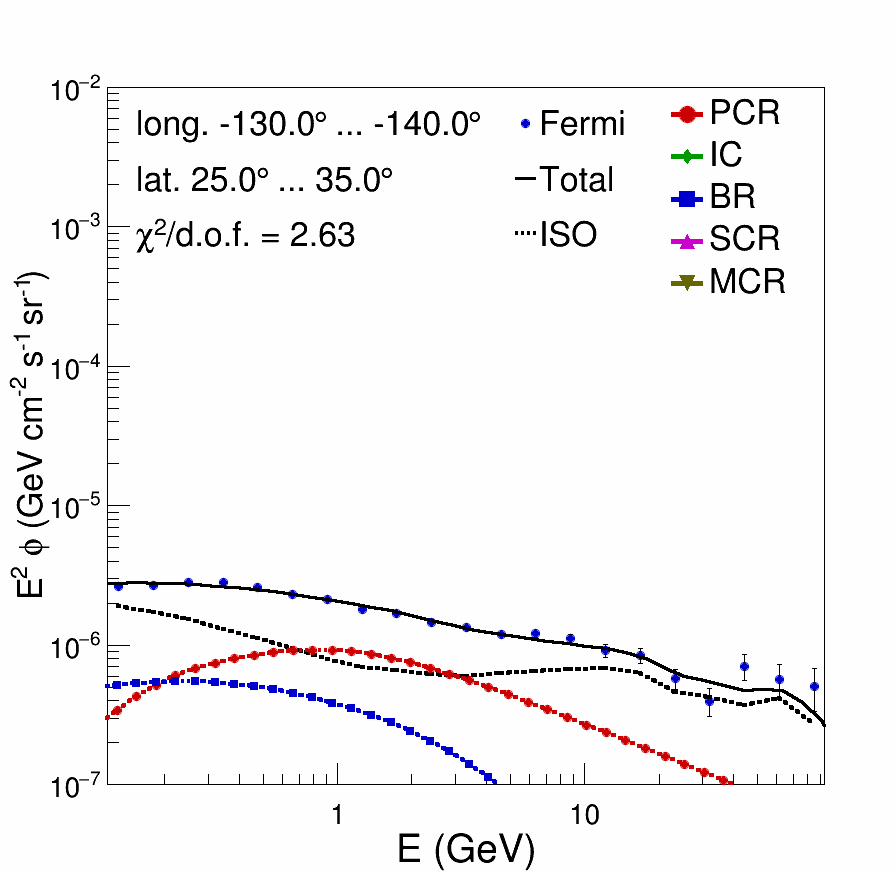}
\includegraphics[width=0.16\textwidth,height=0.16\textwidth,clip]{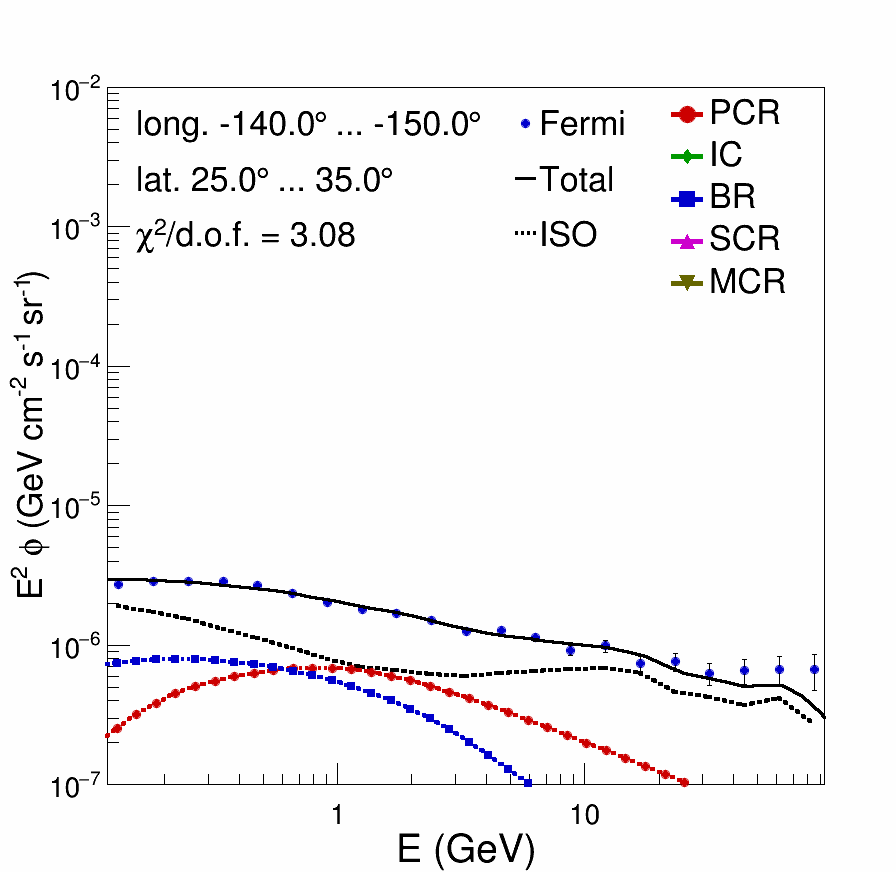}
\includegraphics[width=0.16\textwidth,height=0.16\textwidth,clip]{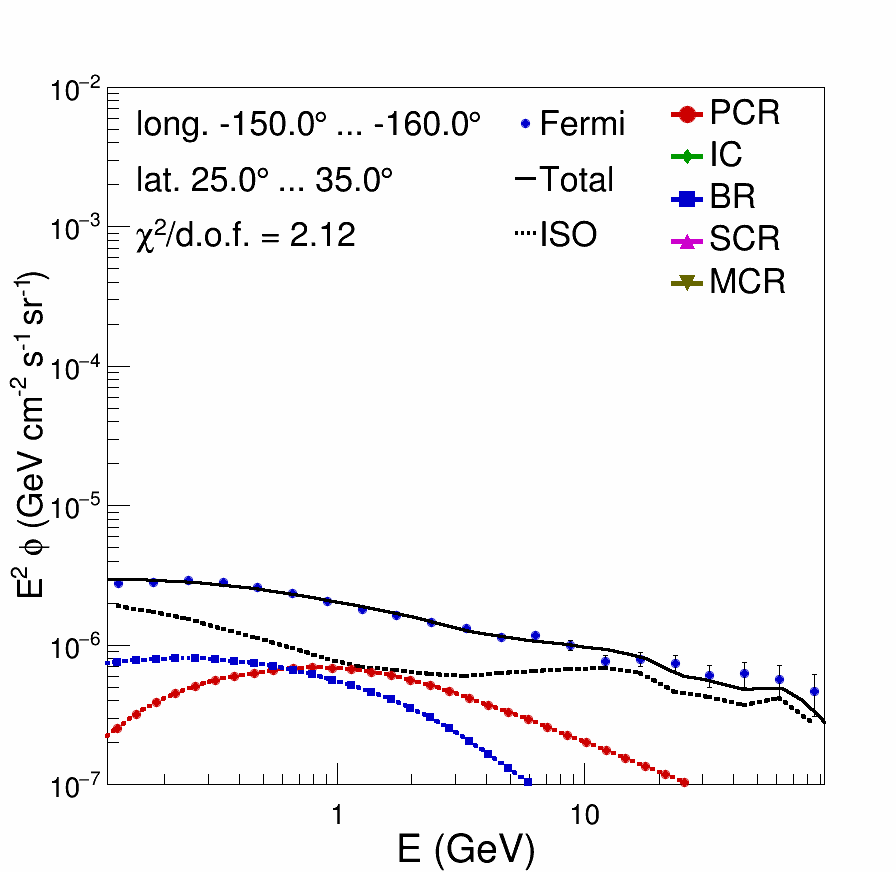}
\includegraphics[width=0.16\textwidth,height=0.16\textwidth,clip]{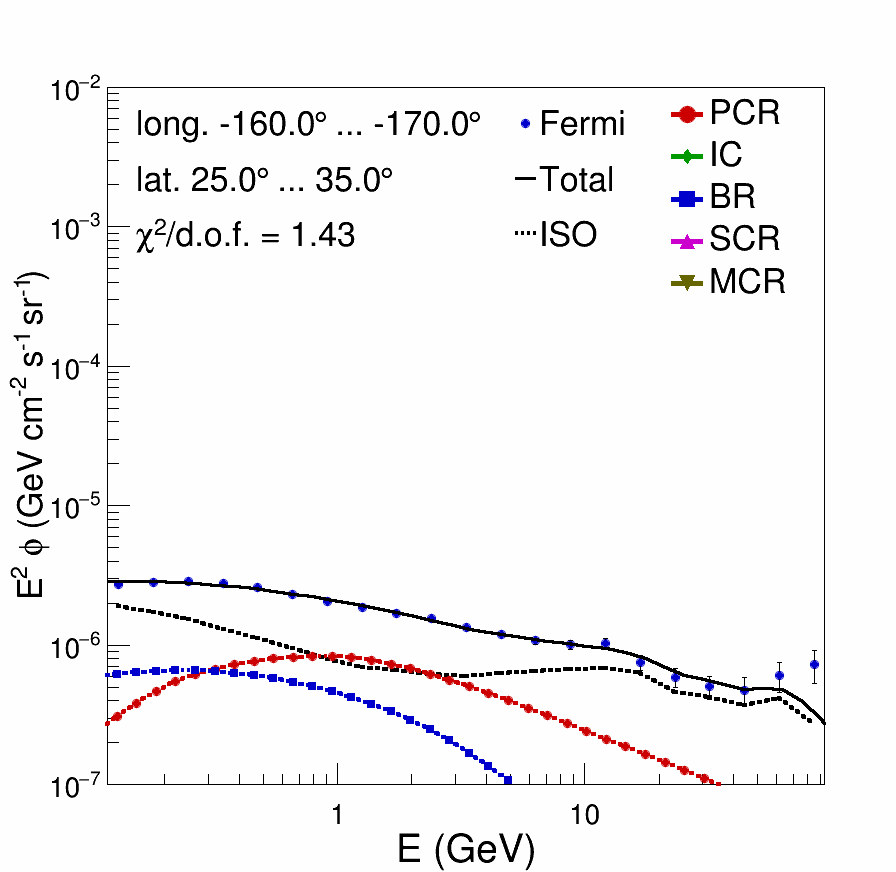}
\includegraphics[width=0.16\textwidth,height=0.16\textwidth,clip]{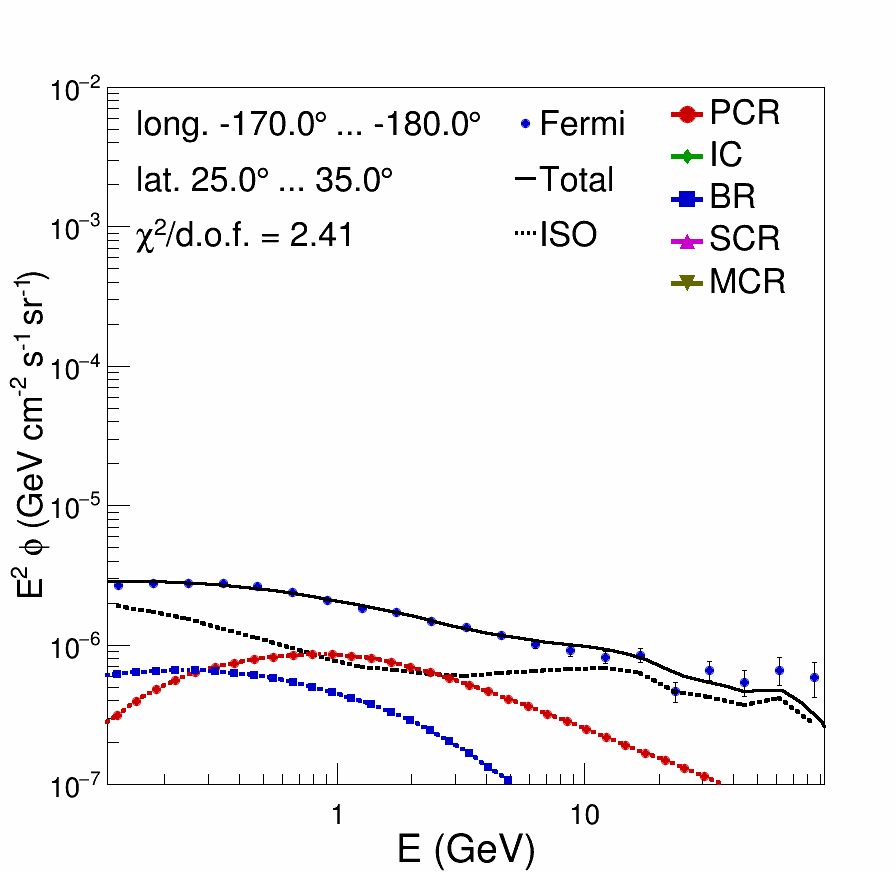}
\caption[]{Template fits for latitudes  with $25.0^\circ<b<35.0^\circ$ and longitudes decreasing from 180$^\circ$ to -180$^\circ$. \label{F15}
}
\end{figure}
\begin{figure}
\centering
\includegraphics[width=0.16\textwidth,height=0.16\textwidth,clip]{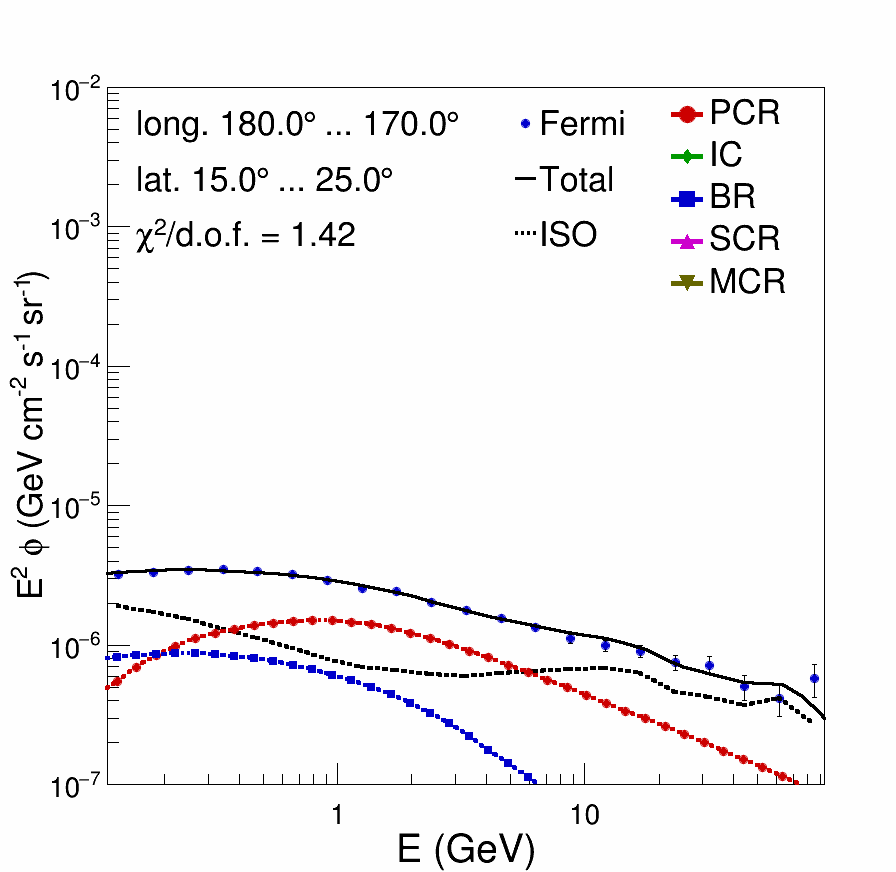}
\includegraphics[width=0.16\textwidth,height=0.16\textwidth,clip]{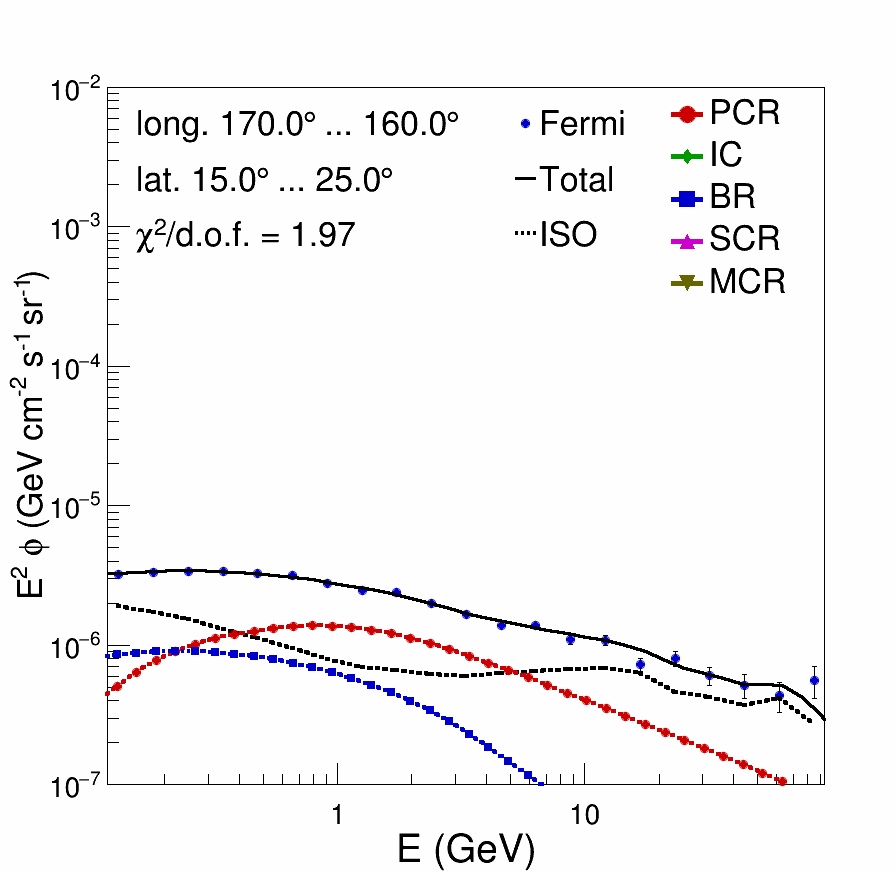}
\includegraphics[width=0.16\textwidth,height=0.16\textwidth,clip]{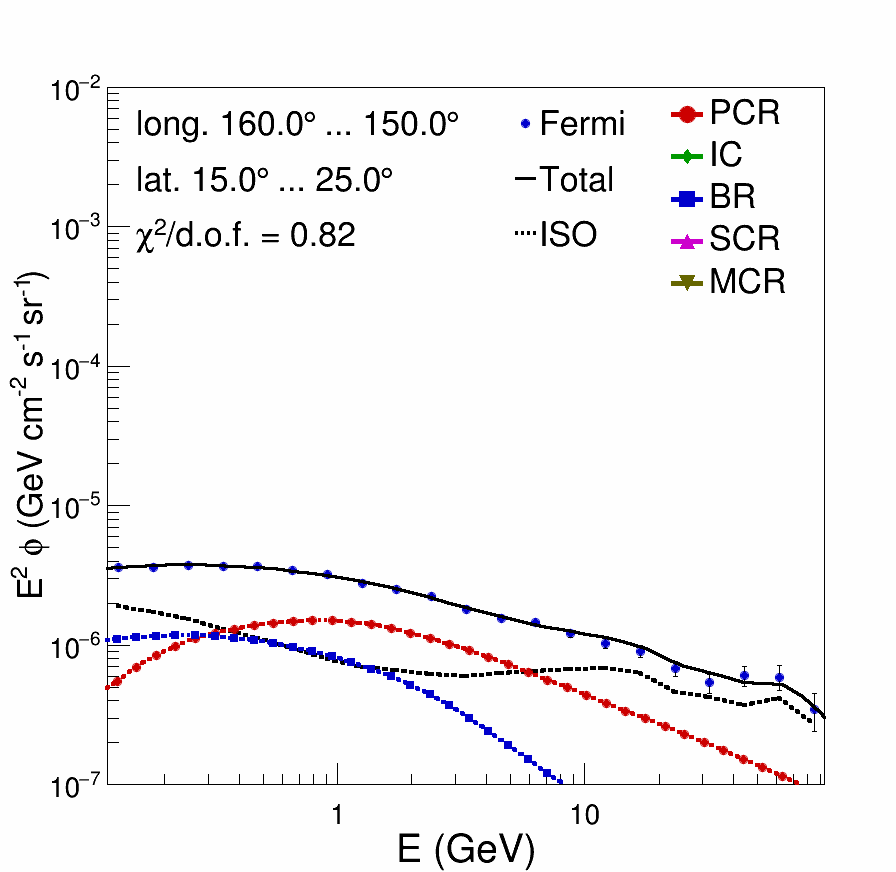}
\includegraphics[width=0.16\textwidth,height=0.16\textwidth,clip]{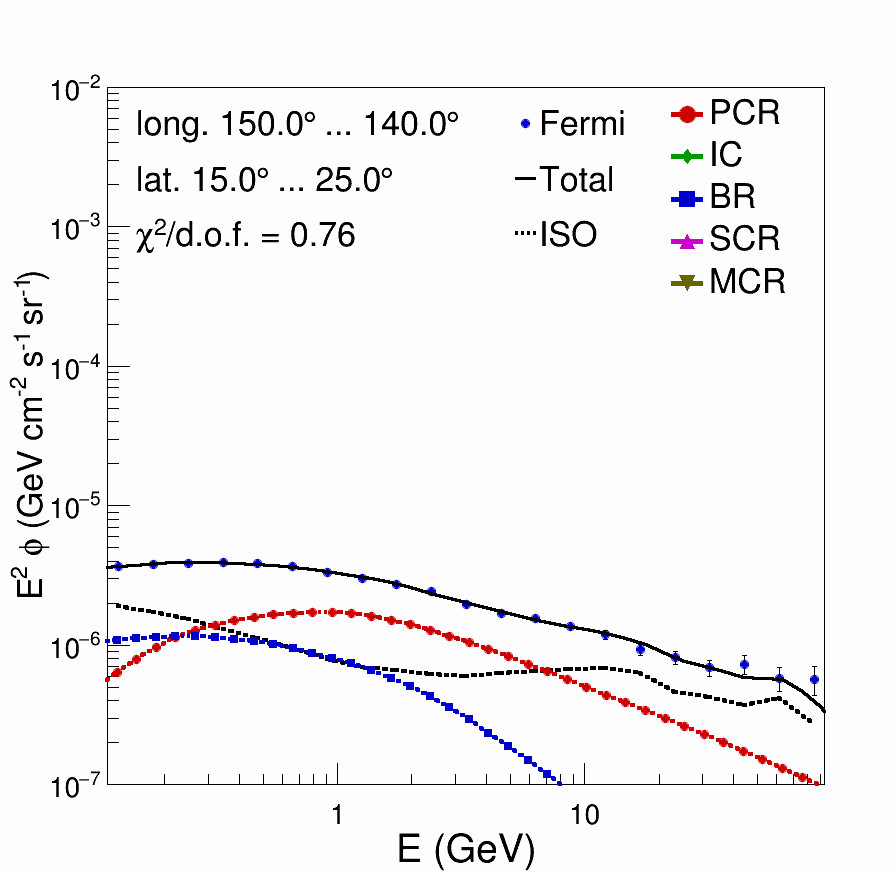}
\includegraphics[width=0.16\textwidth,height=0.16\textwidth,clip]{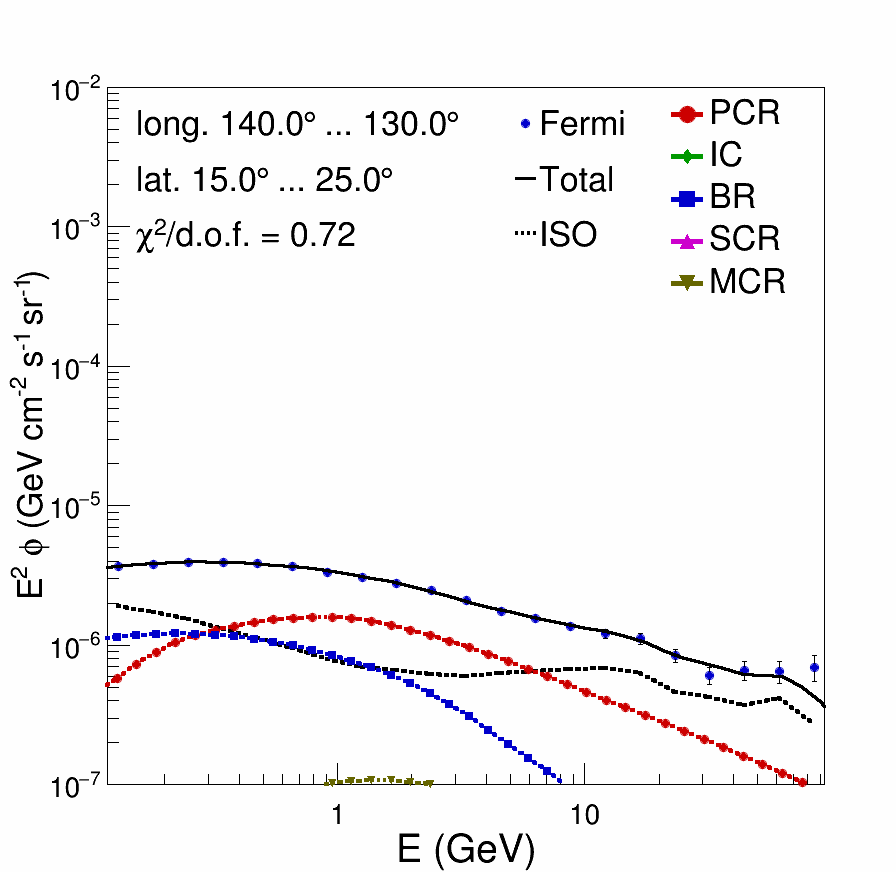}
\includegraphics[width=0.16\textwidth,height=0.16\textwidth,clip]{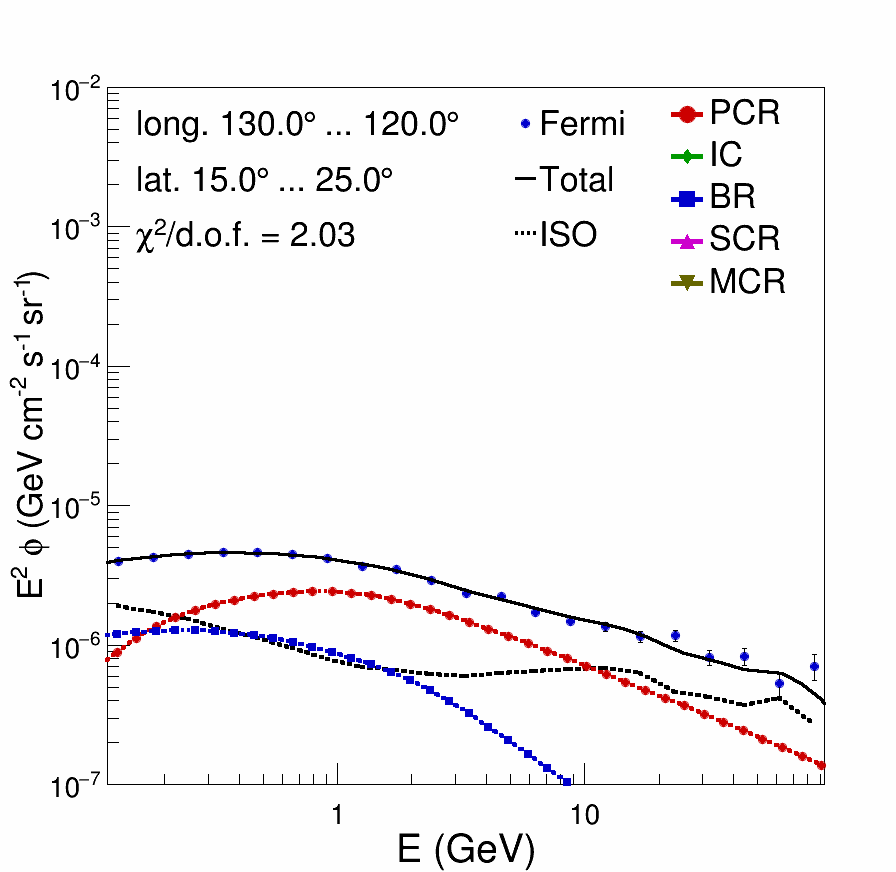}
\includegraphics[width=0.16\textwidth,height=0.16\textwidth,clip]{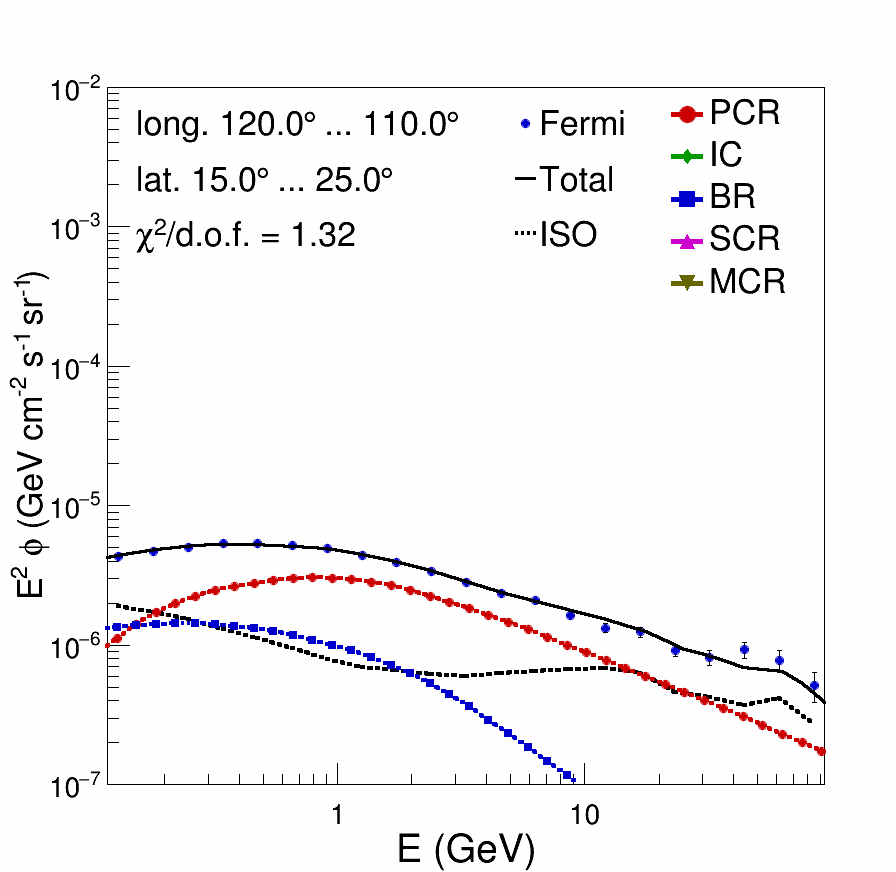}
\includegraphics[width=0.16\textwidth,height=0.16\textwidth,clip]{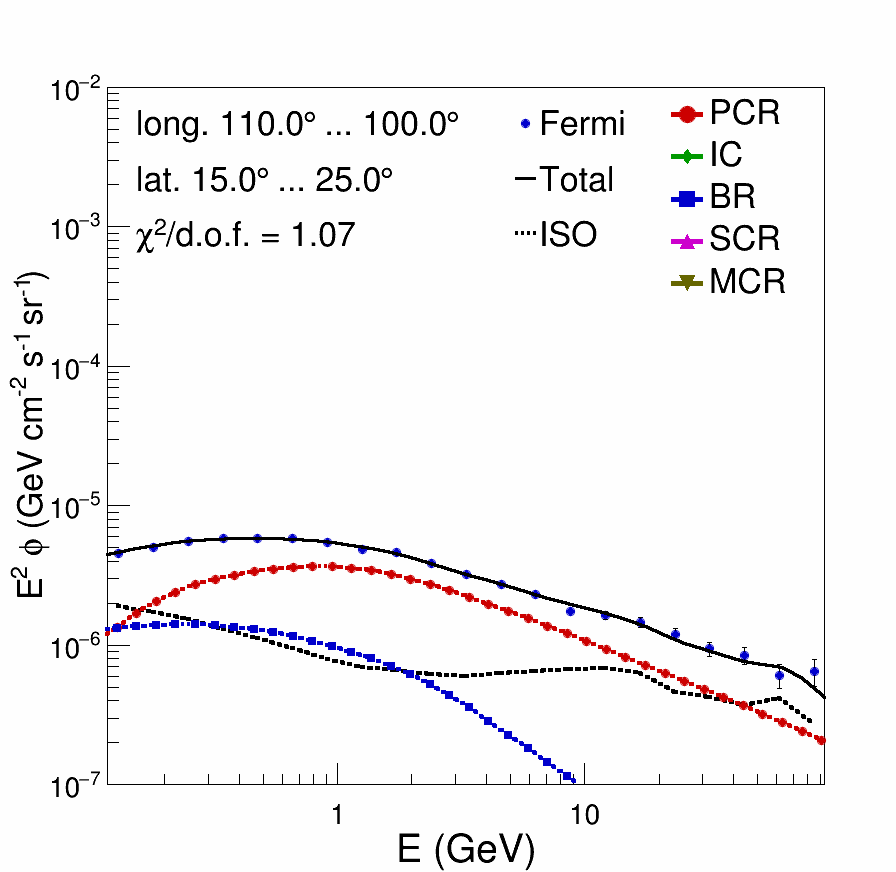}
\includegraphics[width=0.16\textwidth,height=0.16\textwidth,clip]{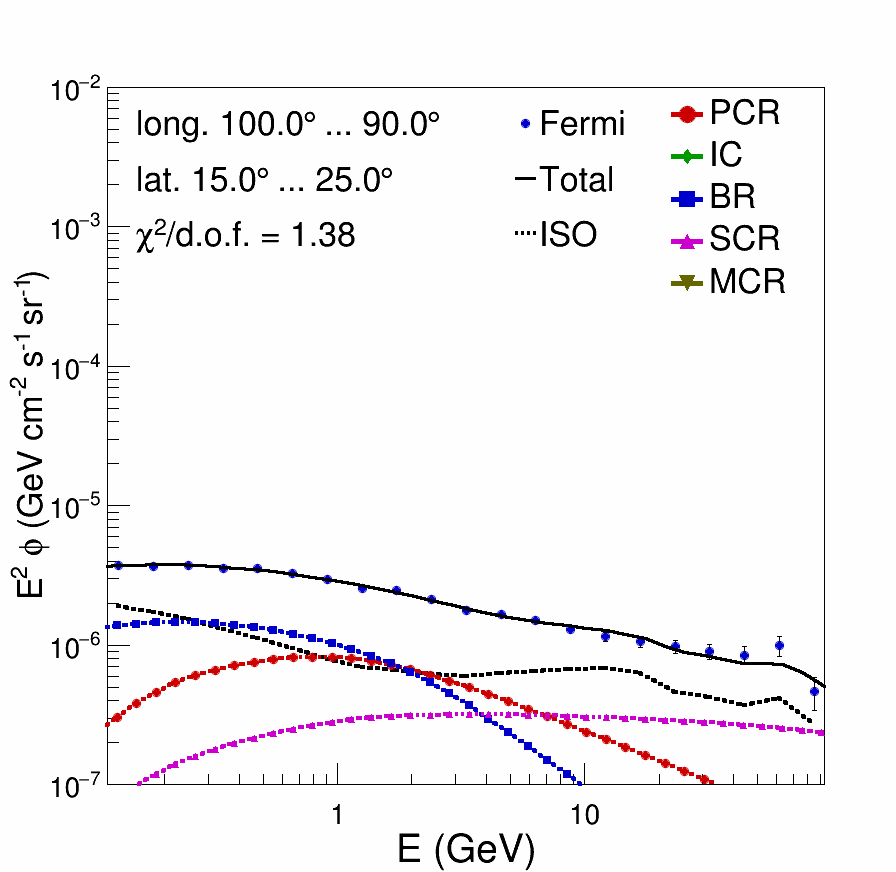}
\includegraphics[width=0.16\textwidth,height=0.16\textwidth,clip]{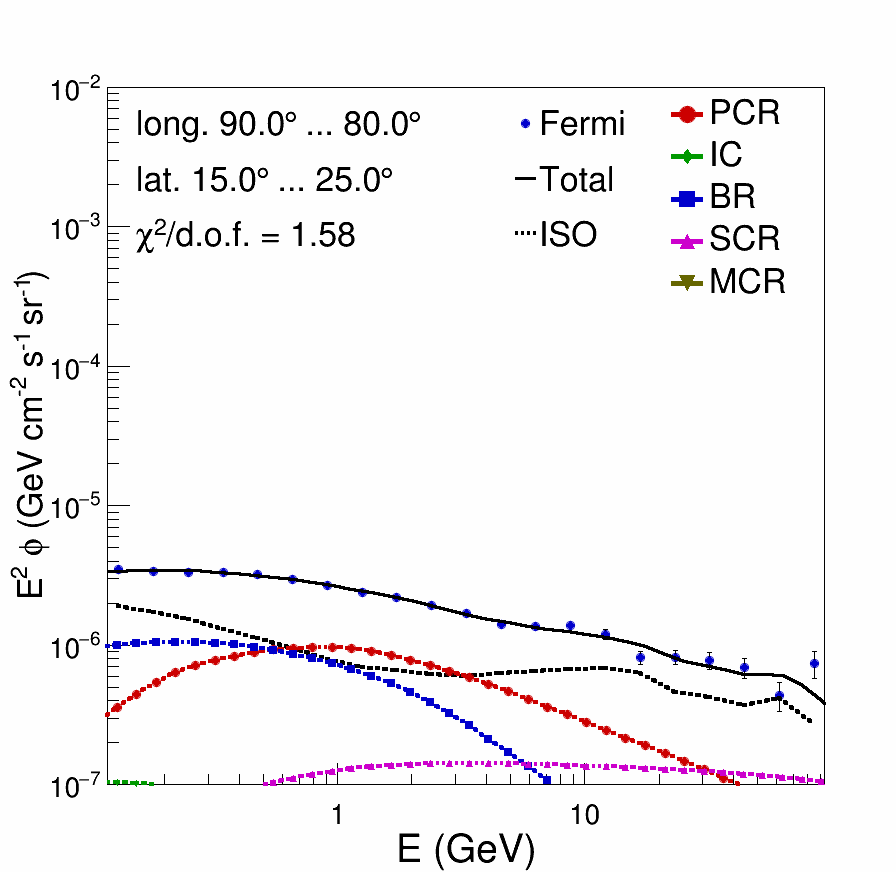}
\includegraphics[width=0.16\textwidth,height=0.16\textwidth,clip]{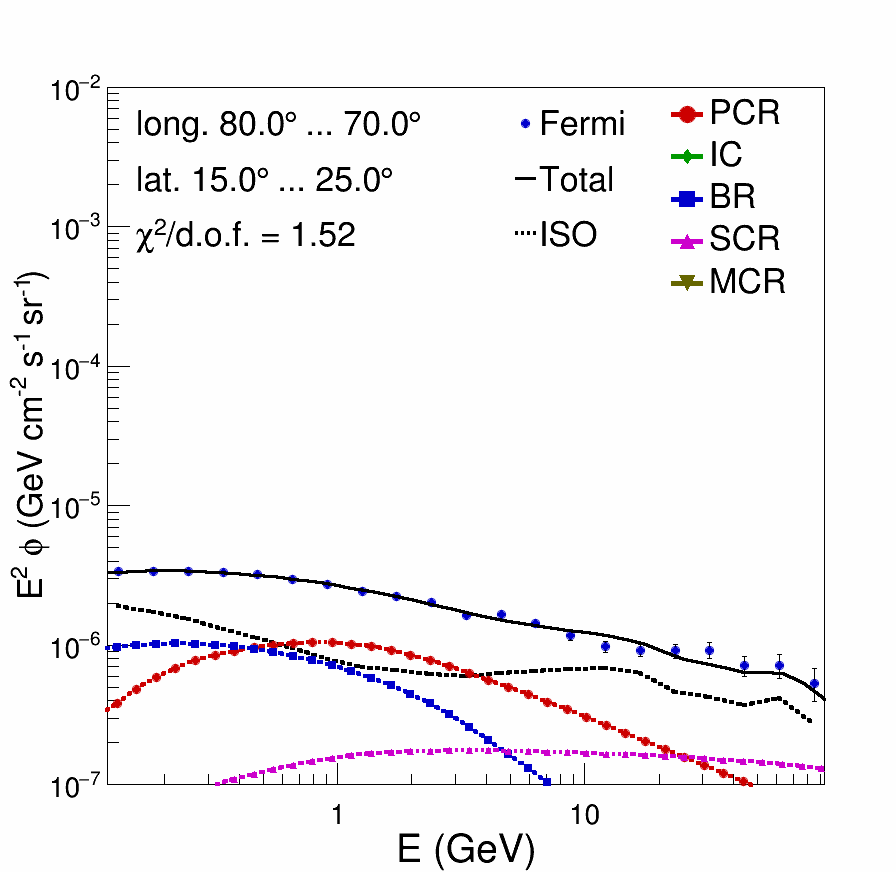}
\includegraphics[width=0.16\textwidth,height=0.16\textwidth,clip]{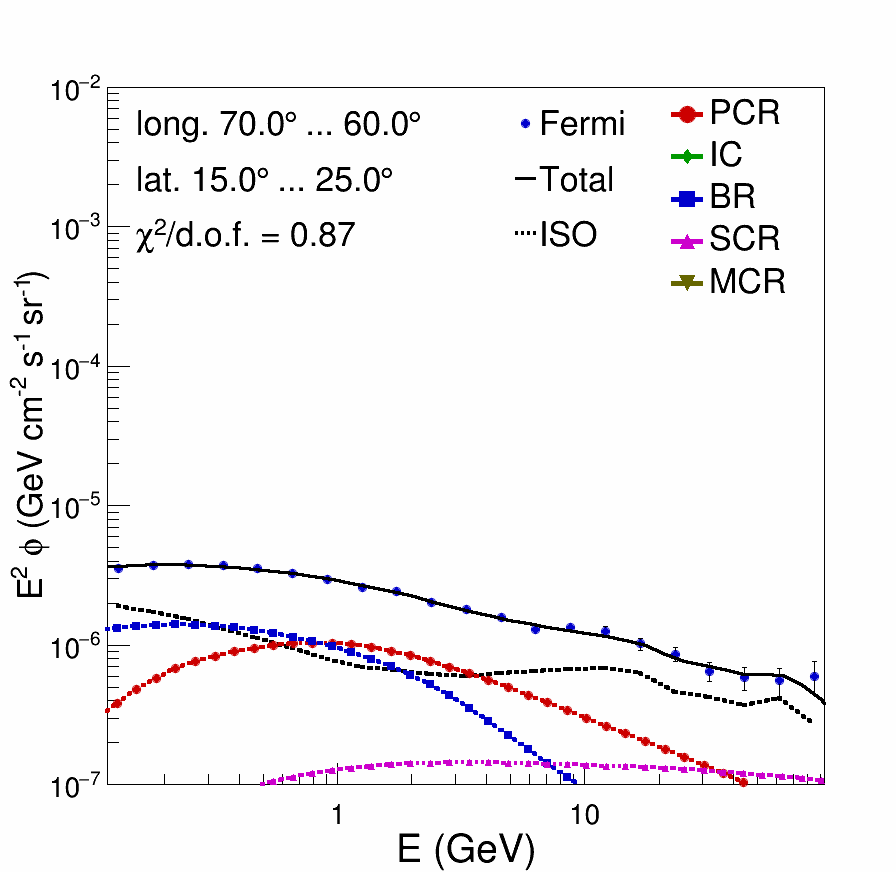}
\includegraphics[width=0.16\textwidth,height=0.16\textwidth,clip]{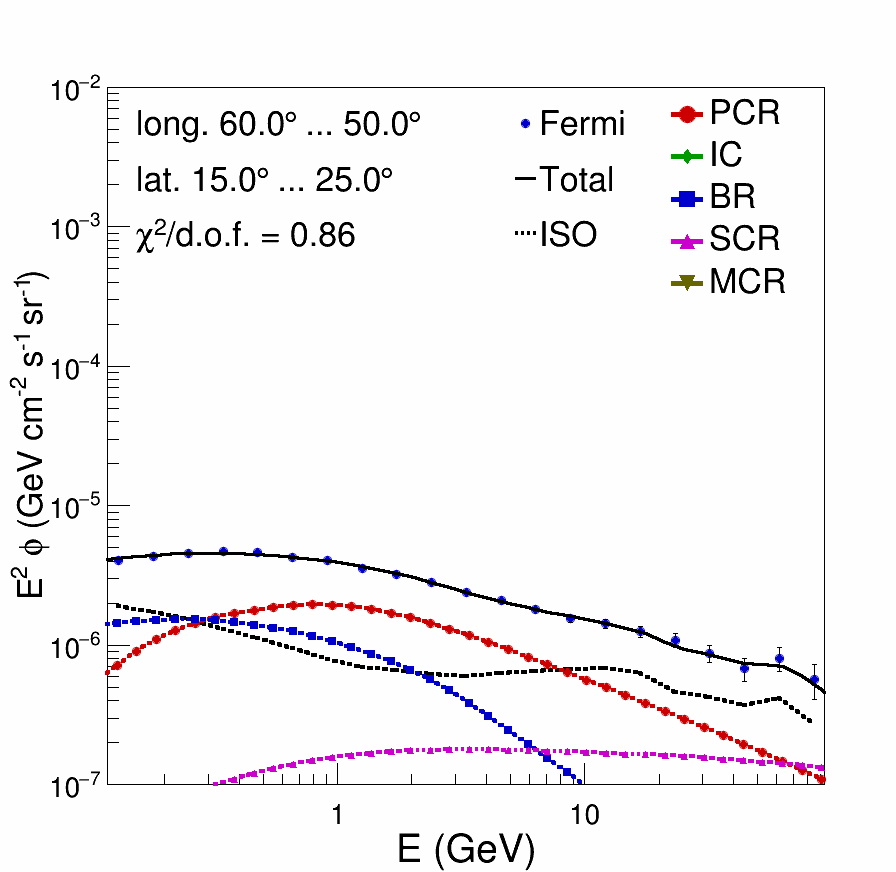}
\includegraphics[width=0.16\textwidth,height=0.16\textwidth,clip]{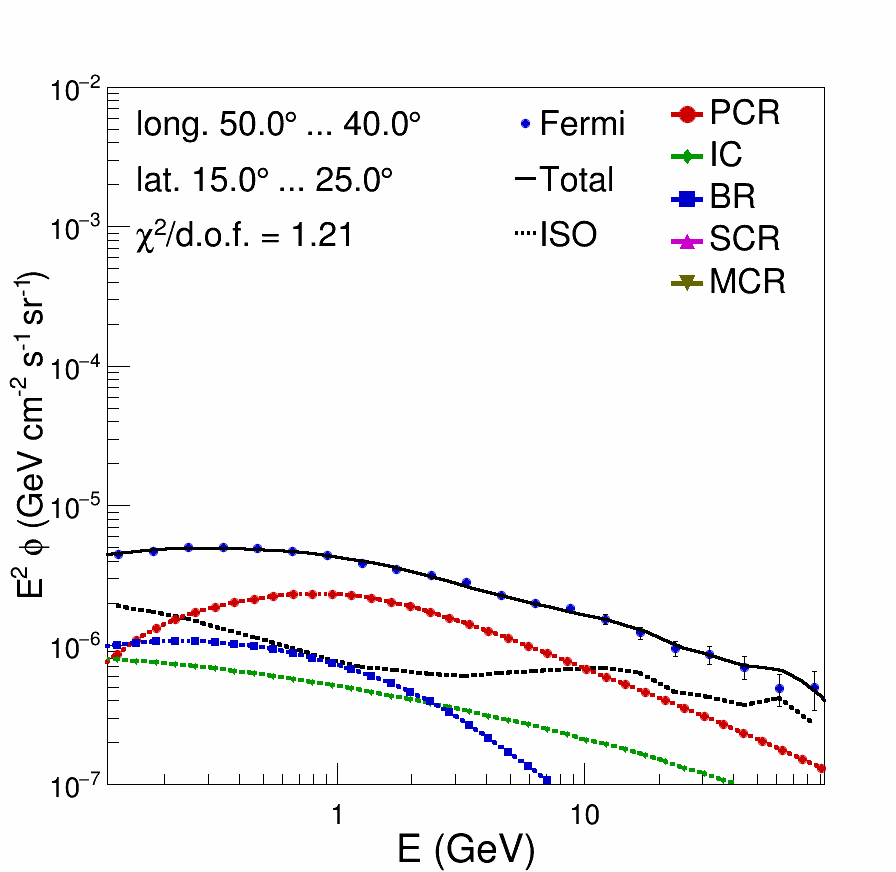}
\includegraphics[width=0.16\textwidth,height=0.16\textwidth,clip]{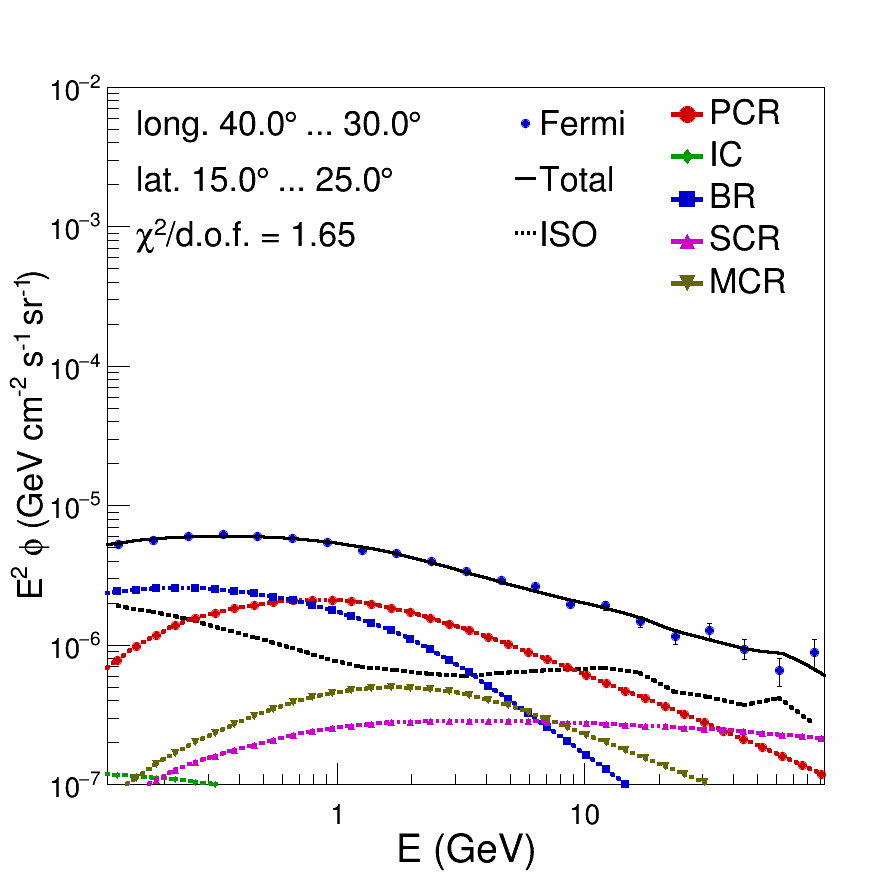}
\includegraphics[width=0.16\textwidth,height=0.16\textwidth,clip]{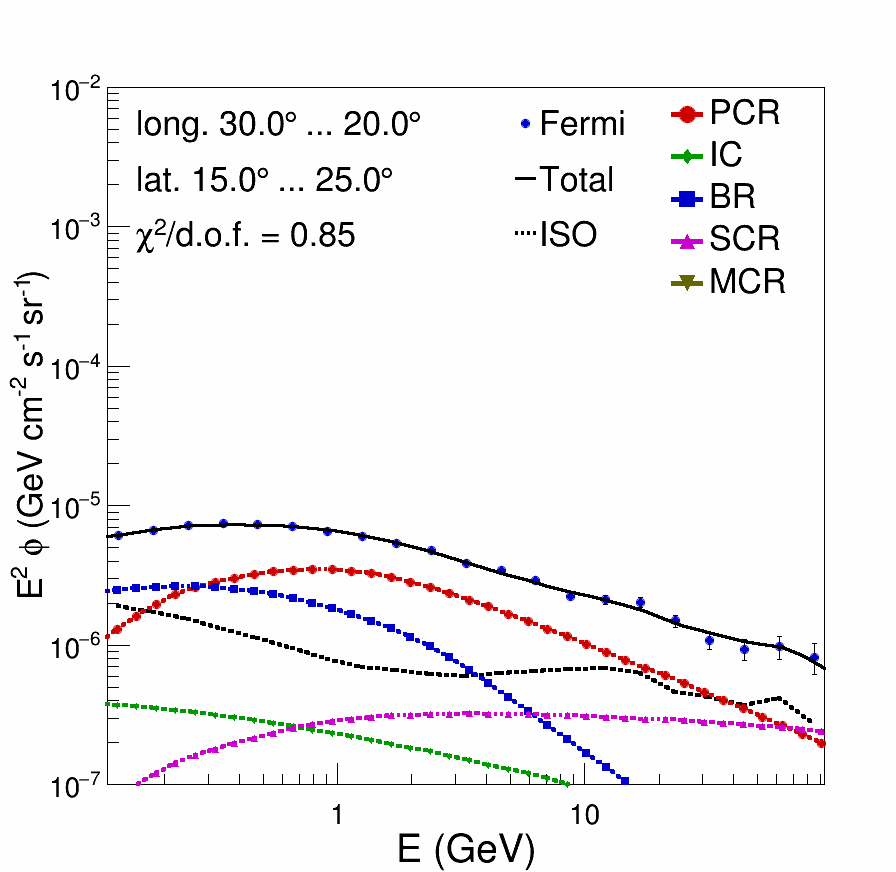}
\includegraphics[width=0.16\textwidth,height=0.16\textwidth,clip]{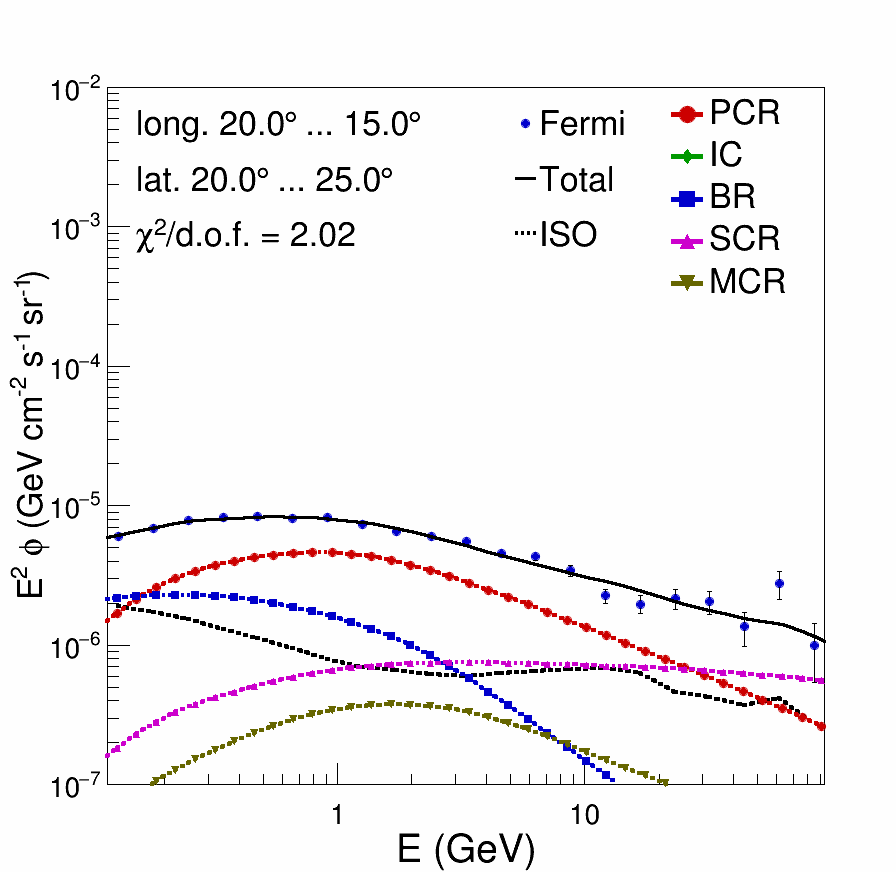}
\includegraphics[width=0.16\textwidth,height=0.16\textwidth,clip]{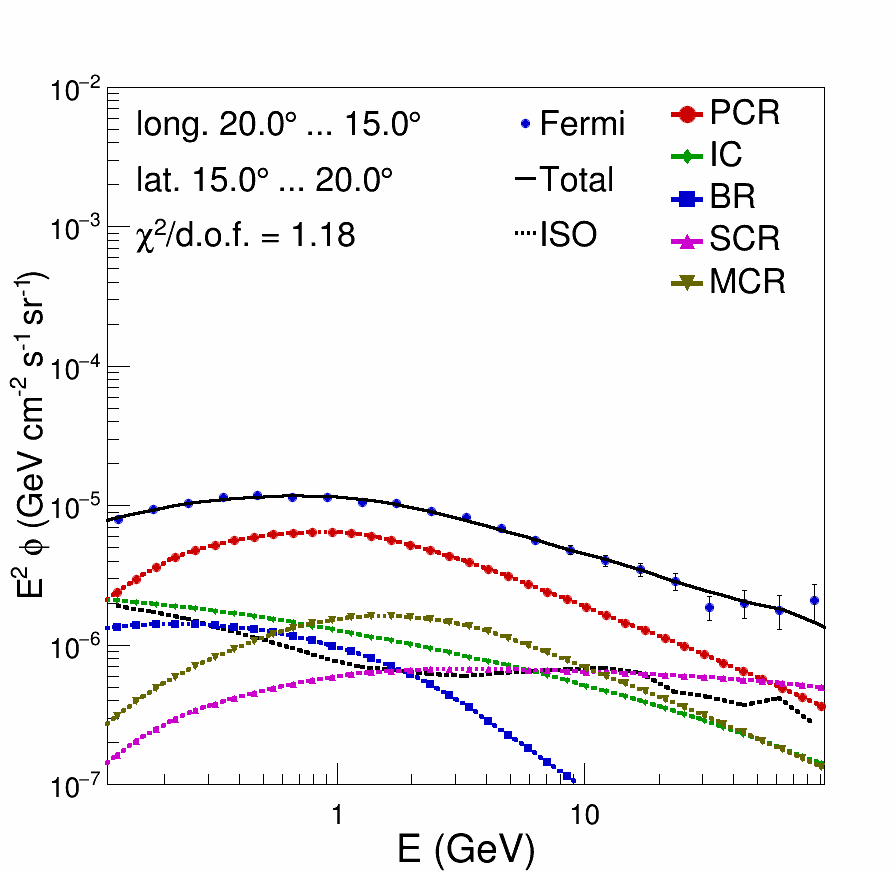}
\includegraphics[width=0.16\textwidth,height=0.16\textwidth,clip]{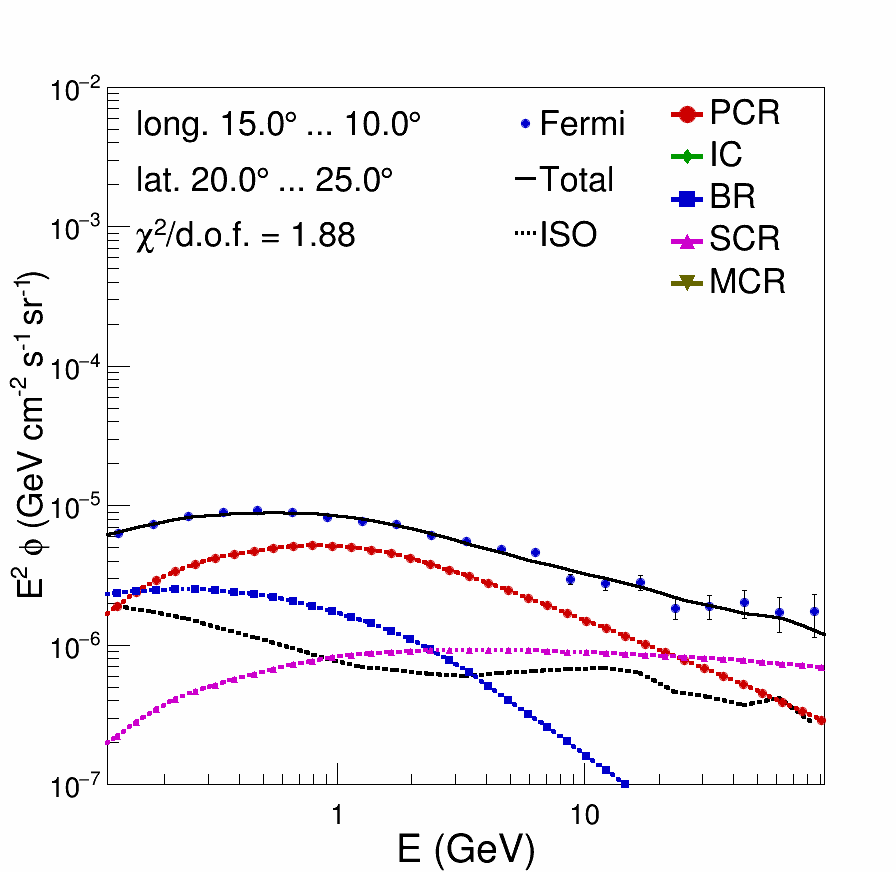}
\includegraphics[width=0.16\textwidth,height=0.16\textwidth,clip]{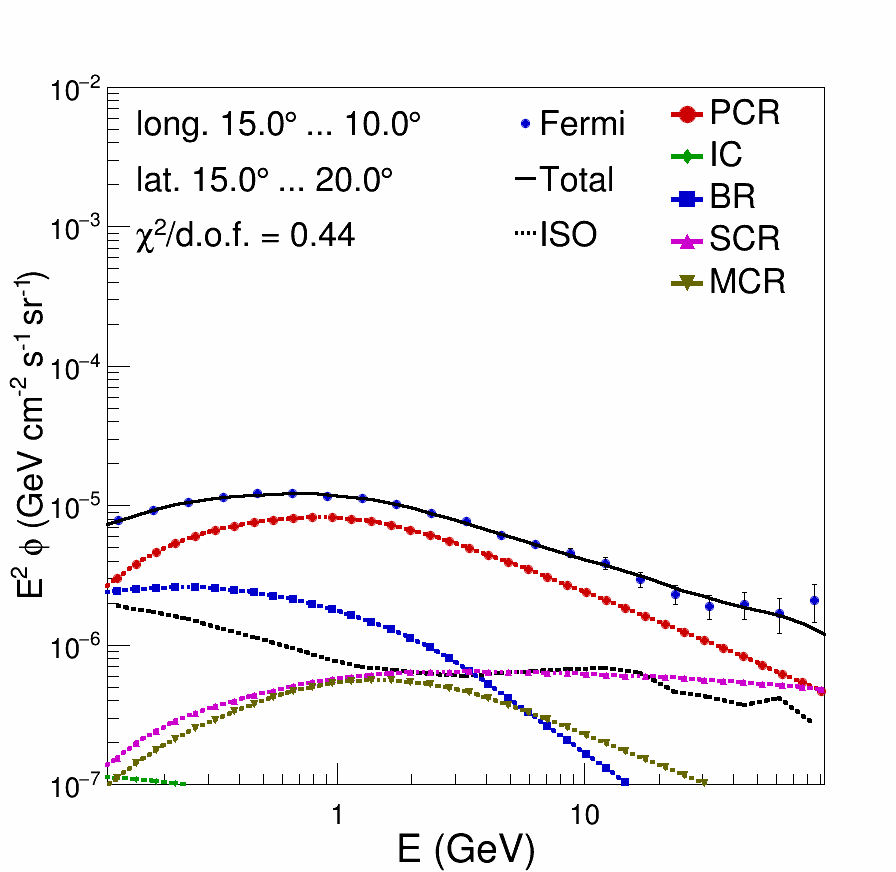}
\includegraphics[width=0.16\textwidth,height=0.16\textwidth,clip]{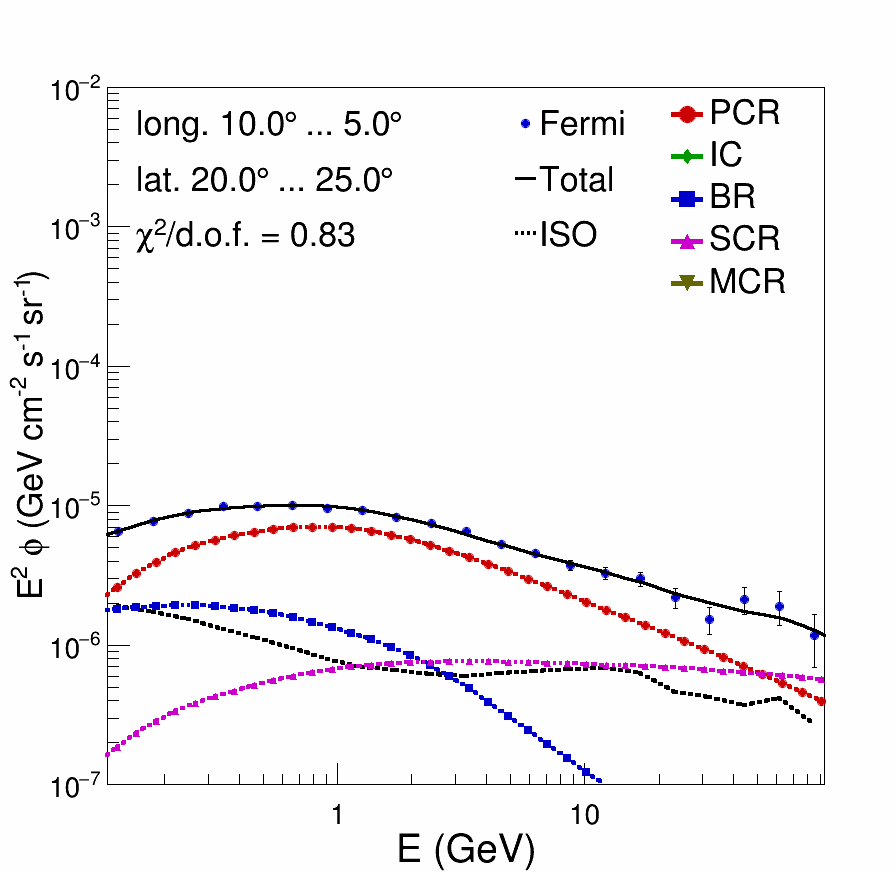}
\includegraphics[width=0.16\textwidth,height=0.16\textwidth,clip]{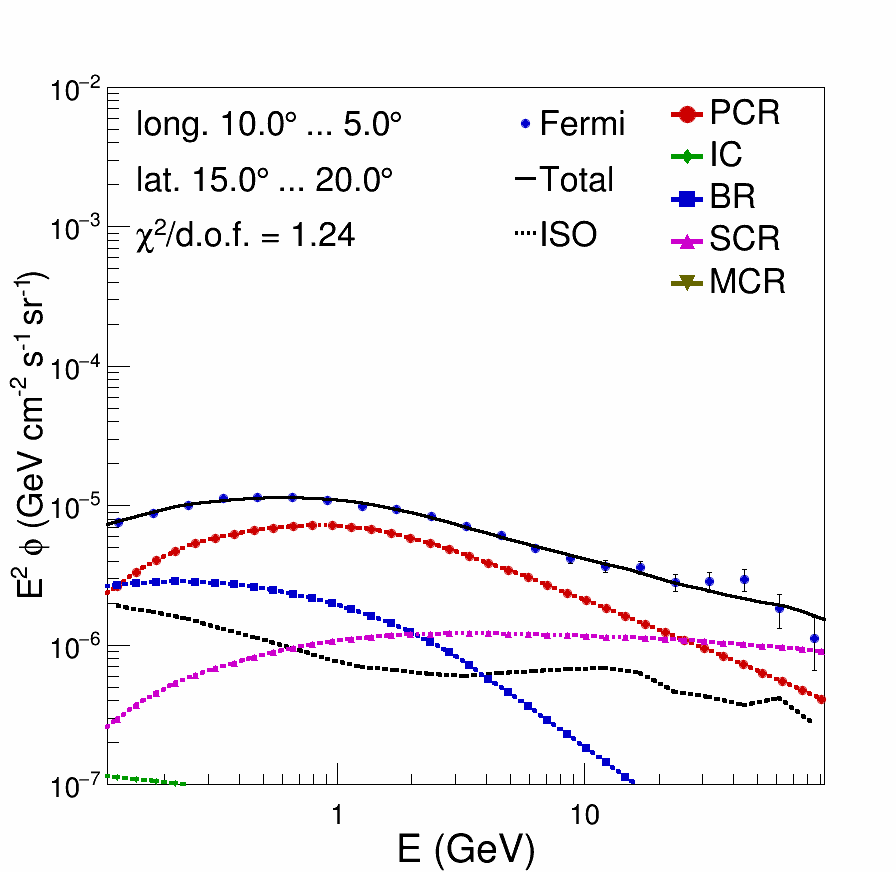}
\includegraphics[width=0.16\textwidth,height=0.16\textwidth,clip]{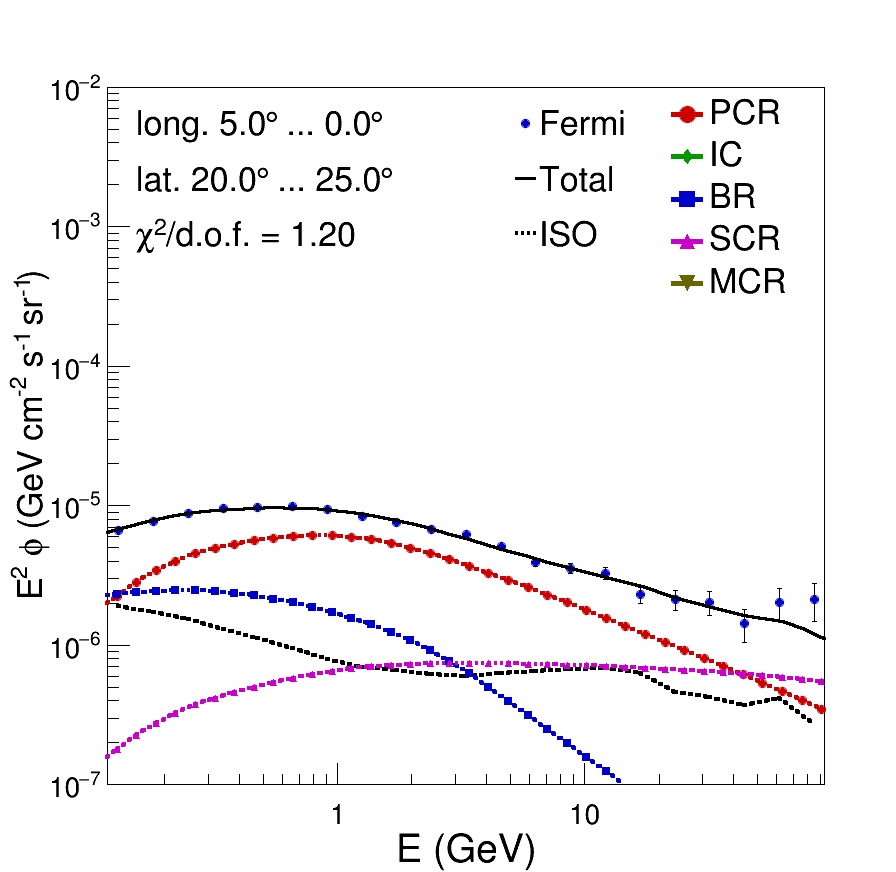}
\includegraphics[width=0.16\textwidth,height=0.16\textwidth,clip]{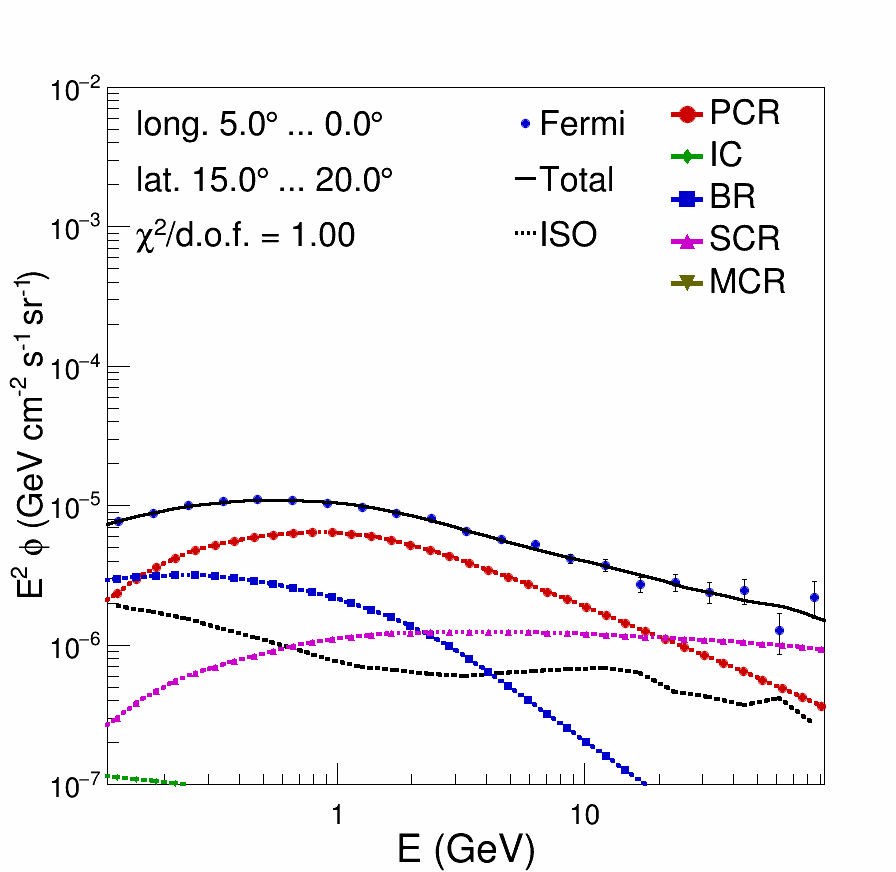}
\includegraphics[width=0.16\textwidth,height=0.16\textwidth,clip]{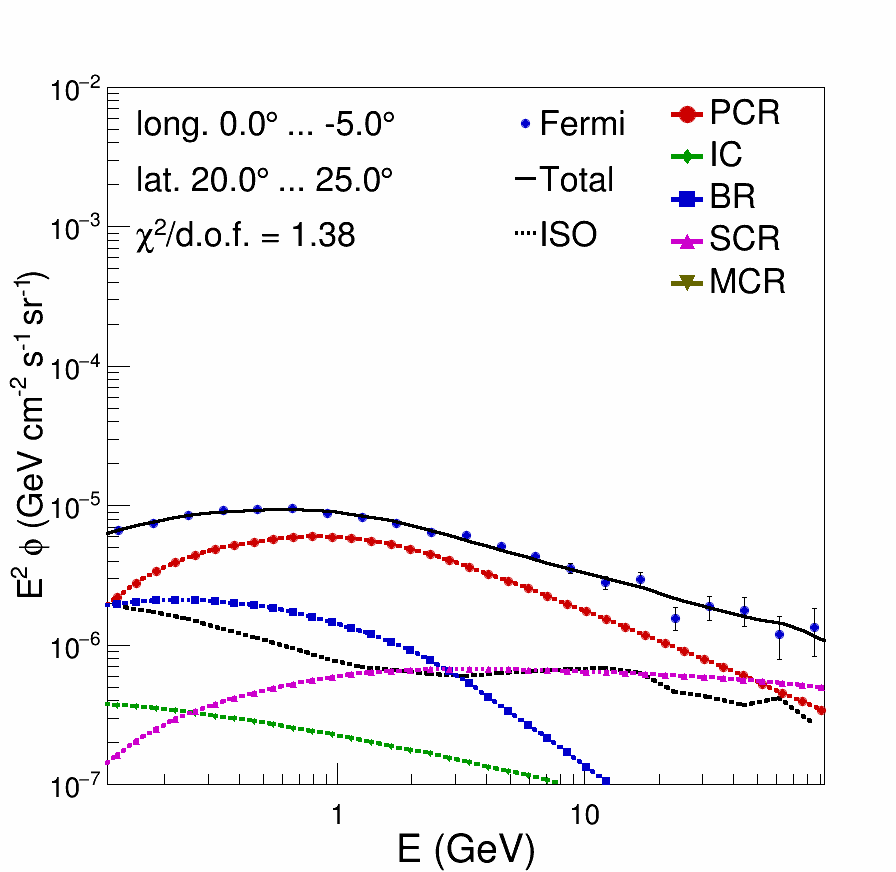}
\includegraphics[width=0.16\textwidth,height=0.16\textwidth,clip]{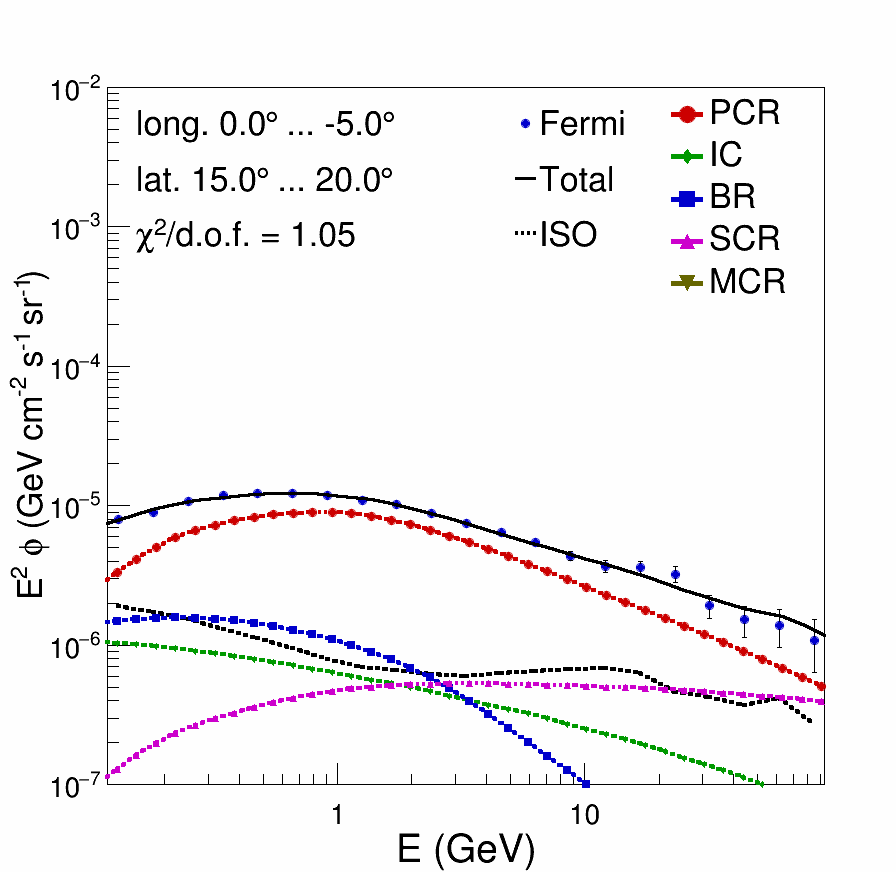}
\includegraphics[width=0.16\textwidth,height=0.16\textwidth,clip]{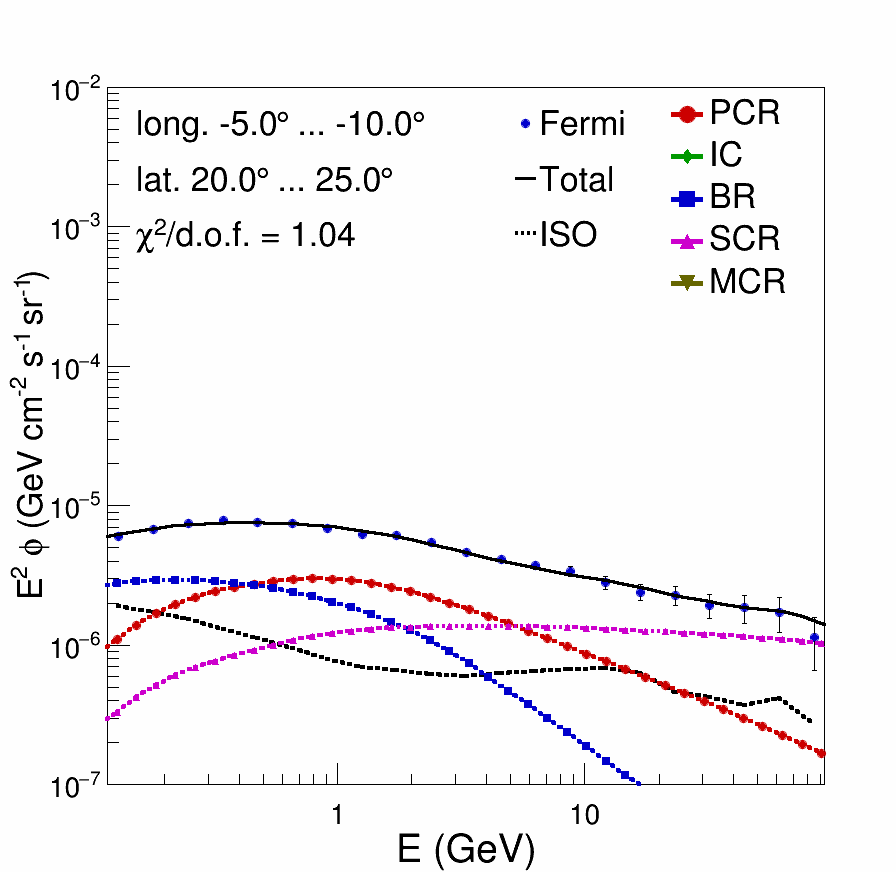}
\includegraphics[width=0.16\textwidth,height=0.16\textwidth,clip]{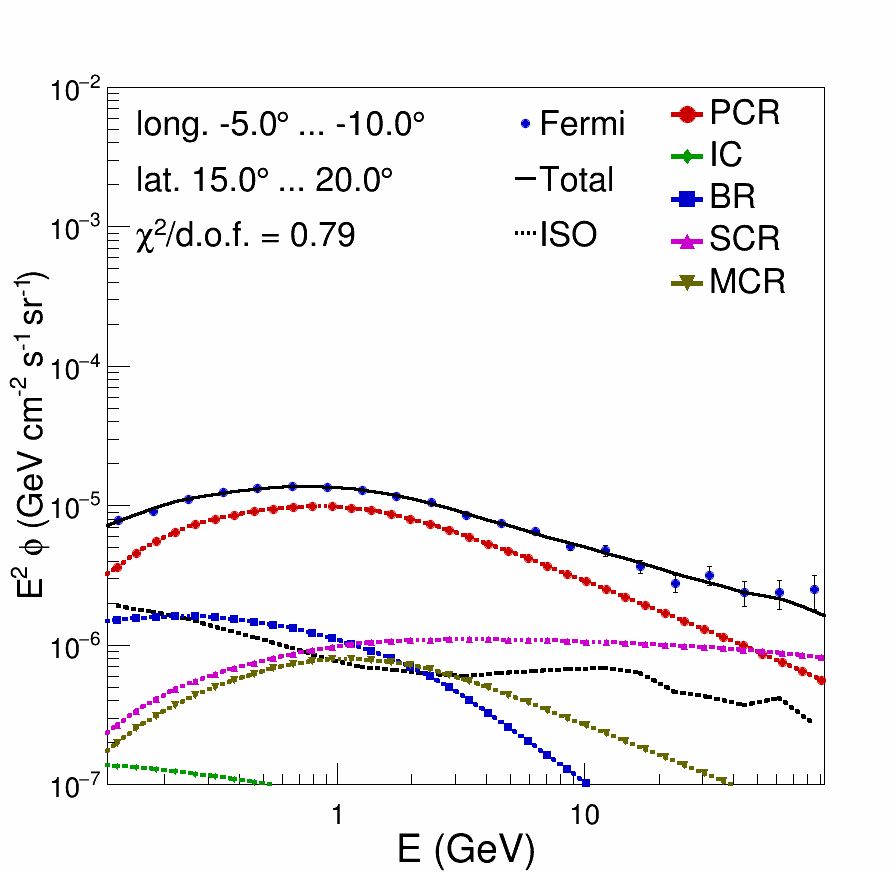}
\includegraphics[width=0.16\textwidth,height=0.16\textwidth,clip]{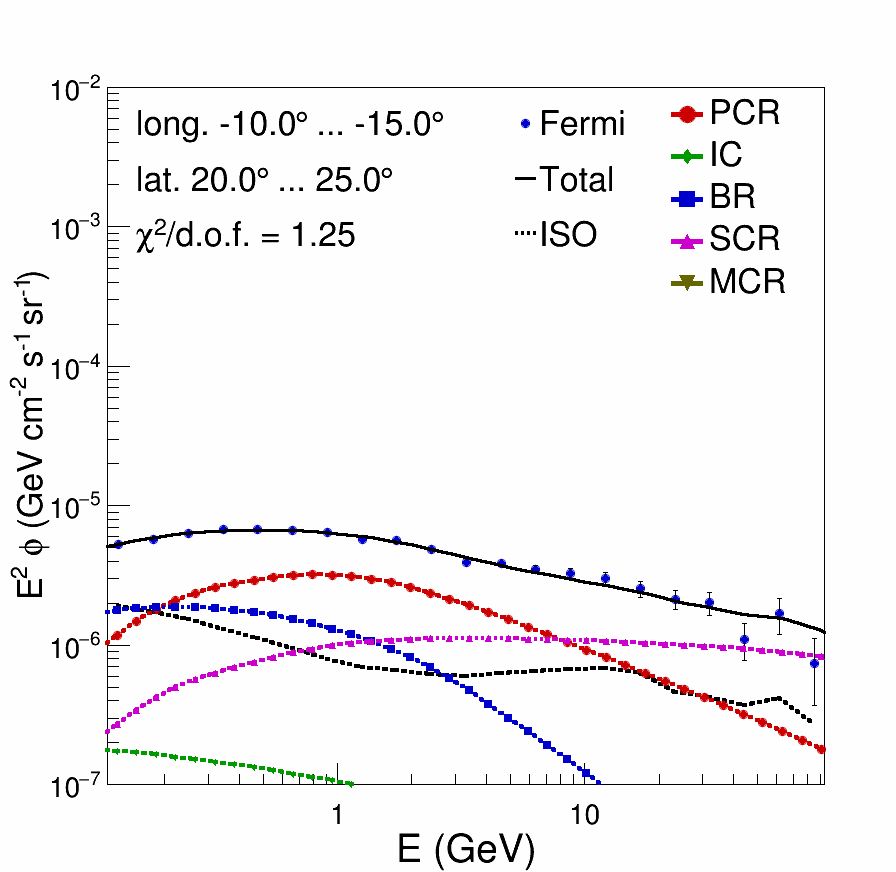}
\includegraphics[width=0.16\textwidth,height=0.16\textwidth,clip]{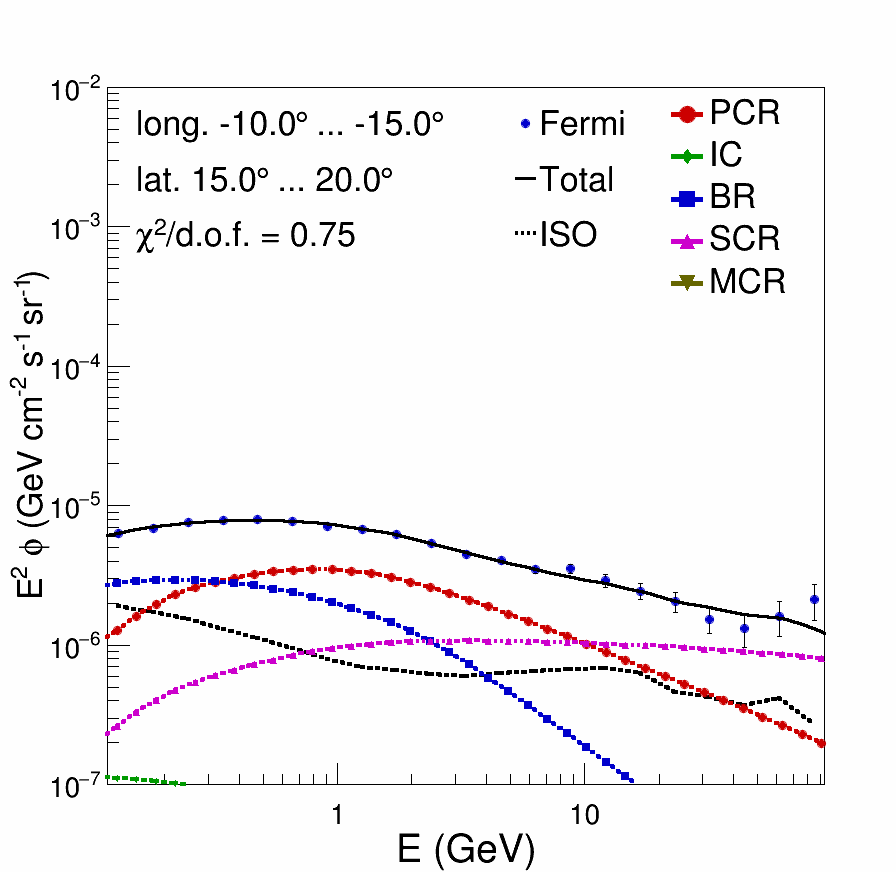}
\includegraphics[width=0.16\textwidth,height=0.16\textwidth,clip]{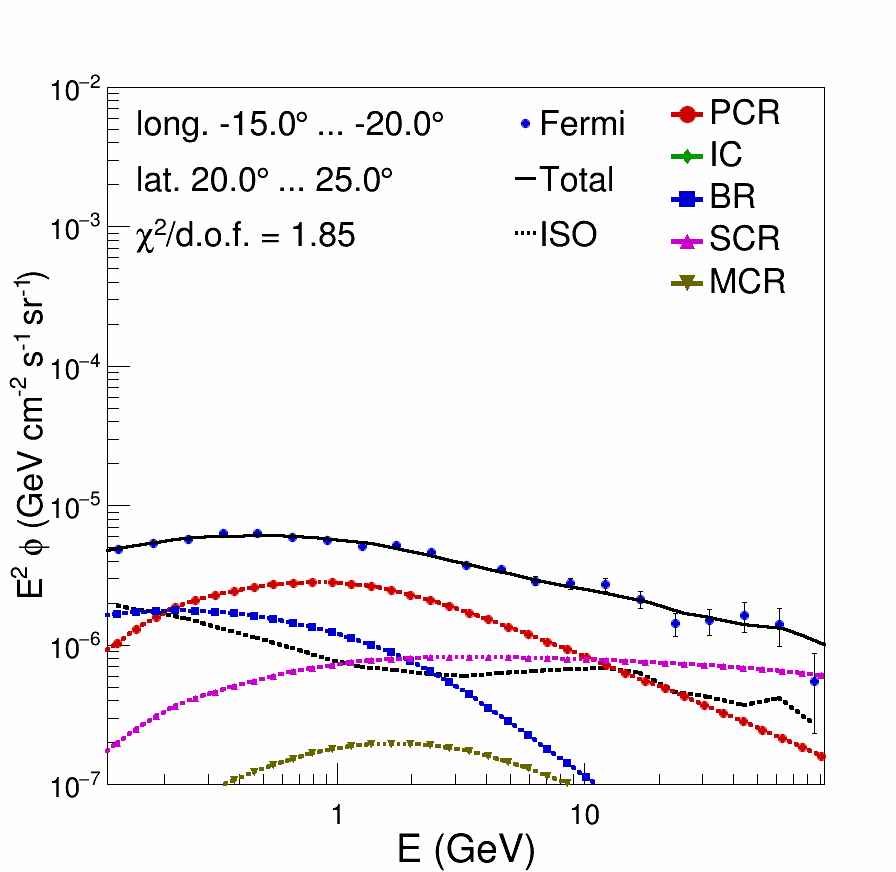}
\includegraphics[width=0.16\textwidth,height=0.16\textwidth,clip]{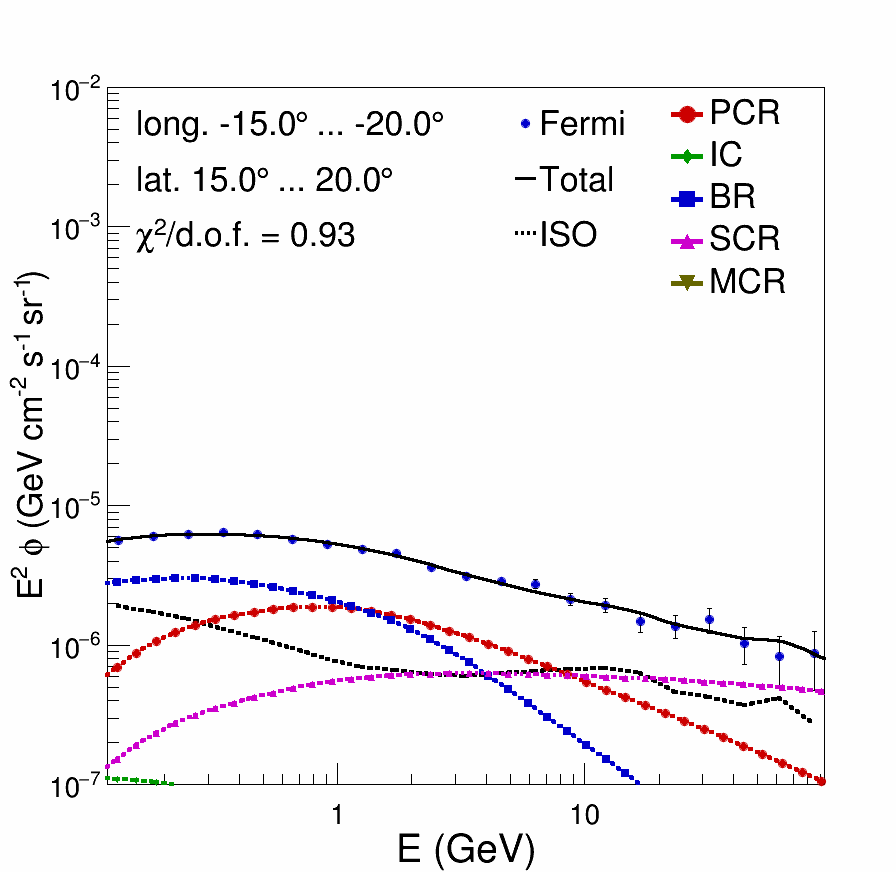}
\includegraphics[width=0.16\textwidth,height=0.16\textwidth,clip]{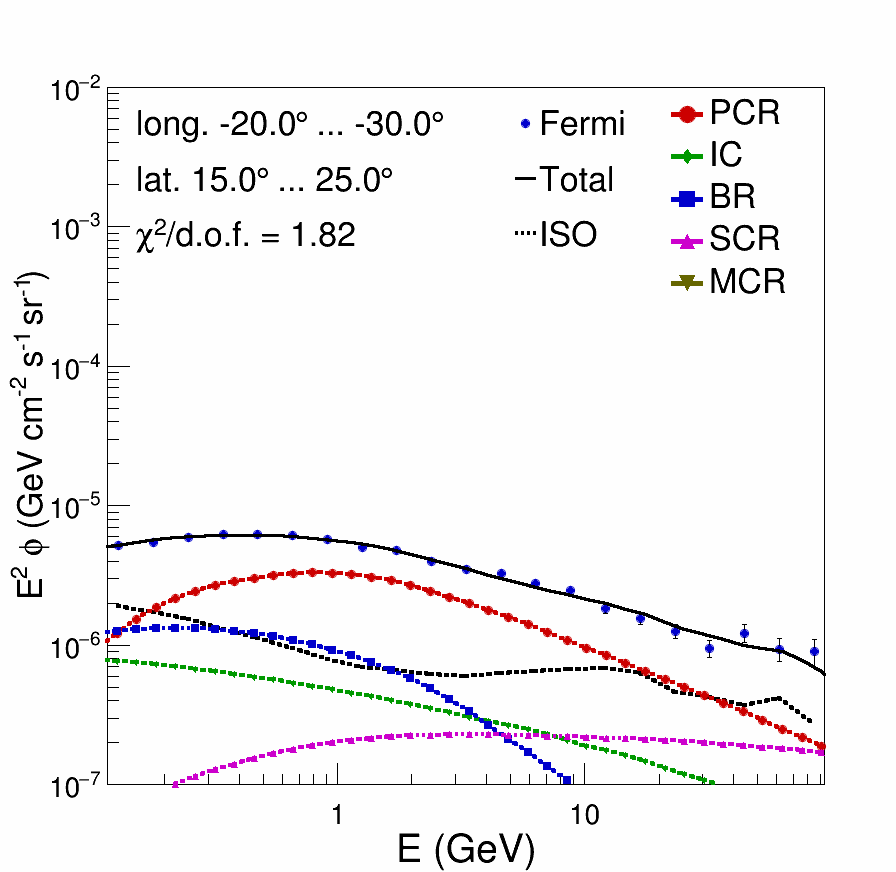}
\includegraphics[width=0.16\textwidth,height=0.16\textwidth,clip]{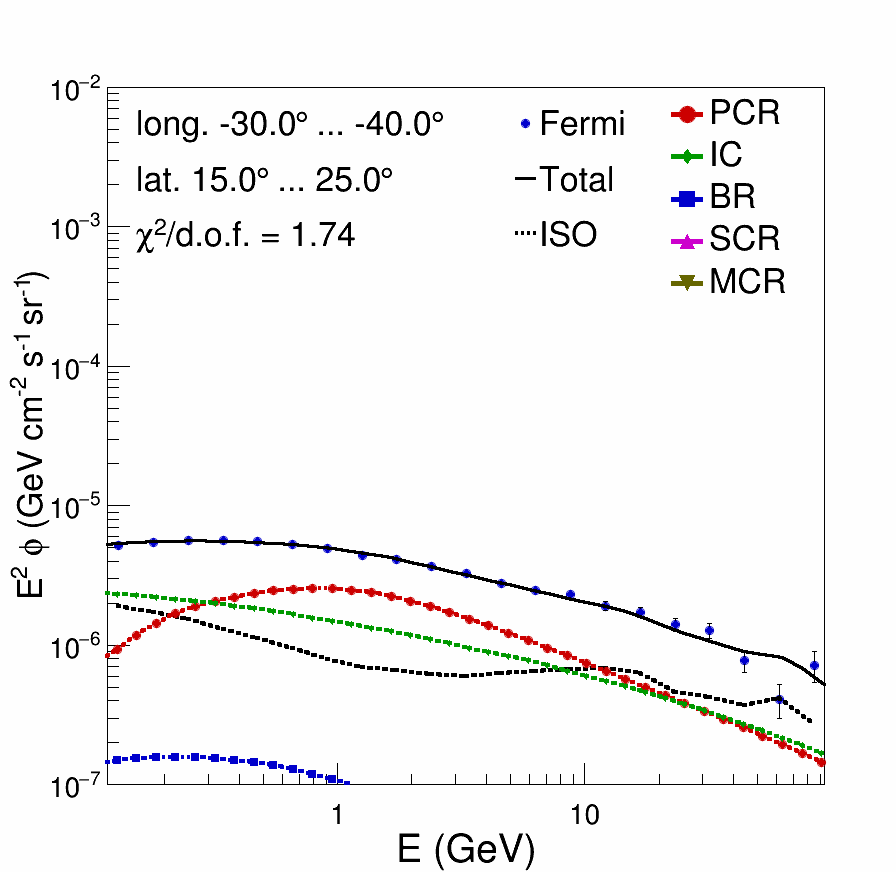}
\includegraphics[width=0.16\textwidth,height=0.16\textwidth,clip]{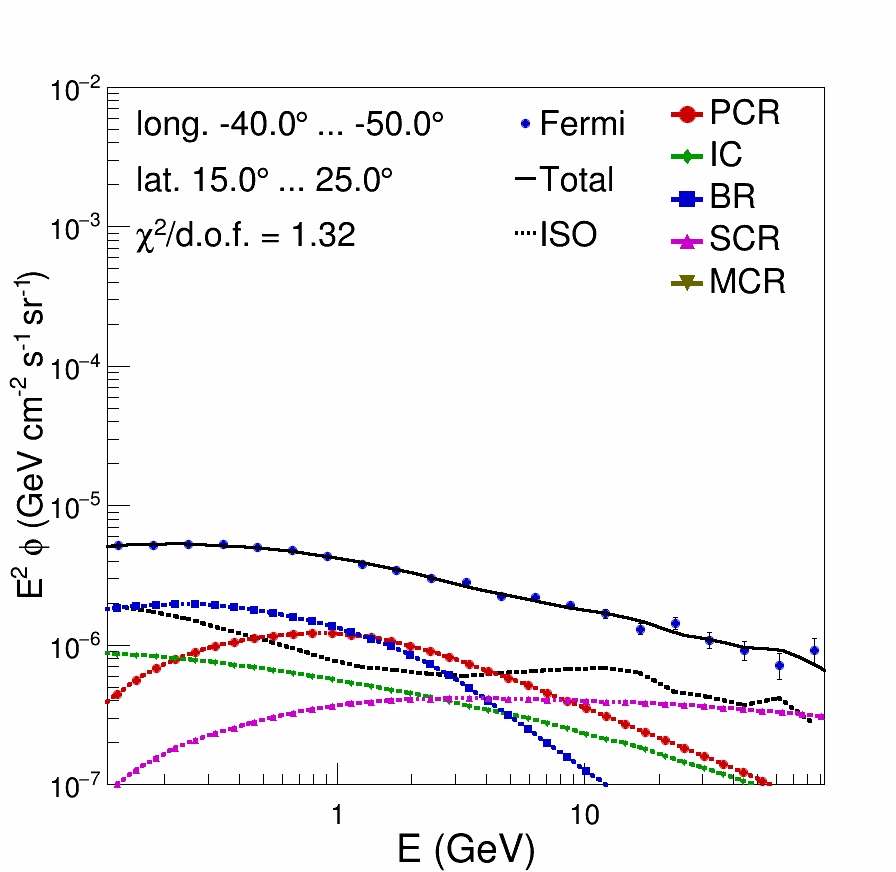}
\includegraphics[width=0.16\textwidth,height=0.16\textwidth,clip]{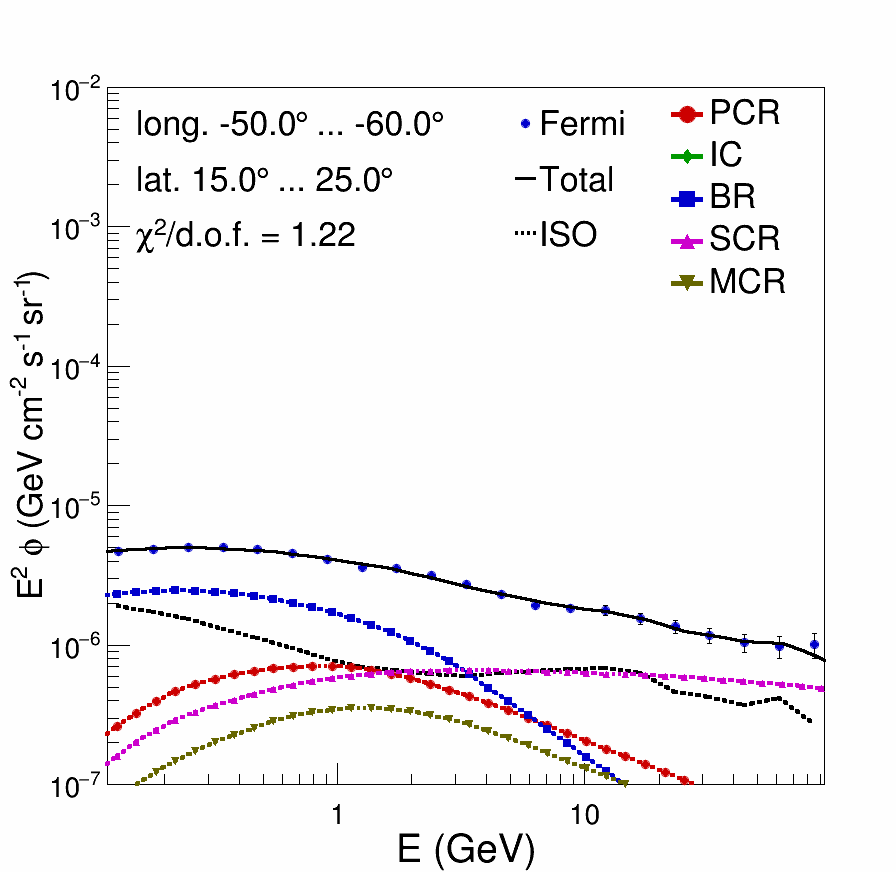}
\includegraphics[width=0.16\textwidth,height=0.16\textwidth,clip]{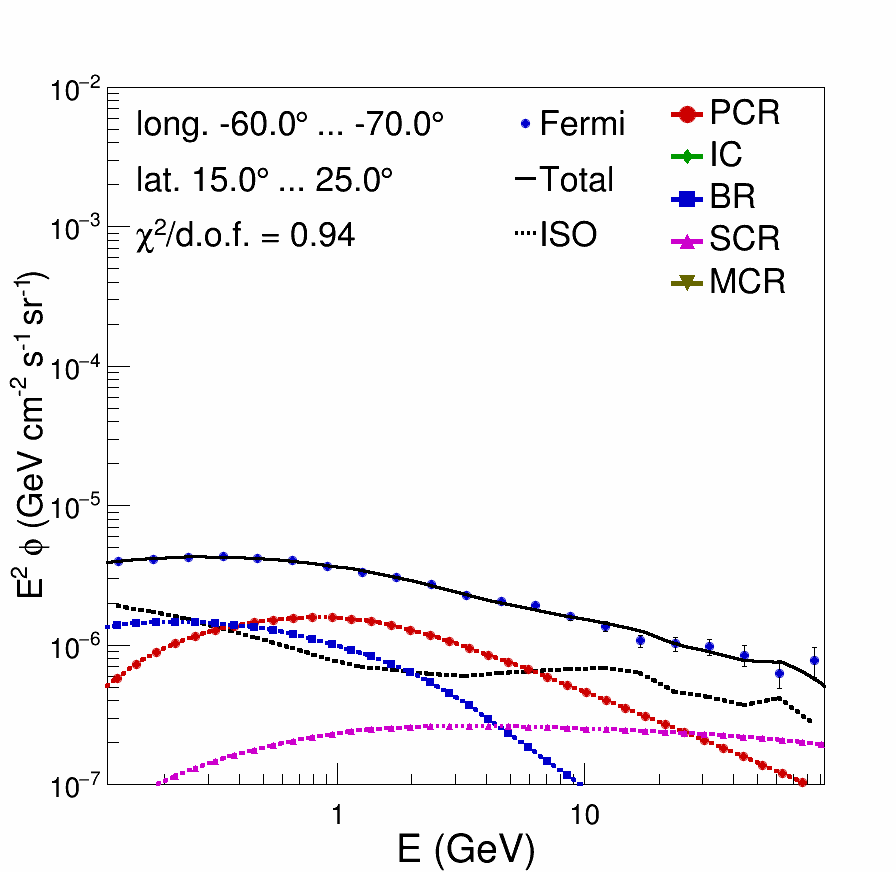}
\includegraphics[width=0.16\textwidth,height=0.16\textwidth,clip]{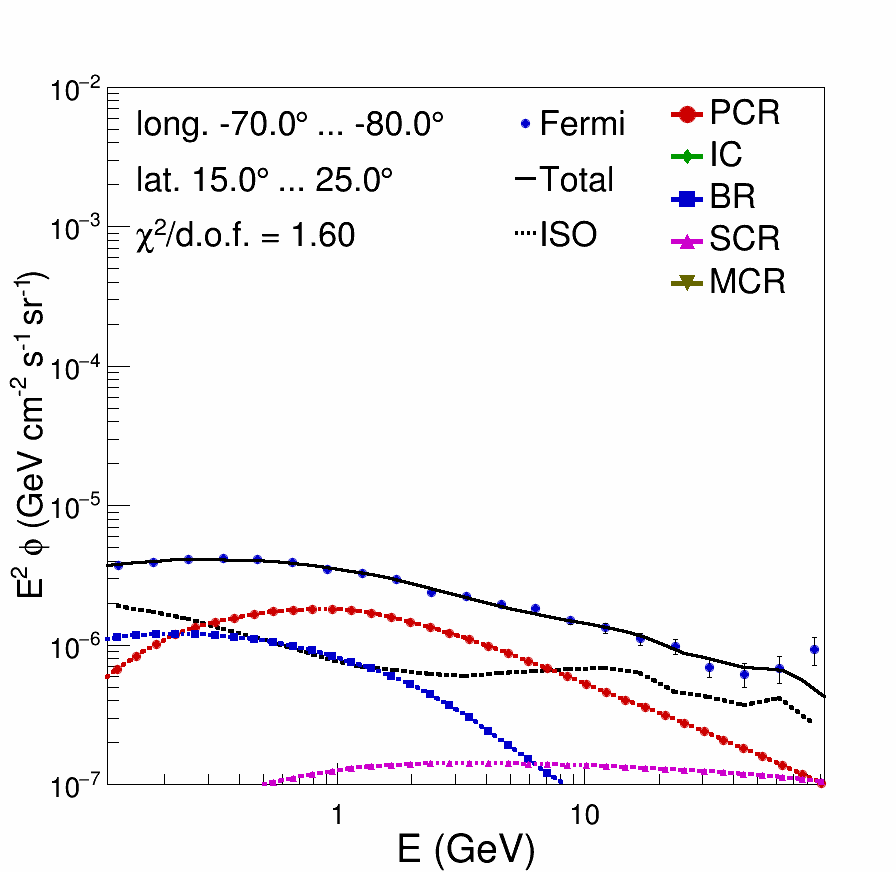}
\includegraphics[width=0.16\textwidth,height=0.16\textwidth,clip]{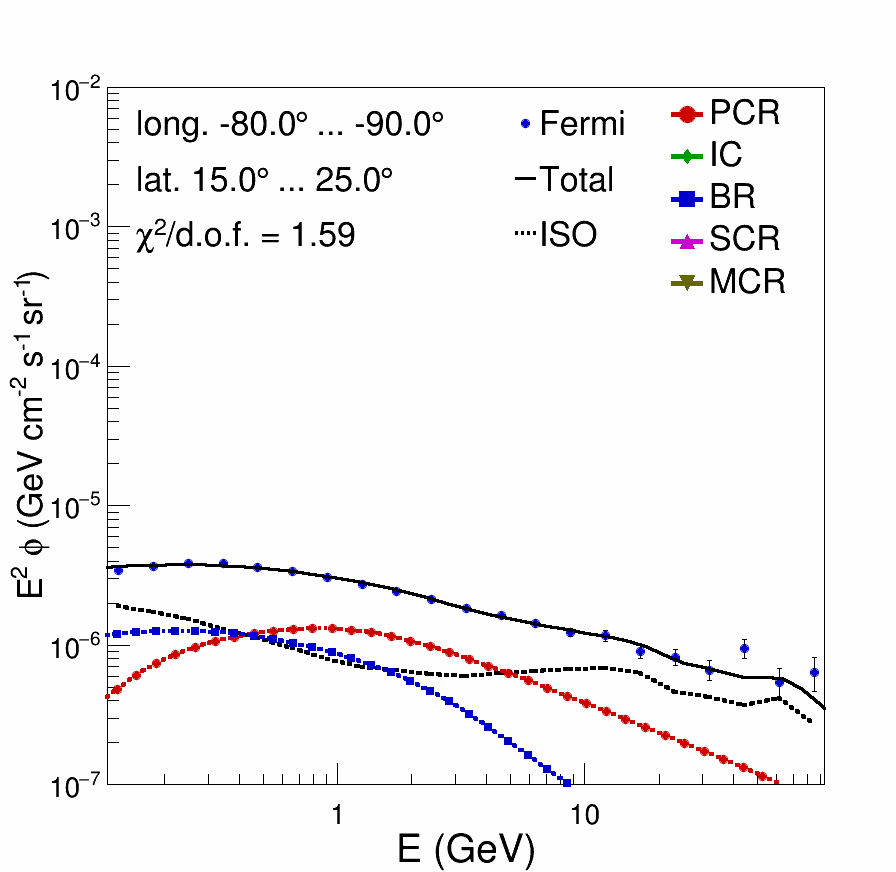}
\includegraphics[width=0.16\textwidth,height=0.16\textwidth,clip]{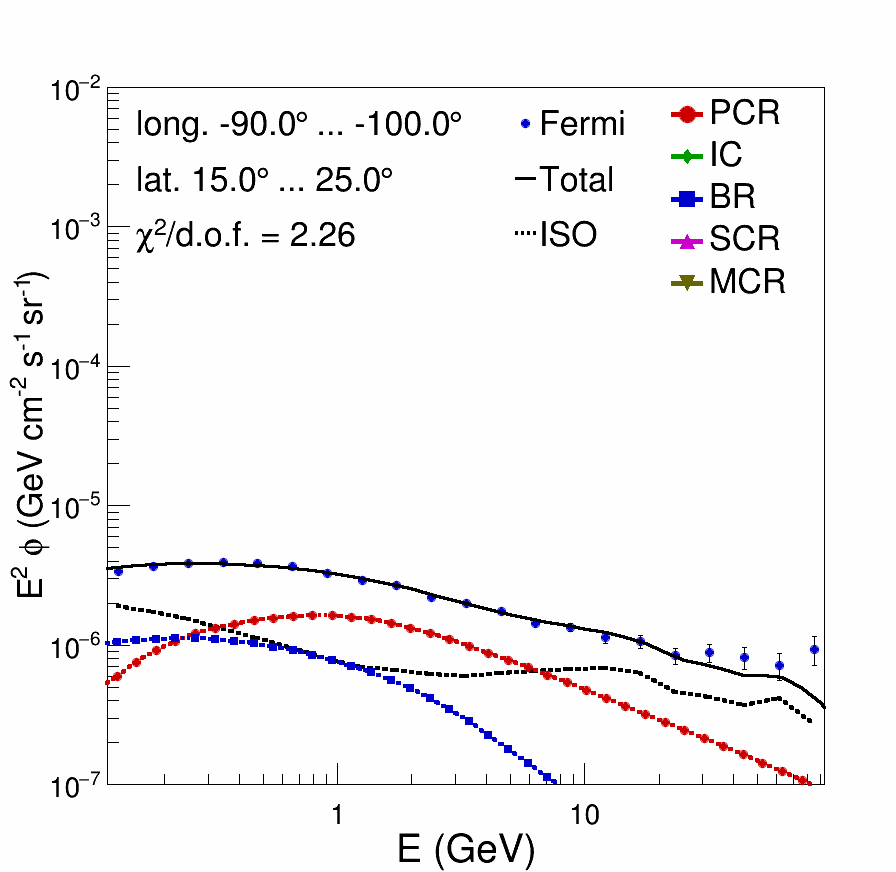}
\includegraphics[width=0.16\textwidth,height=0.16\textwidth,clip]{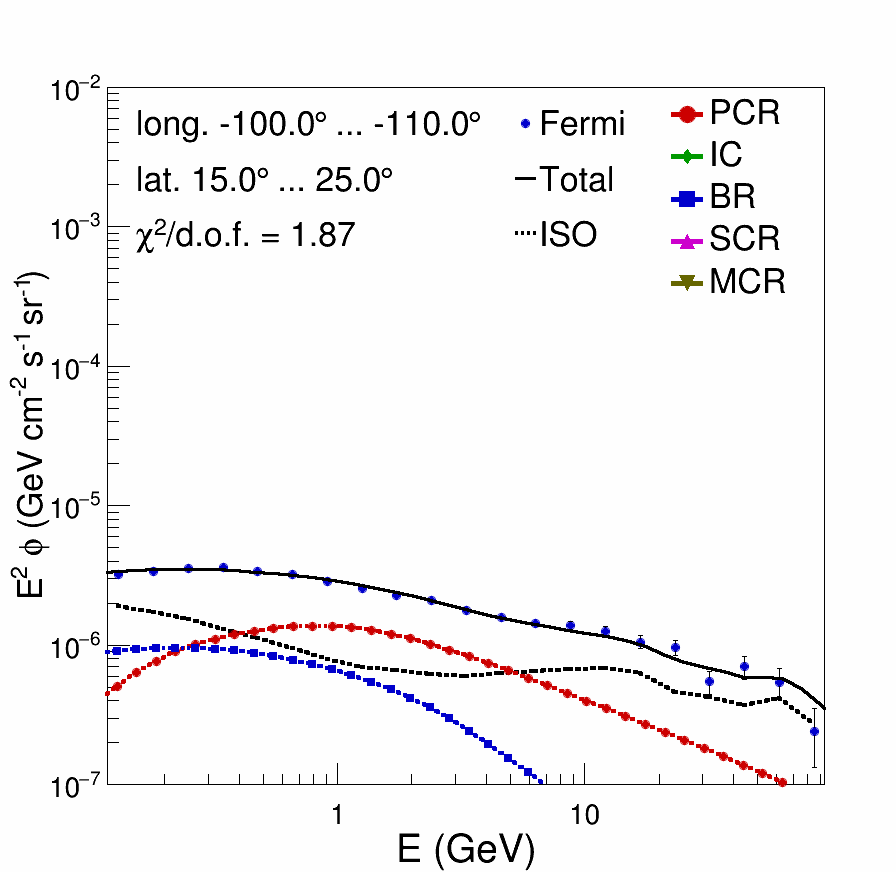}
\includegraphics[width=0.16\textwidth,height=0.16\textwidth,clip]{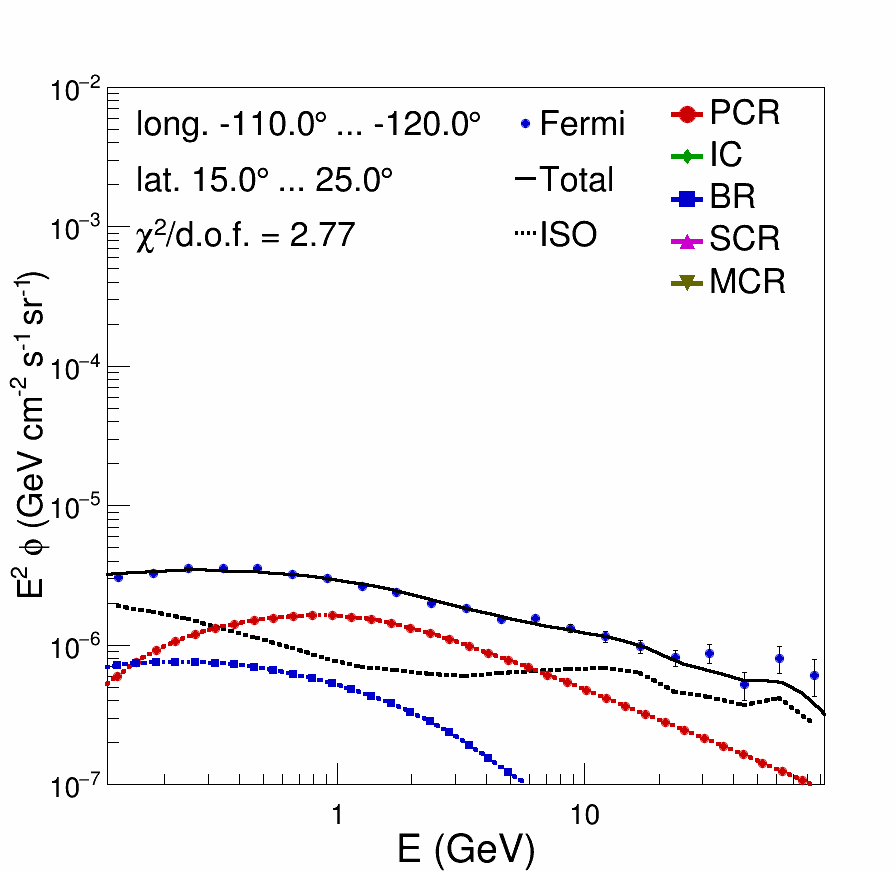}
\includegraphics[width=0.16\textwidth,height=0.16\textwidth,clip]{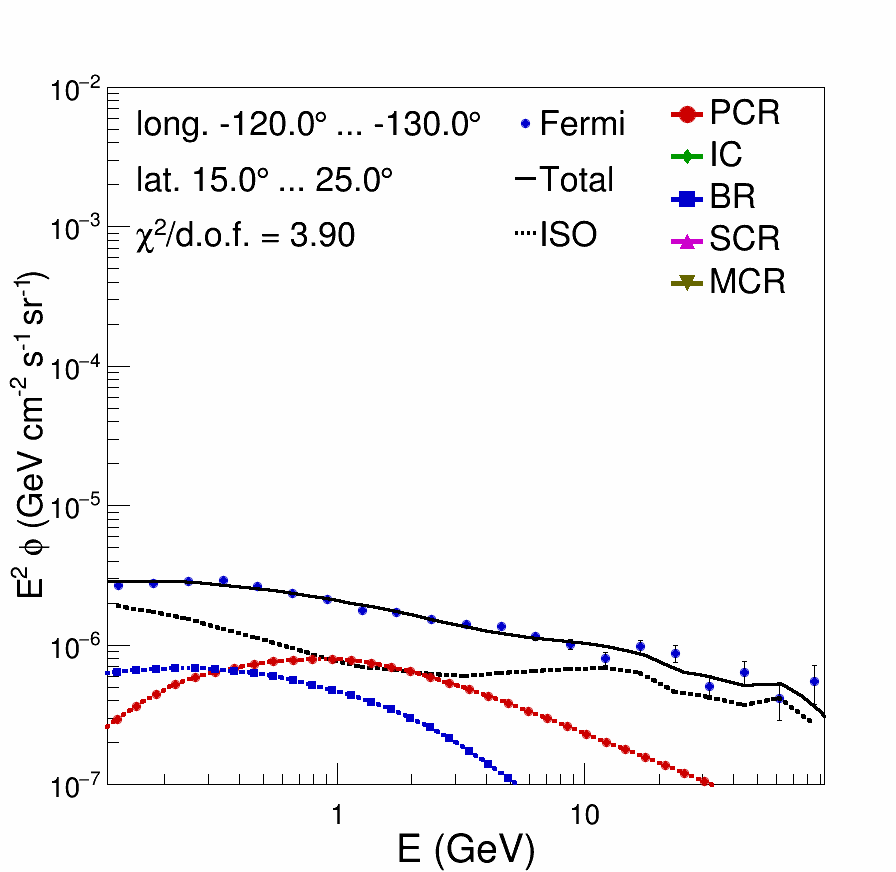}
\includegraphics[width=0.16\textwidth,height=0.16\textwidth,clip]{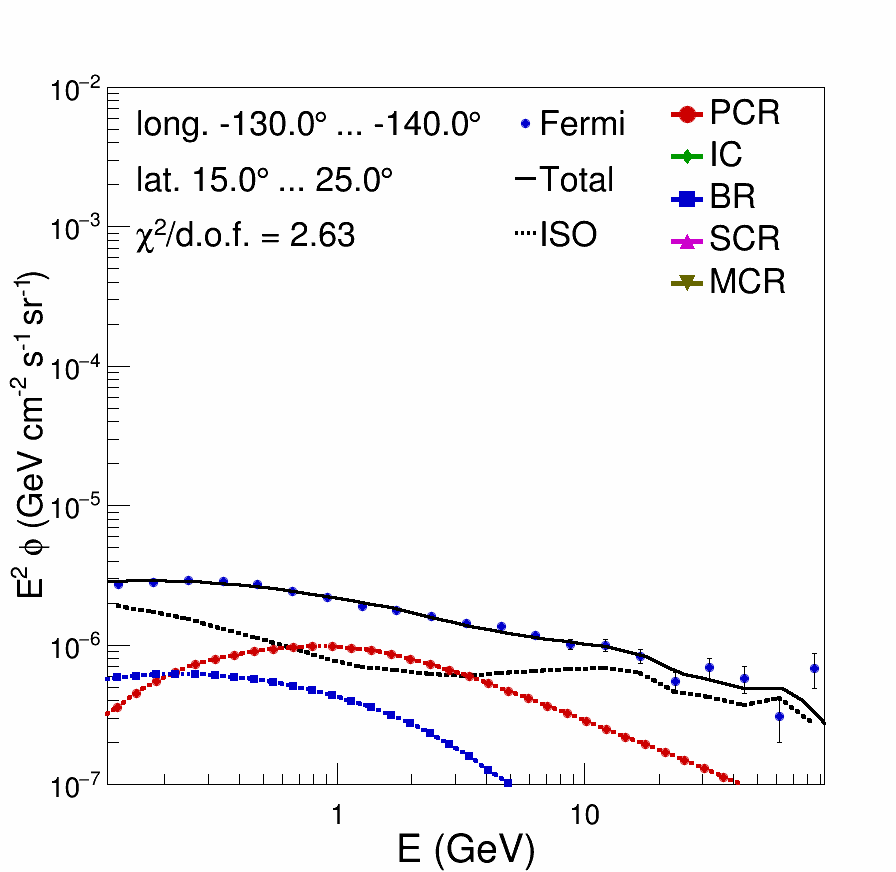}
\includegraphics[width=0.16\textwidth,height=0.16\textwidth,clip]{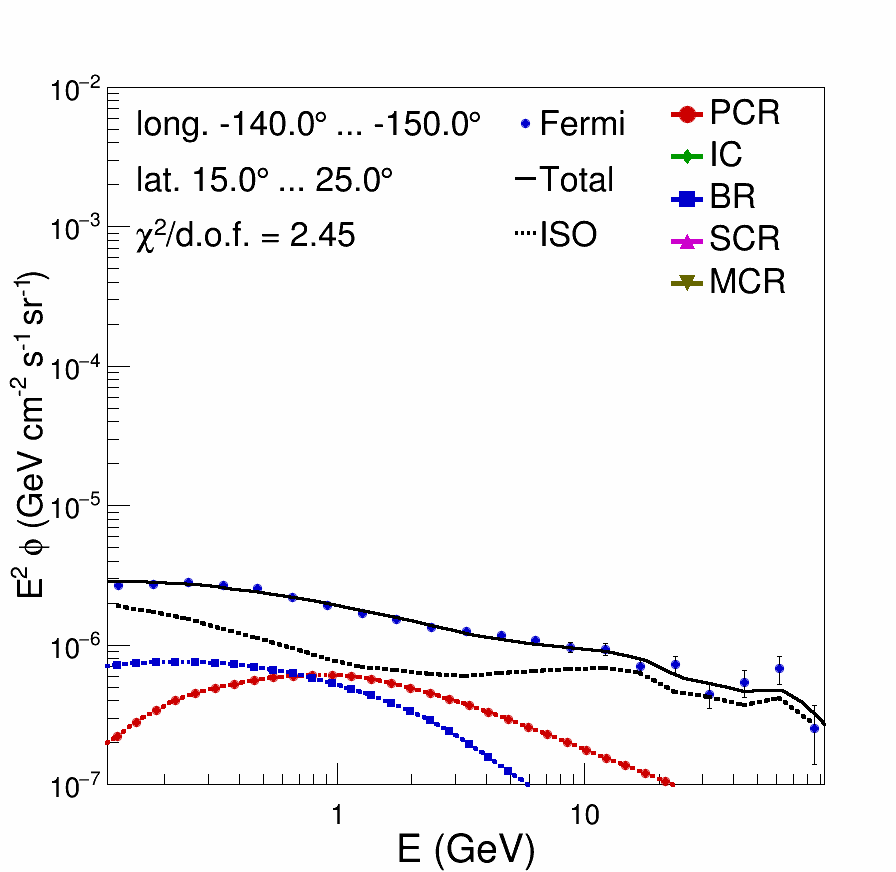}
\includegraphics[width=0.16\textwidth,height=0.16\textwidth,clip]{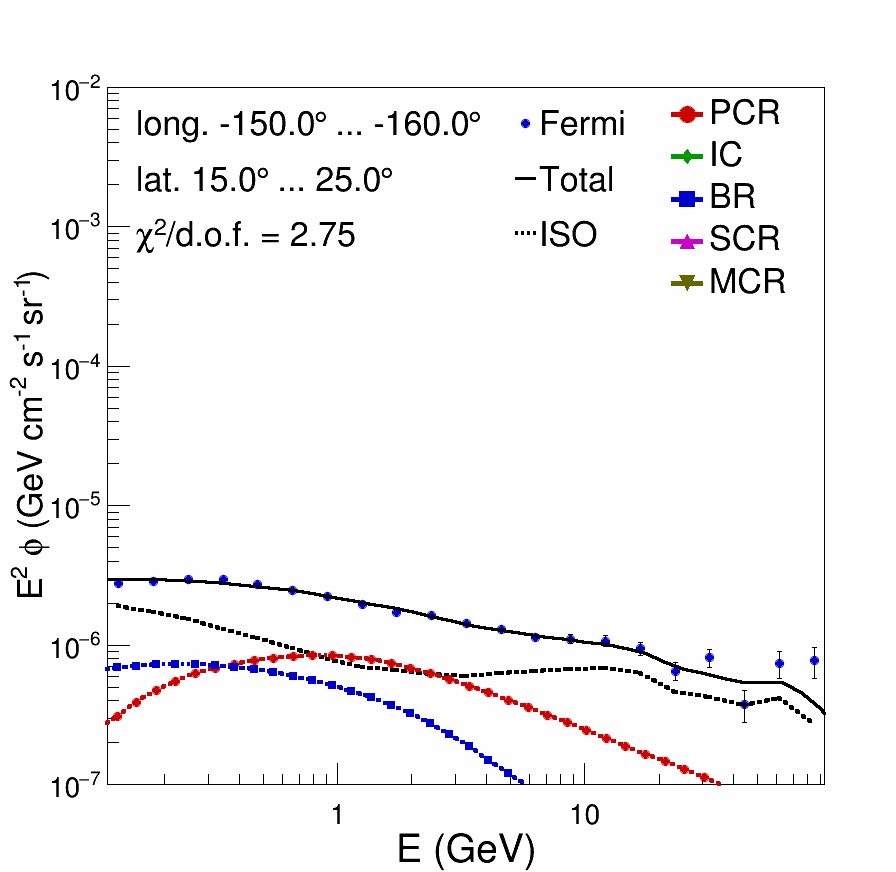}
\includegraphics[width=0.16\textwidth,height=0.16\textwidth,clip]{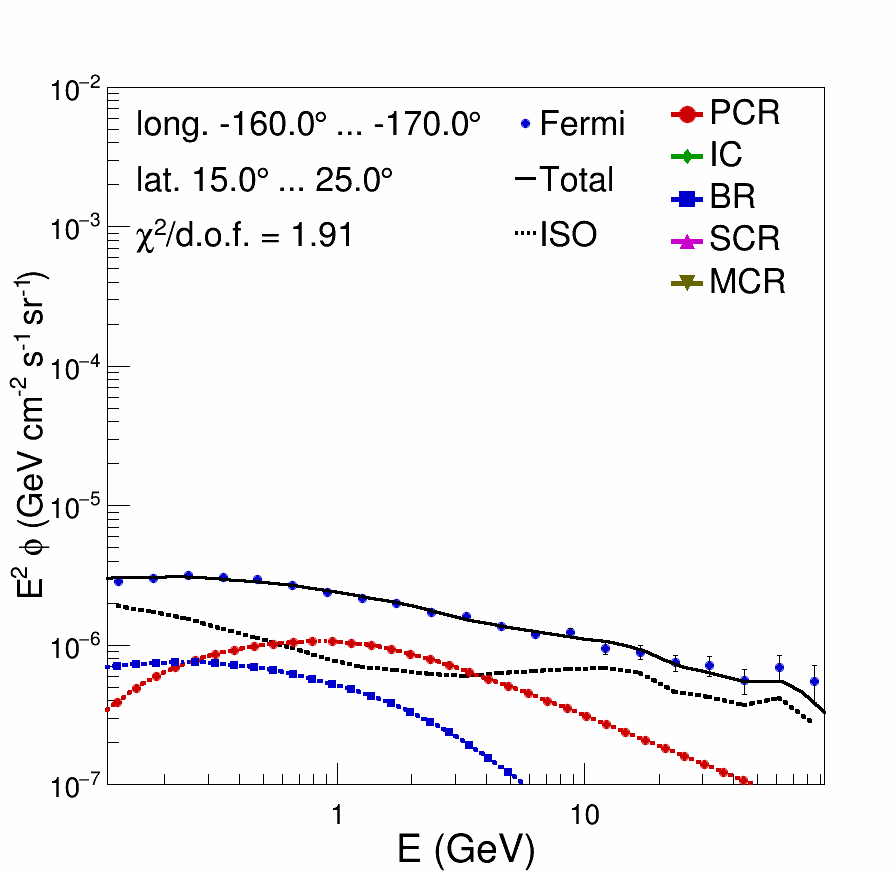}
\includegraphics[width=0.16\textwidth,height=0.16\textwidth,clip]{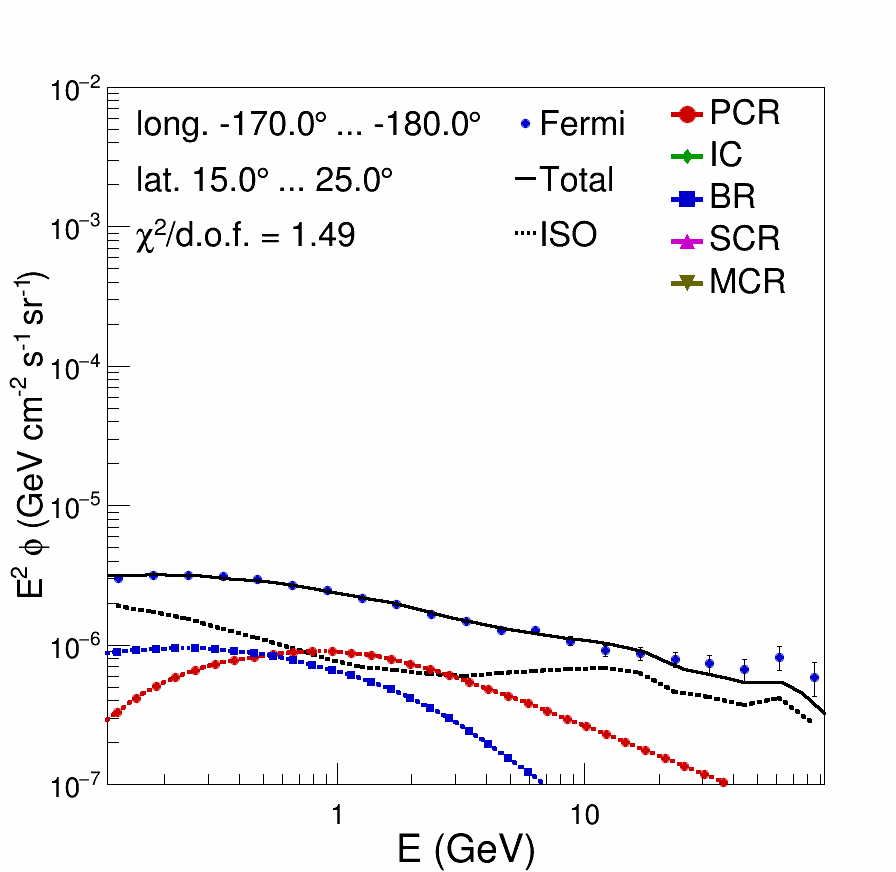}
\caption[]{Template fits for latitudes  with $15.0^\circ<b<25.0^\circ$ and longitudes decreasing from 180$^\circ$ to -180$^\circ$. \label{F16}
}
\end{figure}
\begin{figure}
\centering
\includegraphics[width=0.16\textwidth,height=0.16\textwidth,clip]{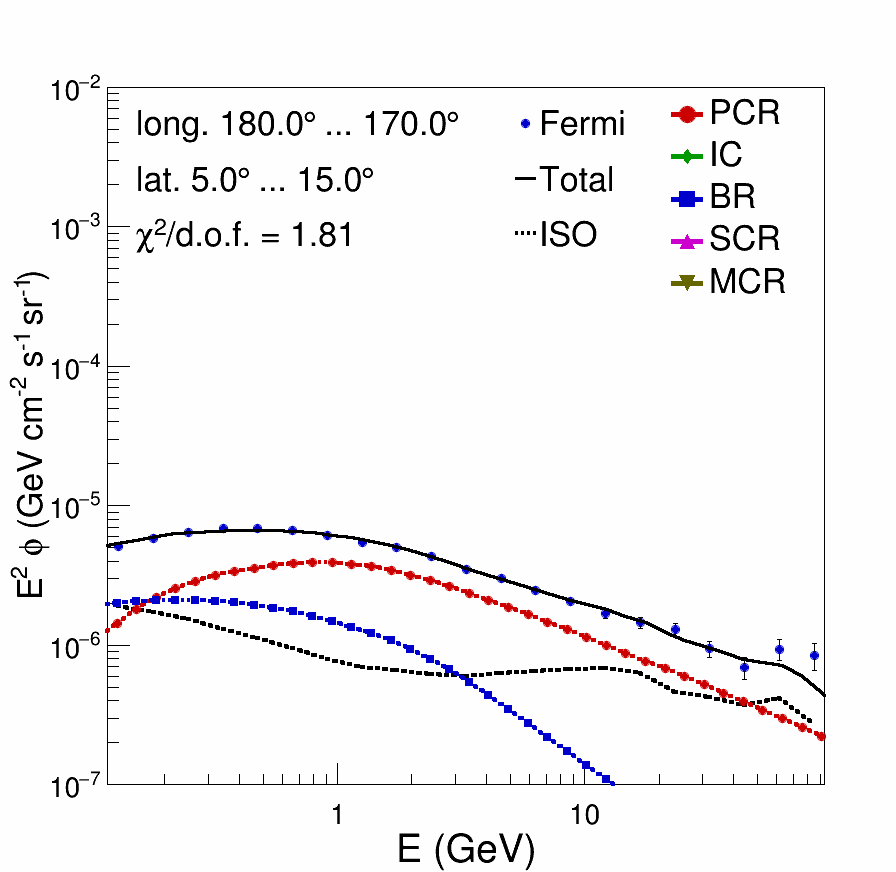}
\includegraphics[width=0.16\textwidth,height=0.16\textwidth,clip]{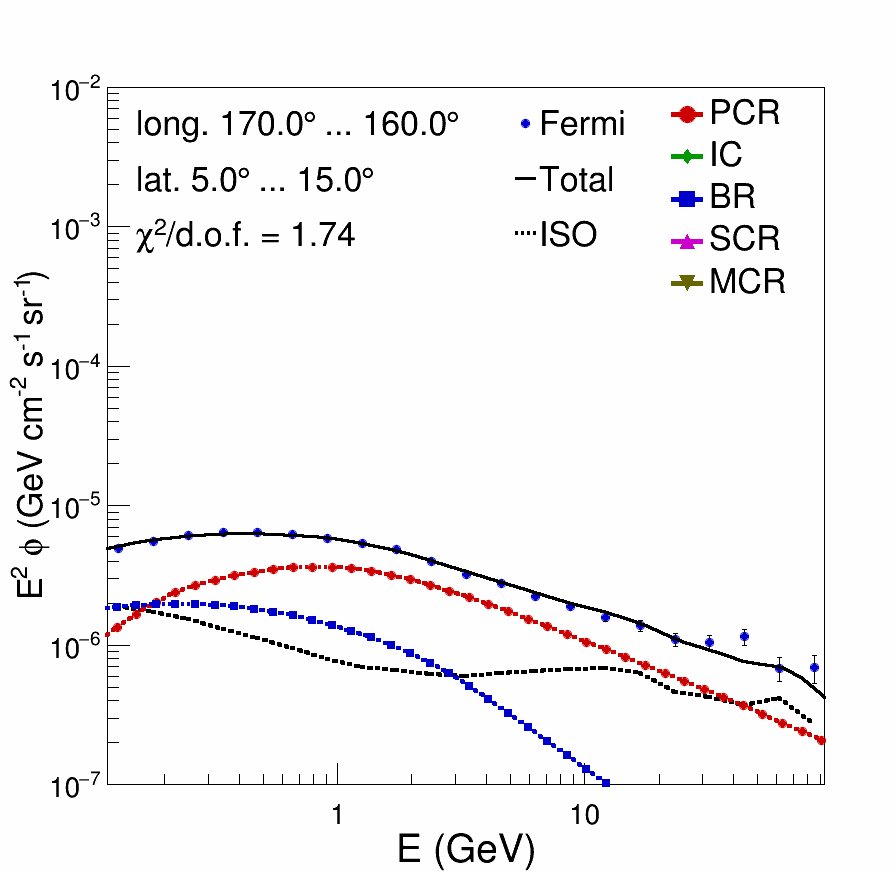}
\includegraphics[width=0.16\textwidth,height=0.16\textwidth,clip]{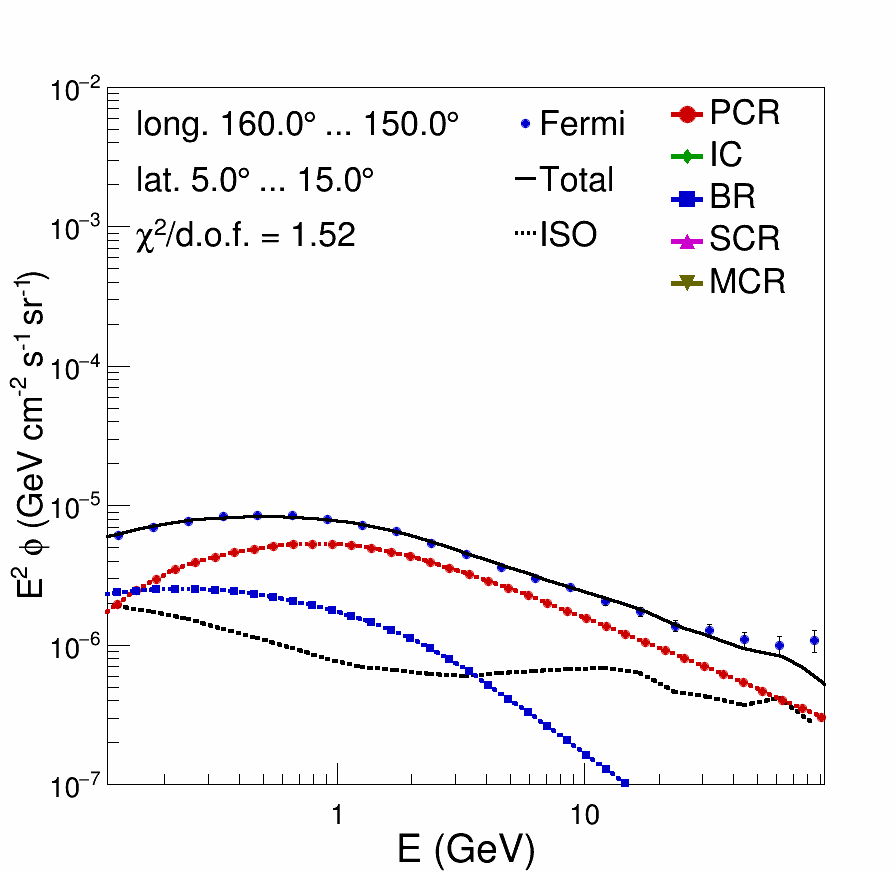}
\includegraphics[width=0.16\textwidth,height=0.16\textwidth,clip]{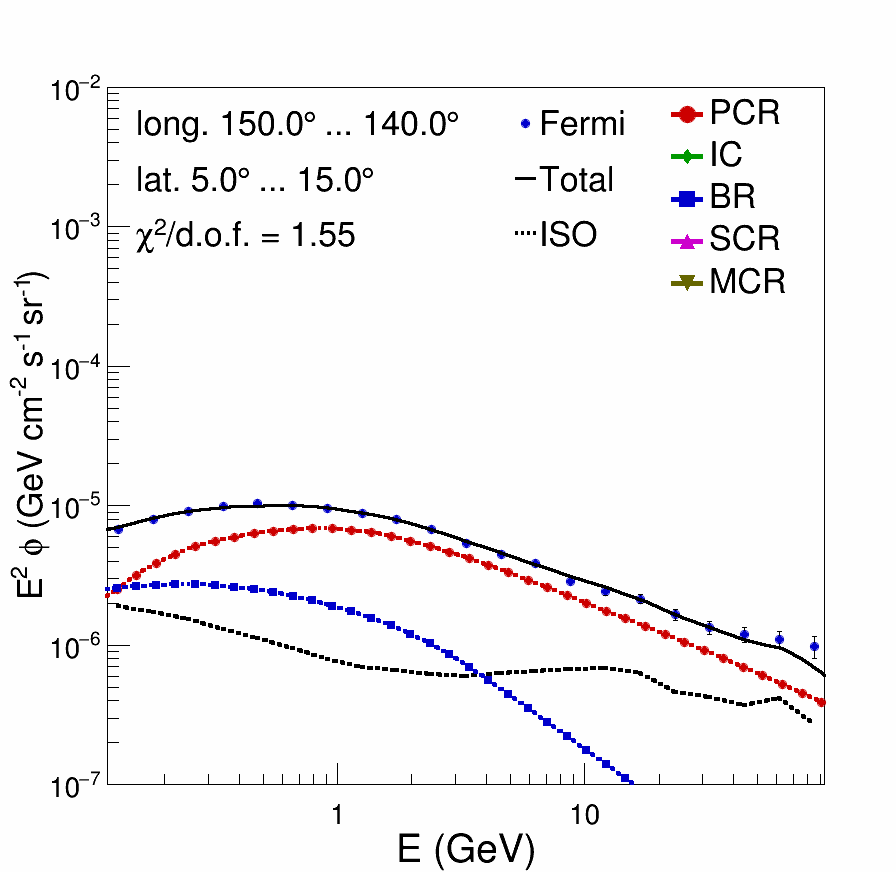}
\includegraphics[width=0.16\textwidth,height=0.16\textwidth,clip]{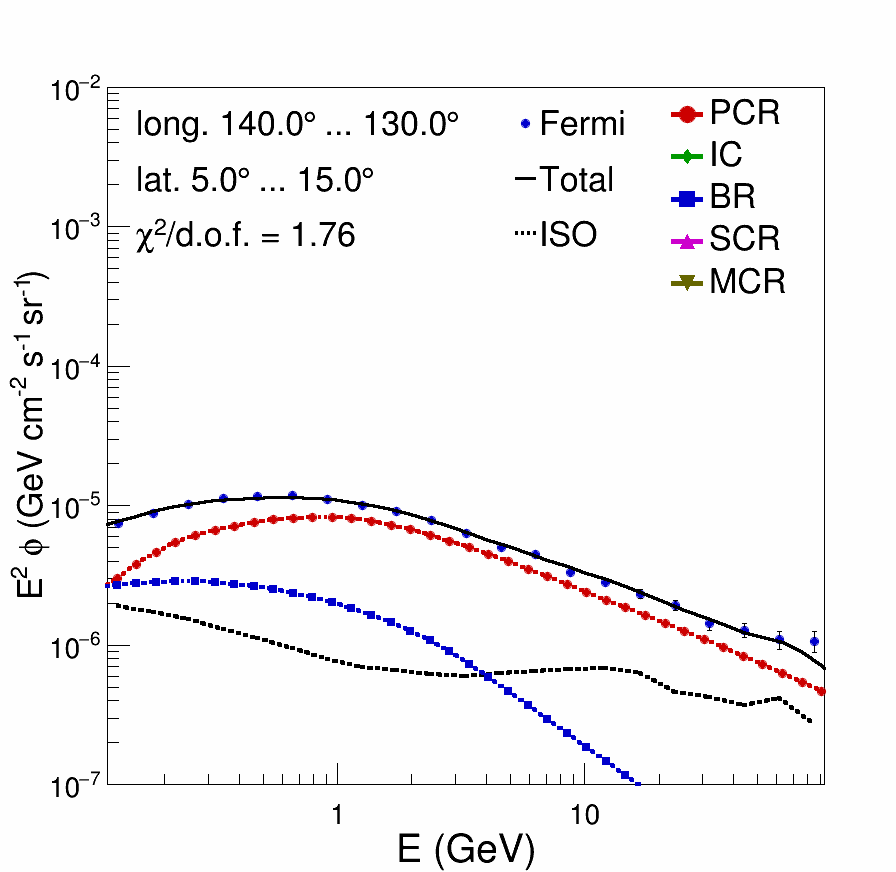}
\includegraphics[width=0.16\textwidth,height=0.16\textwidth,clip]{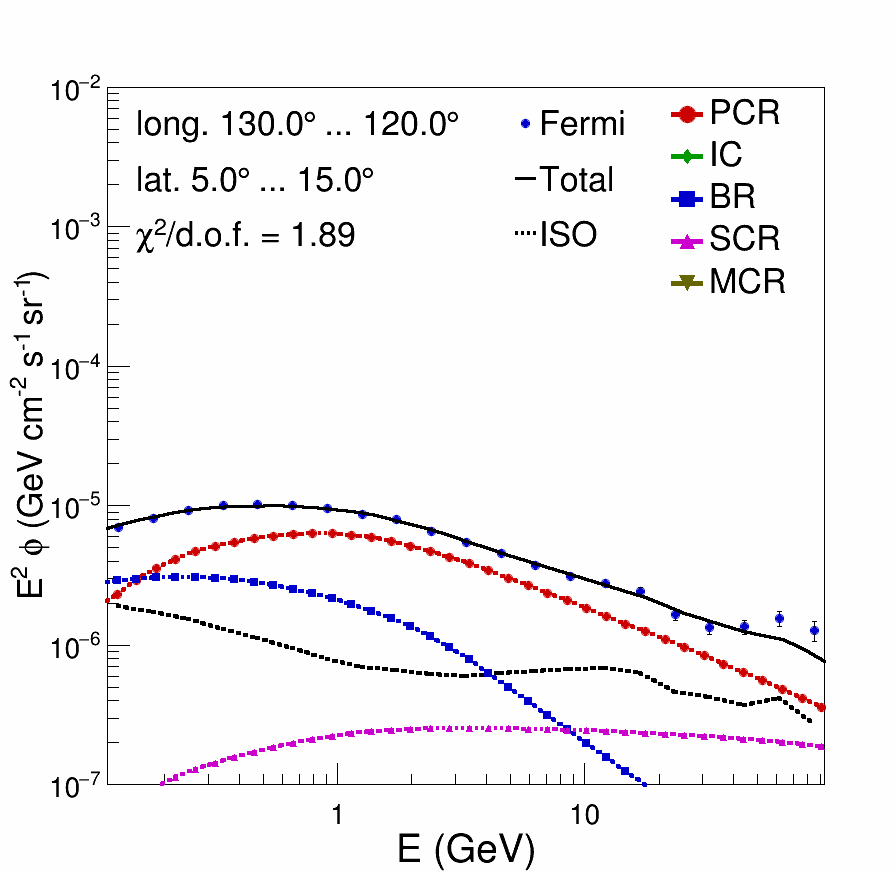}
\includegraphics[width=0.16\textwidth,height=0.16\textwidth,clip]{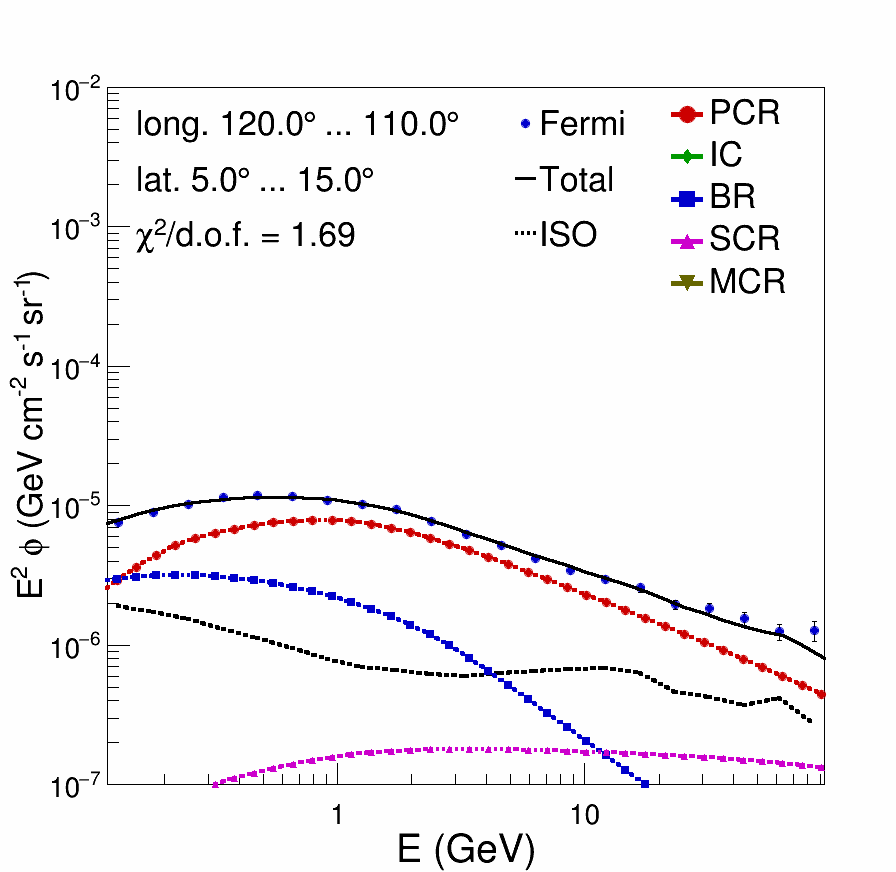}
\includegraphics[width=0.16\textwidth,height=0.16\textwidth,clip]{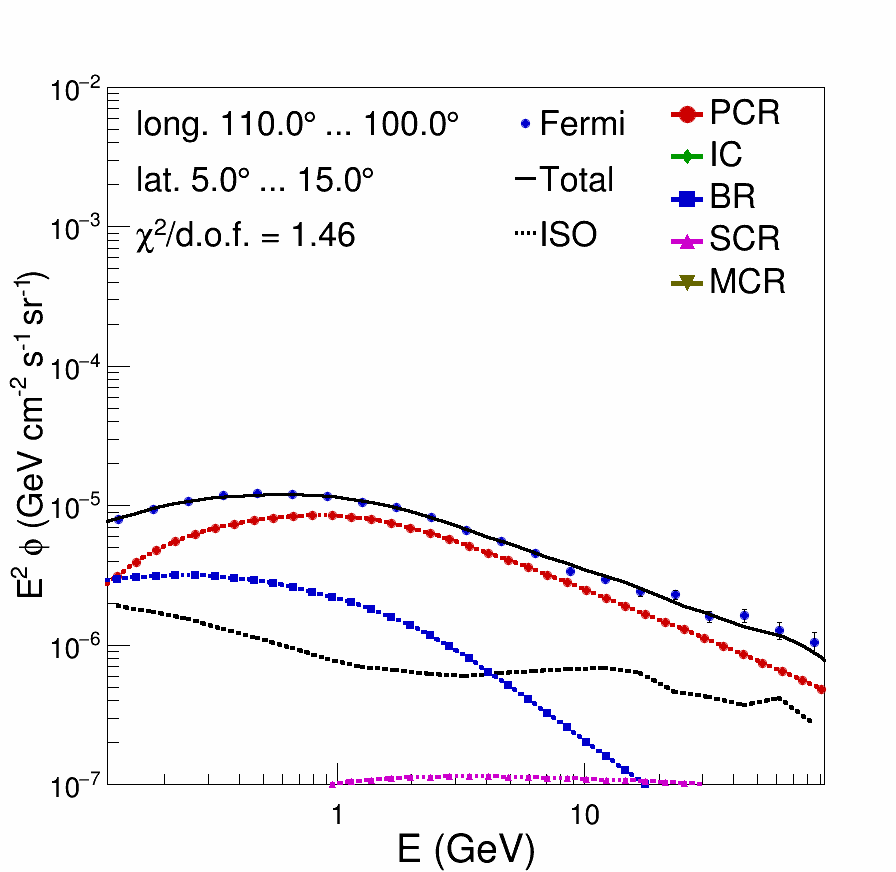}
\includegraphics[width=0.16\textwidth,height=0.16\textwidth,clip]{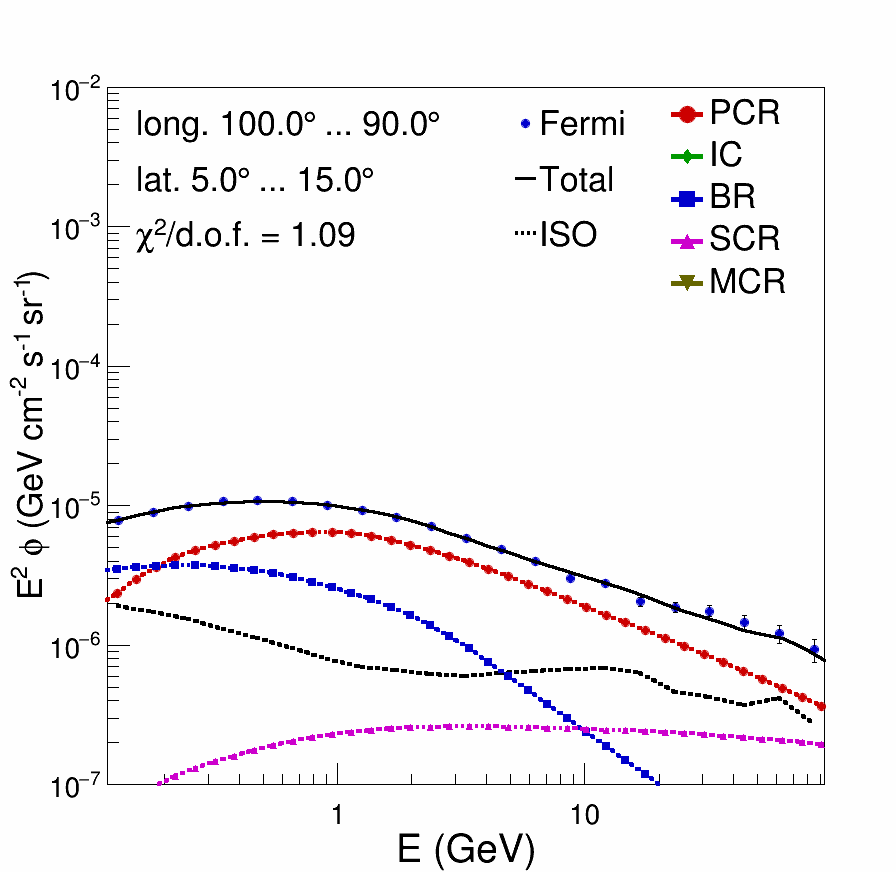}
\includegraphics[width=0.16\textwidth,height=0.16\textwidth,clip]{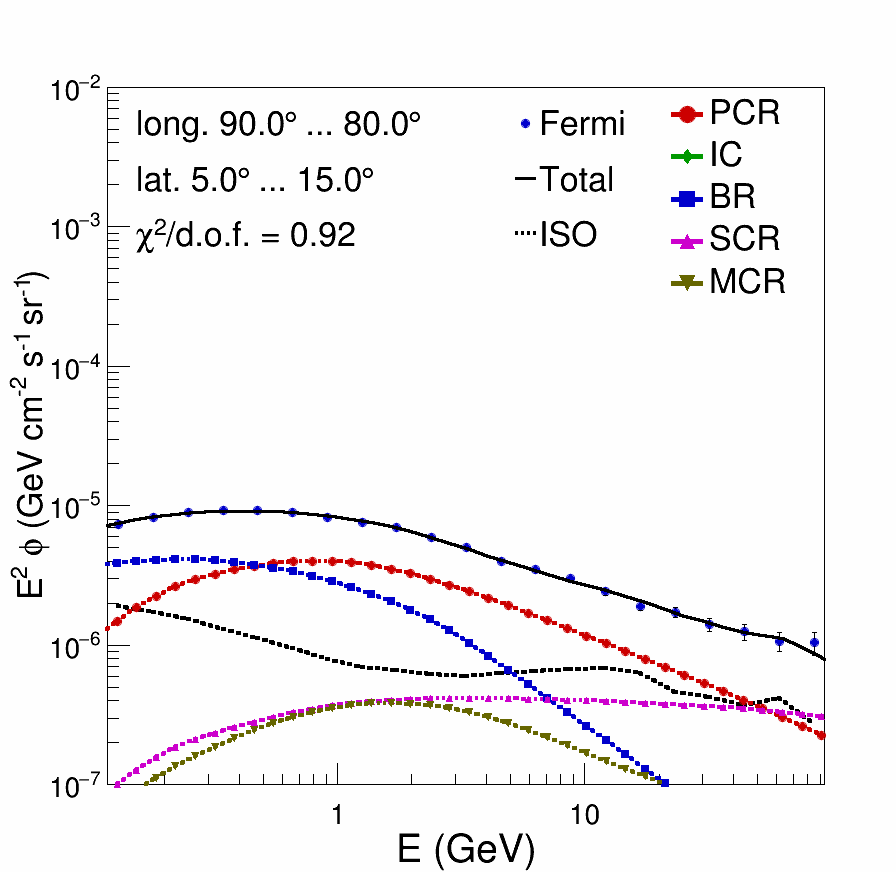}
\includegraphics[width=0.16\textwidth,height=0.16\textwidth,clip]{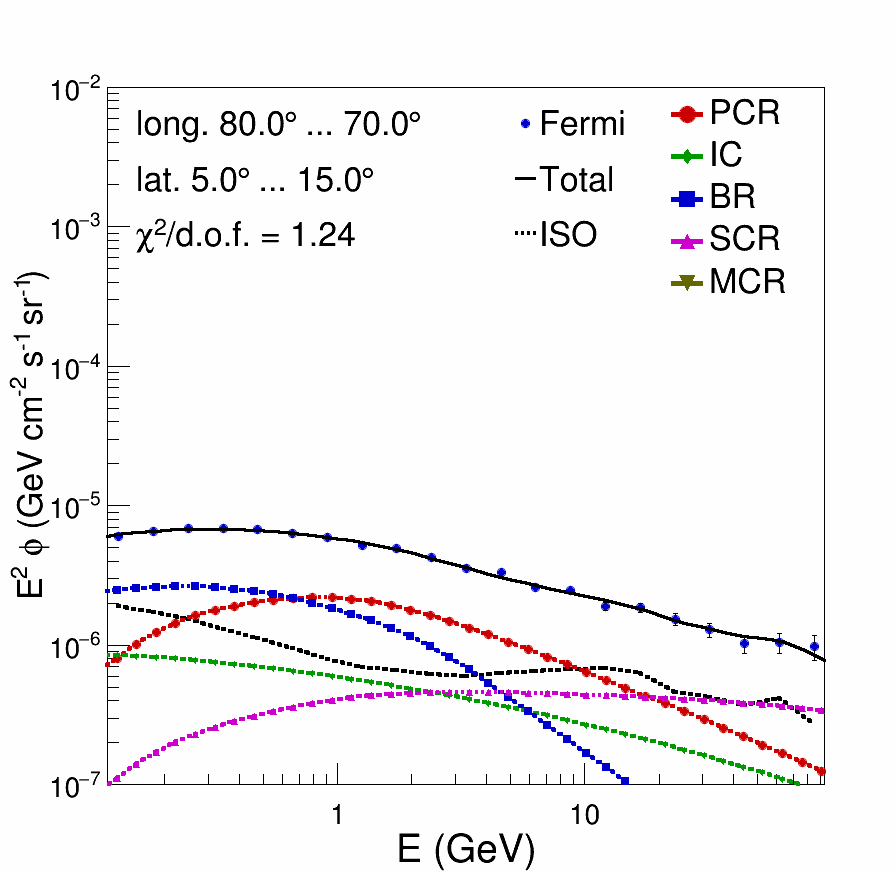}
\includegraphics[width=0.16\textwidth,height=0.16\textwidth,clip]{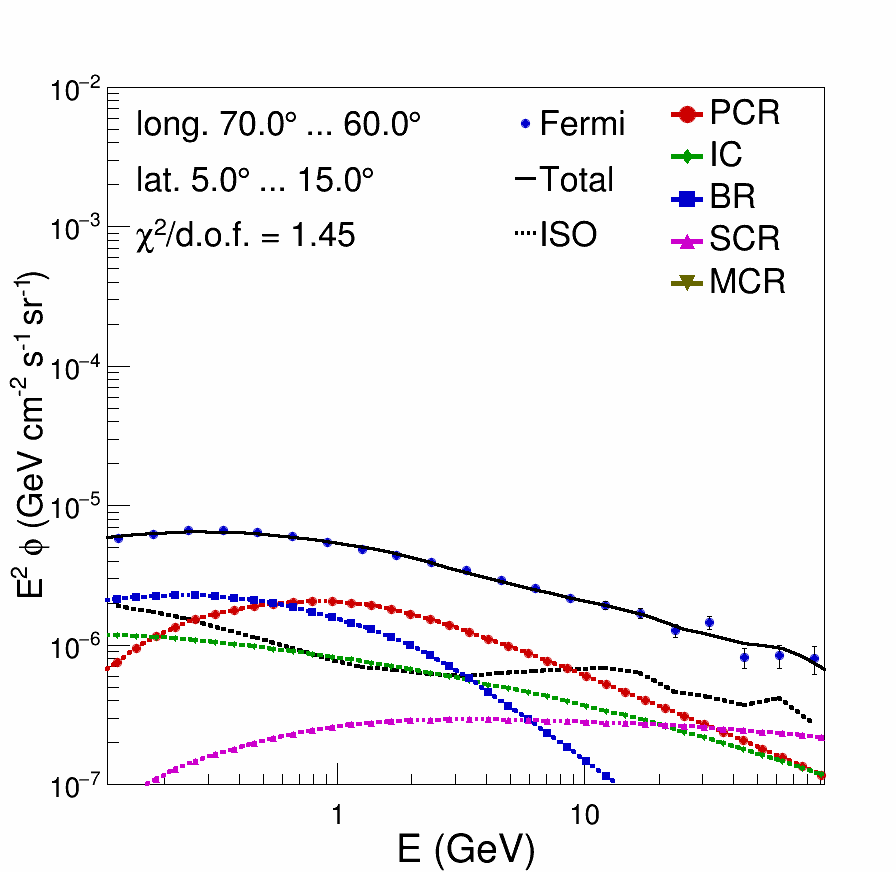}
\includegraphics[width=0.16\textwidth,height=0.16\textwidth,clip]{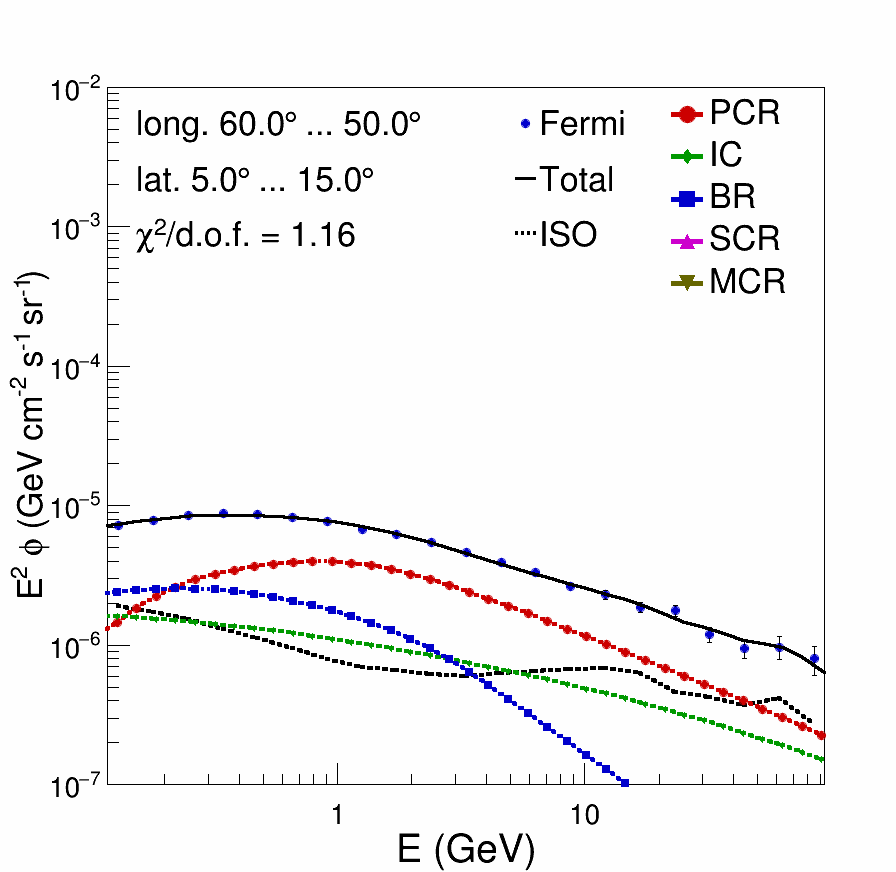}
\includegraphics[width=0.16\textwidth,height=0.16\textwidth,clip]{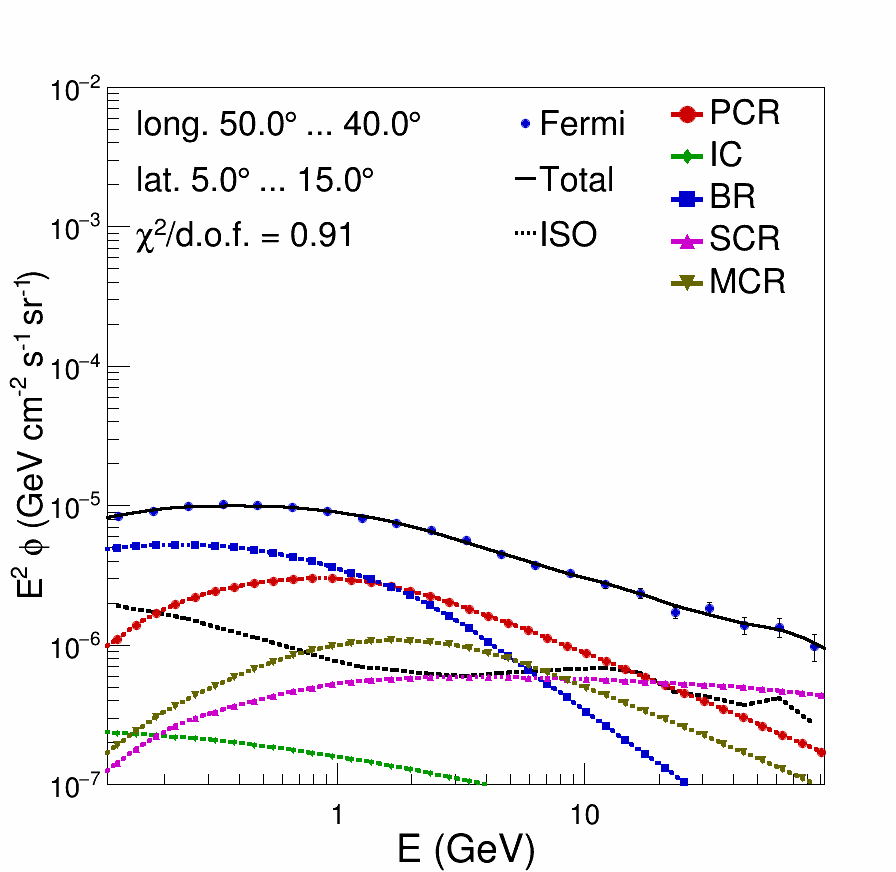}
\includegraphics[width=0.16\textwidth,height=0.16\textwidth,clip]{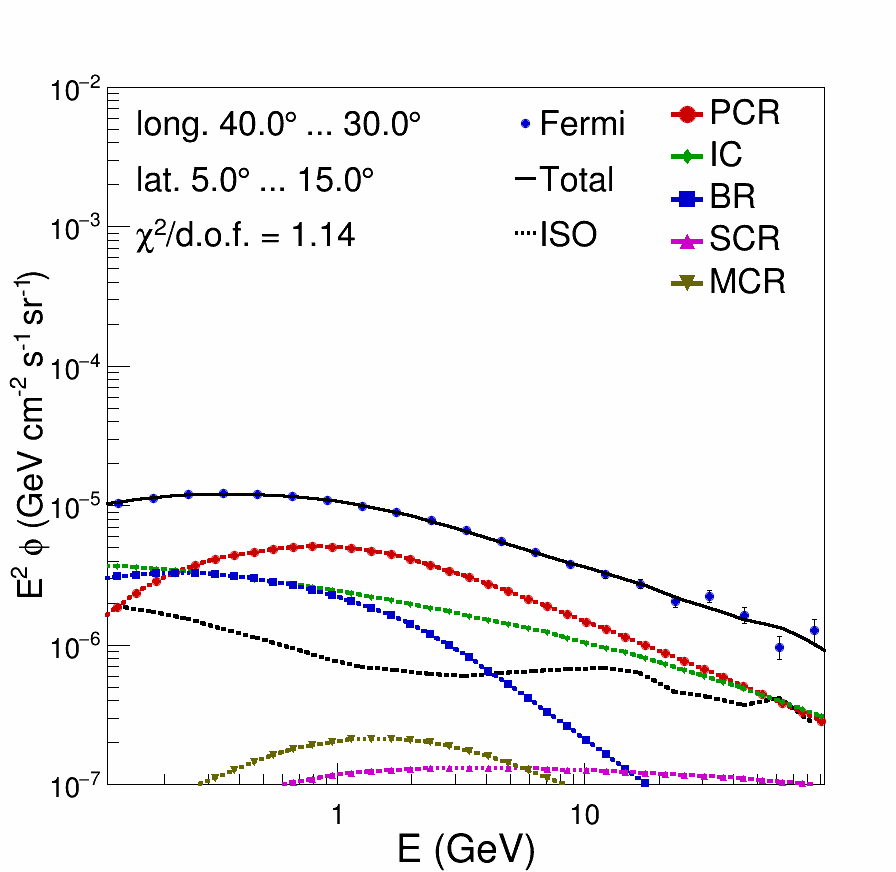}
\includegraphics[width=0.16\textwidth,height=0.16\textwidth,clip]{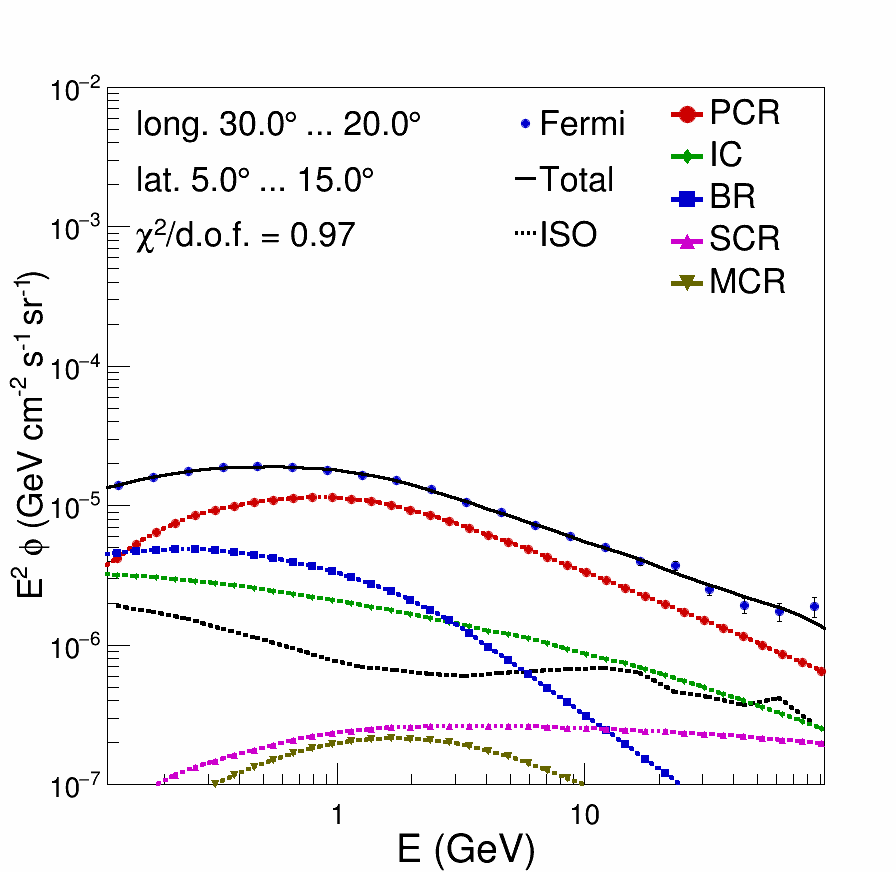}
\includegraphics[width=0.16\textwidth,height=0.16\textwidth,clip]{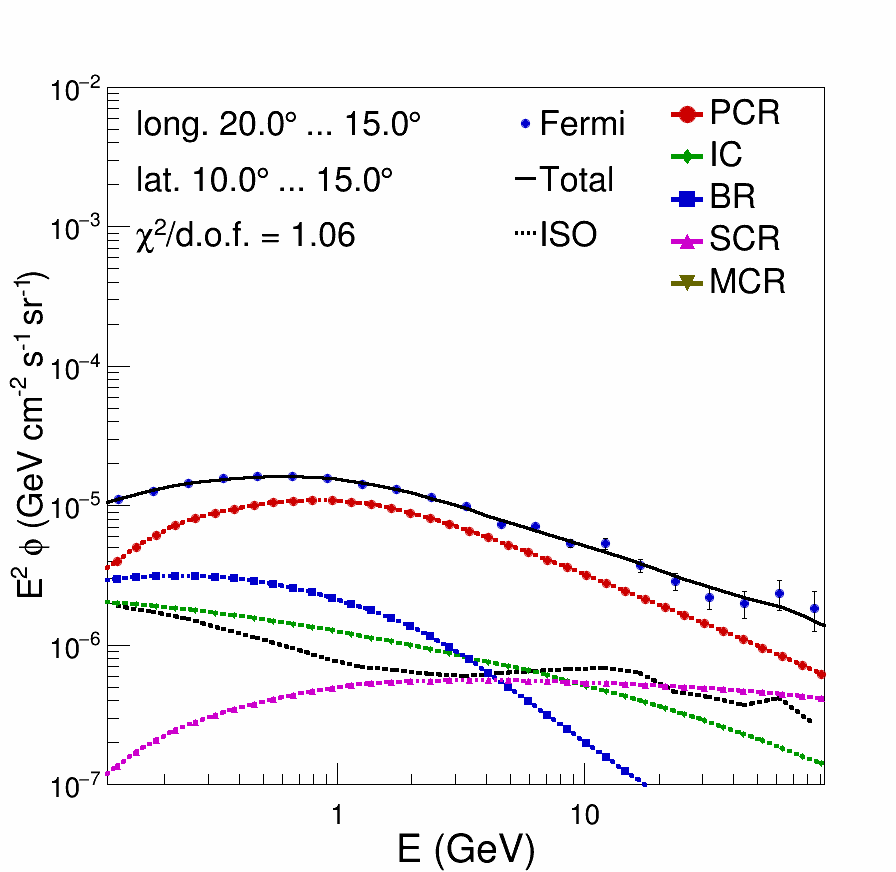}
\includegraphics[width=0.16\textwidth,height=0.16\textwidth,clip]{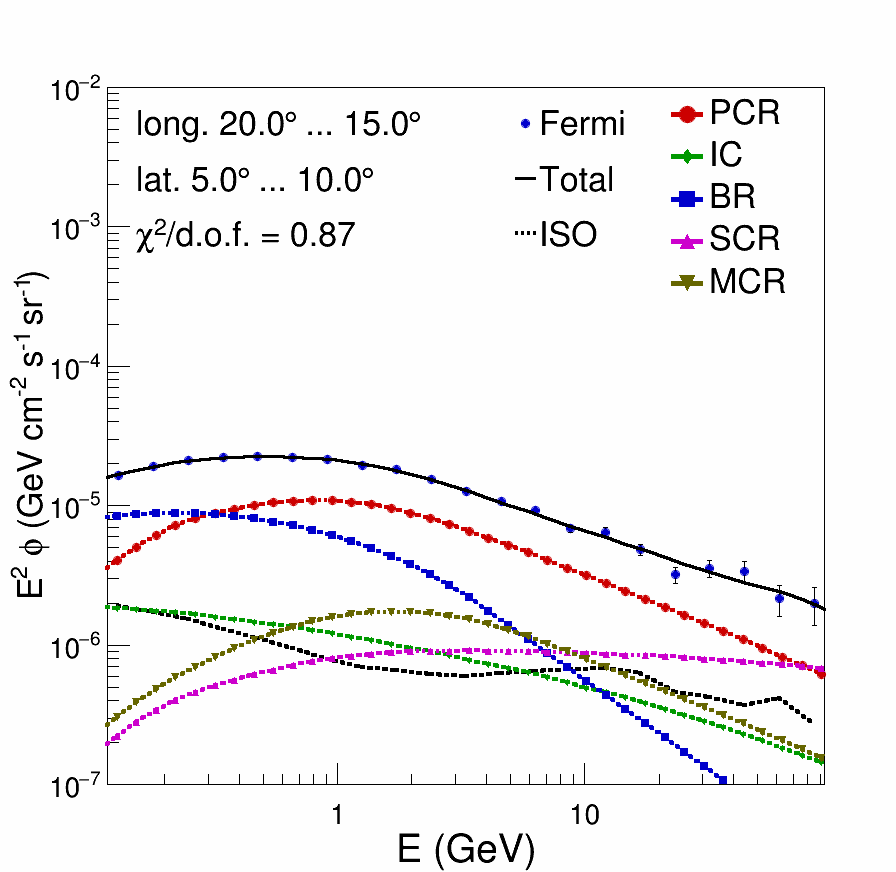}
\includegraphics[width=0.16\textwidth,height=0.16\textwidth,clip]{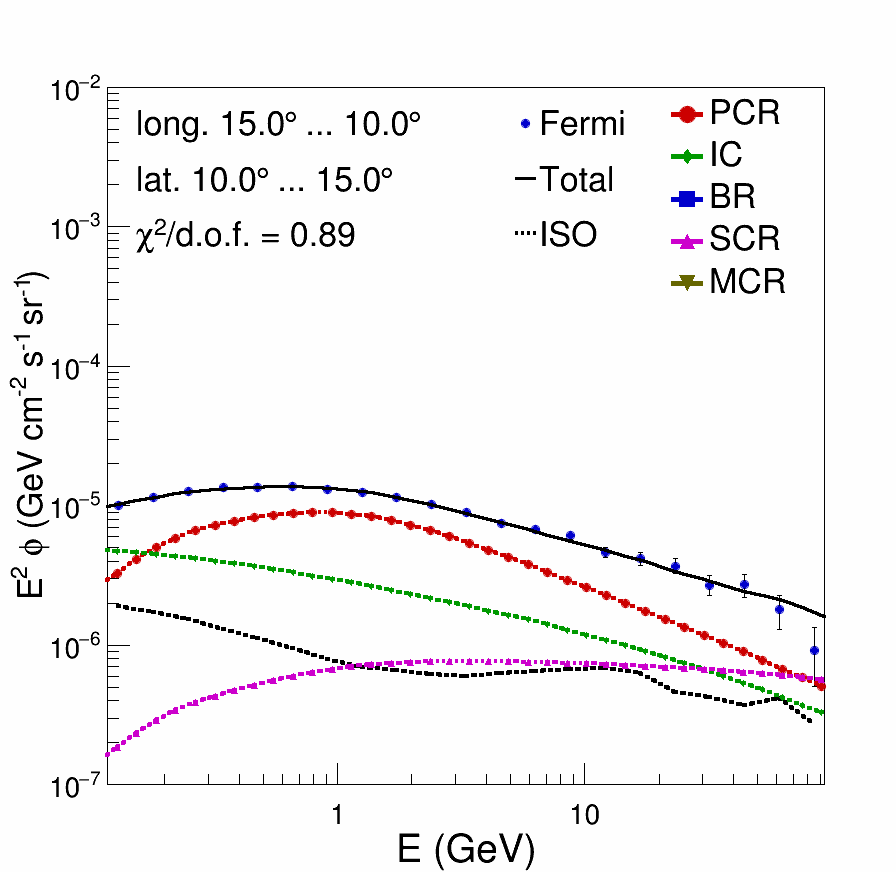}
\includegraphics[width=0.16\textwidth,height=0.16\textwidth,clip]{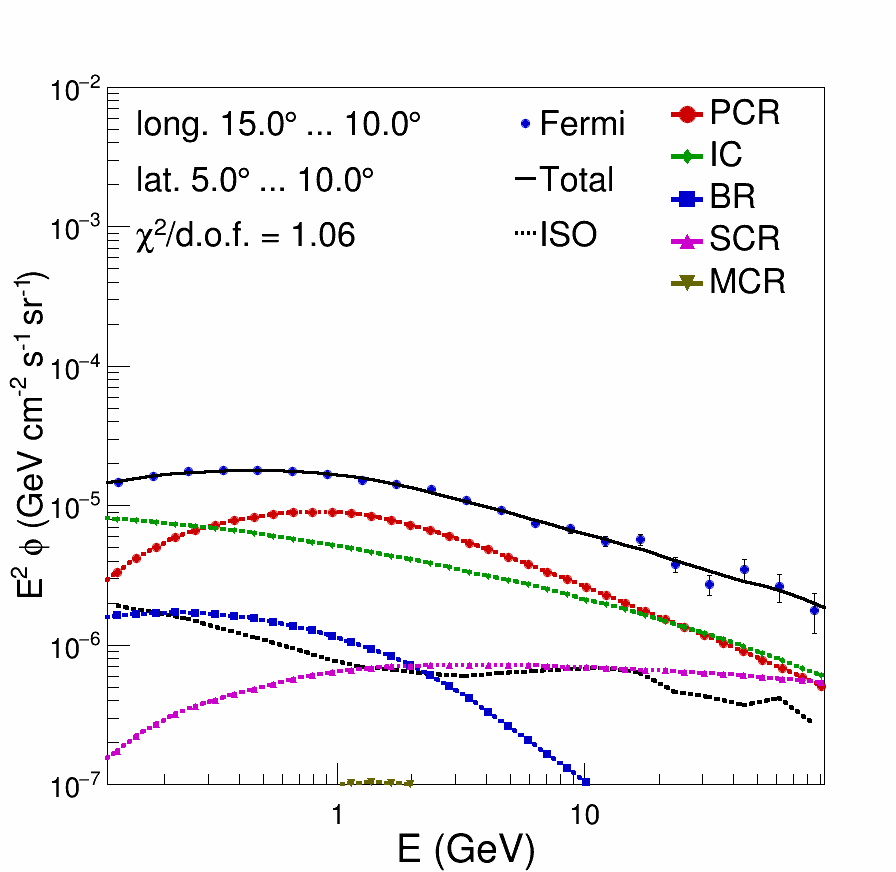}
\includegraphics[width=0.16\textwidth,height=0.16\textwidth,clip]{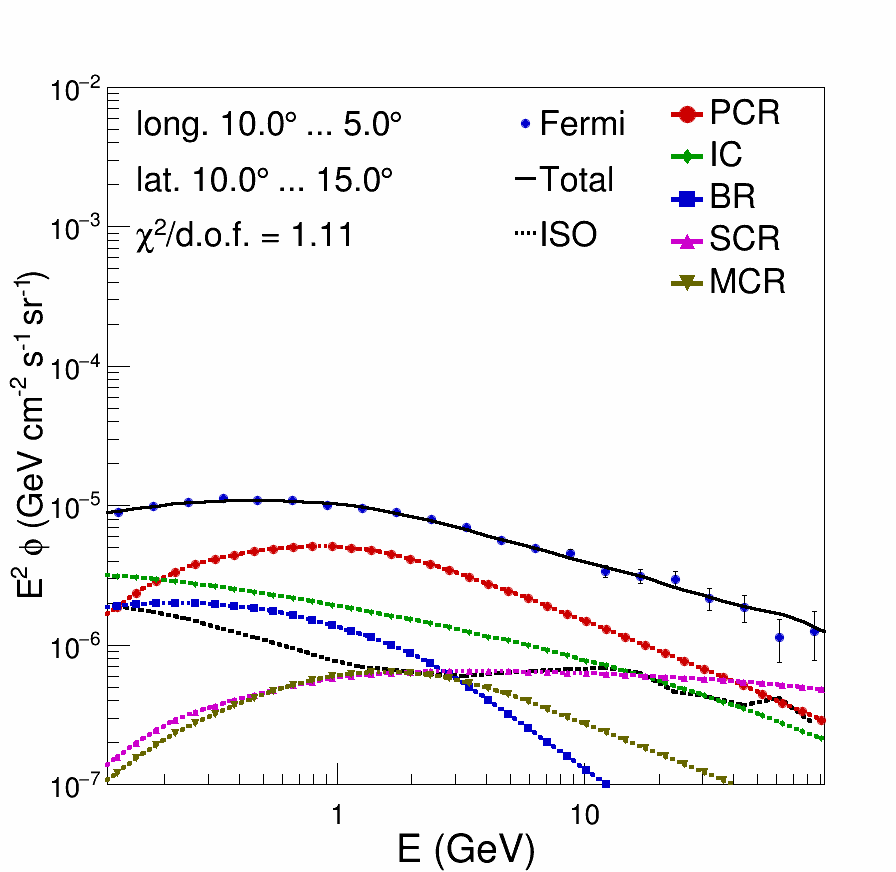}
\includegraphics[width=0.16\textwidth,height=0.16\textwidth,clip]{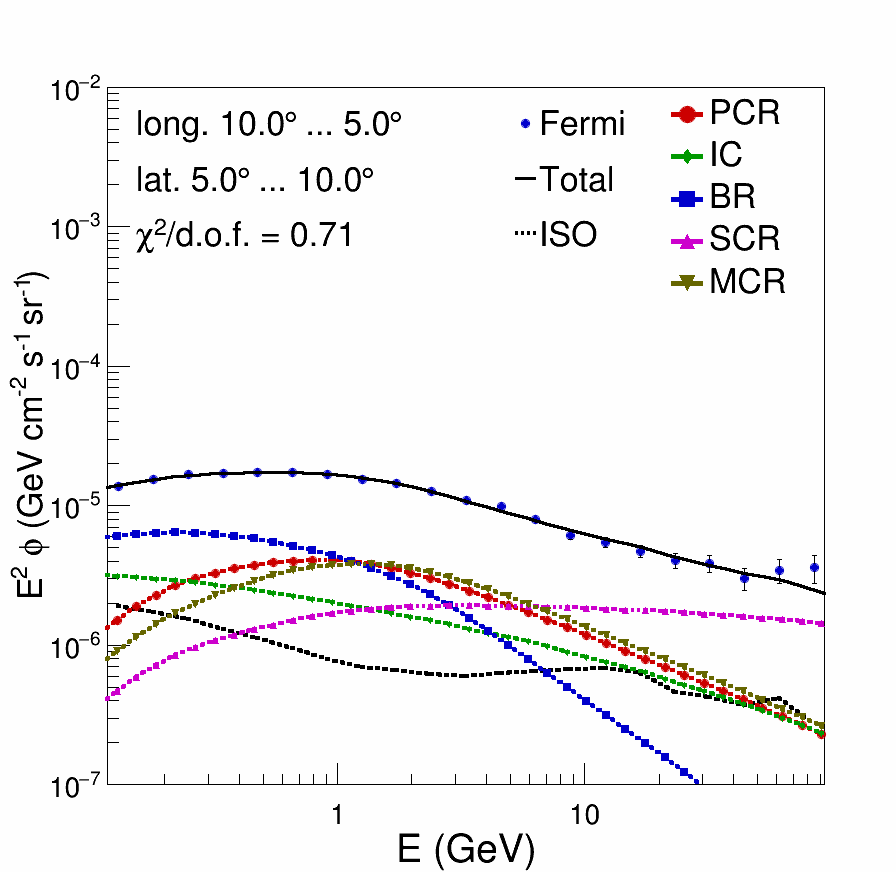}
\includegraphics[width=0.16\textwidth,height=0.16\textwidth,clip]{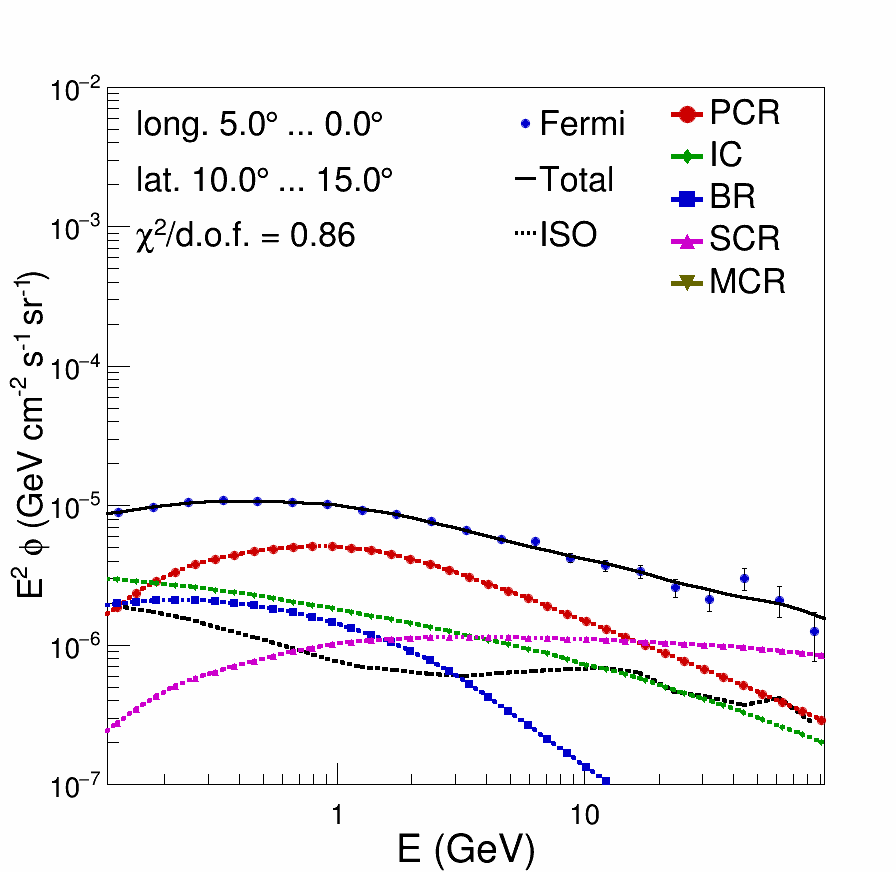}
\includegraphics[width=0.16\textwidth,height=0.16\textwidth,clip]{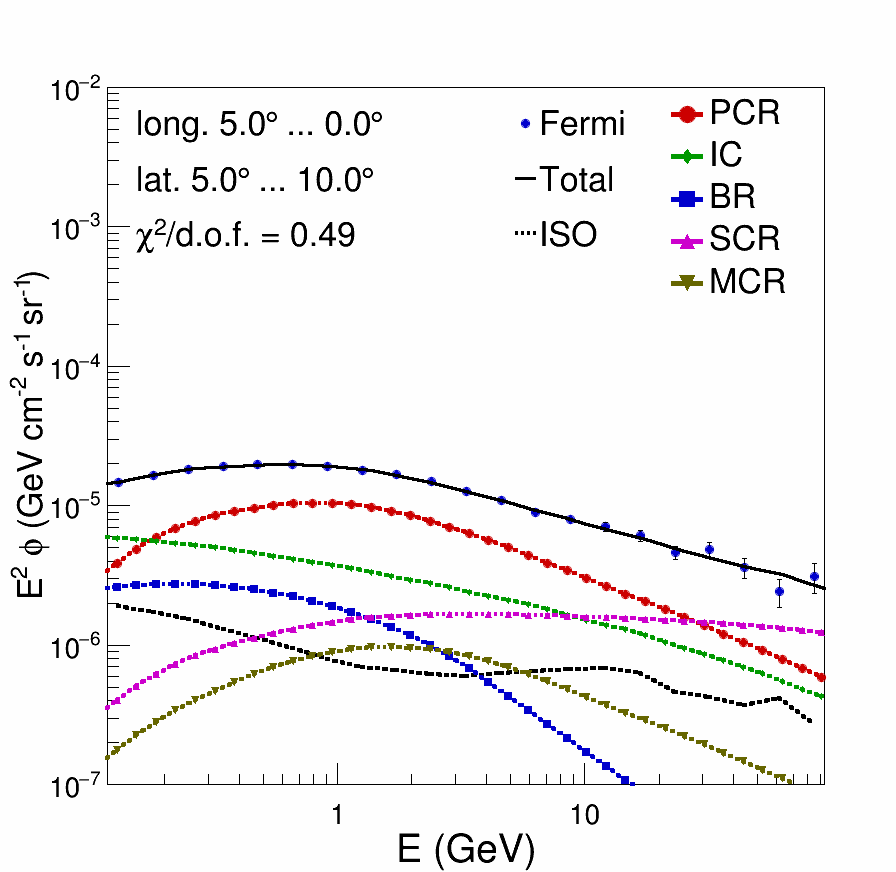}
\includegraphics[width=0.16\textwidth,height=0.16\textwidth,clip]{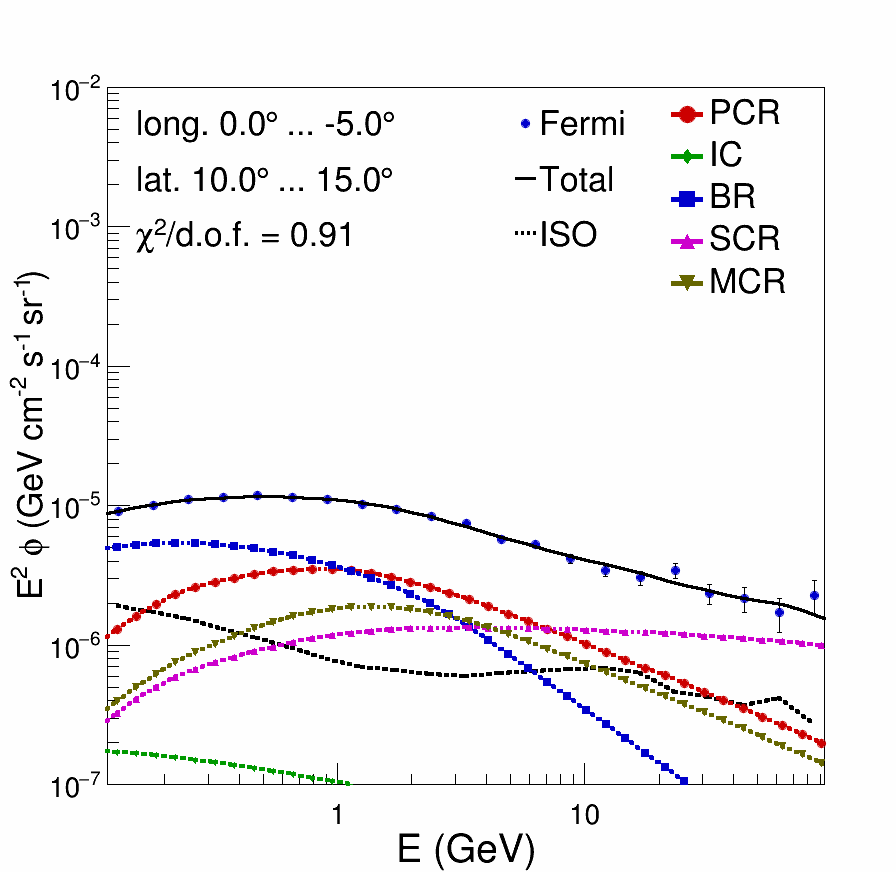}
\includegraphics[width=0.16\textwidth,height=0.16\textwidth,clip]{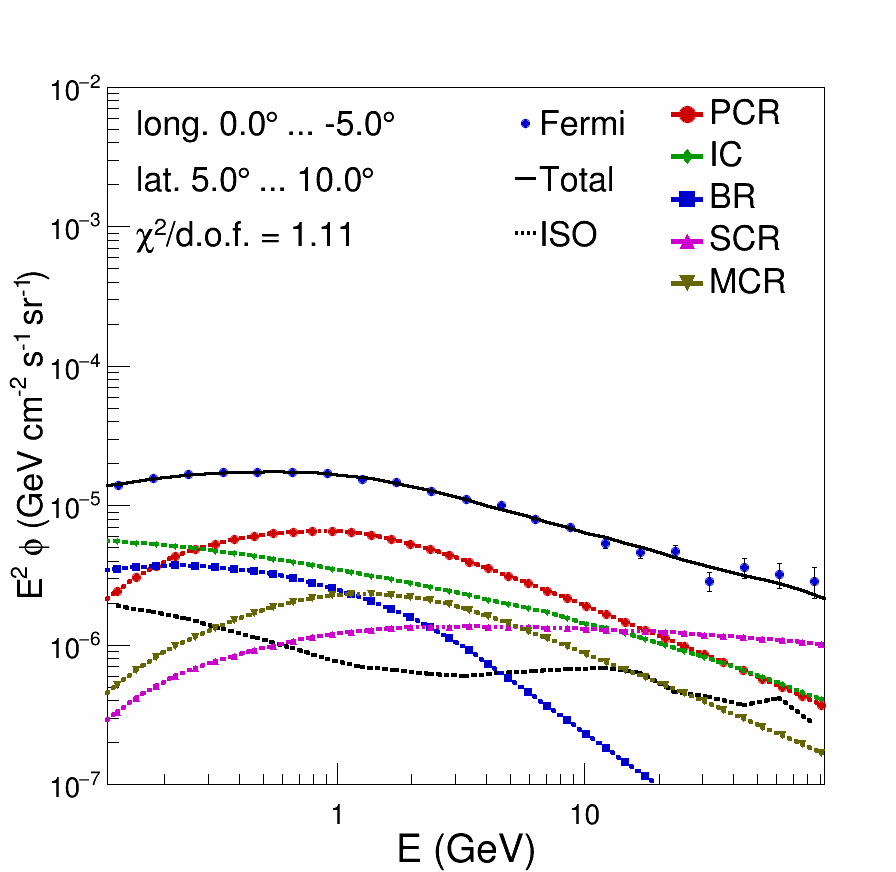}
\includegraphics[width=0.16\textwidth,height=0.16\textwidth,clip]{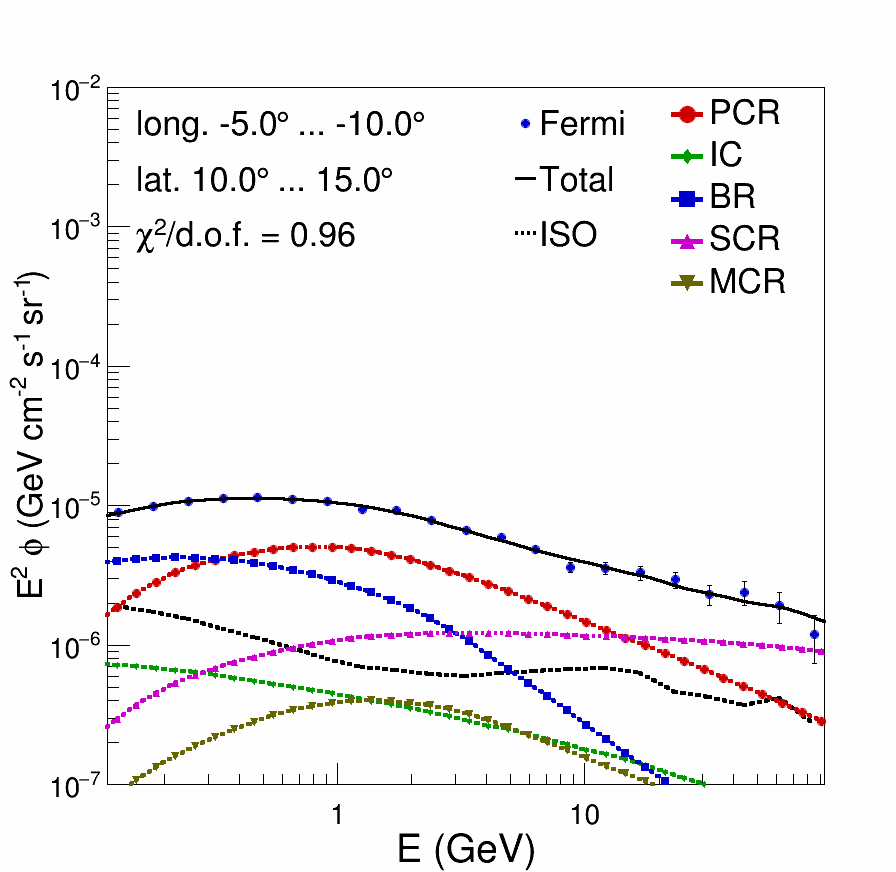}
\includegraphics[width=0.16\textwidth,height=0.16\textwidth,clip]{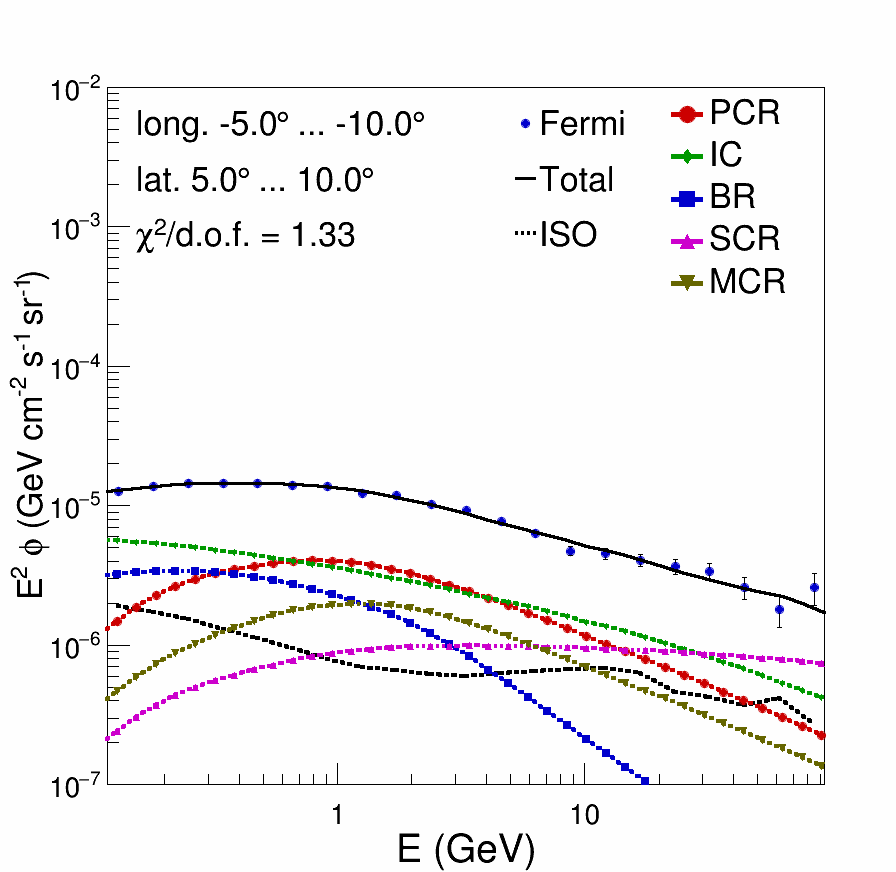}
\includegraphics[width=0.16\textwidth,height=0.16\textwidth,clip]{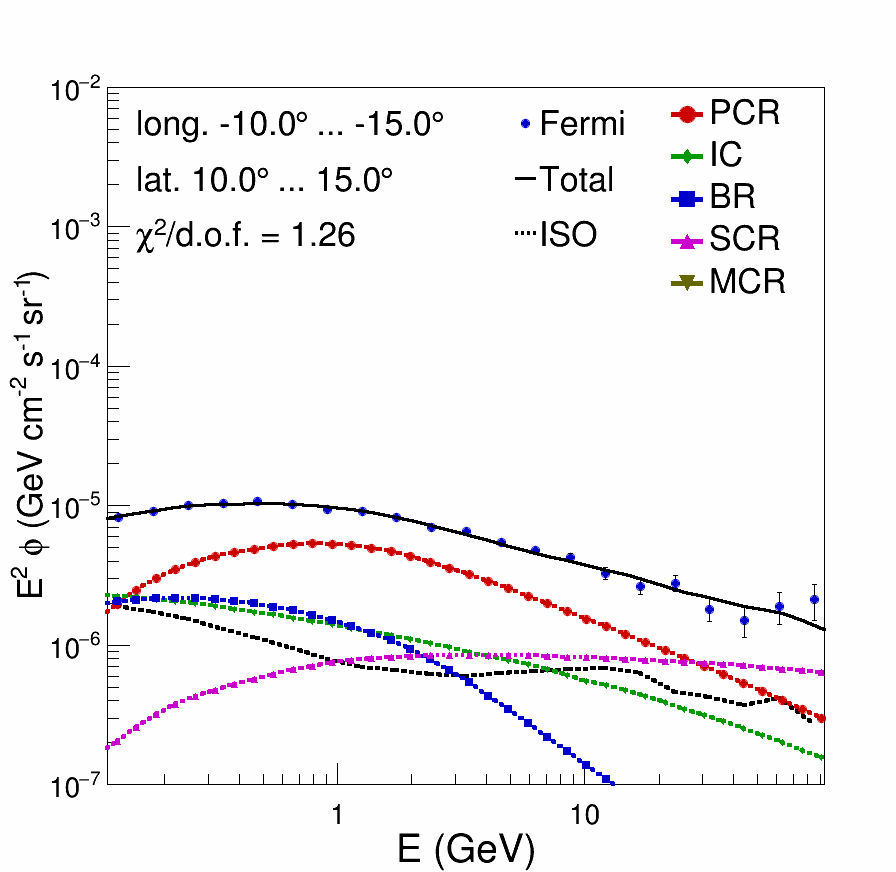}
\includegraphics[width=0.16\textwidth,height=0.16\textwidth,clip]{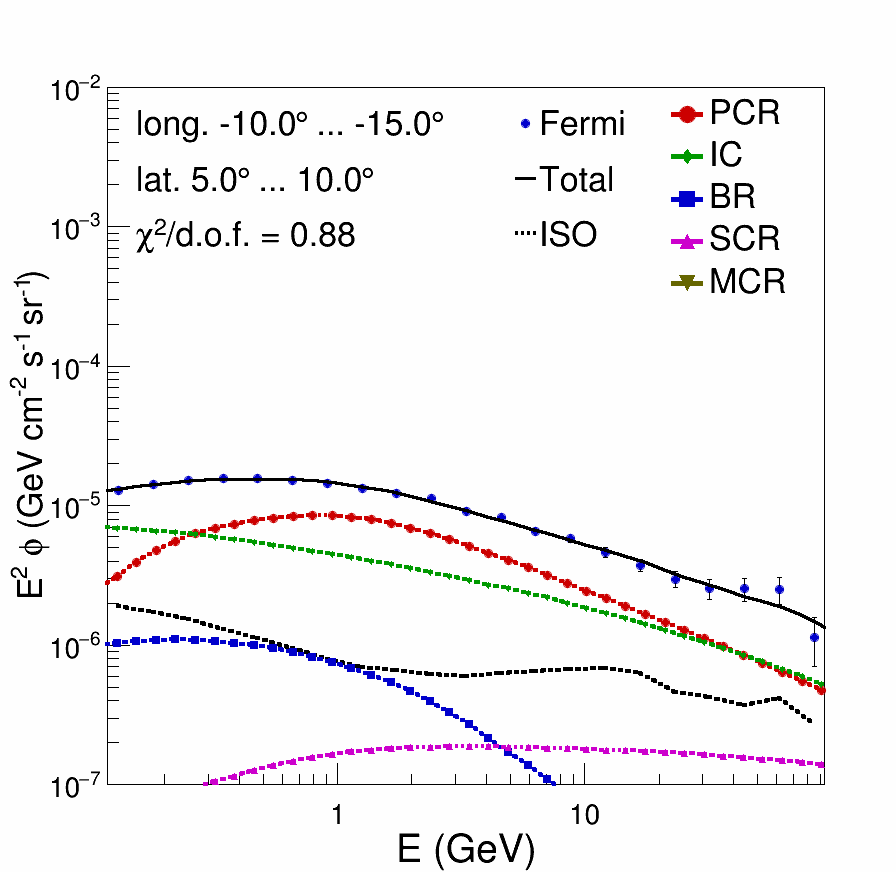}
\includegraphics[width=0.16\textwidth,height=0.16\textwidth,clip]{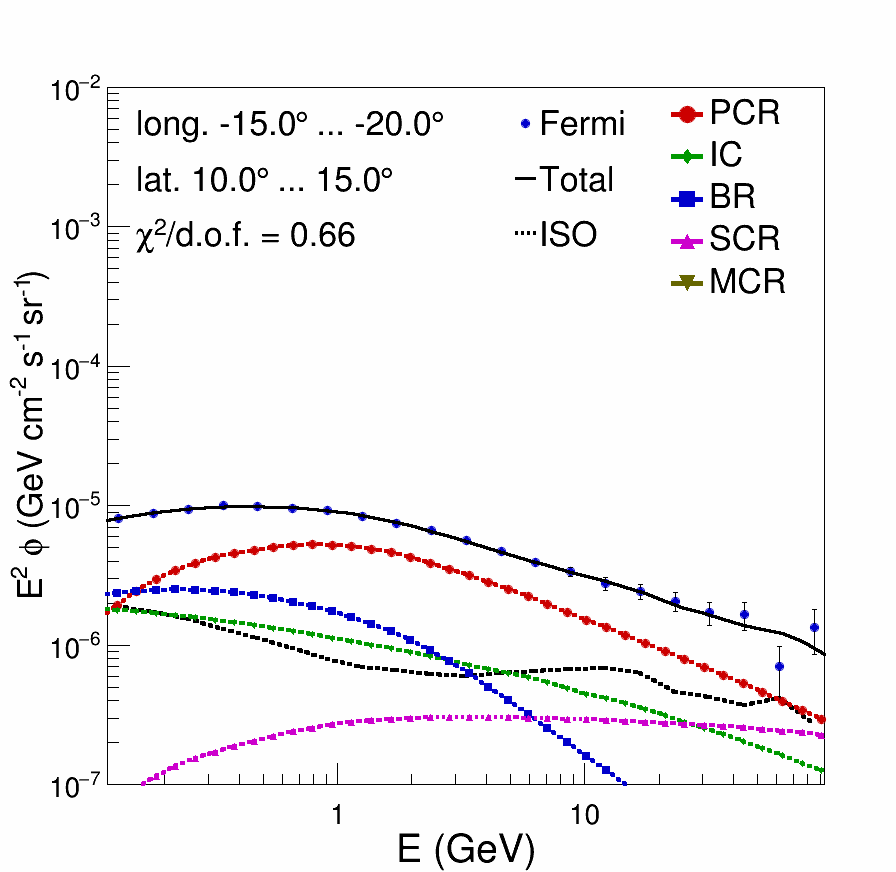}
\includegraphics[width=0.16\textwidth,height=0.16\textwidth,clip]{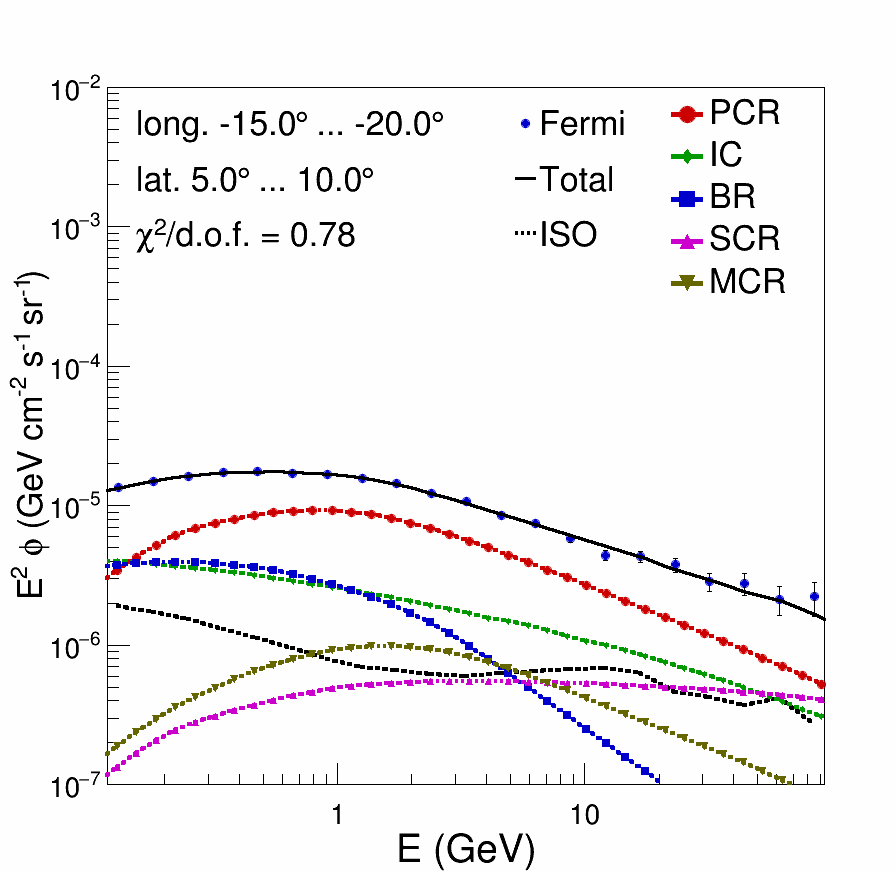}
\includegraphics[width=0.16\textwidth,height=0.16\textwidth,clip]{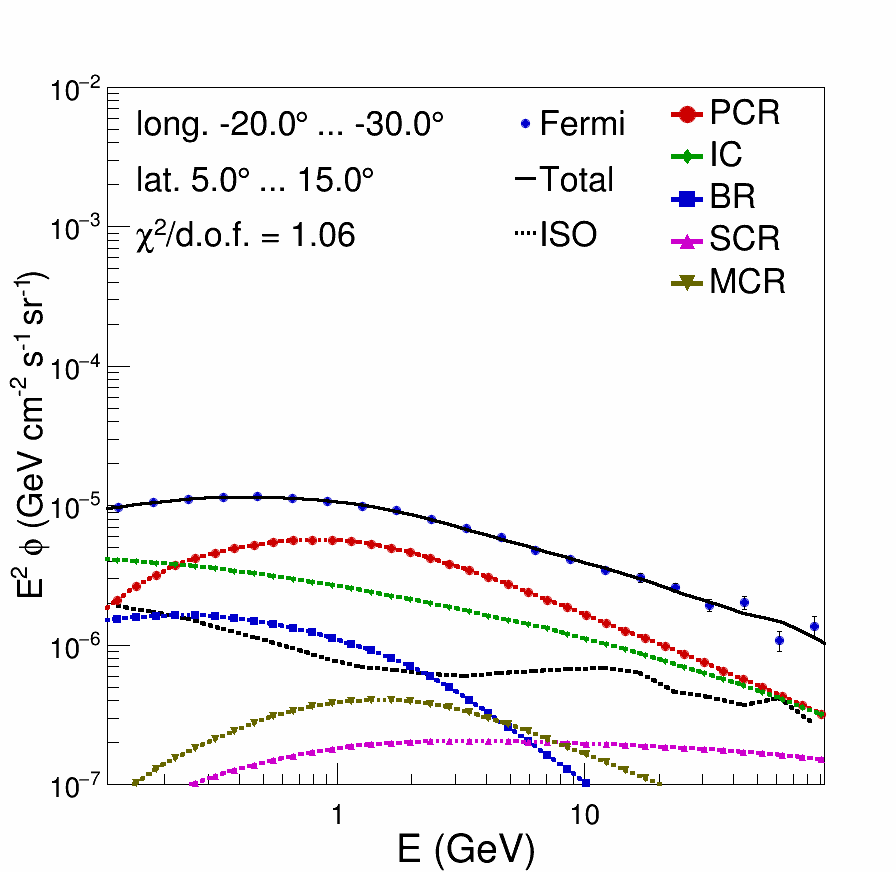}
\includegraphics[width=0.16\textwidth,height=0.16\textwidth,clip]{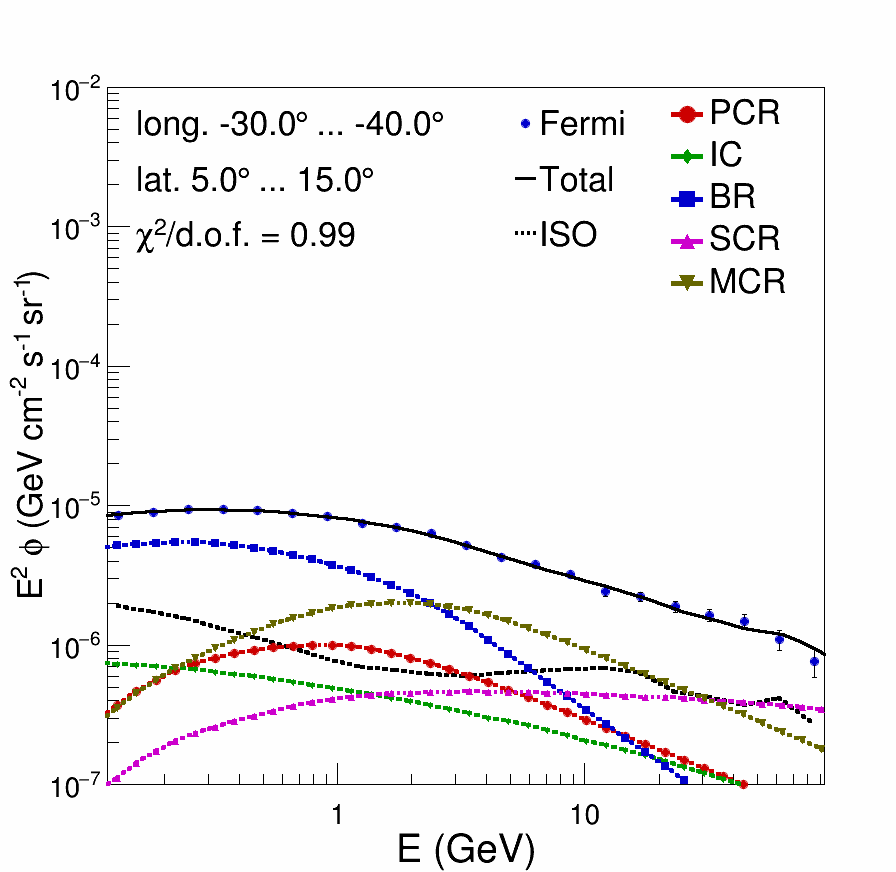}
\includegraphics[width=0.16\textwidth,height=0.16\textwidth,clip]{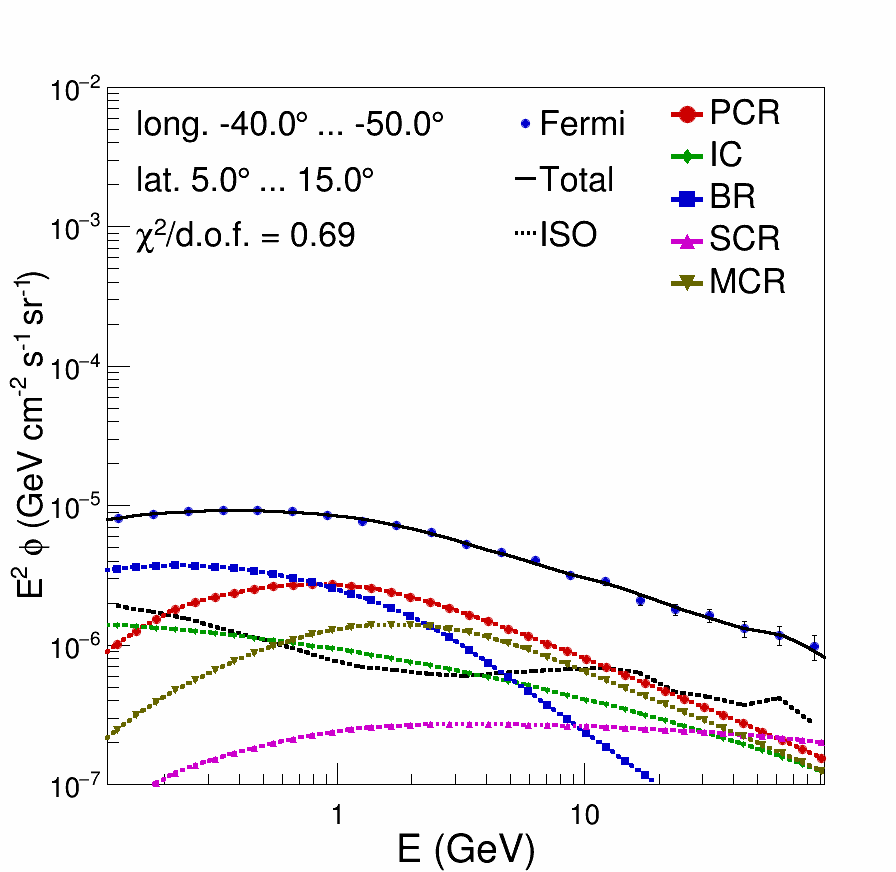}
\includegraphics[width=0.16\textwidth,height=0.16\textwidth,clip]{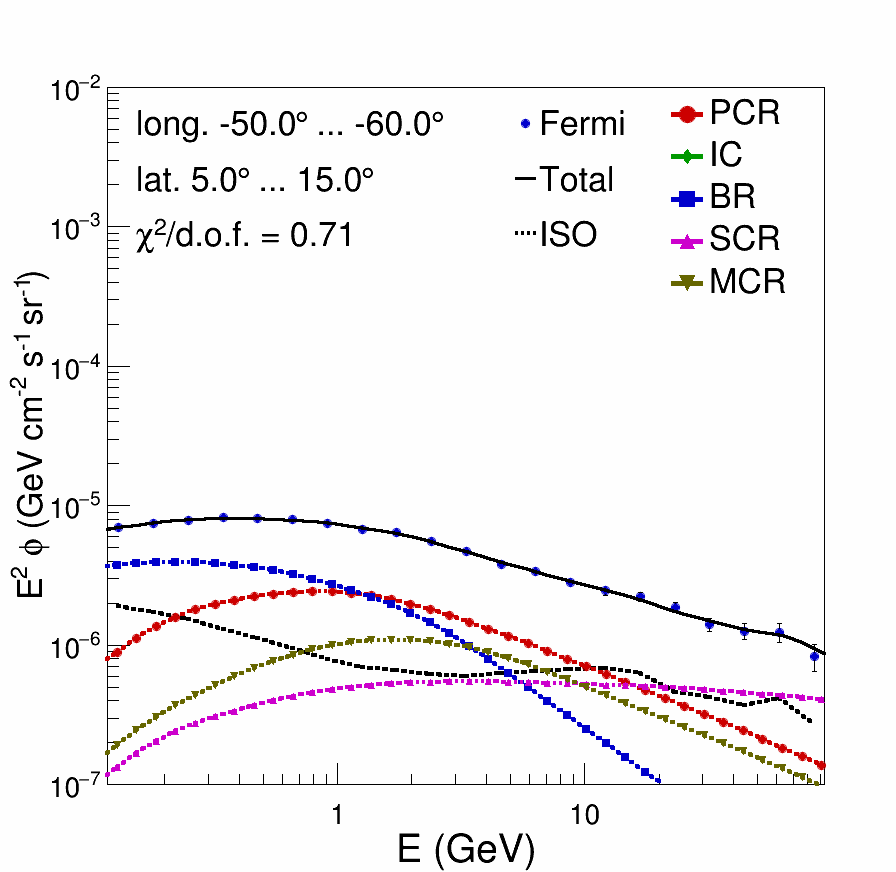}
\includegraphics[width=0.16\textwidth,height=0.16\textwidth,clip]{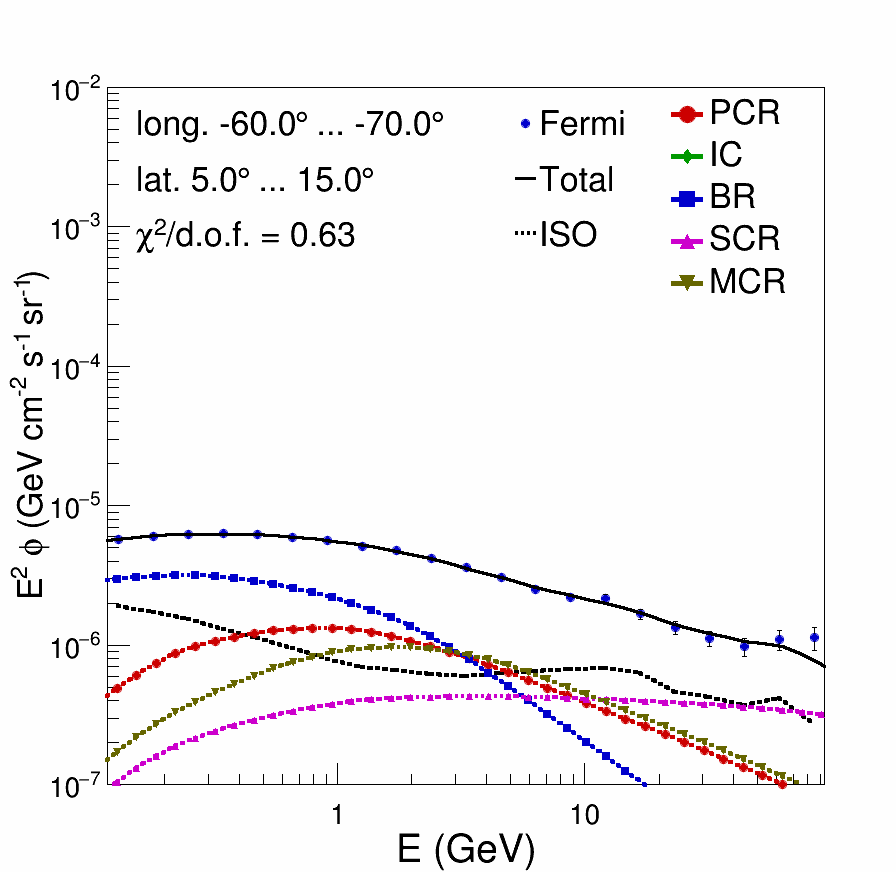}
\includegraphics[width=0.16\textwidth,height=0.16\textwidth,clip]{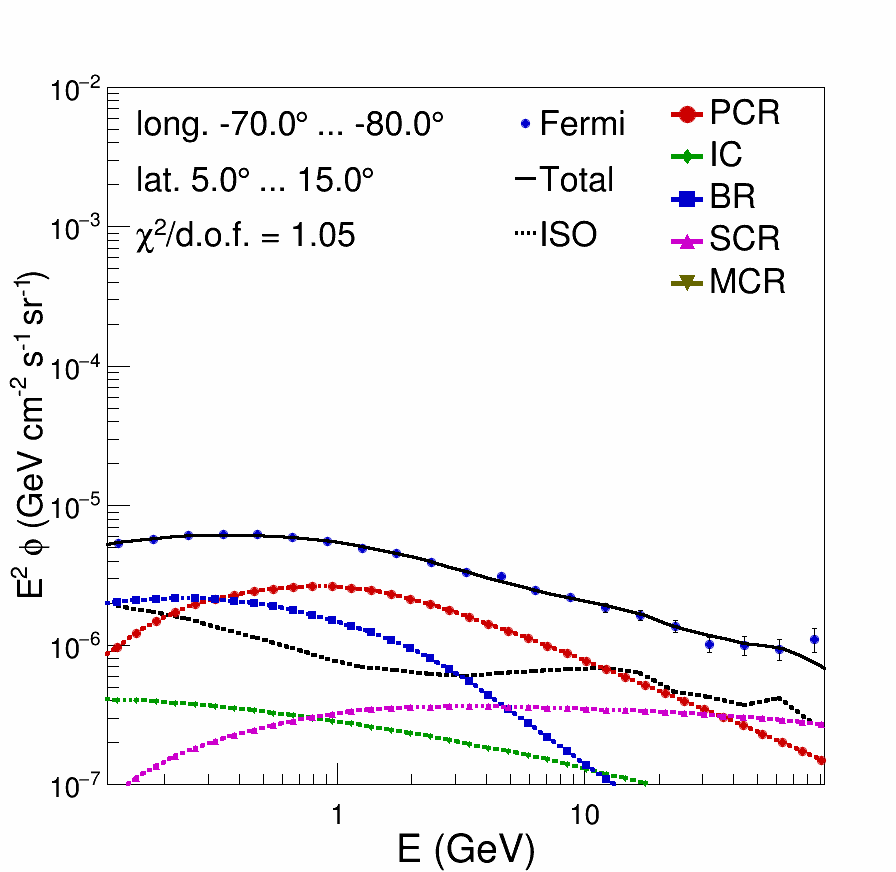}
\includegraphics[width=0.16\textwidth,height=0.16\textwidth,clip]{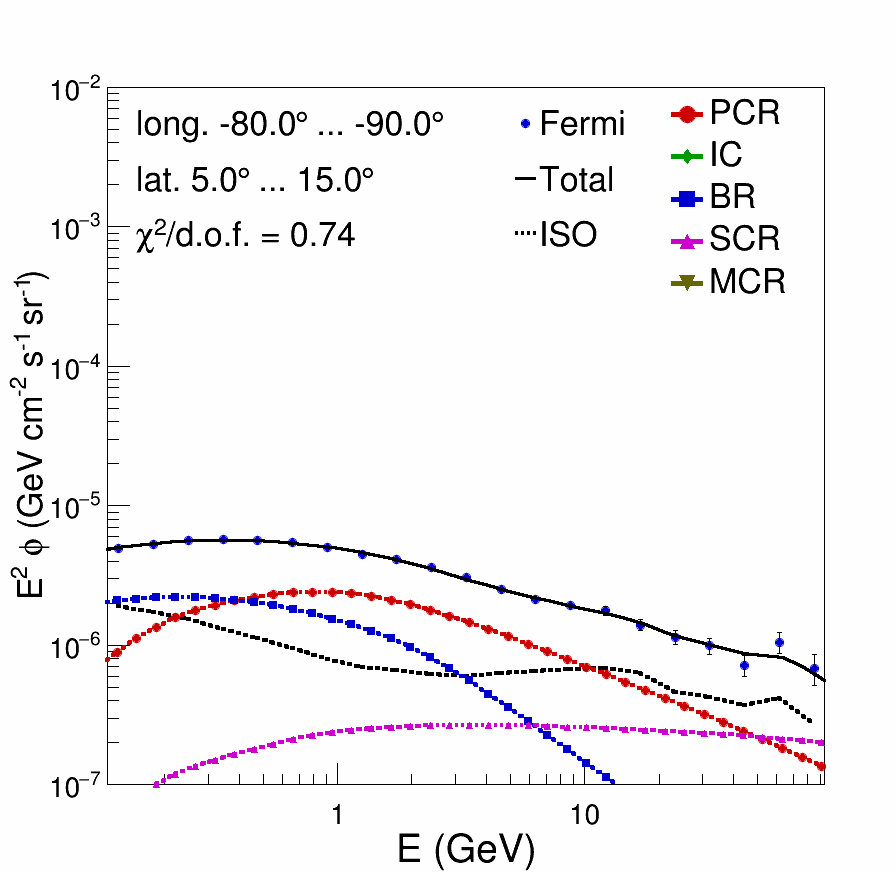}
\includegraphics[width=0.16\textwidth,height=0.16\textwidth,clip]{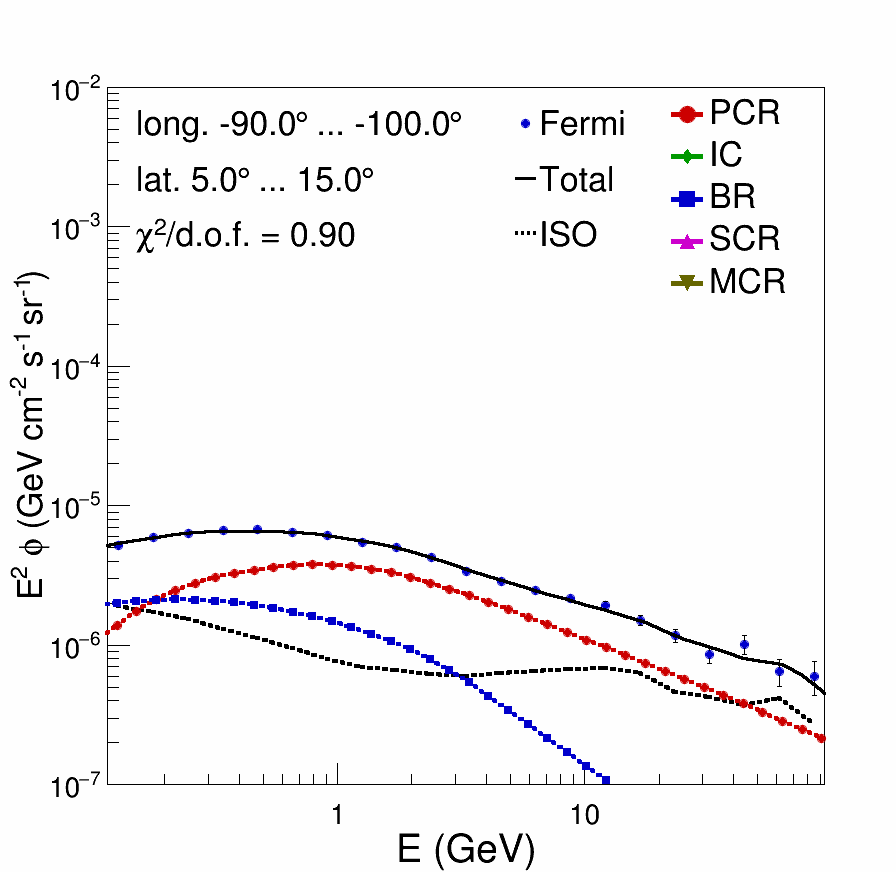}
\includegraphics[width=0.16\textwidth,height=0.16\textwidth,clip]{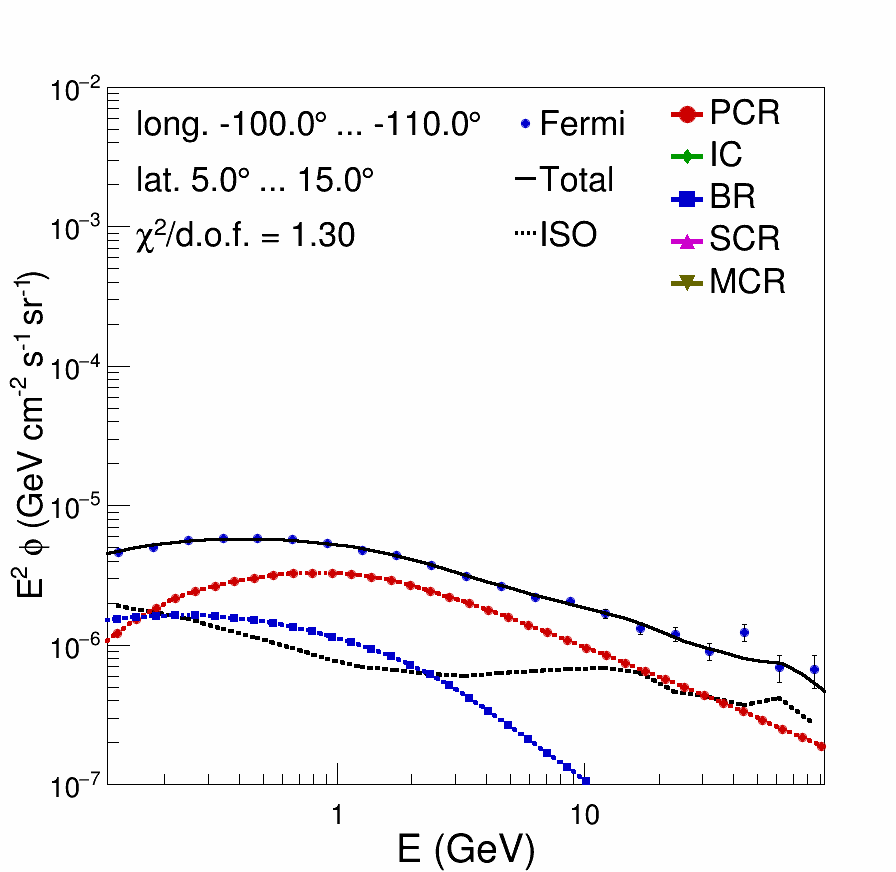}
\includegraphics[width=0.16\textwidth,height=0.16\textwidth,clip]{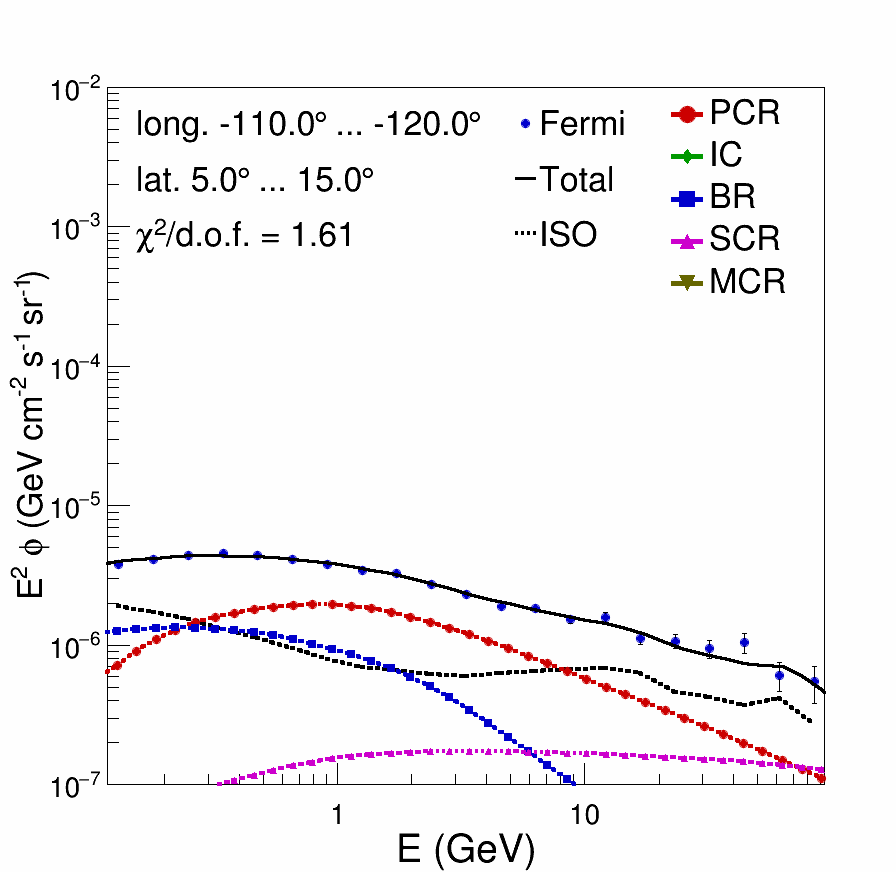}
\includegraphics[width=0.16\textwidth,height=0.16\textwidth,clip]{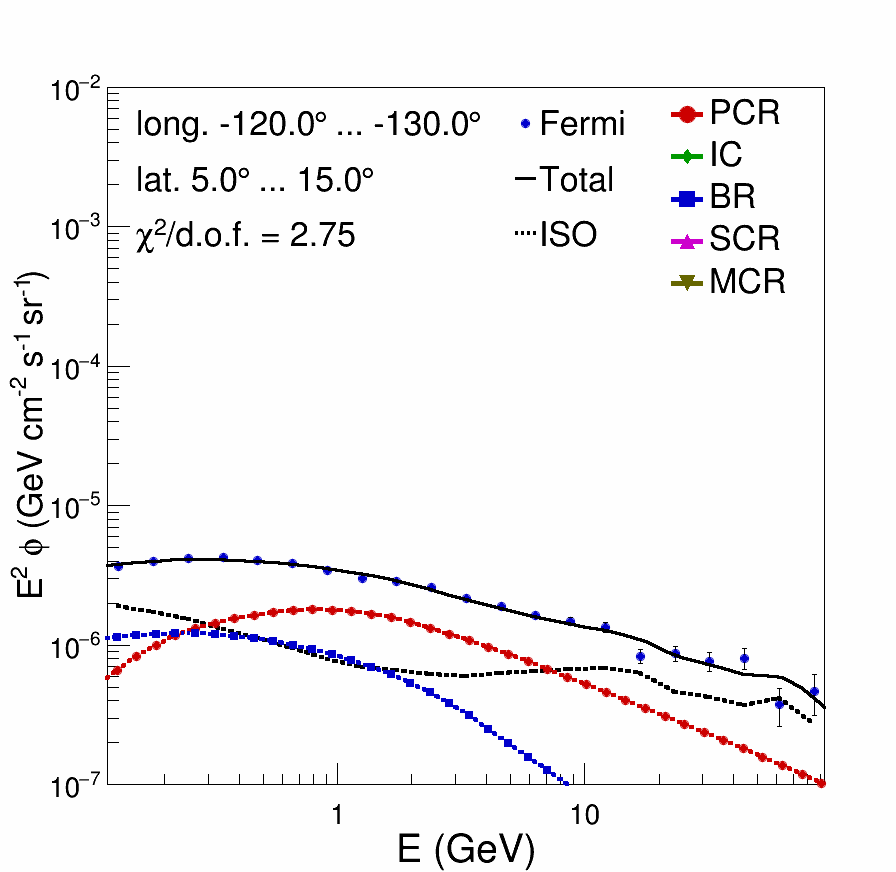}
\includegraphics[width=0.16\textwidth,height=0.16\textwidth,clip]{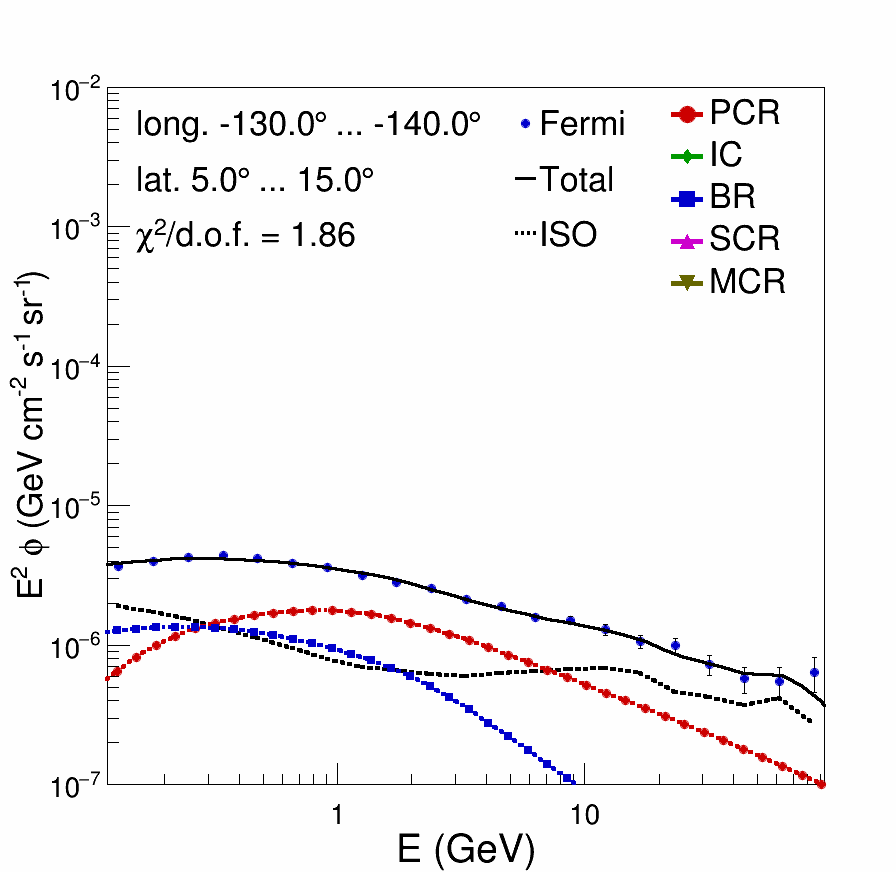}
\includegraphics[width=0.16\textwidth,height=0.16\textwidth,clip]{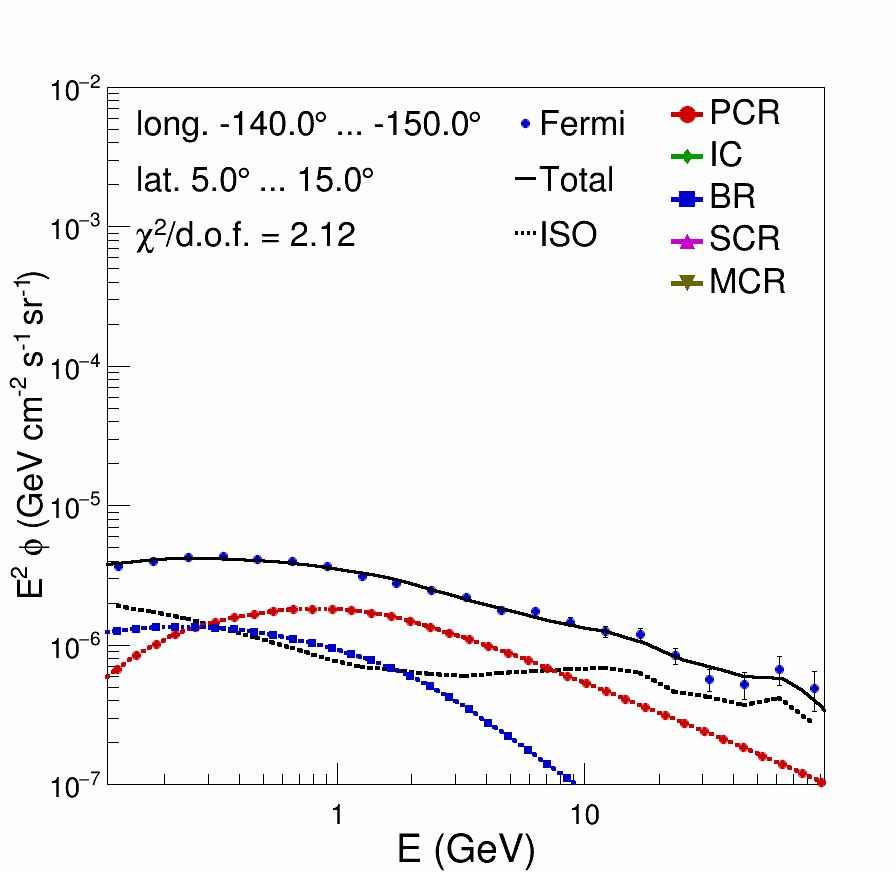}
\includegraphics[width=0.16\textwidth,height=0.16\textwidth,clip]{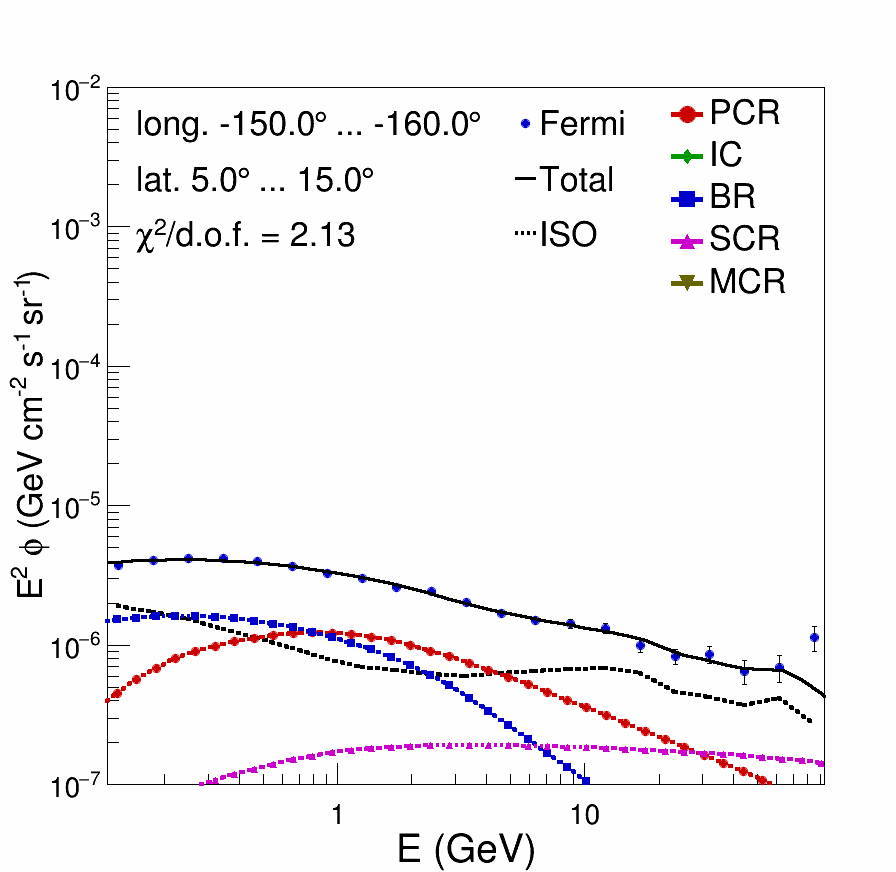}
\includegraphics[width=0.16\textwidth,height=0.16\textwidth,clip]{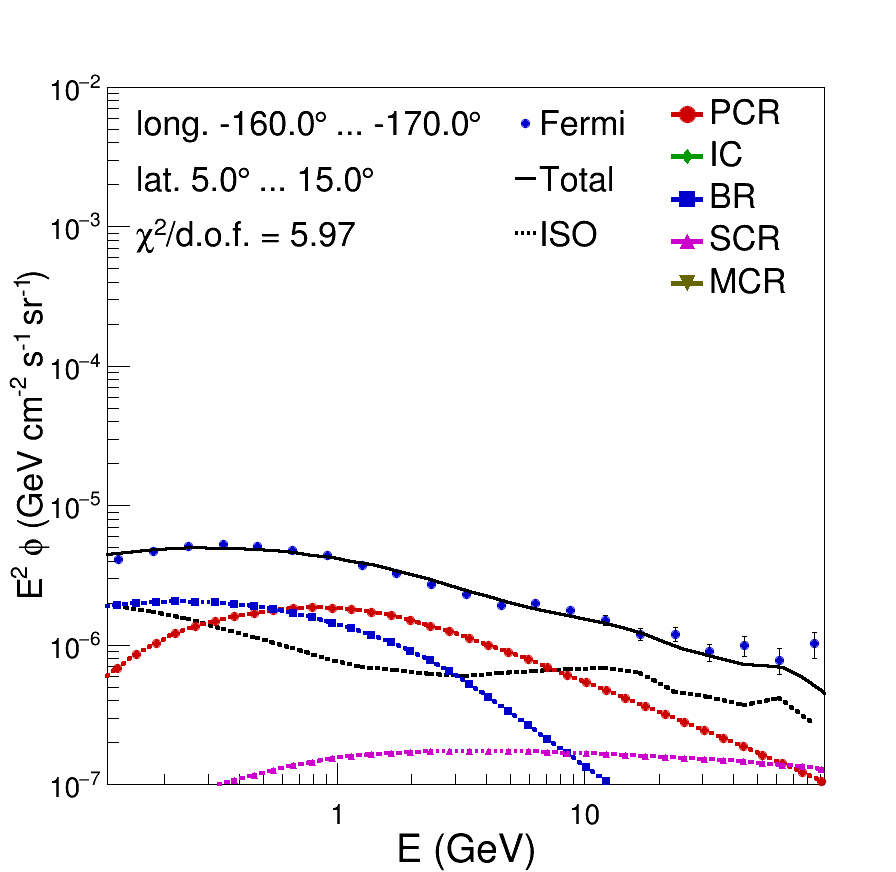}
\includegraphics[width=0.16\textwidth,height=0.16\textwidth,clip]{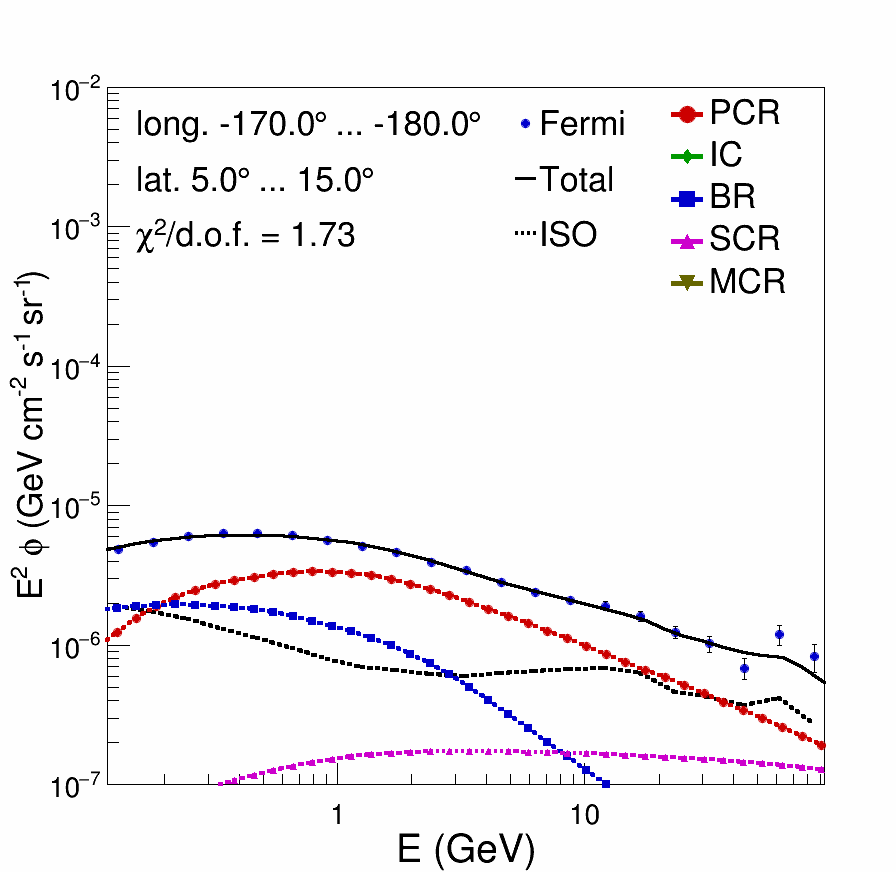}
\caption[]{Template fits for latitudes  with $5.0^\circ<b<15.0^\circ$ and longitudes decreasing from 180$^\circ$ to -180$^\circ$. \label{F17}
}
\end{figure}
\begin{figure}
\centering
\includegraphics[width=0.16\textwidth,height=0.16\textwidth,clip]{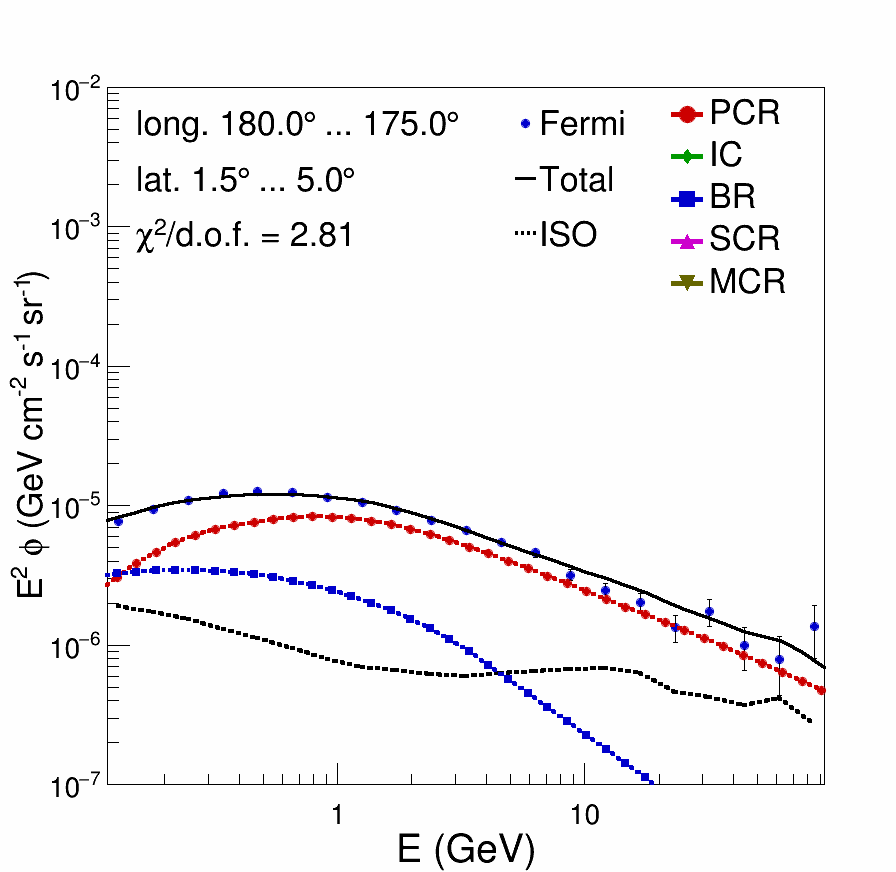}
\includegraphics[width=0.16\textwidth,height=0.16\textwidth,clip]{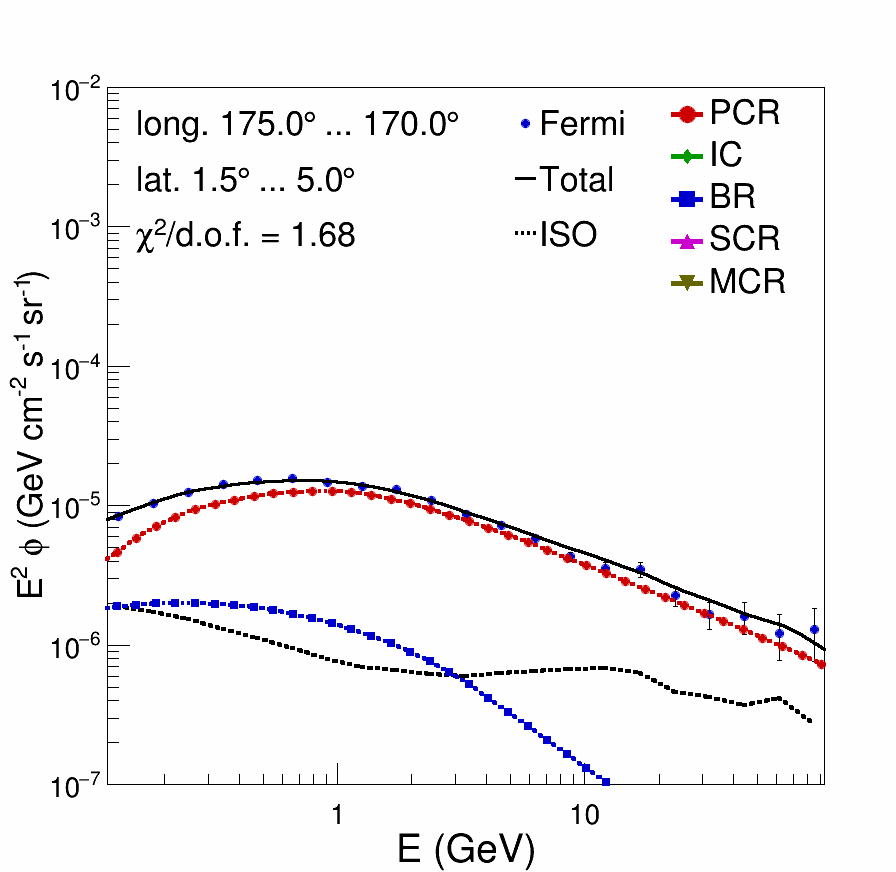}
\includegraphics[width=0.16\textwidth,height=0.16\textwidth,clip]{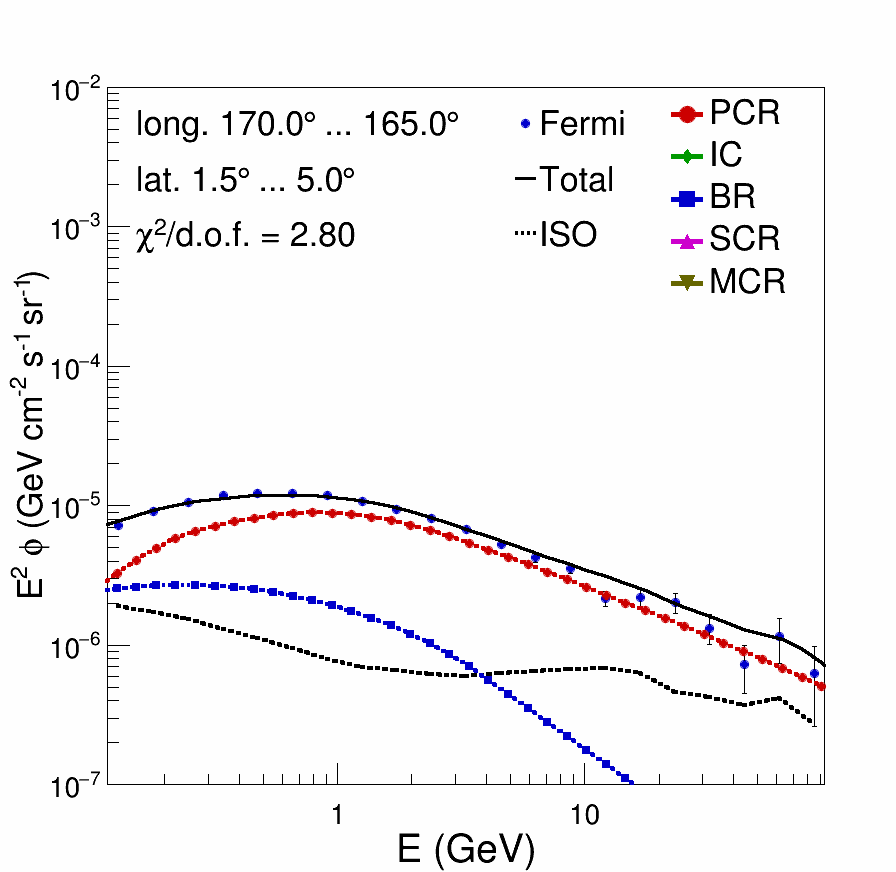}
\includegraphics[width=0.16\textwidth,height=0.16\textwidth,clip]{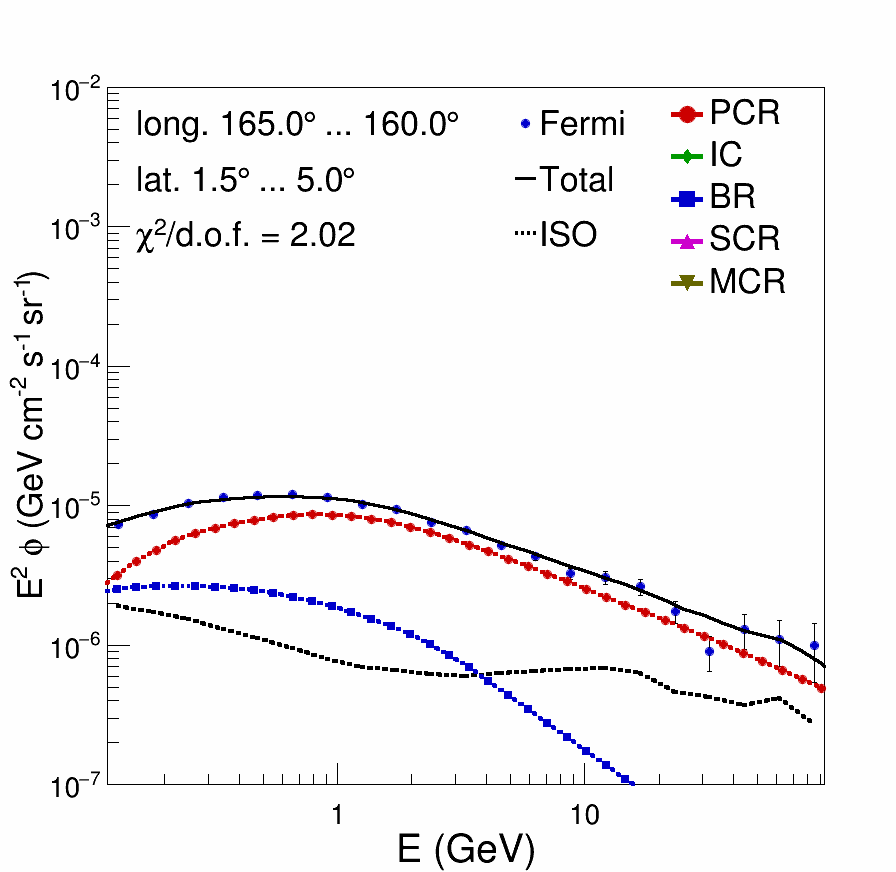}
\includegraphics[width=0.16\textwidth,height=0.16\textwidth,clip]{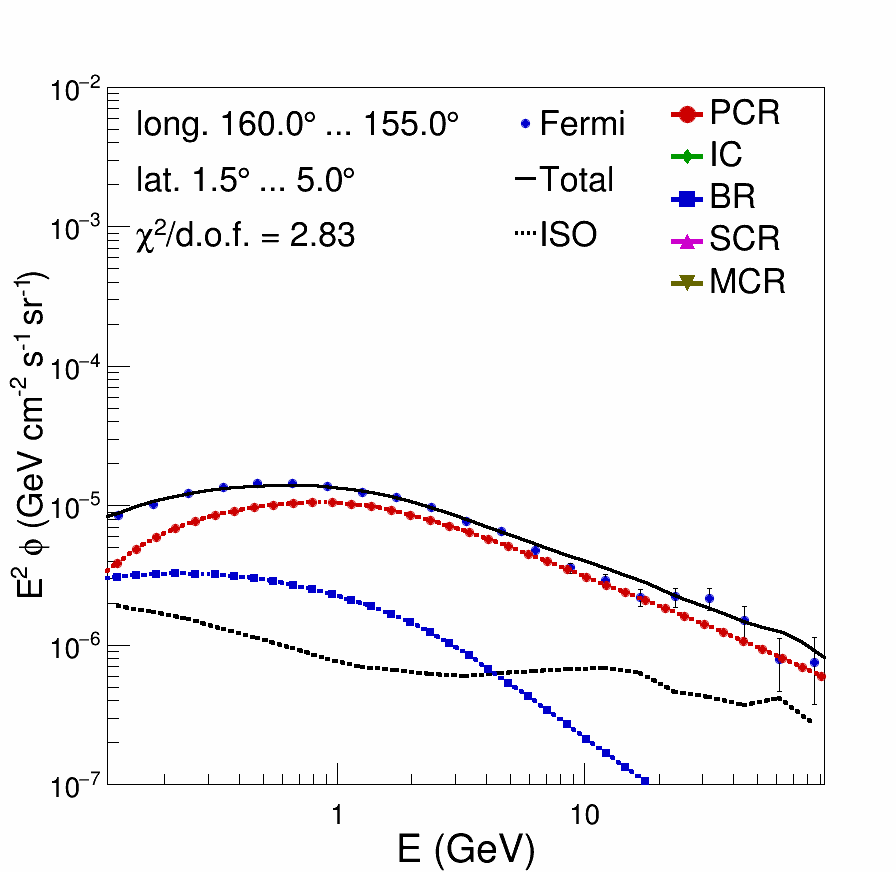}
\includegraphics[width=0.16\textwidth,height=0.16\textwidth,clip]{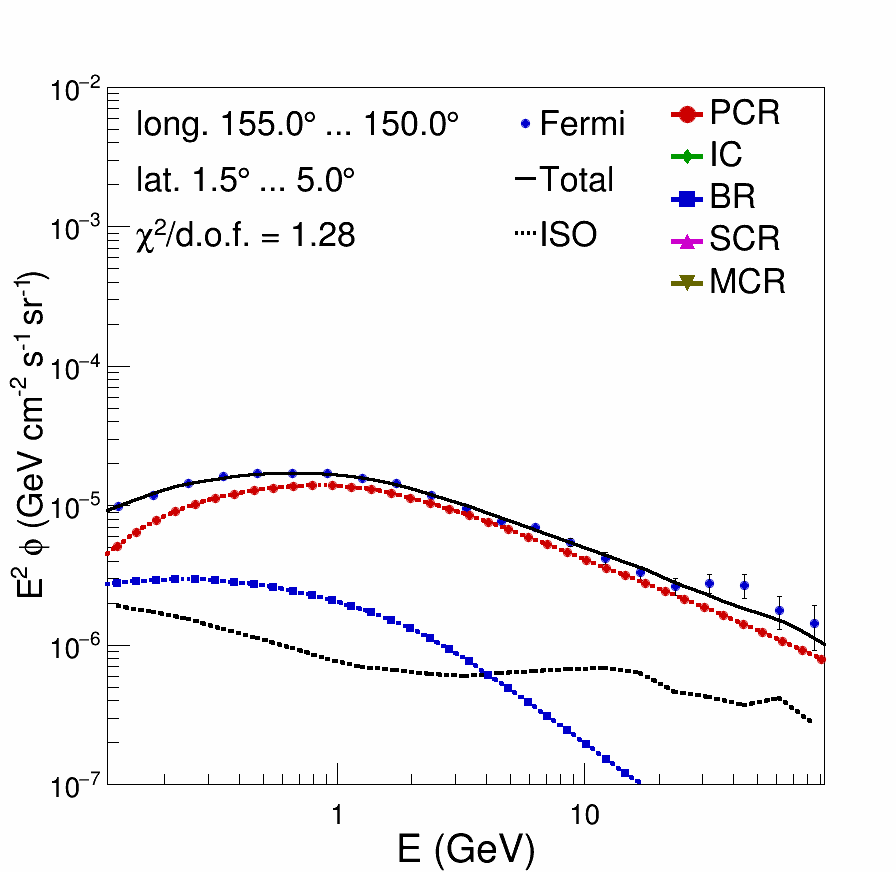}
\includegraphics[width=0.16\textwidth,height=0.16\textwidth,clip]{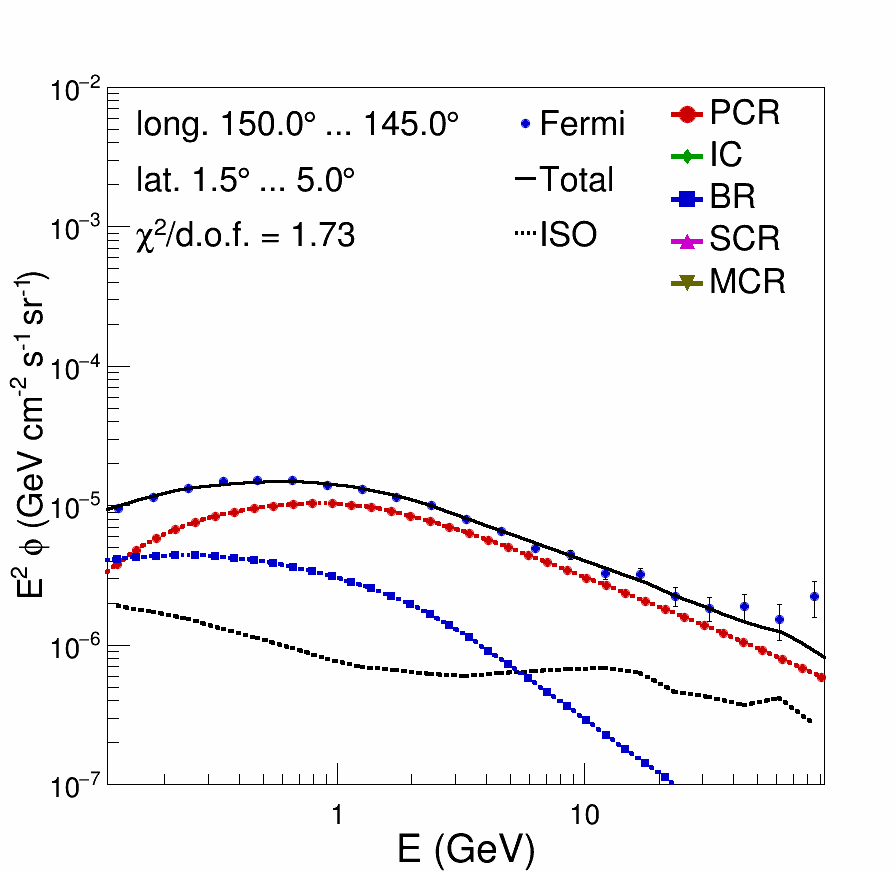}
\includegraphics[width=0.16\textwidth,height=0.16\textwidth,clip]{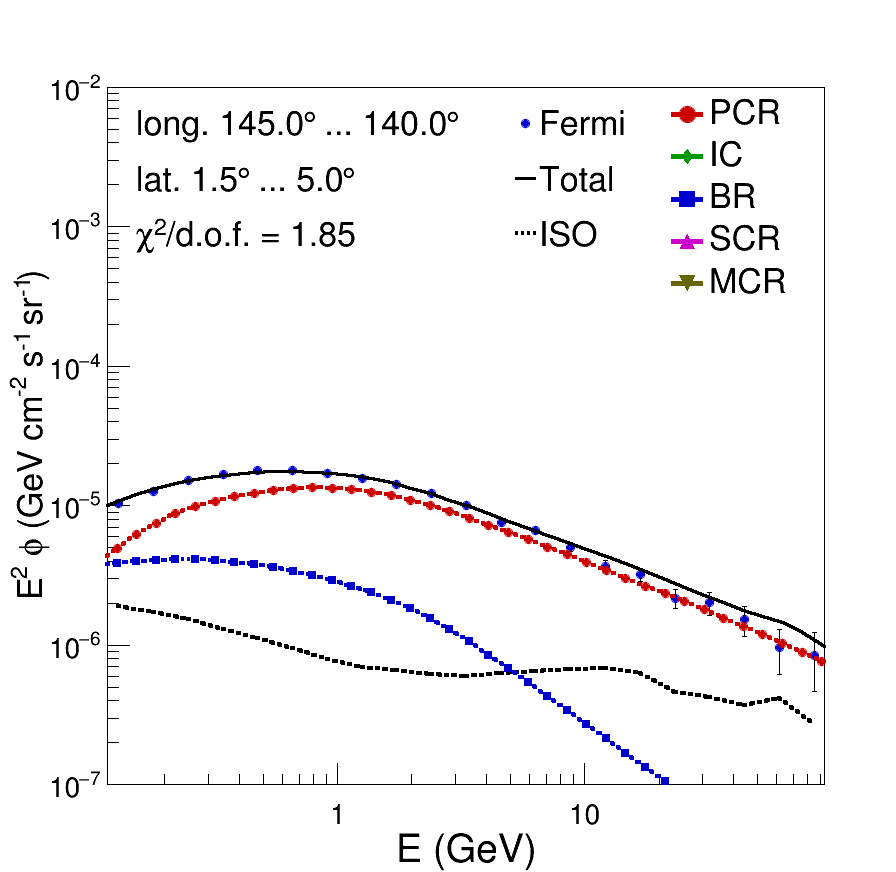}
\includegraphics[width=0.16\textwidth,height=0.16\textwidth,clip]{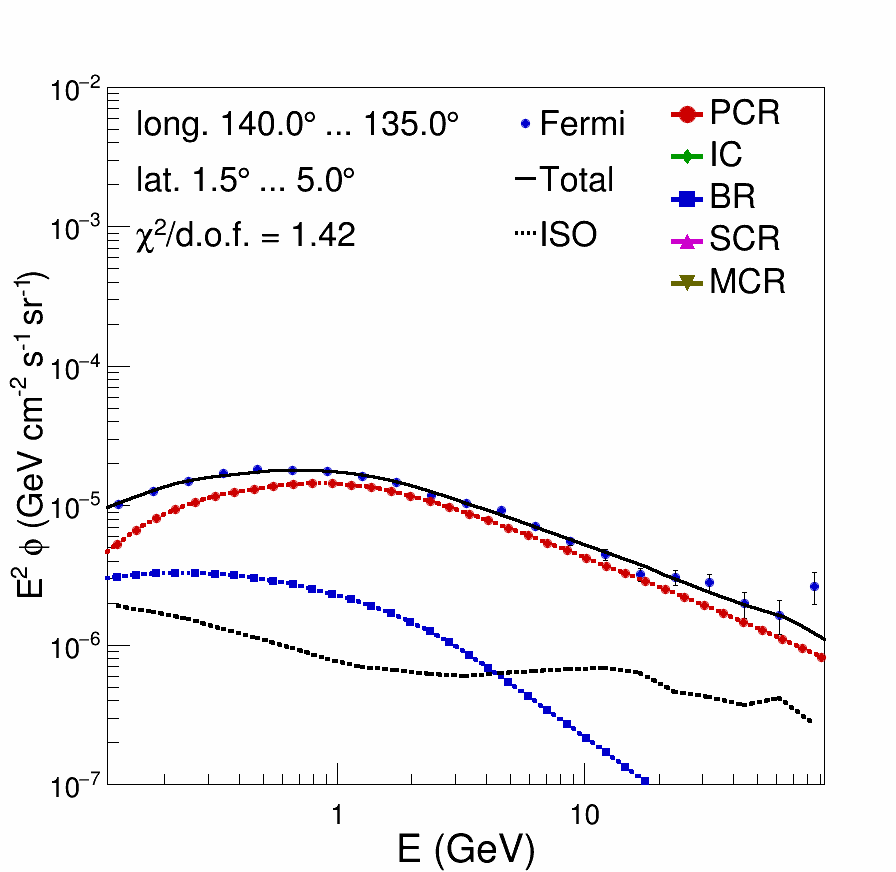}
\includegraphics[width=0.16\textwidth,height=0.16\textwidth,clip]{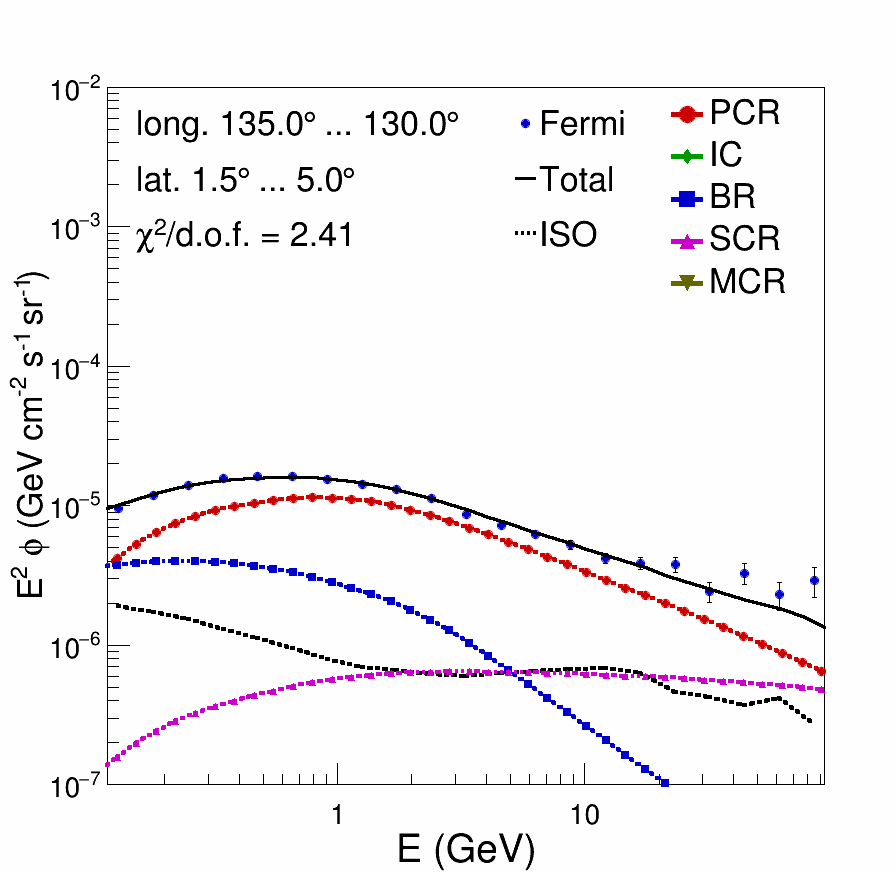}
\includegraphics[width=0.16\textwidth,height=0.16\textwidth,clip]{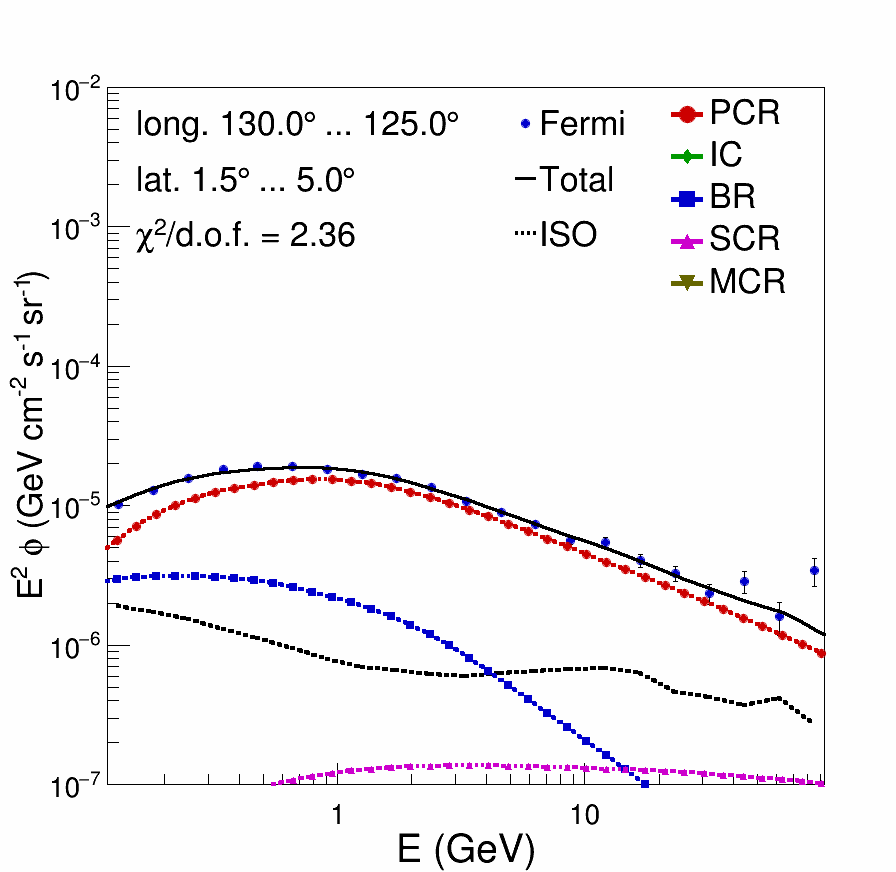}
\includegraphics[width=0.16\textwidth,height=0.16\textwidth,clip]{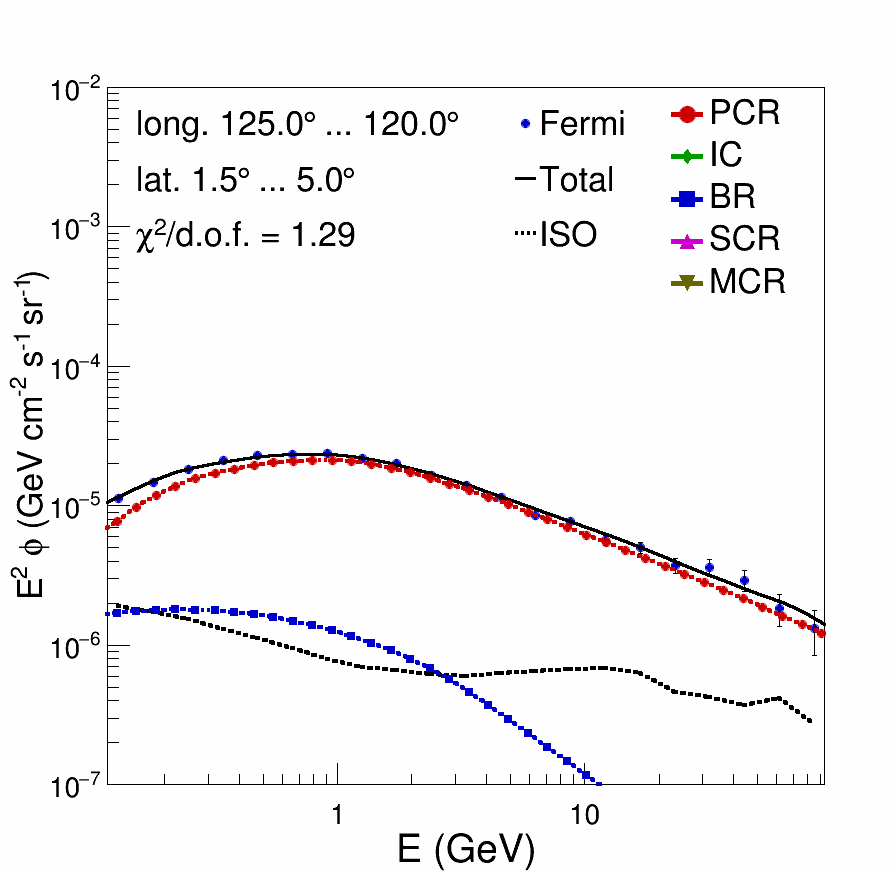}
\includegraphics[width=0.16\textwidth,height=0.16\textwidth,clip]{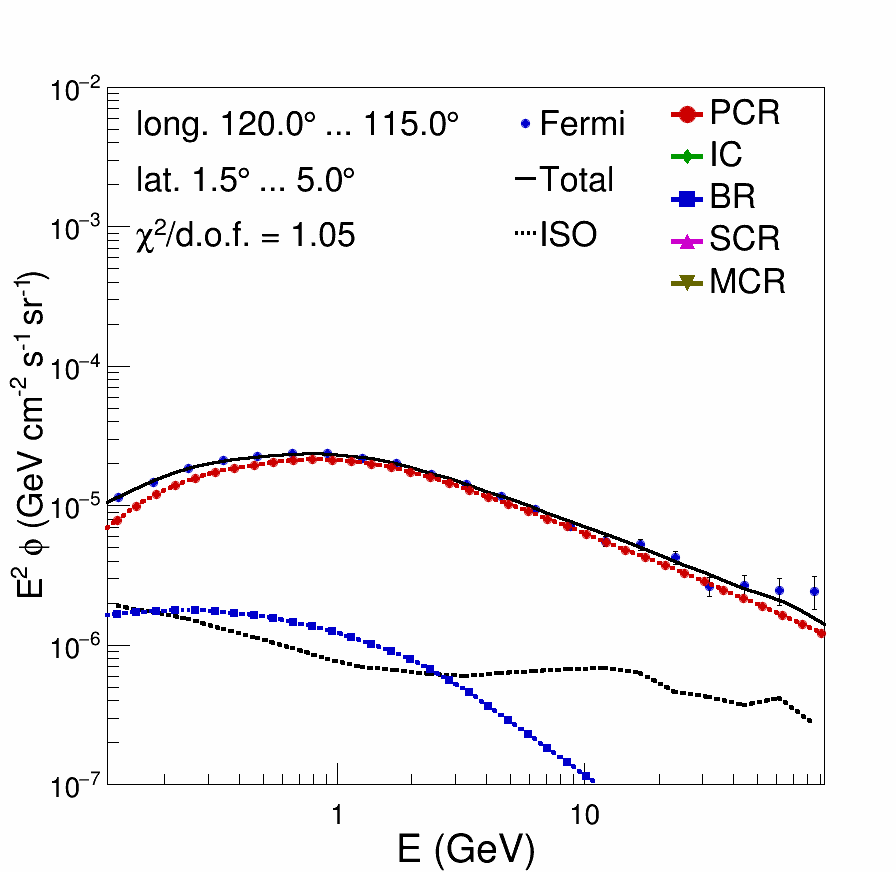}
\includegraphics[width=0.16\textwidth,height=0.16\textwidth,clip]{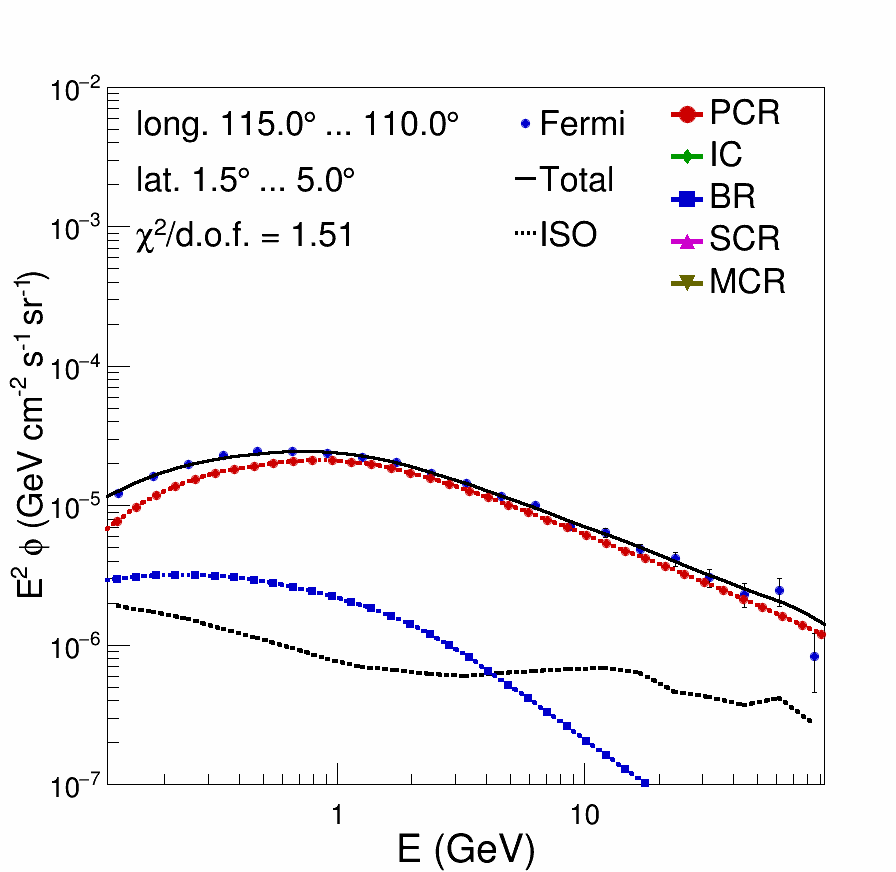}
\includegraphics[width=0.16\textwidth,height=0.16\textwidth,clip]{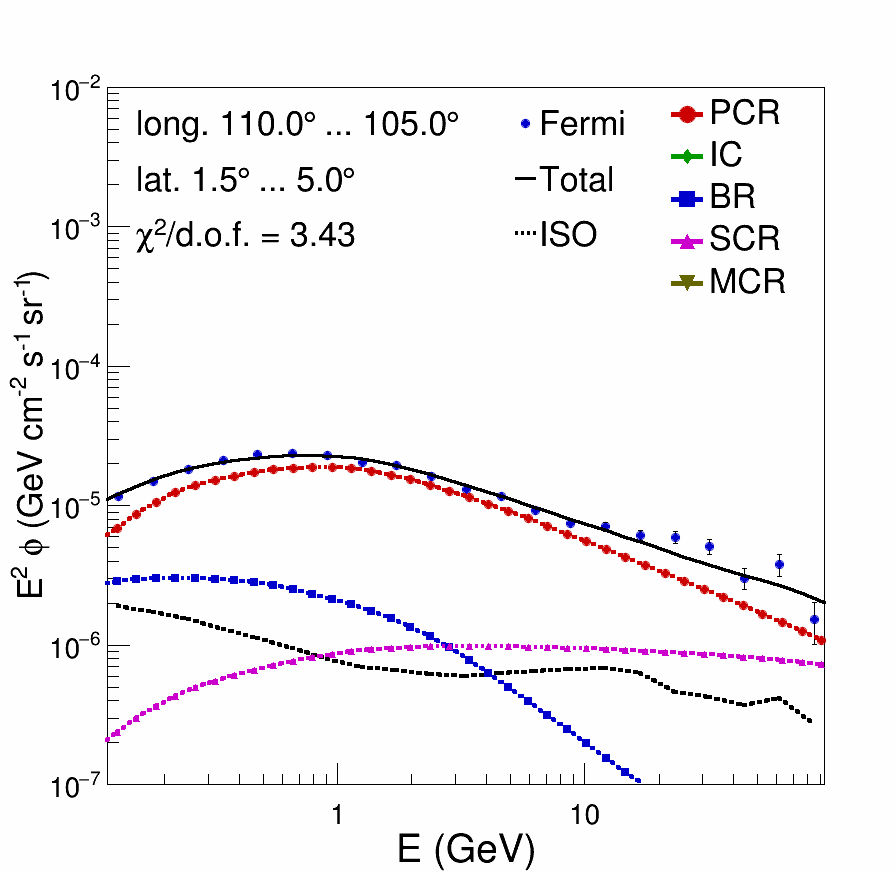}
\includegraphics[width=0.16\textwidth,height=0.16\textwidth,clip]{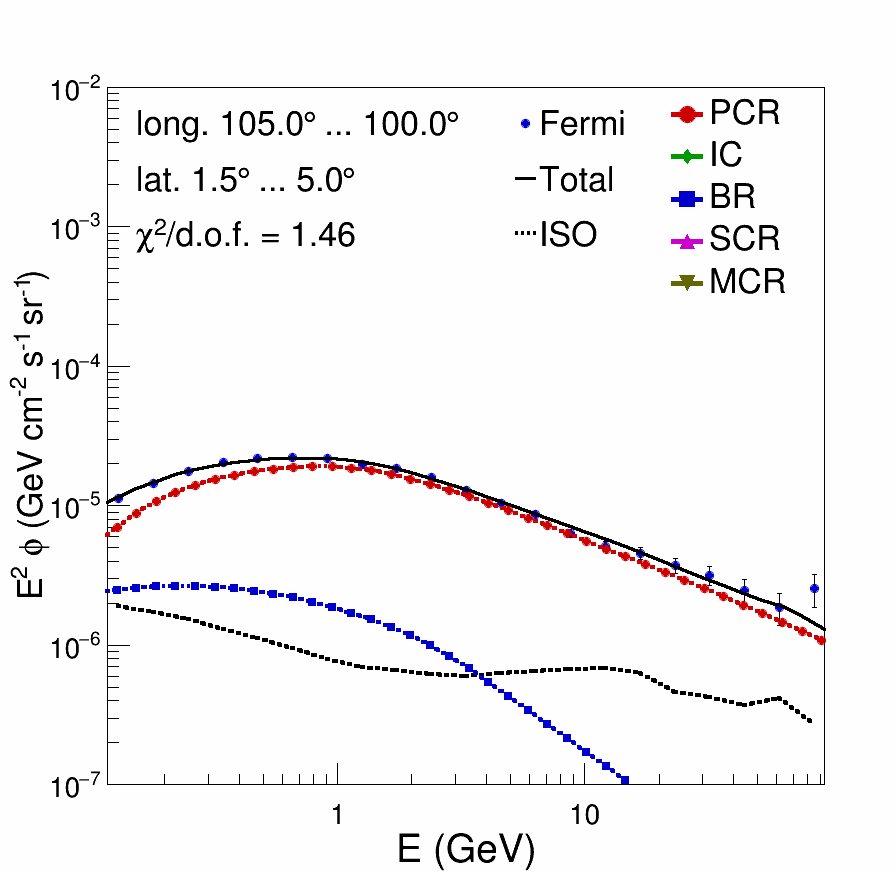}
\includegraphics[width=0.16\textwidth,height=0.16\textwidth,clip]{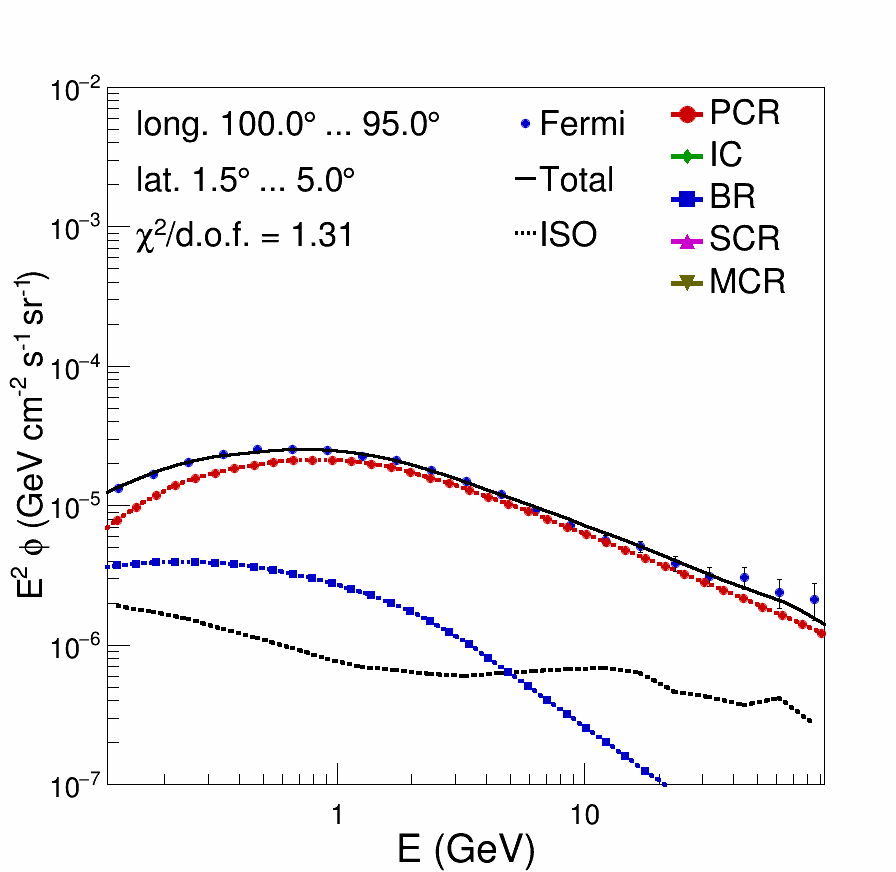}
\includegraphics[width=0.16\textwidth,height=0.16\textwidth,clip]{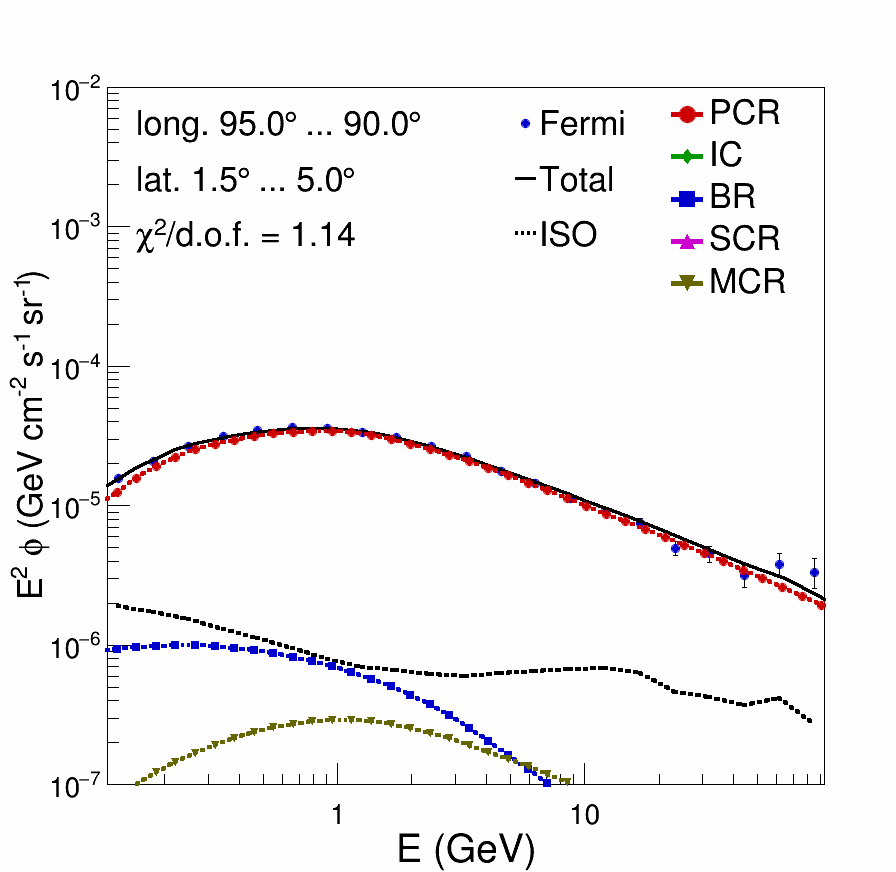}
\includegraphics[width=0.16\textwidth,height=0.16\textwidth,clip]{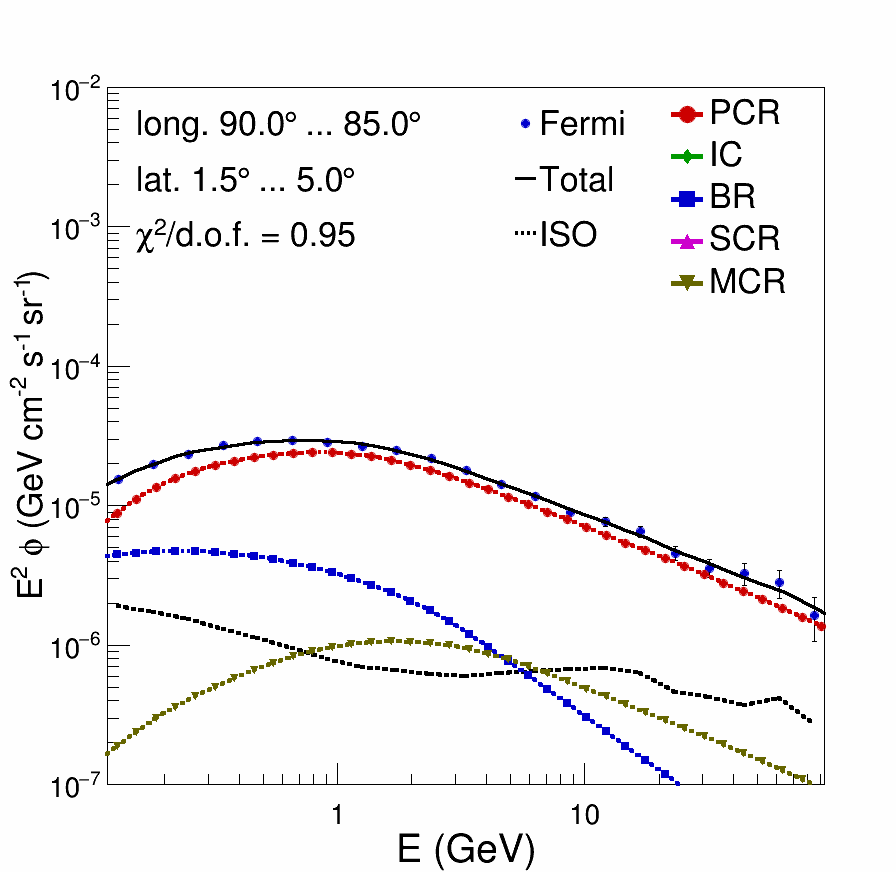}
\includegraphics[width=0.16\textwidth,height=0.16\textwidth,clip]{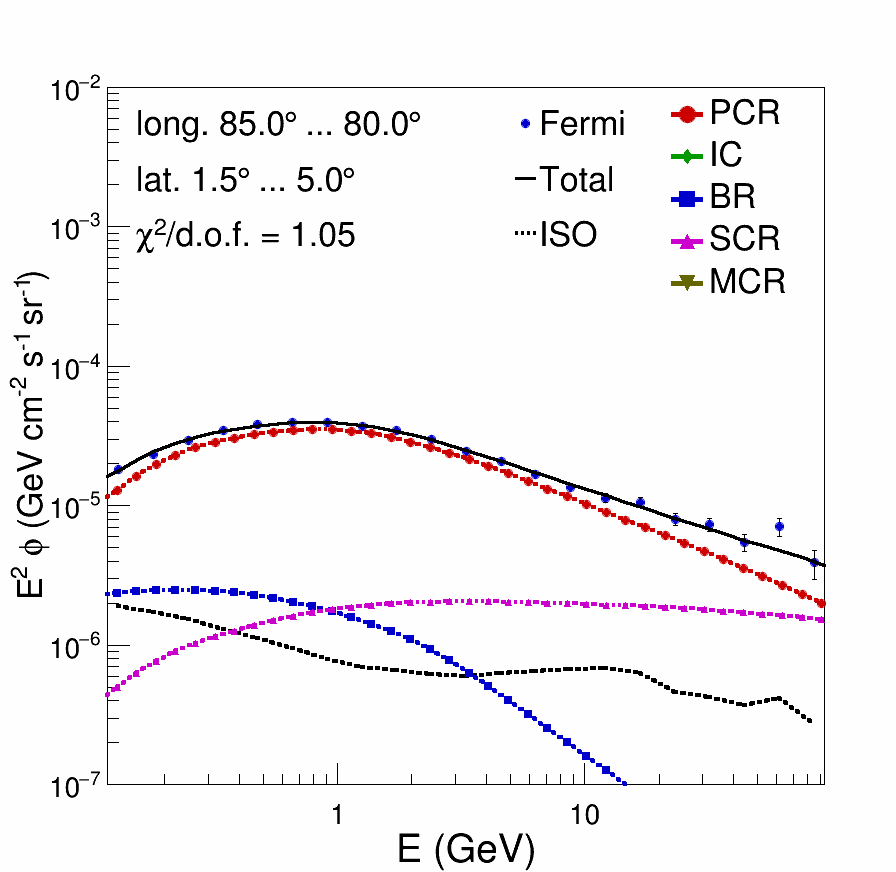}
\includegraphics[width=0.16\textwidth,height=0.16\textwidth,clip]{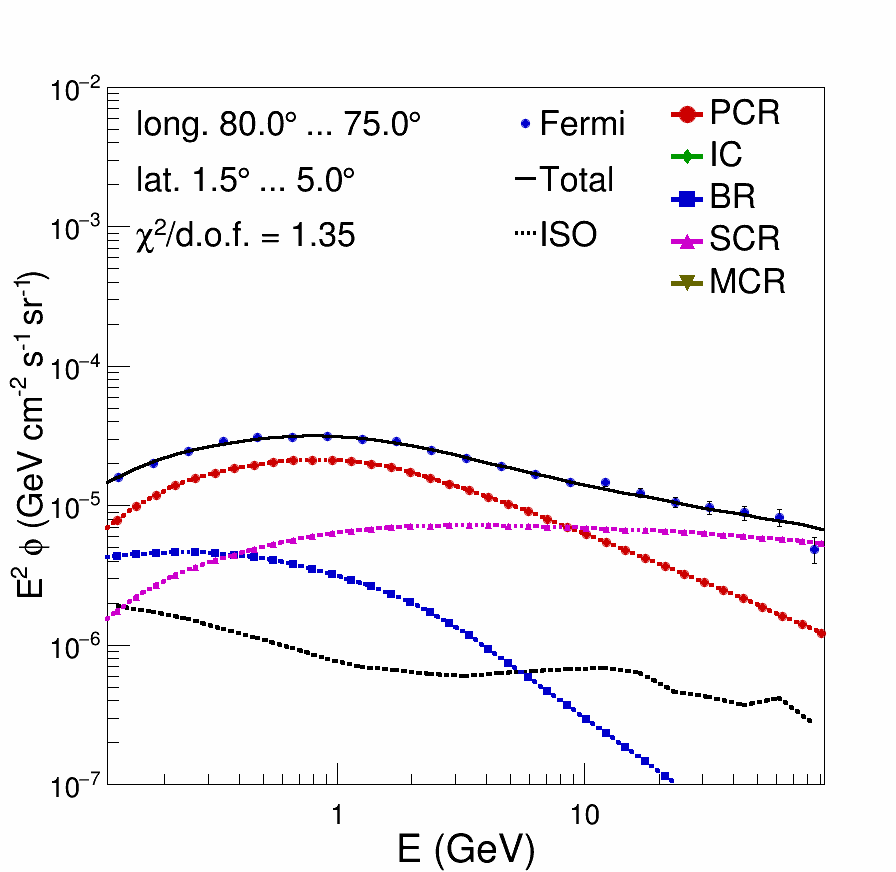}
\includegraphics[width=0.16\textwidth,height=0.16\textwidth,clip]{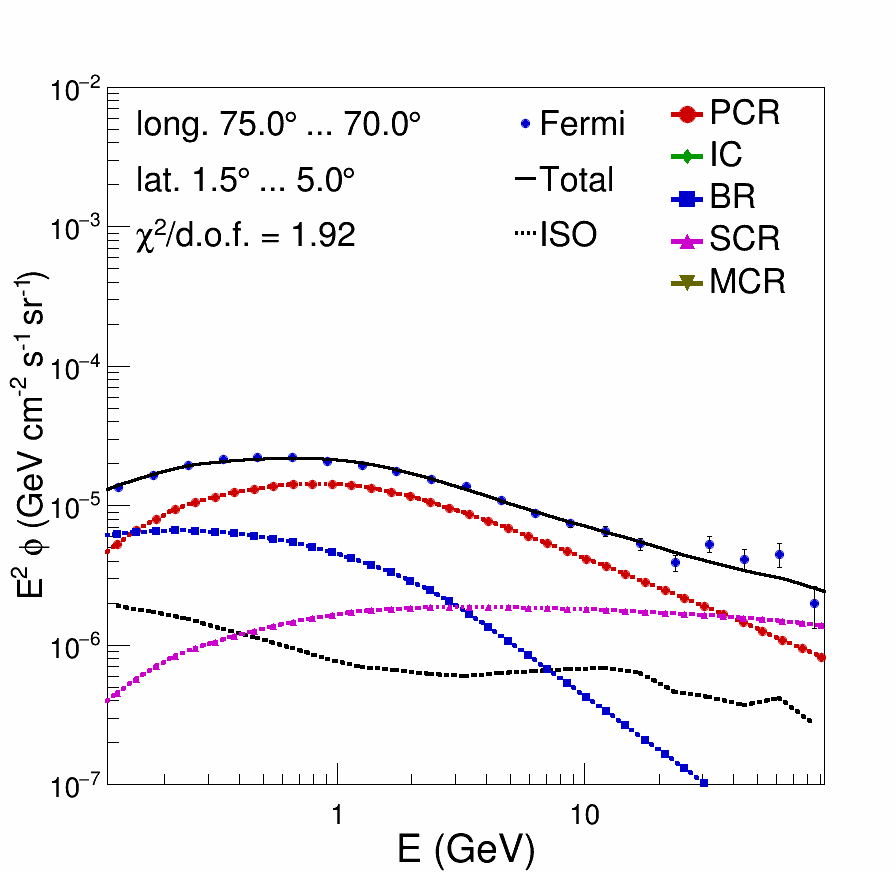}
\includegraphics[width=0.16\textwidth,height=0.16\textwidth,clip]{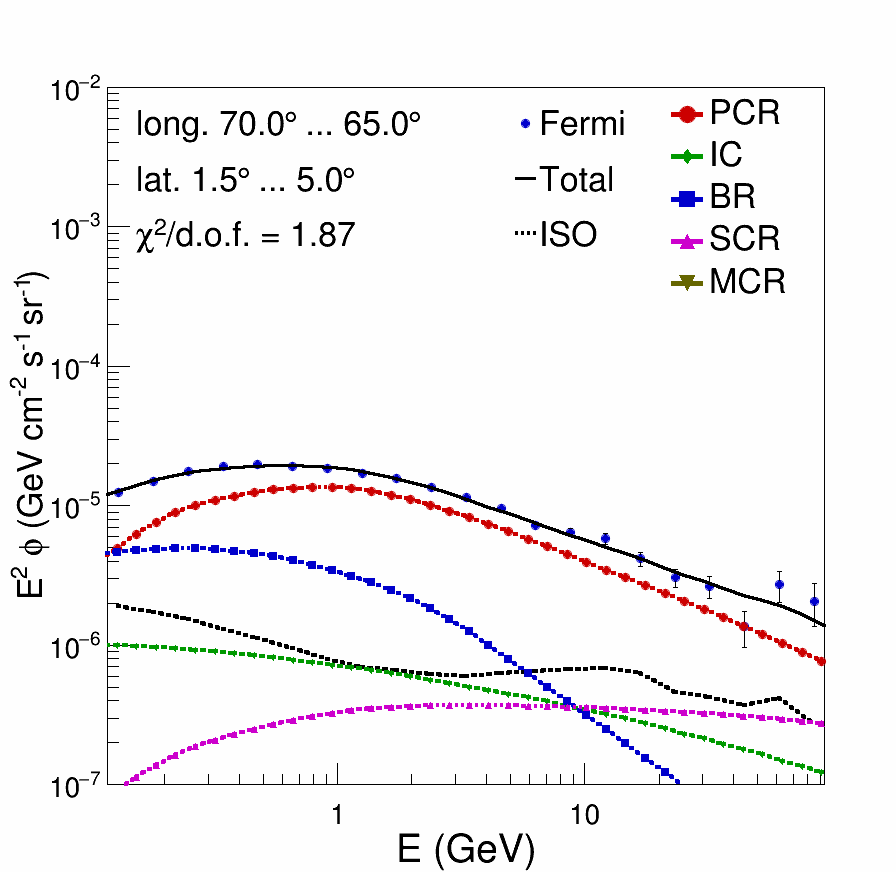}
\includegraphics[width=0.16\textwidth,height=0.16\textwidth,clip]{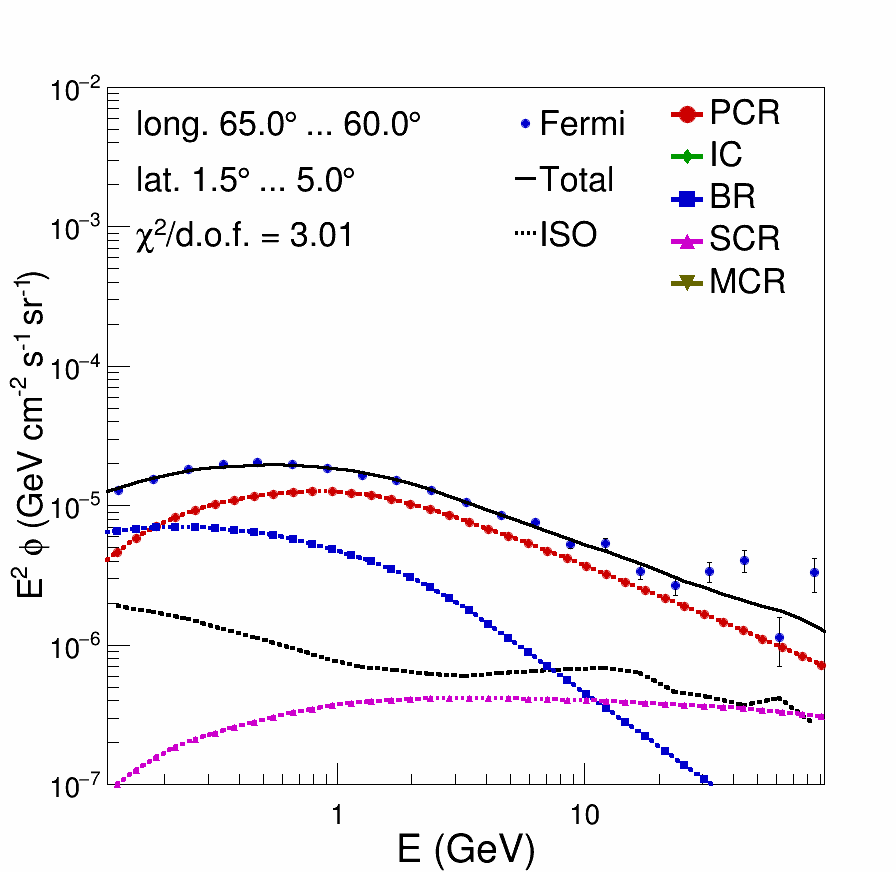}
\includegraphics[width=0.16\textwidth,height=0.16\textwidth,clip]{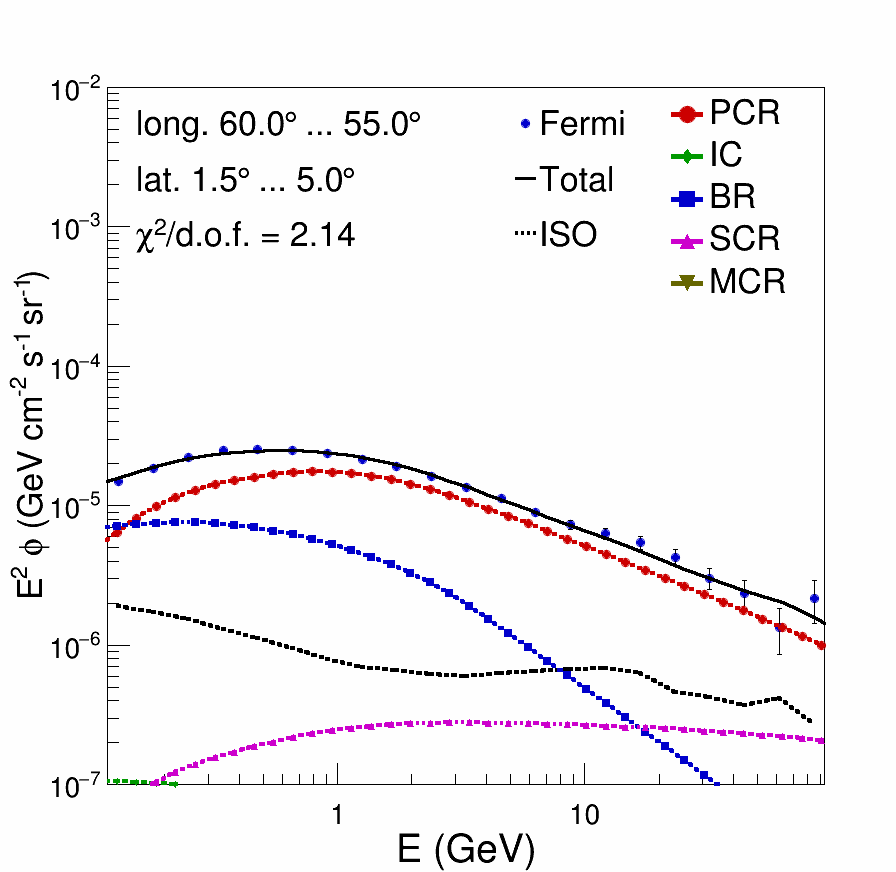}
\includegraphics[width=0.16\textwidth,height=0.16\textwidth,clip]{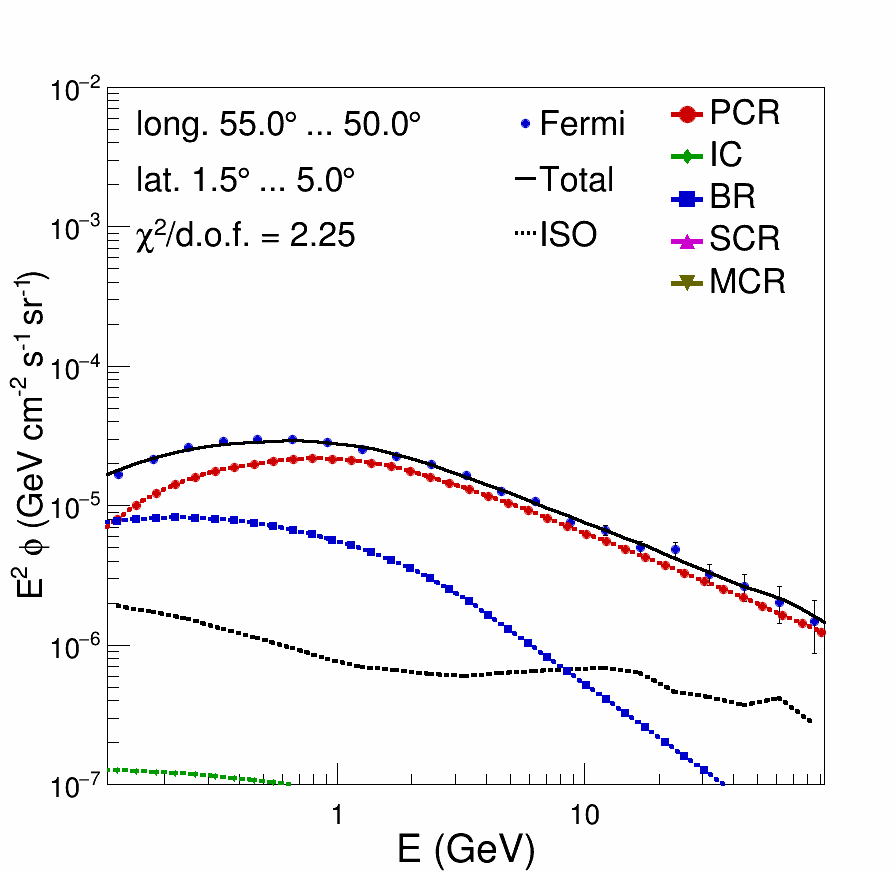}
\includegraphics[width=0.16\textwidth,height=0.16\textwidth,clip]{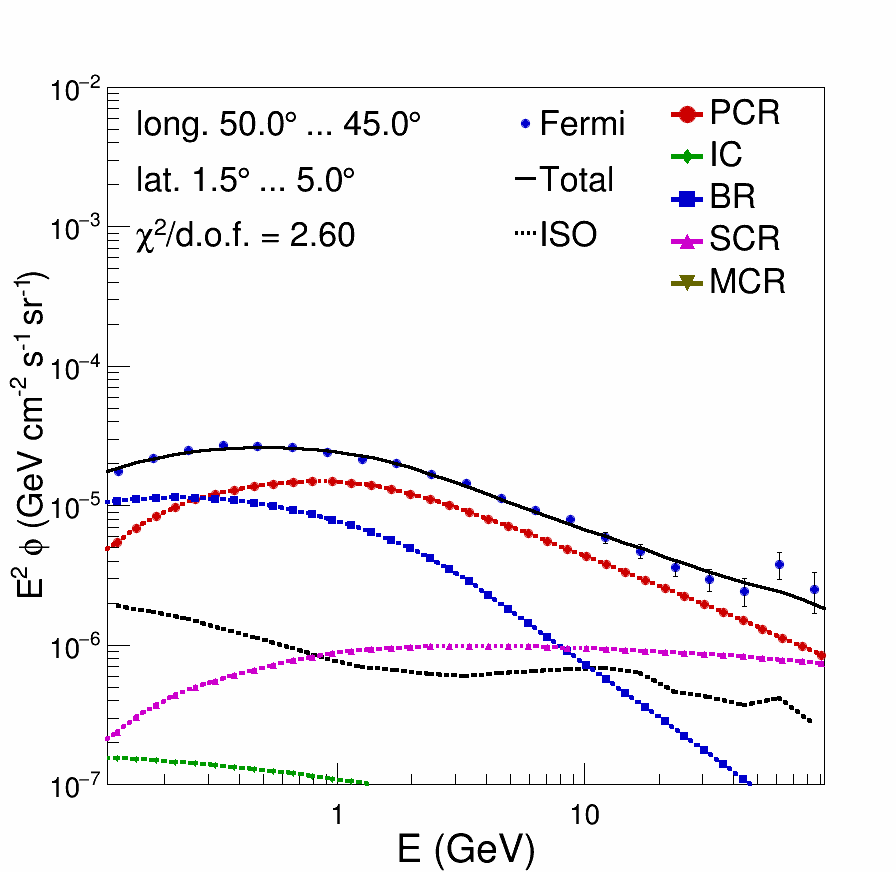}
\includegraphics[width=0.16\textwidth,height=0.16\textwidth,clip]{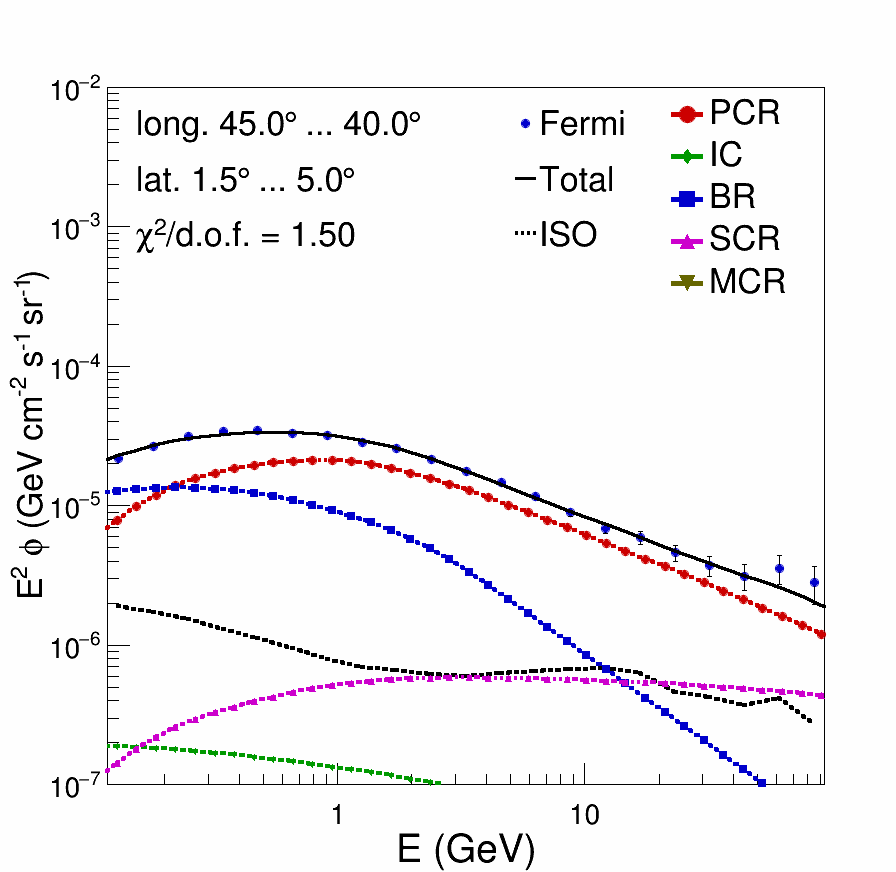}
\includegraphics[width=0.16\textwidth,height=0.16\textwidth,clip]{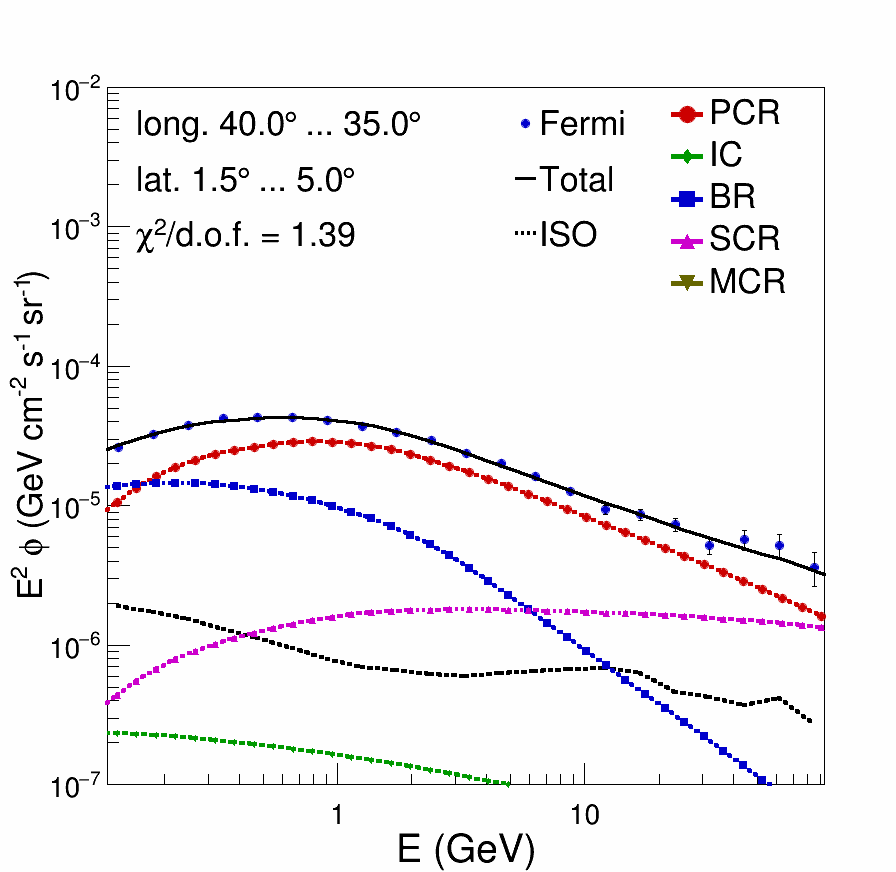}
\includegraphics[width=0.16\textwidth,height=0.16\textwidth,clip]{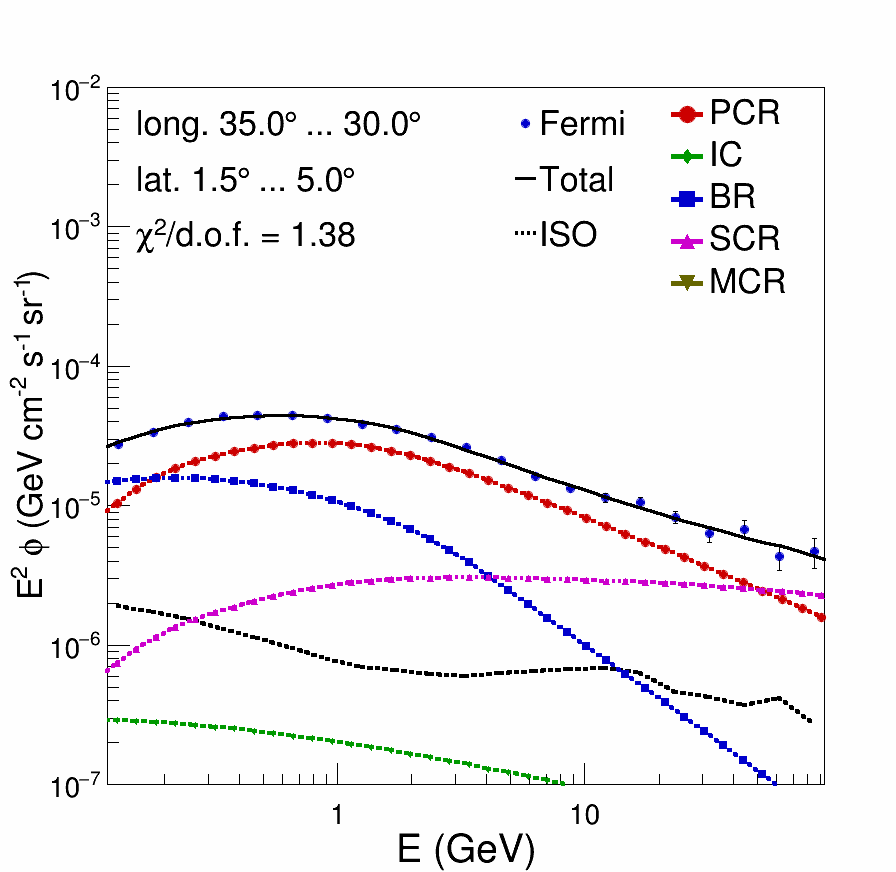}
\includegraphics[width=0.16\textwidth,height=0.16\textwidth,clip]{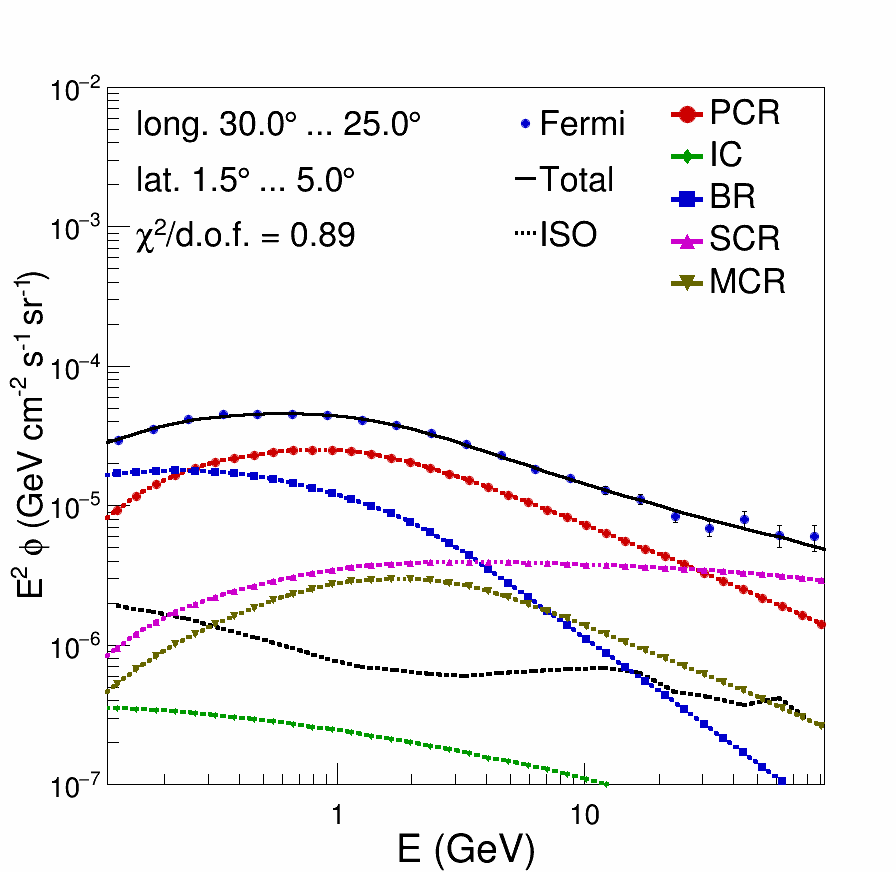}
\includegraphics[width=0.16\textwidth,height=0.16\textwidth,clip]{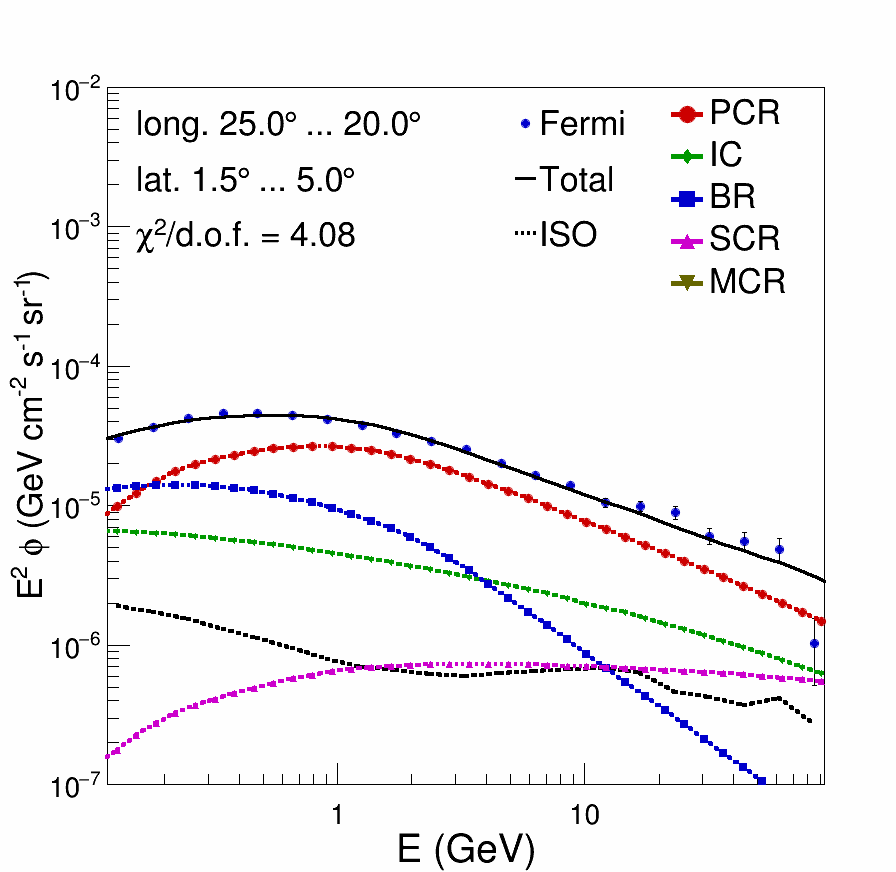}
\includegraphics[width=0.16\textwidth,height=0.16\textwidth,clip]{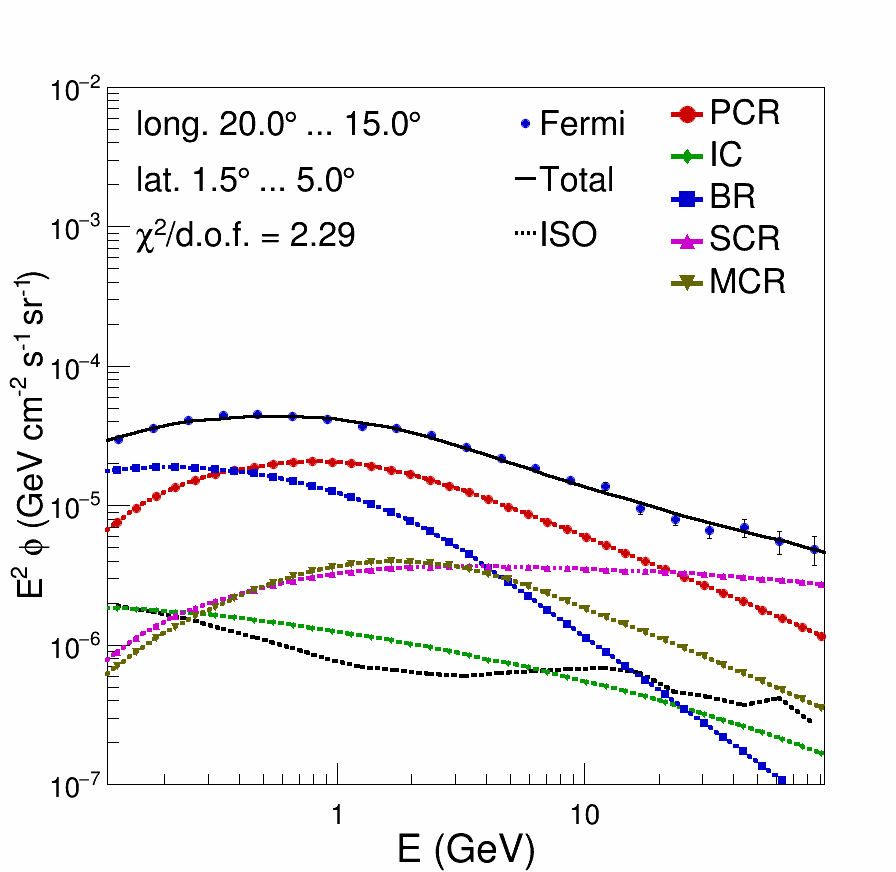}
\includegraphics[width=0.16\textwidth,height=0.16\textwidth,clip]{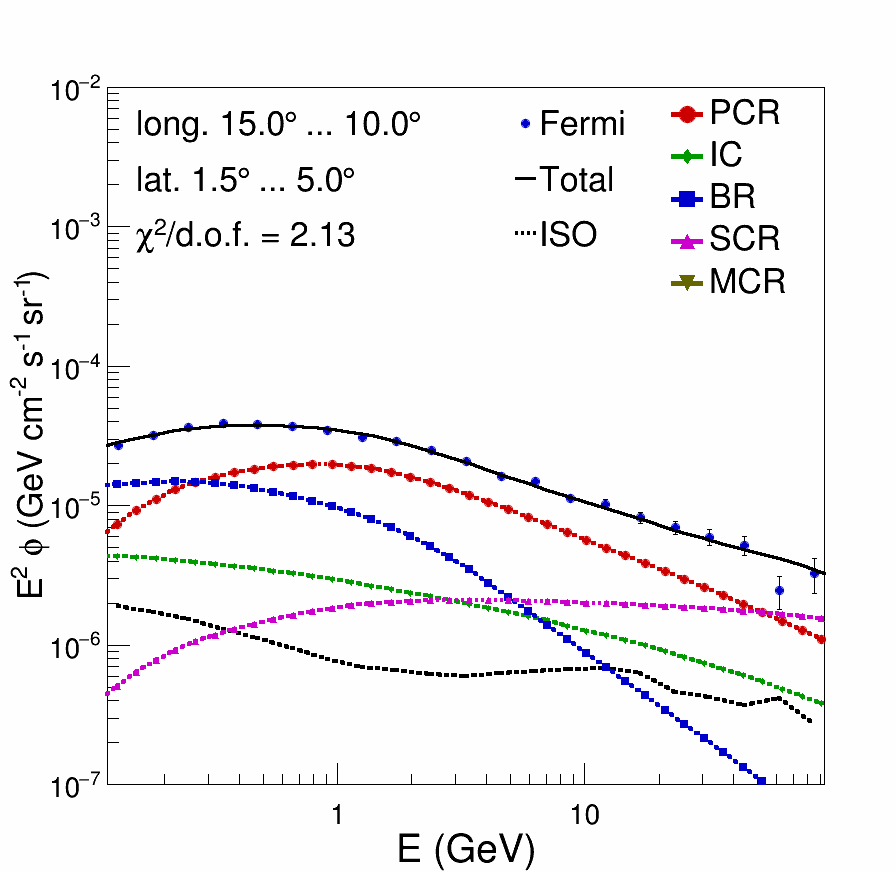}
\includegraphics[width=0.16\textwidth,height=0.16\textwidth,clip]{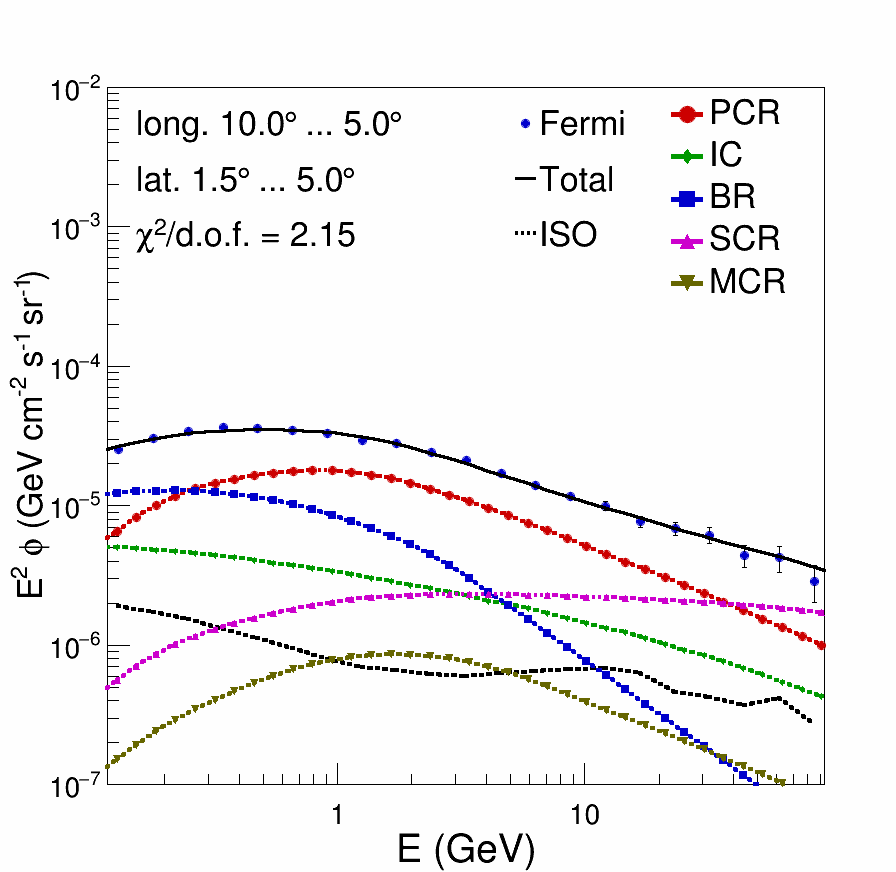}
\includegraphics[width=0.16\textwidth,height=0.16\textwidth,clip]{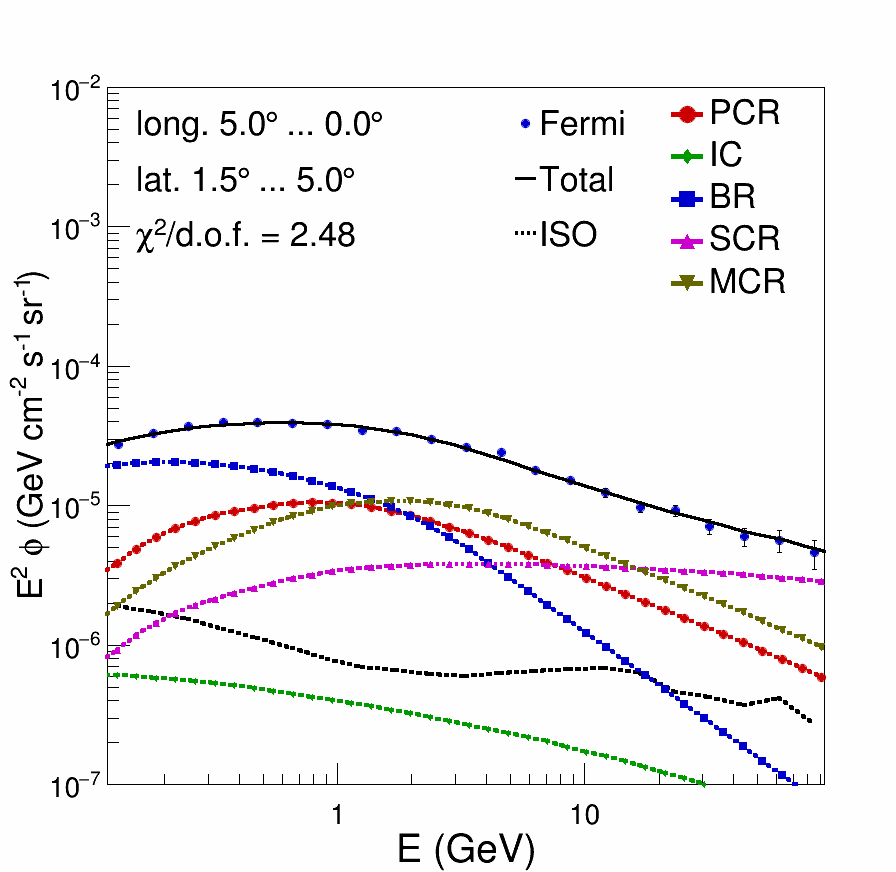}
\caption[]{Template fits for latitudes  with $1.5^\circ<b<5.0^\circ$ and longitudes decreasing from 180$^\circ$ to 0$^\circ$.} \label{F18}
\end{figure}
\begin{figure}
\centering
\includegraphics[width=0.16\textwidth,height=0.16\textwidth,clip]{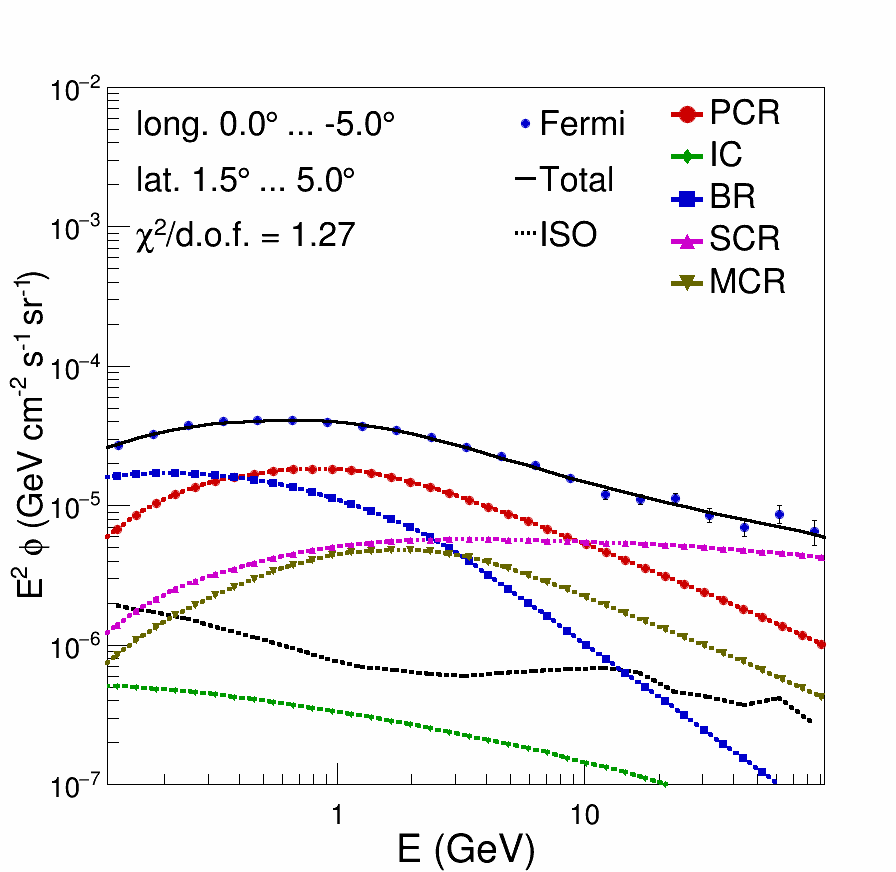}
\includegraphics[width=0.16\textwidth,height=0.16\textwidth,clip]{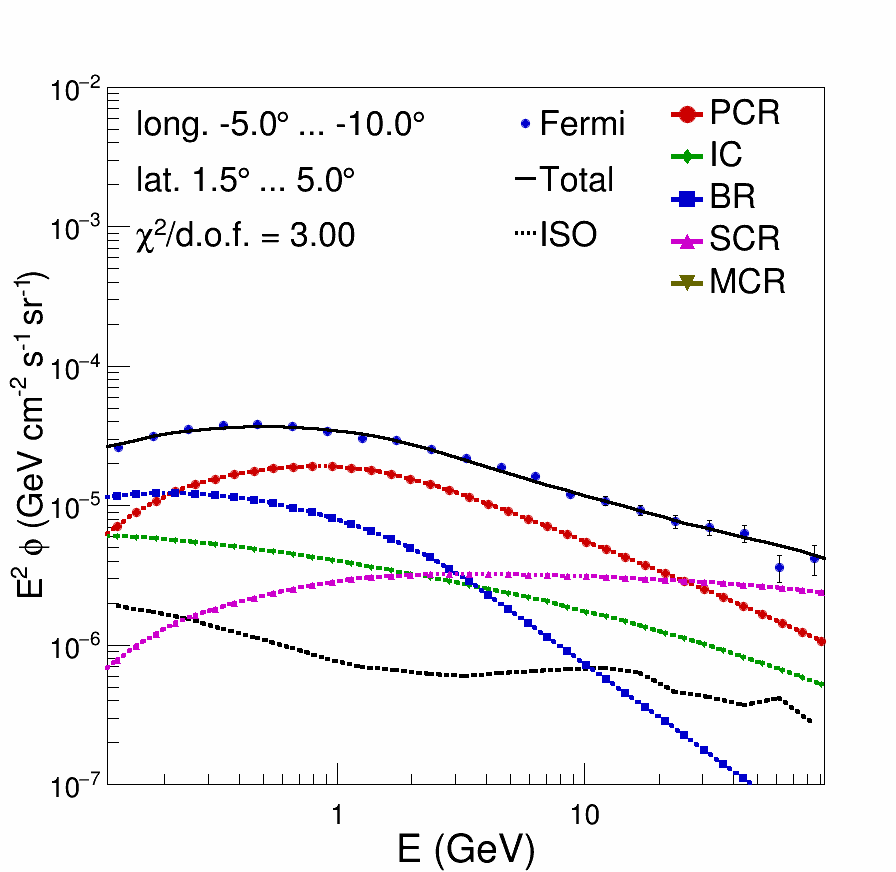}
\includegraphics[width=0.16\textwidth,height=0.16\textwidth,clip]{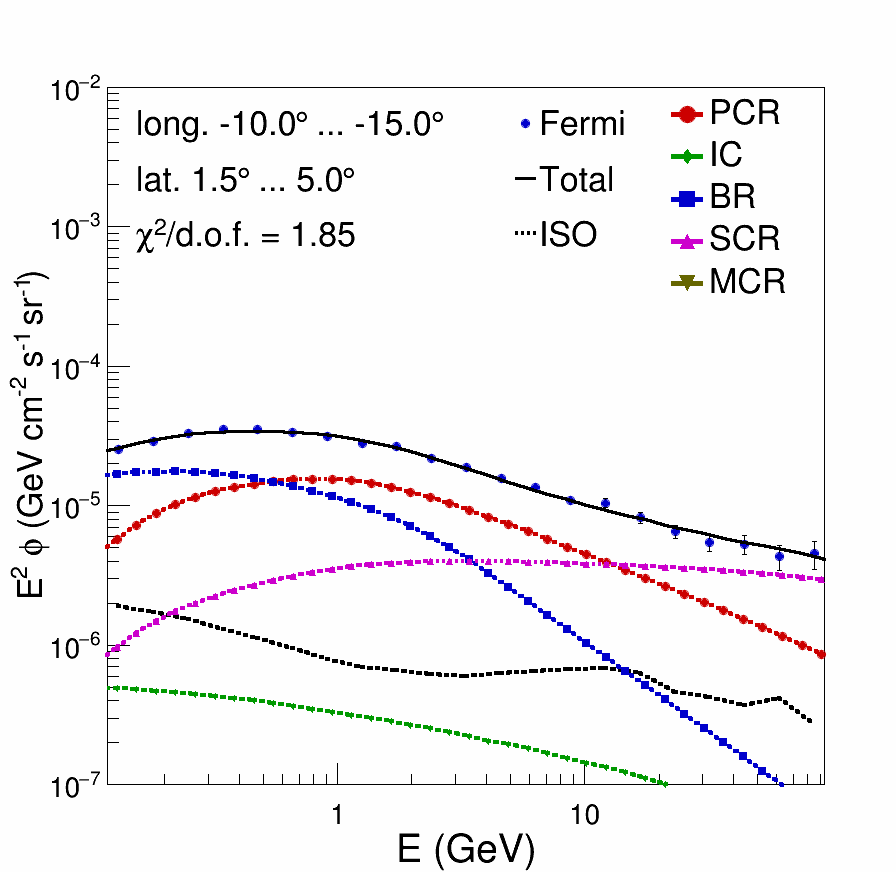}
\includegraphics[width=0.16\textwidth,height=0.16\textwidth,clip]{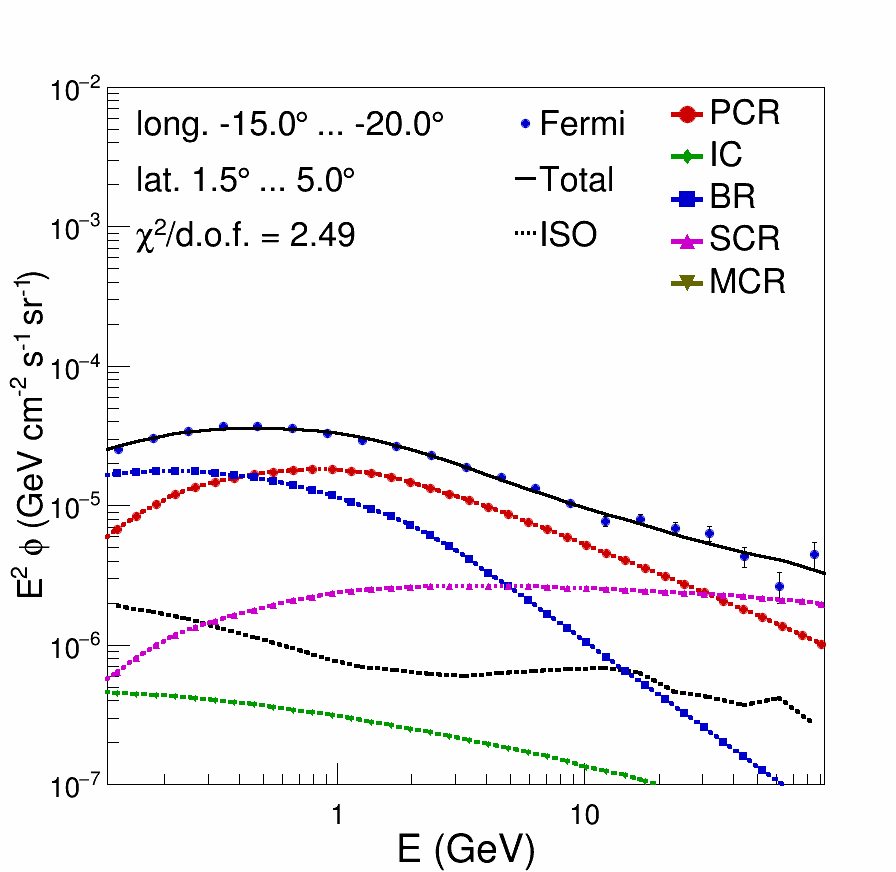}
\includegraphics[width=0.16\textwidth,height=0.16\textwidth,clip]{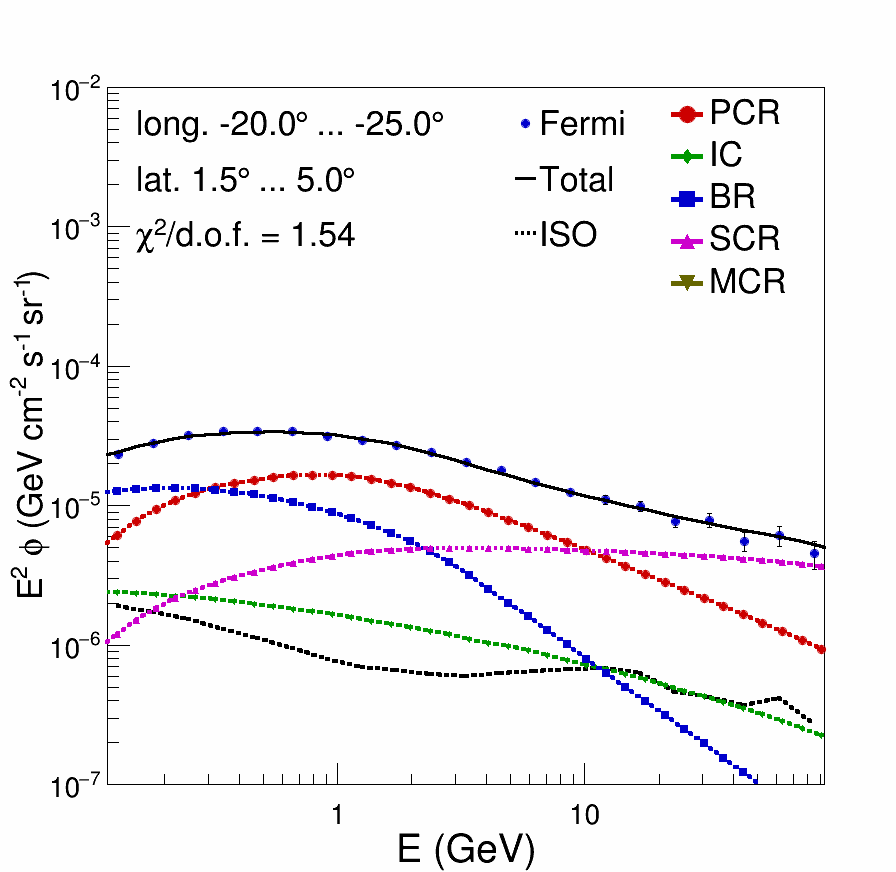}
\includegraphics[width=0.16\textwidth,height=0.16\textwidth,clip]{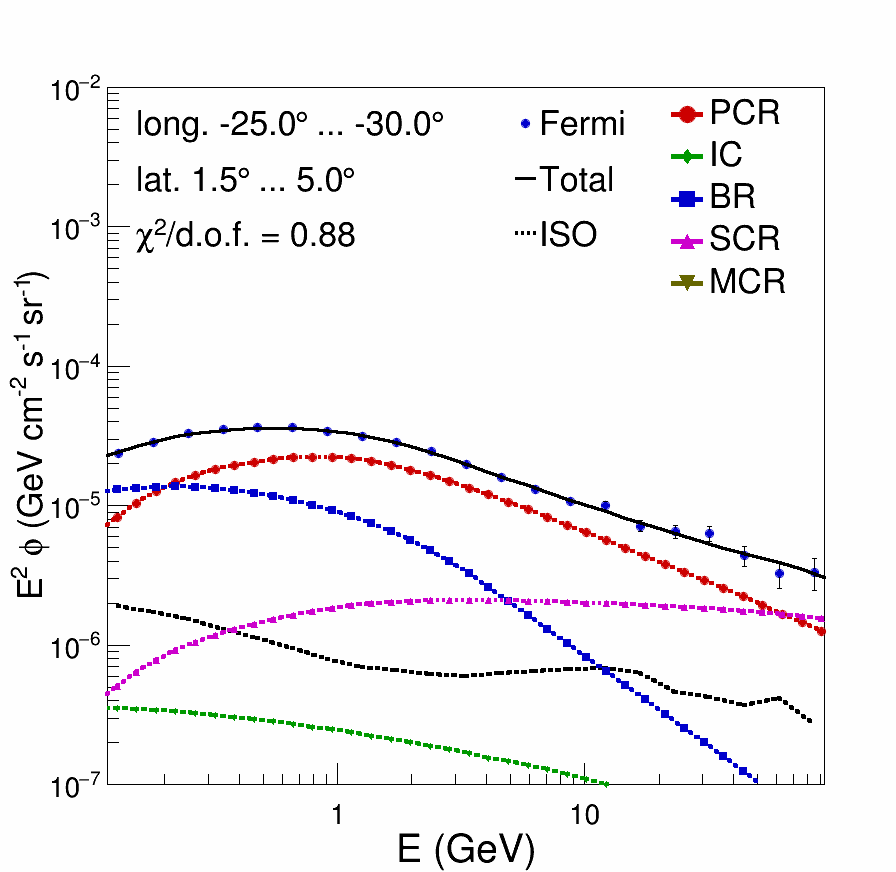}
\includegraphics[width=0.16\textwidth,height=0.16\textwidth,clip]{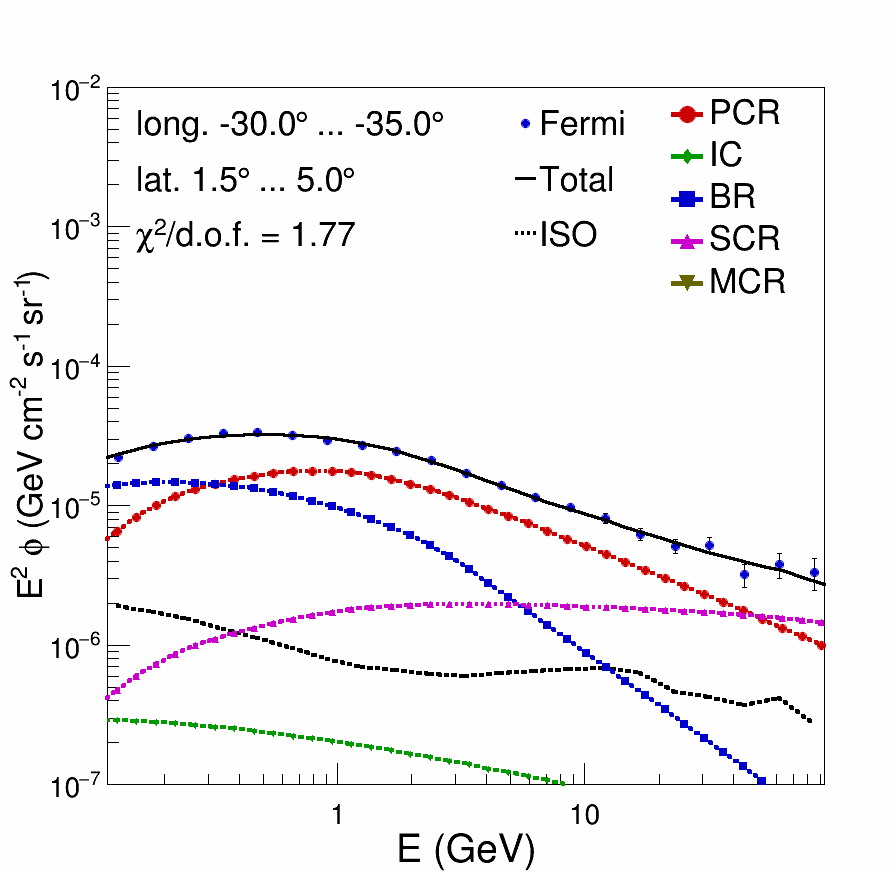}
\includegraphics[width=0.16\textwidth,height=0.16\textwidth,clip]{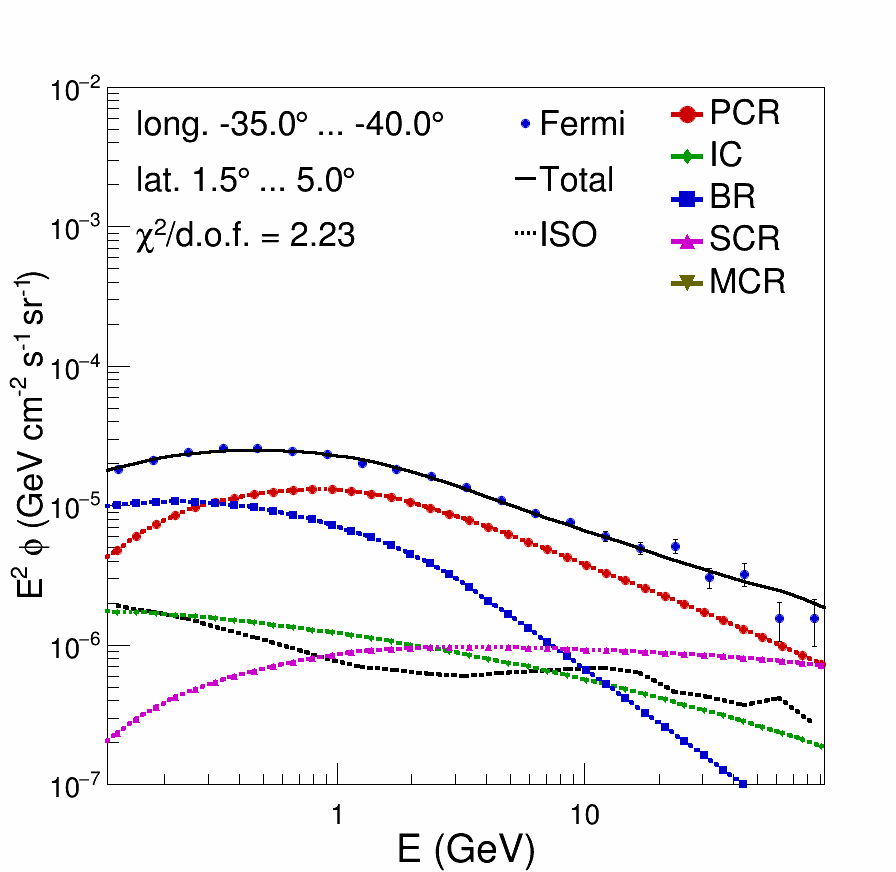}
\includegraphics[width=0.16\textwidth,height=0.16\textwidth,clip]{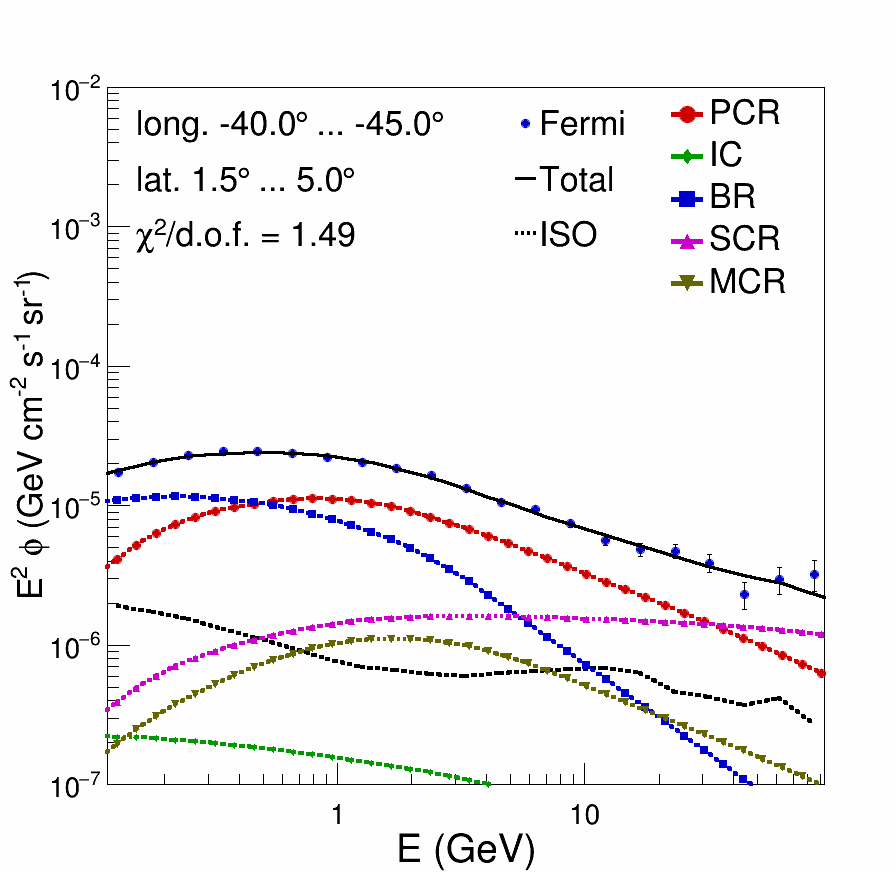}
\includegraphics[width=0.16\textwidth,height=0.16\textwidth,clip]{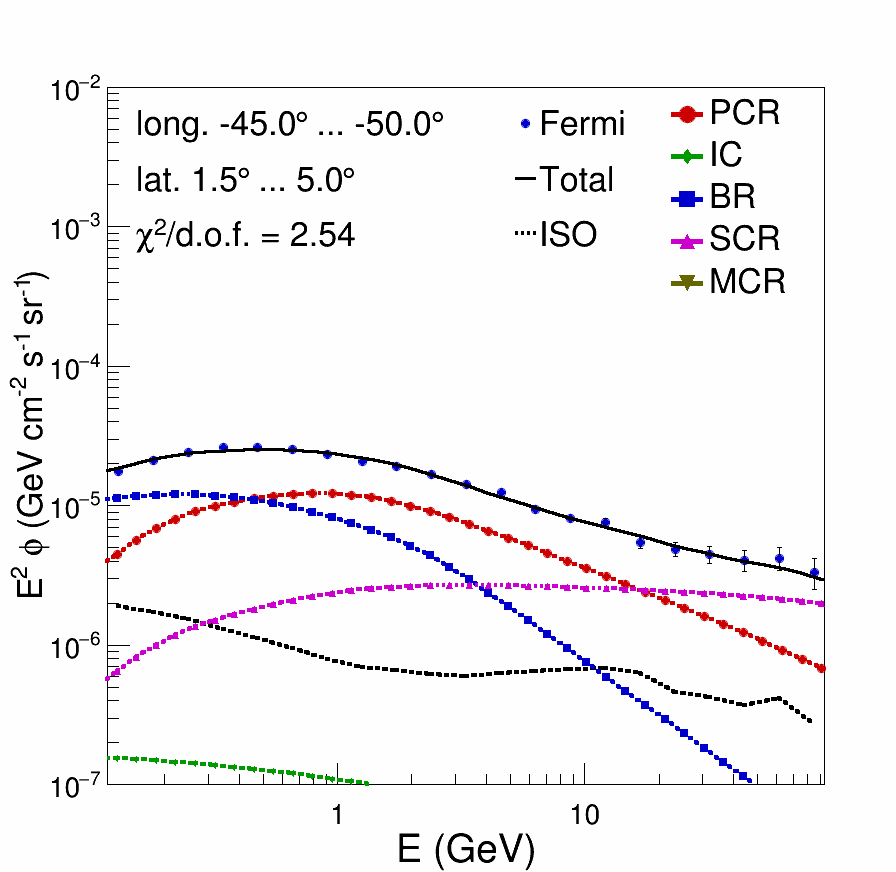}
\includegraphics[width=0.16\textwidth,height=0.16\textwidth,clip]{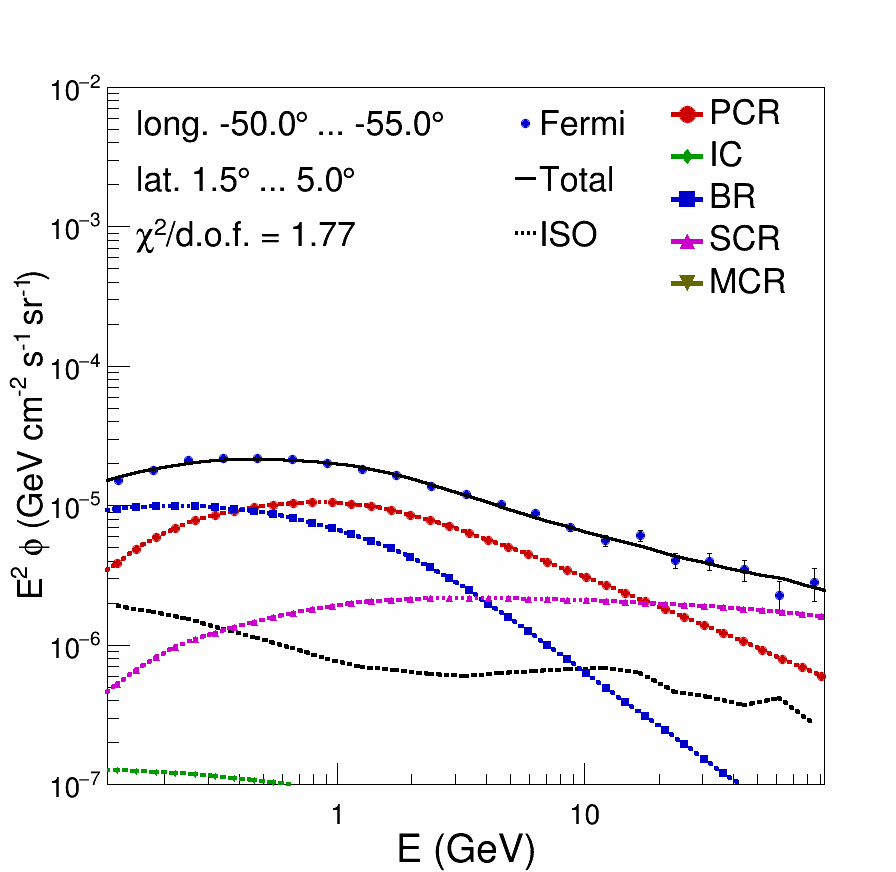}
\includegraphics[width=0.16\textwidth,height=0.16\textwidth,clip]{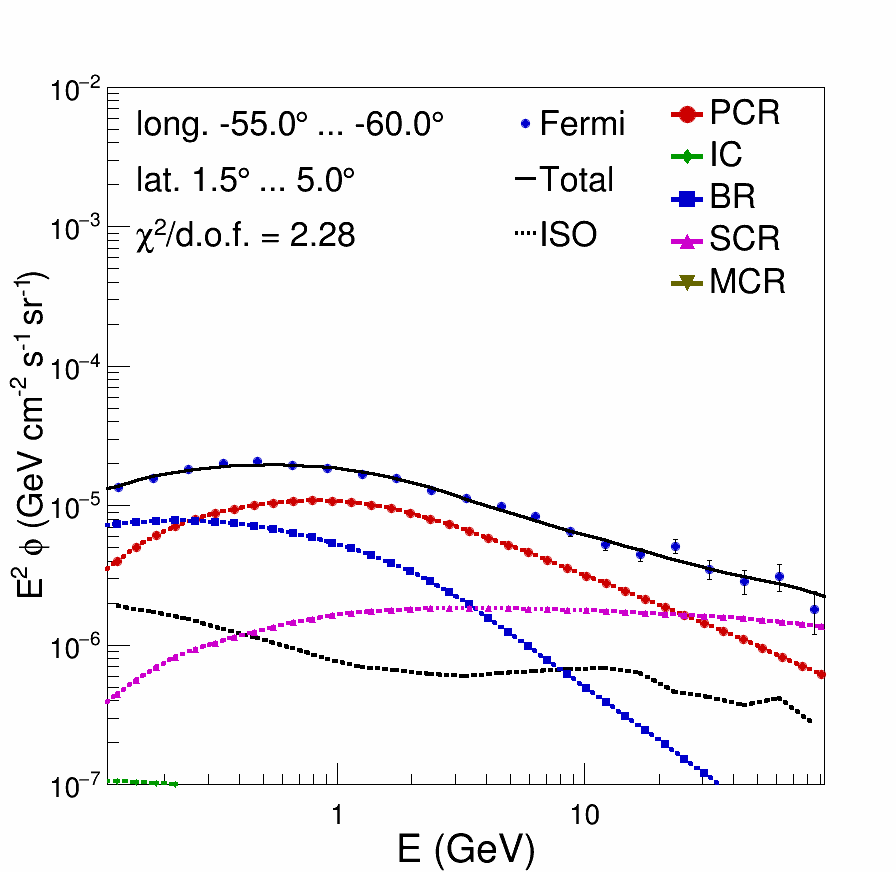}
\includegraphics[width=0.16\textwidth,height=0.16\textwidth,clip]{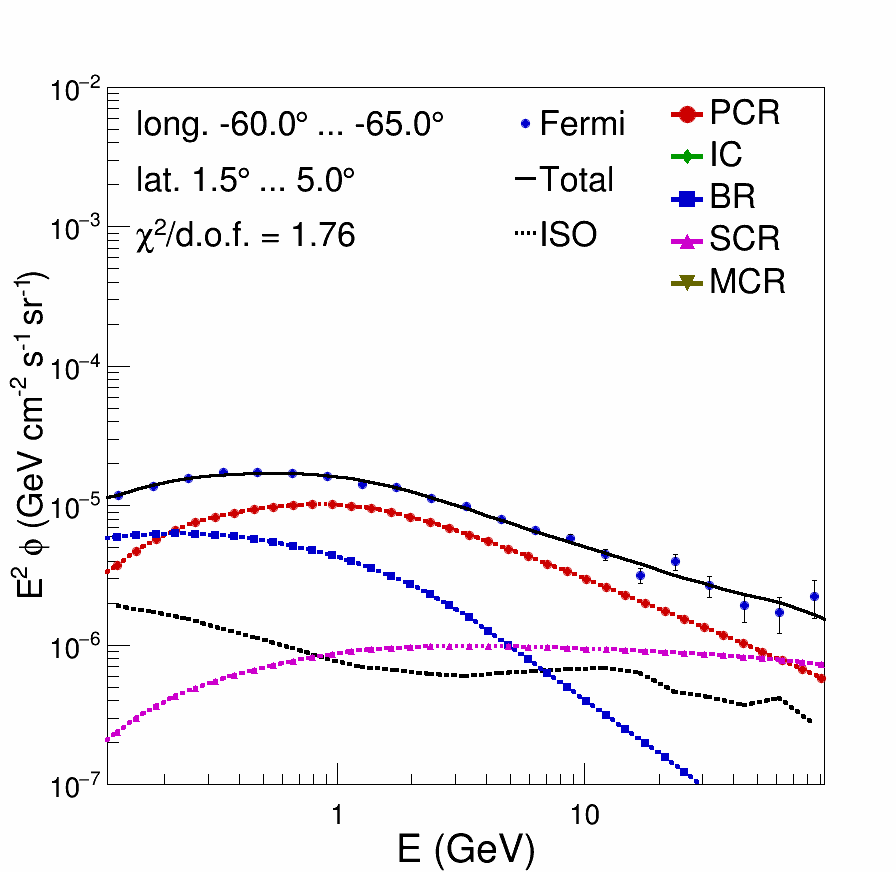}
\includegraphics[width=0.16\textwidth,height=0.16\textwidth,clip]{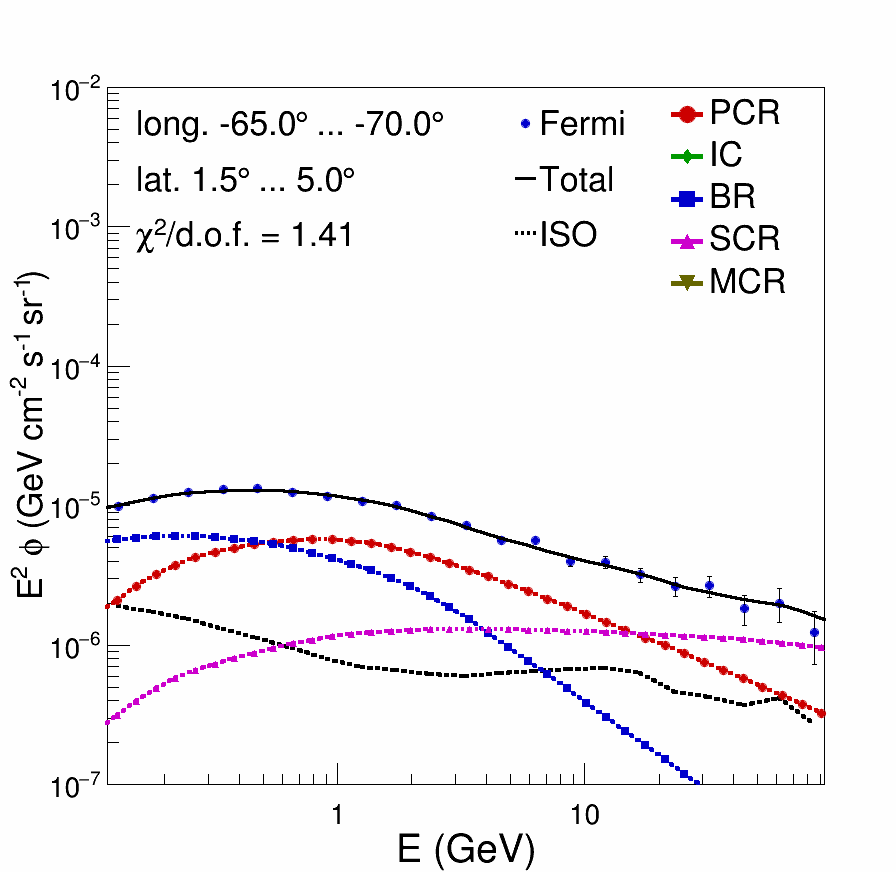}
\includegraphics[width=0.16\textwidth,height=0.16\textwidth,clip]{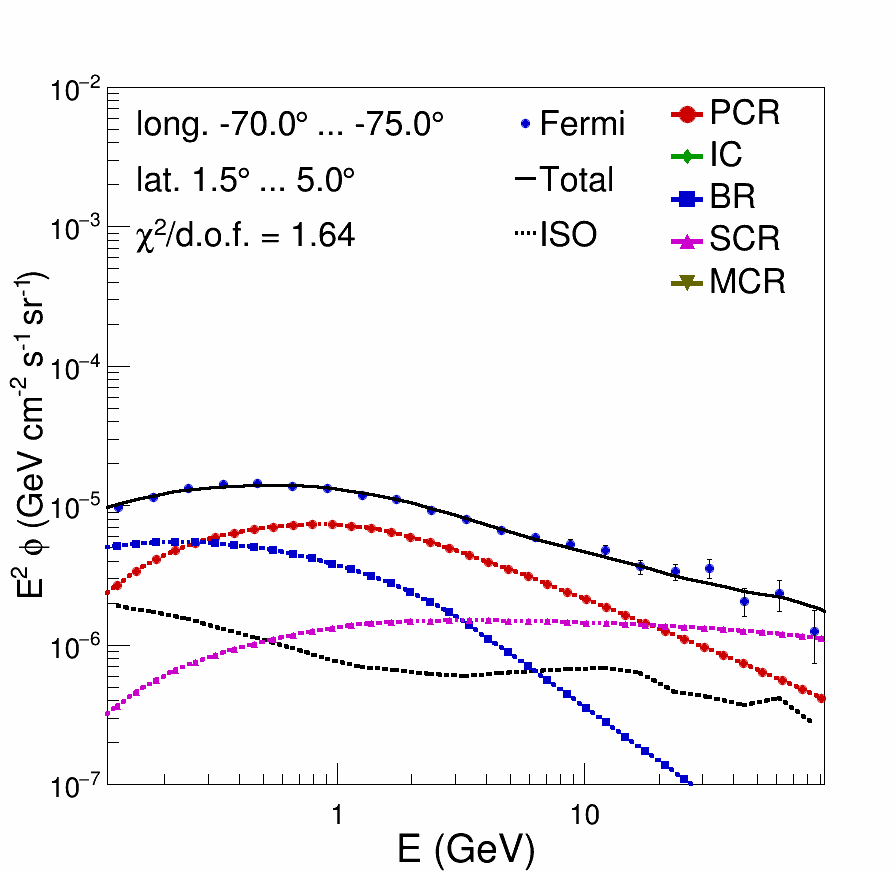}
\includegraphics[width=0.16\textwidth,height=0.16\textwidth,clip]{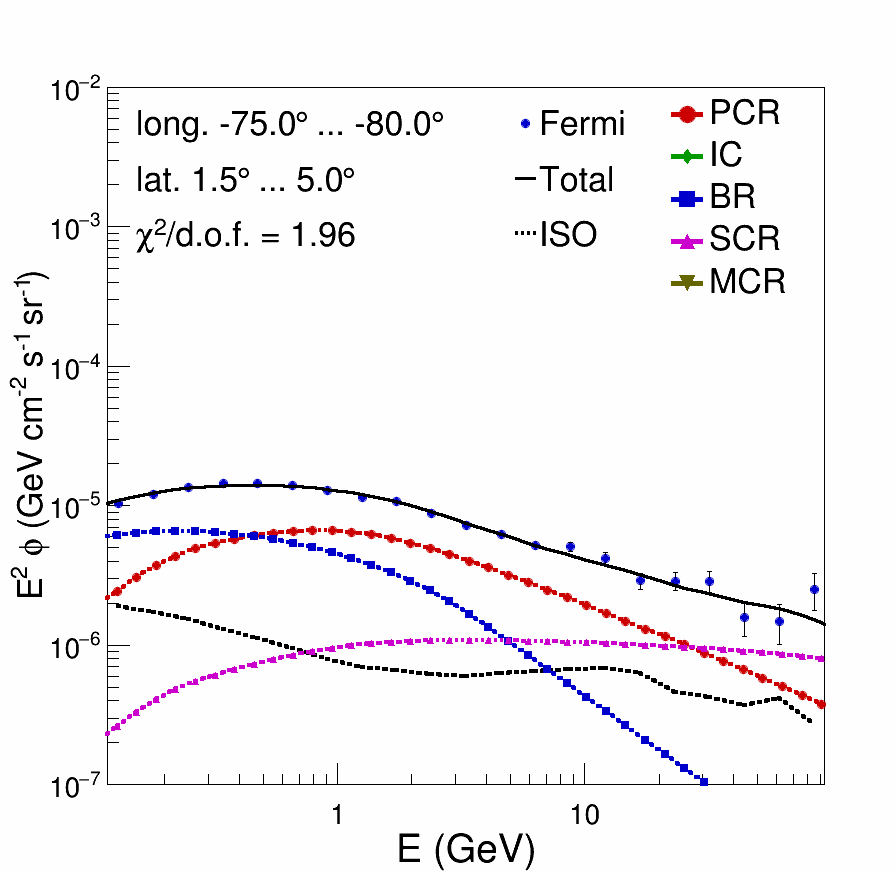}
\includegraphics[width=0.16\textwidth,height=0.16\textwidth,clip]{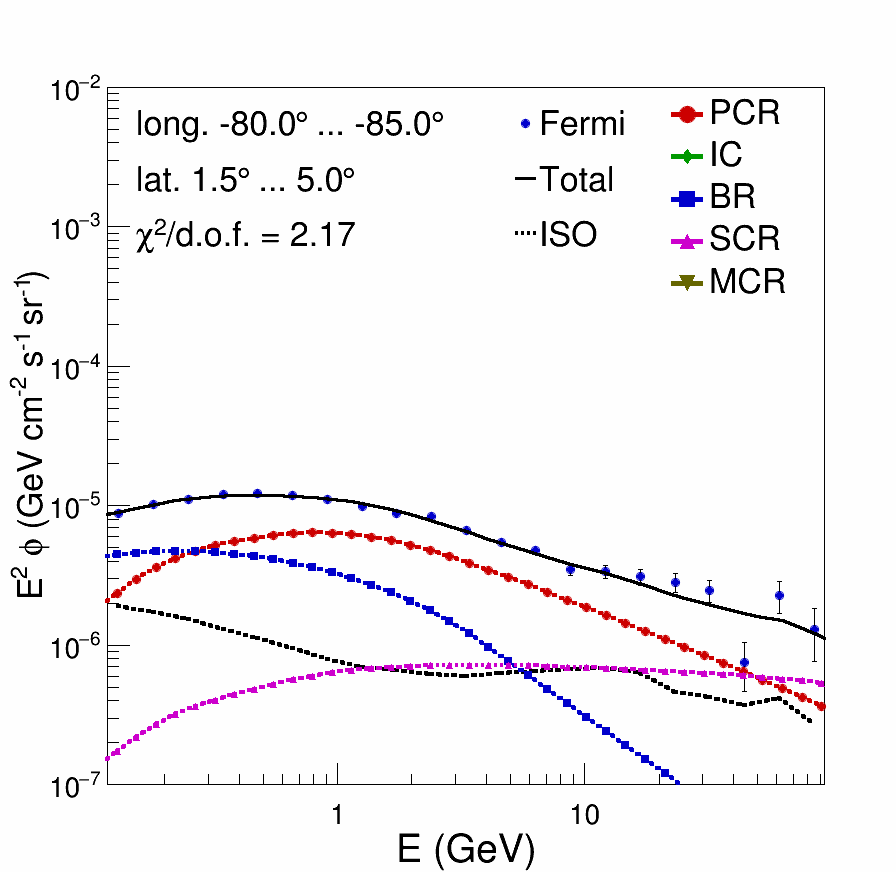}
\includegraphics[width=0.16\textwidth,height=0.16\textwidth,clip]{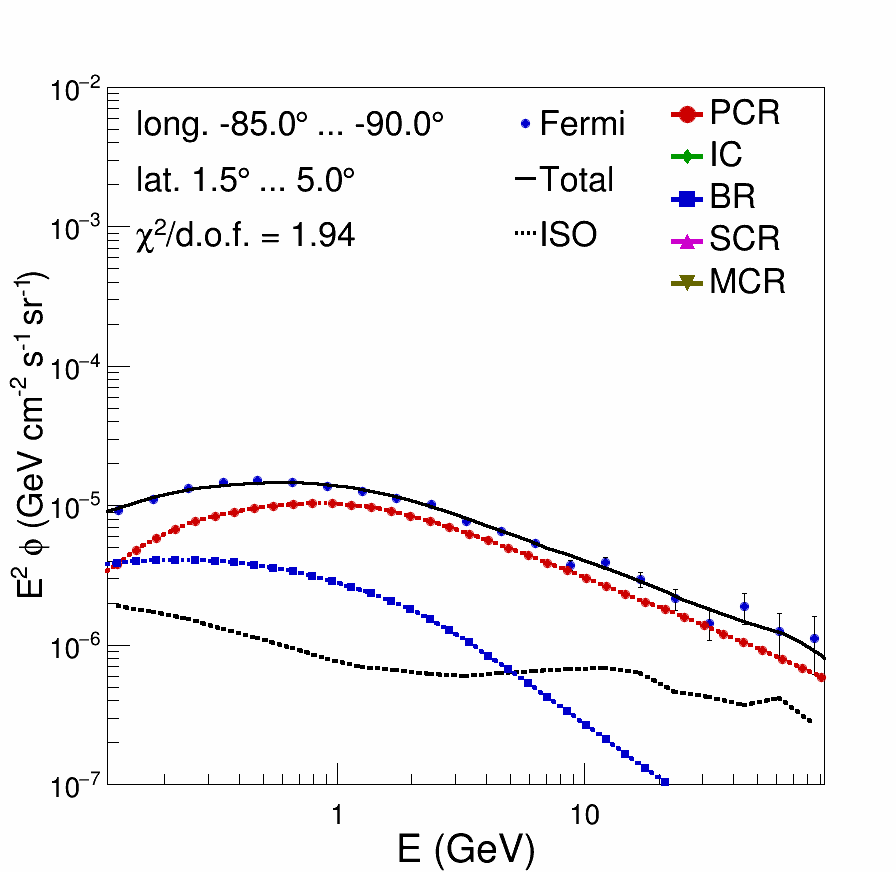}
\includegraphics[width=0.16\textwidth,height=0.16\textwidth,clip]{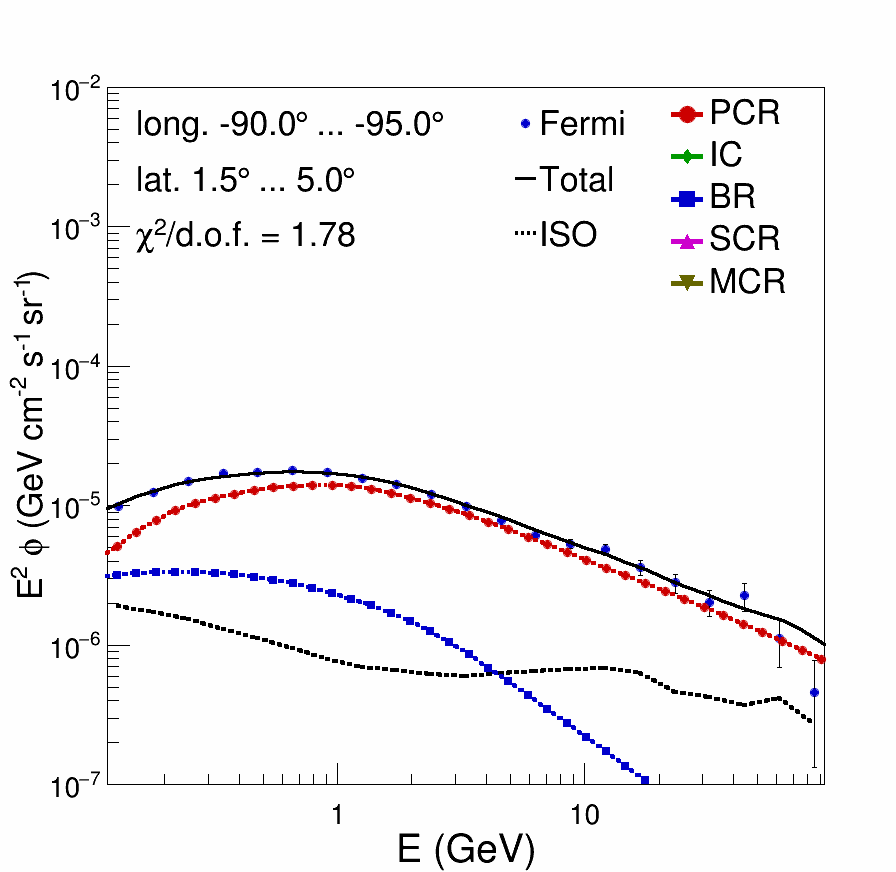}
\includegraphics[width=0.16\textwidth,height=0.16\textwidth,clip]{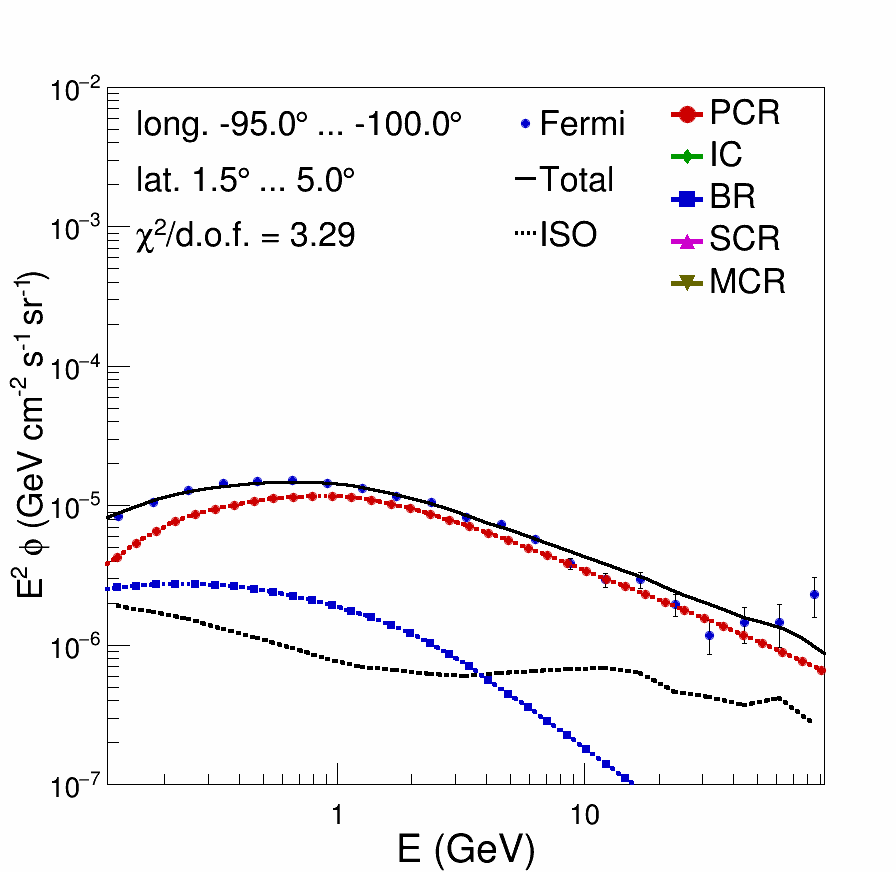}
\includegraphics[width=0.16\textwidth,height=0.16\textwidth,clip]{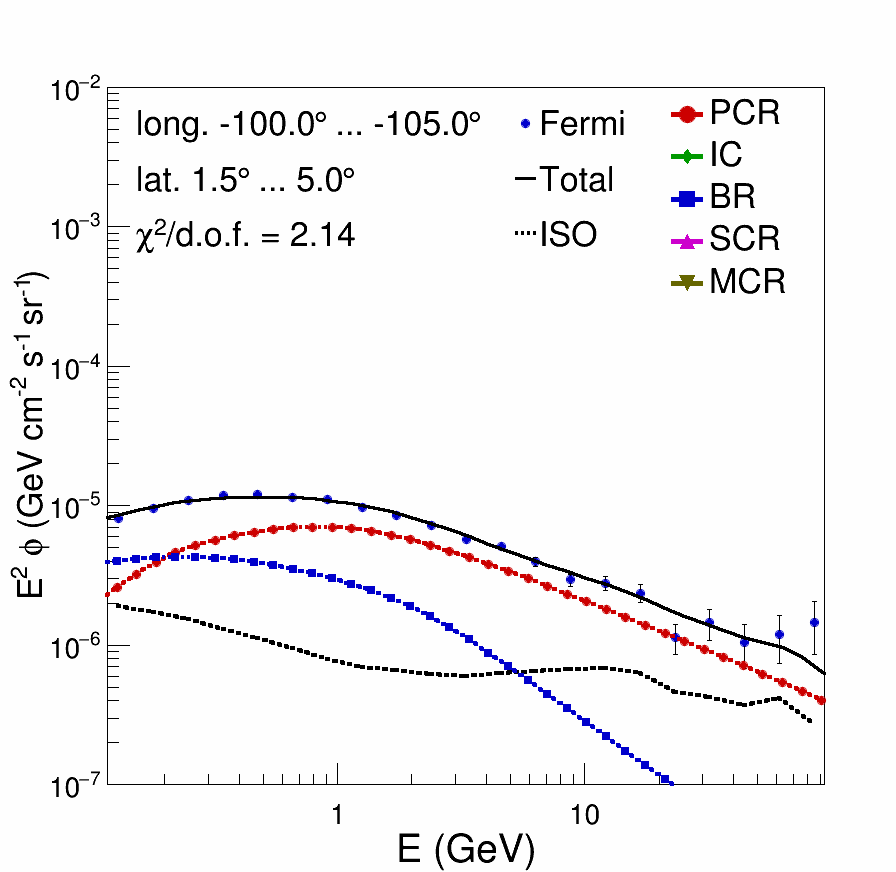}
\includegraphics[width=0.16\textwidth,height=0.16\textwidth,clip]{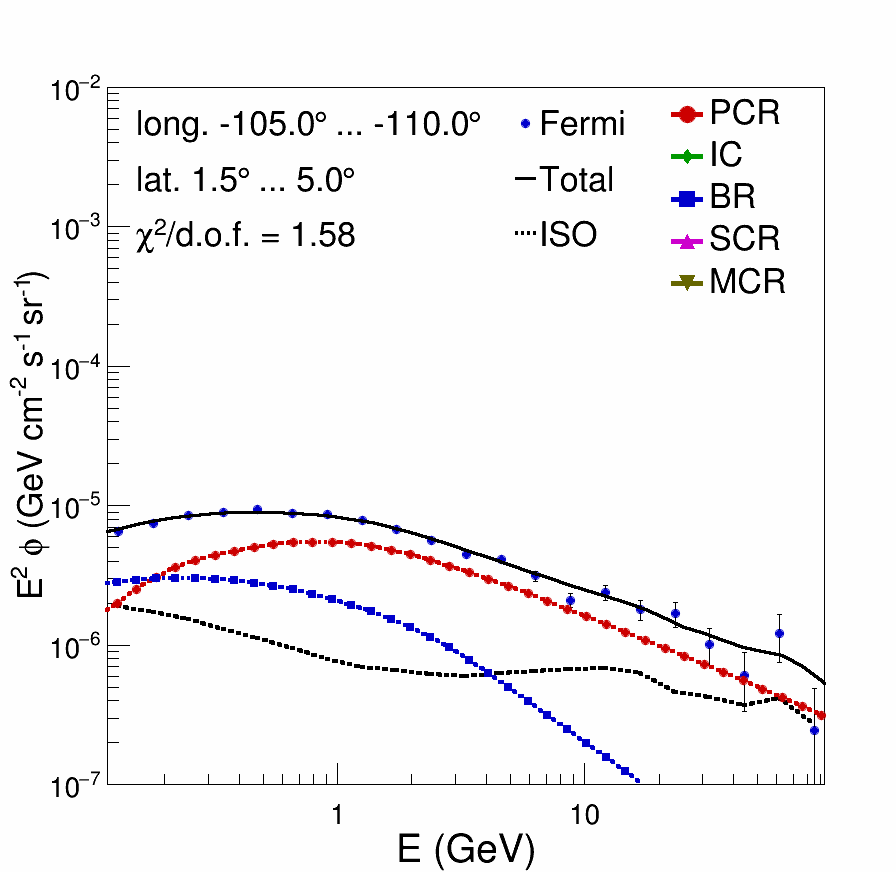}
\includegraphics[width=0.16\textwidth,height=0.16\textwidth,clip]{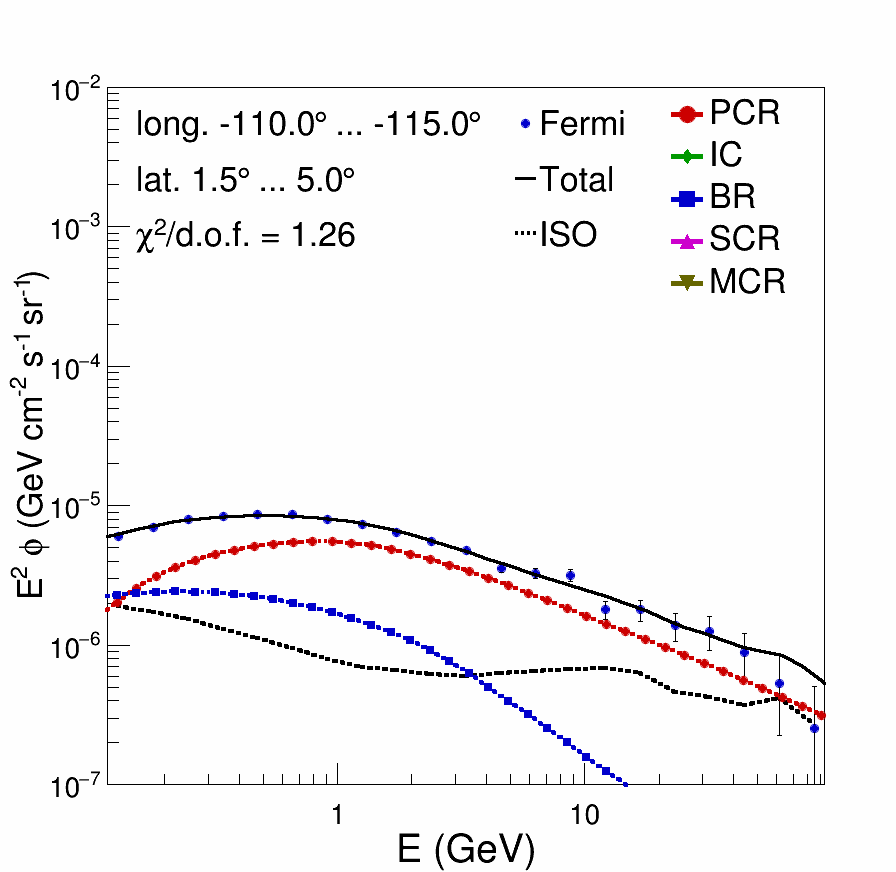}
\includegraphics[width=0.16\textwidth,height=0.16\textwidth,clip]{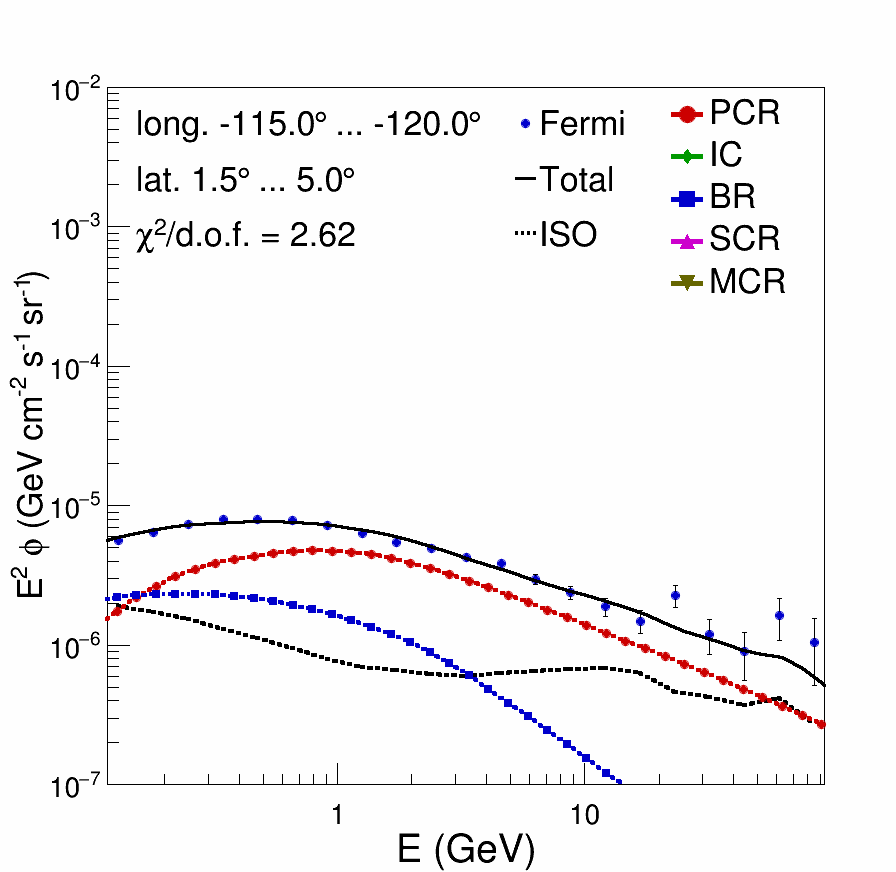}
\includegraphics[width=0.16\textwidth,height=0.16\textwidth,clip]{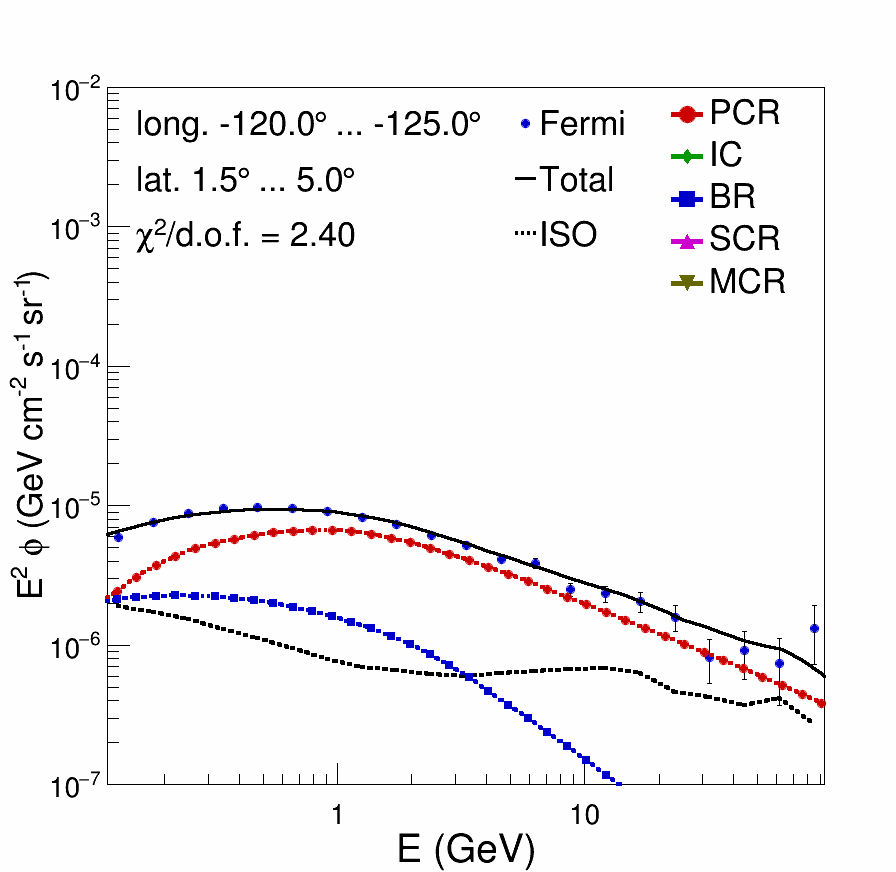}
\includegraphics[width=0.16\textwidth,height=0.16\textwidth,clip]{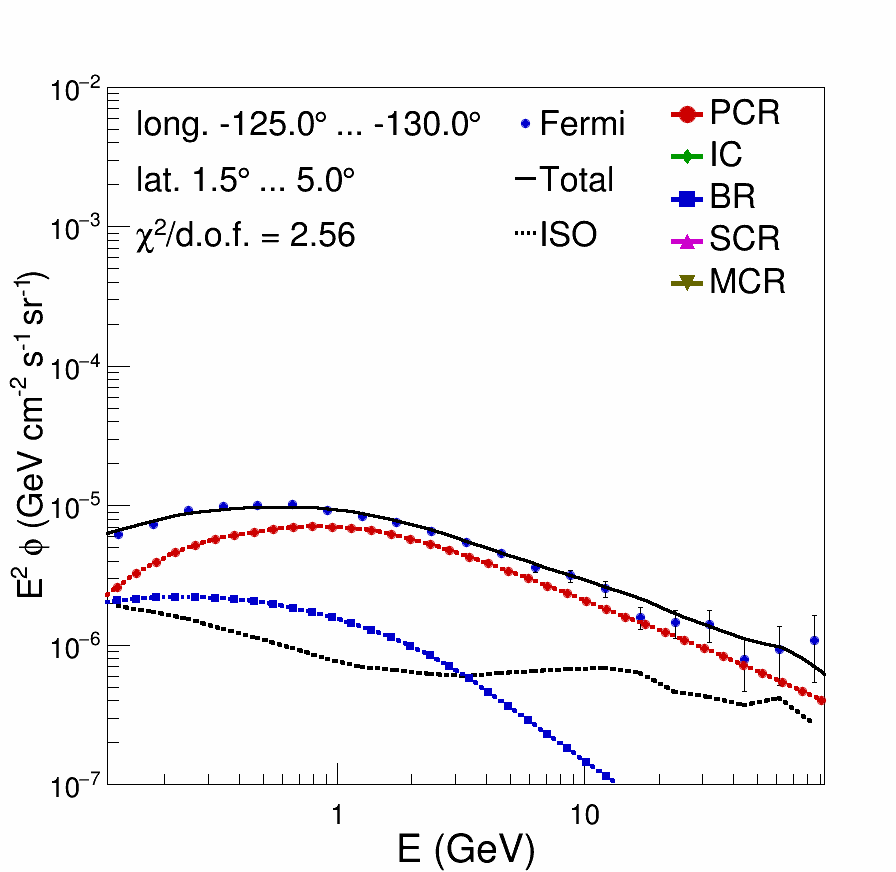}
\includegraphics[width=0.16\textwidth,height=0.16\textwidth,clip]{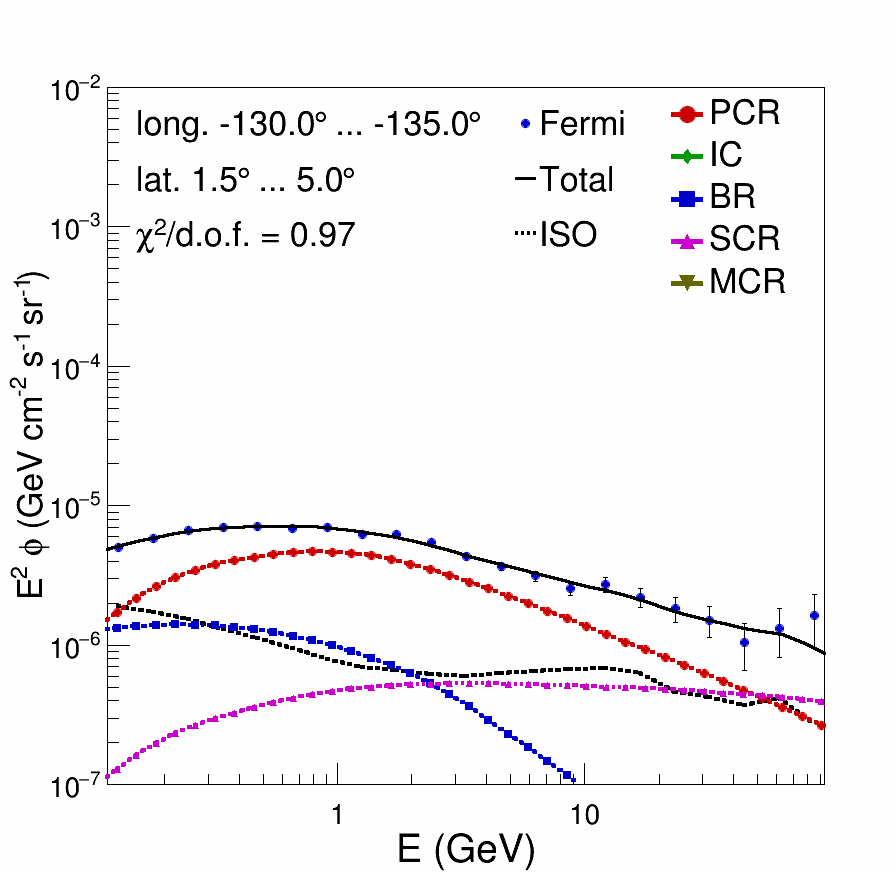}
\includegraphics[width=0.16\textwidth,height=0.16\textwidth,clip]{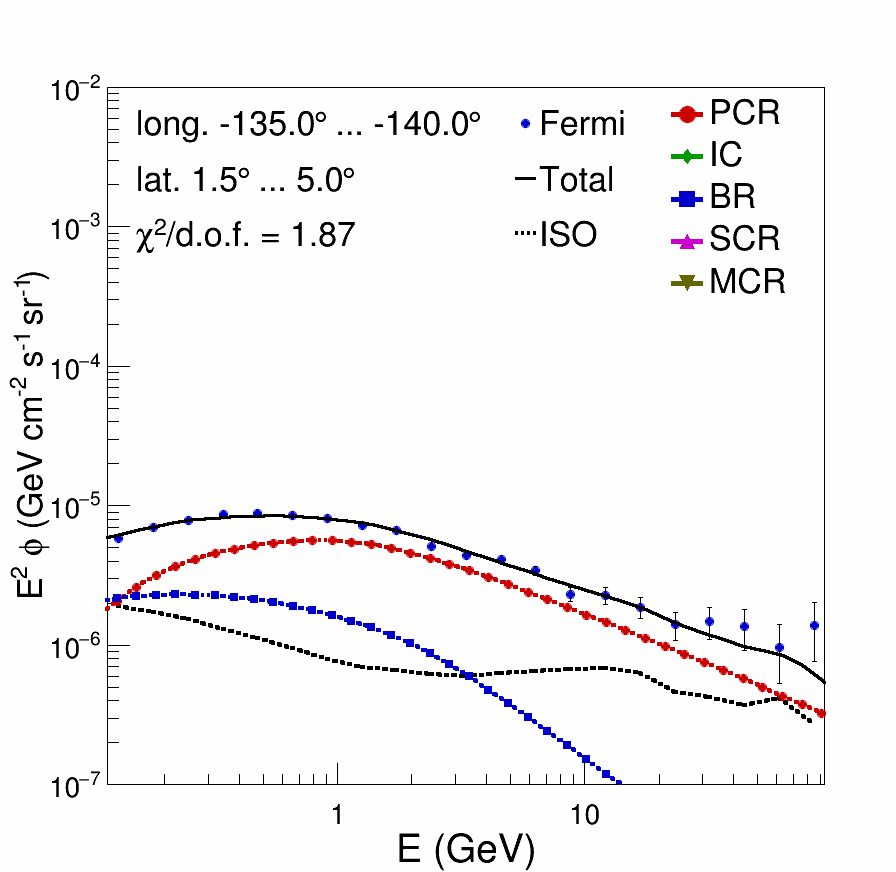}
\includegraphics[width=0.16\textwidth,height=0.16\textwidth,clip]{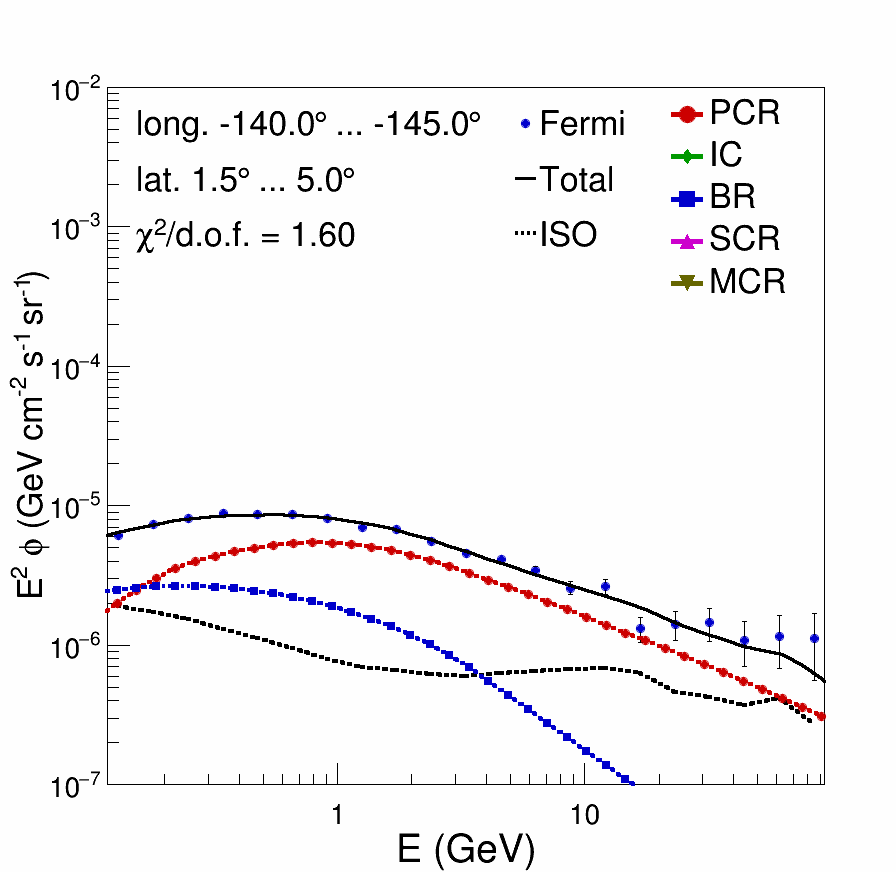}
\includegraphics[width=0.16\textwidth,height=0.16\textwidth,clip]{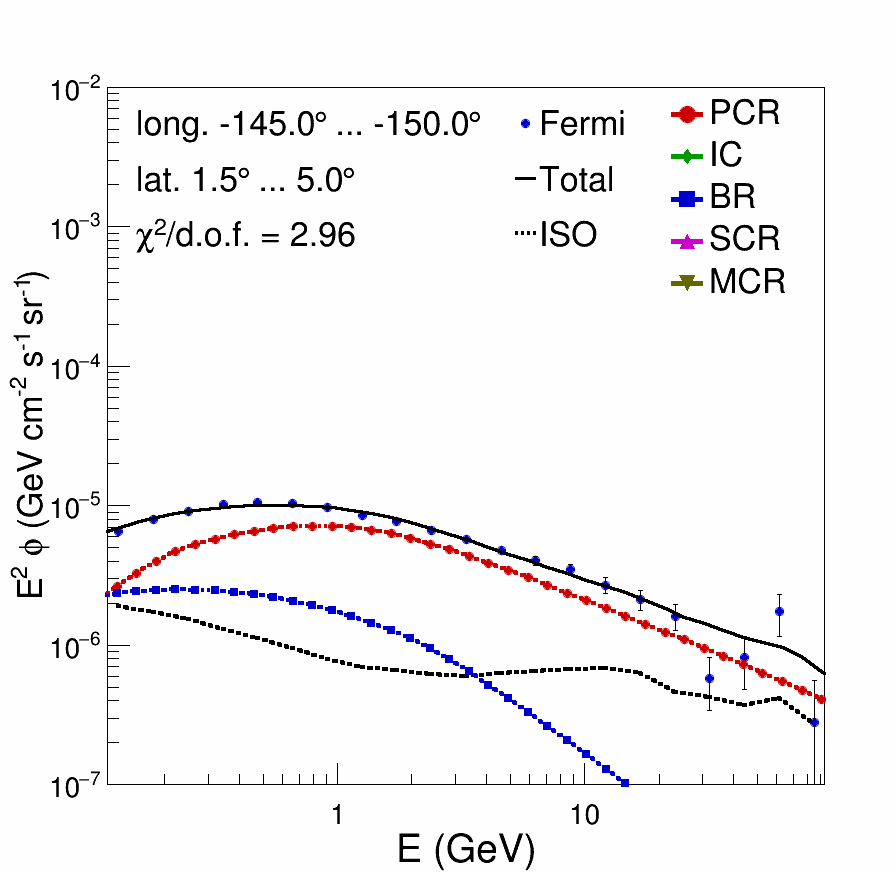}
\includegraphics[width=0.16\textwidth,height=0.16\textwidth,clip]{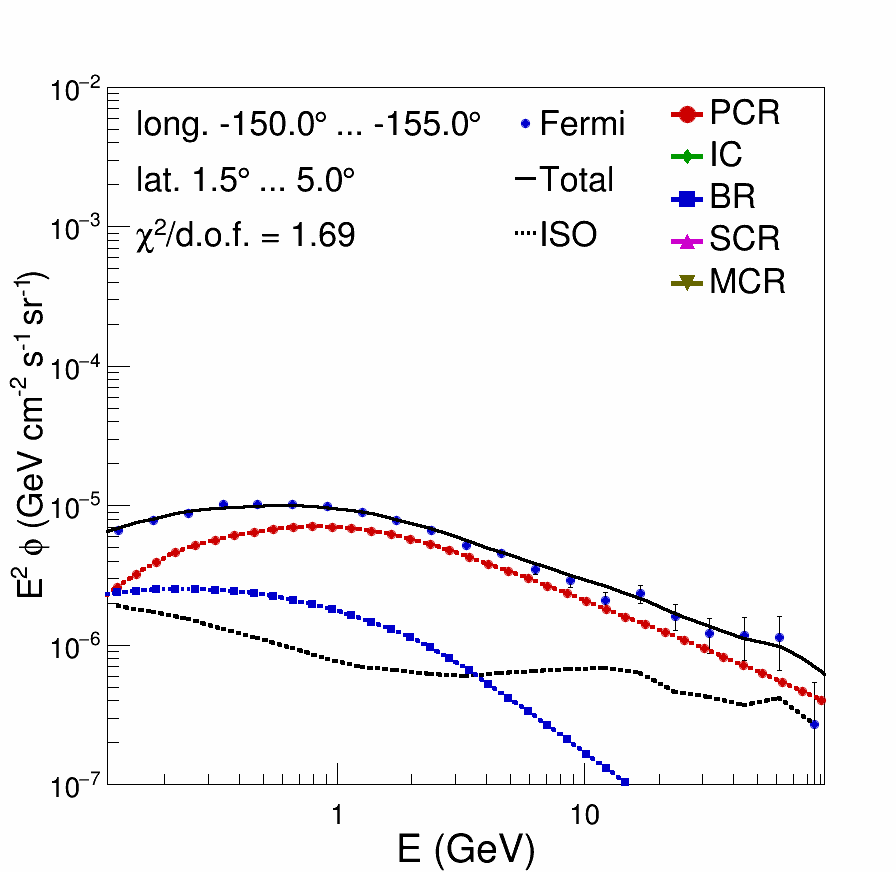}
\includegraphics[width=0.16\textwidth,height=0.16\textwidth,clip]{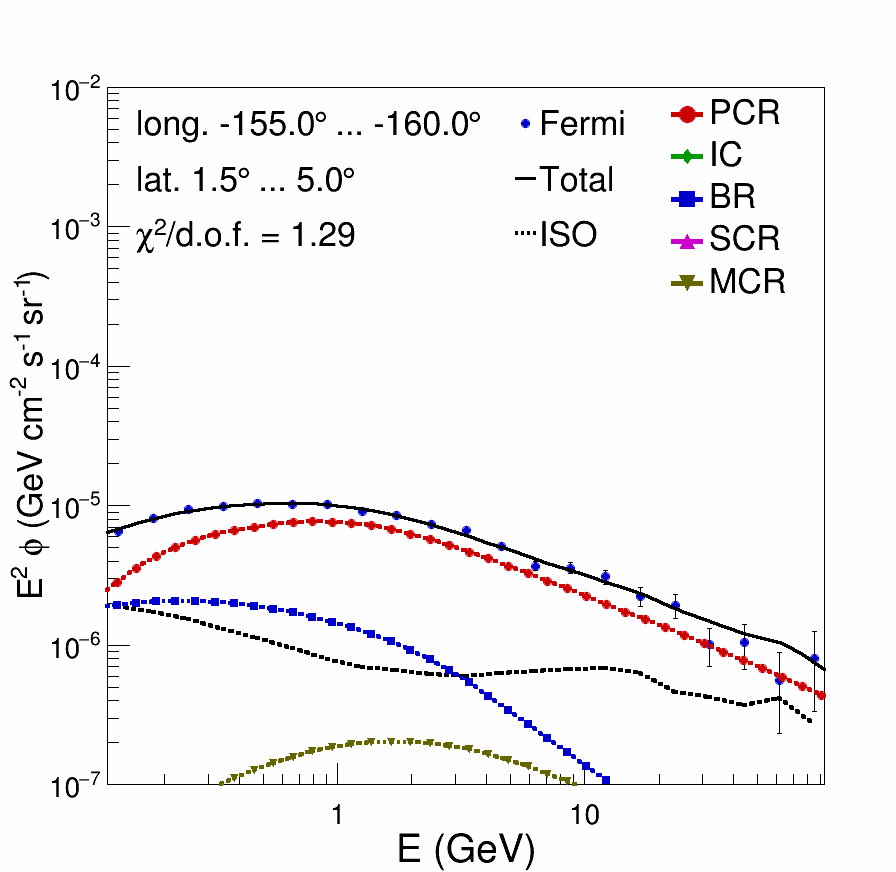}
\includegraphics[width=0.16\textwidth,height=0.16\textwidth,clip]{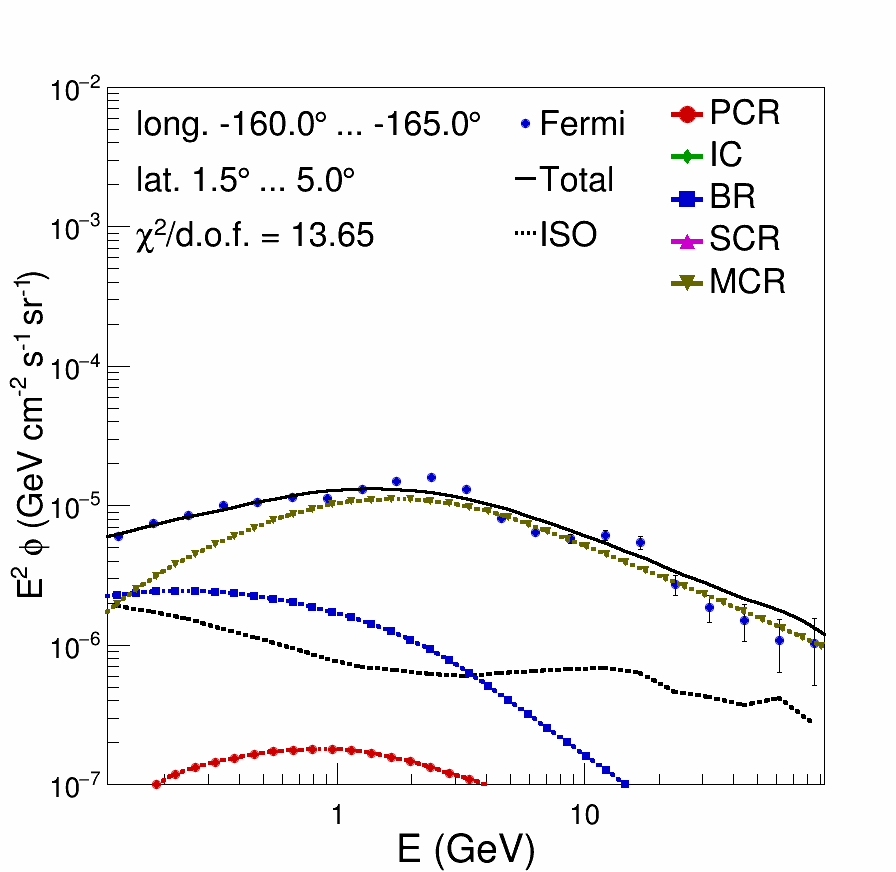}
\includegraphics[width=0.16\textwidth,height=0.16\textwidth,clip]{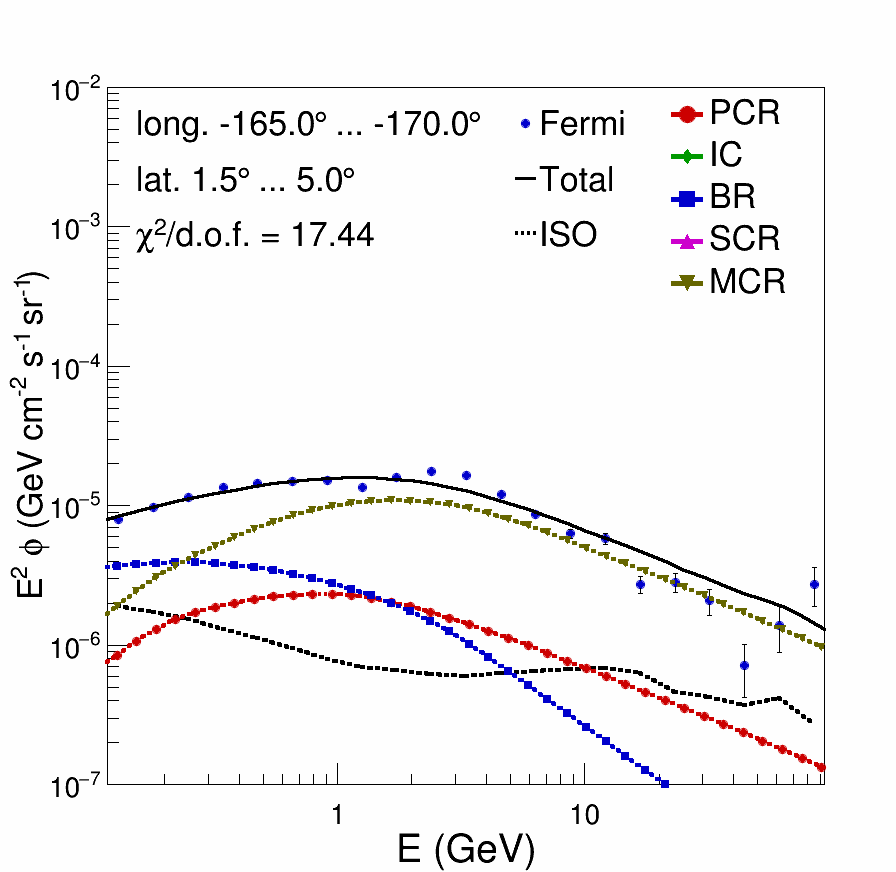}
\includegraphics[width=0.16\textwidth,height=0.16\textwidth,clip]{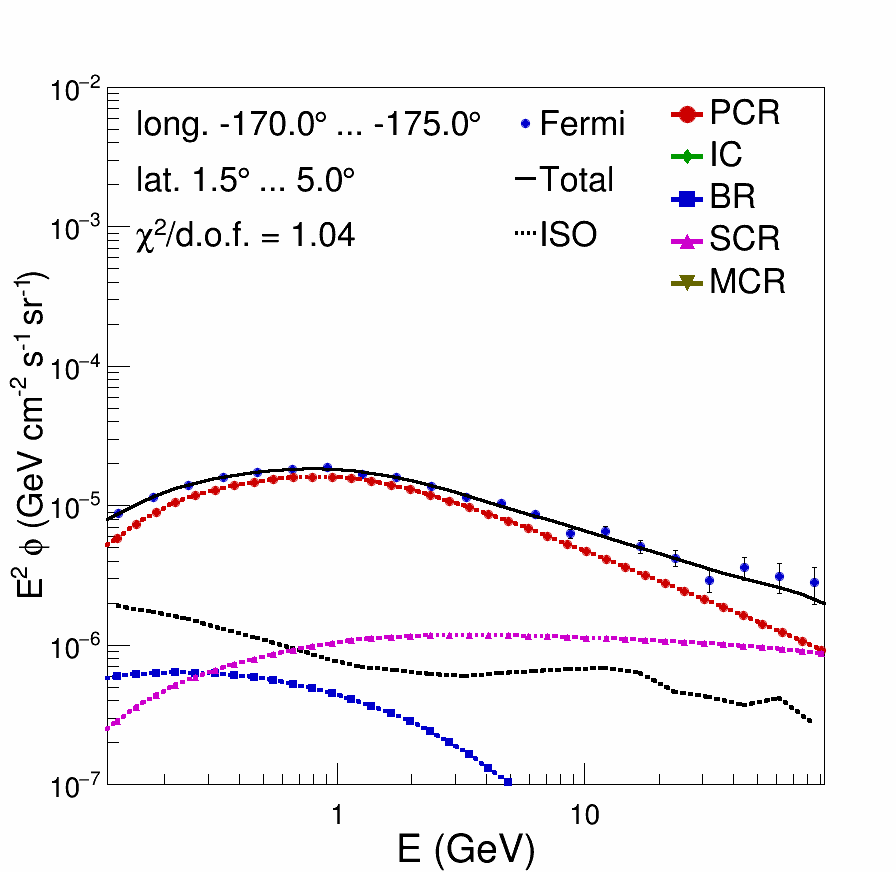}
\includegraphics[width=0.16\textwidth,height=0.16\textwidth,clip]{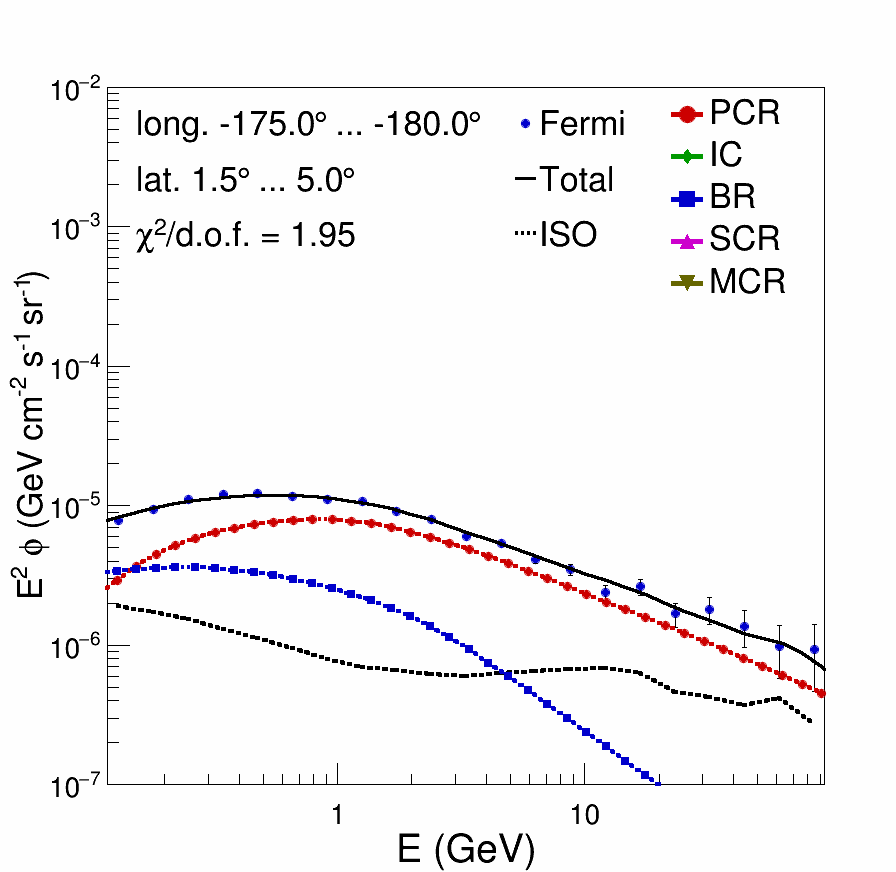}
\caption[]{Template fits for latitudes  with $1.5^\circ<b<5.0^\circ$ and longitudes decreasing from 0$^\circ$ to -180$^\circ$.} \label{F19}
\end{figure}
\begin{figure}
\centering
\includegraphics[width=0.16\textwidth,height=0.16\textwidth,clip]{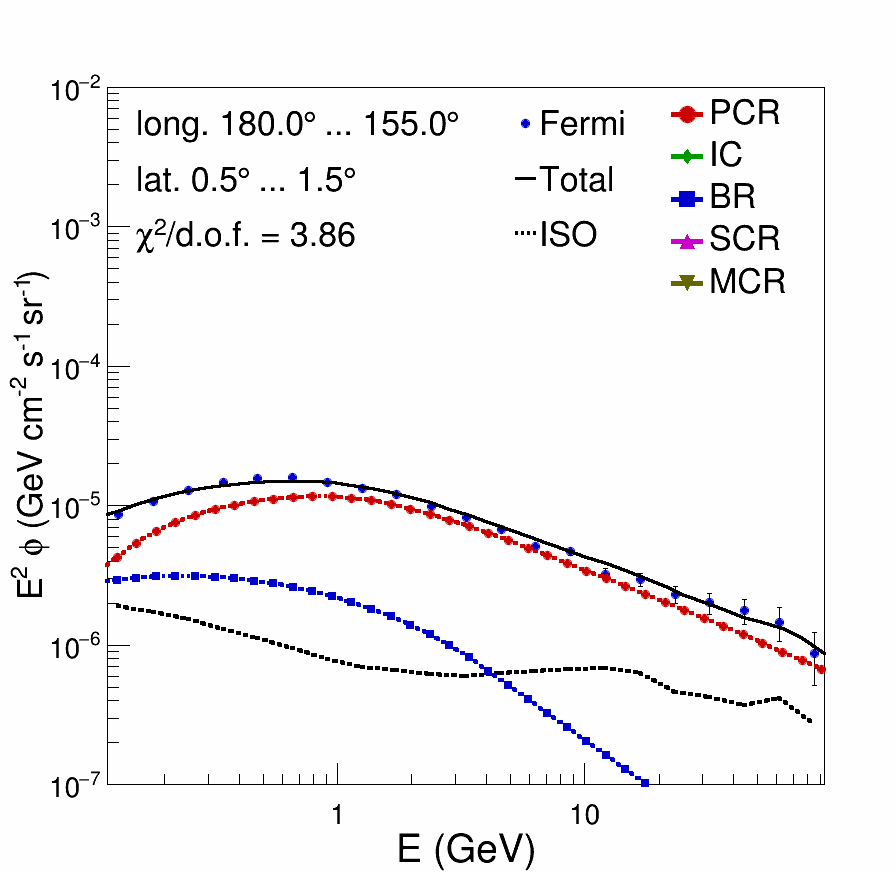}
\includegraphics[width=0.16\textwidth,height=0.16\textwidth,clip]{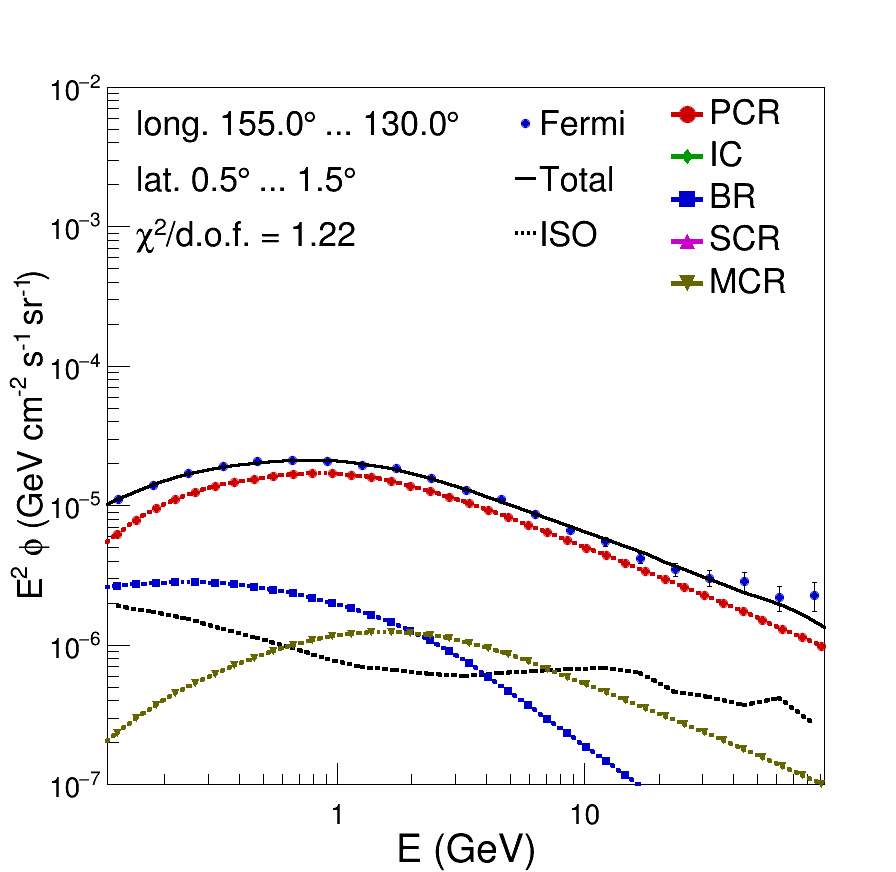}
\includegraphics[width=0.16\textwidth,height=0.16\textwidth,clip]{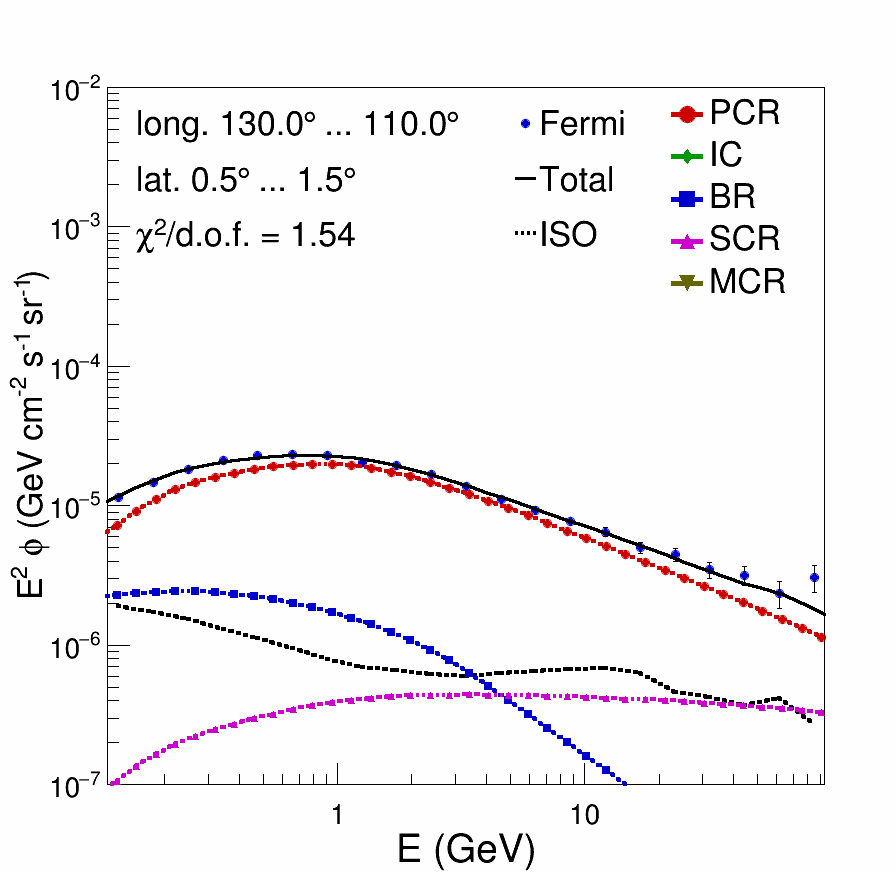}
\includegraphics[width=0.16\textwidth,height=0.16\textwidth,clip]{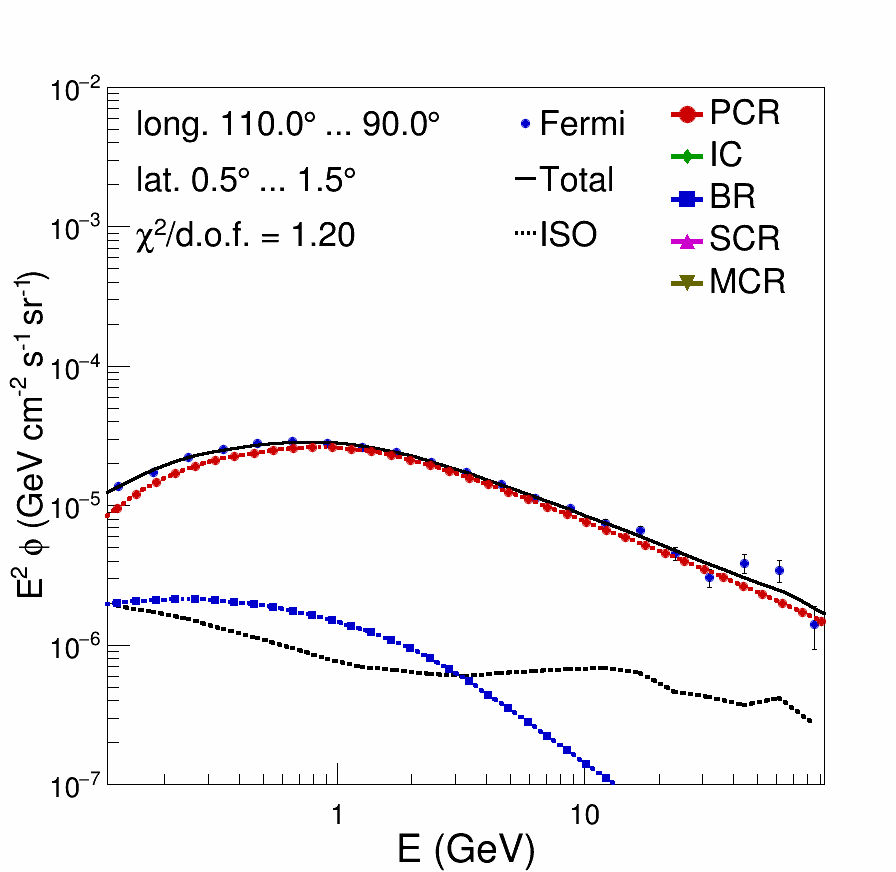}
\includegraphics[width=0.16\textwidth,height=0.16\textwidth,clip]{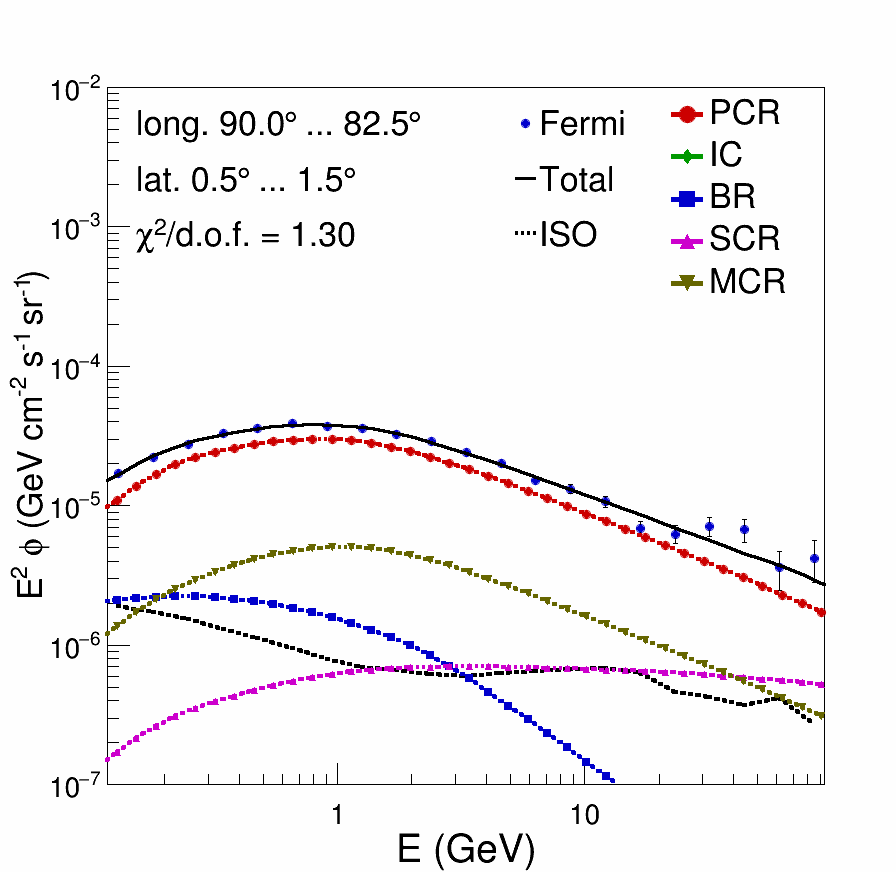}
\includegraphics[width=0.16\textwidth,height=0.16\textwidth,clip]{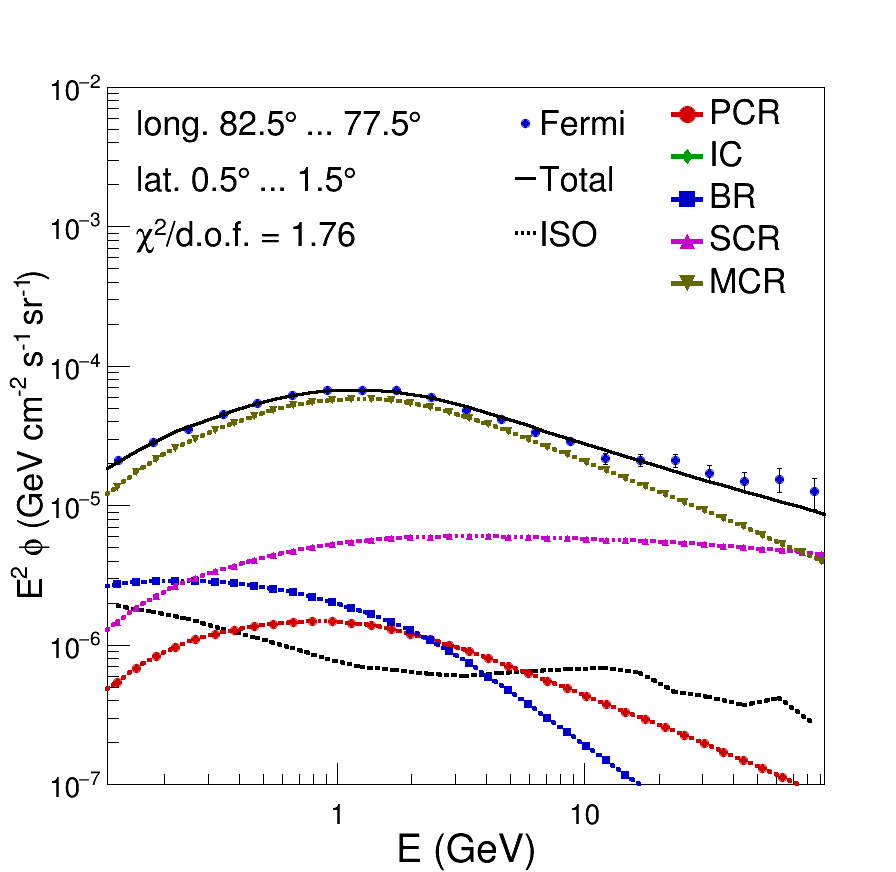}
\includegraphics[width=0.16\textwidth,height=0.16\textwidth,clip]{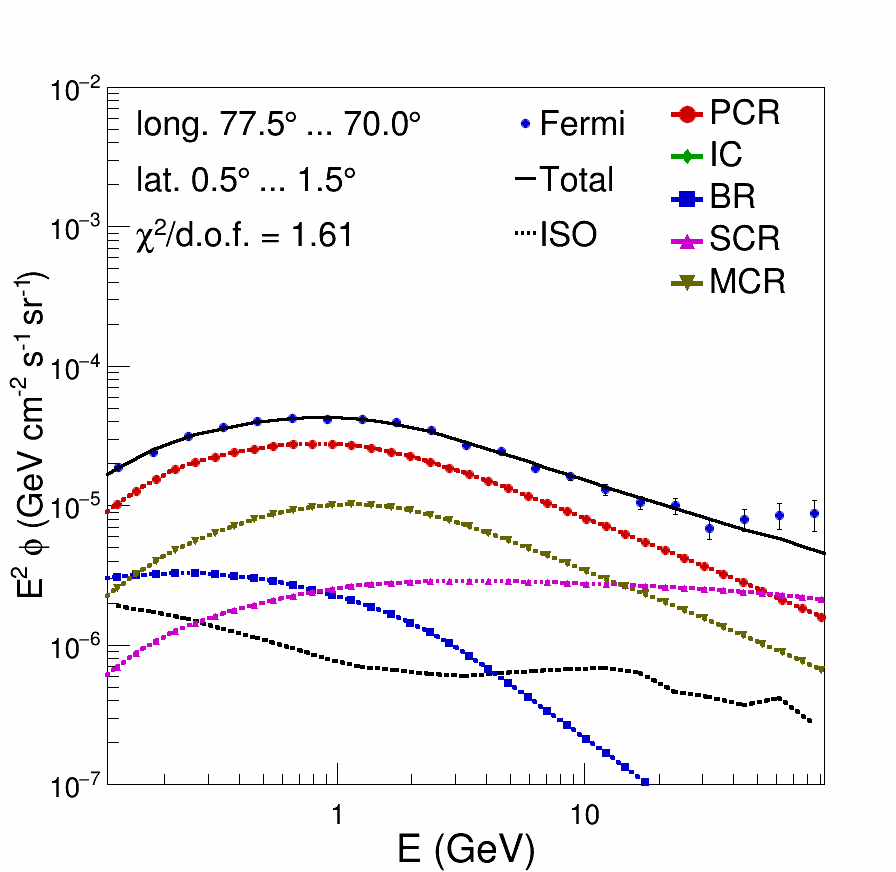}
\includegraphics[width=0.16\textwidth,height=0.16\textwidth,clip]{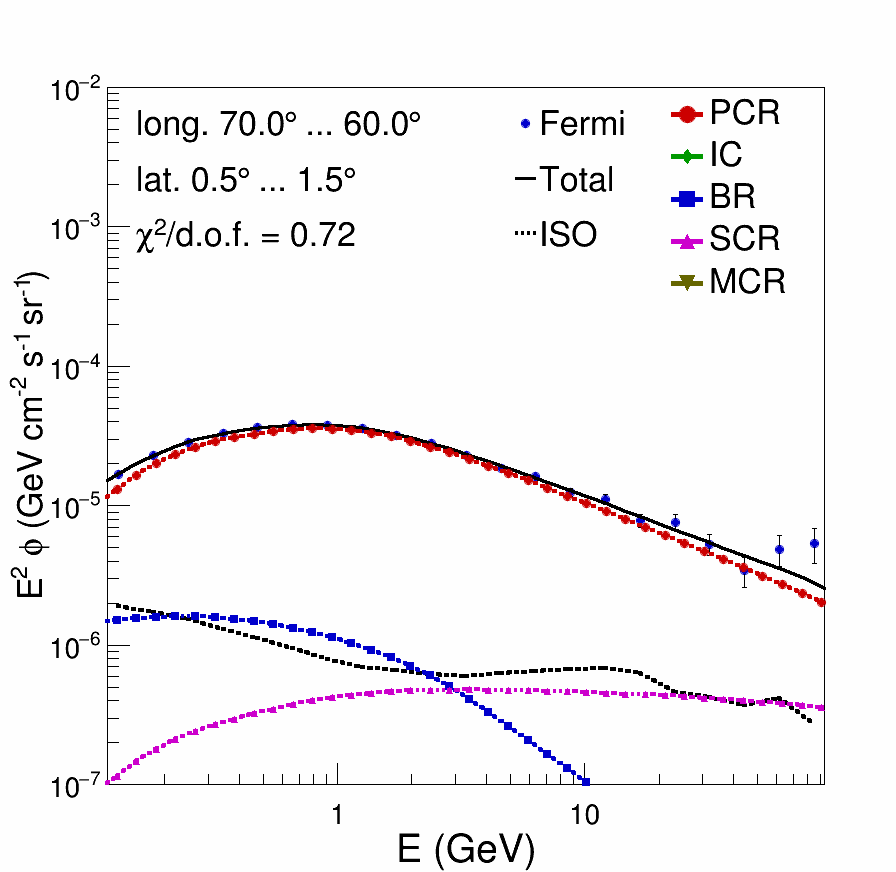}
\includegraphics[width=0.16\textwidth,height=0.16\textwidth,clip]{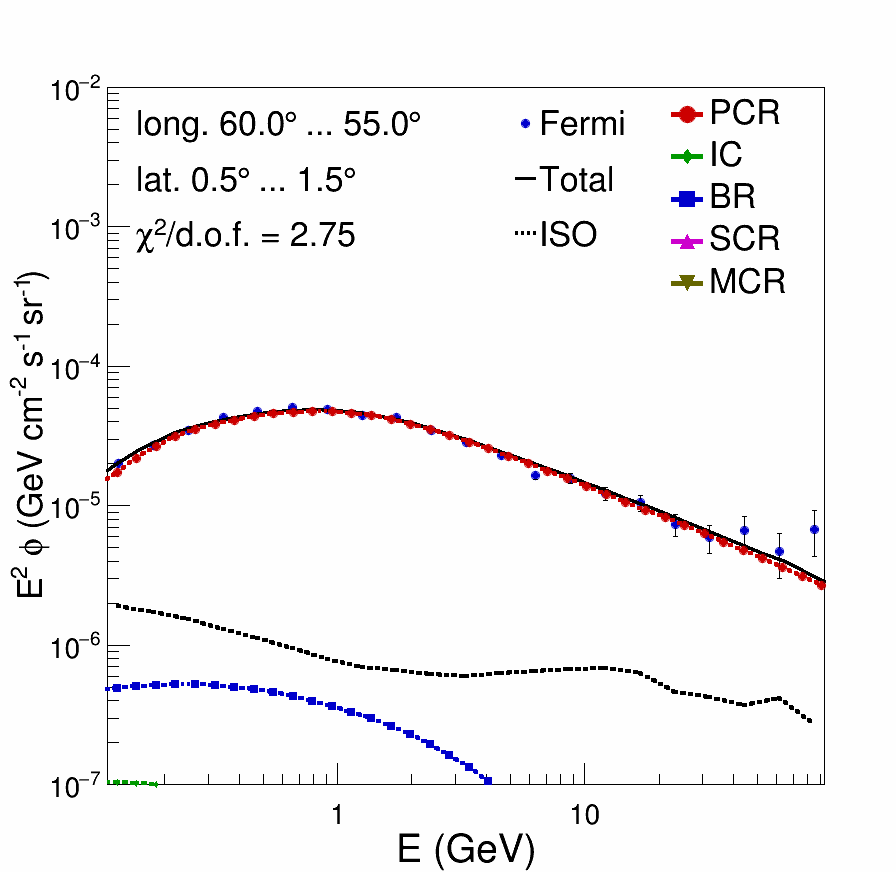}
\includegraphics[width=0.16\textwidth,height=0.16\textwidth,clip]{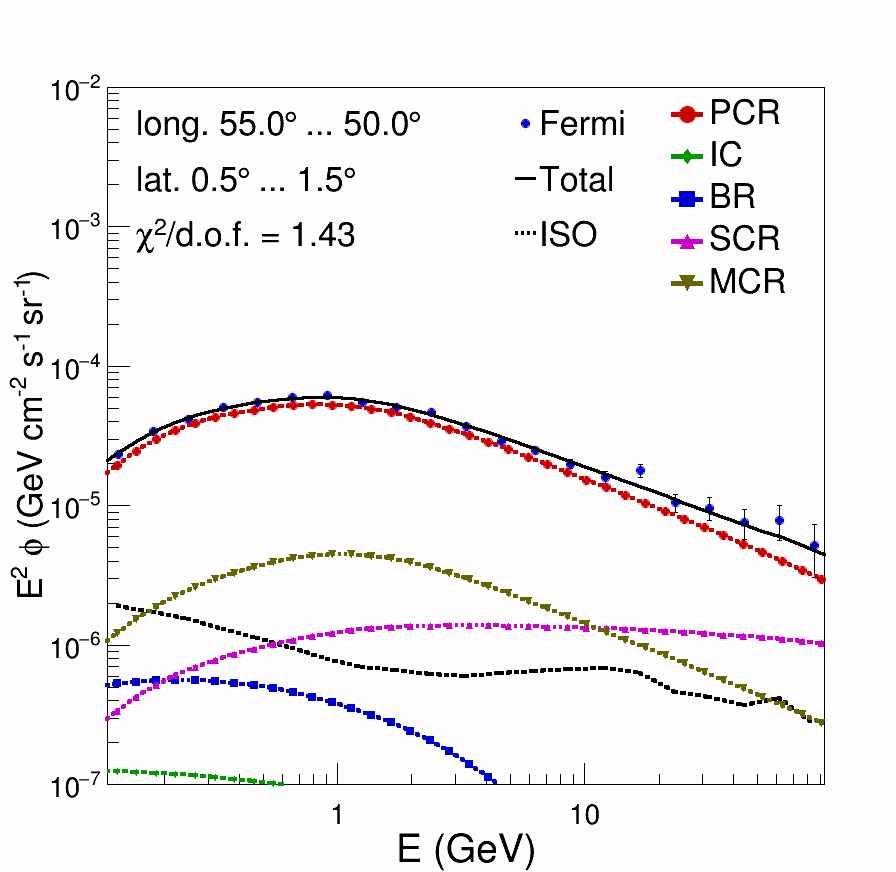}
\includegraphics[width=0.16\textwidth,height=0.16\textwidth,clip]{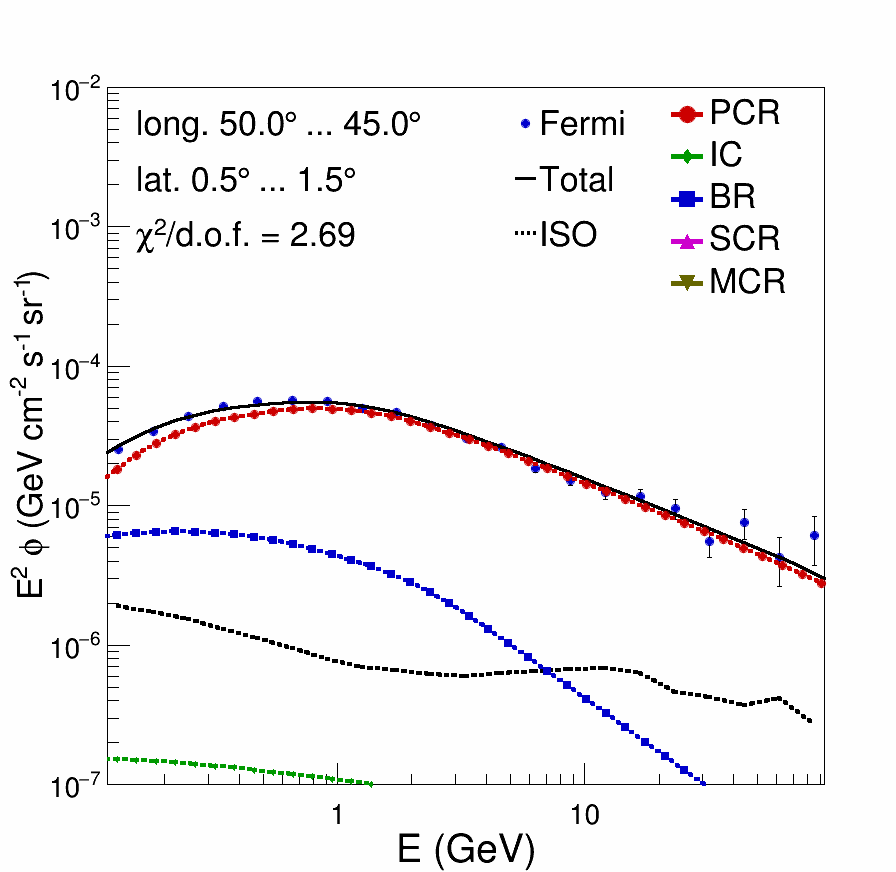}
\includegraphics[width=0.16\textwidth,height=0.16\textwidth,clip]{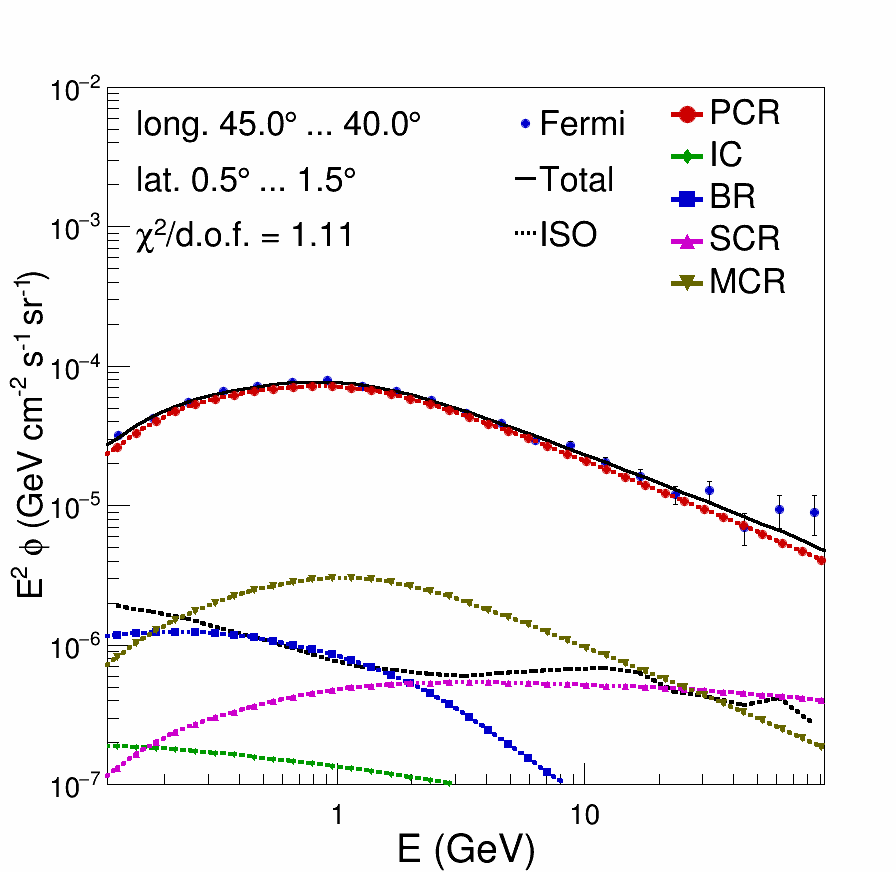}
\includegraphics[width=0.16\textwidth,height=0.16\textwidth,clip]{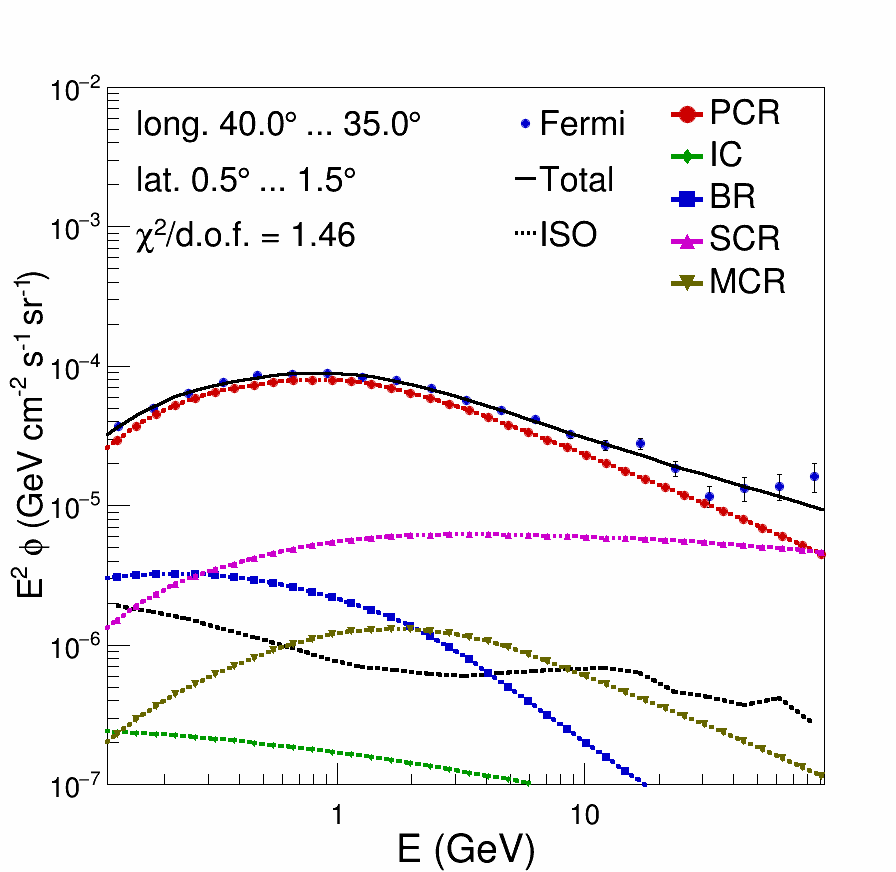}
\includegraphics[width=0.16\textwidth,height=0.16\textwidth,clip]{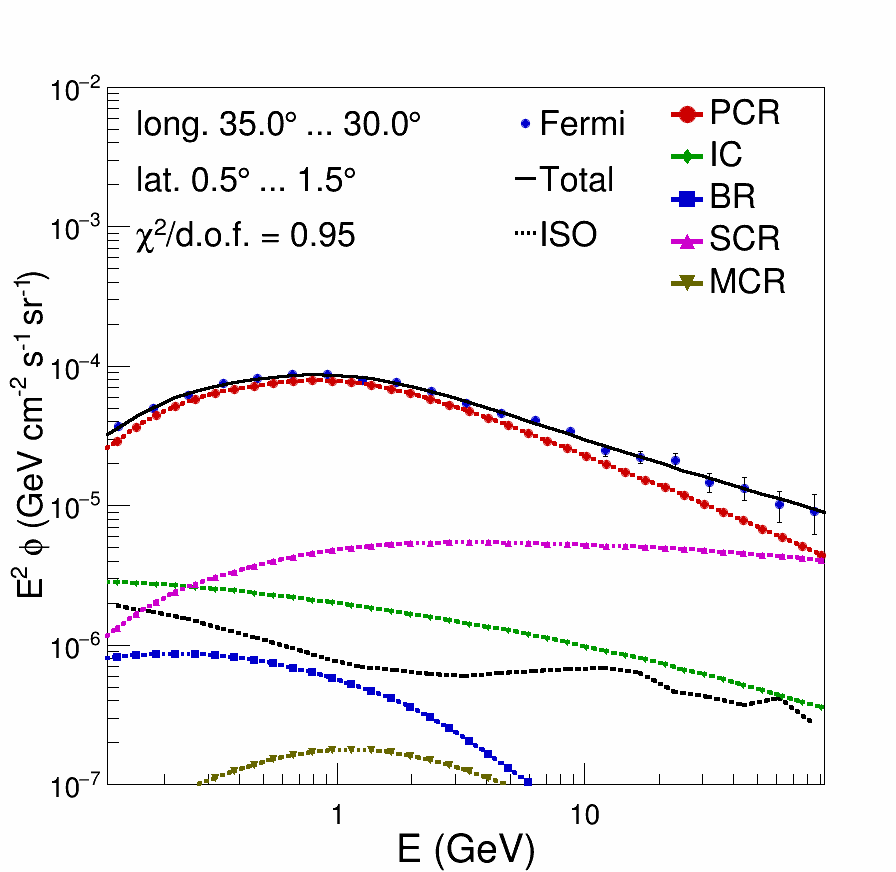}
\includegraphics[width=0.16\textwidth,height=0.16\textwidth,clip]{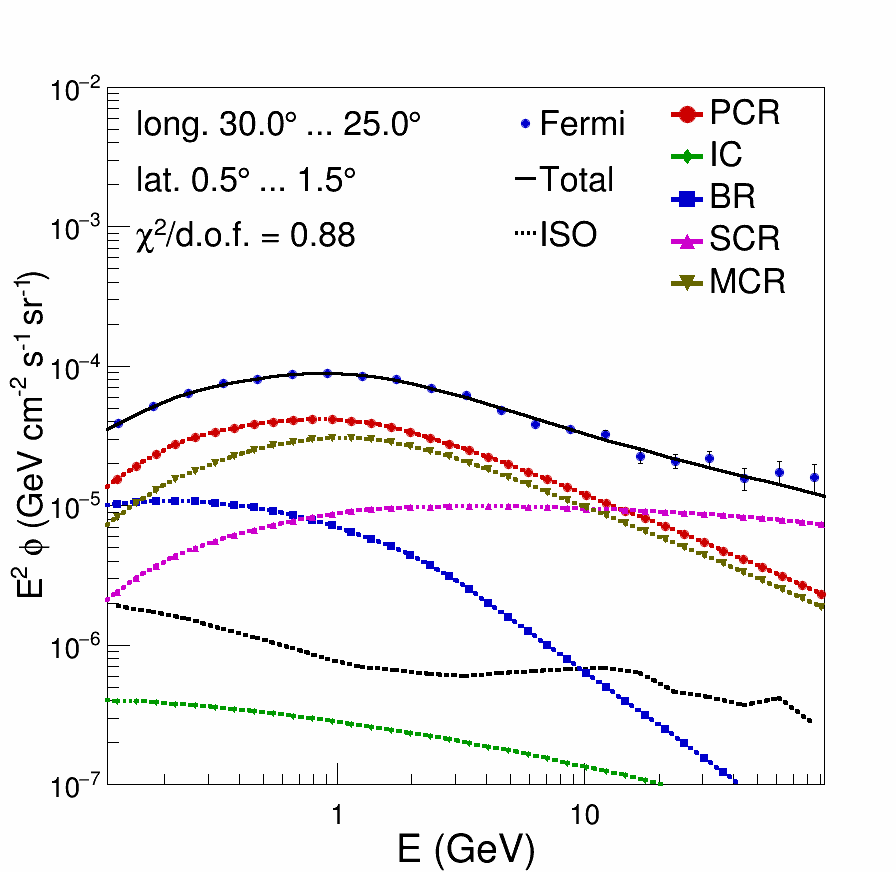}
\includegraphics[width=0.16\textwidth,height=0.16\textwidth,clip]{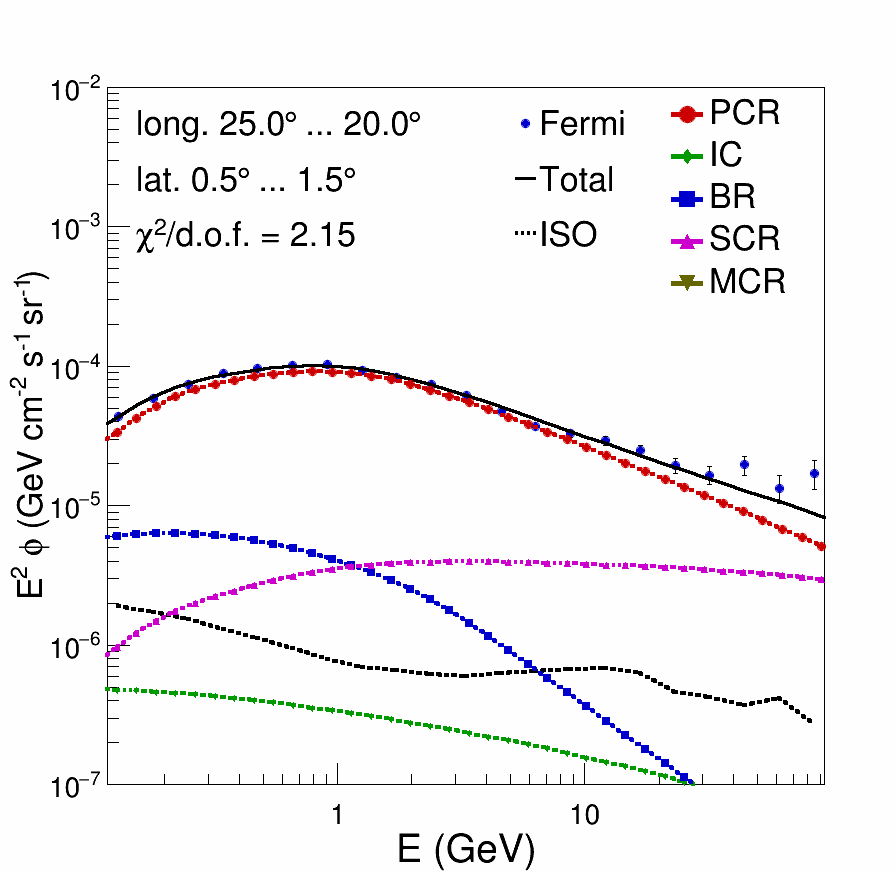}
\includegraphics[width=0.16\textwidth,height=0.16\textwidth,clip]{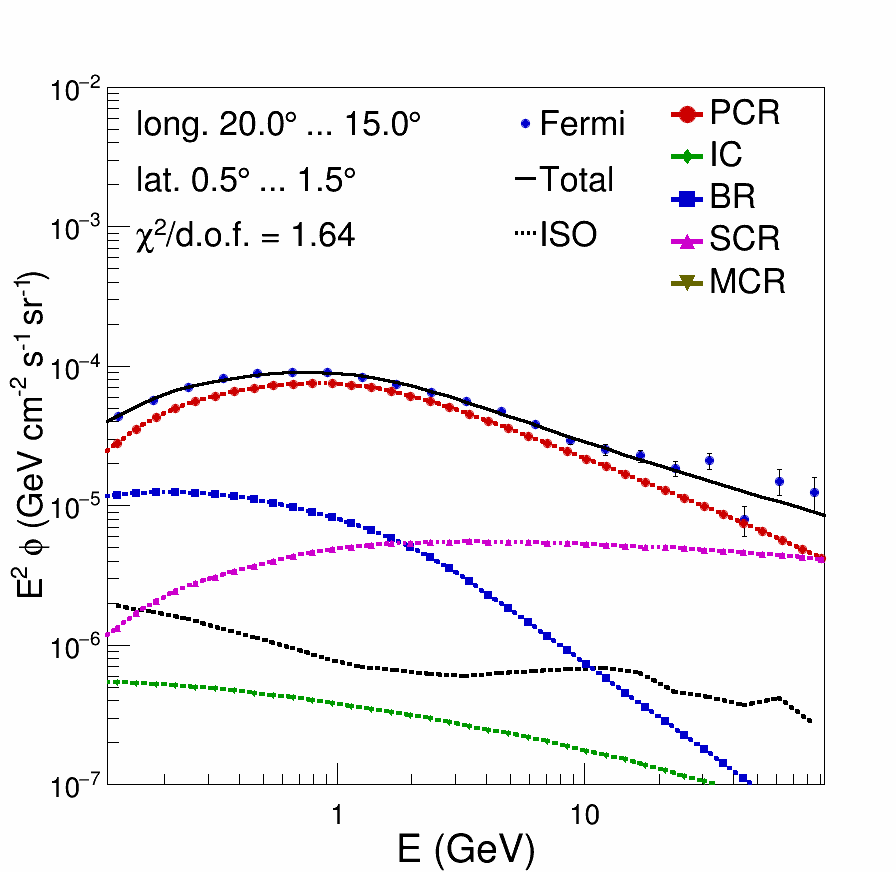}
\includegraphics[width=0.16\textwidth,height=0.16\textwidth,clip]{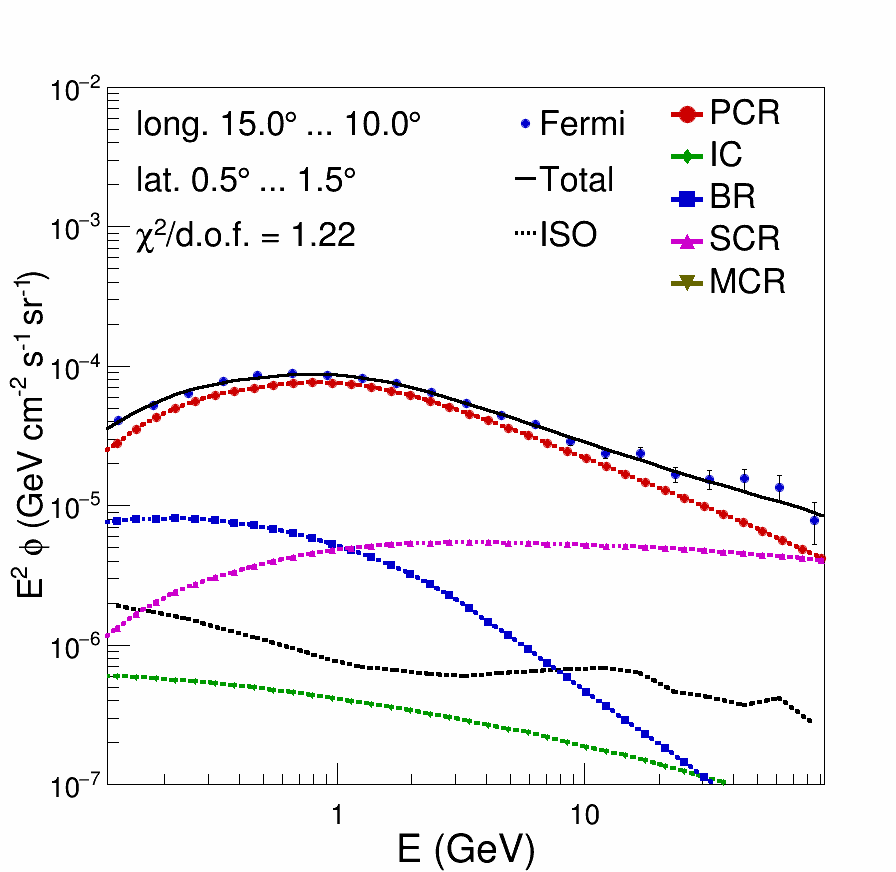}
\includegraphics[width=0.16\textwidth,height=0.16\textwidth,clip]{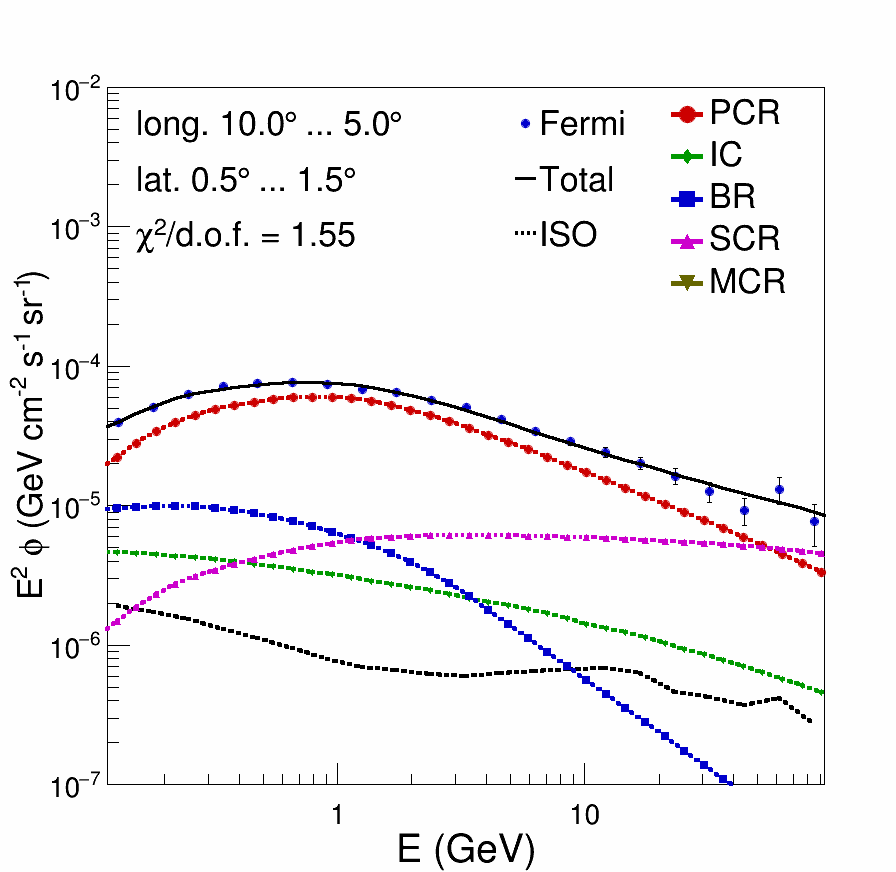}
\includegraphics[width=0.16\textwidth,height=0.16\textwidth,clip]{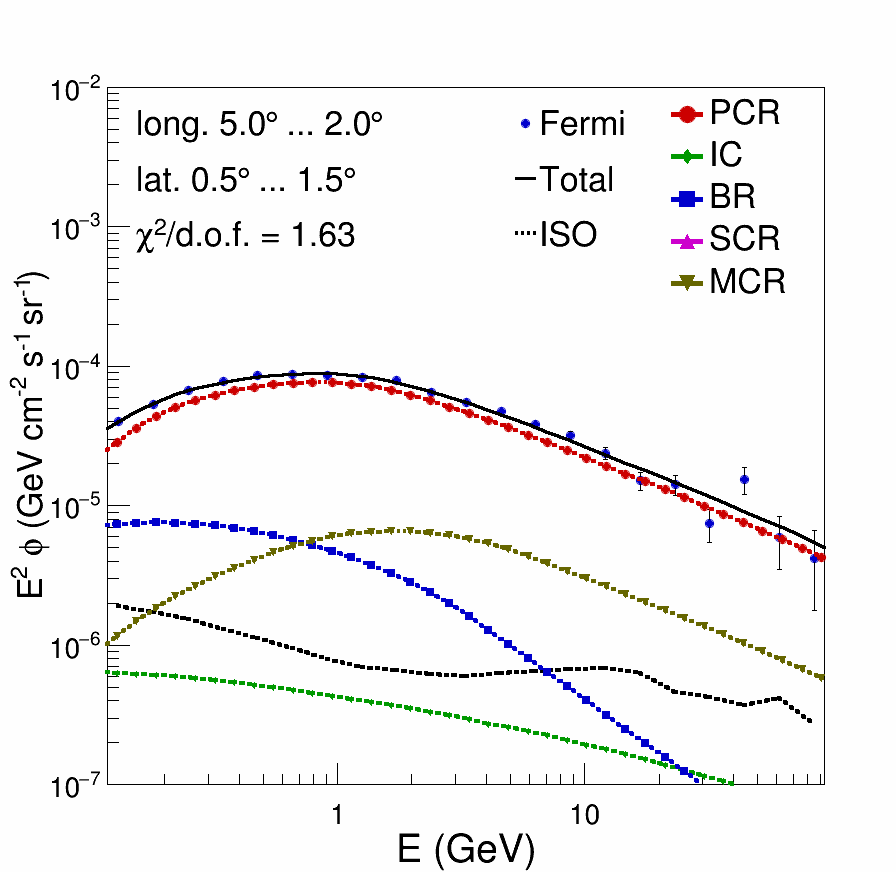}
\includegraphics[width=0.16\textwidth,height=0.16\textwidth,clip]{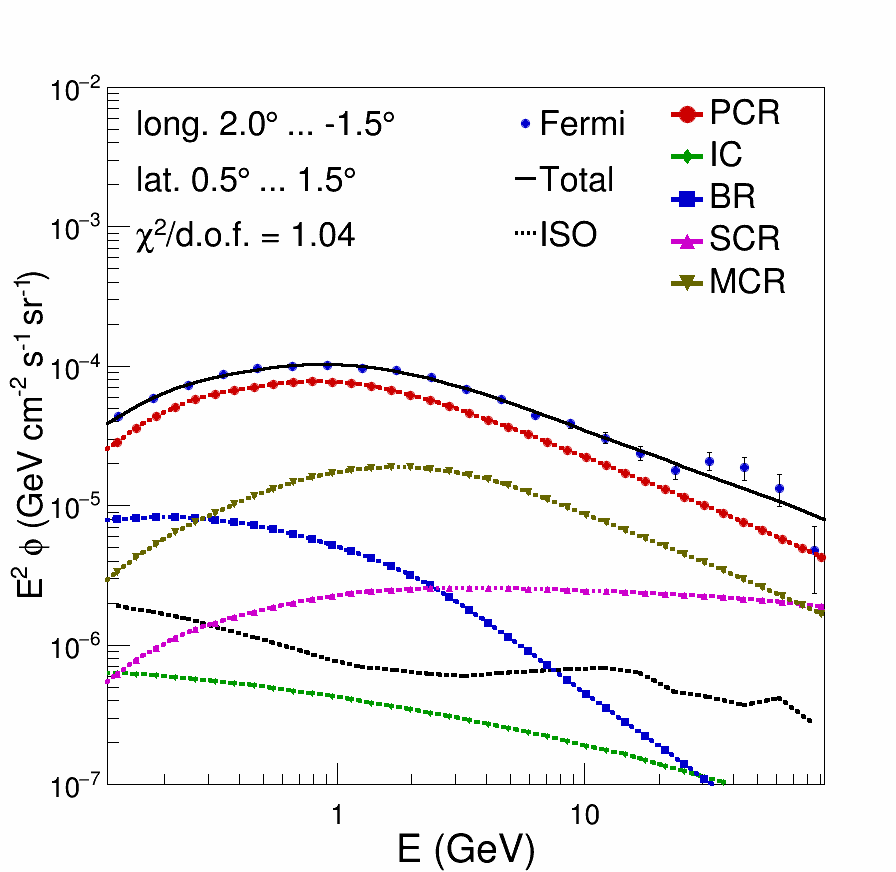}
\includegraphics[width=0.16\textwidth,height=0.16\textwidth,clip]{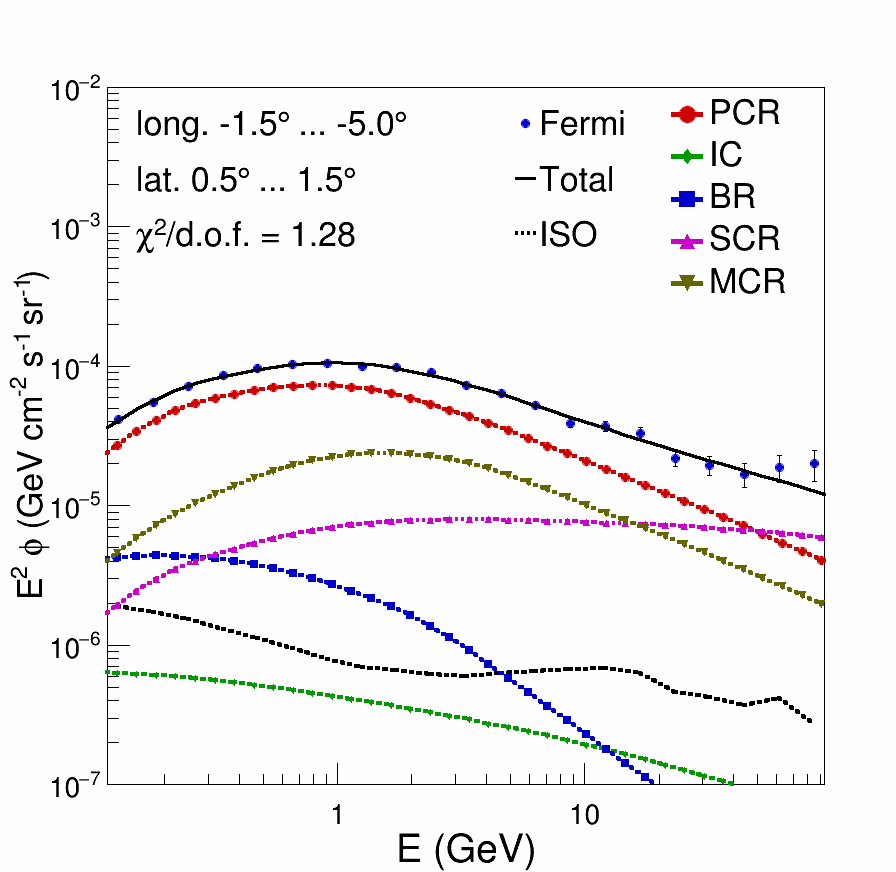}
\includegraphics[width=0.16\textwidth,height=0.16\textwidth,clip]{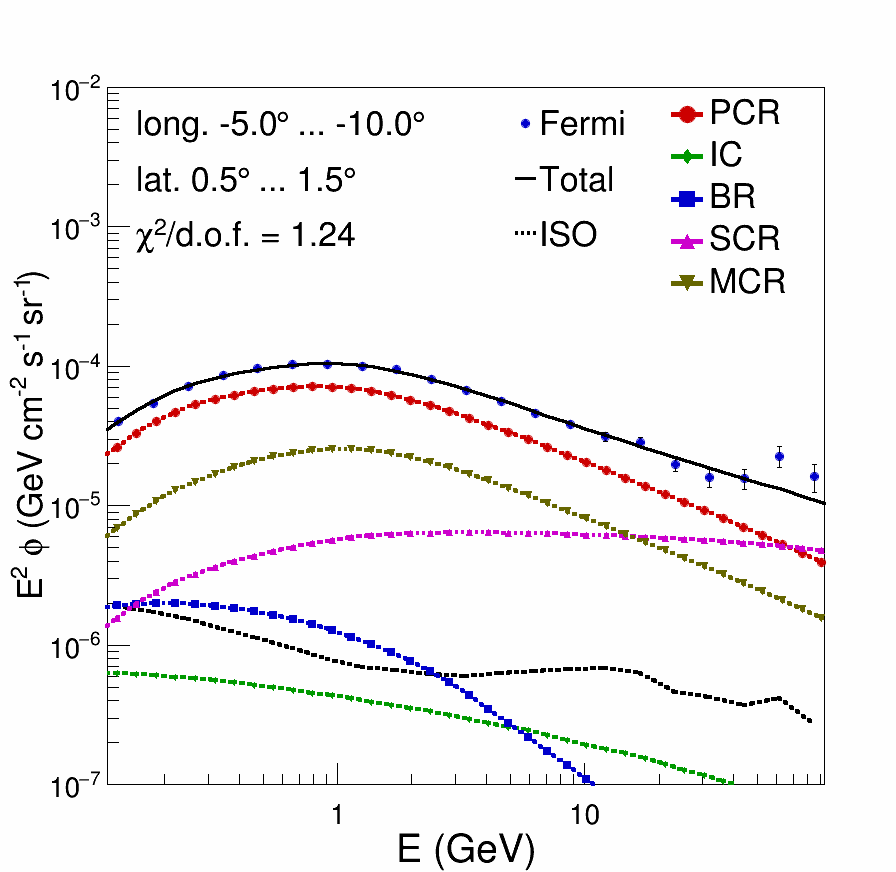}
\includegraphics[width=0.16\textwidth,height=0.16\textwidth,clip]{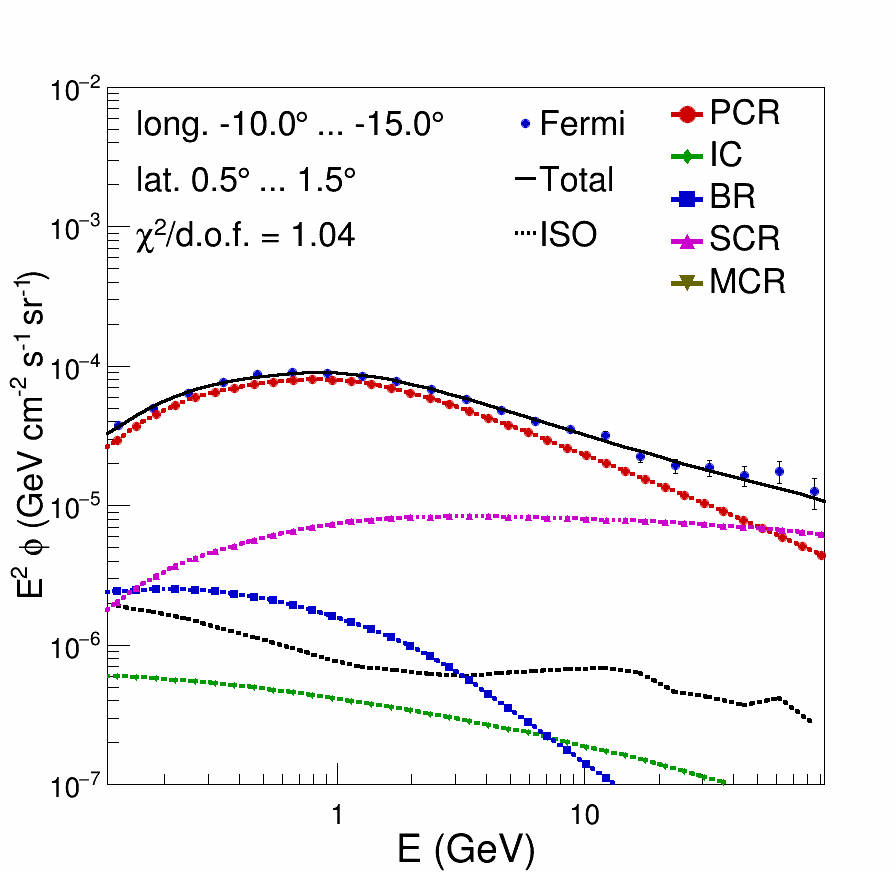}
\includegraphics[width=0.16\textwidth,height=0.16\textwidth,clip]{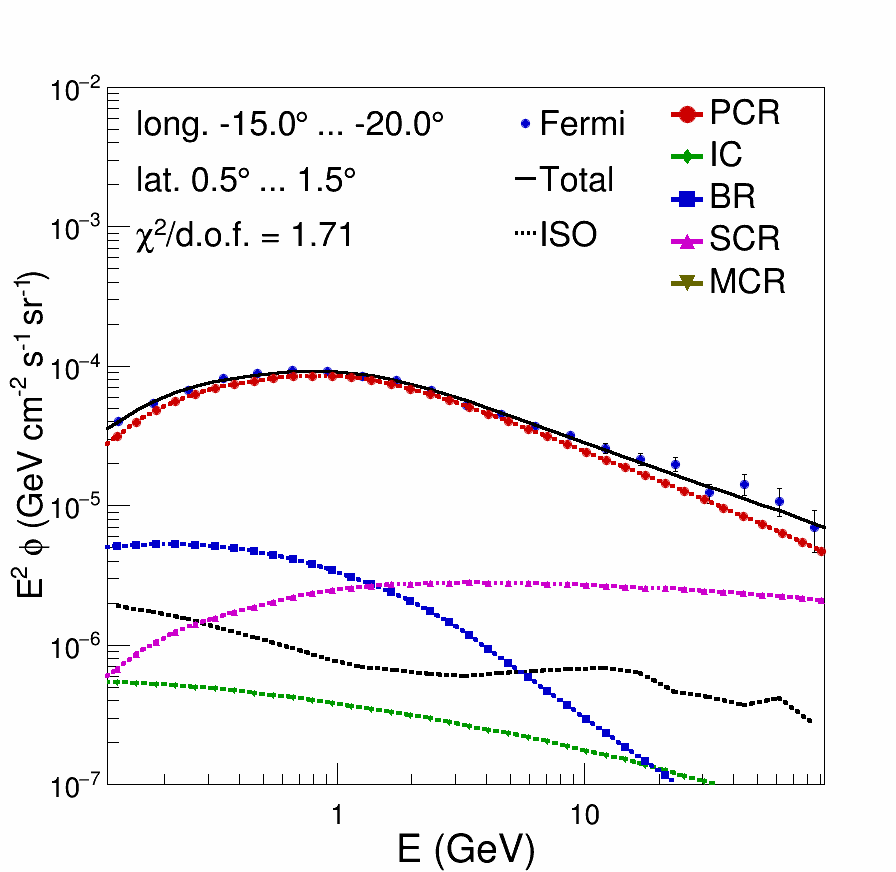}
\includegraphics[width=0.16\textwidth,height=0.16\textwidth,clip]{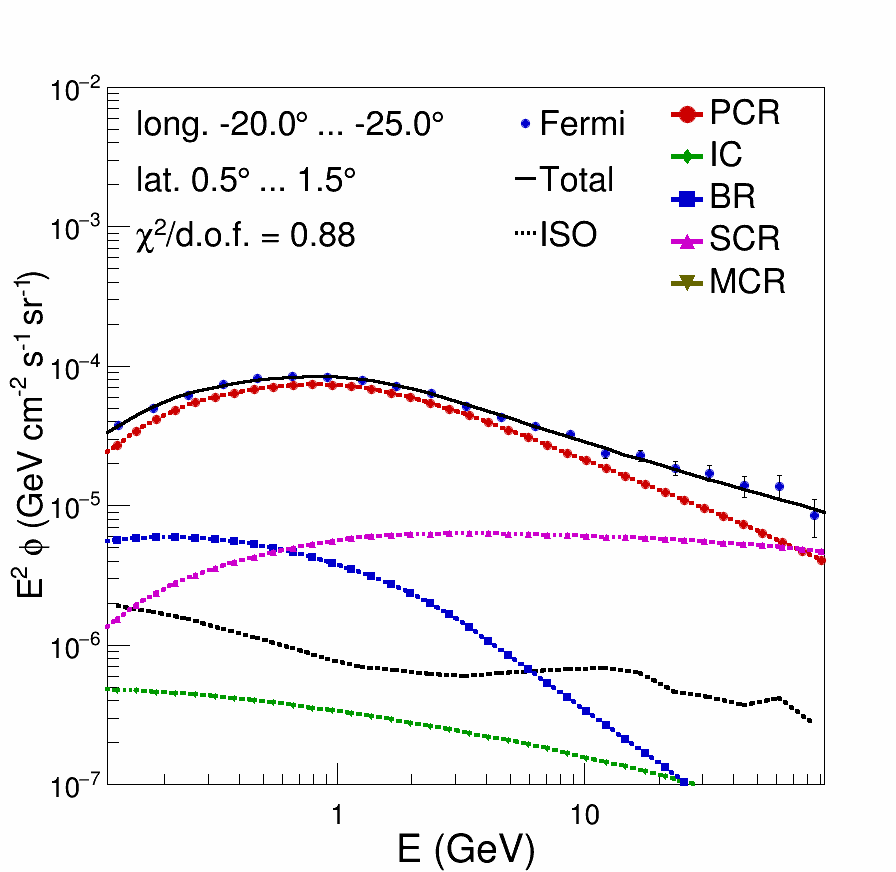}
\includegraphics[width=0.16\textwidth,height=0.16\textwidth,clip]{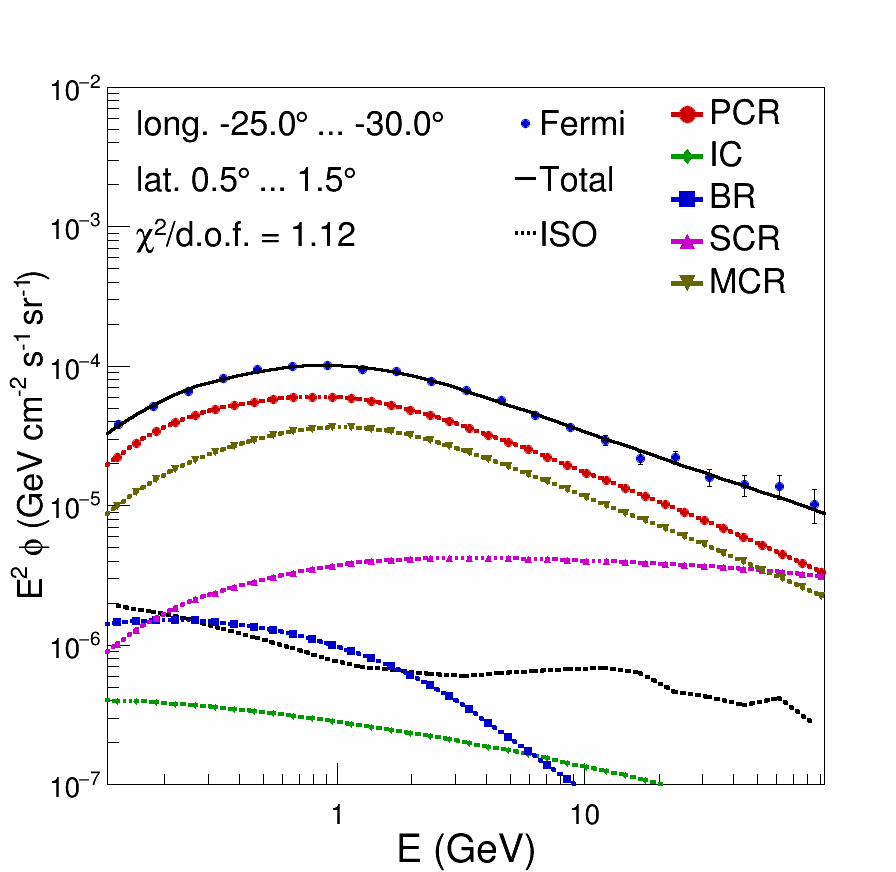}
\includegraphics[width=0.16\textwidth,height=0.16\textwidth,clip]{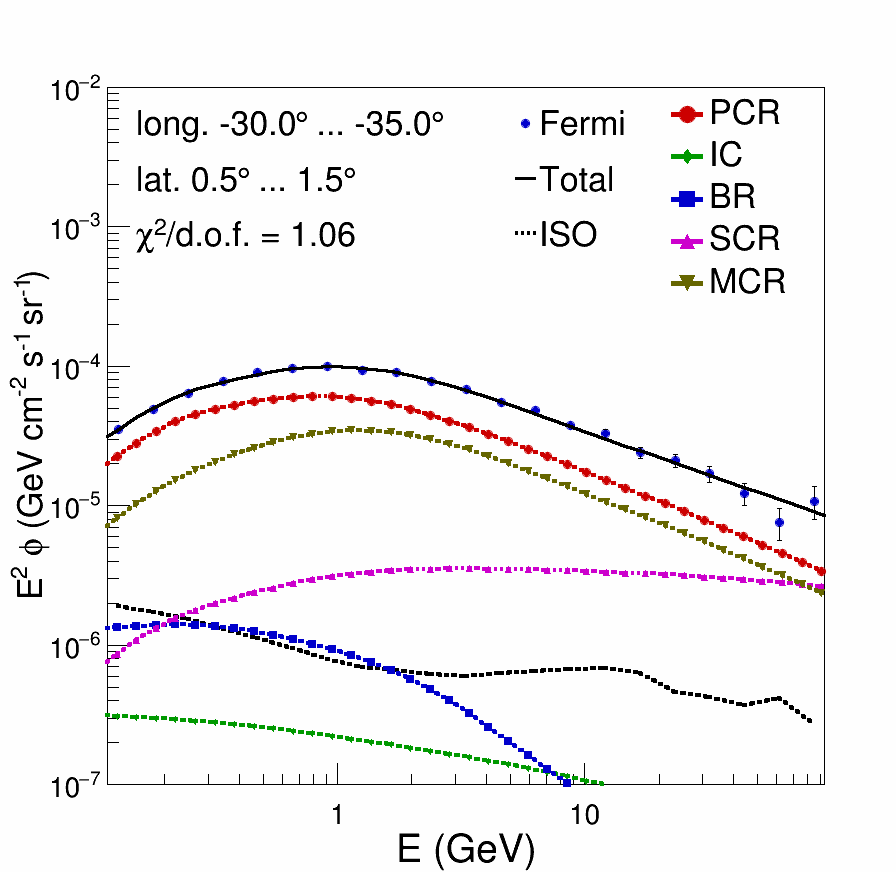}
\includegraphics[width=0.16\textwidth,height=0.16\textwidth,clip]{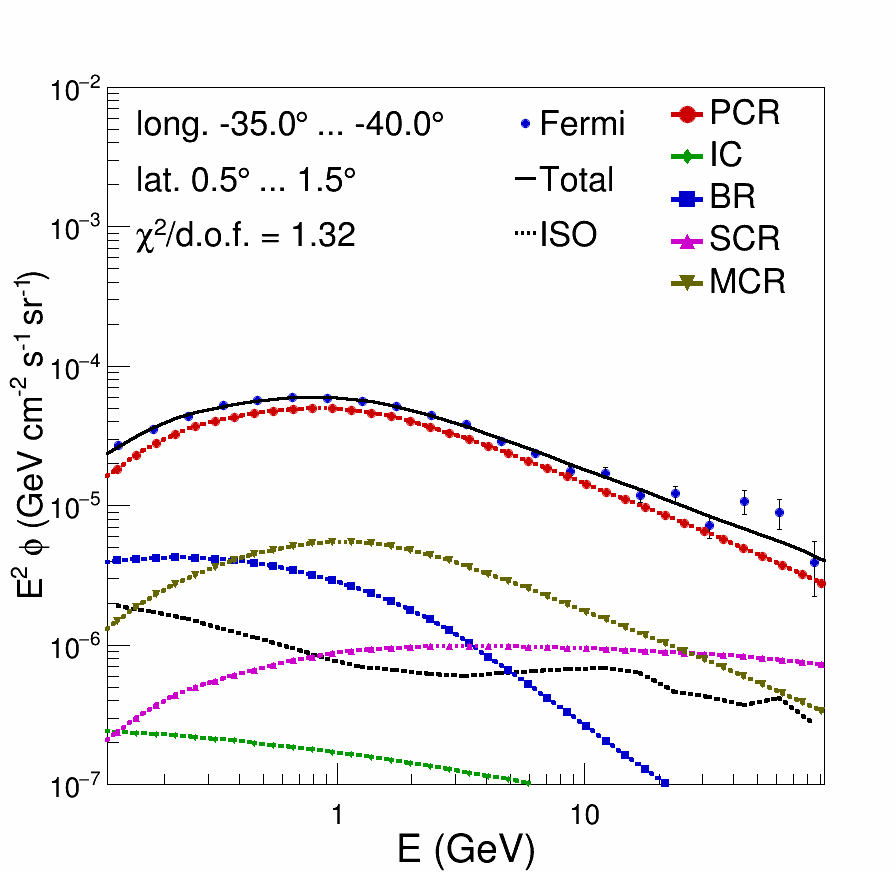}
\includegraphics[width=0.16\textwidth,height=0.16\textwidth,clip]{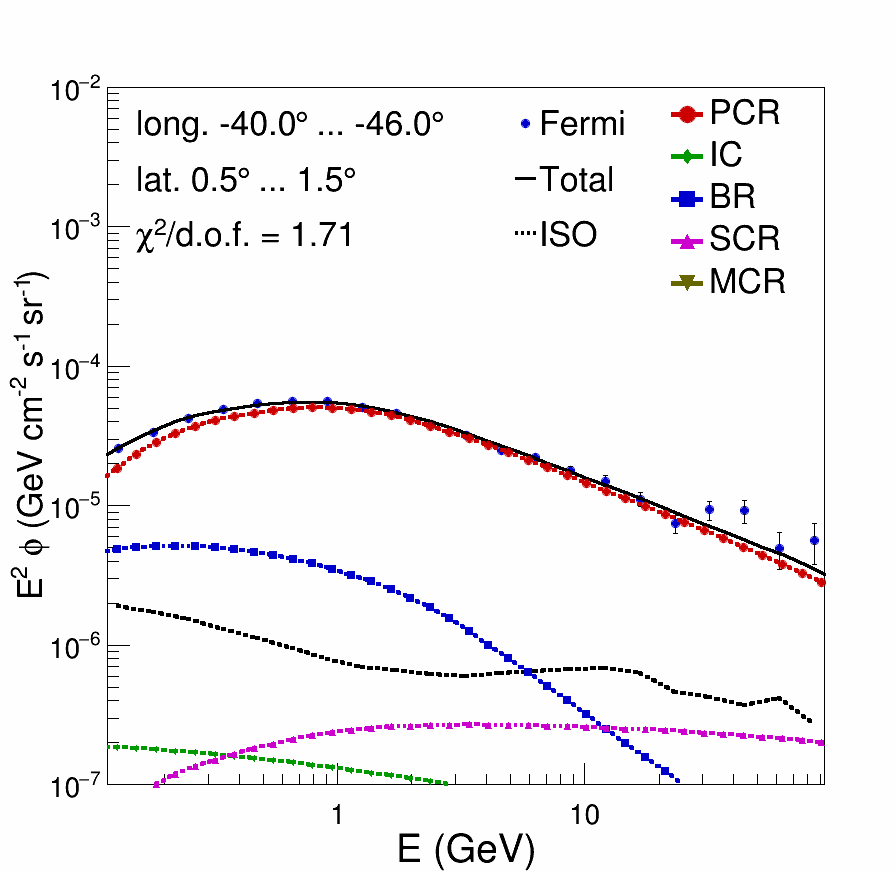}
\includegraphics[width=0.16\textwidth,height=0.16\textwidth,clip]{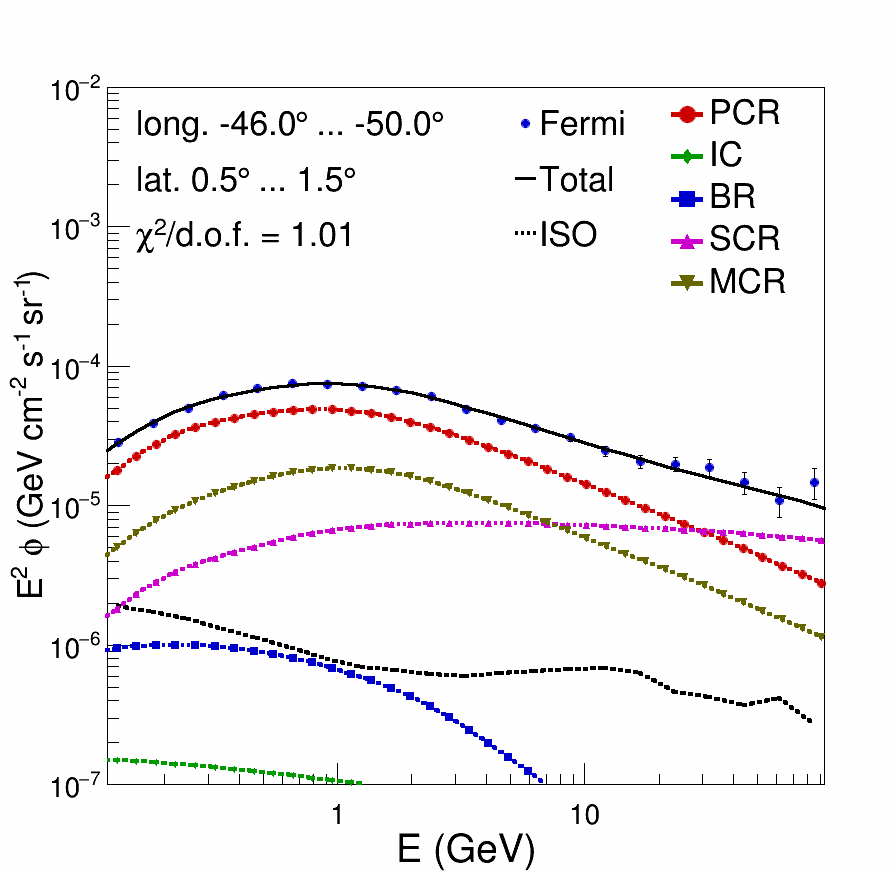}
\includegraphics[width=0.16\textwidth,height=0.16\textwidth,clip]{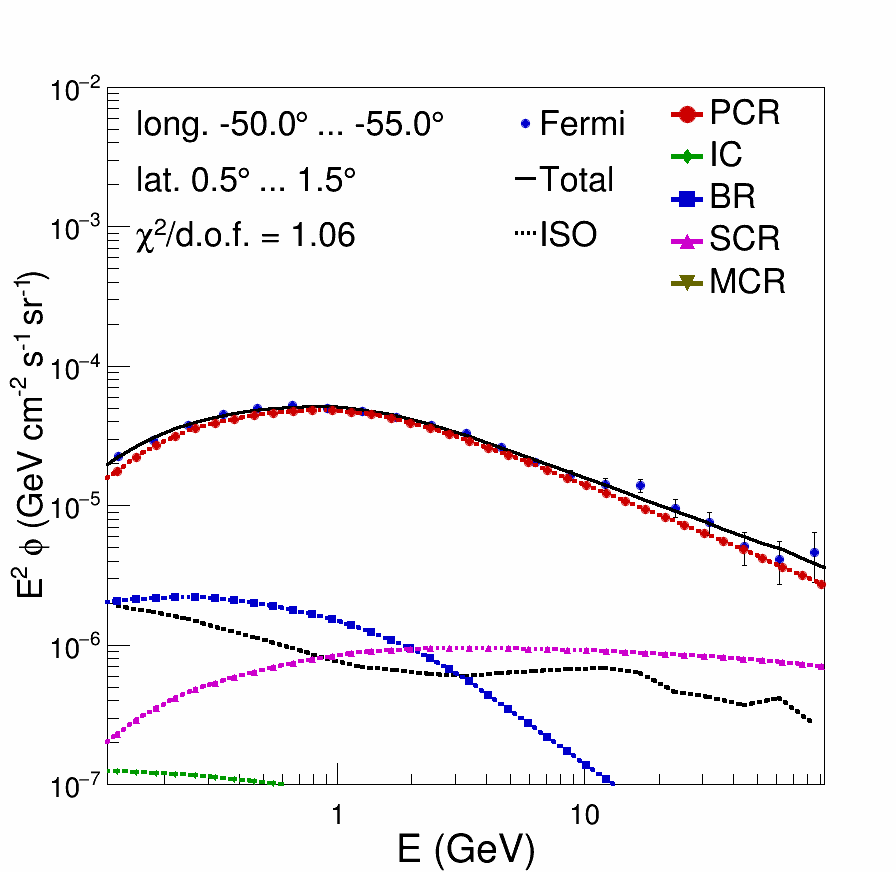}
\includegraphics[width=0.16\textwidth,height=0.16\textwidth,clip]{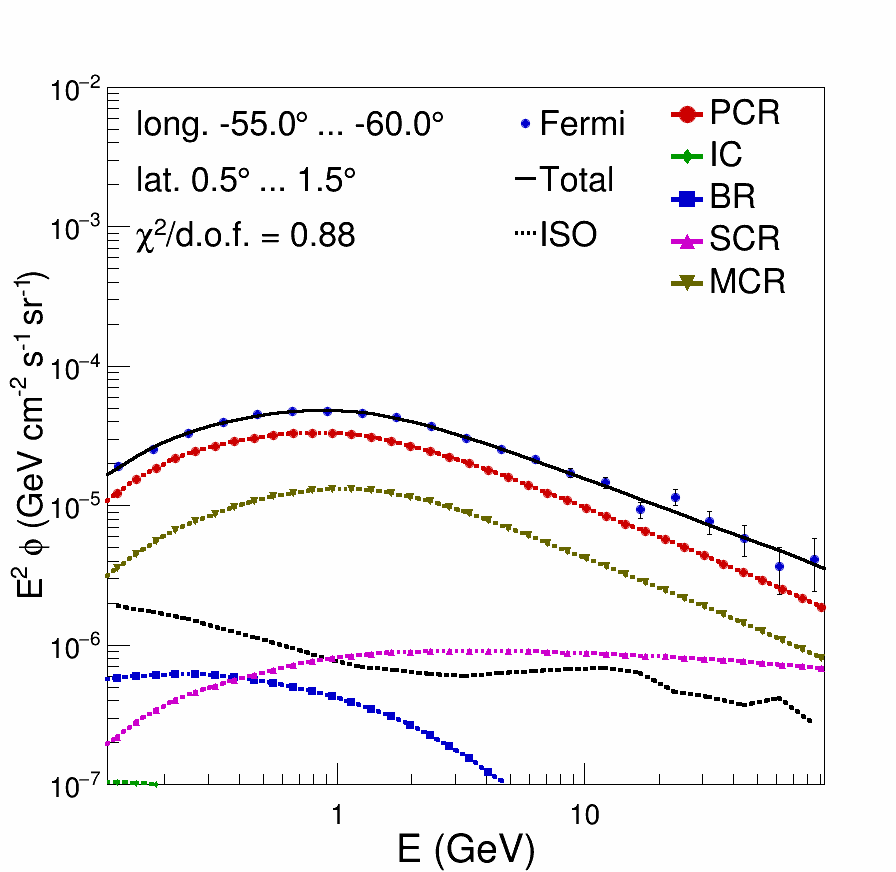}
\includegraphics[width=0.16\textwidth,height=0.16\textwidth,clip]{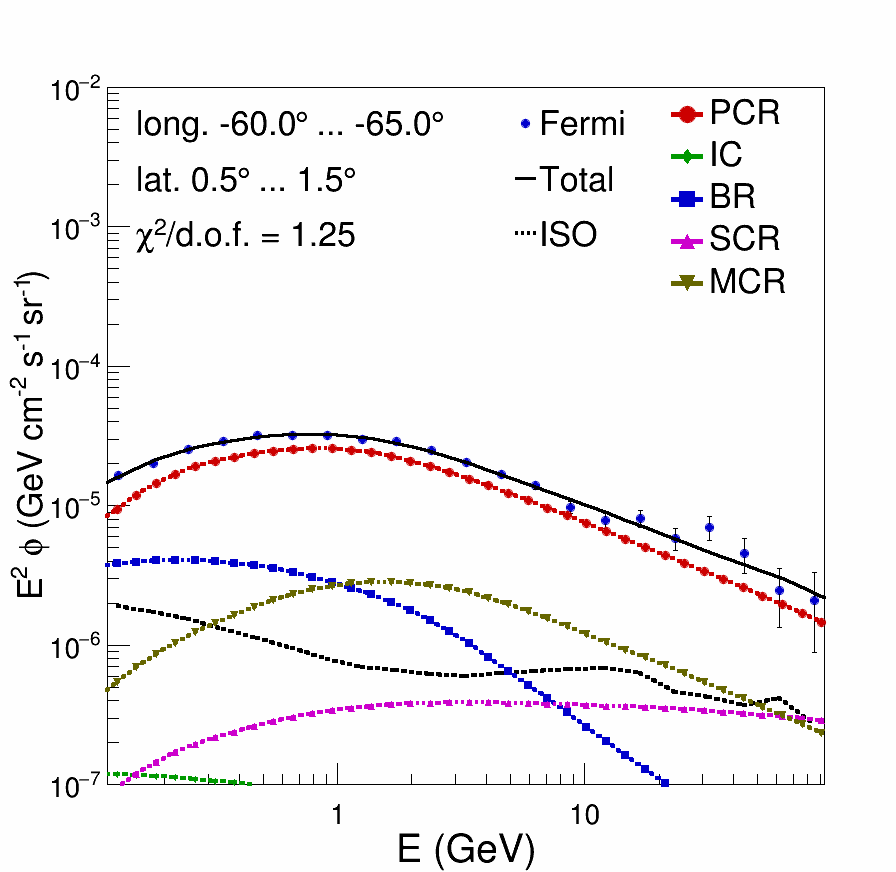}
\includegraphics[width=0.16\textwidth,height=0.16\textwidth,clip]{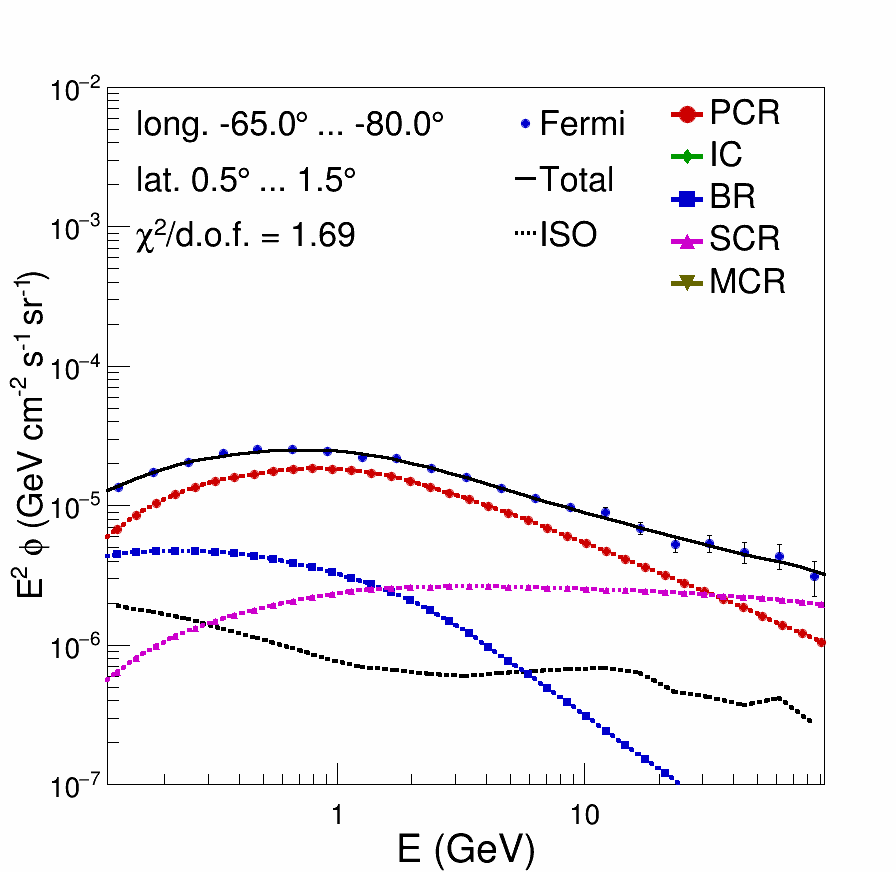}
\includegraphics[width=0.16\textwidth,height=0.16\textwidth,clip]{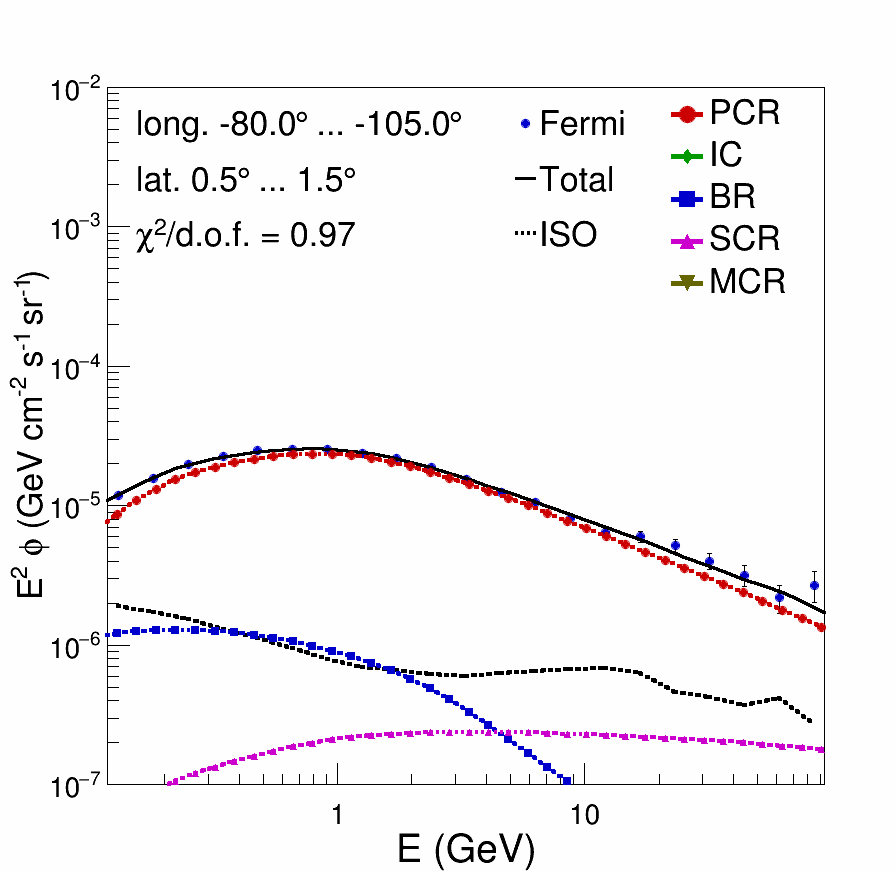}
\includegraphics[width=0.16\textwidth,height=0.16\textwidth,clip]{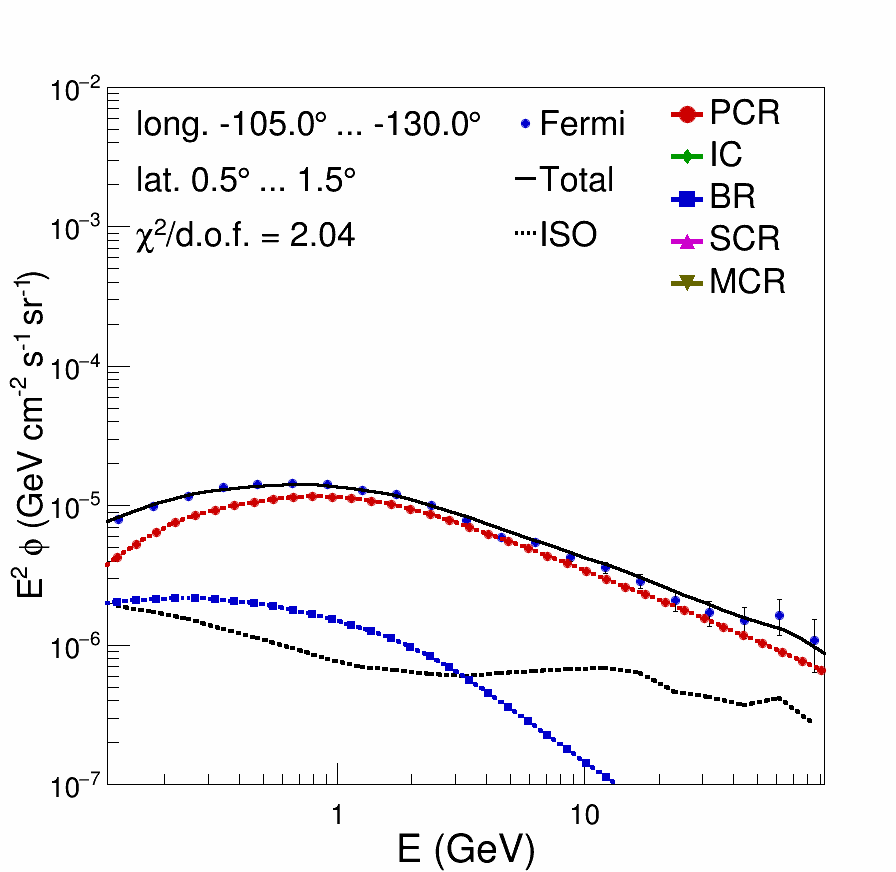}
\includegraphics[width=0.16\textwidth,height=0.16\textwidth,clip]{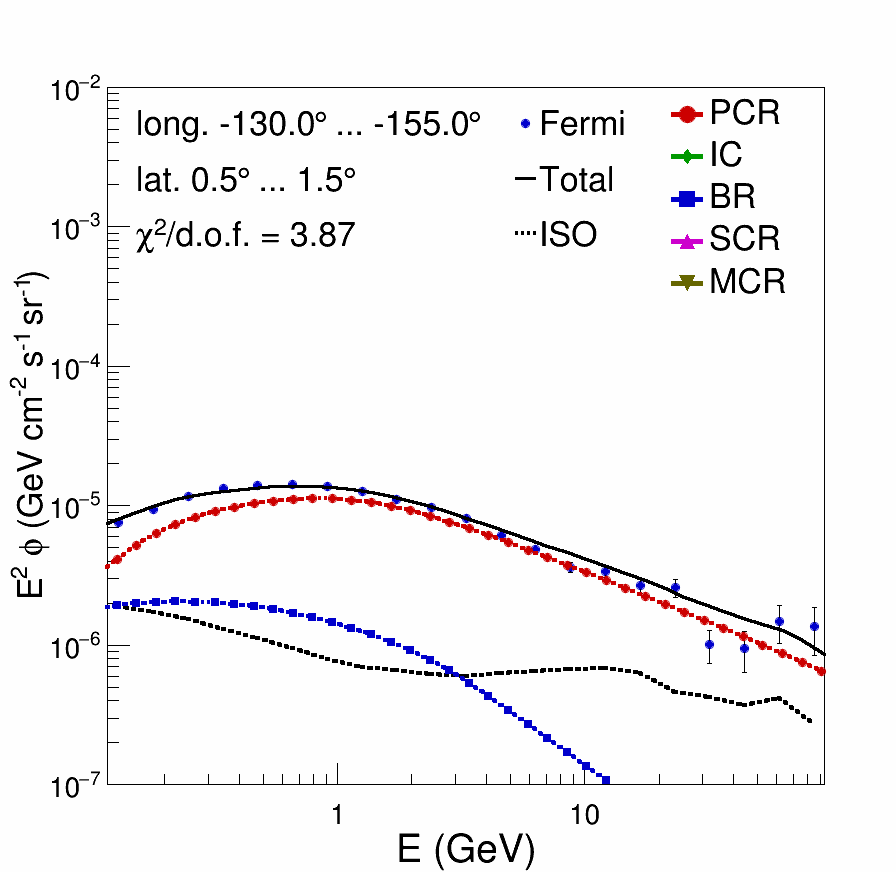}
\includegraphics[width=0.16\textwidth,height=0.16\textwidth,clip]{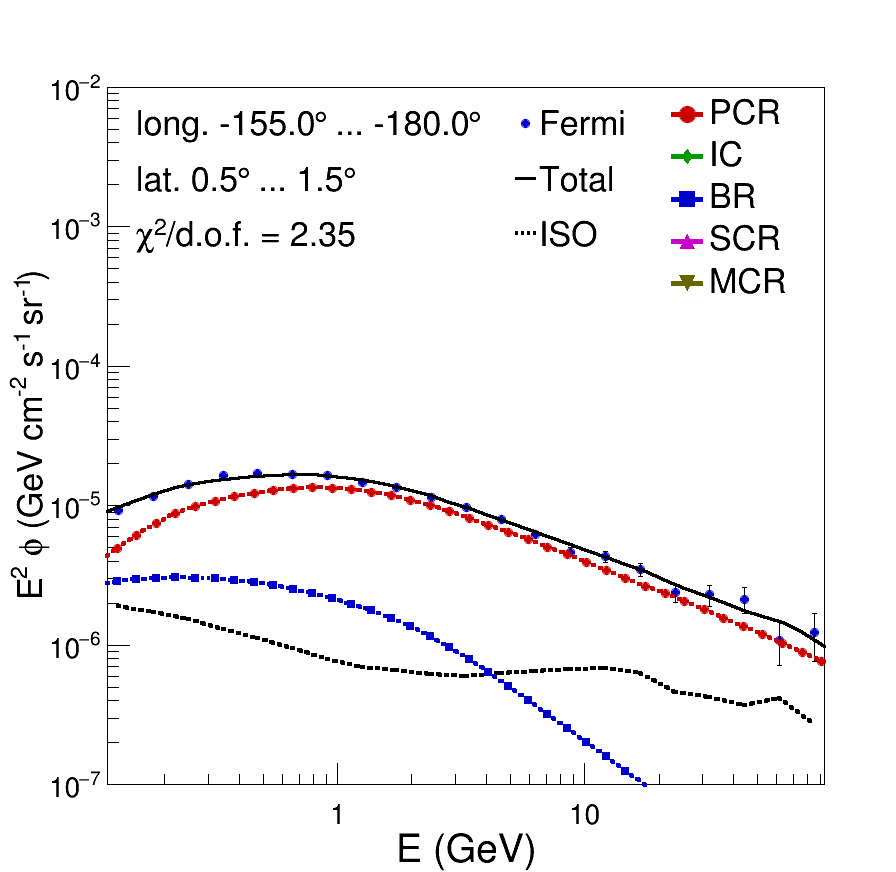}
\caption[]{Template fits for latitudes  with $0.5^\circ<b<1.5^\circ$ and longitudes decreasing from 180$^\circ$ to -180$^\circ$.} \label{F20}
\end{figure}
\clearpage
\begin{figure}
\centering
\includegraphics[width=0.16\textwidth,height=0.16\textwidth,clip]{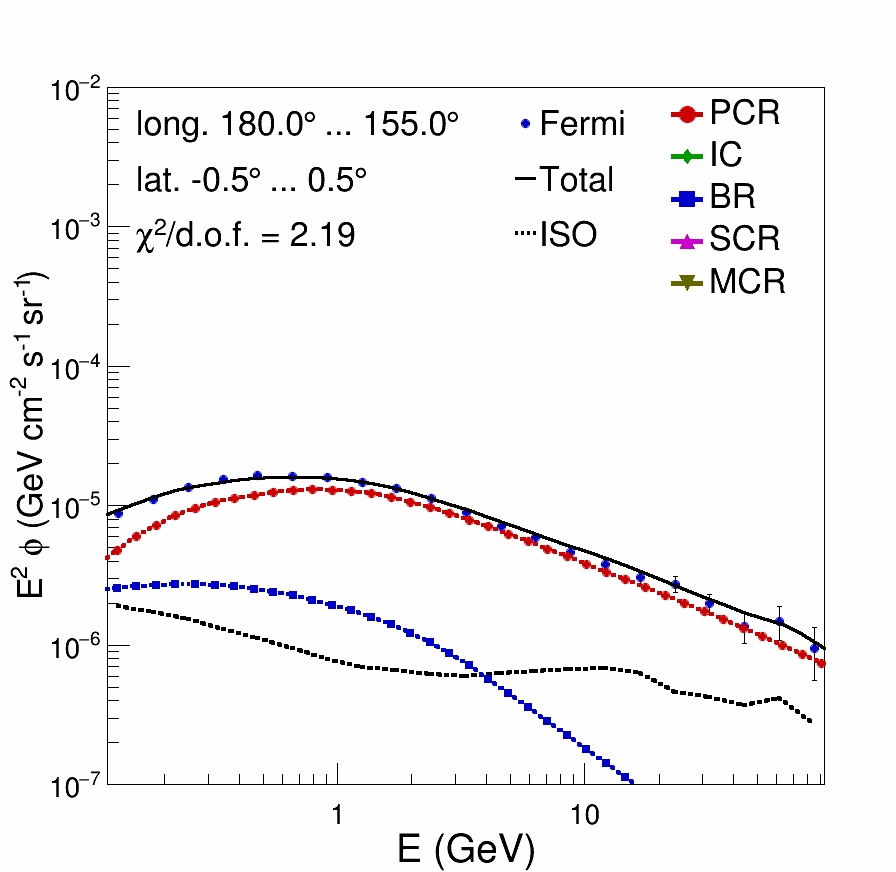}
\includegraphics[width=0.16\textwidth,height=0.16\textwidth,clip]{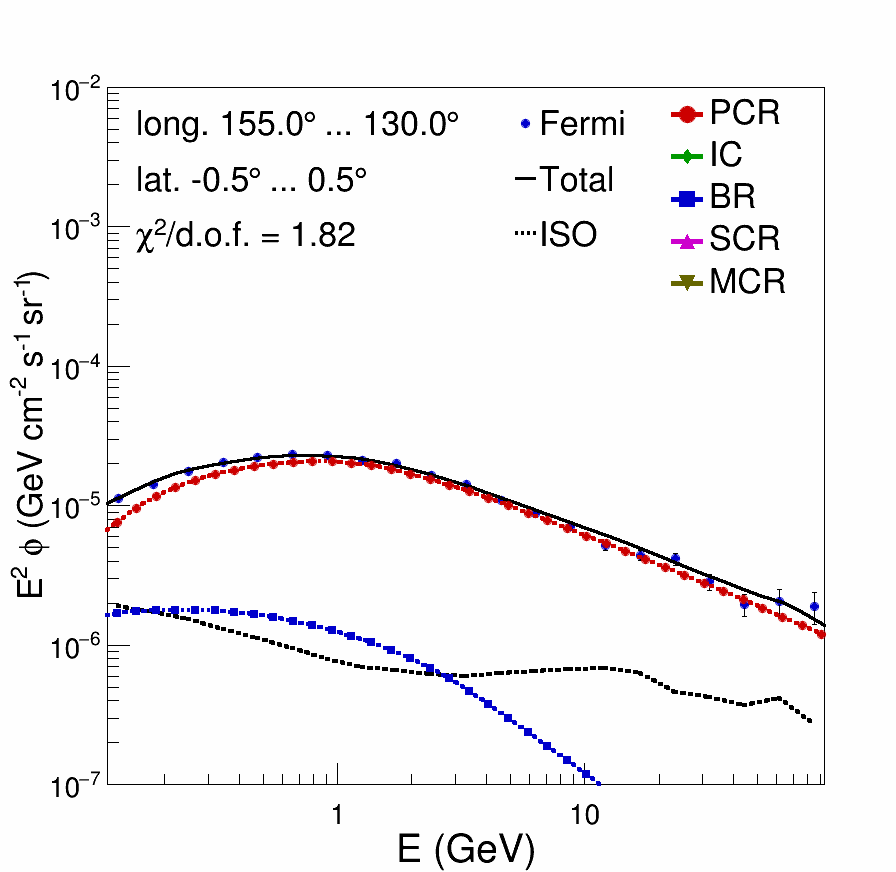}
\includegraphics[width=0.16\textwidth,height=0.16\textwidth,clip]{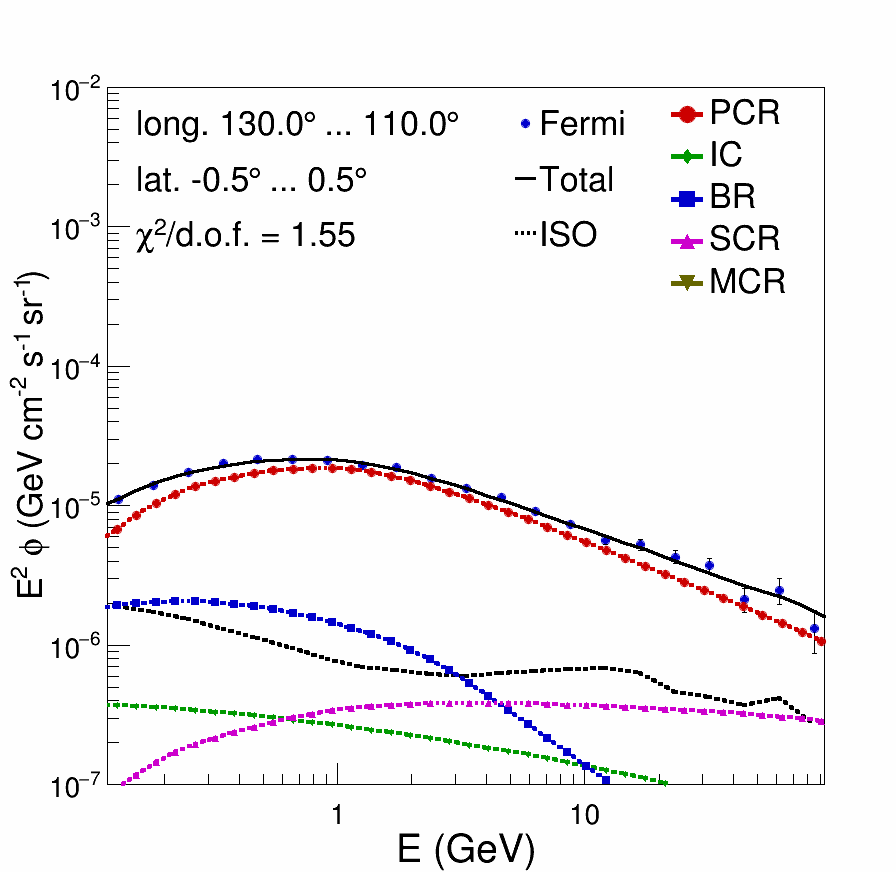}
\includegraphics[width=0.16\textwidth,height=0.16\textwidth,clip]{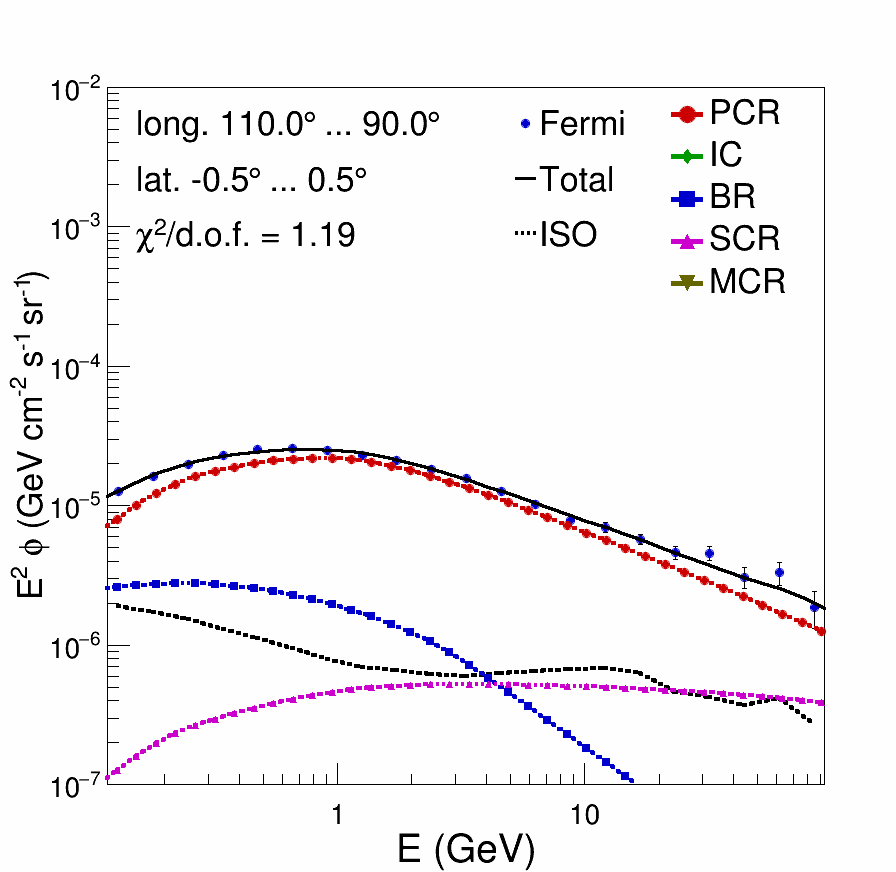}
\includegraphics[width=0.16\textwidth,height=0.16\textwidth,clip]{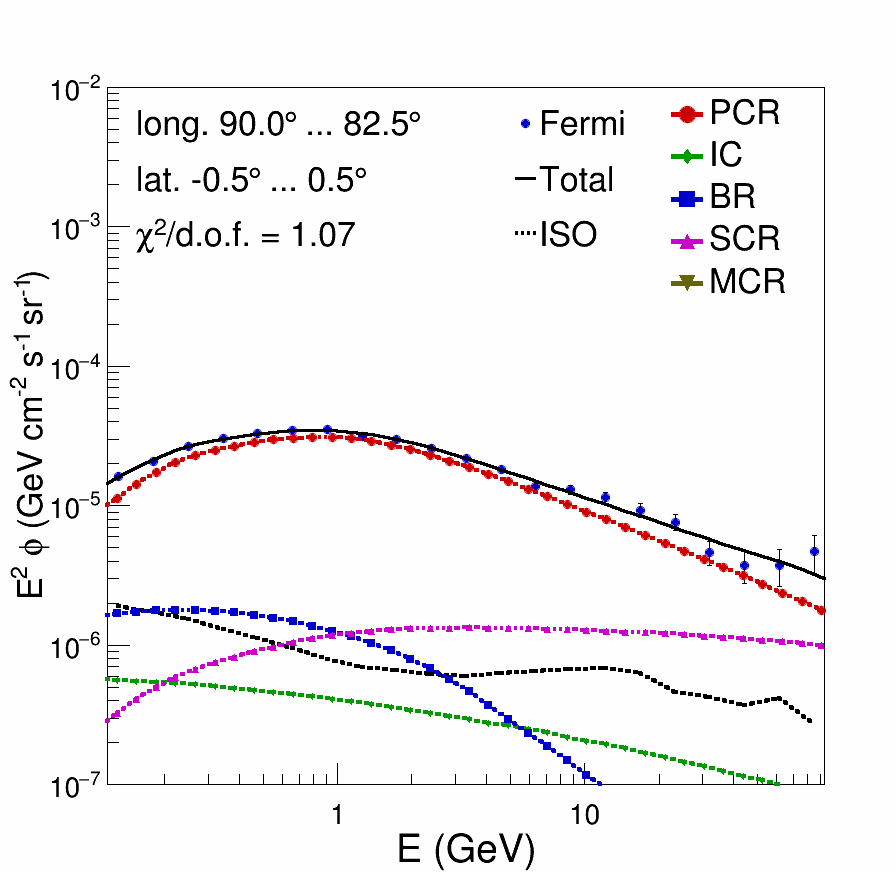}
\includegraphics[width=0.16\textwidth,height=0.16\textwidth,clip]{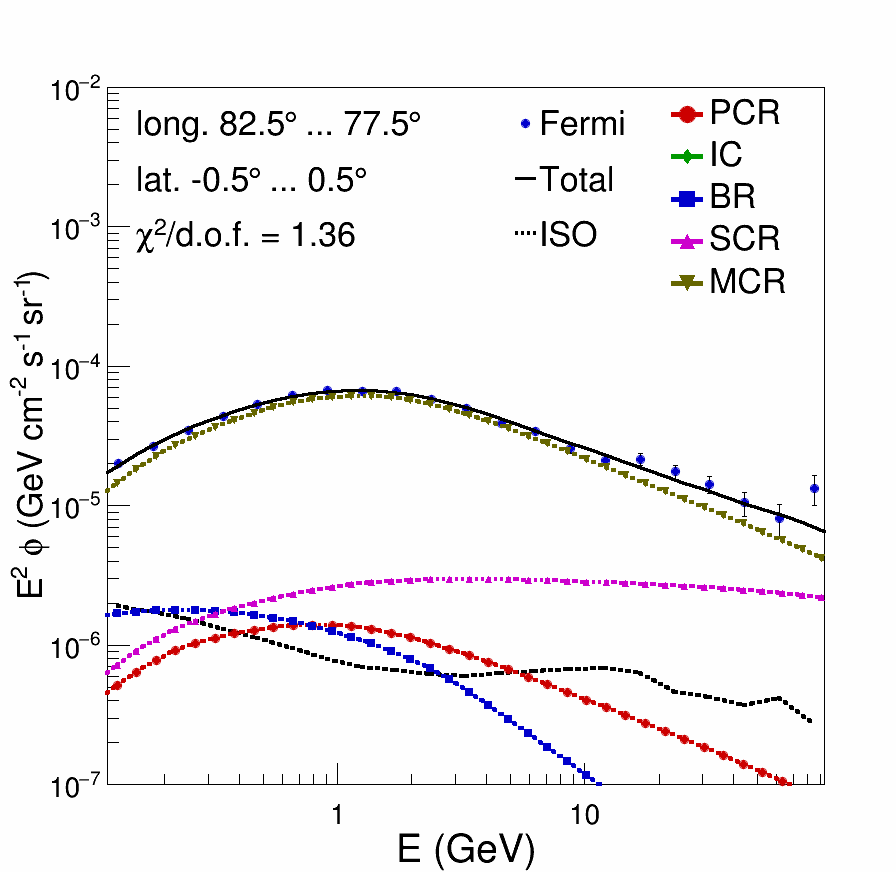}
\includegraphics[width=0.16\textwidth,height=0.16\textwidth,clip]{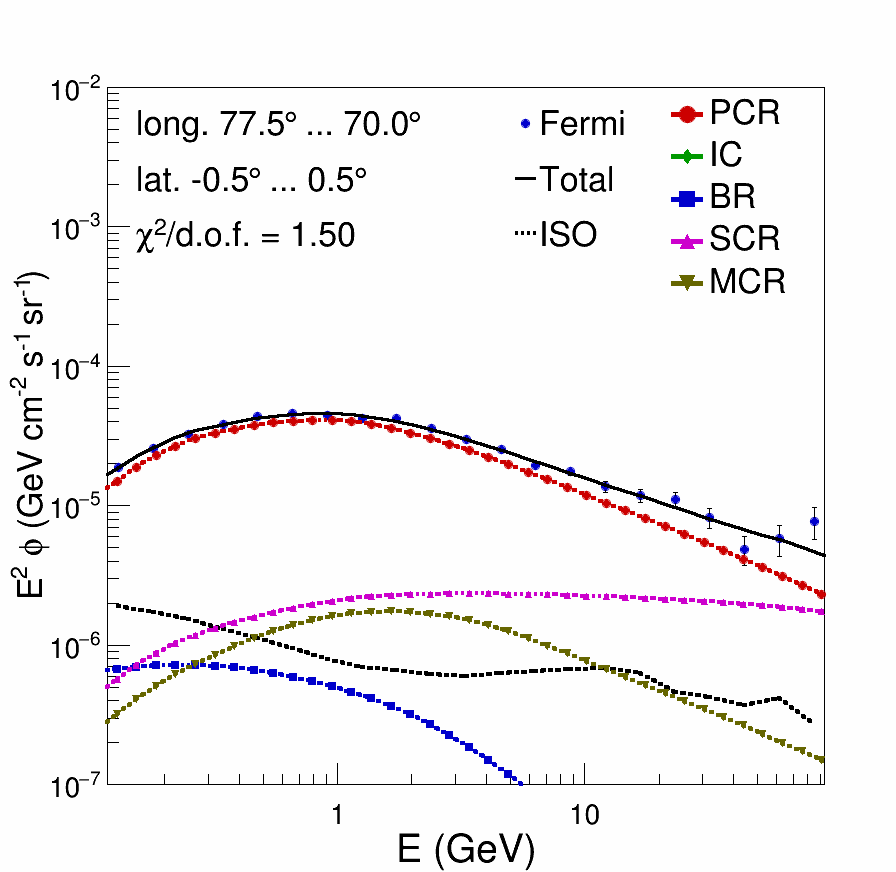}
\includegraphics[width=0.16\textwidth,height=0.16\textwidth,clip]{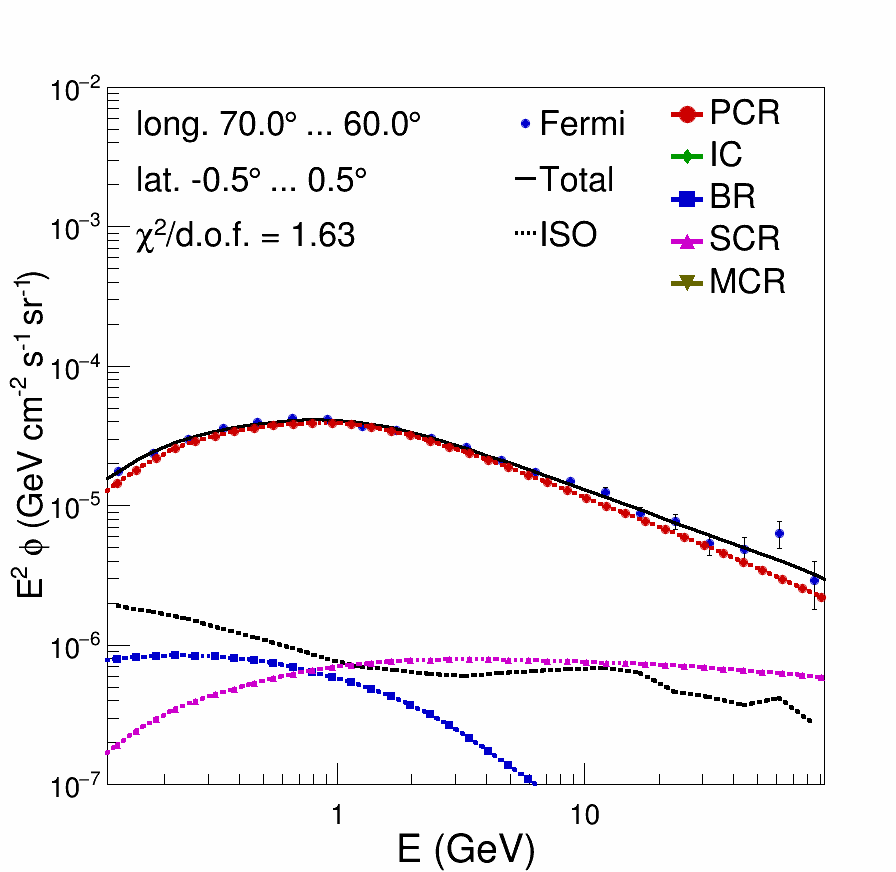}
\includegraphics[width=0.16\textwidth,height=0.16\textwidth,clip]{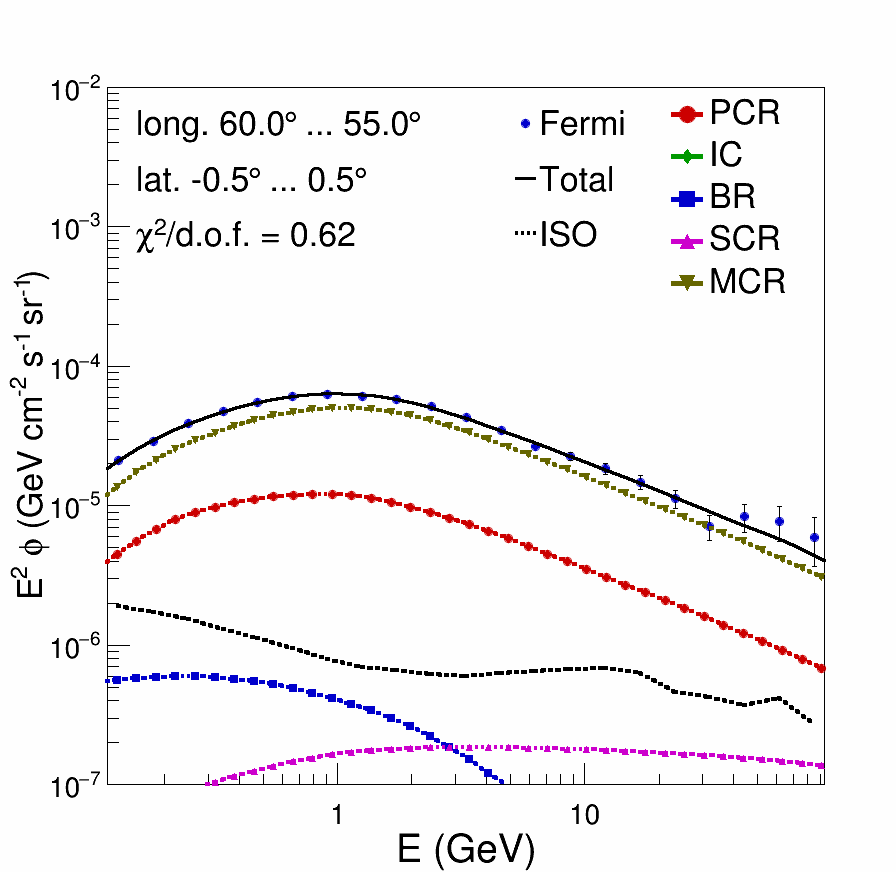}
\includegraphics[width=0.16\textwidth,height=0.16\textwidth,clip]{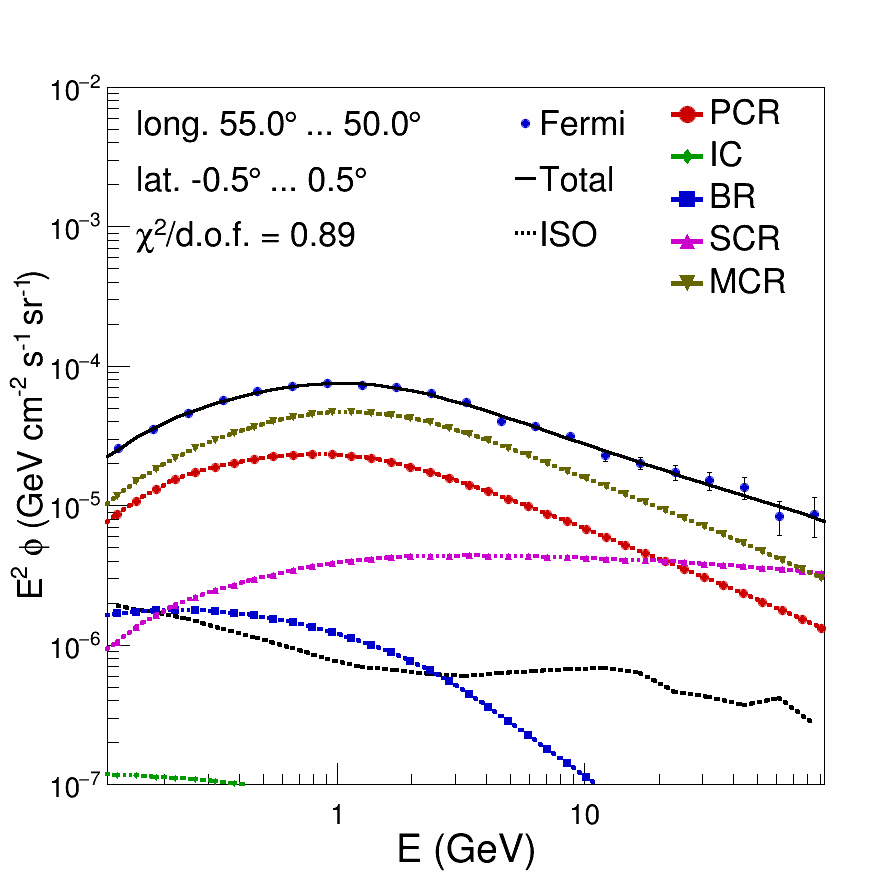}
\includegraphics[width=0.16\textwidth,height=0.16\textwidth,clip]{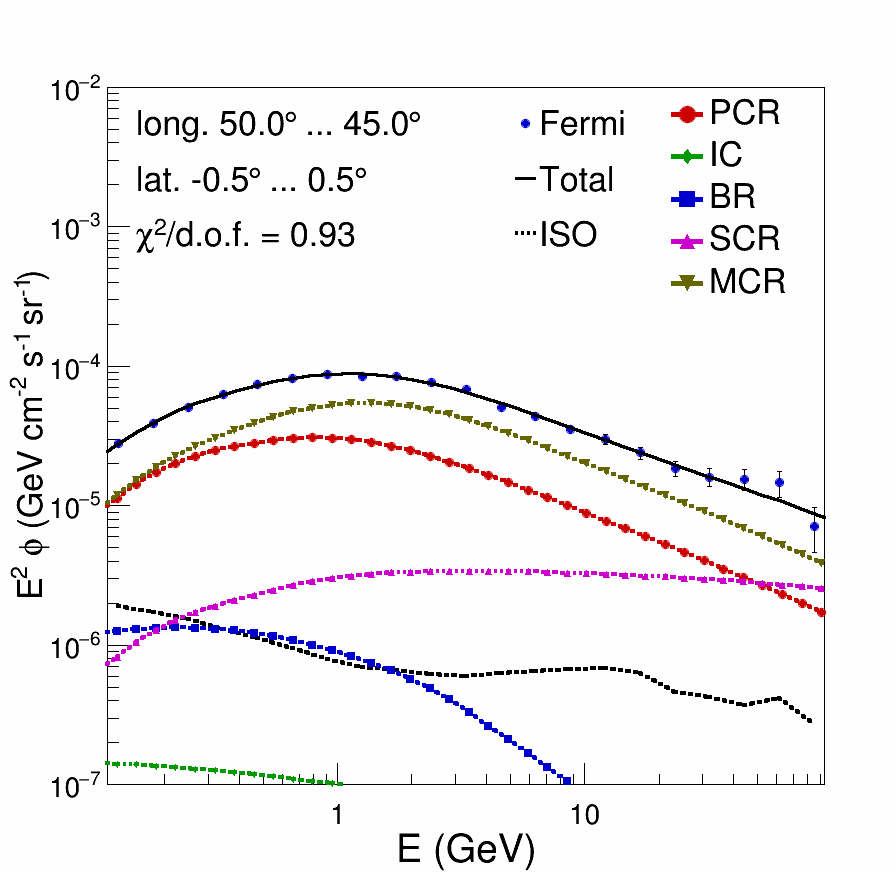}
\includegraphics[width=0.16\textwidth,height=0.16\textwidth,clip]{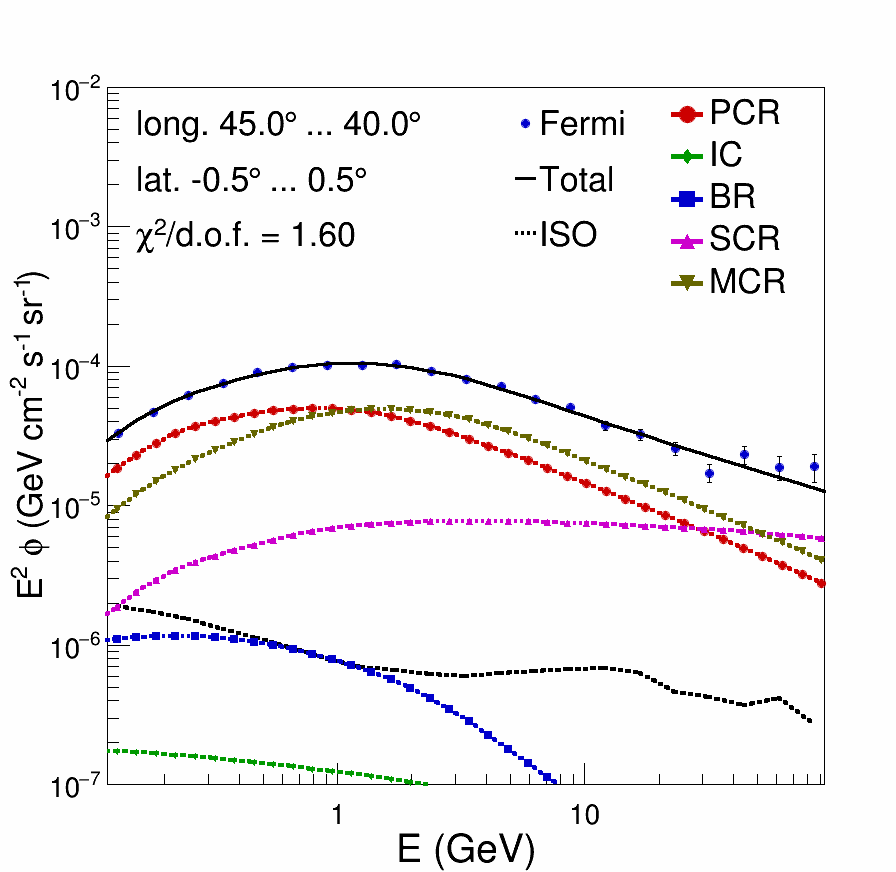}
\includegraphics[width=0.16\textwidth,height=0.16\textwidth,clip]{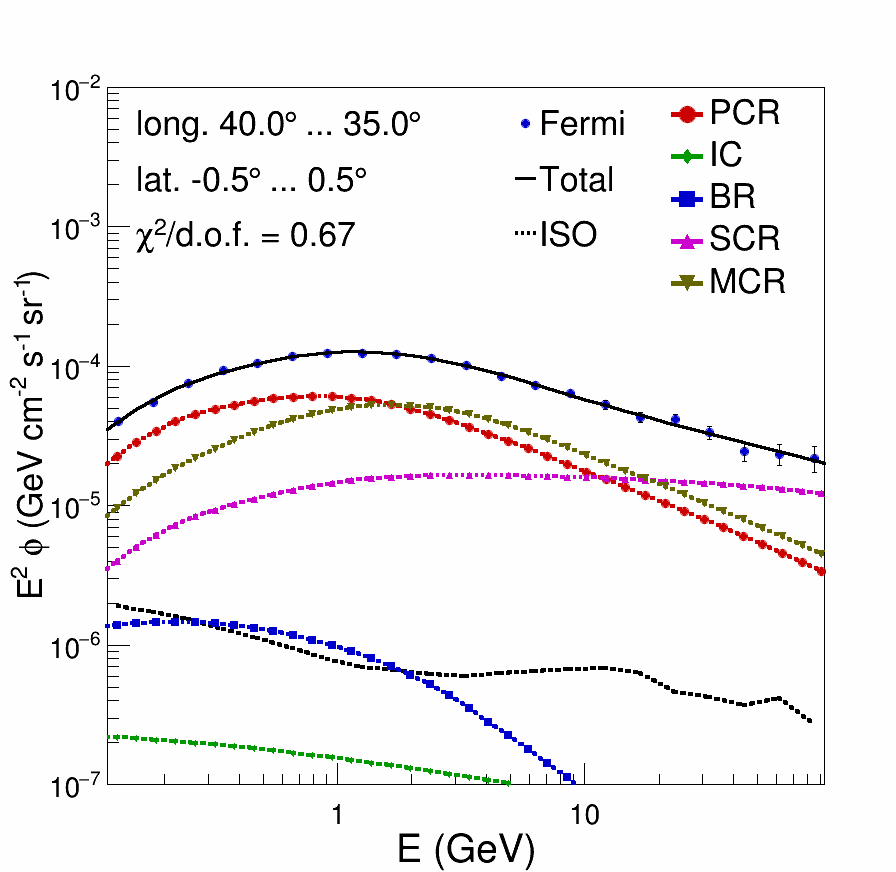}
\includegraphics[width=0.16\textwidth,height=0.16\textwidth,clip]{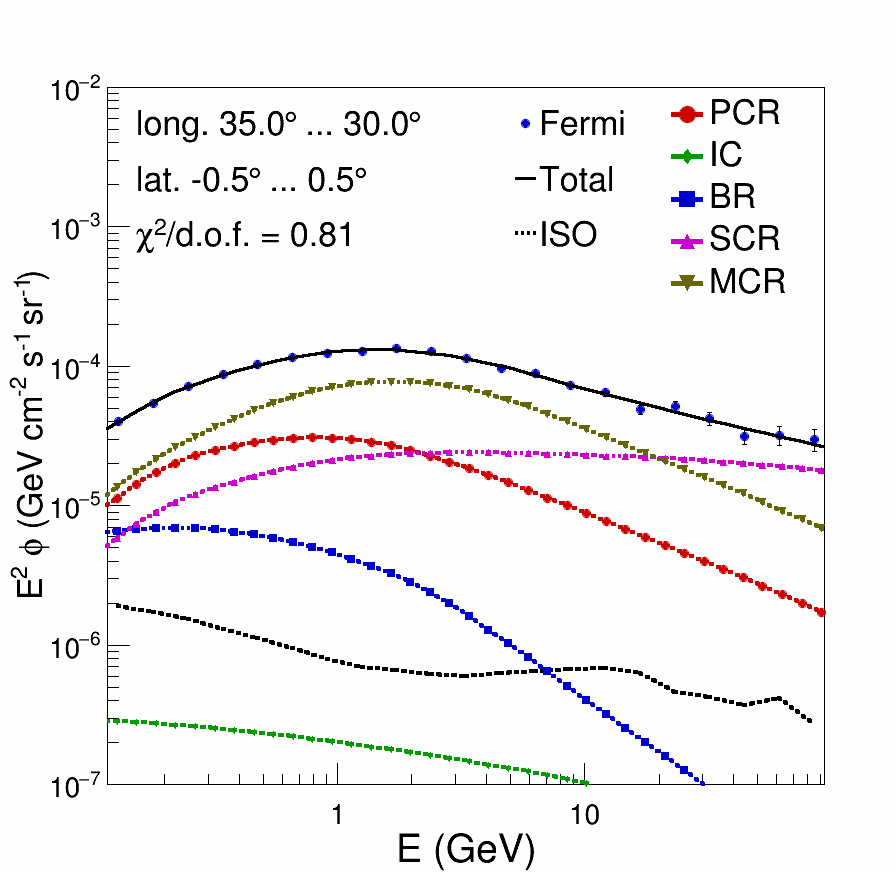}
\includegraphics[width=0.16\textwidth,height=0.16\textwidth,clip]{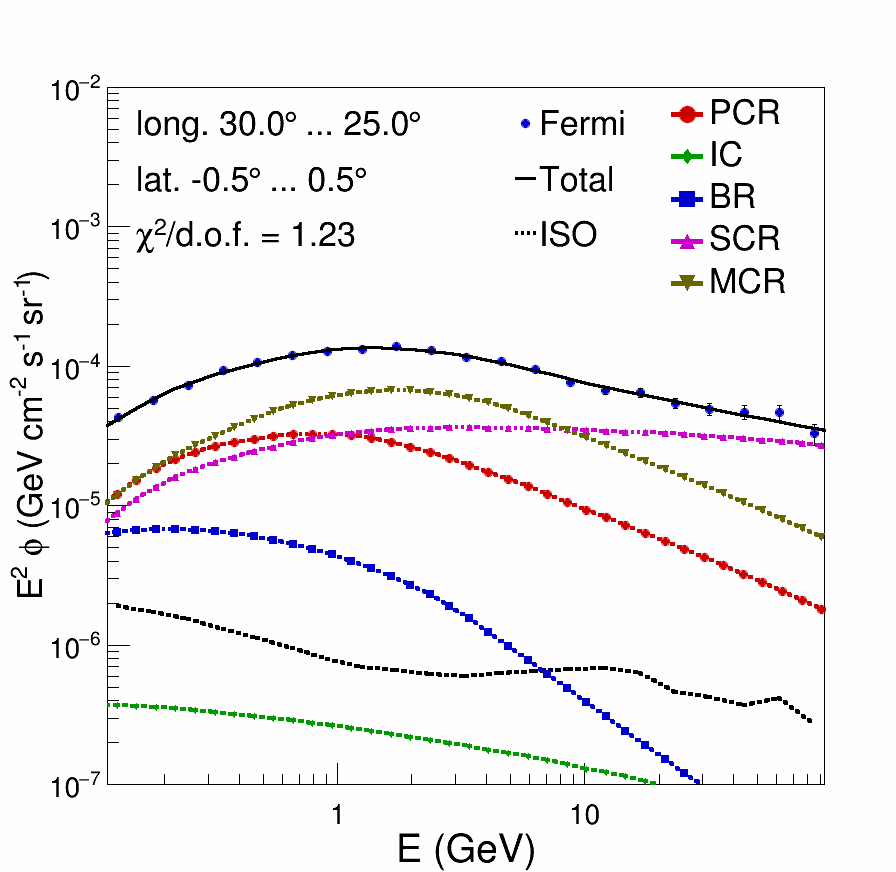}
\includegraphics[width=0.16\textwidth,height=0.16\textwidth,clip]{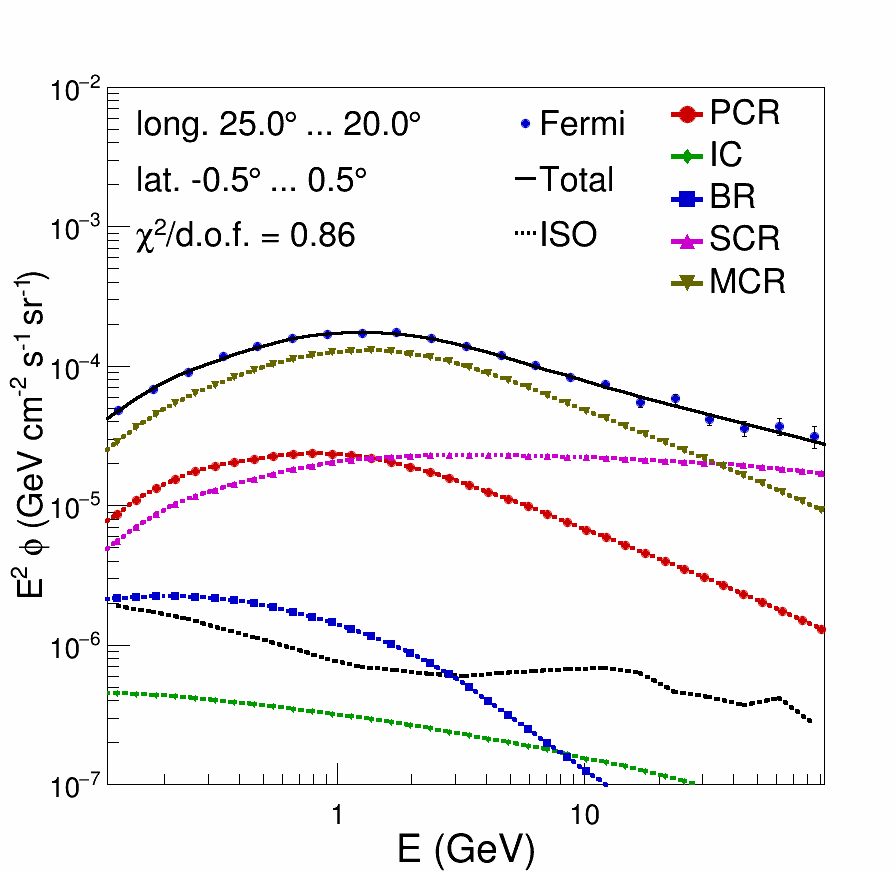}
\includegraphics[width=0.16\textwidth,height=0.16\textwidth,clip]{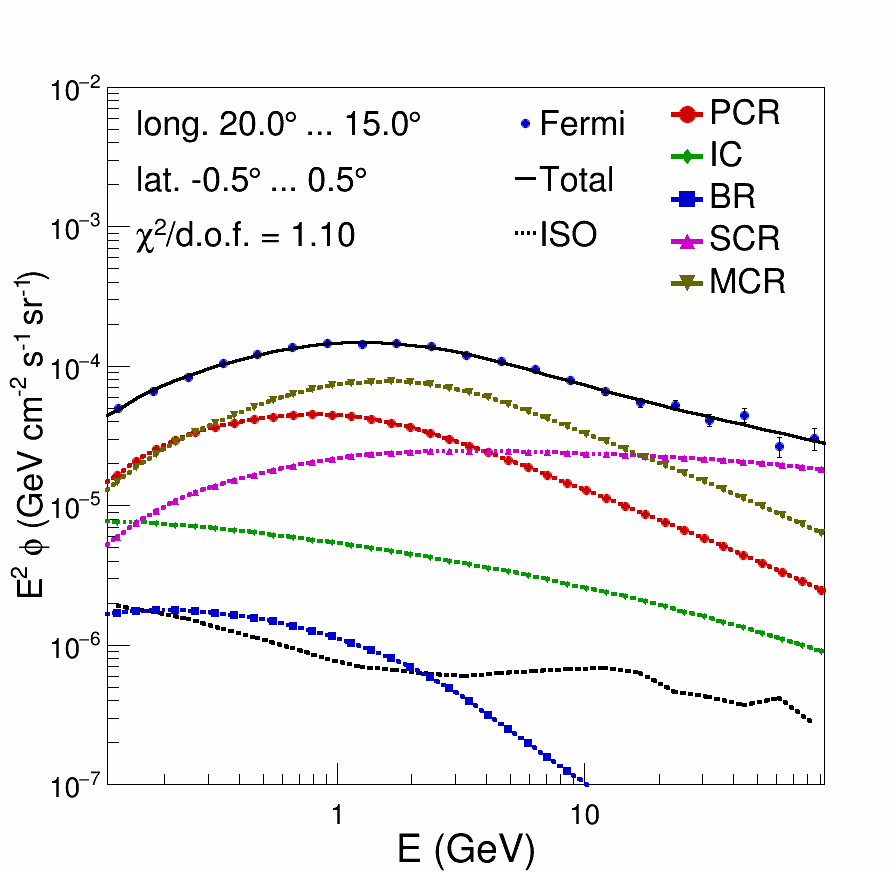}
\includegraphics[width=0.16\textwidth,height=0.16\textwidth,clip]{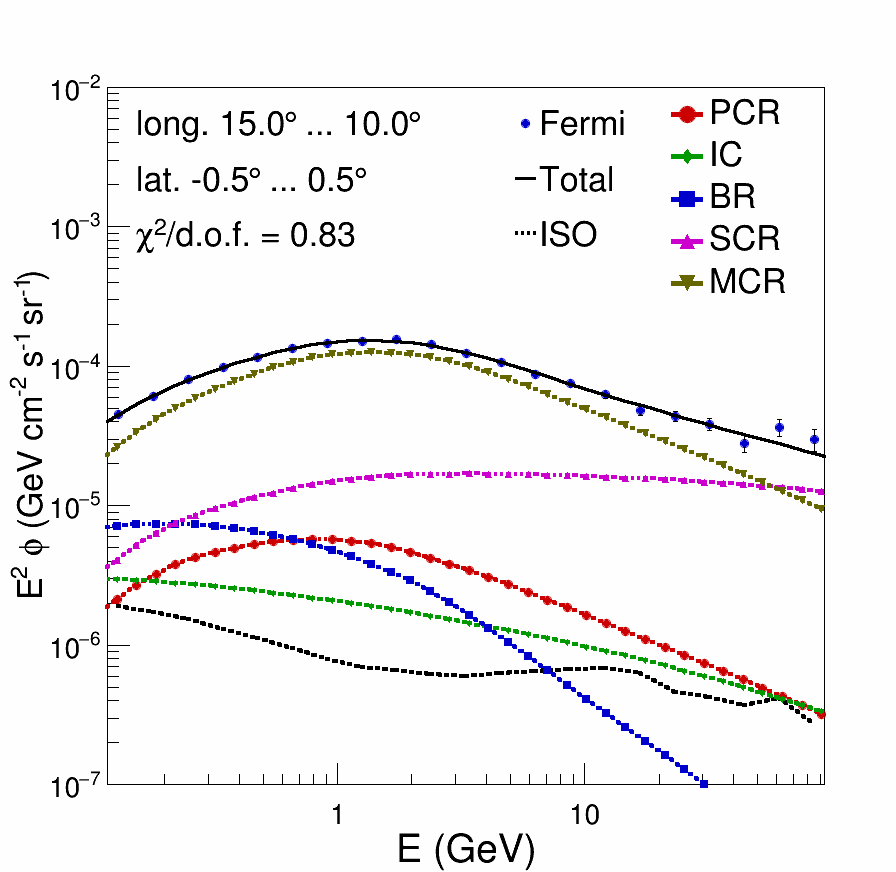}
\includegraphics[width=0.16\textwidth,height=0.16\textwidth,clip]{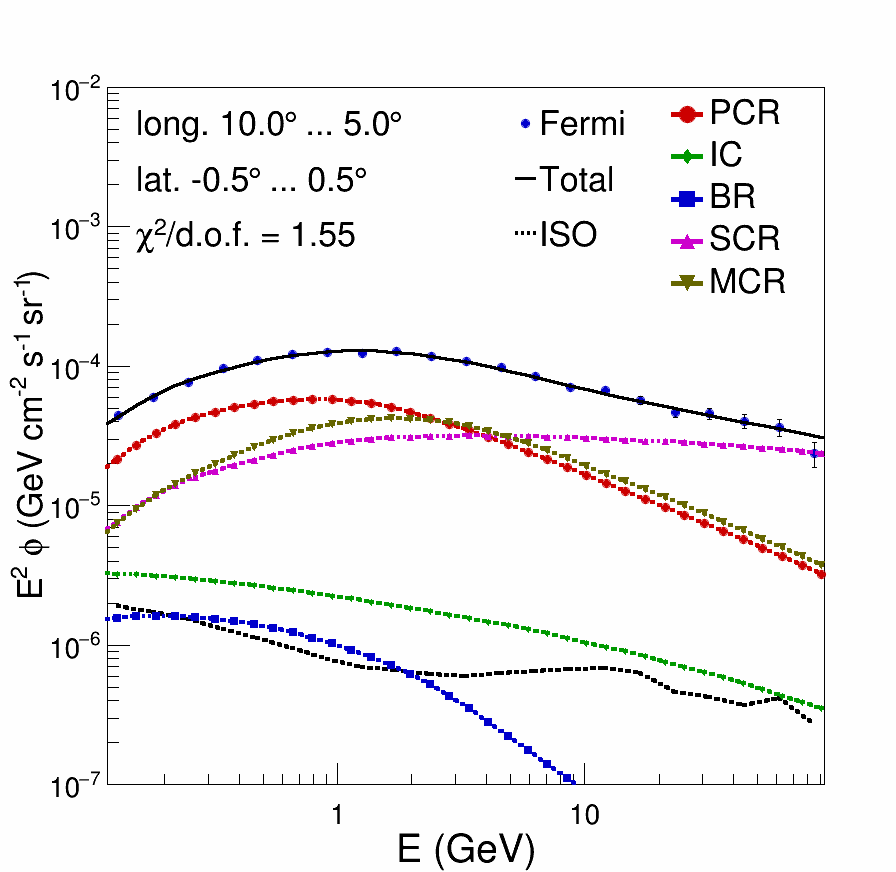}
\includegraphics[width=0.16\textwidth,height=0.16\textwidth,clip]{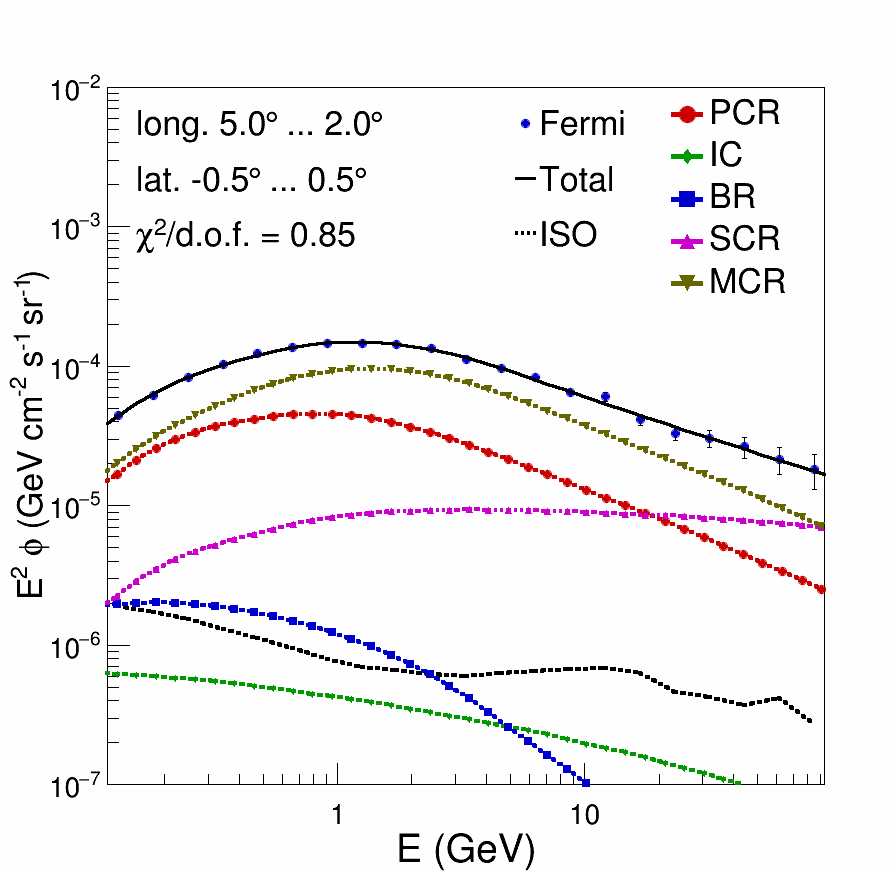}
\includegraphics[width=0.16\textwidth,height=0.16\textwidth,clip]{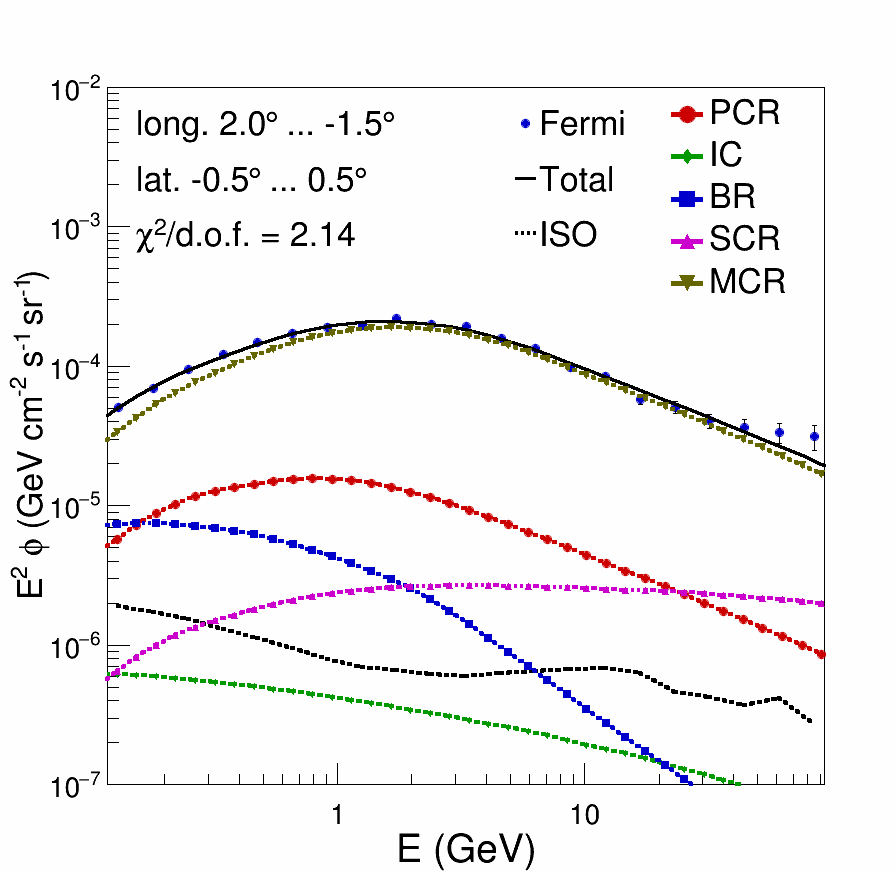}
\includegraphics[width=0.16\textwidth,height=0.16\textwidth,clip]{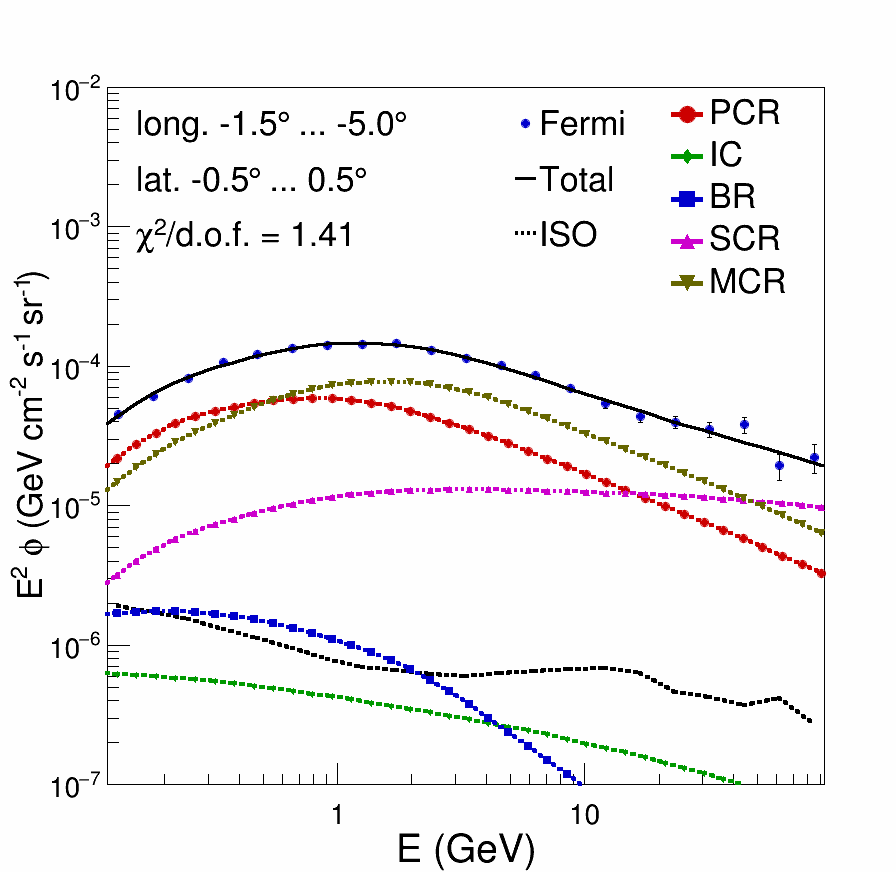}
\includegraphics[width=0.16\textwidth,height=0.16\textwidth,clip]{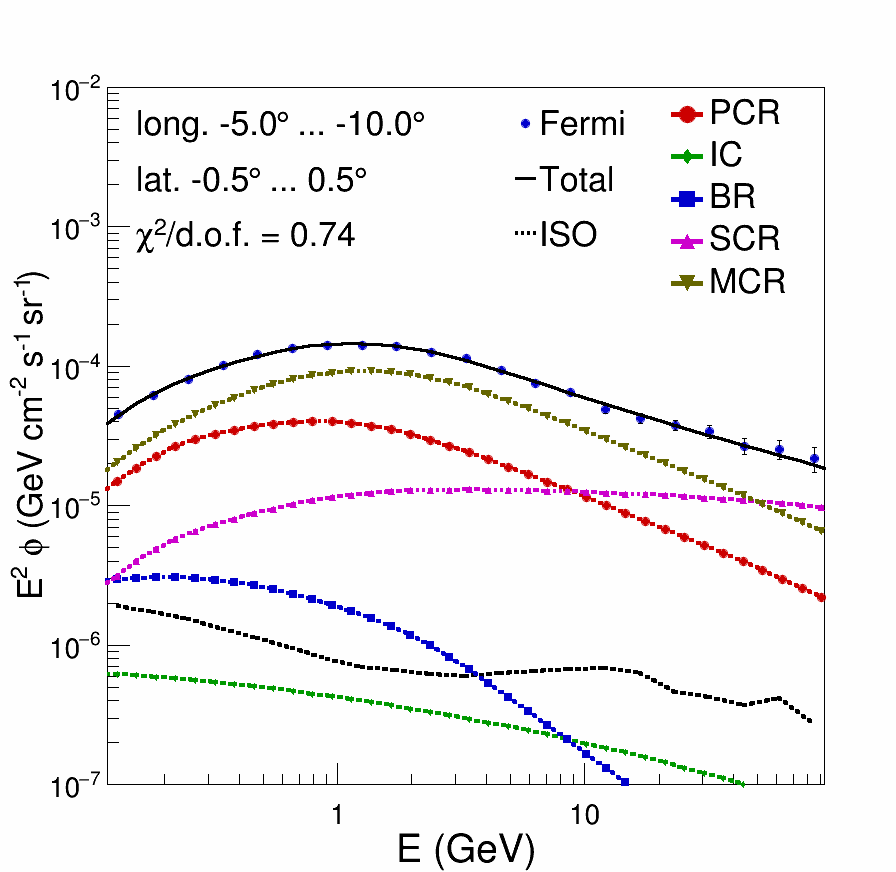}
\includegraphics[width=0.16\textwidth,height=0.16\textwidth,clip]{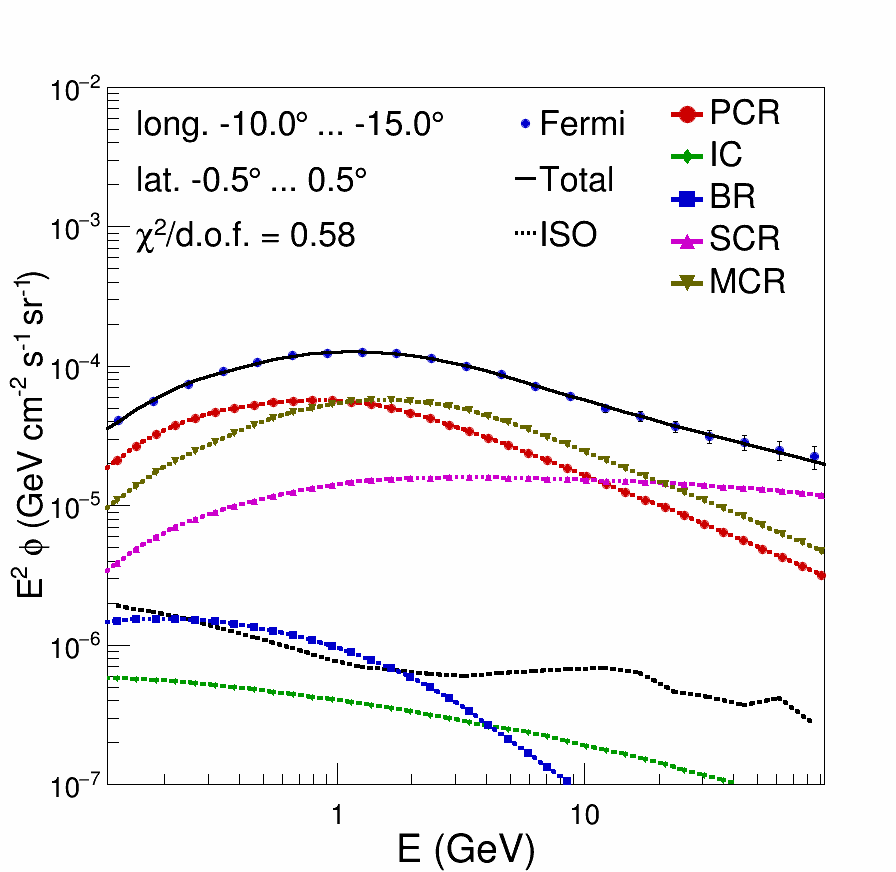}
\includegraphics[width=0.16\textwidth,height=0.16\textwidth,clip]{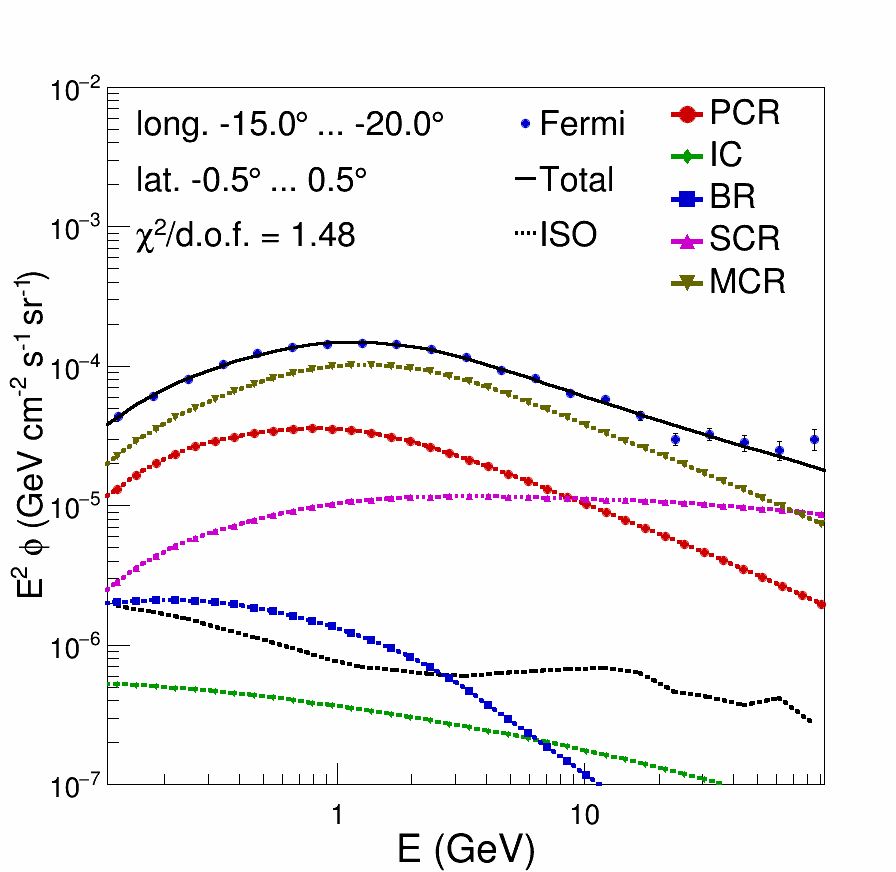}
\includegraphics[width=0.16\textwidth,height=0.16\textwidth,clip]{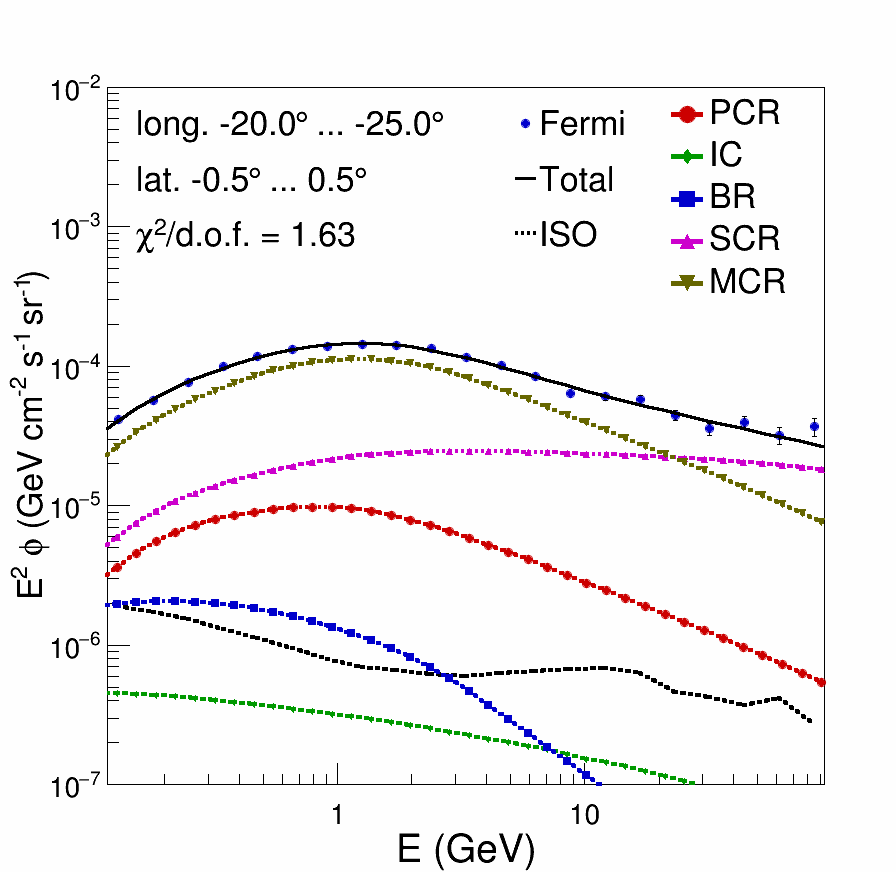}
\includegraphics[width=0.16\textwidth,height=0.16\textwidth,clip]{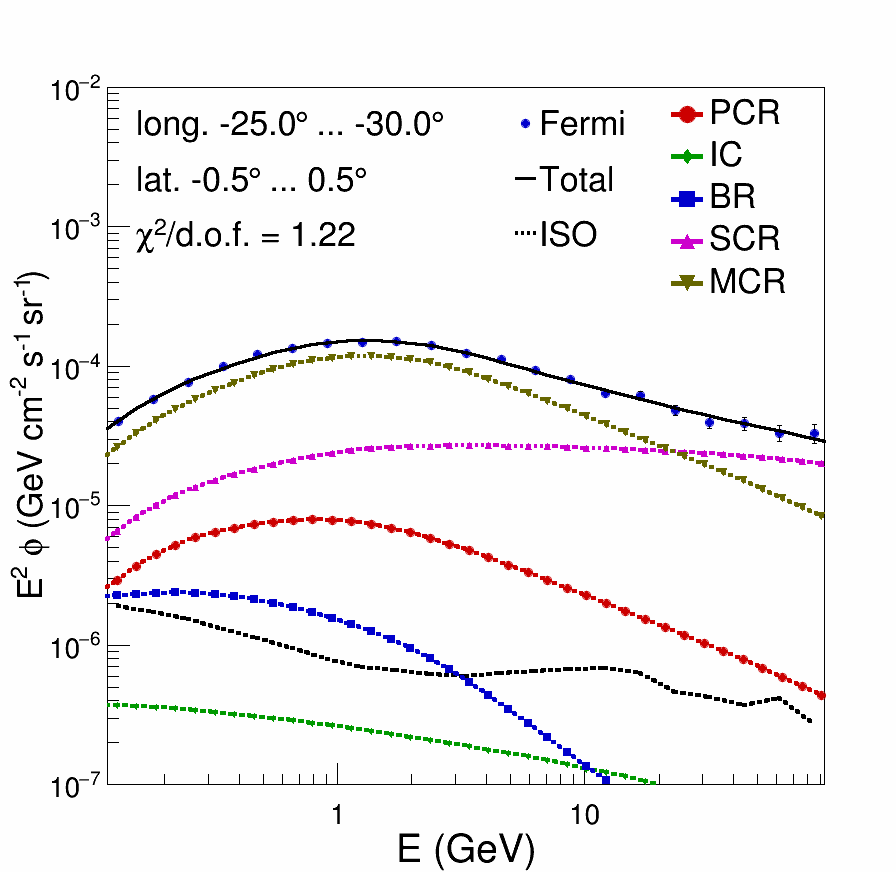}
\includegraphics[width=0.16\textwidth,height=0.16\textwidth,clip]{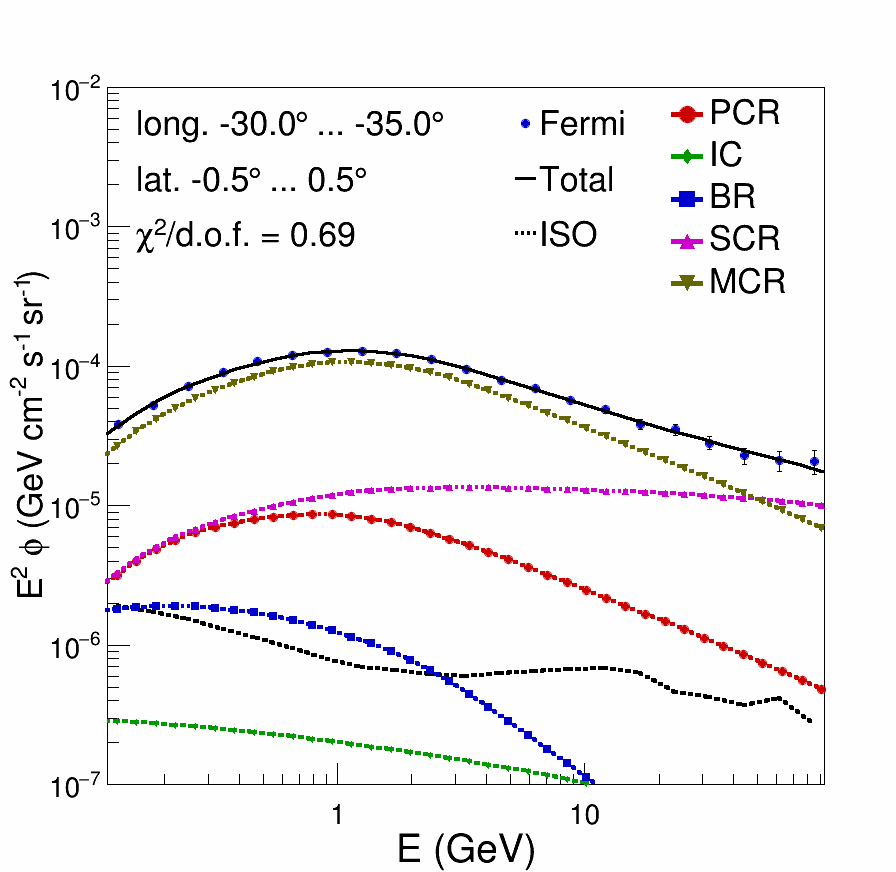}
\includegraphics[width=0.16\textwidth,height=0.16\textwidth,clip]{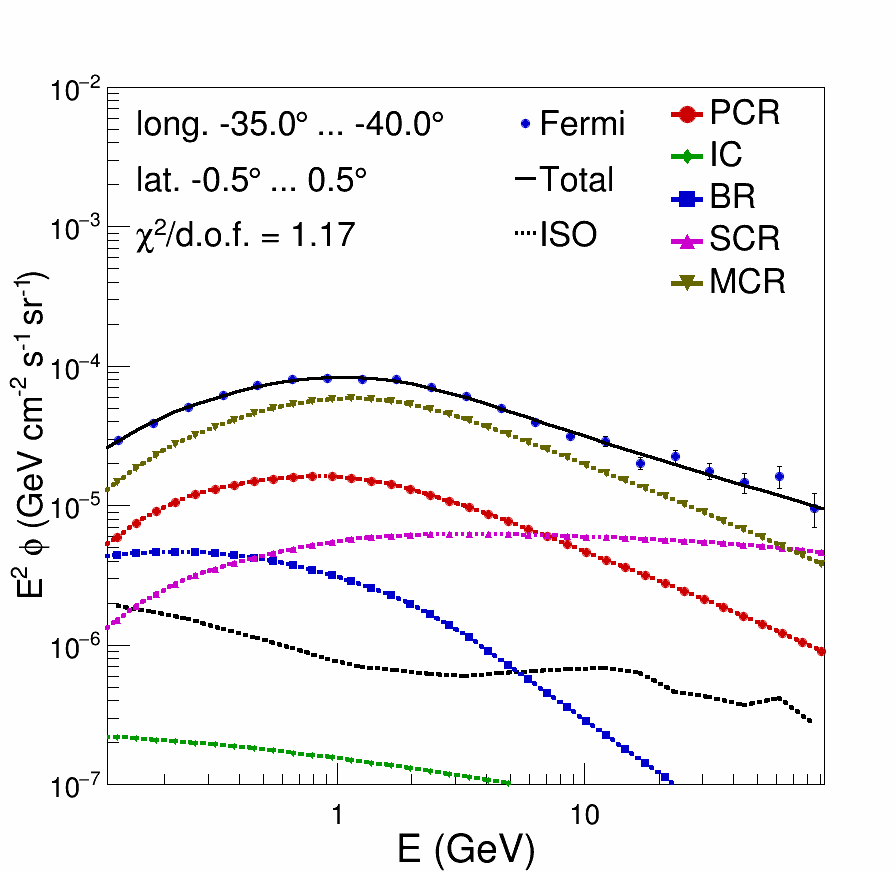}
\includegraphics[width=0.16\textwidth,height=0.16\textwidth,clip]{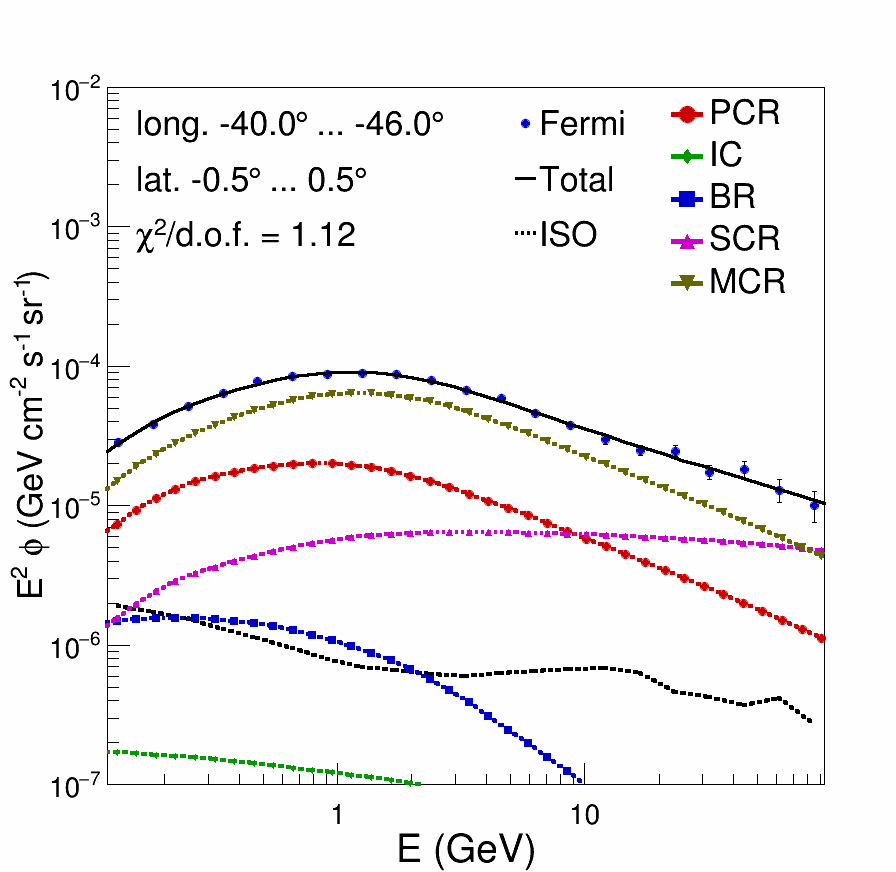}
\includegraphics[width=0.16\textwidth,height=0.16\textwidth,clip]{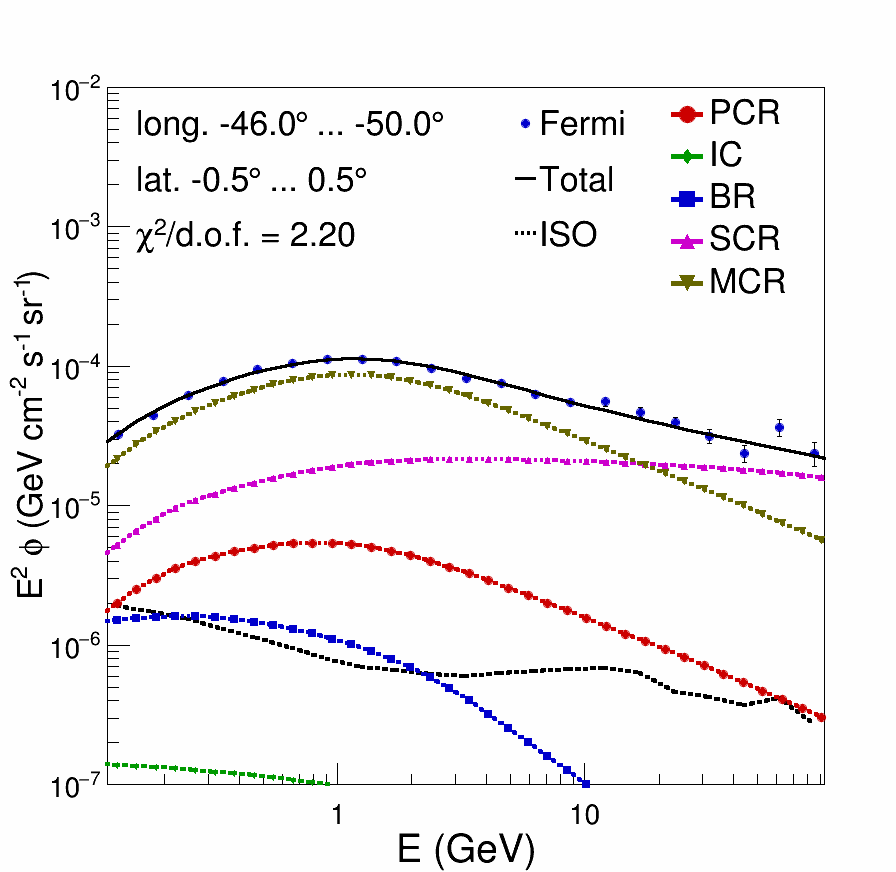}
\includegraphics[width=0.16\textwidth,height=0.16\textwidth,clip]{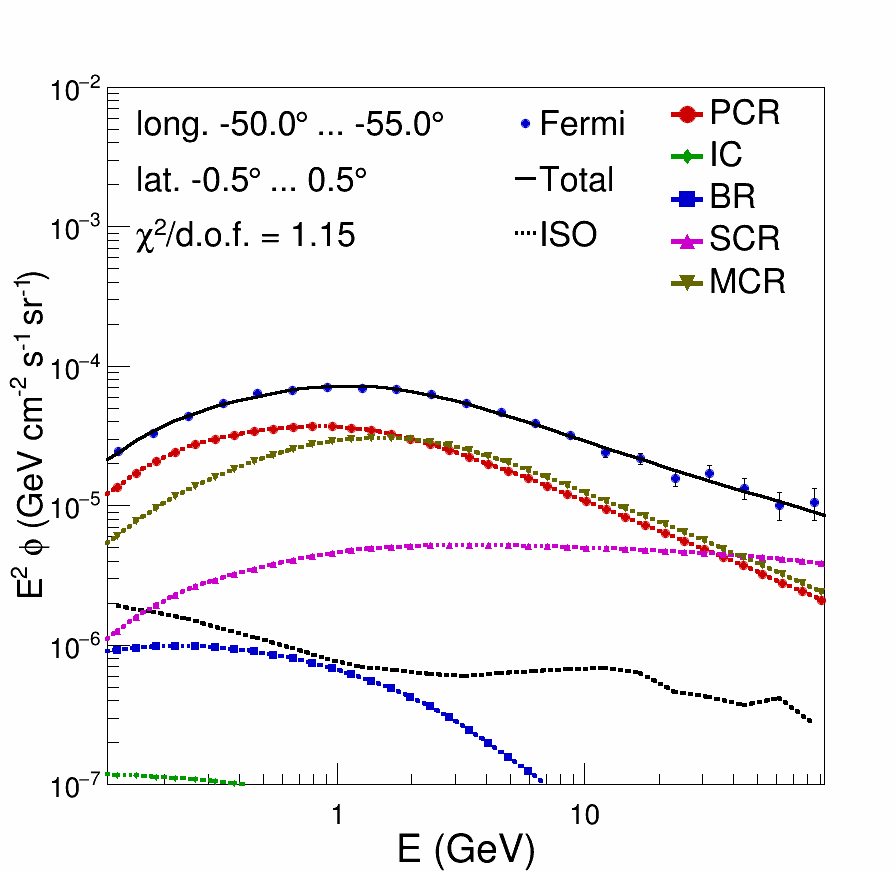}
\includegraphics[width=0.16\textwidth,height=0.16\textwidth,clip]{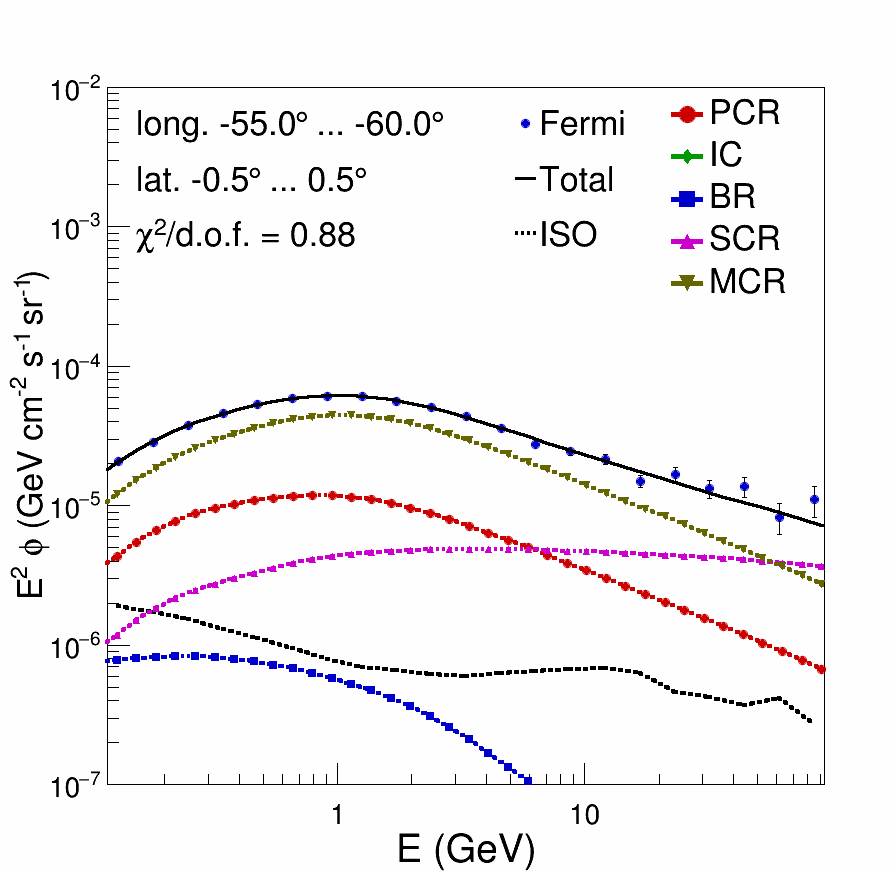}
\includegraphics[width=0.16\textwidth,height=0.16\textwidth,clip]{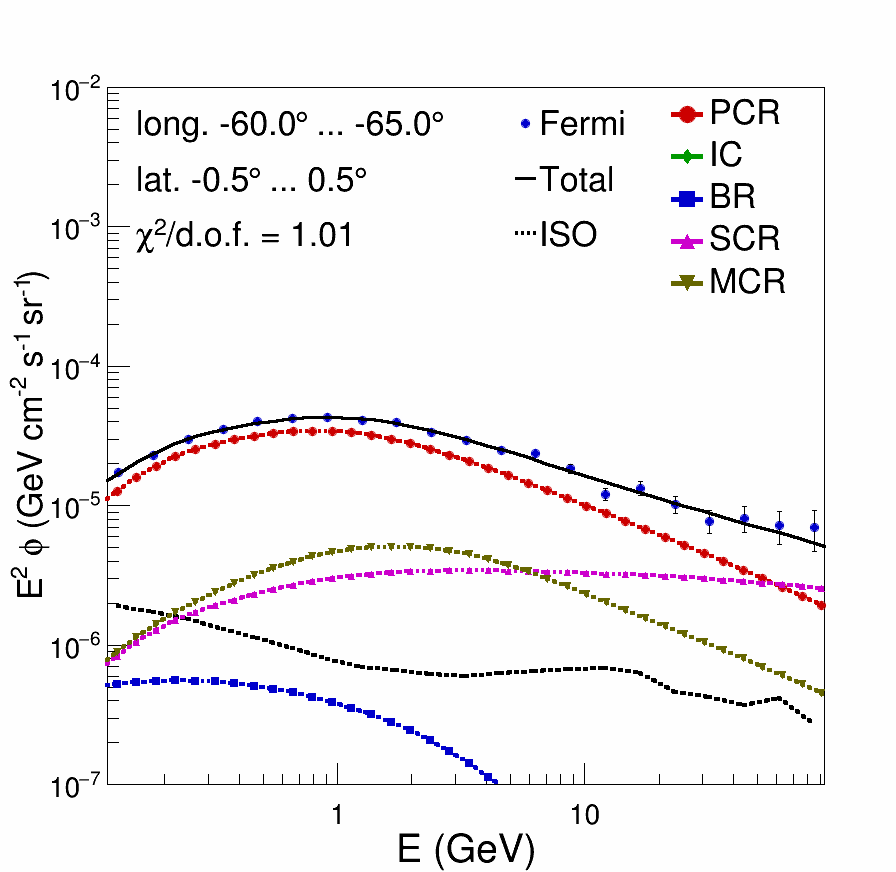}
\includegraphics[width=0.16\textwidth,height=0.16\textwidth,clip]{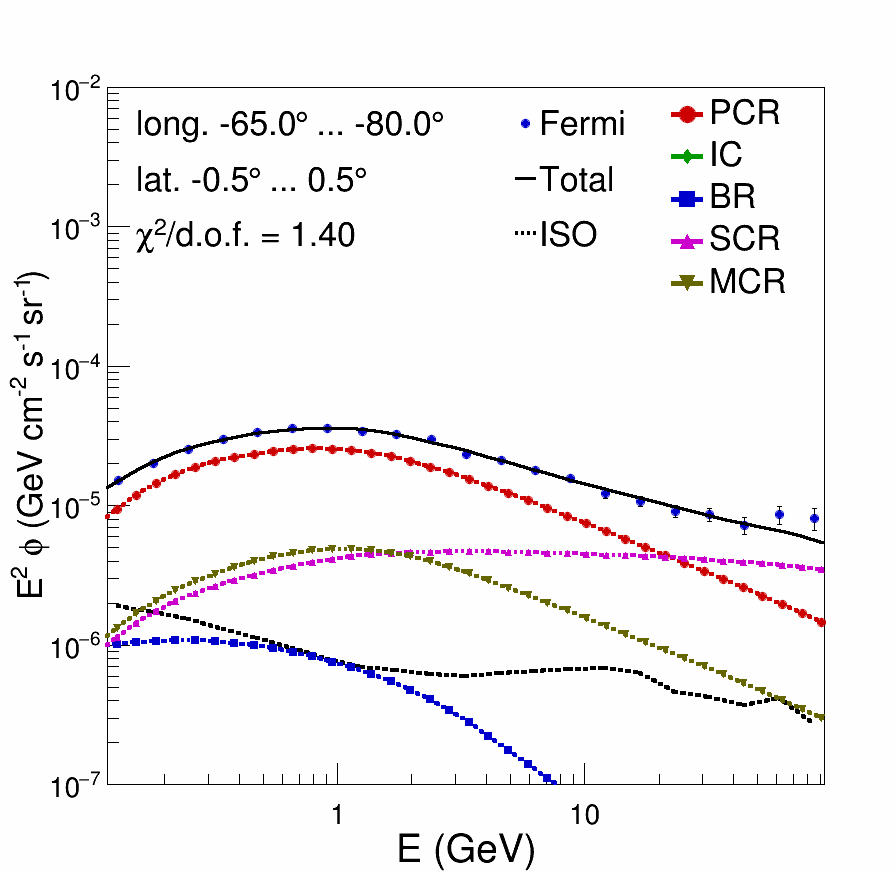}
\includegraphics[width=0.16\textwidth,height=0.16\textwidth,clip]{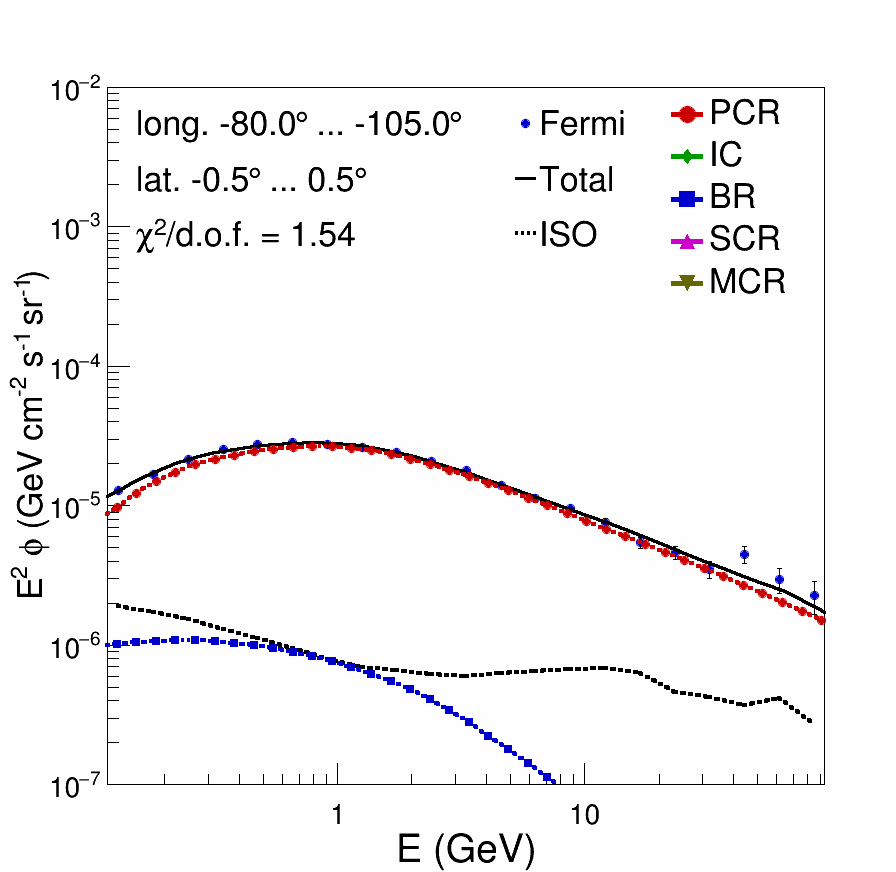}
\includegraphics[width=0.16\textwidth,height=0.16\textwidth,clip]{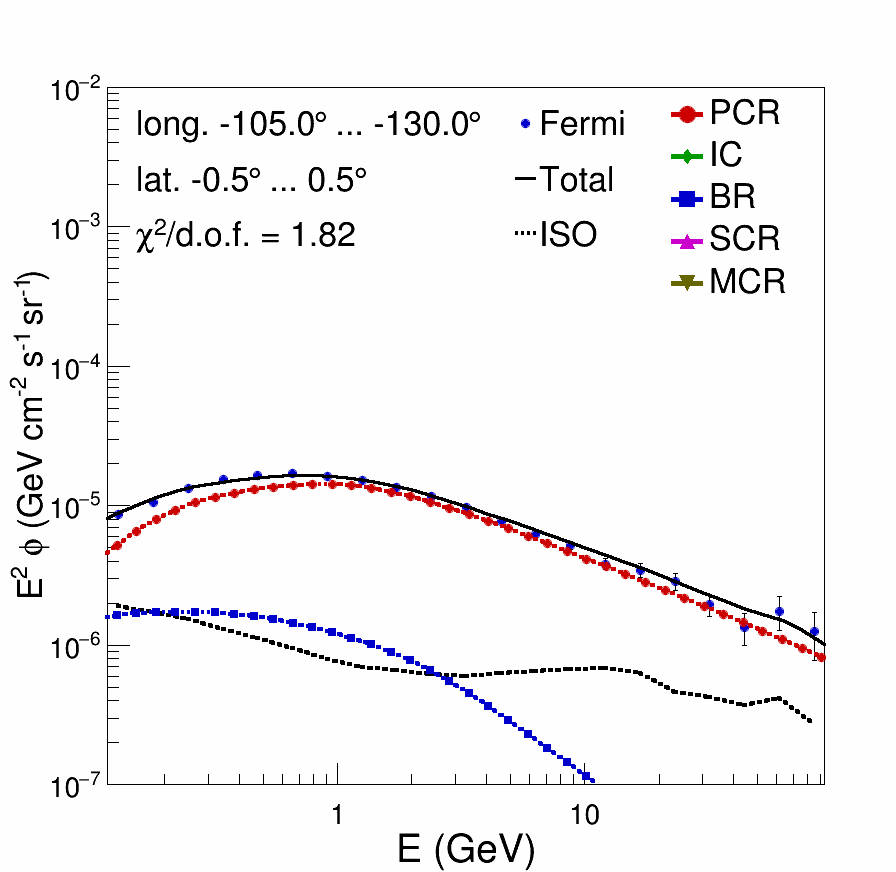}
\includegraphics[width=0.16\textwidth,height=0.16\textwidth,clip]{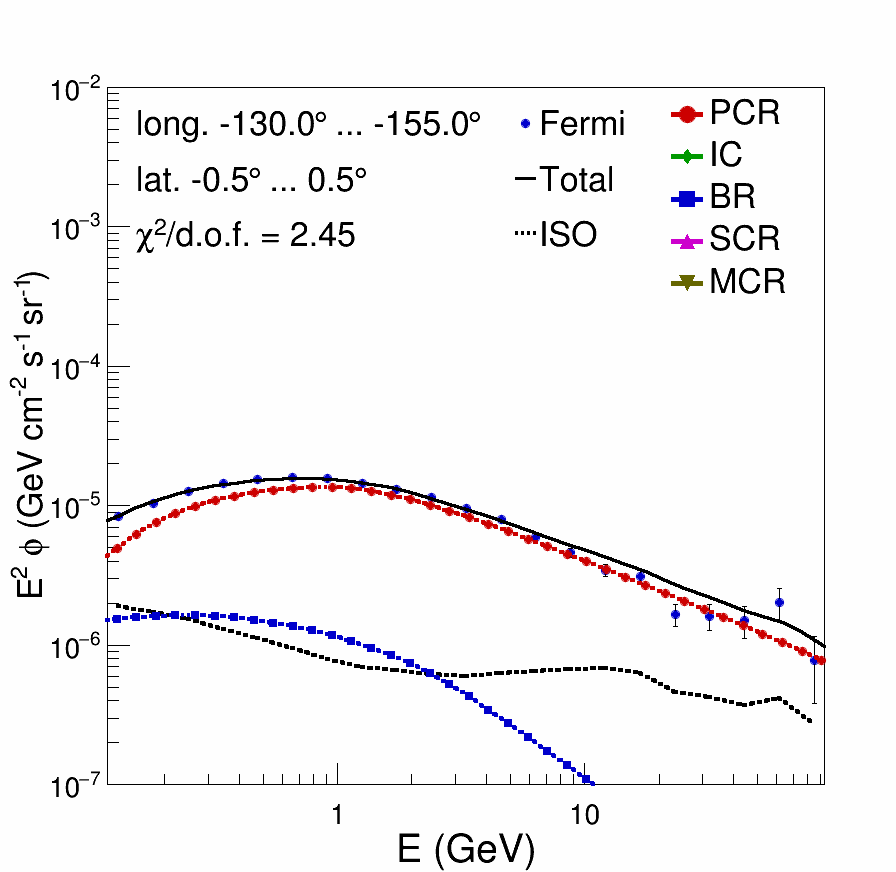}
\includegraphics[width=0.16\textwidth,height=0.16\textwidth,clip]{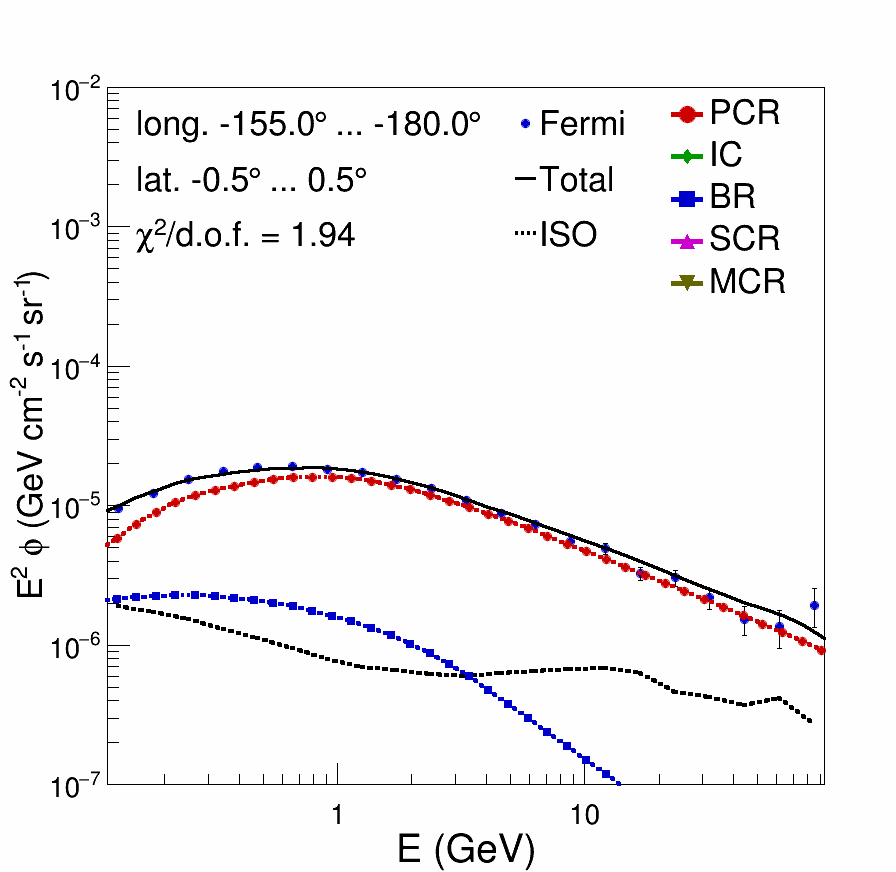}
\caption[]{Template fits for latitudes  with $-0.5^\circ<b<0.5^\circ$ and longitudes decreasing from 180$^\circ$ to -180$^\circ$.} \label{F21}
\end{figure}
\begin{figure}
\centering
	\includegraphics[width=0.16\textwidth,height=0.16\textwidth,clip]{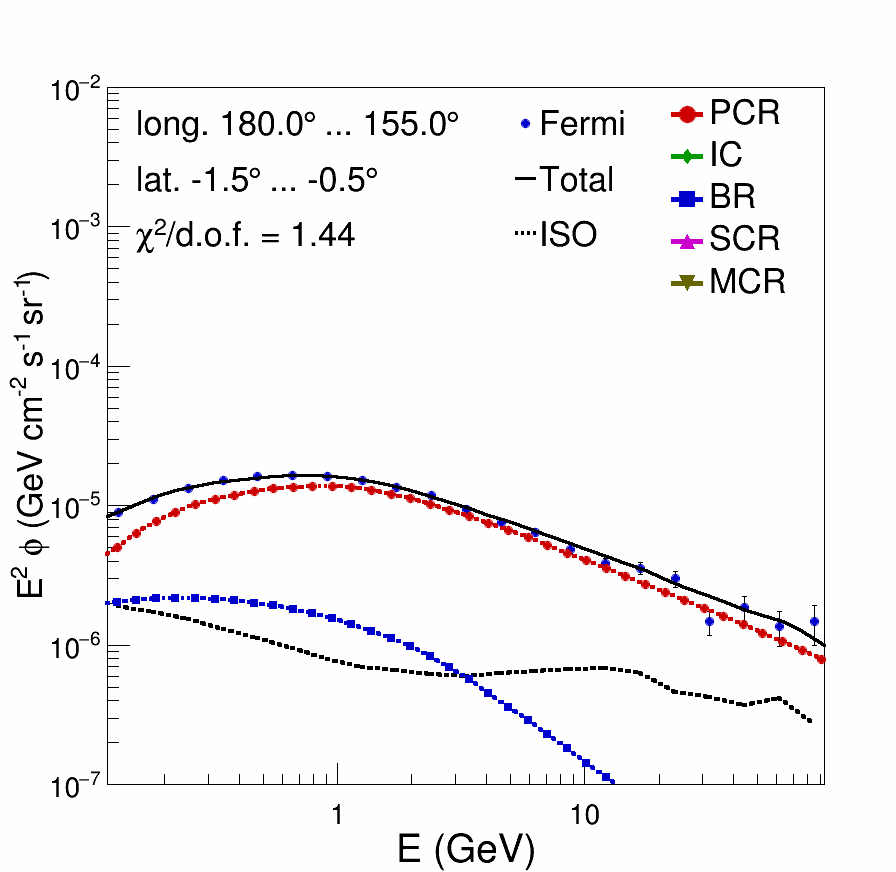}
	\includegraphics[width=0.16\textwidth,height=0.16\textwidth,clip]{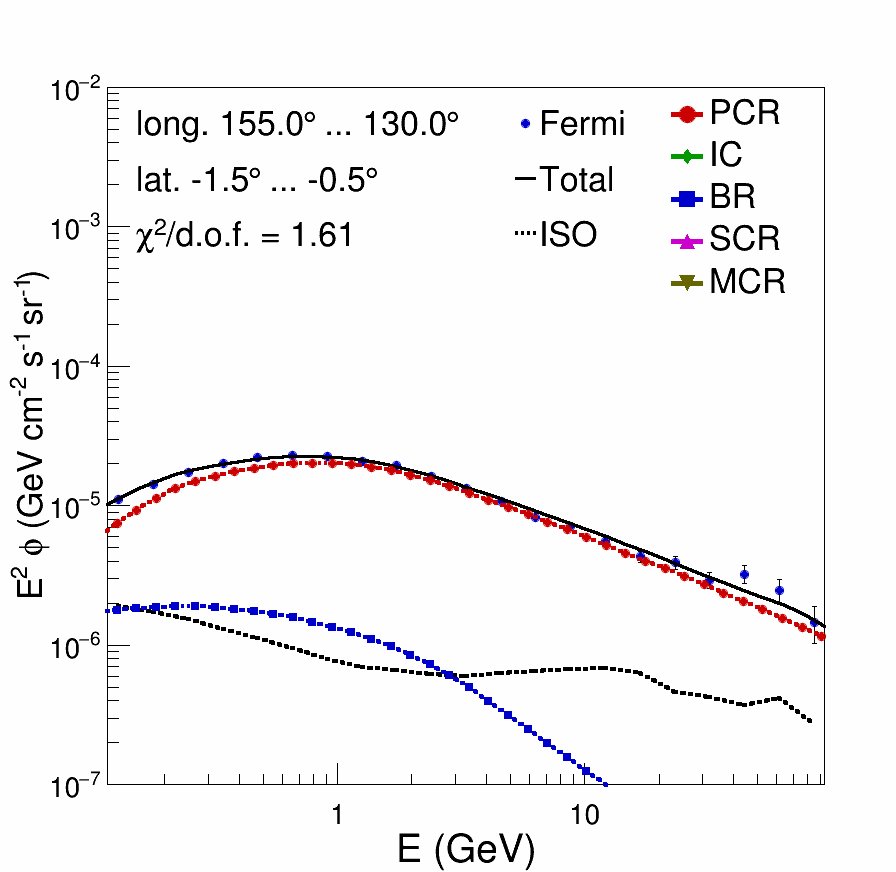}
	\includegraphics[width=0.16\textwidth,height=0.16\textwidth,clip]{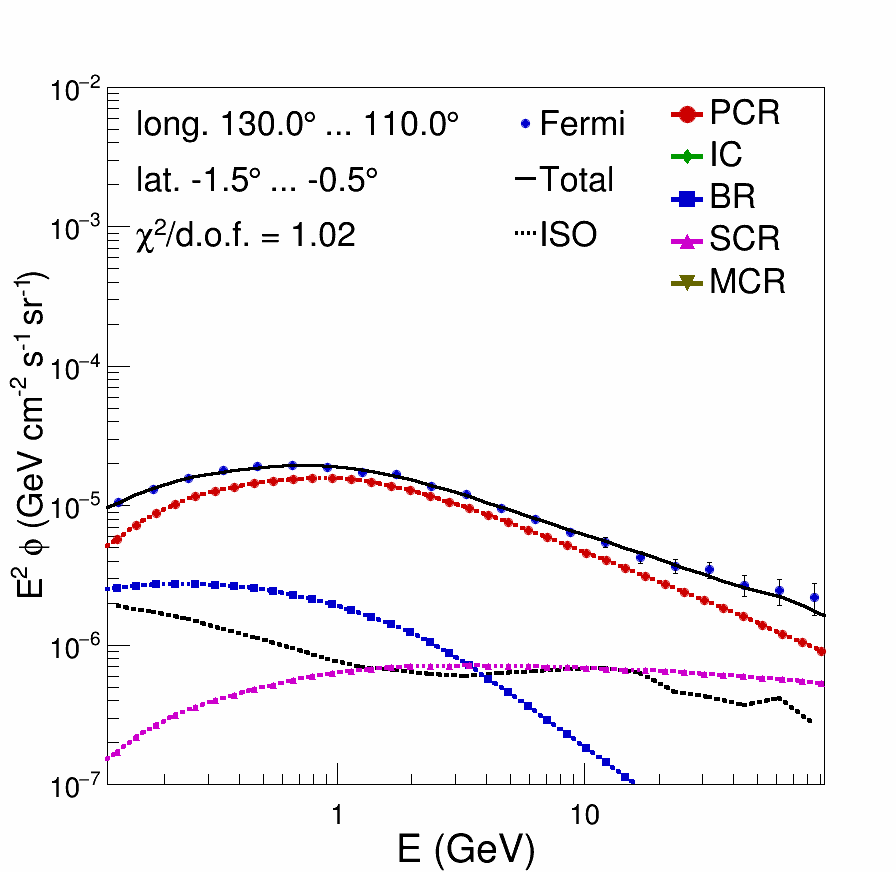}
	\includegraphics[width=0.16\textwidth,height=0.16\textwidth,clip]{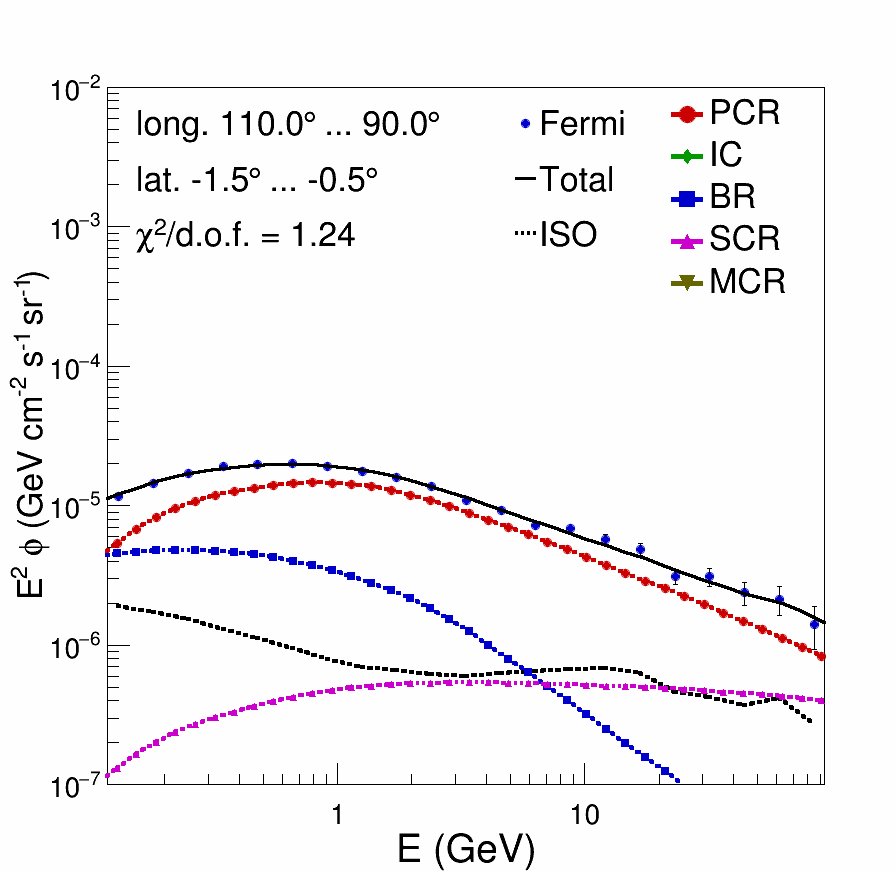}
	\includegraphics[width=0.16\textwidth,height=0.16\textwidth,clip]{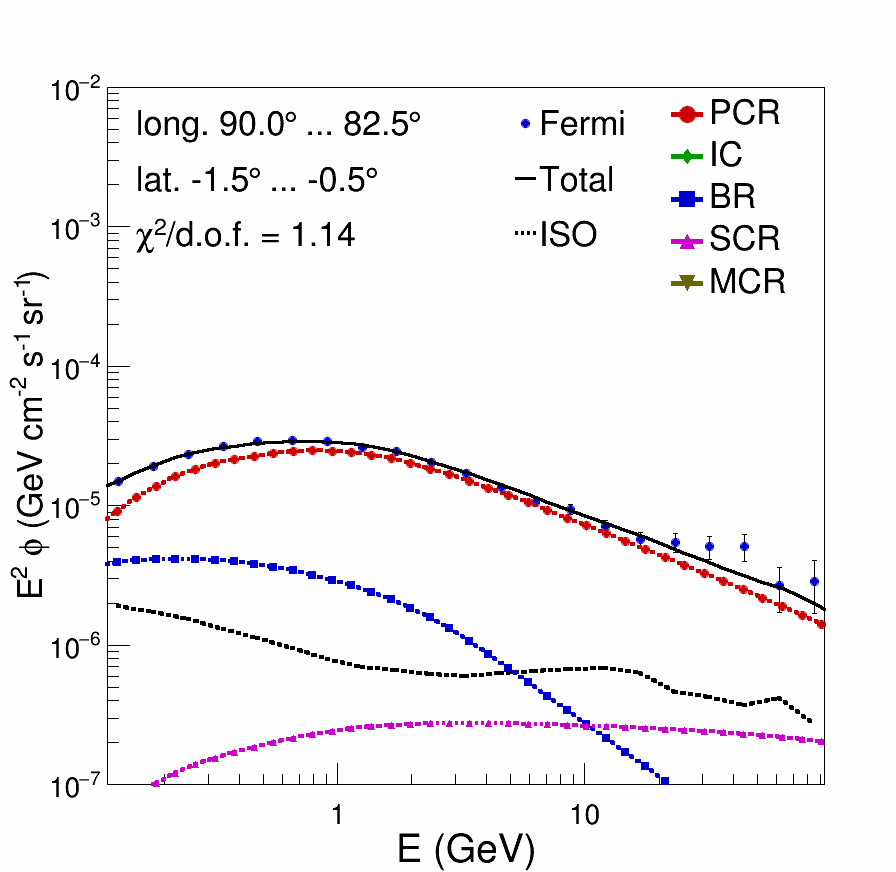}
	\includegraphics[width=0.16\textwidth,height=0.16\textwidth,clip]{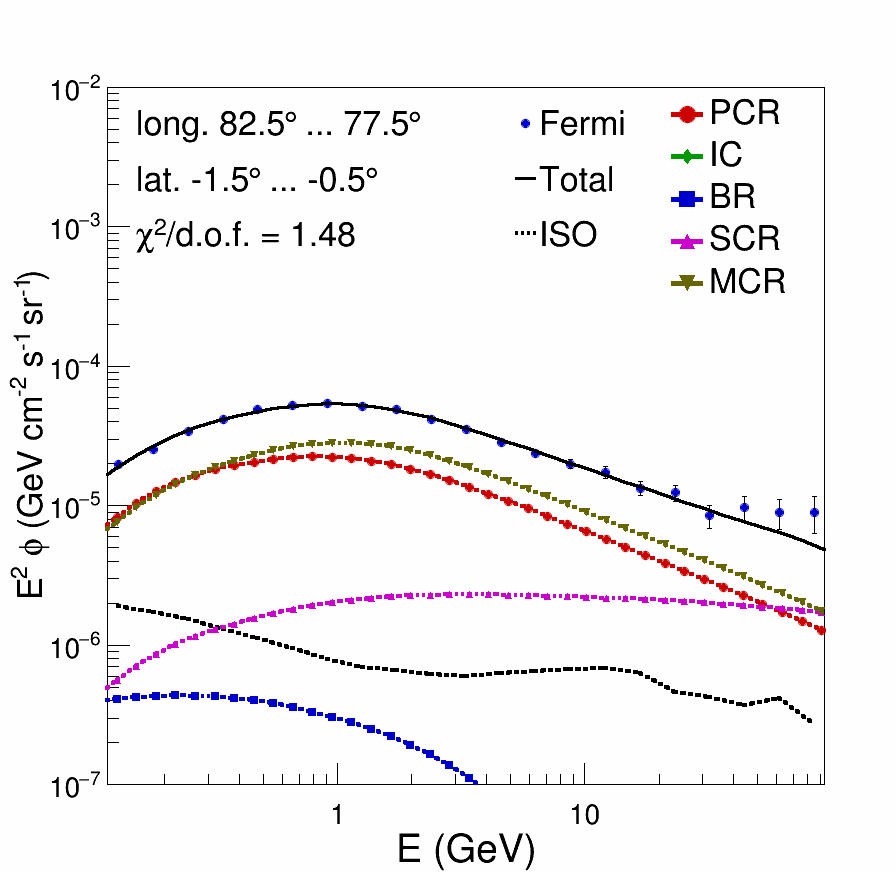}
	\includegraphics[width=0.16\textwidth,height=0.16\textwidth,clip]{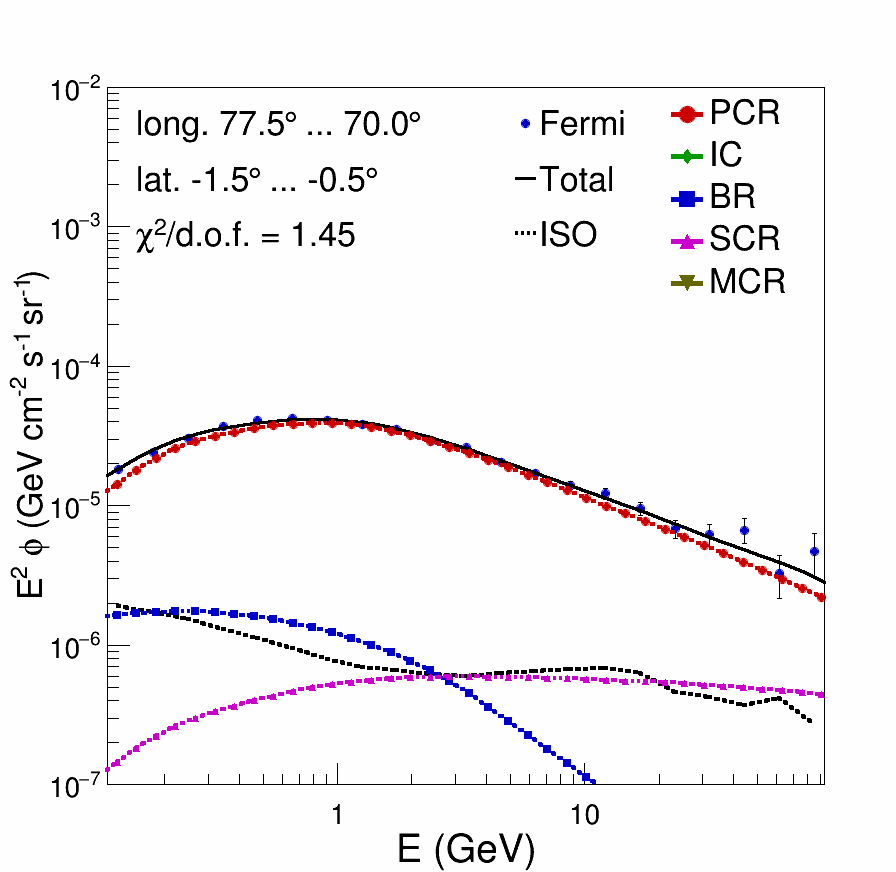}
	\includegraphics[width=0.16\textwidth,height=0.16\textwidth,clip]{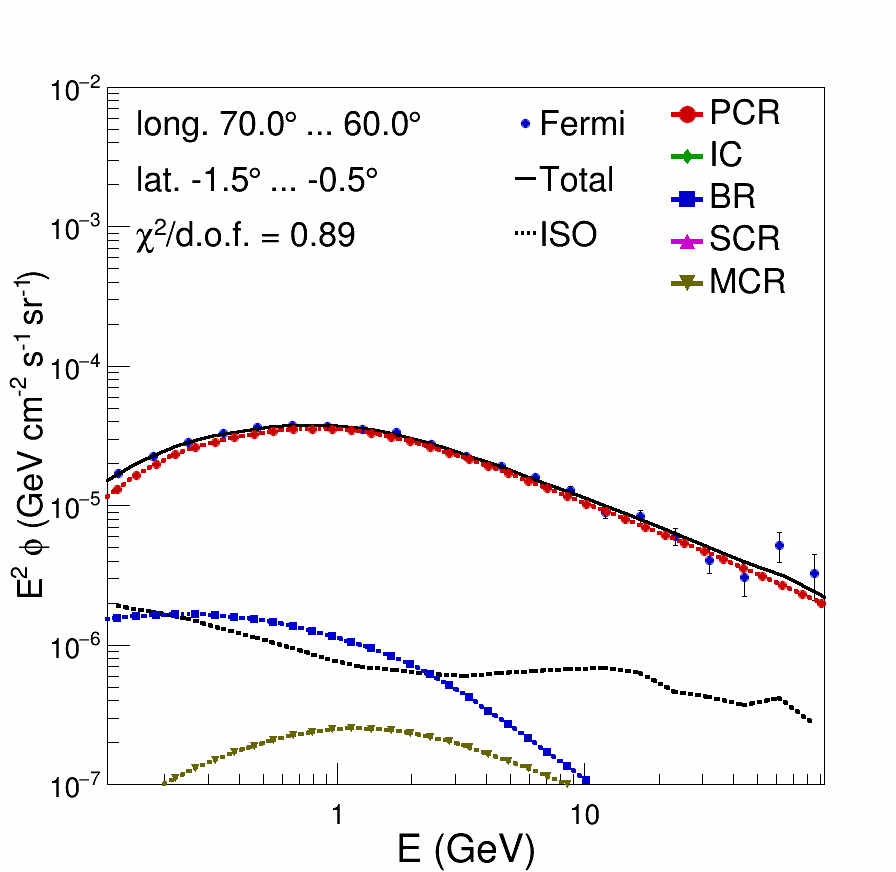}
	\includegraphics[width=0.16\textwidth,height=0.16\textwidth,clip]{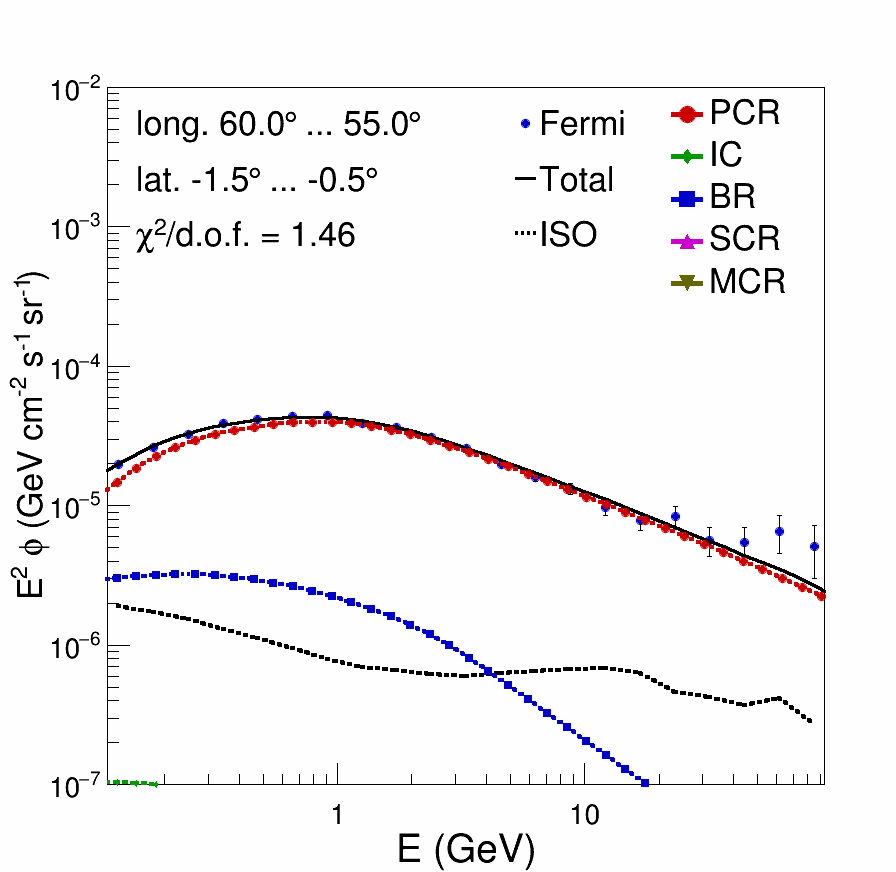}
	\includegraphics[width=0.16\textwidth,height=0.16\textwidth,clip]{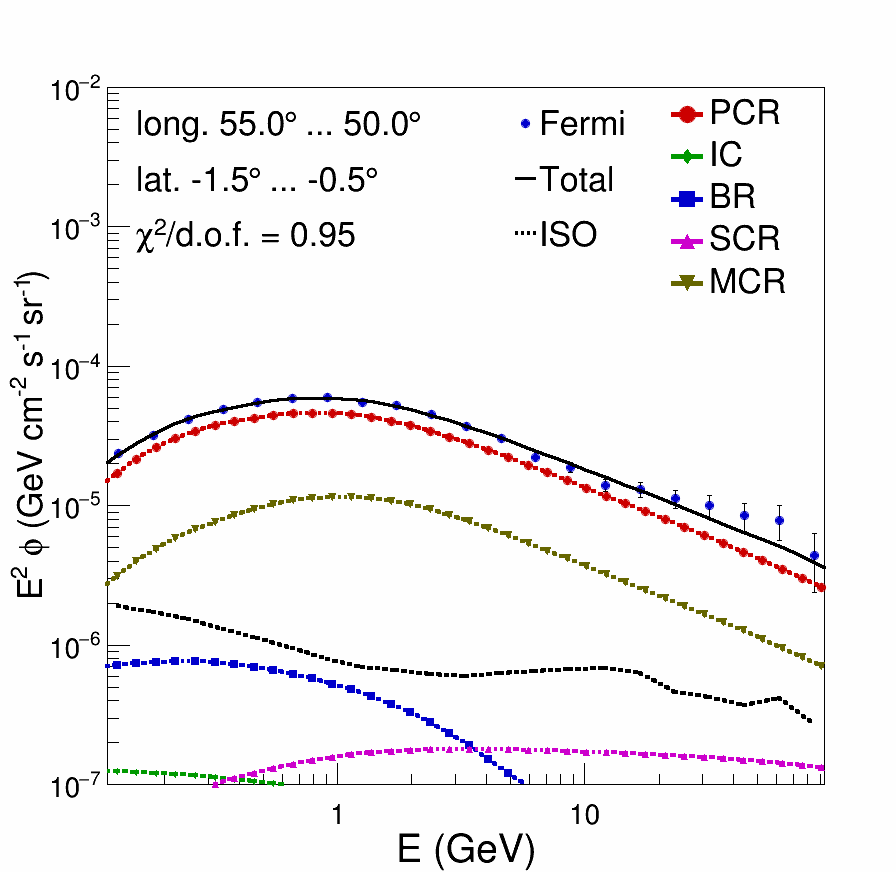}
	\includegraphics[width=0.16\textwidth,height=0.16\textwidth,clip]{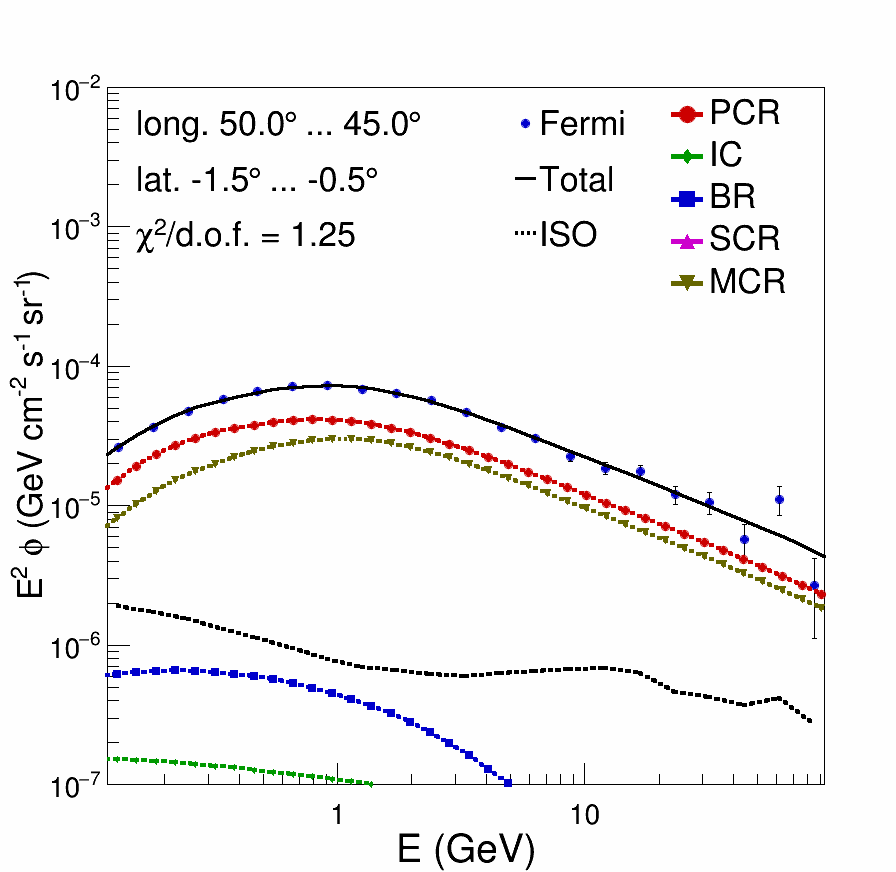}			    \includegraphics[width=0.16\textwidth,height=0.16\textwidth,clip]{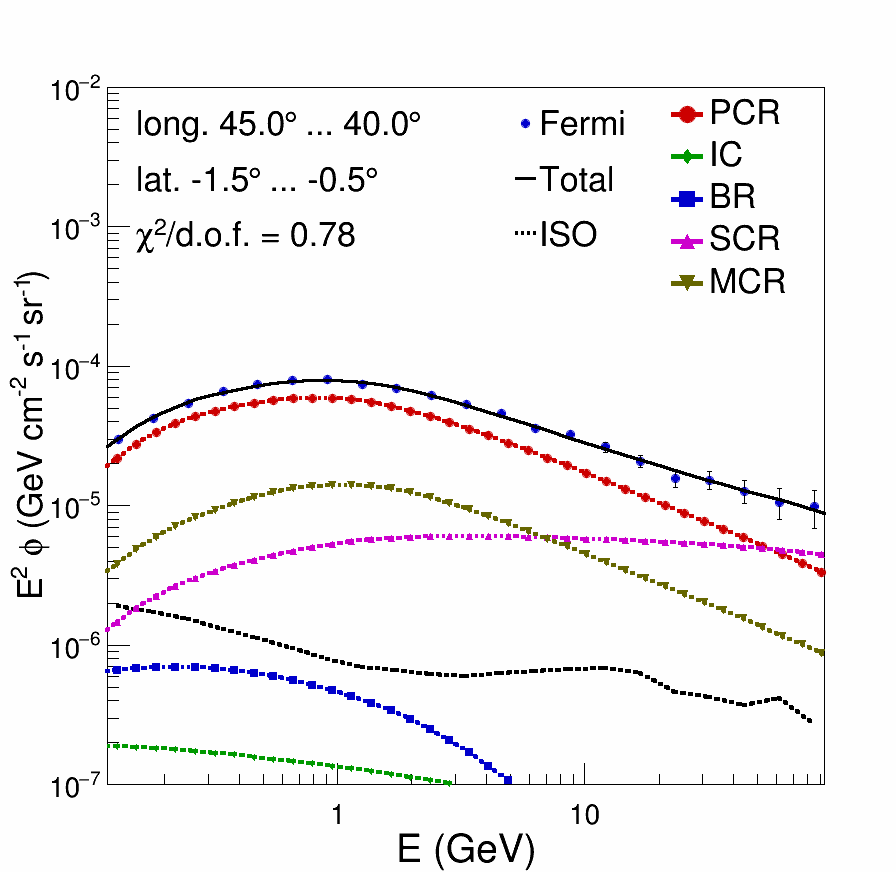}
	\includegraphics[width=0.16\textwidth,height=0.16\textwidth,clip]{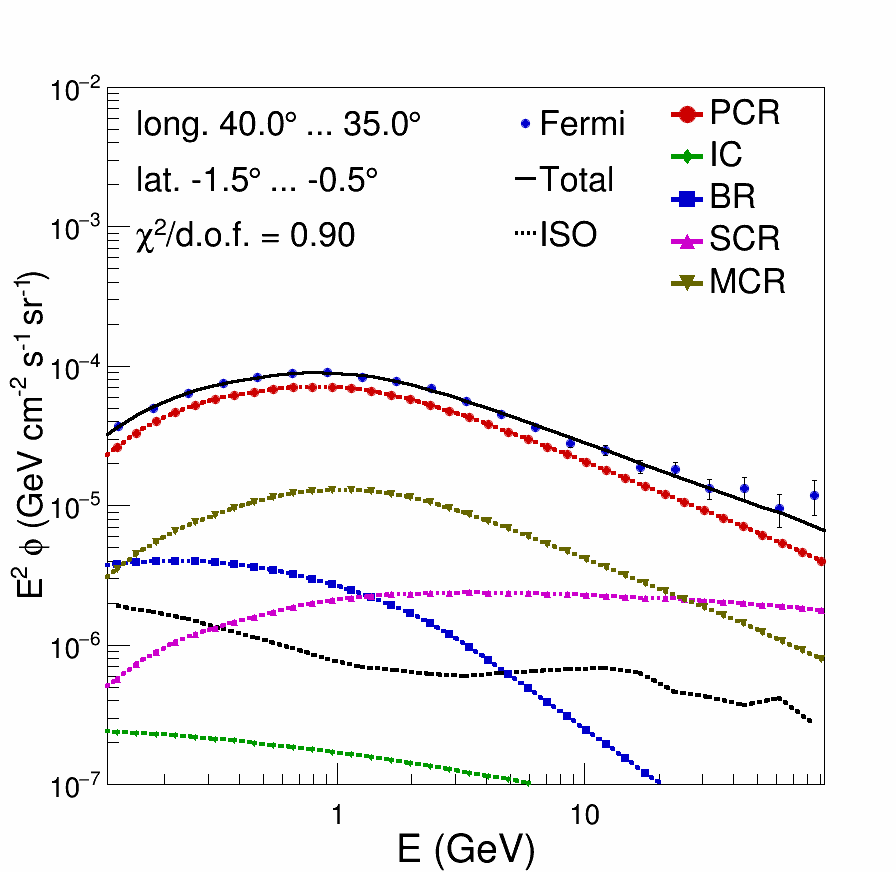}
	\includegraphics[width=0.16\textwidth,height=0.16\textwidth,clip]{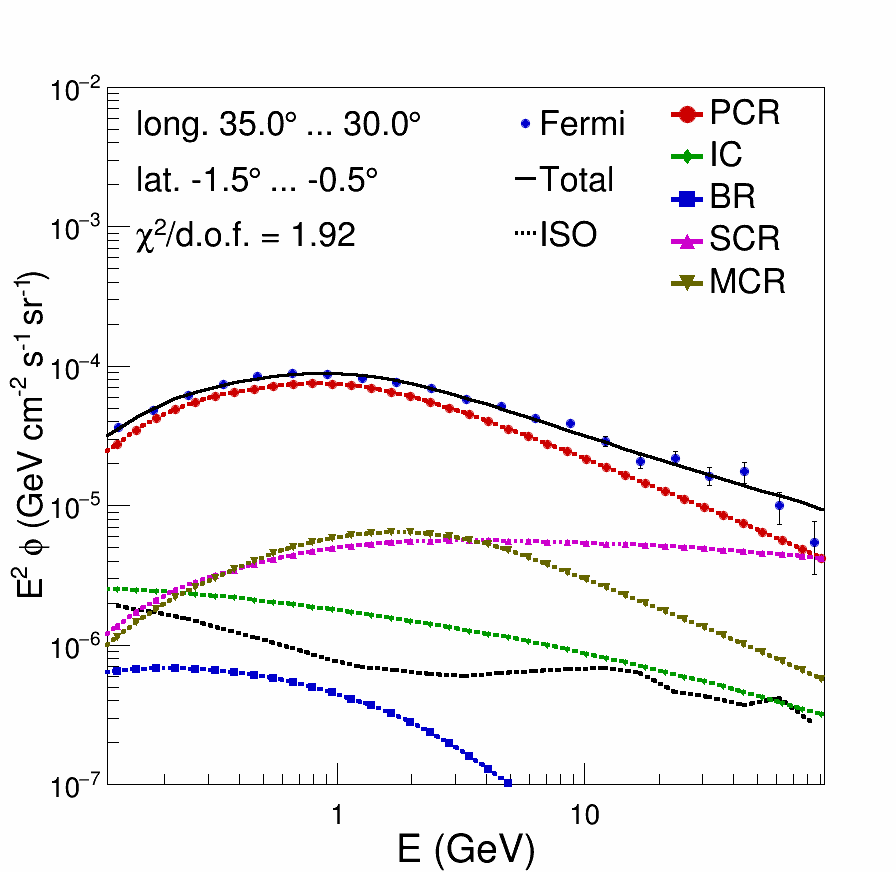}
	\includegraphics[width=0.16\textwidth,height=0.16\textwidth,clip]{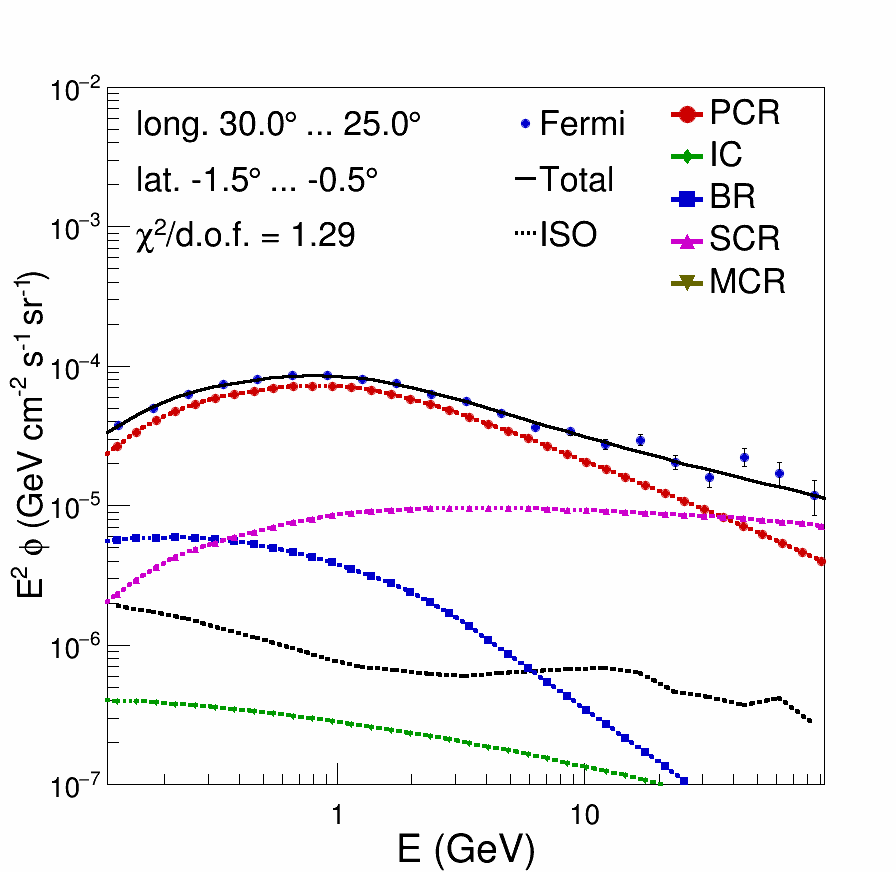}
	\includegraphics[width=0.16\textwidth,height=0.16\textwidth,clip]{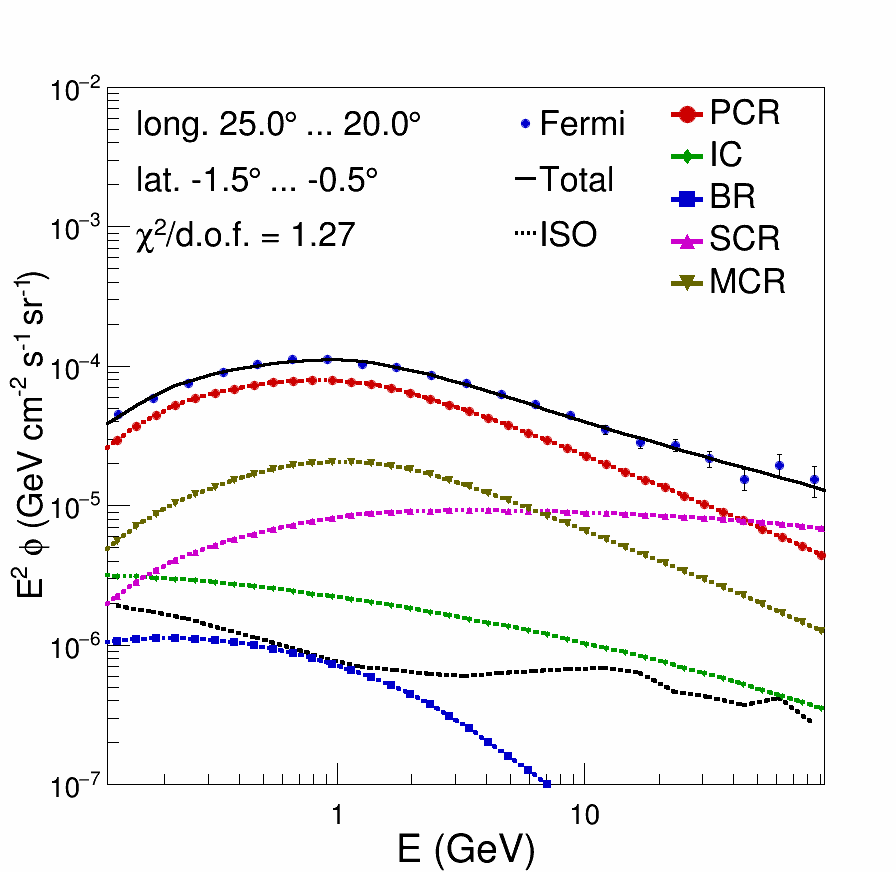}
	\includegraphics[width=0.16\textwidth,height=0.16\textwidth,clip]{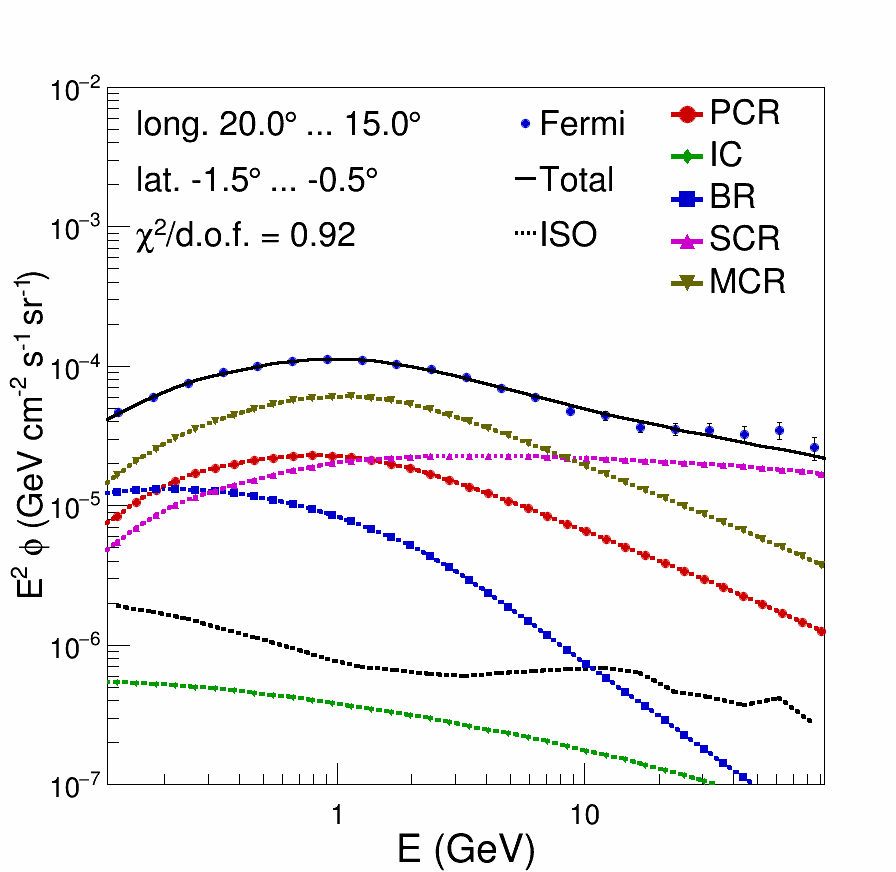}
	\includegraphics[width=0.16\textwidth,height=0.16\textwidth,clip]{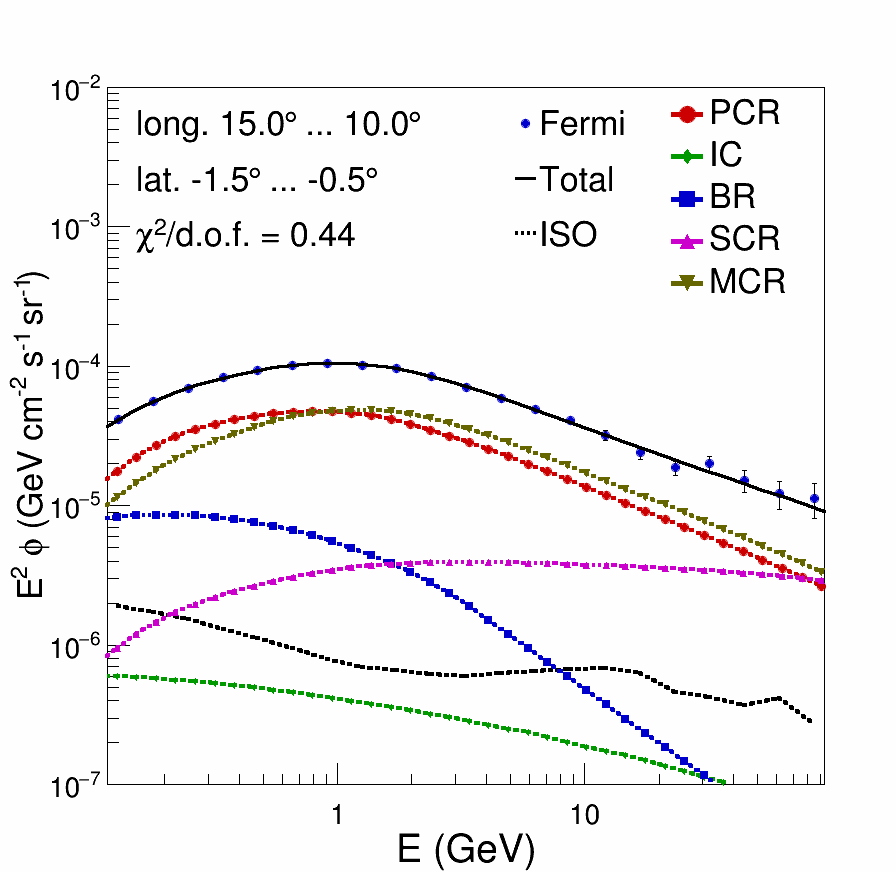}
	\includegraphics[width=0.16\textwidth,height=0.16\textwidth,clip]{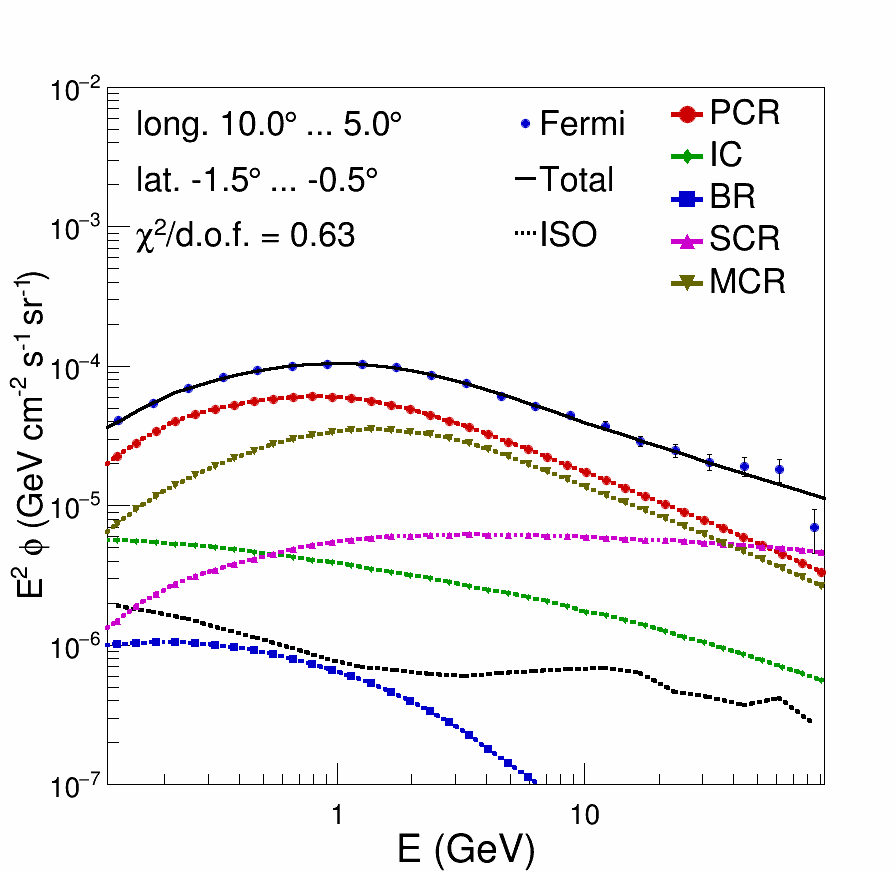}
	\includegraphics[width=0.16\textwidth,height=0.16\textwidth,clip]{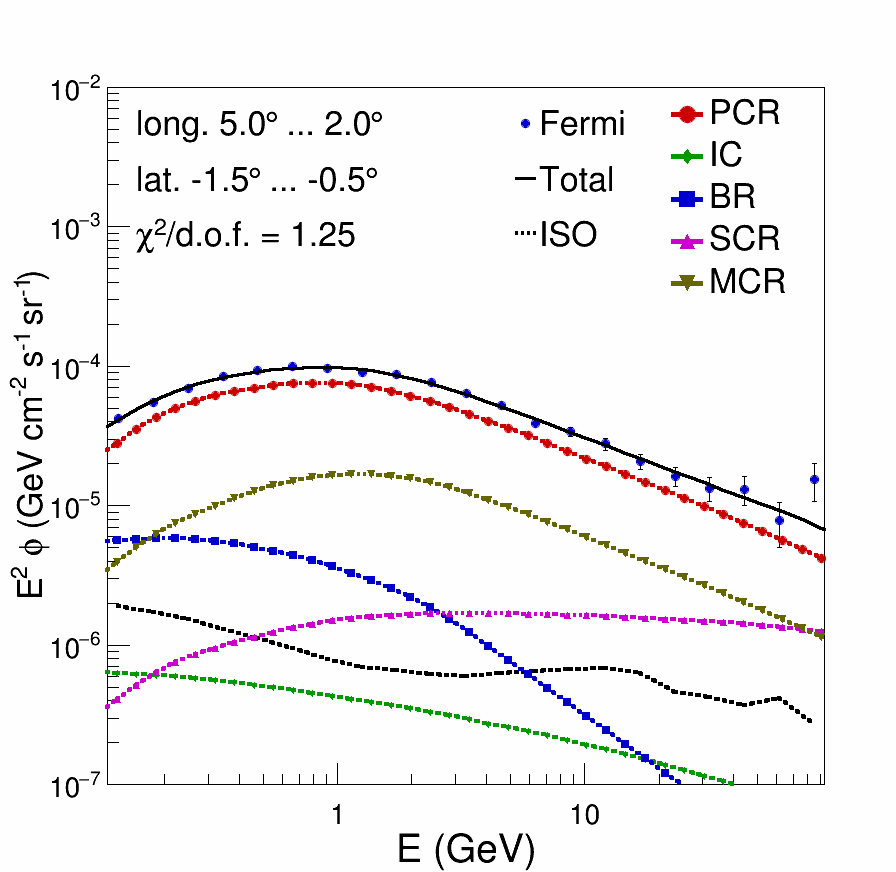}
	\includegraphics[width=0.16\textwidth,height=0.16\textwidth,clip]{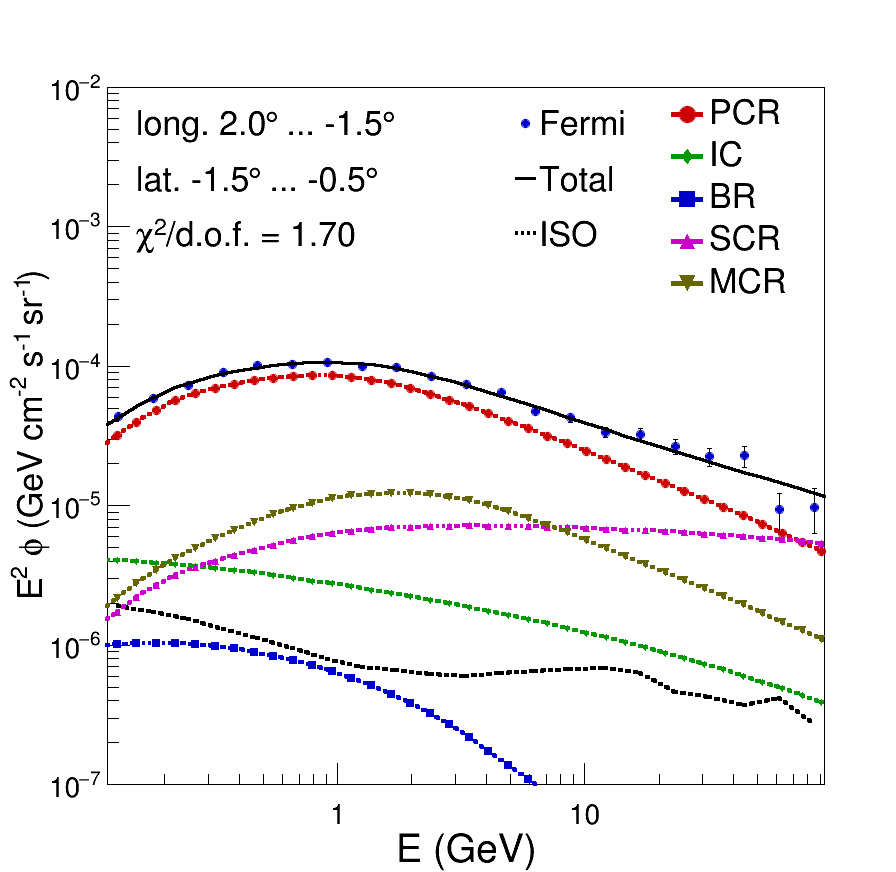}
	\includegraphics[width=0.16\textwidth,height=0.16\textwidth,clip]{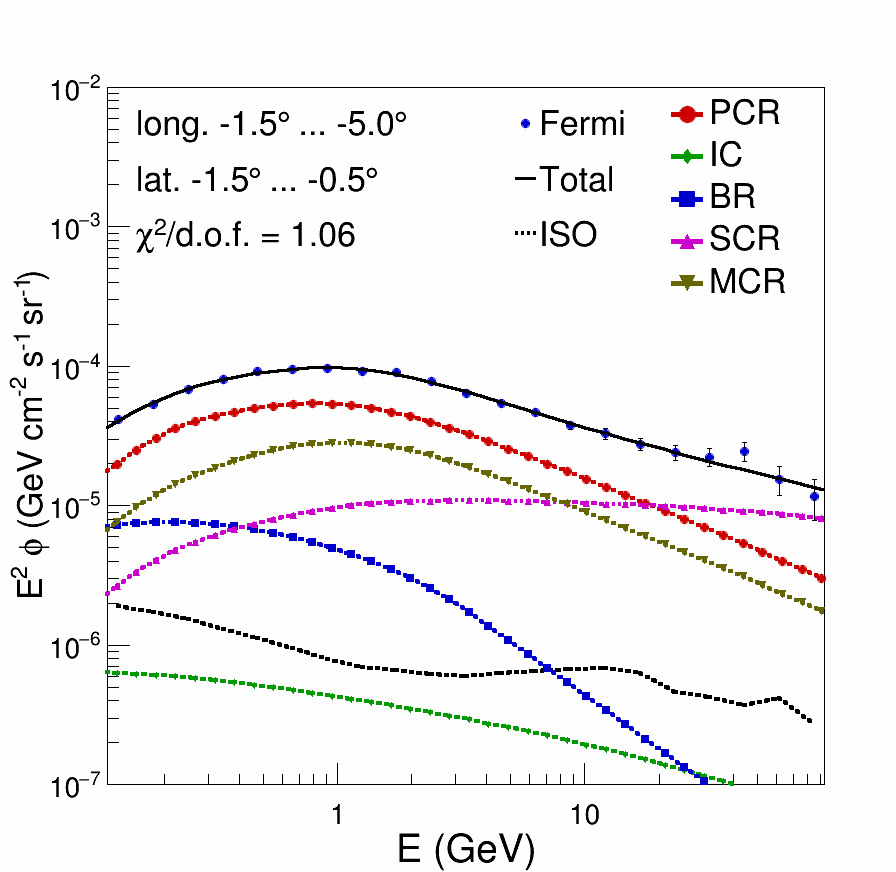}
	\includegraphics[width=0.16\textwidth,height=0.16\textwidth,clip]{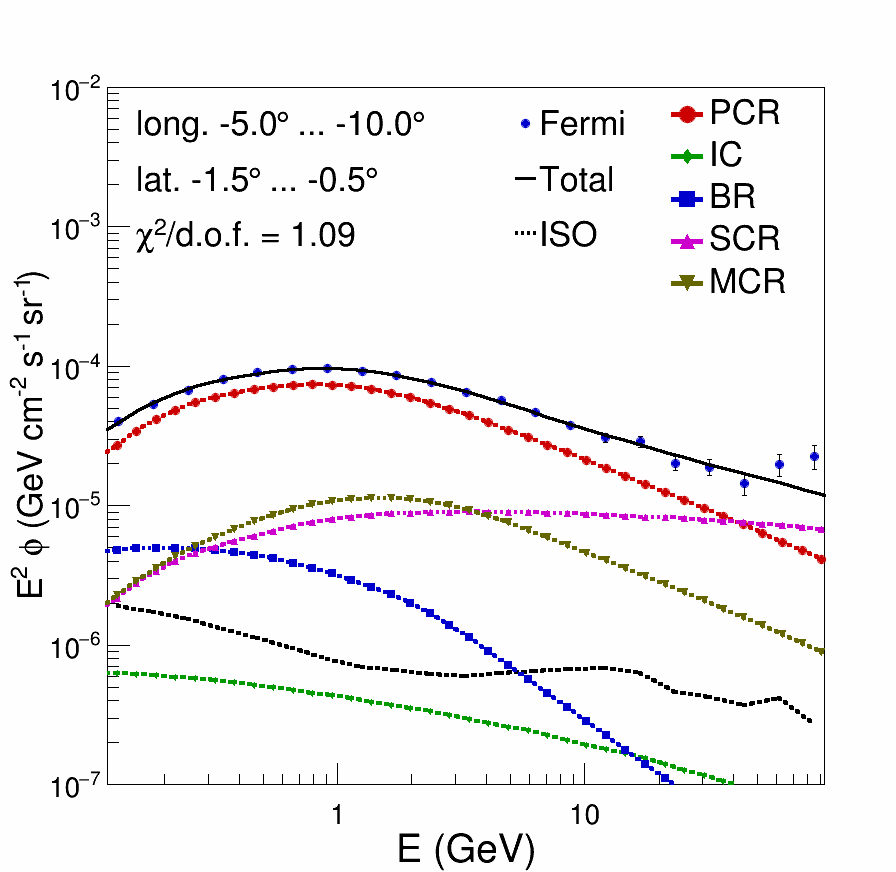}
	\includegraphics[width=0.16\textwidth,height=0.16\textwidth,clip]{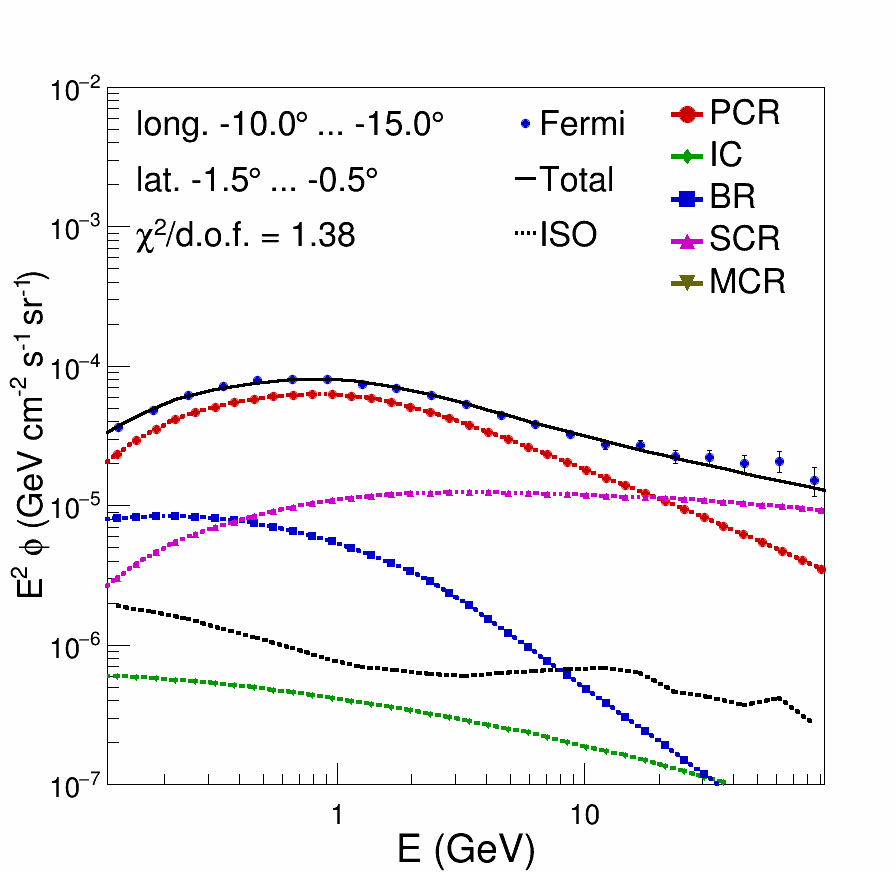}
	\includegraphics[width=0.16\textwidth,height=0.16\textwidth,clip]{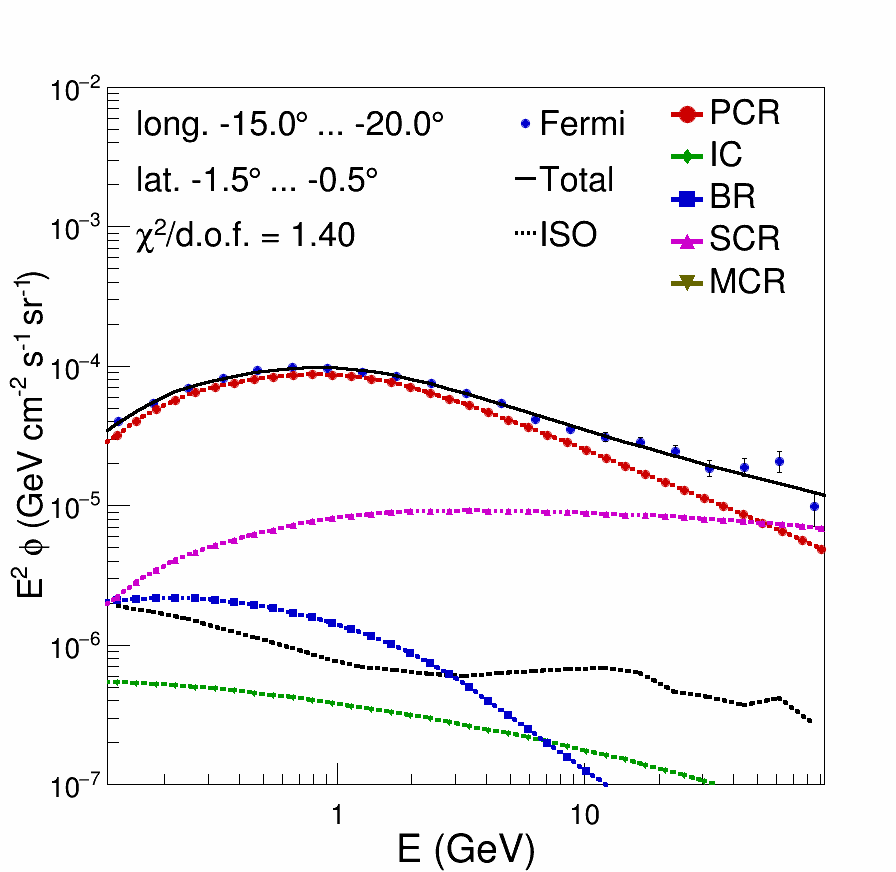}
	\includegraphics[width=0.16\textwidth,height=0.16\textwidth,clip]{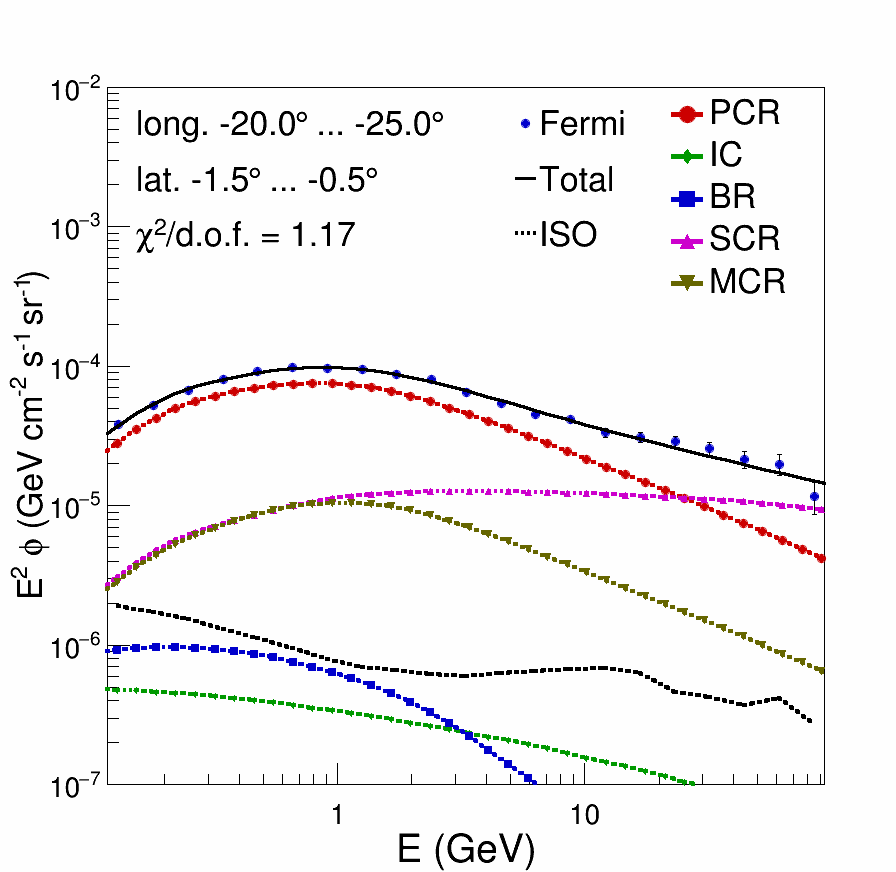}
	\includegraphics[width=0.16\textwidth,height=0.16\textwidth,clip]{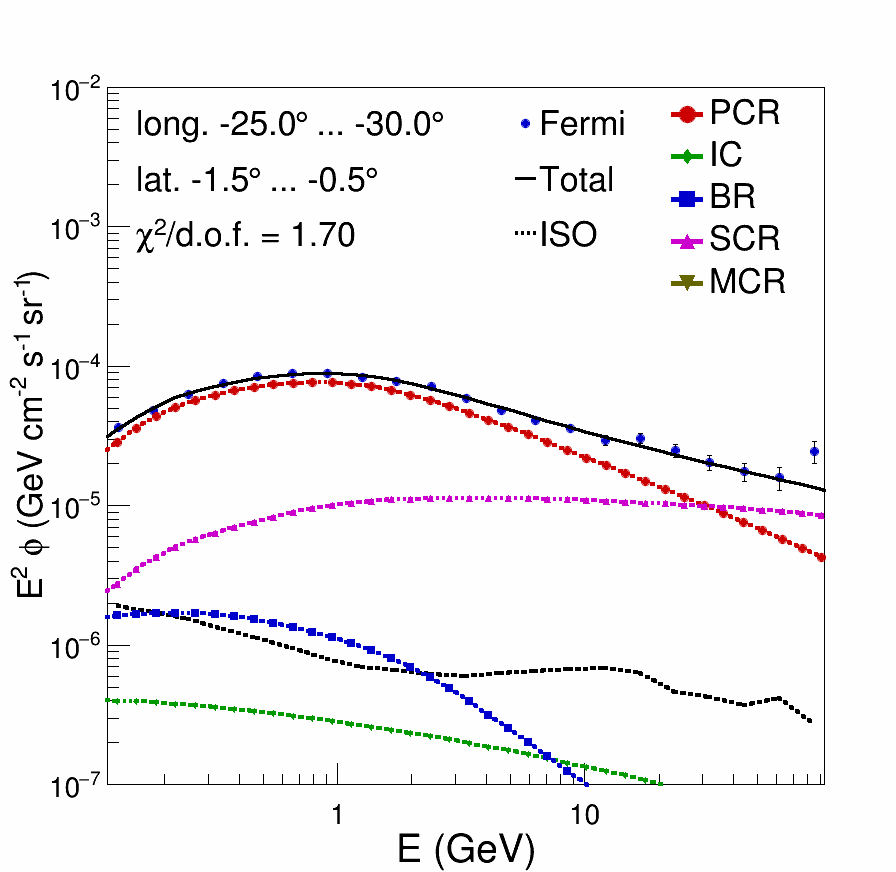}
	\includegraphics[width=0.16\textwidth,height=0.16\textwidth,clip]{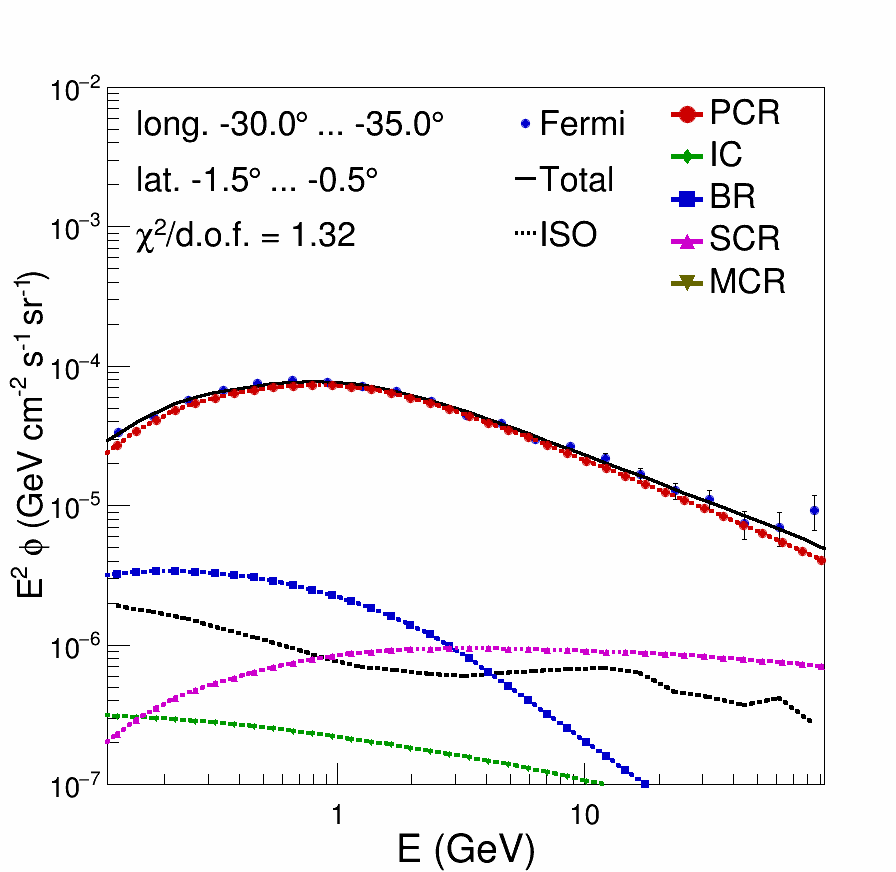}
	\includegraphics[width=0.16\textwidth,height=0.16\textwidth,clip]{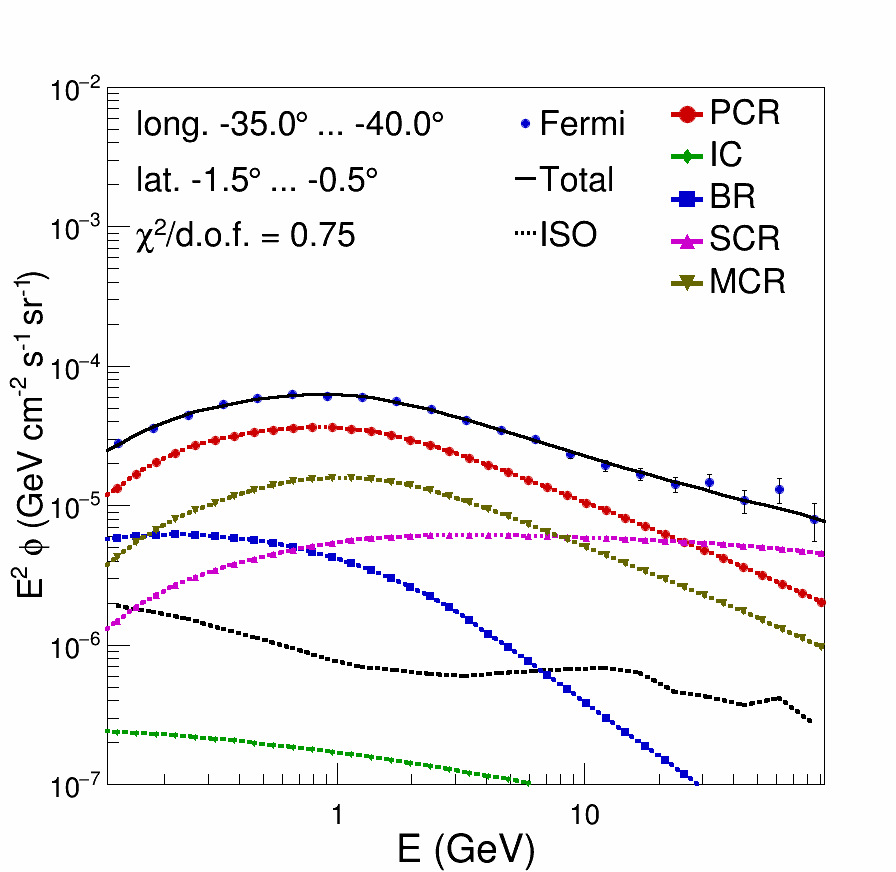}
	\includegraphics[width=0.16\textwidth,height=0.16\textwidth,clip]{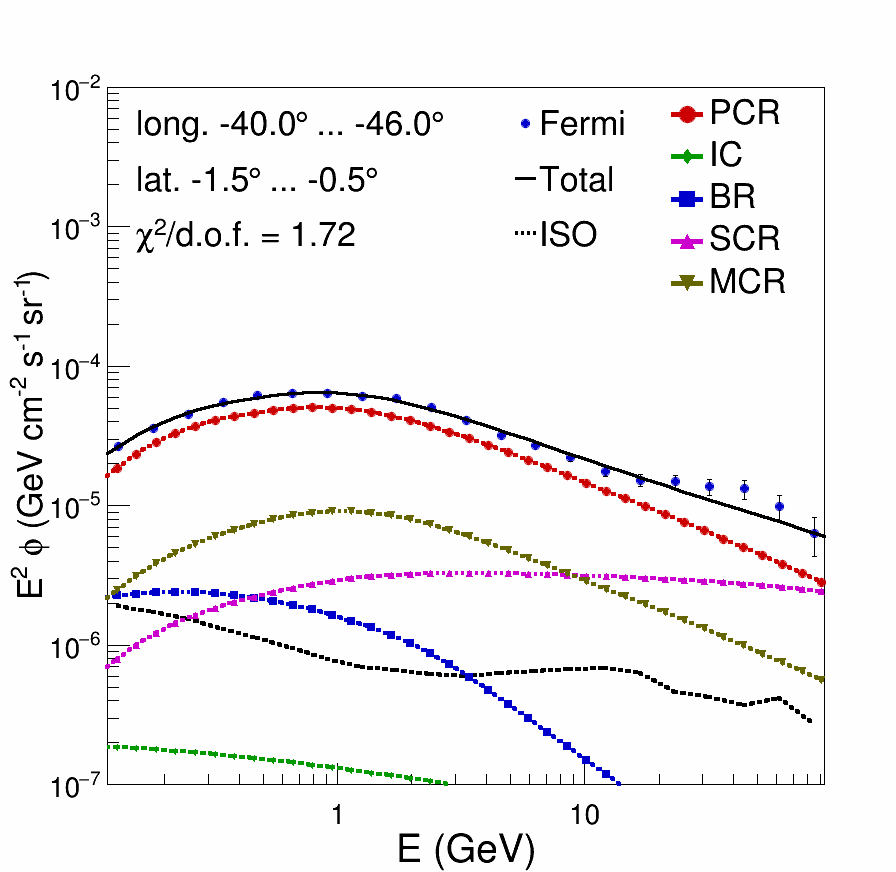}
	\includegraphics[width=0.16\textwidth,height=0.16\textwidth,clip]{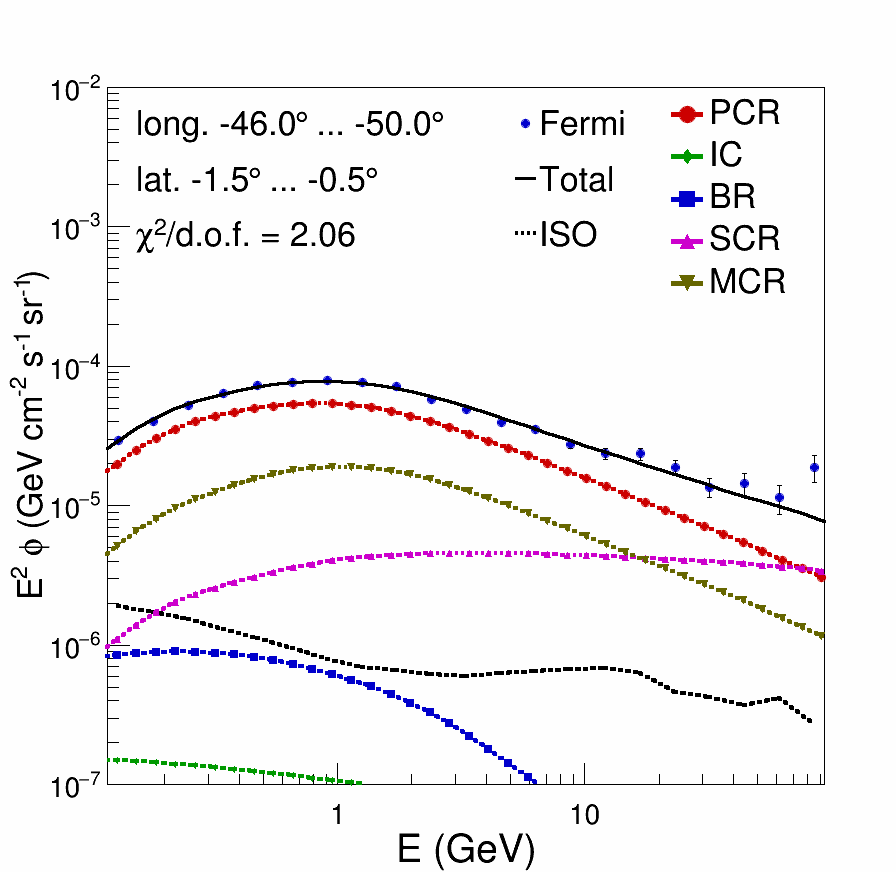}
	\includegraphics[width=0.16\textwidth,height=0.16\textwidth,clip]{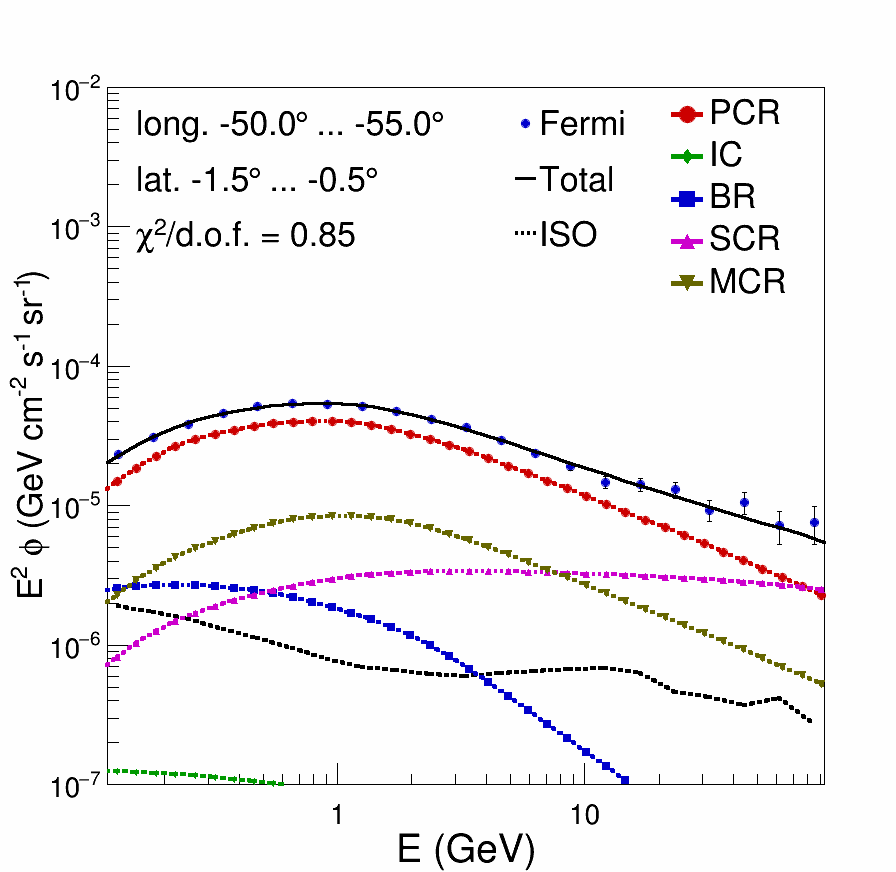}
	\includegraphics[width=0.16\textwidth,height=0.16\textwidth,clip]{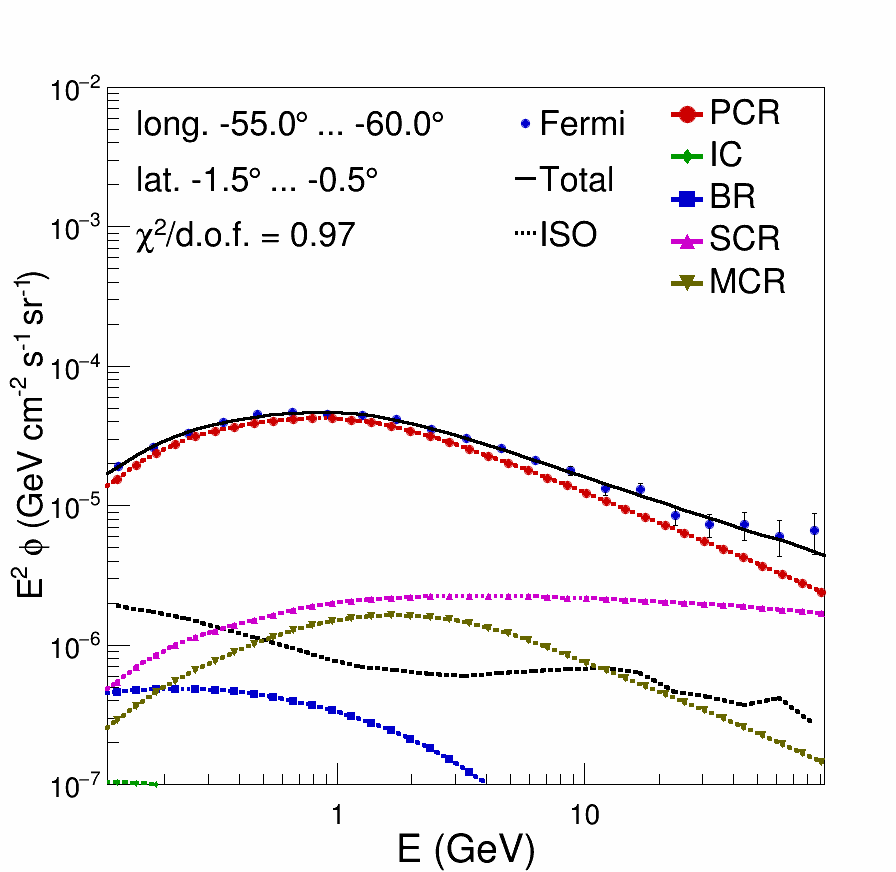}
	\includegraphics[width=0.16\textwidth,height=0.16\textwidth,clip]{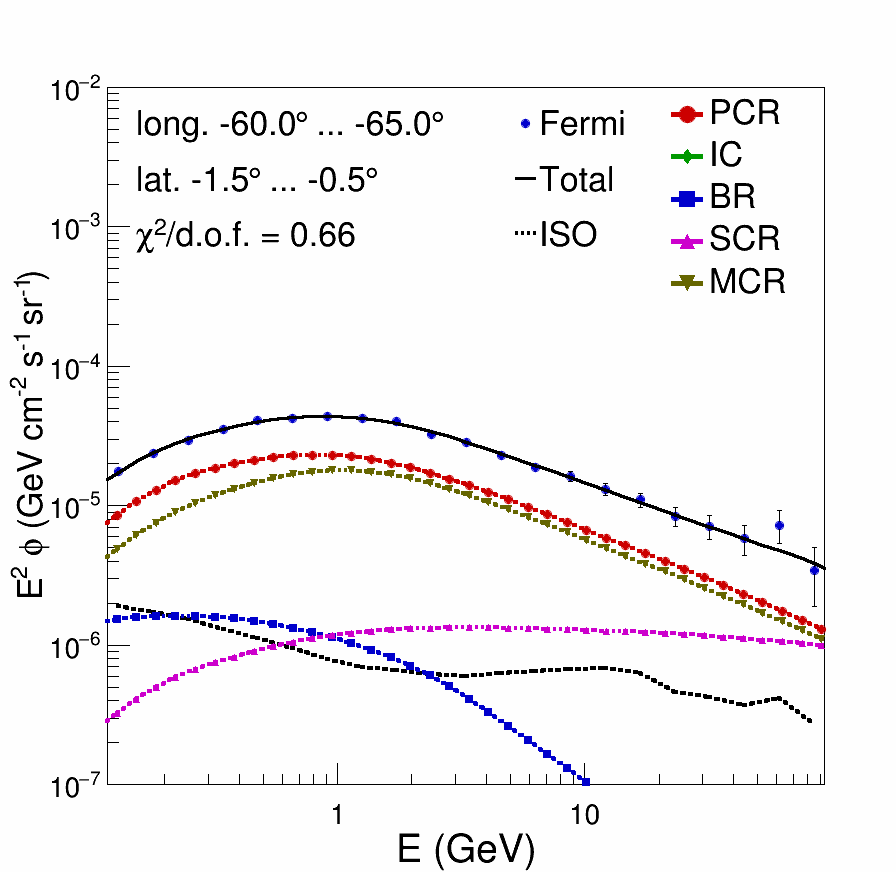}
	\includegraphics[width=0.16\textwidth,height=0.16\textwidth,clip]{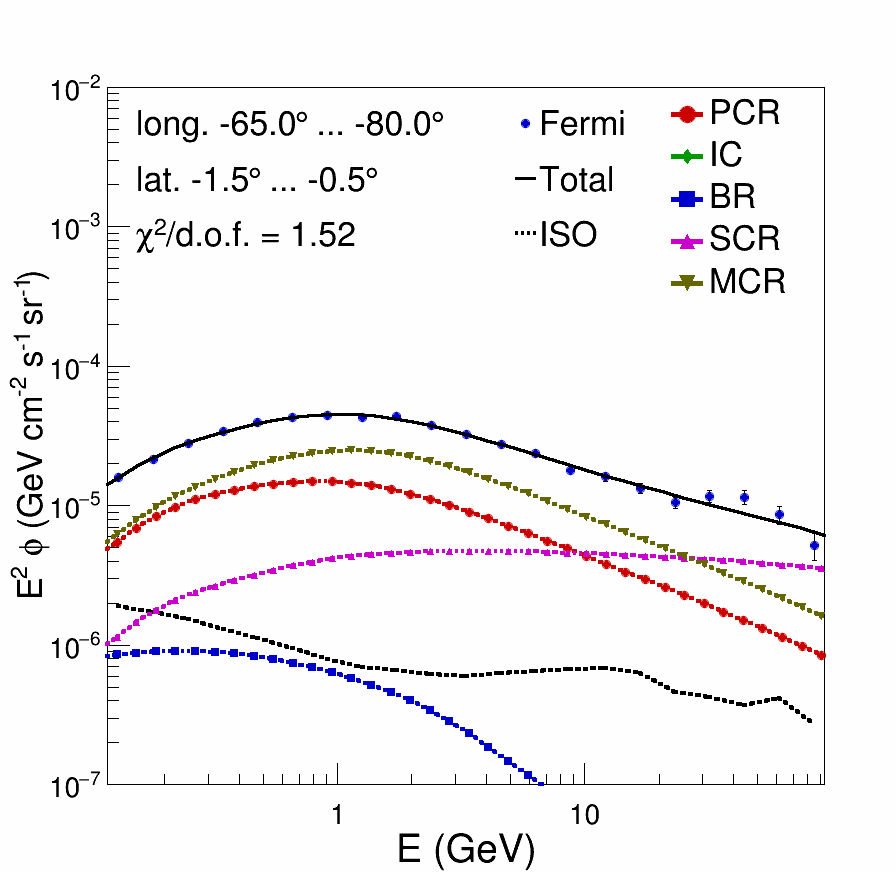}
	\includegraphics[width=0.16\textwidth,height=0.16\textwidth,clip]{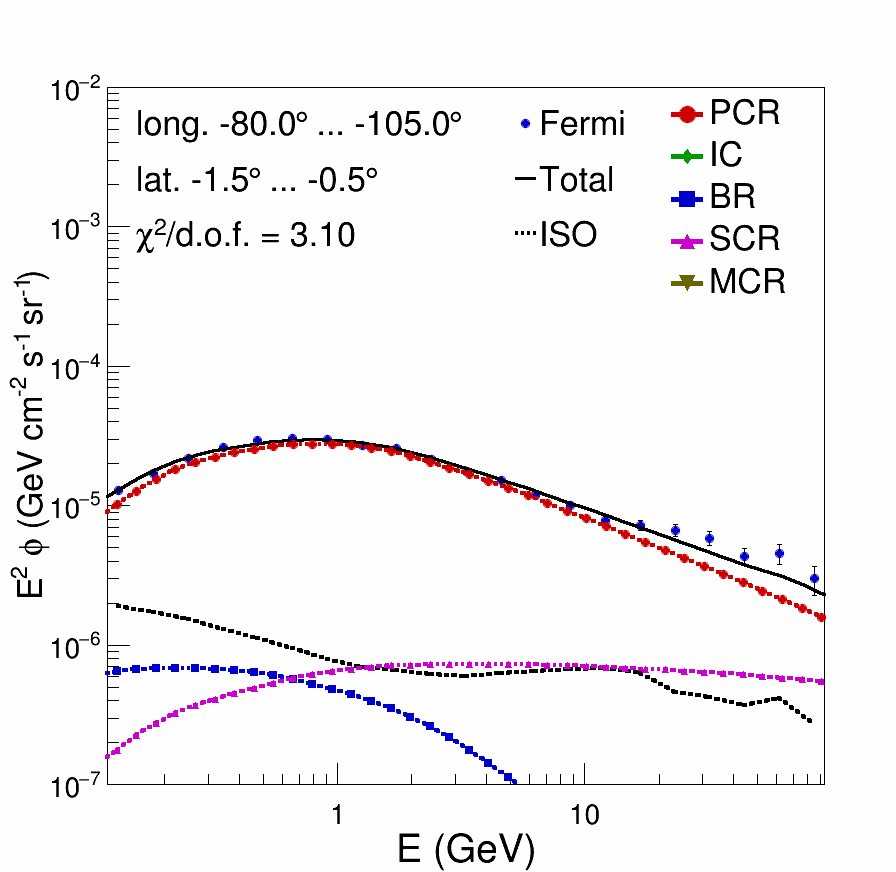}
	\includegraphics[width=0.16\textwidth,height=0.16\textwidth,clip]{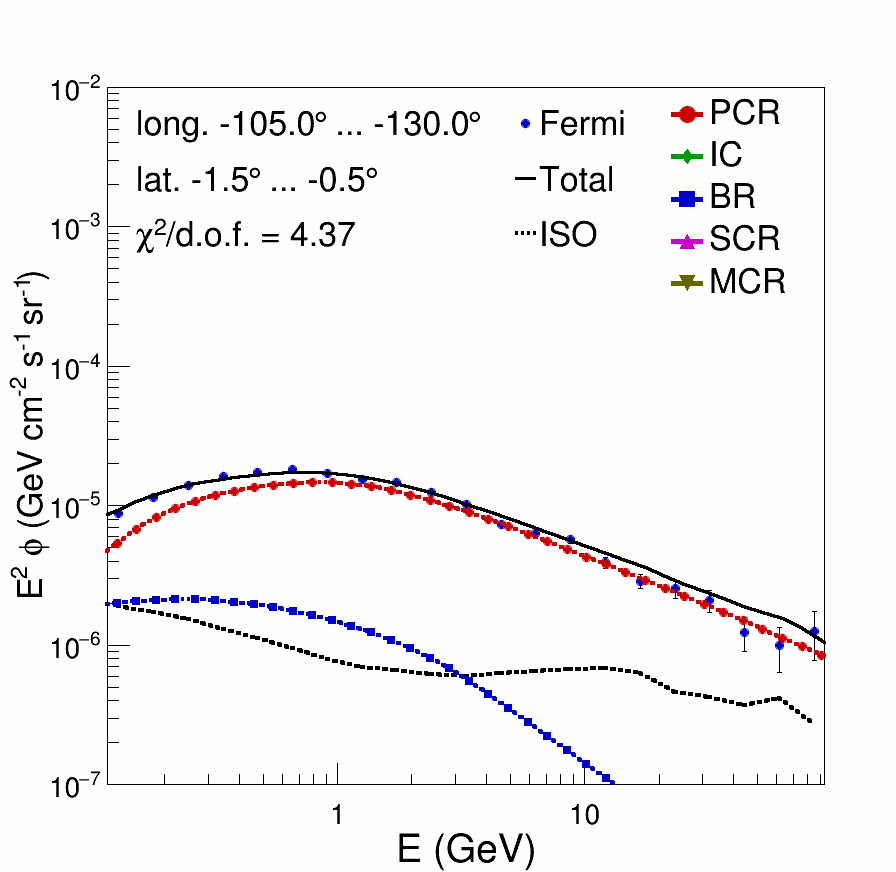}
	\includegraphics[width=0.16\textwidth,height=0.16\textwidth,clip]{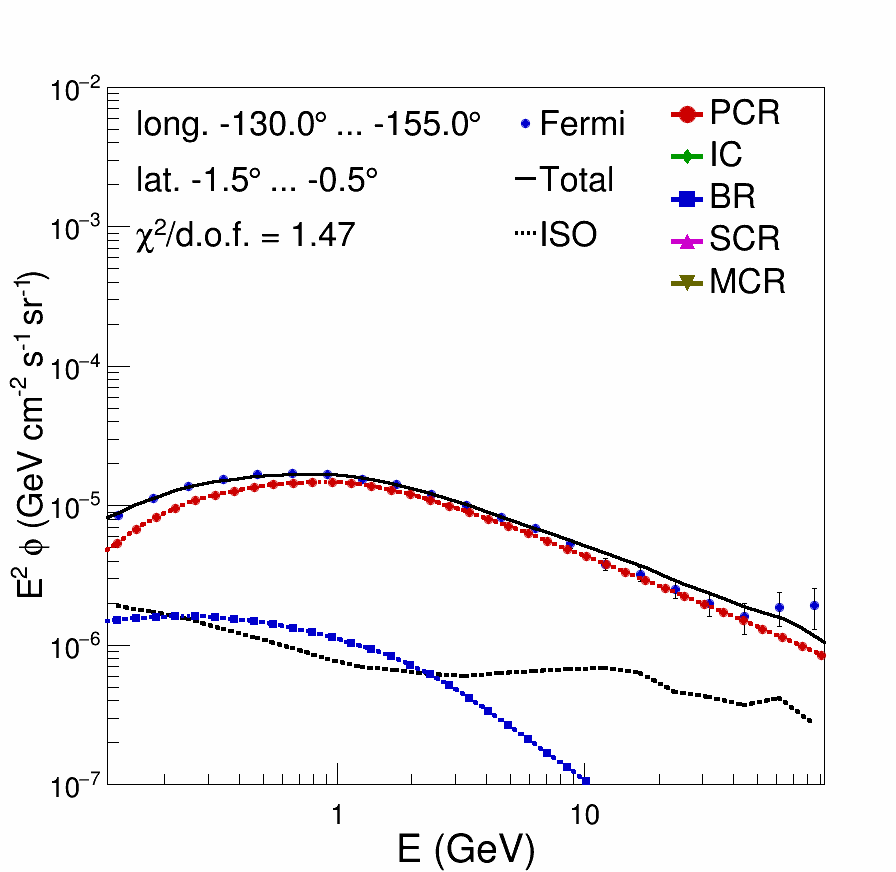}
	\includegraphics[width=0.16\textwidth,height=0.16\textwidth,clip]{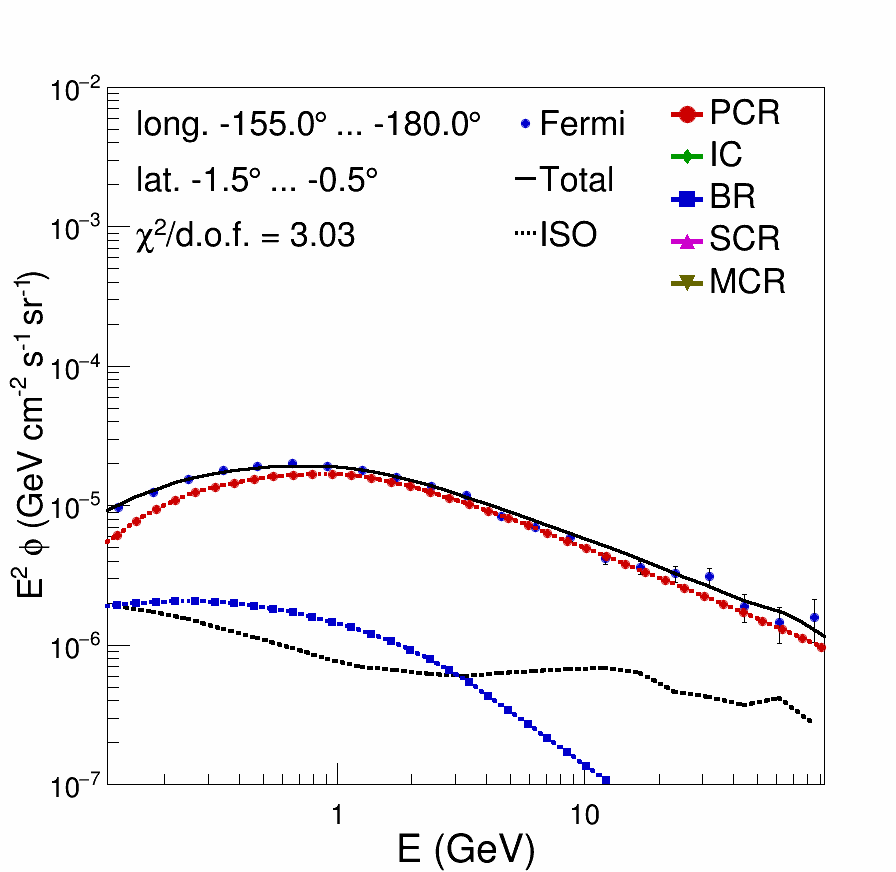}
\caption[]{Template fits for latitudes  with $-1.5^\circ<b<-0.5^\circ$ and longitudes decreasing from 180$^\circ$ to -180$^\circ$.} \label{F22}
\end{figure}
\begin{figure}
\centering
\includegraphics[width=0.16\textwidth,height=0.16\textwidth,clip]{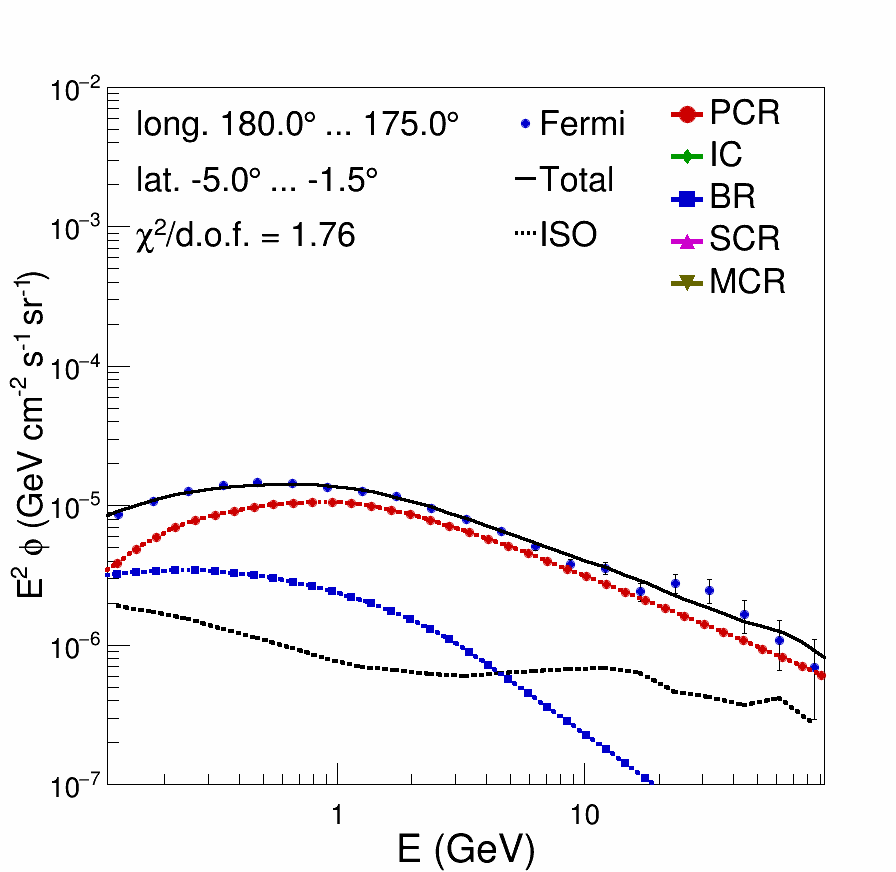}
\includegraphics[width=0.16\textwidth,height=0.16\textwidth,clip]{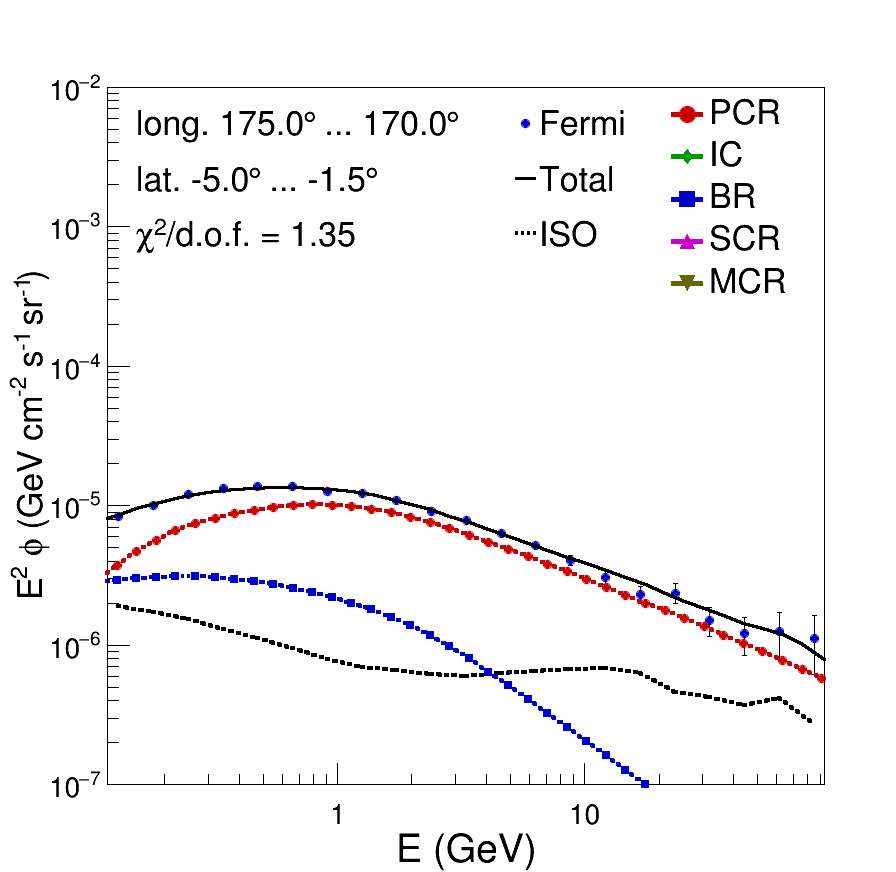}
\includegraphics[width=0.16\textwidth,height=0.16\textwidth,clip]{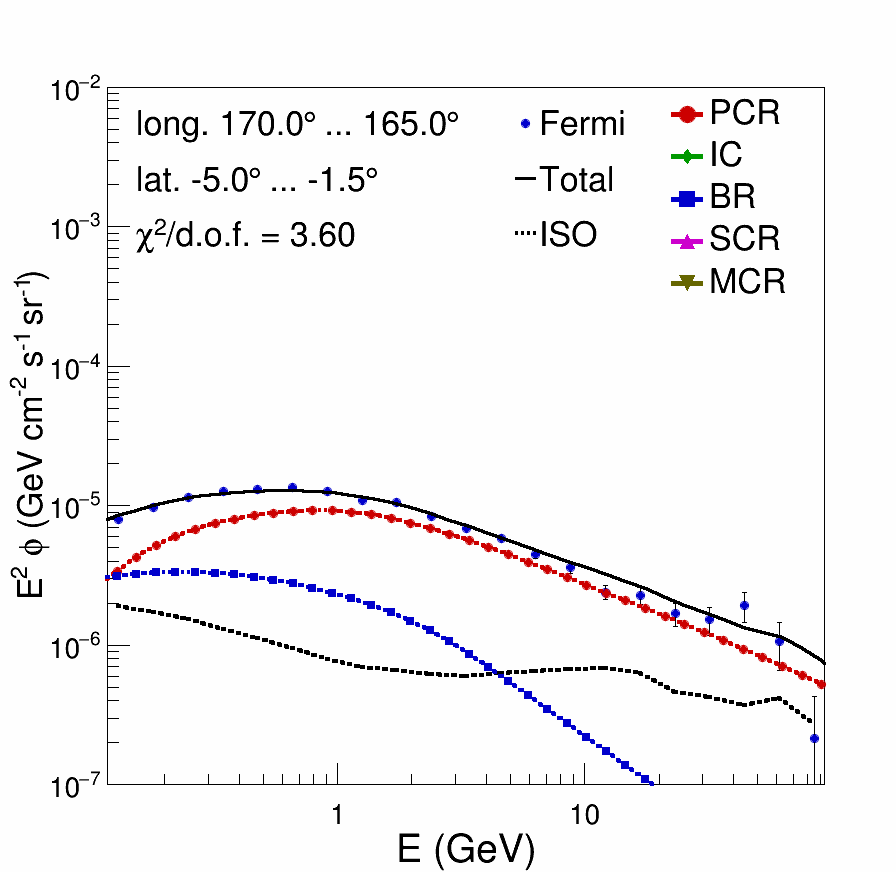}
\includegraphics[width=0.16\textwidth,height=0.16\textwidth,clip]{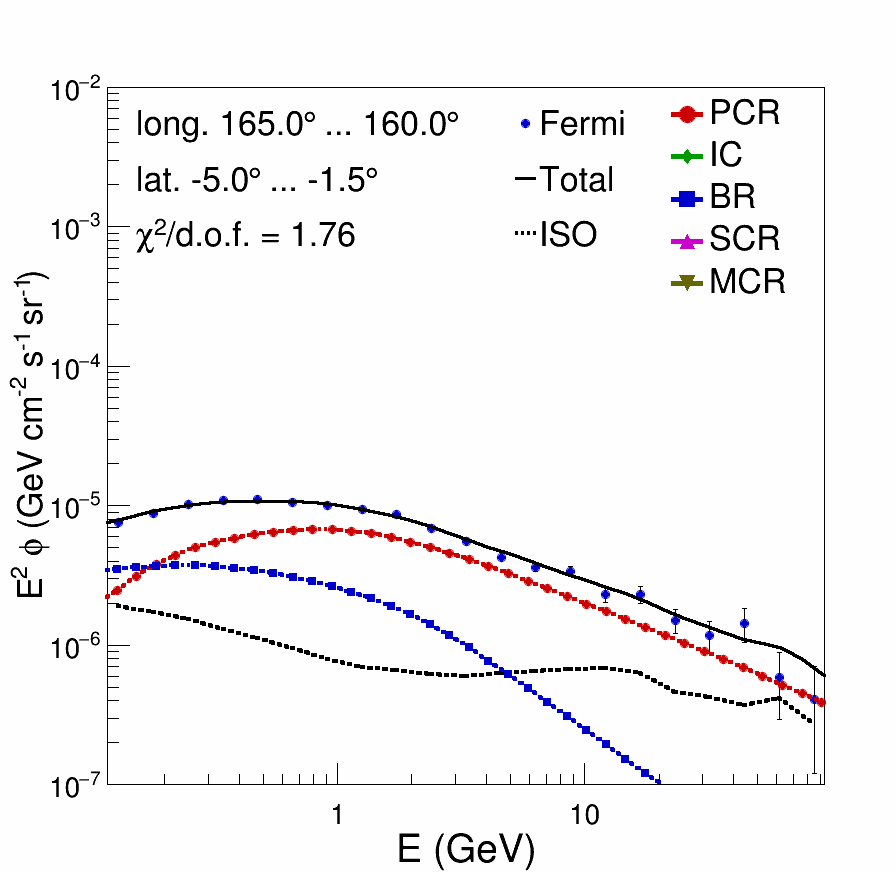}
\includegraphics[width=0.16\textwidth,height=0.16\textwidth,clip]{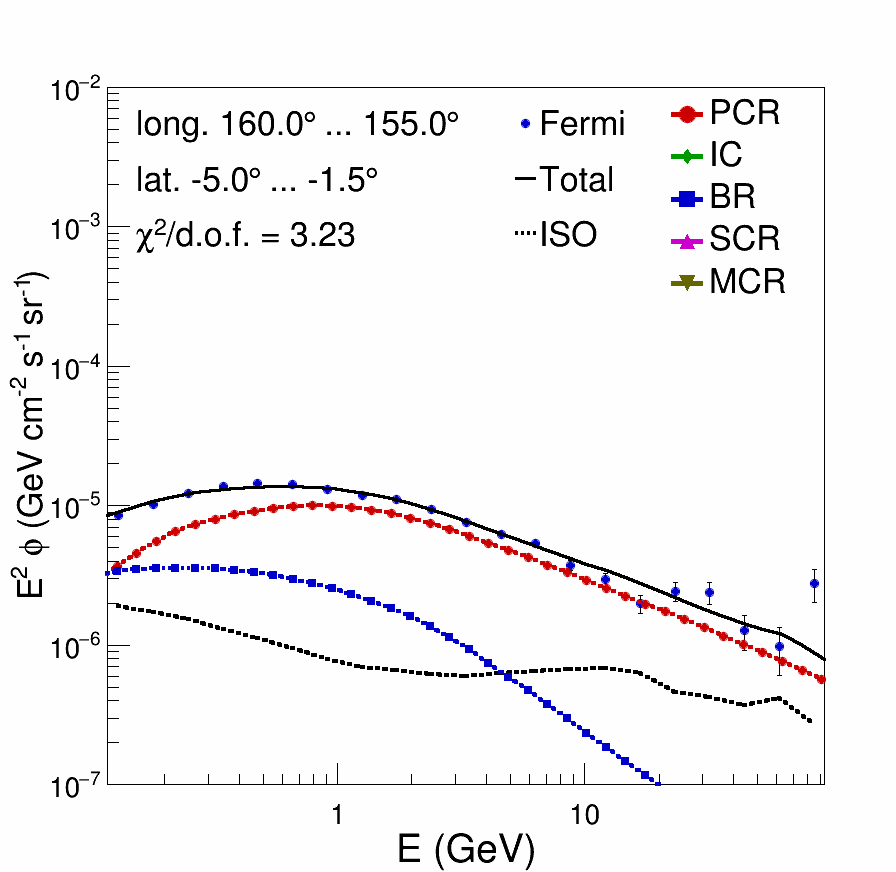}
\includegraphics[width=0.16\textwidth,height=0.16\textwidth,clip]{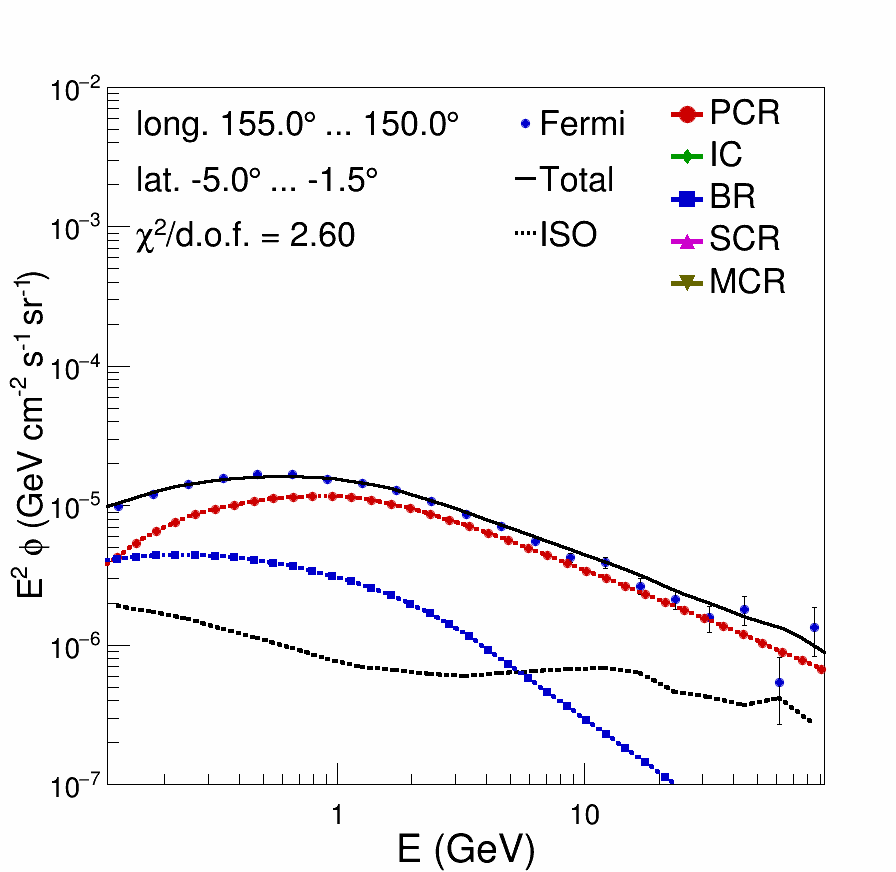}
\includegraphics[width=0.16\textwidth,height=0.16\textwidth,clip]{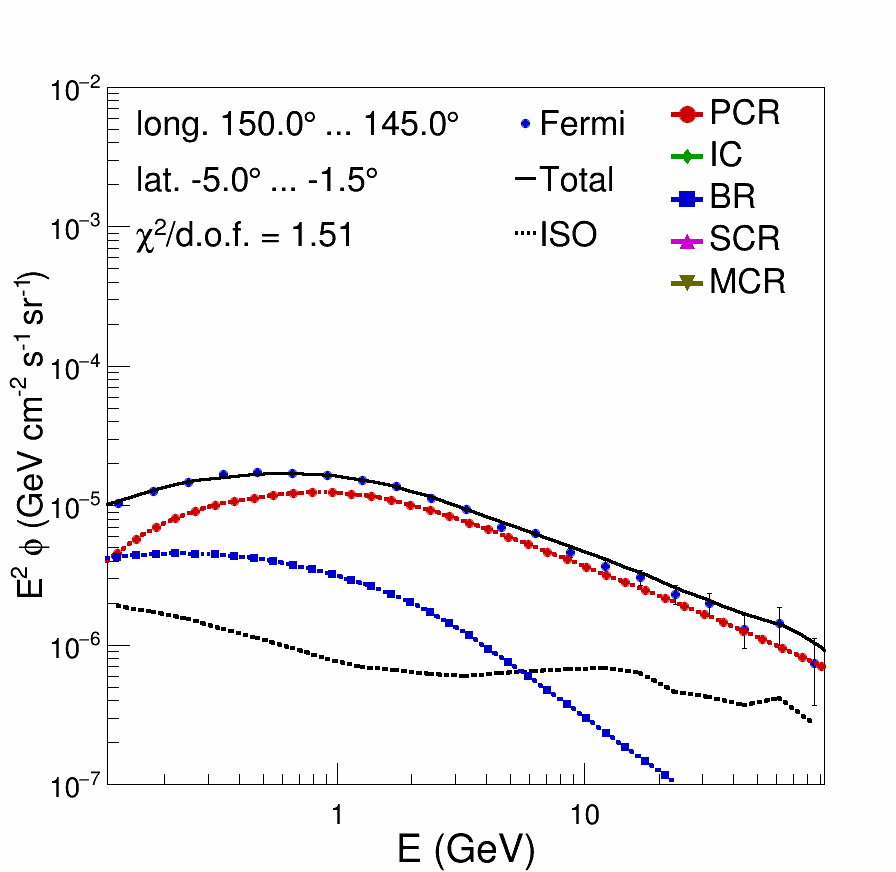}
\includegraphics[width=0.16\textwidth,height=0.16\textwidth,clip]{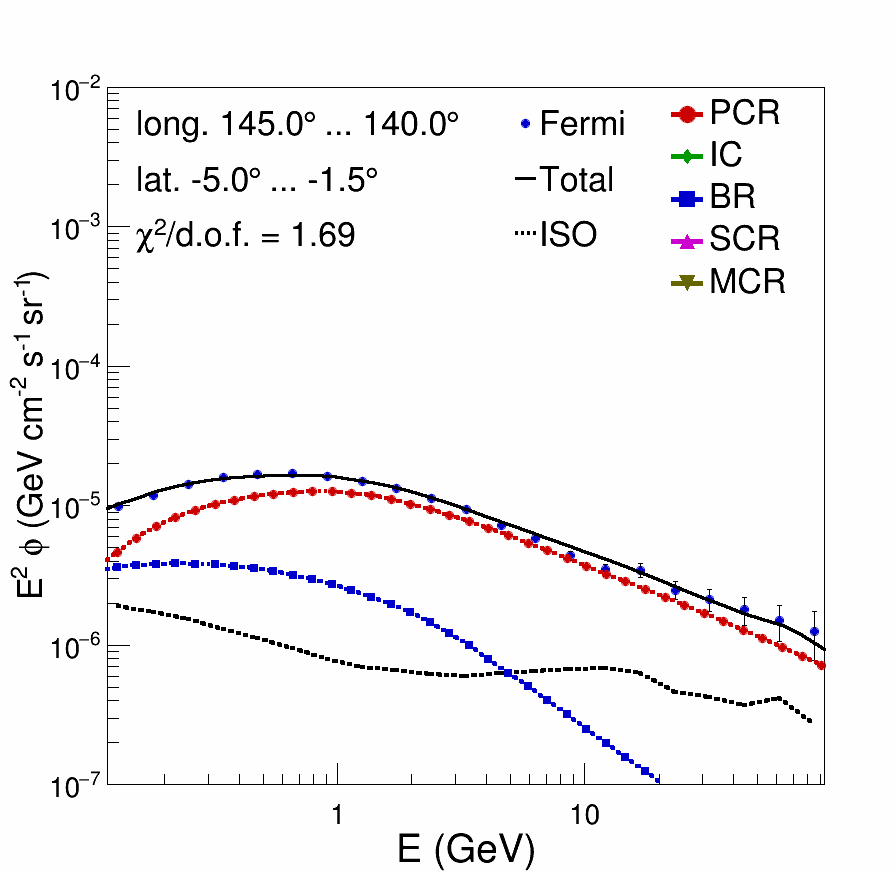}
\includegraphics[width=0.16\textwidth,height=0.16\textwidth,clip]{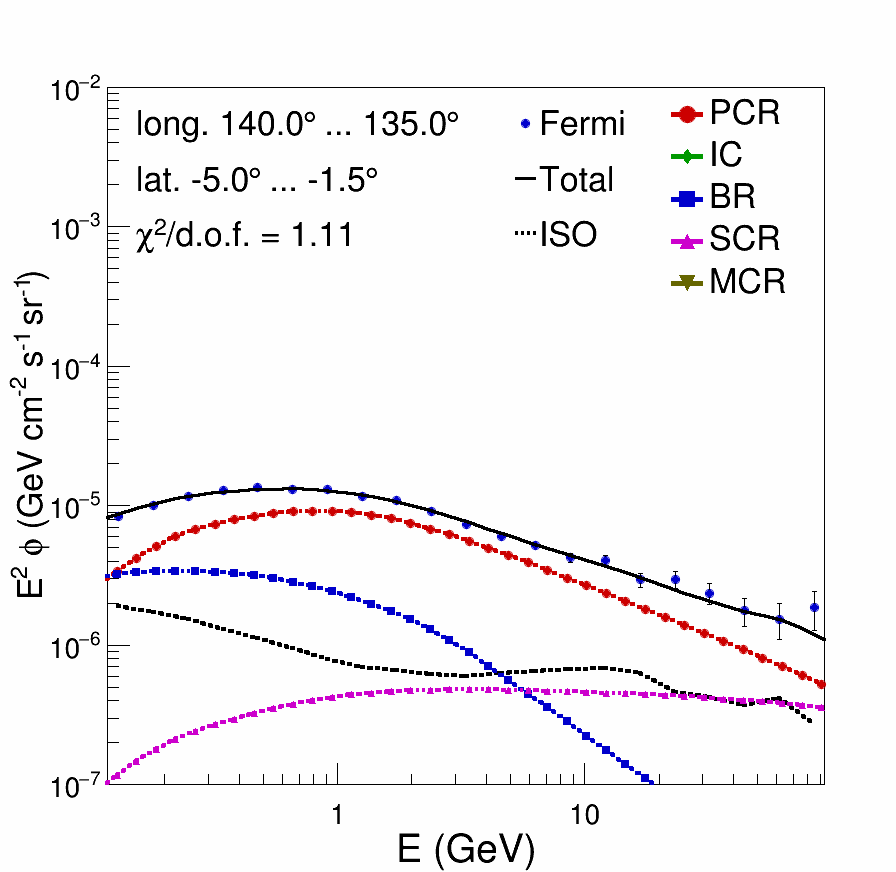}
\includegraphics[width=0.16\textwidth,height=0.16\textwidth,clip]{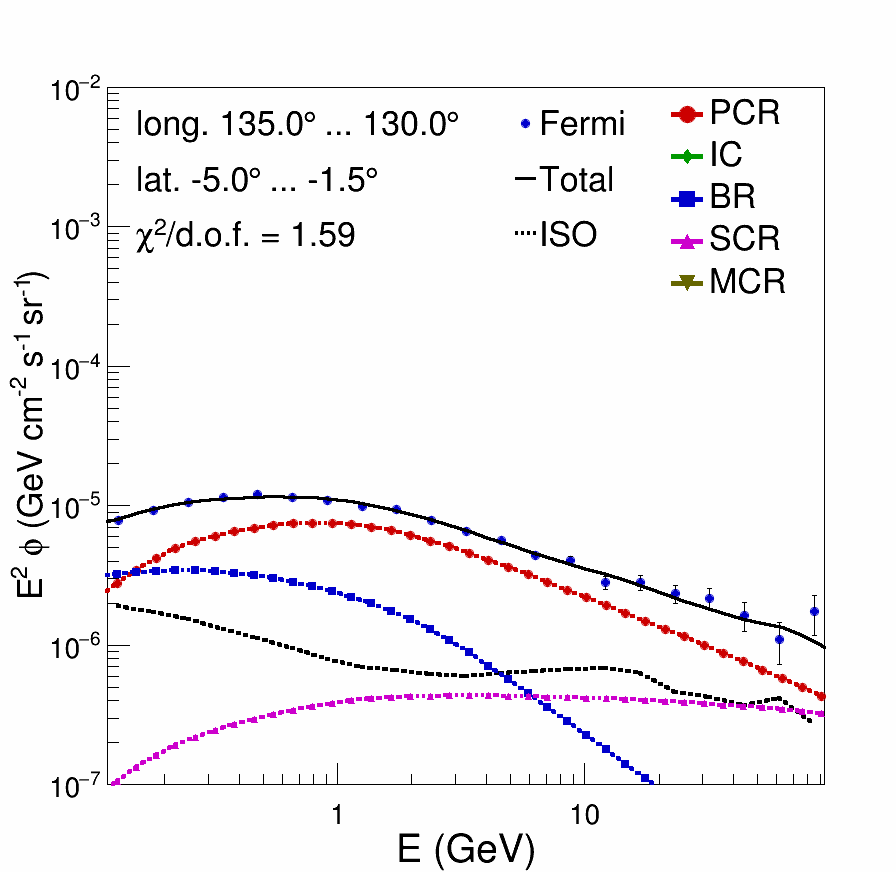}
\includegraphics[width=0.16\textwidth,height=0.16\textwidth,clip]{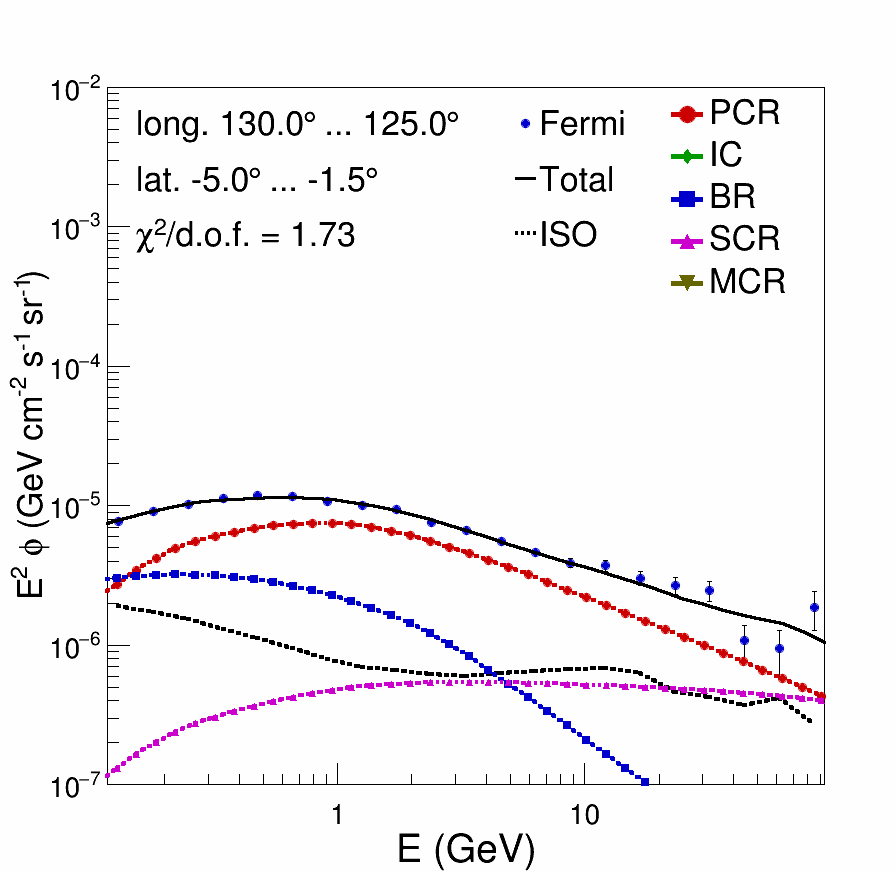}
\includegraphics[width=0.16\textwidth,height=0.16\textwidth,clip]{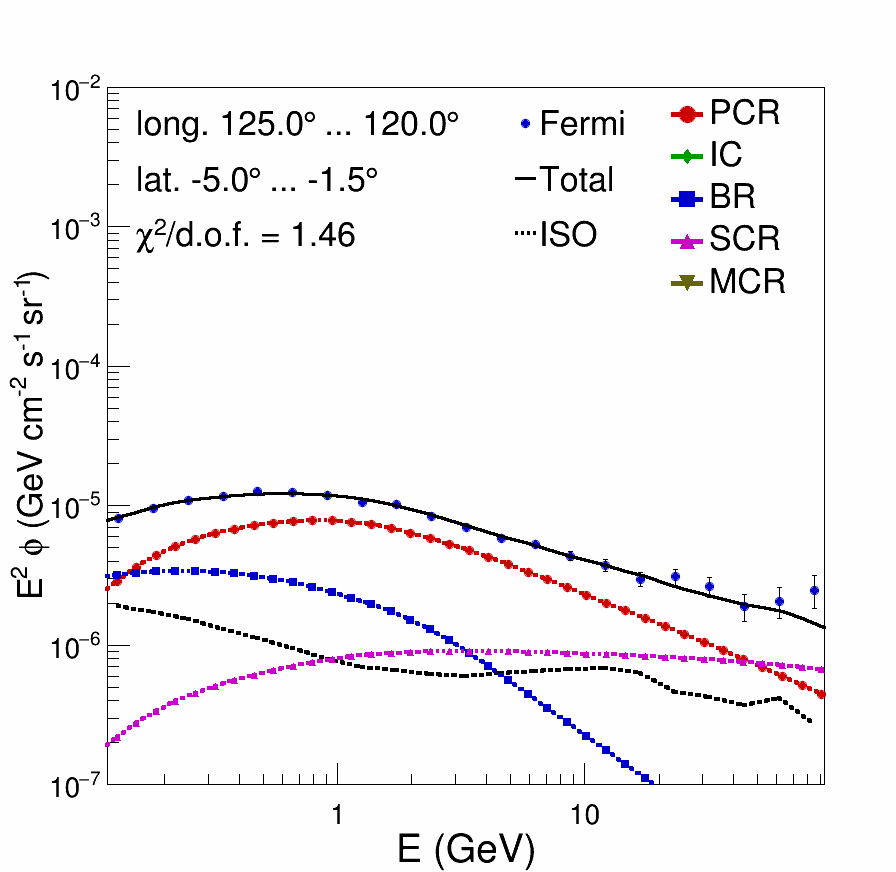}
\includegraphics[width=0.16\textwidth,height=0.16\textwidth,clip]{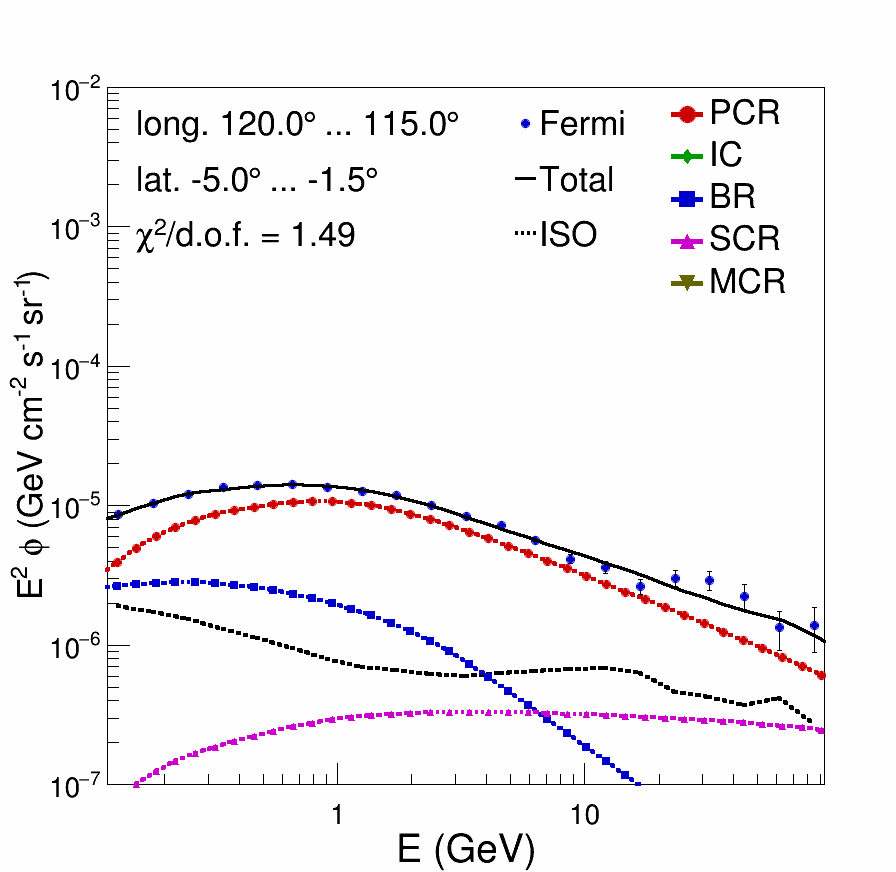}
\includegraphics[width=0.16\textwidth,height=0.16\textwidth,clip]{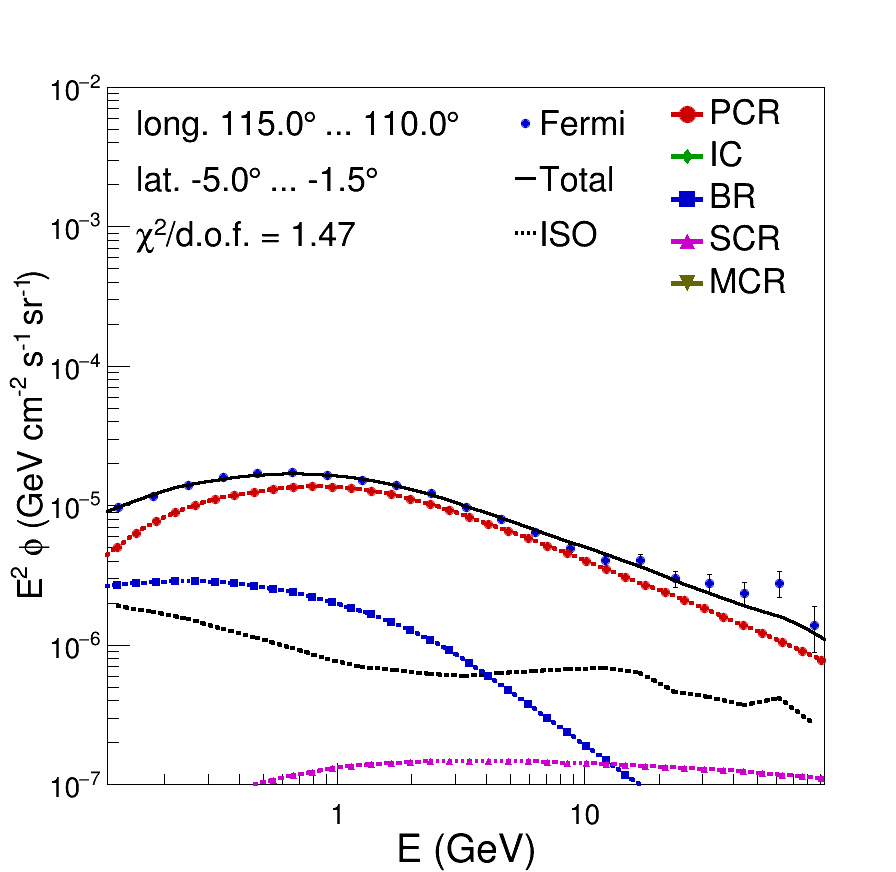}
\includegraphics[width=0.16\textwidth,height=0.16\textwidth,clip]{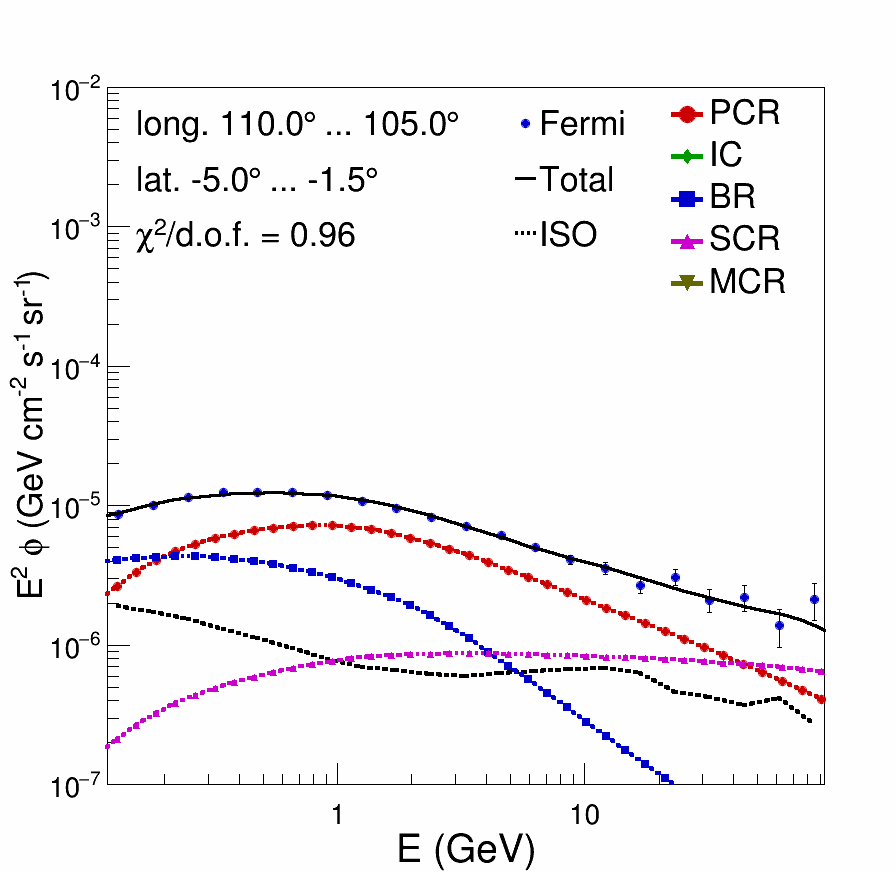}
\includegraphics[width=0.16\textwidth,height=0.16\textwidth,clip]{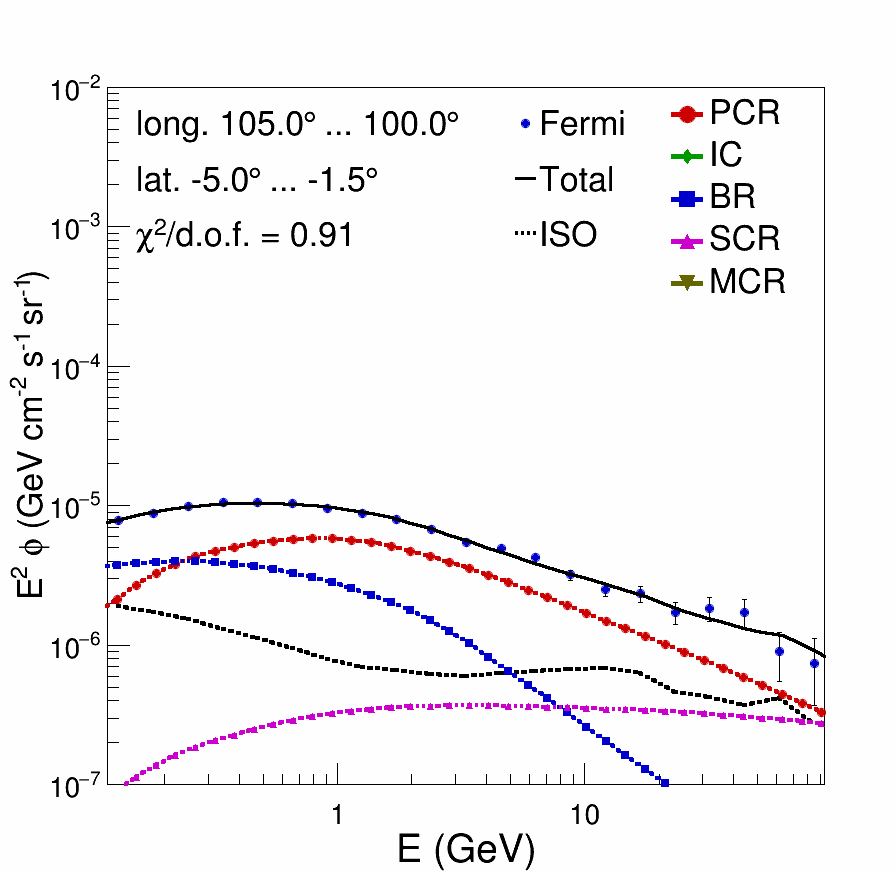}
\includegraphics[width=0.16\textwidth,height=0.16\textwidth,clip]{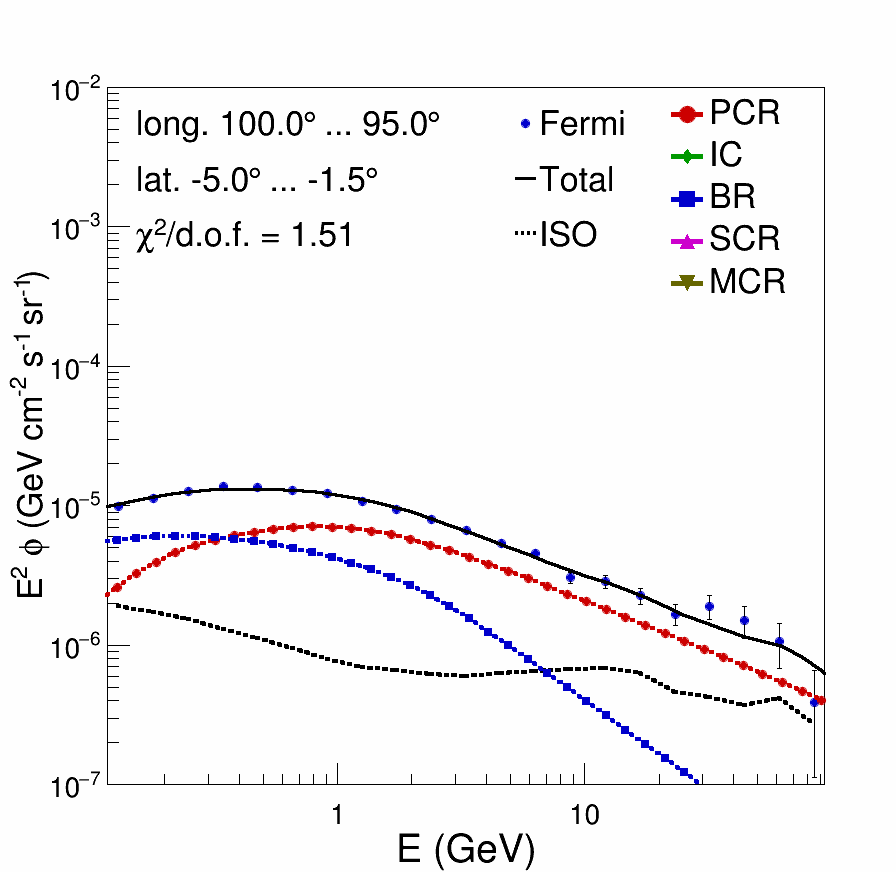}
\includegraphics[width=0.16\textwidth,height=0.16\textwidth,clip]{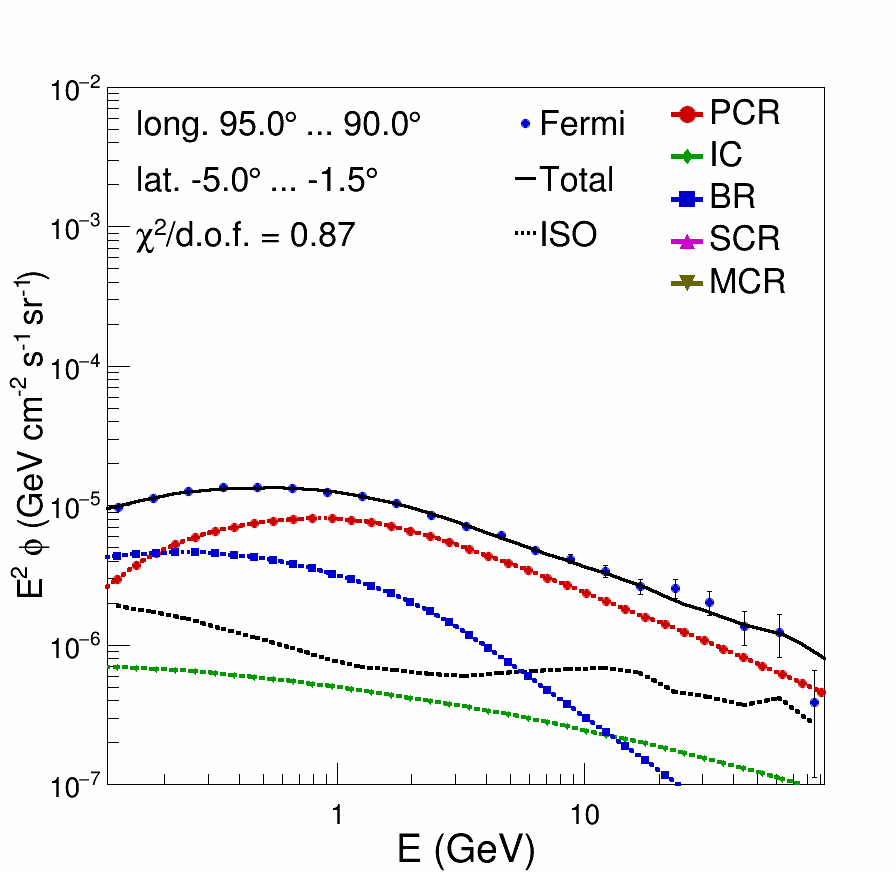}
\includegraphics[width=0.16\textwidth,height=0.16\textwidth,clip]{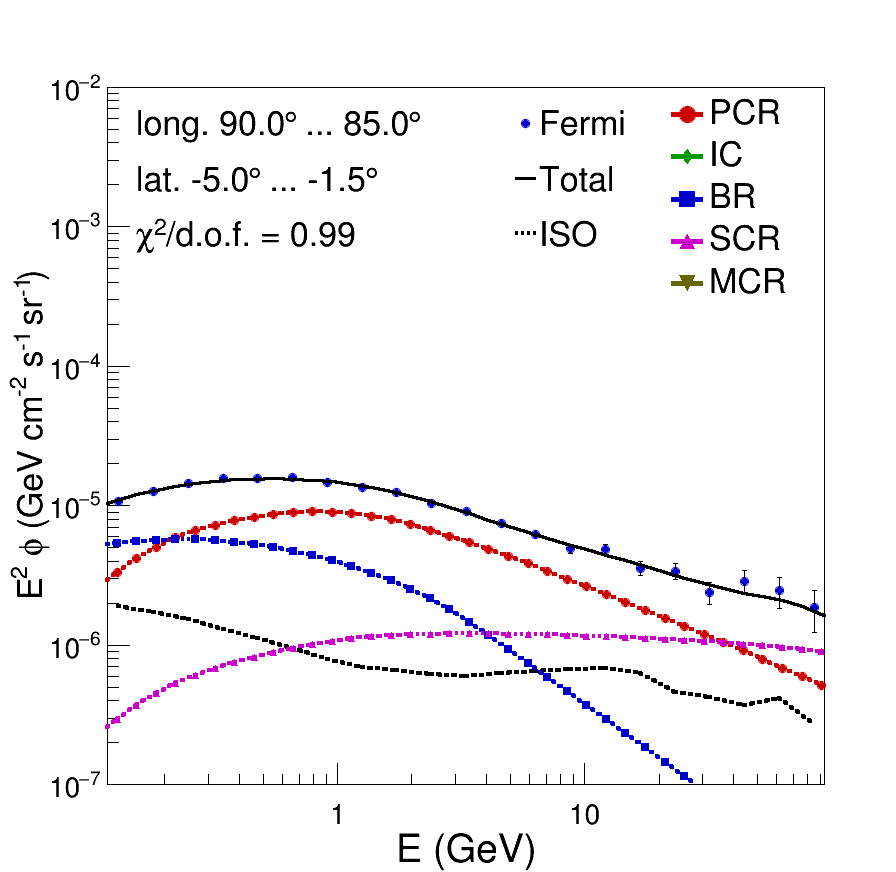}
\includegraphics[width=0.16\textwidth,height=0.16\textwidth,clip]{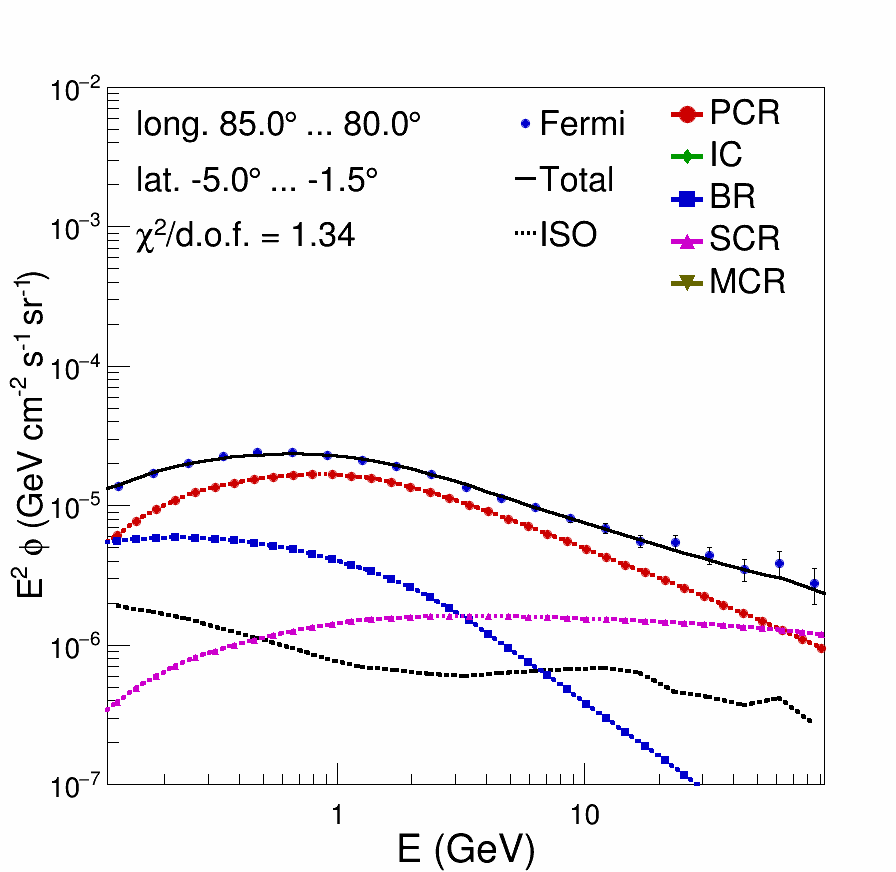}
\includegraphics[width=0.16\textwidth,height=0.16\textwidth,clip]{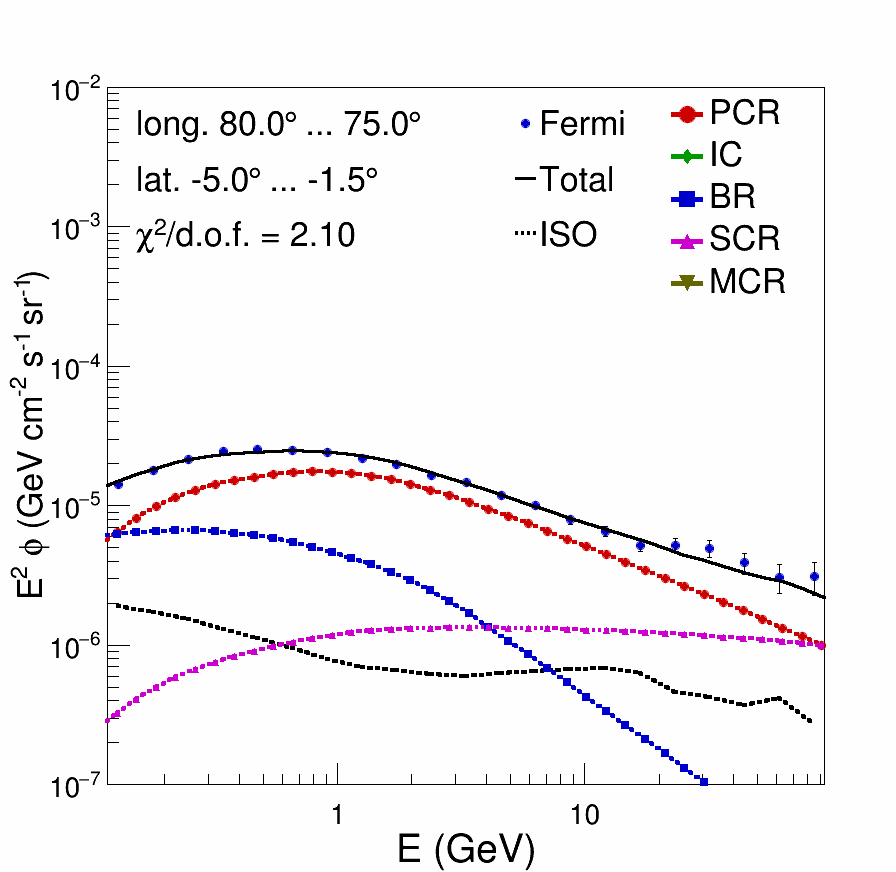}
\includegraphics[width=0.16\textwidth,height=0.16\textwidth,clip]{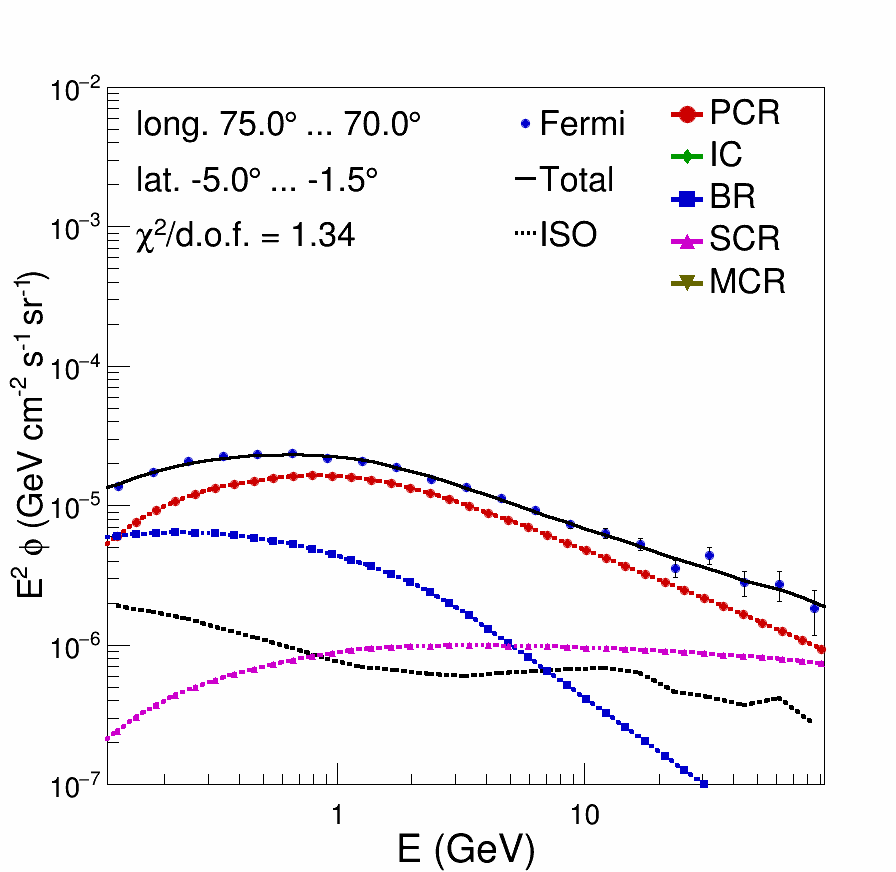}
\includegraphics[width=0.16\textwidth,height=0.16\textwidth,clip]{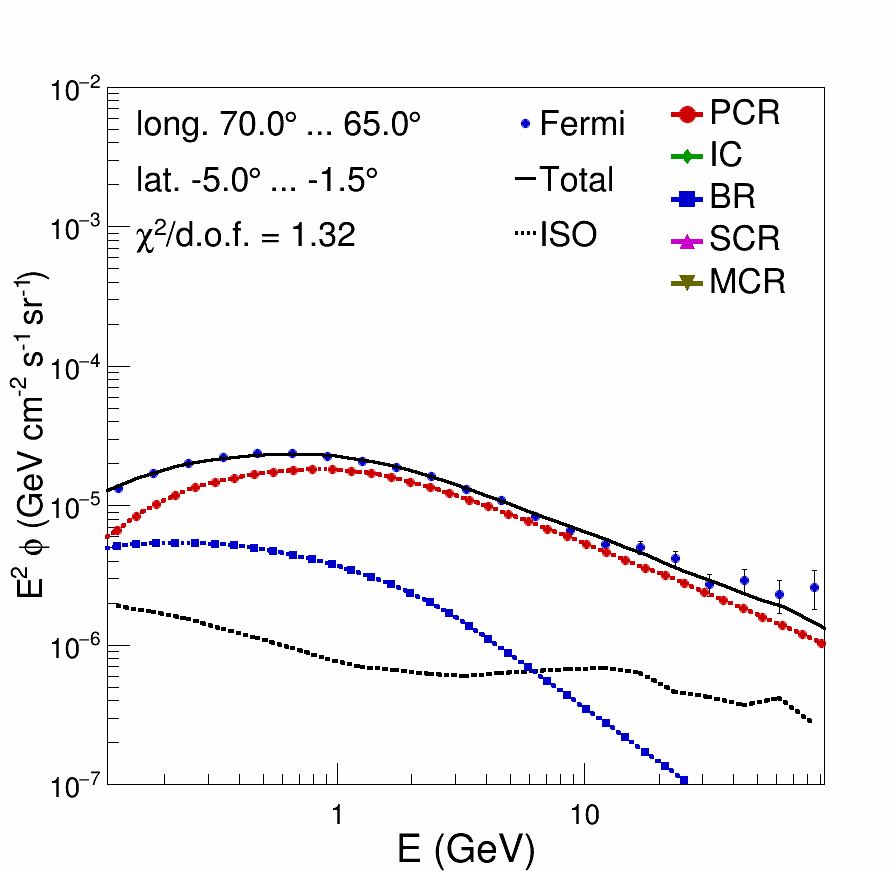}
\includegraphics[width=0.16\textwidth,height=0.16\textwidth,clip]{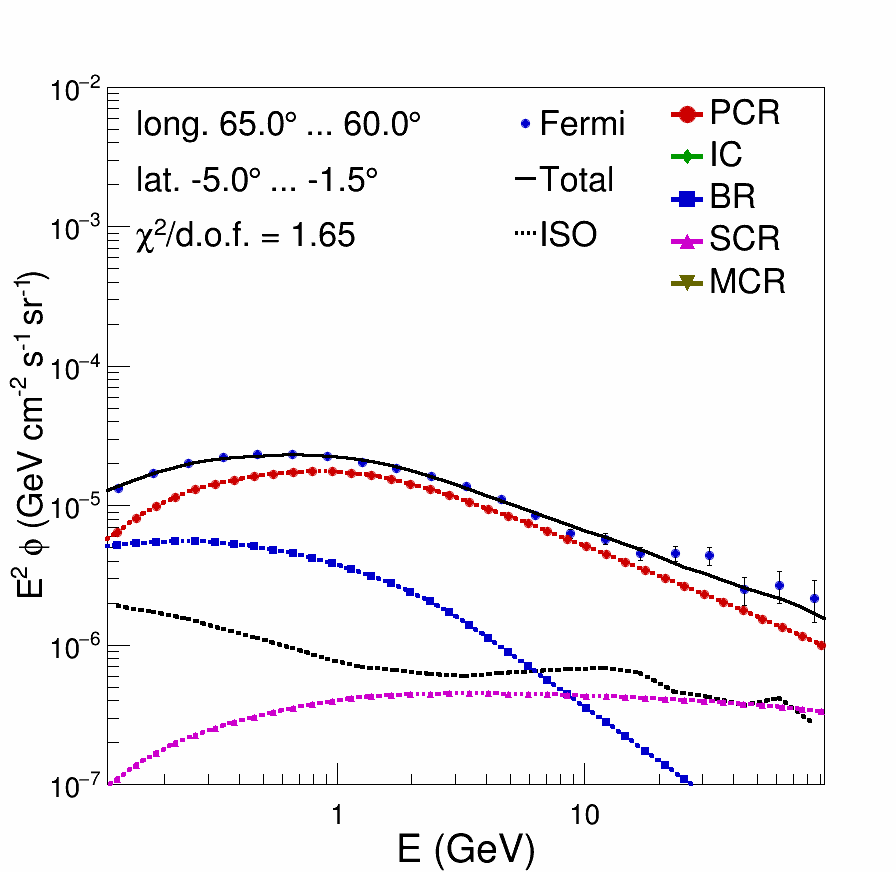}
\includegraphics[width=0.16\textwidth,height=0.16\textwidth,clip]{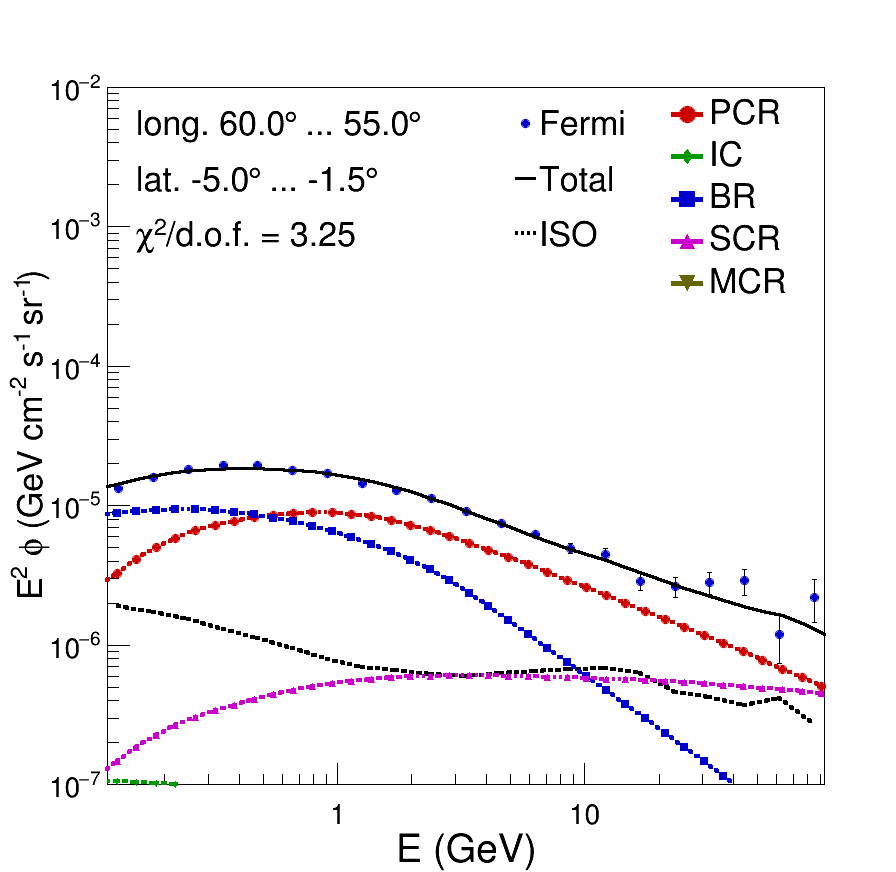}
\includegraphics[width=0.16\textwidth,height=0.16\textwidth,clip]{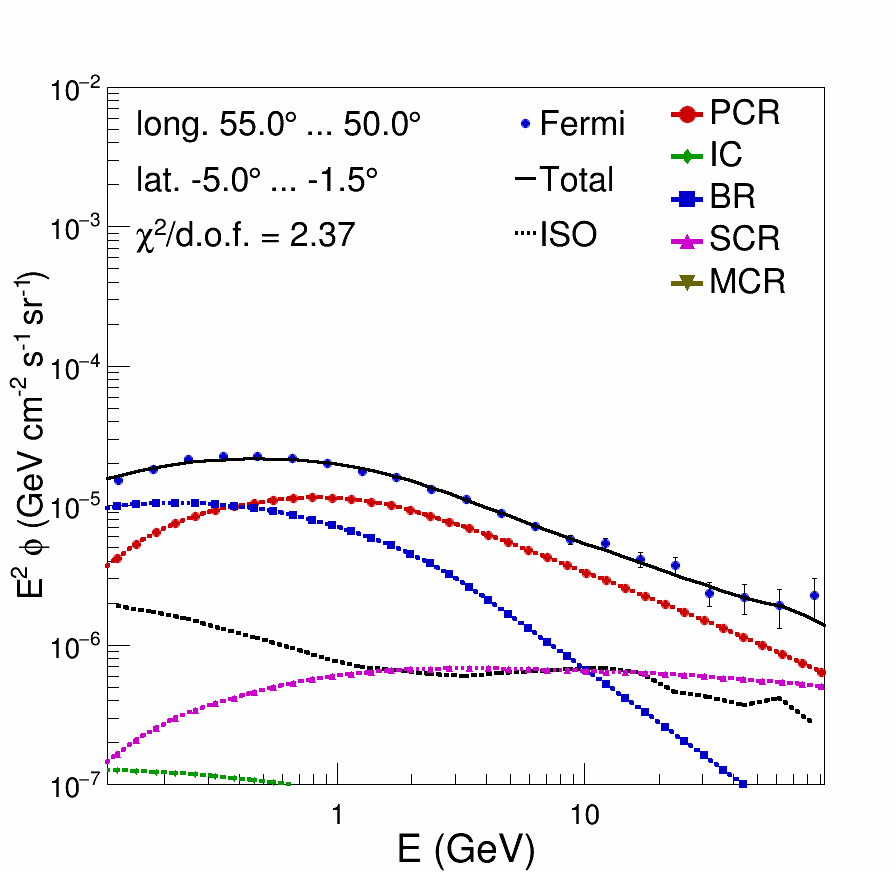}
\includegraphics[width=0.16\textwidth,height=0.16\textwidth,clip]{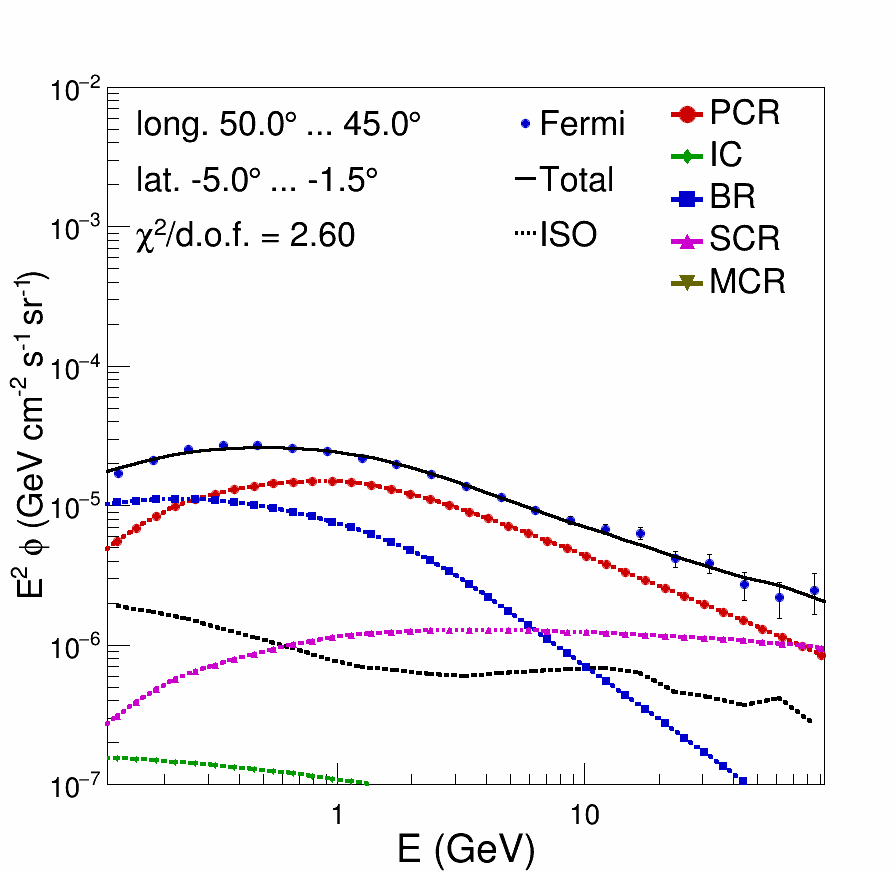}
\includegraphics[width=0.16\textwidth,height=0.16\textwidth,clip]{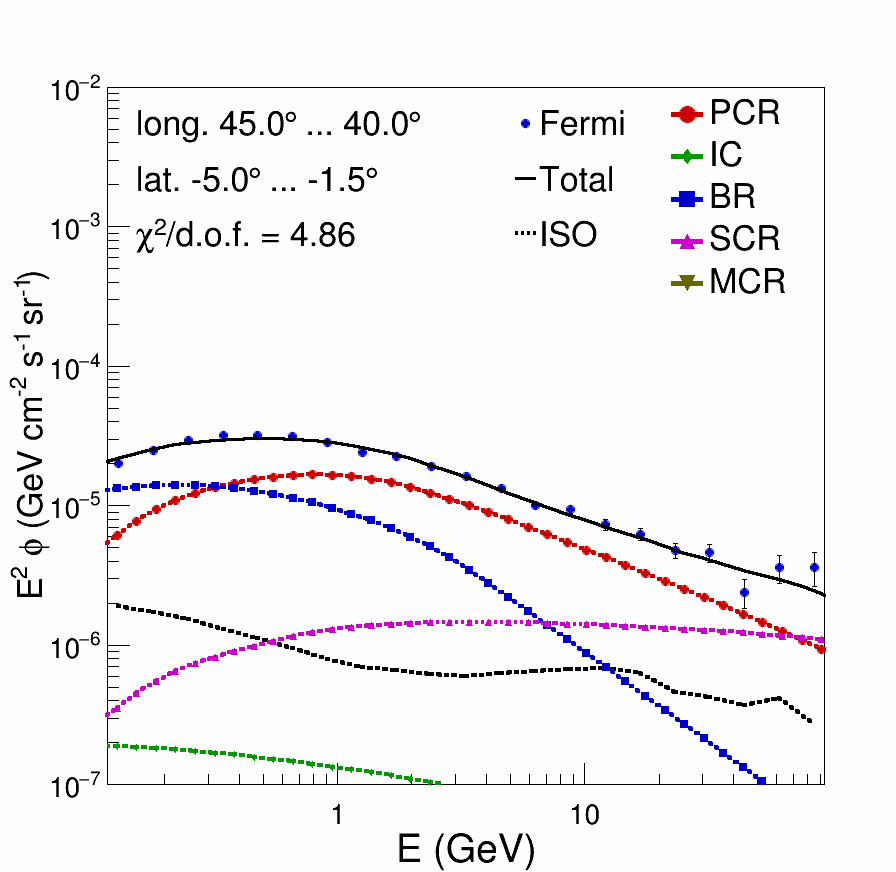}
\includegraphics[width=0.16\textwidth,height=0.16\textwidth,clip]{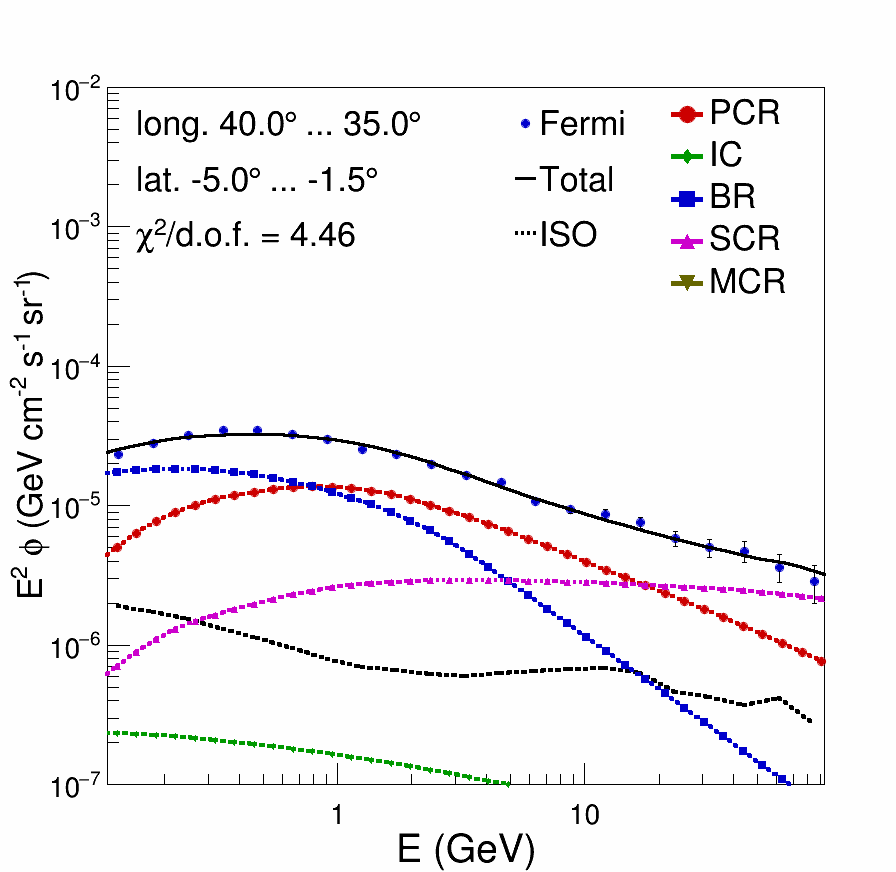}
\includegraphics[width=0.16\textwidth,height=0.16\textwidth,clip]{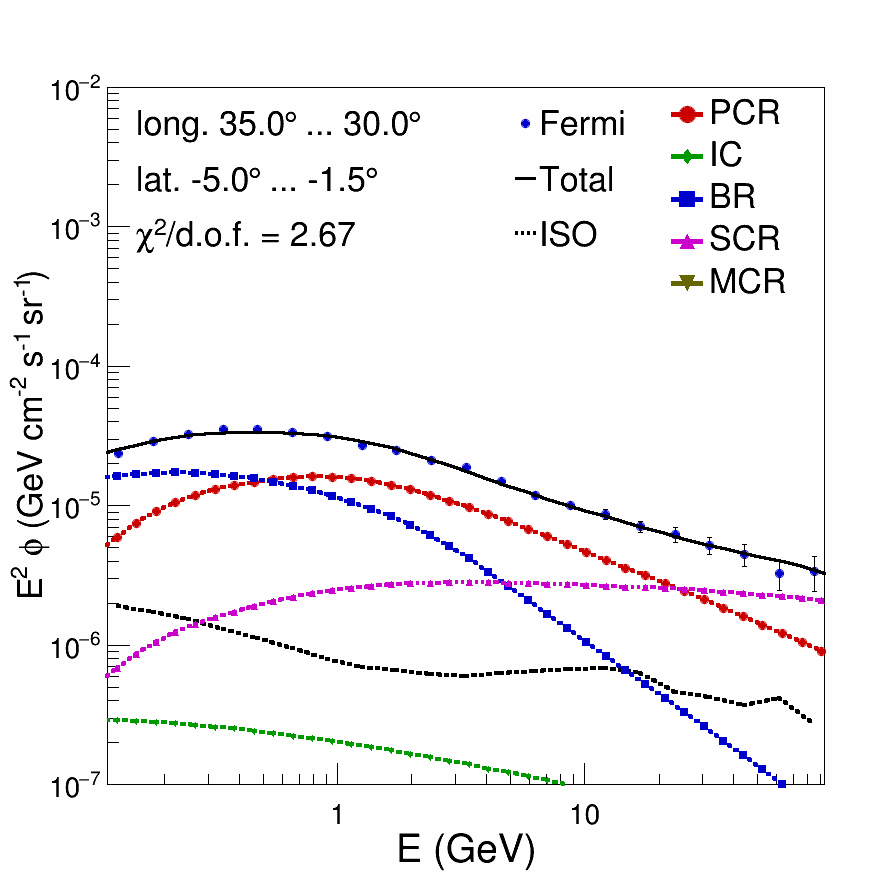}
\includegraphics[width=0.16\textwidth,height=0.16\textwidth,clip]{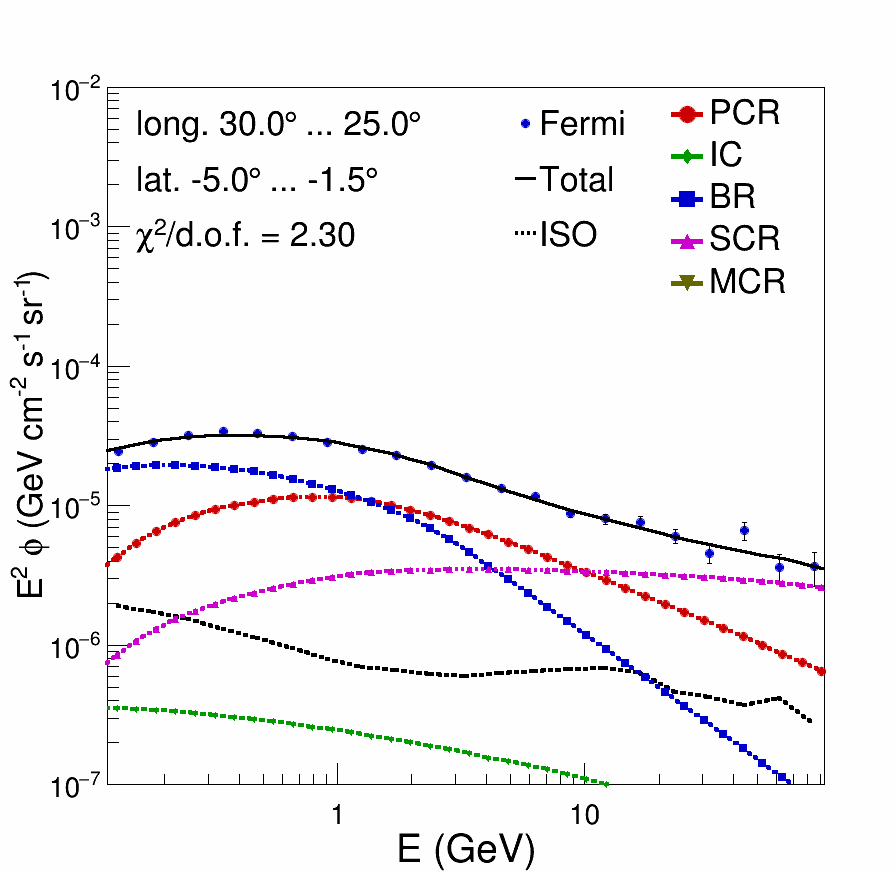}
\includegraphics[width=0.16\textwidth,height=0.16\textwidth,clip]{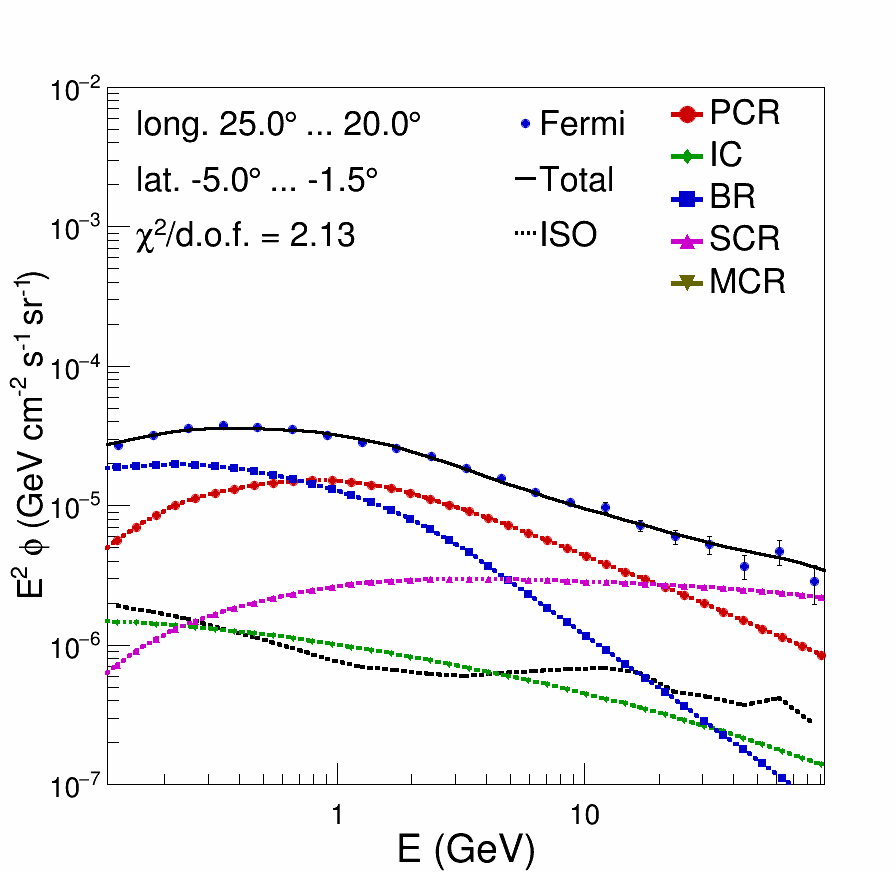}
\includegraphics[width=0.16\textwidth,height=0.16\textwidth,clip]{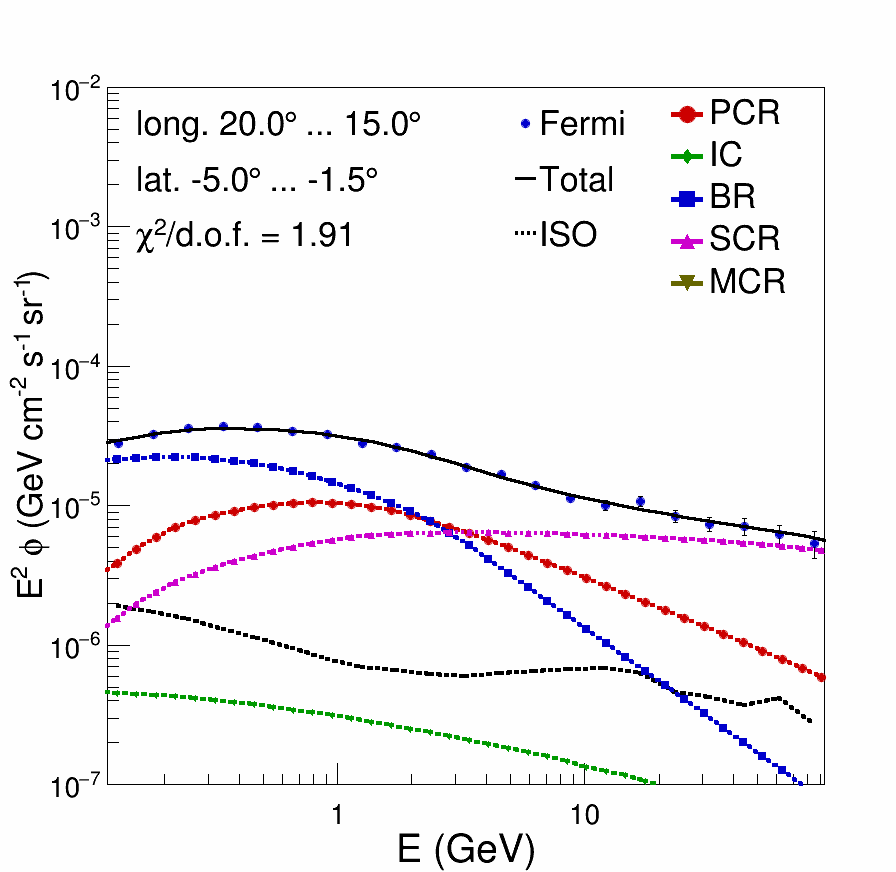}
\includegraphics[width=0.16\textwidth,height=0.16\textwidth,clip]{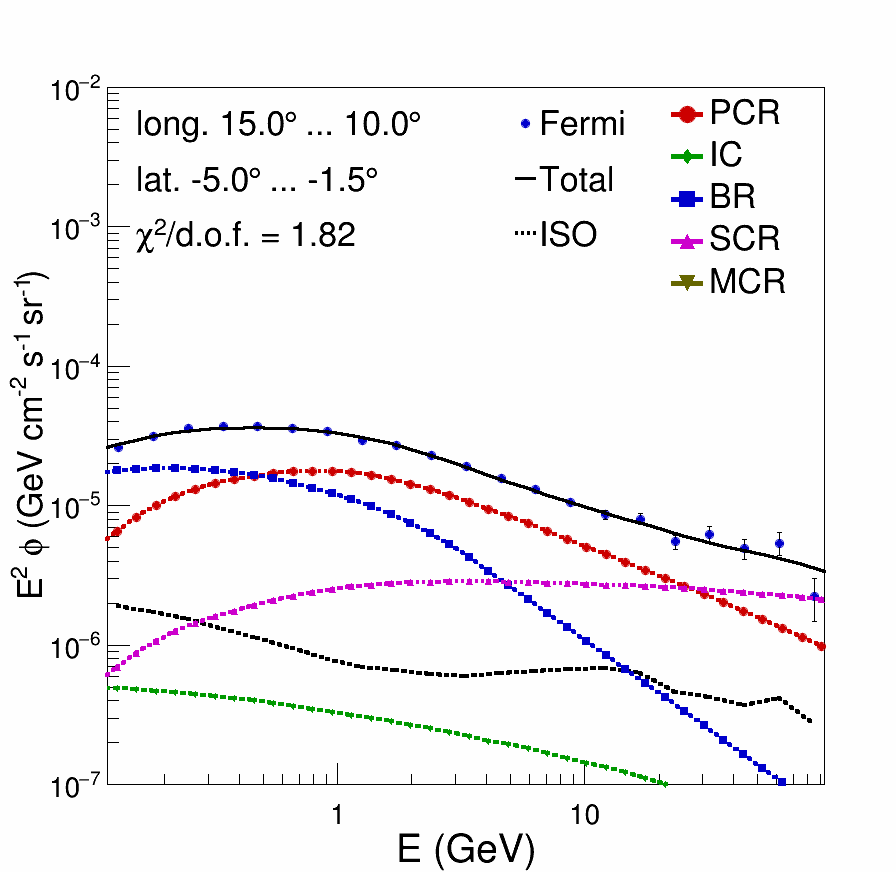}
\includegraphics[width=0.16\textwidth,height=0.16\textwidth,clip]{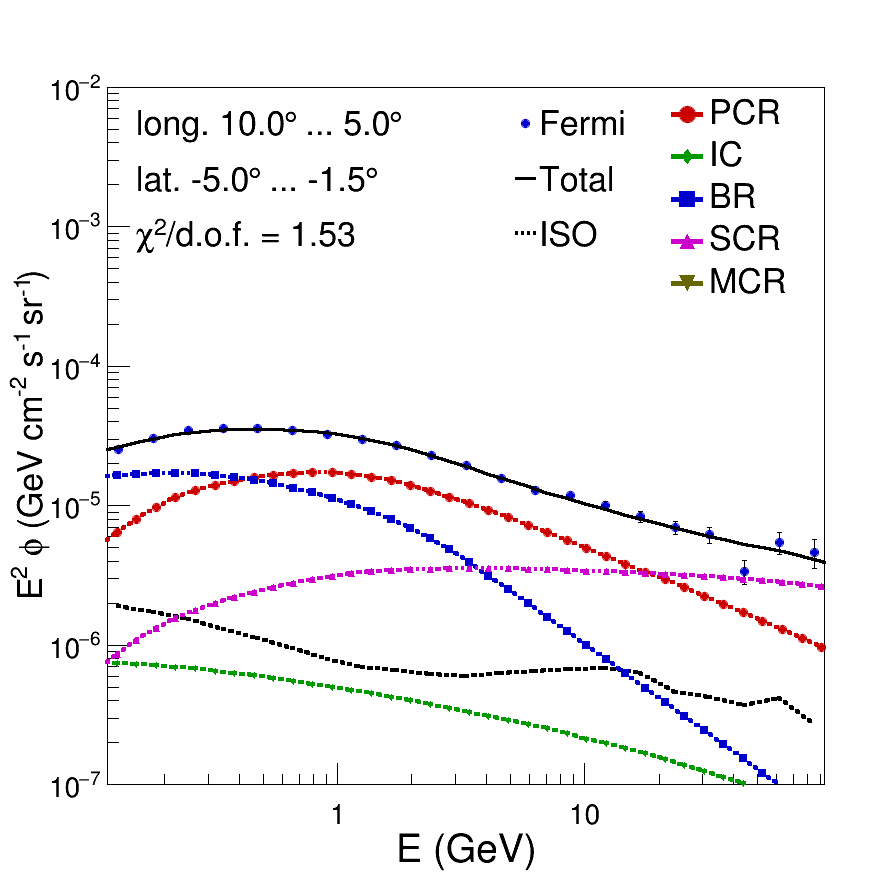}
\includegraphics[width=0.16\textwidth,height=0.16\textwidth,clip]{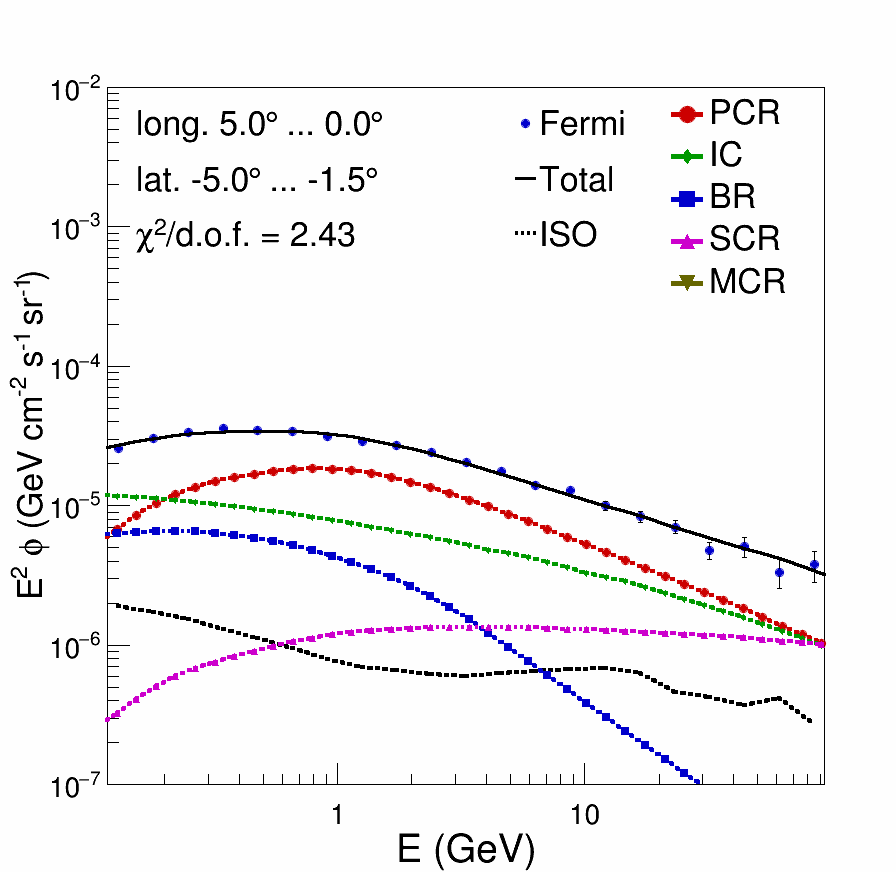}
\caption[]{Template fits for latitudes  with $-5.0^\circ<b<-1.5^\circ$ and longitudes decreasing from 180$^\circ$ to 0$^\circ$.} \label{F23}
\end{figure}
\begin{figure}
\centering
\includegraphics[width=0.16\textwidth,height=0.16\textwidth,clip]{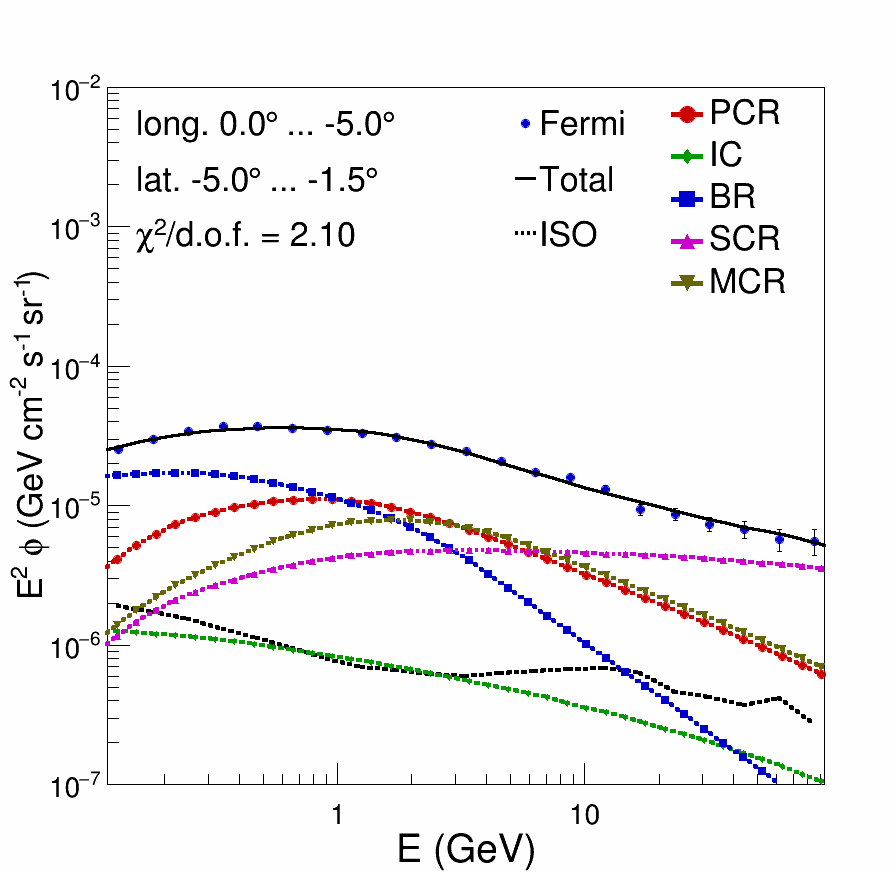}
\includegraphics[width=0.16\textwidth,height=0.16\textwidth,clip]{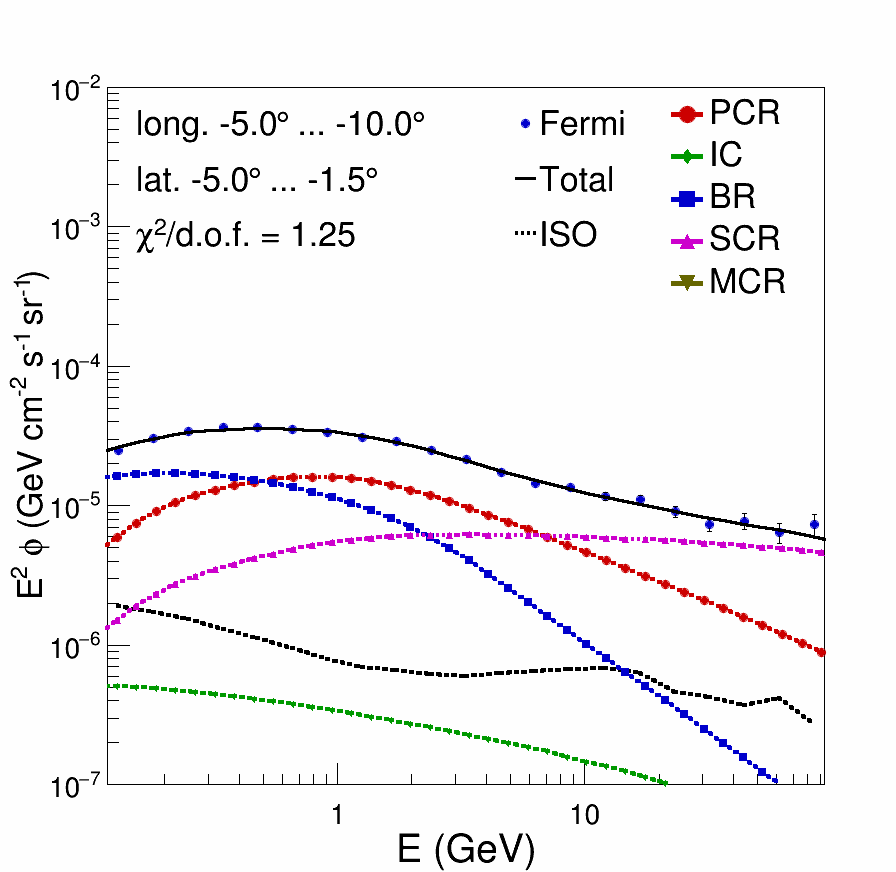}
\includegraphics[width=0.16\textwidth,height=0.16\textwidth,clip]{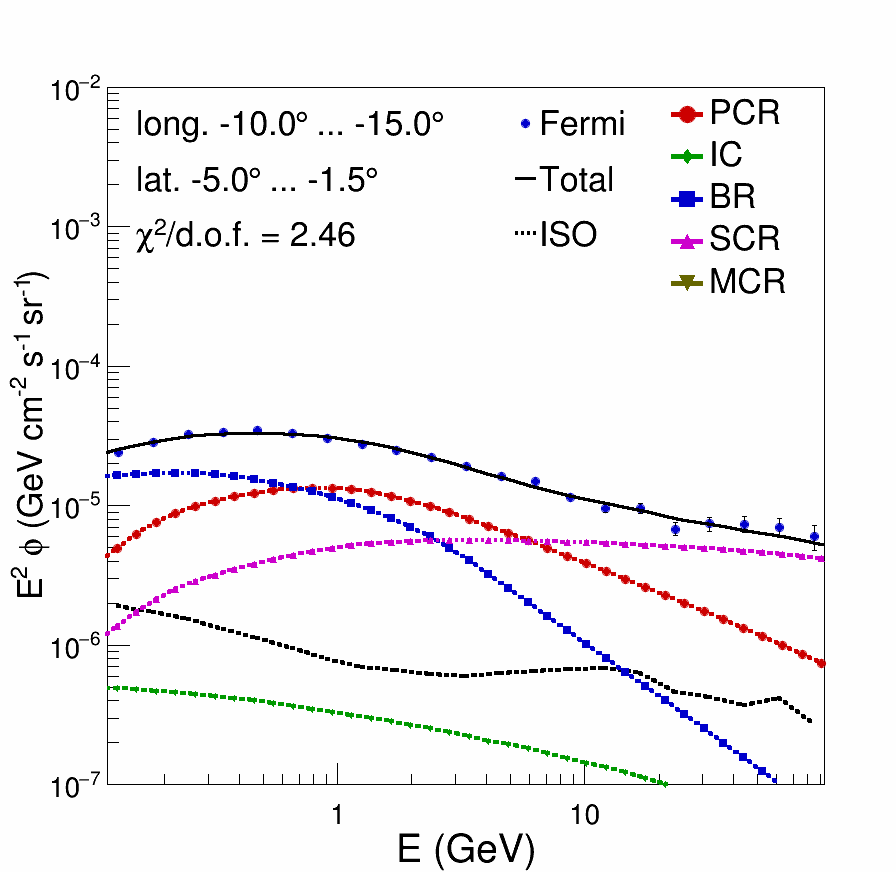}
\includegraphics[width=0.16\textwidth,height=0.16\textwidth,clip]{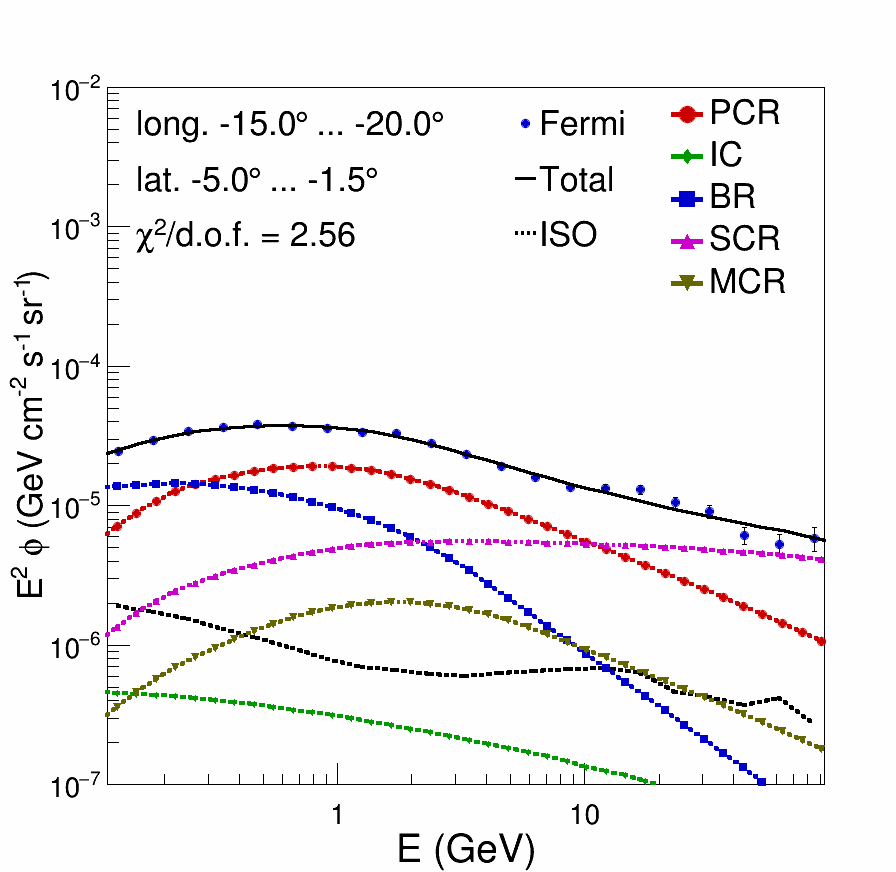}
\includegraphics[width=0.16\textwidth,height=0.16\textwidth,clip]{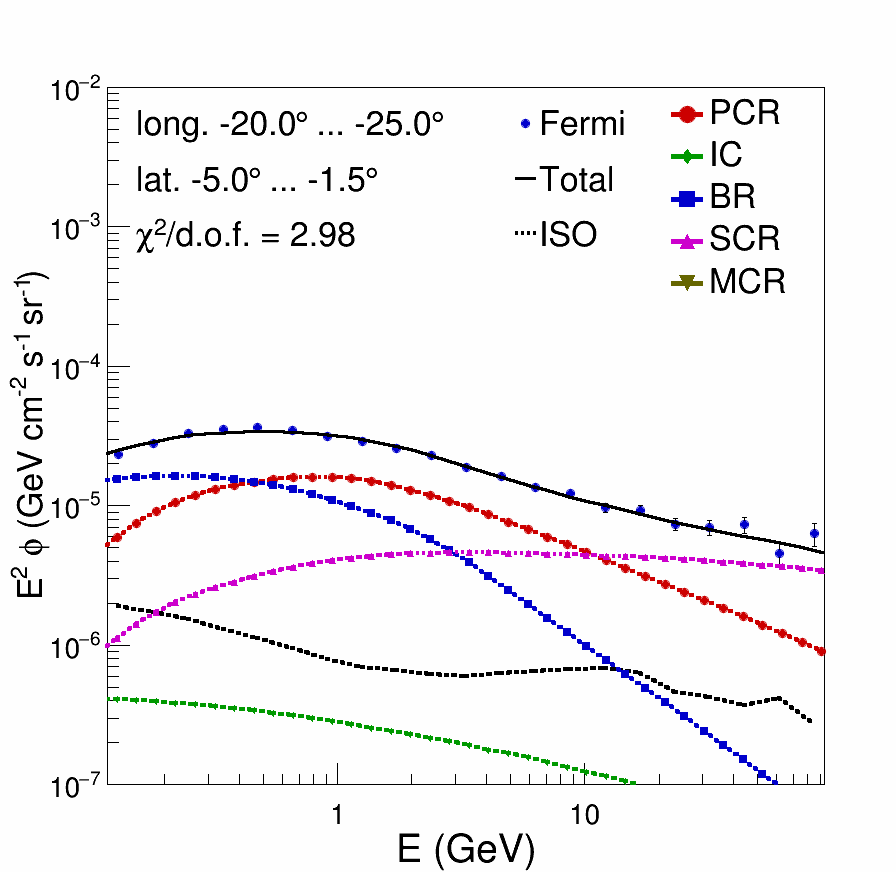}
\includegraphics[width=0.16\textwidth,height=0.16\textwidth,clip]{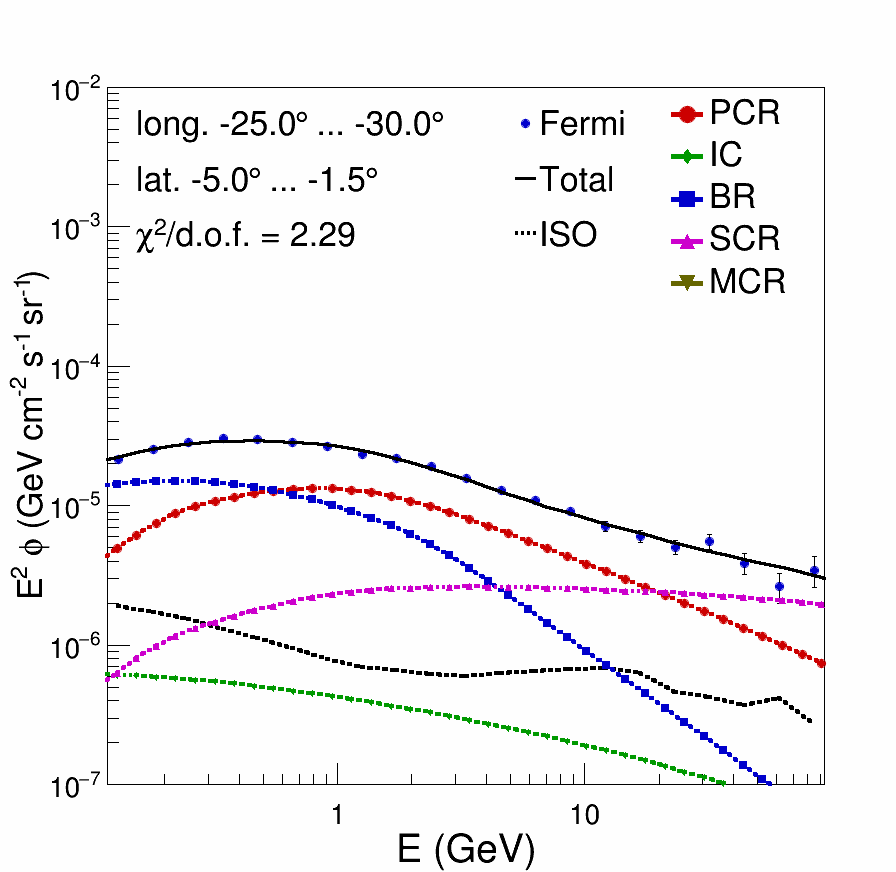}
\includegraphics[width=0.16\textwidth,height=0.16\textwidth,clip]{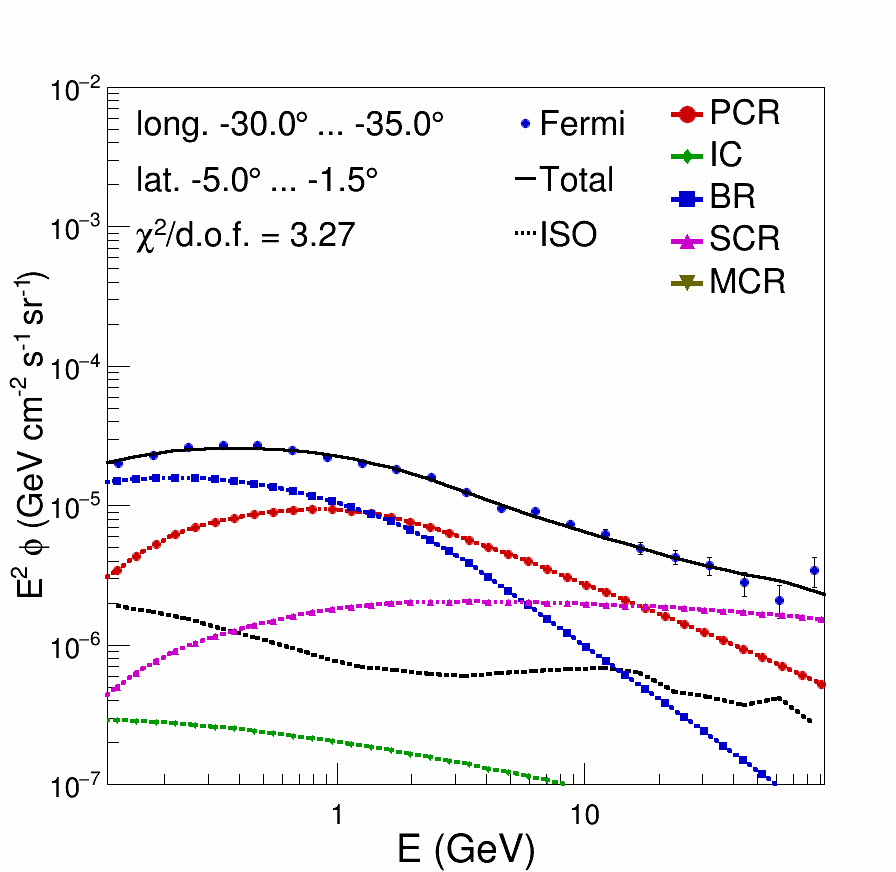}
\includegraphics[width=0.16\textwidth,height=0.16\textwidth,clip]{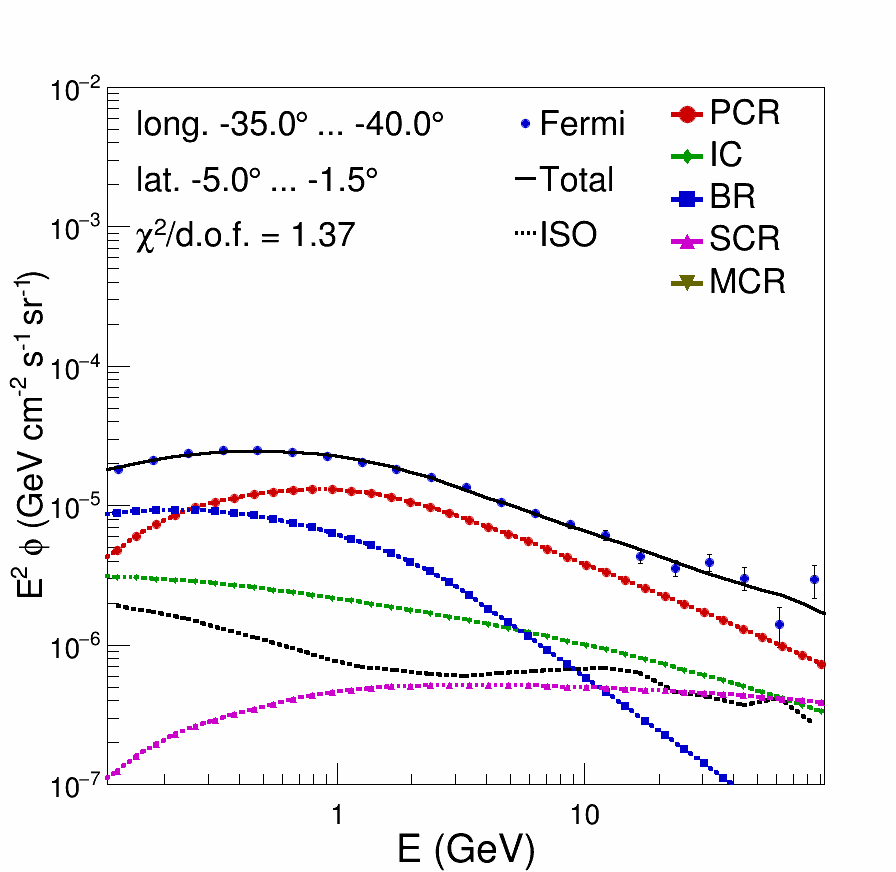}
\includegraphics[width=0.16\textwidth,height=0.16\textwidth,clip]{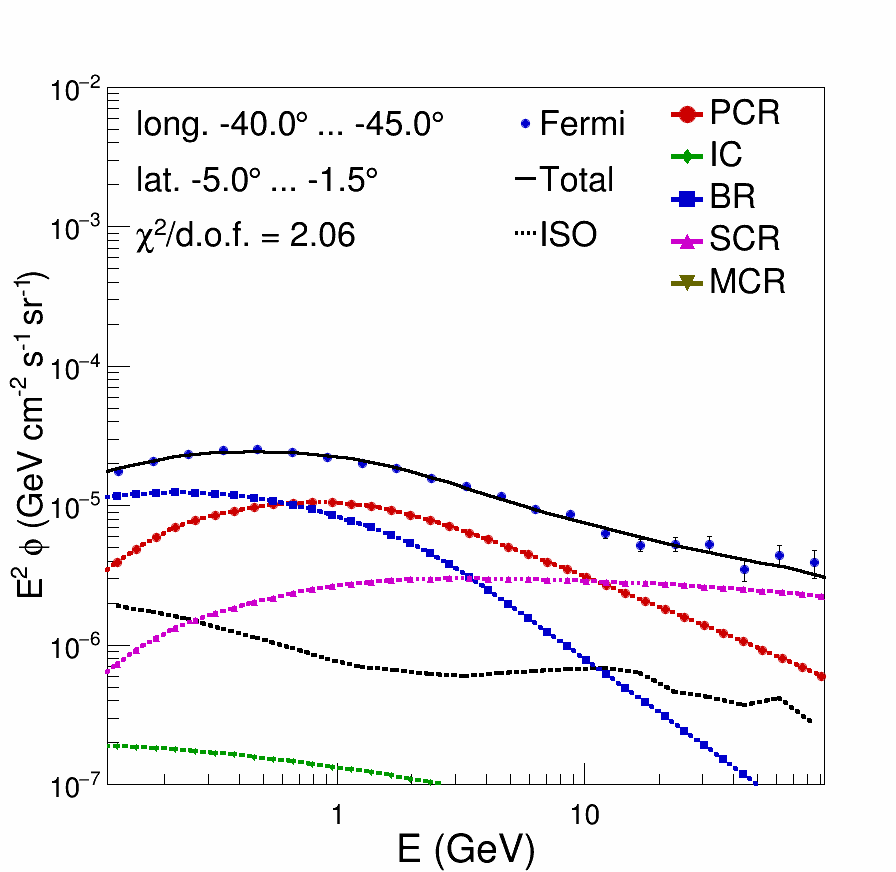}
\includegraphics[width=0.16\textwidth,height=0.16\textwidth,clip]{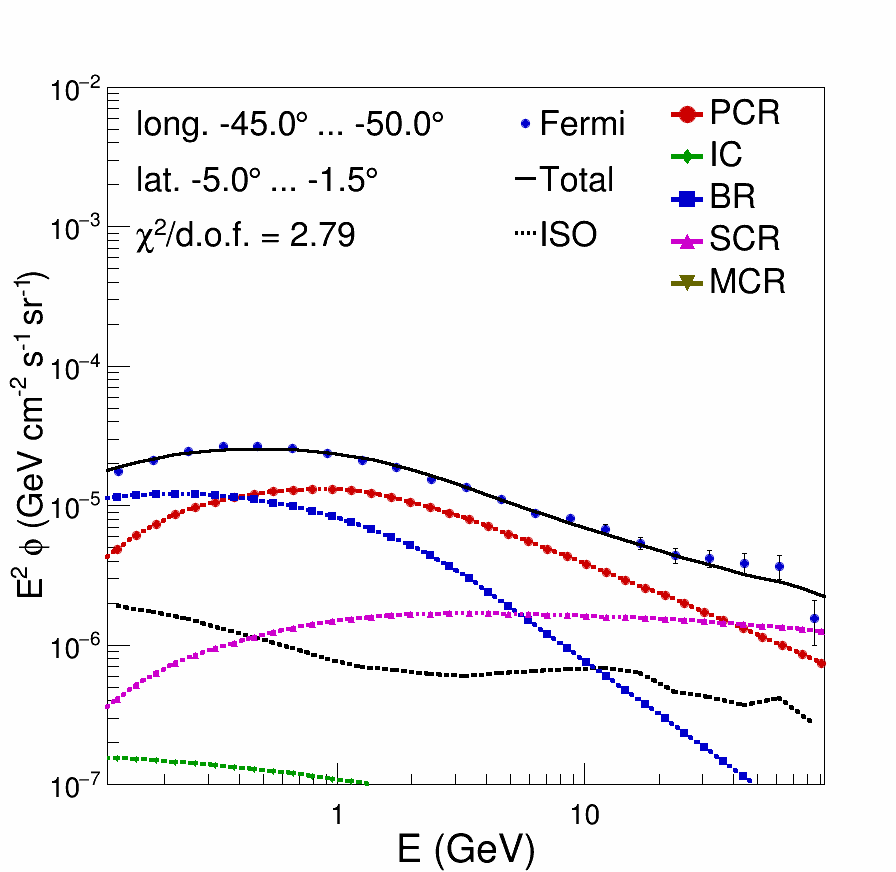}
\includegraphics[width=0.16\textwidth,height=0.16\textwidth,clip]{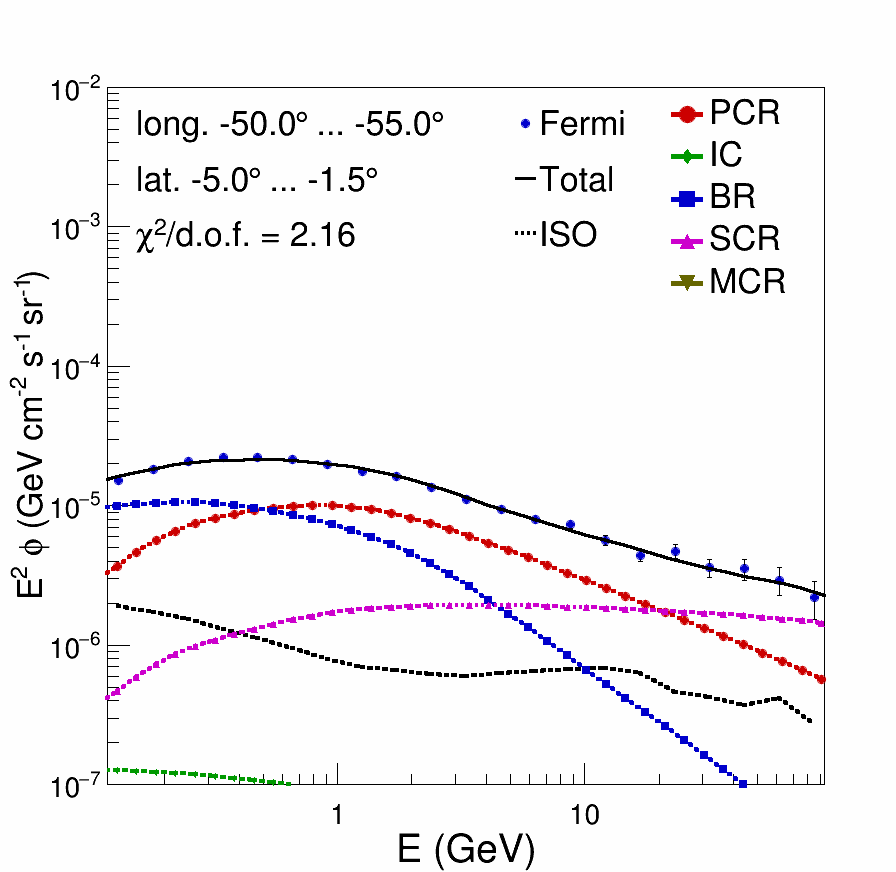}
\includegraphics[width=0.16\textwidth,height=0.16\textwidth,clip]{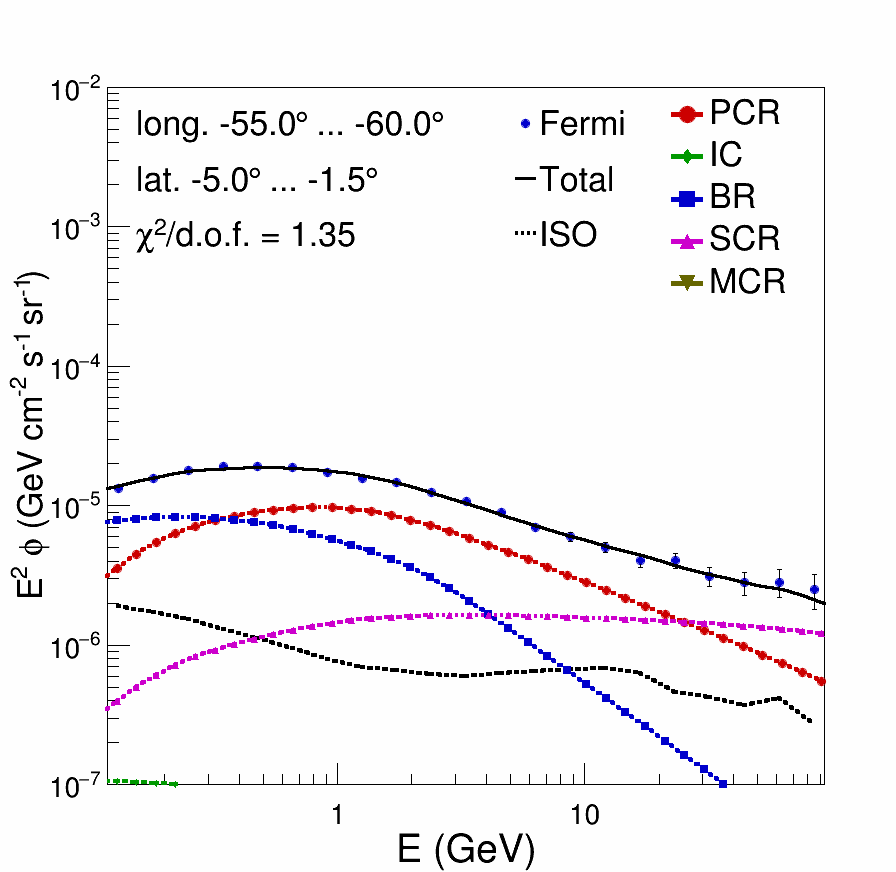}
\includegraphics[width=0.16\textwidth,height=0.16\textwidth,clip]{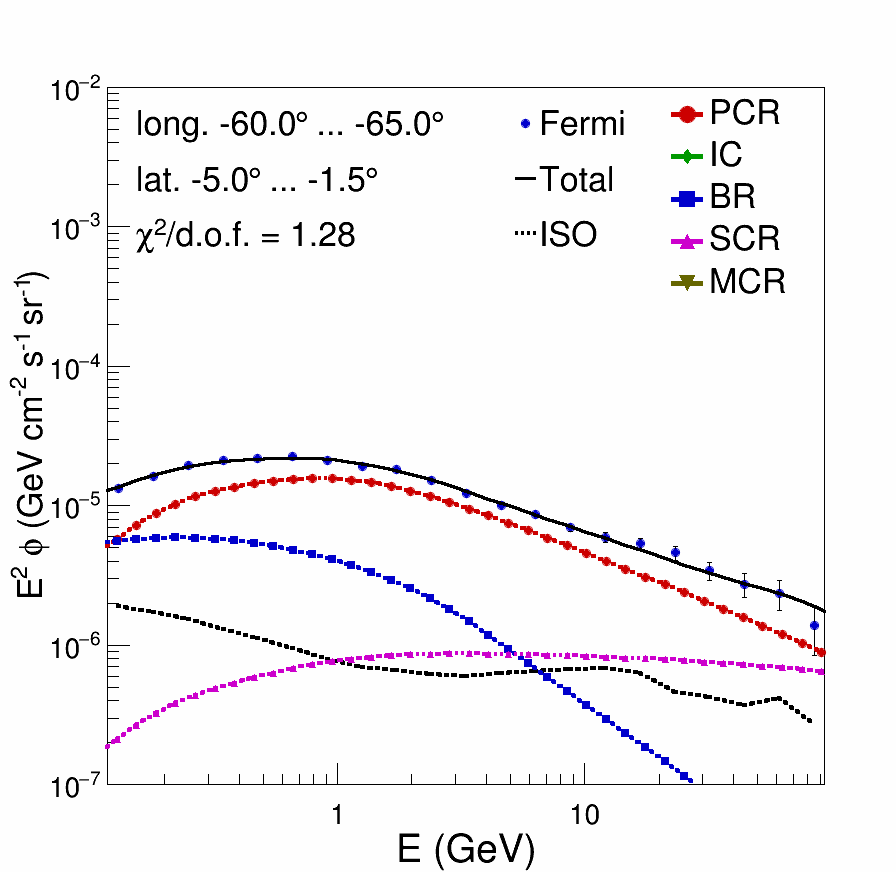}
\includegraphics[width=0.16\textwidth,height=0.16\textwidth,clip]{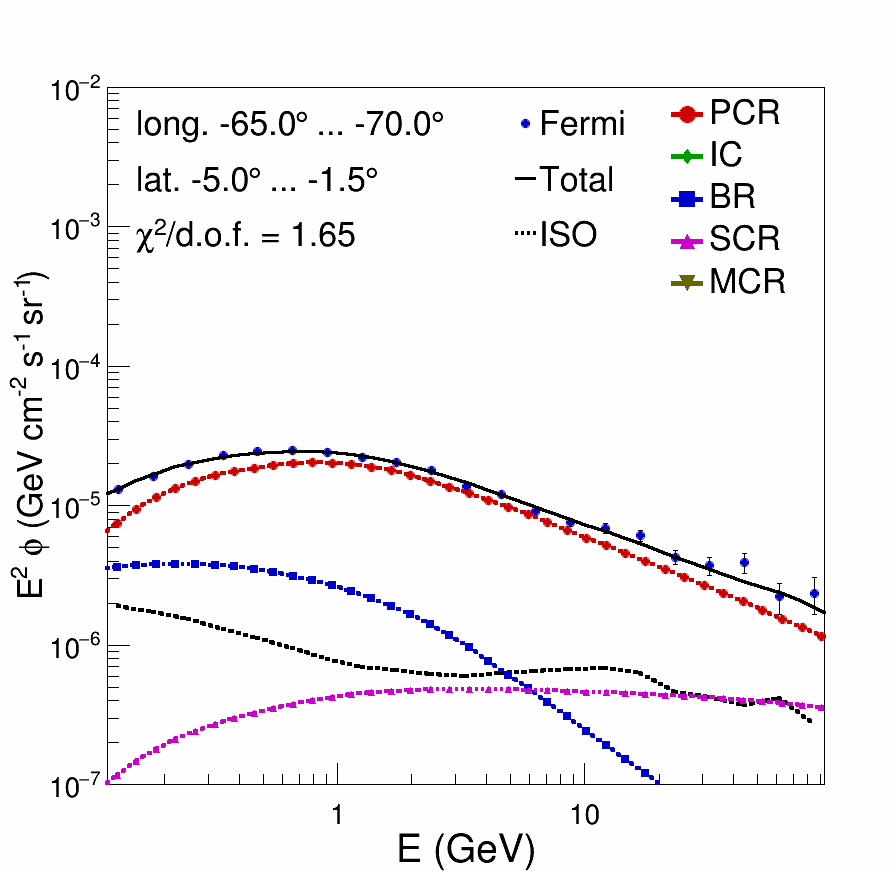}
\includegraphics[width=0.16\textwidth,height=0.16\textwidth,clip]{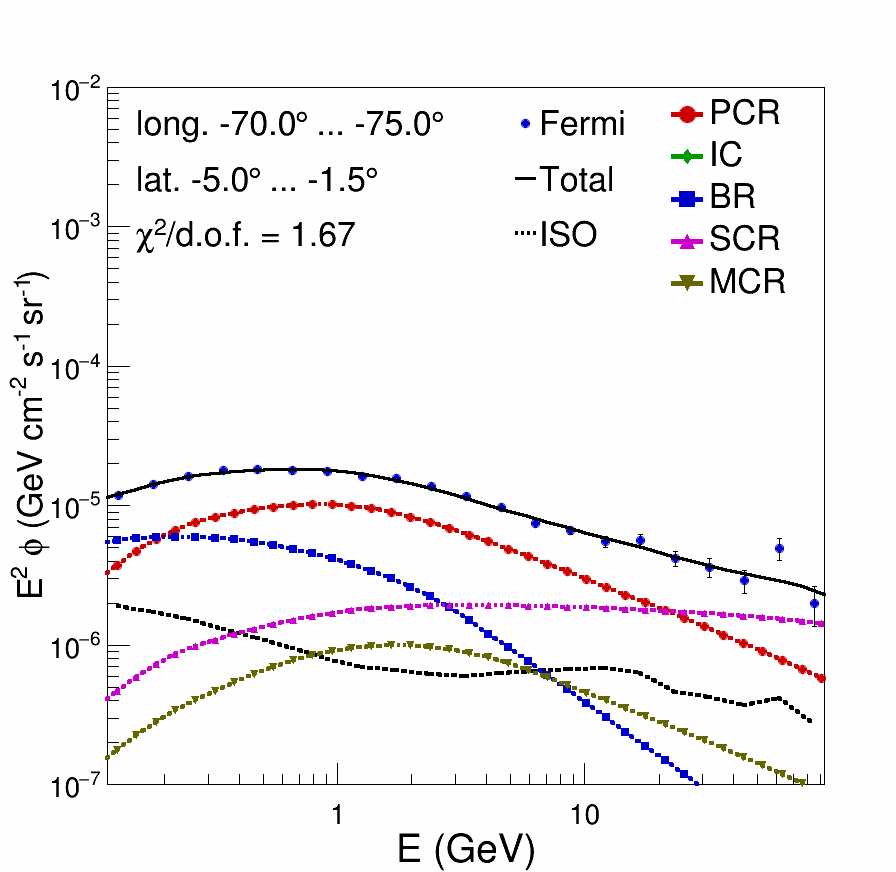}
\includegraphics[width=0.16\textwidth,height=0.16\textwidth,clip]{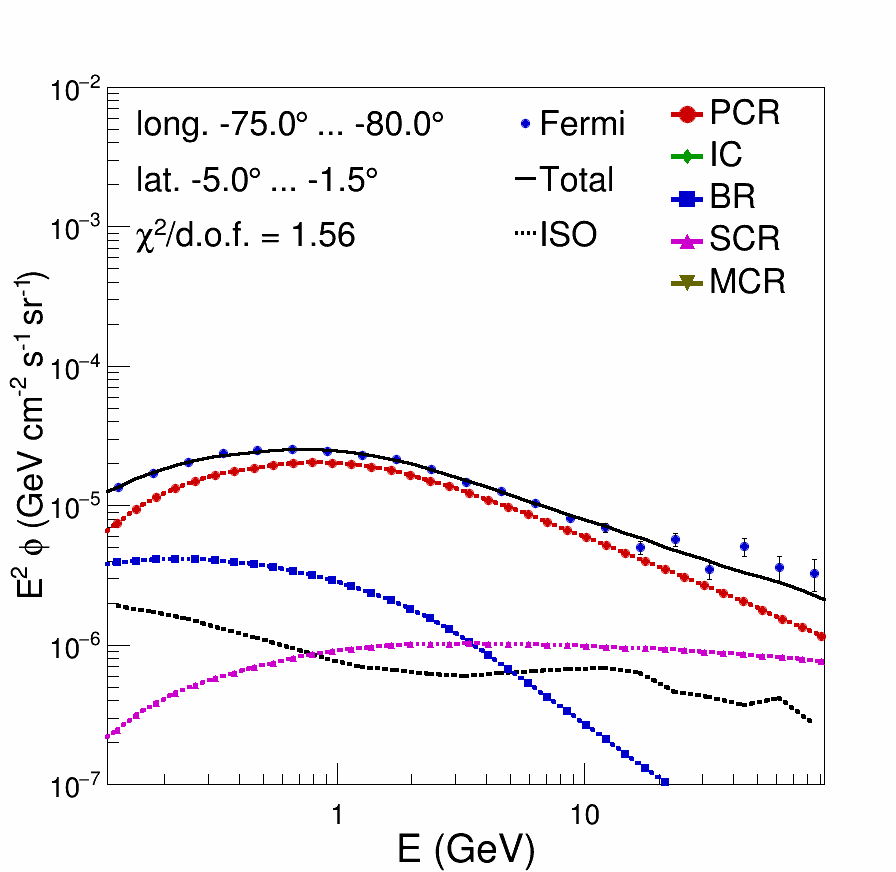}
\includegraphics[width=0.16\textwidth,height=0.16\textwidth,clip]{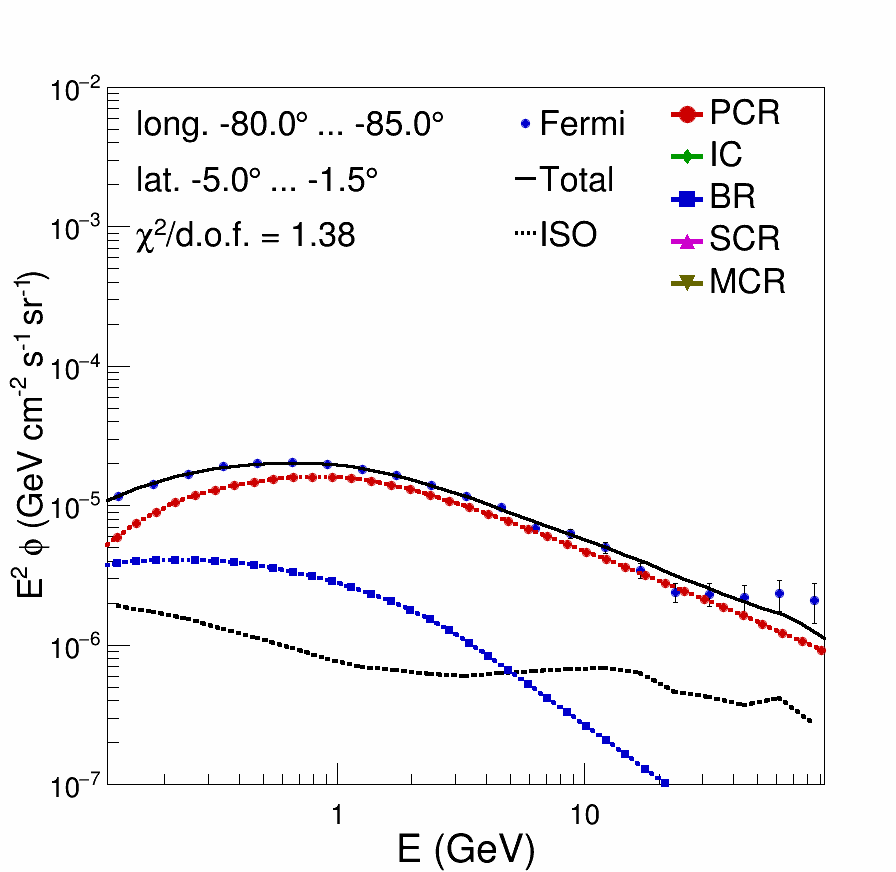}
\includegraphics[width=0.16\textwidth,height=0.16\textwidth,clip]{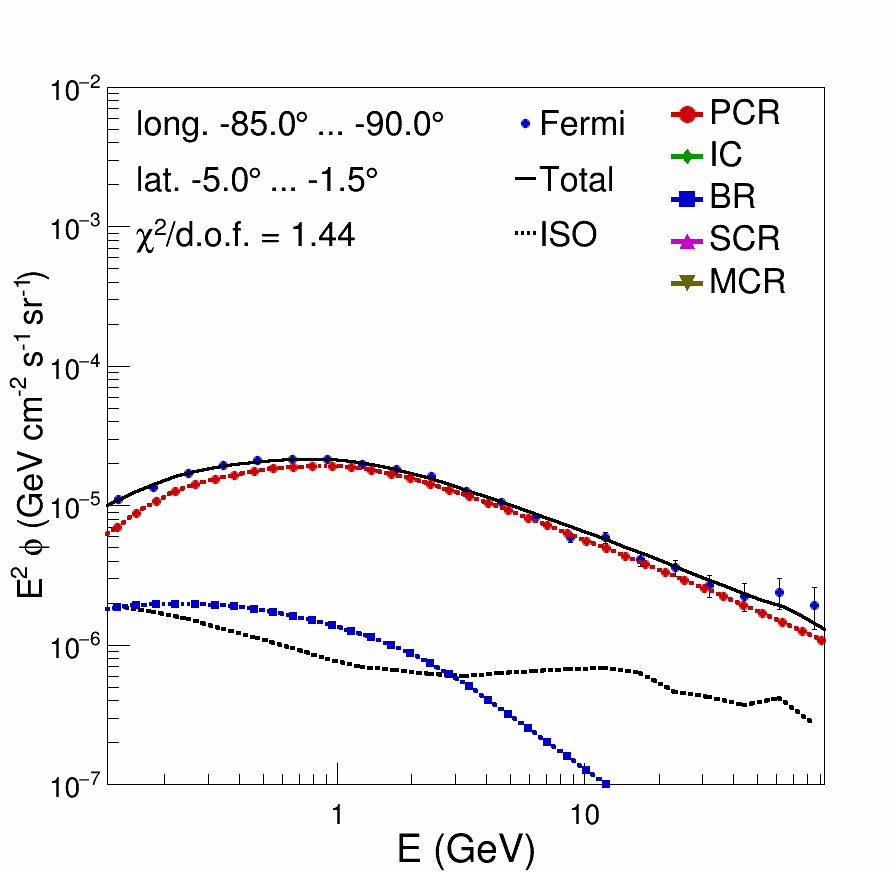}
\includegraphics[width=0.16\textwidth,height=0.16\textwidth,clip]{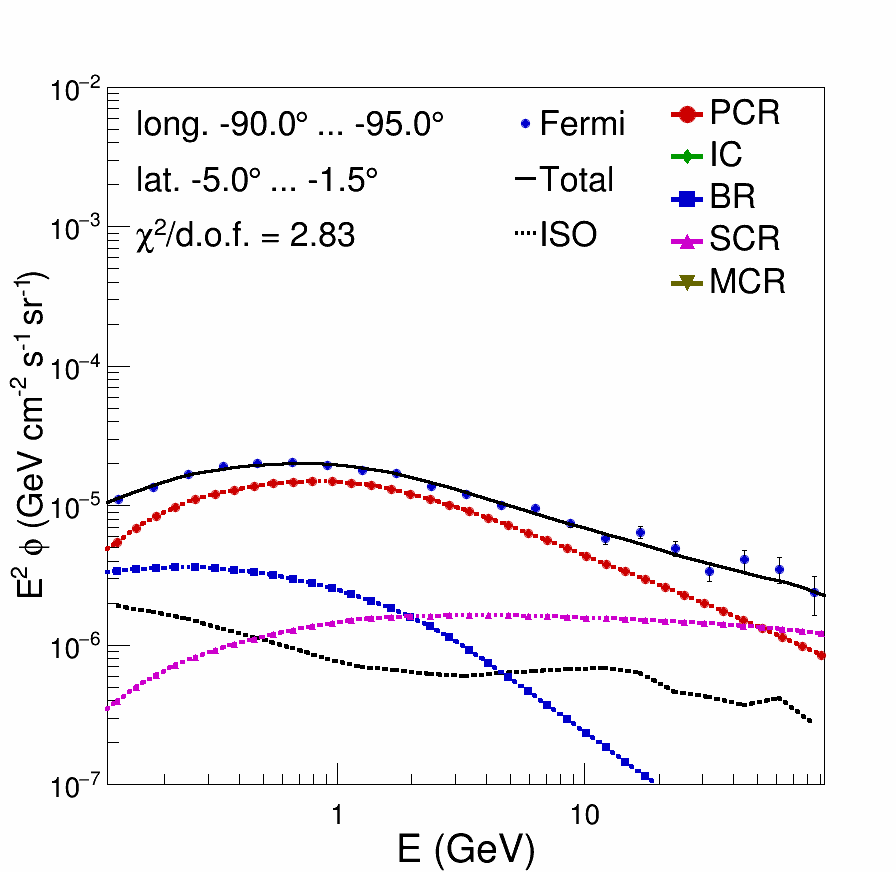}
\includegraphics[width=0.16\textwidth,height=0.16\textwidth,clip]{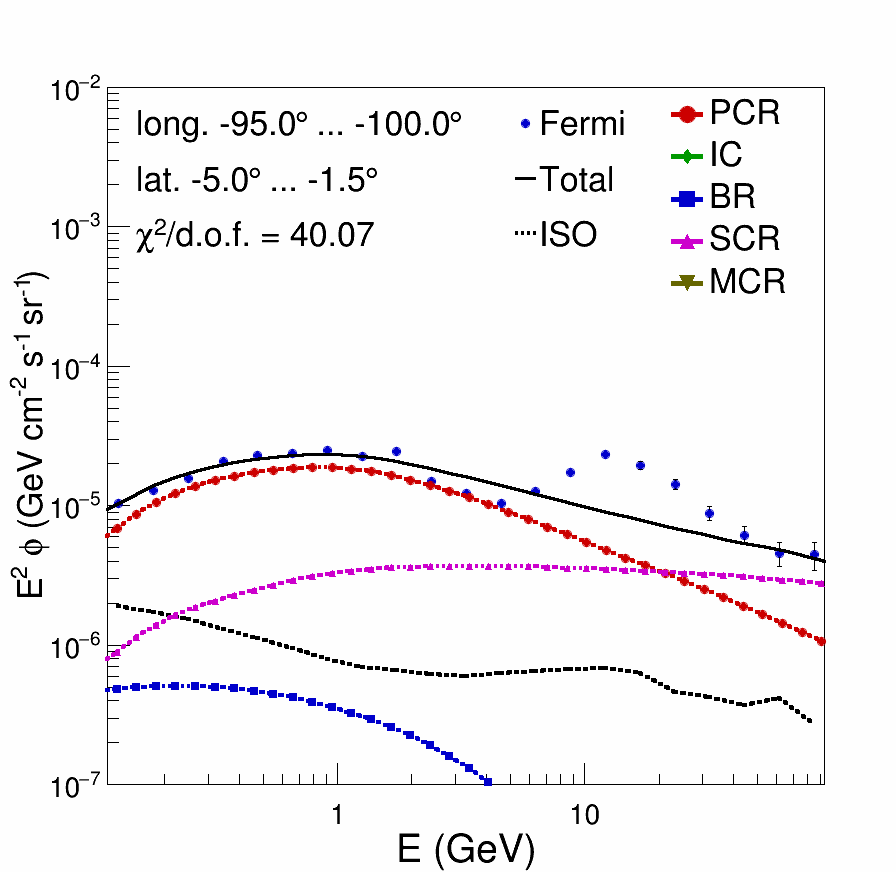}
\includegraphics[width=0.16\textwidth,height=0.16\textwidth,clip]{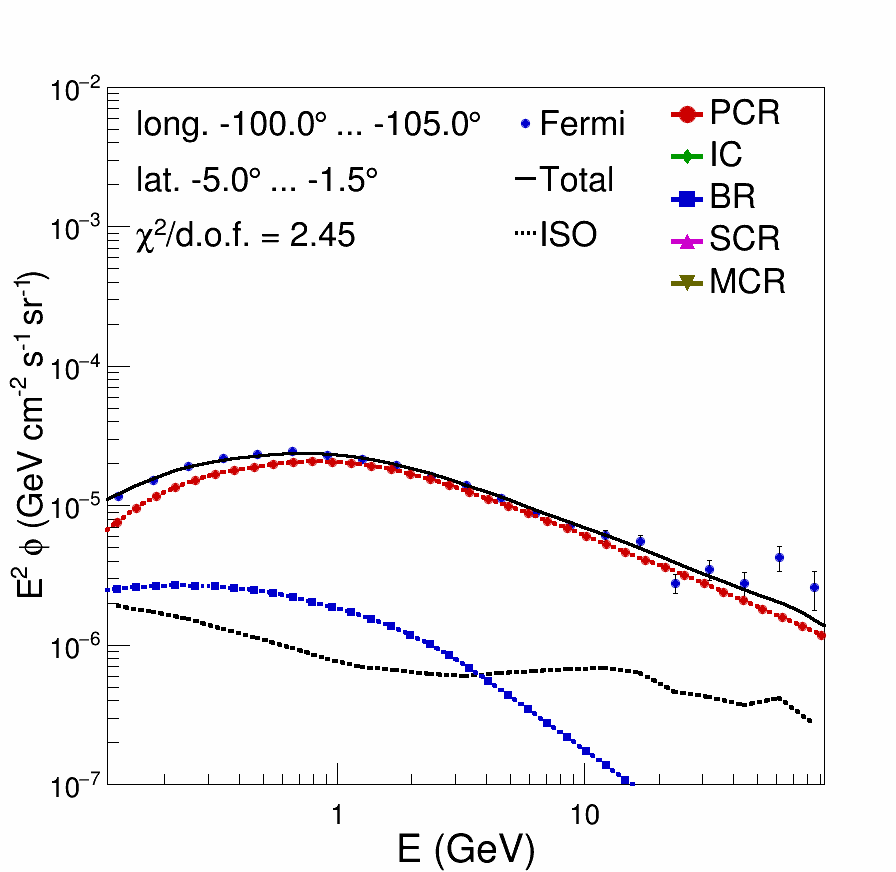}
\includegraphics[width=0.16\textwidth,height=0.16\textwidth,clip]{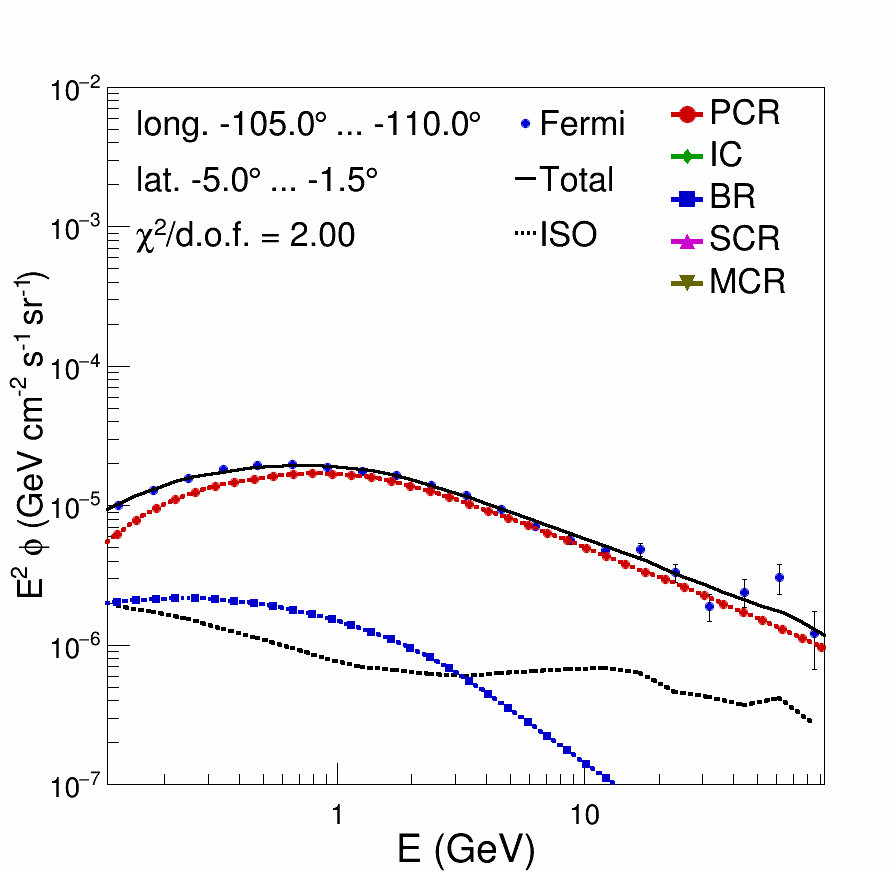}
\includegraphics[width=0.16\textwidth,height=0.16\textwidth,clip]{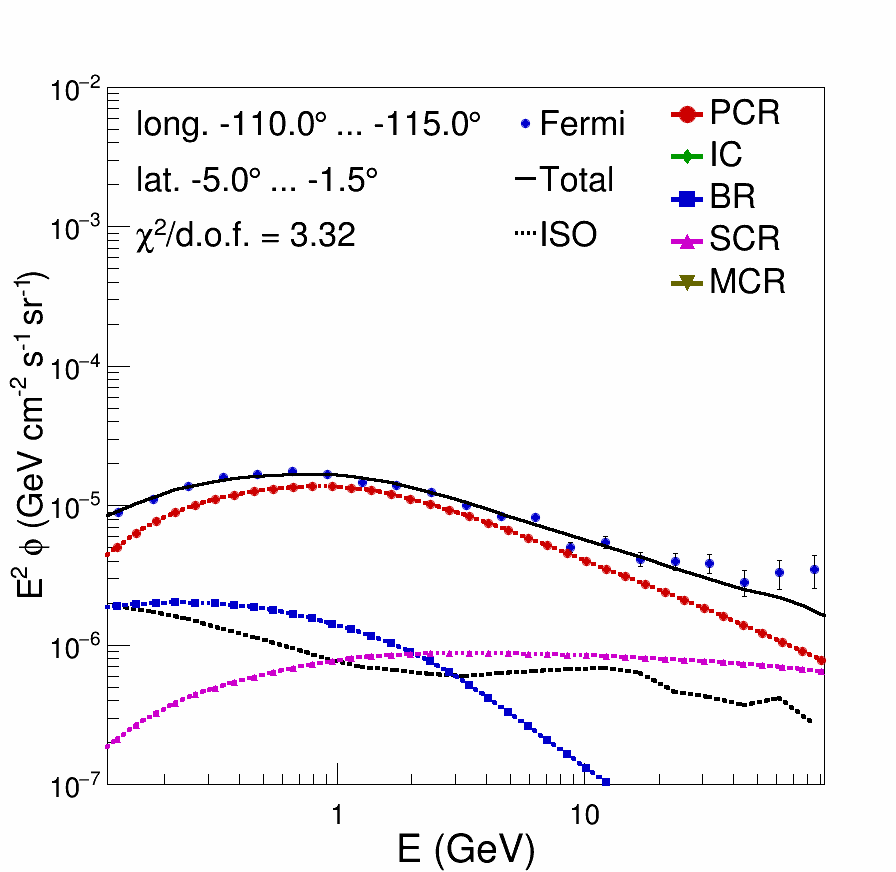}
\includegraphics[width=0.16\textwidth,height=0.16\textwidth,clip]{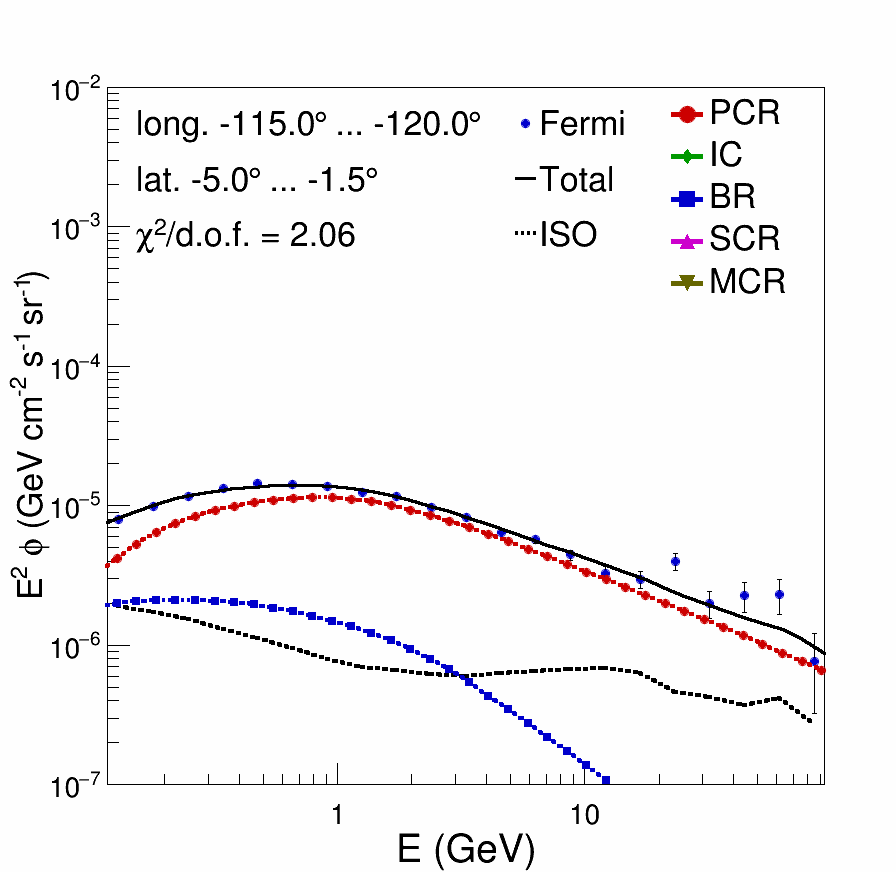}
\includegraphics[width=0.16\textwidth,height=0.16\textwidth,clip]{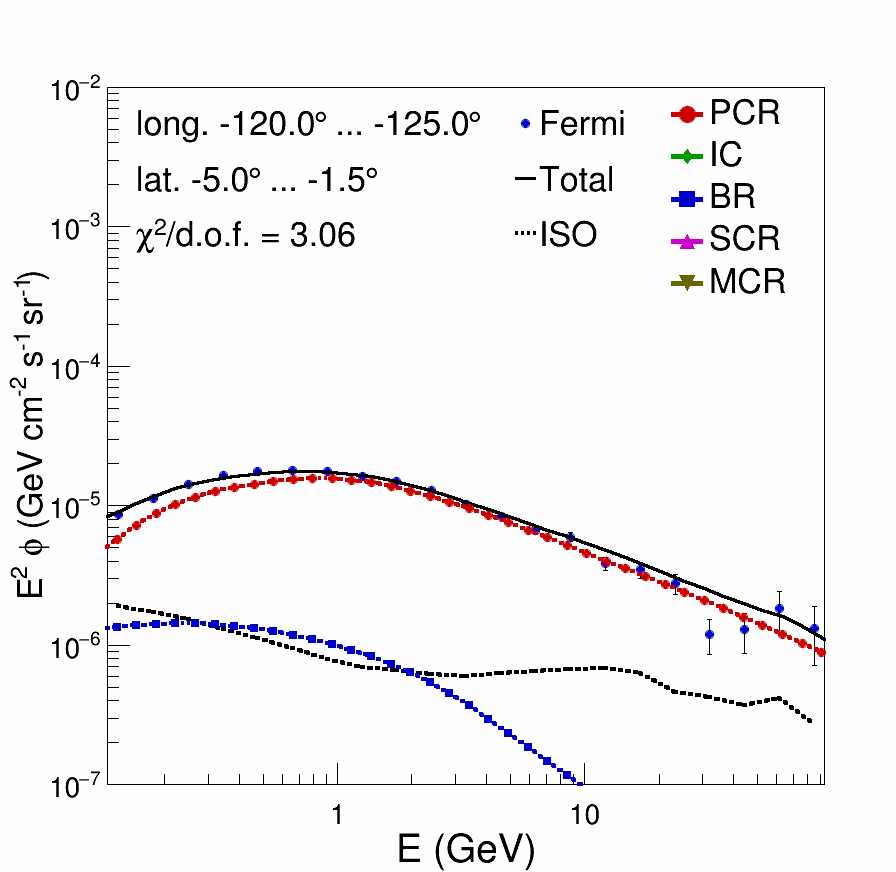}
\includegraphics[width=0.16\textwidth,height=0.16\textwidth,clip]{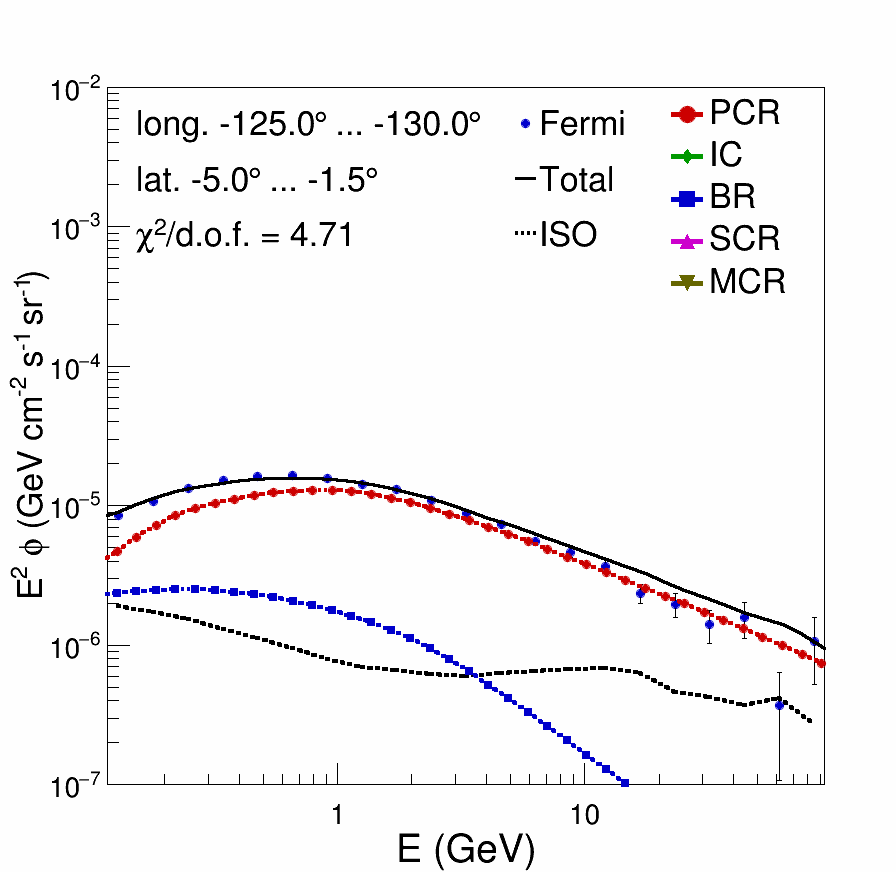}
\includegraphics[width=0.16\textwidth,height=0.16\textwidth,clip]{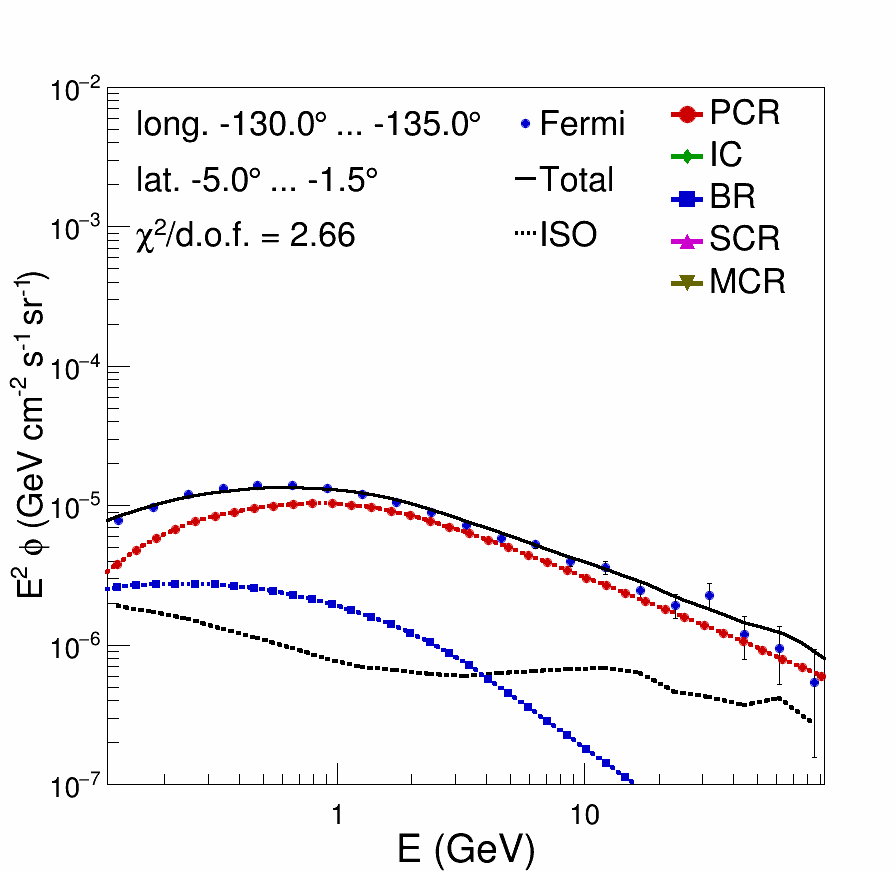}
\includegraphics[width=0.16\textwidth,height=0.16\textwidth,clip]{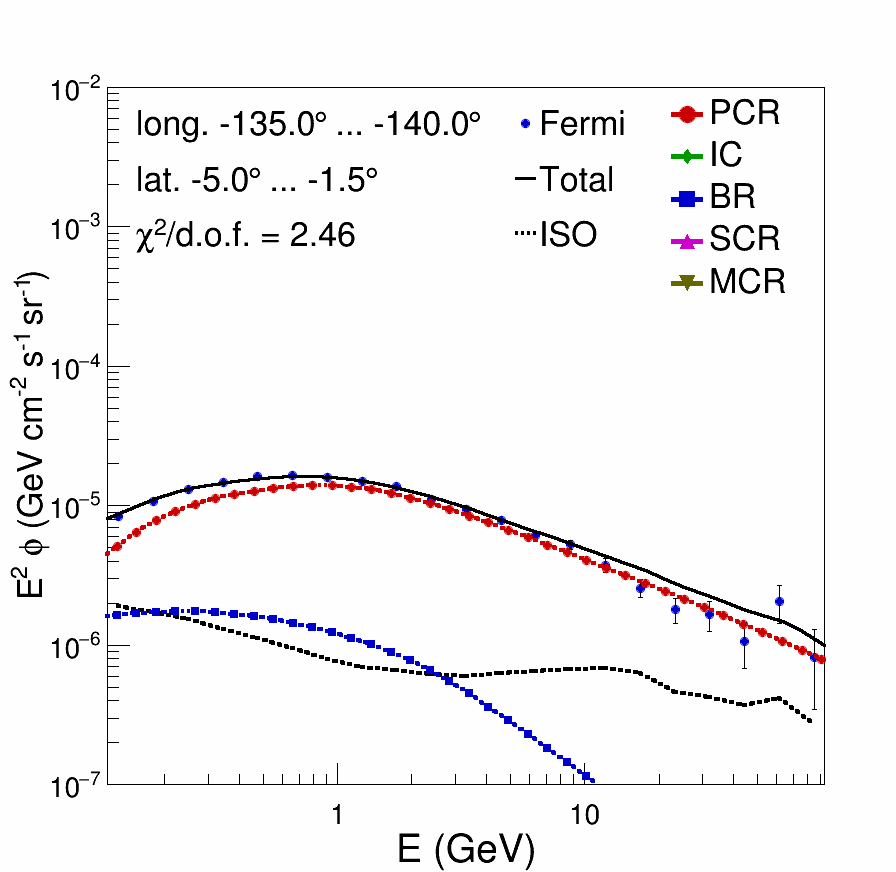}
\includegraphics[width=0.16\textwidth,height=0.16\textwidth,clip]{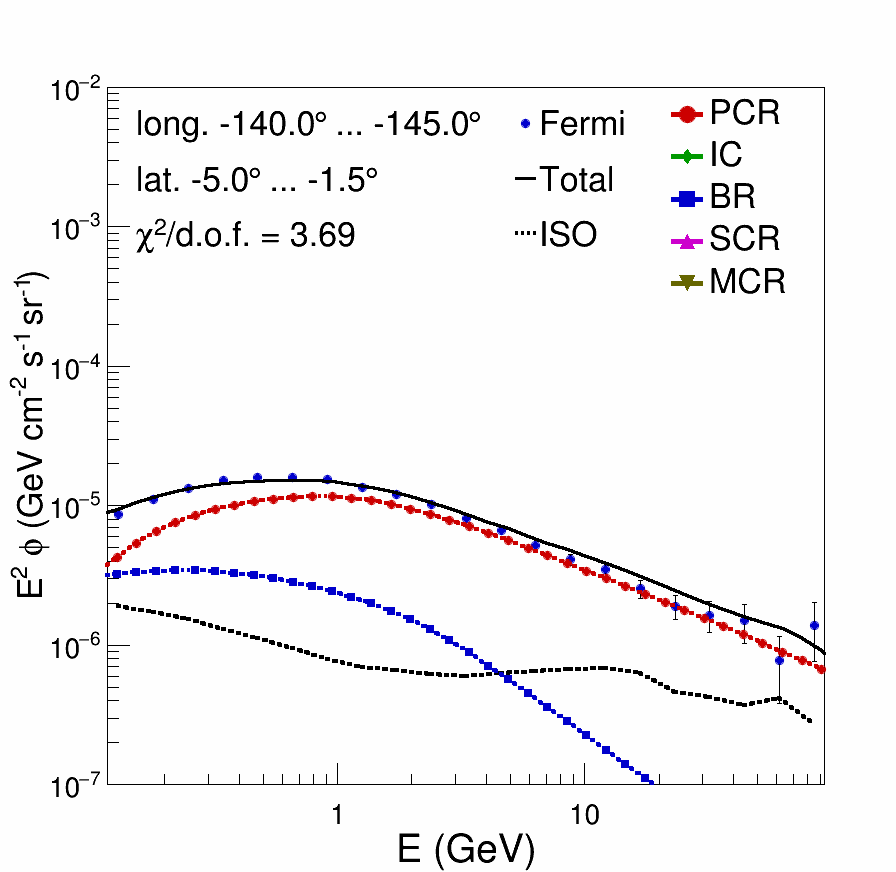}
\includegraphics[width=0.16\textwidth,height=0.16\textwidth,clip]{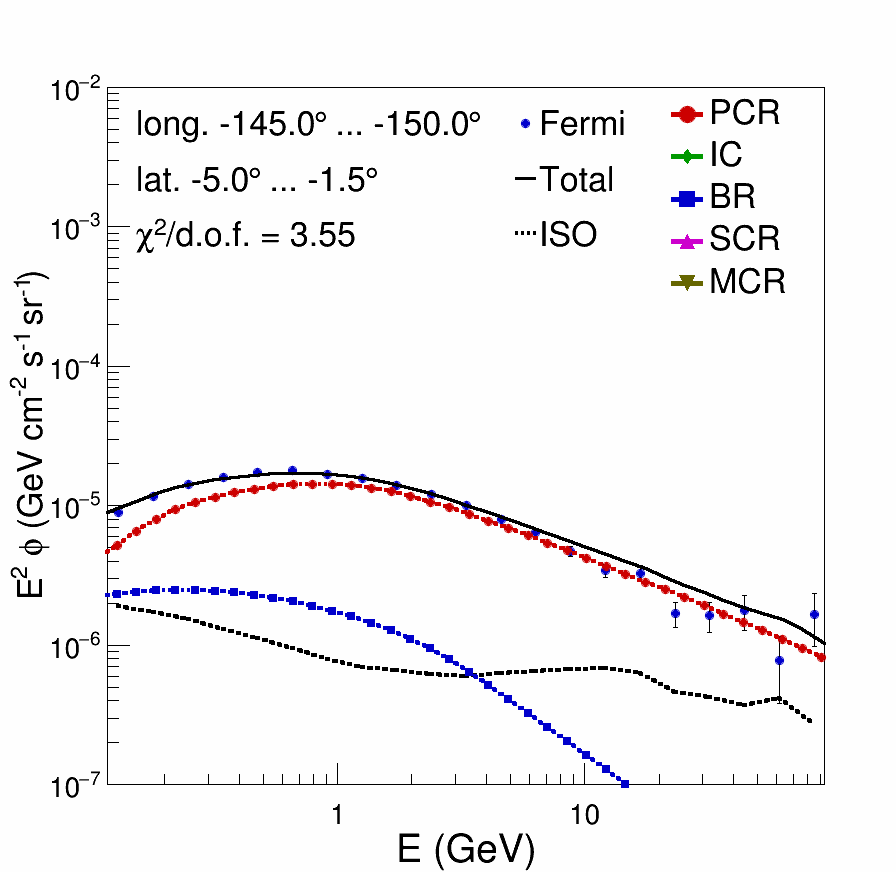}
\includegraphics[width=0.16\textwidth,height=0.16\textwidth,clip]{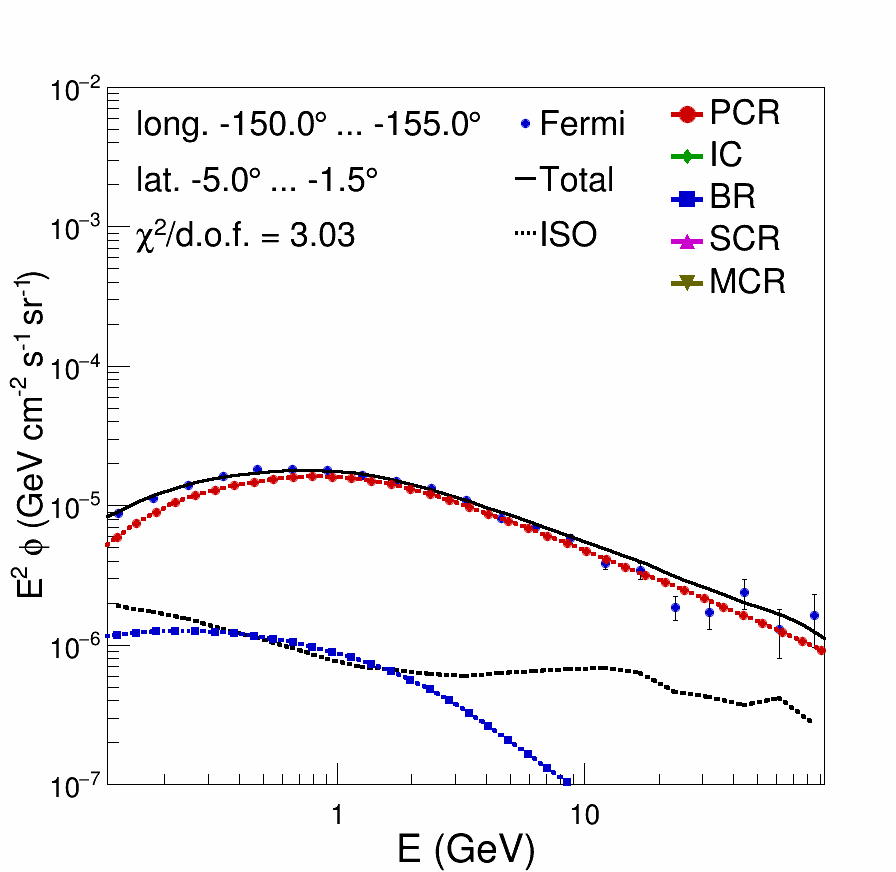}
\includegraphics[width=0.16\textwidth,height=0.16\textwidth,clip]{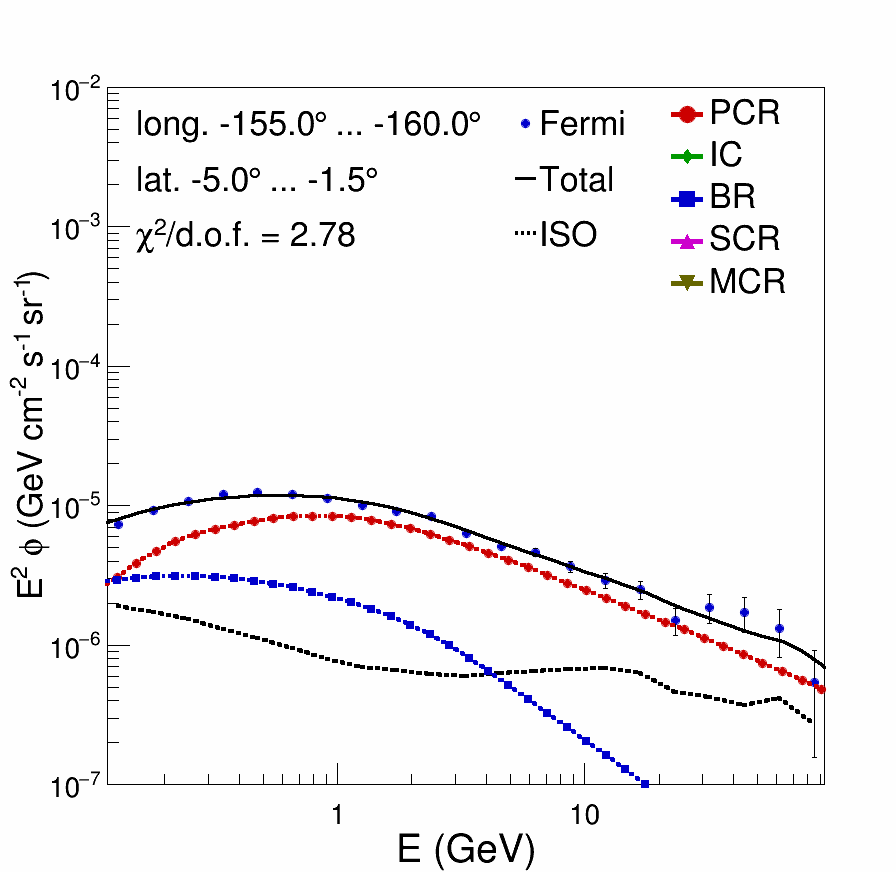}
\includegraphics[width=0.16\textwidth,height=0.16\textwidth,clip]{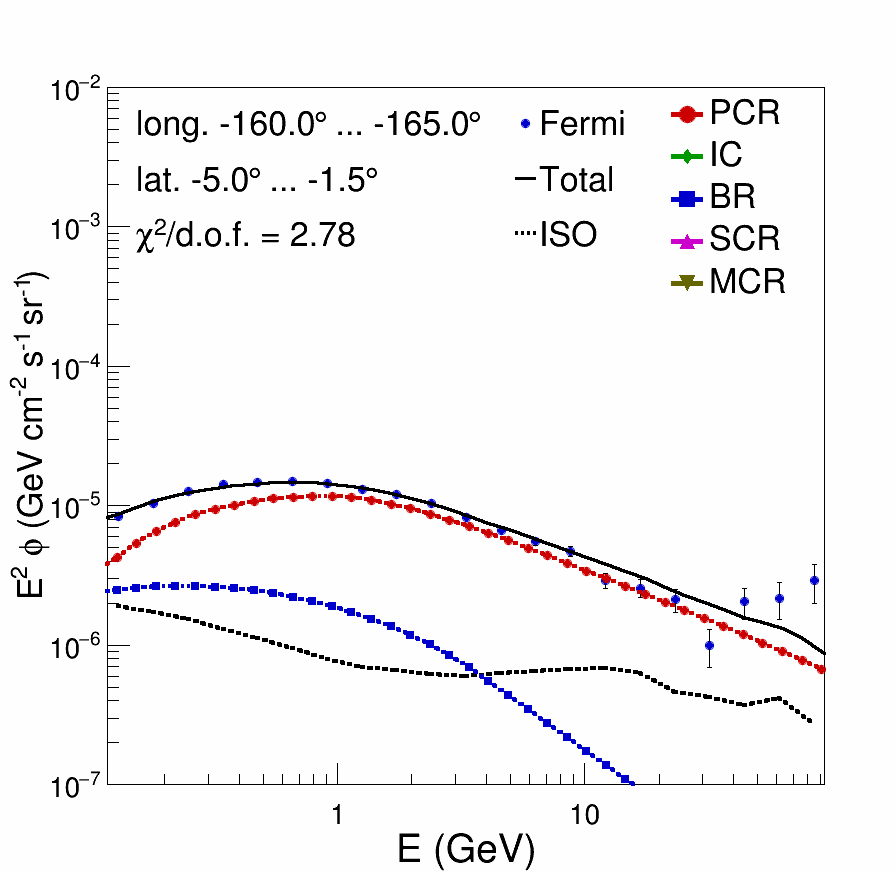}
\includegraphics[width=0.16\textwidth,height=0.16\textwidth,clip]{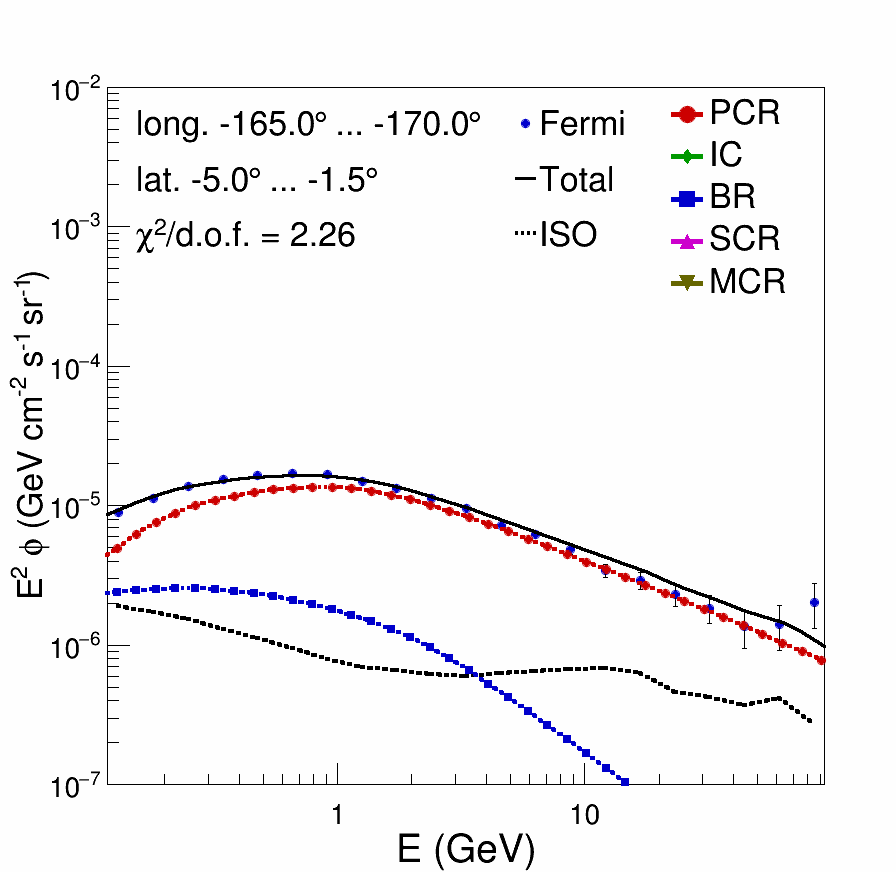}
\includegraphics[width=0.16\textwidth,height=0.16\textwidth,clip]{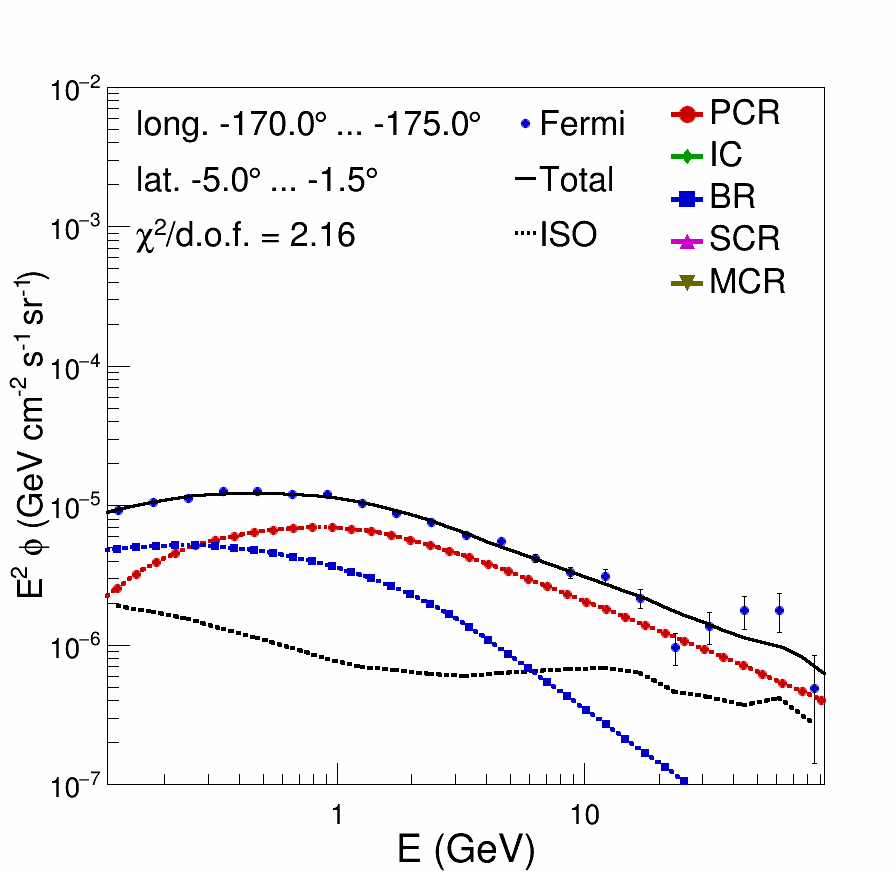}
\includegraphics[width=0.16\textwidth,height=0.16\textwidth,clip]{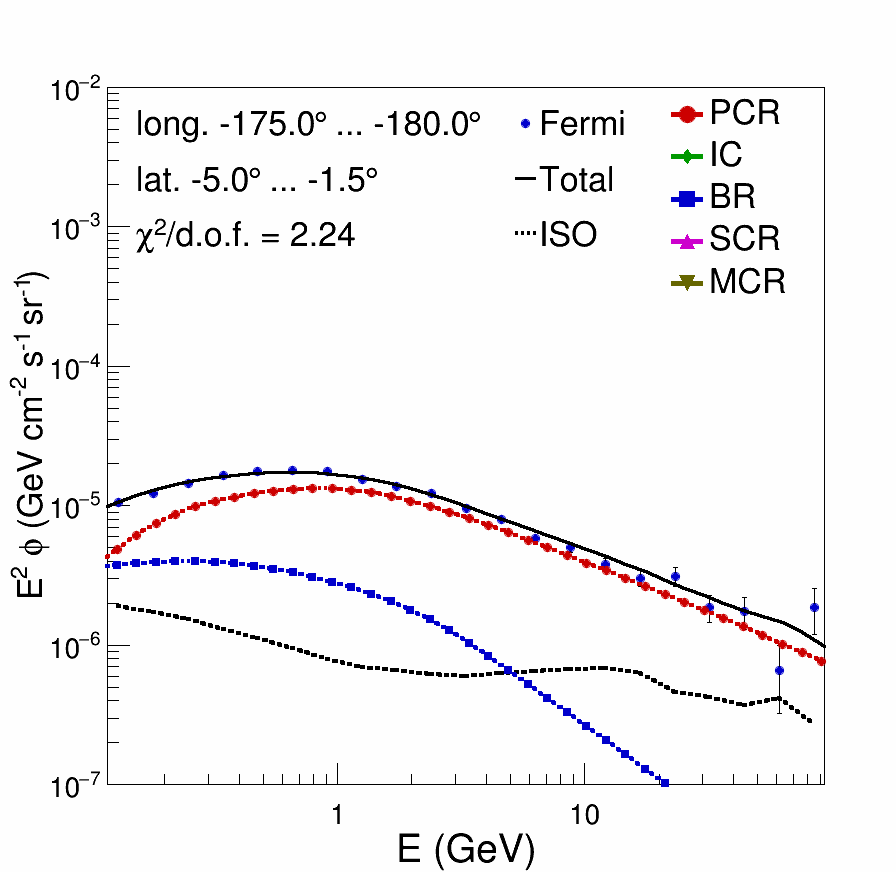}
\caption[]{Template fits for latitudes  with $-5.0^\circ<b<-1.5^\circ$ and longitudes decreasing from 0$^\circ$ to -180$^\circ$.} \label{F24}
\end{figure}
\begin{figure}
\centering
\includegraphics[width=0.16\textwidth,height=0.16\textwidth,clip]{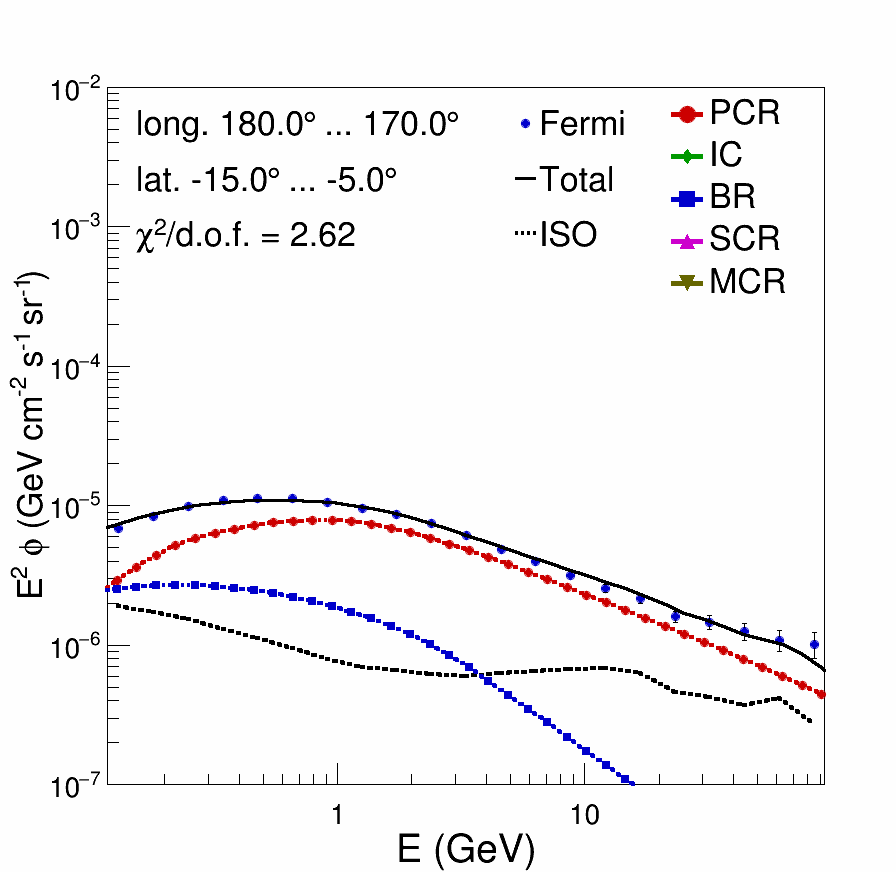}
\includegraphics[width=0.16\textwidth,height=0.16\textwidth,clip]{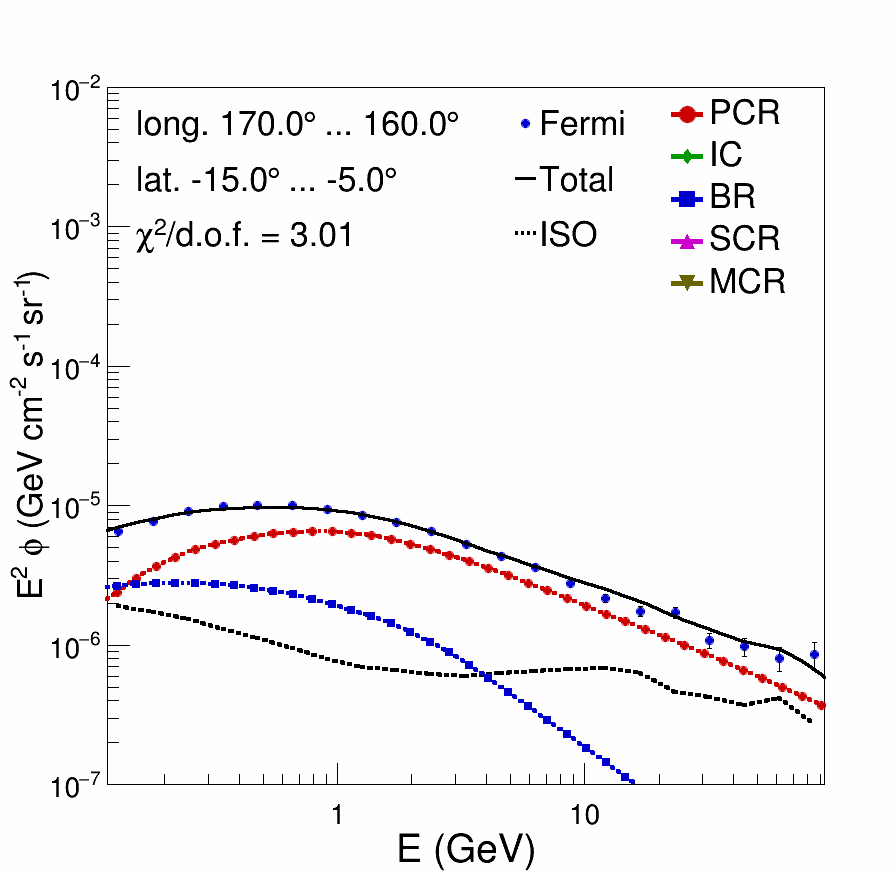}
\includegraphics[width=0.16\textwidth,height=0.16\textwidth,clip]{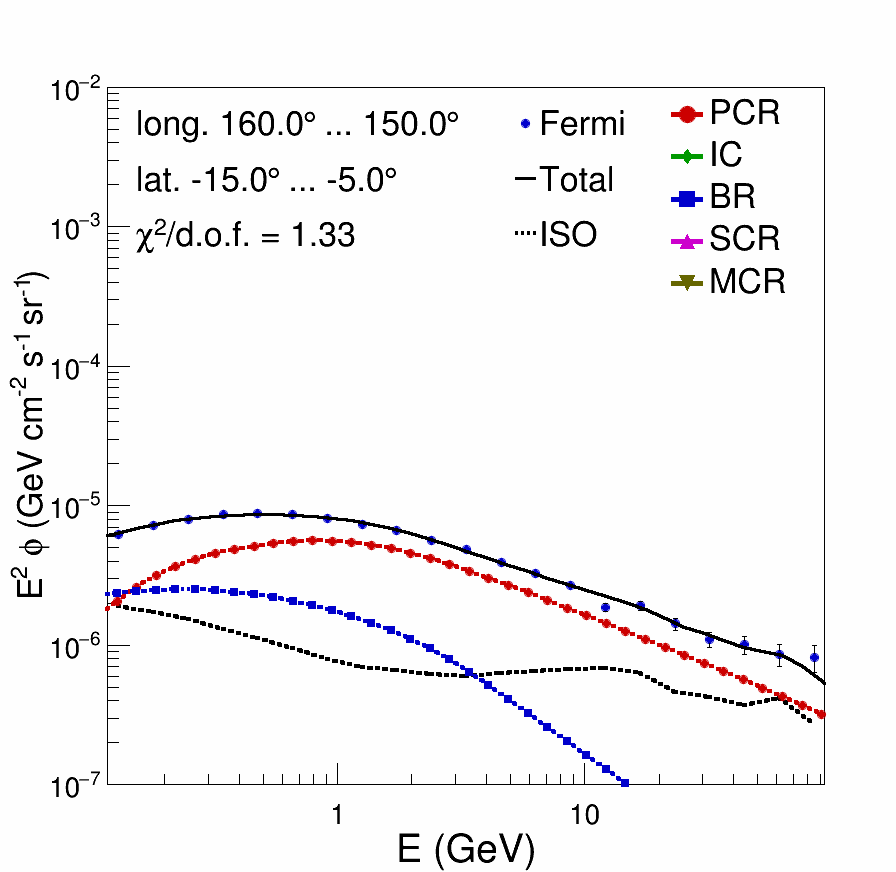}
\includegraphics[width=0.16\textwidth,height=0.16\textwidth,clip]{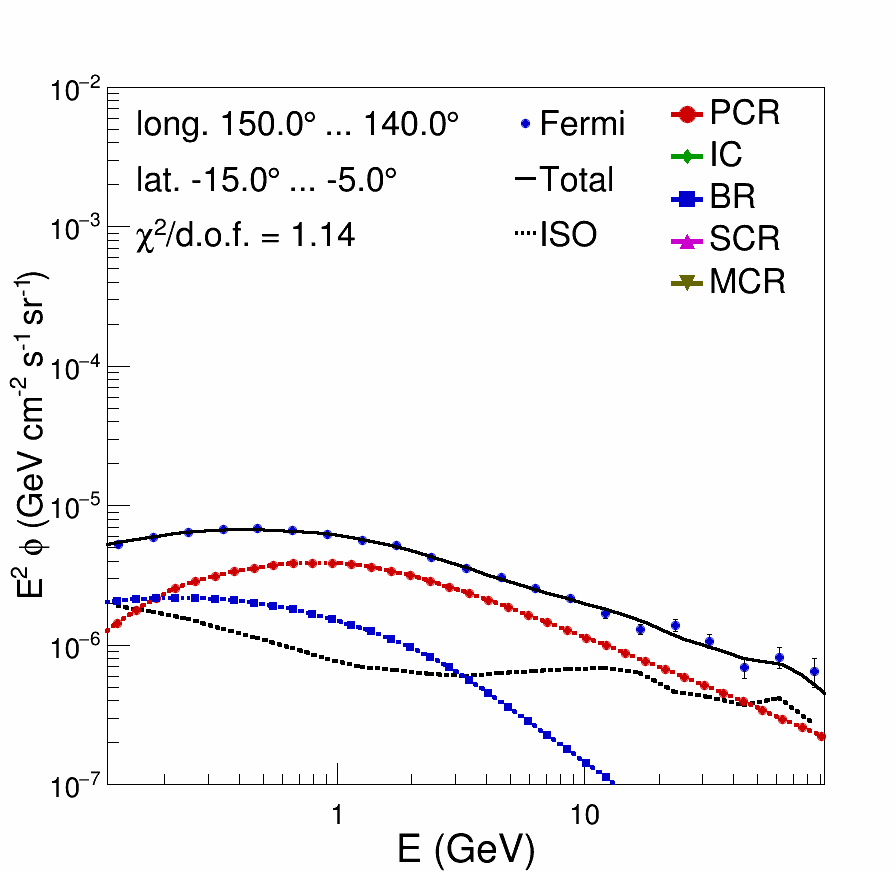}
\includegraphics[width=0.16\textwidth,height=0.16\textwidth,clip]{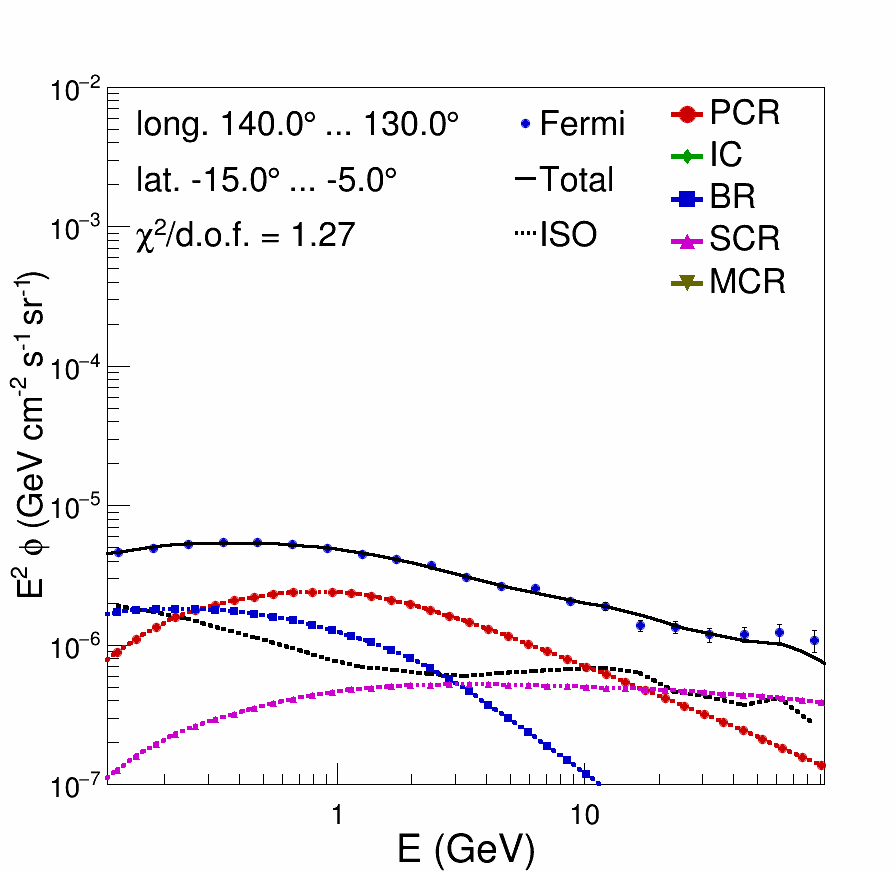}
\includegraphics[width=0.16\textwidth,height=0.16\textwidth,clip]{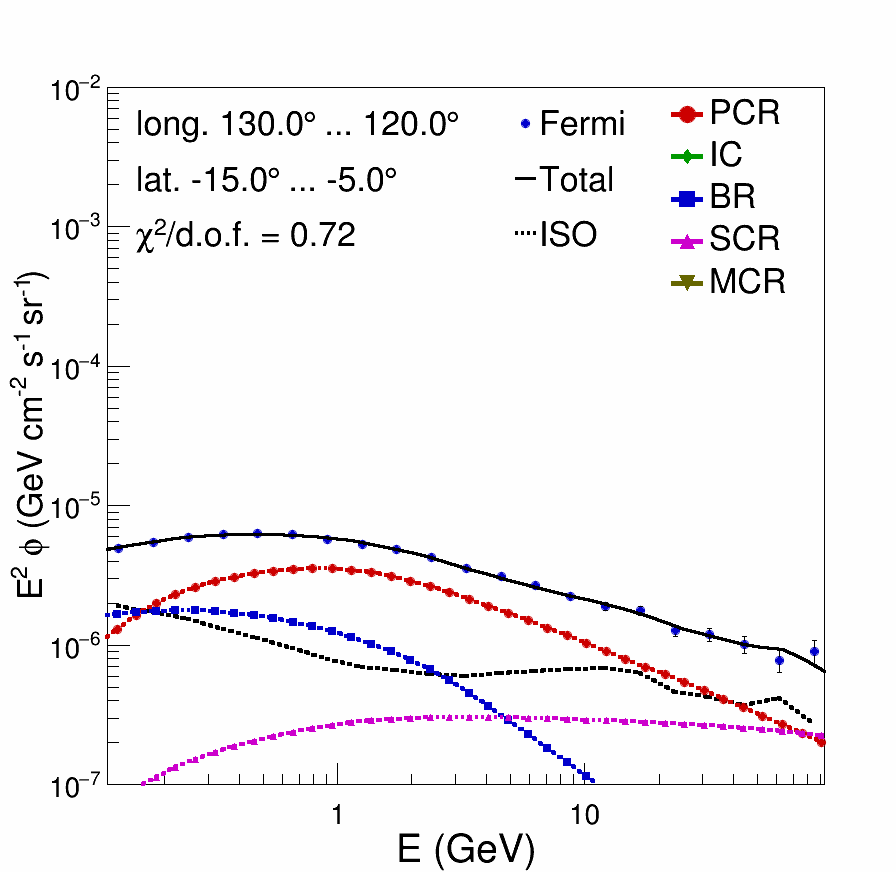}
\includegraphics[width=0.16\textwidth,height=0.16\textwidth,clip]{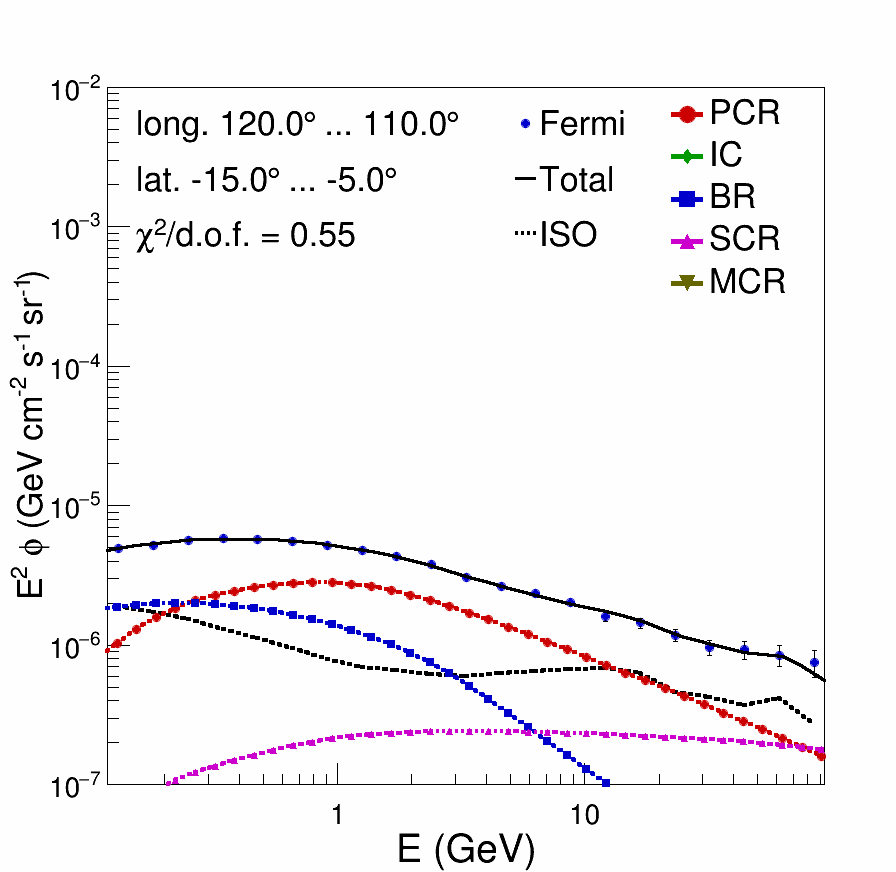}
\includegraphics[width=0.16\textwidth,height=0.16\textwidth,clip]{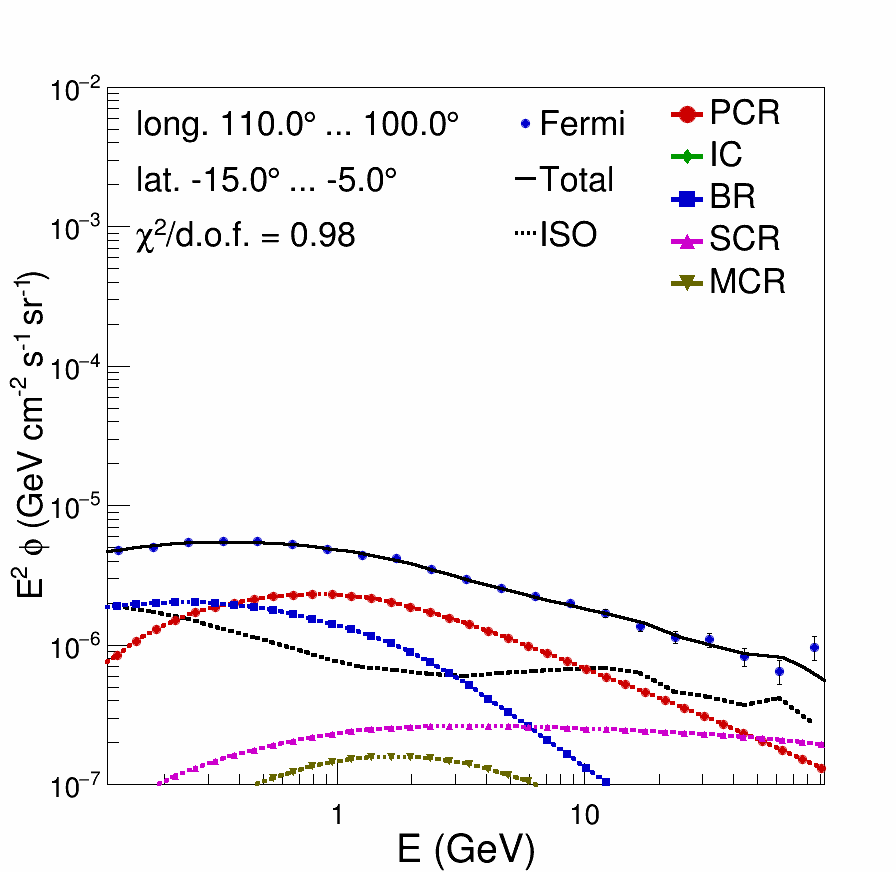}
\includegraphics[width=0.16\textwidth,height=0.16\textwidth,clip]{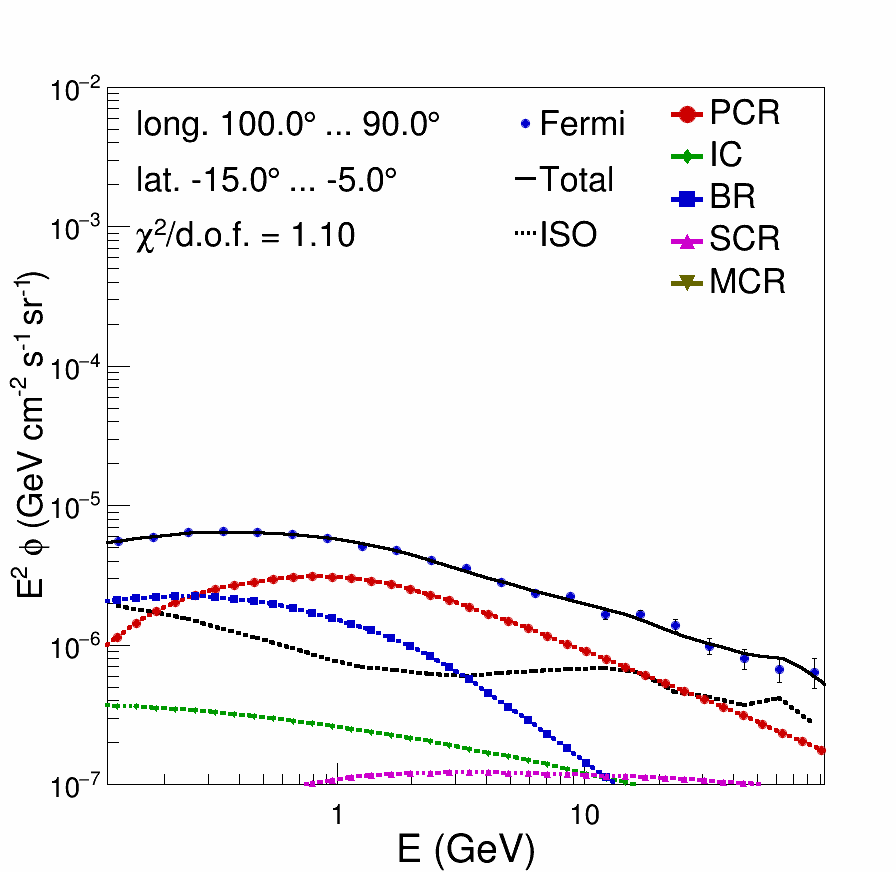}
\includegraphics[width=0.16\textwidth,height=0.16\textwidth,clip]{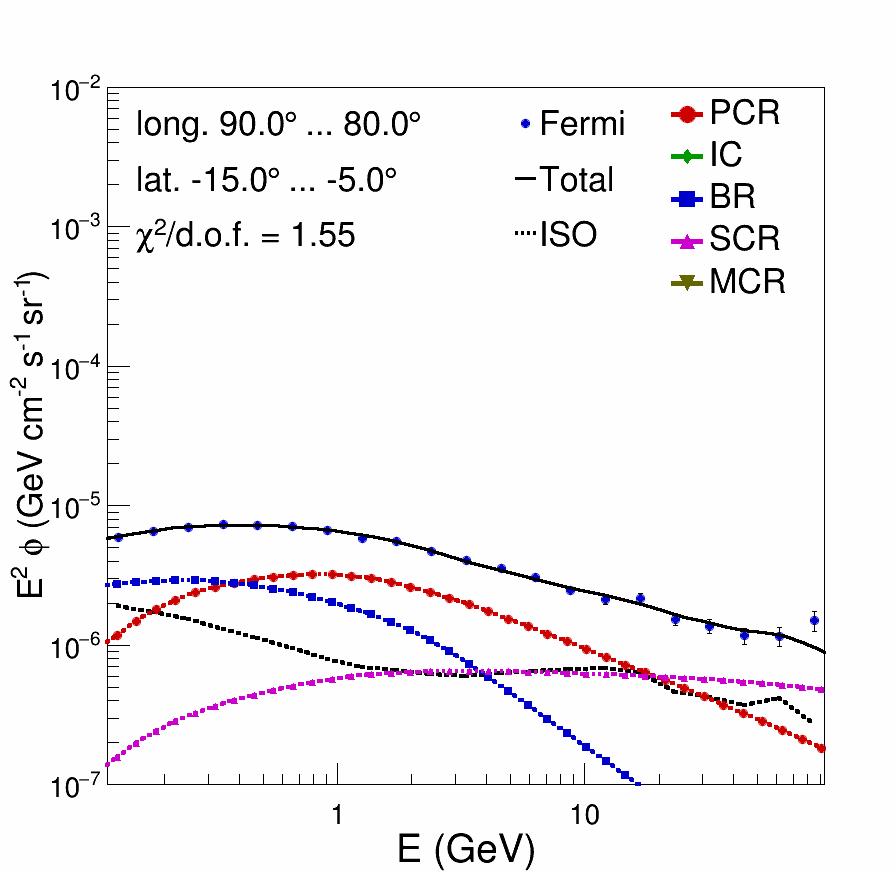}
\includegraphics[width=0.16\textwidth,height=0.16\textwidth,clip]{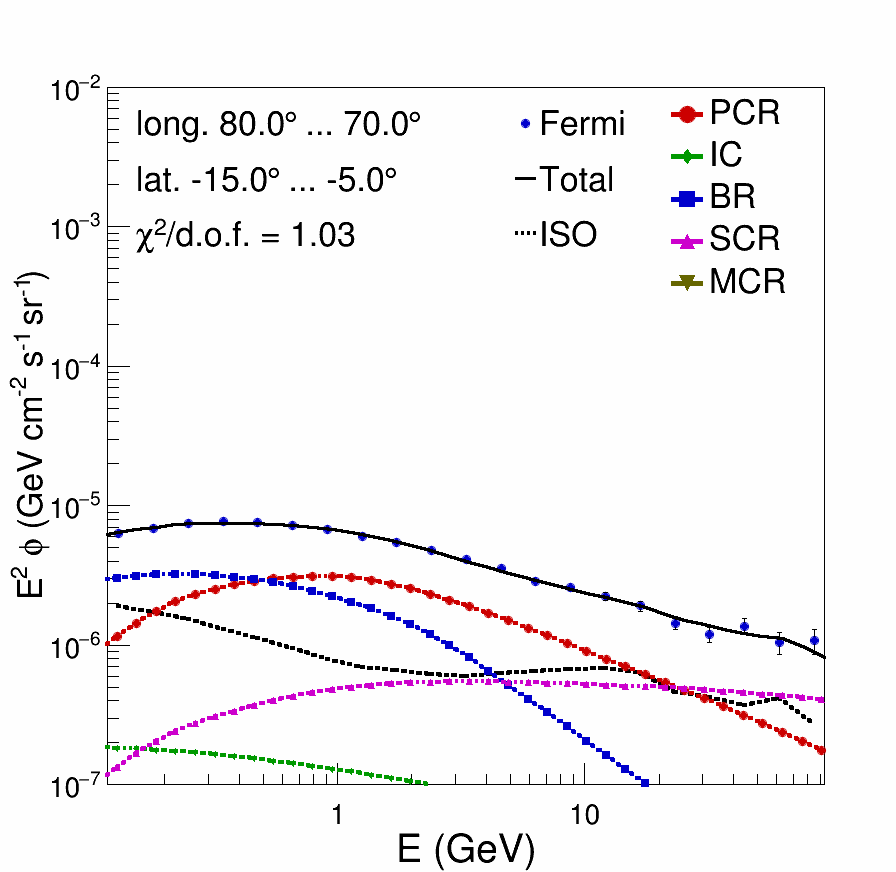}
\includegraphics[width=0.16\textwidth,height=0.16\textwidth,clip]{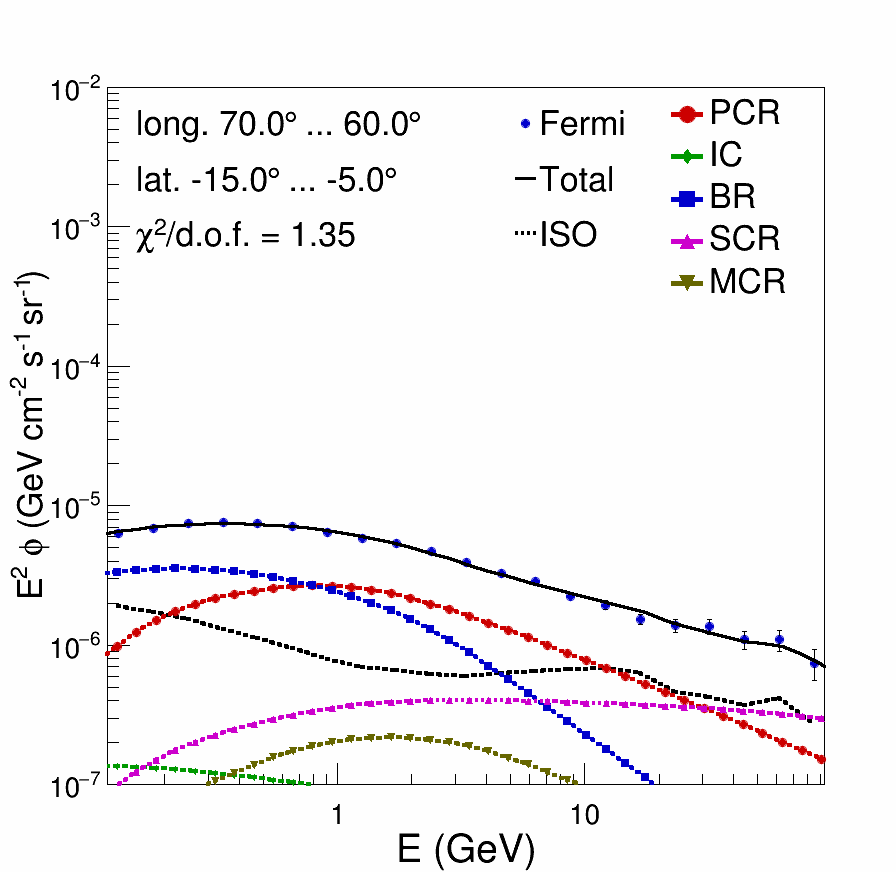}
\includegraphics[width=0.16\textwidth,height=0.16\textwidth,clip]{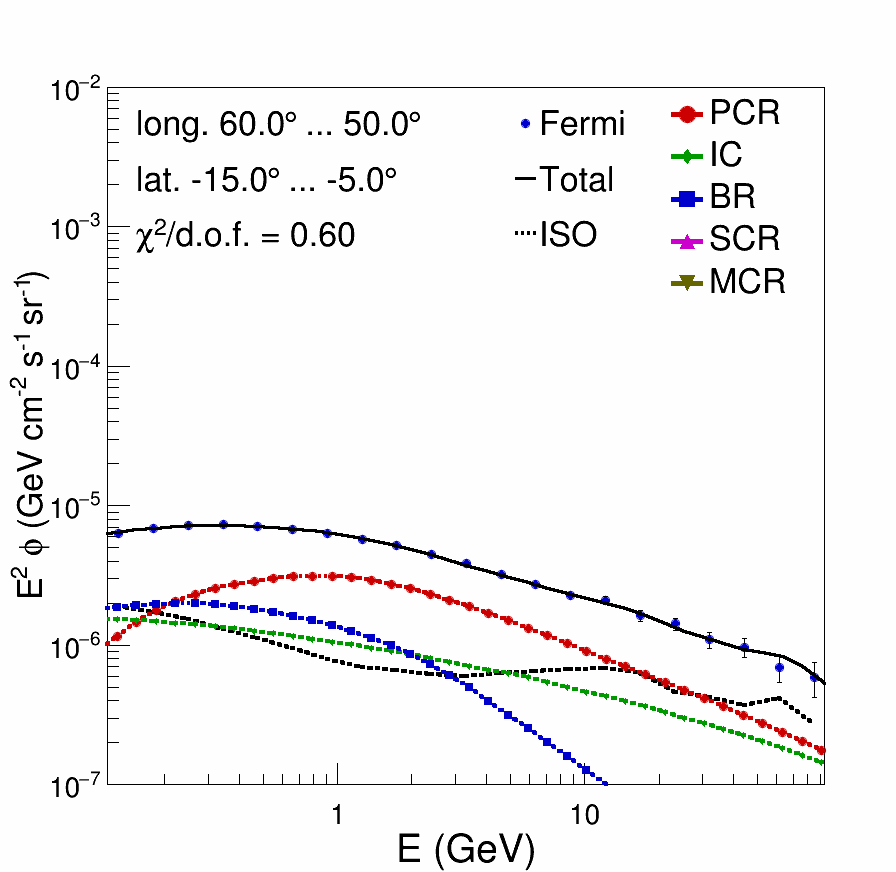}
\includegraphics[width=0.16\textwidth,height=0.16\textwidth,clip]{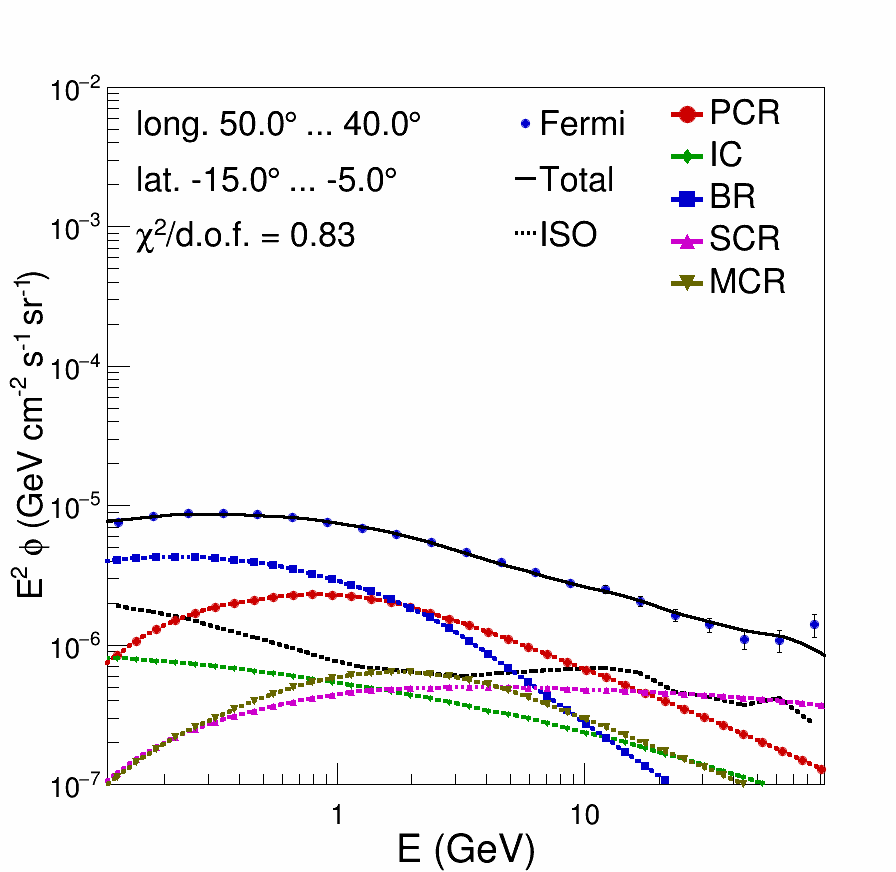}
\includegraphics[width=0.16\textwidth,height=0.16\textwidth,clip]{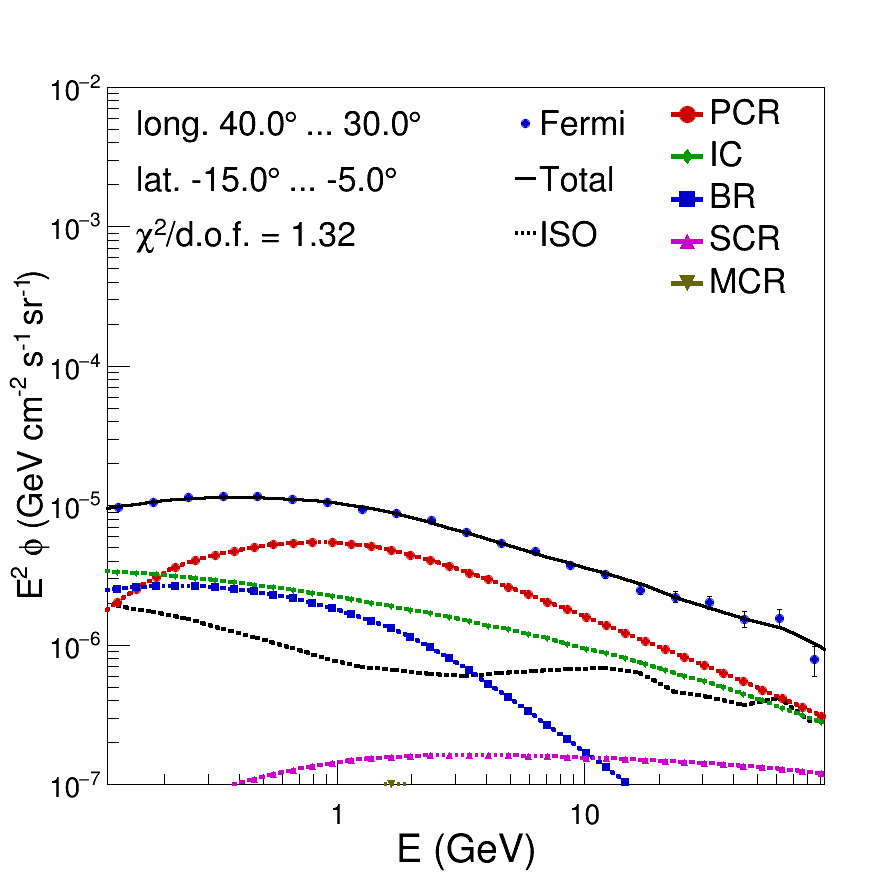}
\includegraphics[width=0.16\textwidth,height=0.16\textwidth,clip]{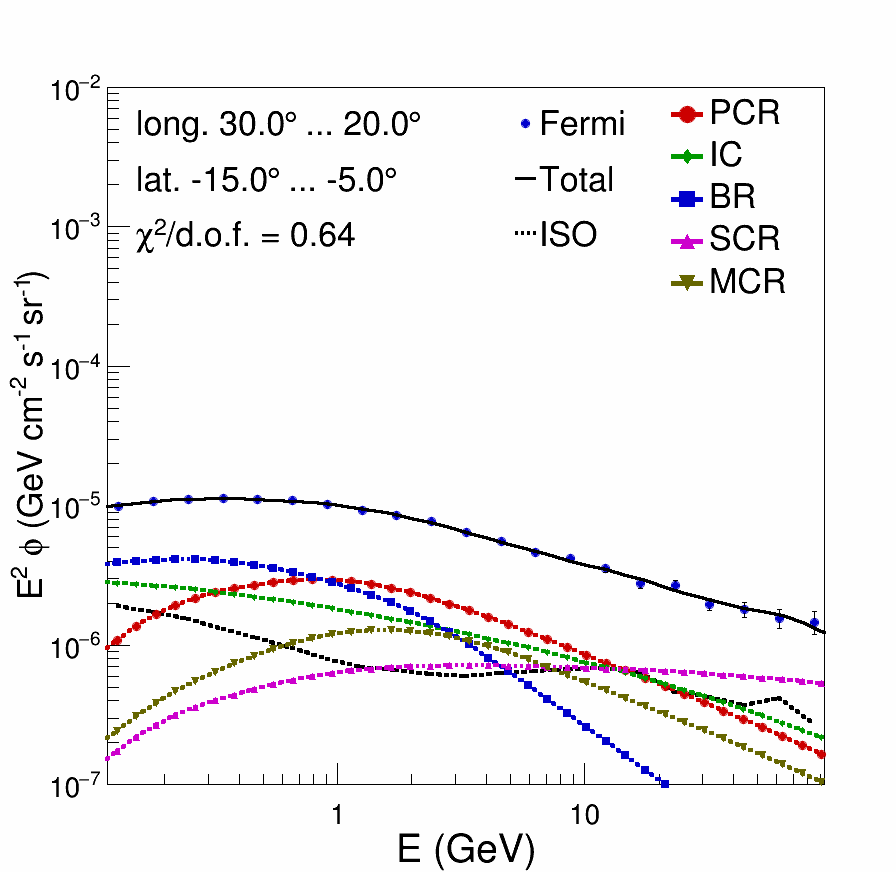}
\includegraphics[width=0.16\textwidth,height=0.16\textwidth,clip]{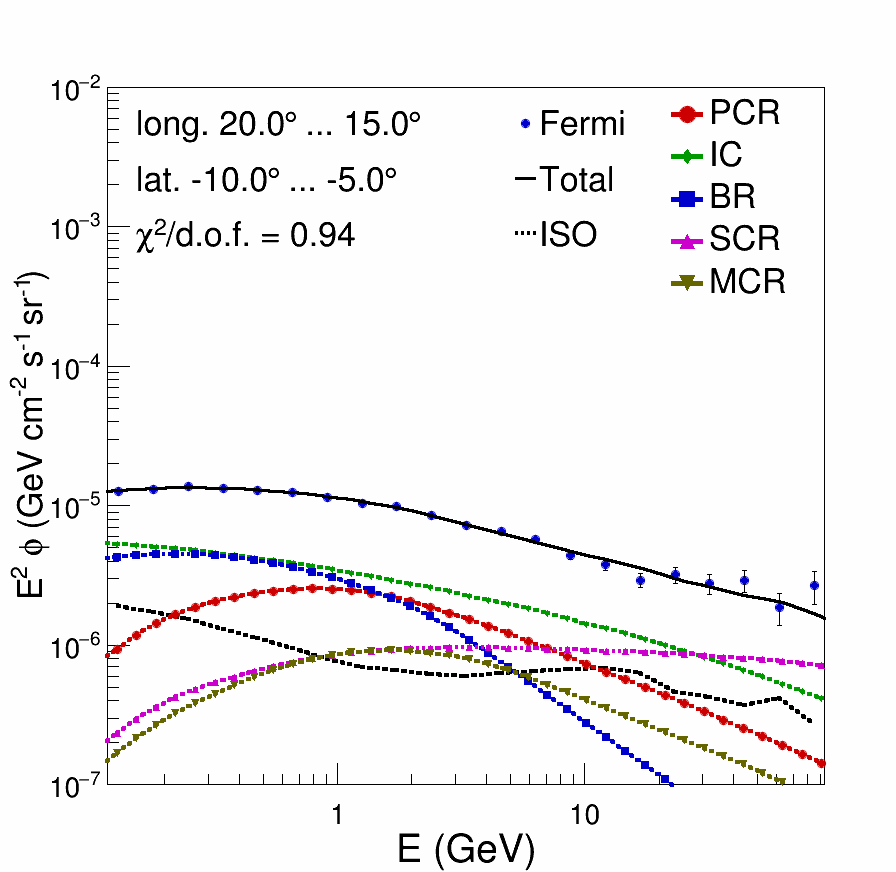}
\includegraphics[width=0.16\textwidth,height=0.16\textwidth,clip]{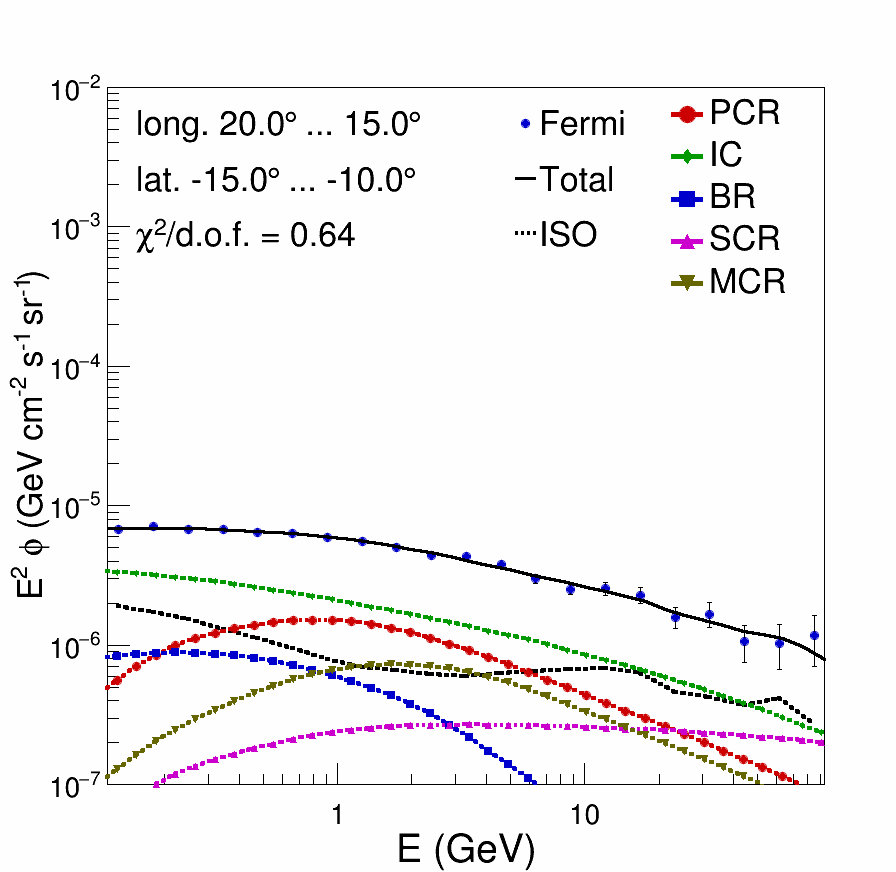}
\includegraphics[width=0.16\textwidth,height=0.16\textwidth,clip]{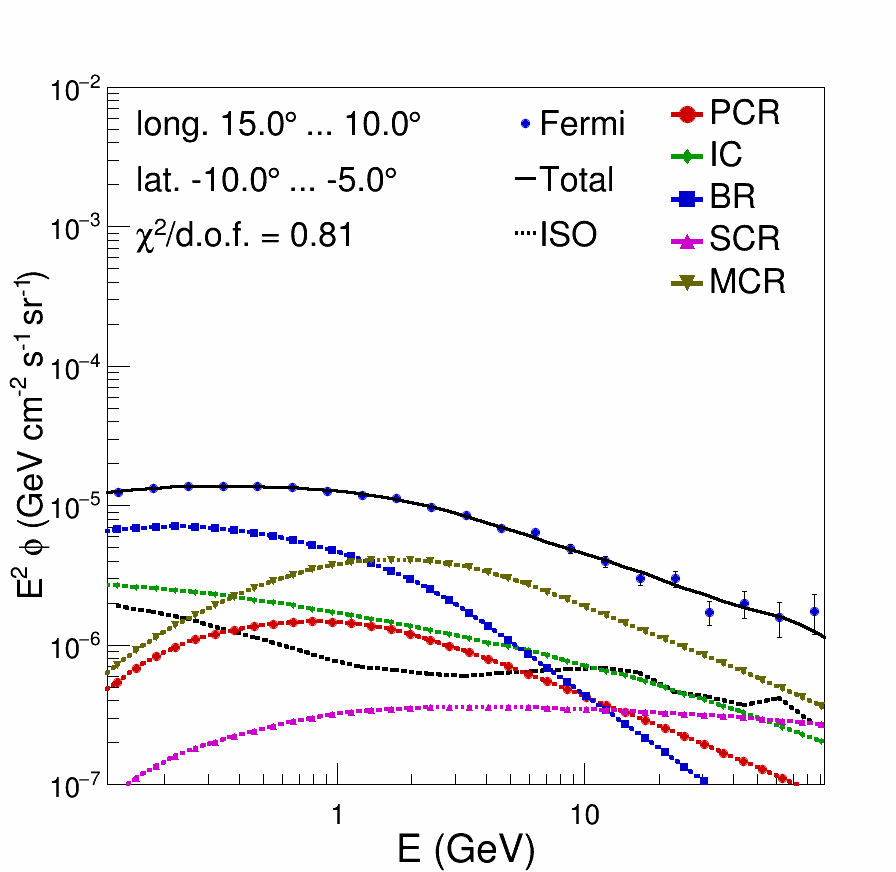}
\includegraphics[width=0.16\textwidth,height=0.16\textwidth,clip]{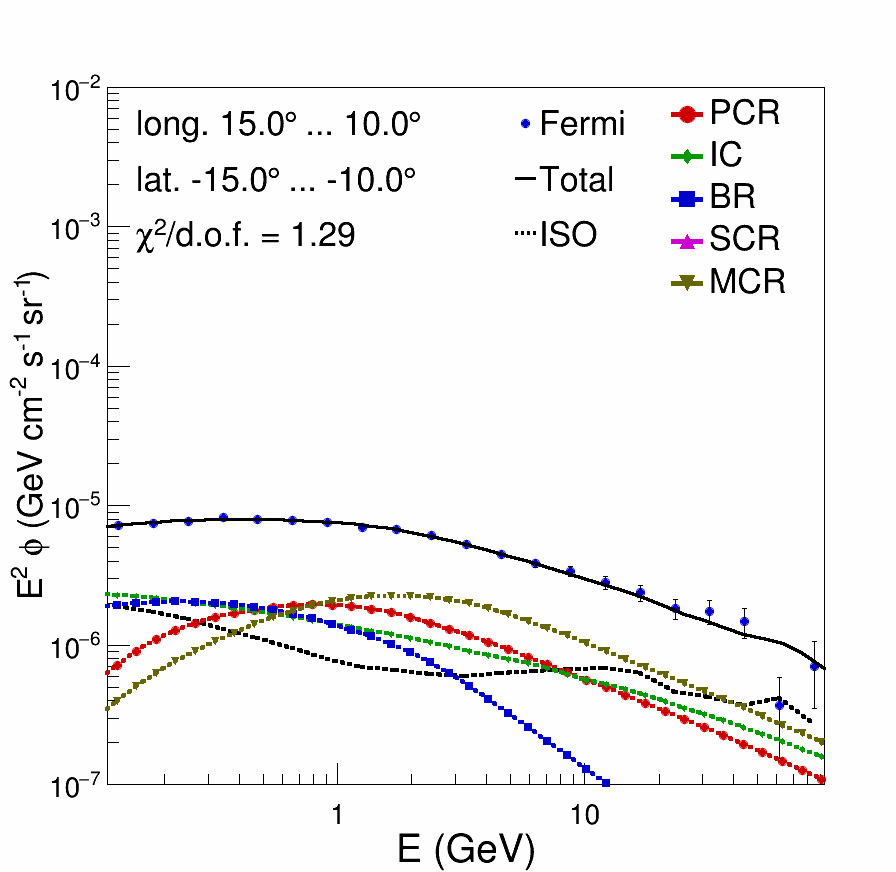}
\includegraphics[width=0.16\textwidth,height=0.16\textwidth,clip]{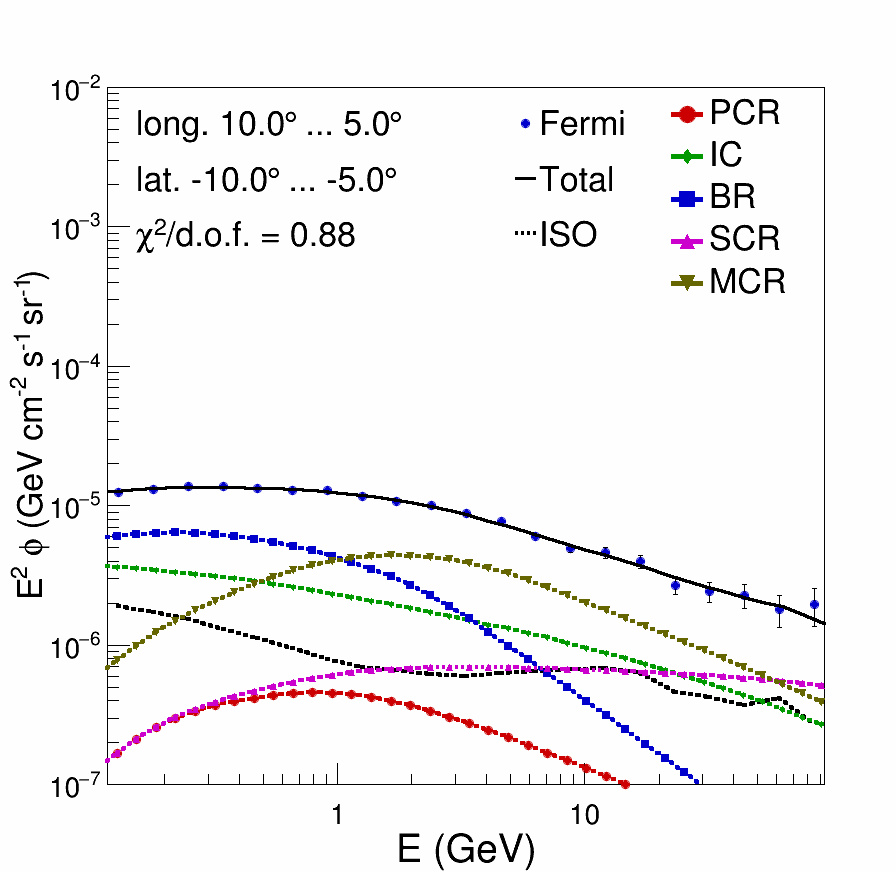}
\includegraphics[width=0.16\textwidth,height=0.16\textwidth,clip]{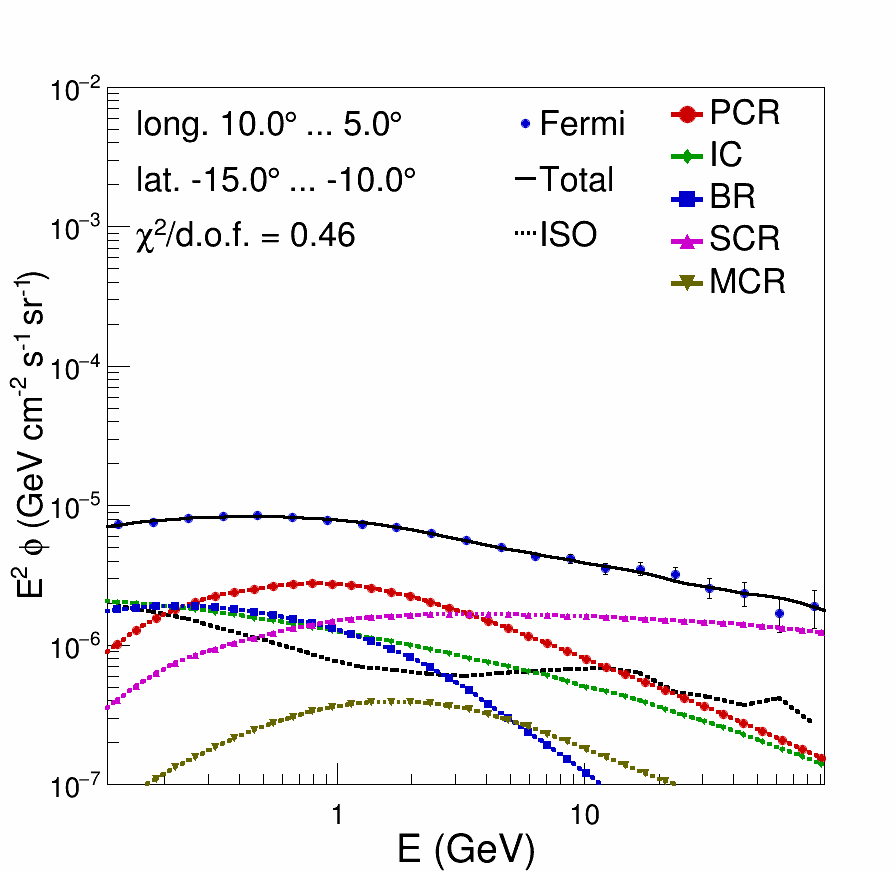}
\includegraphics[width=0.16\textwidth,height=0.16\textwidth,clip]{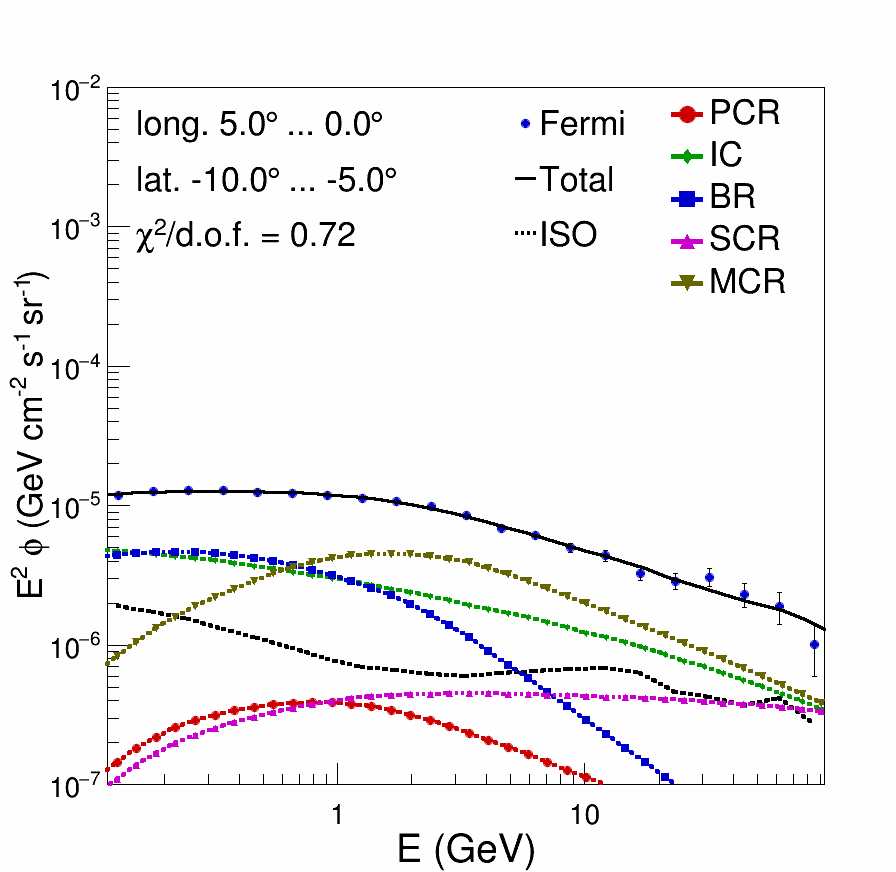}
\includegraphics[width=0.16\textwidth,height=0.16\textwidth,clip]{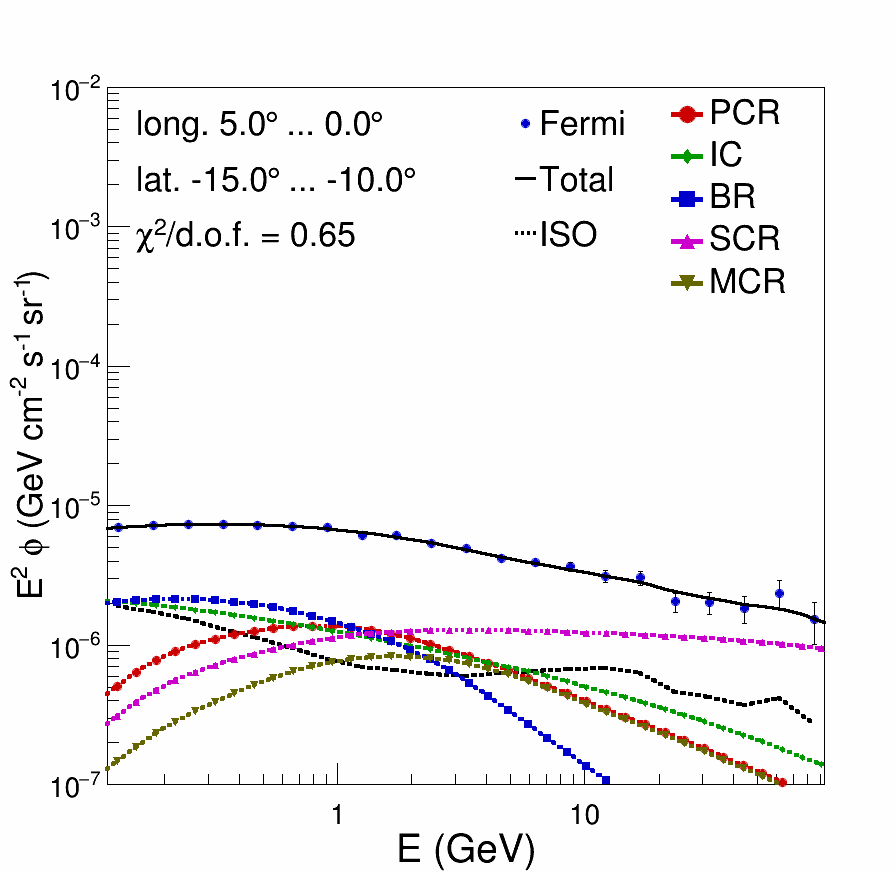}
\includegraphics[width=0.16\textwidth,height=0.16\textwidth,clip]{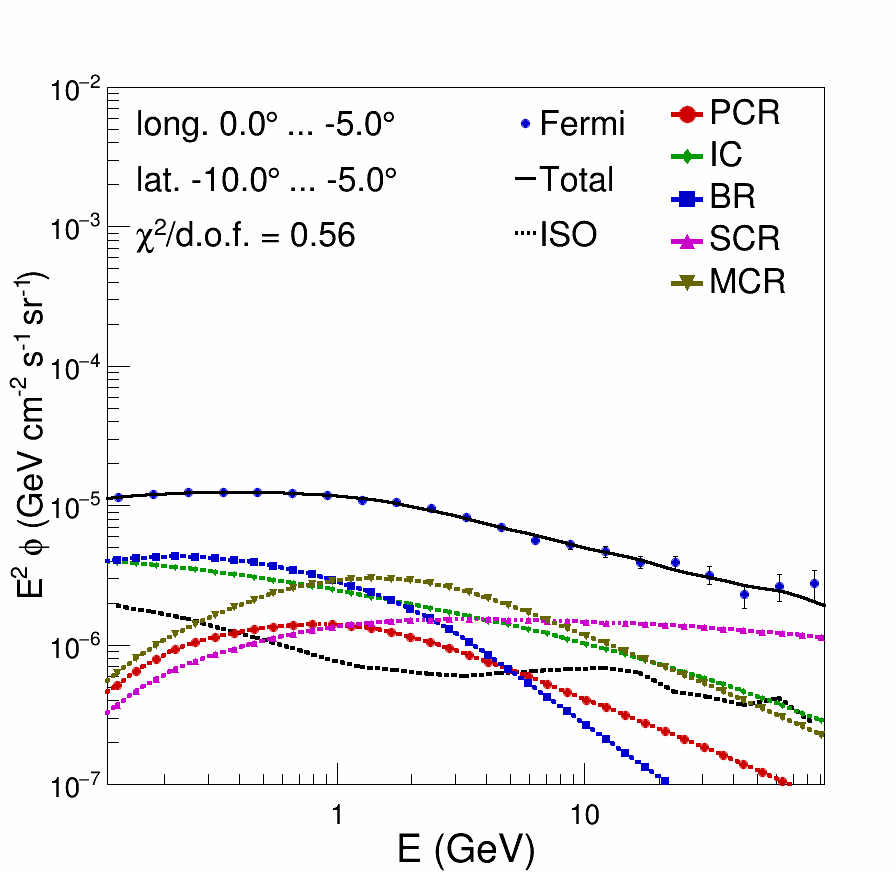}
\includegraphics[width=0.16\textwidth,height=0.16\textwidth,clip]{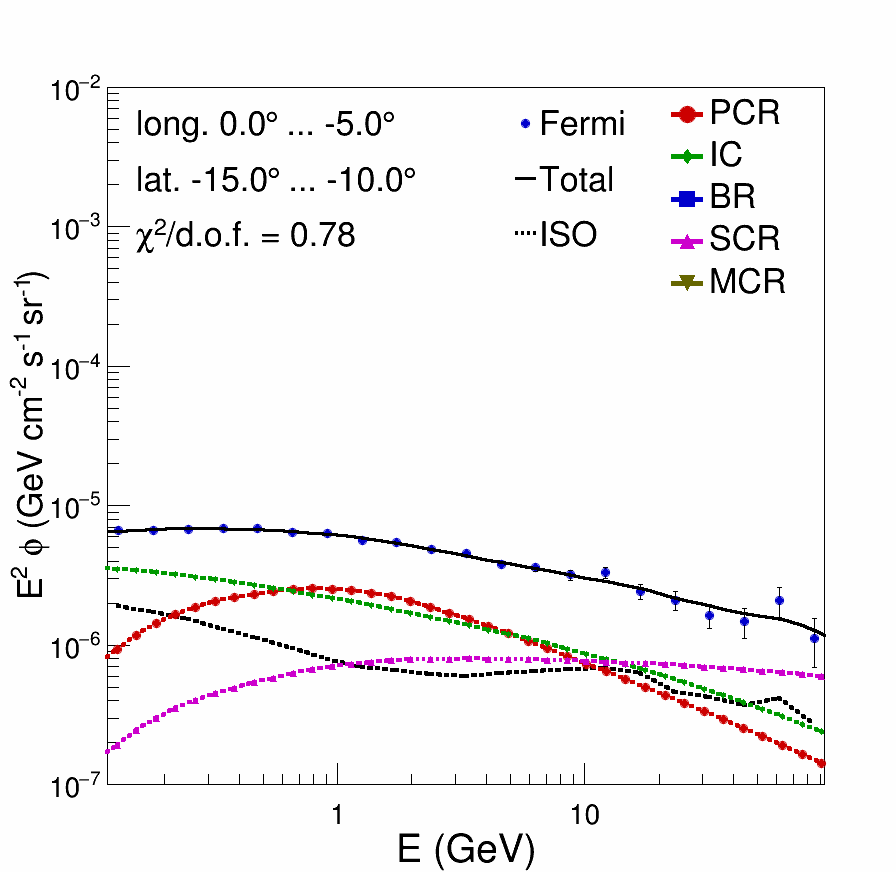}
\includegraphics[width=0.16\textwidth,height=0.16\textwidth,clip]{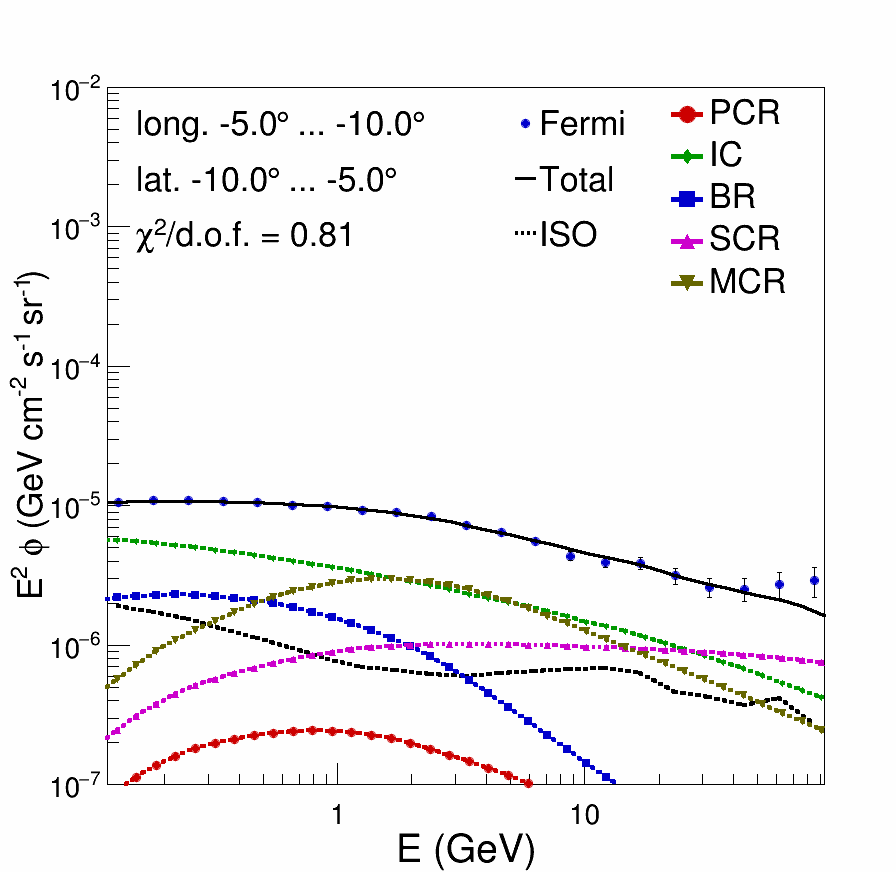}
\includegraphics[width=0.16\textwidth,height=0.16\textwidth,clip]{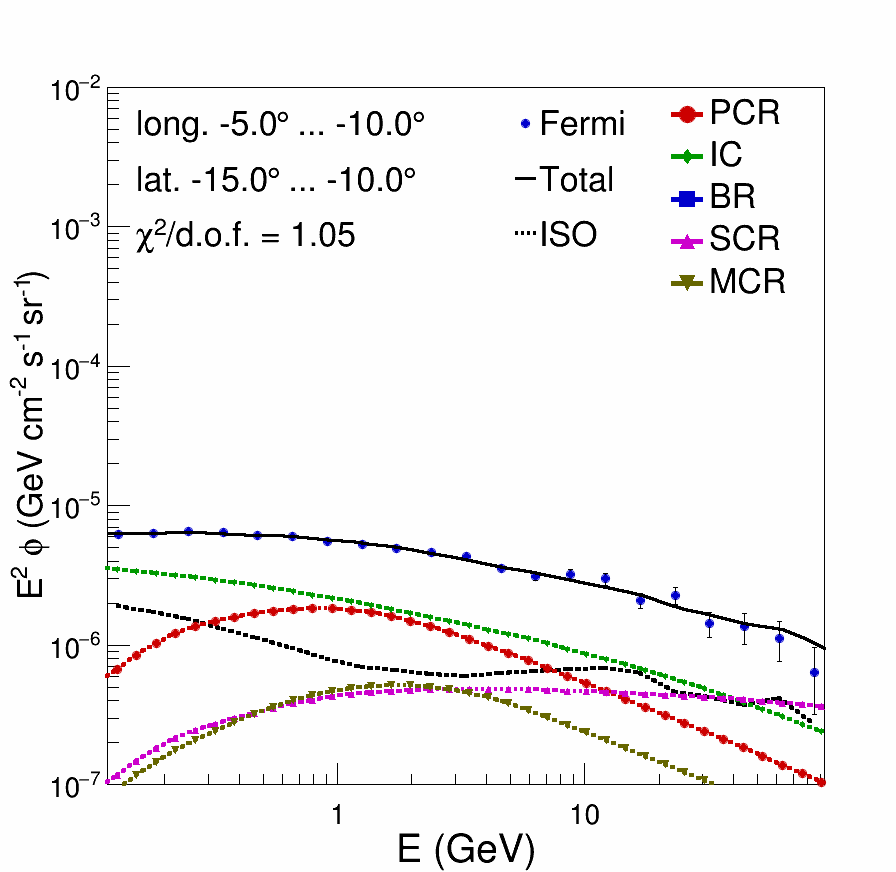}
\includegraphics[width=0.16\textwidth,height=0.16\textwidth,clip]{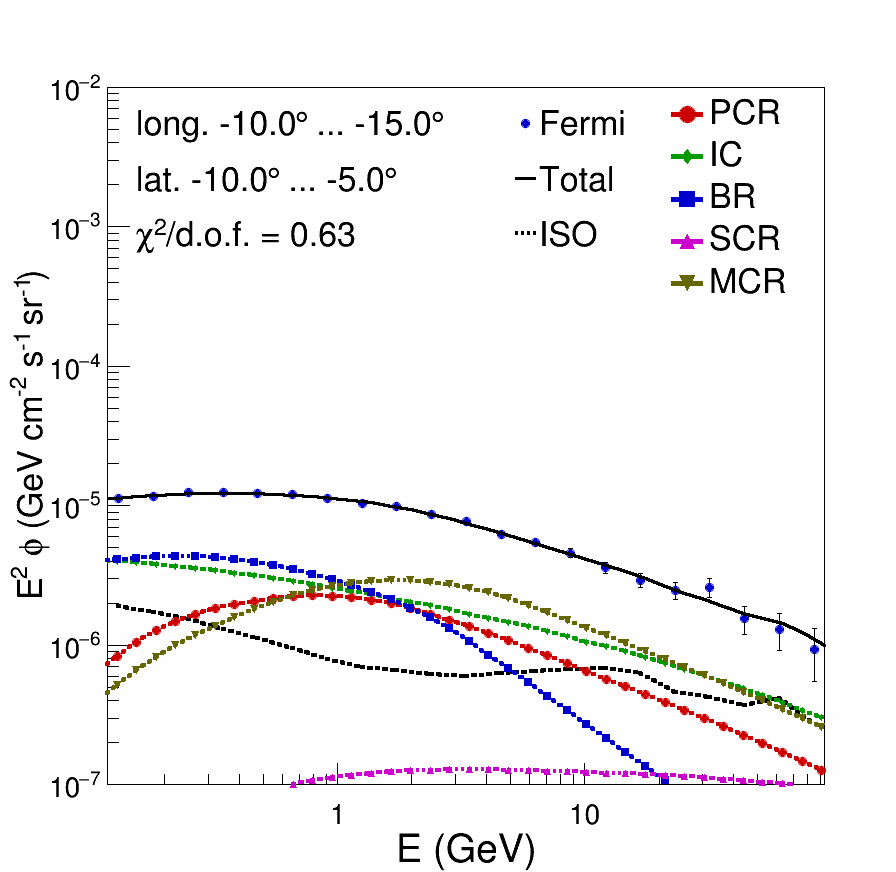}
\includegraphics[width=0.16\textwidth,height=0.16\textwidth,clip]{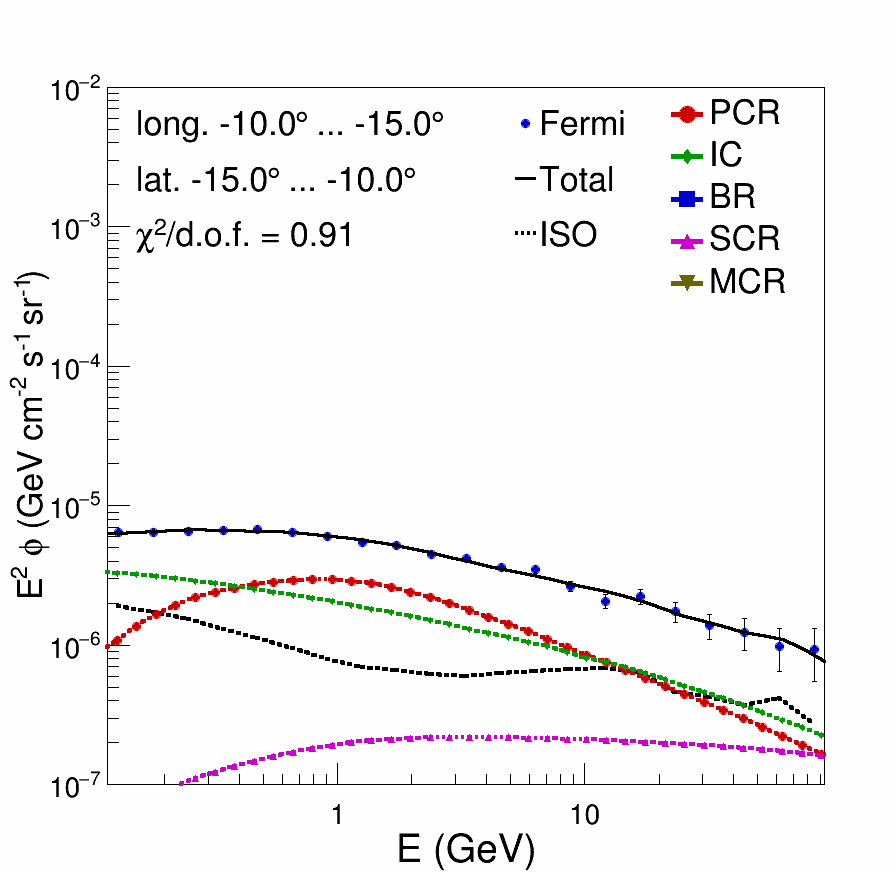}
\includegraphics[width=0.16\textwidth,height=0.16\textwidth,clip]{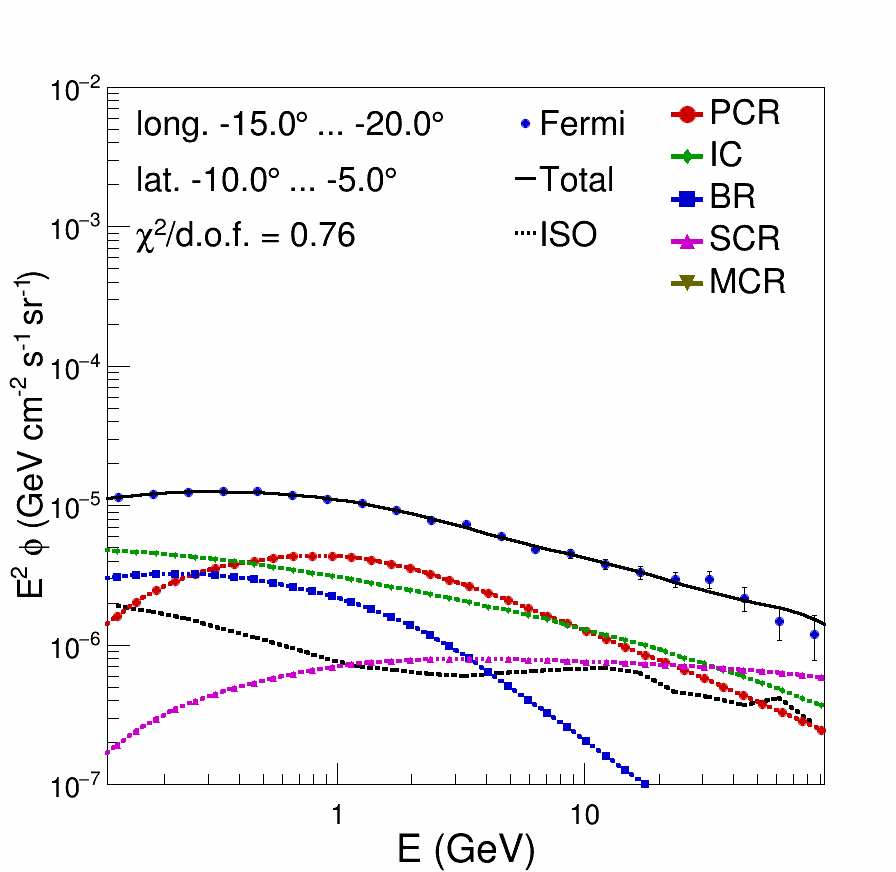}
\includegraphics[width=0.16\textwidth,height=0.16\textwidth,clip]{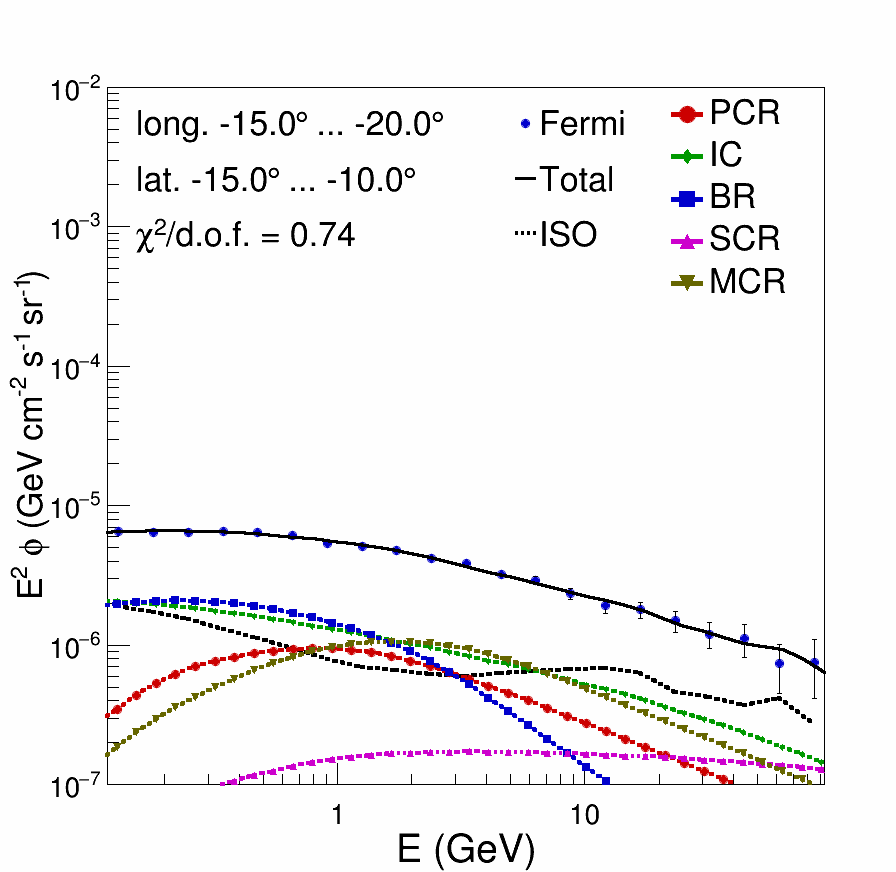}
\includegraphics[width=0.16\textwidth,height=0.16\textwidth,clip]{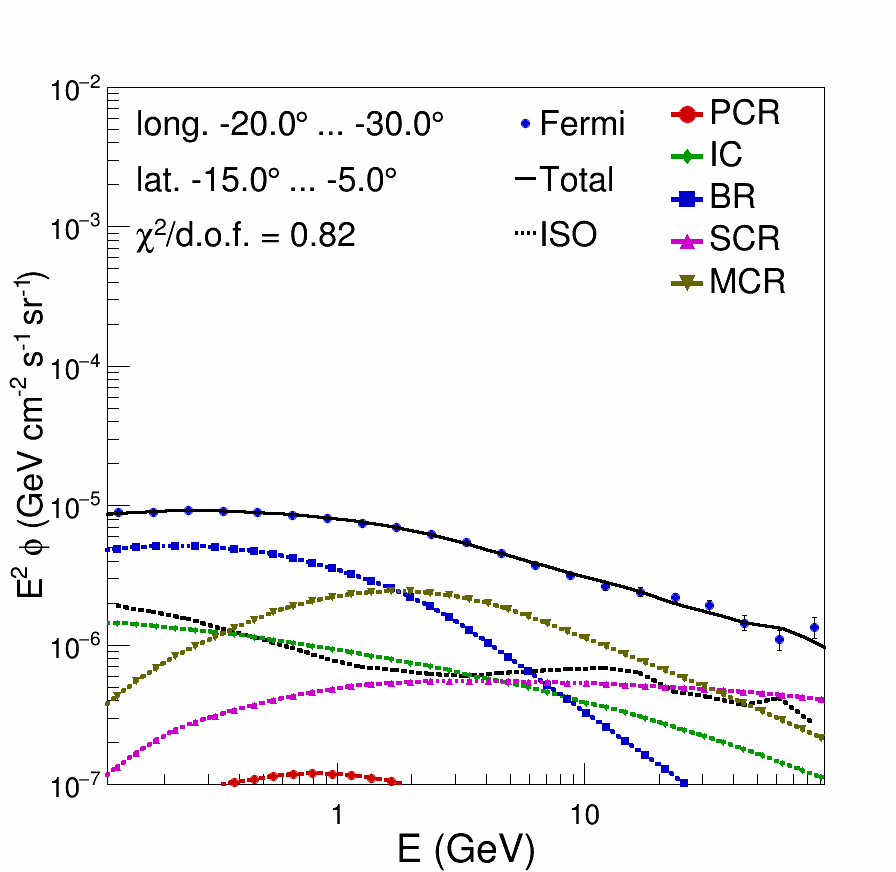}
\includegraphics[width=0.16\textwidth,height=0.16\textwidth,clip]{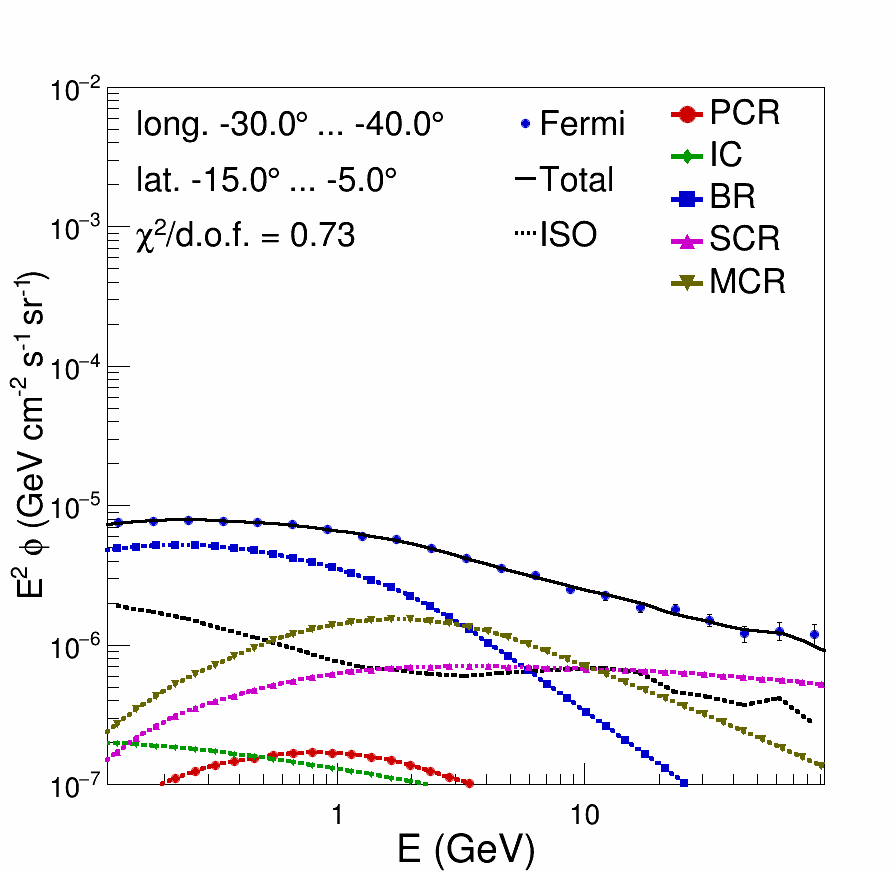}
\includegraphics[width=0.16\textwidth,height=0.16\textwidth,clip]{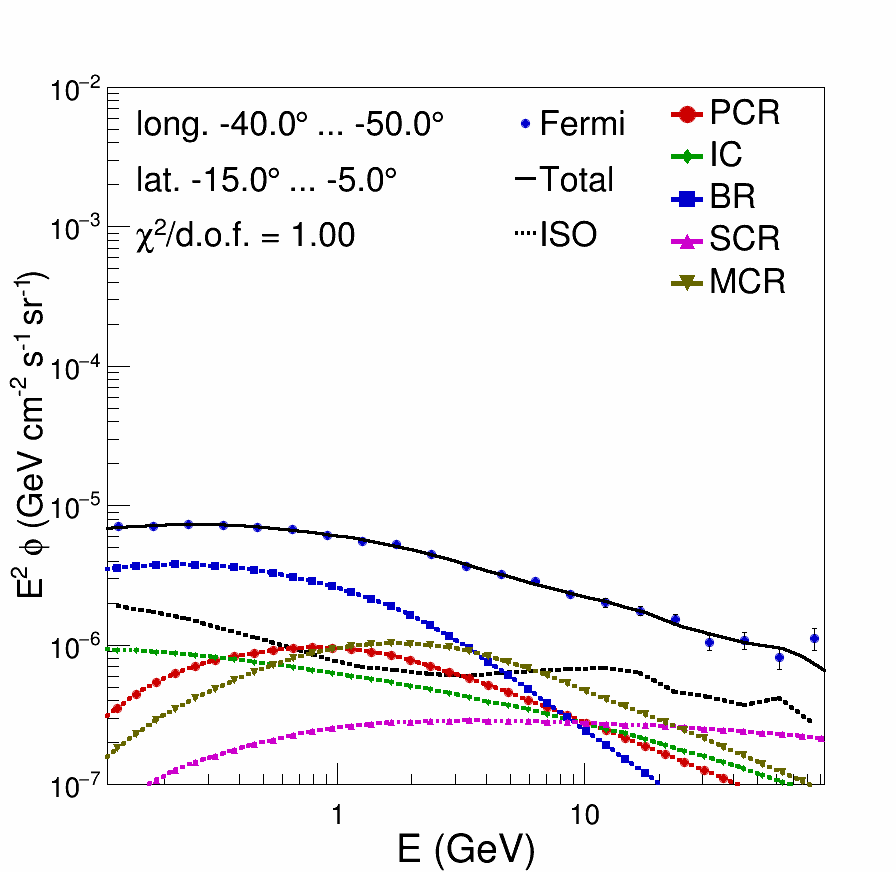}
\includegraphics[width=0.16\textwidth,height=0.16\textwidth,clip]{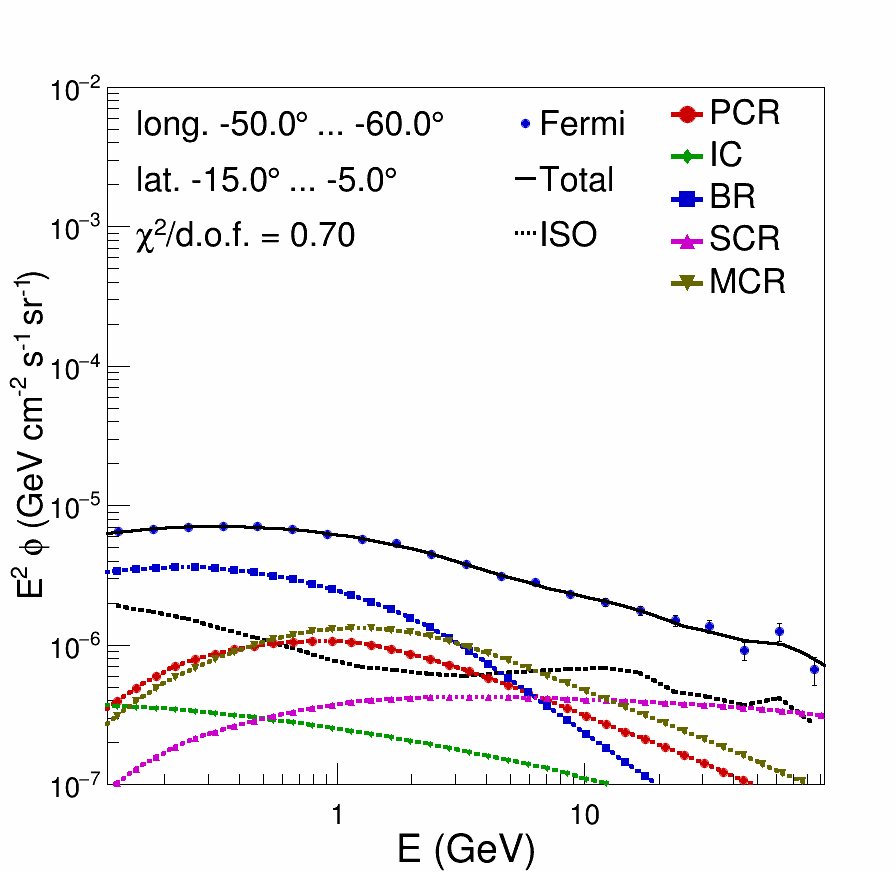}
\includegraphics[width=0.16\textwidth,height=0.16\textwidth,clip]{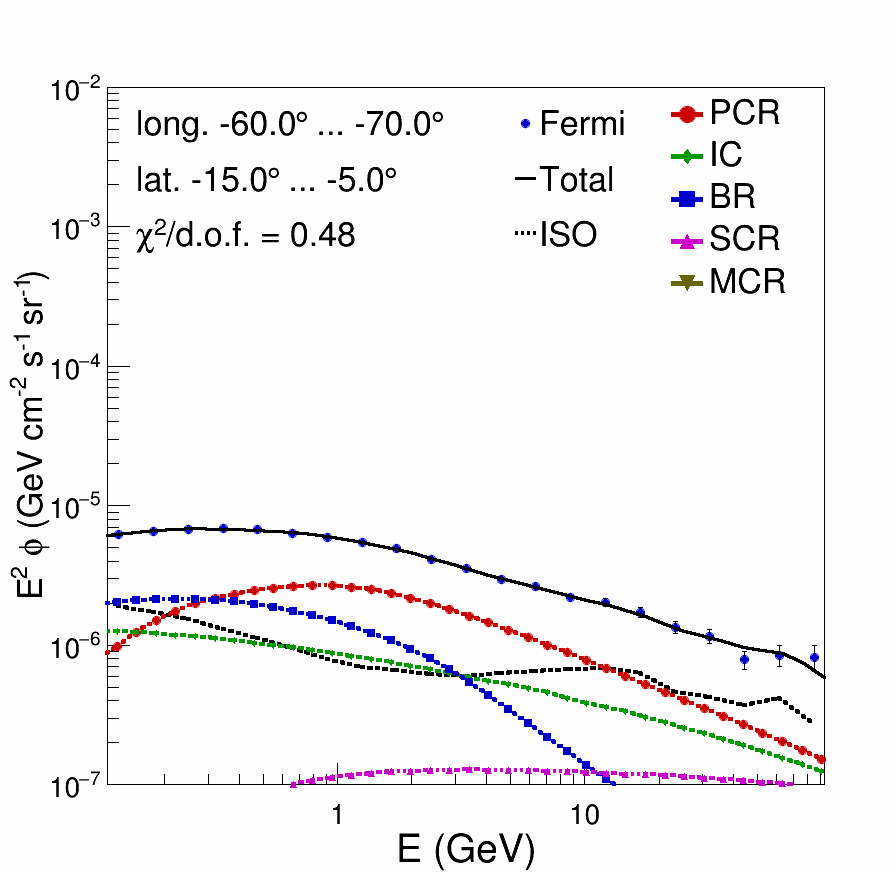}
\includegraphics[width=0.16\textwidth,height=0.16\textwidth,clip]{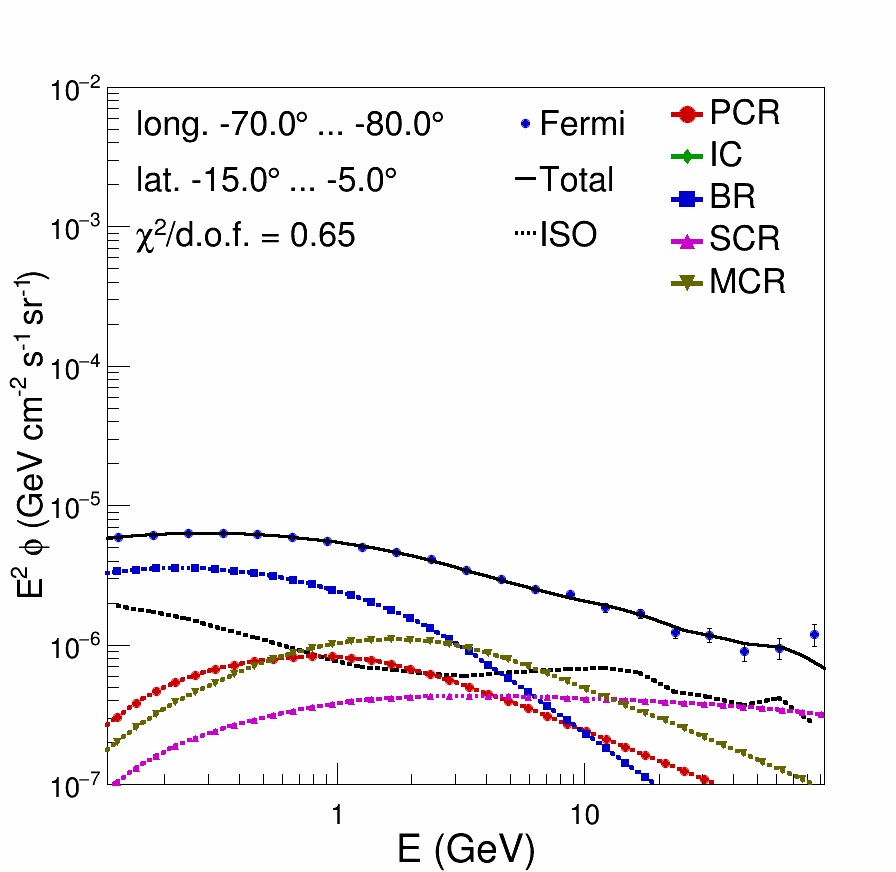}
\includegraphics[width=0.16\textwidth,height=0.16\textwidth,clip]{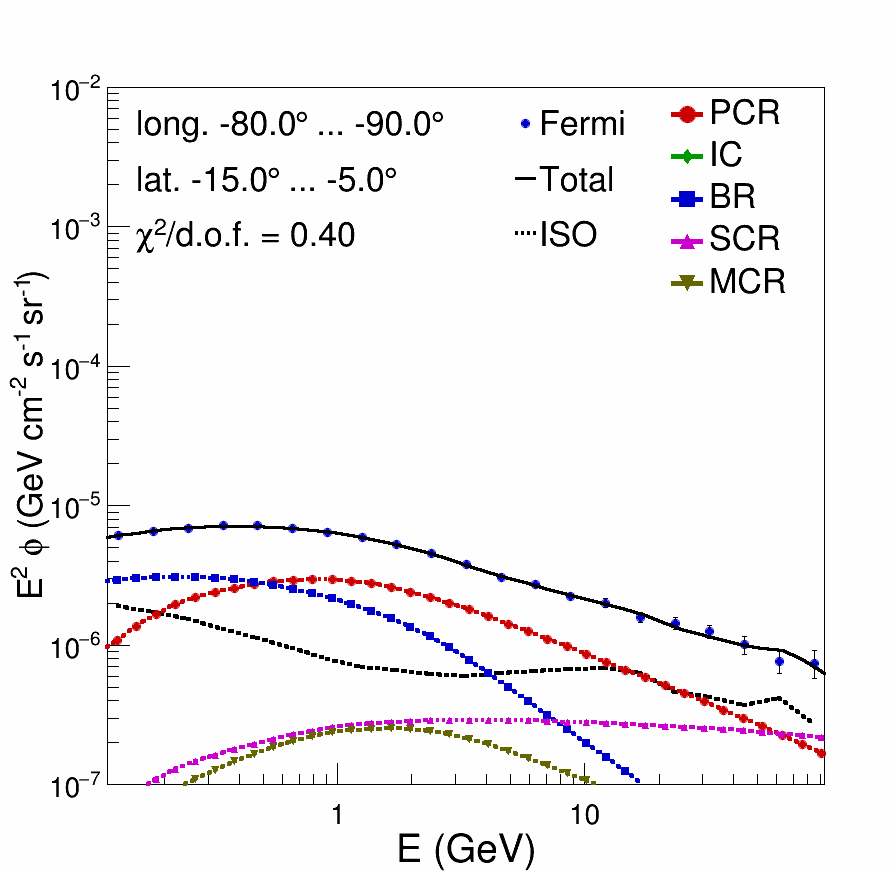}
\includegraphics[width=0.16\textwidth,height=0.16\textwidth,clip]{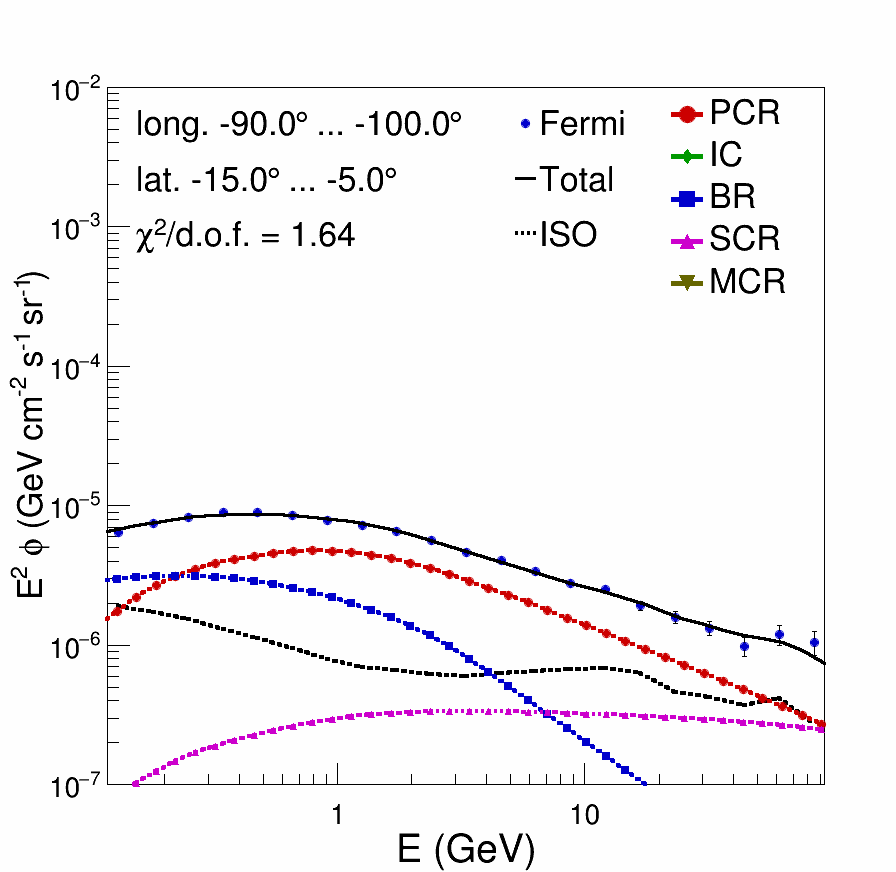}
\includegraphics[width=0.16\textwidth,height=0.16\textwidth,clip]{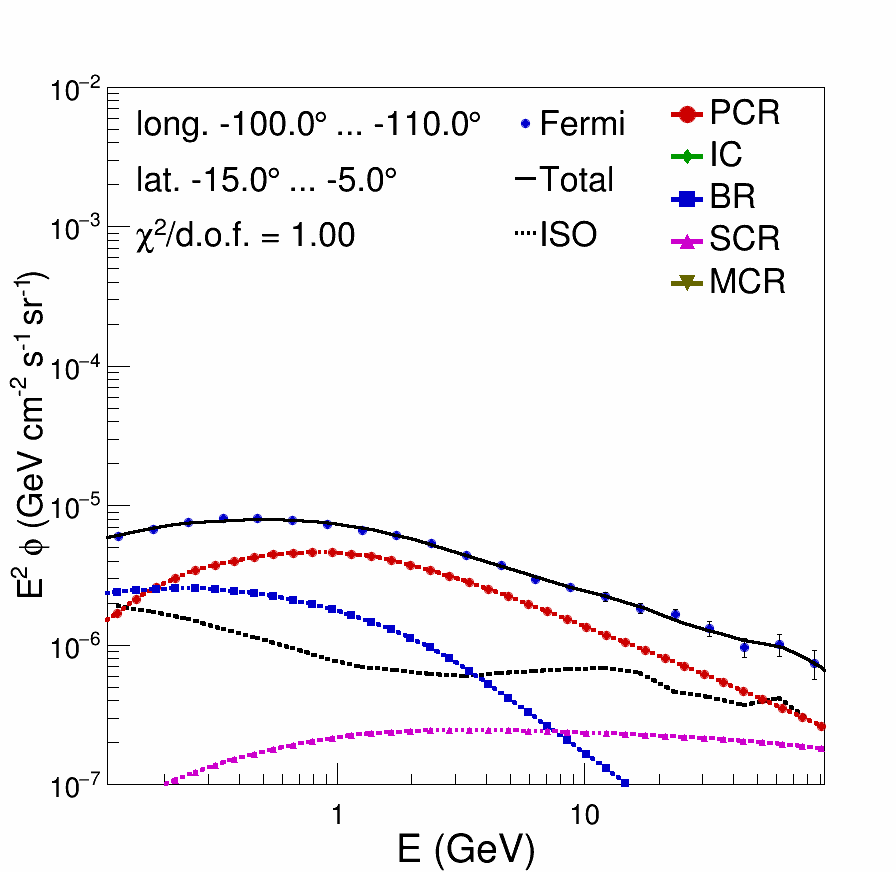}
\includegraphics[width=0.16\textwidth,height=0.16\textwidth,clip]{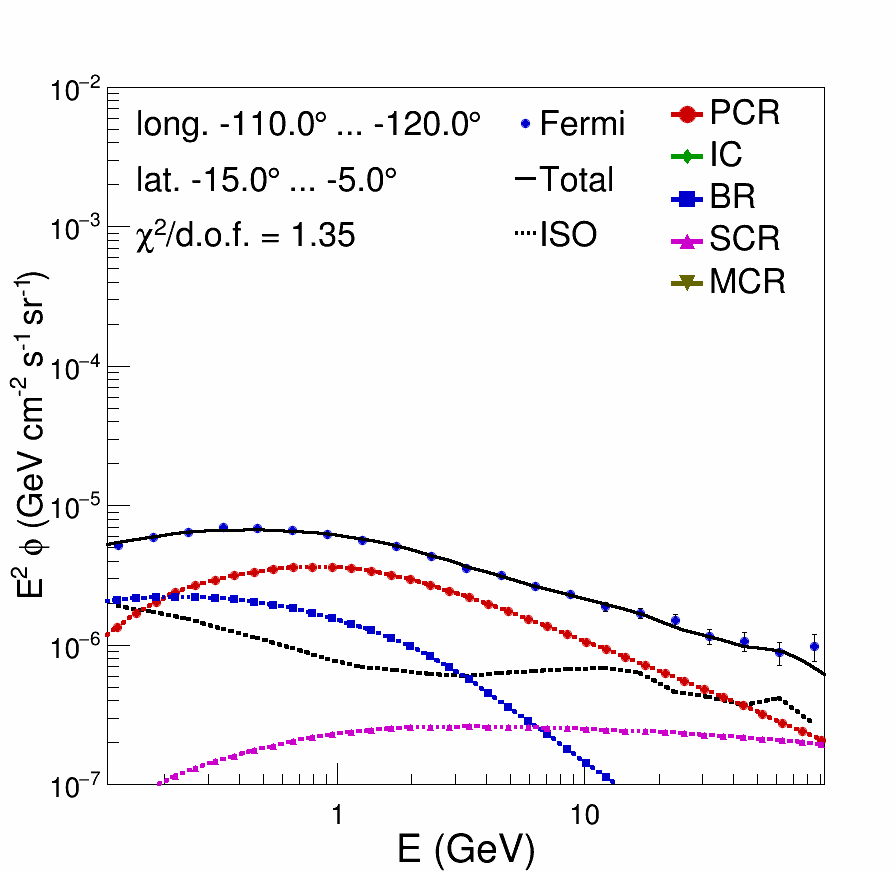}
\includegraphics[width=0.16\textwidth,height=0.16\textwidth,clip]{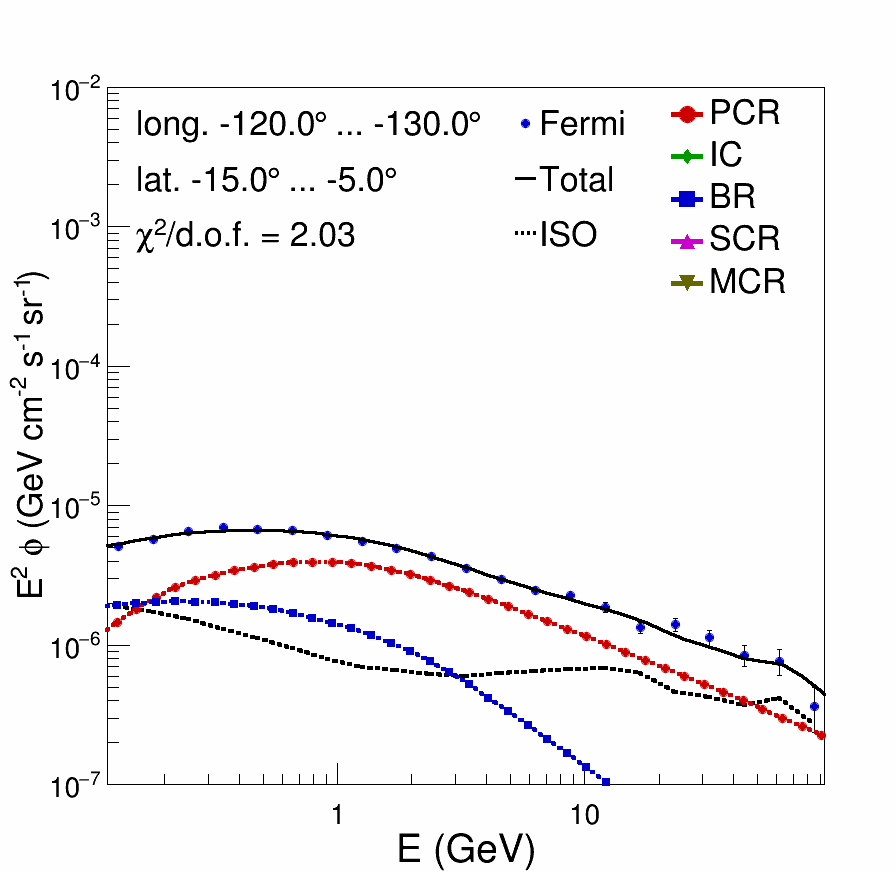}
\includegraphics[width=0.16\textwidth,height=0.16\textwidth,clip]{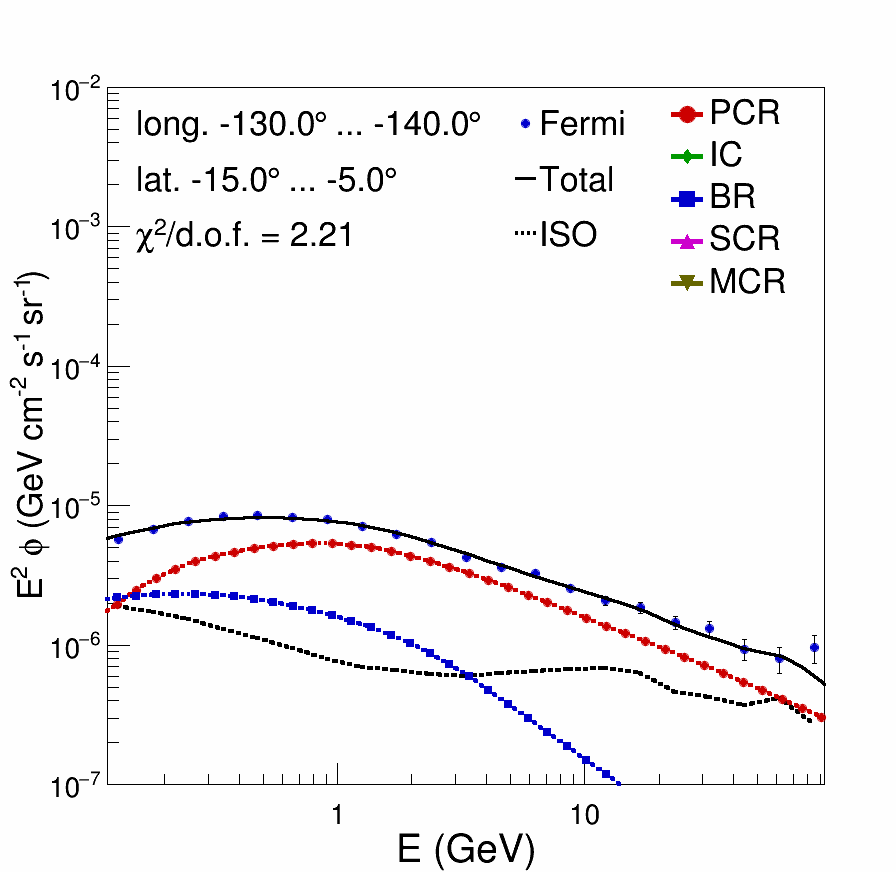}
\includegraphics[width=0.16\textwidth,height=0.16\textwidth,clip]{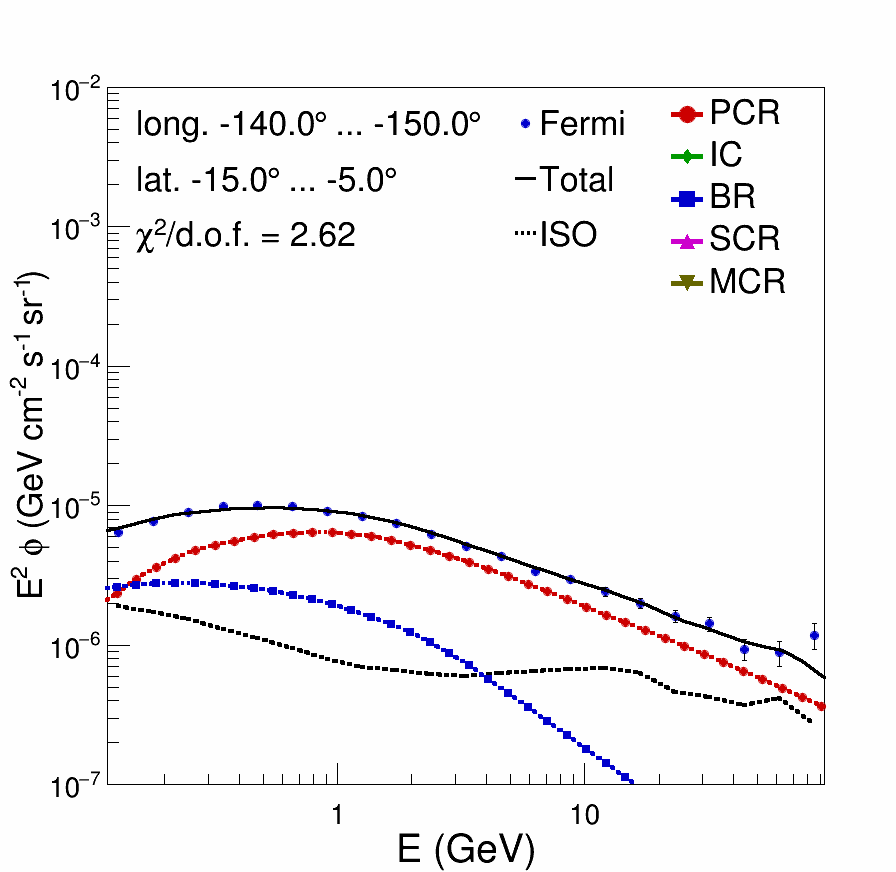}
\includegraphics[width=0.16\textwidth,height=0.16\textwidth,clip]{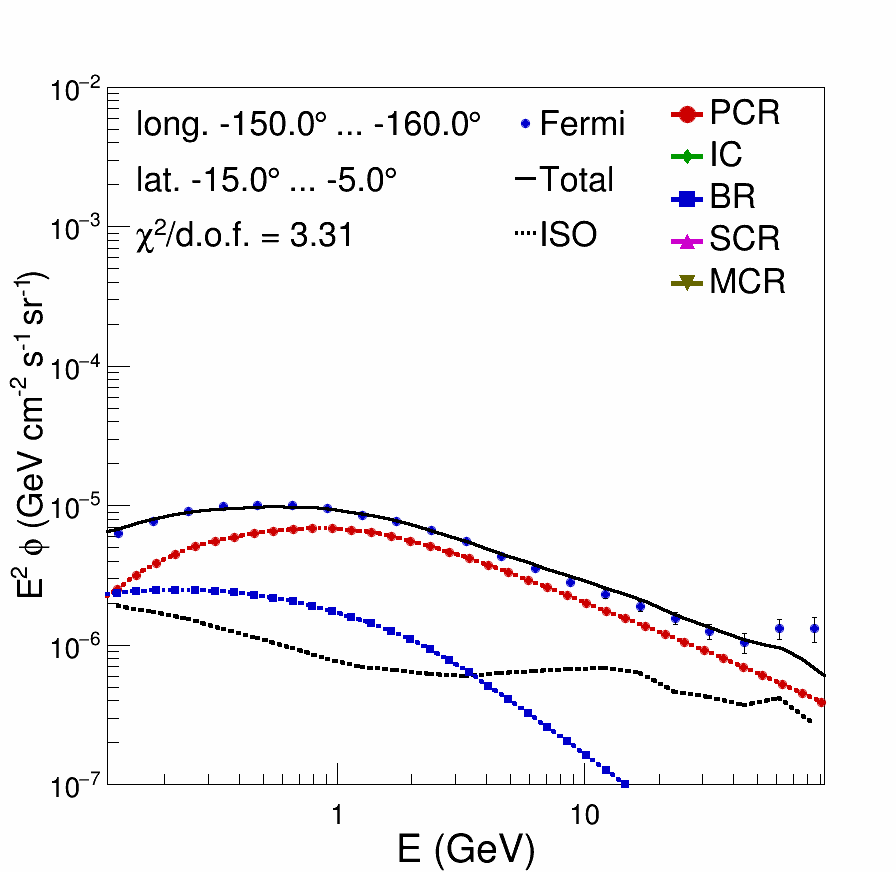}
\includegraphics[width=0.16\textwidth,height=0.16\textwidth,clip]{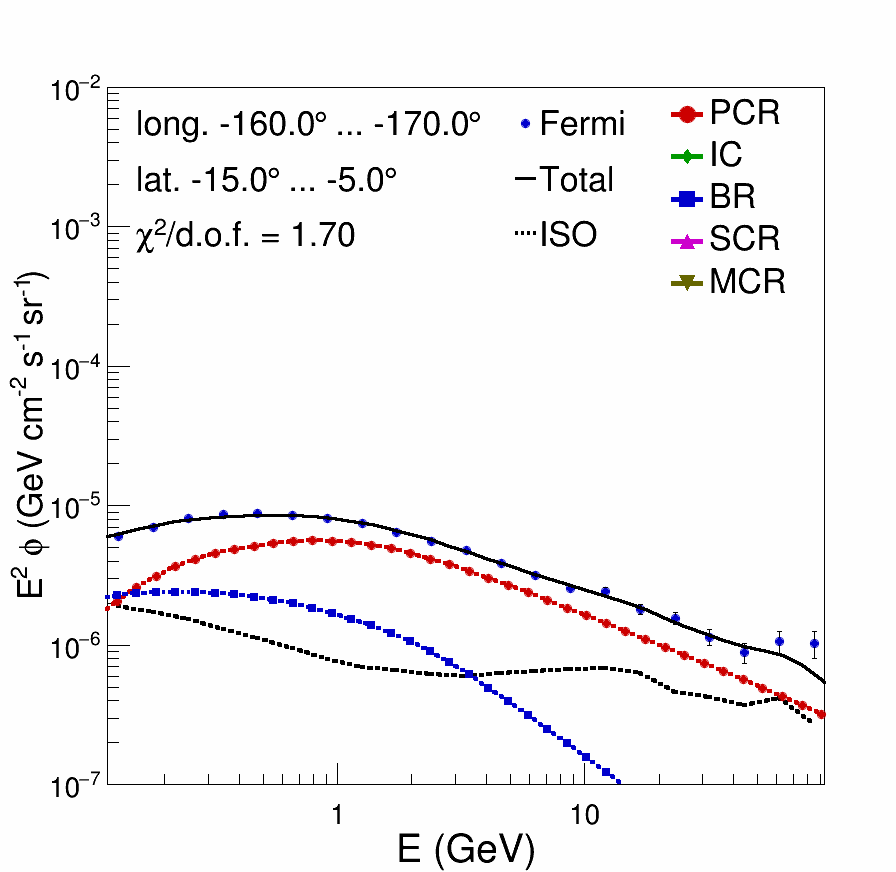}
\includegraphics[width=0.16\textwidth,height=0.16\textwidth,clip]{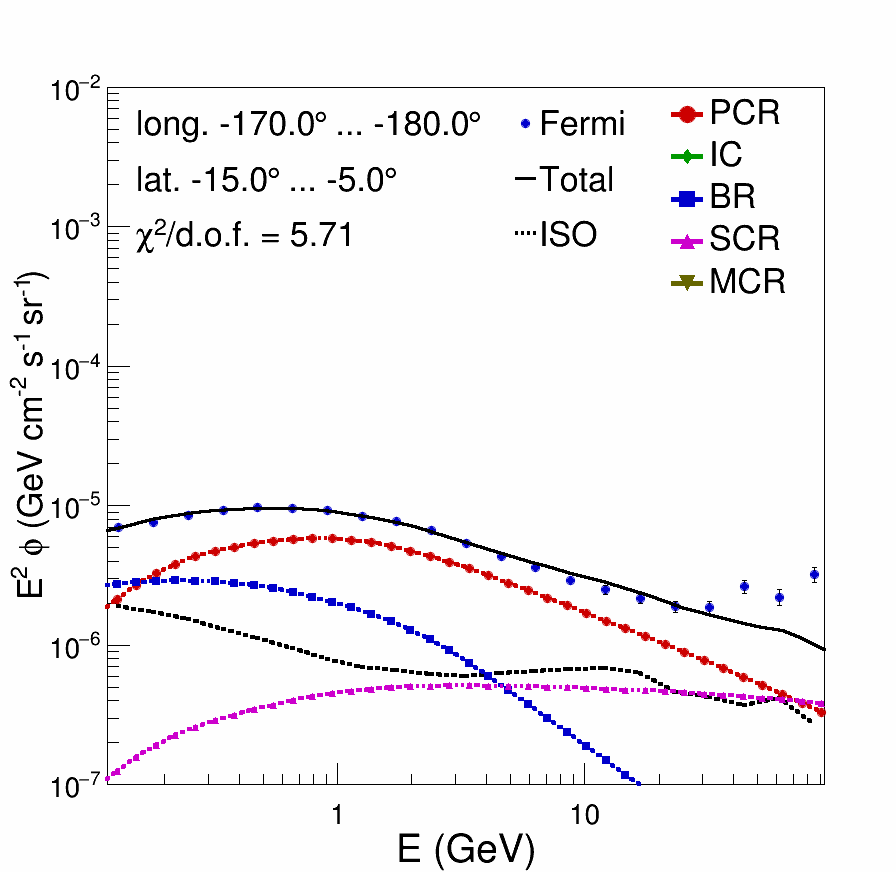}
\caption[]{Template fits for latitudes  with $-15.0^\circ<b<-5.0^\circ$ and longitudes decreasing from 180$^\circ$ to -180$^\circ$.} \label{F25}
\end{figure}
\begin{figure}
\centering
\includegraphics[width=0.16\textwidth,height=0.16\textwidth,clip]{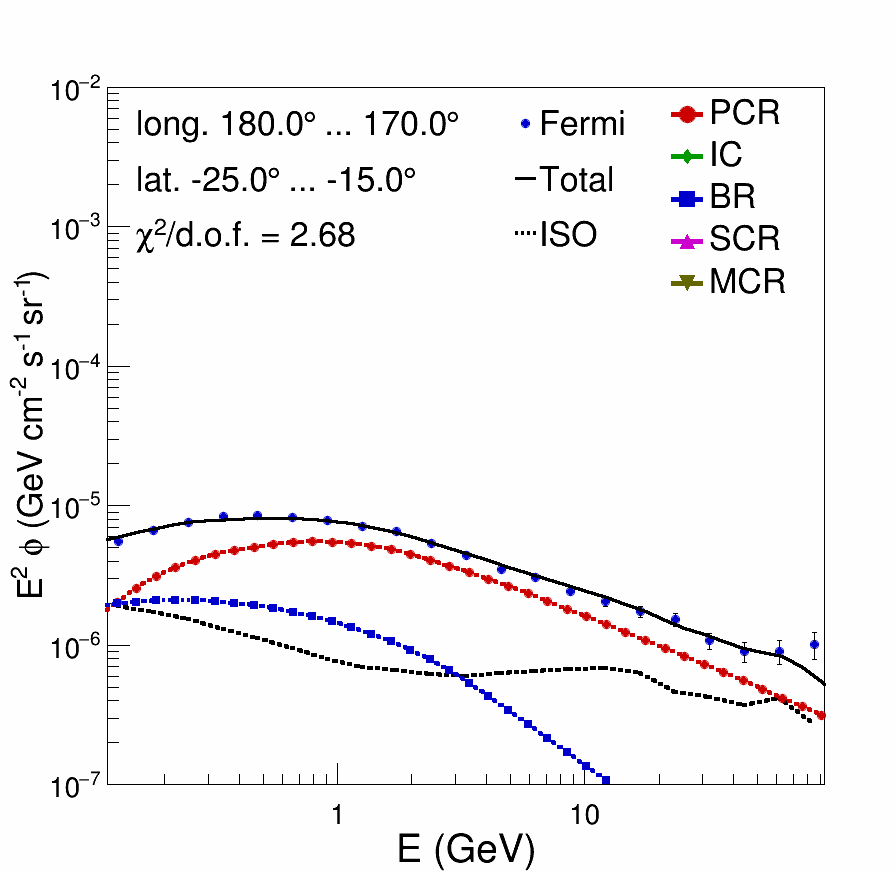}
\includegraphics[width=0.16\textwidth,height=0.16\textwidth,clip]{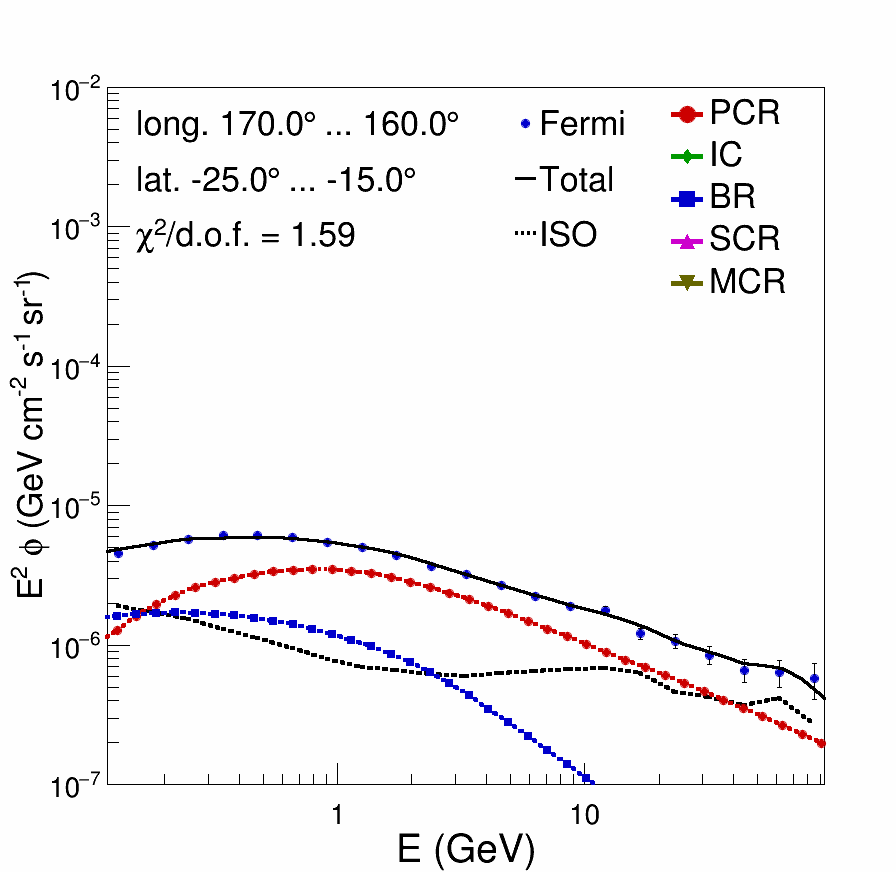}
\includegraphics[width=0.16\textwidth,height=0.16\textwidth,clip]{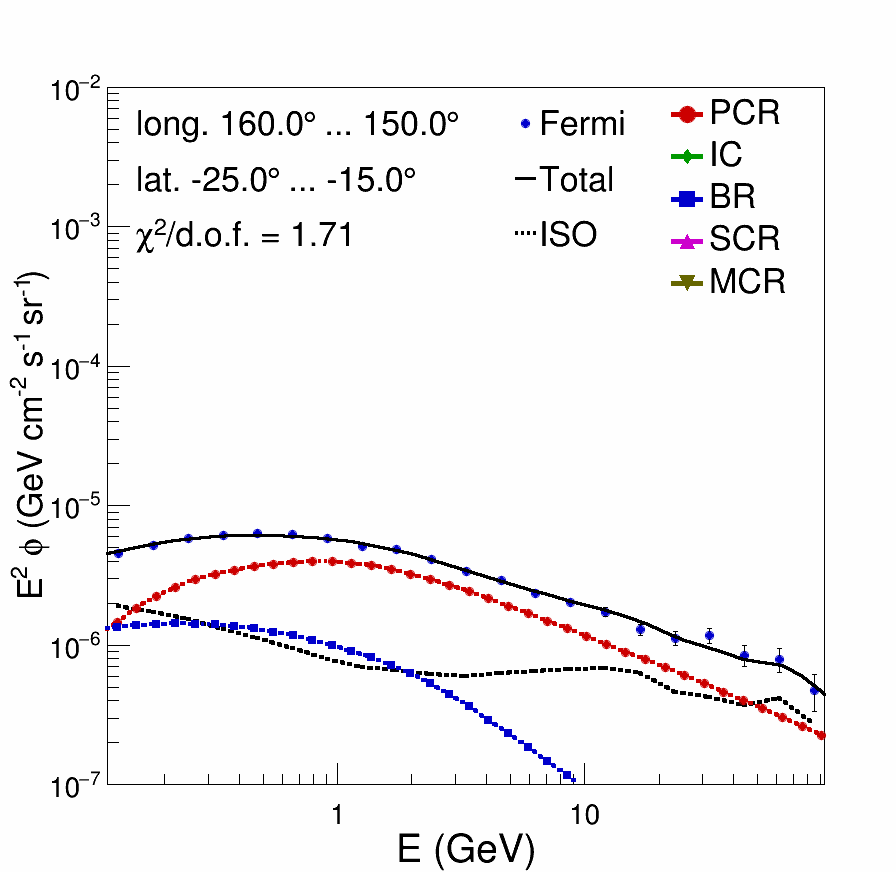}
\includegraphics[width=0.16\textwidth,height=0.16\textwidth,clip]{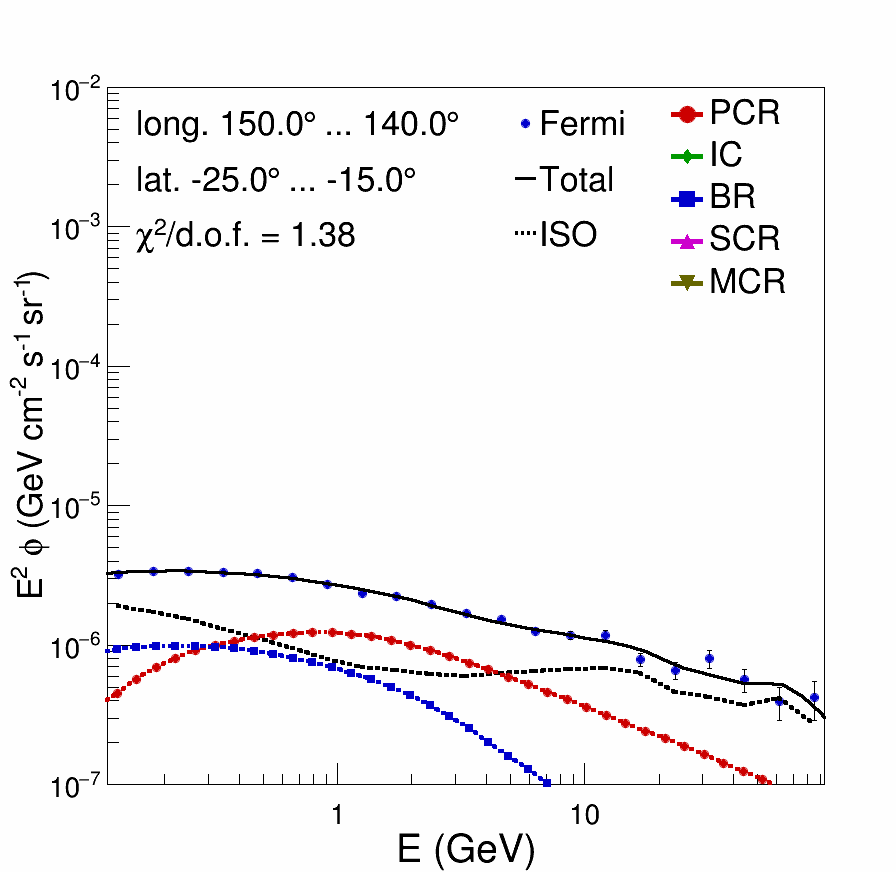}
\includegraphics[width=0.16\textwidth,height=0.16\textwidth,clip]{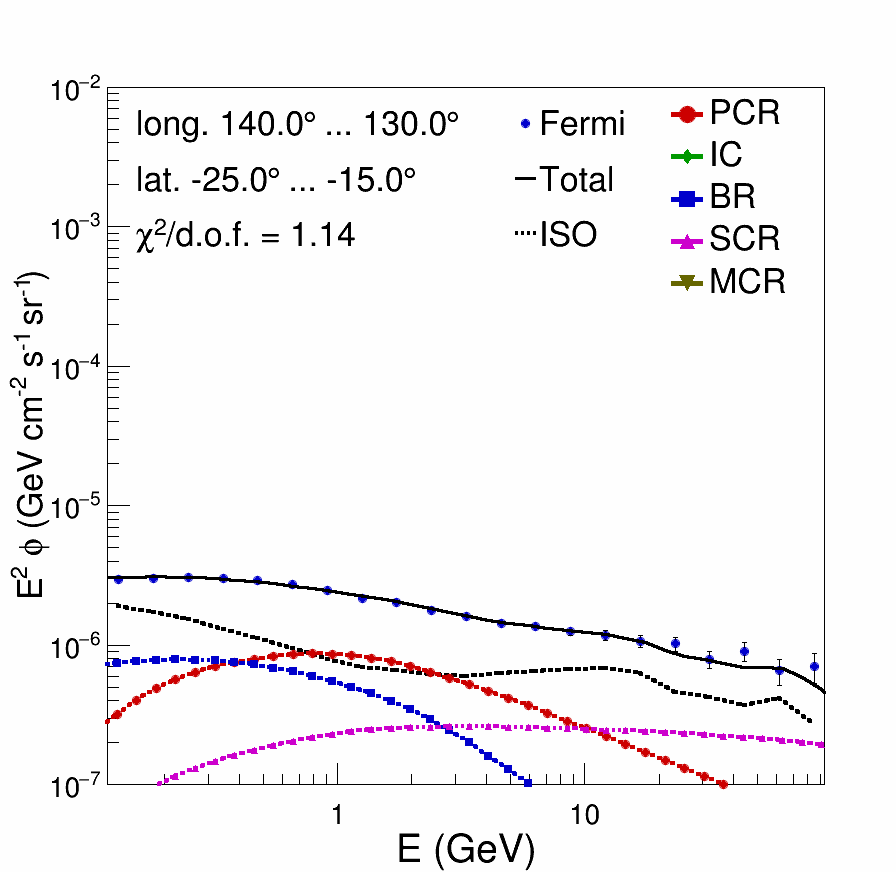}
\includegraphics[width=0.16\textwidth,height=0.16\textwidth,clip]{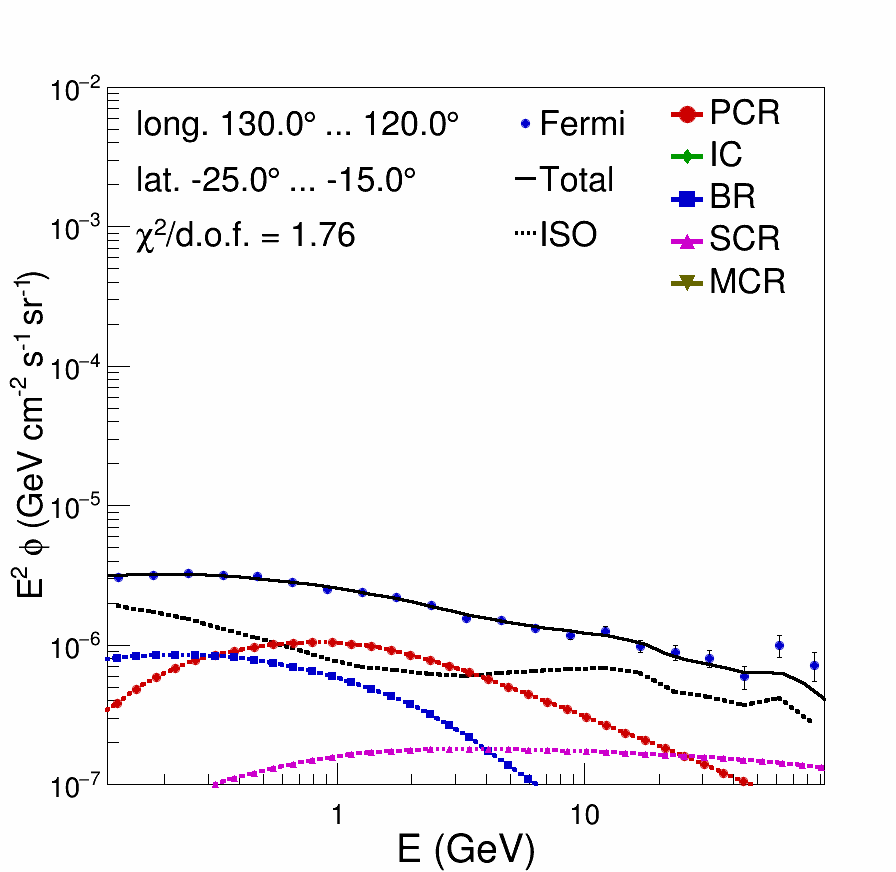}
\includegraphics[width=0.16\textwidth,height=0.16\textwidth,clip]{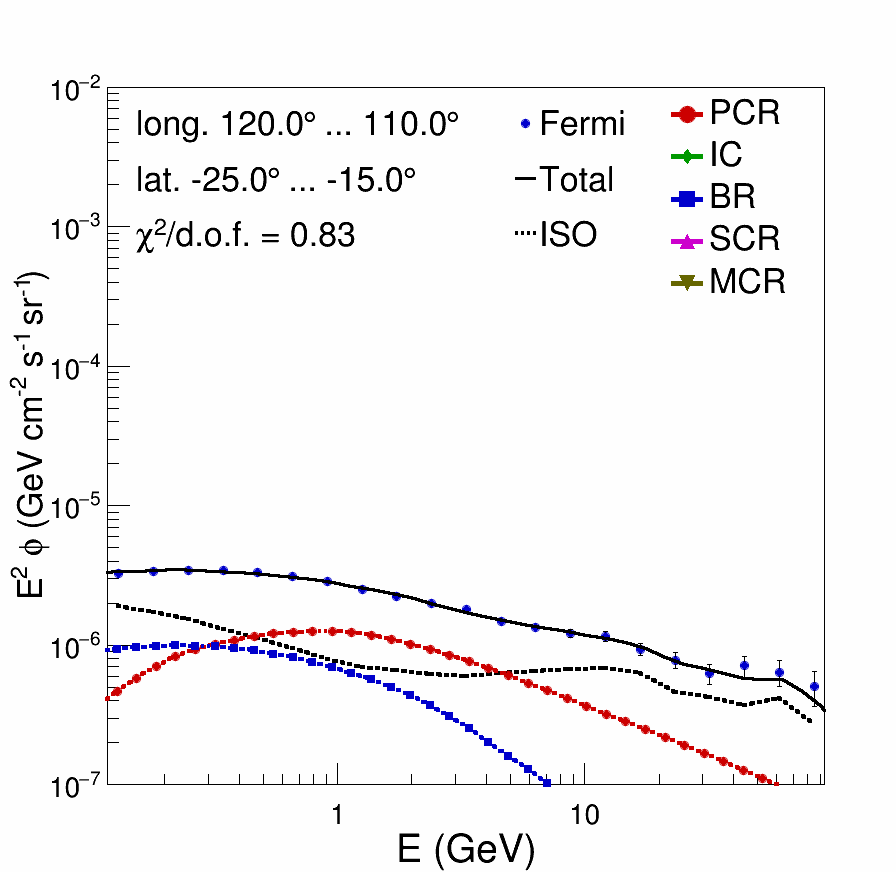}
\includegraphics[width=0.16\textwidth,height=0.16\textwidth,clip]{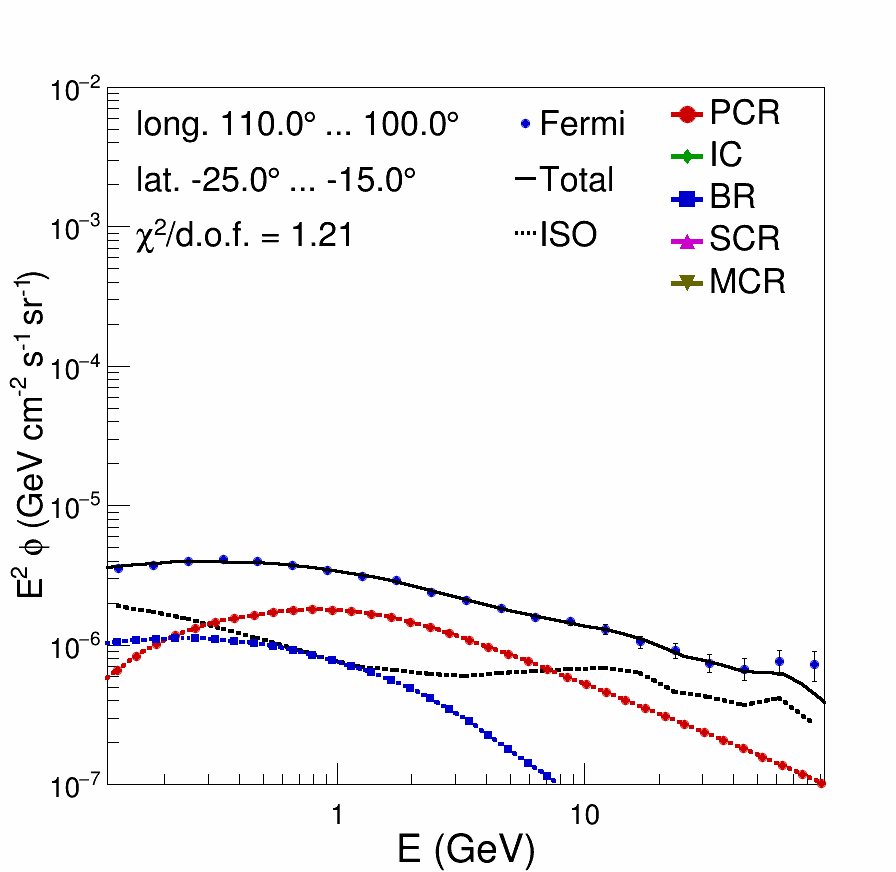}
\includegraphics[width=0.16\textwidth,height=0.16\textwidth,clip]{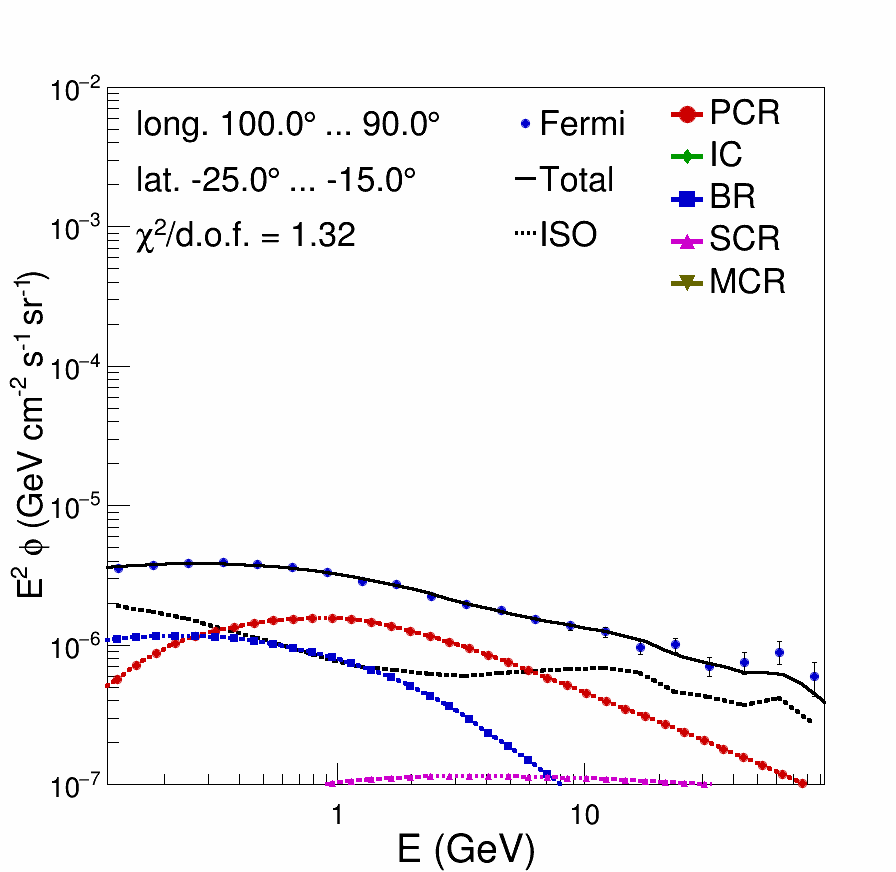}
\includegraphics[width=0.16\textwidth,height=0.16\textwidth,clip]{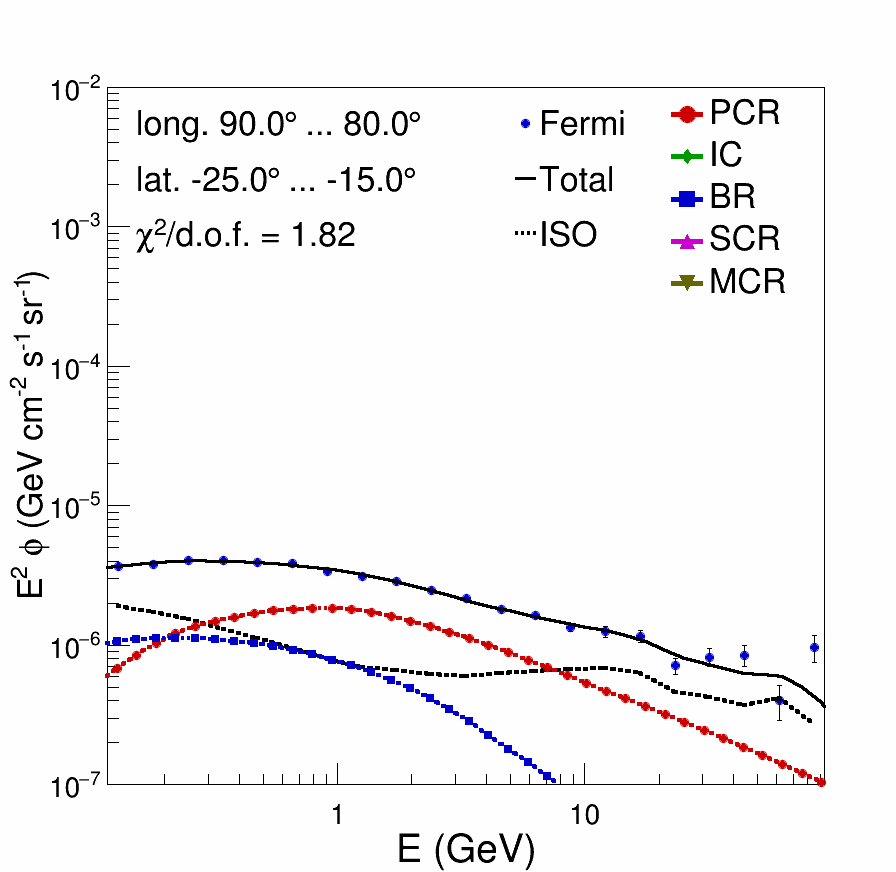}
\includegraphics[width=0.16\textwidth,height=0.16\textwidth,clip]{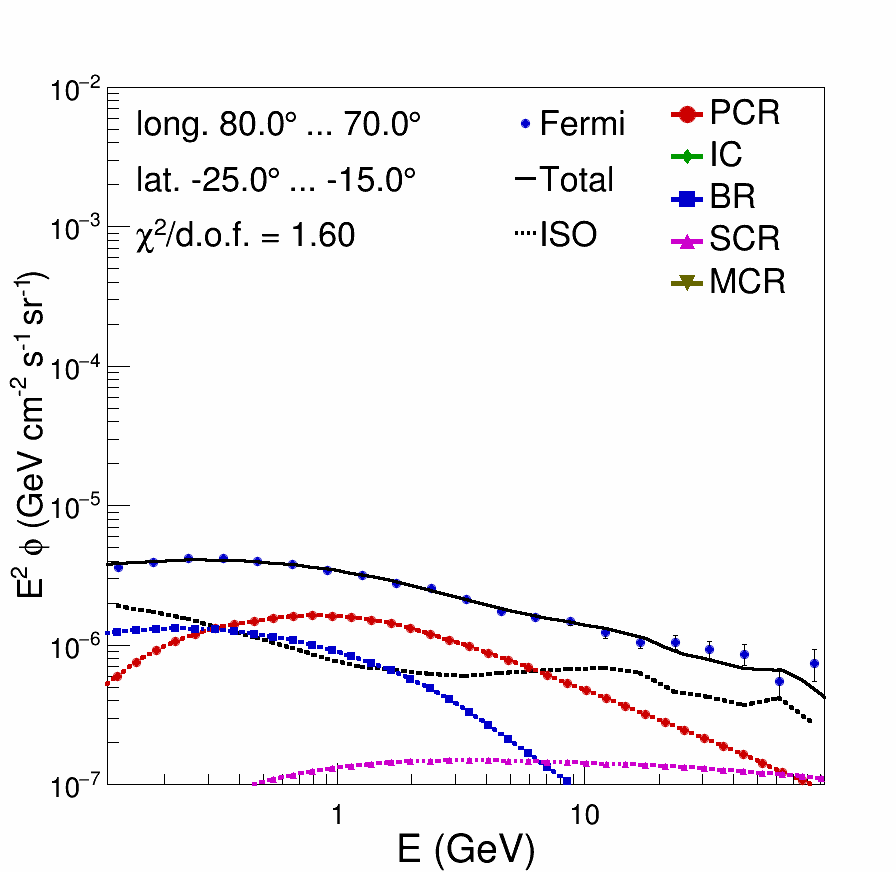}
\includegraphics[width=0.16\textwidth,height=0.16\textwidth,clip]{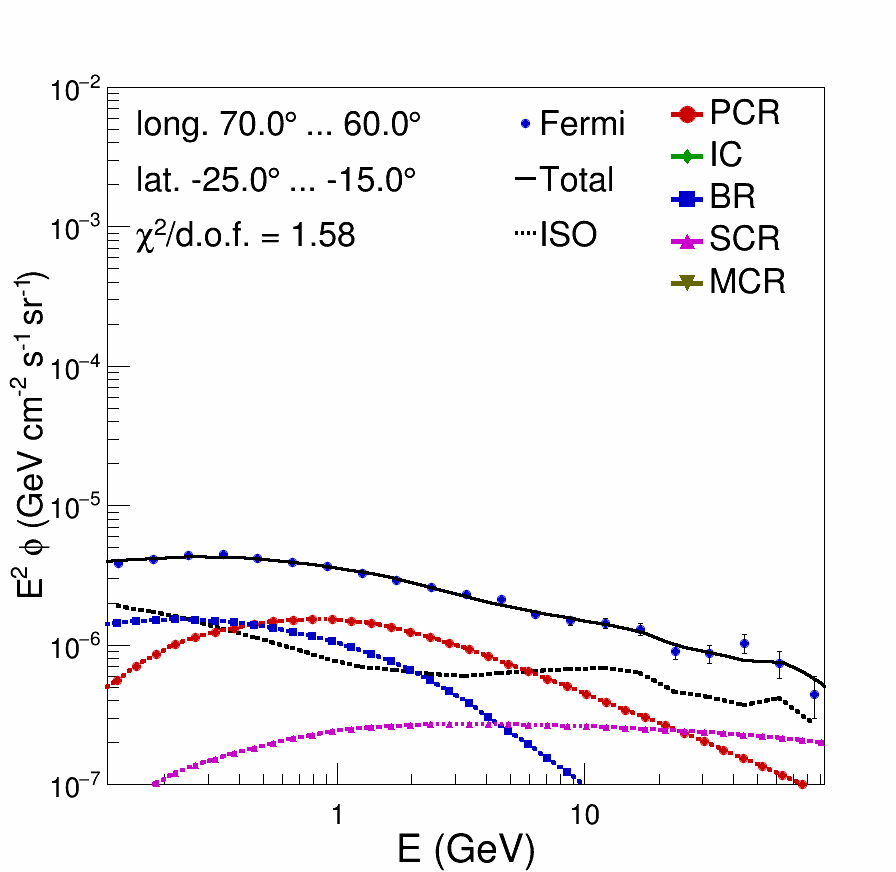}
\includegraphics[width=0.16\textwidth,height=0.16\textwidth,clip]{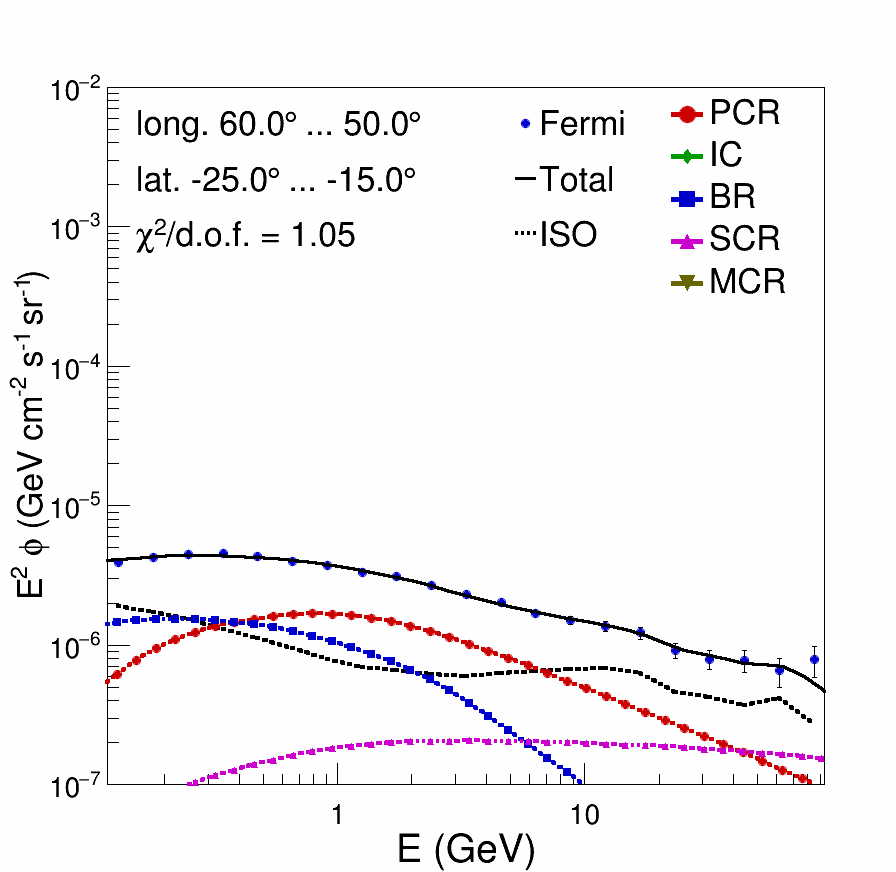}
\includegraphics[width=0.16\textwidth,height=0.16\textwidth,clip]{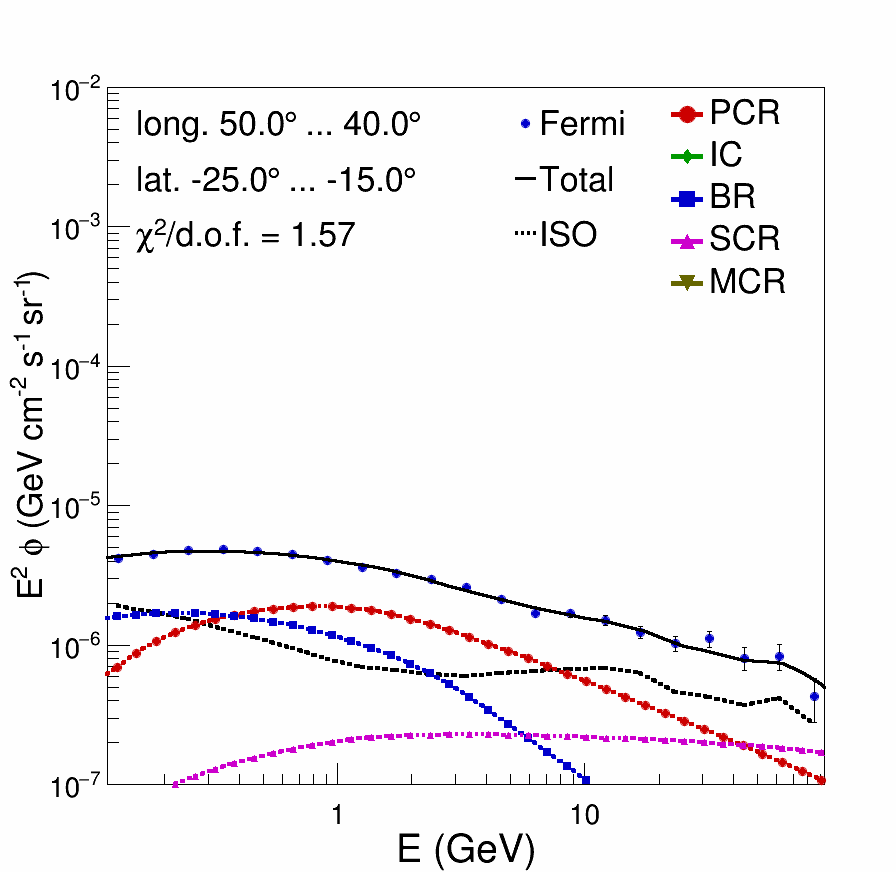}
\includegraphics[width=0.16\textwidth,height=0.16\textwidth,clip]{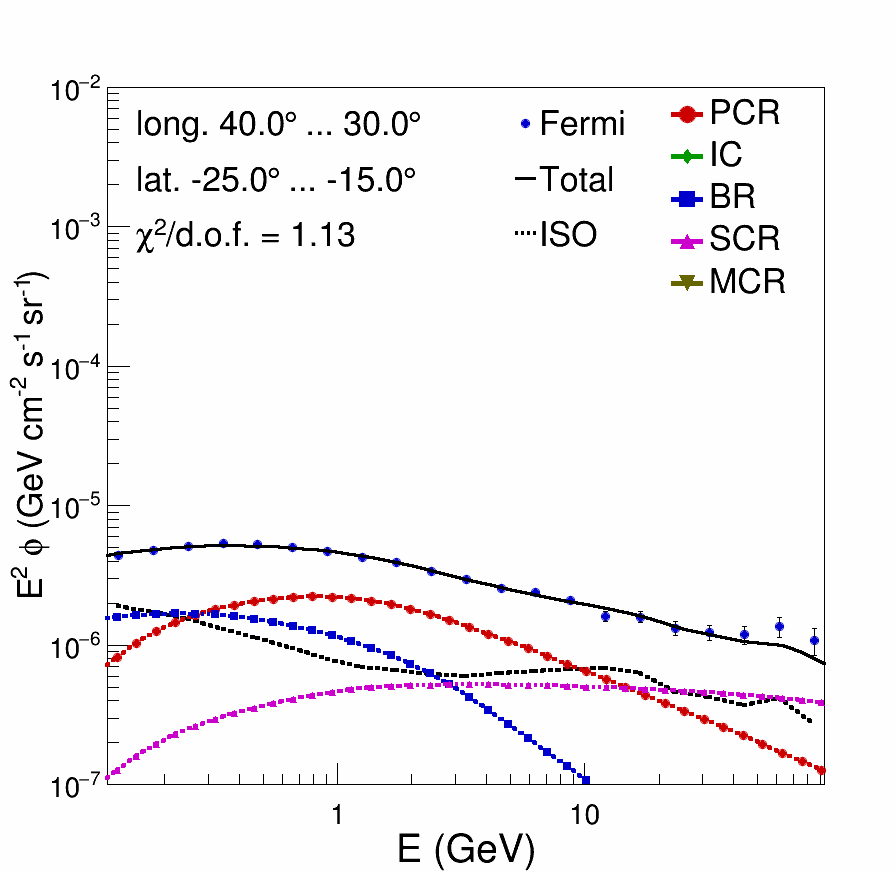}
\includegraphics[width=0.16\textwidth,height=0.16\textwidth,clip]{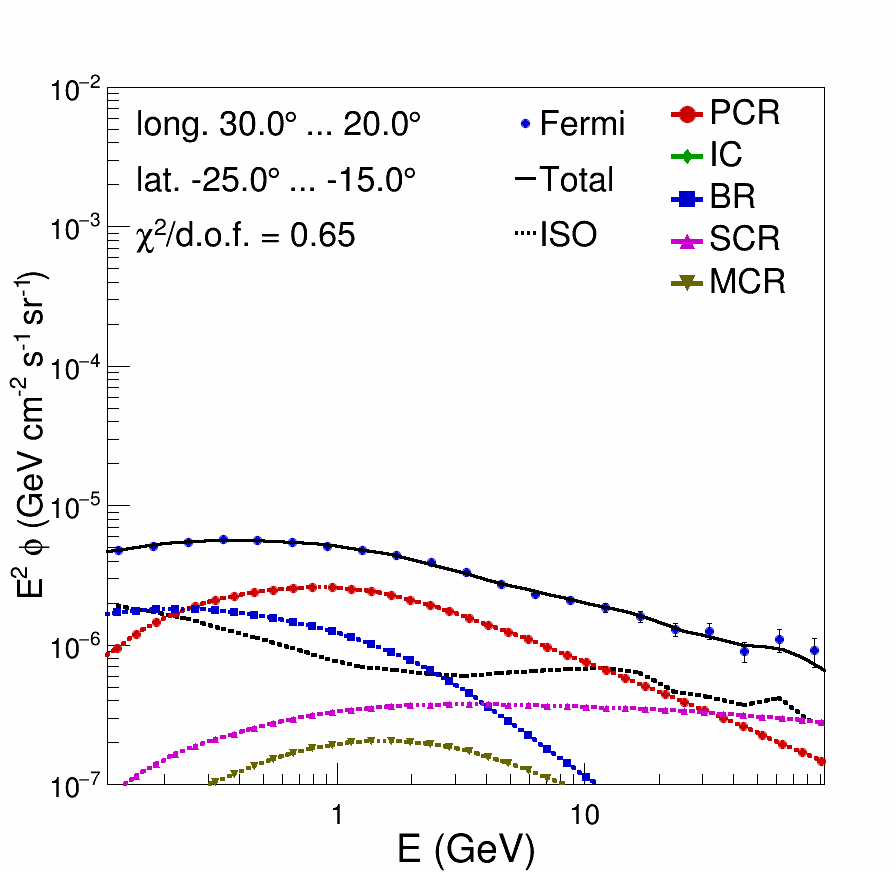}
\includegraphics[width=0.16\textwidth,height=0.16\textwidth,clip]{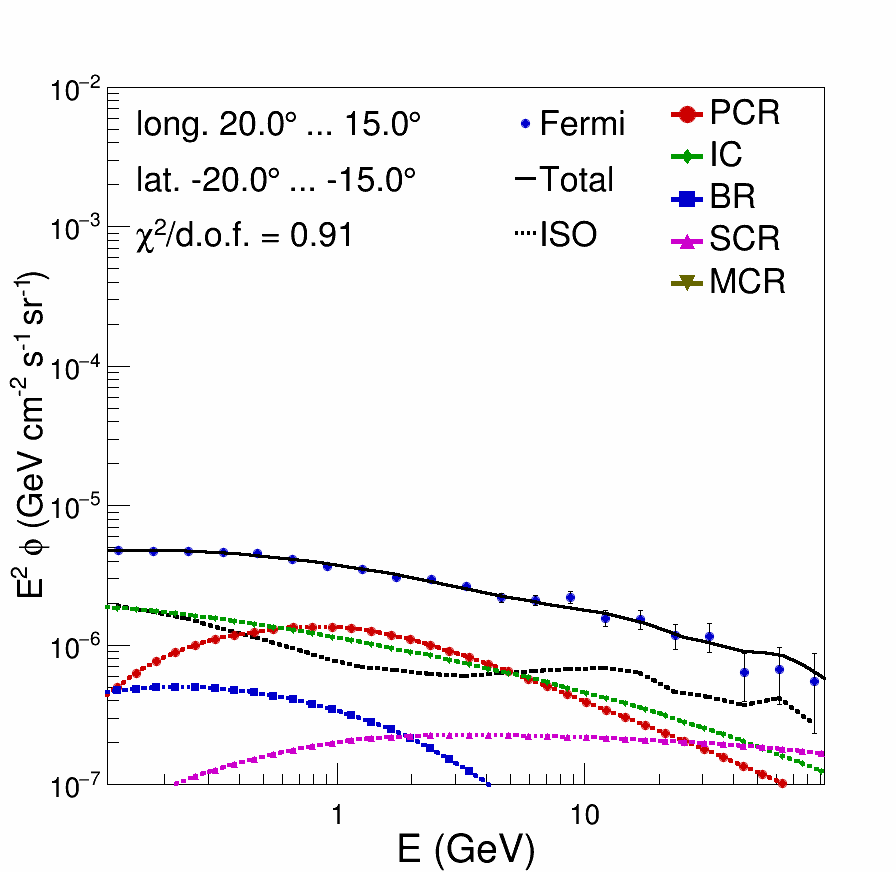}
\includegraphics[width=0.16\textwidth,height=0.16\textwidth,clip]{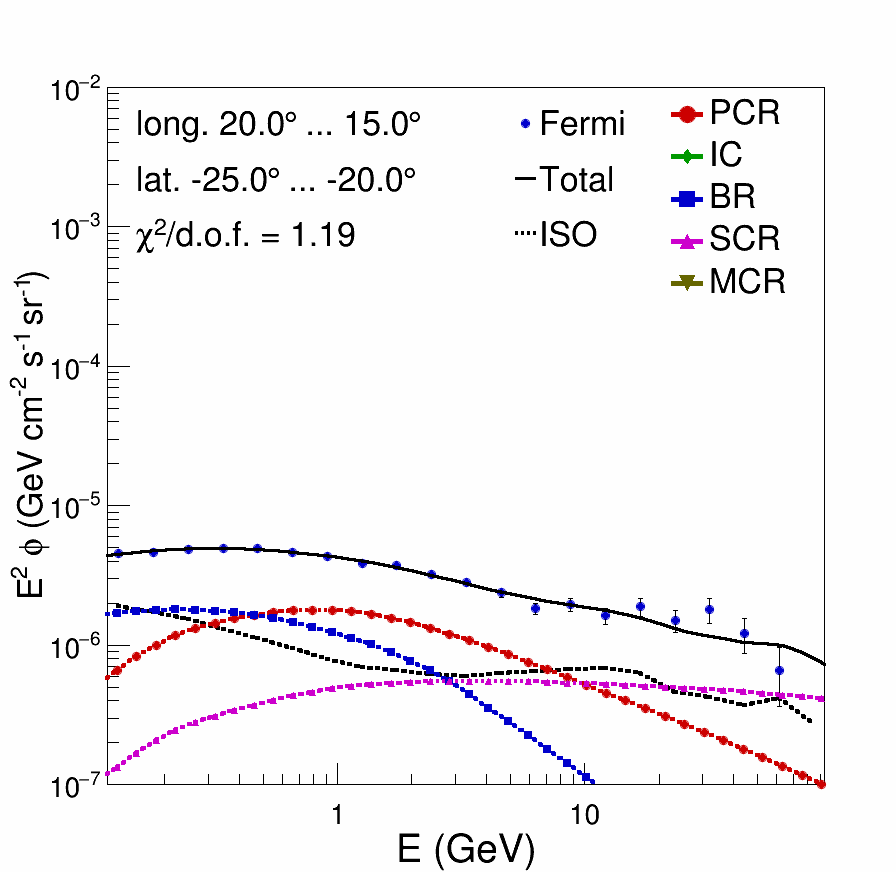}
\includegraphics[width=0.16\textwidth,height=0.16\textwidth,clip]{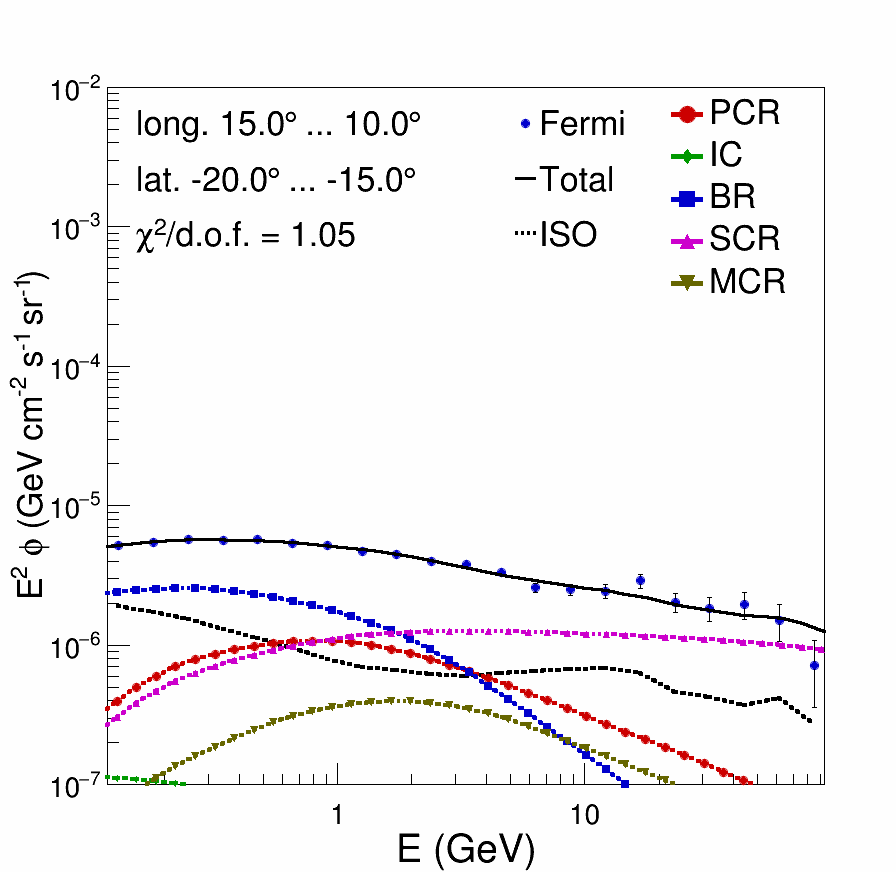}
\includegraphics[width=0.16\textwidth,height=0.16\textwidth,clip]{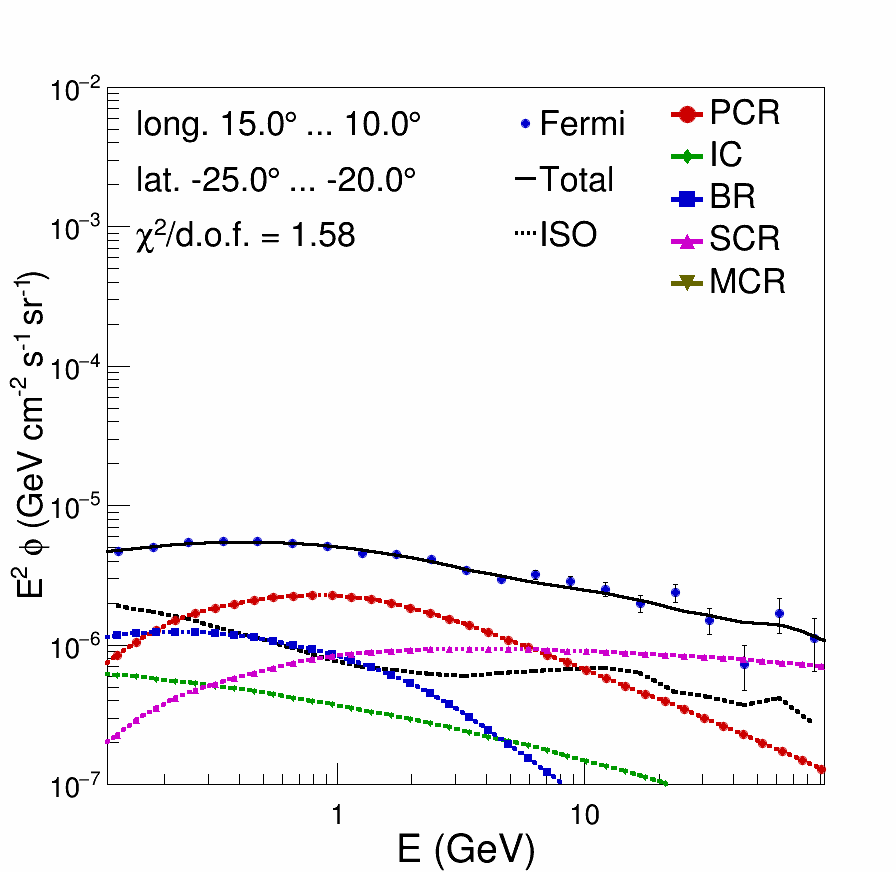}
\includegraphics[width=0.16\textwidth,height=0.16\textwidth,clip]{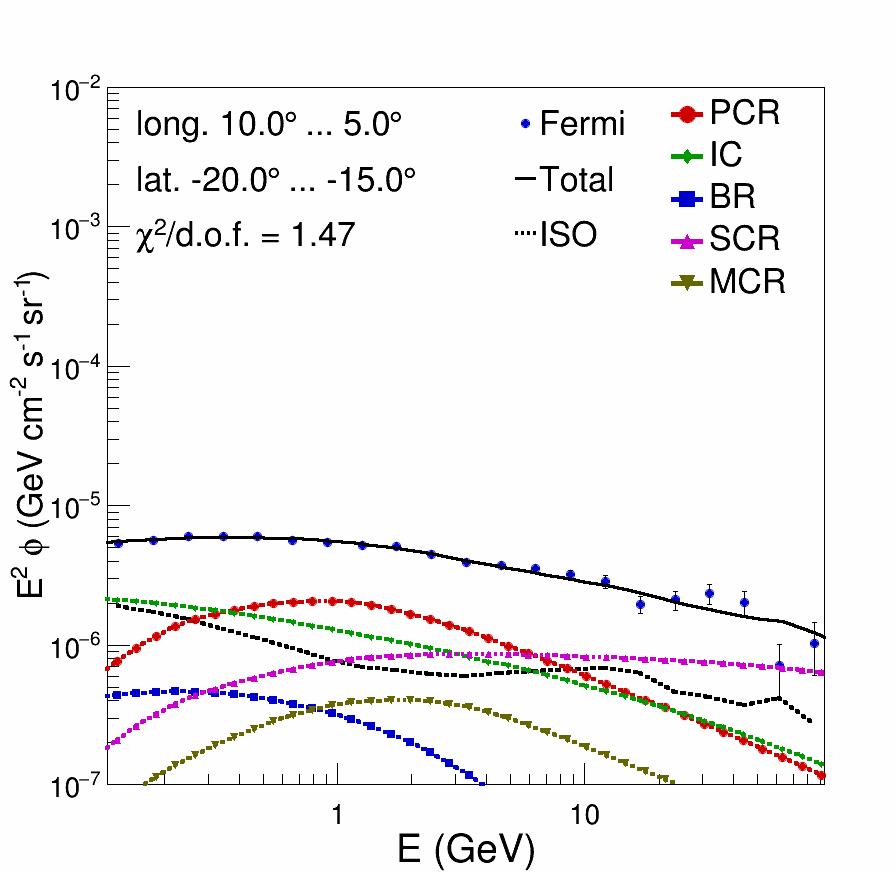}
\includegraphics[width=0.16\textwidth,height=0.16\textwidth,clip]{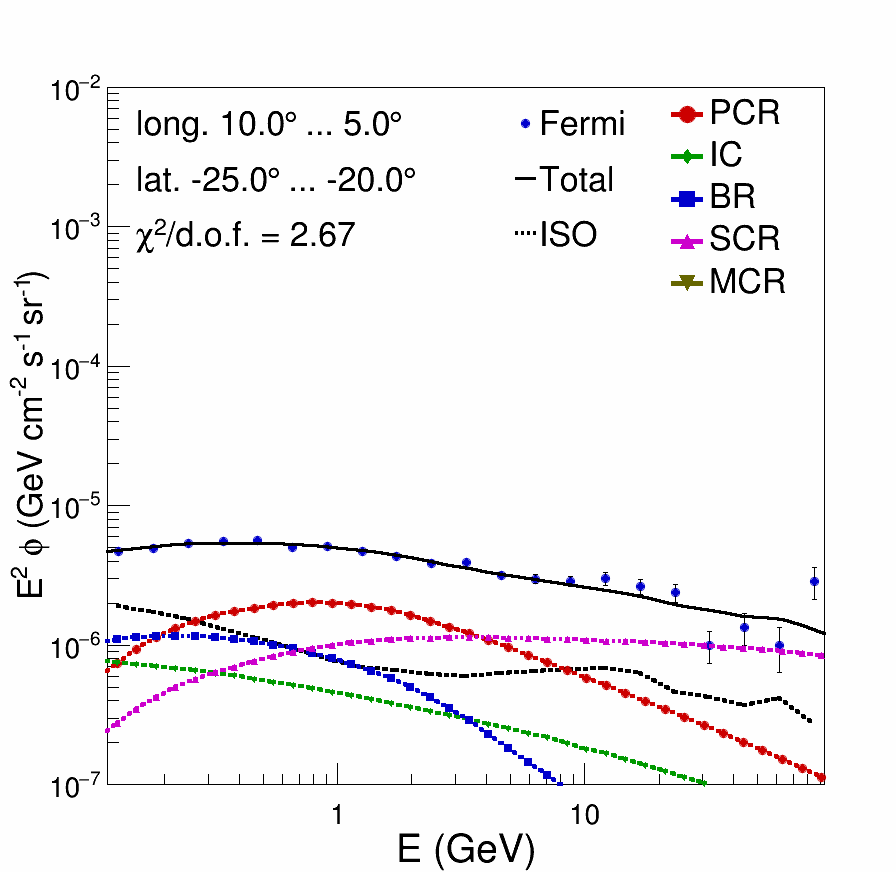}
\includegraphics[width=0.16\textwidth,height=0.16\textwidth,clip]{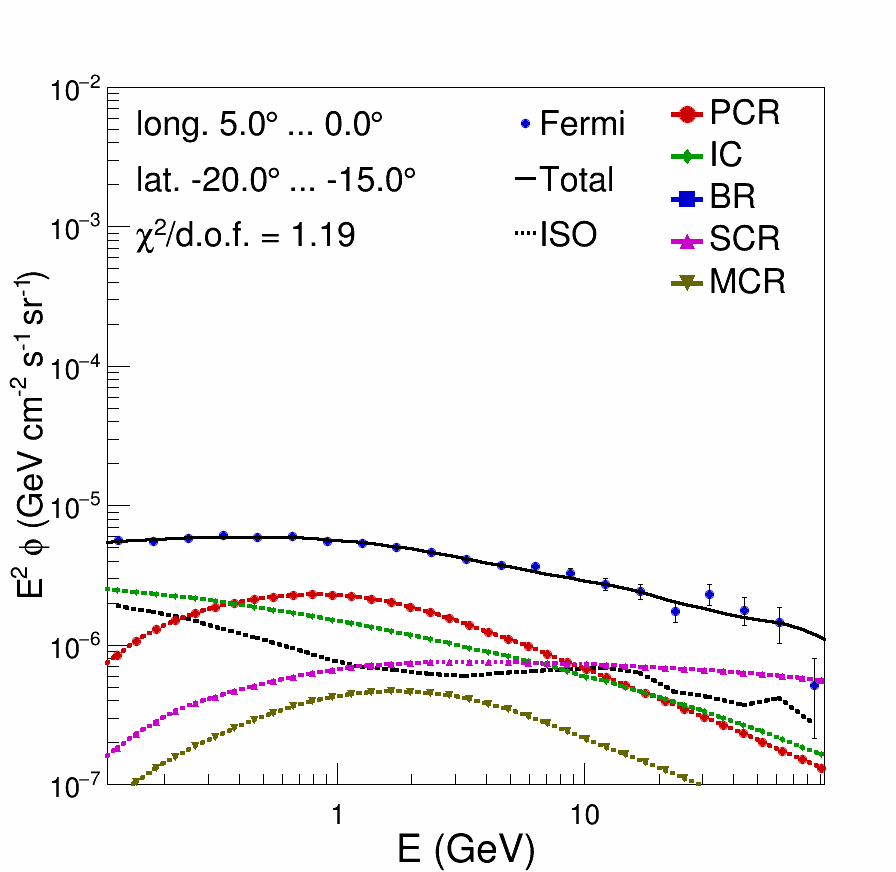}
\includegraphics[width=0.16\textwidth,height=0.16\textwidth,clip]{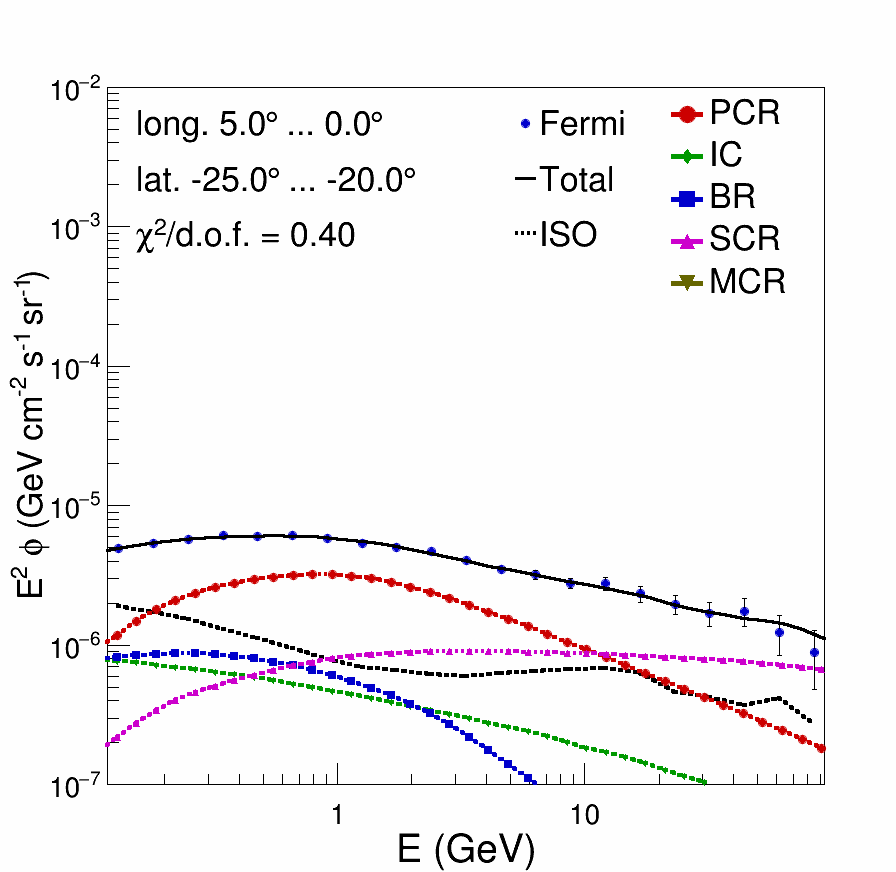}
\includegraphics[width=0.16\textwidth,height=0.16\textwidth,clip]{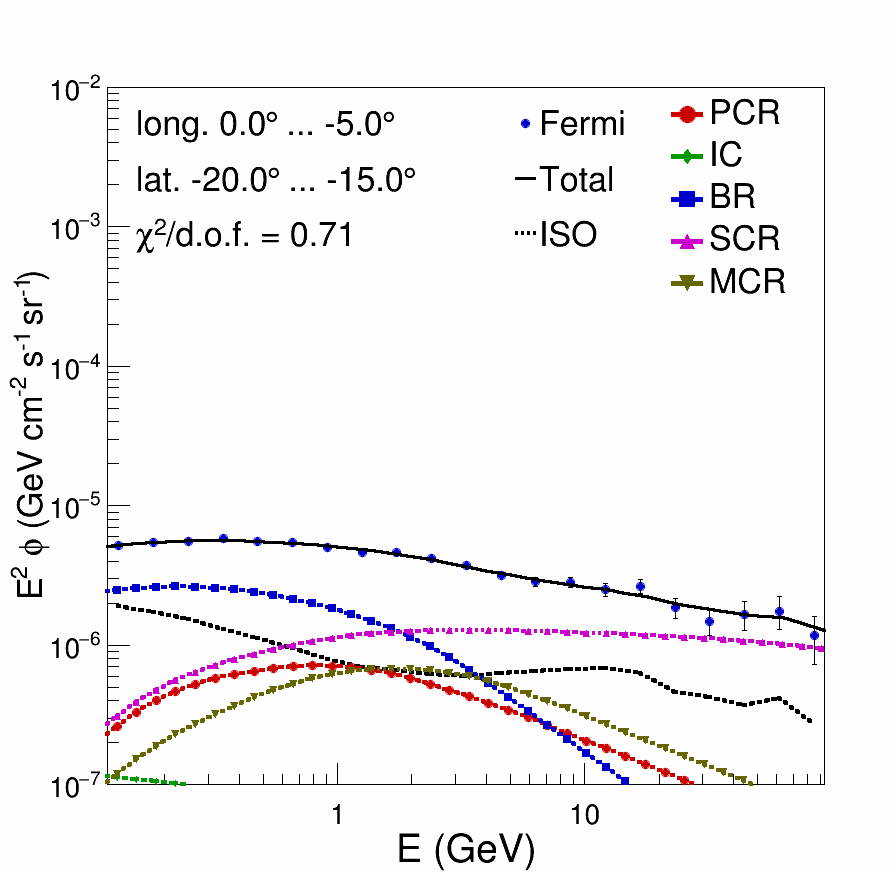}
\includegraphics[width=0.16\textwidth,height=0.16\textwidth,clip]{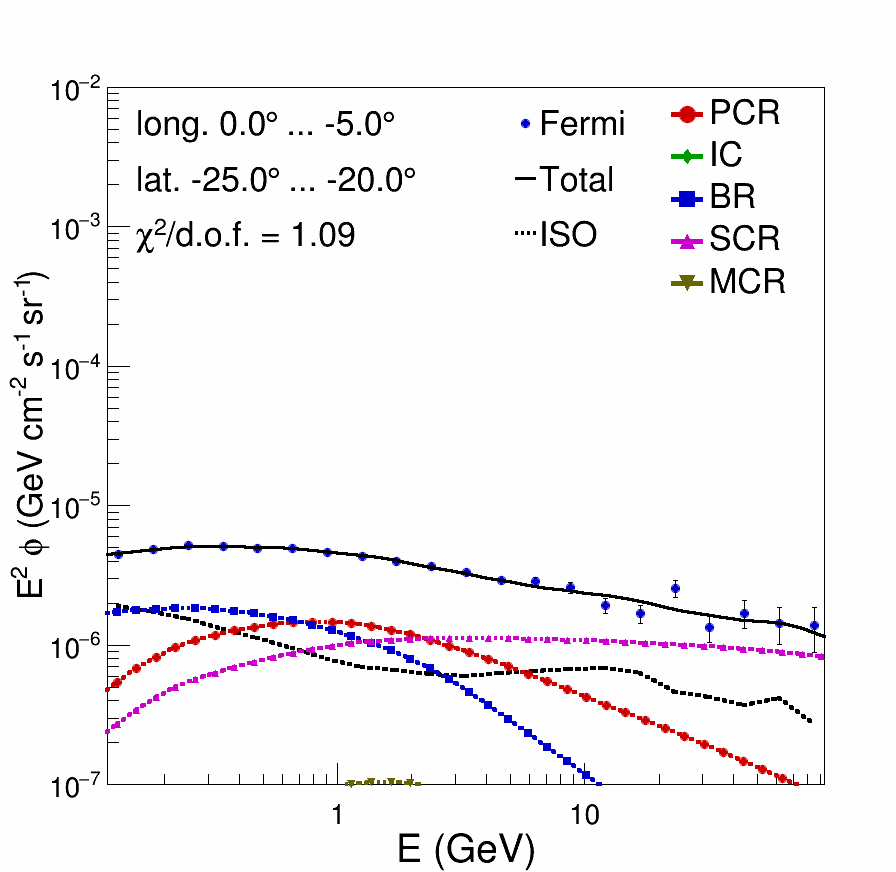}
\includegraphics[width=0.16\textwidth,height=0.16\textwidth,clip]{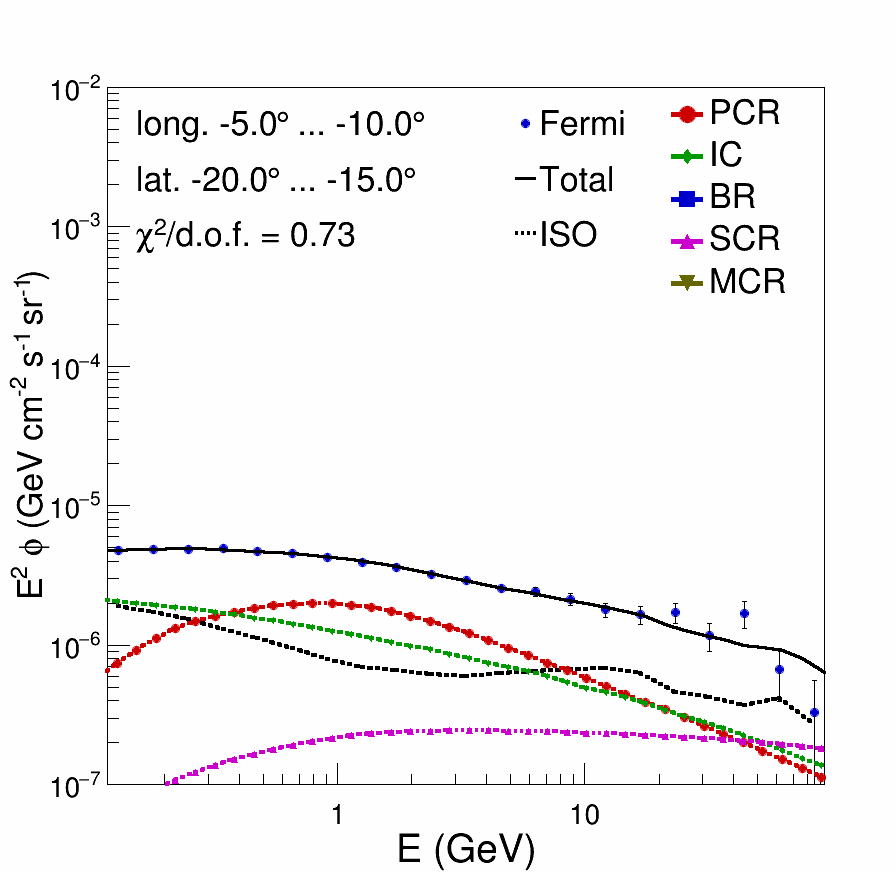}
\includegraphics[width=0.16\textwidth,height=0.16\textwidth,clip]{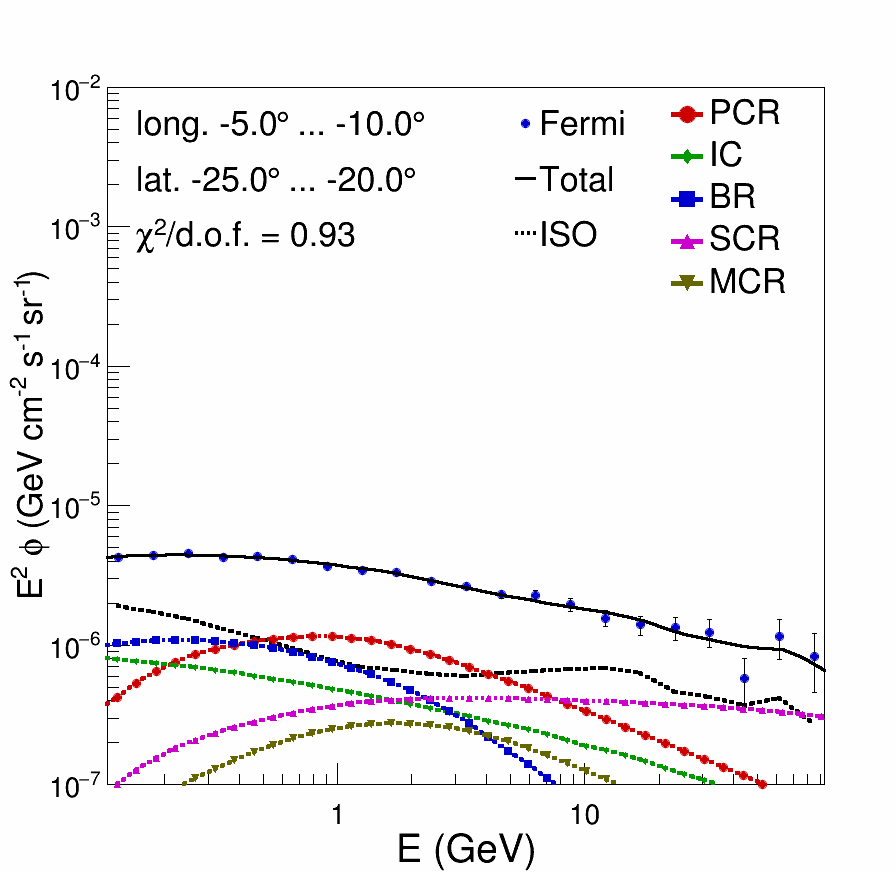}
\includegraphics[width=0.16\textwidth,height=0.16\textwidth,clip]{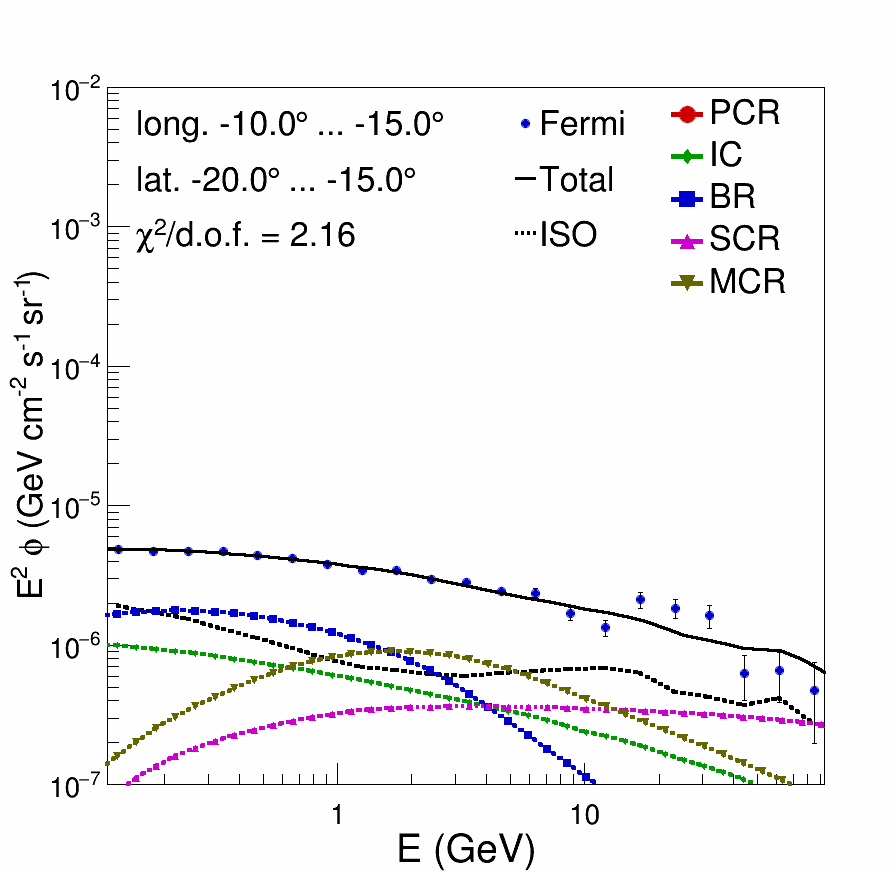}
\includegraphics[width=0.16\textwidth,height=0.16\textwidth,clip]{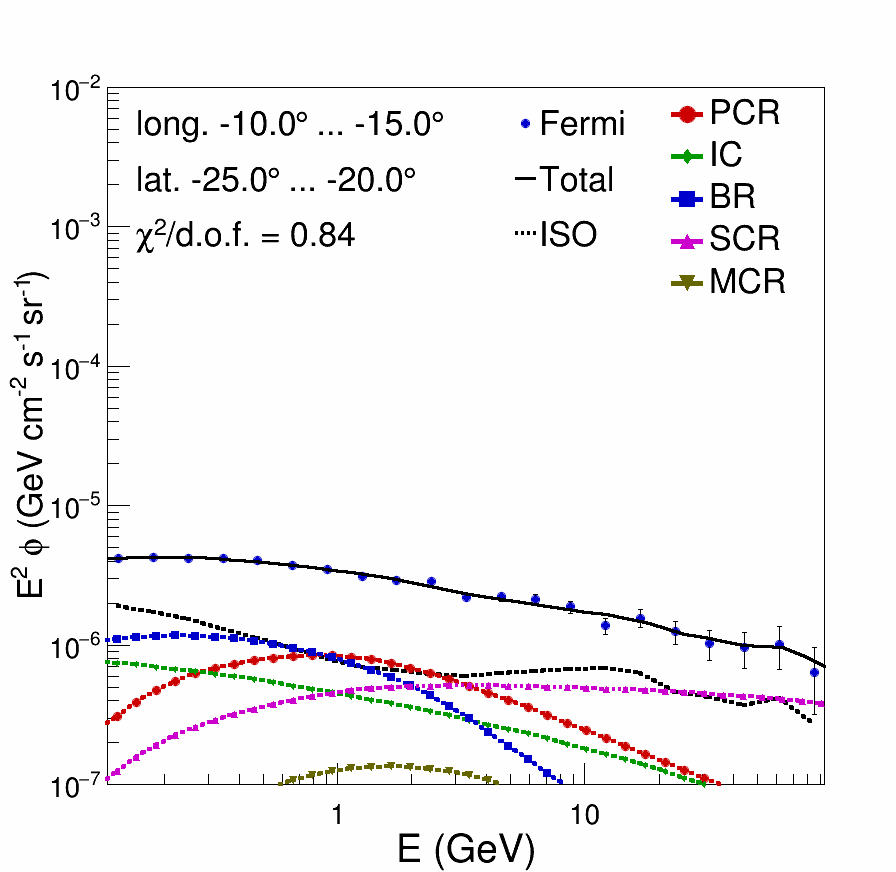}
\includegraphics[width=0.16\textwidth,height=0.16\textwidth,clip]{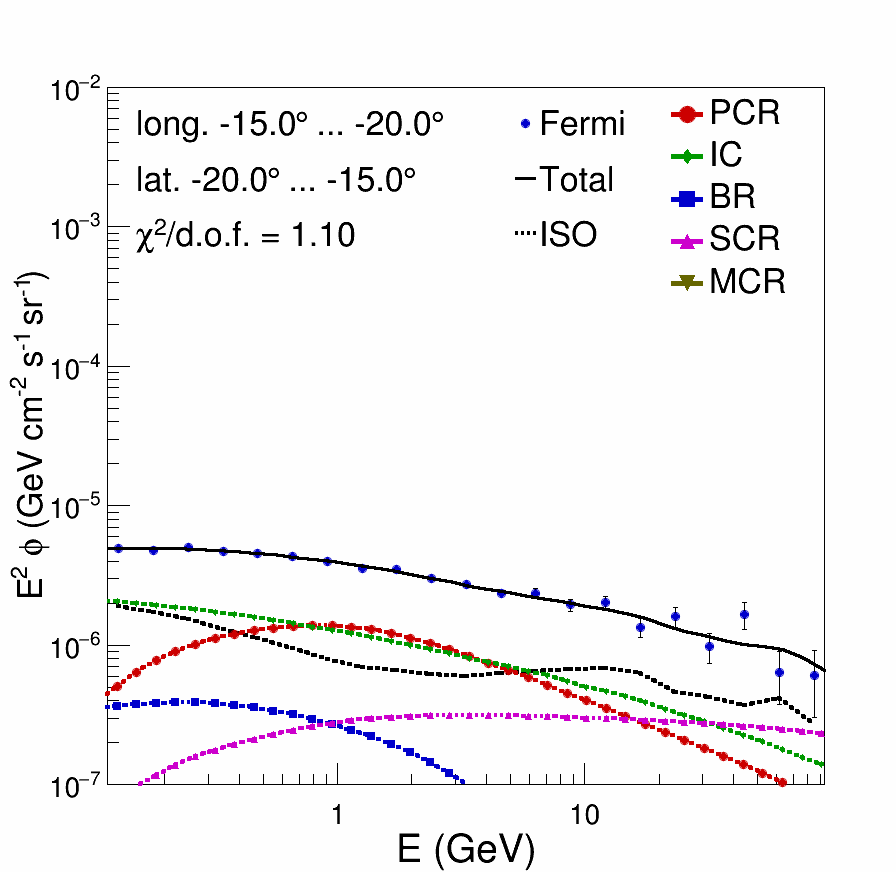}
\includegraphics[width=0.16\textwidth,height=0.16\textwidth,clip]{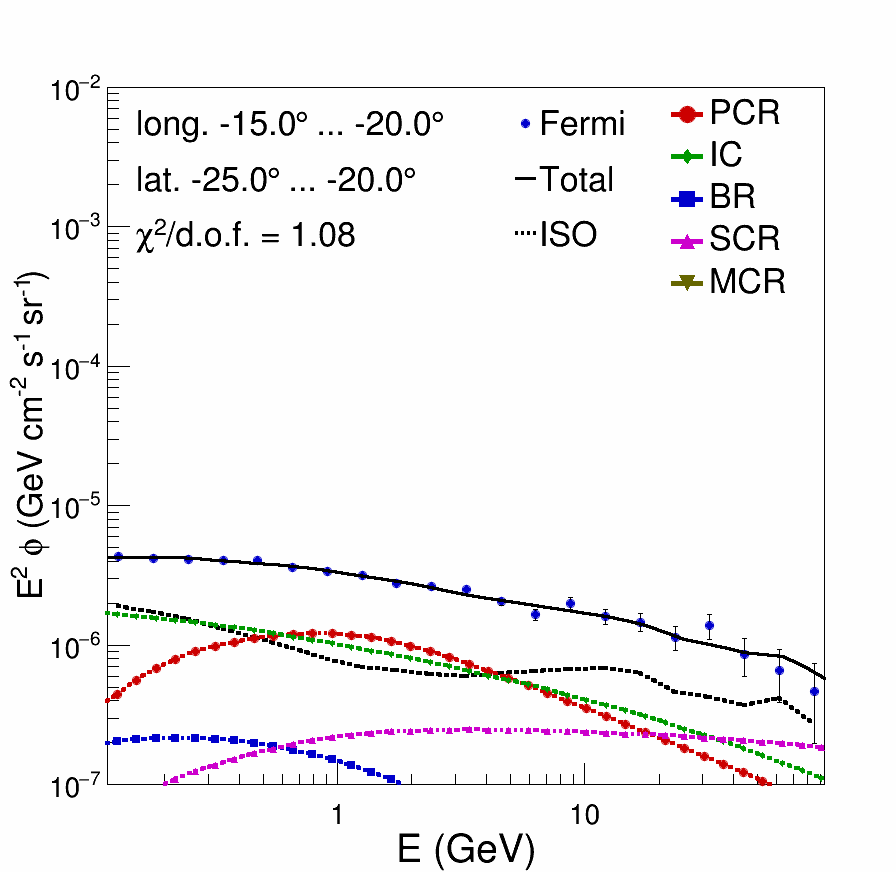}
\includegraphics[width=0.16\textwidth,height=0.16\textwidth,clip]{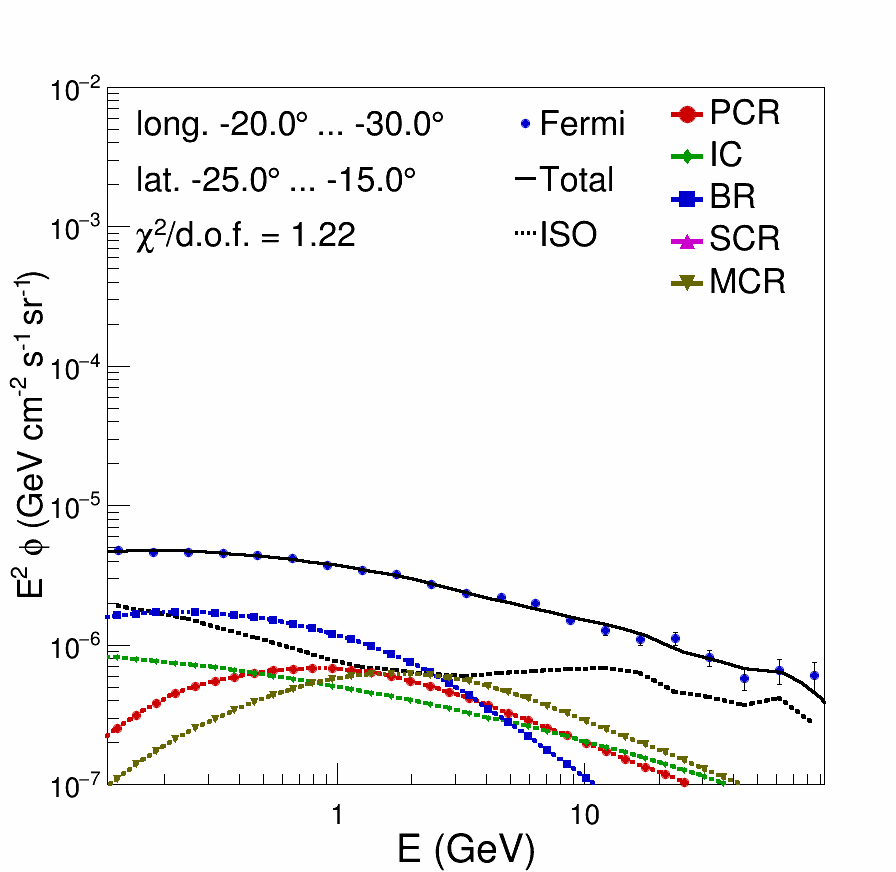}
\includegraphics[width=0.16\textwidth,height=0.16\textwidth,clip]{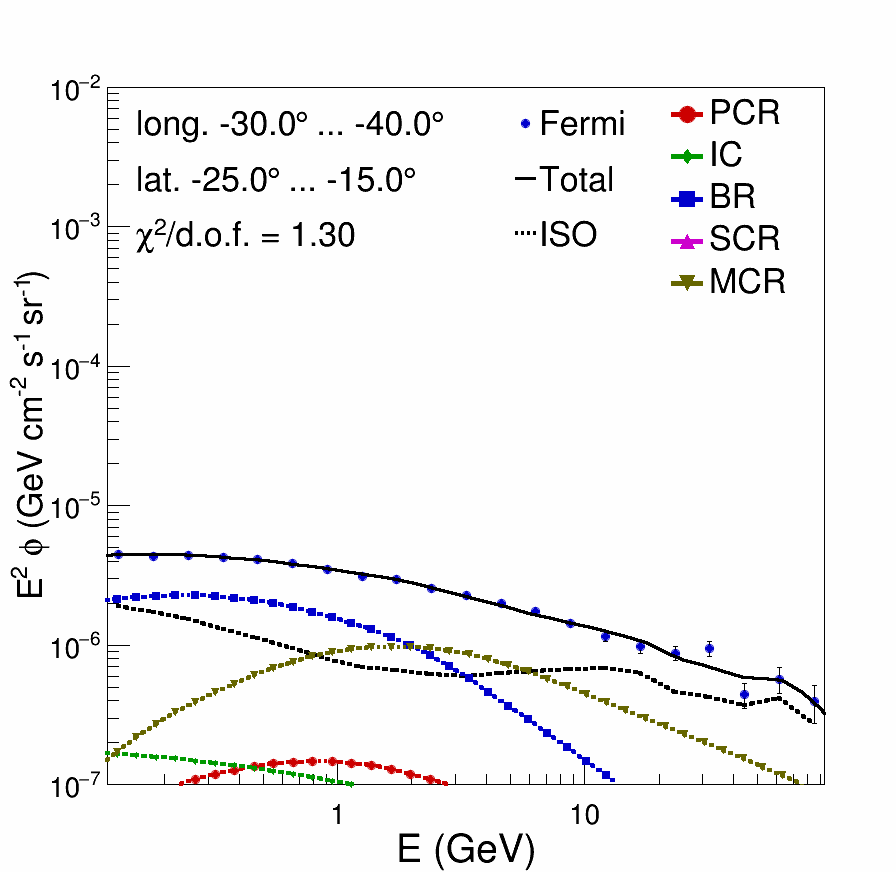}
\includegraphics[width=0.16\textwidth,height=0.16\textwidth,clip]{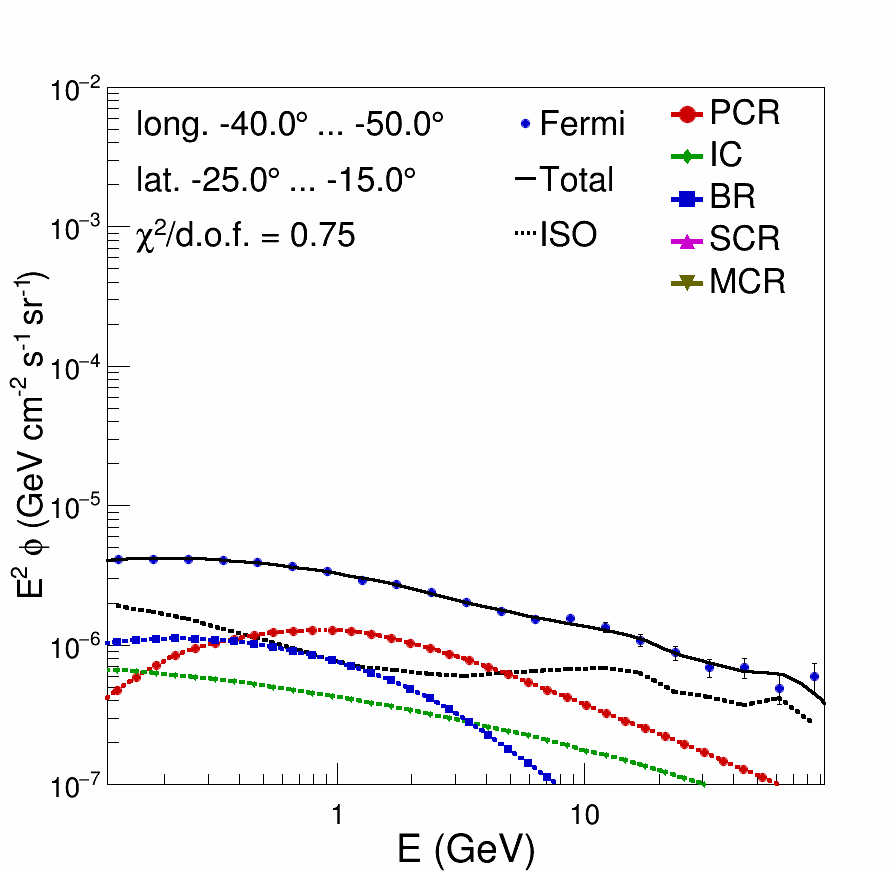}
\includegraphics[width=0.16\textwidth,height=0.16\textwidth,clip]{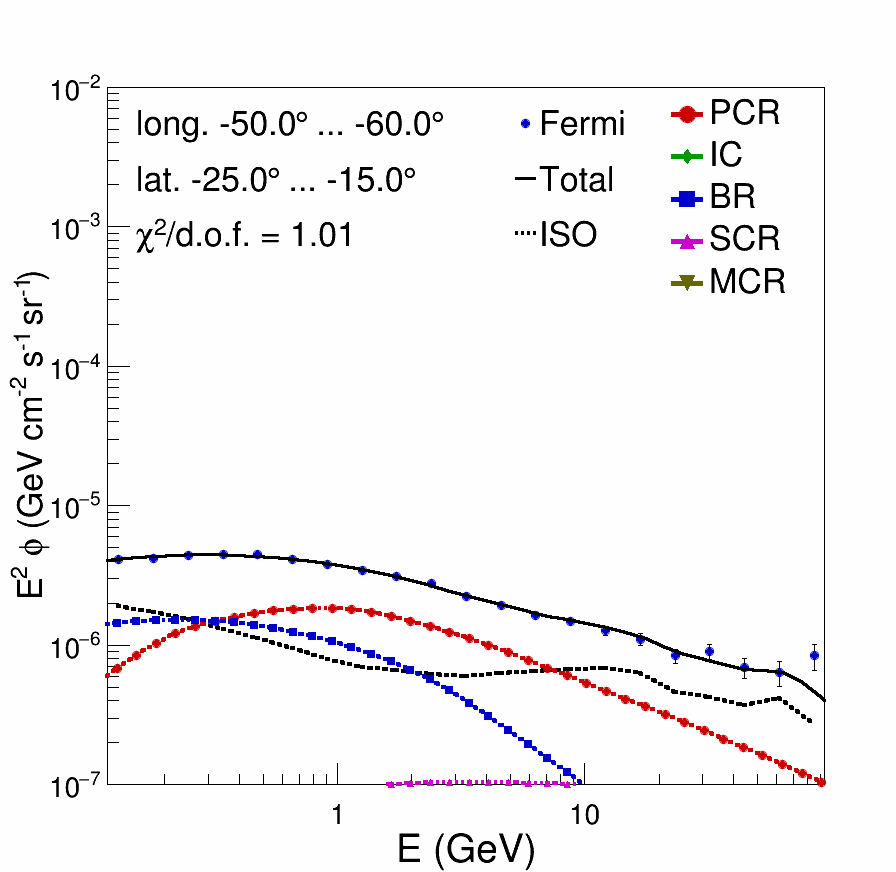}
\includegraphics[width=0.16\textwidth,height=0.16\textwidth,clip]{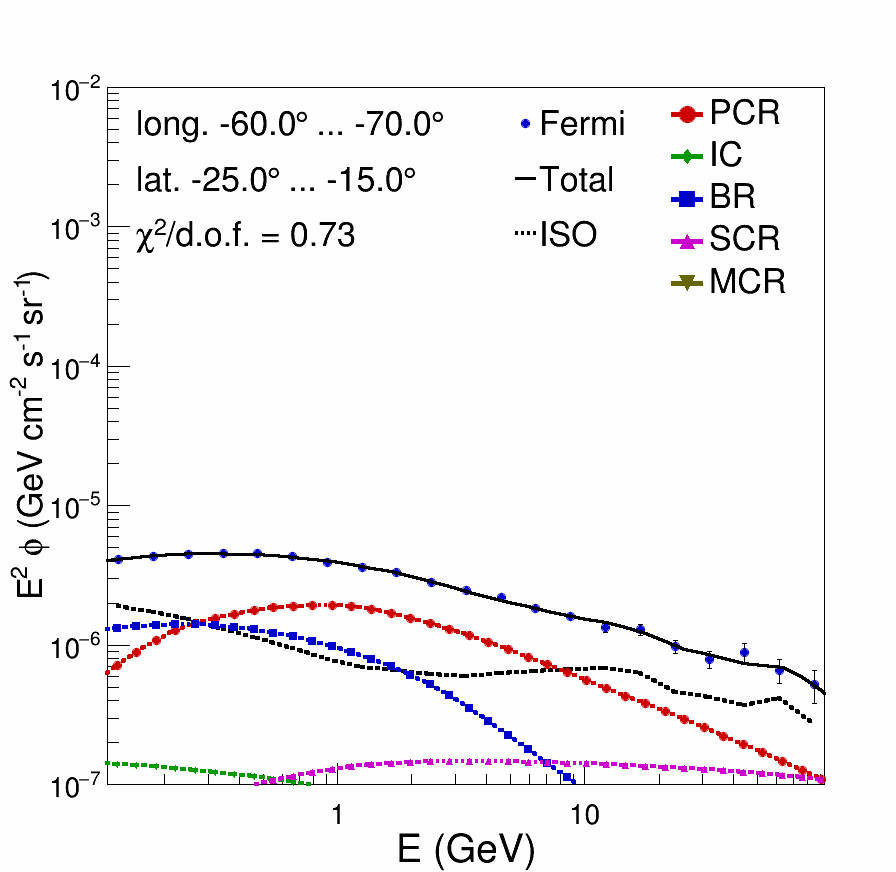}
\includegraphics[width=0.16\textwidth,height=0.16\textwidth,clip]{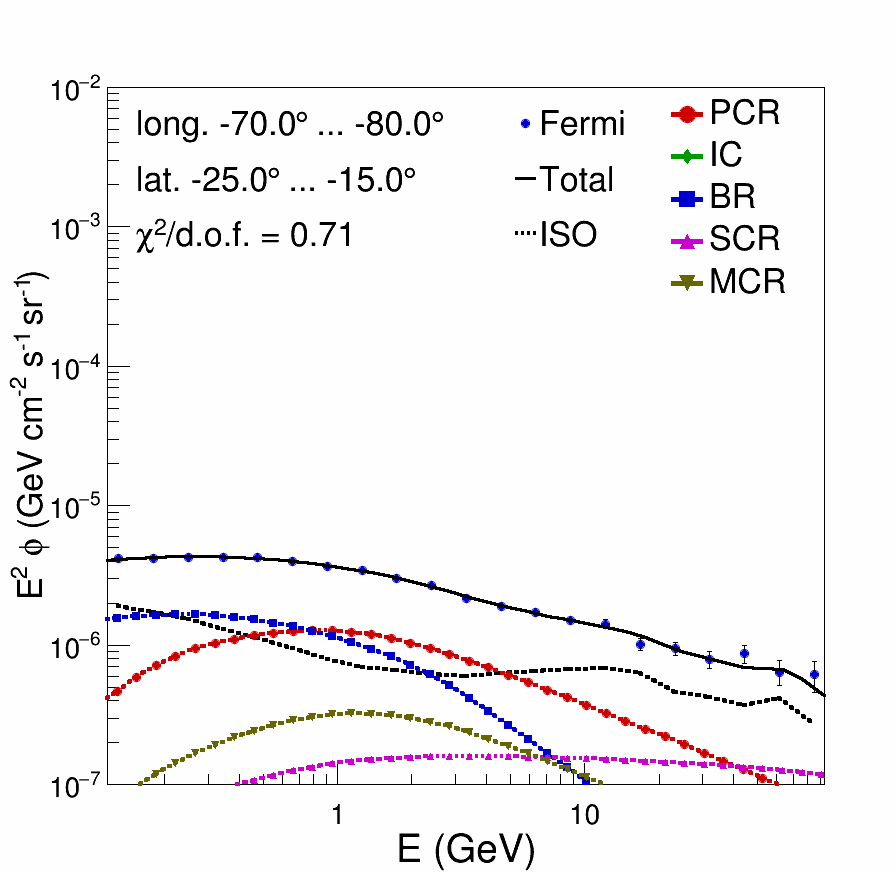}
\includegraphics[width=0.16\textwidth,height=0.16\textwidth,clip]{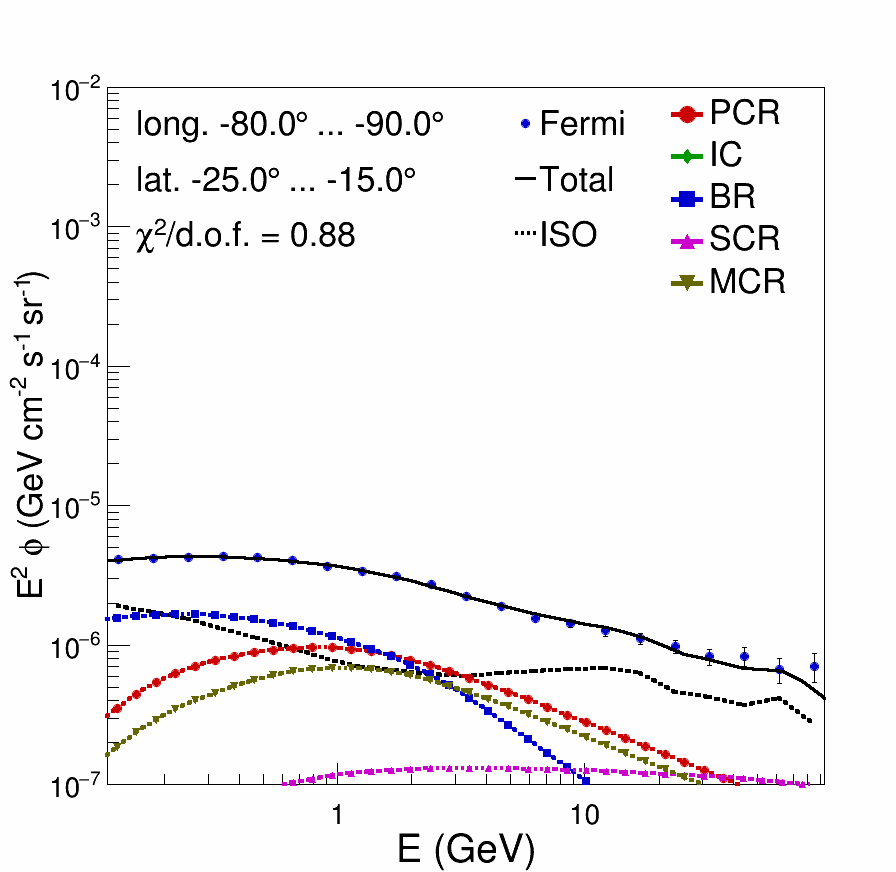}
\includegraphics[width=0.16\textwidth,height=0.16\textwidth,clip]{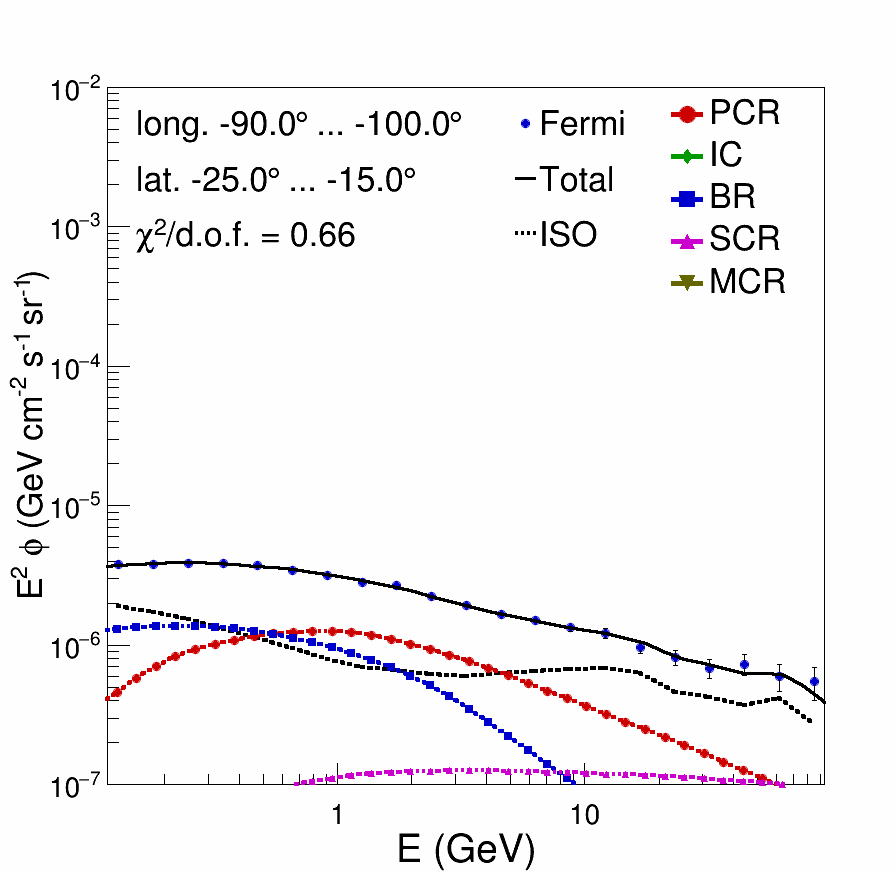}
\includegraphics[width=0.16\textwidth,height=0.16\textwidth,clip]{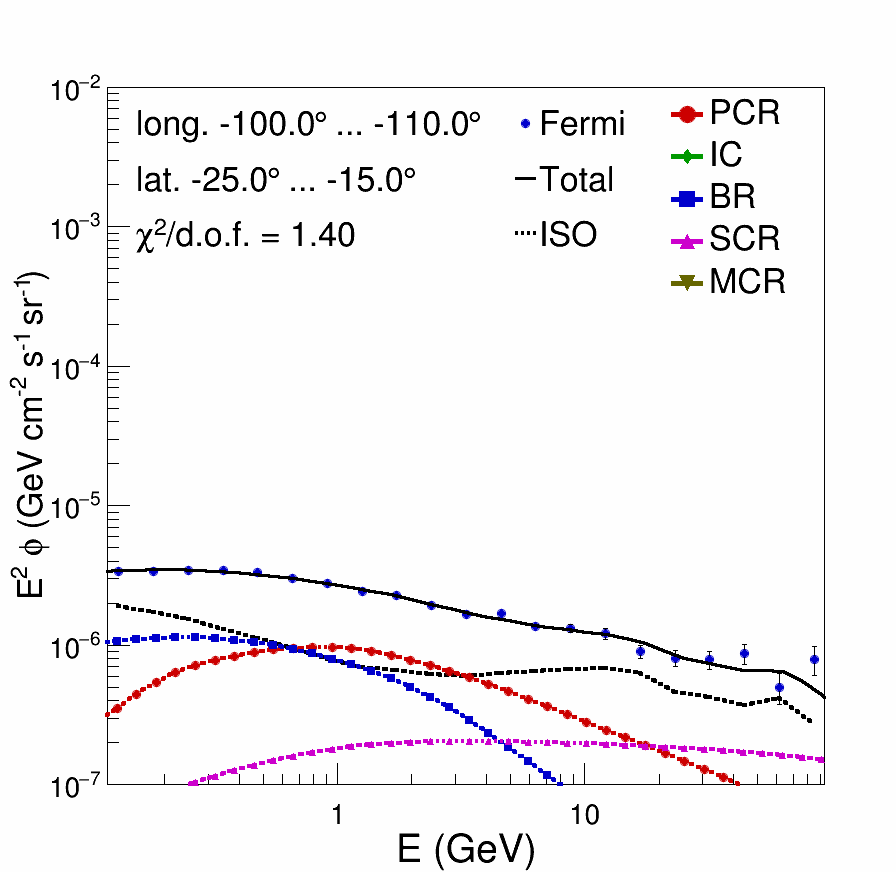}
\includegraphics[width=0.16\textwidth,height=0.16\textwidth,clip]{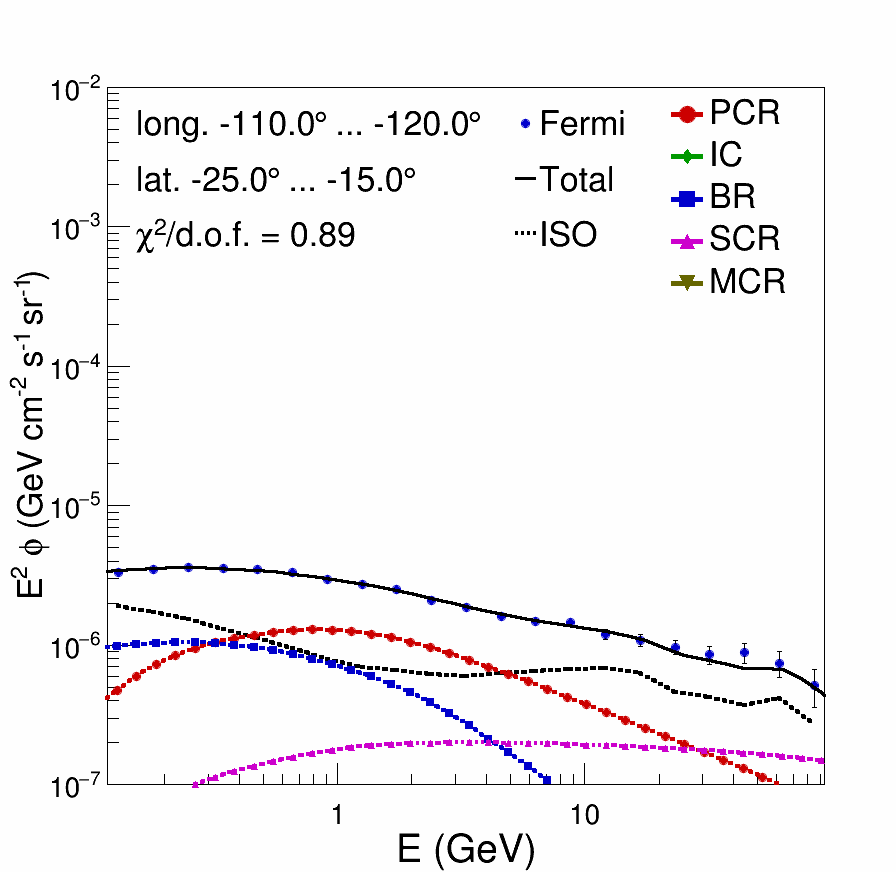}
\includegraphics[width=0.16\textwidth,height=0.16\textwidth,clip]{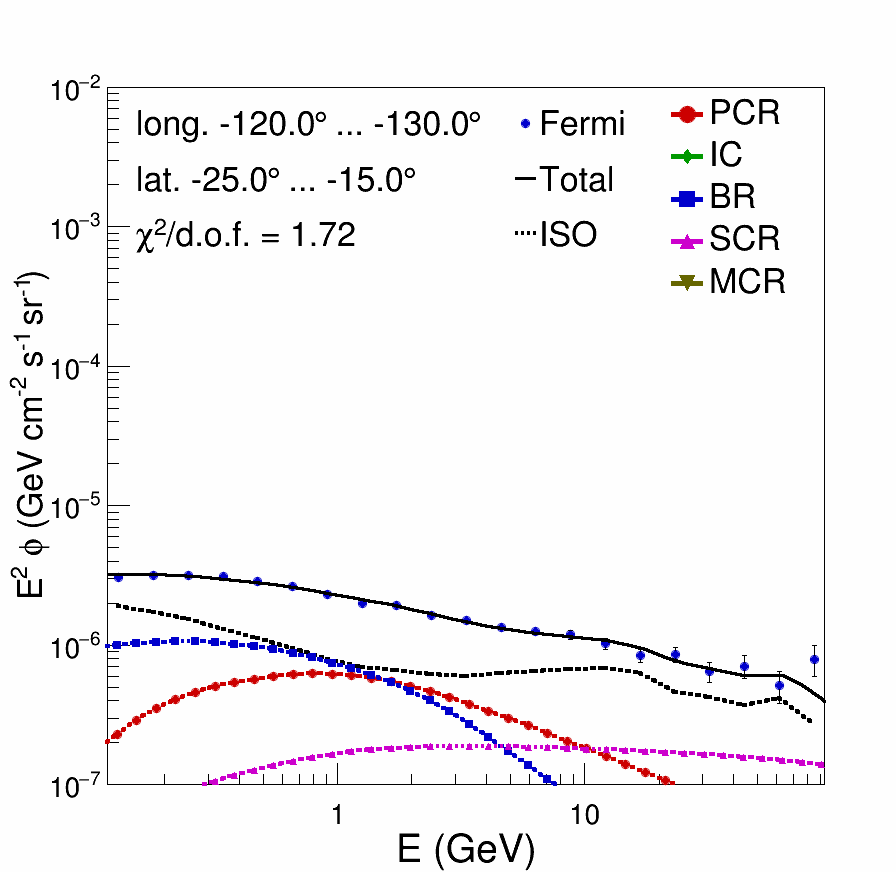}
\includegraphics[width=0.16\textwidth,height=0.16\textwidth,clip]{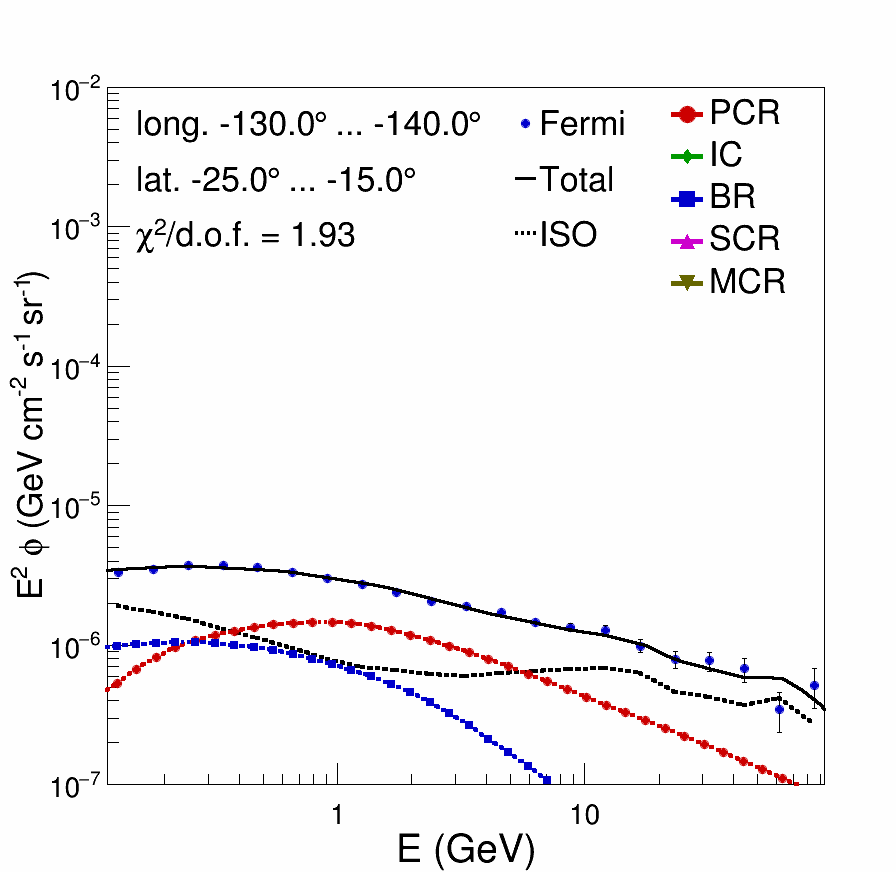}
\includegraphics[width=0.16\textwidth,height=0.16\textwidth,clip]{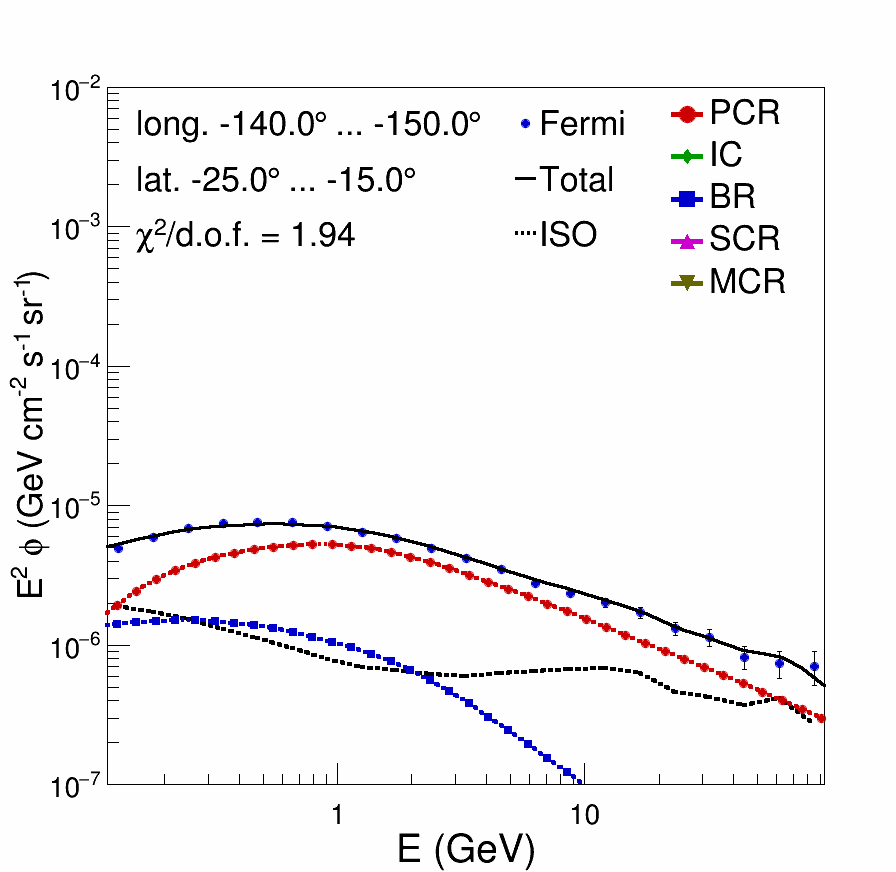}
\includegraphics[width=0.16\textwidth,height=0.16\textwidth,clip]{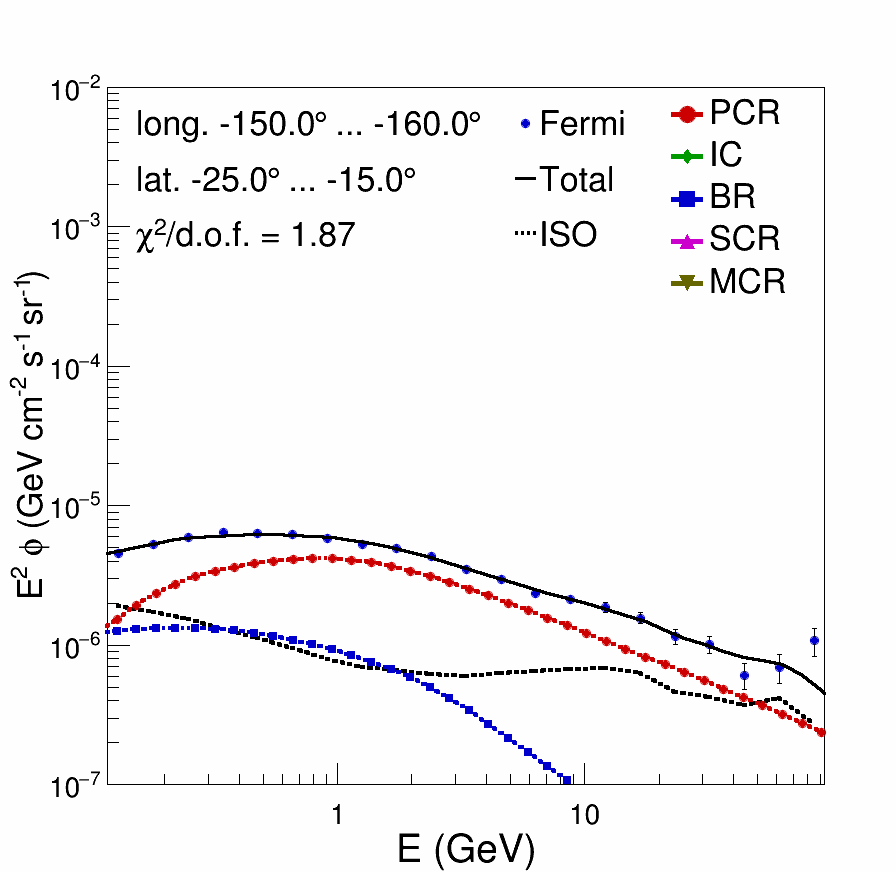}
\includegraphics[width=0.16\textwidth,height=0.16\textwidth,clip]{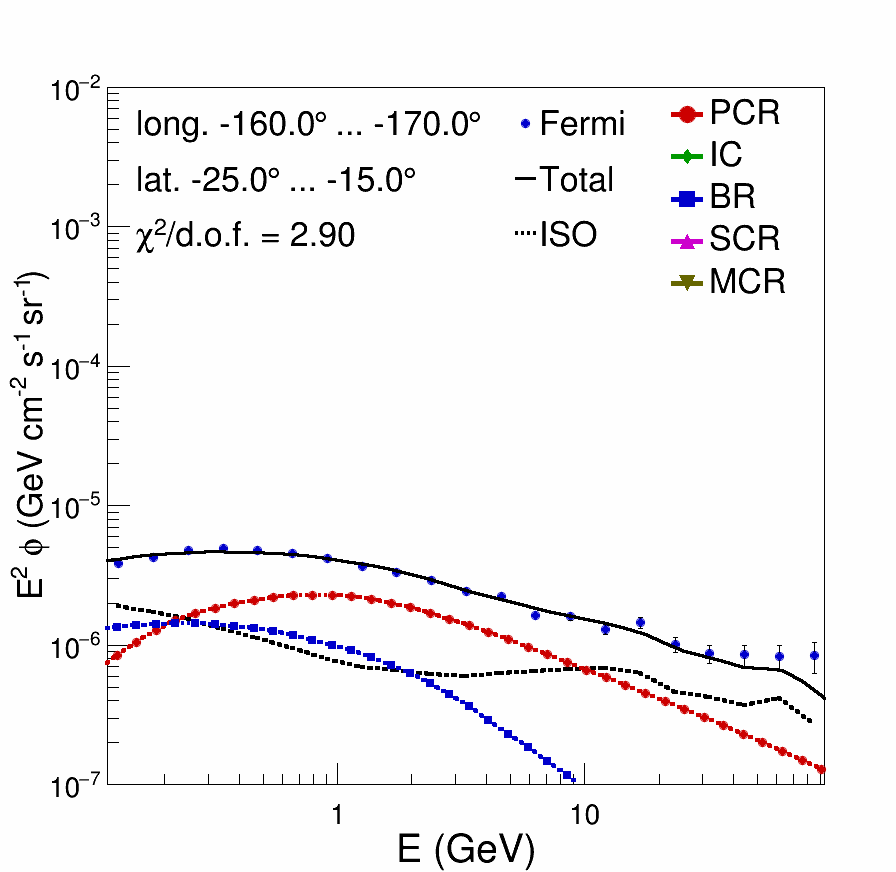}
\includegraphics[width=0.16\textwidth,height=0.16\textwidth,clip]{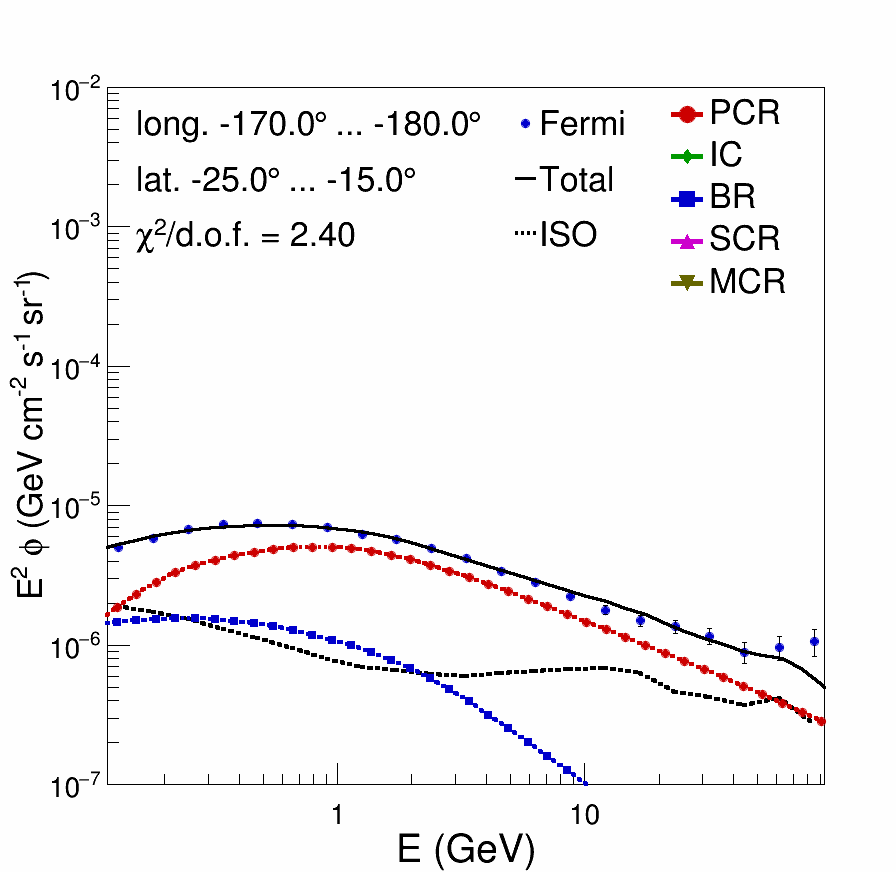}
\caption[]{Template fits for latitudes  with $-25.0^\circ<b<-15.0^\circ$ and longitudes decreasing from 180$^\circ$ to -180$^\circ$.} \label{F26}
\end{figure}
\clearpage
\begin{figure}
\centering
\includegraphics[width=0.16\textwidth,height=0.16\textwidth,clip]{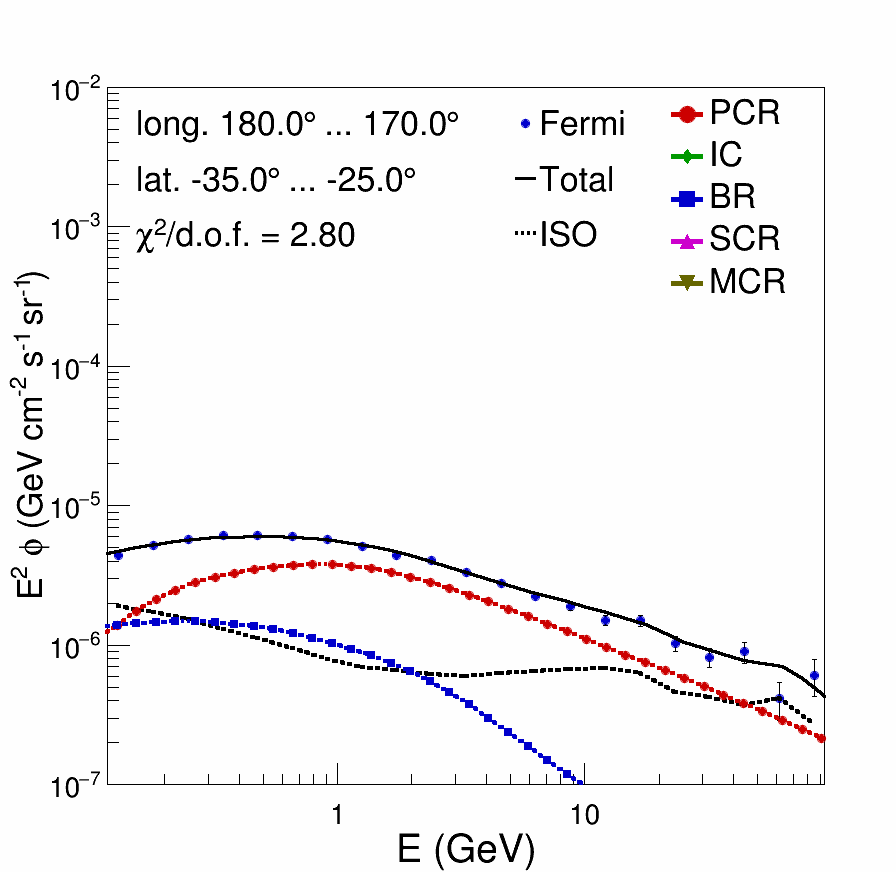}
\includegraphics[width=0.16\textwidth,height=0.16\textwidth,clip]{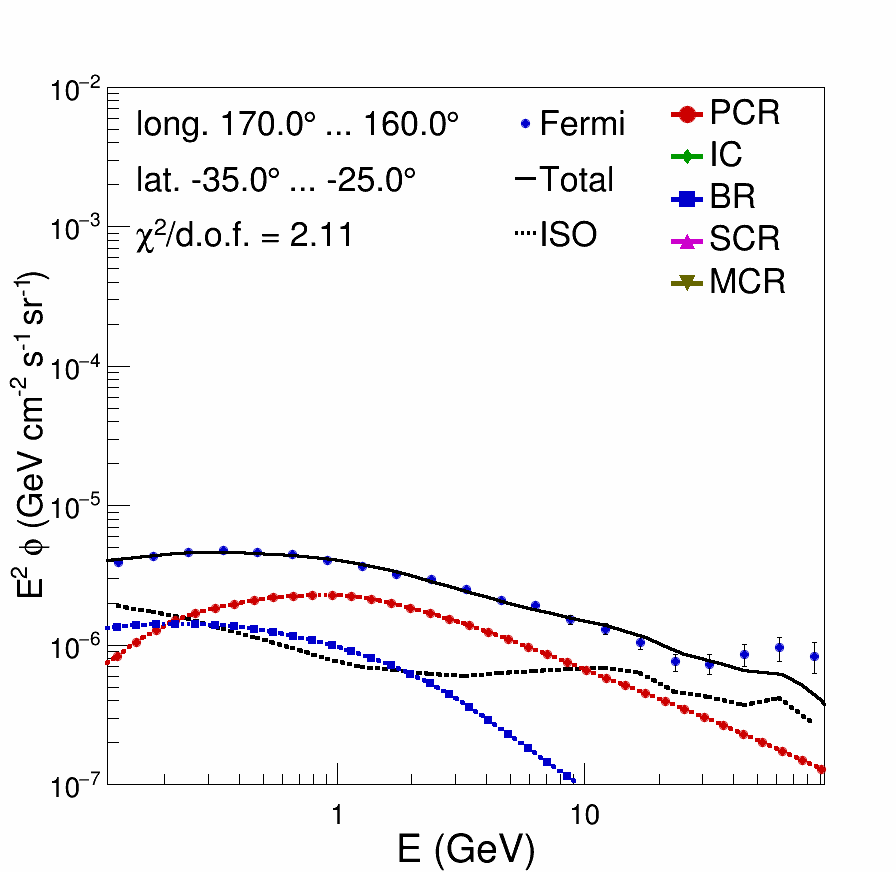}
\includegraphics[width=0.16\textwidth,height=0.16\textwidth,clip]{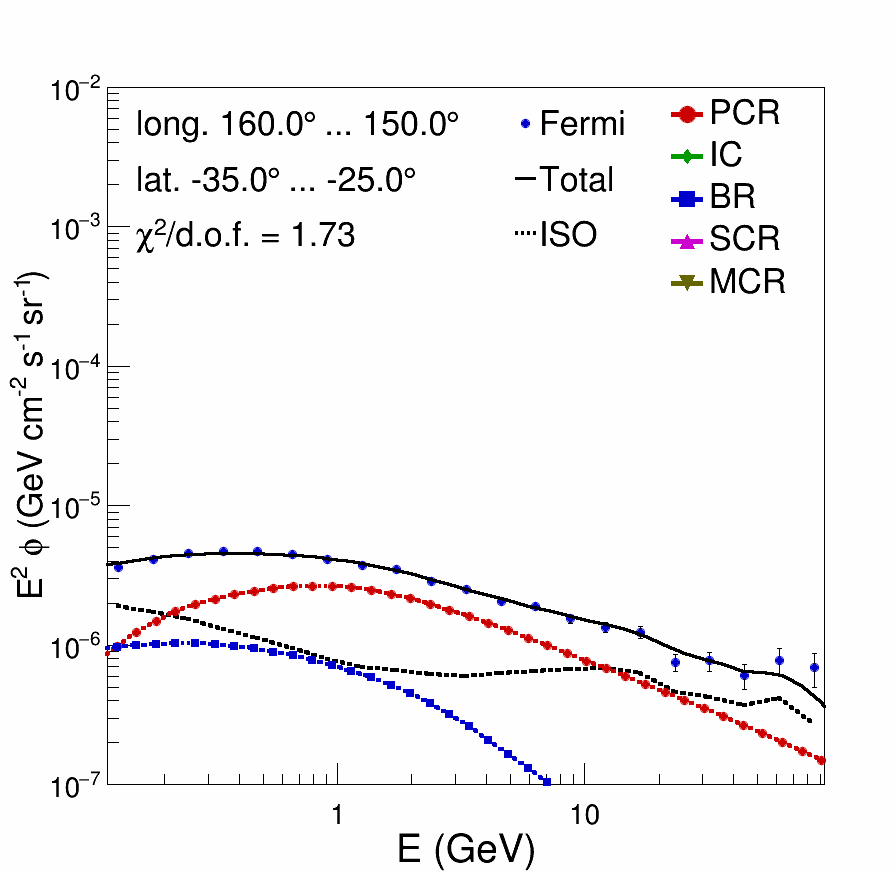}
\includegraphics[width=0.16\textwidth,height=0.16\textwidth,clip]{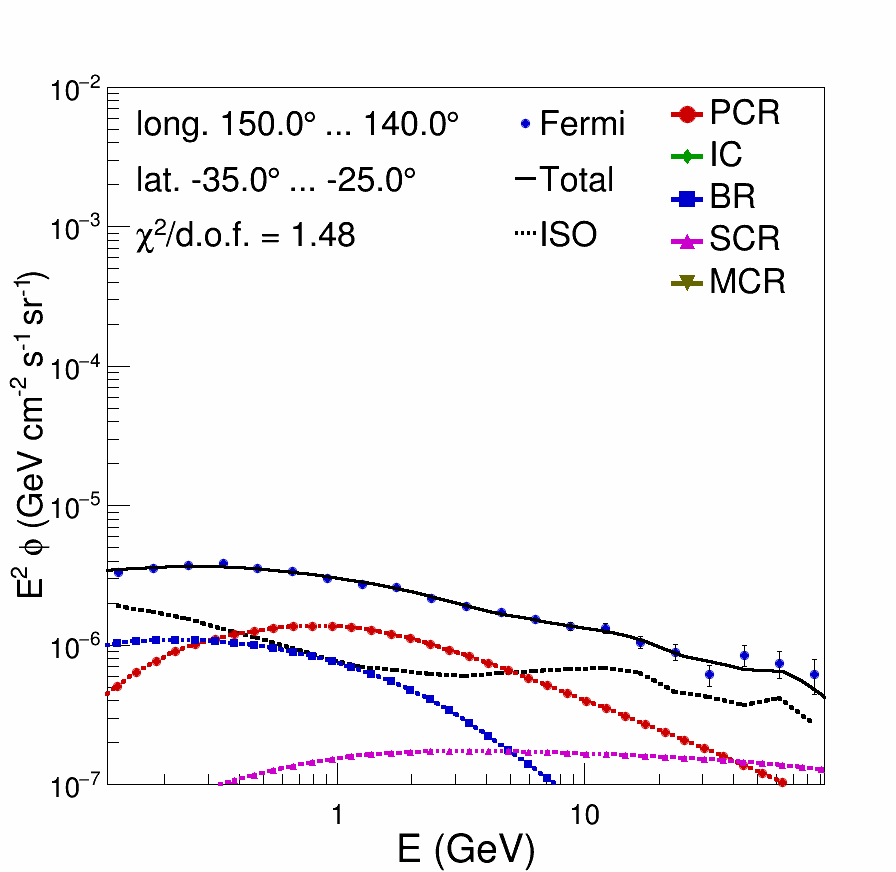}
\includegraphics[width=0.16\textwidth,height=0.16\textwidth,clip]{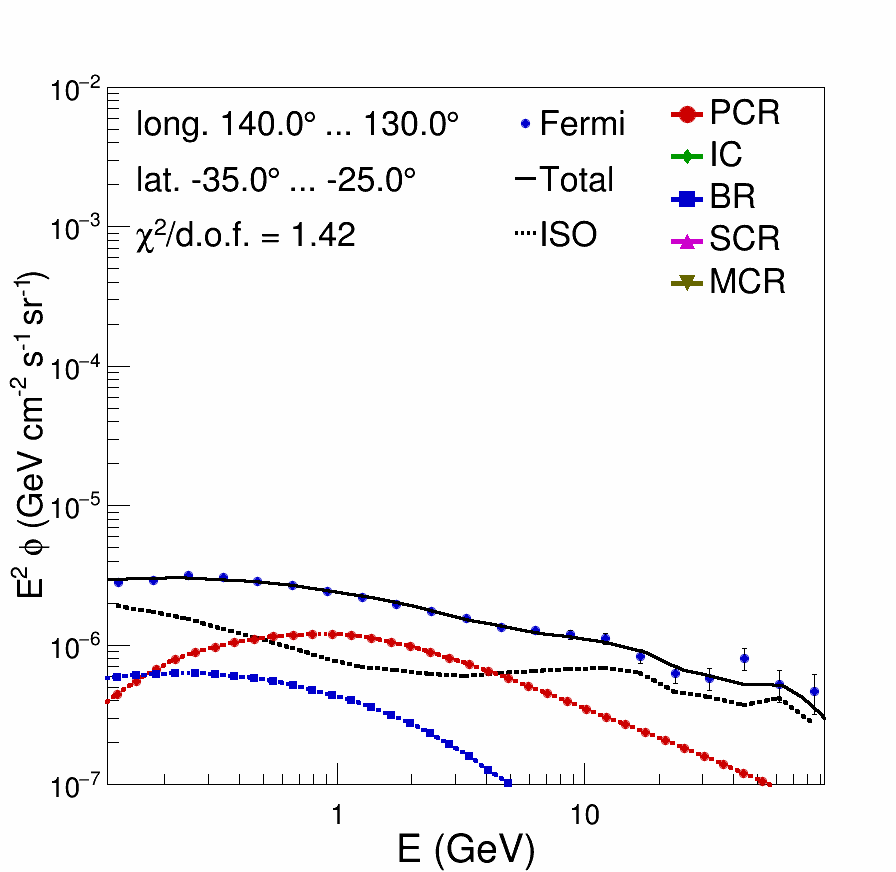}
\includegraphics[width=0.16\textwidth,height=0.16\textwidth,clip]{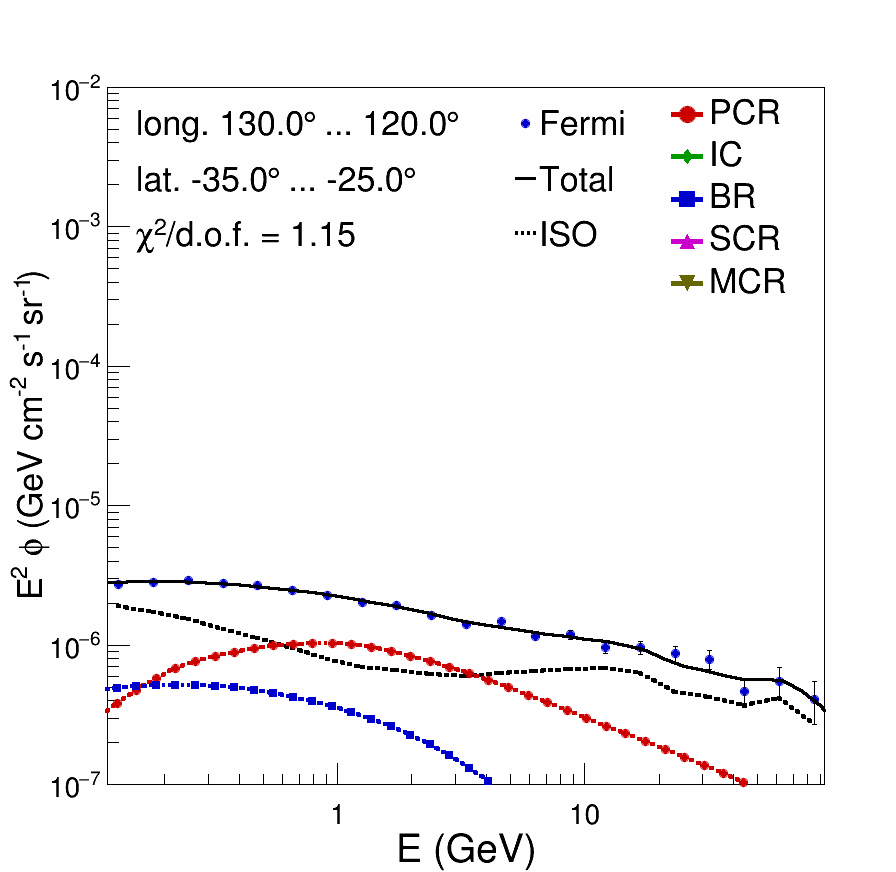}
\includegraphics[width=0.16\textwidth,height=0.16\textwidth,clip]{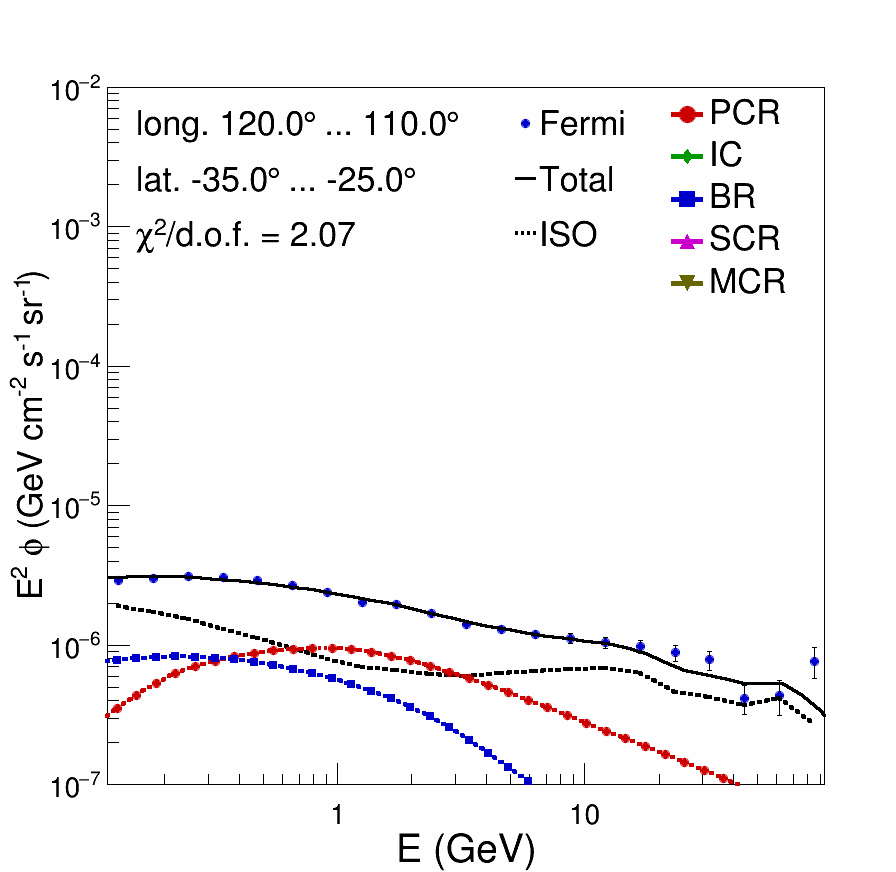}
\includegraphics[width=0.16\textwidth,height=0.16\textwidth,clip]{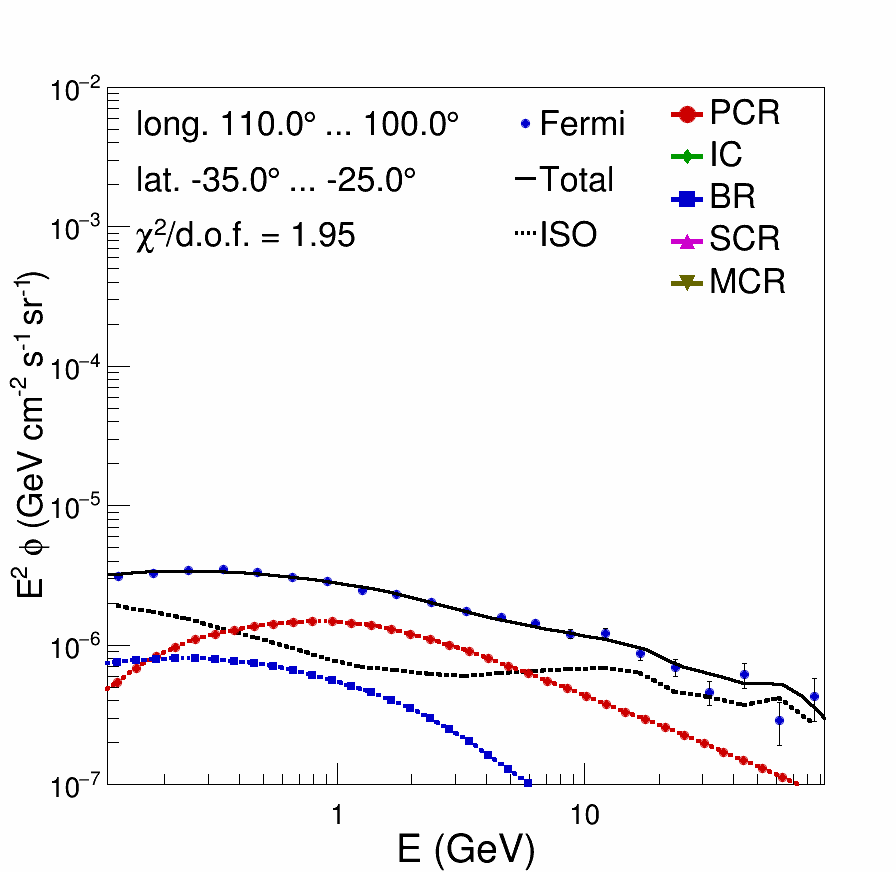}
\includegraphics[width=0.16\textwidth,height=0.16\textwidth,clip]{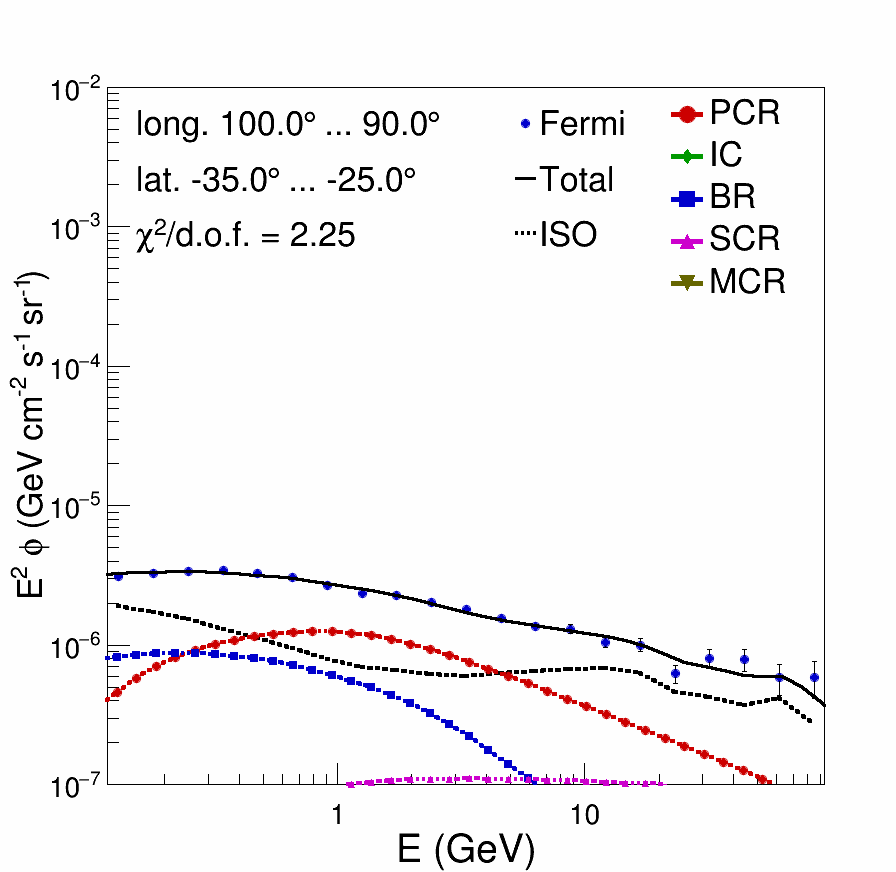}
\includegraphics[width=0.16\textwidth,height=0.16\textwidth,clip]{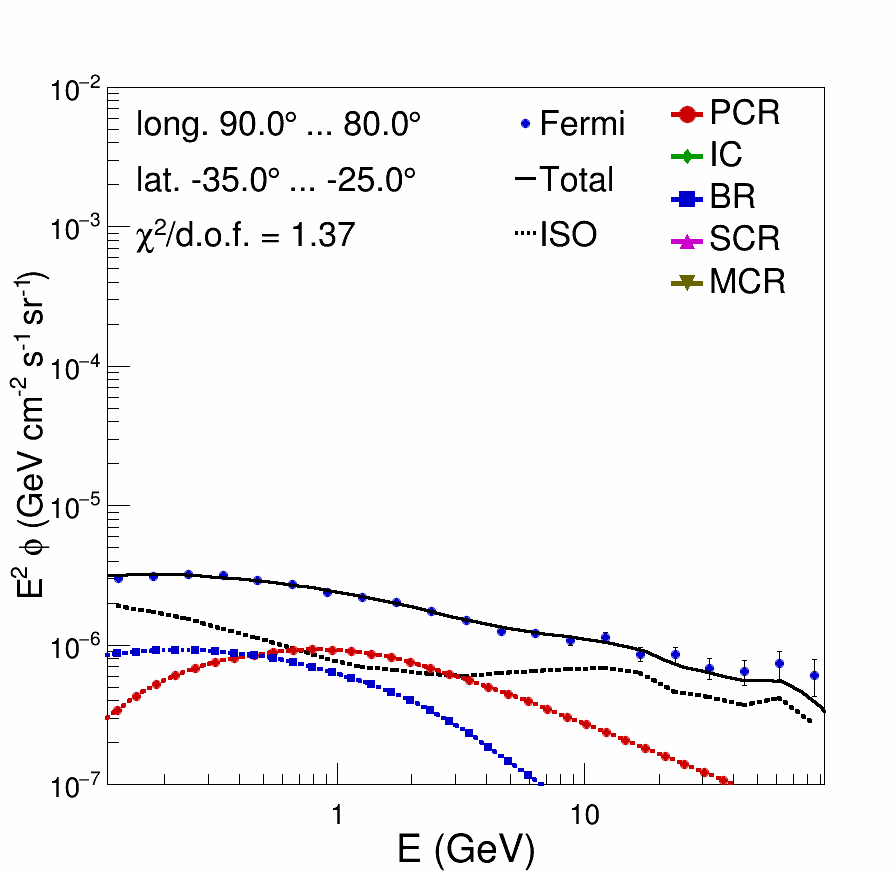}
\includegraphics[width=0.16\textwidth,height=0.16\textwidth,clip]{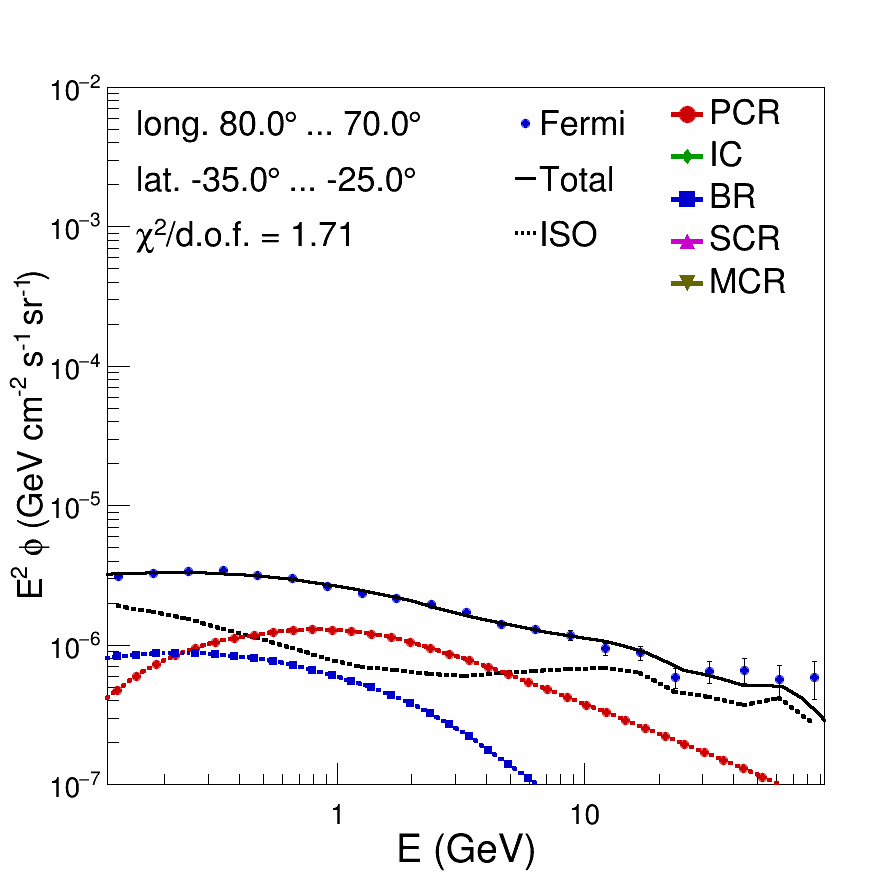}
\includegraphics[width=0.16\textwidth,height=0.16\textwidth,clip]{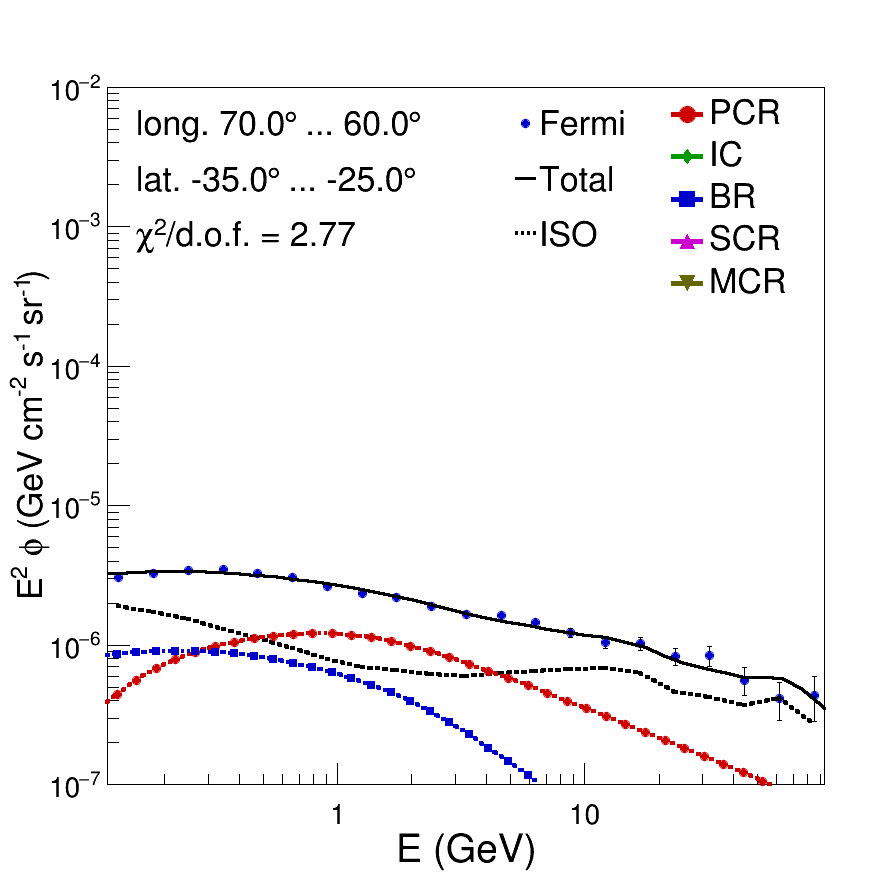}
\includegraphics[width=0.16\textwidth,height=0.16\textwidth,clip]{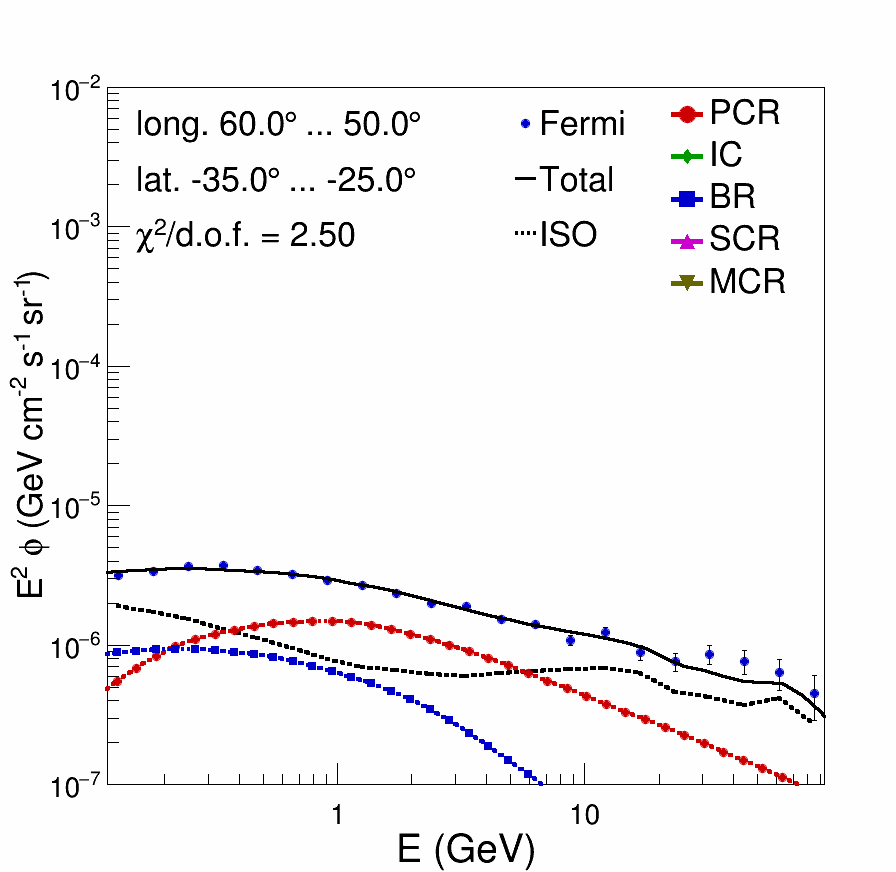}
\includegraphics[width=0.16\textwidth,height=0.16\textwidth,clip]{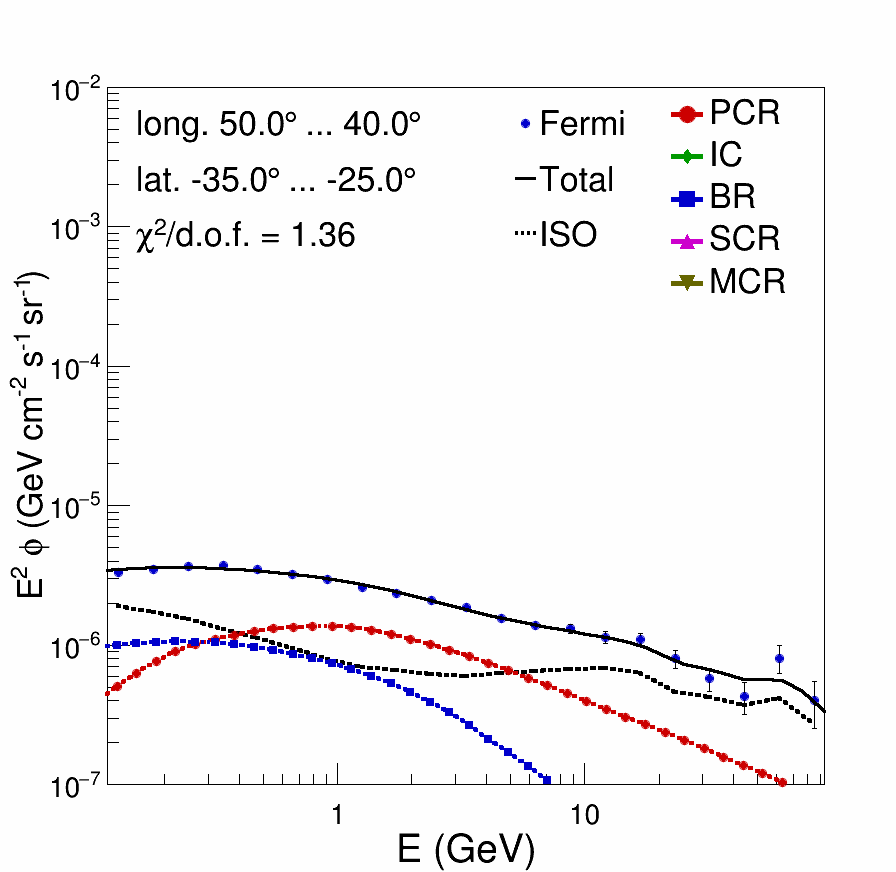}
\includegraphics[width=0.16\textwidth,height=0.16\textwidth,clip]{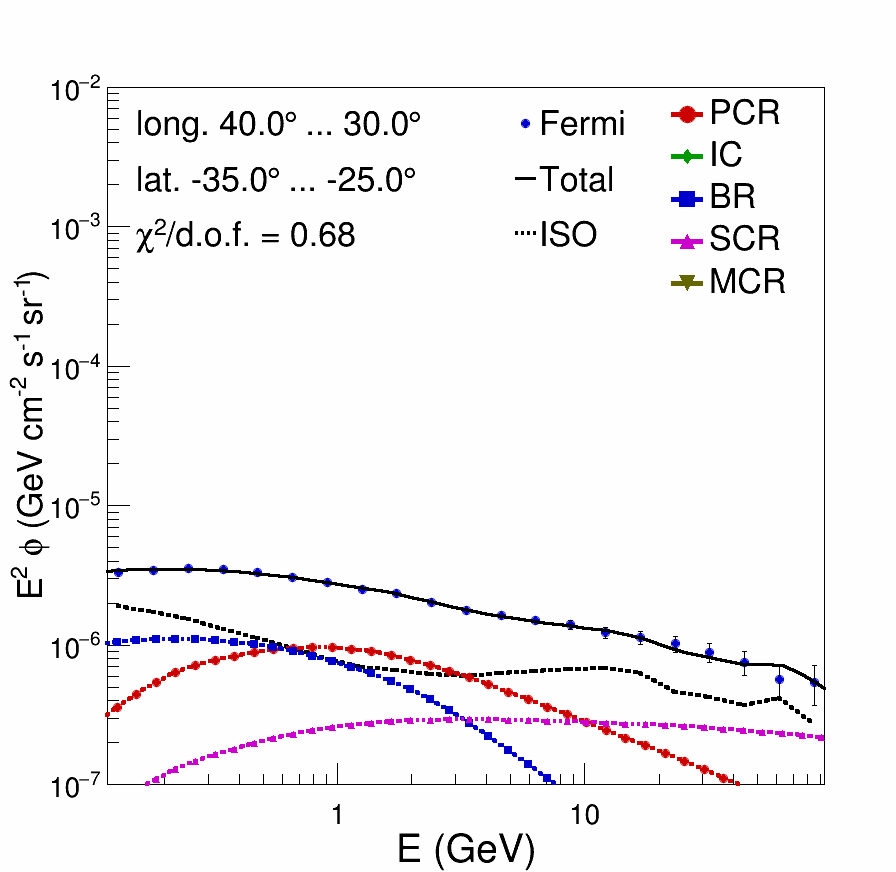}
\includegraphics[width=0.16\textwidth,height=0.16\textwidth,clip]{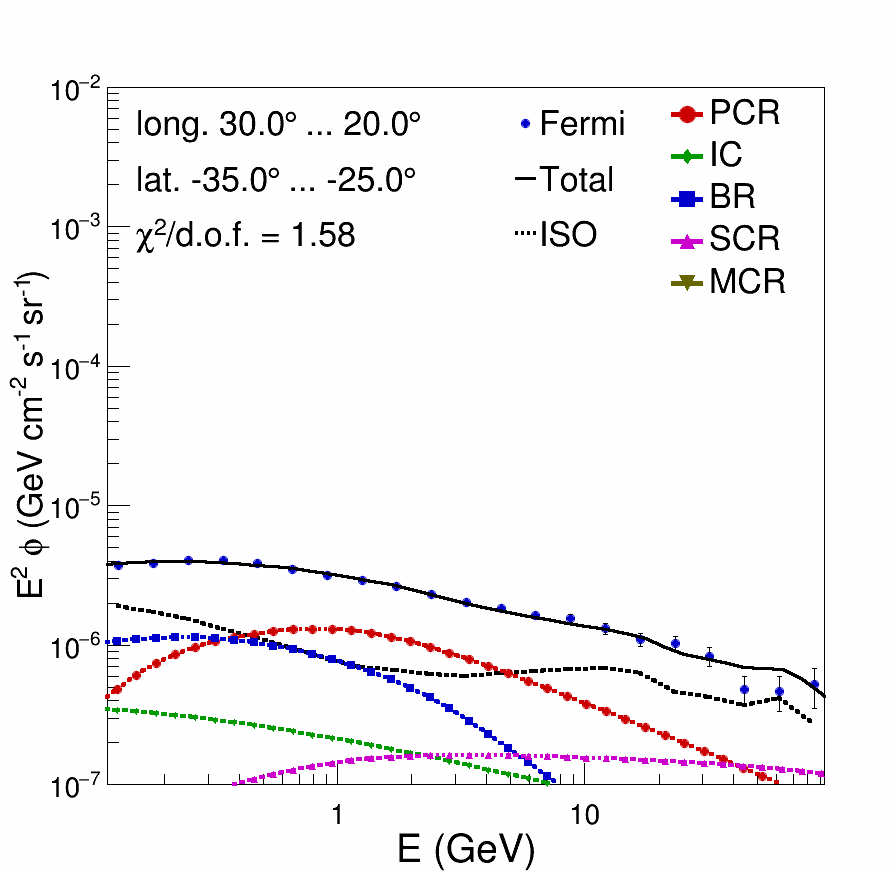}
\includegraphics[width=0.16\textwidth,height=0.16\textwidth,clip]{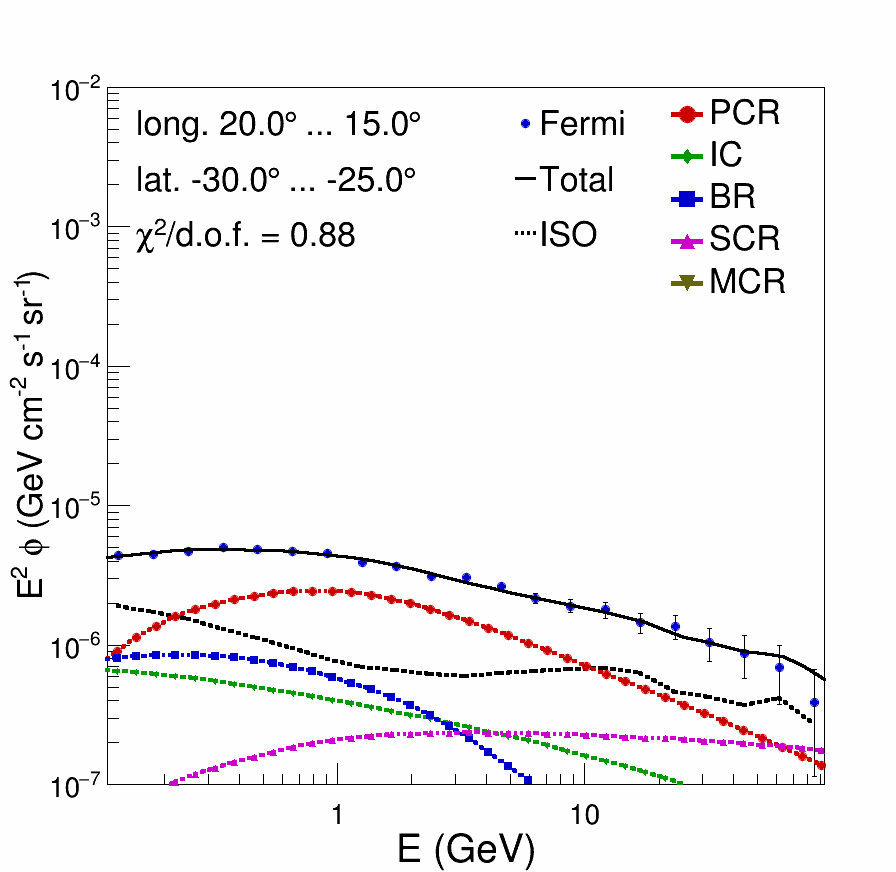}
\includegraphics[width=0.16\textwidth,height=0.16\textwidth,clip]{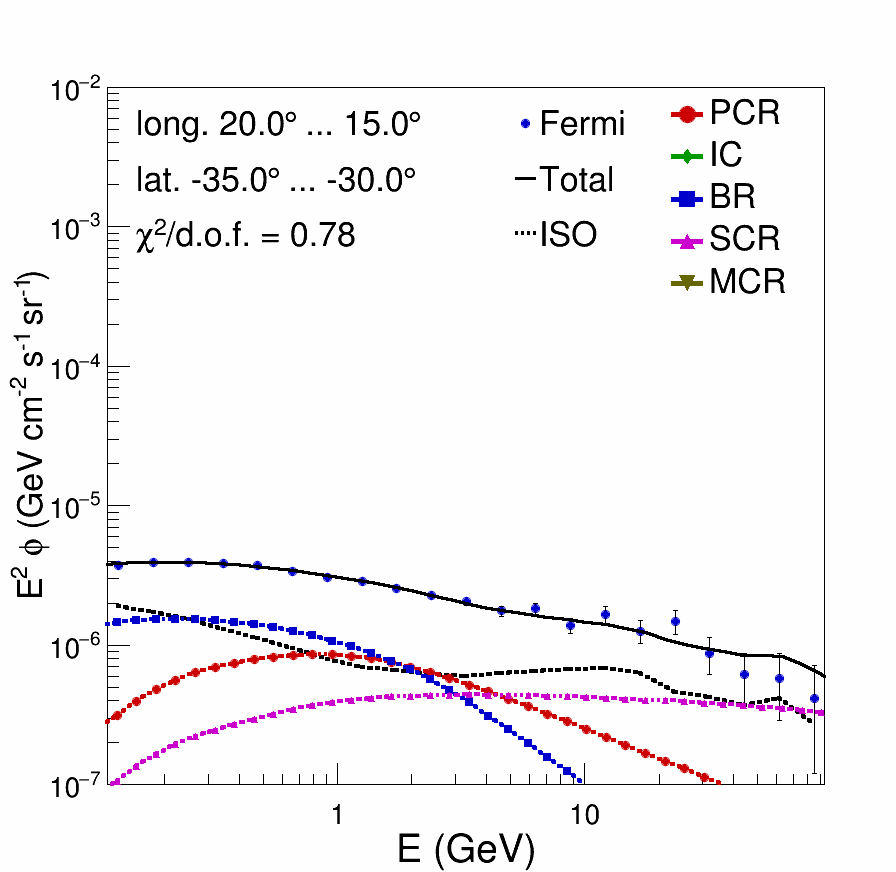}
\includegraphics[width=0.16\textwidth,height=0.16\textwidth,clip]{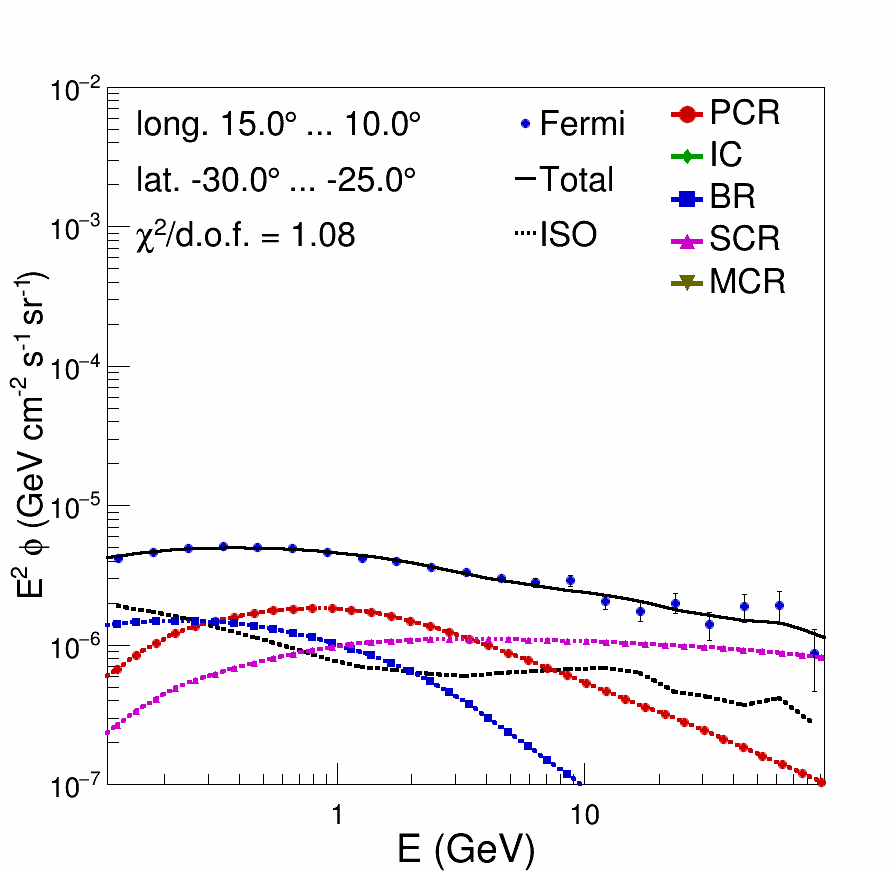}
\includegraphics[width=0.16\textwidth,height=0.16\textwidth,clip]{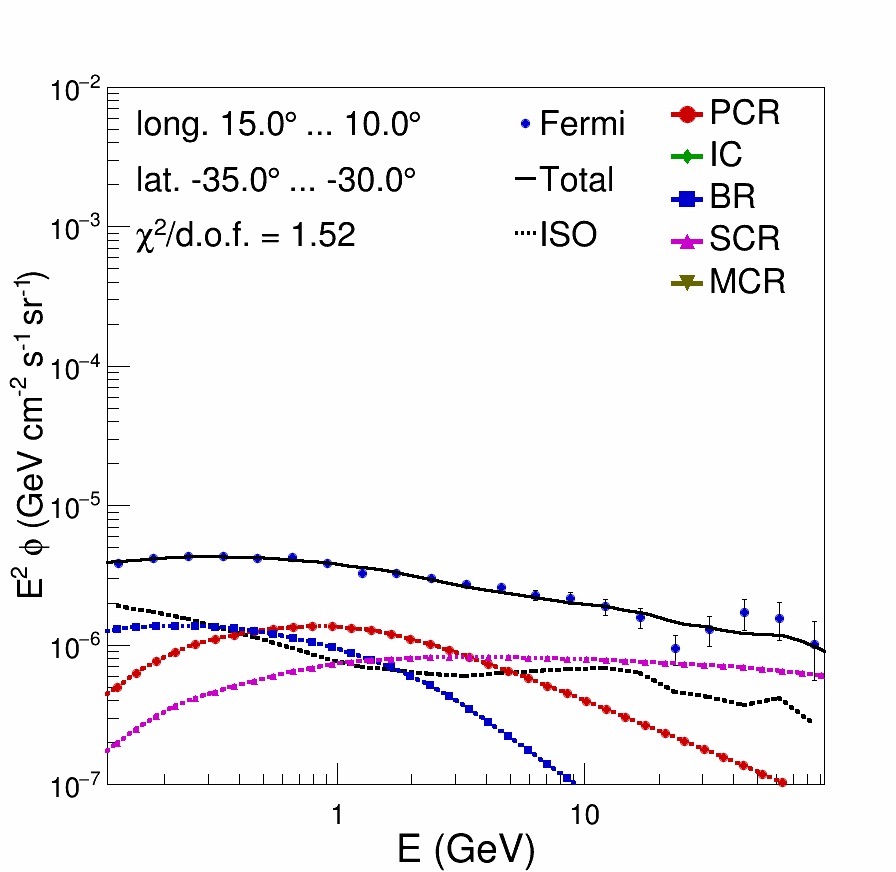}
\includegraphics[width=0.16\textwidth,height=0.16\textwidth,clip]{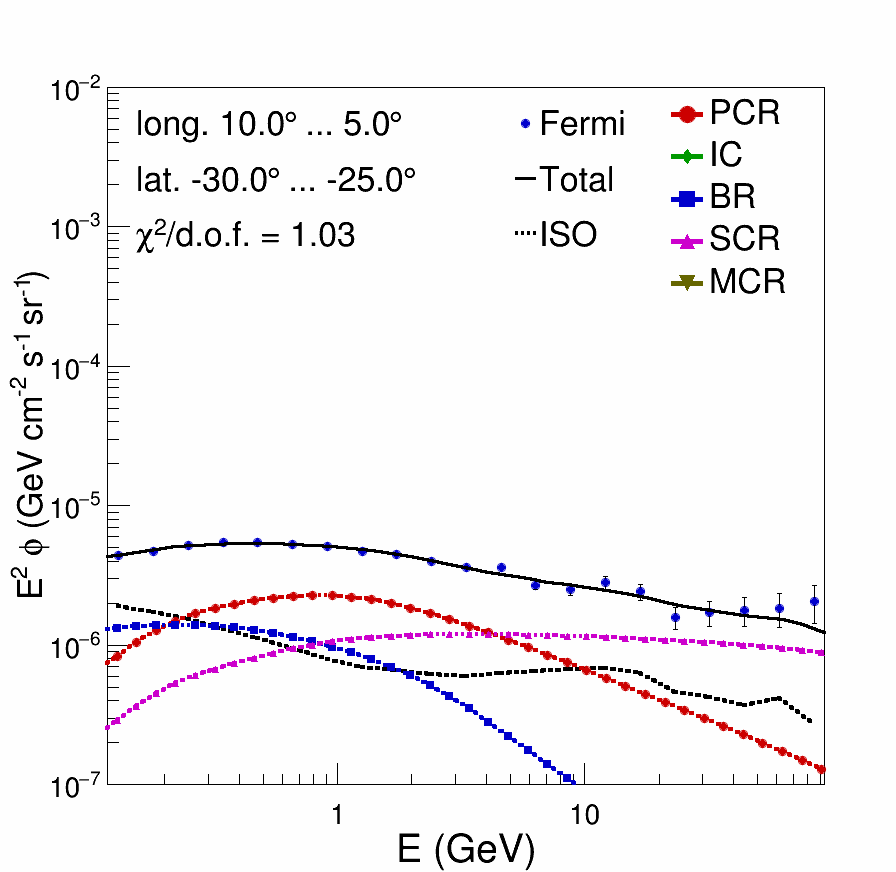}
\includegraphics[width=0.16\textwidth,height=0.16\textwidth,clip]{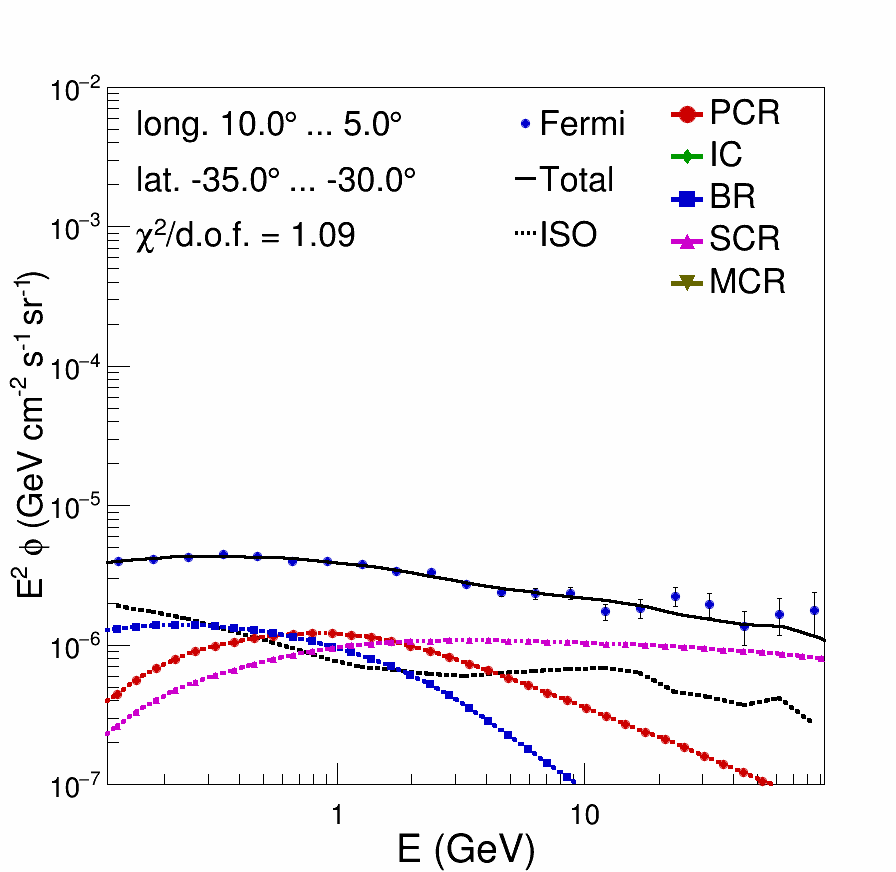}
\includegraphics[width=0.16\textwidth,height=0.16\textwidth,clip]{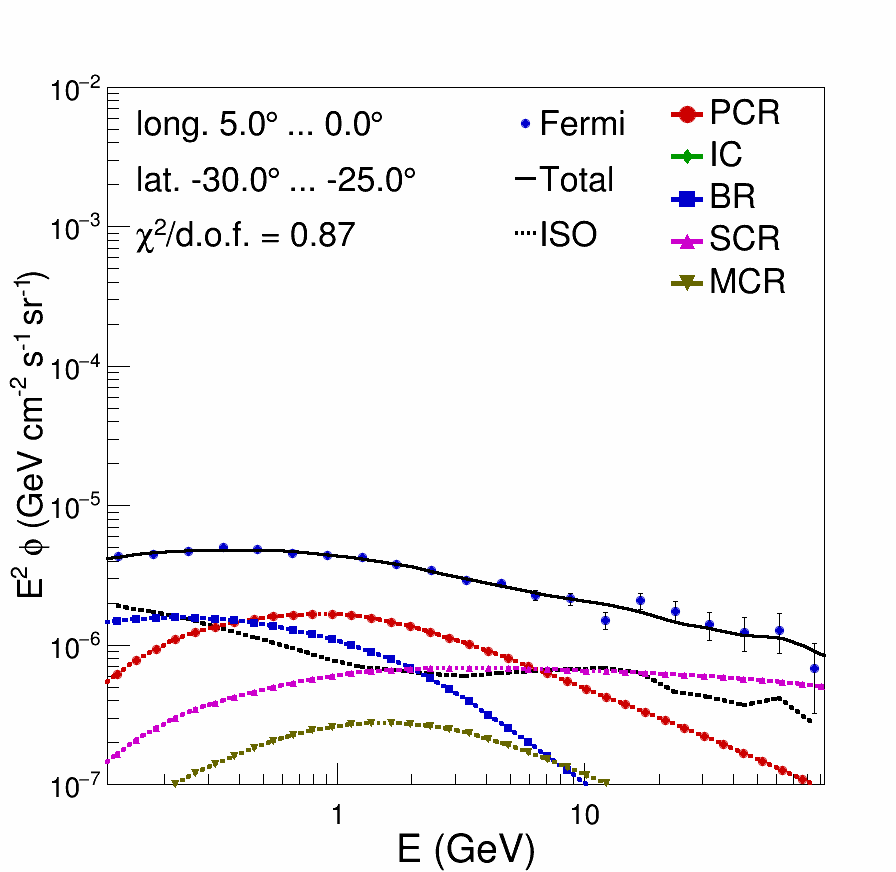}
\includegraphics[width=0.16\textwidth,height=0.16\textwidth,clip]{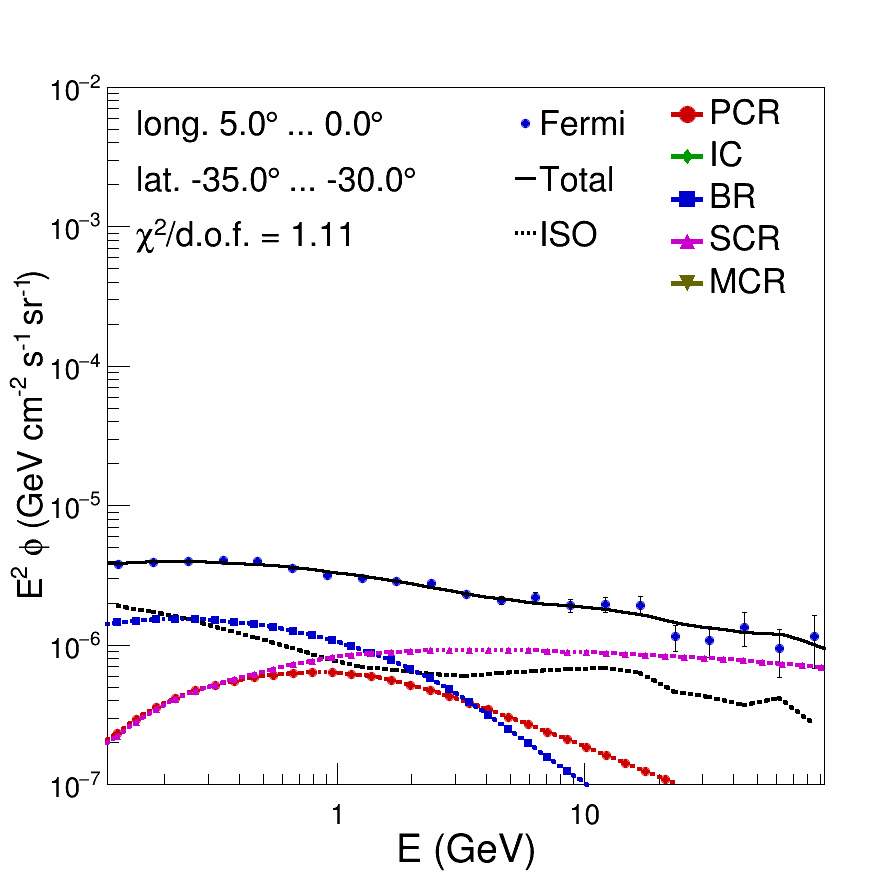}
\includegraphics[width=0.16\textwidth,height=0.16\textwidth,clip]{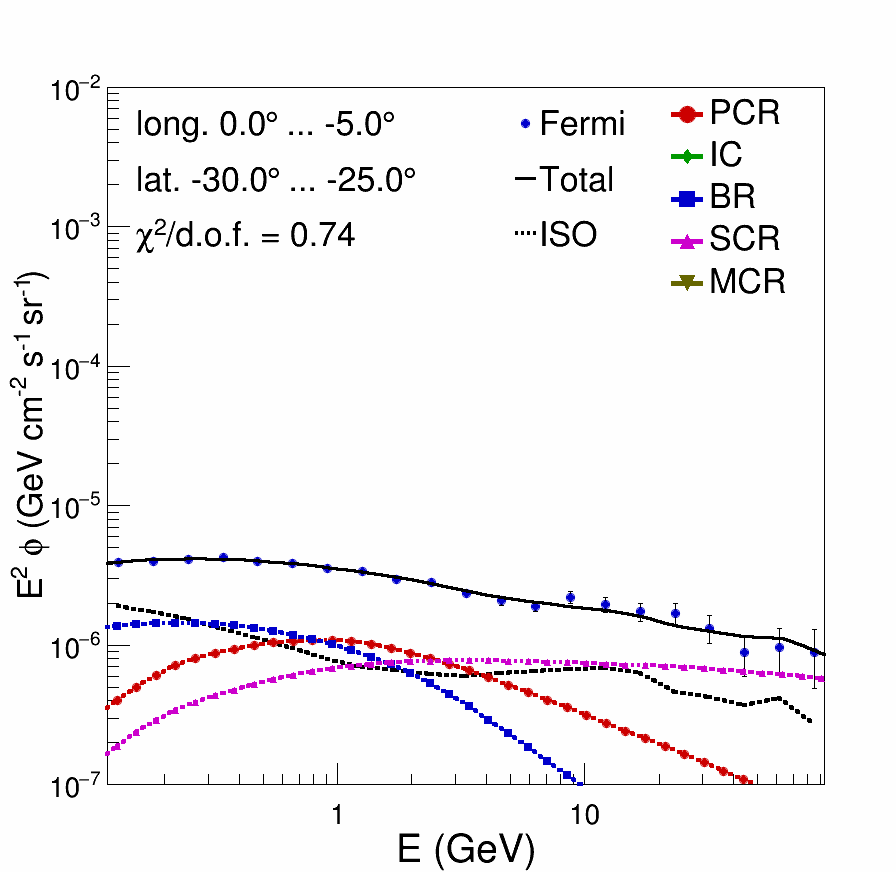}
\includegraphics[width=0.16\textwidth,height=0.16\textwidth,clip]{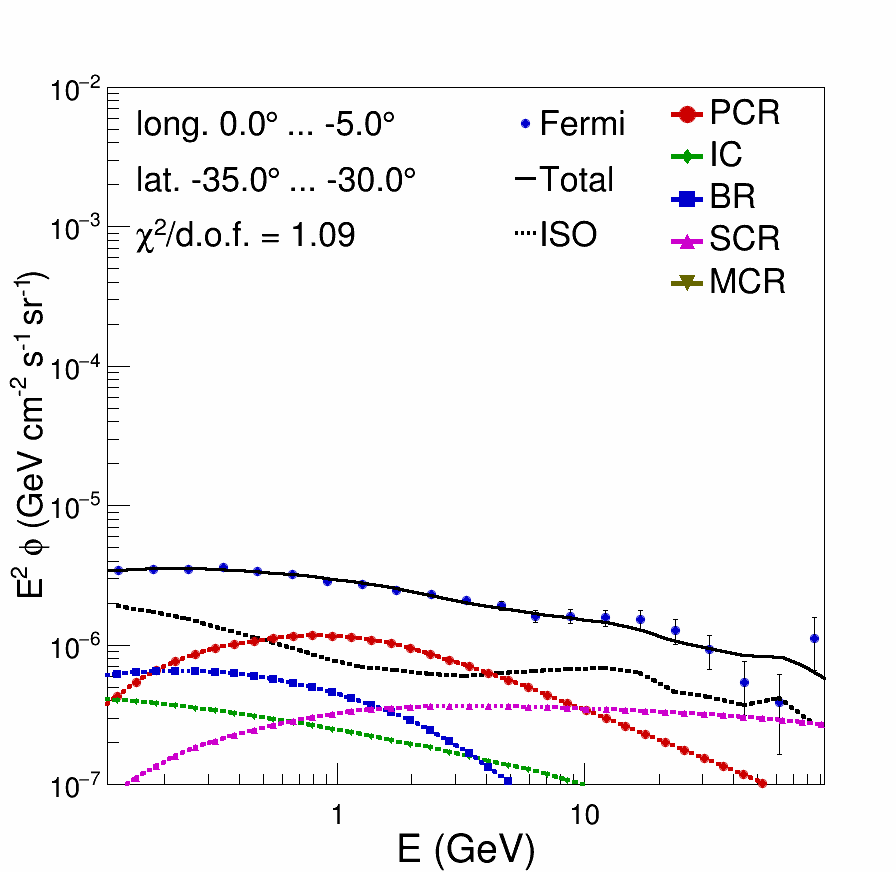}
\includegraphics[width=0.16\textwidth,height=0.16\textwidth,clip]{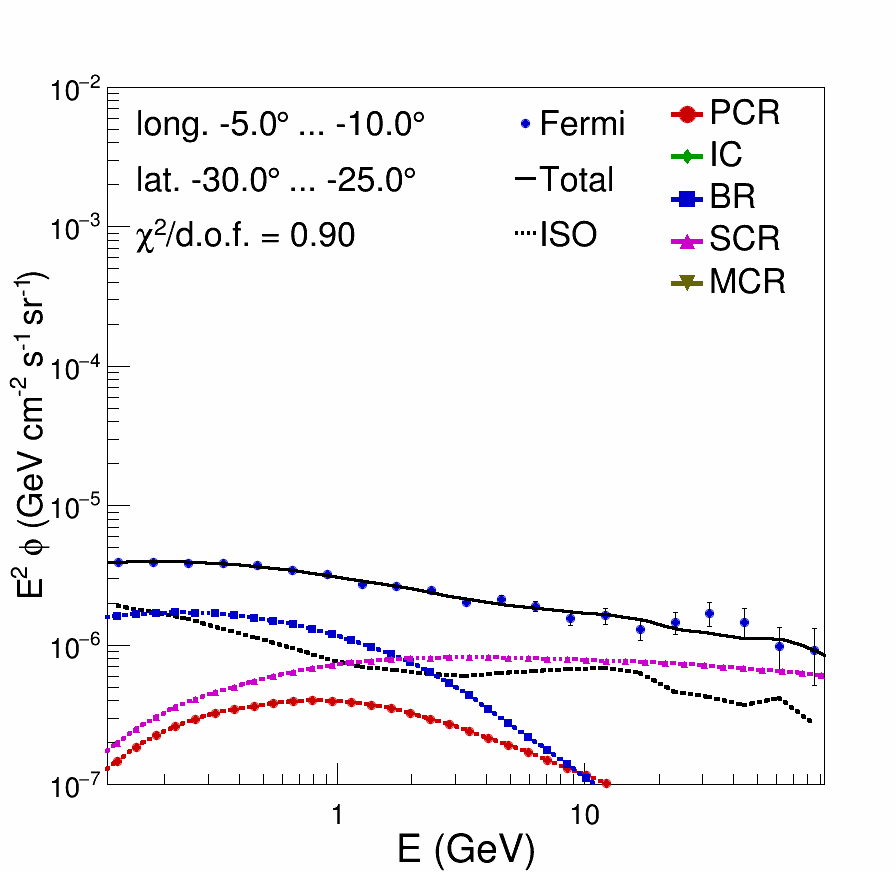}
\includegraphics[width=0.16\textwidth,height=0.16\textwidth,clip]{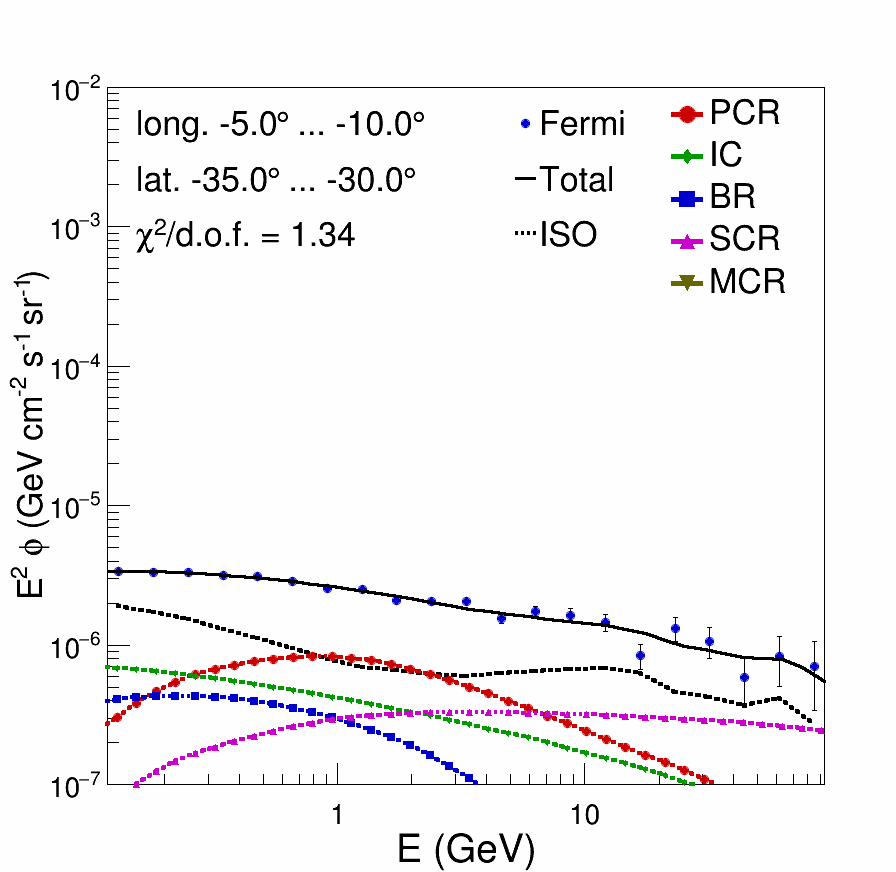}
\includegraphics[width=0.16\textwidth,height=0.16\textwidth,clip]{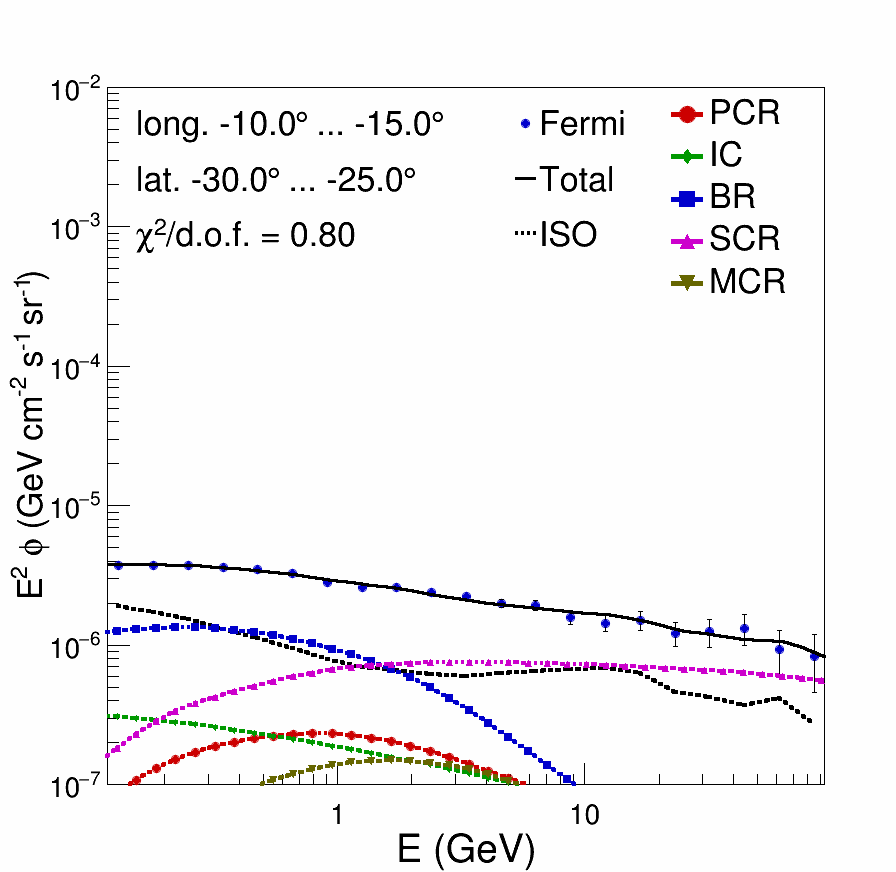}
\includegraphics[width=0.16\textwidth,height=0.16\textwidth,clip]{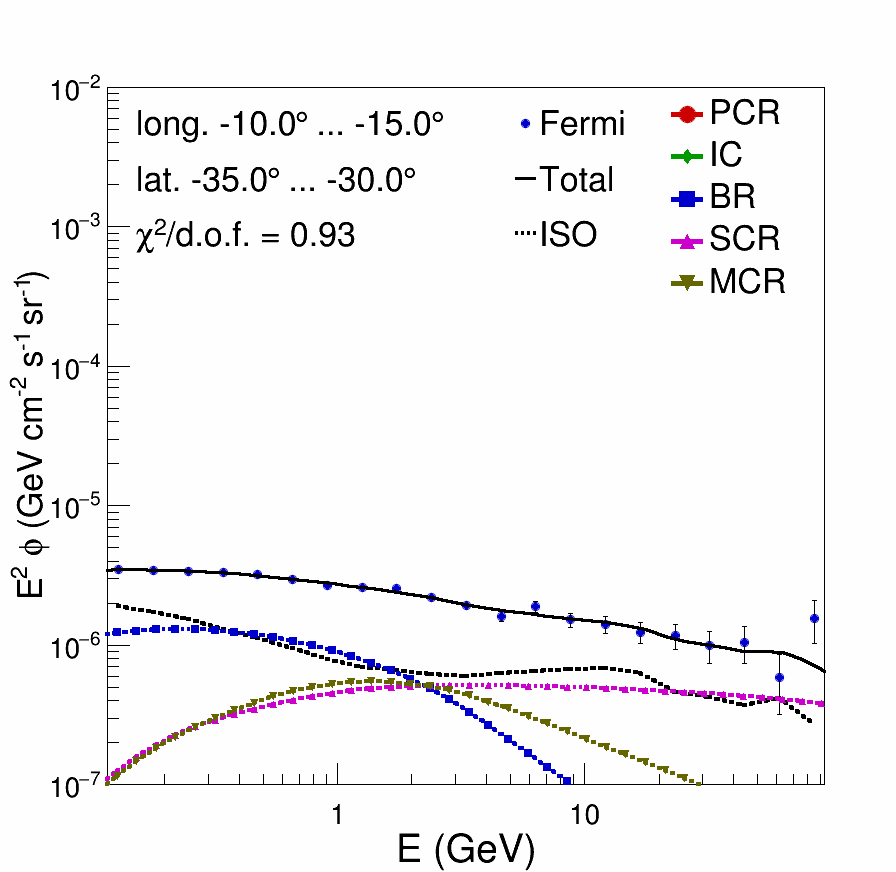}
\includegraphics[width=0.16\textwidth,height=0.16\textwidth,clip]{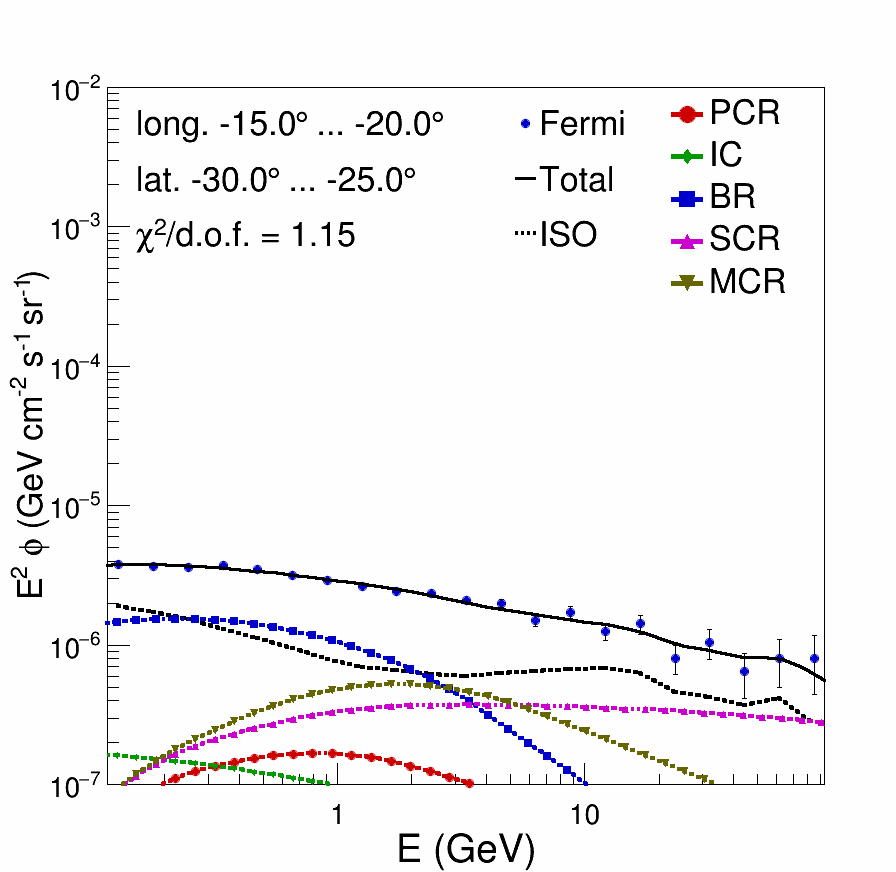}
\includegraphics[width=0.16\textwidth,height=0.16\textwidth,clip]{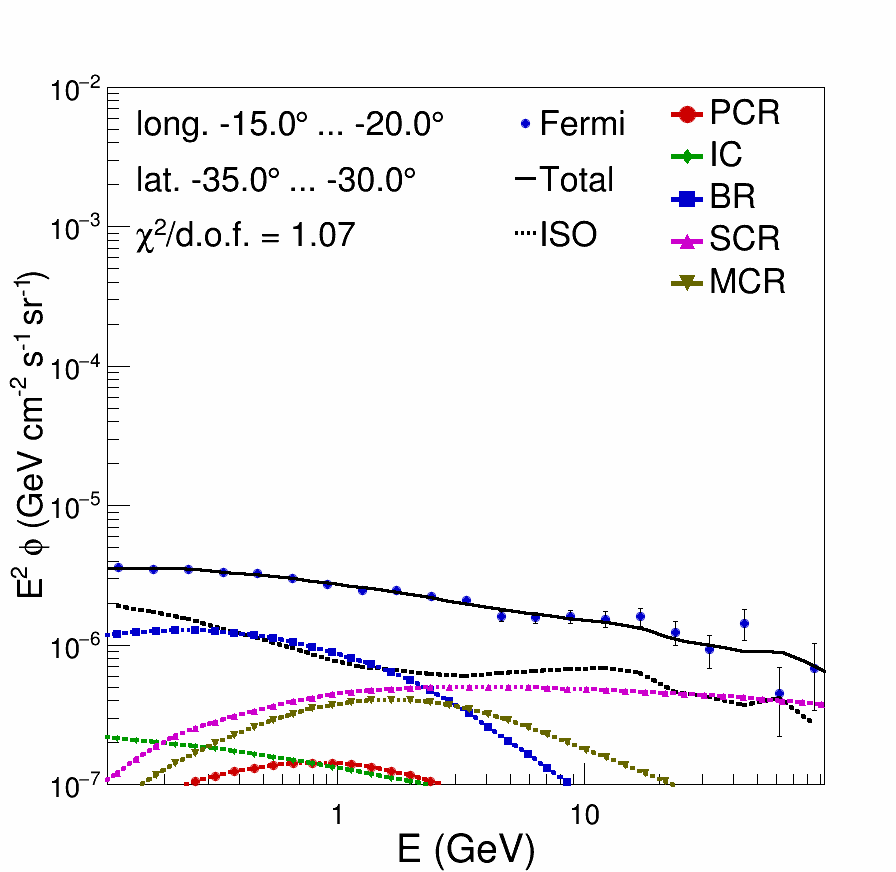}
\includegraphics[width=0.16\textwidth,height=0.16\textwidth,clip]{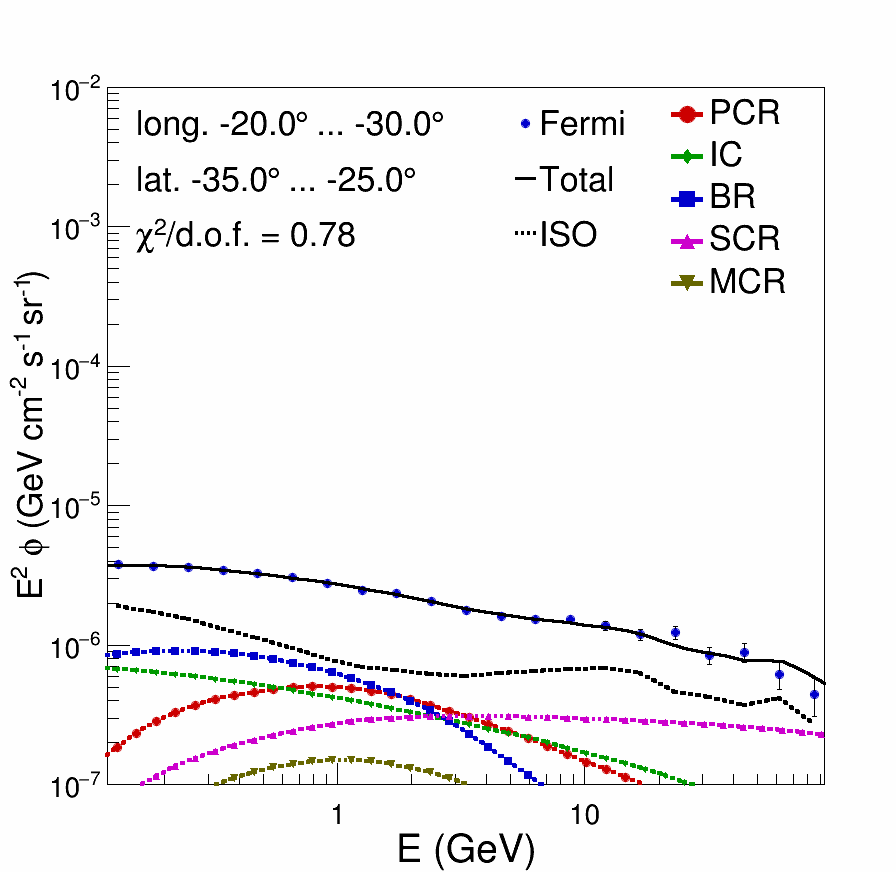}
\includegraphics[width=0.16\textwidth,height=0.16\textwidth,clip]{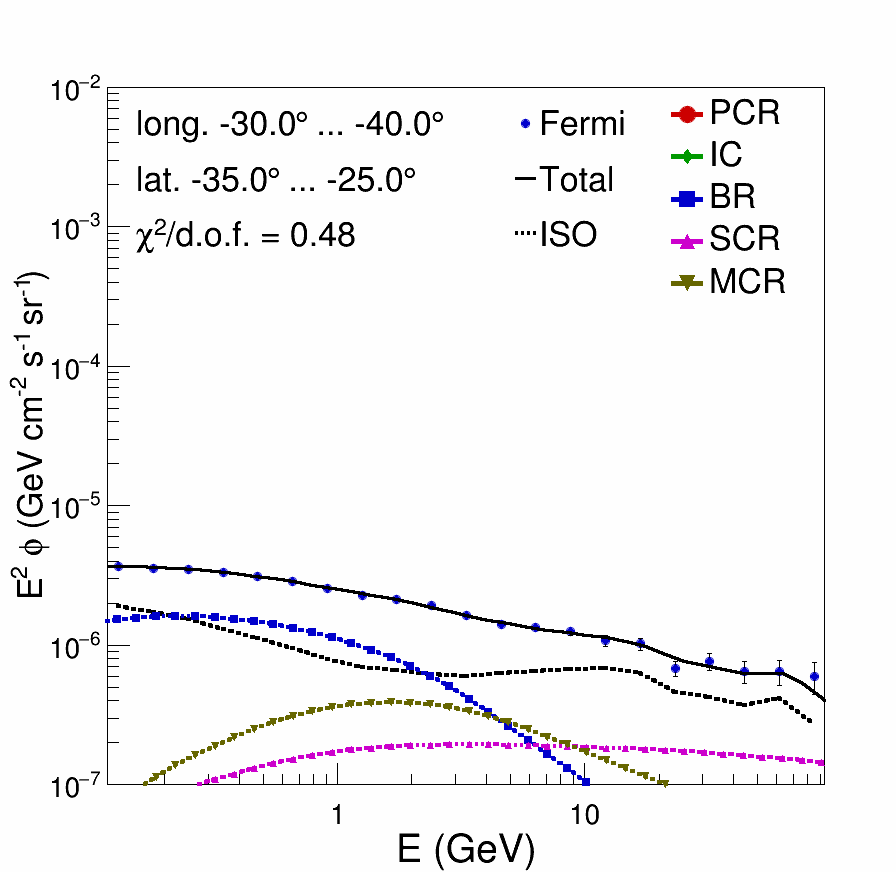}
\includegraphics[width=0.16\textwidth,height=0.16\textwidth,clip]{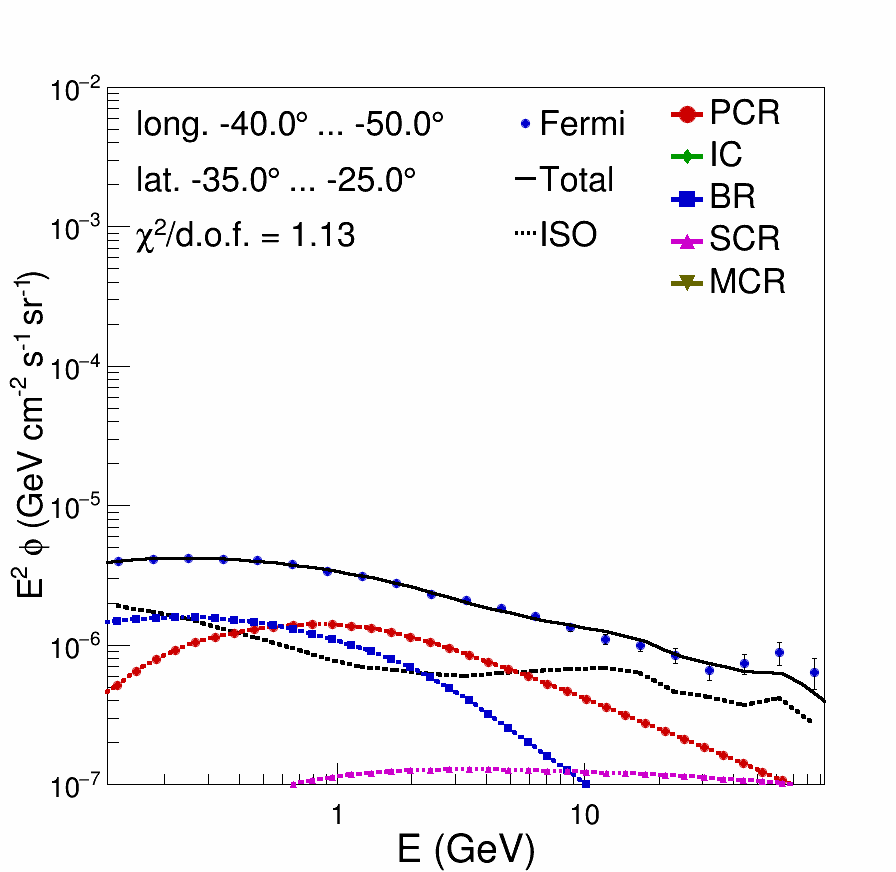}
\includegraphics[width=0.16\textwidth,height=0.16\textwidth,clip]{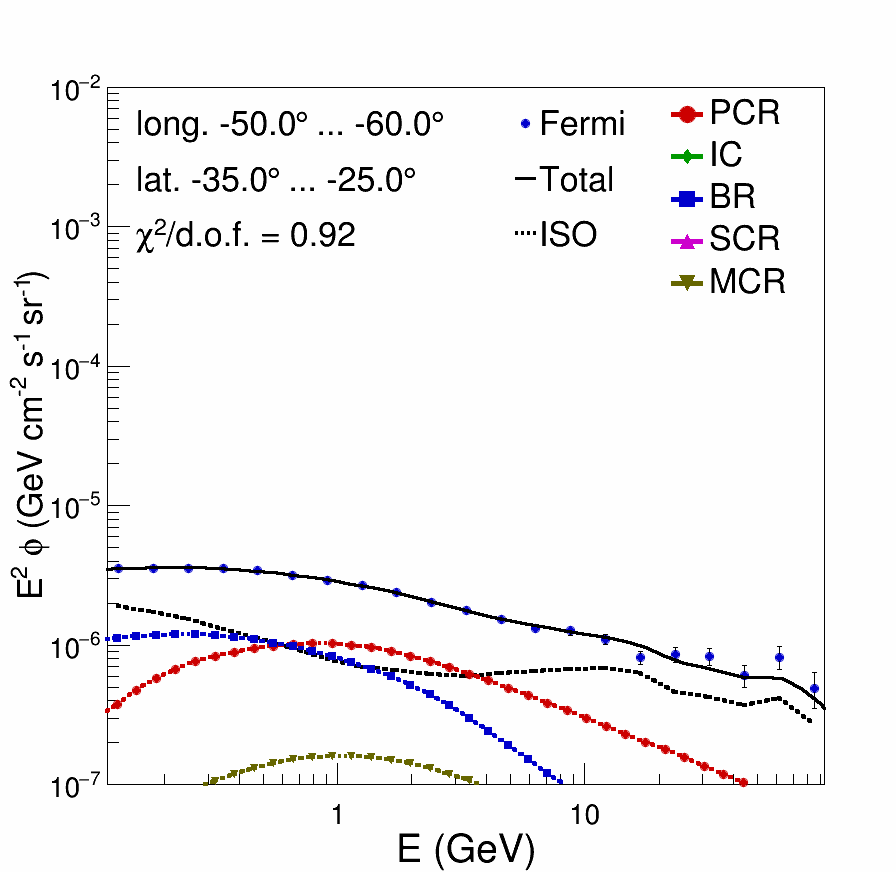}
\includegraphics[width=0.16\textwidth,height=0.16\textwidth,clip]{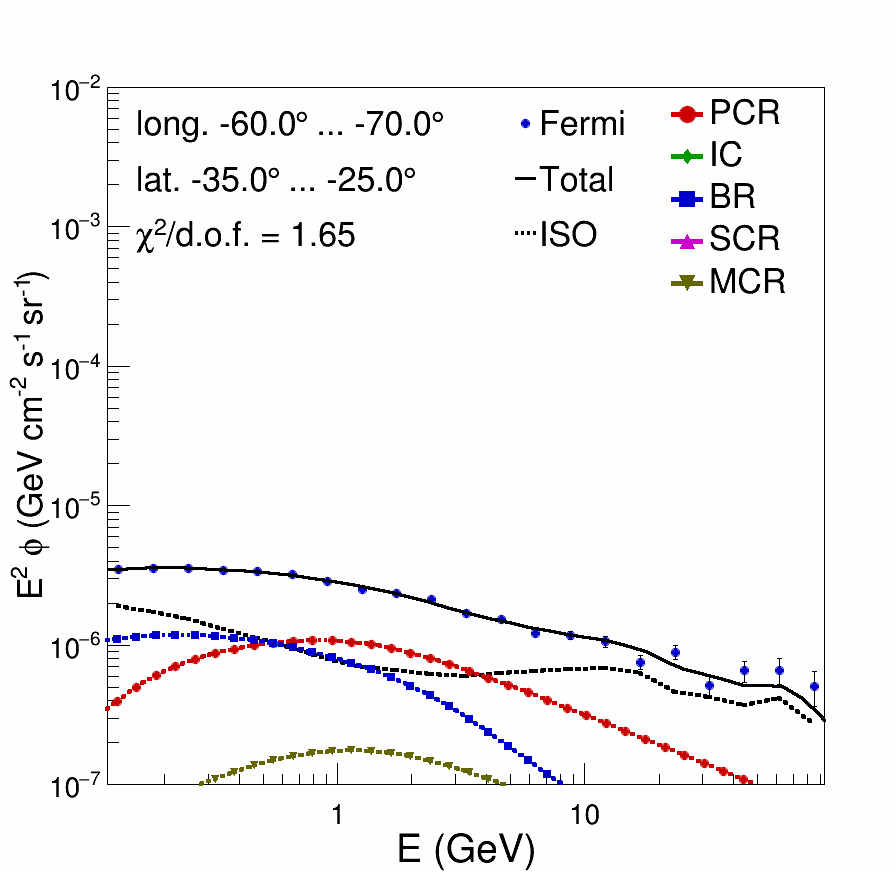}
\includegraphics[width=0.16\textwidth,height=0.16\textwidth,clip]{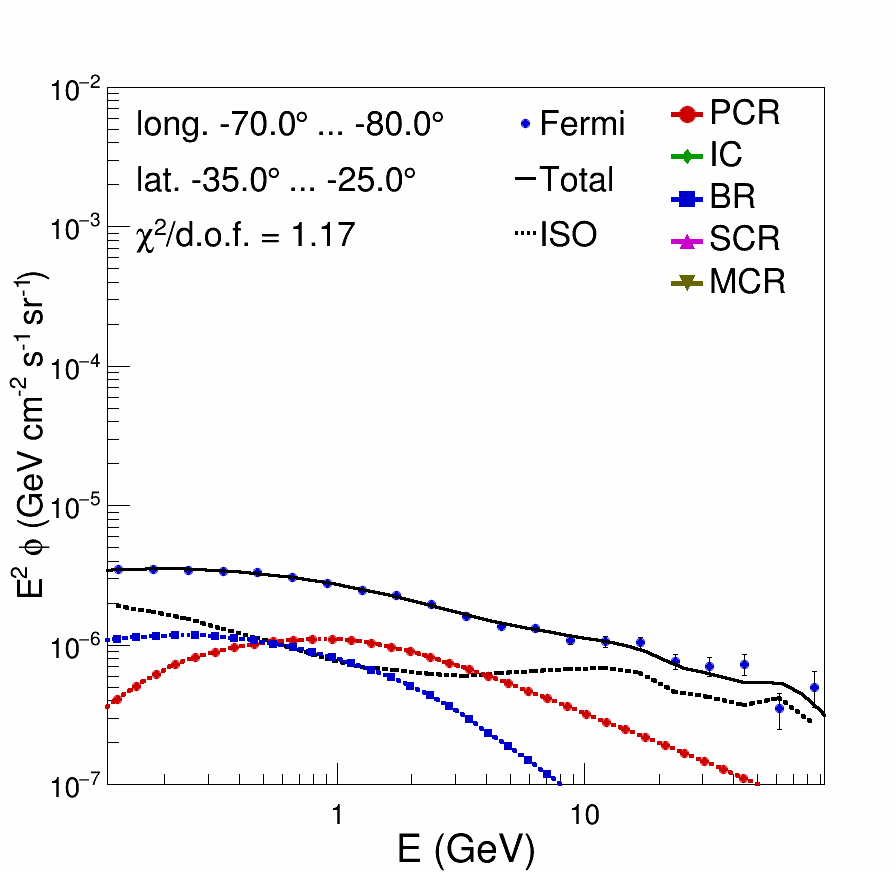}
\includegraphics[width=0.16\textwidth,height=0.16\textwidth,clip]{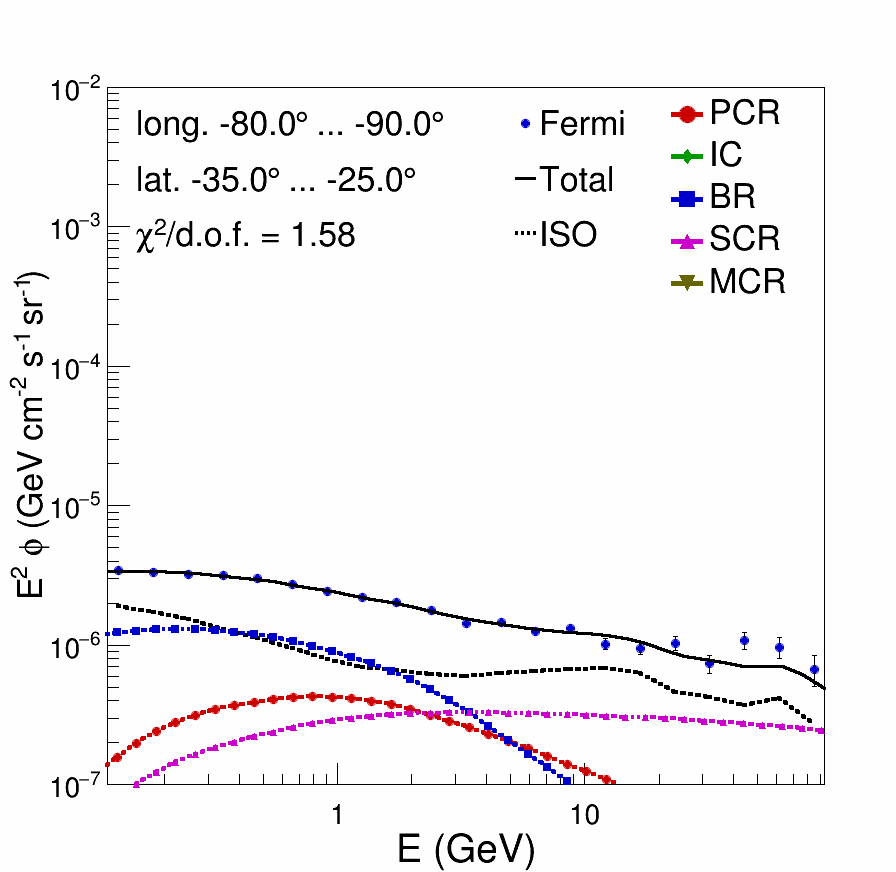}
\includegraphics[width=0.16\textwidth,height=0.16\textwidth,clip]{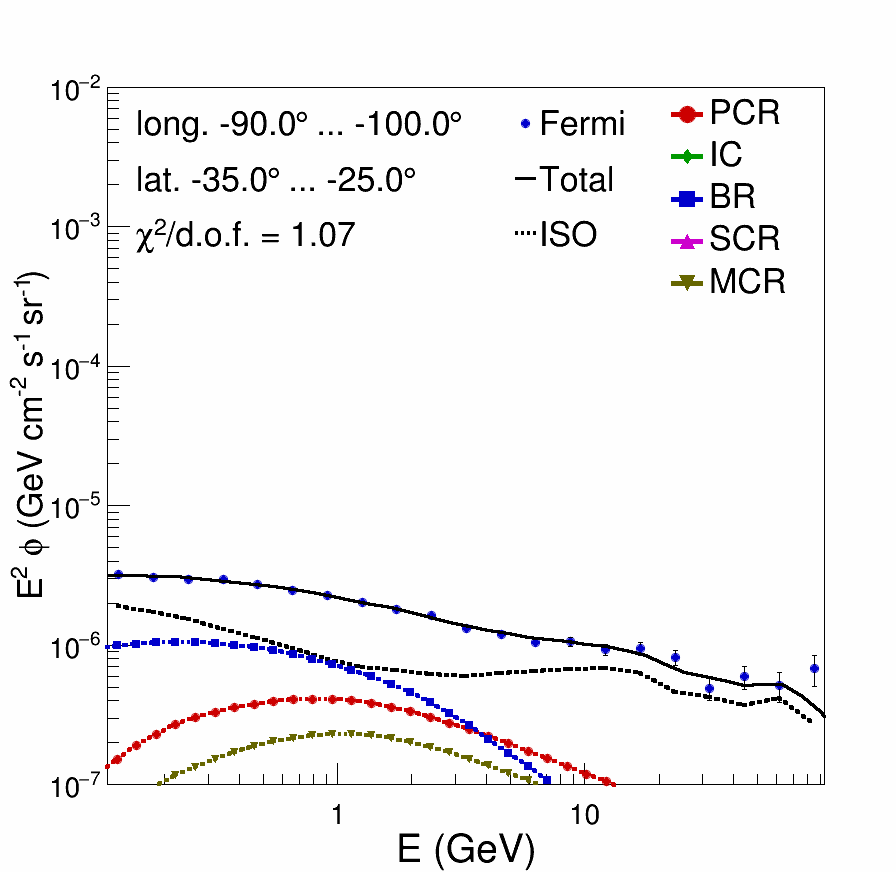}
\includegraphics[width=0.16\textwidth,height=0.16\textwidth,clip]{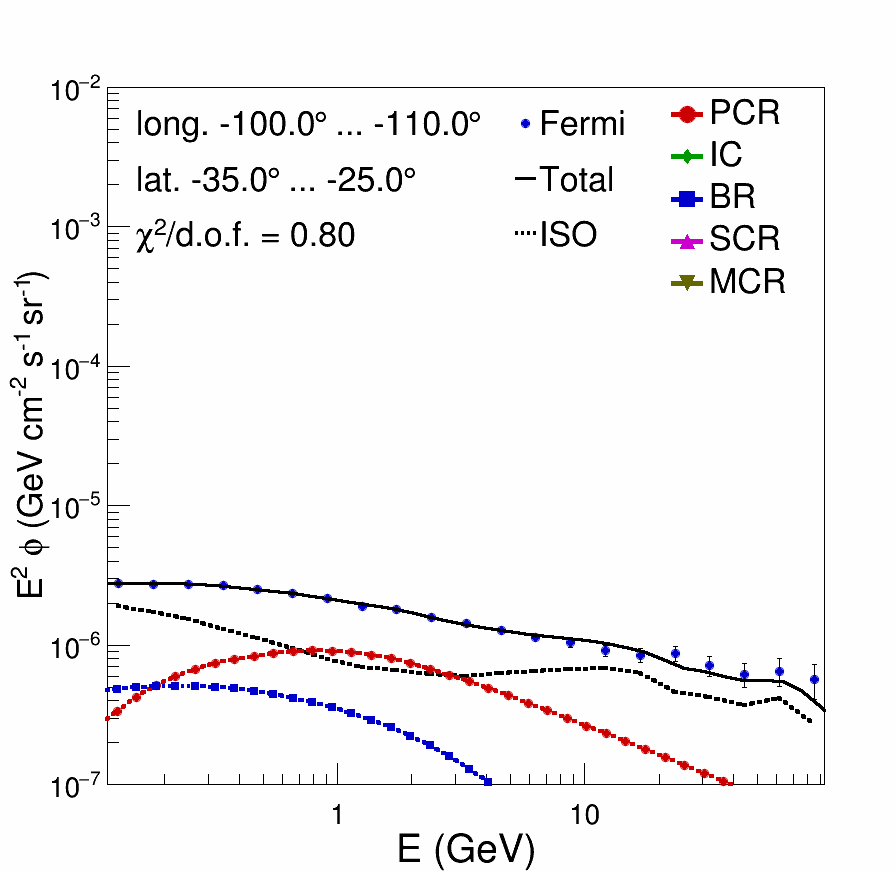}
\includegraphics[width=0.16\textwidth,height=0.16\textwidth,clip]{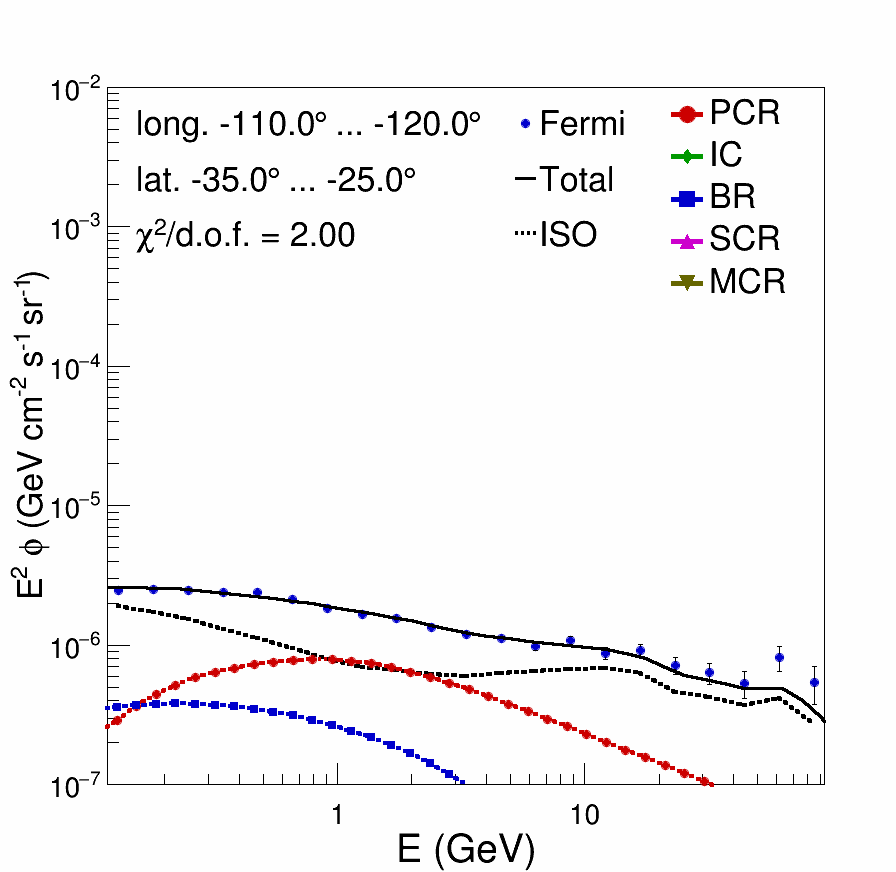}
\includegraphics[width=0.16\textwidth,height=0.16\textwidth,clip]{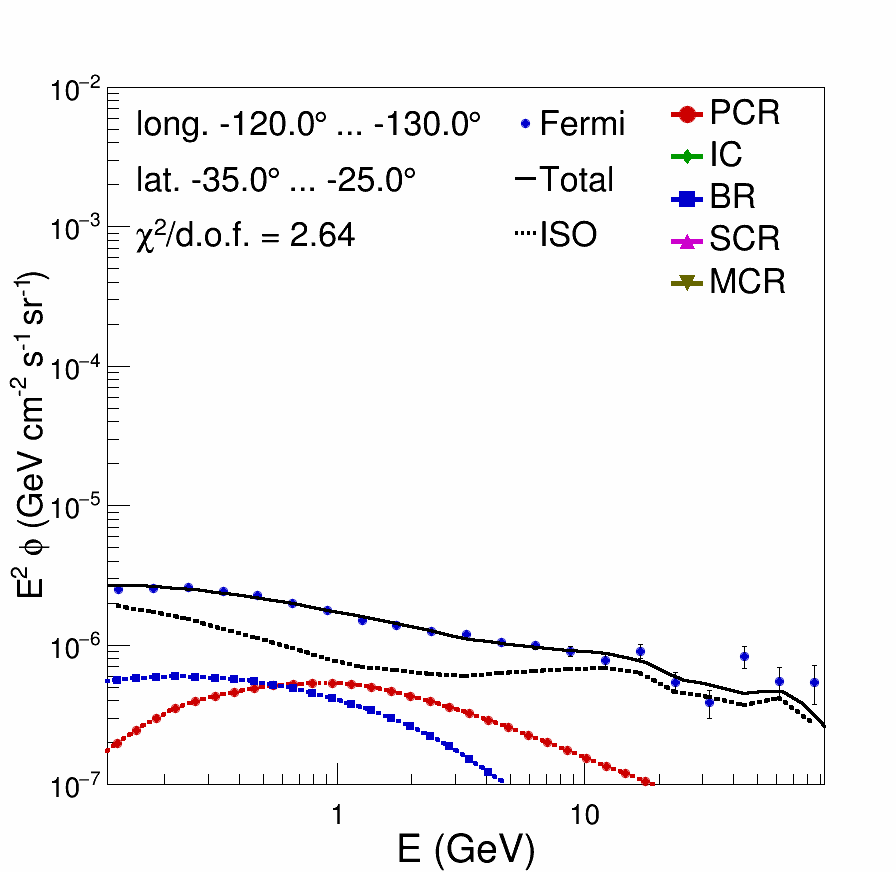}
\includegraphics[width=0.16\textwidth,height=0.16\textwidth,clip]{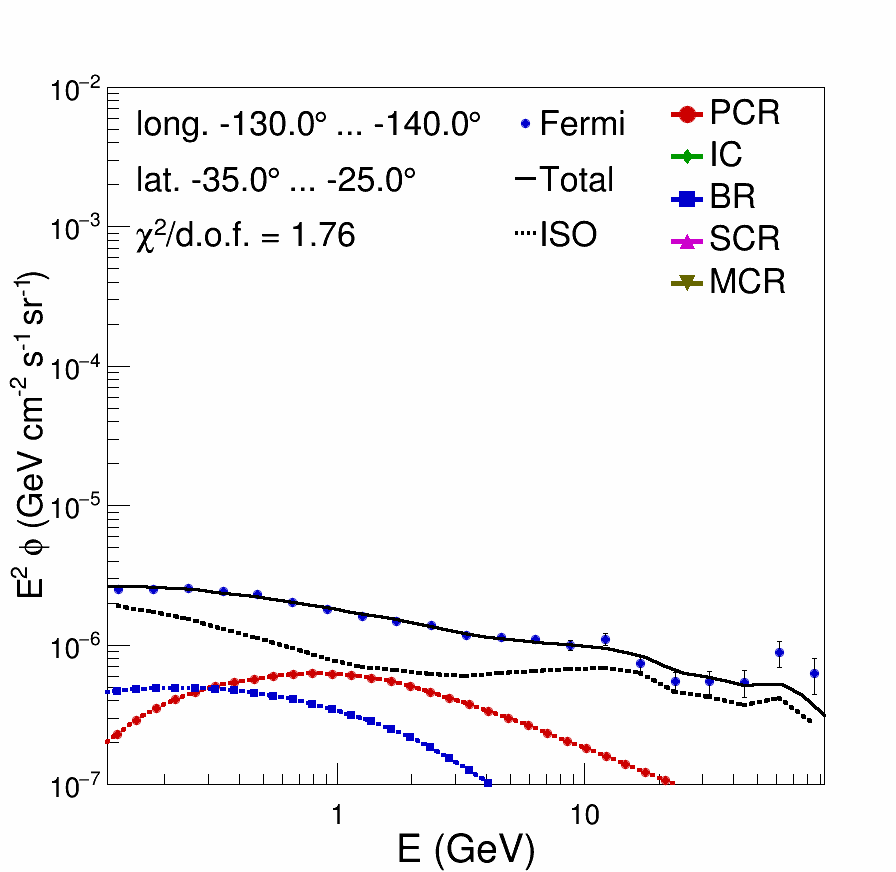}
\includegraphics[width=0.16\textwidth,height=0.16\textwidth,clip]{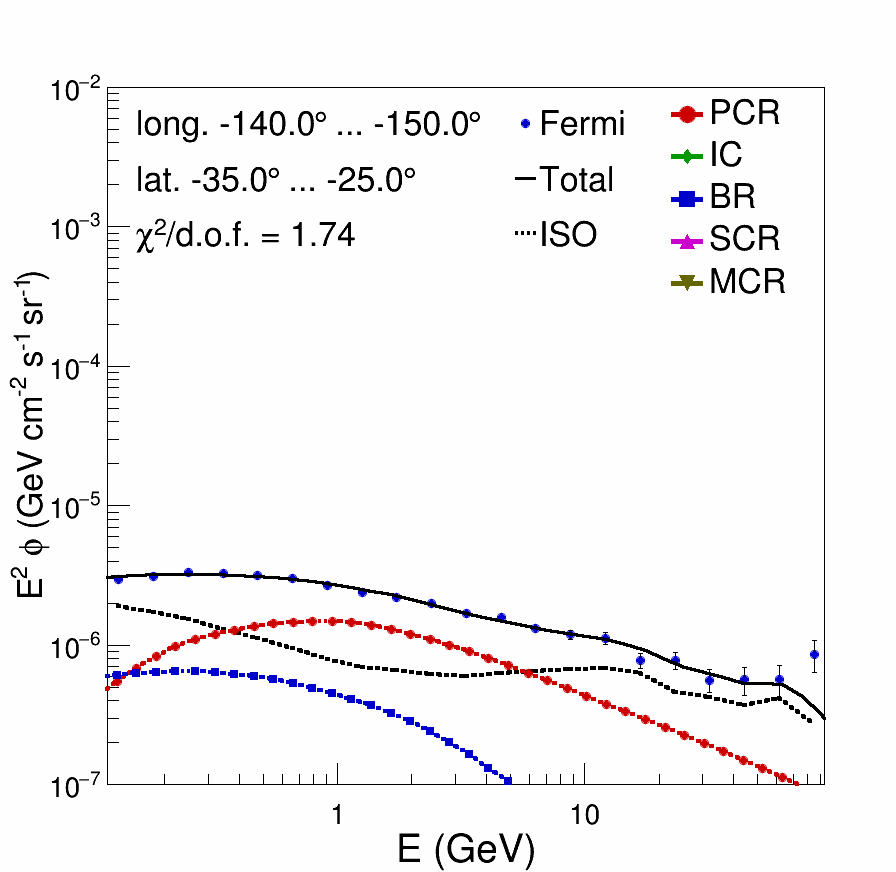}
\includegraphics[width=0.16\textwidth,height=0.16\textwidth,clip]{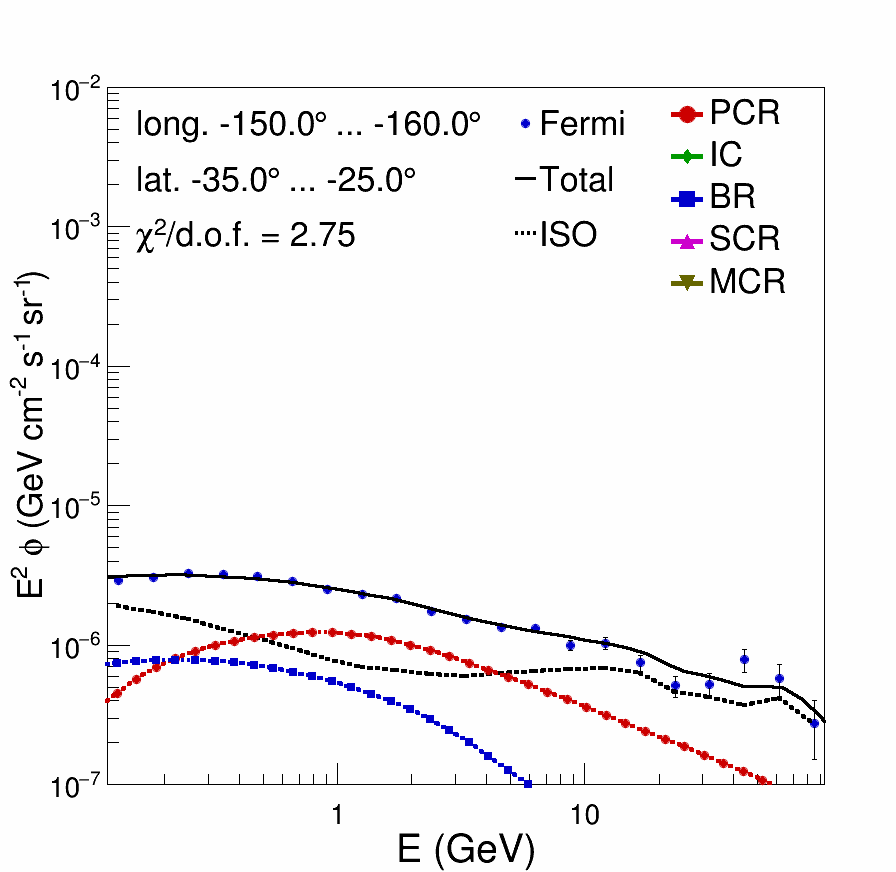}
\includegraphics[width=0.16\textwidth,height=0.16\textwidth,clip]{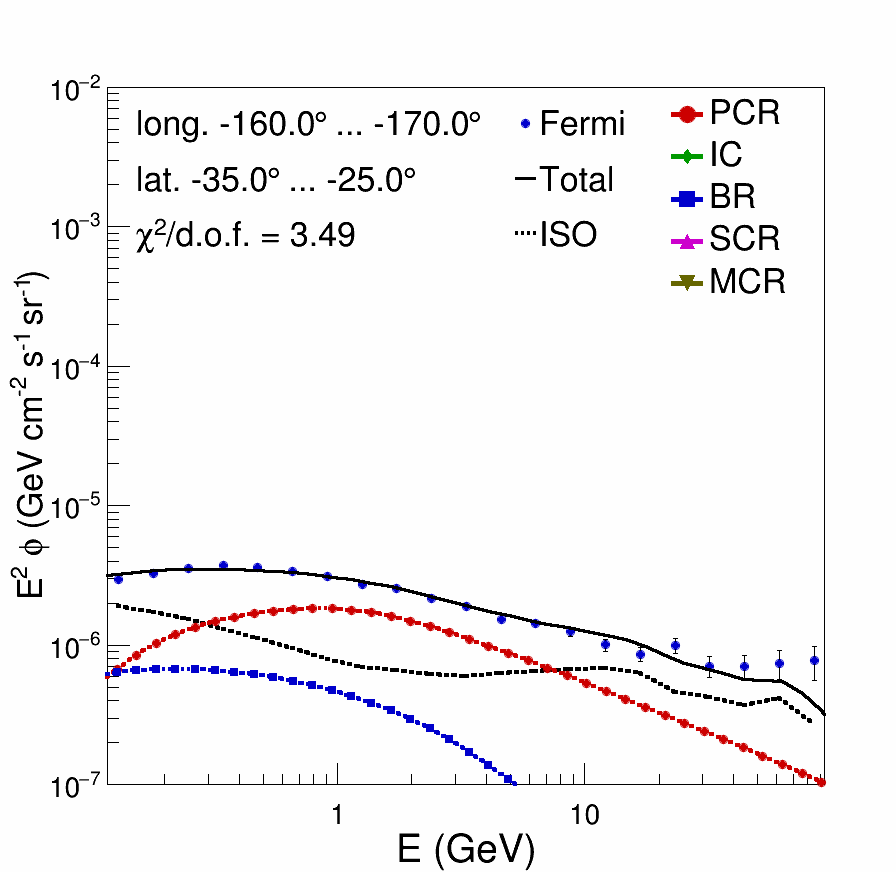}
\includegraphics[width=0.16\textwidth,height=0.16\textwidth,clip]{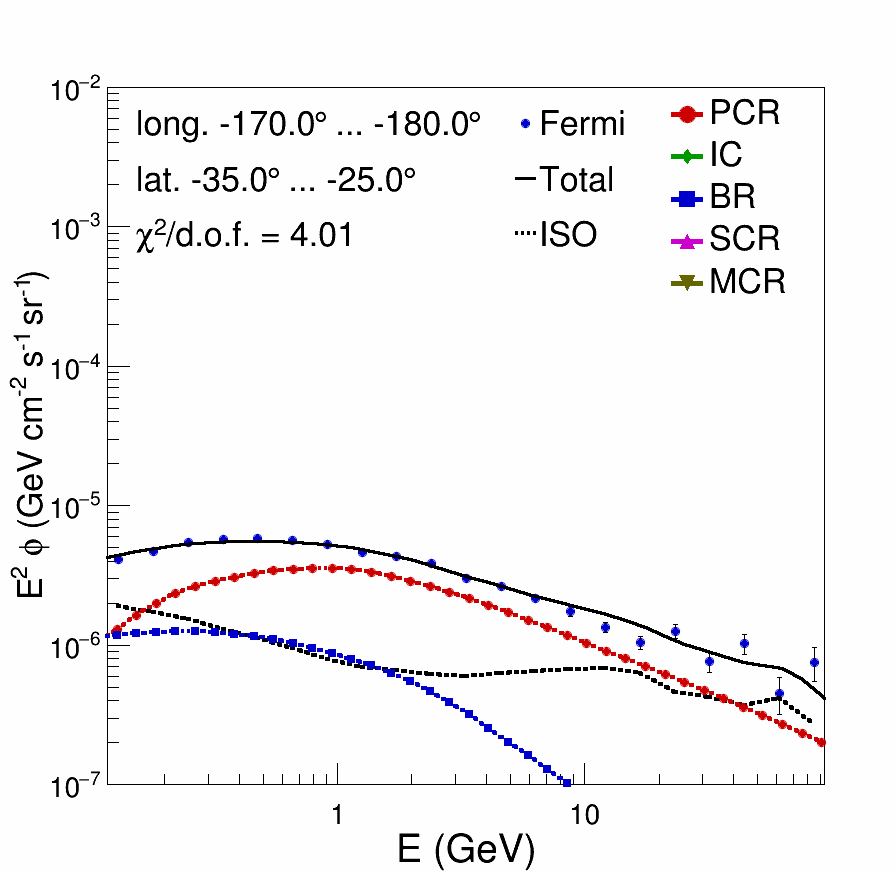}
\caption[]{Template fits for latitudes  with $-35.0^\circ<b<-25.0^\circ$ and longitudes decreasing from 180$^\circ$ to -180$^\circ$.} \label{F27}
\end{figure}
\begin{figure}
\centering
\includegraphics[width=0.16\textwidth,height=0.16\textwidth,clip]{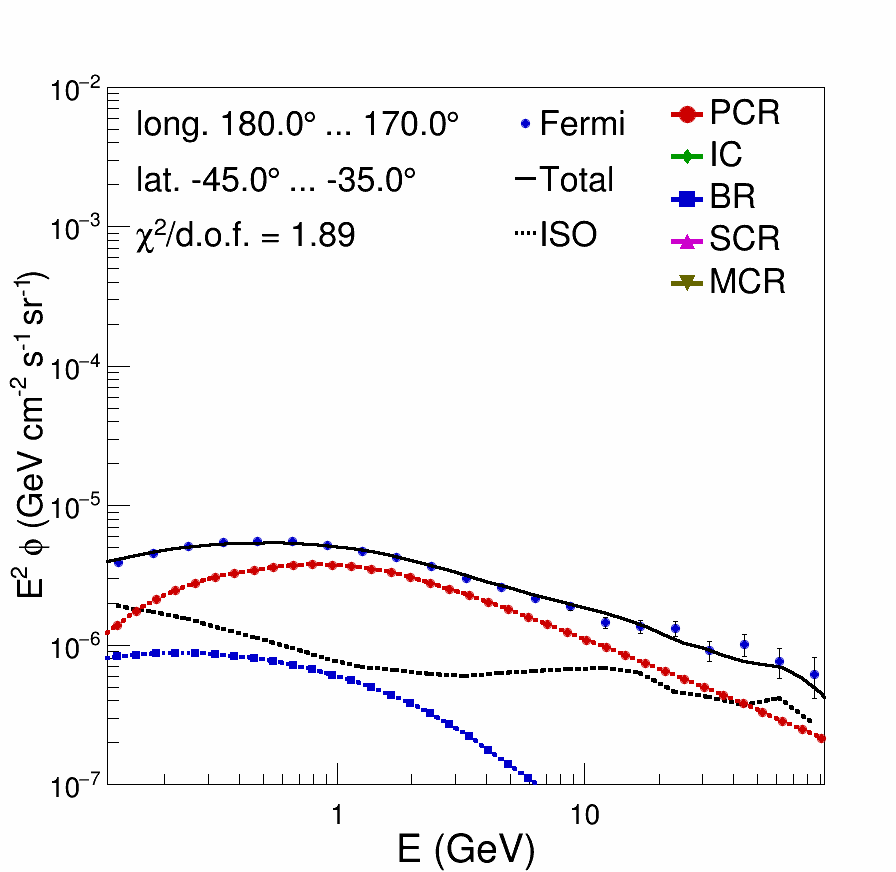}
\includegraphics[width=0.16\textwidth,height=0.16\textwidth,clip]{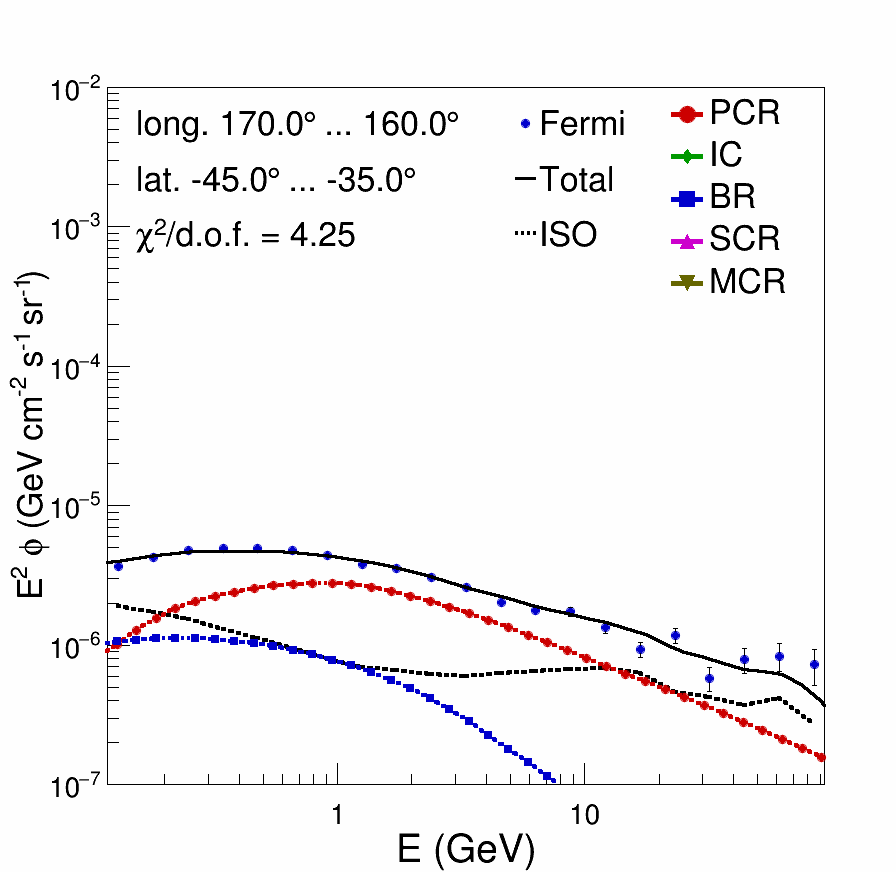}
\includegraphics[width=0.16\textwidth,height=0.16\textwidth,clip]{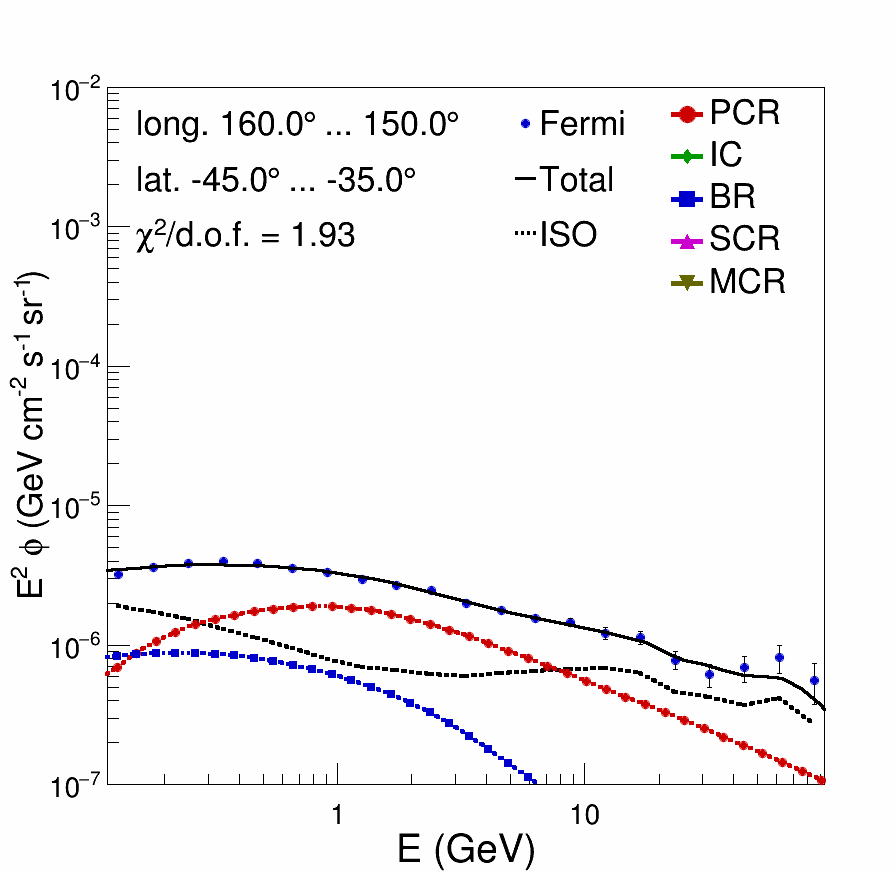}
\includegraphics[width=0.16\textwidth,height=0.16\textwidth,clip]{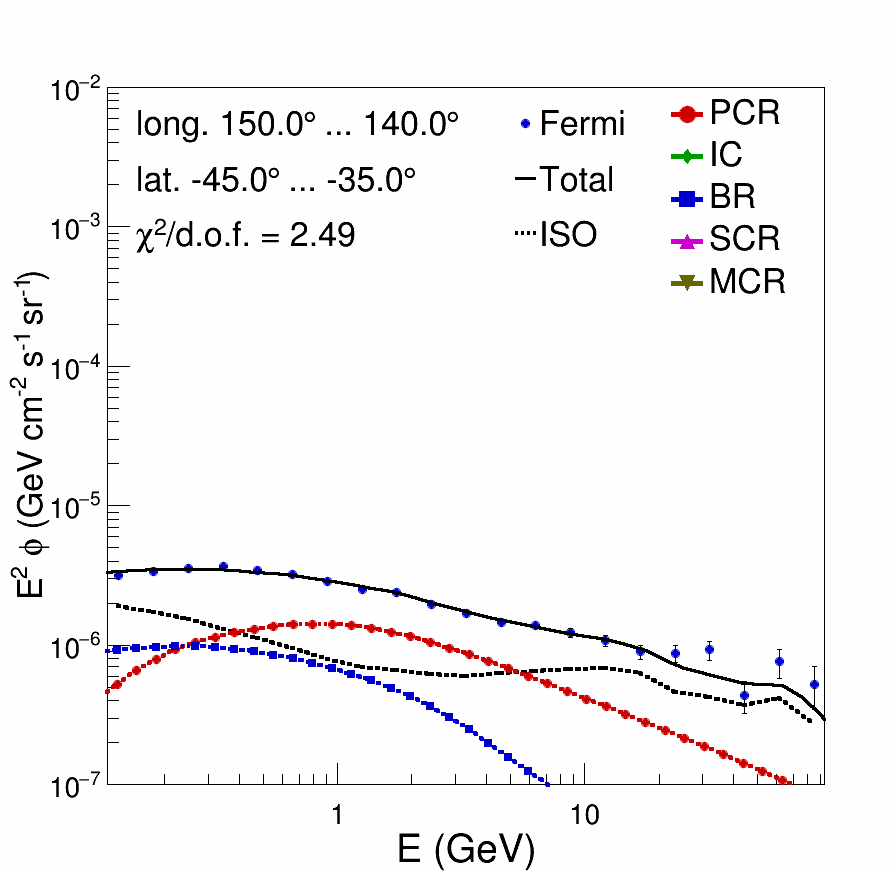}
\includegraphics[width=0.16\textwidth,height=0.16\textwidth,clip]{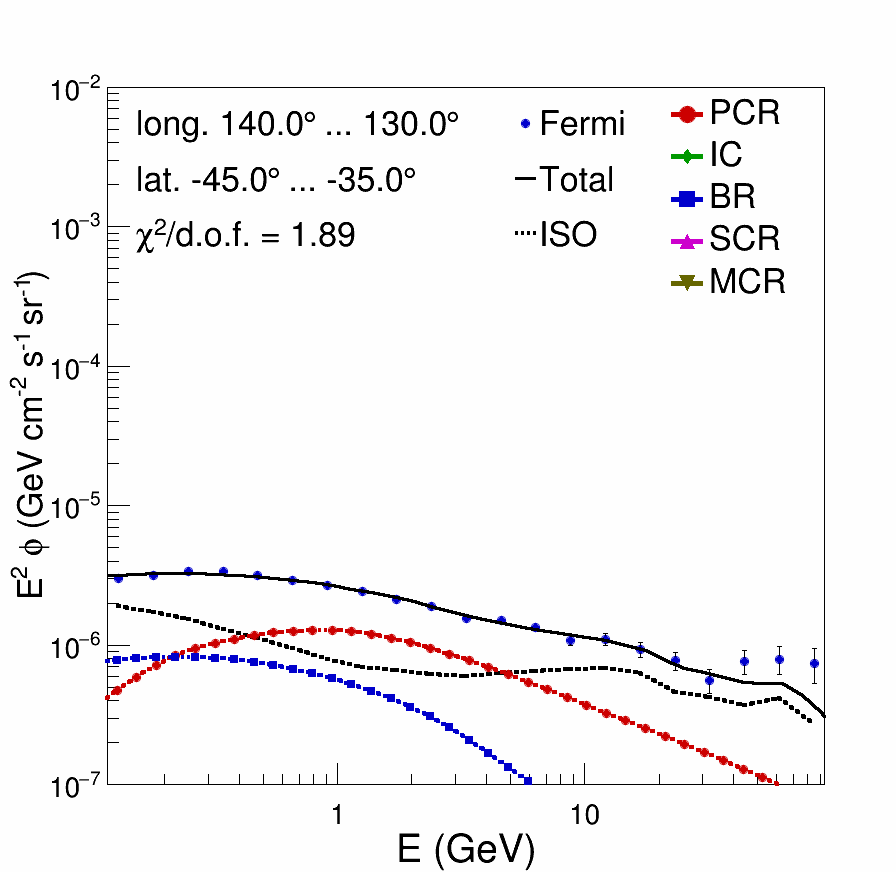}
\includegraphics[width=0.16\textwidth,height=0.16\textwidth,clip]{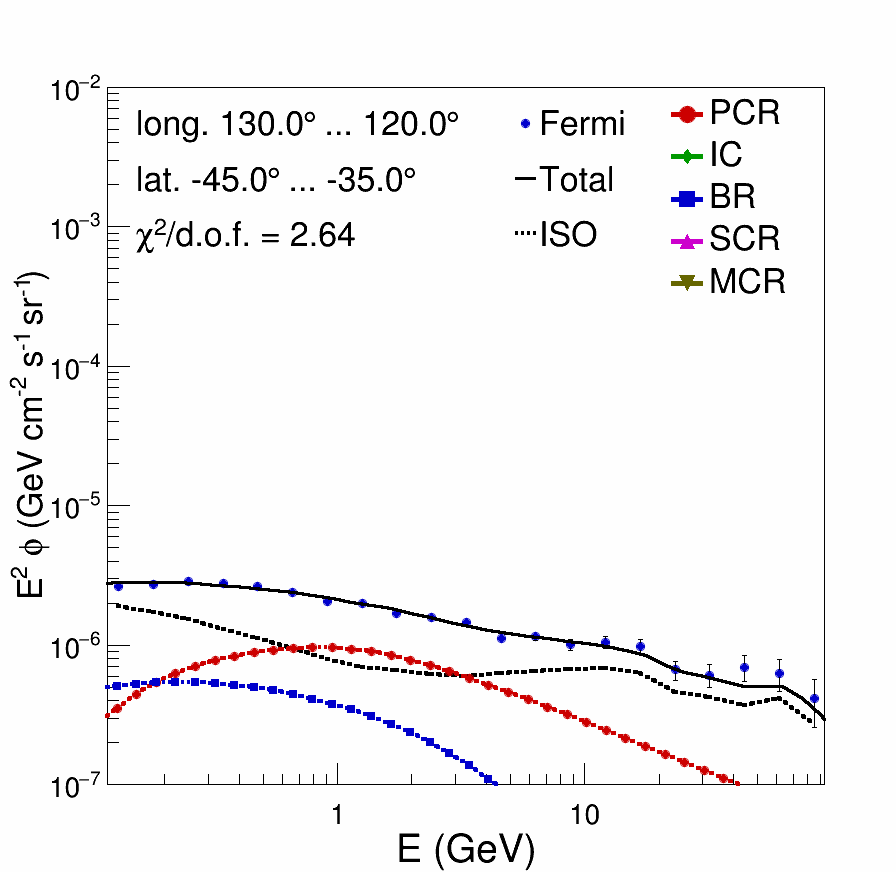}
\includegraphics[width=0.16\textwidth,height=0.16\textwidth,clip]{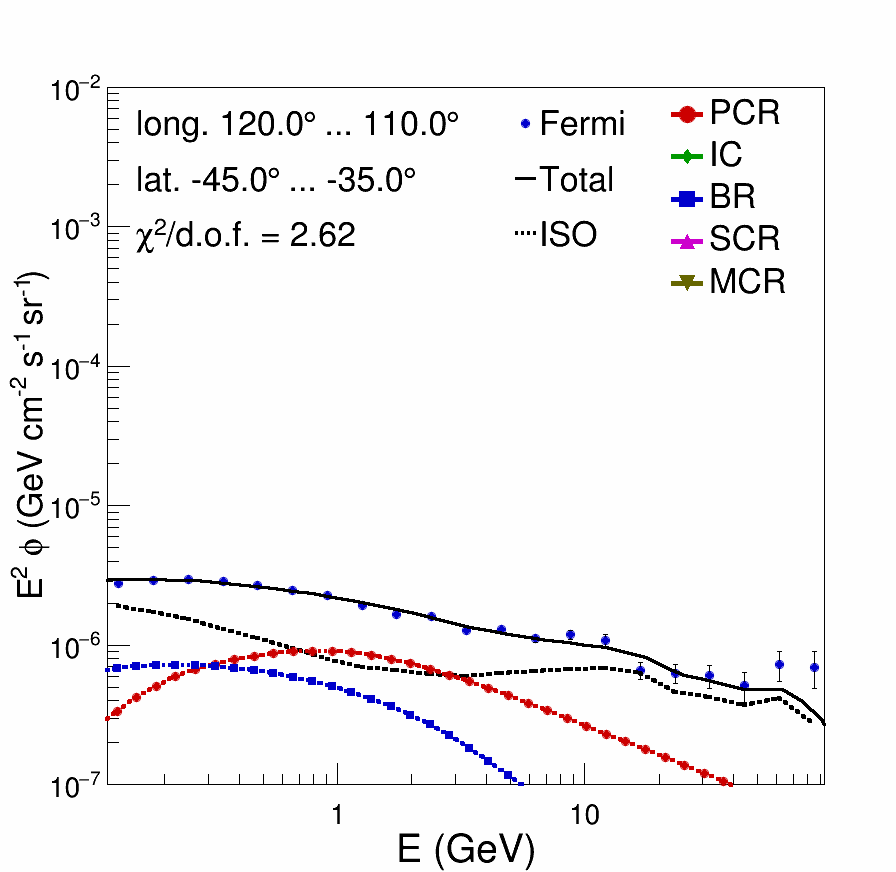}
\includegraphics[width=0.16\textwidth,height=0.16\textwidth,clip]{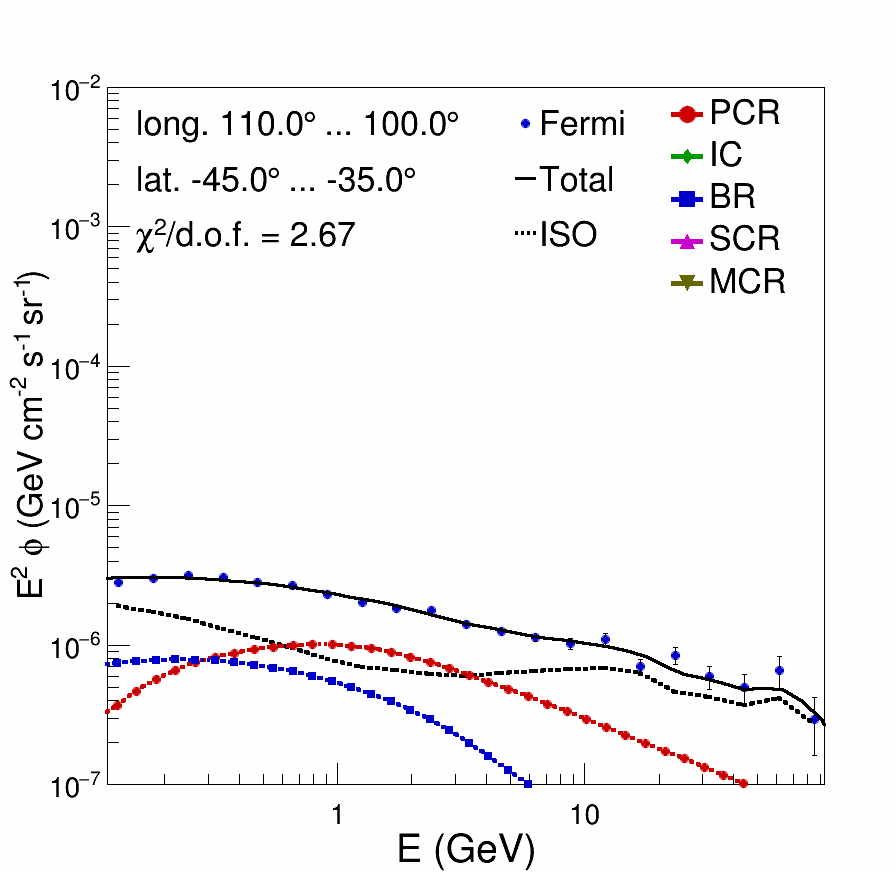}
\includegraphics[width=0.16\textwidth,height=0.16\textwidth,clip]{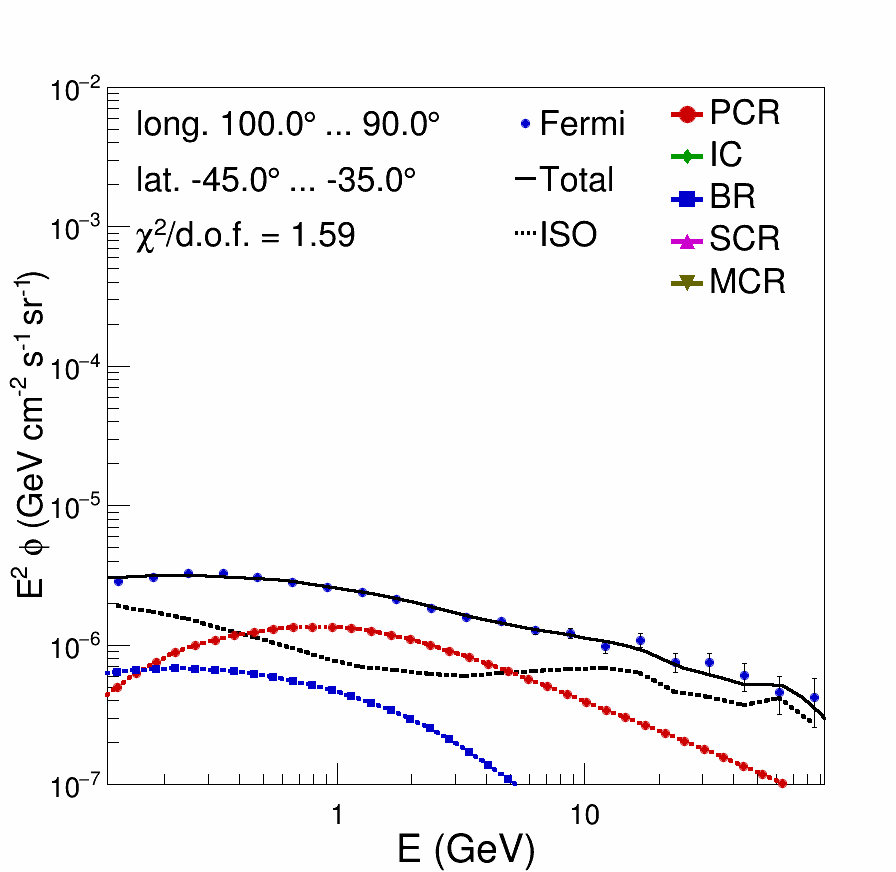}
\includegraphics[width=0.16\textwidth,height=0.16\textwidth,clip]{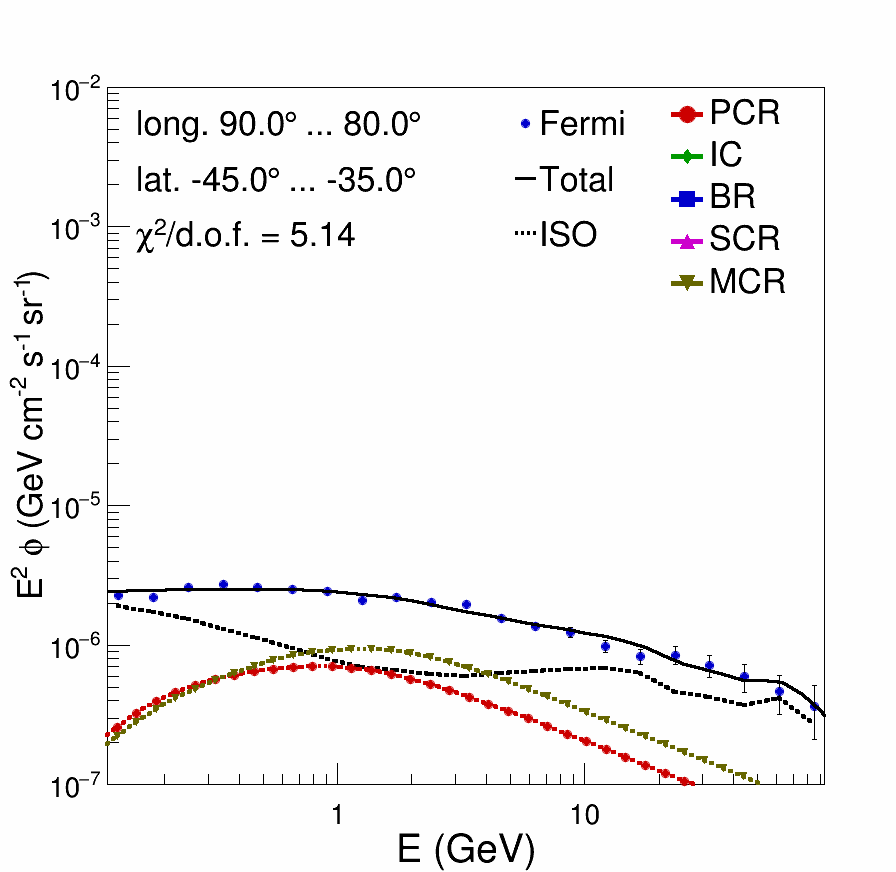}
\includegraphics[width=0.16\textwidth,height=0.16\textwidth,clip]{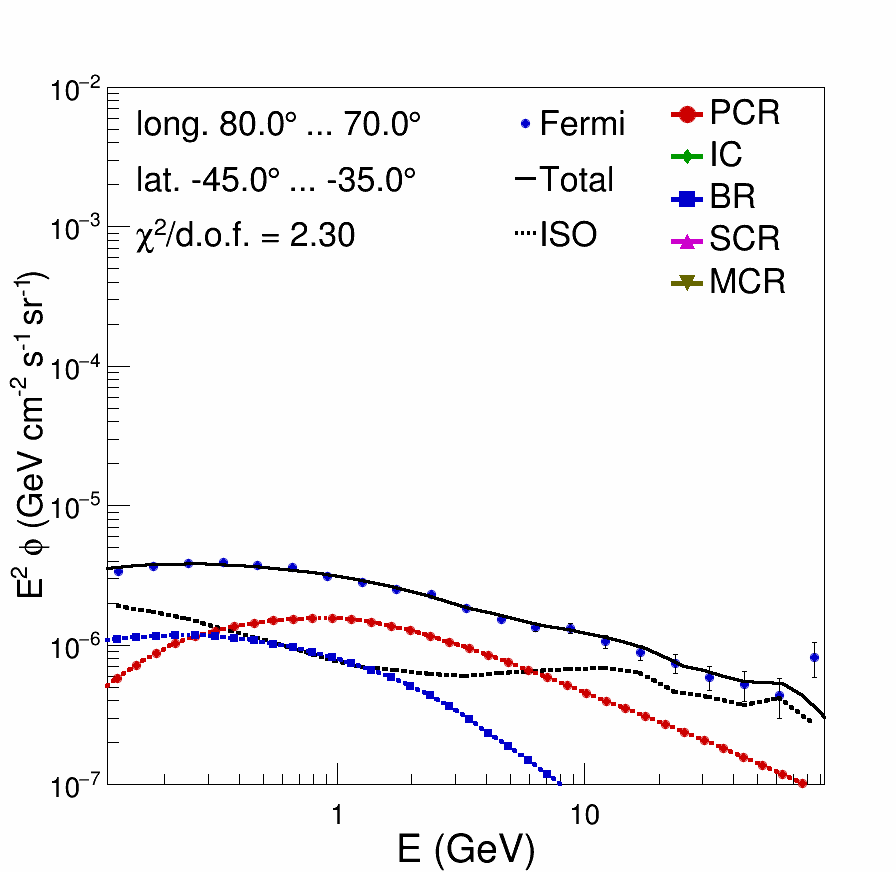}
\includegraphics[width=0.16\textwidth,height=0.16\textwidth,clip]{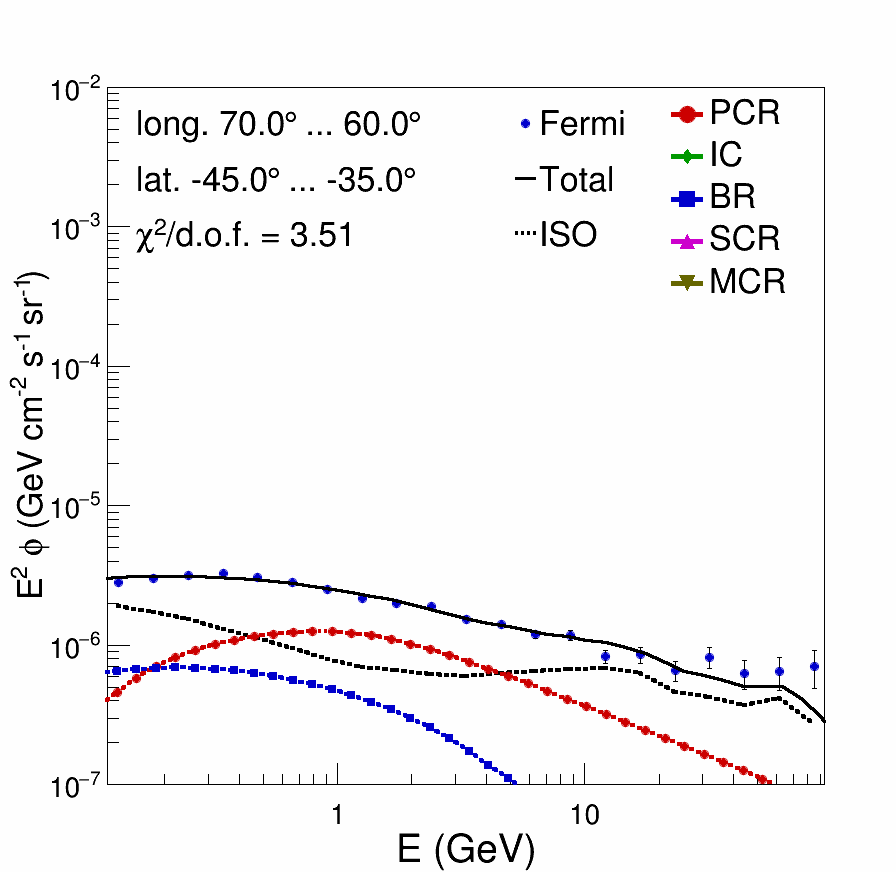}
\includegraphics[width=0.16\textwidth,height=0.16\textwidth,clip]{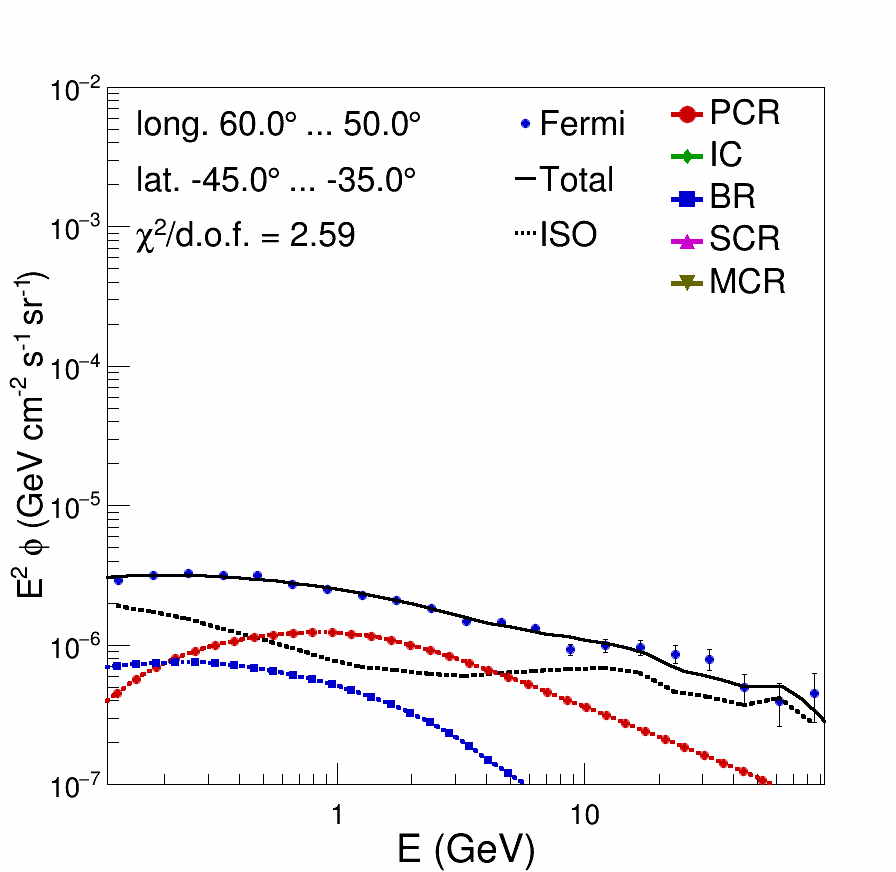}
\includegraphics[width=0.16\textwidth,height=0.16\textwidth,clip]{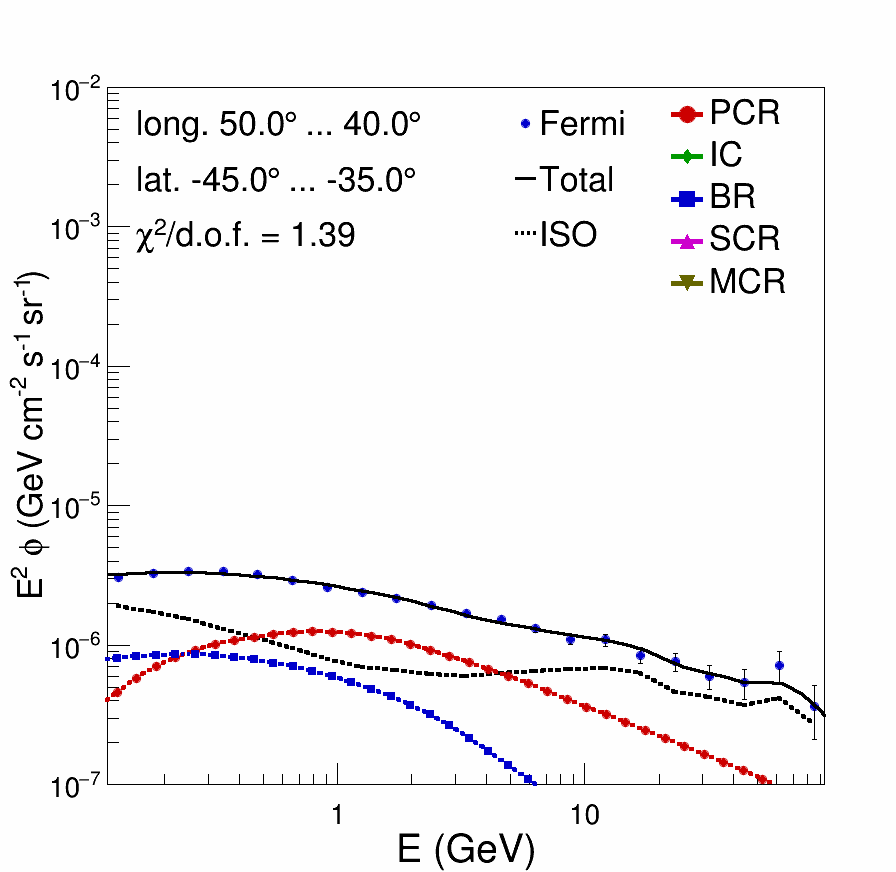}
\includegraphics[width=0.16\textwidth,height=0.16\textwidth,clip]{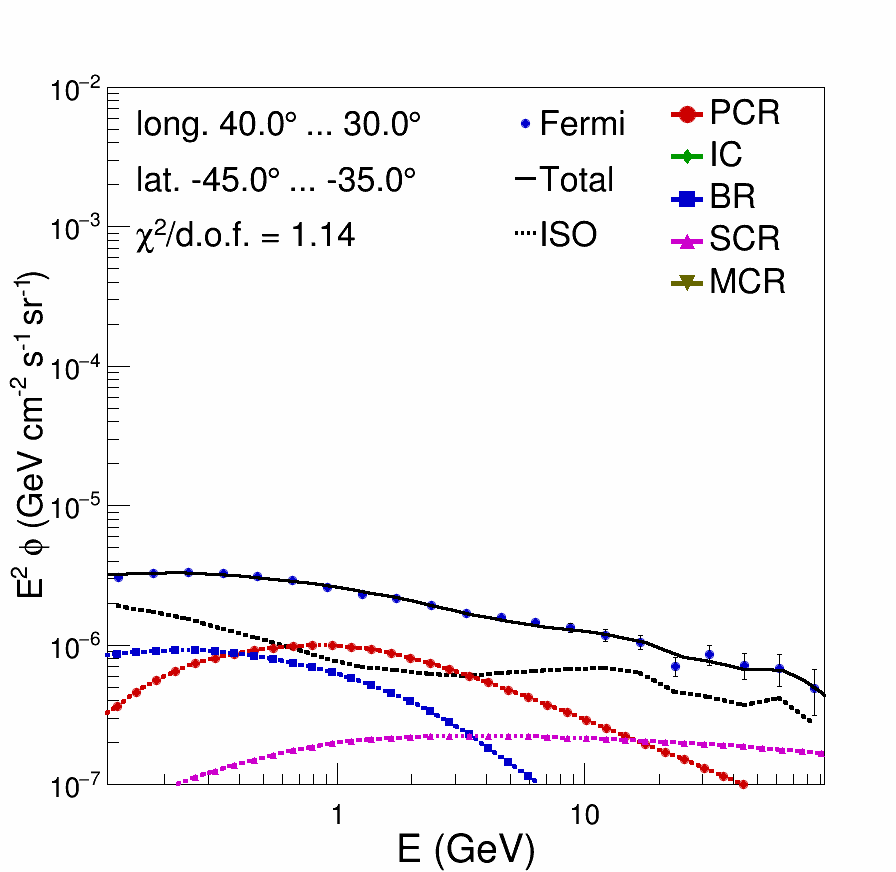}
\includegraphics[width=0.16\textwidth,height=0.16\textwidth,clip]{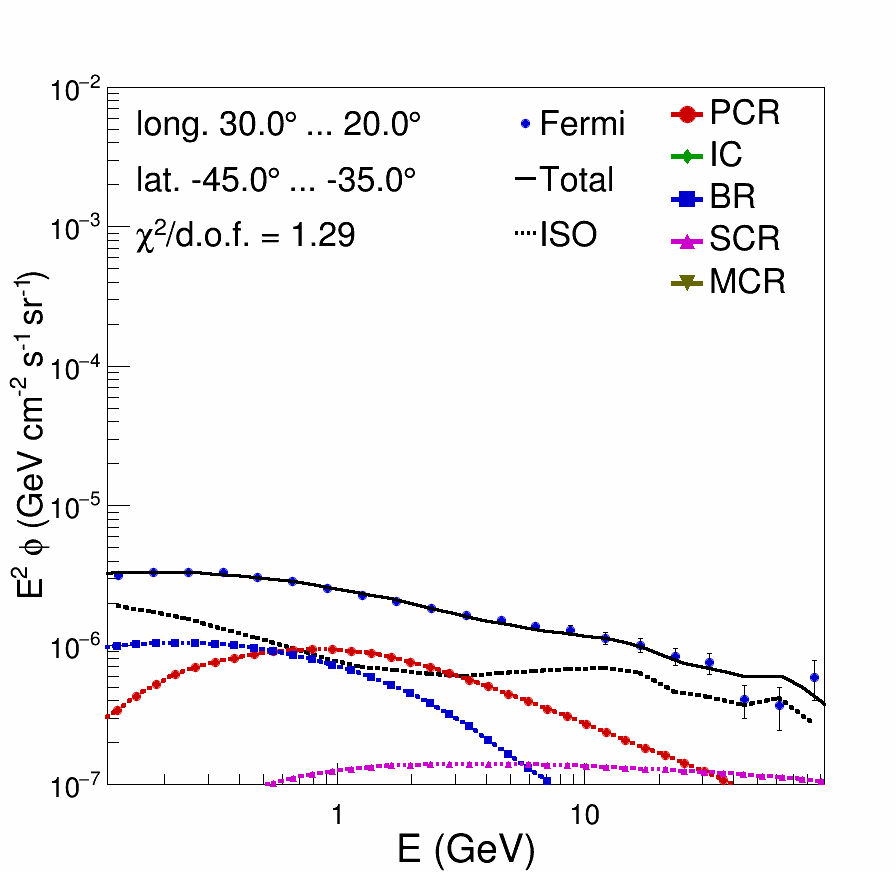}
\includegraphics[width=0.16\textwidth,height=0.16\textwidth,clip]{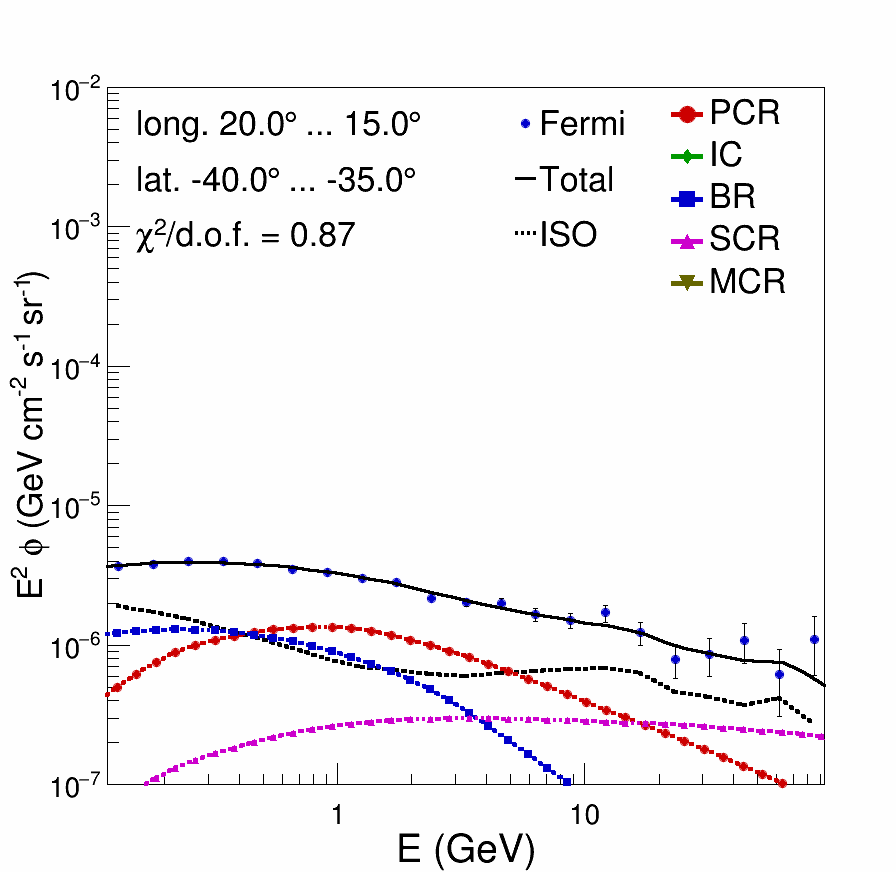}
\includegraphics[width=0.16\textwidth,height=0.16\textwidth,clip]{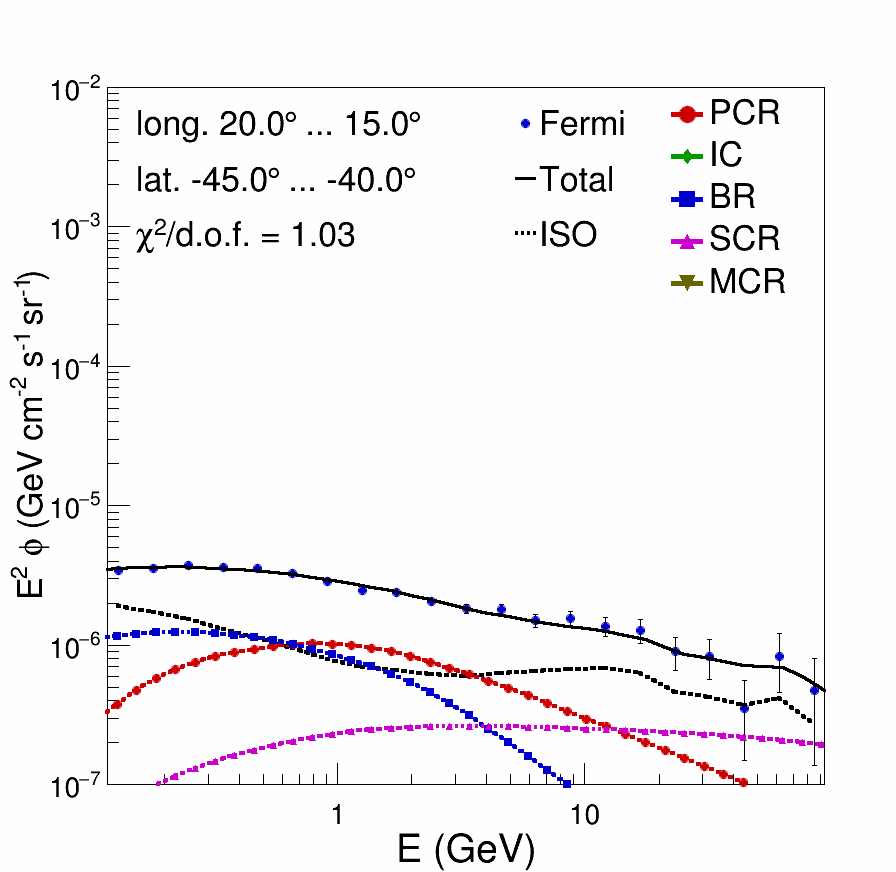}
\includegraphics[width=0.16\textwidth,height=0.16\textwidth,clip]{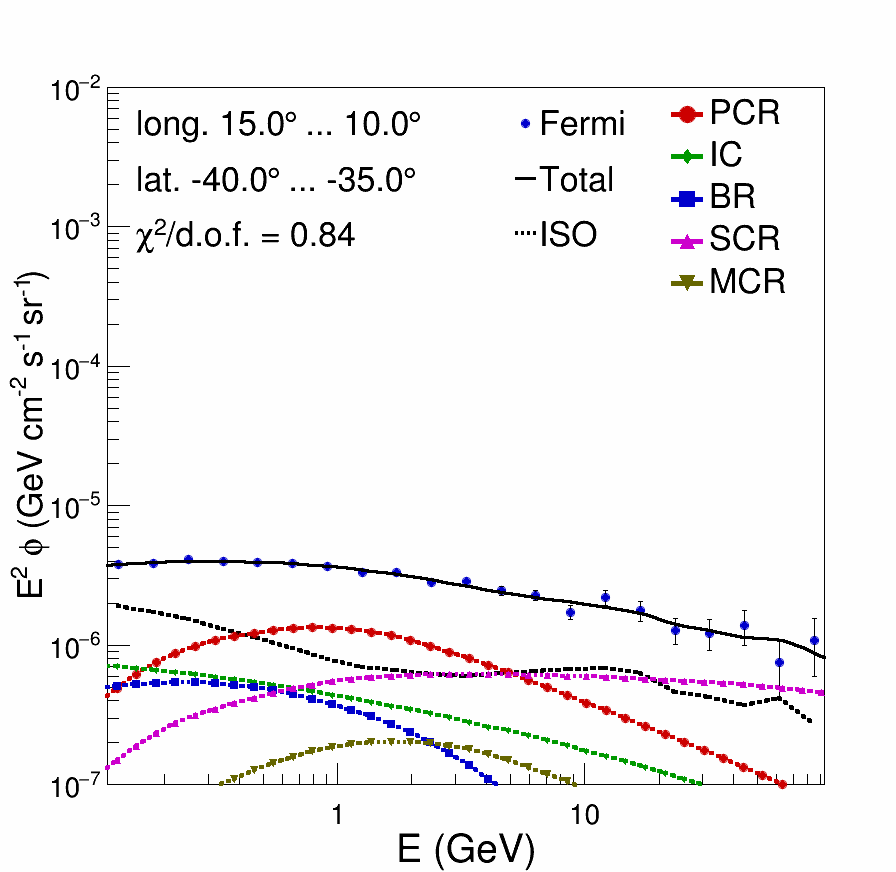}
\includegraphics[width=0.16\textwidth,height=0.16\textwidth,clip]{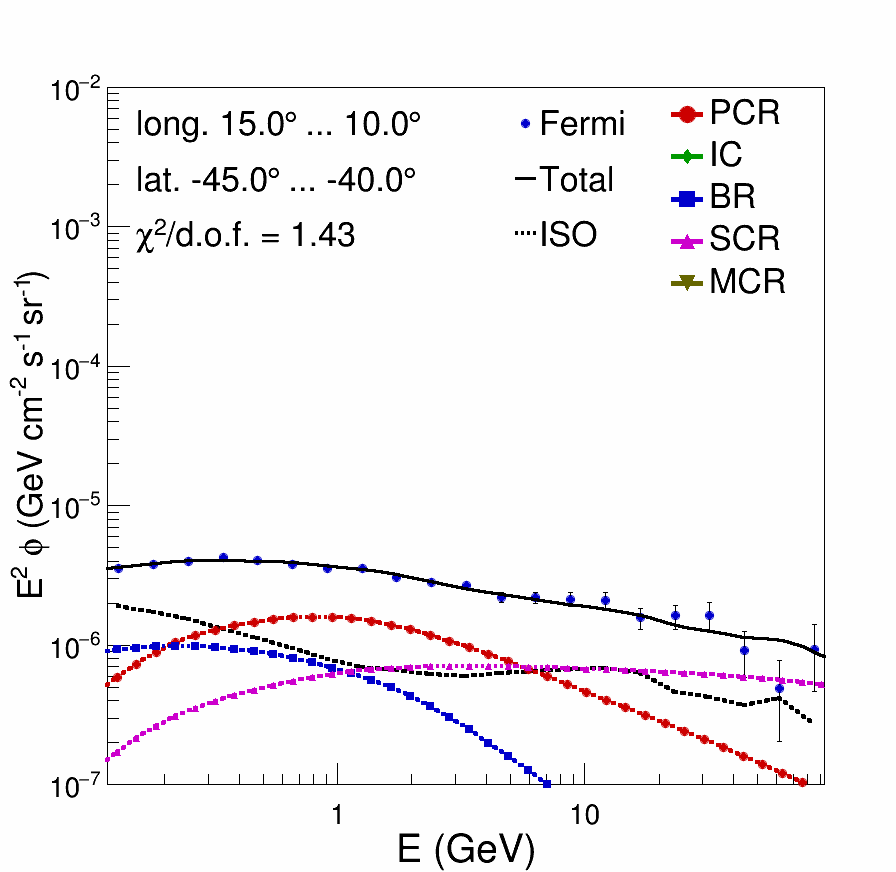}
\includegraphics[width=0.16\textwidth,height=0.16\textwidth,clip]{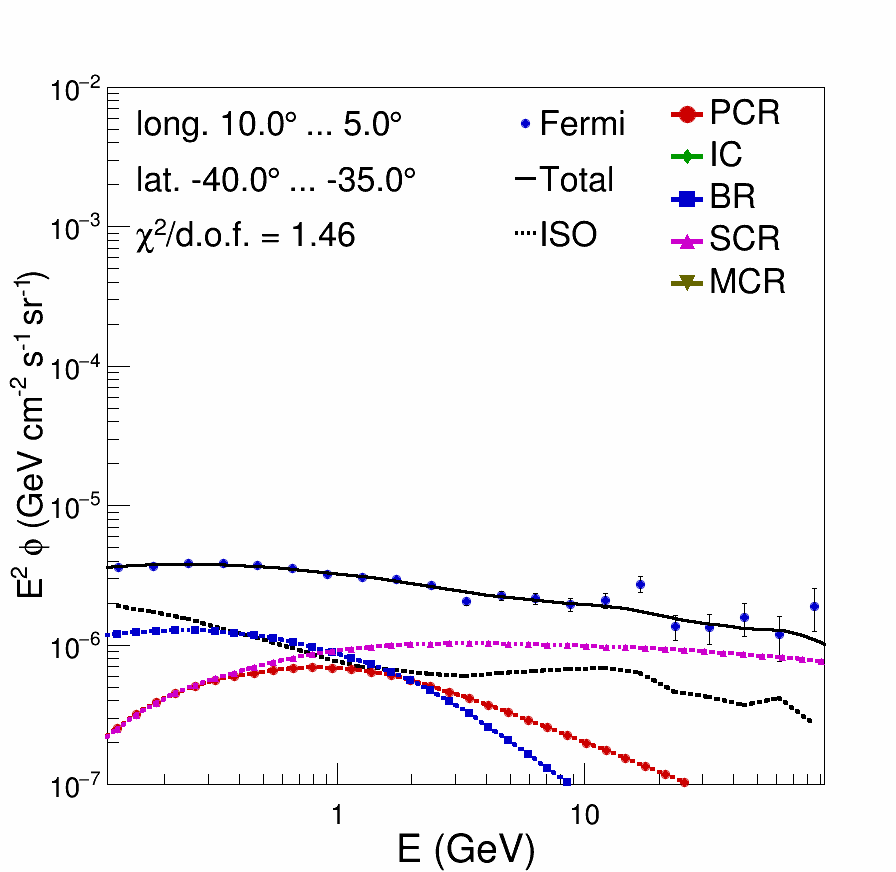}
\includegraphics[width=0.16\textwidth,height=0.16\textwidth,clip]{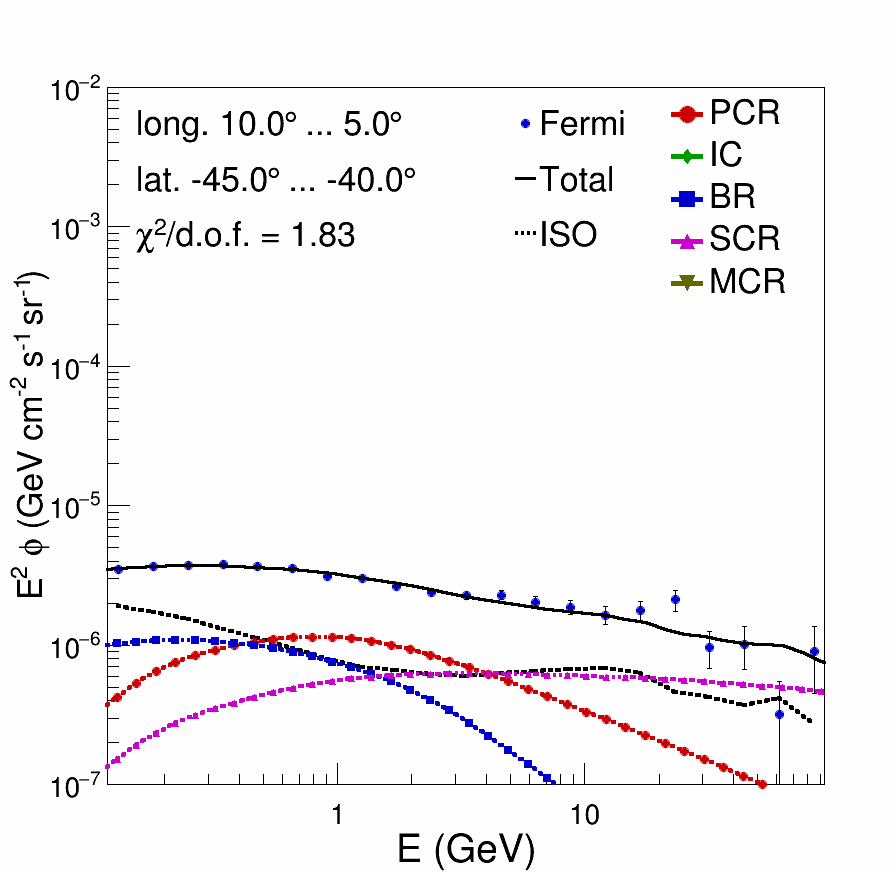}
\includegraphics[width=0.16\textwidth,height=0.16\textwidth,clip]{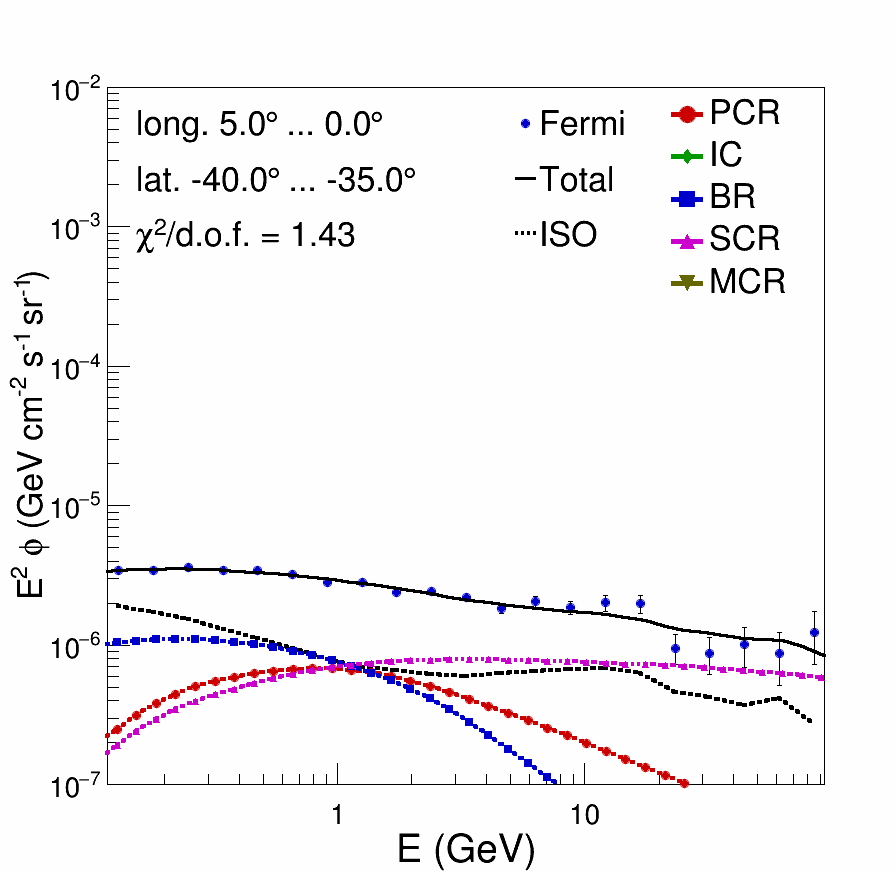}
\includegraphics[width=0.16\textwidth,height=0.16\textwidth,clip]{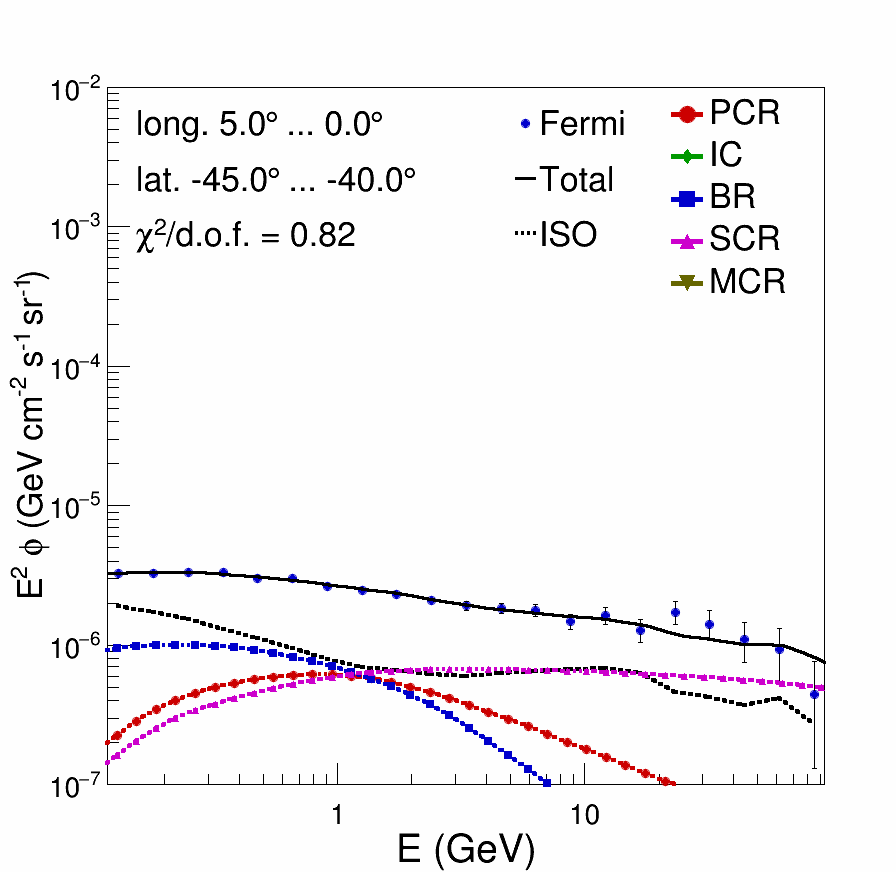}
\includegraphics[width=0.16\textwidth,height=0.16\textwidth,clip]{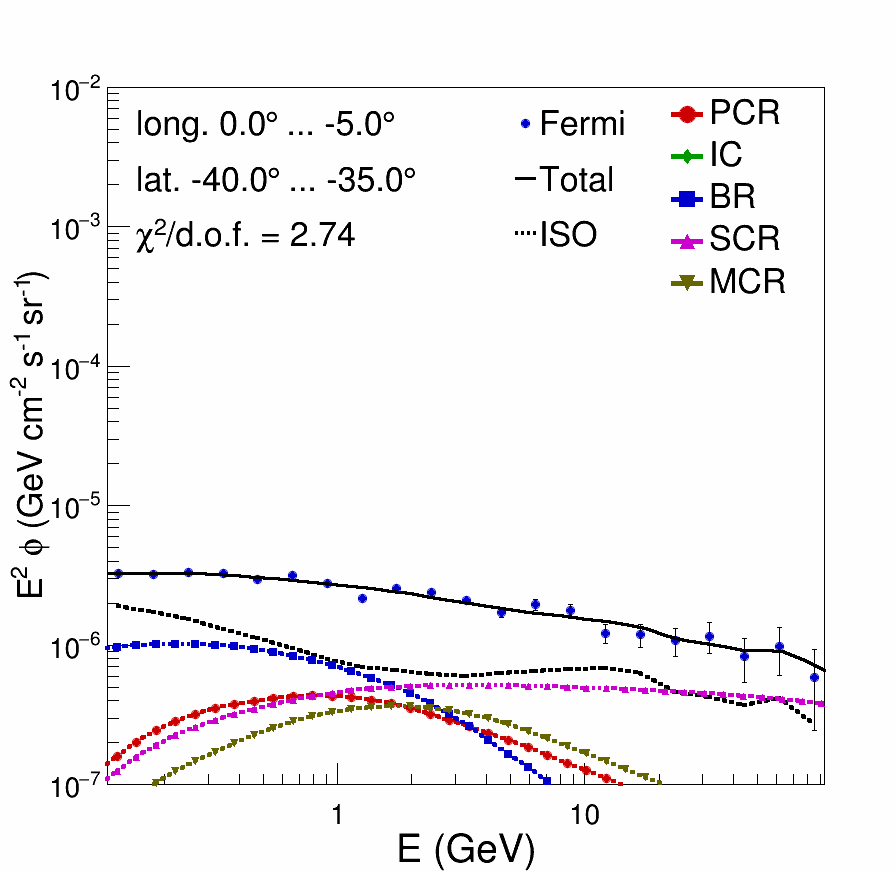}
\includegraphics[width=0.16\textwidth,height=0.16\textwidth,clip]{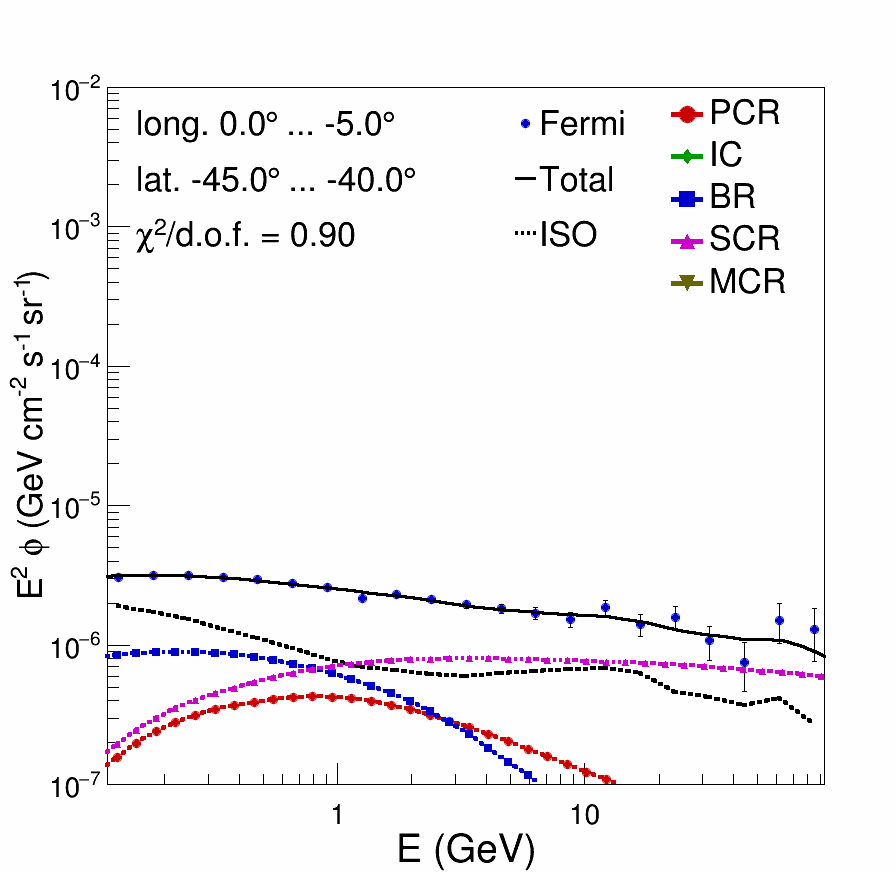}
\includegraphics[width=0.16\textwidth,height=0.16\textwidth,clip]{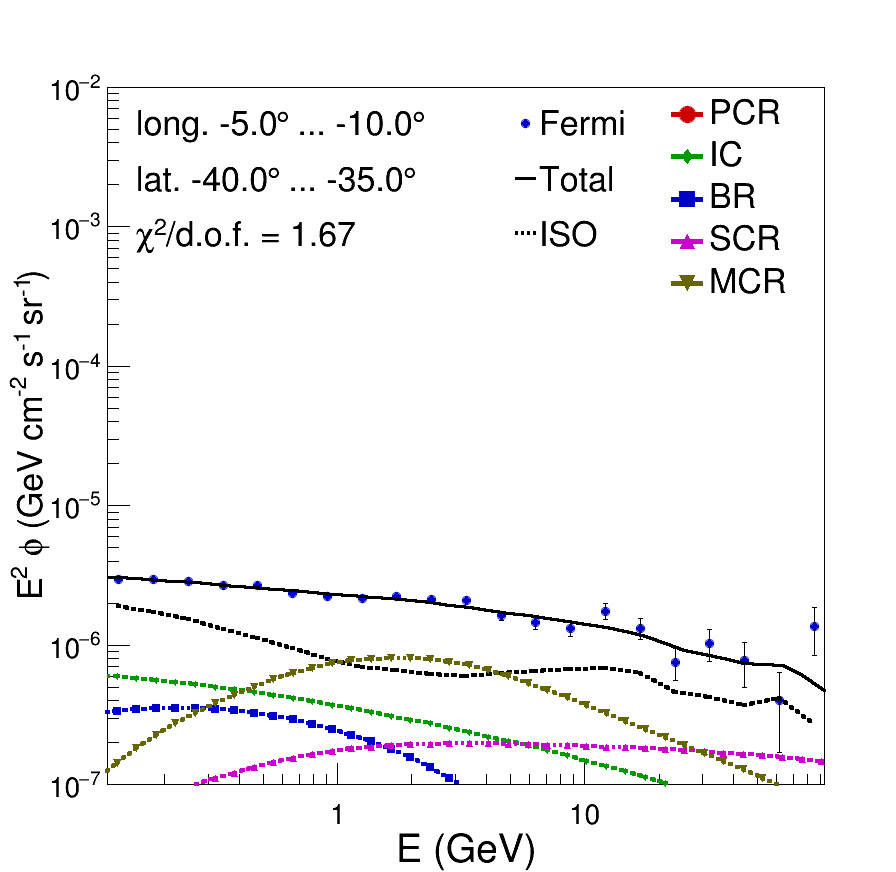}
\includegraphics[width=0.16\textwidth,height=0.16\textwidth,clip]{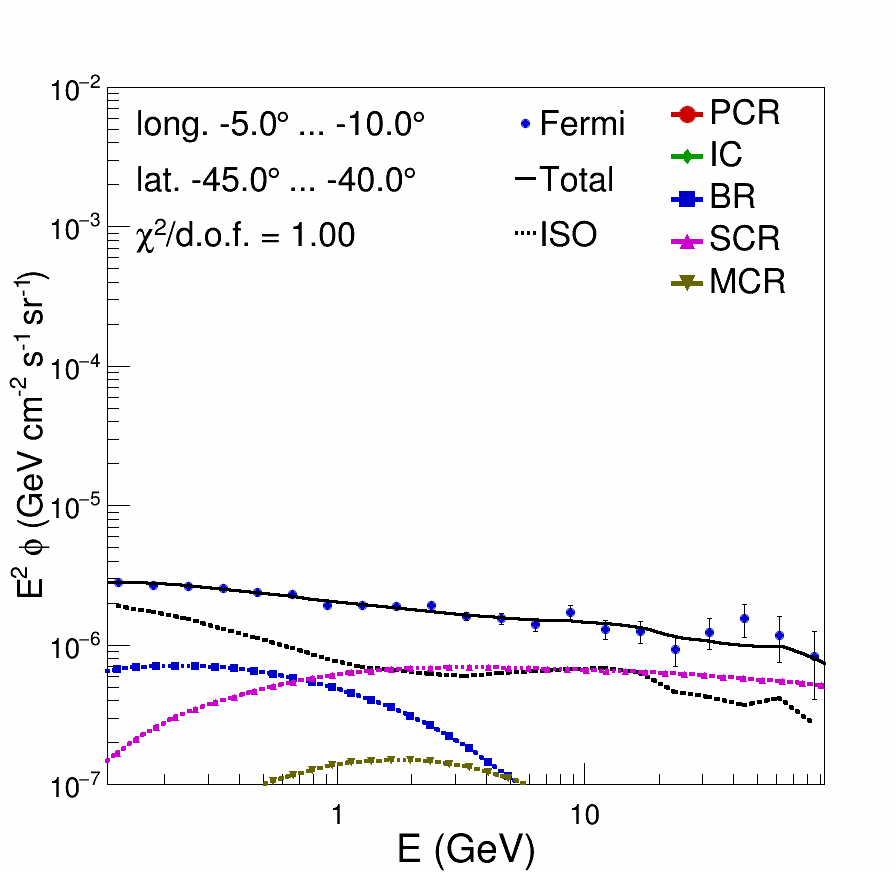}
\includegraphics[width=0.16\textwidth,height=0.16\textwidth,clip]{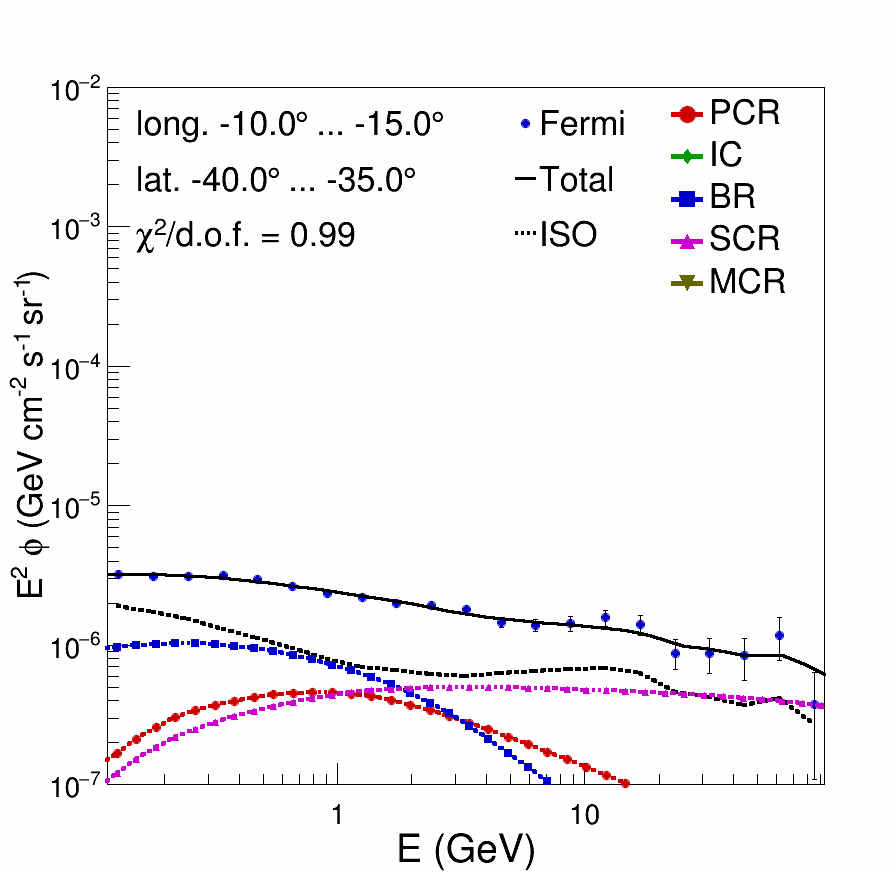}
\includegraphics[width=0.16\textwidth,height=0.16\textwidth,clip]{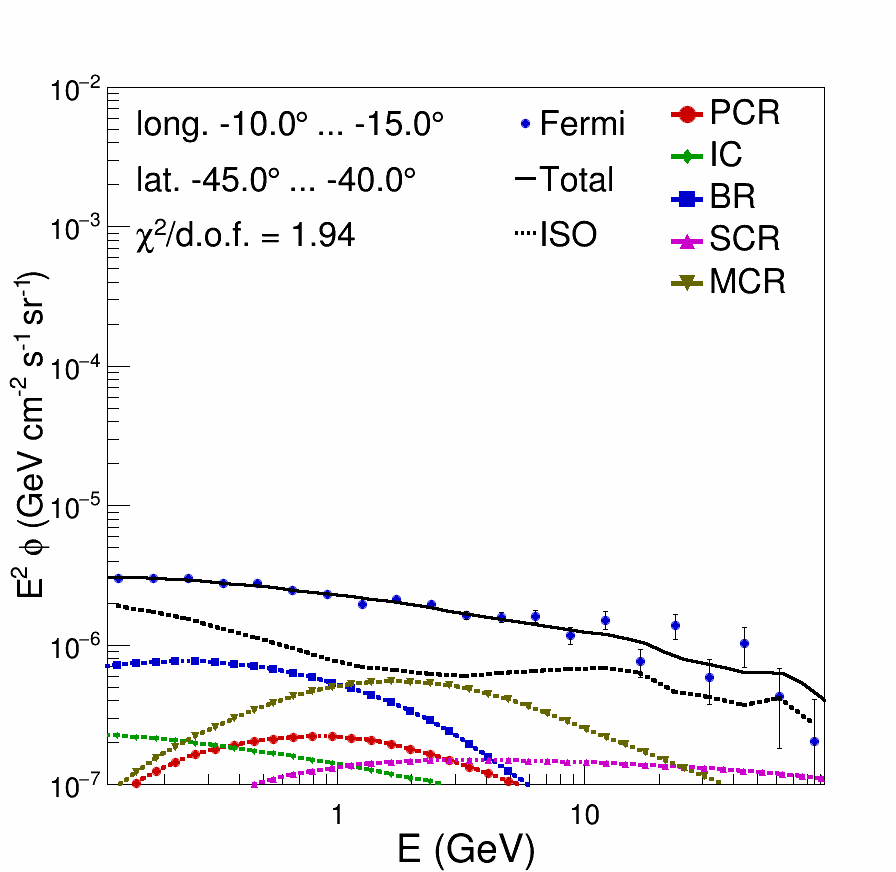}
\includegraphics[width=0.16\textwidth,height=0.16\textwidth,clip]{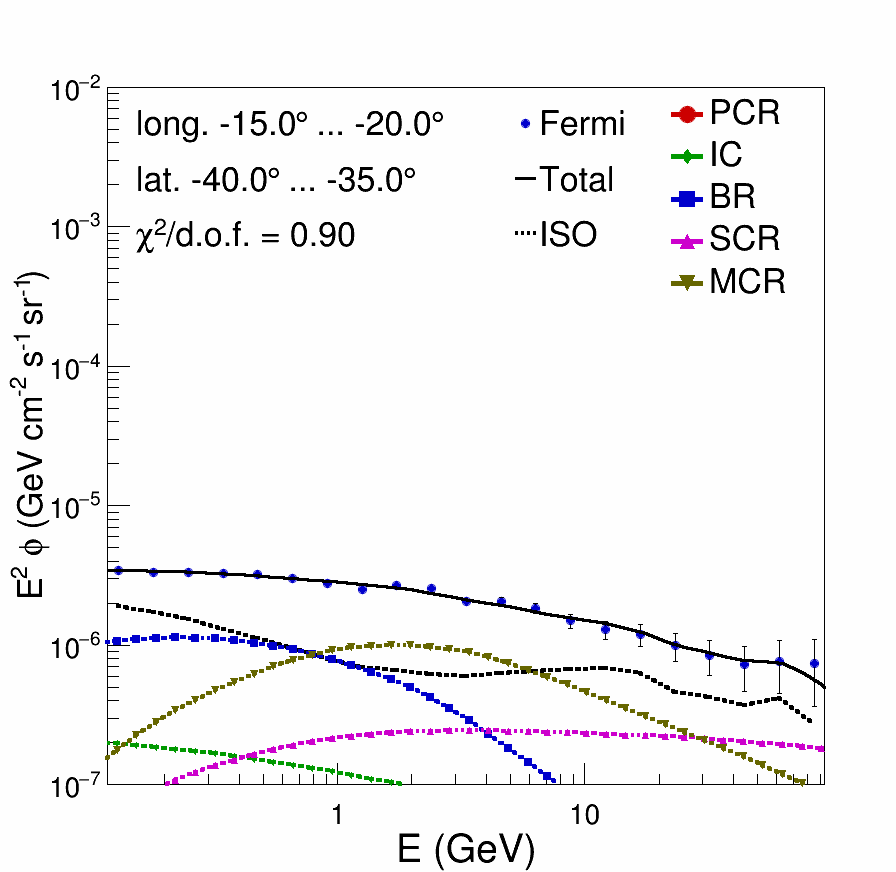}
\includegraphics[width=0.16\textwidth,height=0.16\textwidth,clip]{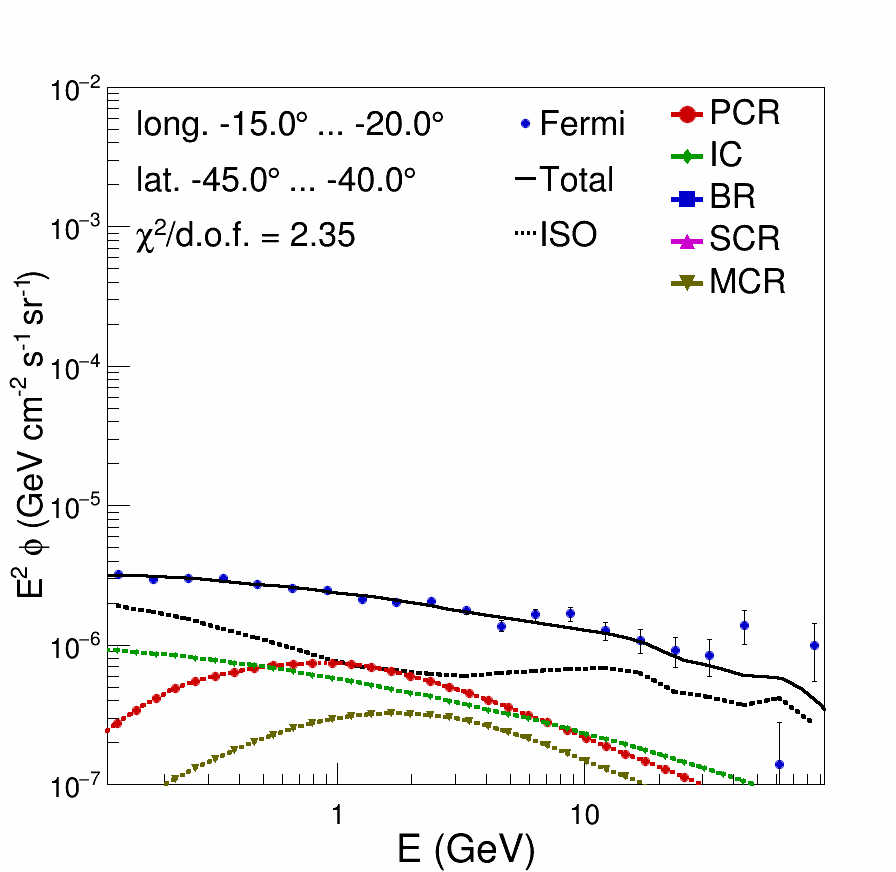}
\includegraphics[width=0.16\textwidth,height=0.16\textwidth,clip]{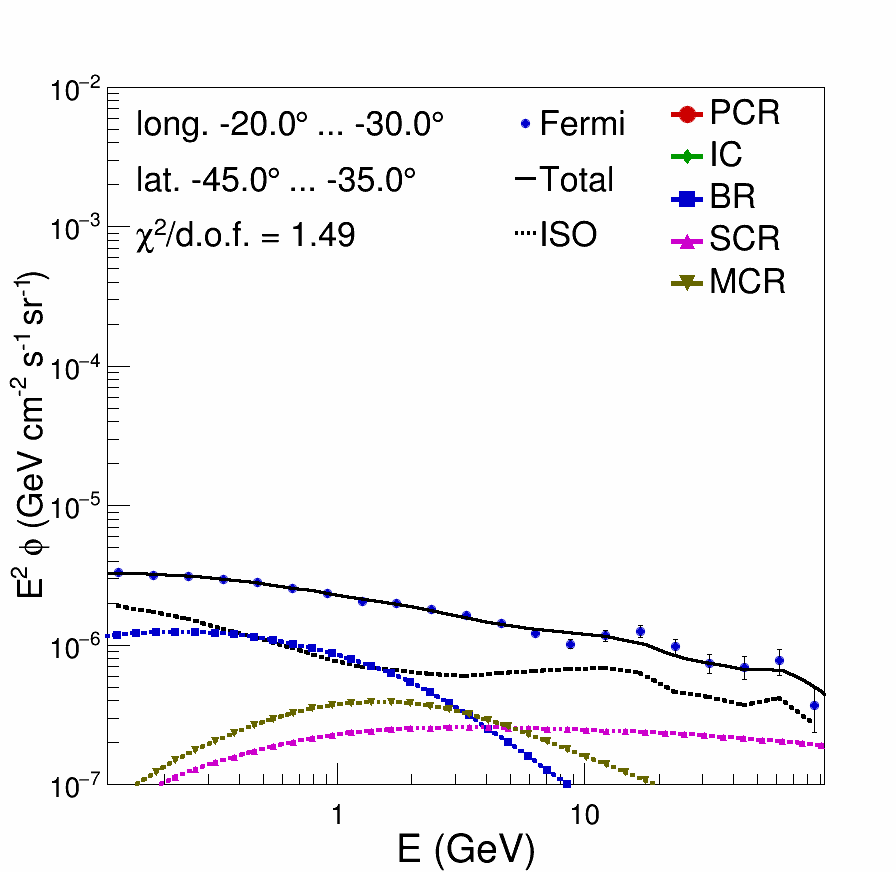}
\includegraphics[width=0.16\textwidth,height=0.16\textwidth,clip]{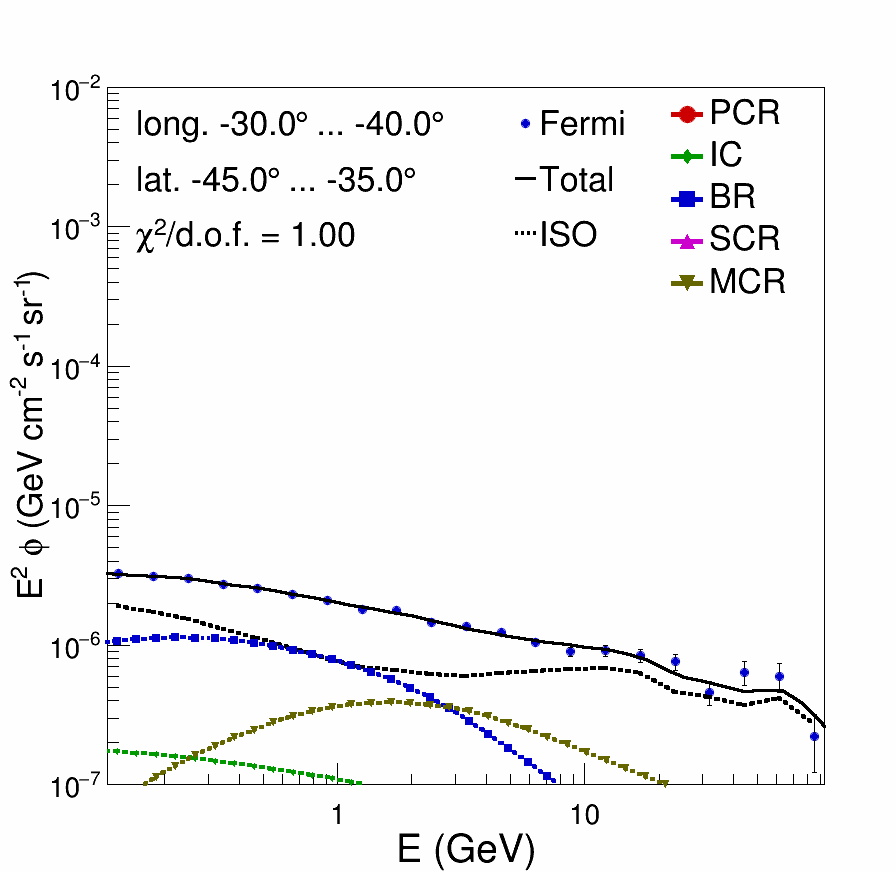}
\includegraphics[width=0.16\textwidth,height=0.16\textwidth,clip]{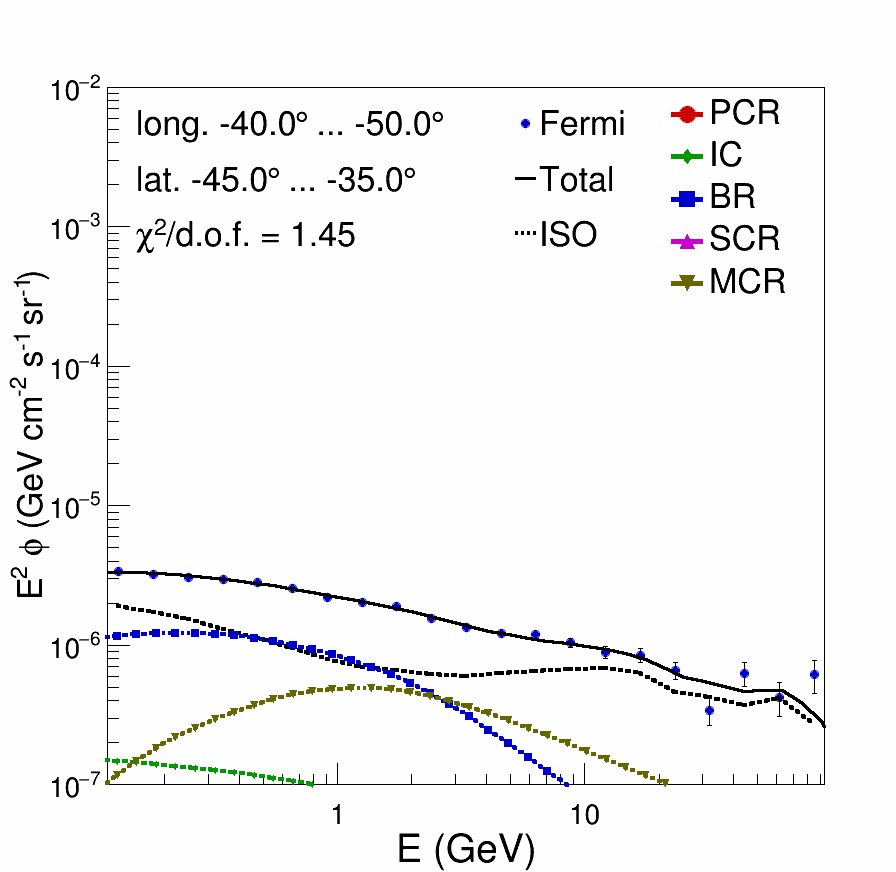}
\includegraphics[width=0.16\textwidth,height=0.16\textwidth,clip]{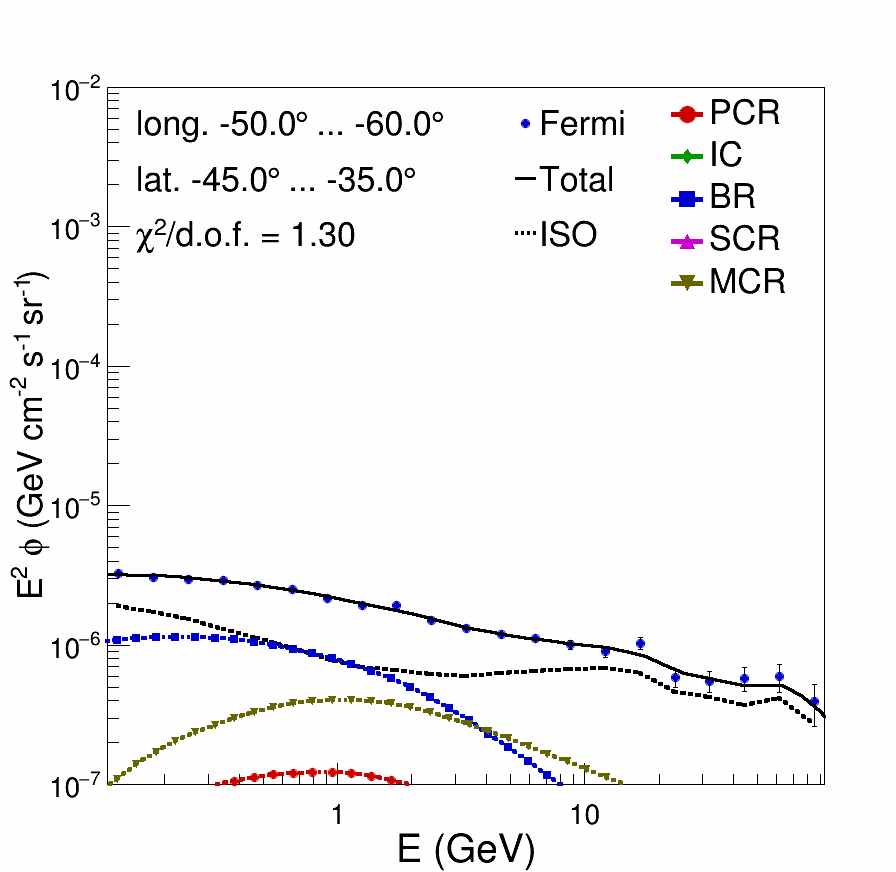}
\includegraphics[width=0.16\textwidth,height=0.16\textwidth,clip]{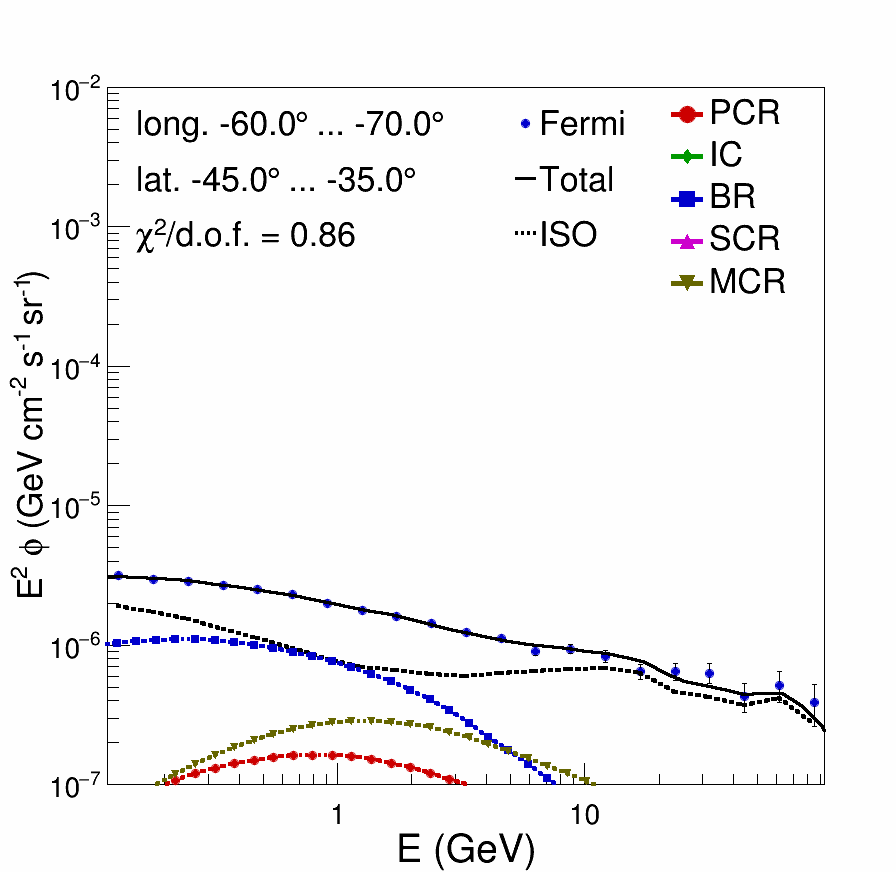}
\includegraphics[width=0.16\textwidth,height=0.16\textwidth,clip]{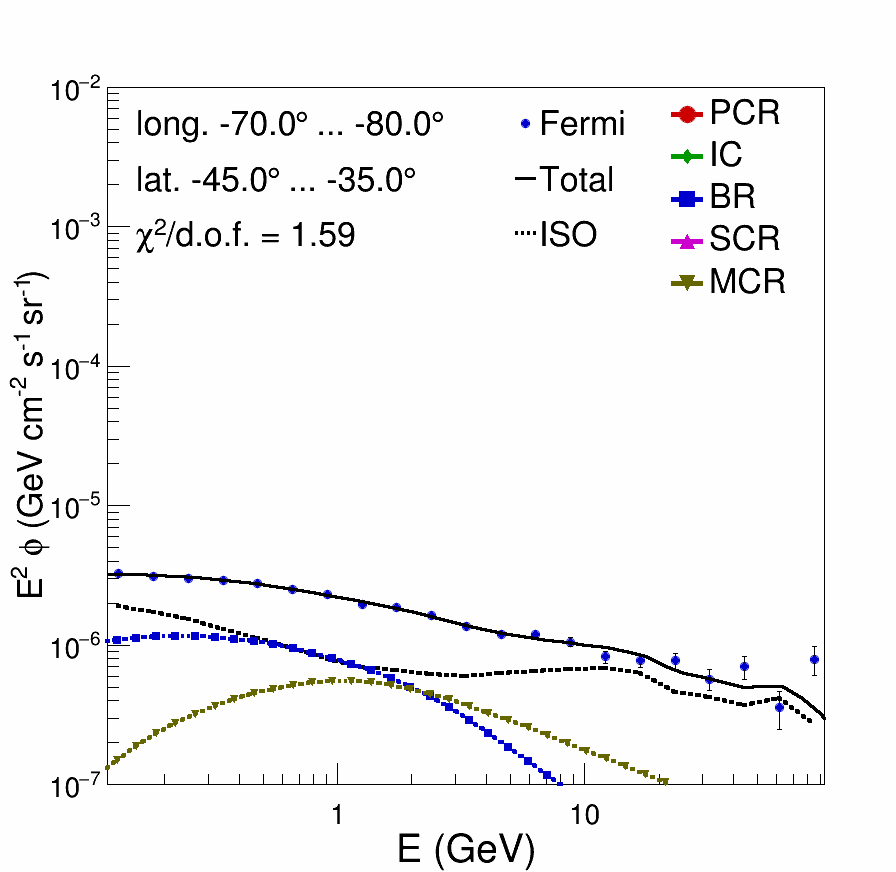}
\includegraphics[width=0.16\textwidth,height=0.16\textwidth,clip]{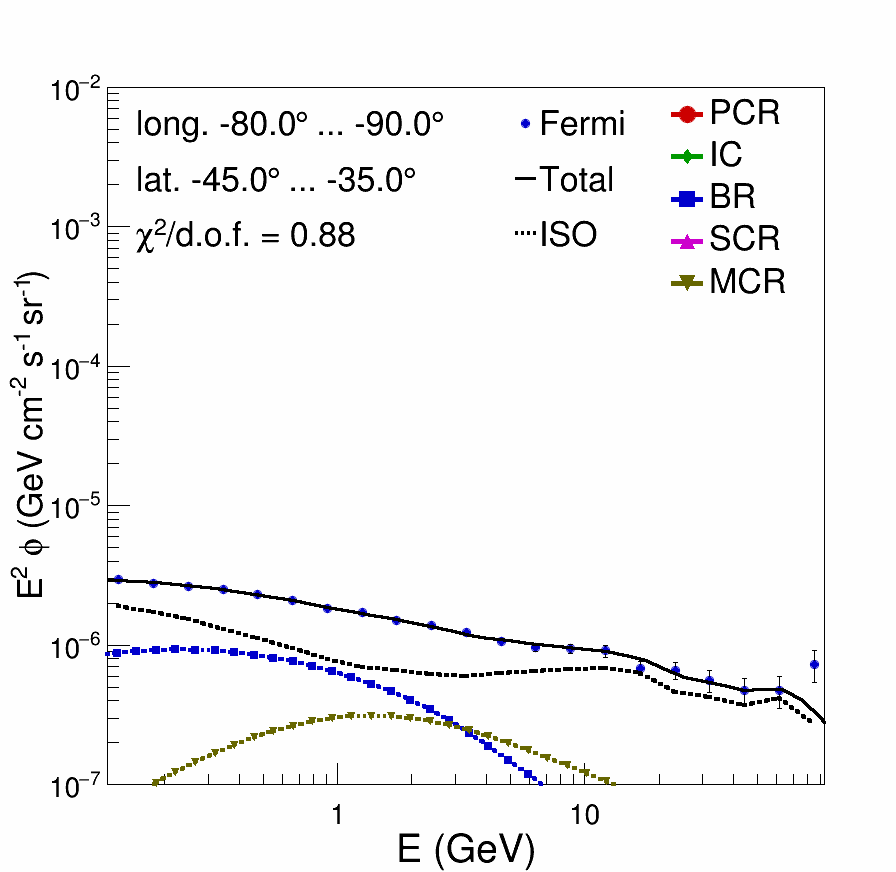}
\includegraphics[width=0.16\textwidth,height=0.16\textwidth,clip]{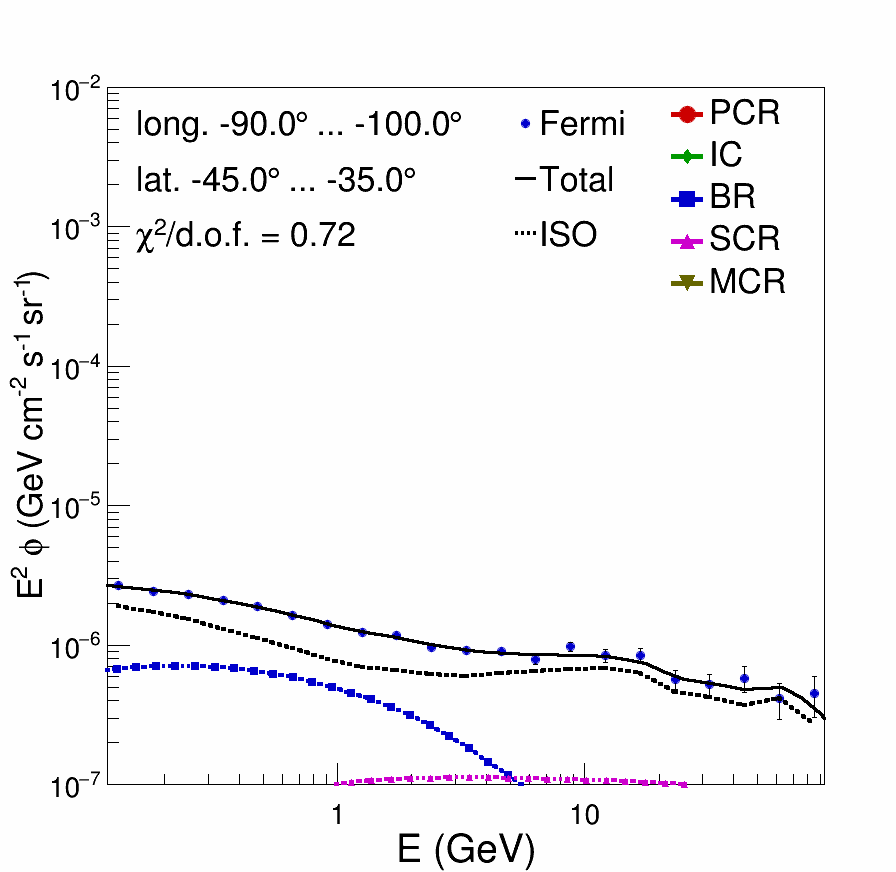}
\includegraphics[width=0.16\textwidth,height=0.16\textwidth,clip]{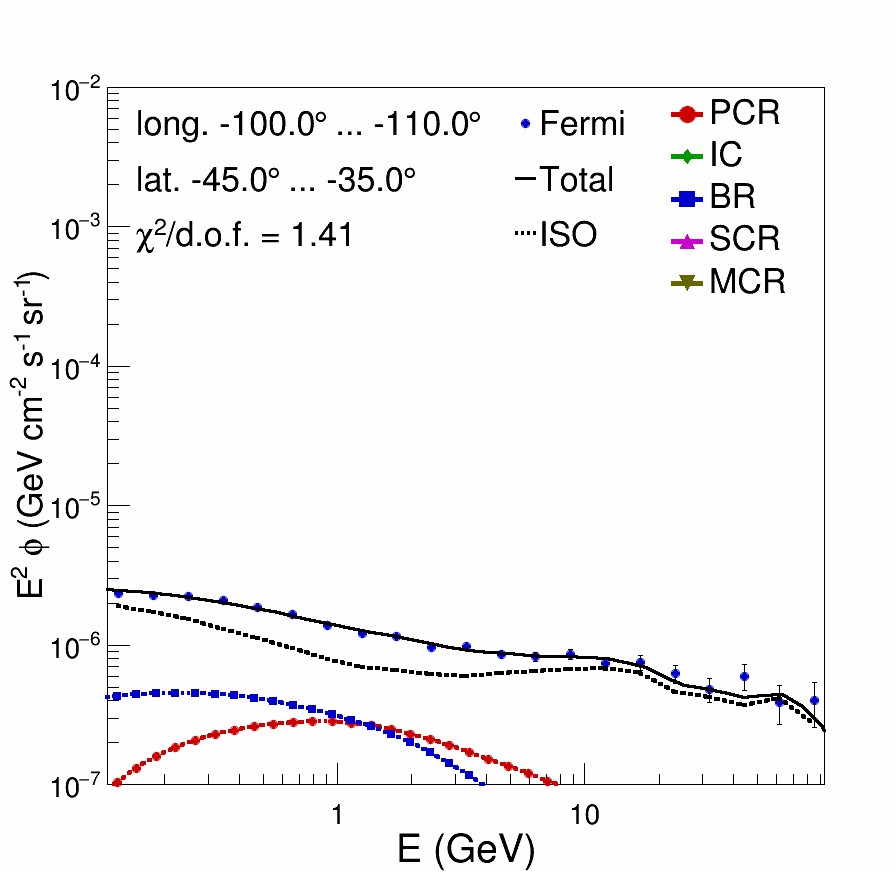}
\includegraphics[width=0.16\textwidth,height=0.16\textwidth,clip]{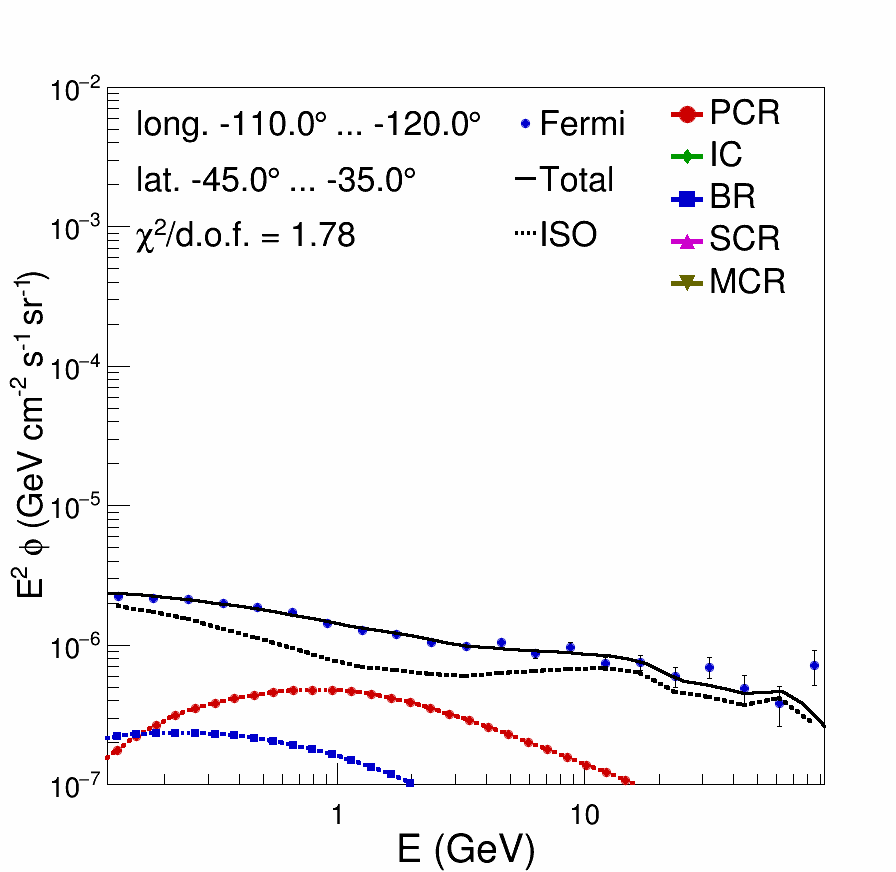}
\includegraphics[width=0.16\textwidth,height=0.16\textwidth,clip]{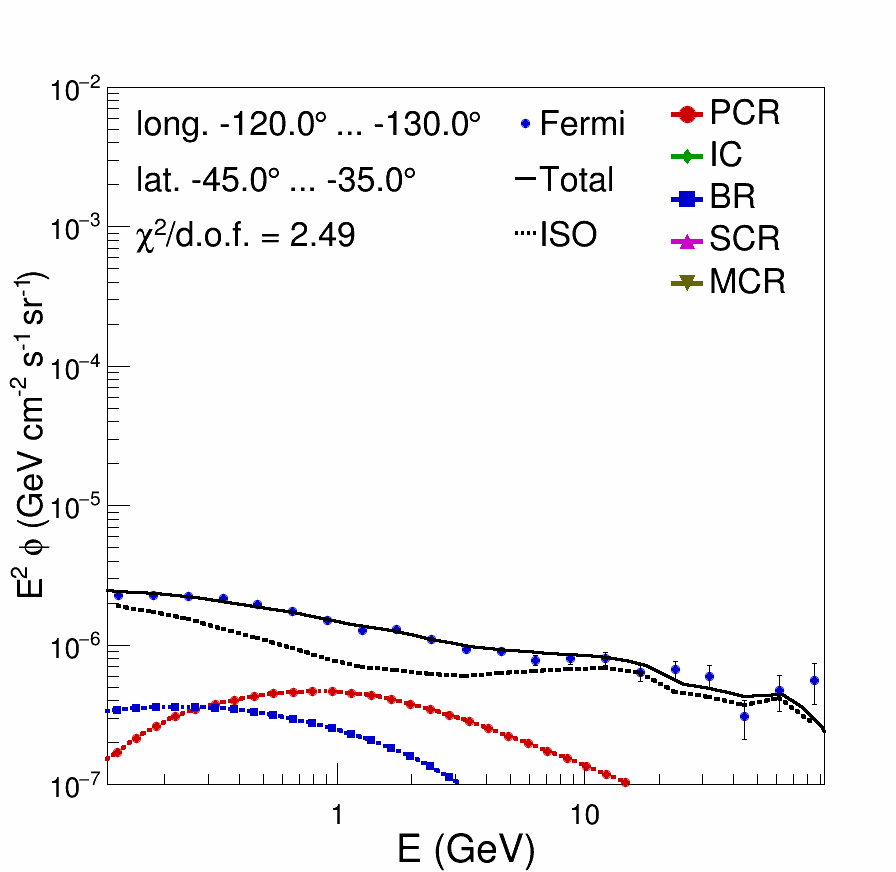}
\includegraphics[width=0.16\textwidth,height=0.16\textwidth,clip]{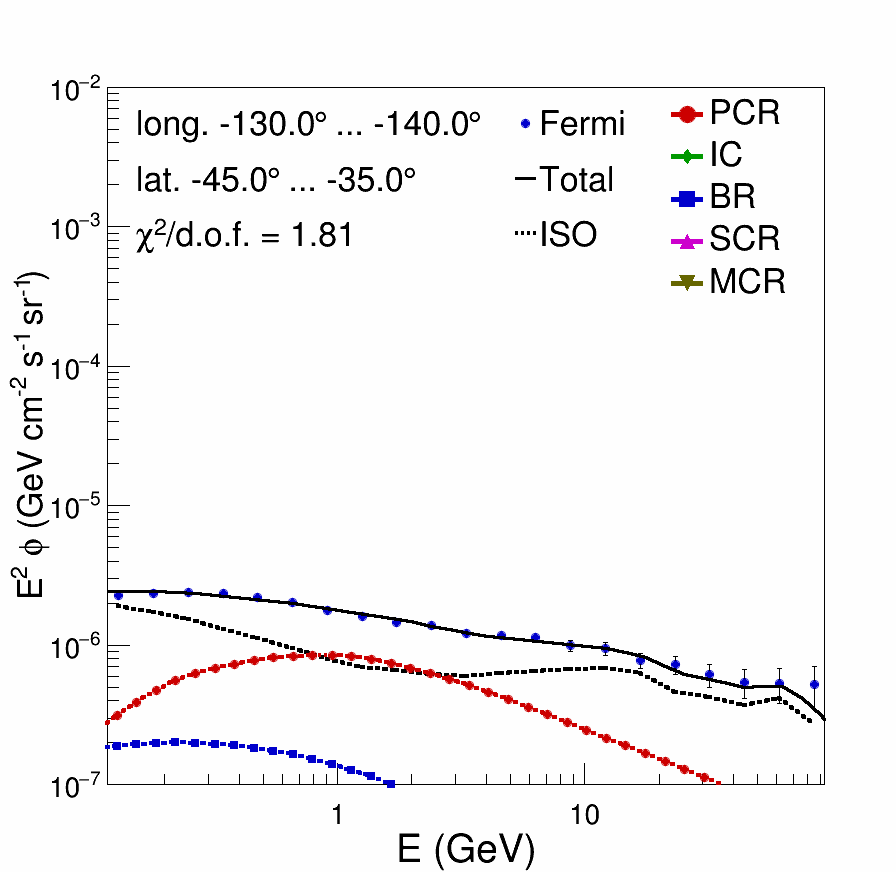}
\includegraphics[width=0.16\textwidth,height=0.16\textwidth,clip]{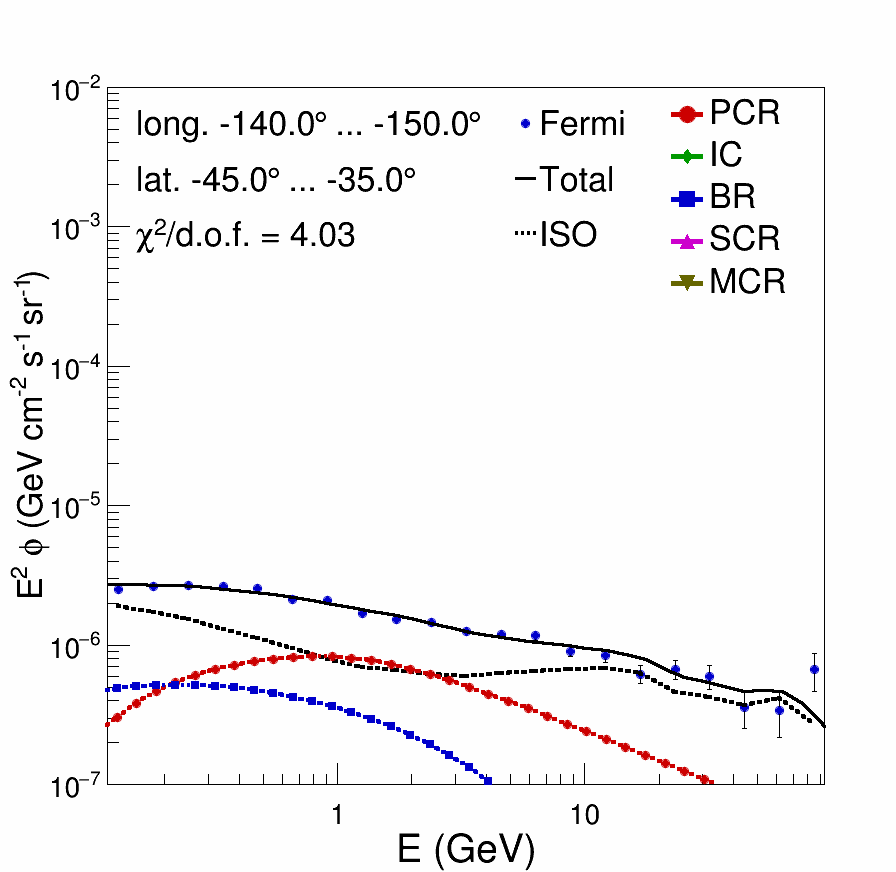}
\includegraphics[width=0.16\textwidth,height=0.16\textwidth,clip]{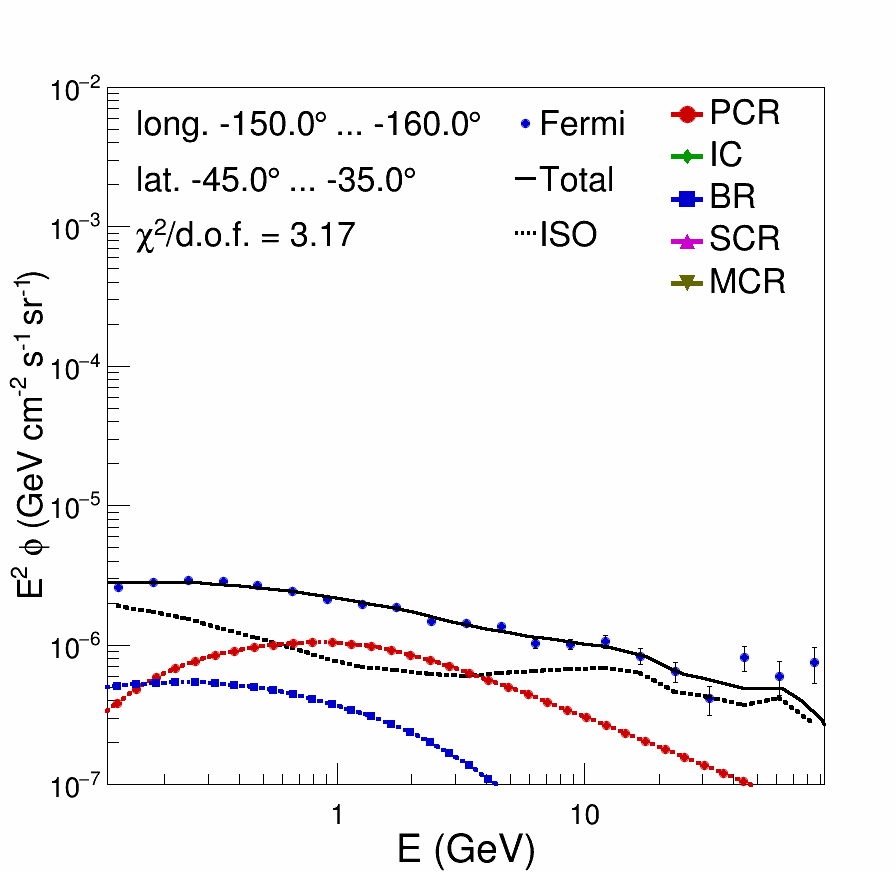}
\includegraphics[width=0.16\textwidth,height=0.16\textwidth,clip]{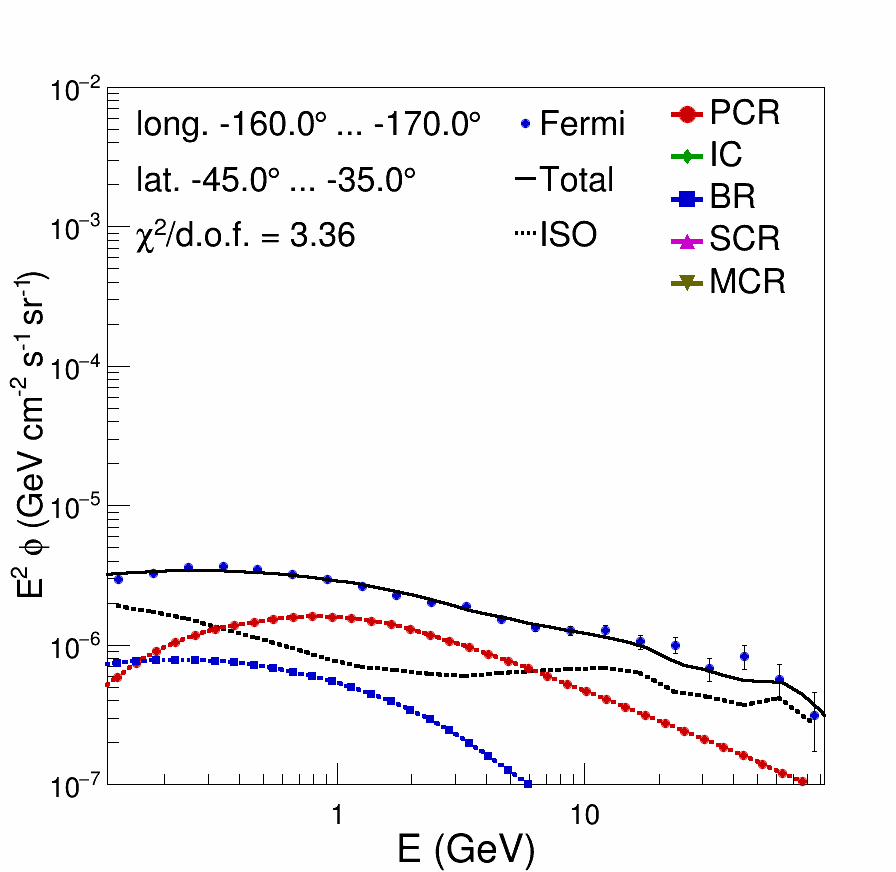}
\includegraphics[width=0.16\textwidth,height=0.16\textwidth,clip]{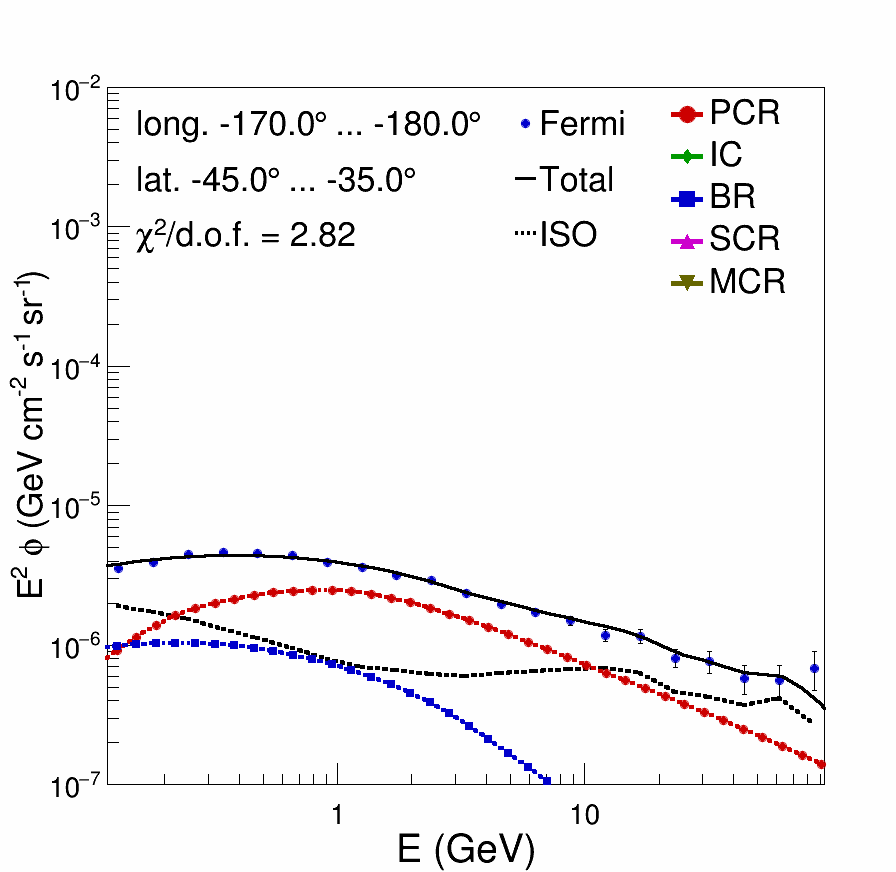}
\caption[]{Template fits for latitudes  with $-45.0^\circ<b<-35.0^\circ$ and longitudes decreasing from 180$^\circ$ to -180$^\circ$.} \label{F28}
\end{figure}
\begin{figure}
\centering
\includegraphics[width=0.16\textwidth,height=0.16\textwidth,clip]{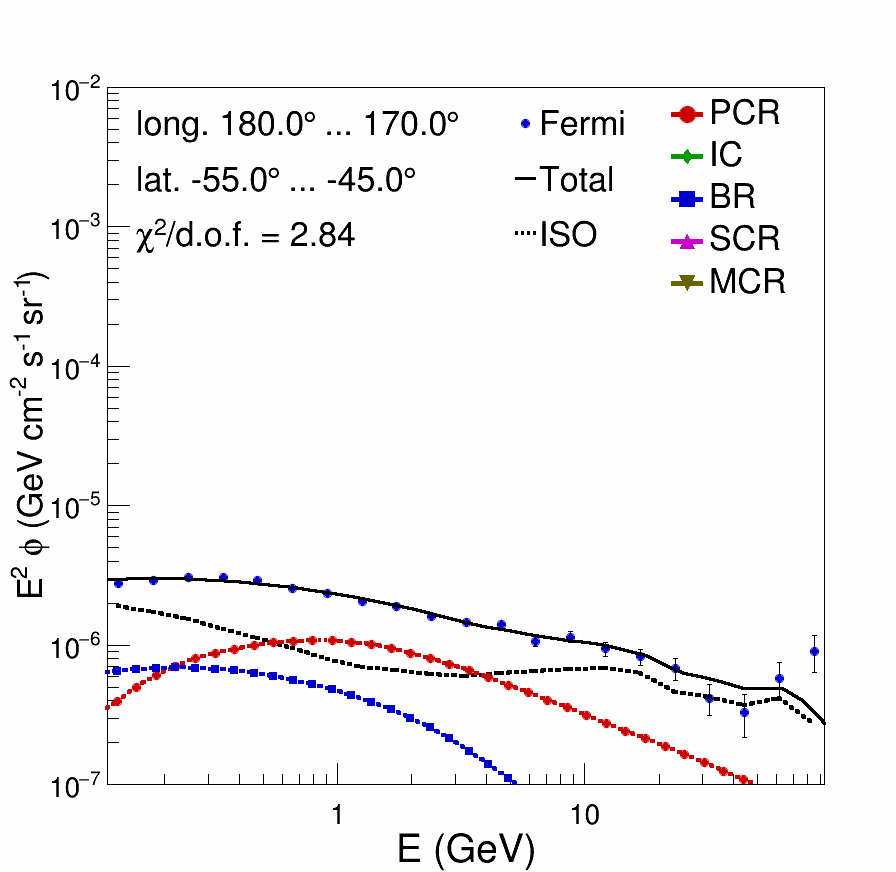}
\includegraphics[width=0.16\textwidth,height=0.16\textwidth,clip]{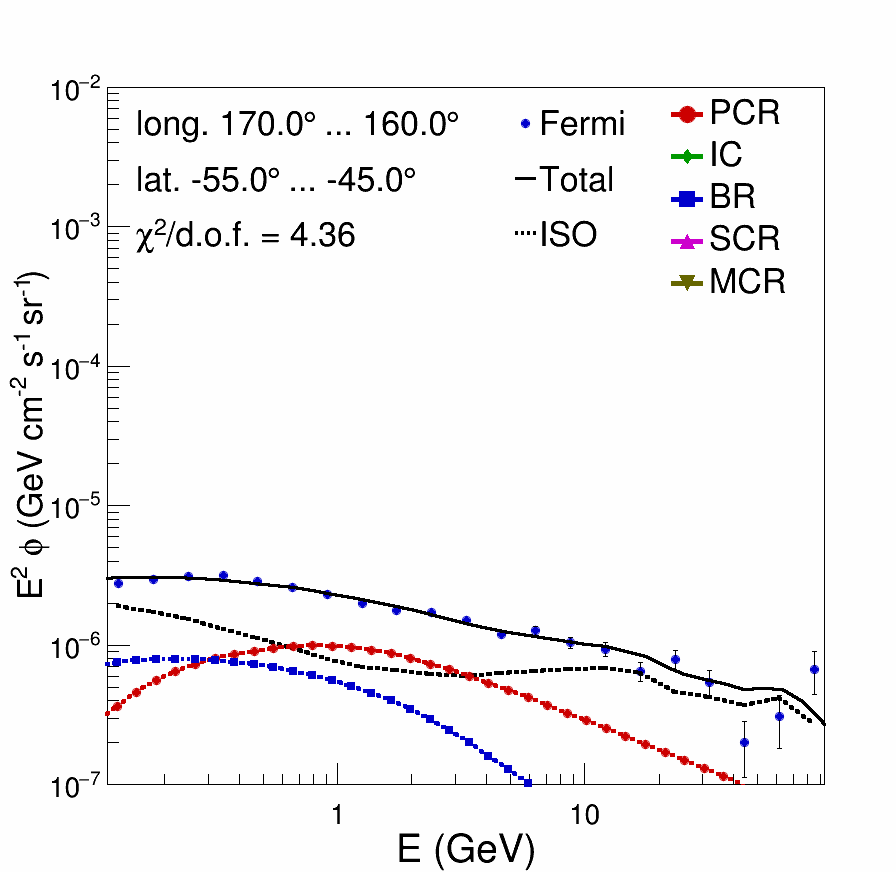}
\includegraphics[width=0.16\textwidth,height=0.16\textwidth,clip]{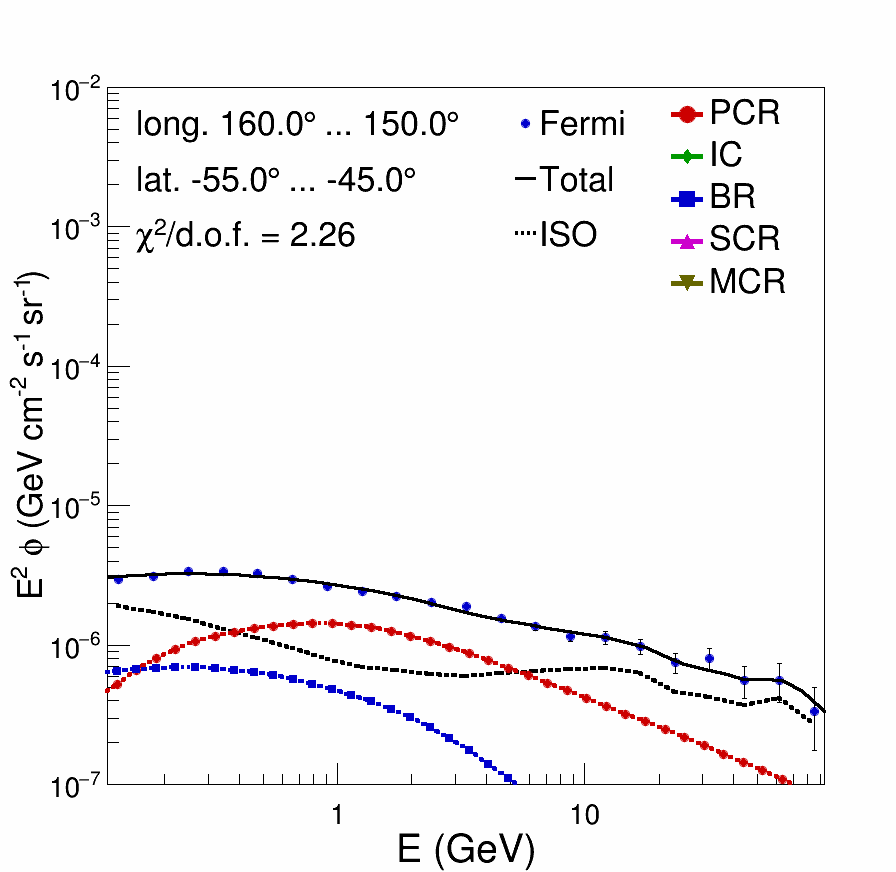}
\includegraphics[width=0.16\textwidth,height=0.16\textwidth,clip]{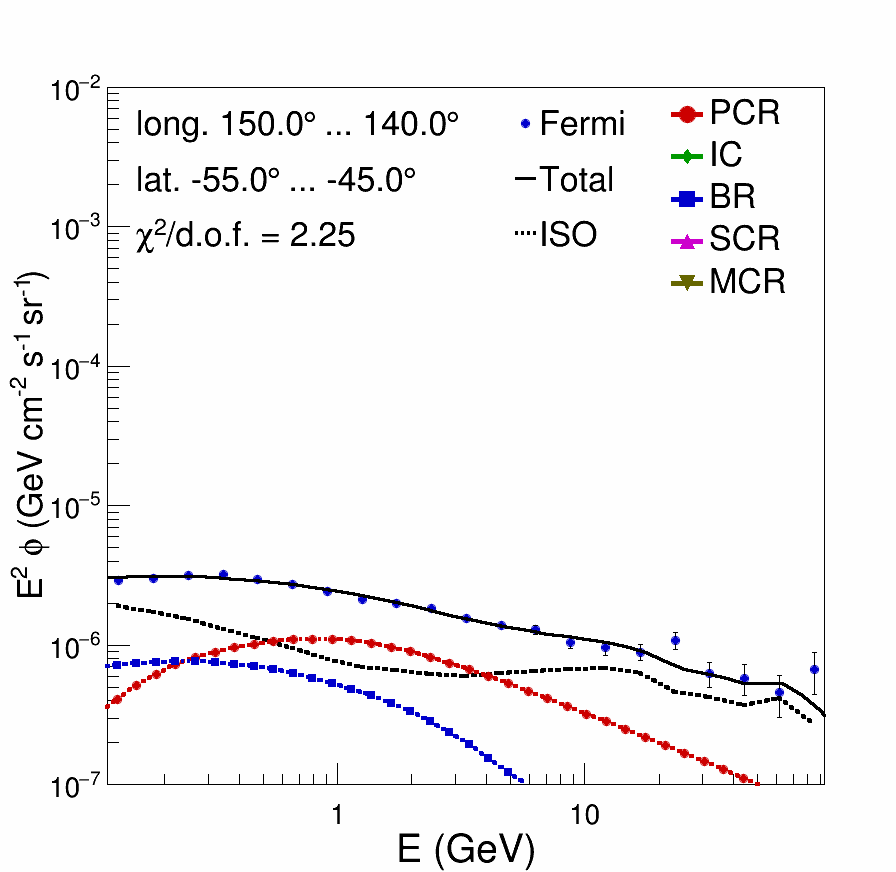}
\includegraphics[width=0.16\textwidth,height=0.16\textwidth,clip]{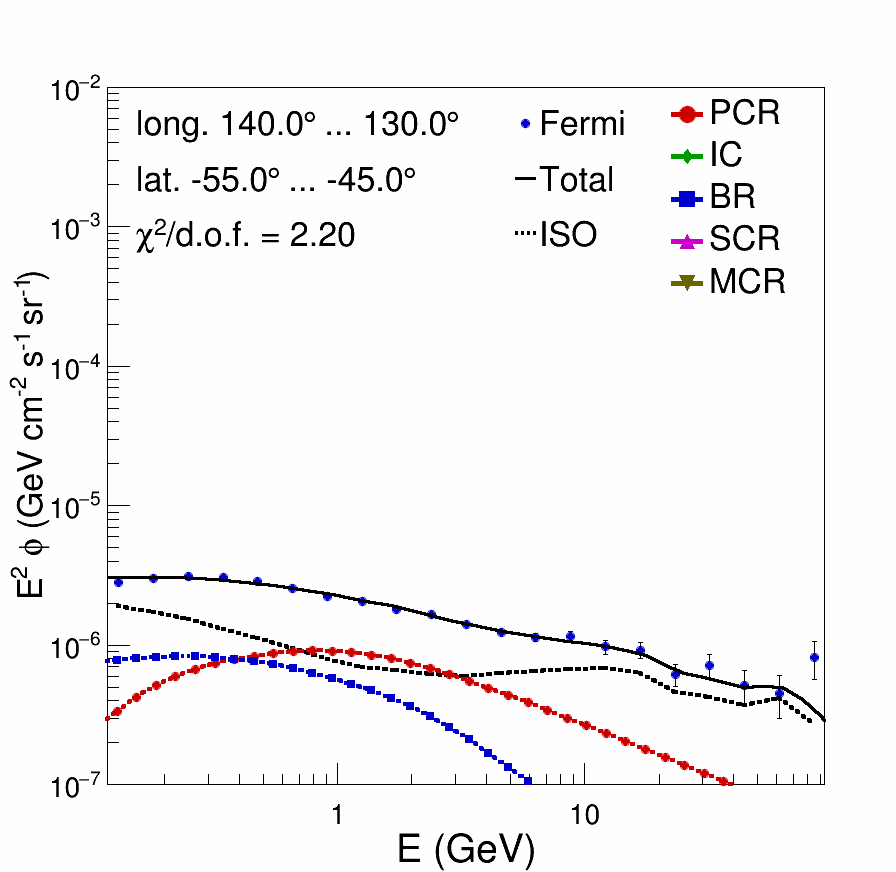}
\includegraphics[width=0.16\textwidth,height=0.16\textwidth,clip]{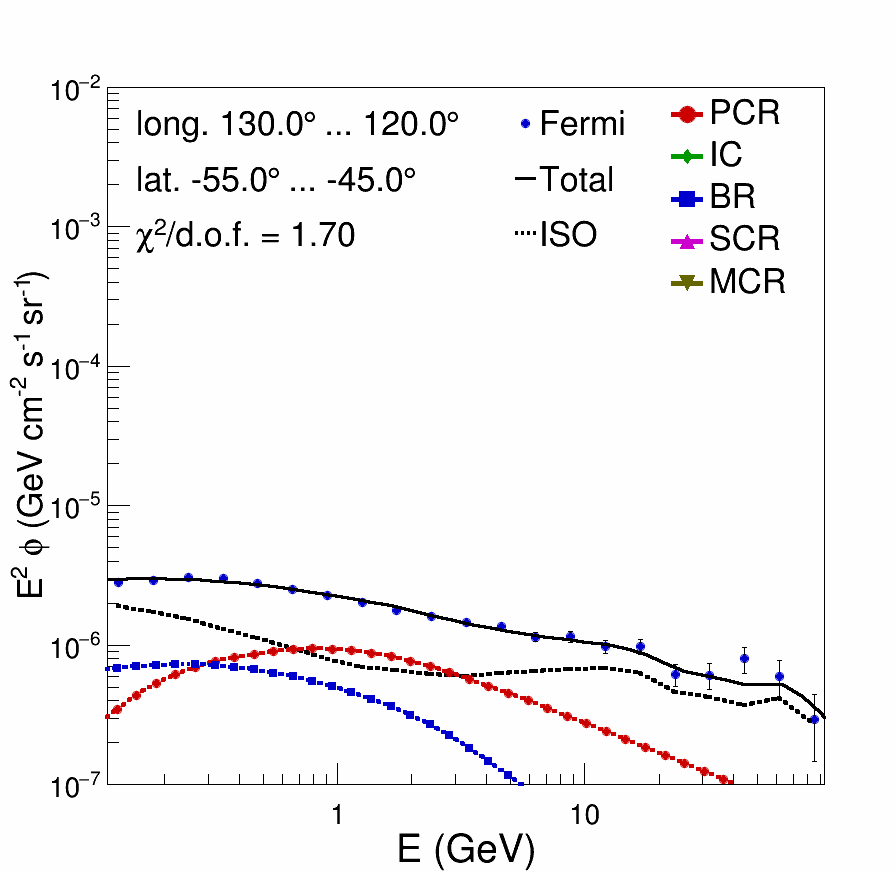}
\includegraphics[width=0.16\textwidth,height=0.16\textwidth,clip]{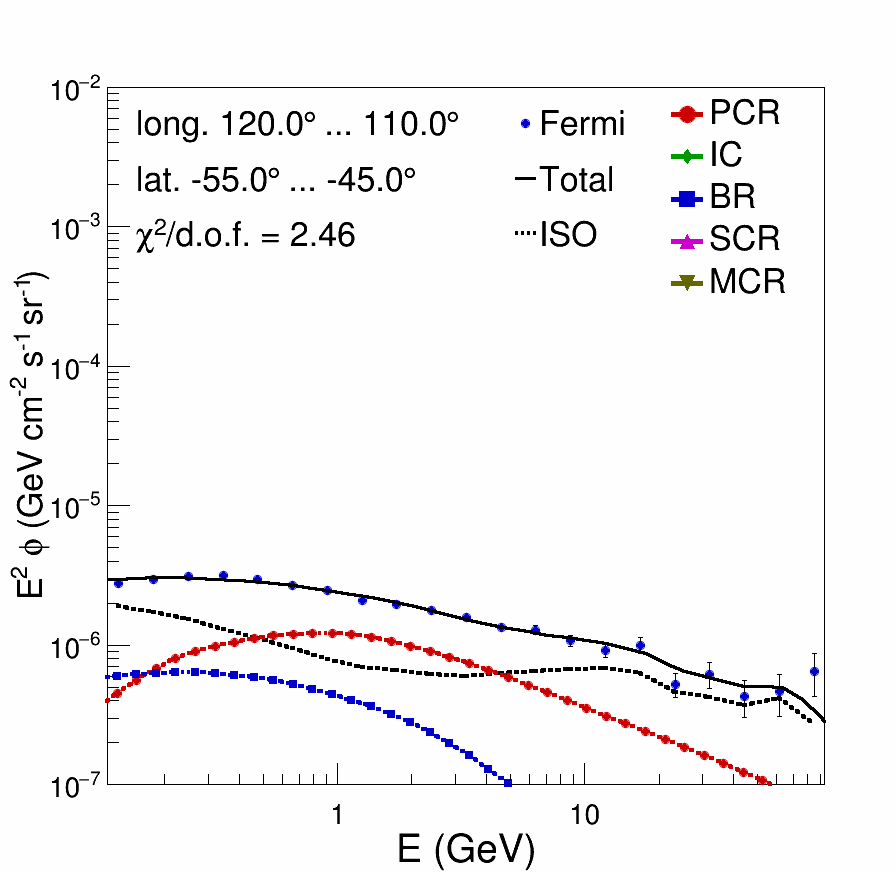}
\includegraphics[width=0.16\textwidth,height=0.16\textwidth,clip]{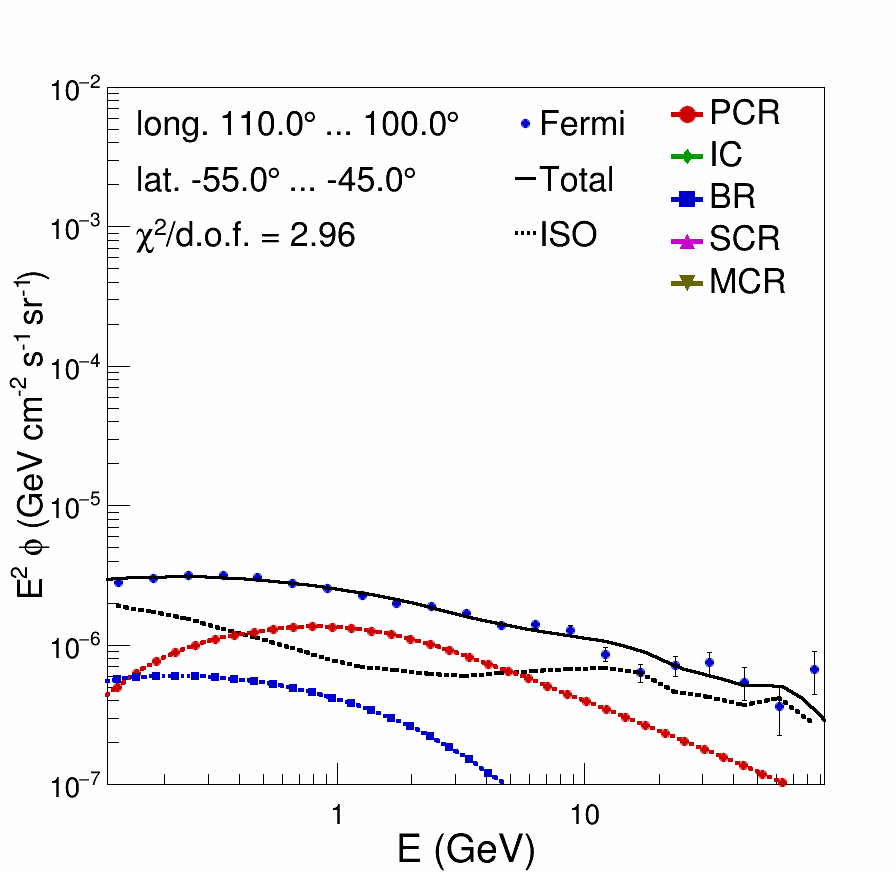}
\includegraphics[width=0.16\textwidth,height=0.16\textwidth,clip]{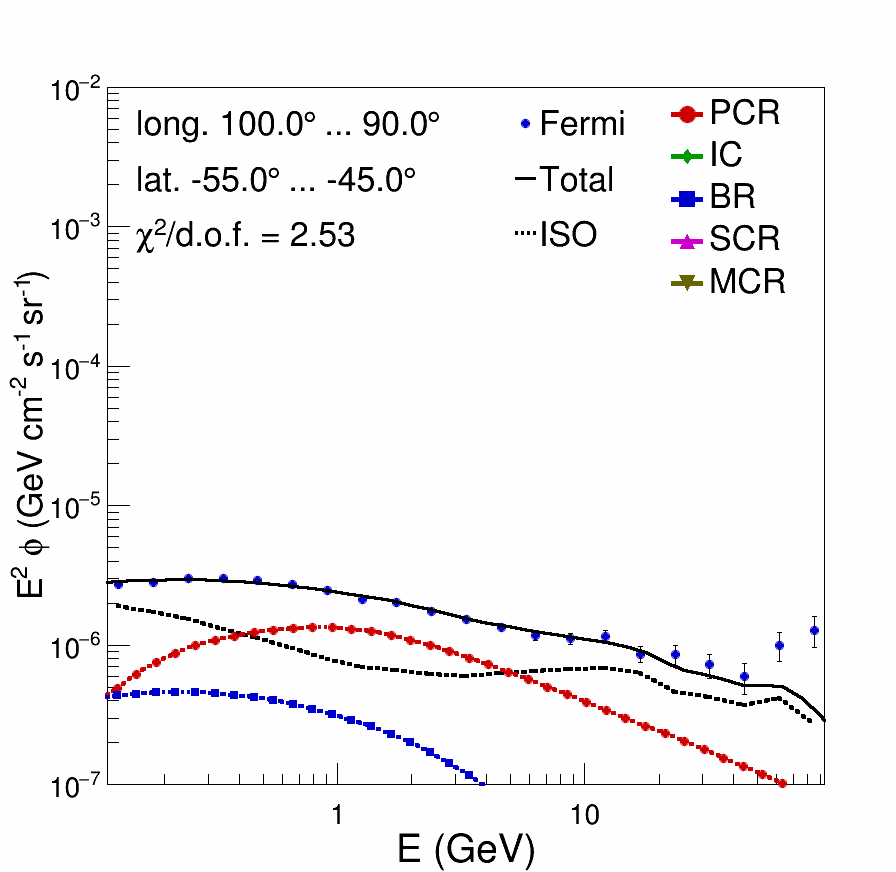}
\includegraphics[width=0.16\textwidth,height=0.16\textwidth,clip]{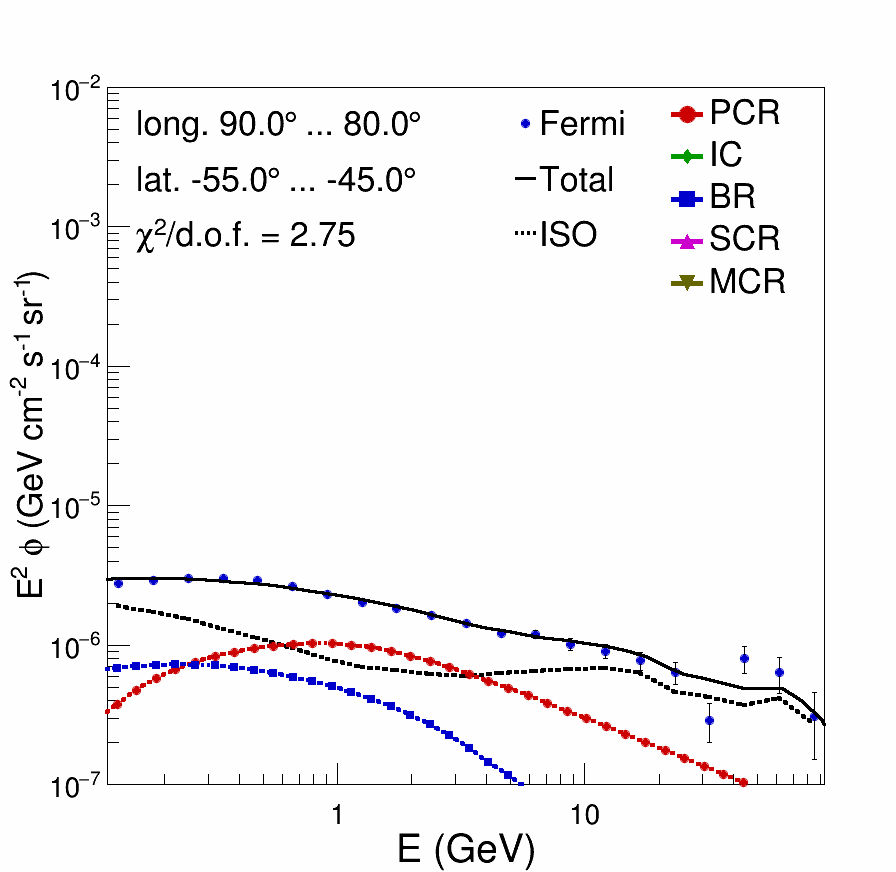}
\includegraphics[width=0.16\textwidth,height=0.16\textwidth,clip]{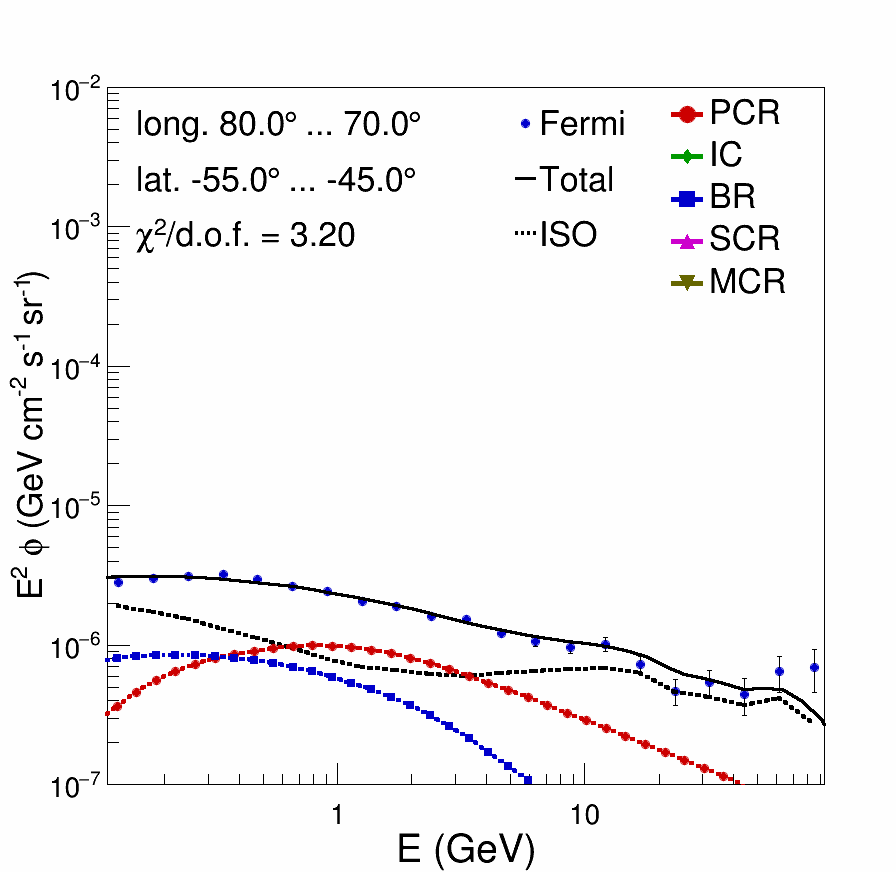}
\includegraphics[width=0.16\textwidth,height=0.16\textwidth,clip]{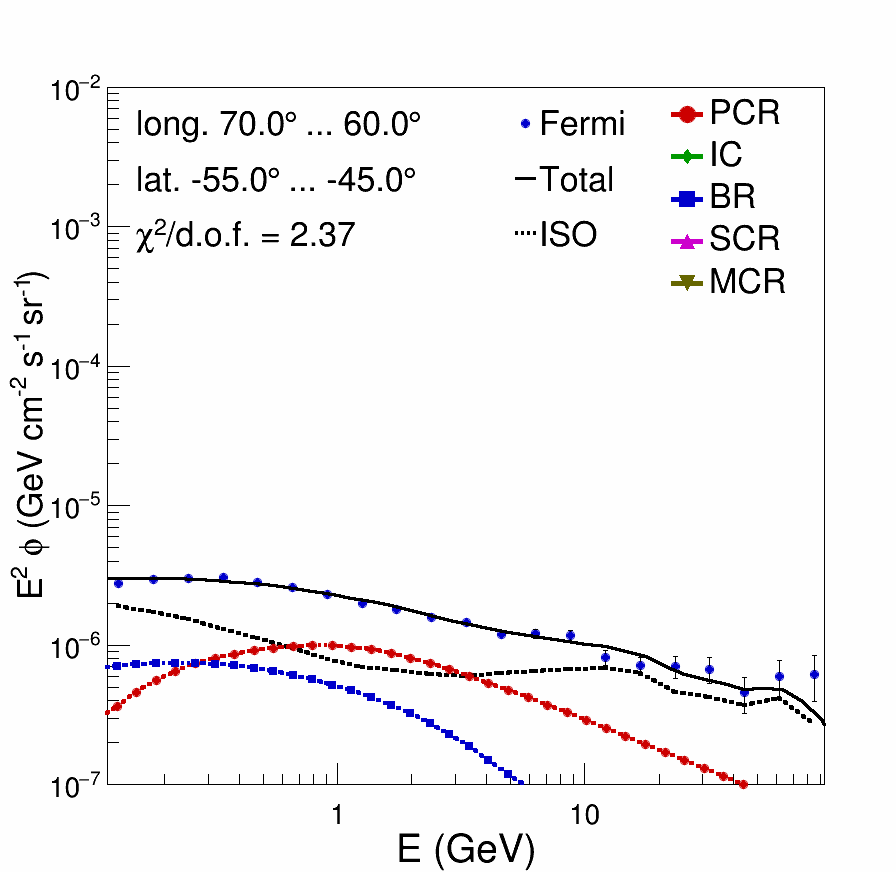}
\includegraphics[width=0.16\textwidth,height=0.16\textwidth,clip]{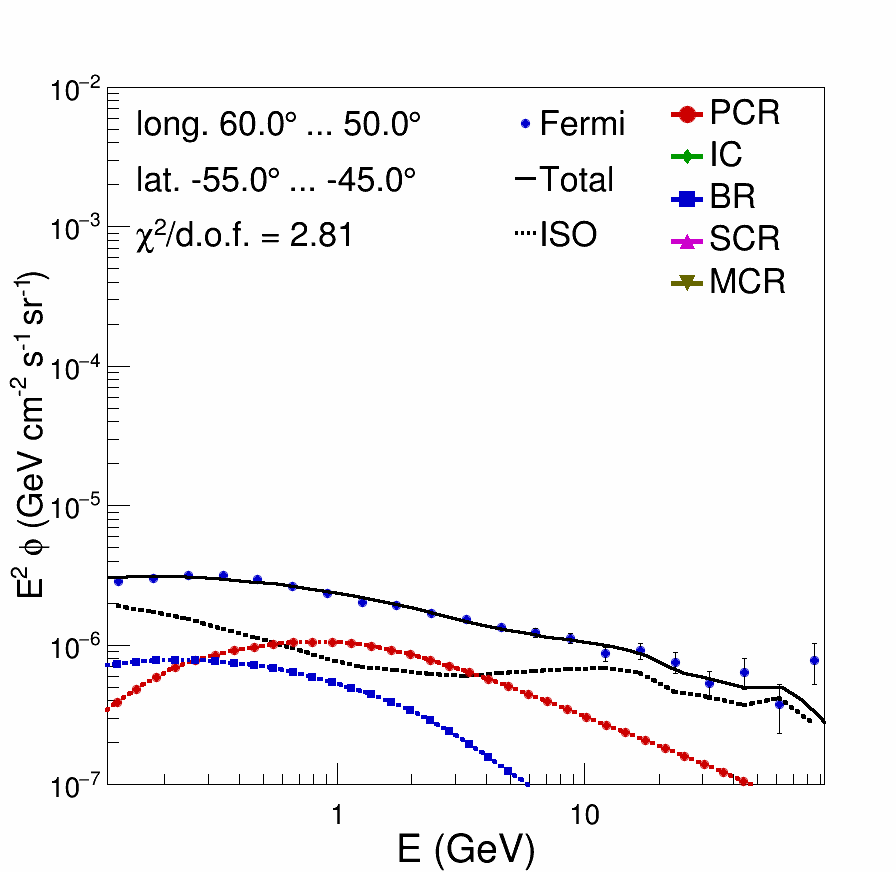}
\includegraphics[width=0.16\textwidth,height=0.16\textwidth,clip]{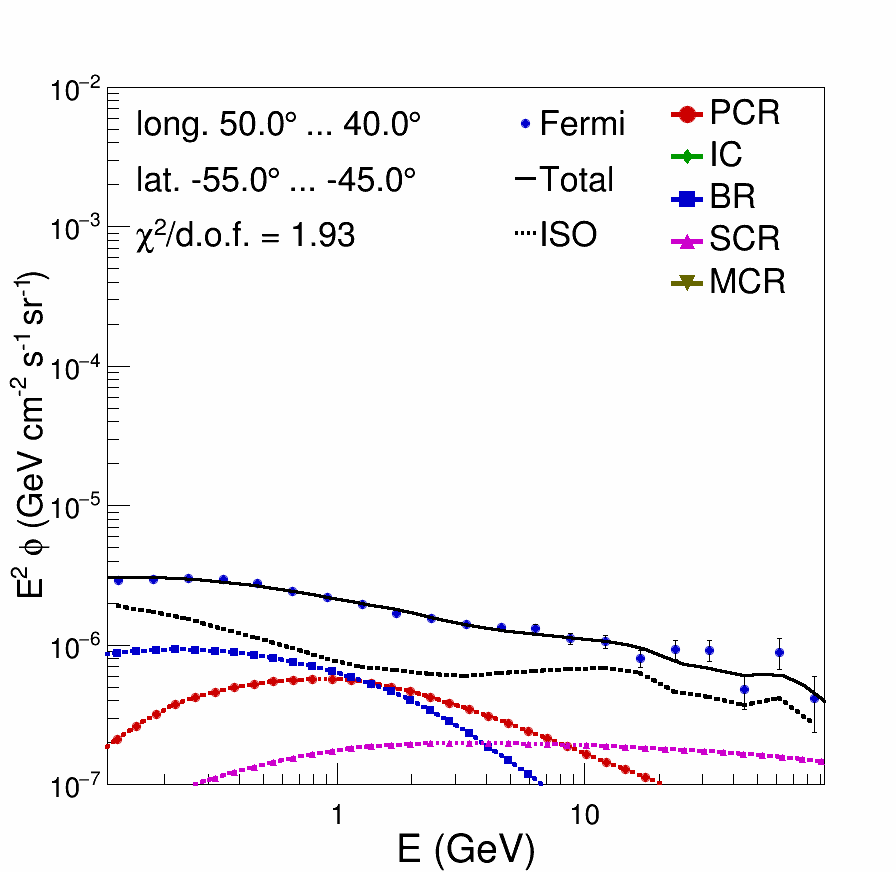}
\includegraphics[width=0.16\textwidth,height=0.16\textwidth,clip]{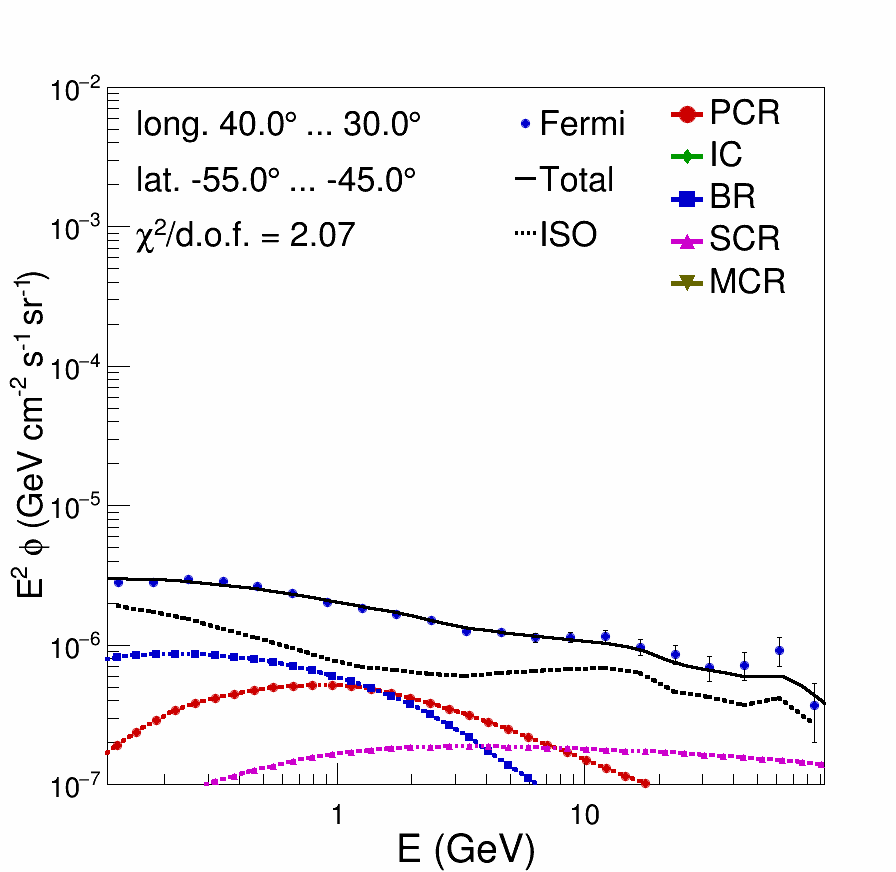}
\includegraphics[width=0.16\textwidth,height=0.16\textwidth,clip]{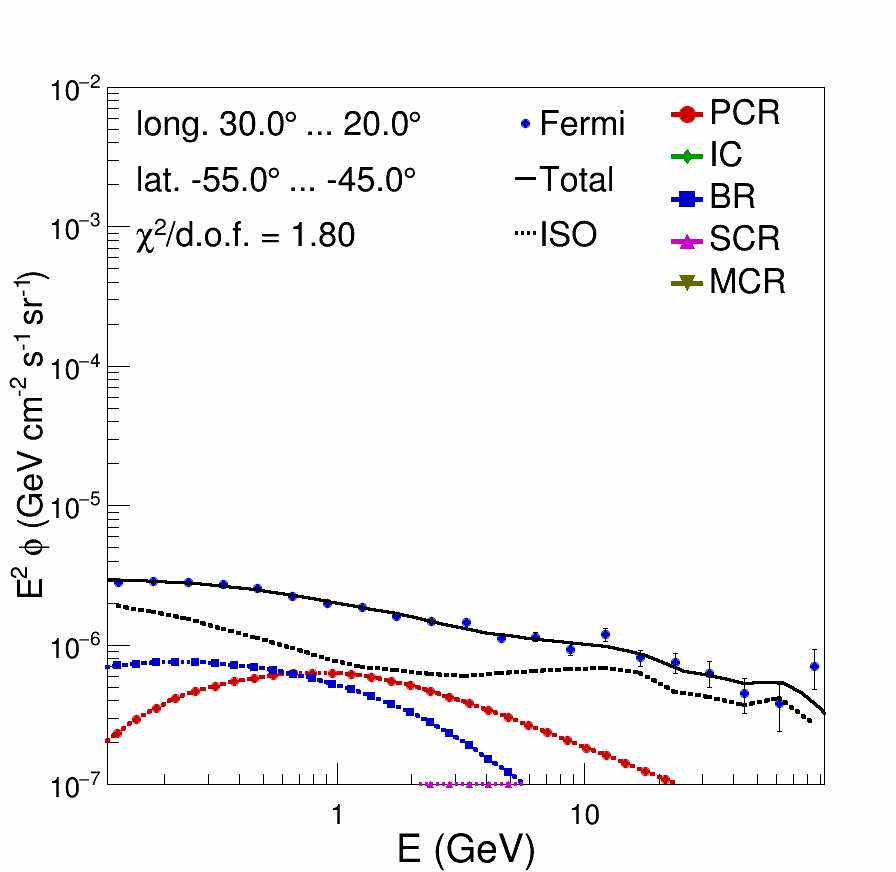}
\includegraphics[width=0.16\textwidth,height=0.16\textwidth,clip]{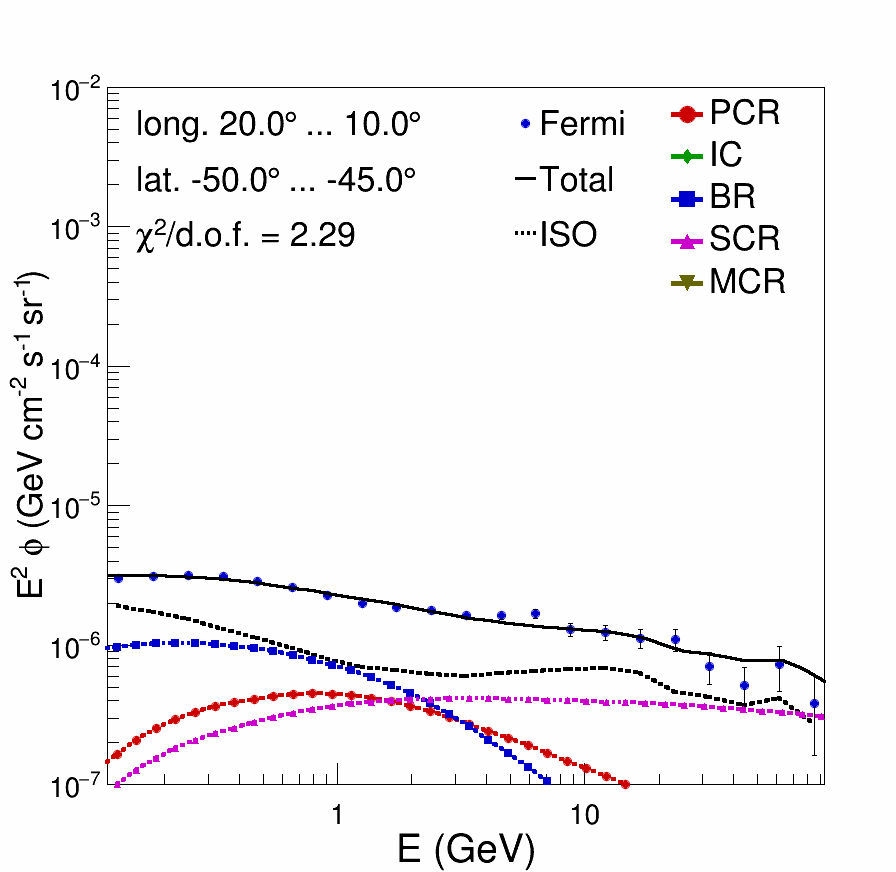}
\includegraphics[width=0.16\textwidth,height=0.16\textwidth,clip]{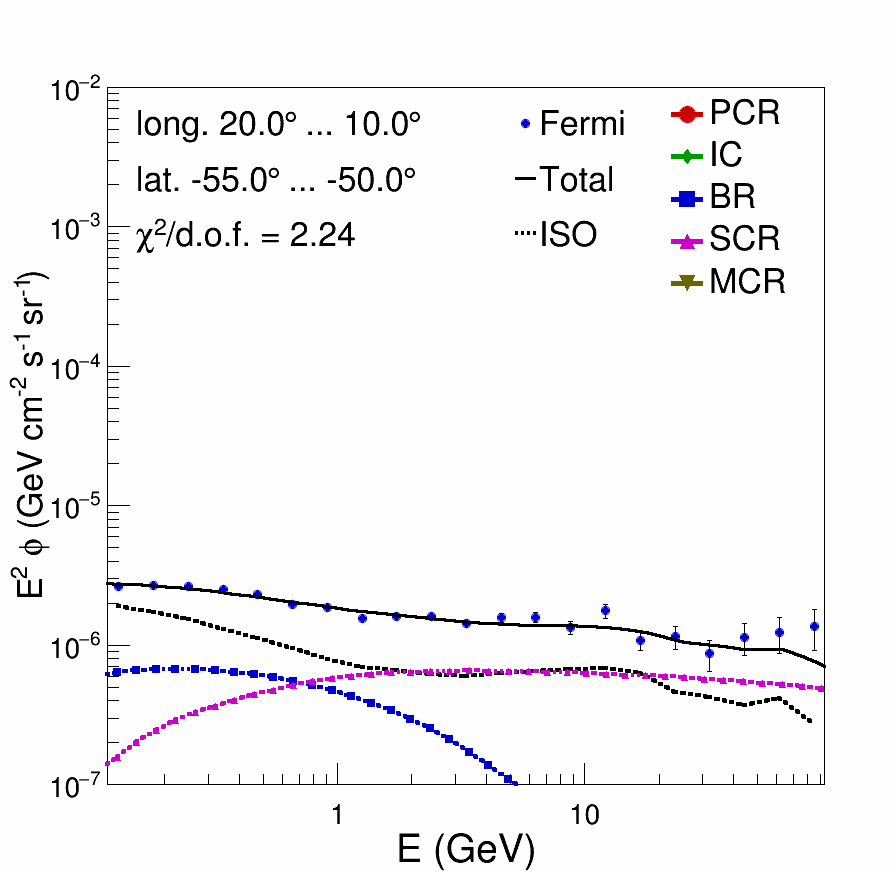}
\includegraphics[width=0.16\textwidth,height=0.16\textwidth,clip]{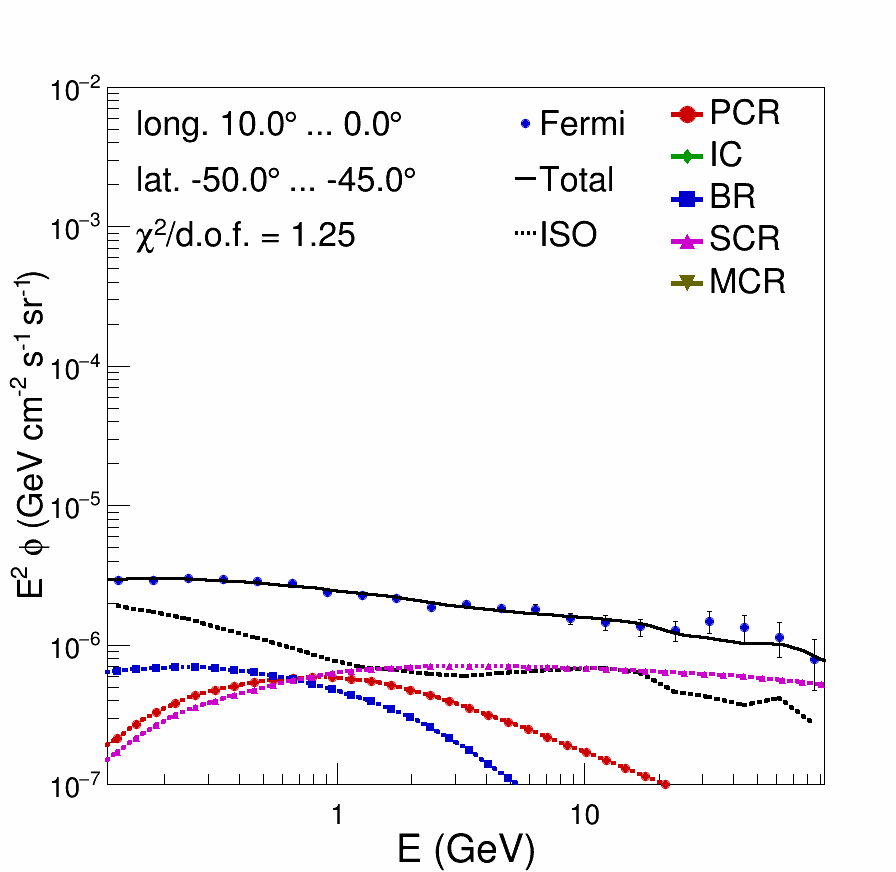}
\includegraphics[width=0.16\textwidth,height=0.16\textwidth,clip]{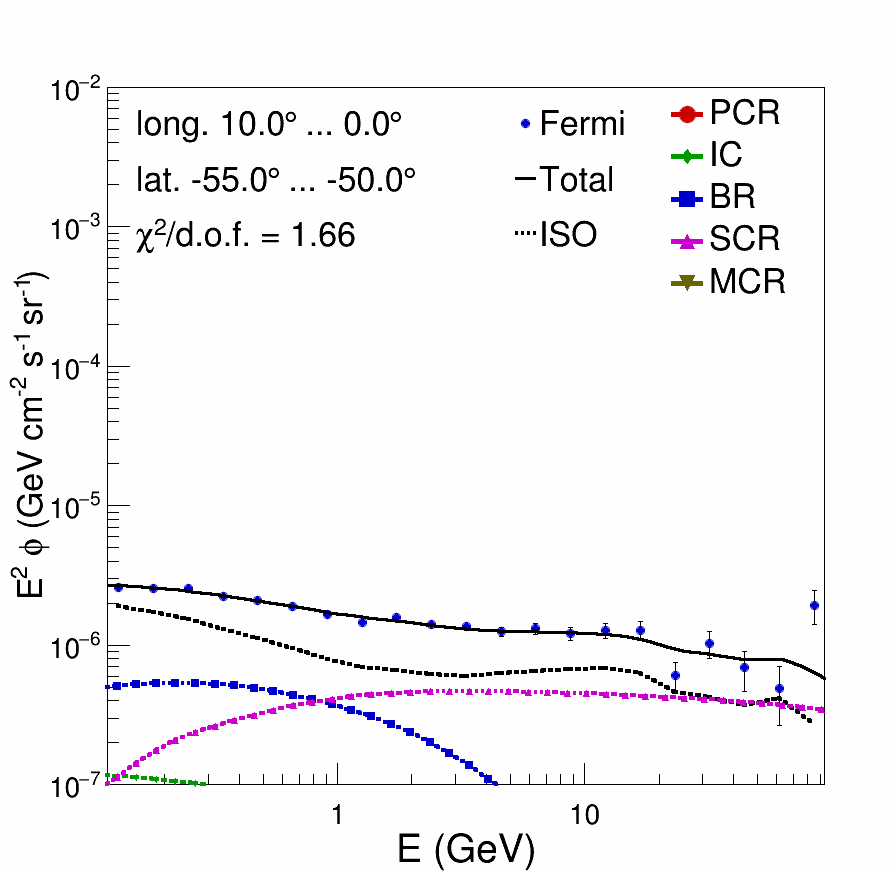}
\includegraphics[width=0.16\textwidth,height=0.16\textwidth,clip]{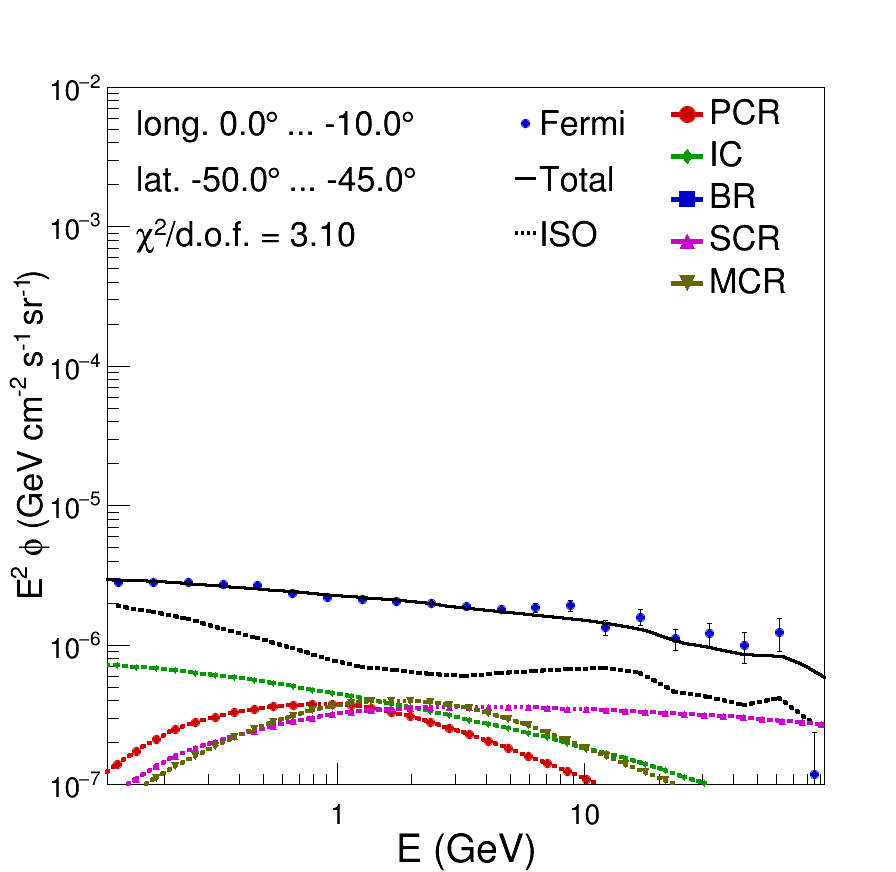}
\includegraphics[width=0.16\textwidth,height=0.16\textwidth,clip]{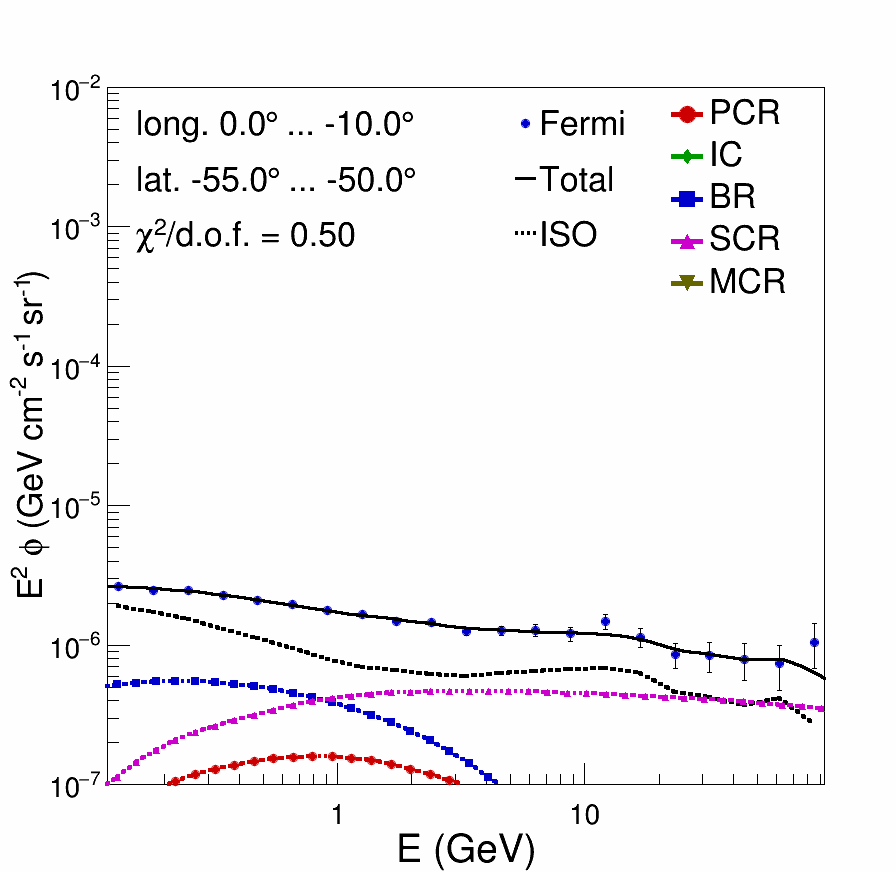}
\includegraphics[width=0.16\textwidth,height=0.16\textwidth,clip]{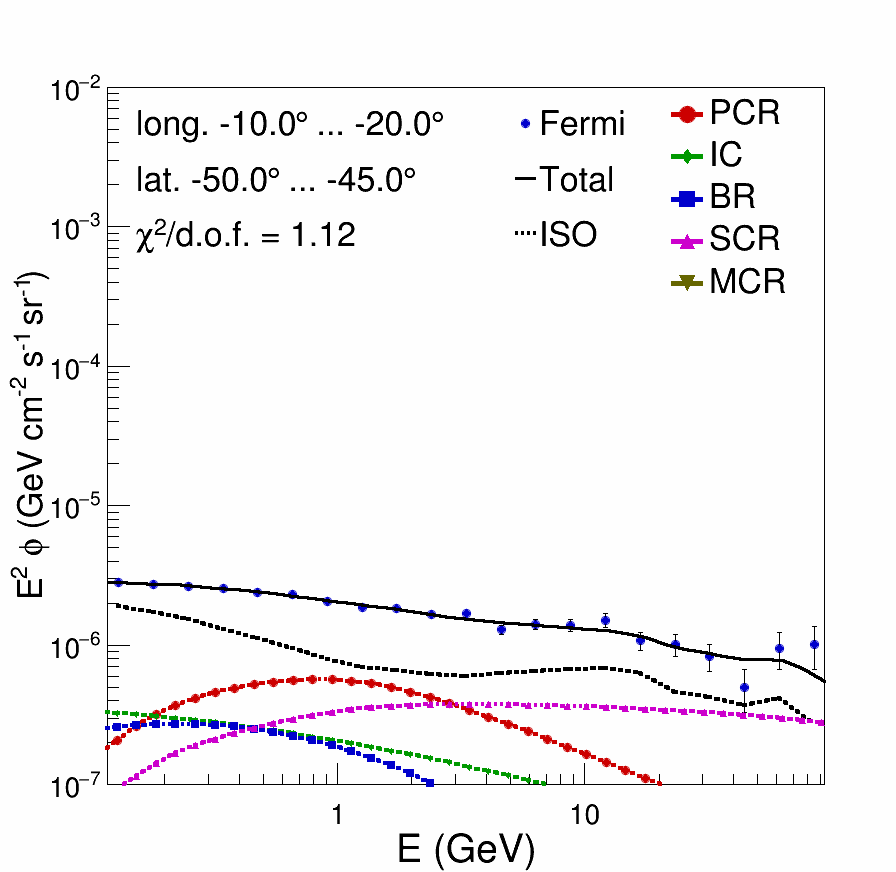}
\includegraphics[width=0.16\textwidth,height=0.16\textwidth,clip]{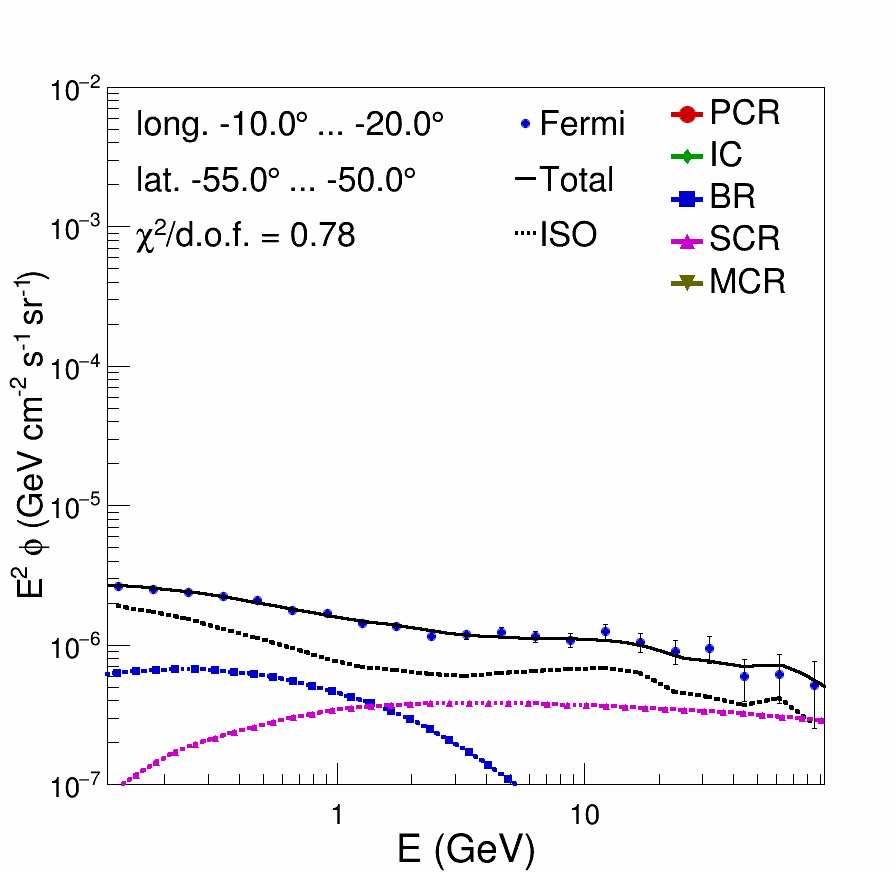}
\includegraphics[width=0.16\textwidth,height=0.16\textwidth,clip]{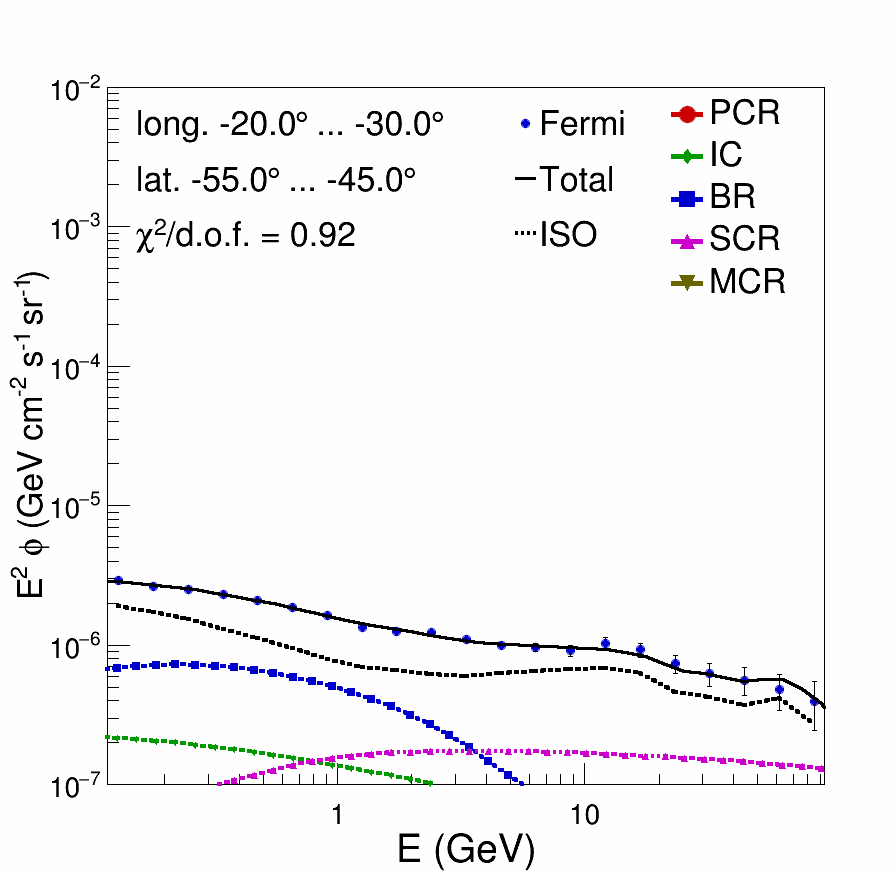}
\includegraphics[width=0.16\textwidth,height=0.16\textwidth,clip]{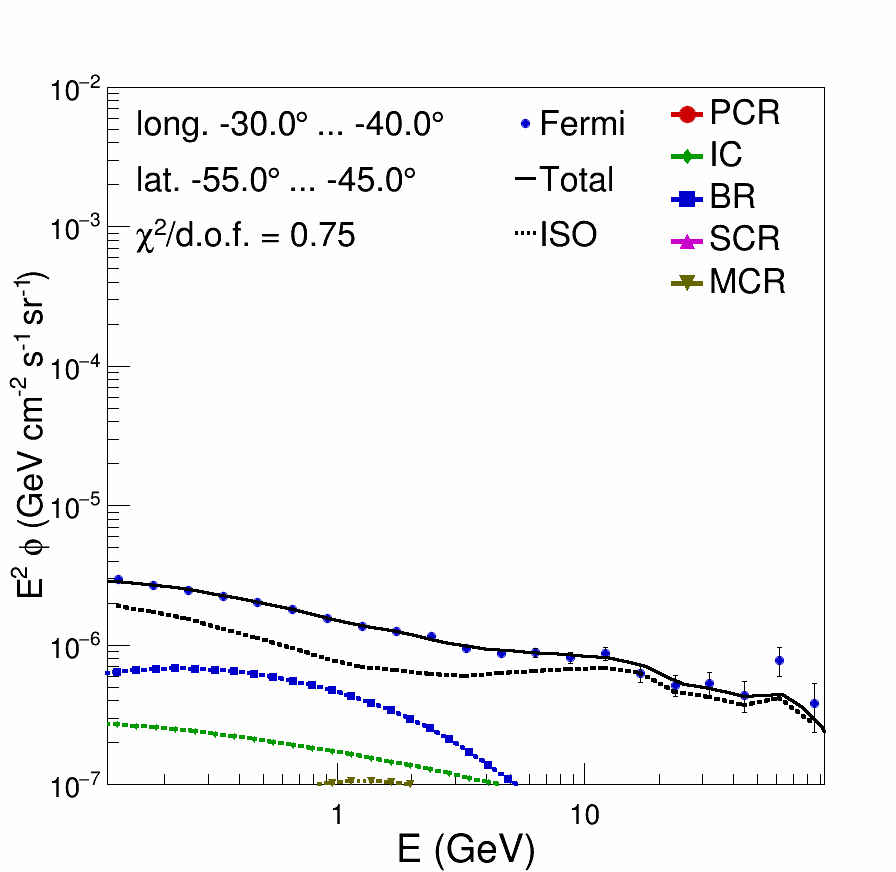}
\includegraphics[width=0.16\textwidth,height=0.16\textwidth,clip]{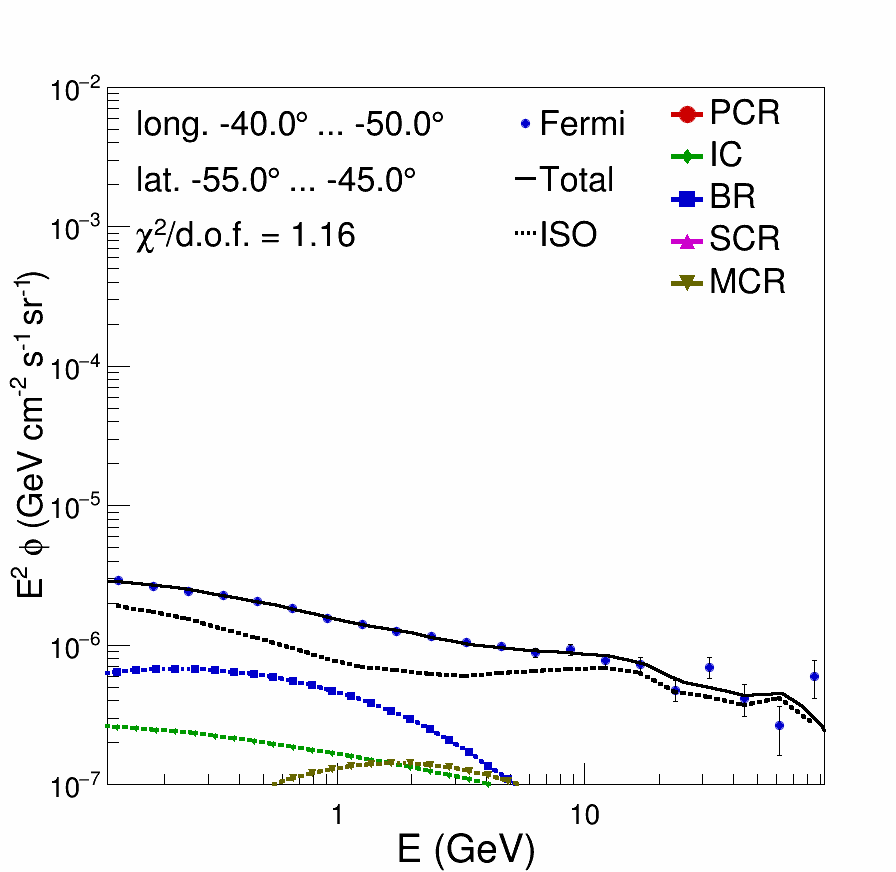}
\includegraphics[width=0.16\textwidth,height=0.16\textwidth,clip]{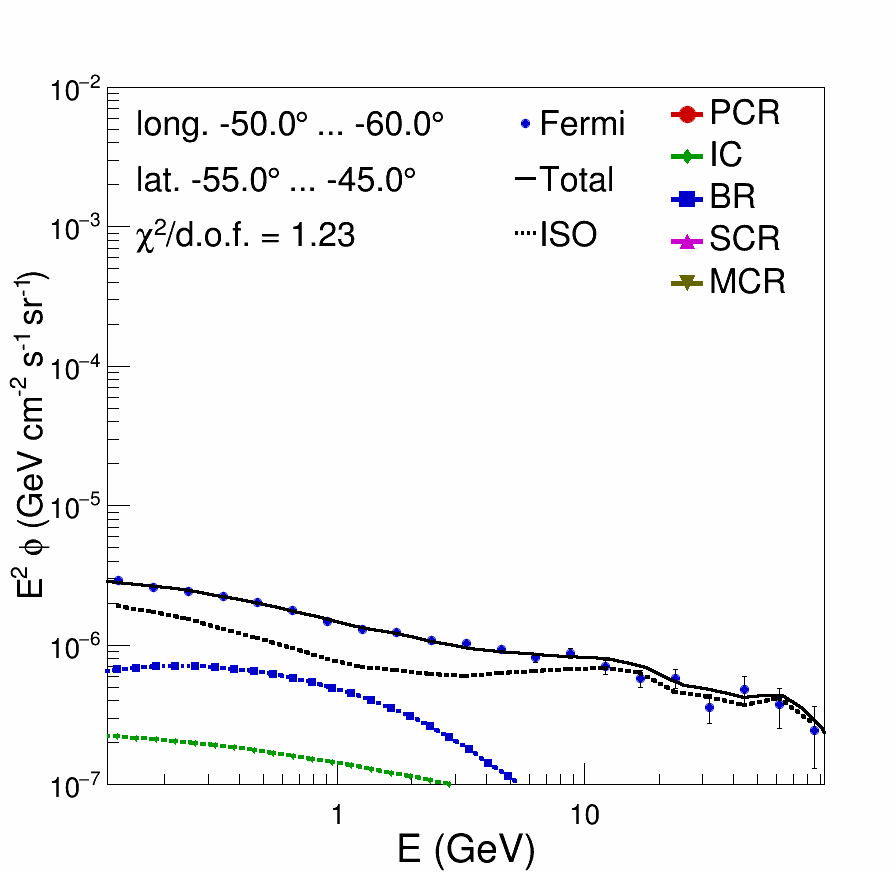}
\includegraphics[width=0.16\textwidth,height=0.16\textwidth,clip]{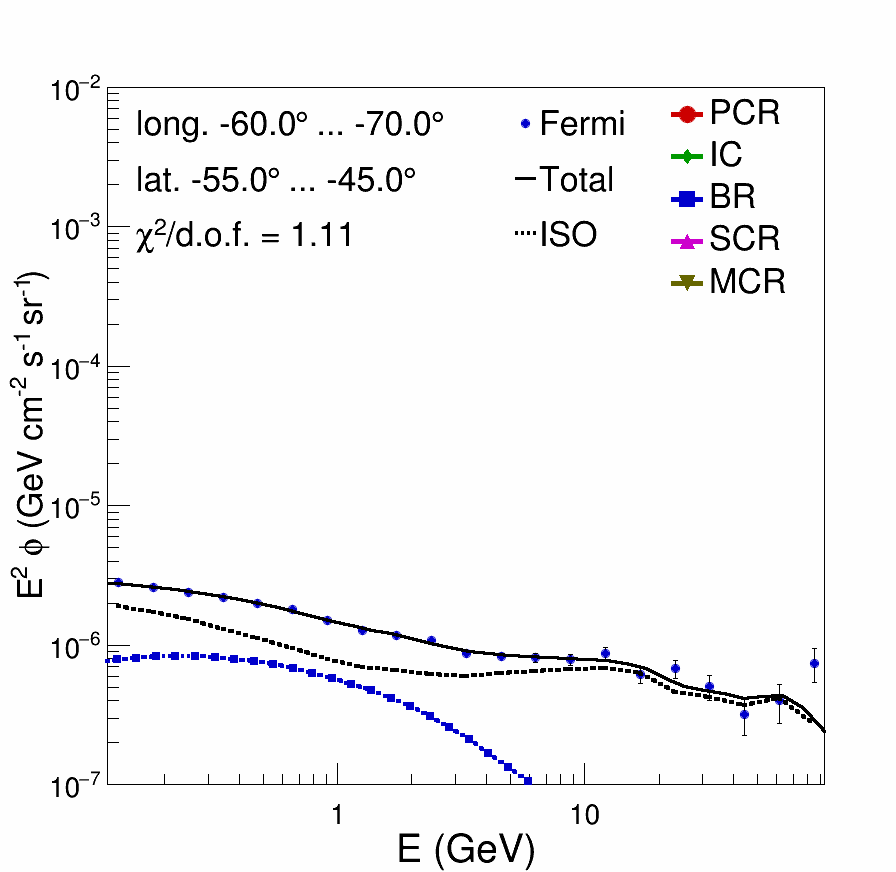}
\includegraphics[width=0.16\textwidth,height=0.16\textwidth,clip]{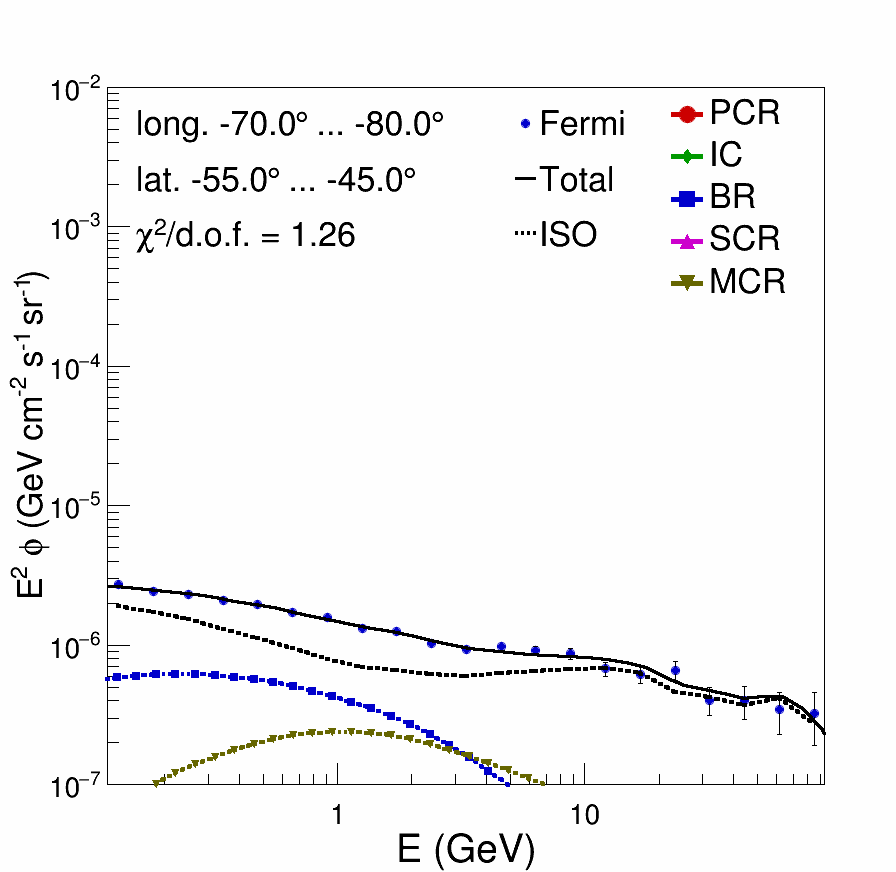}
\includegraphics[width=0.16\textwidth,height=0.16\textwidth,clip]{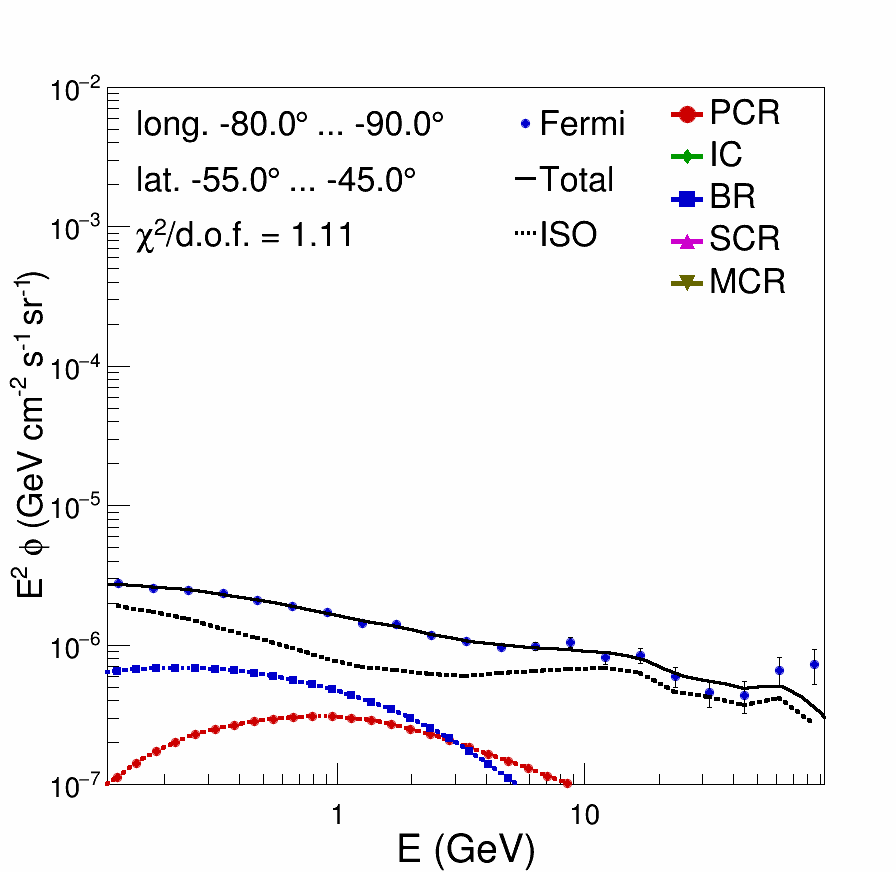}
\includegraphics[width=0.16\textwidth,height=0.16\textwidth,clip]{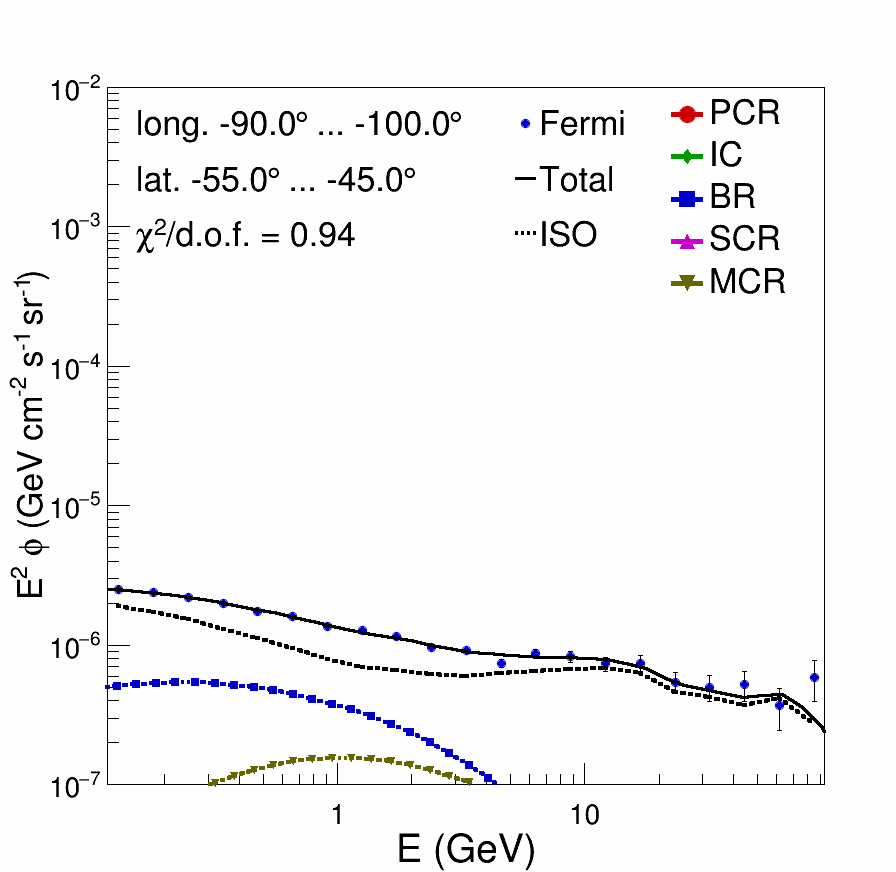}
\includegraphics[width=0.16\textwidth,height=0.16\textwidth,clip]{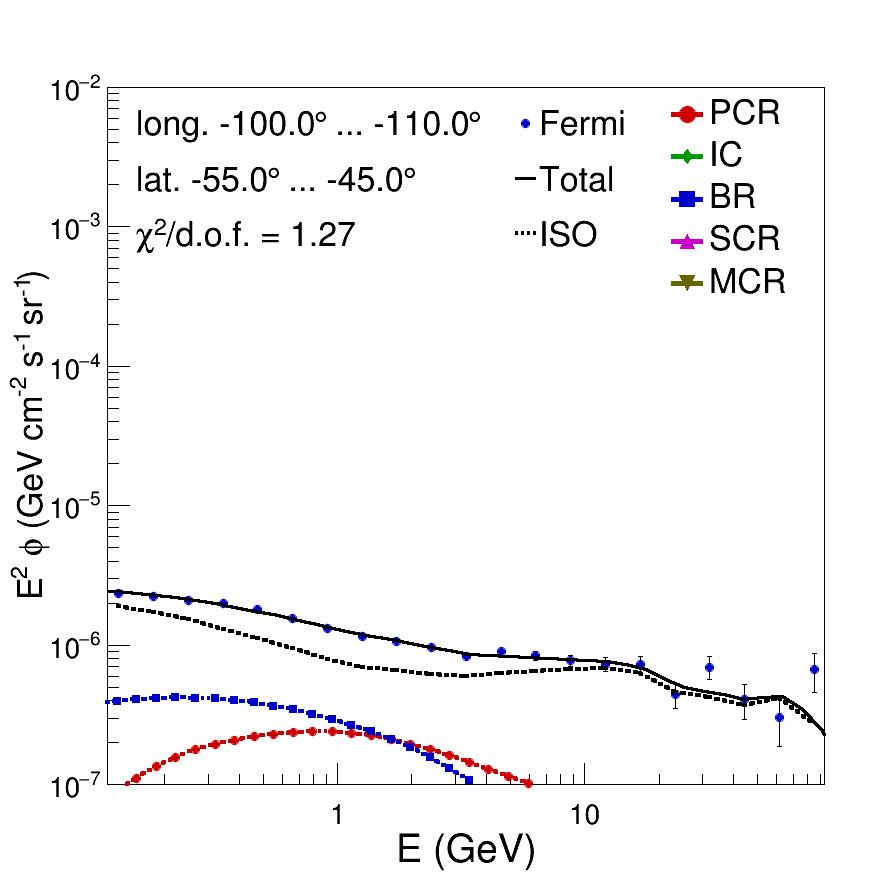}
\includegraphics[width=0.16\textwidth,height=0.16\textwidth,clip]{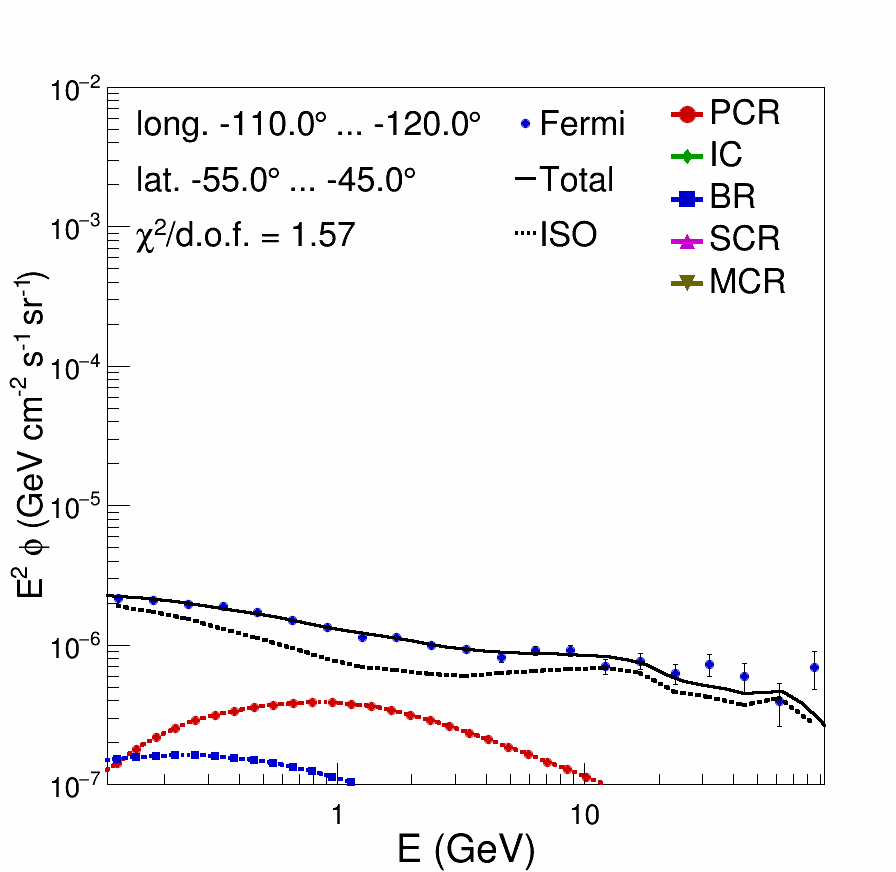}
\includegraphics[width=0.16\textwidth,height=0.16\textwidth,clip]{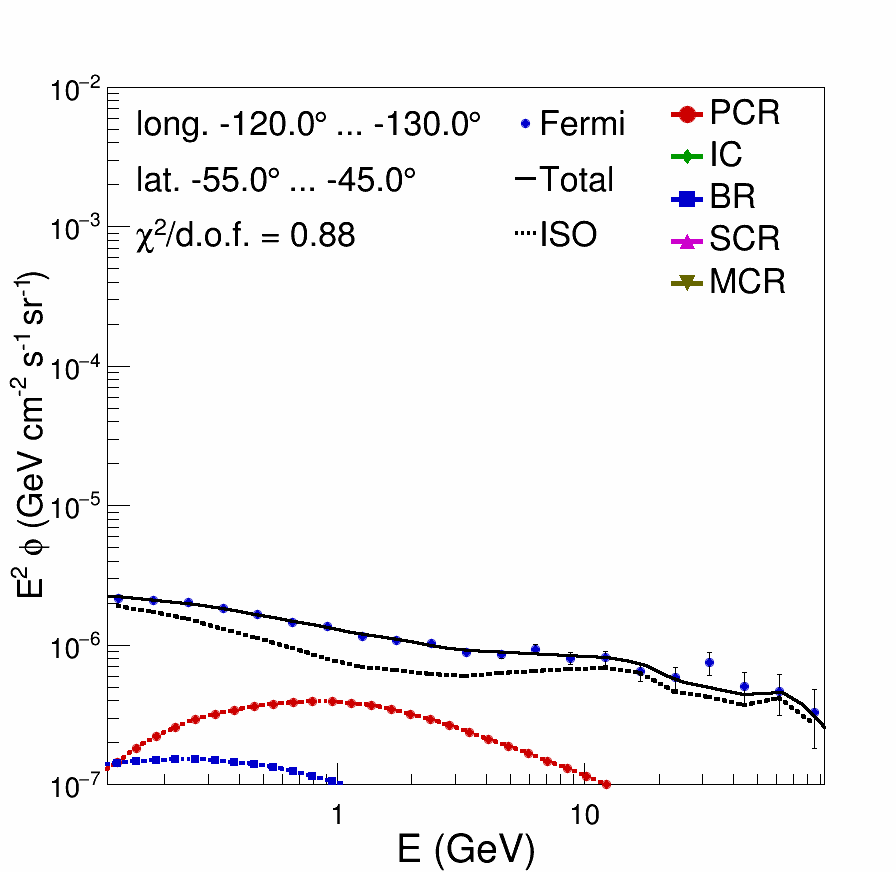}
\includegraphics[width=0.16\textwidth,height=0.16\textwidth,clip]{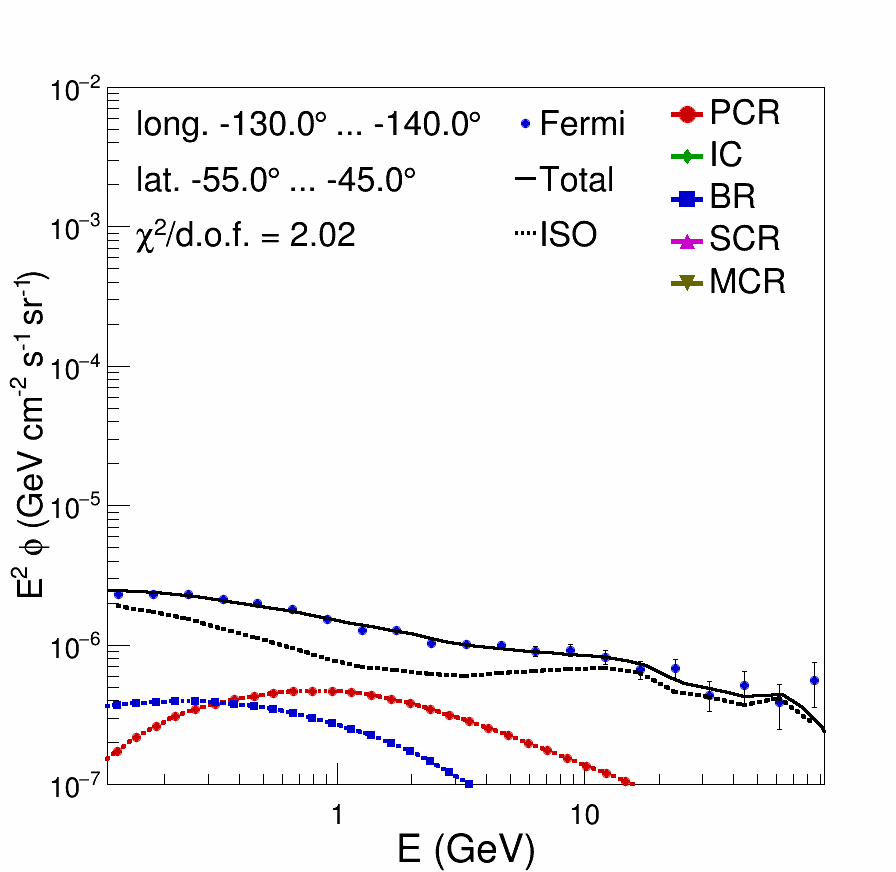}
\includegraphics[width=0.16\textwidth,height=0.16\textwidth,clip]{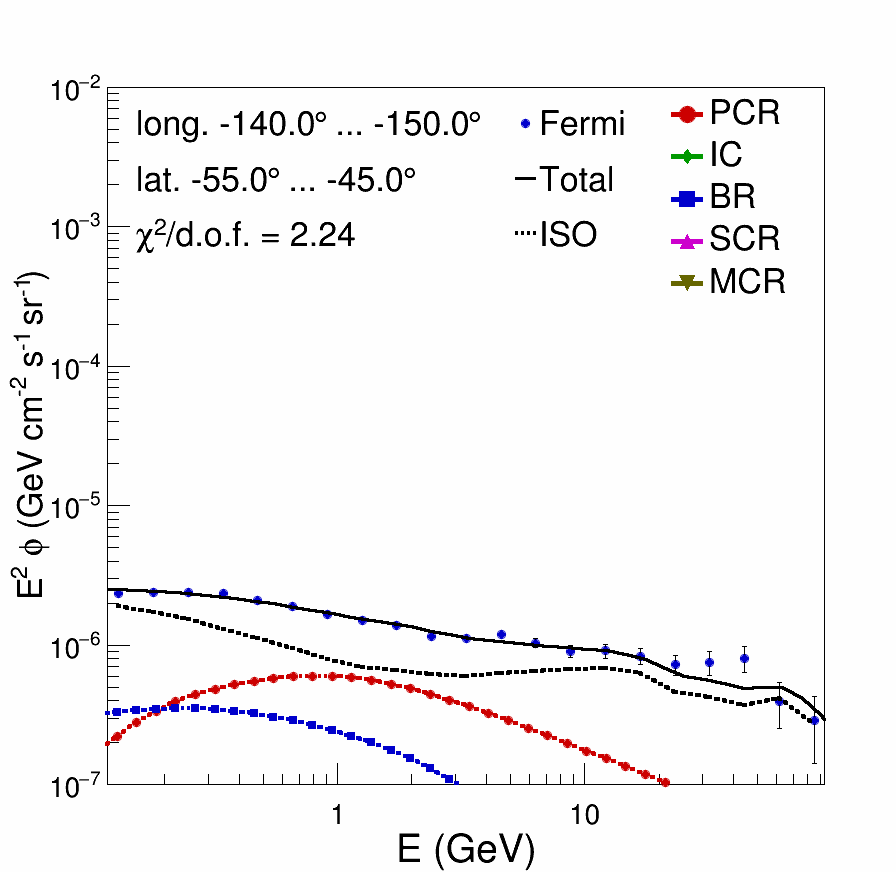}
\includegraphics[width=0.16\textwidth,height=0.16\textwidth,clip]{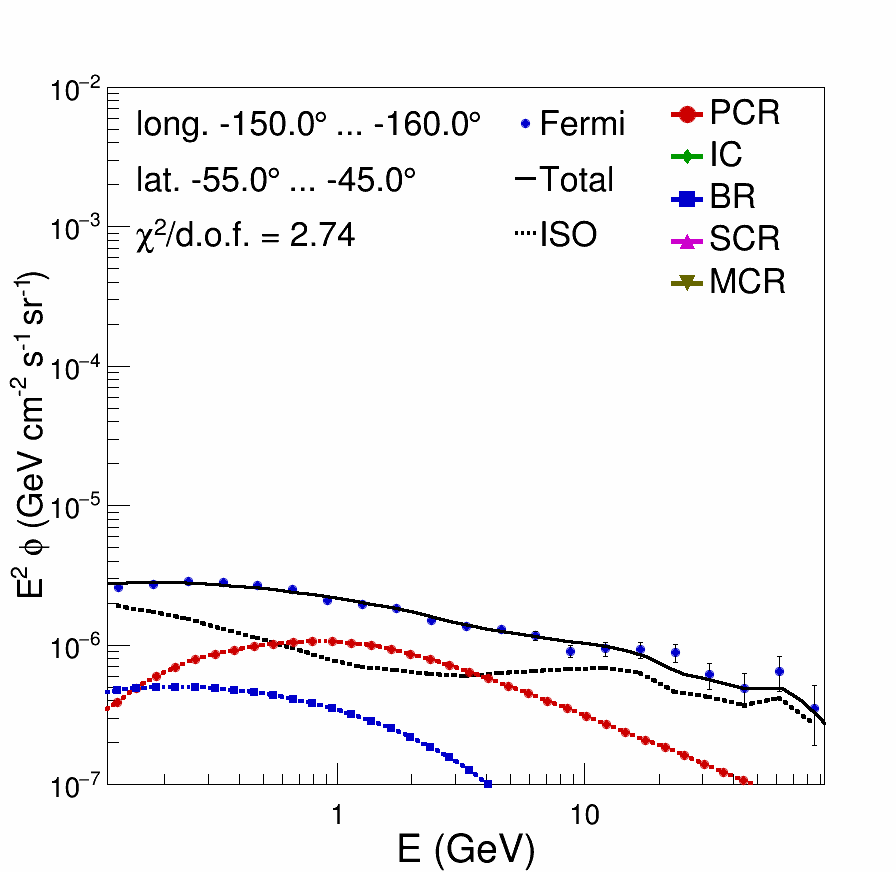}
\includegraphics[width=0.16\textwidth,height=0.16\textwidth,clip]{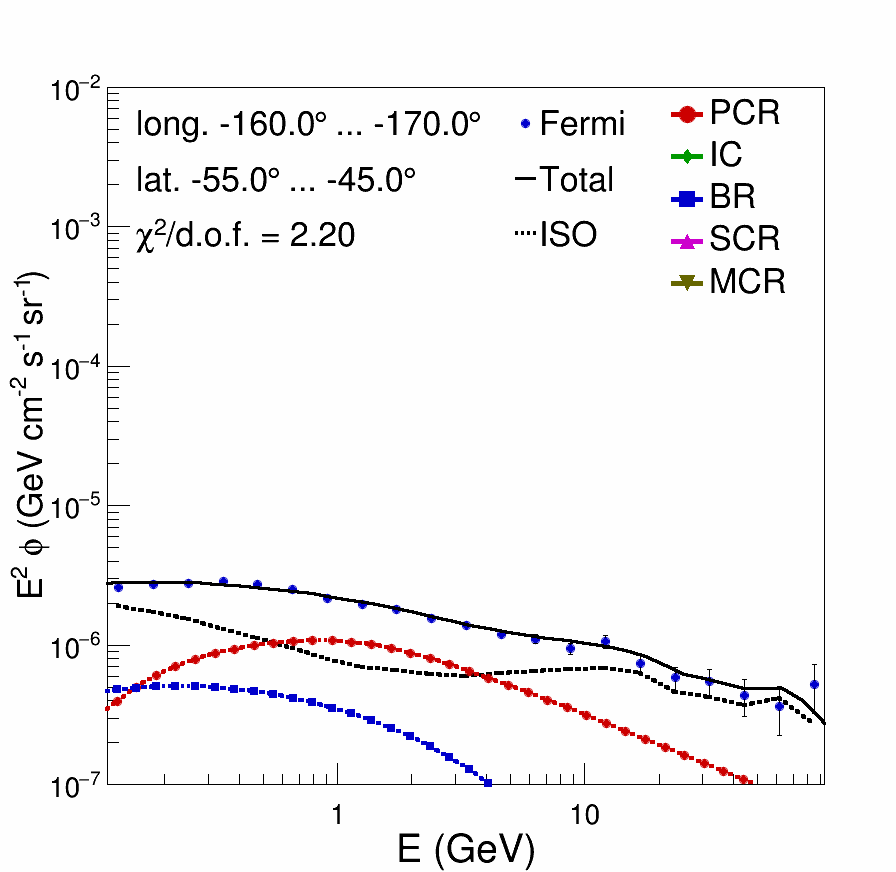}
\includegraphics[width=0.16\textwidth,height=0.16\textwidth,clip]{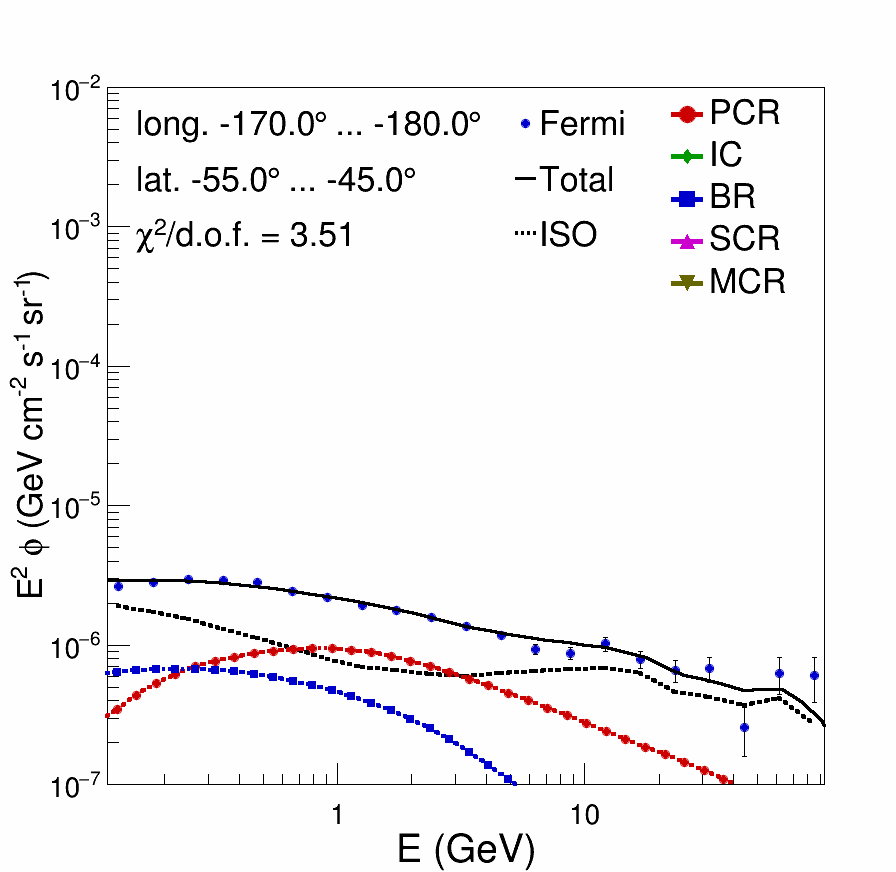}
\caption[]{Template fits for latitudes  with $-55.0^\circ<b<-45.0^\circ$ and longitudes decreasing from 180$^\circ$ to -180$^\circ$.} \label{F29}
\end{figure}
\begin{figure}
\centering
\includegraphics[width=0.16\textwidth,height=0.16\textwidth,clip]{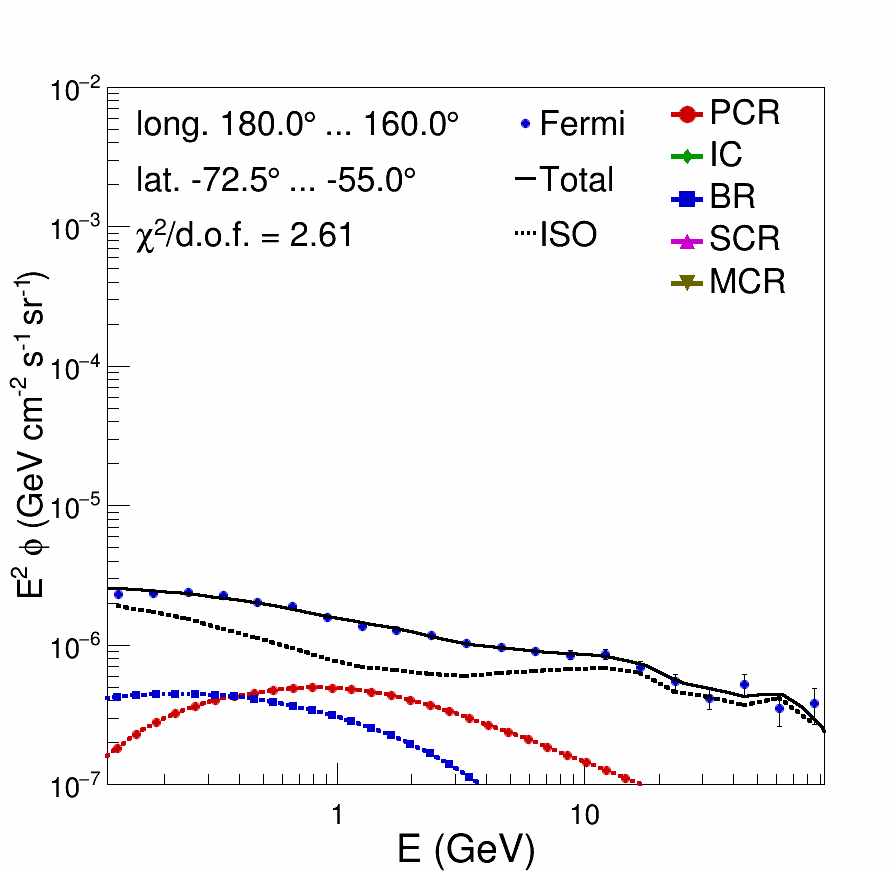}
\includegraphics[width=0.16\textwidth,height=0.16\textwidth,clip]{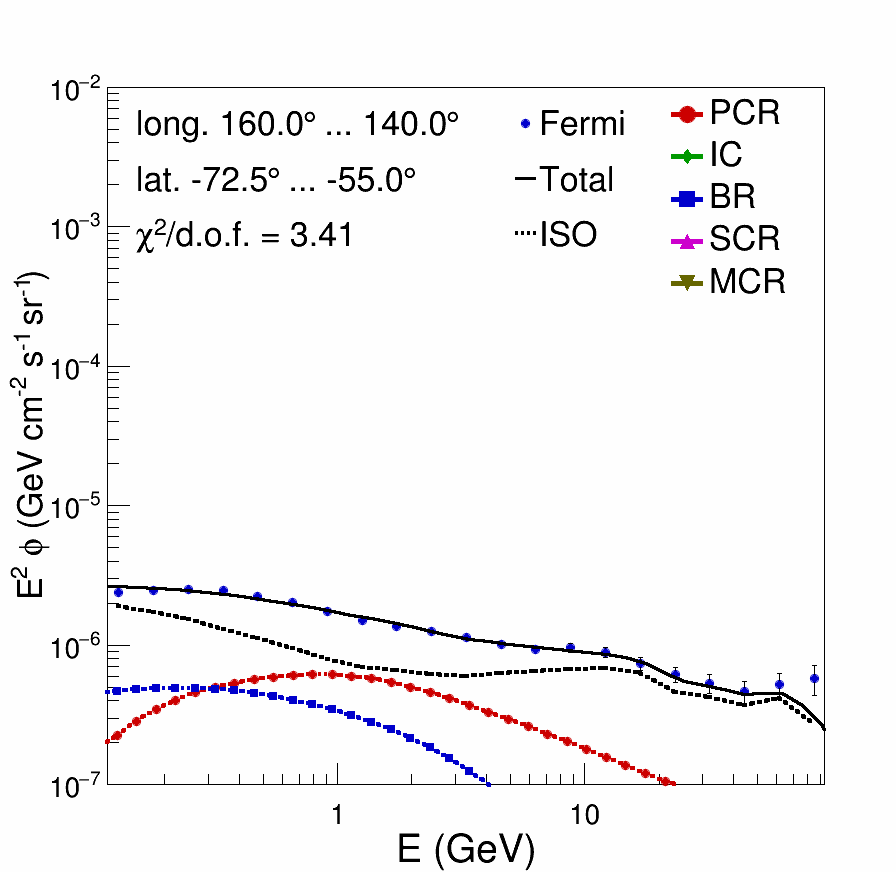}
\includegraphics[width=0.16\textwidth,height=0.16\textwidth,clip]{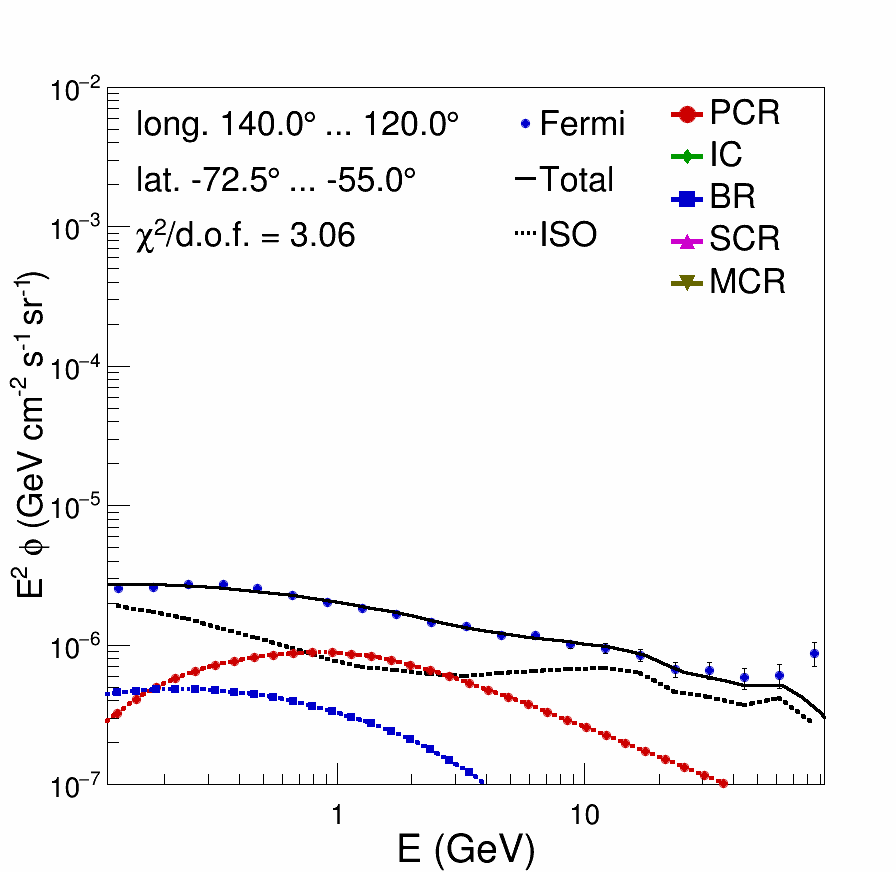}
\includegraphics[width=0.16\textwidth,height=0.16\textwidth,clip]{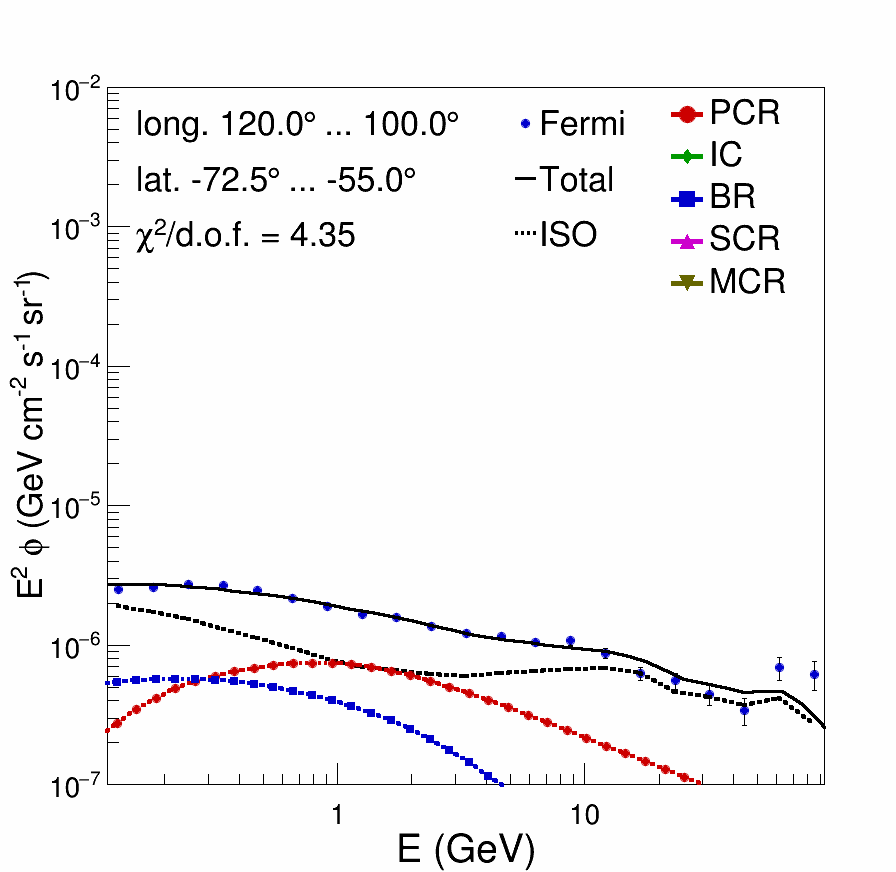}
\includegraphics[width=0.16\textwidth,height=0.16\textwidth,clip]{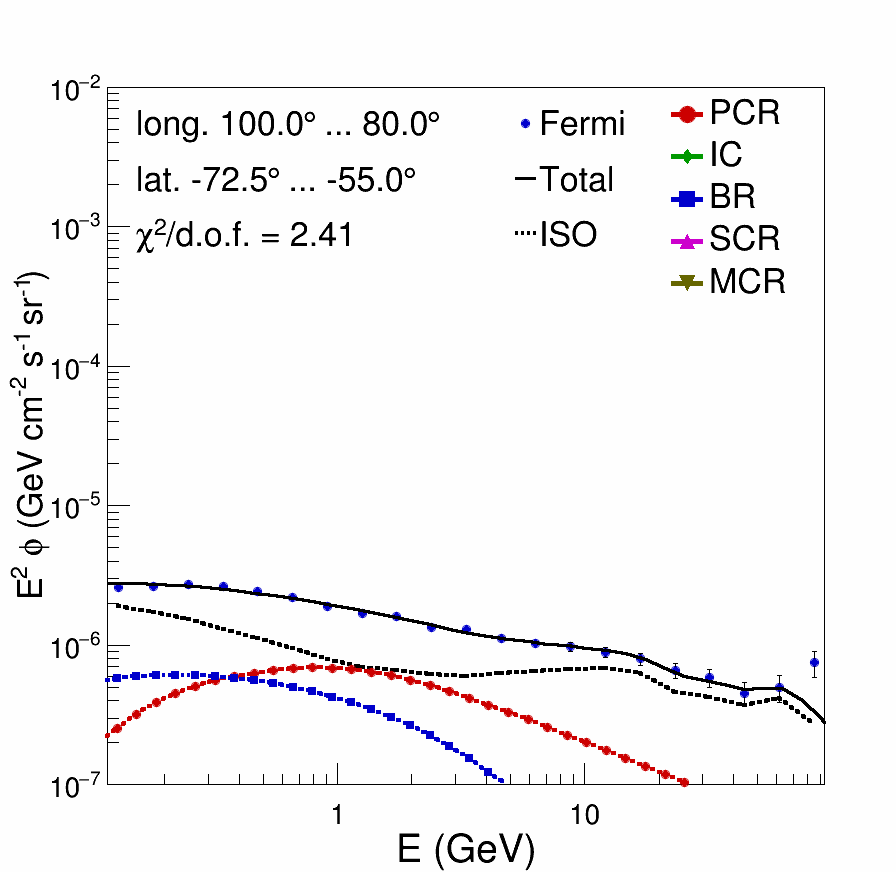}
\includegraphics[width=0.16\textwidth,height=0.16\textwidth,clip]{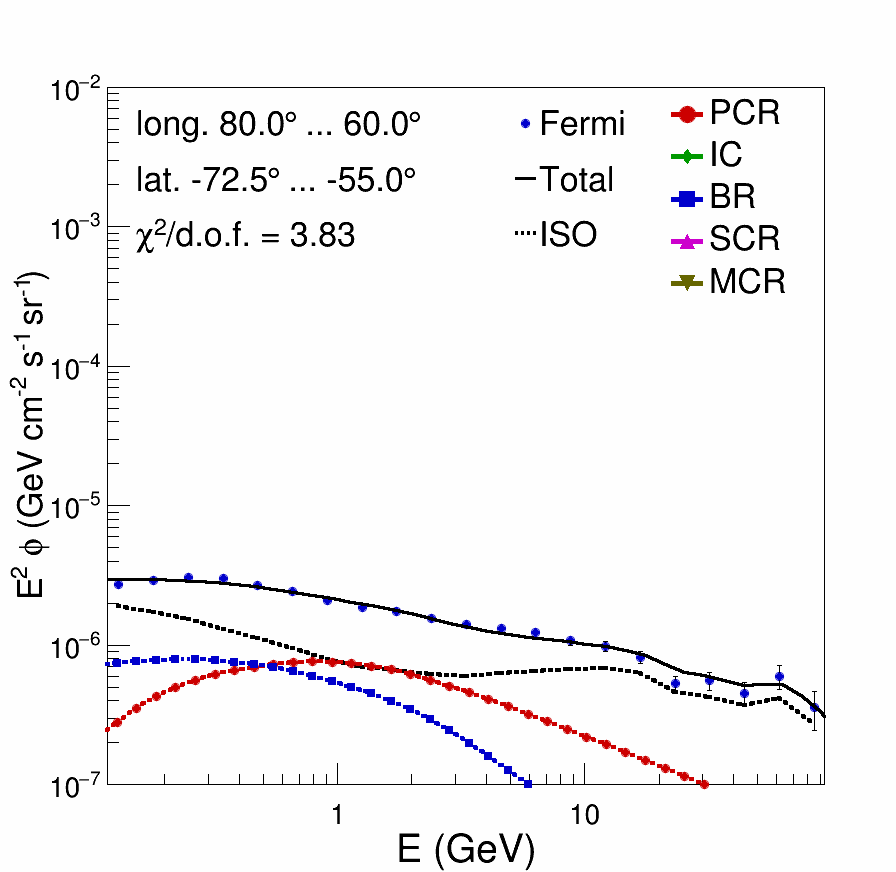}
\includegraphics[width=0.16\textwidth,height=0.16\textwidth,clip]{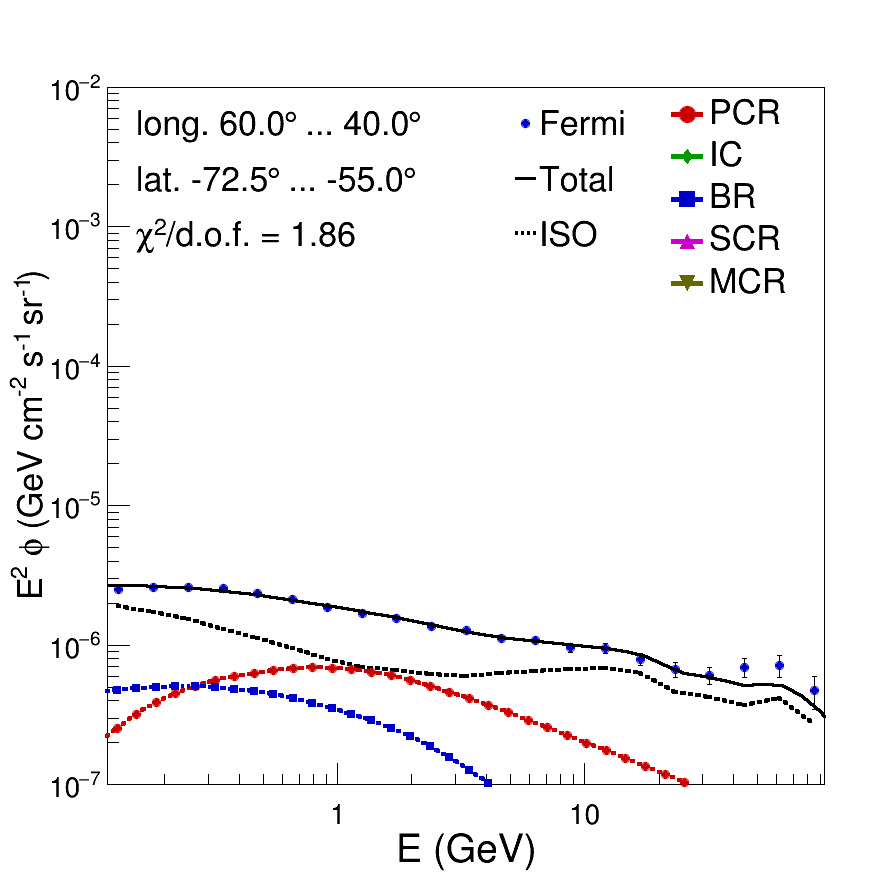}
\includegraphics[width=0.16\textwidth,height=0.16\textwidth,clip]{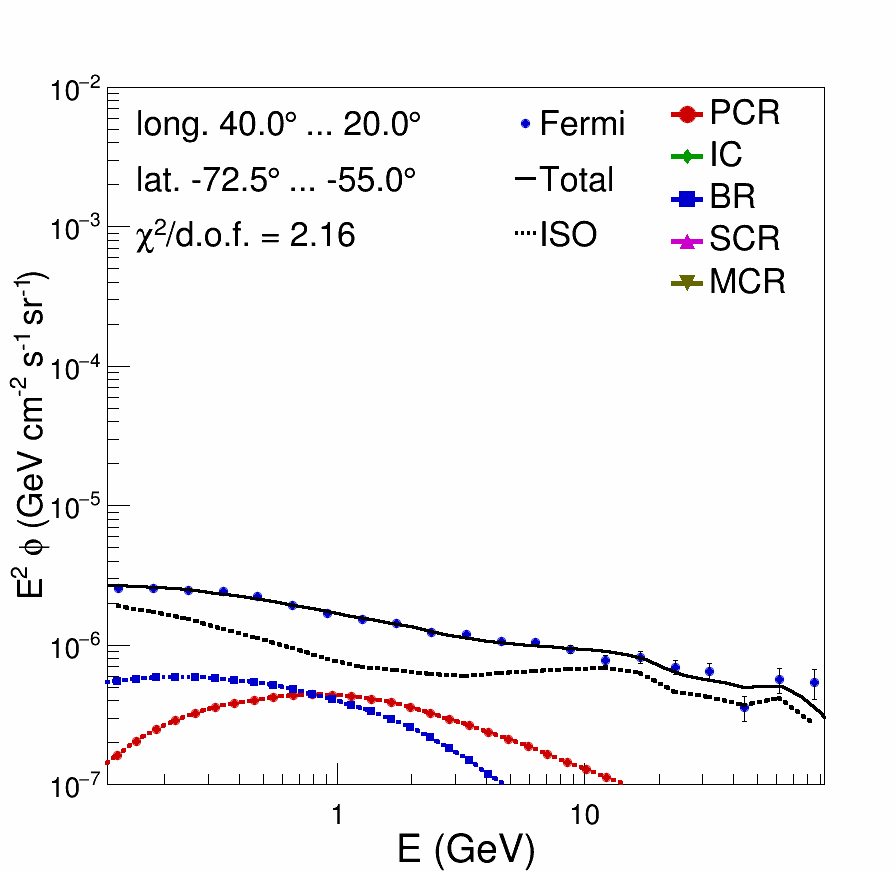}
\includegraphics[width=0.16\textwidth,height=0.16\textwidth,clip]{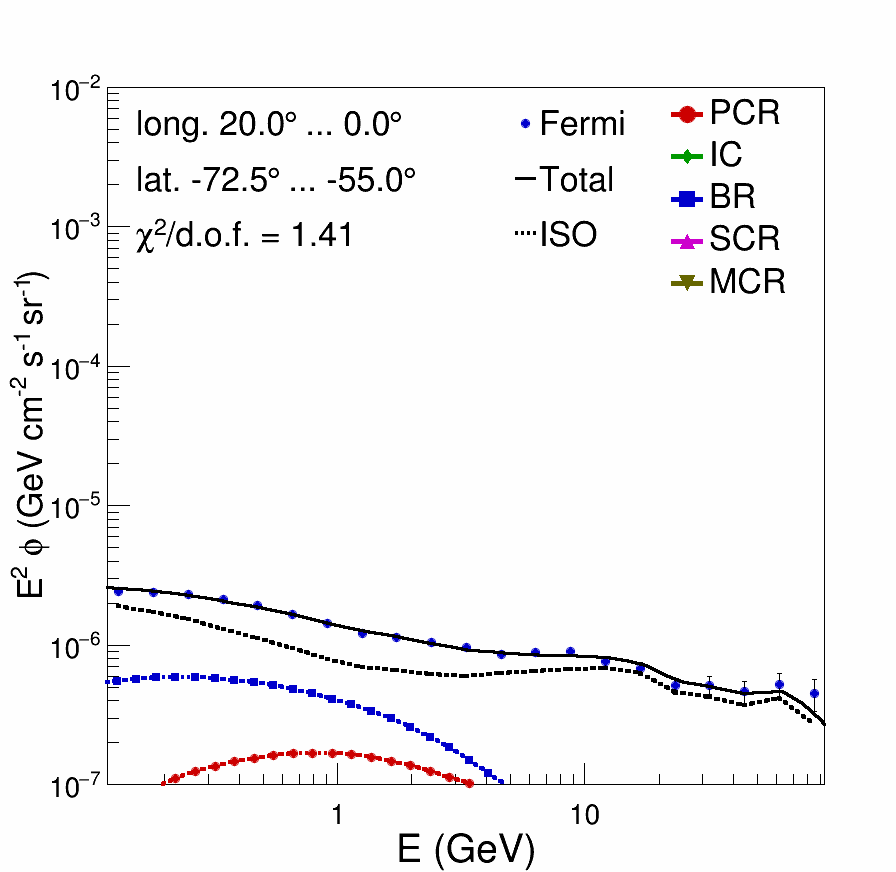}
\includegraphics[width=0.16\textwidth,height=0.16\textwidth,clip]{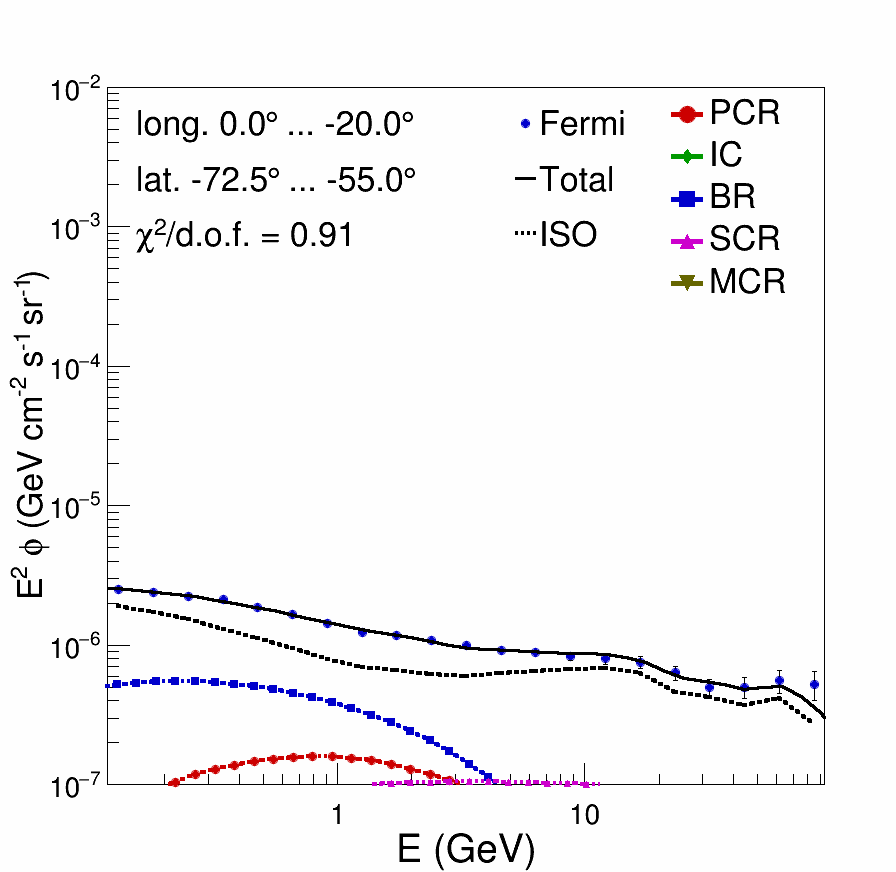}
\includegraphics[width=0.16\textwidth,height=0.16\textwidth,clip]{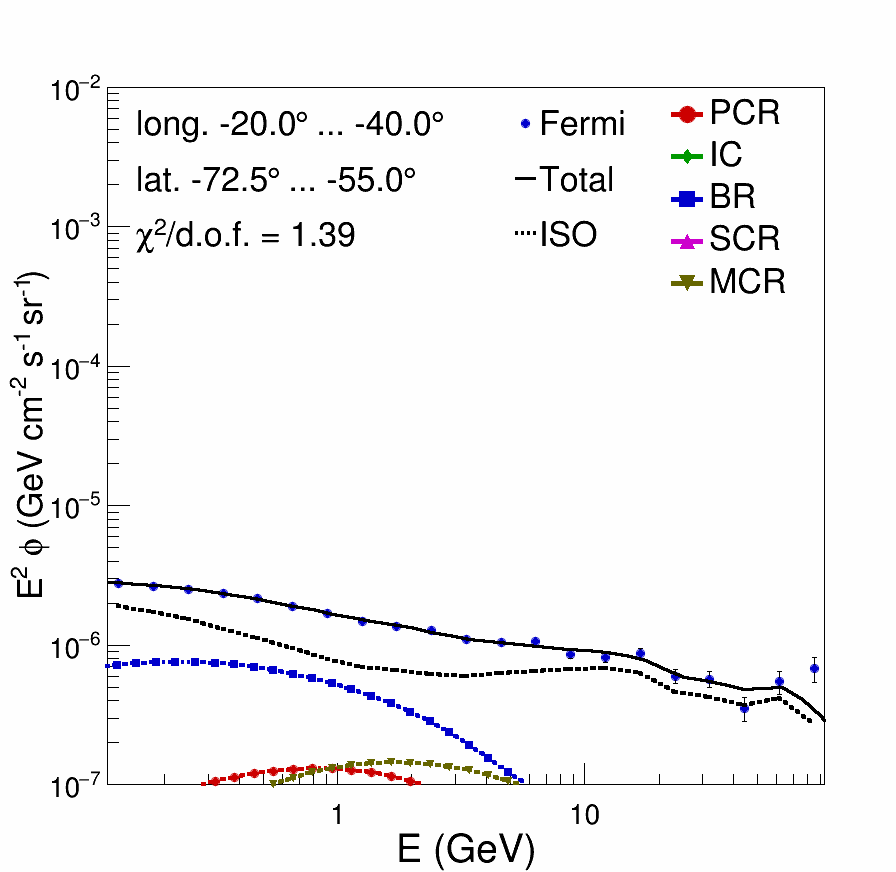}
\includegraphics[width=0.16\textwidth,height=0.16\textwidth,clip]{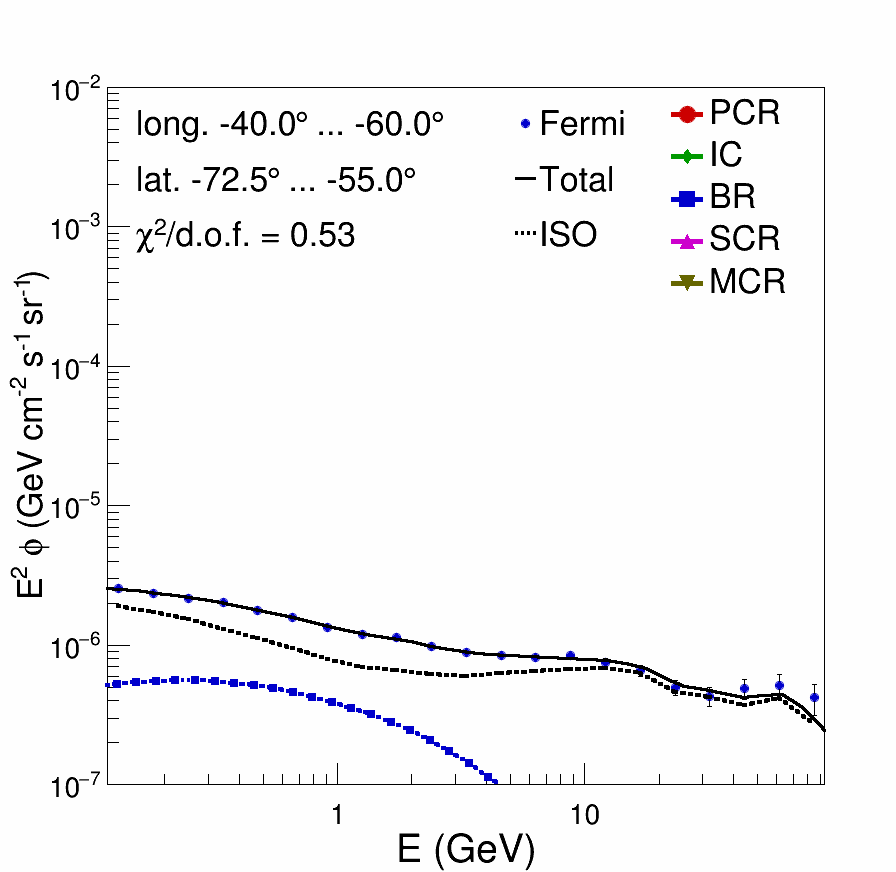}
\includegraphics[width=0.16\textwidth,height=0.16\textwidth,clip]{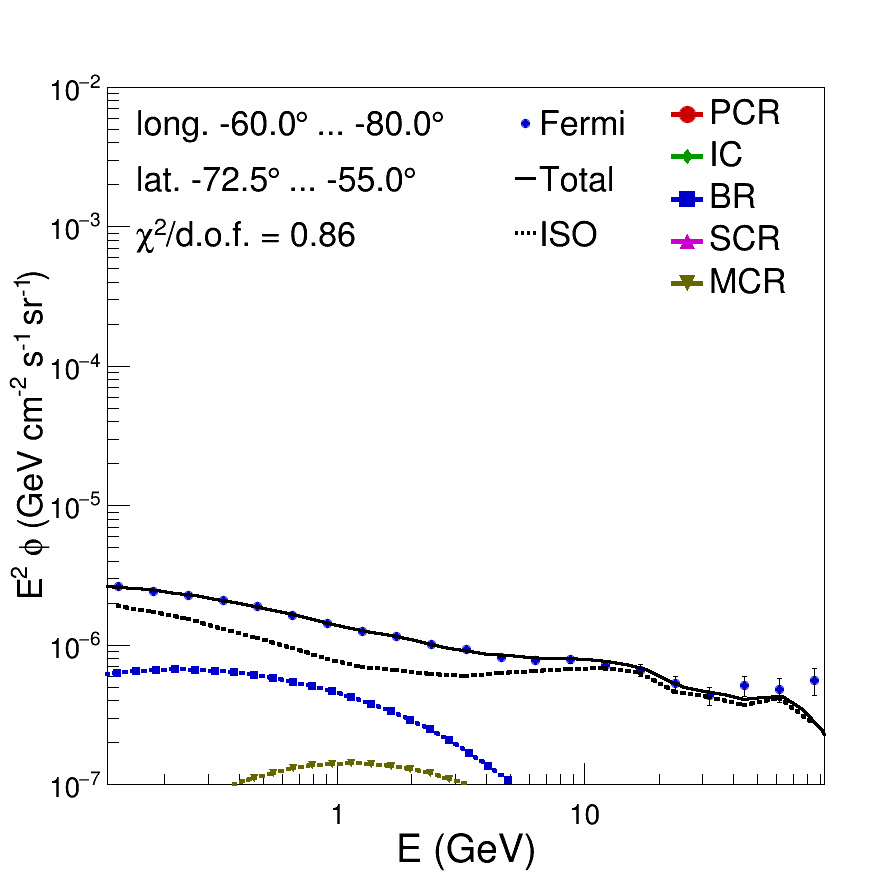}
\includegraphics[width=0.16\textwidth,height=0.16\textwidth,clip]{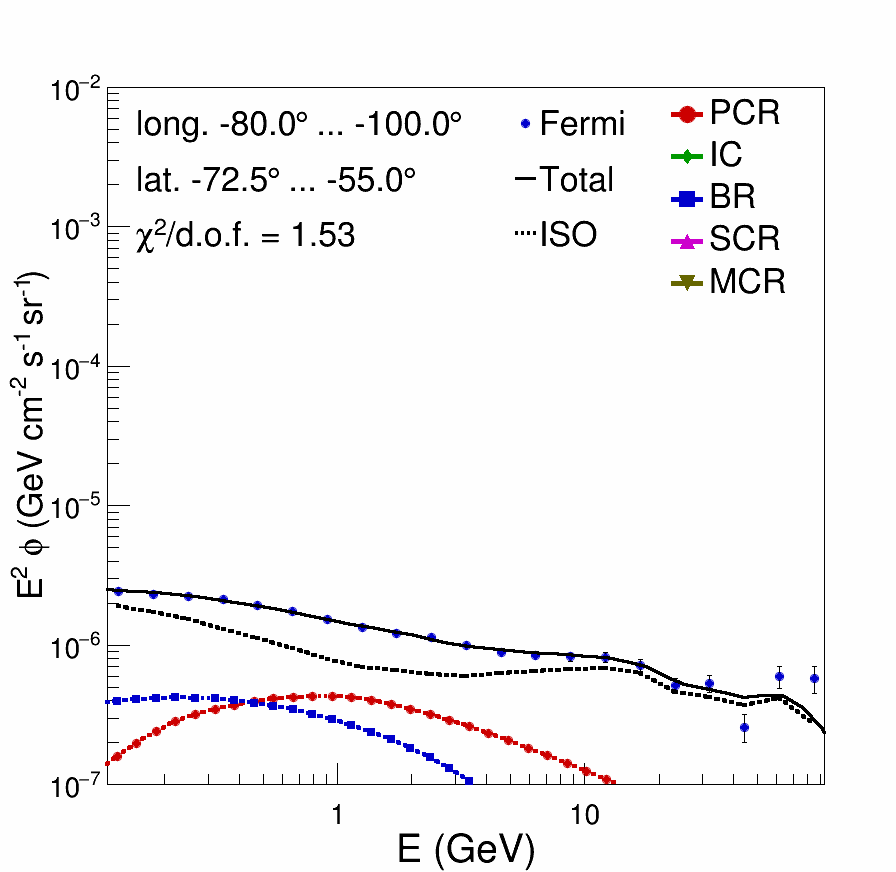}
\includegraphics[width=0.16\textwidth,height=0.16\textwidth,clip]{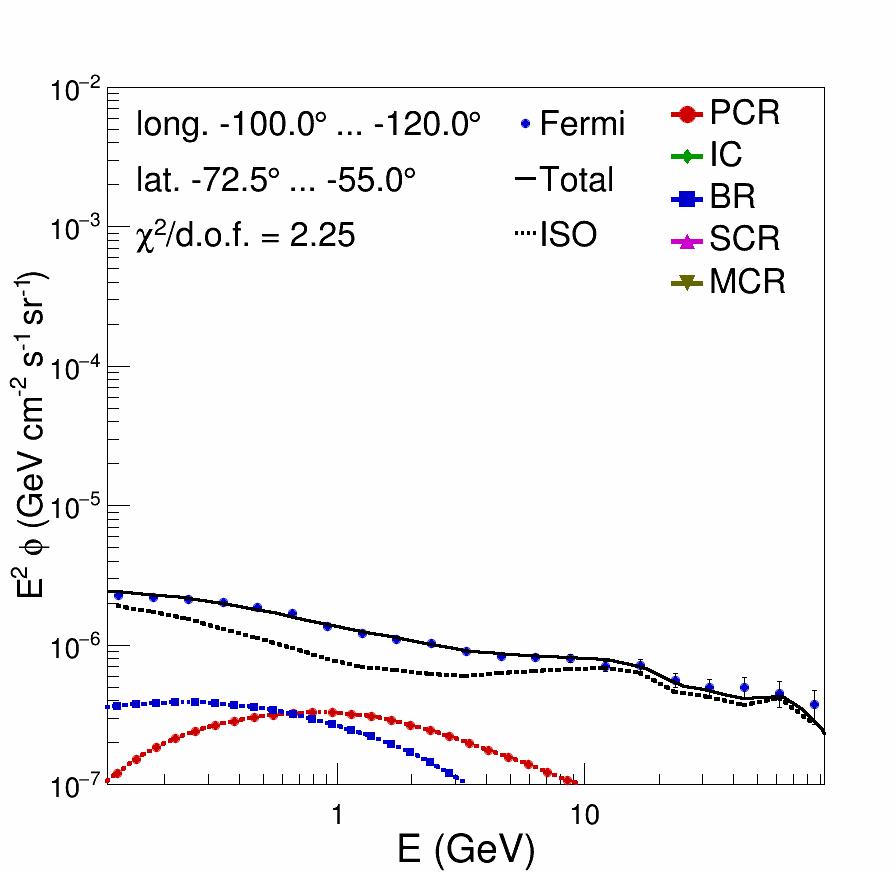}
\includegraphics[width=0.16\textwidth,height=0.16\textwidth,clip]{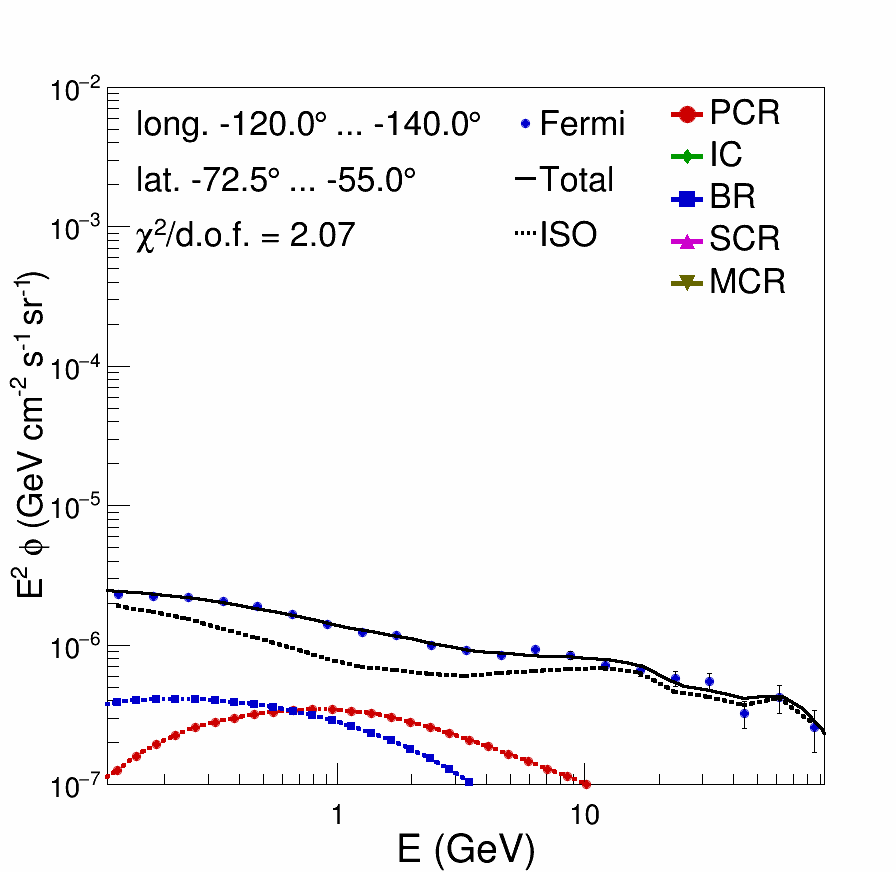}
\includegraphics[width=0.16\textwidth,height=0.16\textwidth,clip]{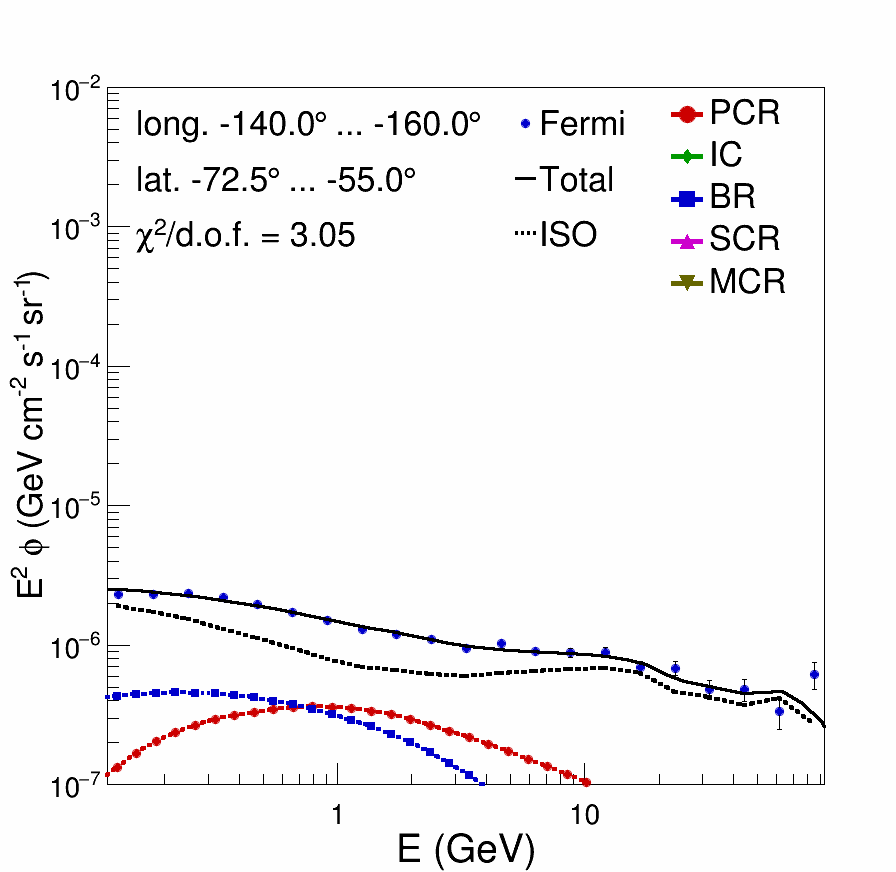}
\includegraphics[width=0.16\textwidth,height=0.16\textwidth,clip]{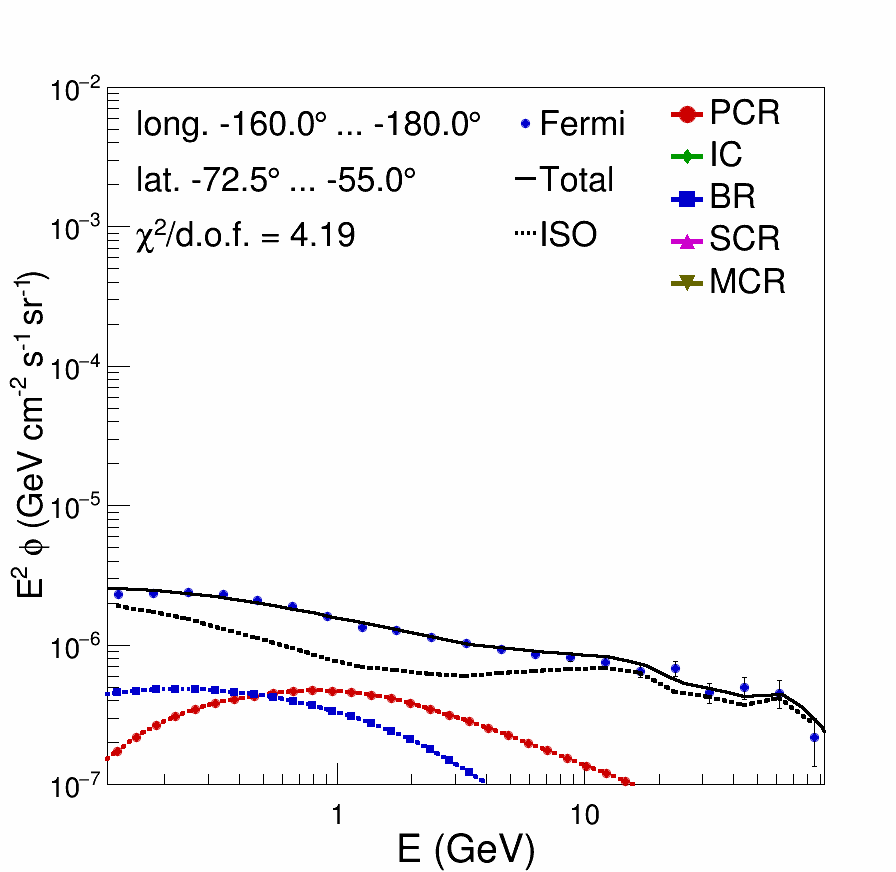}
\caption[]{Template fits for latitudes  with $-72.5^\circ<b<-55.0^\circ$ and longitudes decreasing from 180$^\circ$ to -180$^\circ$.} \label{F30}
\end{figure}
\begin{figure}
\includegraphics[width=0.16\textwidth,height=0.16\textwidth,clip]{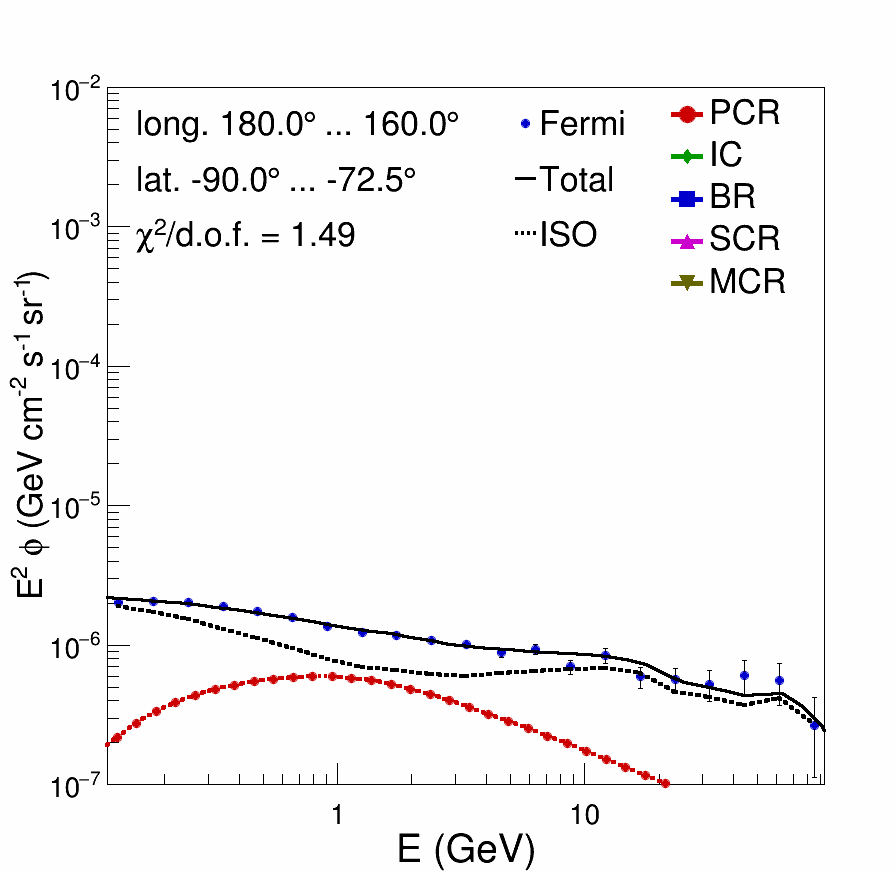}
\includegraphics[width=0.16\textwidth,height=0.16\textwidth,clip]{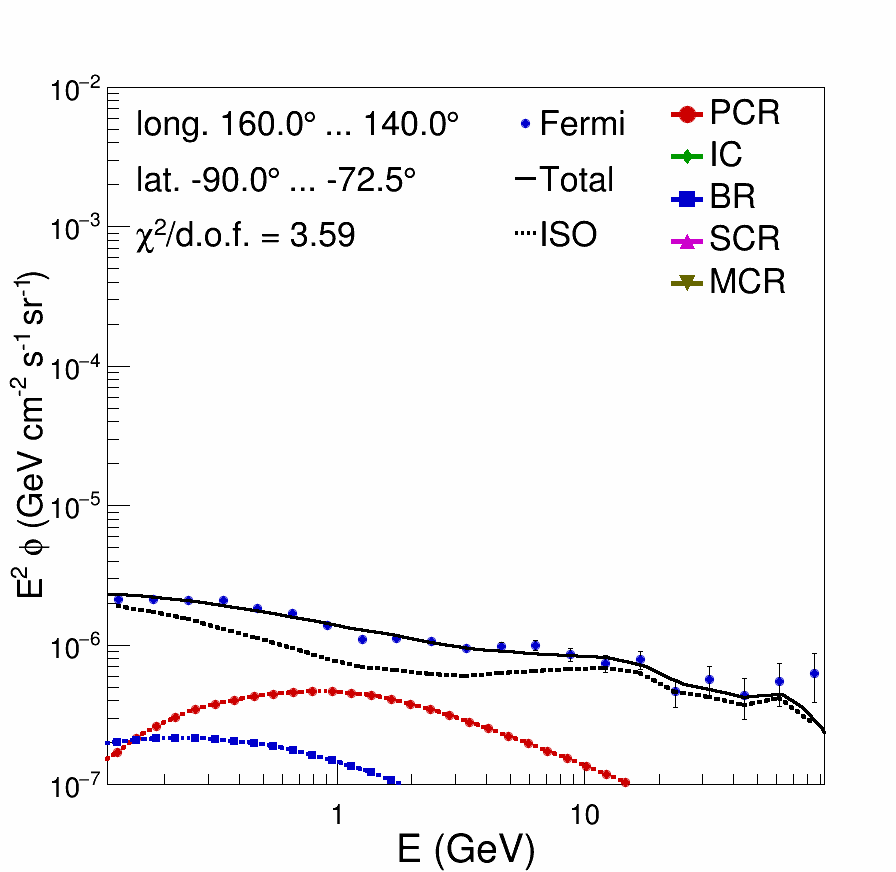}
\includegraphics[width=0.16\textwidth,height=0.16\textwidth,clip]{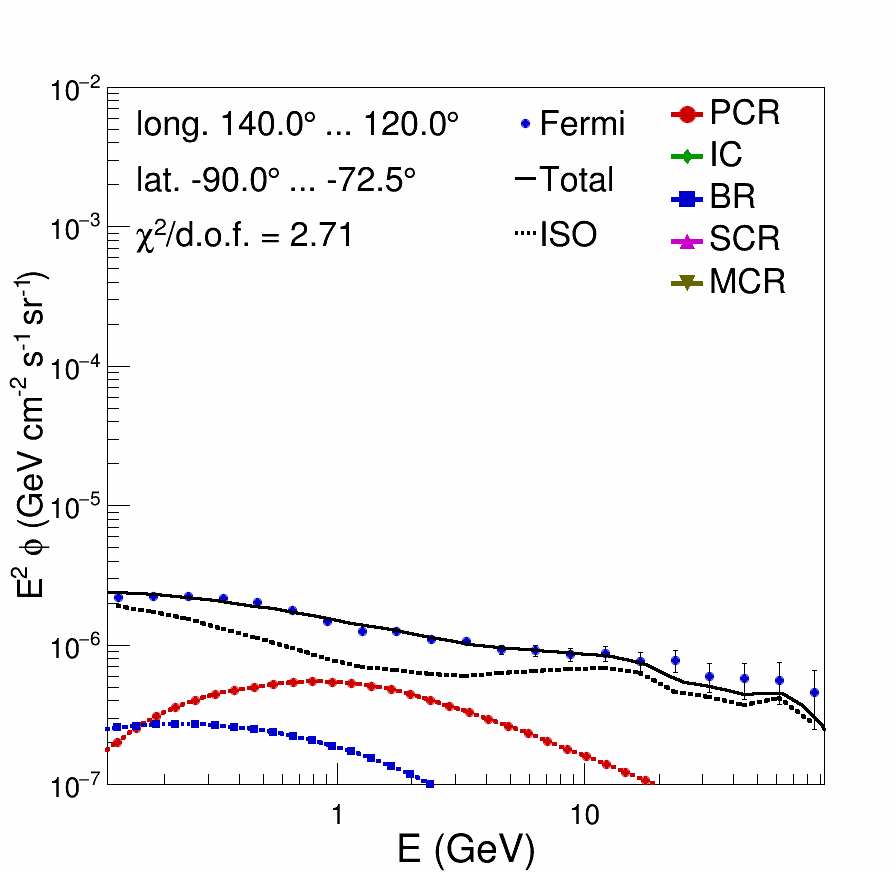}
\includegraphics[width=0.16\textwidth,height=0.16\textwidth,clip]{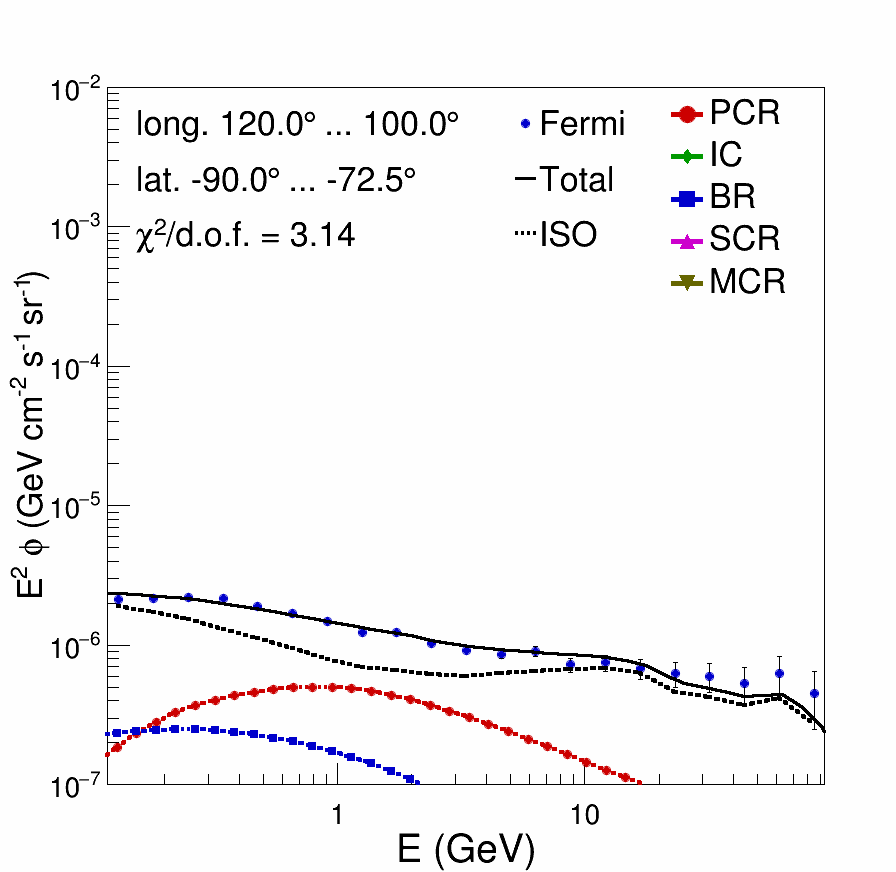}
\includegraphics[width=0.16\textwidth,height=0.16\textwidth,clip]{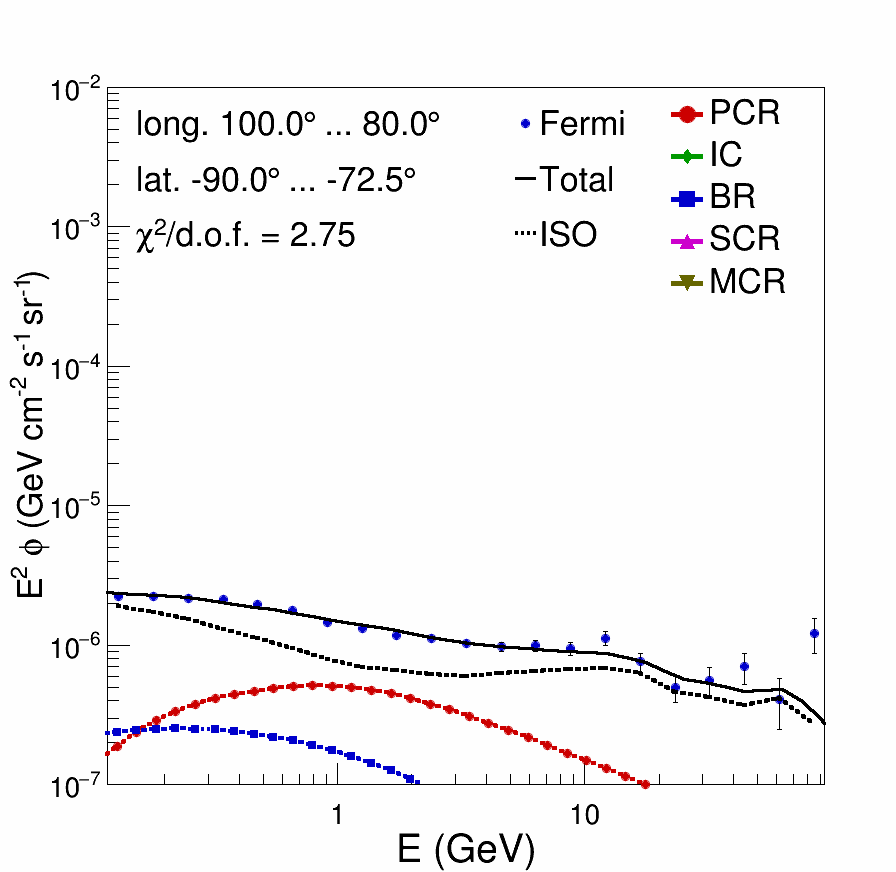}
\includegraphics[width=0.16\textwidth,height=0.16\textwidth,clip]{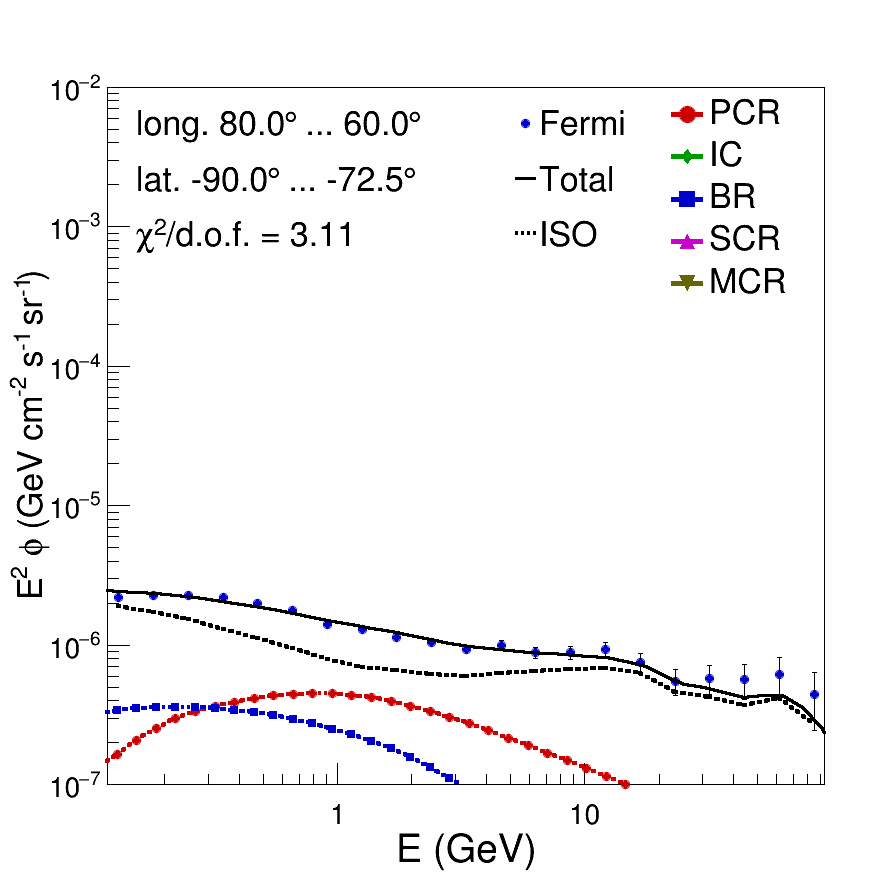}
\includegraphics[width=0.16\textwidth,height=0.16\textwidth,clip]{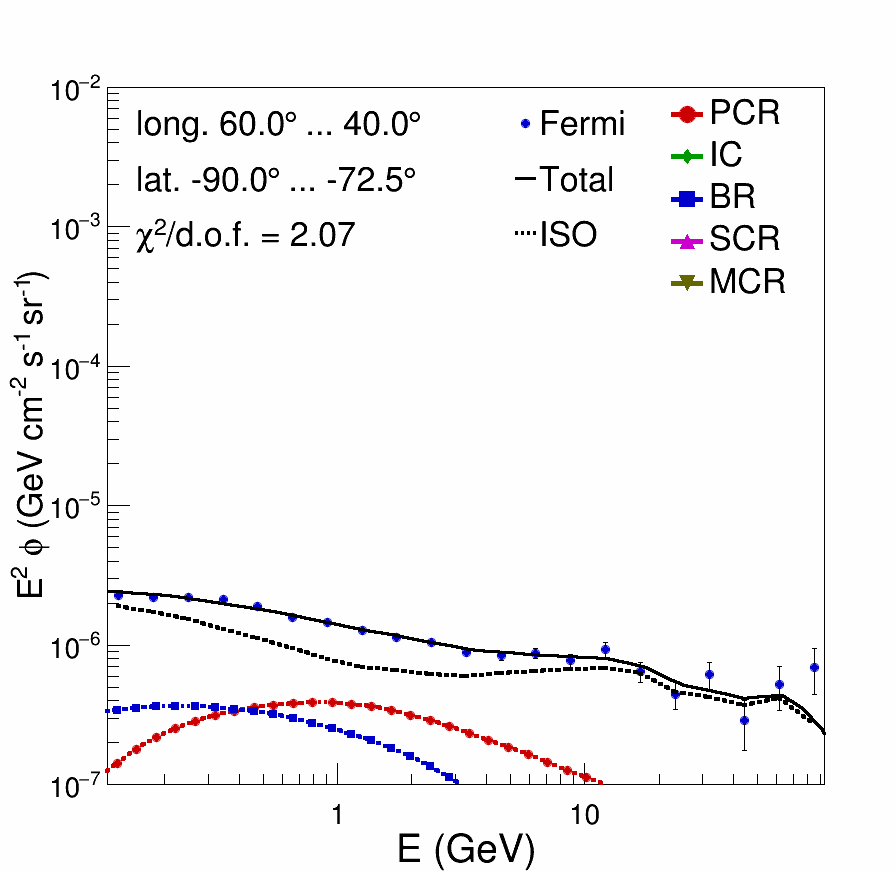}
\includegraphics[width=0.16\textwidth,height=0.16\textwidth,clip]{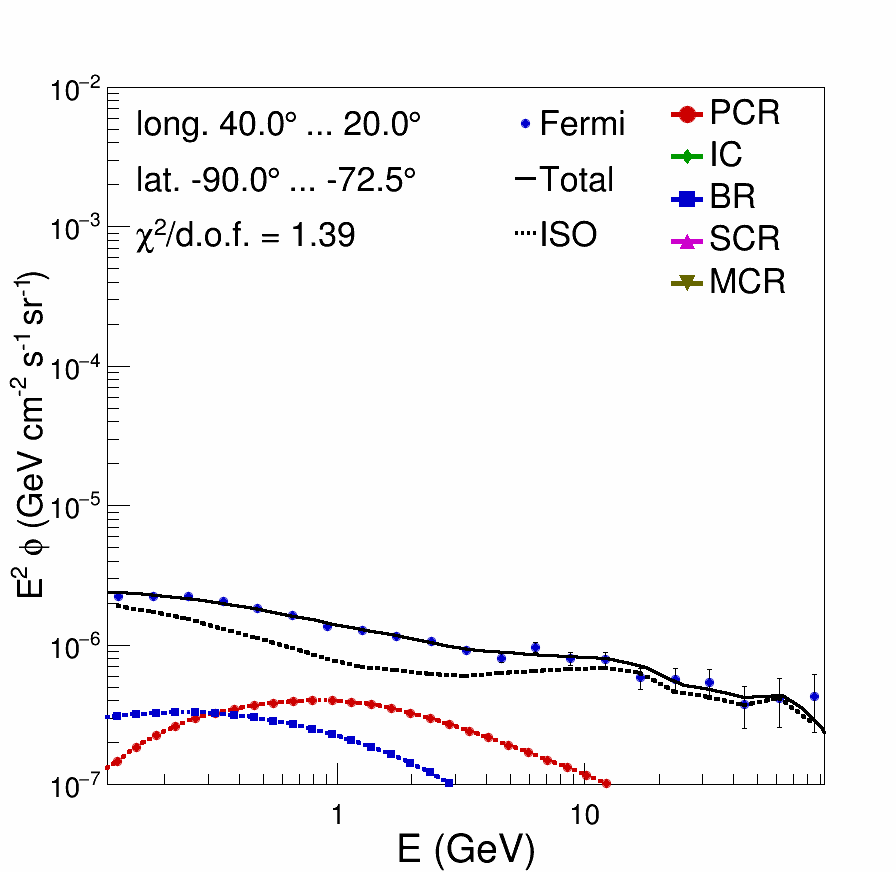}
\includegraphics[width=0.16\textwidth,height=0.16\textwidth,clip]{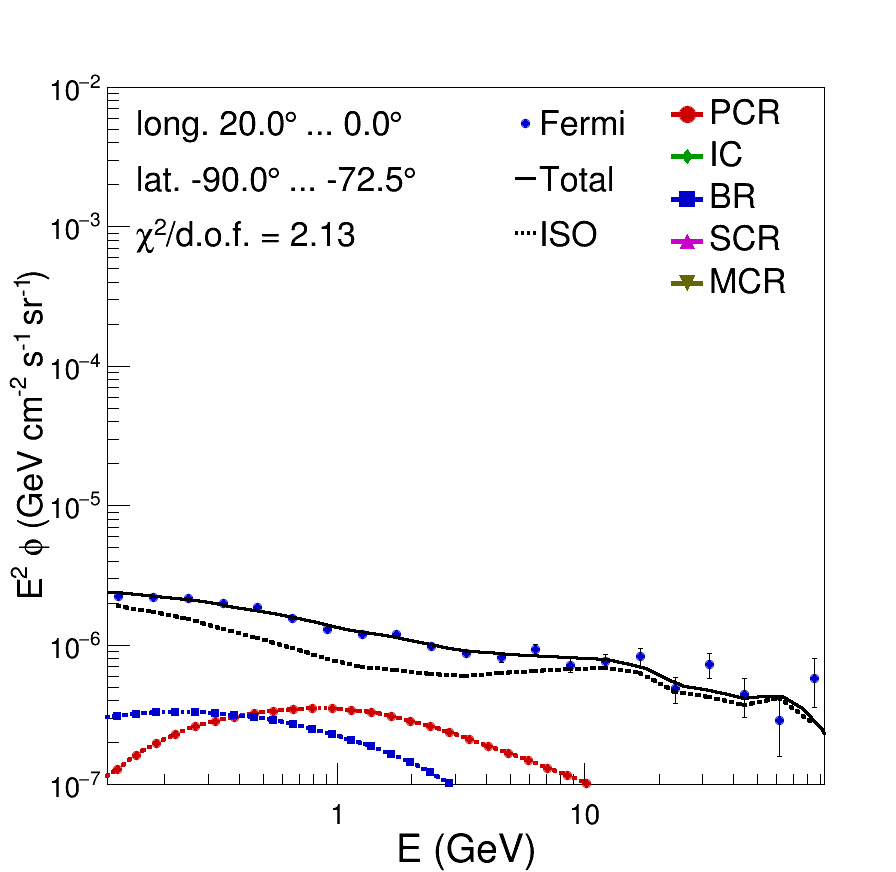}
\includegraphics[width=0.16\textwidth,height=0.16\textwidth,clip]{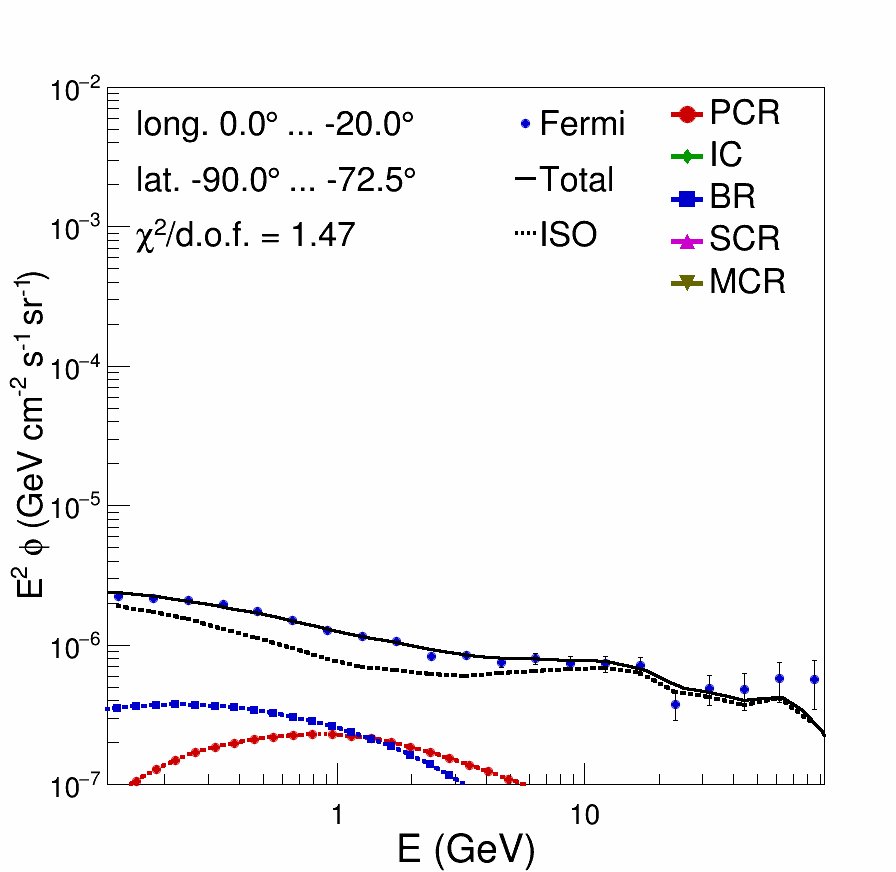}
\includegraphics[width=0.16\textwidth,height=0.16\textwidth,clip]{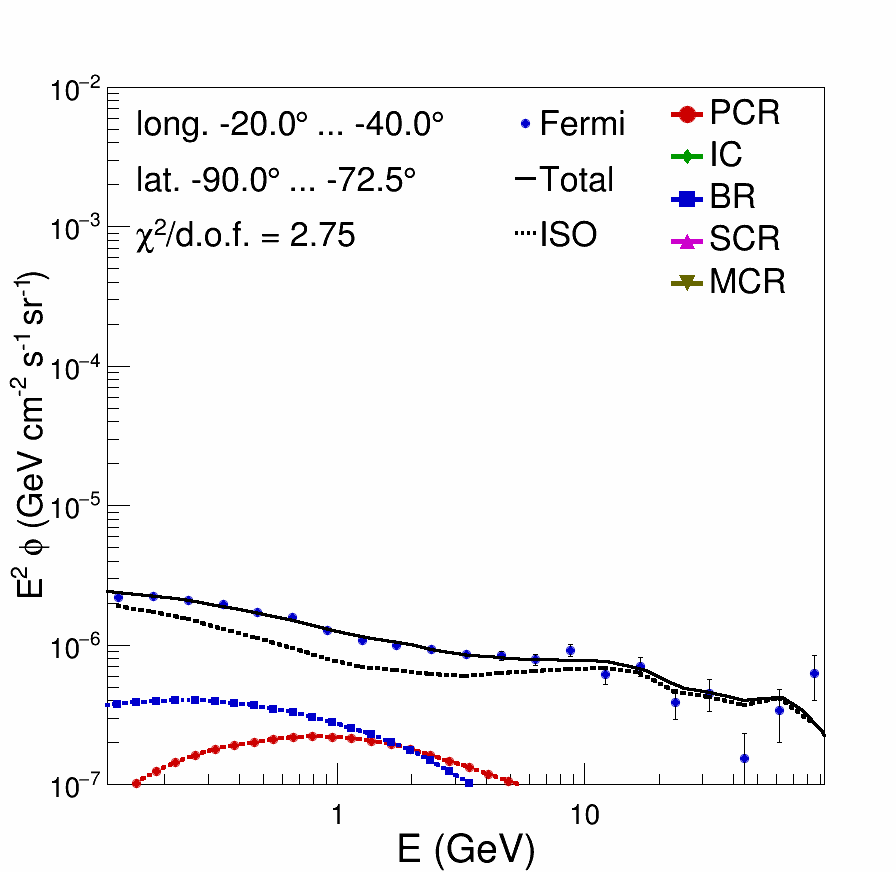}
\includegraphics[width=0.16\textwidth,height=0.16\textwidth,clip]{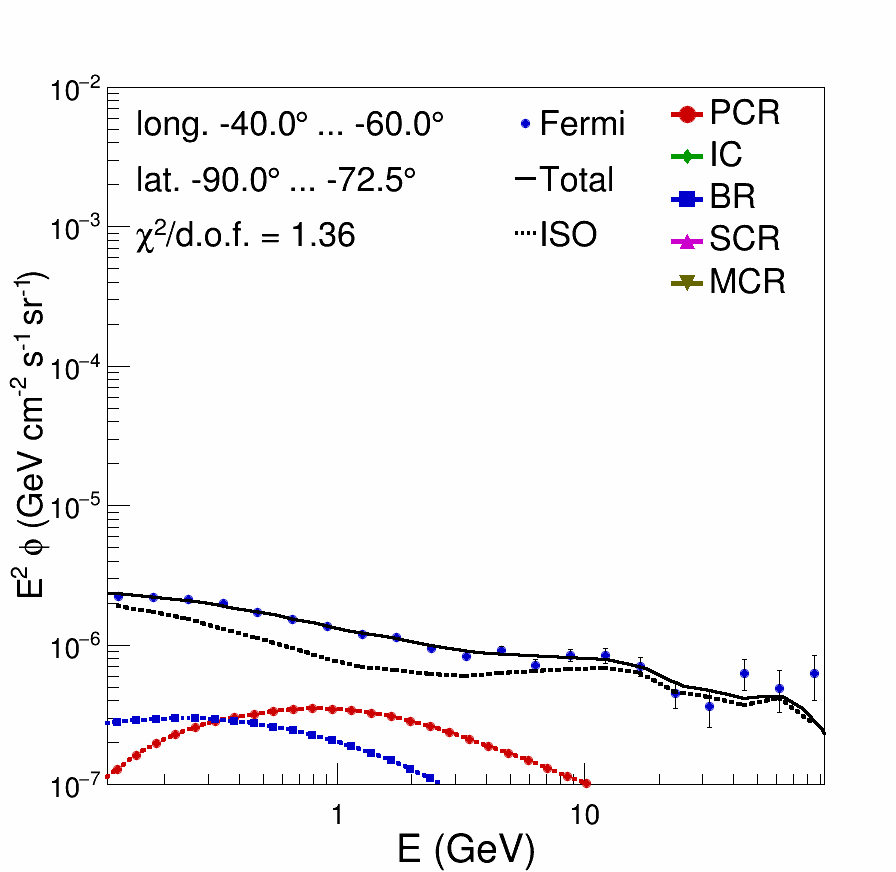}
\includegraphics[width=0.16\textwidth,height=0.16\textwidth,clip]{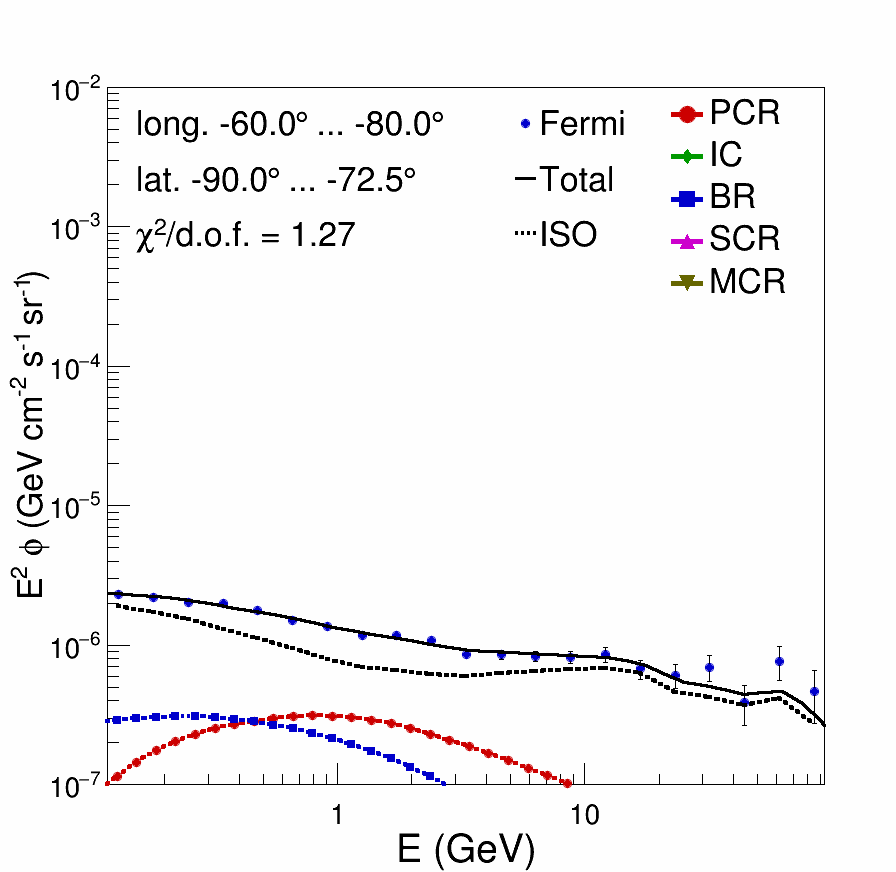}
\includegraphics[width=0.16\textwidth,height=0.16\textwidth,clip]{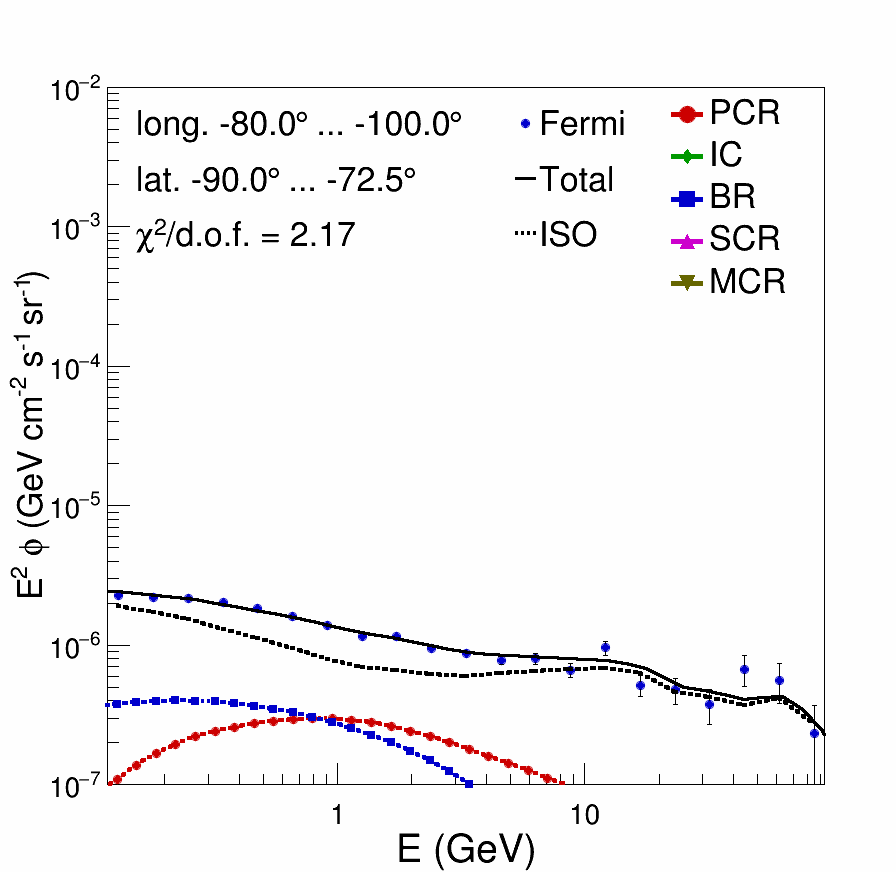}
\includegraphics[width=0.16\textwidth,height=0.16\textwidth,clip]{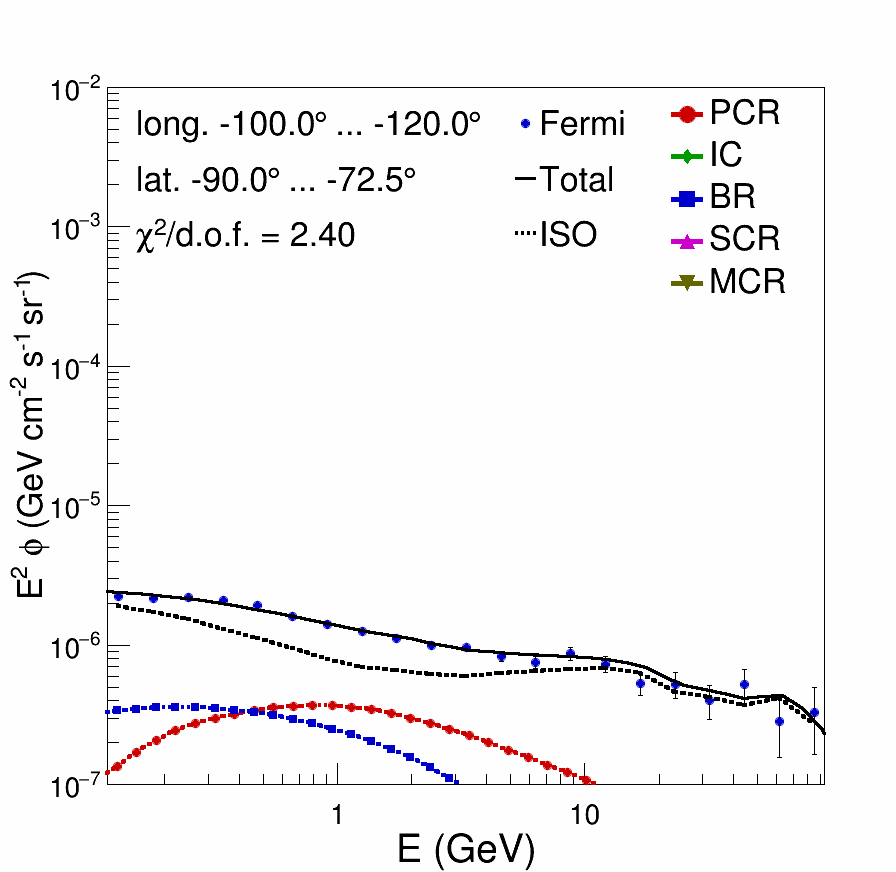}
\includegraphics[width=0.16\textwidth,height=0.16\textwidth,clip]{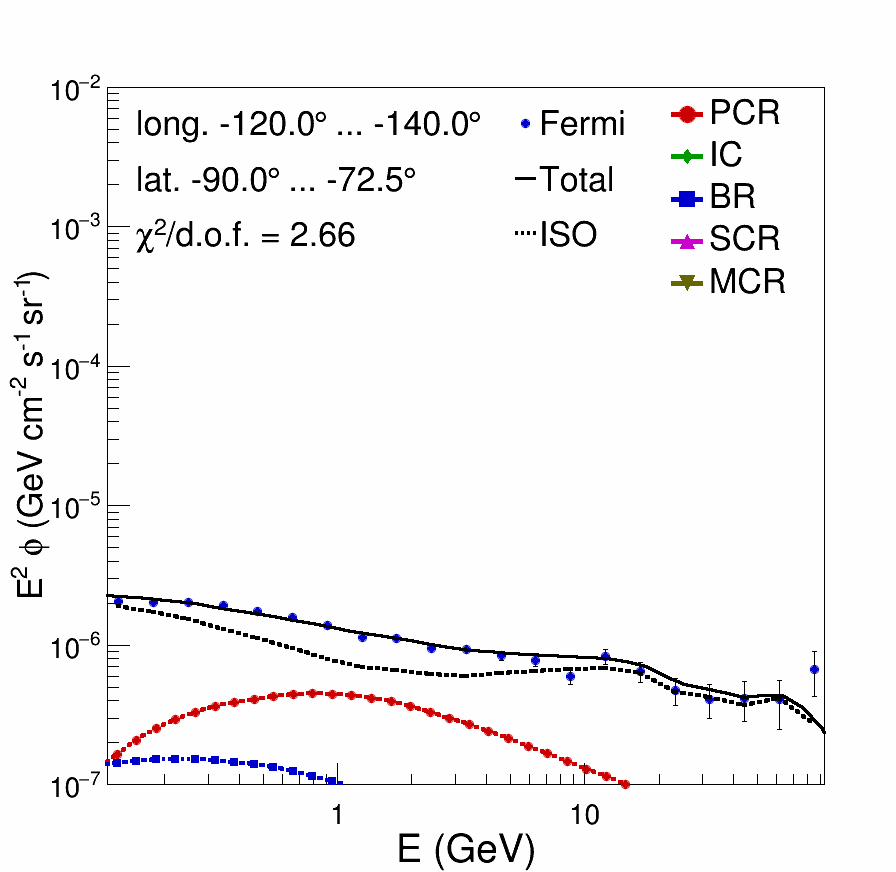}
\includegraphics[width=0.16\textwidth,height=0.16\textwidth,clip]{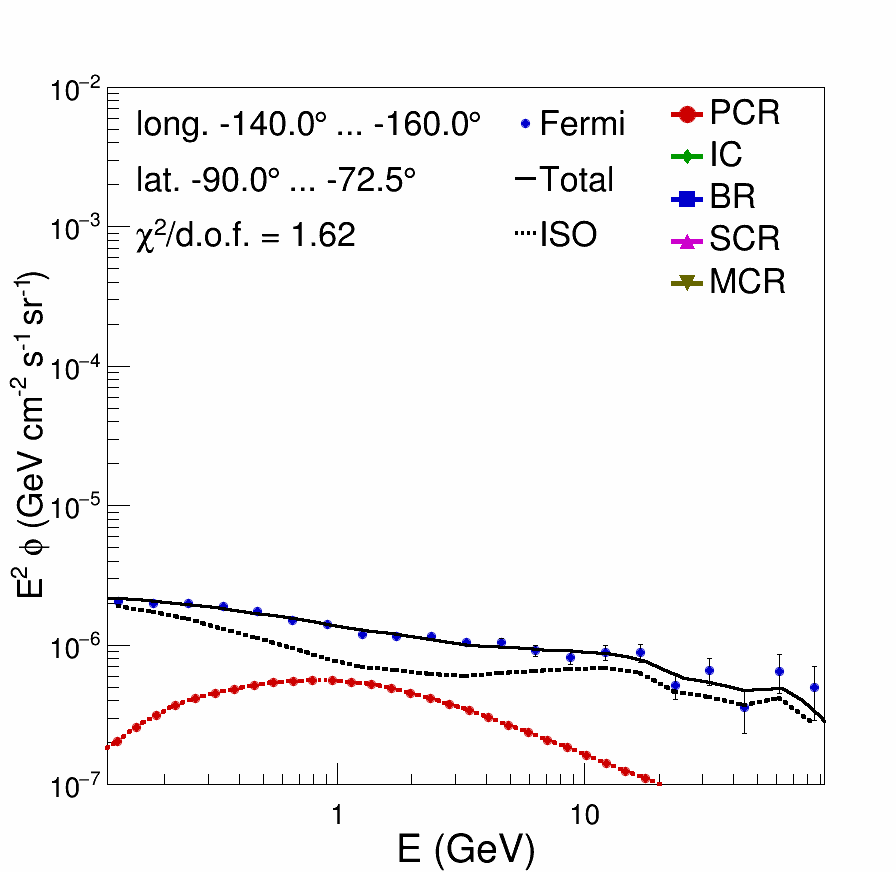}
\includegraphics[width=0.16\textwidth,height=0.16\textwidth,clip]{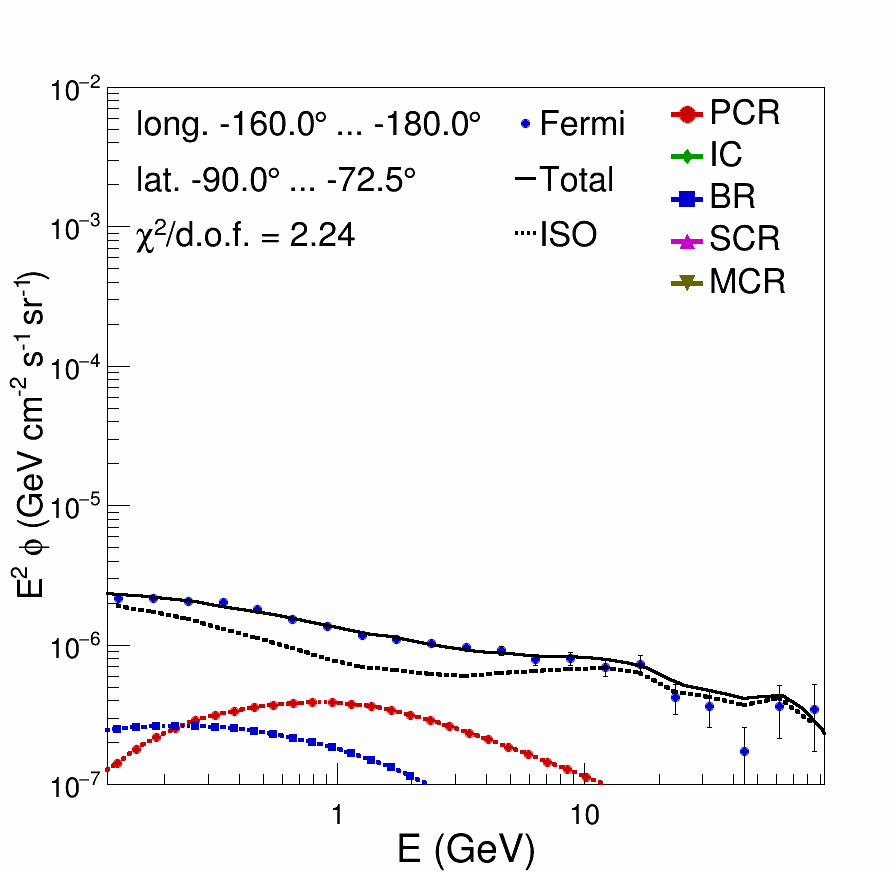}
\caption[]{Template fits for latitudes  with $-90.0^\circ<b<-72.5^\circ$ and longitudes decreasing from 180$^\circ$ to -180$^\circ$.} \label{F31}
\end{figure}

\clearpage

 \providecommand{\href}[2]{#2}\begingroup\raggedright\endgroup
\end{document}